\documentclass[structabstract]{aa}  
\usepackage{amsmath}
\usepackage{graphicx} 
\usepackage{txfonts}
\usepackage{natbib}
\usepackage{lscape}
\usepackage{hyperref}
\usepackage{siunitx}  
\usepackage{xcolor}   
\newcommand{\GG}{\mbox{$G$}}
\newcommand{\GBP}{\mbox{$G_{\rm BP}$}}
\newcommand{\GRP}{\mbox{$G_{\rm RP}$}}
\newcommand{\GBPmGRP}{\mbox{$G_{\rm BP}-G_{\rm RP}$}}
\newcommand{\GBPmGRPmin}{\mbox{$(G_{\rm BP}-G_{\rm RP})_{\rm min}$}}
\newcommand{\GBPmGRPmax}{\mbox{$(G_{\rm BP}-G_{\rm RP})_{\rm max}$}}

\newcommand{\sGz}{\mbox{$s_{G{\rm,0}}$}}

\newcommand{\sGc}{\mbox{$s_{G{\rm,c}}$}}
\newcommand{\sG}{\mbox{$s_G$}}

\newcommand{\sXz}{\mbox{$s_{X{\rm,0}}$}}
\newcommand{\sXi}{\mbox{$s_{X{\rm,i}}$}}
\newcommand{\sXins}{\mbox{$s_{X{\rm,ins}}$}}
\newcommand{\sXc}{\mbox{$s_{X{\rm,c}}$}}
\newcommand{\sX}{\mbox{$s_X$}}
\newcommand{\sigmasX}{\mbox{$\sigma_{s_X}$}}
\newcommand{\sigmasXi}{\mbox{$\sigma_{X{\rm,i}}$}}
\newcommand{\sGBP}{\mbox{$s_{G_{\rm BP}}$}}
\newcommand{\sGRP}{\mbox{$s_{G_{\rm RP}}$}}
\newcommand{\sGBPz}{\mbox{$s_{G_{\rm BP}{\rm,0}}$}}
\newcommand{\sGRPz}{\mbox{$s_{G_{\rm RP}{\rm,0}}$}}

\newcommand{\Gabs}{\mbox{$G_{\rm abs}$}}

\newcommand{\mci}[1]{\multicolumn{1}{c}{#1}}
\newcommand{\mcii}[1]{\multicolumn{2}{c}{#1}}
\newcommand{\mciii}[1]{\multicolumn{3}{c}{#1}}

\newcommand{\mcv}[1]{\multicolumn{5}{c}{#1}}

\newcommand{\pic}{\mbox{$\varpi_{\rm c}$}}
\newcommand{\spic}{\mbox{$\sigma_{\varpi_{\rm c}}$}}

\newcommand{\VO}[1]{Villafranca~O-{#1}}

\renewcommand{\arraystretch}{1.2}

\begin{document}

   \title{Stellar variability in \textit{Gaia} DR3 \linebreak
          I. Three-band photometric dispersions for 145 million sources}
   \titlerunning{Stellar variability in \textit{Gaia} DR3}

   \author{J. Ma\'{\i}z Apell\'aniz \inst{1}
           \and
           G. Holgado\inst{1}
           \and
           M. Pantaleoni Gonz\'alez\inst{1,2}
%           D. J. Lennon\inst{3}
           \and
           J. A. Caballero\inst{1}
           }
   \authorrunning{J. Ma\'{\i}z Apell\'aniz et al.}

   \institute{Centro de Astrobiolog\'{\i}a. CSIC-INTA. Campus ESAC. 
              C. bajo del castillo s/n. 
              E-\num{28692} Villanueva de la Ca\~nada, Madrid, Spain.\linebreak
              \email{jmaiz@cab.inta-csic.es} 
              \and
              Departamento de Astrof{\'\i}sica y F{\'\i}sica de la Atm\'osfera. Universidad Complutense de Madrid. 
              E-\num{28040} Madrid, Spain. 
%              \and
%              Instituto de Astrof{\'\i}sica de Canarias.
%              C/ V\'{\i}a L\'actea s/n.
%              E-\num{38200} La Laguna, Santa Cruz de Tenerife, Spain.
              }

   \date{Received 27 March 2023 / Accepted XX XXX 2023}

% \abstract{}{}{}{}{} 
% 5 {} token are mandatory
 
  \abstract
   {The unparalleled characteristics of \textit{Gaia} photometry in terms of calibration, stability, time span, dynamic range, 
    full-sky coverage, and complementary information make it an excellent choice to study stellar variability.}
   {To measure the photometric dispersion in the \GG+\GBP+\GRP\ bands of the \num{145677450} third \textit{Gaia} data release (DR3)
    five-parameter sources with $\GG \le 17$ mag and \GBPmGRP\ between $-1.0$ and $8.0$ mag. To use that unbiased sample to analyze 
    stellar variability in the Milky Way (MW), LMC, and SMC.}
   {For each band we convert from magnitude uncertainties to observed photometric dispersions, calculate the 
    instrumental component as a function of apparent magnitude and color, and use it to transform the observed dispersions into the 
    astrophysical ones: \sG, \sGBP, and \sGRP. We give variability indices in the three bands for the whole
    sample indicating whether the objects are non-variable, marginally variable, or clearly so. We 
    use the subsample of Rimoldini~et~al. with light curves and variability types to calibrate our results and establish
    their limitations.}
   {The position of an object in the dispersion-dispersion planes can be used to constrain 
    its variability type, a direct application of these results. We use information from the MW, LMC, and SMC 
    color-absolute magnitude diagrams (CAMDs) to discuss variability across the Hertzsprung-Russell diagram. White dwarfs and 
    B-type subdwarfs are more variable than main sequence (MS) or red clump (RC) stars, with a flat distribution in \sG\ up to 
    10~mmag and with variability decreasing for the former 
    with age. The MS region in the \textit{Gaia} CAMD includes a mixture of populations
    from the MS itself and from other evolutionary phases. Its \sG\ distribution peaks at low values ($\sim 1-2$~mmag) but it 
    has a large tail dominated by eclipsing binaries, RR~Lyr stars, and young stellar objects. RC stars
    are characterized by little variability, with their \sG\ distribution peaking at 1~mmag or less. The 
    stars in the pre-main-sequence (PMS) region are highly variable, with a power law distribution in \sG\ with slope 2.75 and a 
    cutoff for values lower than 7~mmag. The luminous red stars region of the \textit{Gaia}~CAMD has the highest 
    variability, with its extreme dominated by AGB stars and with a power law in \sG\ with slope $\sim$2.2 that extends from 
    there to a cutoff of 7~mmag. We show that our method can be used to search for LMC Cepheids. We analyze four 
    stellar clusters with O stars (\VO{016},~O-021,~O-024,~and~O-026) and detect a strong difference in \sG\ between 
    stars that are already in the MS and those that are still in the PMS.}
   {}
   \maketitle
%
%________________________________________________________________

\section{Introduction}

$\,\!$\indent The \textit{Gaia} mission \citep{Prusetal16} is arguably the best example of the paradigm change that large-scale 
surveys have introduced in astronomy. Not long ago, most astronomical research was carried on by individuals or small groups
going on observing runs and conducting analyses of one to several hundreds of objects using their own reduction. 
Such programs are still carried on today and are useful and necessary, but the weight in astronomy has shifted towards 
large-scale surveys that cover the whole sky or a significant fraction of it, are conducted by large teams, involve samples several 
orders of magnitude larger, and are reduced using common procedures that ease calibration.

The (full) third \textit{Gaia} data release (DR3) provides astrometric, photometric, and spectroscopic data for $1.8\times 10^9$
sources \citep{Valletal22}. One of the novelties of DR3 is the inclusion of photometric variability information for over 12 million
sources (less than 1\% of the full sample) in the form of epoch photometry and variability classifications \citep{Eyeretal22}. The 
classifications based on the light curves are given by \citet{Rimoetal22}, from now on R22. For the specific case of the 2+ million
eclipsing binaries, see \citet{Mowletal22}. The R22 sample was selected from sources expected to be variable, so it is
biased in that sense, something we will verify in this paper.

The quality and quantity of information provided by \textit{Gaia} allows for a limited analysis of photometric variability without the
use of light curves but based instead on the observed photometric dispersions. Previous papers have done that 
\citep{Vioqetal20,Guidetal21,Mowletal21,Andretal21,Barletal22,BarbMann23} but they have not used the full potential provided by 
using a large fraction of the \textit{Gaia}~DR3 sample. That is the purpose of this paper: to calculate astrophysical photometric 
dispersions for the majority of \textit{Gaia}~DR3 objects with $\GG \le 17$~mag. We will follow it up with papers on short-timescale
variables (Ma{\'\i}z Apell\'aniz et al. in prep.) and with an analysis of the \textit{Gaia}~DR3 light curves for Galactic O~stars 
(Holgado et al. in prep.). Possible topics for future papers include the photometric variability of massive stars, the study of 
low-mass stars targeted by radial-velocity planet surveys, and an analysis of luminous red stars in the Magellanic Clouds.

In Section~2 we present our sample selection and methodology. In Section~3 we give our general results for variability across the 
\textit{Gaia} color-absolute magnitude 
diagram (CAMD), in Section~4 we discuss some examples of applications of the data presented in this paper, and we end with a summary.

\section{Data and methods}

\subsection{Sample selection}

\begin{figure}
\centerline{\includegraphics[width=\linewidth]{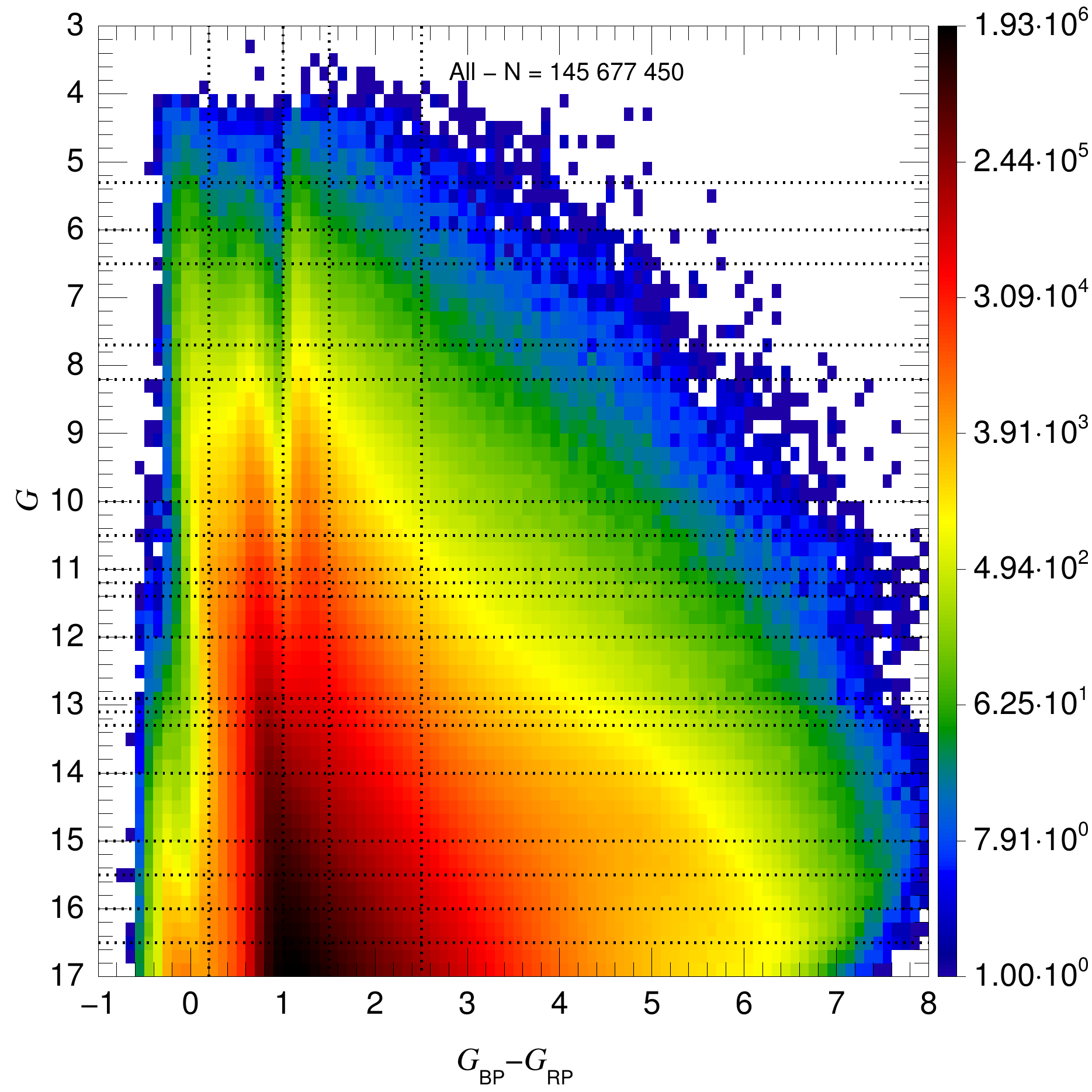}}
\caption{Source density (logarithmic scale) in the color-magnitude plane for the sample in this paper. Dotted black lines 
         indicate the magnitude-color bins used to determine the characteristics of the photometric dispersion.}
\label{CMD_all}
\end{figure}

$\,\!$\indent We started by querying the \textit{Gaia}~DR3 catalog for all sources with (a) $\GG \le 17$~mag, (b) five-parameter 
astrometric solutions, and (c) \GBPmGRP\ between $-$1.0 and 8.0~mag. The reason for constraining the sample in magnitude is that 
fainter sources have larger instrumental photometric dispersions (see below), thus limiting the usefulness of a variability analysis. 
In addition, the primary interest of this analysis is stellar variability and the fainter range sampled by {\it Gaia} has a larger 
fraction of QSOs and galaxies. The exclusion of six-parameter solutions is due to their generally poorer photometric 
quality and our desire to keep a sample with uniform properties. In any case, the proportion of six-parameter solutions for 
$\GG \le 17$~mag is small except for the brightest stars (See Fig.~12 in \citealt{Lindetal21b}). The third condition only removes a 
small number of stars, as bluer stars do not exist in theory (one object excluded) and redder stars are scarce and have in general
poor astrometry (74 objects excluded). The query resulted in \num{145677450} objects (8\% of the total \textit{Gaia}~DR3 sample) and 
their density distribution in \GG\ vs. \GBPmGRP\ is shown in Fig.~\ref{CMD_all}.

\begin{table*}
\caption{Statistics for the stars in the sample with R22 variability classifications. See subsection 3.1.1 in that
         paper for additional information on the comments column.}
\label{vartypestats}
\centerline{
\begin{tabular}{llrrl}
\hline
Variability type   & Acronym                              & Number        & Perc.   & Comments                                 \\
\hline
Solar-like         & SOLAR\_LIKE                          & \num{1659858} &  37.028 & Includes several related types           \\
LPV                & LPV                                  &  \num{904937} &  20.188 & Includes several related types           \\
$\delta$ Sct+      & DSCT$|$GDOR$|$SXPHE                  &  \num{687083} &  15.328 & 3 types of pulsating AF stars            \\
RS CVn             & RS                                   &  \num{539356} &  12.032 &                                          \\
Eclpsing binary    & ECL                                  &  \num{513359} &  11.452 &                                          \\
RR Lyr             & RR                                   &   \num{71859} &   1.603 & Includes different modes                 \\
YSO                & YSO                                  &   \num{38301} &   0.854 & Includes several related types           \\
Ellipsoidal        & ELL                                  &   \num{26951} &   0.601 &                                          \\
Cepheid            & CEP                                  &   \num{10687} &   0.238 & Includes several related types           \\
$\alpha^2$ CVn+    & ACV$|$CP$|$MCP$|$ROAM$|$ROAP$|$SXARI &    \num{9100} &   0.203 & Several types of magnetic variables      \\
Be+                & BE$|$GCAS$|$SDOR$|$WR                &    \num{7599} &   0.170 & 4 types of irregular high-mass variables \\
Short timescale    & S                                    &    \num{4185} &   0.093 &                                          \\
AGN                & AGN                                  &    \num{3578} &   0.080 &                                          \\
$\beta$ Cep        & BCEP                                 &    \num{1361} &   0.030 &                                          \\
Slowly pulsating B & SPB                                  &    \num{1153} &   0.026 &                                          \\
sdB                & SDB                                  &     \num{887} &   0.020 &                                          \\
CV                 & CV                                   &     \num{651} &   0.015 & Excludes SNe and symbiotic stars         \\
WD                 & WD                                   &     \num{575} &   0.013 & Includes several related types           \\
Symbiotic          & SYST                                 &     \num{526} &   0.012 &                                          \\
$\alpha$ Cyg       & ACYG                                 &     \num{318} &   0.007 &                                          \\
Exoplanet transit  & EP                                   &     \num{210} &   0.005 &                                          \\
R CrB              & RCB                                  &      \num{87} &   0.002 &                                          \\
Microlensing       & MICROLENSING                         &      \num{34} &   0.001 &                                          \\
\hline
Total              &                                      & \num{4482655} & 100.000 &                                          \\
\hline
\end{tabular}
}
\end{table*}

In addition, we also queried the \textit{Gaia}~DR3 catalog for light curve and variability information on the sample. 
\num{4991335} objects (3.426\%) have epoch photometry, \num{4795431} (3.292\%) are tagged as variable, and \num{4482655} (3.077\%)
have a variability classification by R22. We list in Table~\ref{vartypestats} the breakdown by variable type and we 
refer to that table and to R22 for the meaning of the acronyms used for each type.

\begin{figure*}
\centerline{$\!\!\!$\includegraphics[width=0.35\linewidth]{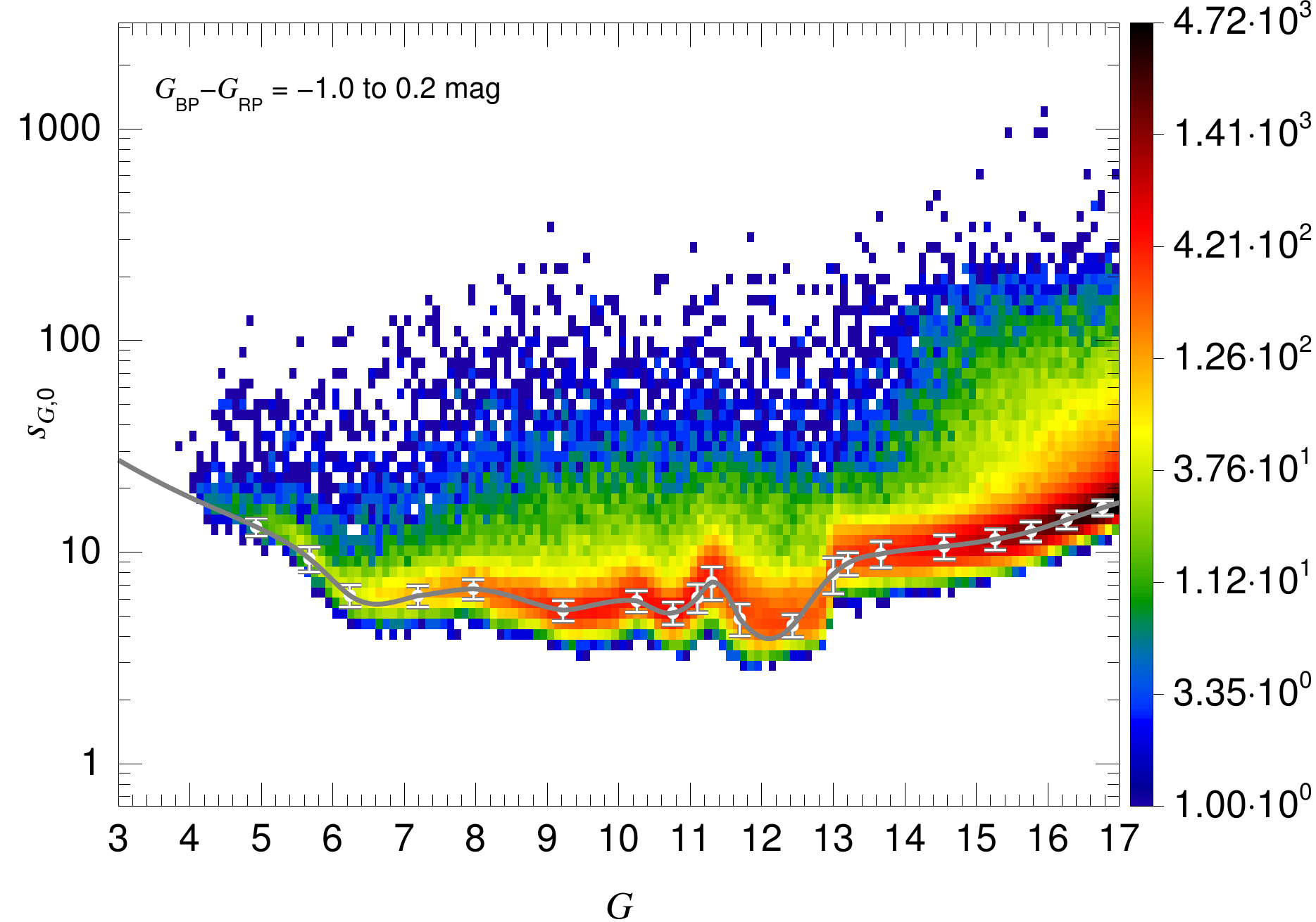}$\!\!\!$
                    \includegraphics[width=0.35\linewidth]{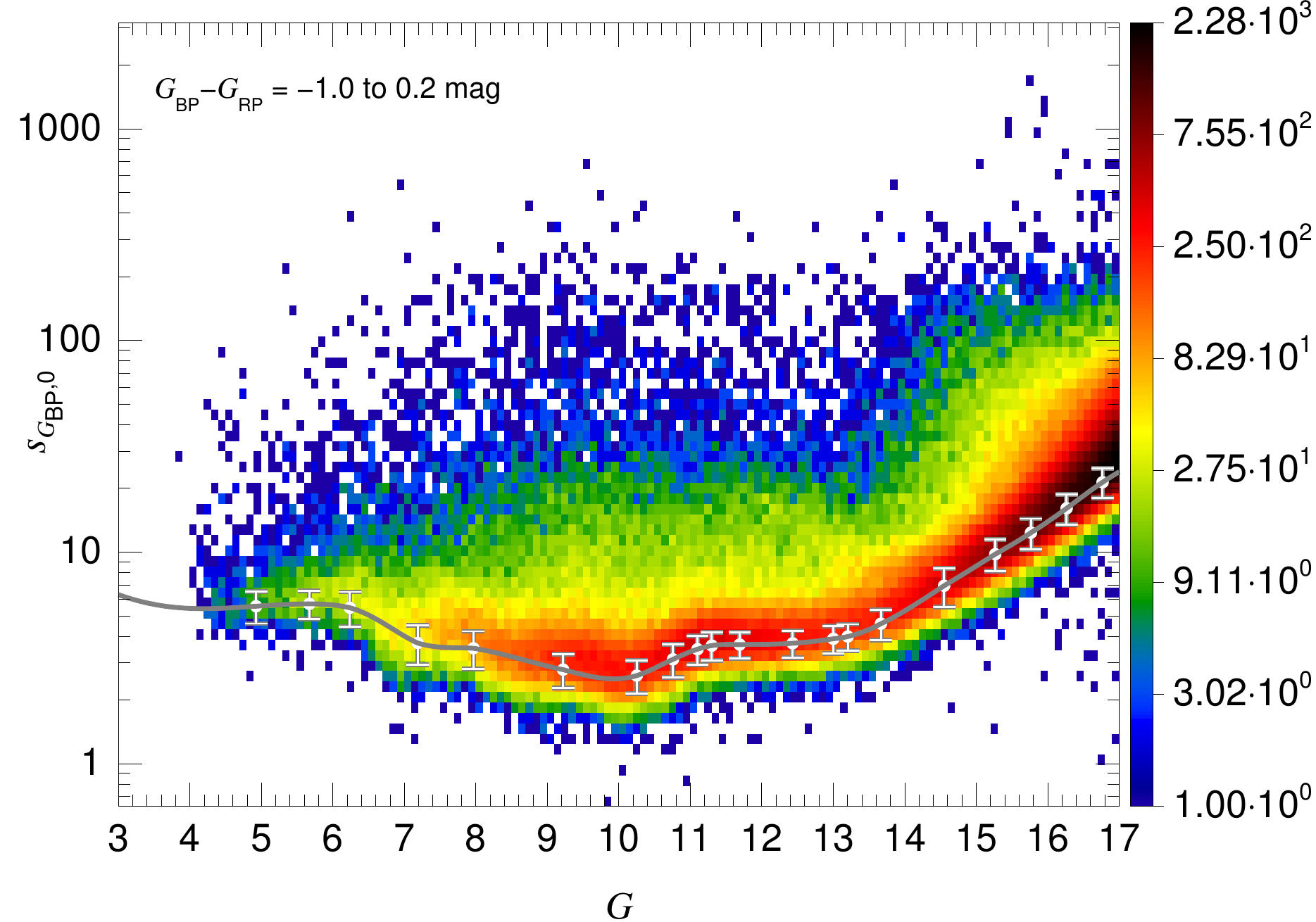}$\!\!\!$
                    \includegraphics[width=0.35\linewidth]{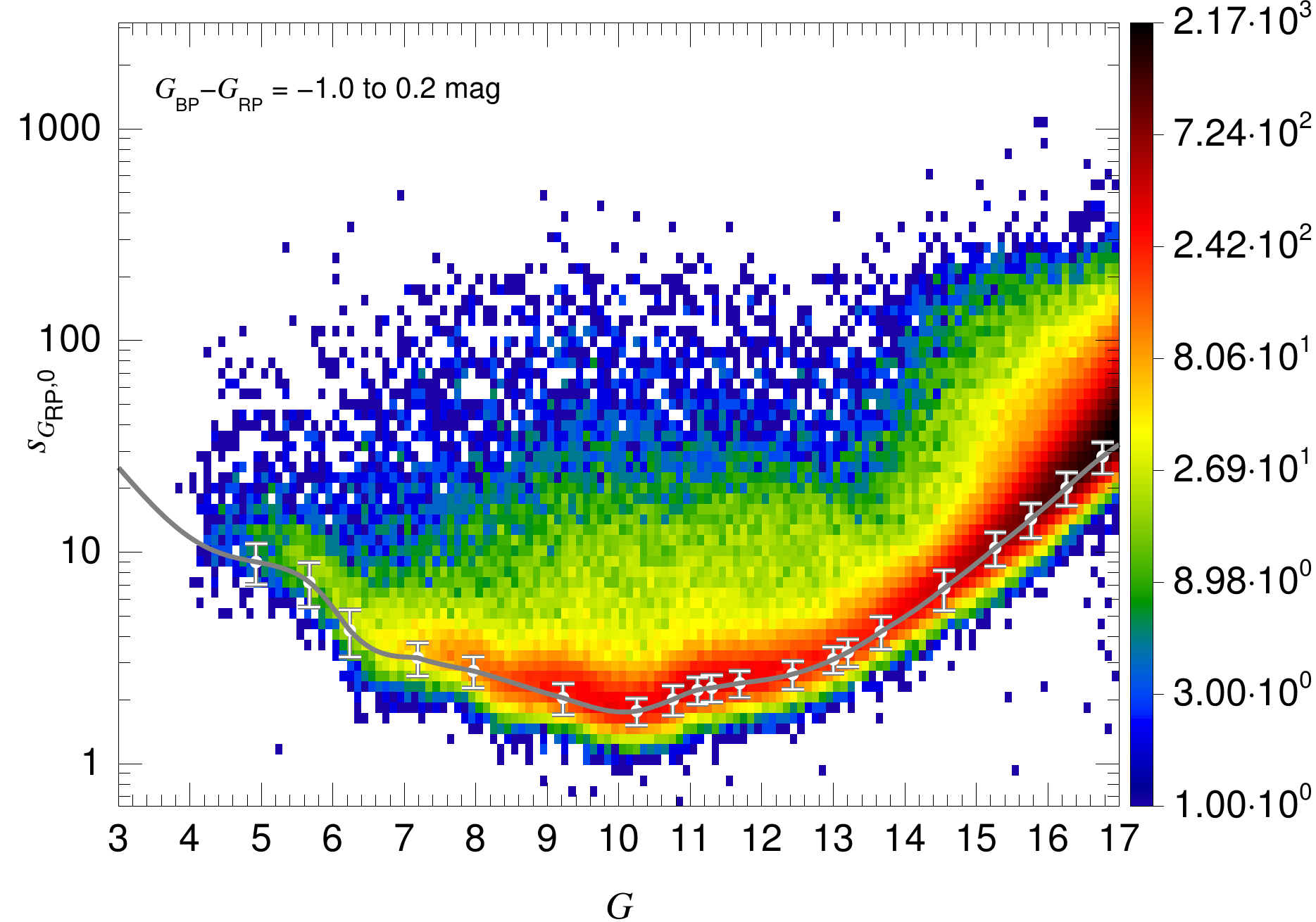}}
\centerline{$\!\!\!$\includegraphics[width=0.35\linewidth]{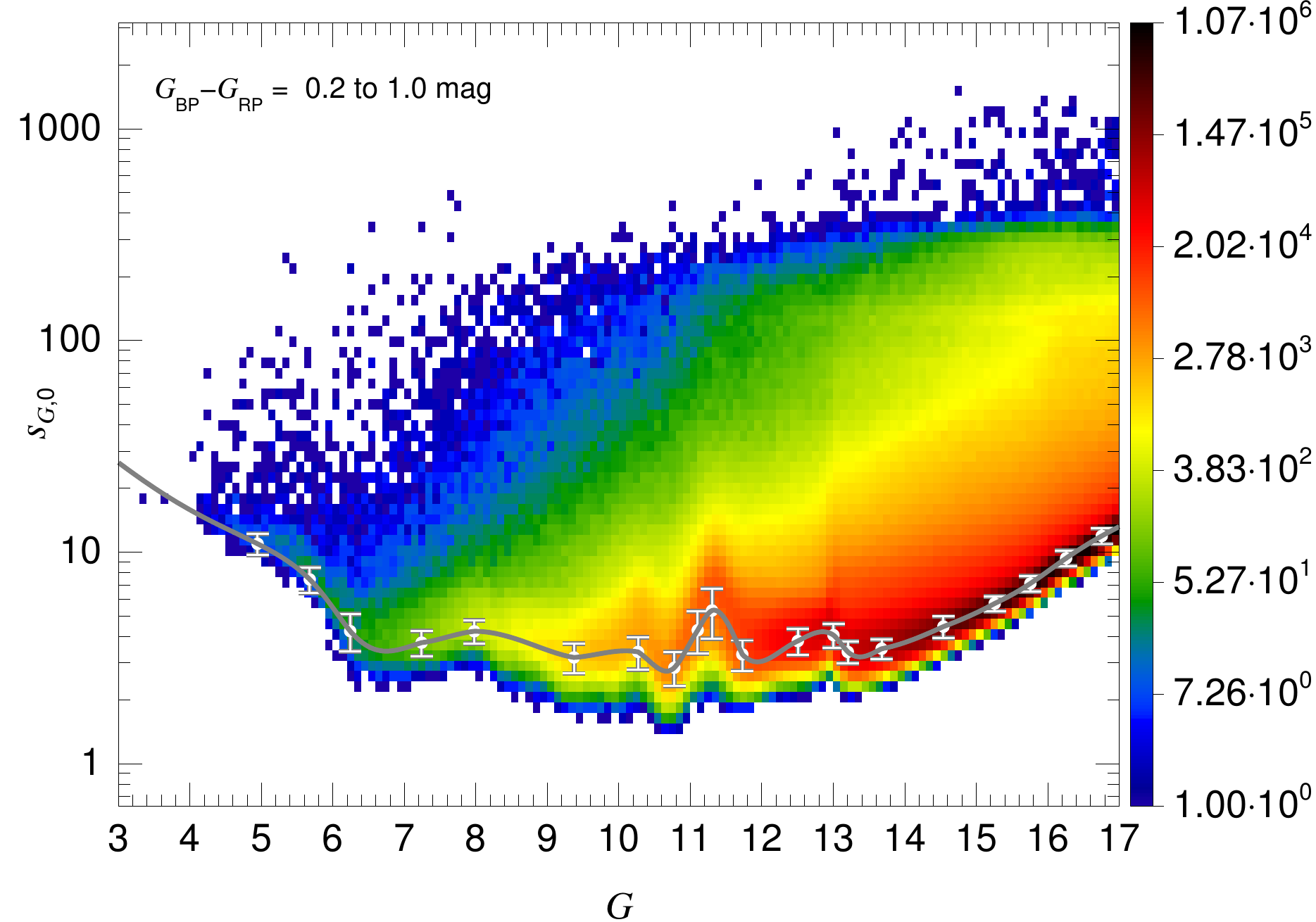}$\!\!\!$
                    \includegraphics[width=0.35\linewidth]{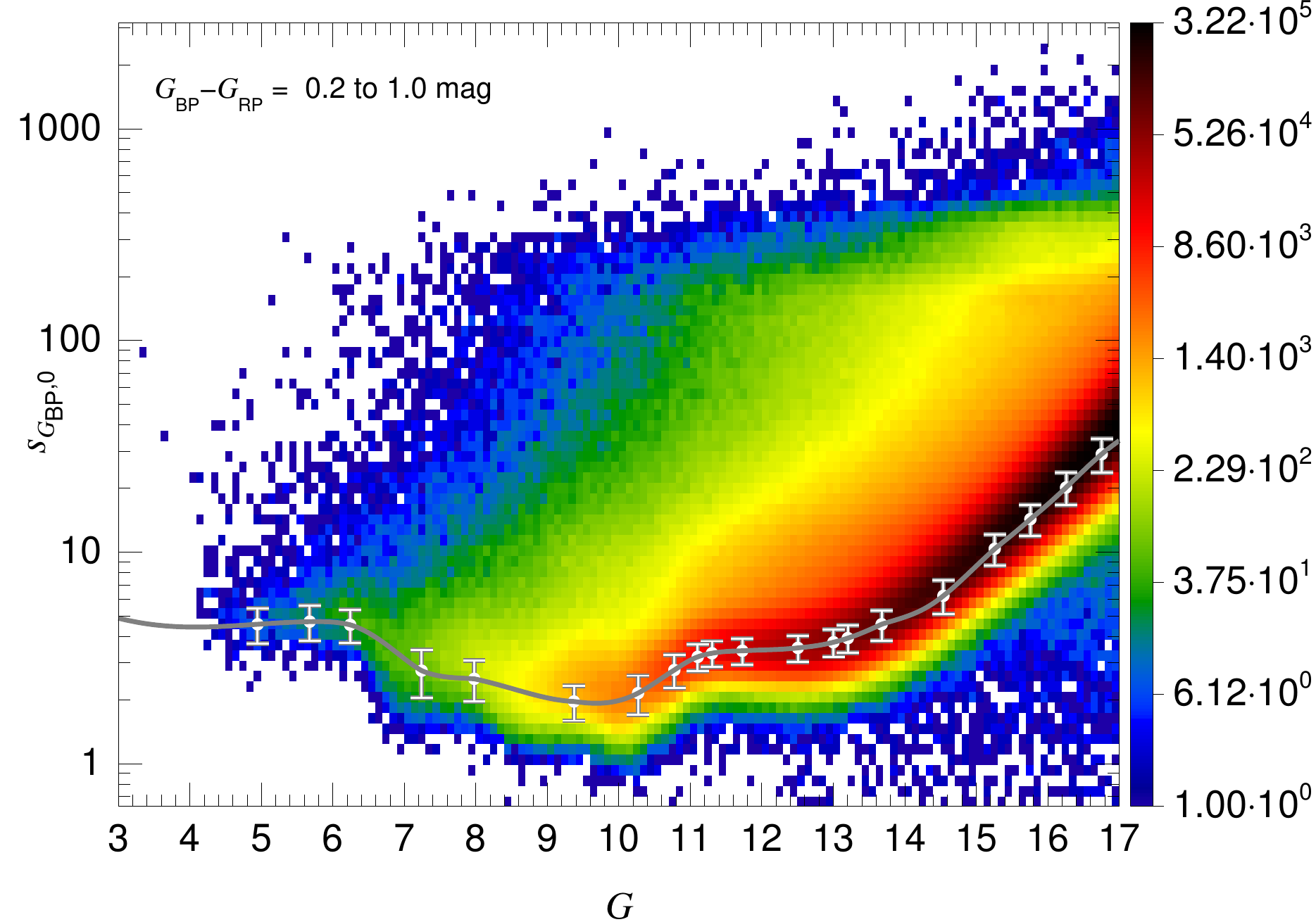}$\!\!\!$
                    \includegraphics[width=0.35\linewidth]{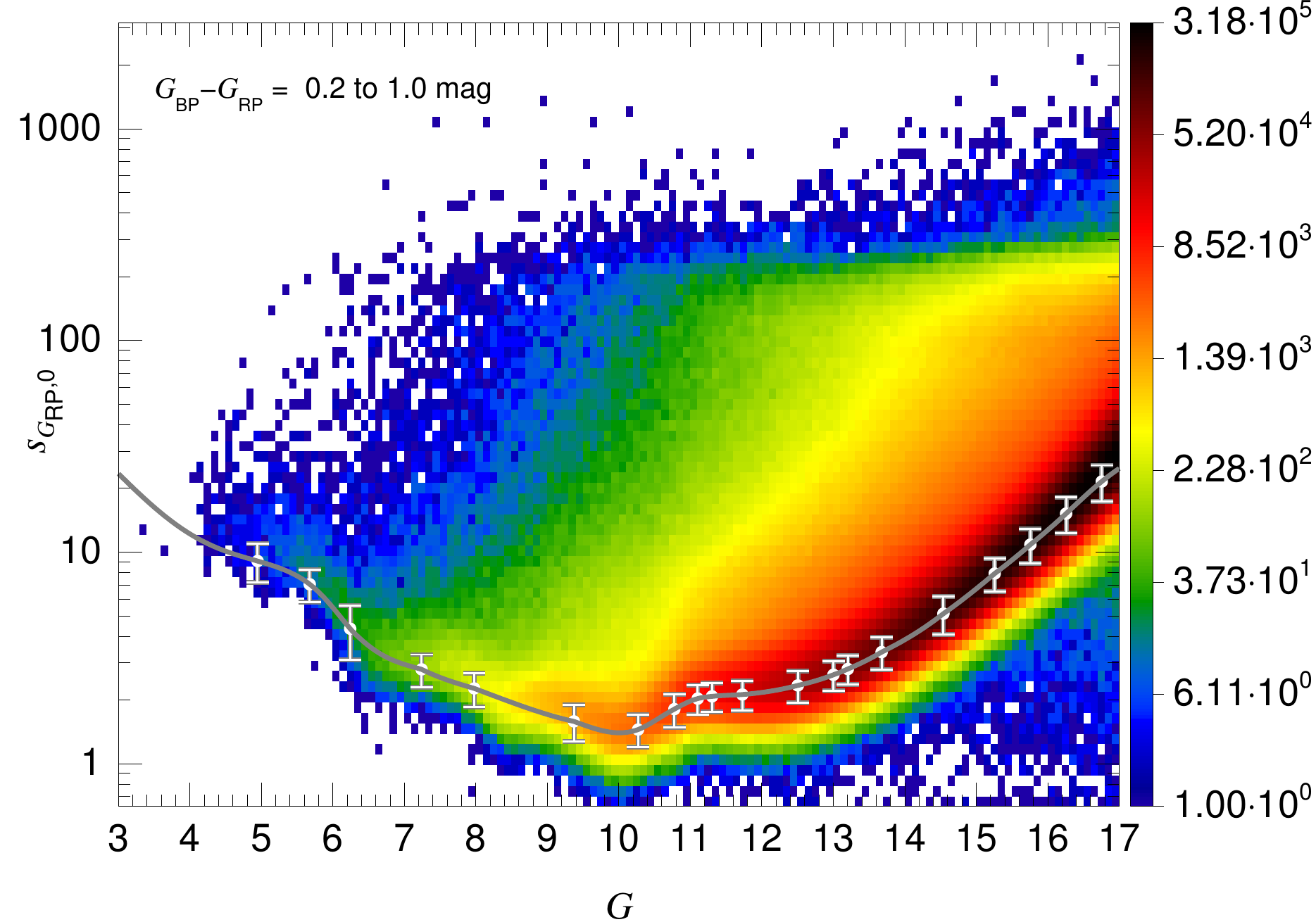}}
\centerline{$\!\!\!$\includegraphics[width=0.35\linewidth]{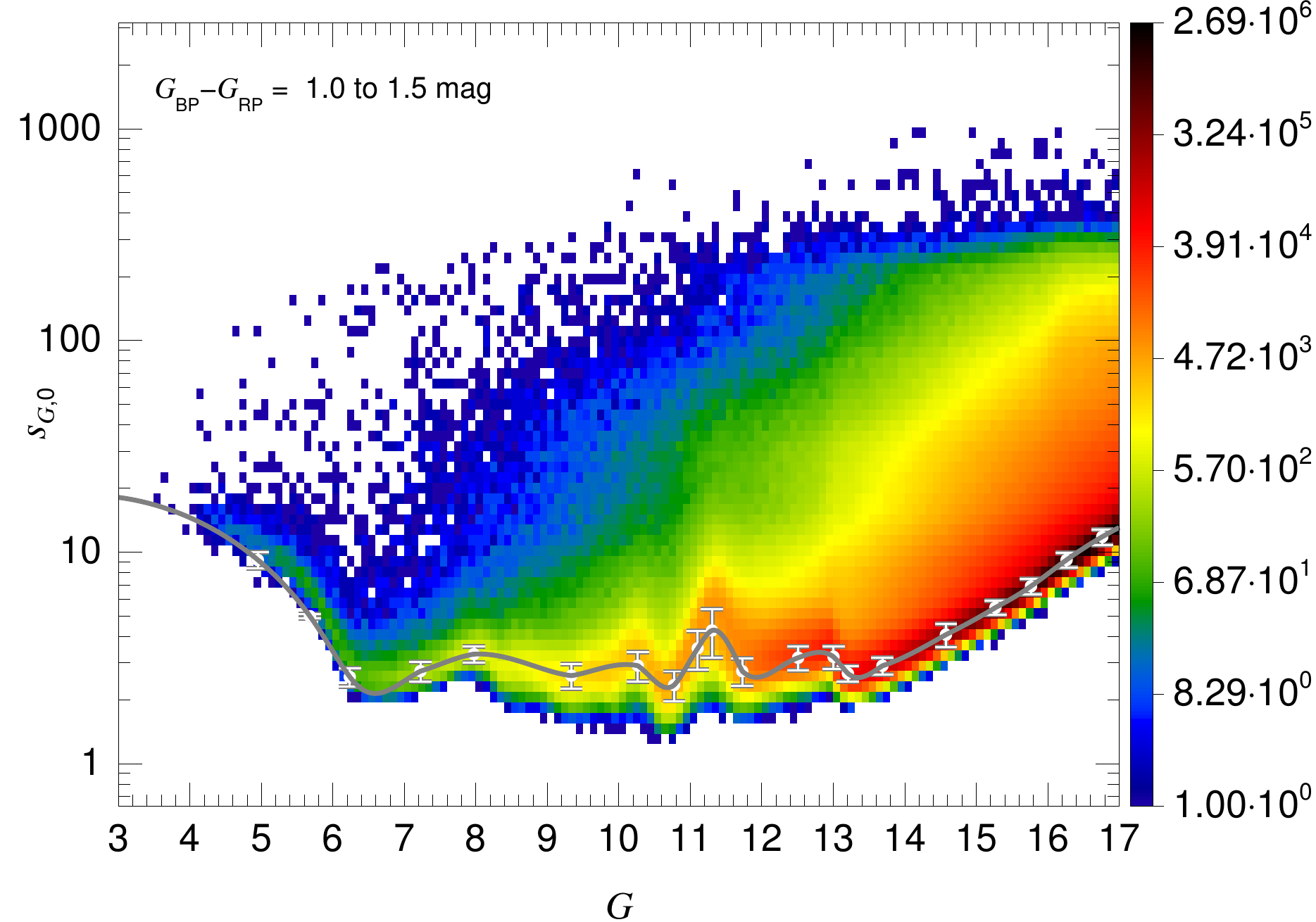}$\!\!\!$
                    \includegraphics[width=0.35\linewidth]{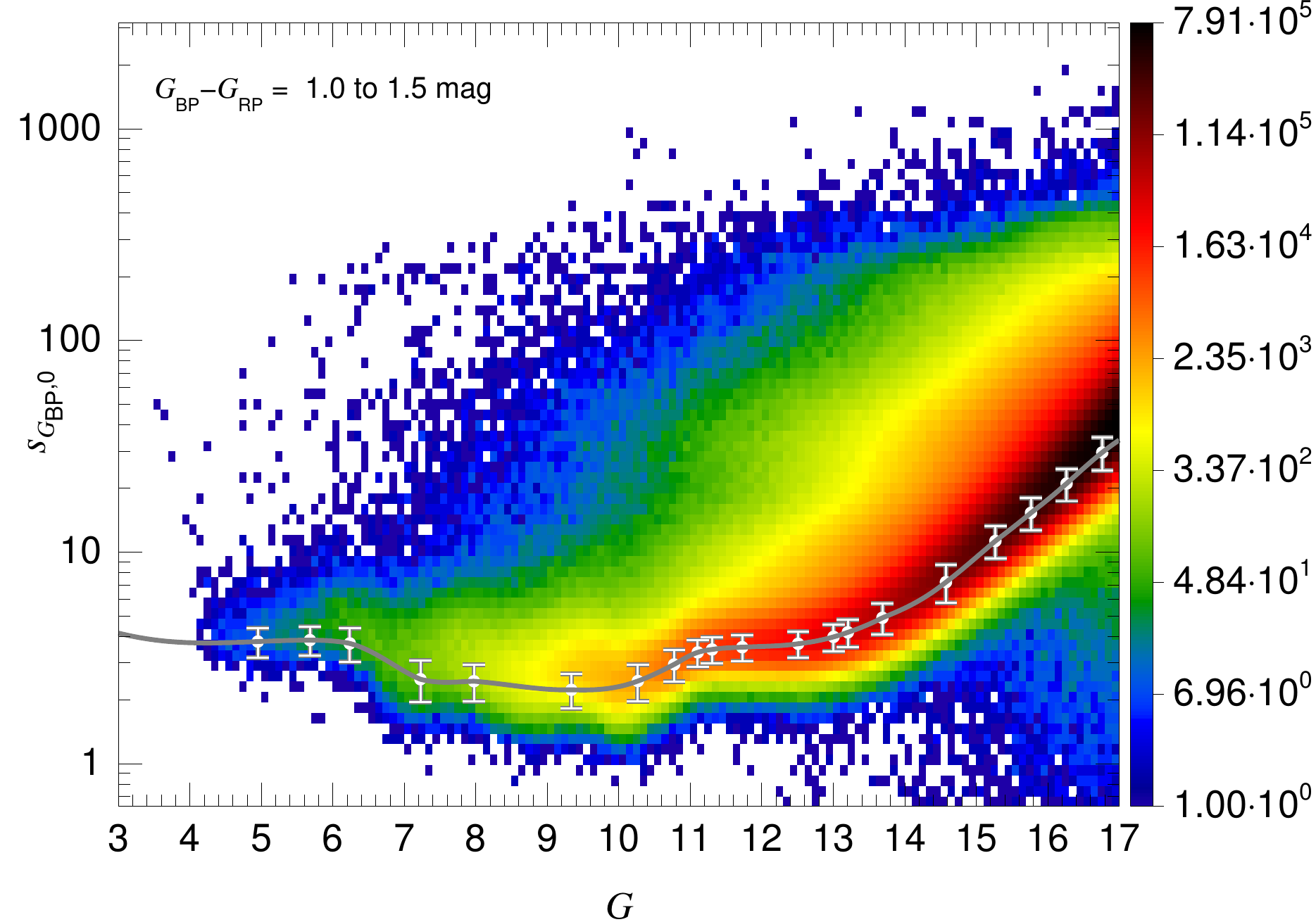}$\!\!\!$
                    \includegraphics[width=0.35\linewidth]{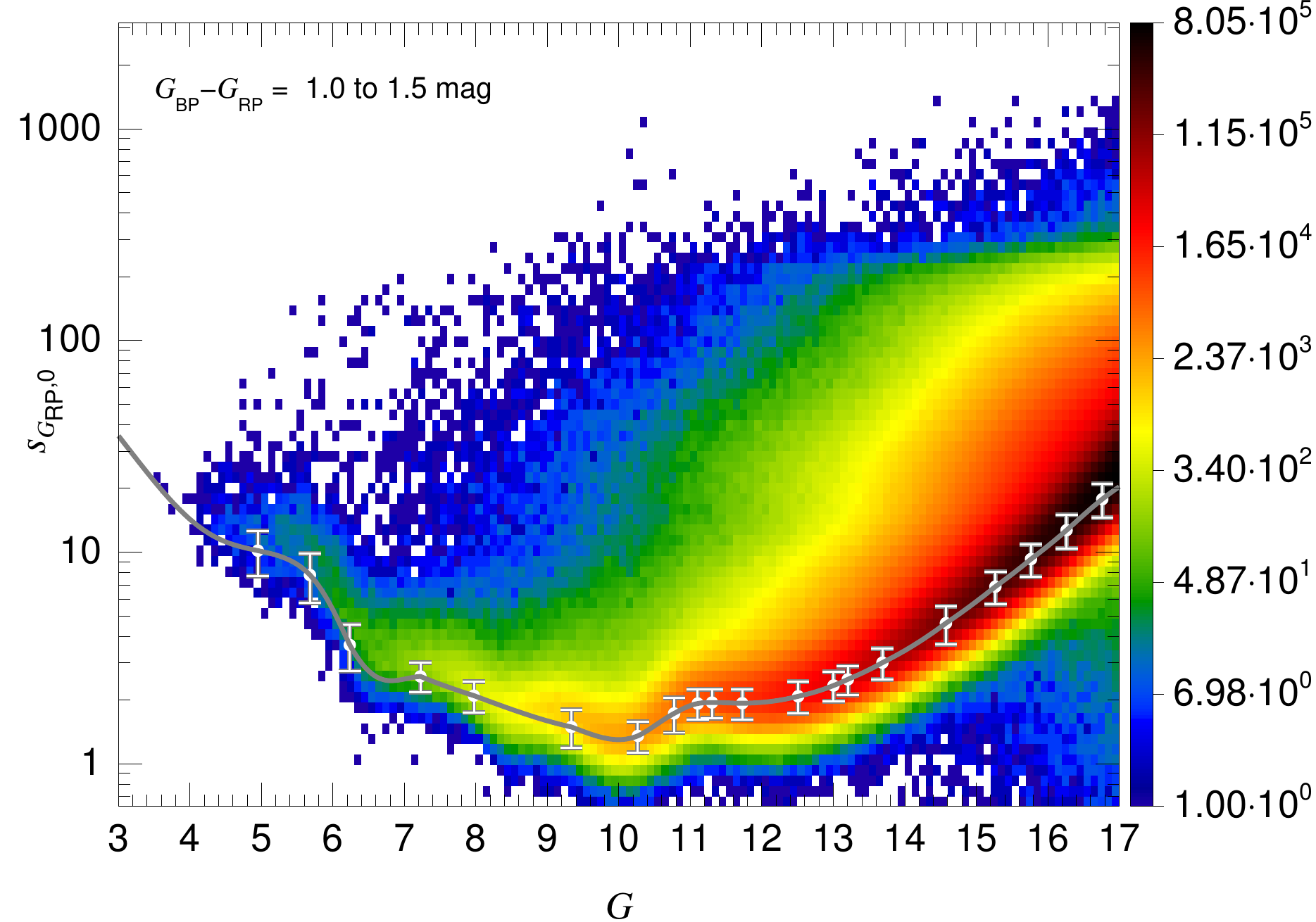}}
\centerline{$\!\!\!$\includegraphics[width=0.35\linewidth]{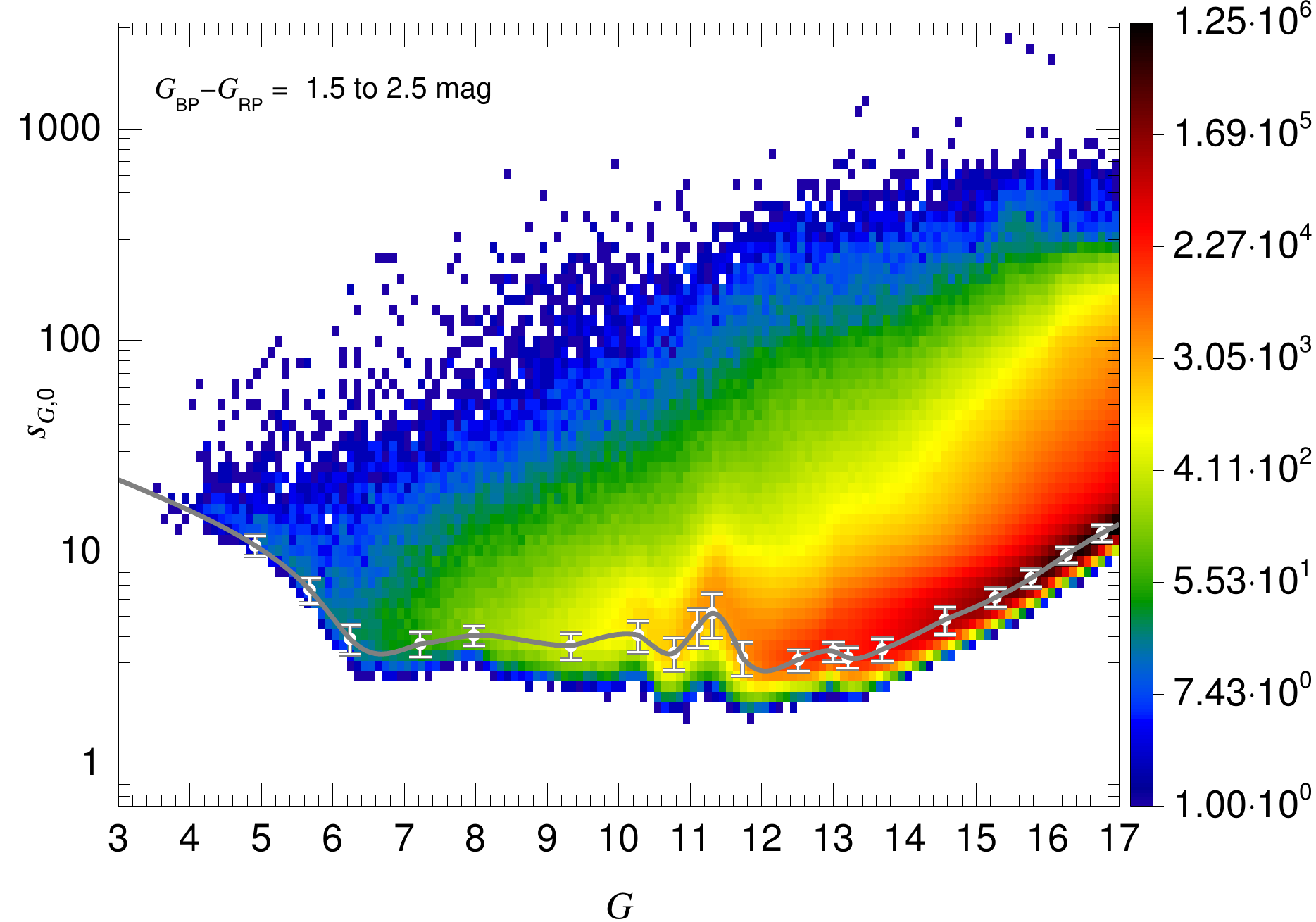}$\!\!\!$
                    \includegraphics[width=0.35\linewidth]{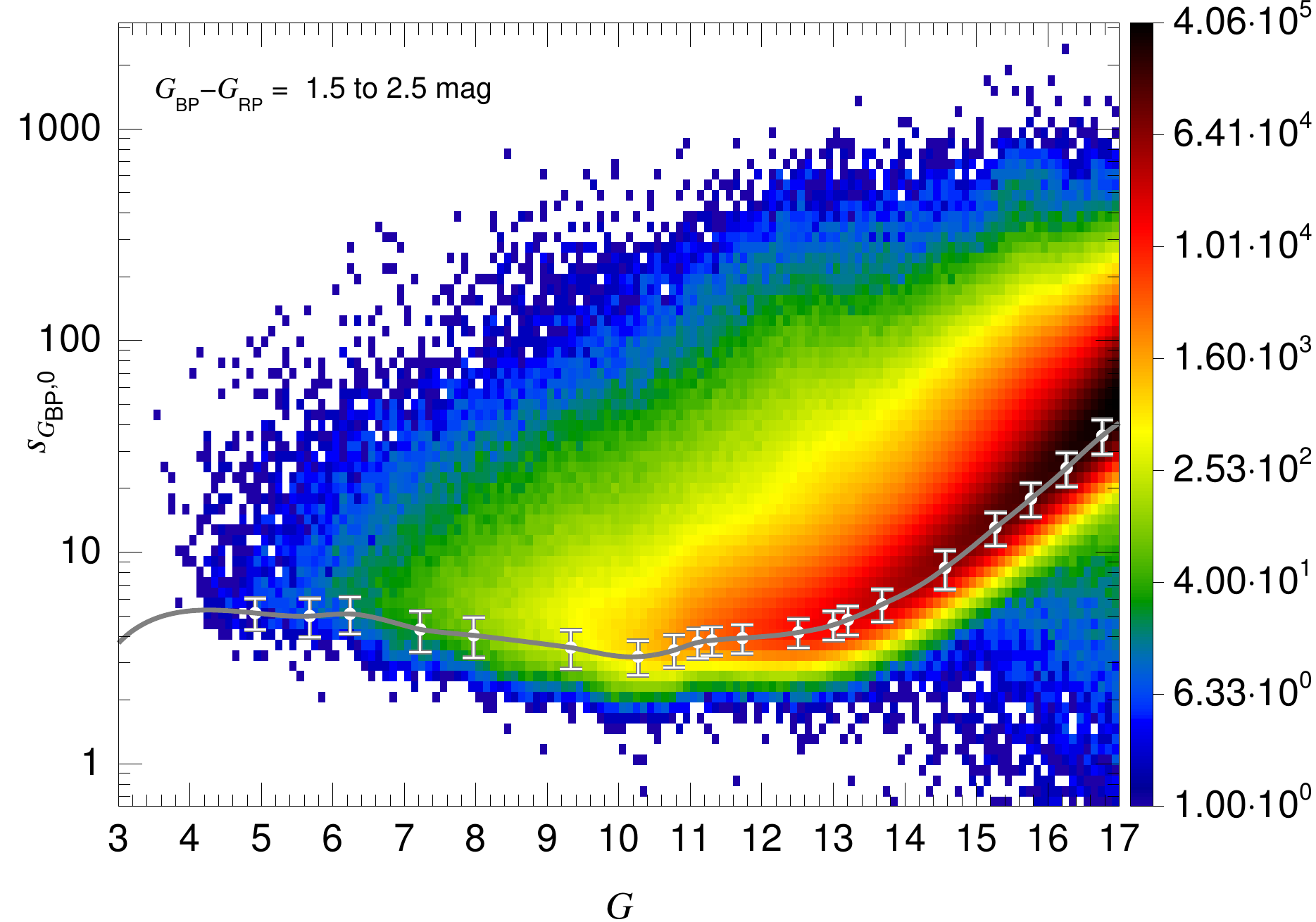}$\!\!\!$
                    \includegraphics[width=0.35\linewidth]{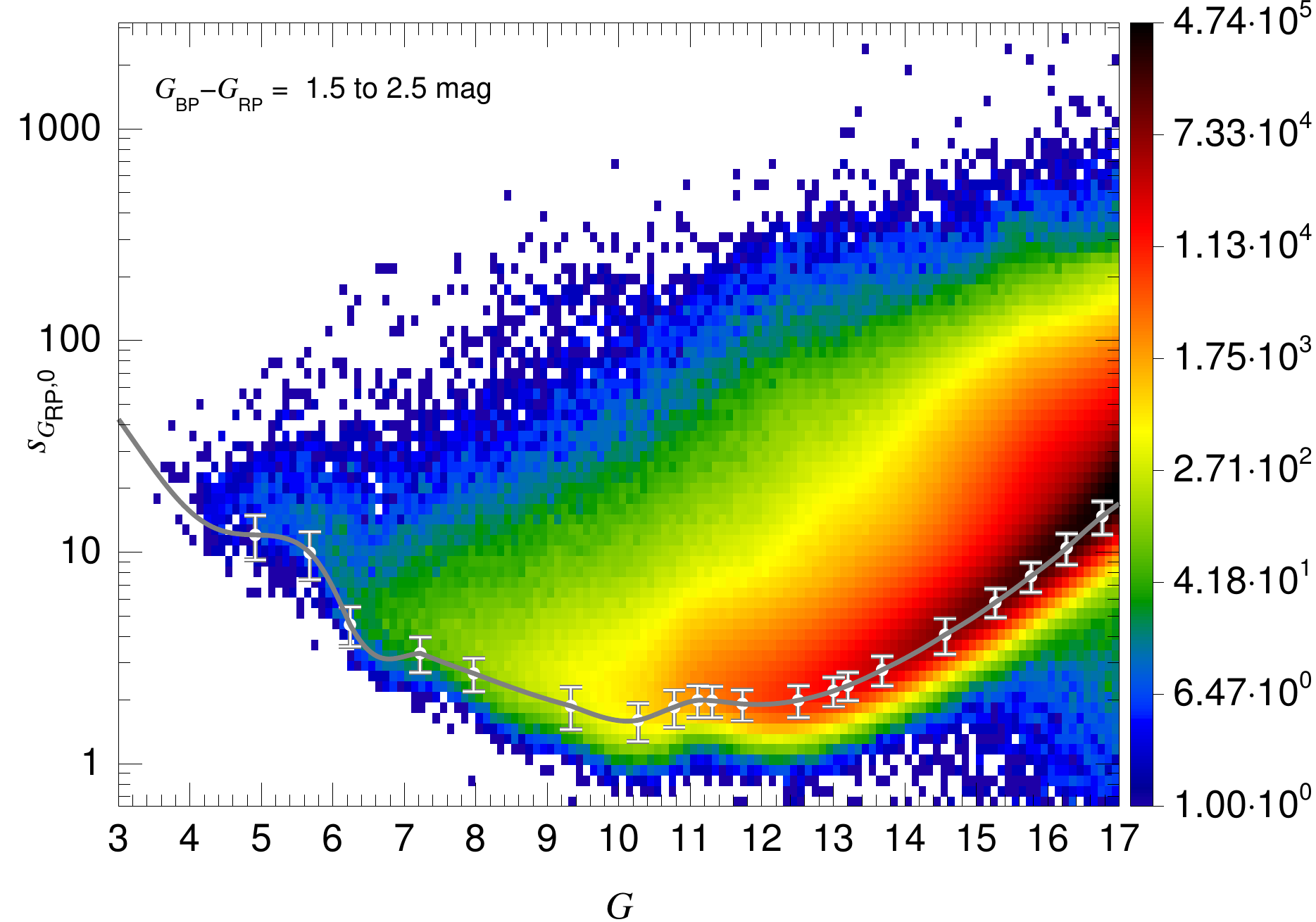}}
\centerline{$\!\!\!$\includegraphics[width=0.35\linewidth]{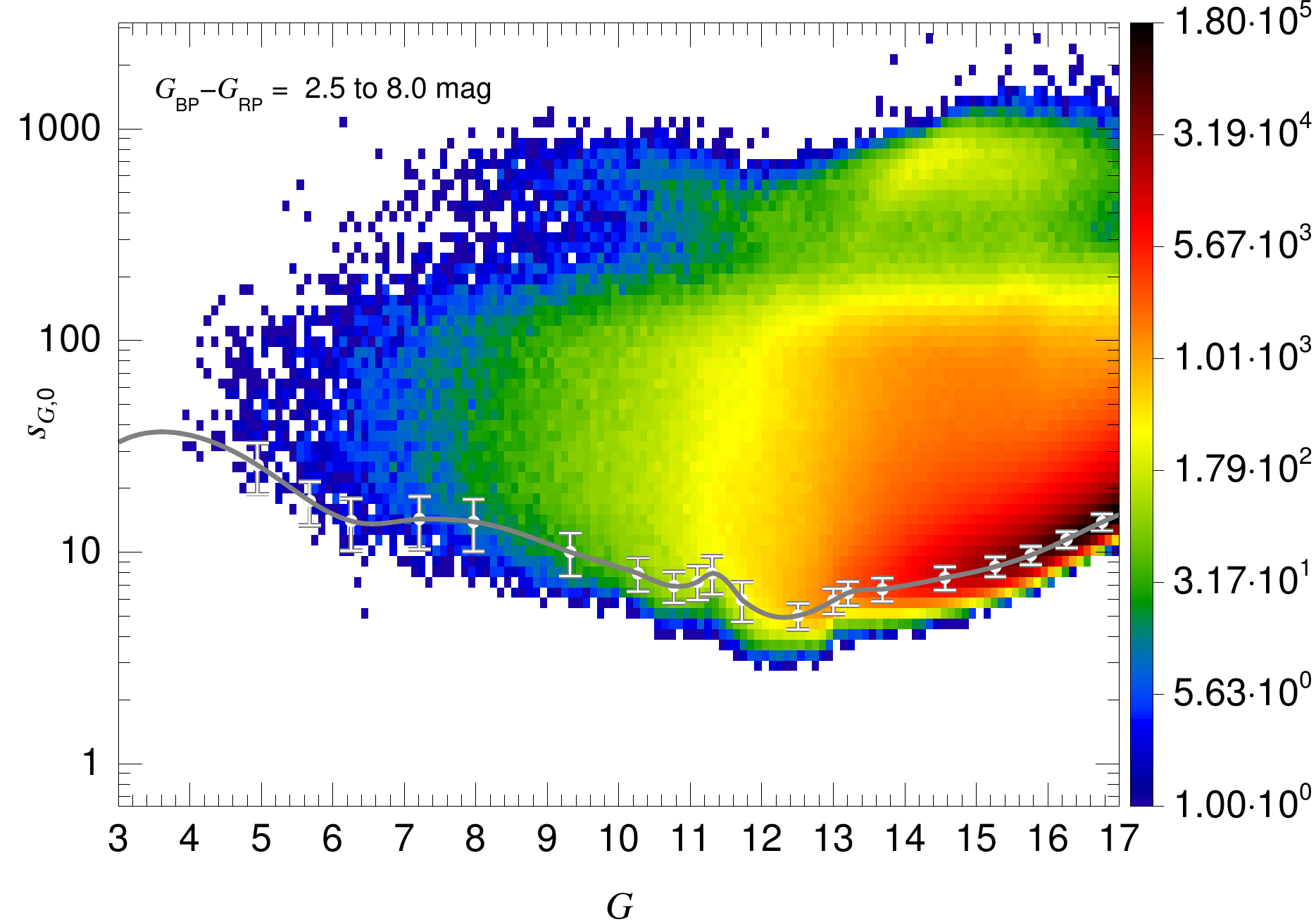}$\!\!\!$
                    \includegraphics[width=0.35\linewidth]{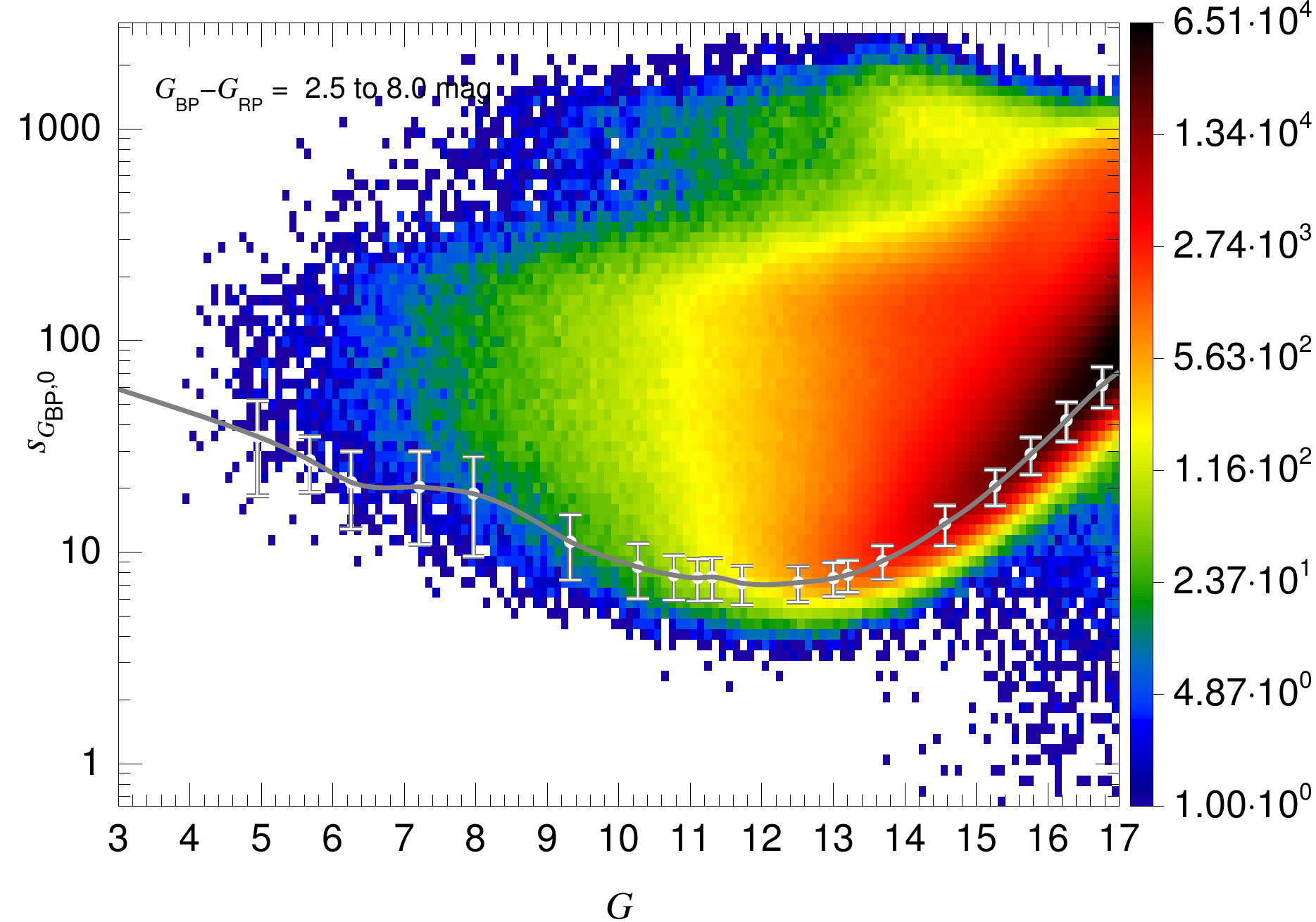}$\!\!\!$
                    \includegraphics[width=0.35\linewidth]{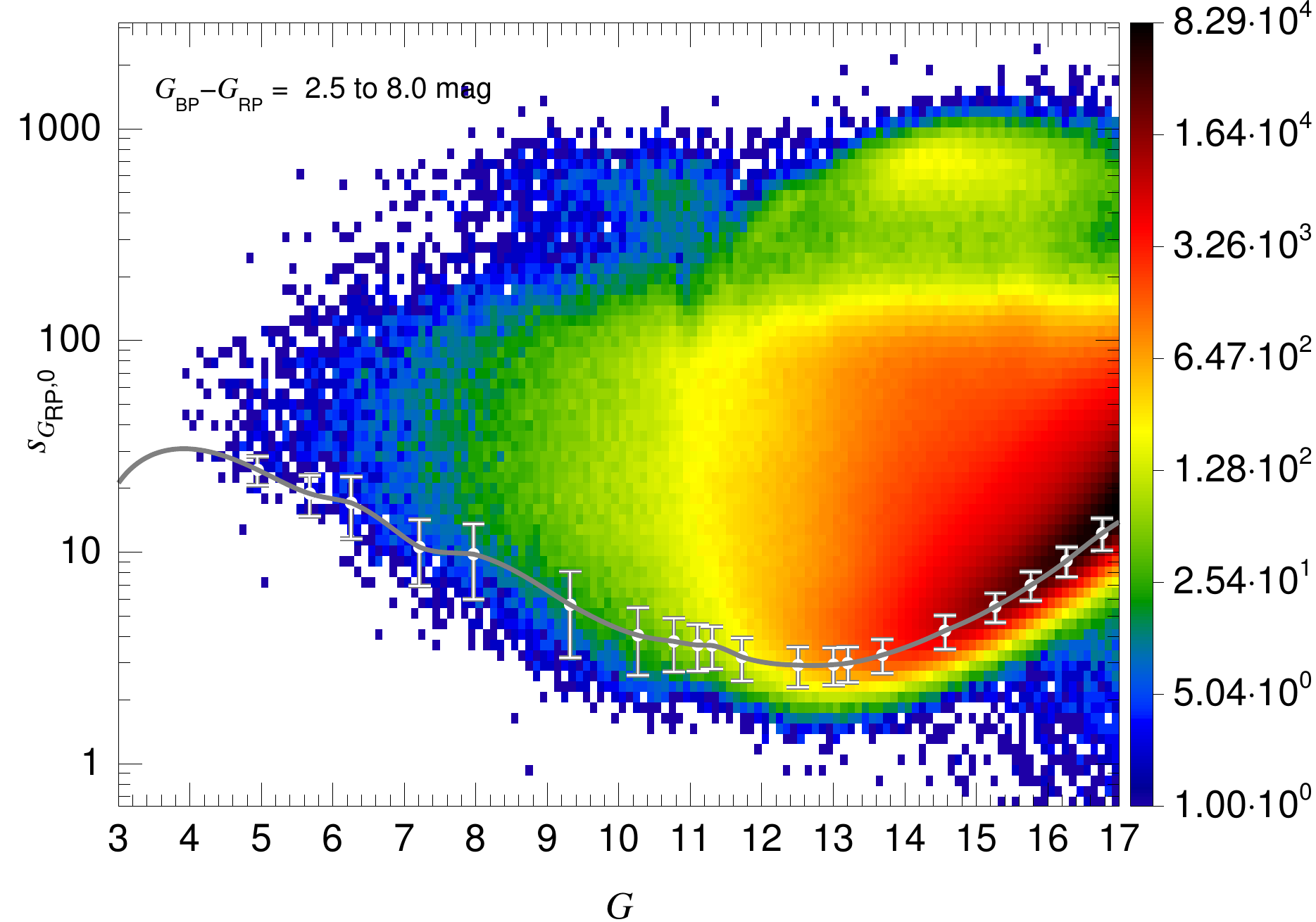}}
\caption{Observed photometric dispersions for \GG\ ({\it left column}), \GBP\ ({\it center column}), and \GRP\ ({\it right column}) 
         for the five \GBPmGRP\ ranges used in this work, from top to bottom: $-$1.0-0.2, 0.2-1.0, 1.0-1.5, 1.5-2.5, and 2.5-8.0 (in 
         mag). The plotted points and error bars are the measurements of the instrumental dispersions \sXi\ and their widths 
         \sigmasXi\ for each magnitude bin and the grey line joining them is a cubic spline fit. The horizontal axes are in mag and 
         the vertical ones in mmag.}
\label{hist_mag_sigma0}
\end{figure*}

\subsection{Parameter calculation}

$\,\!$\indent The published \textit{Gaia}~DR3 magnitudes are the weighted averages of a large number of individual measurements. 
Two measurements of the same star are not identical due to a combination of instrumental and astrophysical (or intrinsic)
variability. The first one can be approximated as a Gaussian distribution, keeping in mind the possibility of outliers (an issue we 
come back to below). As
for the astrophysical variability, the measured magnitudes can have quite different distributions, from symmetrical but
non-Gaussian in cases such as ellipsoidal variables to asymmetrical in most eclipsing binaries or bursting sources. 
However, under the reasonable assumption that both are independent, the relationship between the observed dispersion \sXz\ and the 
astrophysical \sX\ and instrumental \sXins\ dispersions for a band $X$ (\GG, \GBP, or \GRP) is given by:

\begin{equation}
s_{X{\rm,0}}^2 = s_X^2 + s_{X{\rm,ins}}^2,
\label{sX}
\end{equation}

\noindent where \sXz\ itself is the product of the listed magnitude uncertainty $\sigma_X$ by $\sqrt{N_X-1}$. $N_X$ is the 
number of observations used for photometry, as $\sigma_X$ is the (weighted) standard deviation of the mean of the distribution 
(Eqn.~5 in \citealt{Rieletal21}).

The obvious interest is the determination of the astrophysical dispersions \sG, \sGBP, and \sGRP, which requires the prior calculation 
of the instrumental component to be subtracted in quadrature. Those components are primarily a function of magnitude 
\citep{Rieletal21} but can also depend on other parameters such as color. Given the behavior seen in Fig.~14 of \citet{Rieletal21} 
and our equivalent Fig.~\ref{hist_mag_sigma0} (see also the plots in Appendix~E of R22), we proceed in the following way:

\begin{enumerate}
 \item We divide the sample in \GG-magnitude bins between 3~mag and 17~mag in order to capture the trends seen in 
       Fig.~\ref{hist_mag_sigma0}. In particular, the observed dispersions decrease as \GG\ increases for bright stars, then become 
       relatively flat (but with some fine structure) for a range of intermediate values of \GG\ (between 7~mag and 13~mag, 
       approximately), and finally increase between \GG\ of 13~mag and 17~mag. The behavior in the last section is the
       typical one caused by an increase in Poisson noise as stars become fainter. The behavior in the first two is caused by the
       complexities of the \textit{Gaia} CCD observations, including the use of different gate configurations as a function of
       magnitude \citep{Carretal16}. The 20 magnitude bins are selected to trace the structures seen in Fig.~\ref{hist_mag_sigma0}.
 \item We also divide the sample in five \GBPmGRP\ bins to capture the differences as a function of color, yielding a 
       total of 100 magnitude-color bins (Fig.~\ref{CMD_all}). We select the limits at \GBPmGRP\ of 0.2, 1.0, 1.5, and 2.5~magnitudes
       for several reasons, the main one being that we tested several options and those where the ones that better captured the 
       changes in behavior as a function of color while being representative of the different values of the effective wavenumber that 
       is used for astrometry (see Fig.~2 in \citealt{Lindetal21a}). In addition, the five ranges sample different astrophysical 
       populations that are mixed to a degree by extinction. The first range includes OBA stars, subdwarfs, and WDs. We note that 
       most Galactic OB stars are driven out of it by extinction, leaving mostly massive stars in the Magellanic Clouds in it. The 
       second range is dominated by FG stars in the main sequence (MS) and the third one by somewhat later (or more extinguished)
       MS stars but also a significant fraction of red clump (RC) stars, The fourth and fifth include a mixed bag of extinguished 
       versions of the above, M dwarfs, red giants, AGB stars, and YSOs, with the main difference between the two ranges being 
       caused by extinction.
 \item For each of those magnitude-color bins we assume that the distribution of astrophysical dispersions is given by a functional 
       form:
\begin{equation}
f(s_X) = \frac{A}{1 + (s_X/s_{X{\rm,c}})^\alpha},
\label{fsX}
\end{equation}
       where $A$ is a normalizing constant, $\alpha$ is the exponent that characterizes a power-law behavior at large values of
       the dispersion, and \sGc\ is a characteristic value where the distribution flattens to adjust the proportion between
       low-dispersion and high-dispersion values. Below we discuss how well this model describes reality.
 \item The instrumental dispersion \sXins\ in a given magnitude-color bin can be characterized by a Gaussian average value \sXi\ and a 
       width \sigmasXi. The explanation for such a width is that even if we make the bins relatively small, a given sample is 
       bound to contain objects with a range of magnitudes, colors, and number of observations (among other characteristics) that 
       generate slightly different instrumental effects. We use \sigmasXi\ as the uncertainty on \sXi\ when 
       computing \sX\ using Eqn.~\ref{sX} and we verify that $\sigmasXi \ll \sXi$ for all magnitude-color bins. The
       distribution of total dispersions is formed then by applying Eqn.~\ref{sX} in Eqn.~\ref{fsX} and convolving the result
       with the Gaussian distribution in \sXins.
\end{enumerate}

The above procedure leads us to fit the distribution of observed dispersions in each magnitude-color bin with a function of five 
parameters: $A$, \sXc, $\alpha$, \sXi\, and \sigmasXi. We wrote an IDL procedure to do the fits and the results are 
listed in Tables~\ref{fit_results_G},~\ref{fit_results_BP},~and~\ref{fit_results_RP}, where we give the total number of stars $N$ 
instead of $A$ as a normalizing constant to ease interpretation. The corresponding plots are shown in 
Fig.~\ref{hist_sigma0} (each one of them is a vertical slice of one of the plots in Fig.~\ref{hist_mag_sigma0}), where we also 
show the histograms for the three most common variable types and the rest of the variables from R22 in each case, 
noting that only a fraction of the sample (given in each plot) has a variability type from that paper. The plots are 
useful to estimate how many of the variable stars are in R22 and which are the dominant types of variables across 
the \textit{Gaia} color-magnitude diagram (CMD). 
Further below we analyze the distribution of variable stars across the \textit{Gaia} CAMD but for the
time being we stick to the CMD, as our initial purpose is to analyze instrumental effects that depend on magnitude, not on absolute
magnitude. We note that the histograms in Fig.~\ref{hist_sigma0} have a constant logarithmic bin size in the $x$ axis. 
This is done for display purposes but
the fits themselves were performed using variable bin sizes that reduce the differences in the number of stars per bin, as 
using a constant bin size can lead to biases in the fit \citep{MaizUbed05}, which in this case would affect mostly the value of
$\alpha$.

\subsection{Cataloging}

$\,\!$\indent The main result of this paper is a catalog with information on the \num{145677450} \textit{Gaia}~DR3 sources
described above that will be made available through the CDS. Here we describe its content divided in categories:

\paragraph{\textit{Gaia DR3} main catalog.} This is the basic information for each object copied directly from the catalog:
ID, coordinates, \GG\ magnitude, and \GBPmGRP\ color.

\paragraph{Parallaxes and absolute magnitudes.} We list the corrected \textit{Gaia}~DR3 parallax \pic\ and parallax uncertainty \spic\
applying the procedure in \citet{Maizetal21c} and \citet{Maiz22}. In particular, the zero point for objects with $\GG < 13$~mag
is significatively different than the one from \citet{Lindetal21b}, uncertainties are considerably larger (by up to a factor of 3)
than the ones listed in the \textit{Gaia} catalog, and an additional correction factor is applied to those objects with RUWE larger
than 1.4. In addition, for objects within 10\degr\ of the center of the LMC or SMC according to \citet{Lurietal21}, we substitute
their parallaxes and uncertainties by $20.2\pm 0.2$~mas and $15.9\pm 0.6$~mas, using the values from \citet{Pietetal19} and 
\citet{Cionetal00}, respectively\footnote{Subsequent to \citet{Lurietal21}, \citet{JimAetal23b,JimAetal23a} have produced new
membership lists for the LMC and SMC, but for stars within 10\degr\ of the respective galactic centers and with $\GG \le 17$ the
differences are minor.}. To simplify the differentiation between the Milky Way (MW), LMC, and SMC we also generate a
column labelled M, L, or S according to the membership to each galaxy, assigning to the MW anything outside of the 10\degr\ radius of 
the \citet{Lurietal21} sample. The MW sample defined in that way actually includes some distant members of the Magellanic Clouds and 
objects from other galaxies such as M31, but those are a small part of the {\it Gaia}~DR3 sample with $\GG \le 17$~mag. Finally, for 
objects with $\pic/\spic > 5$ (a significantly more restrictive criterion that its equivalent with uncorrected values) we compute the
absolute magnitude \Gabs\ using as distance the inverse of the parallax, which may introduce small biases
\citep{Maiz01a,Maiz05c,Lurietal18}. There are \num{105576353} such objects (72.47\% of the sample), of which \num{413462} are in the 
LMC and \num{83025} in the SMC.

\begin{figure*}
\centerline{$\!\!\!$\includegraphics[width=0.35\linewidth]{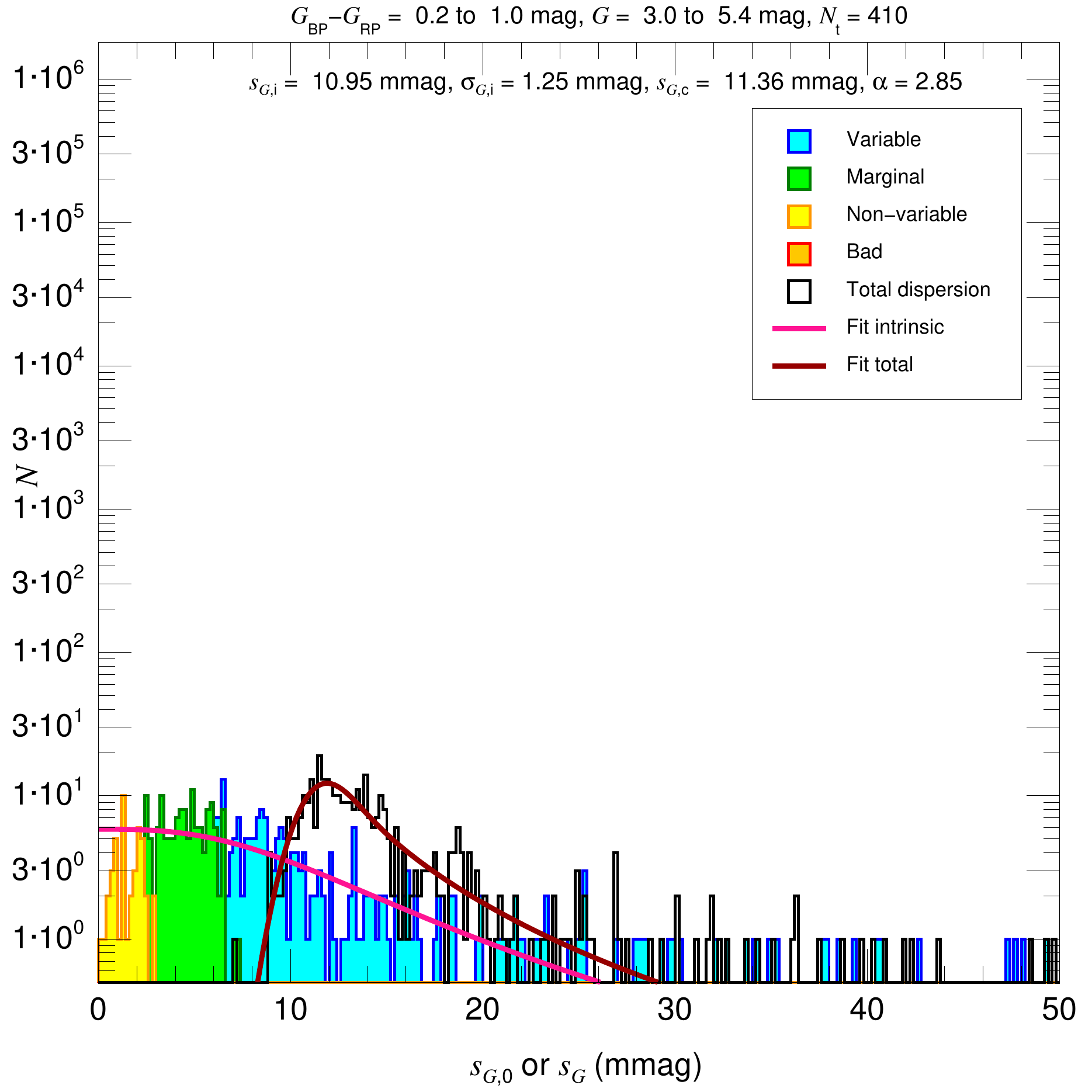}$\!\!\!$
                    \includegraphics[width=0.35\linewidth]{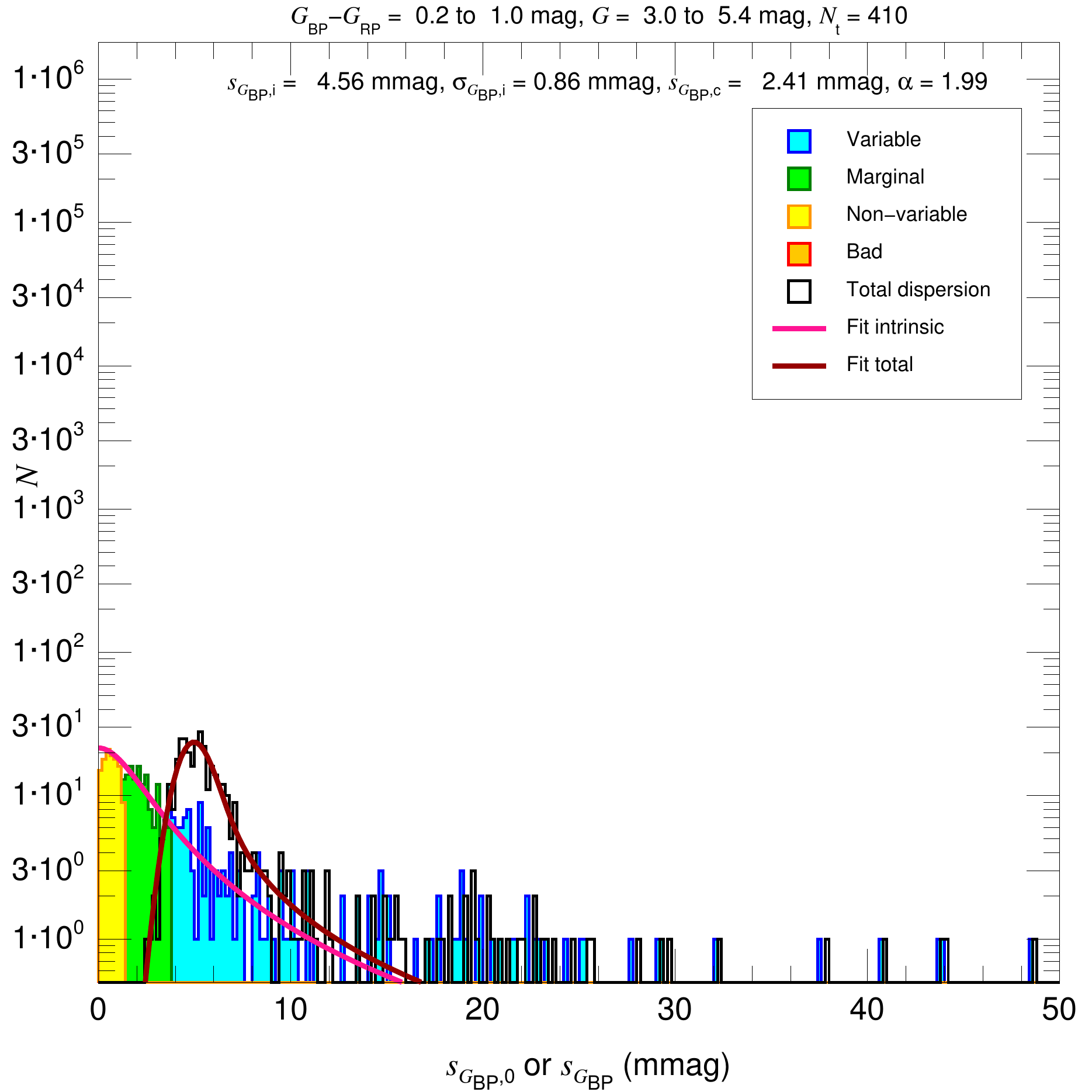}$\!\!\!$
                    \includegraphics[width=0.35\linewidth]{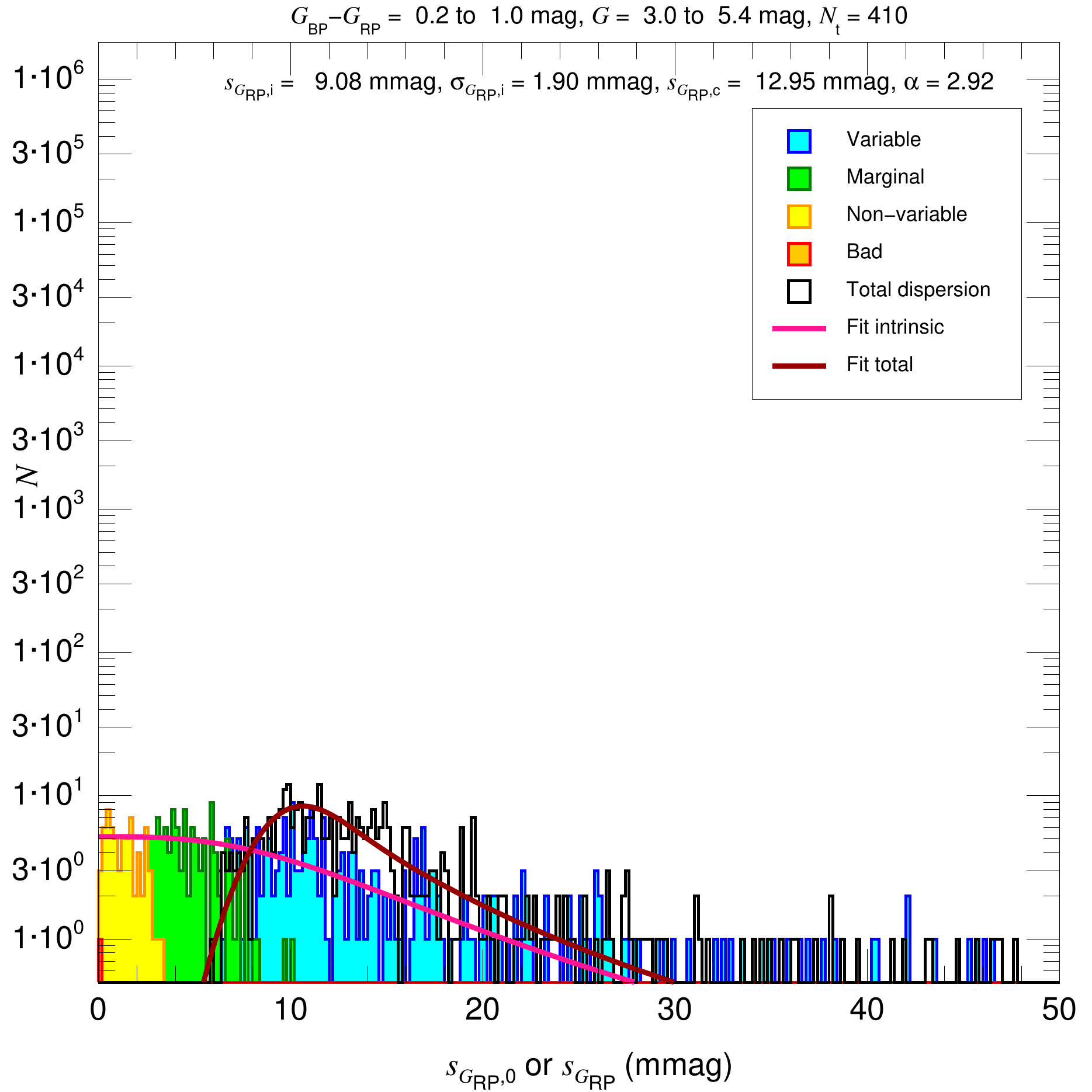}}
\centerline{$\!\!\!$\includegraphics[width=0.35\linewidth]{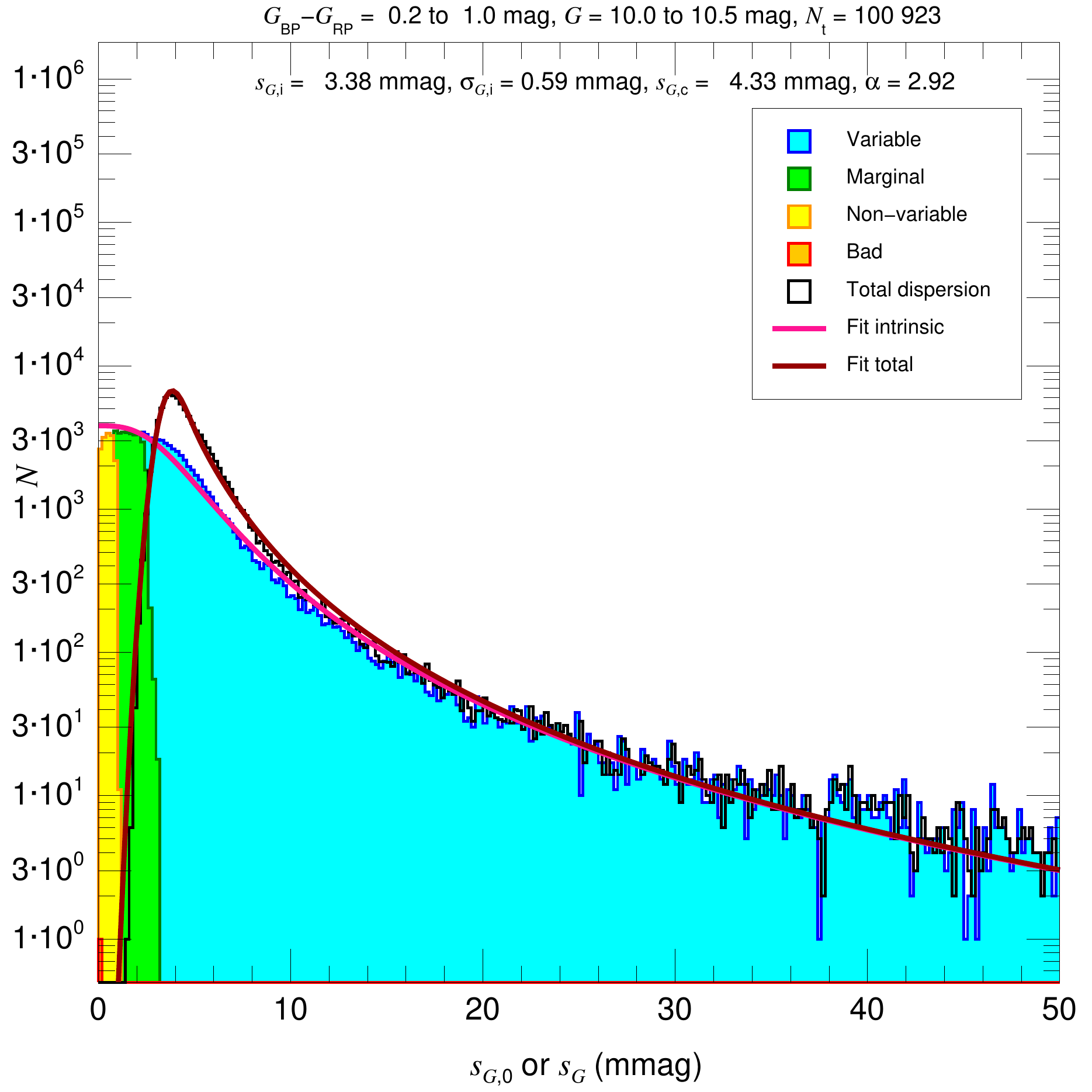}$\!\!\!$
                    \includegraphics[width=0.35\linewidth]{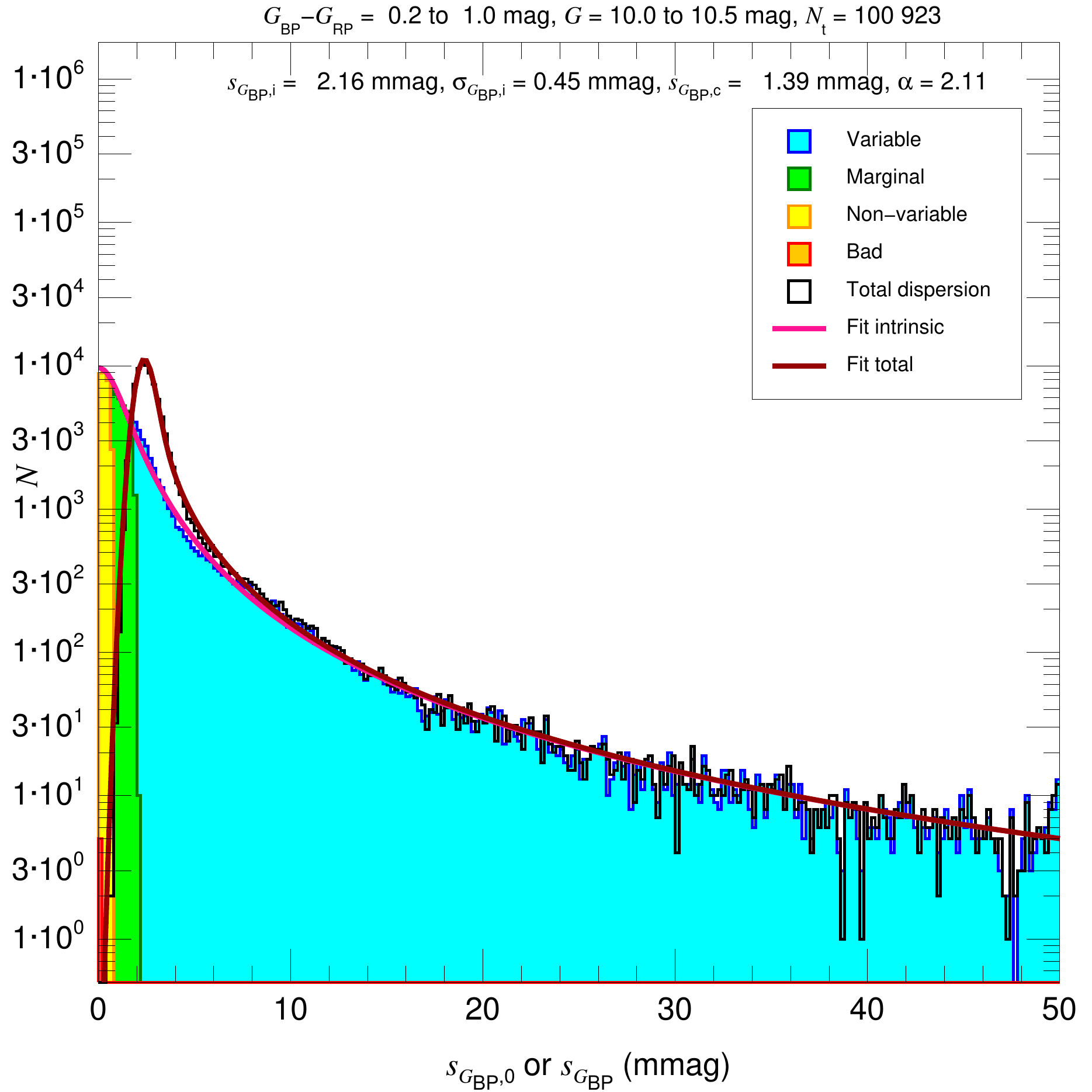}$\!\!\!$
                    \includegraphics[width=0.35\linewidth]{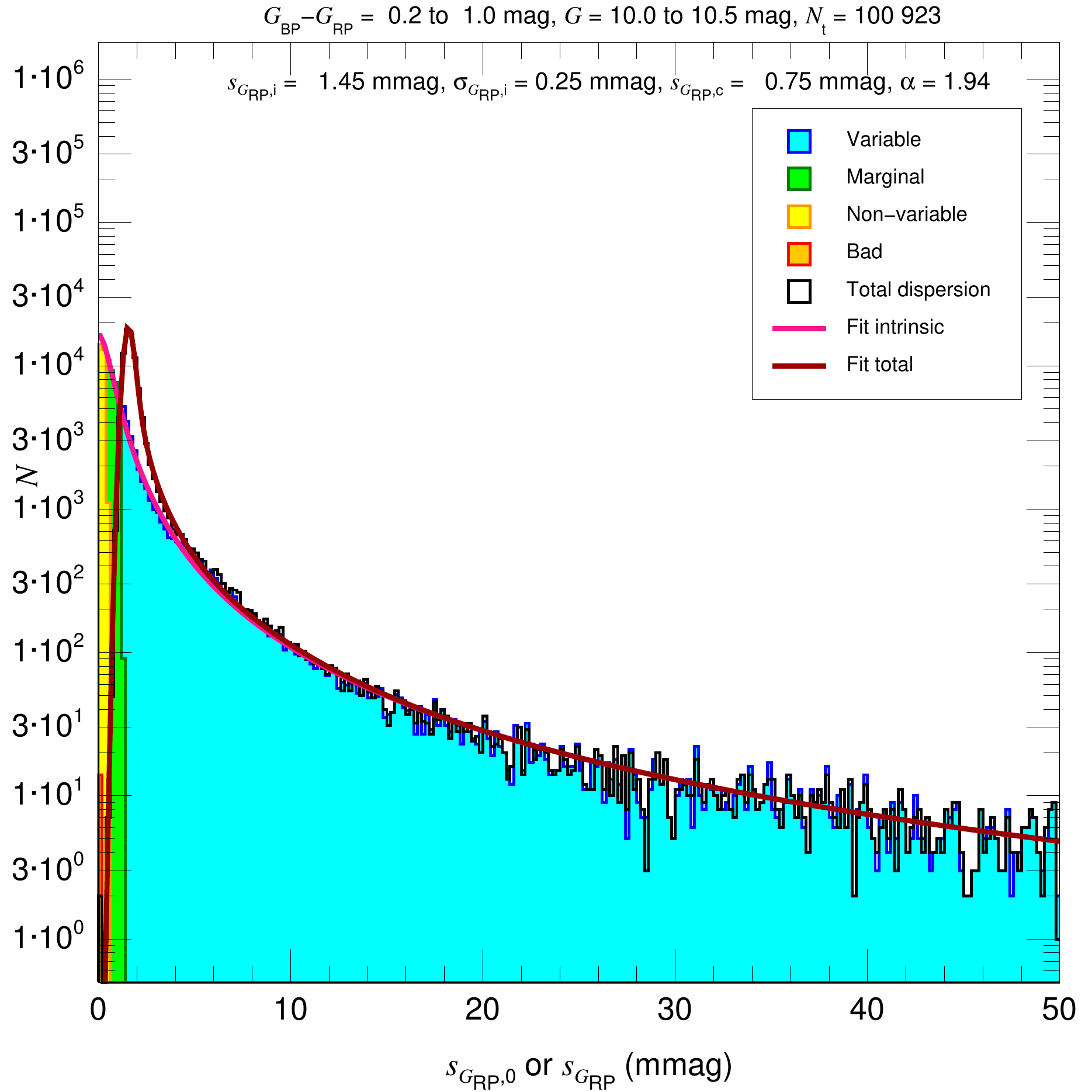}}
\centerline{$\!\!\!$\includegraphics[width=0.35\linewidth]{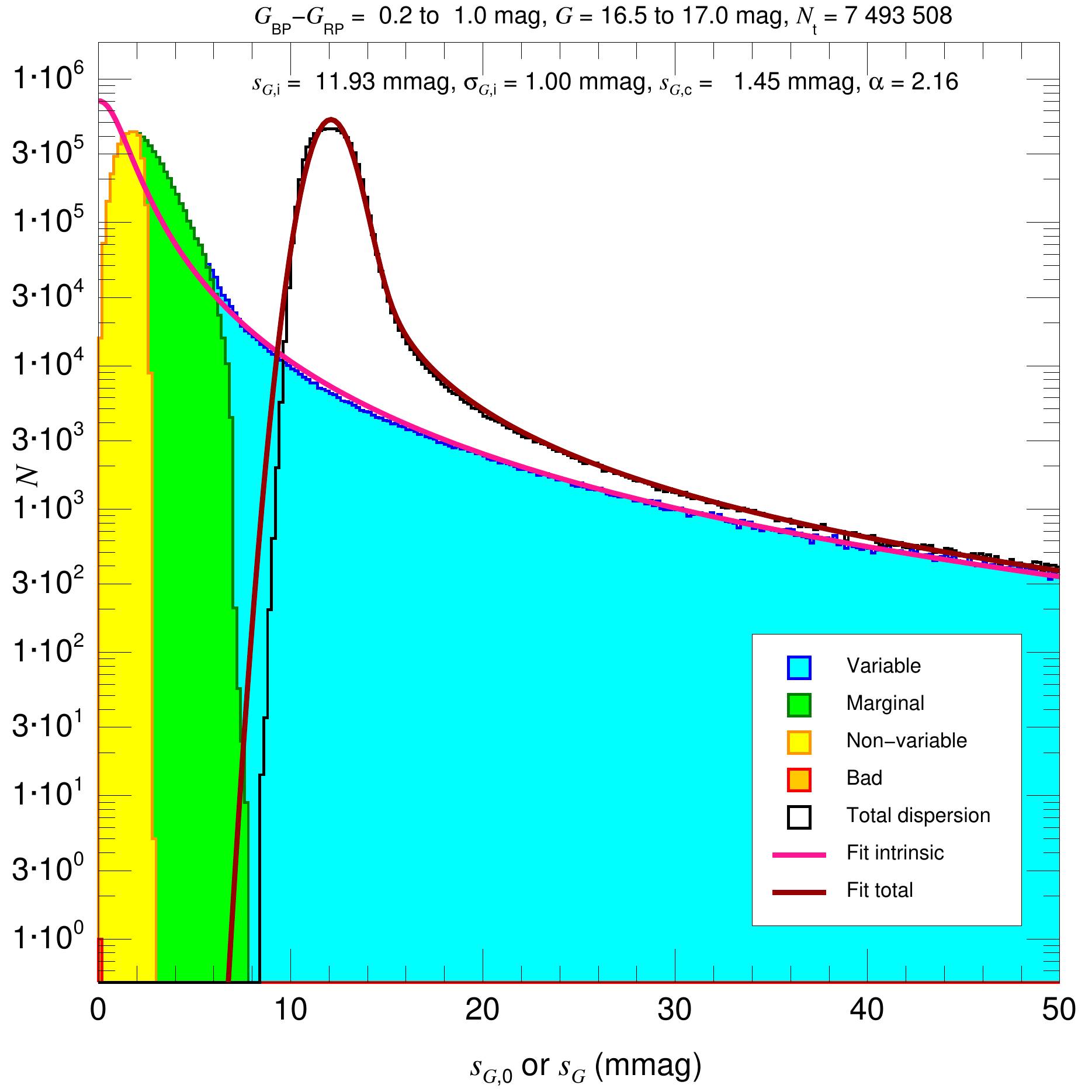}$\!\!\!$
                    \includegraphics[width=0.35\linewidth]{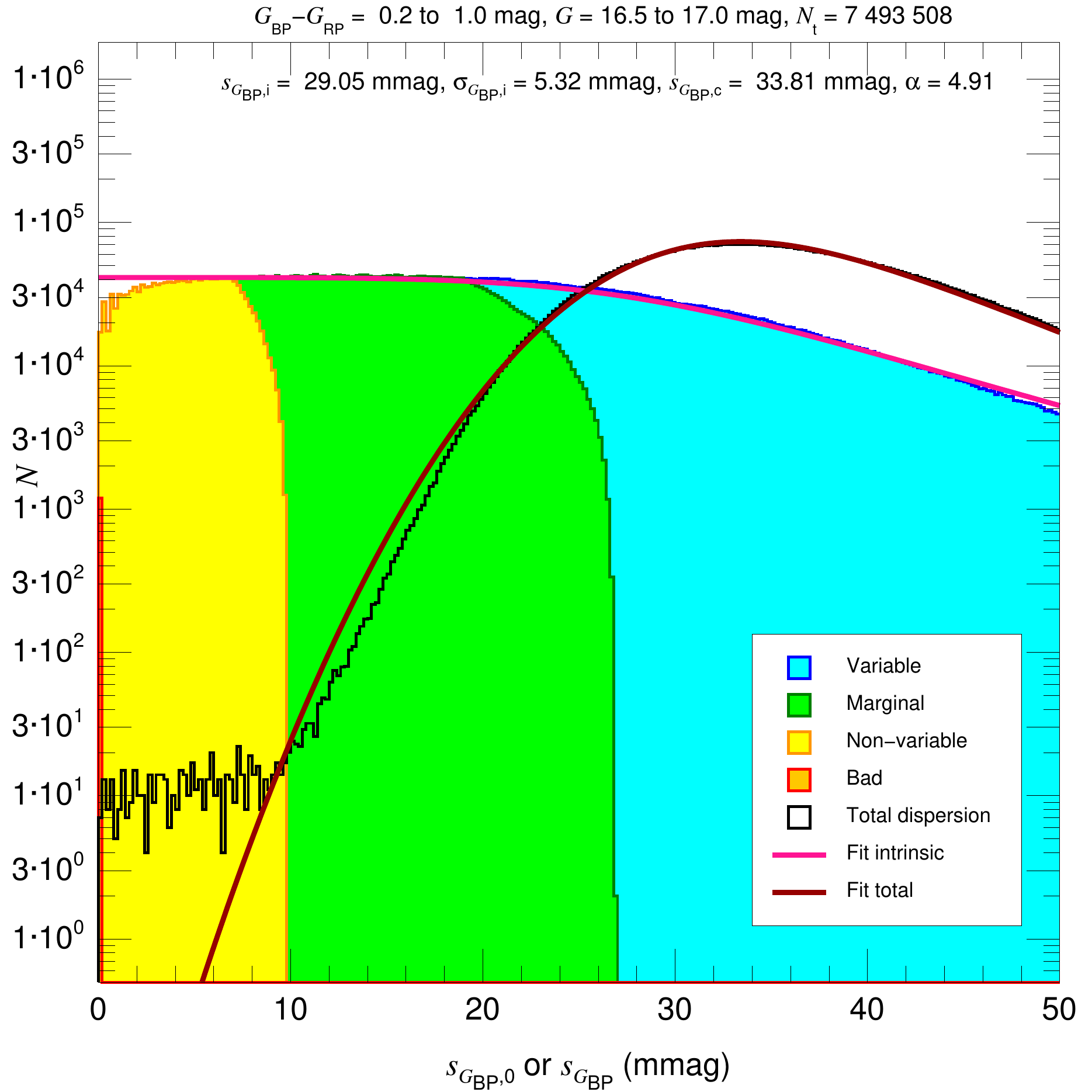}$\!\!\!$
                    \includegraphics[width=0.35\linewidth]{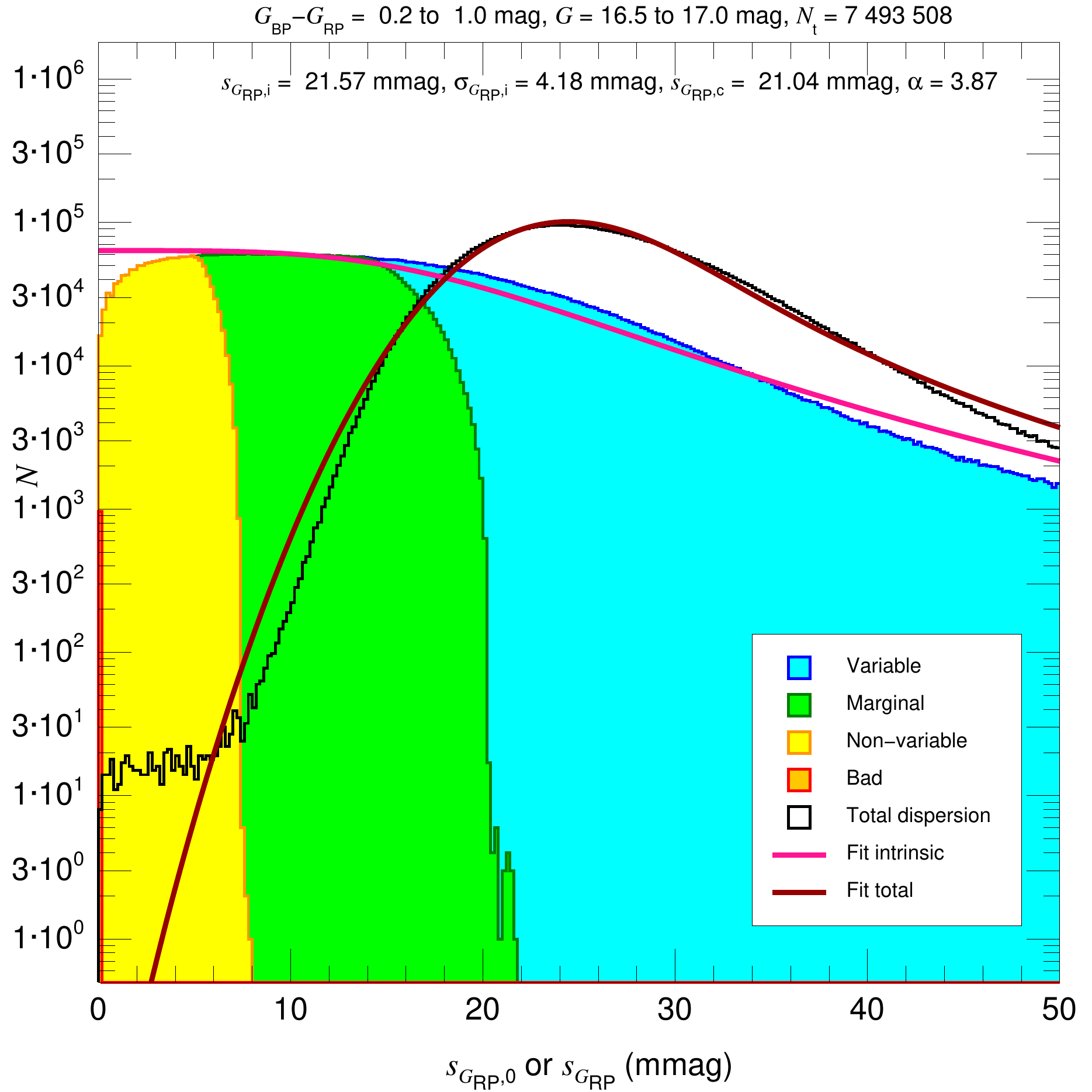}}
\caption{Selected astrophysical dispersion histograms. The three columns show the histograms for \GG\ ({\it left}), \GBP\ 
         ({\it center}) and \GRP\ ({\it right}) and the rows show different magnitude ranges: 3.5-5.4 ({\it top}), 10.0-10.5 
         ({\it middle}), and 16.5-17.0 ({\it bottom}). The \GBPmGRP\ range is always 0.2-1.0 mag. Different colors are 
         used to fill each astrophysical dispersion histogram to show the values of the variability flag for each 
         filter (see text), note that there is one orange bin at most per panel. In addition, an empty histogram shows the 
         corresponding total dispersion histogram. Finally, two lines of different color indicate the fitted function in the form of 
         the astrophysical (pink) and total (brown) dispersions, respectively. The text at the top of each panel gives the results of 
         the fits.}
\label{hist_sigma}
\end{figure*}

\paragraph{Astrophysical dispersions.} For each object in each band $X$ (\GG, \GBP, or \GRP) we give the \sXz\ directly 
computed from the catalog data and \sX, \sigmasX, and \sXi\ obtained from the previous subsection. More specifically, we
use cubic splines in \GG\ for each of the five \GBPmGRP\ bins to calculate the value of \sXi\ and \sigmasXi\
for each of the objects in the sample (see Fig.~\ref{hist_mag_sigma0} for a graphical representation of how this is done). For the 
regions around the \GBPmGRP\ boundaries (0.2, 1.0, 1.5, and 2.5~mag) we linearly interpolate between the cubic splines for the two color
bins, noting that the dependence on \GG\ is significantly stronger than that on \GBPmGRP. We finally use Eqn.~\ref{sX} to obtain \sX\ and
its uncertainty \sigmasX\ by assuming a Gaussian distribution of instrumental uncertainties with mean \sXi\ and width \sigmasXi. The 
error propagation has to be done using the whole distribution, as the standard formula that uses only the first derivative breaks down 
for small values of the dispersion.

\paragraph{Variability information from \textit{Gaia DR3}.} We list three variability-related columns collected from 
\textit{Gaia}~DR3: Two flags from the main catalog, the epoch photometry and the variability tag, and the variable type from R22.

\begin{figure*}
\centerline{$\!\!\!$\includegraphics[width=0.35\linewidth]{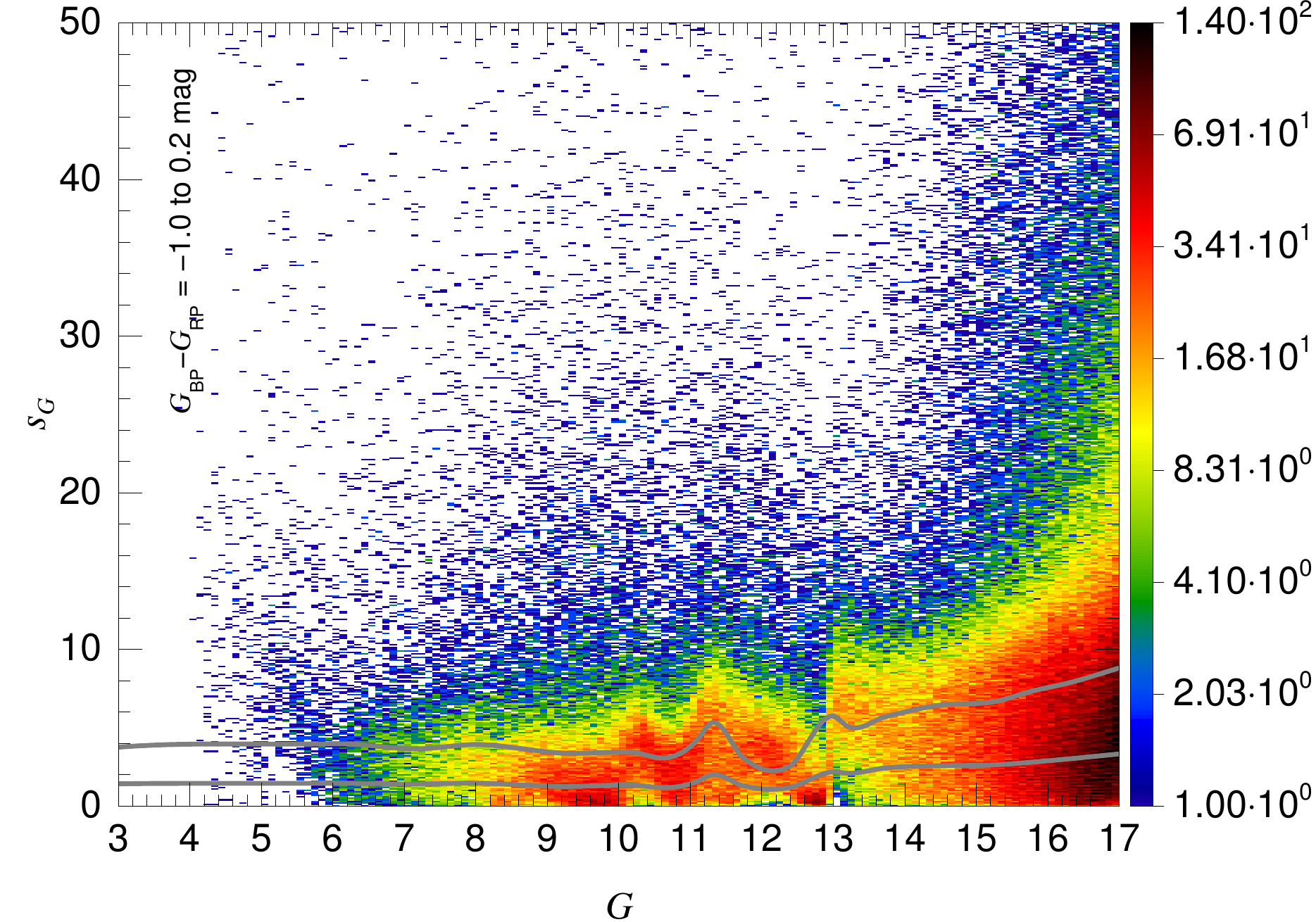}$\!\!\!$
                    \includegraphics[width=0.35\linewidth]{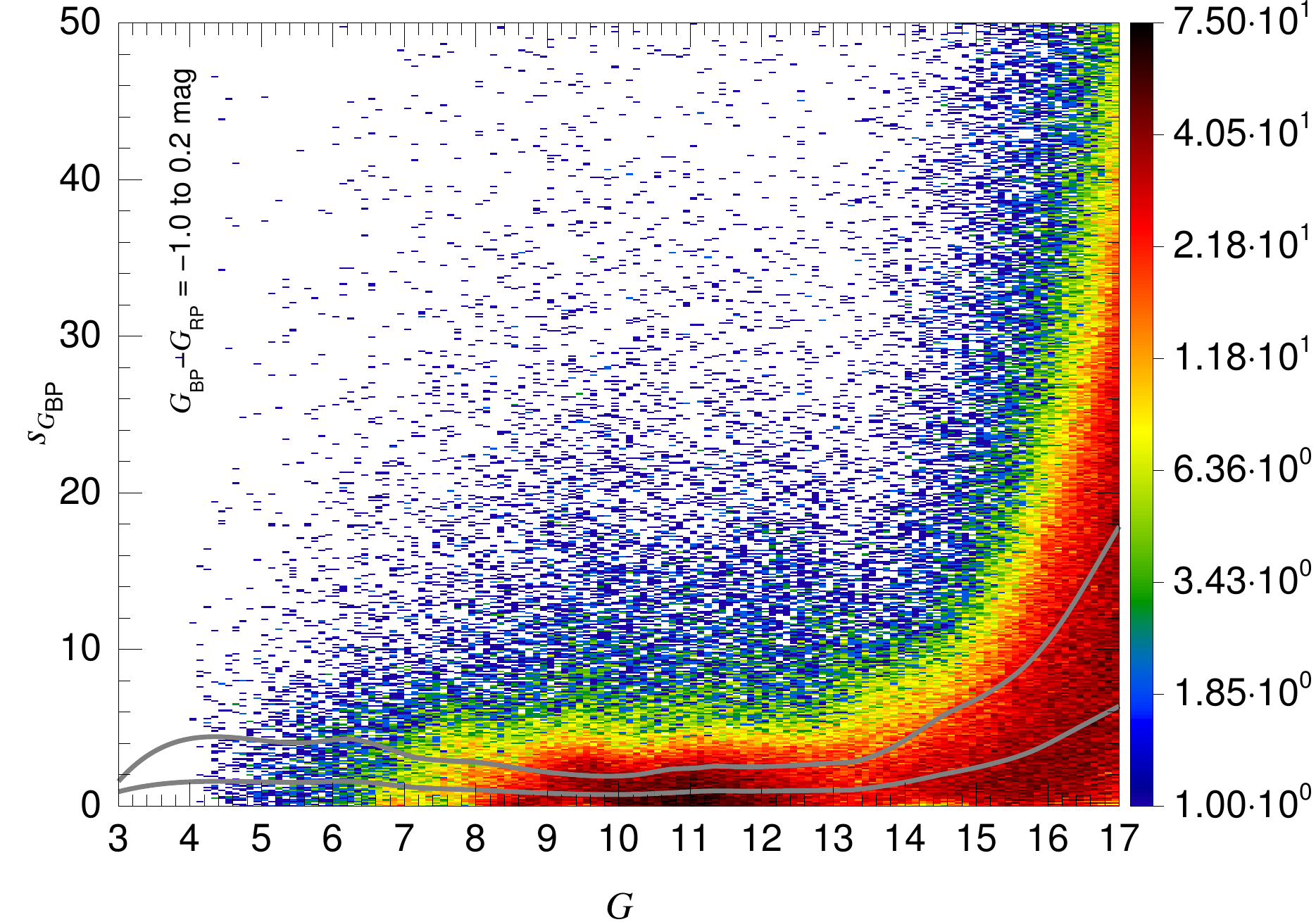}$\!\!\!$
                    \includegraphics[width=0.35\linewidth]{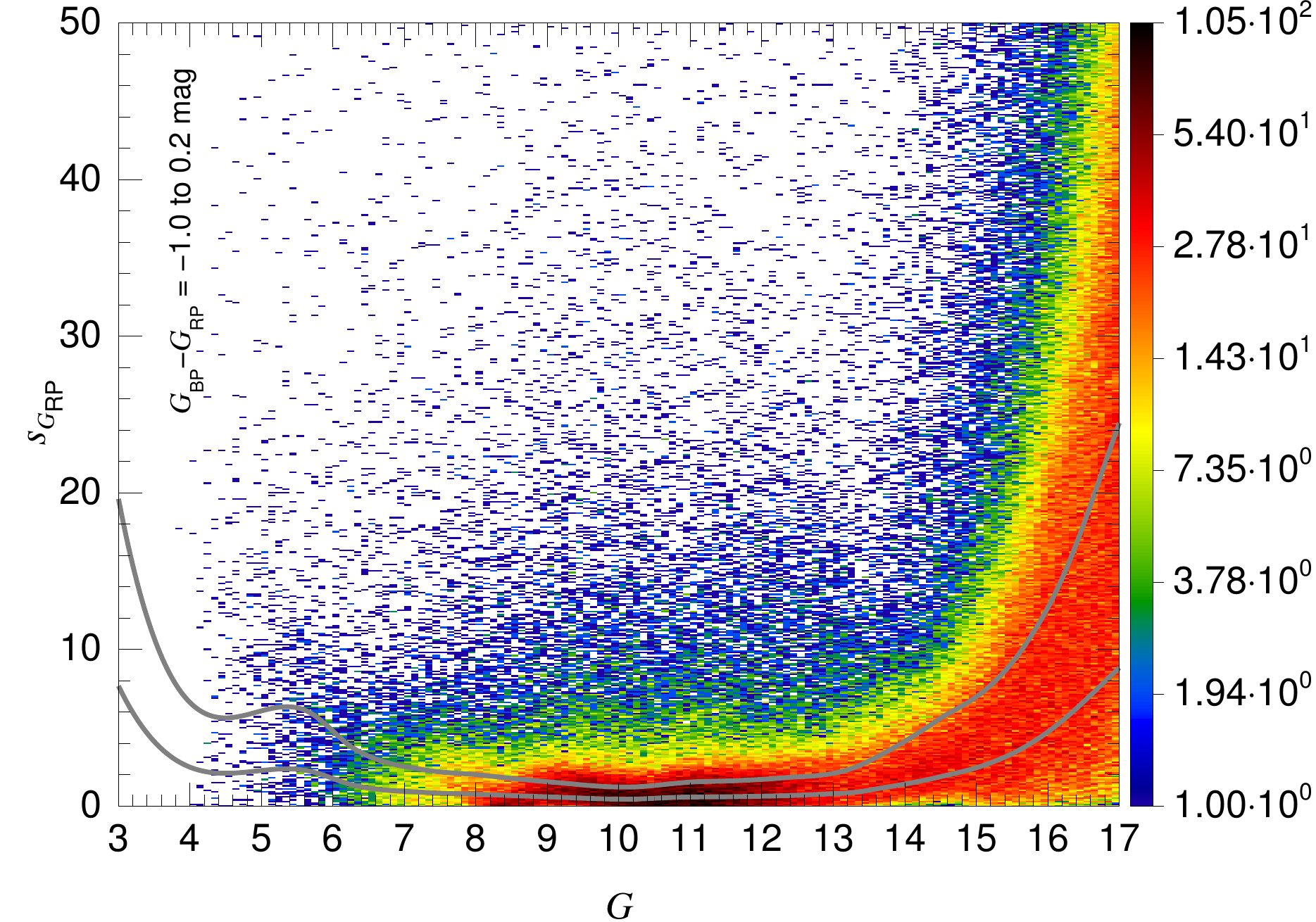}}
\centerline{$\!\!\!$\includegraphics[width=0.35\linewidth]{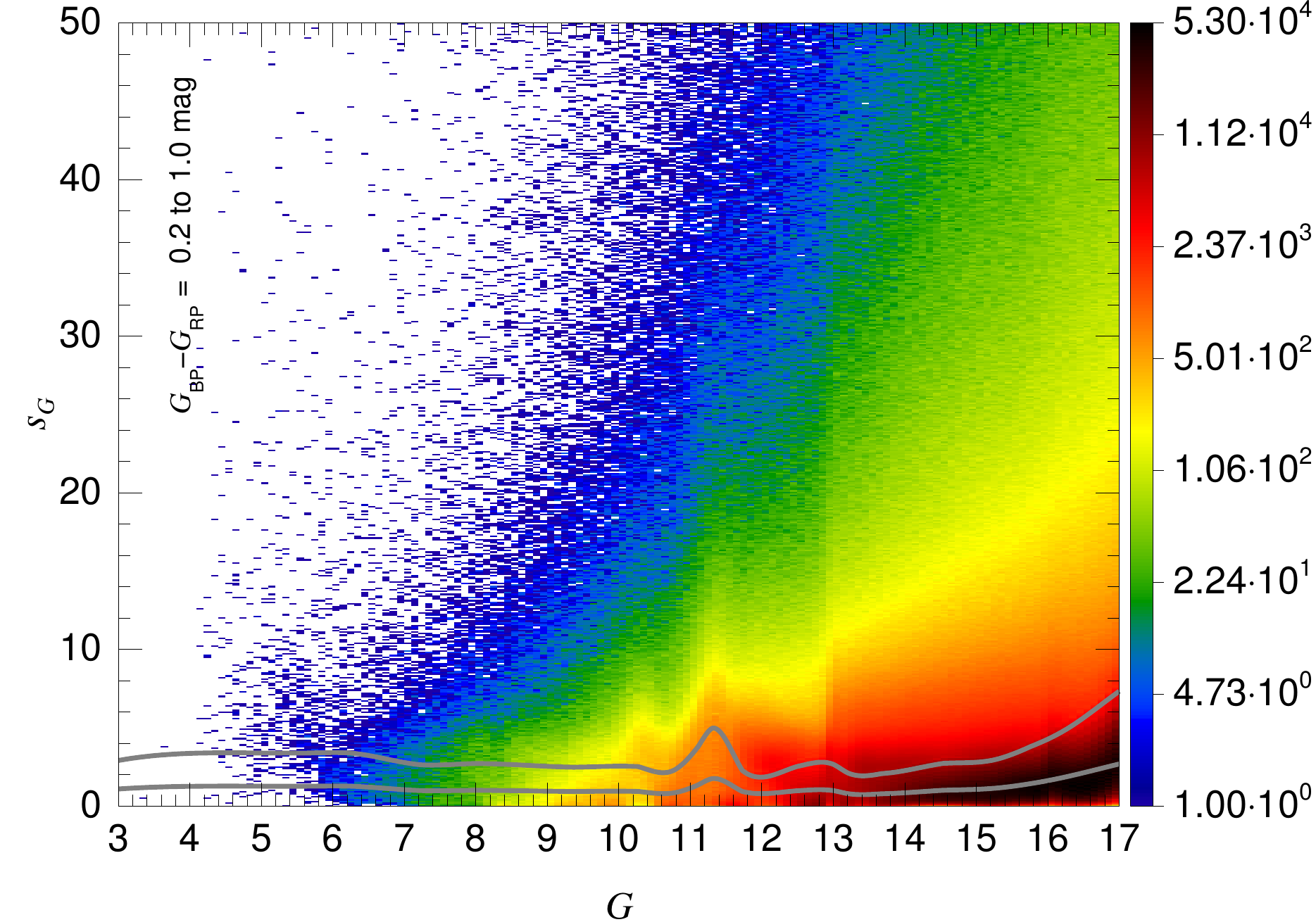}$\!\!\!$
                    \includegraphics[width=0.35\linewidth]{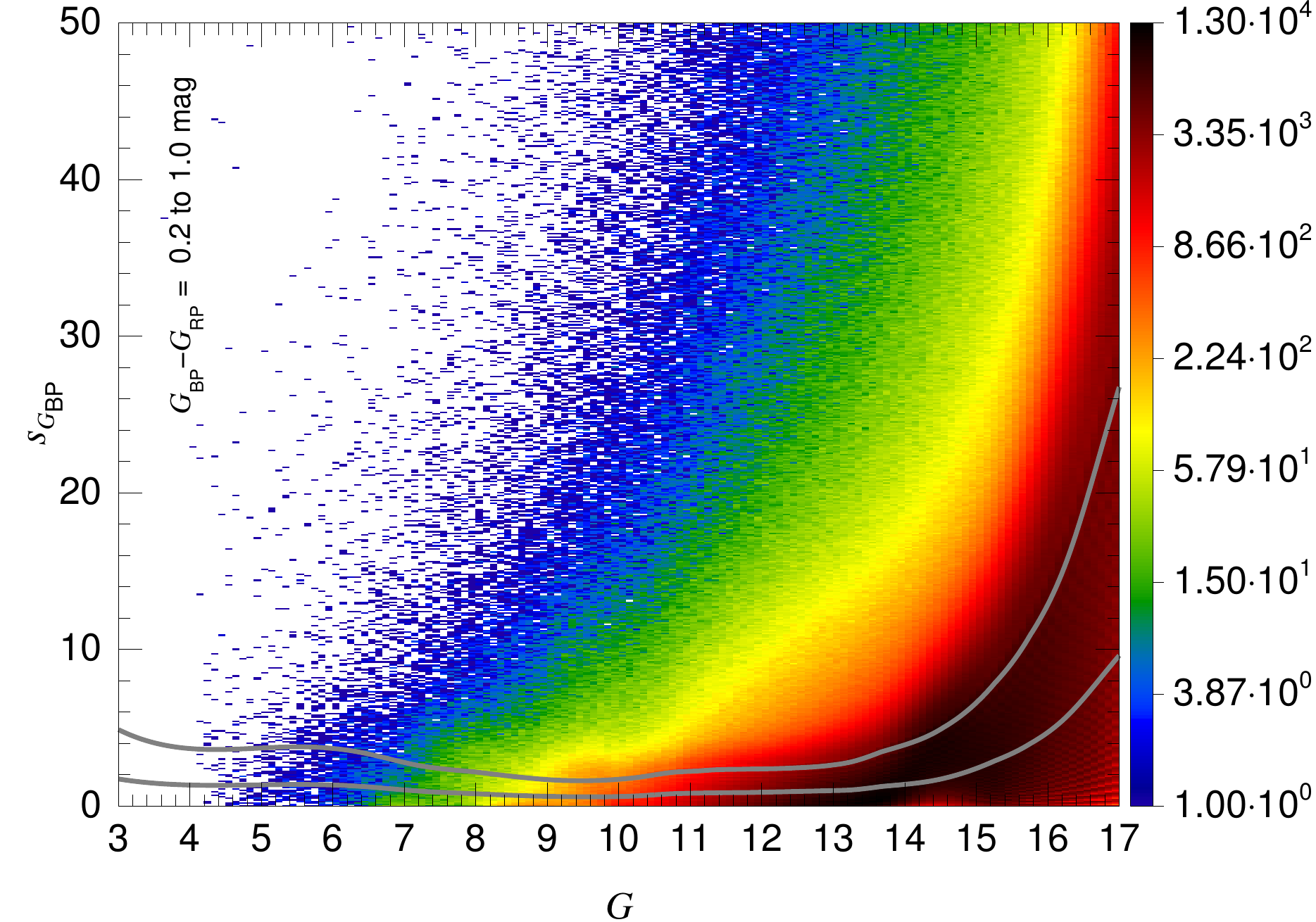}$\!\!\!$
                    \includegraphics[width=0.35\linewidth]{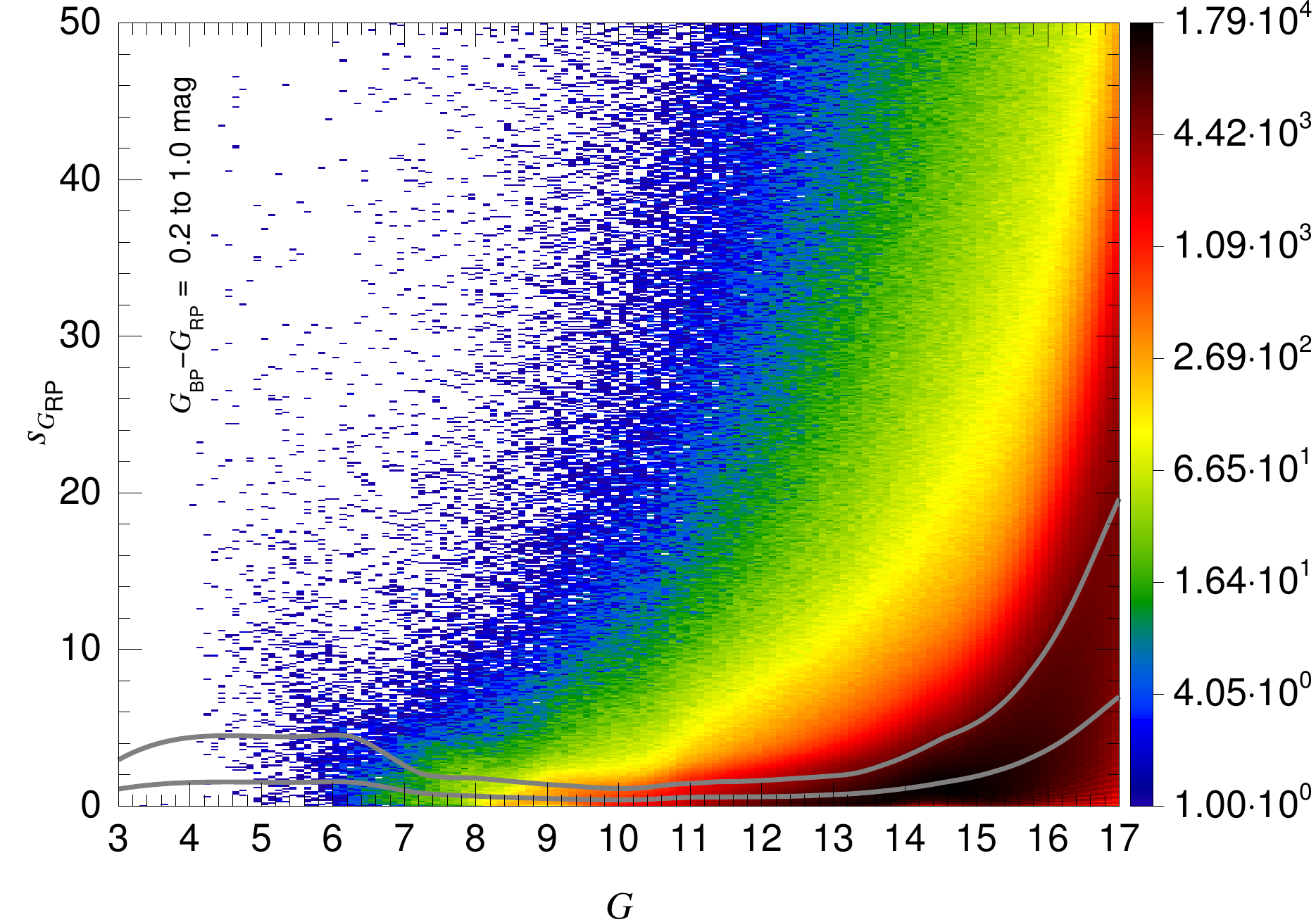}}
\centerline{$\!\!\!$\includegraphics[width=0.35\linewidth]{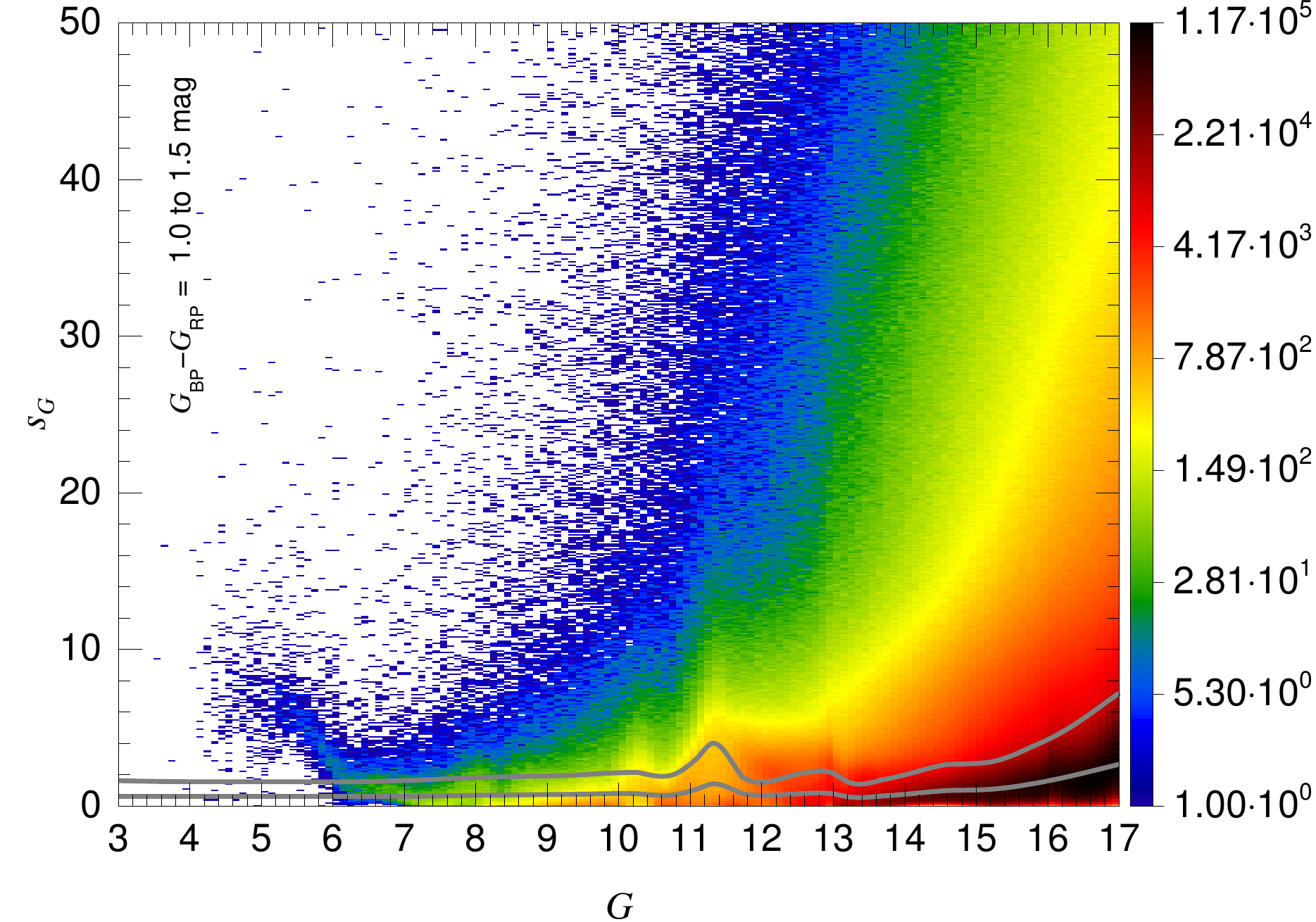}$\!\!\!$
                    \includegraphics[width=0.35\linewidth]{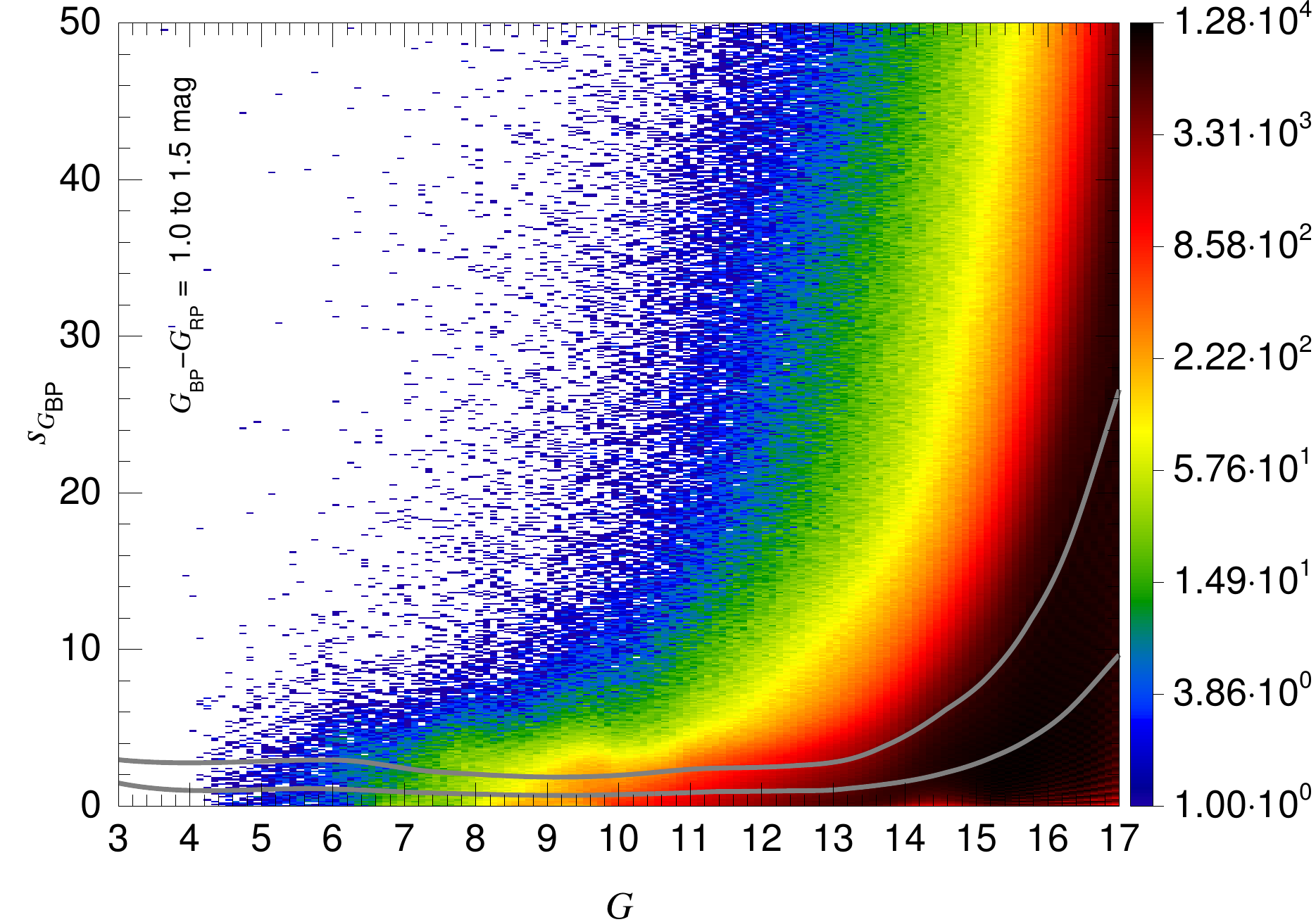}$\!\!\!$
                    \includegraphics[width=0.35\linewidth]{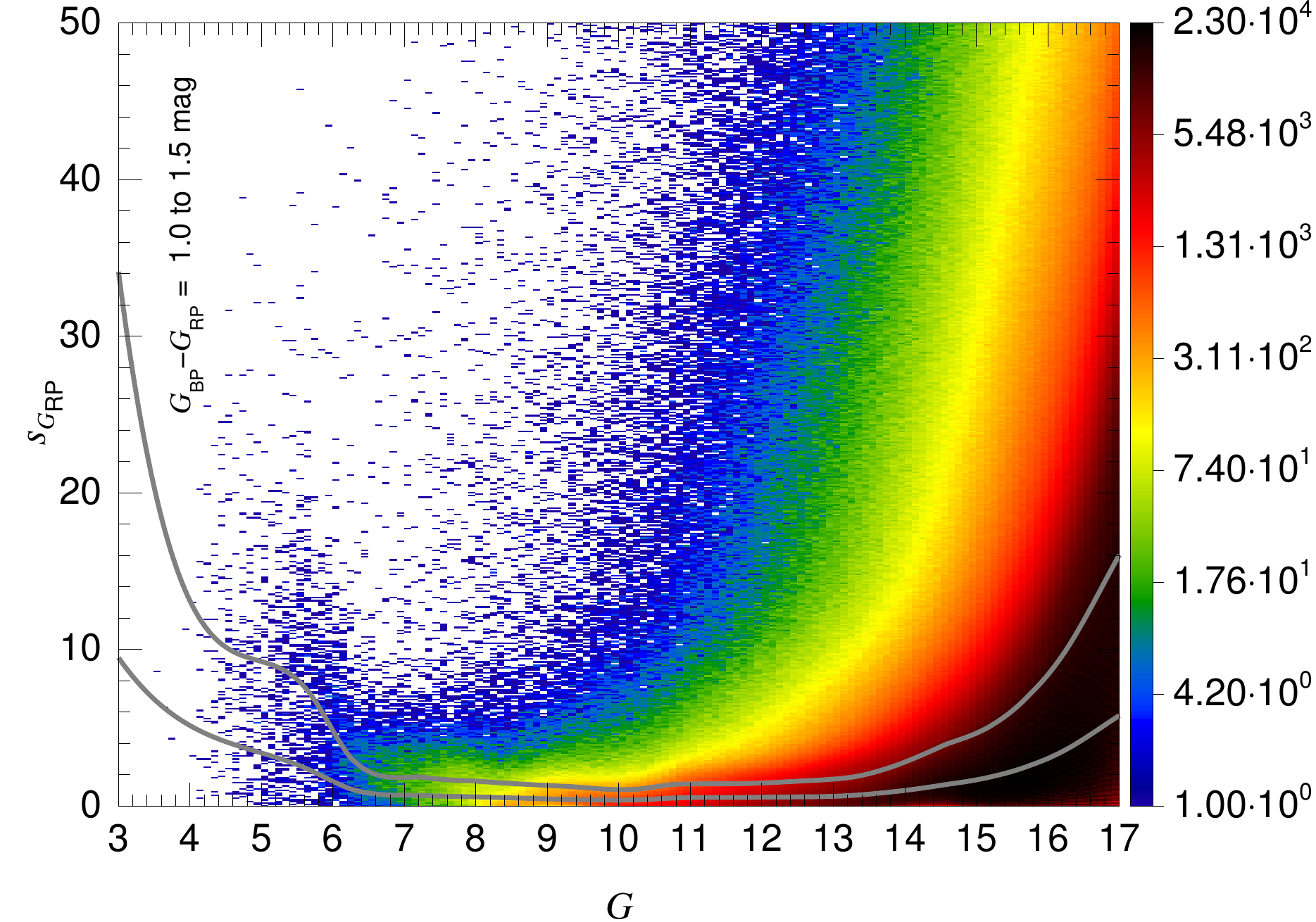}}
\centerline{$\!\!\!$\includegraphics[width=0.35\linewidth]{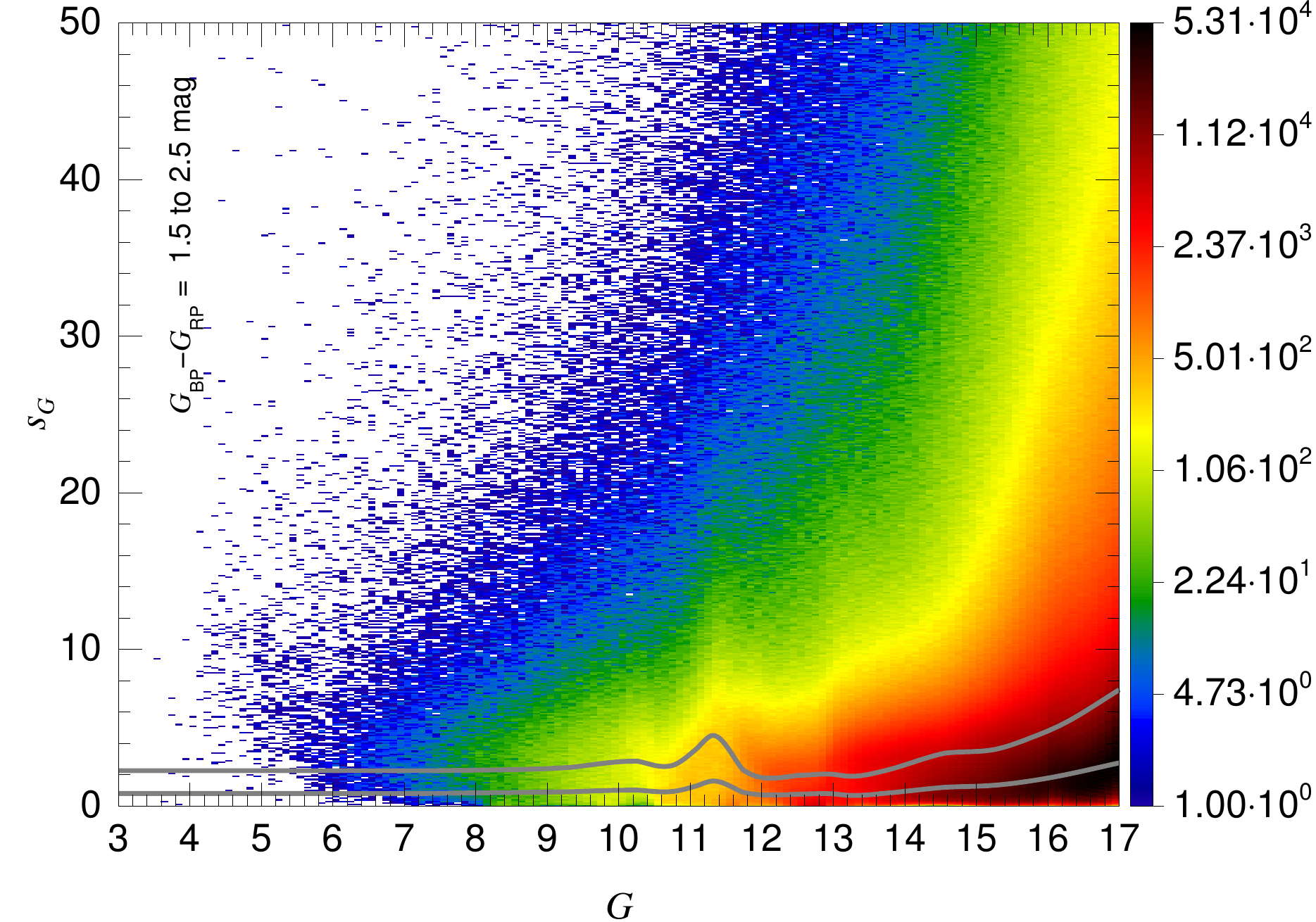}$\!\!\!$
                    \includegraphics[width=0.35\linewidth]{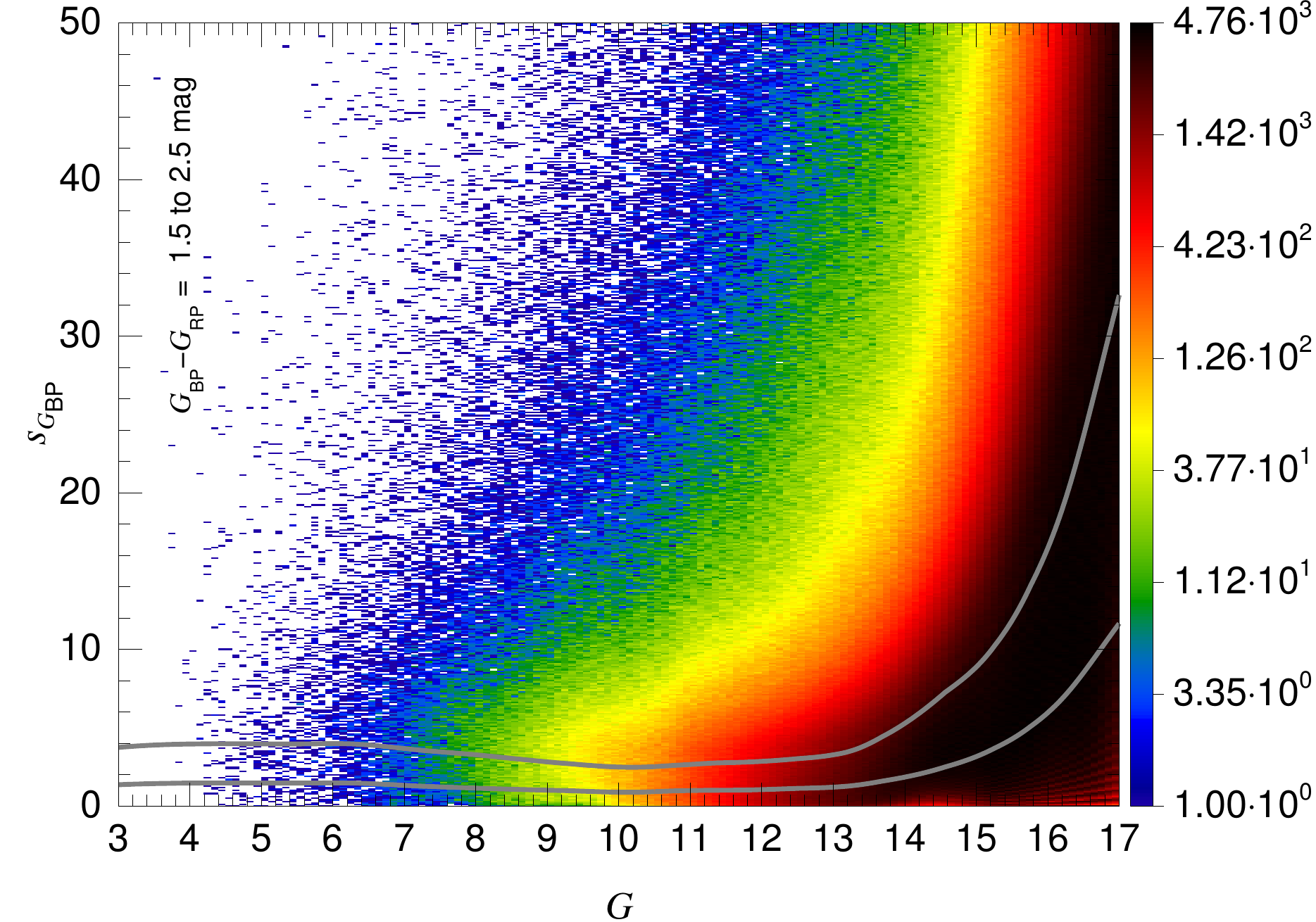}$\!\!\!$
                    \includegraphics[width=0.35\linewidth]{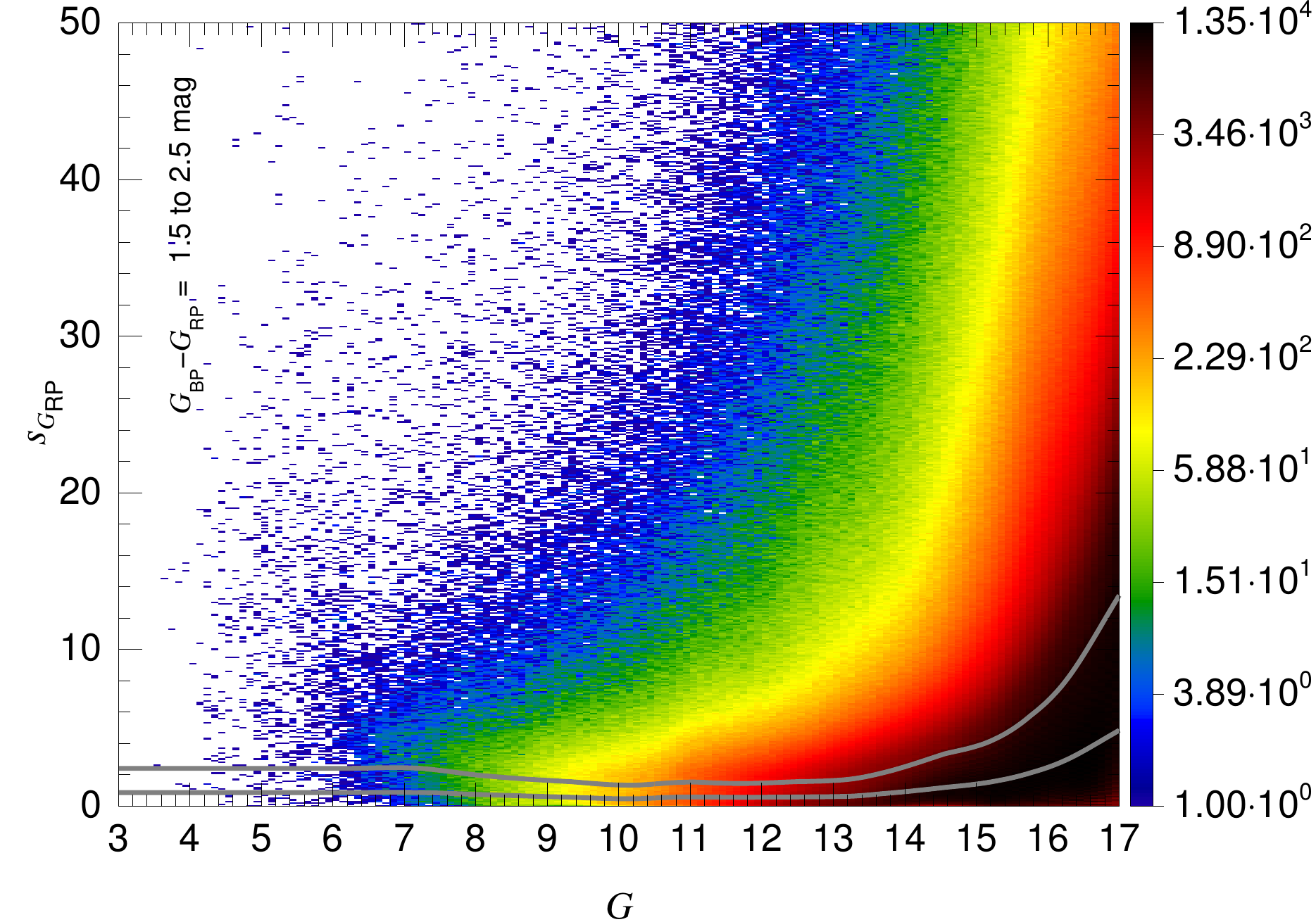}}
\centerline{$\!\!\!$\includegraphics[width=0.35\linewidth]{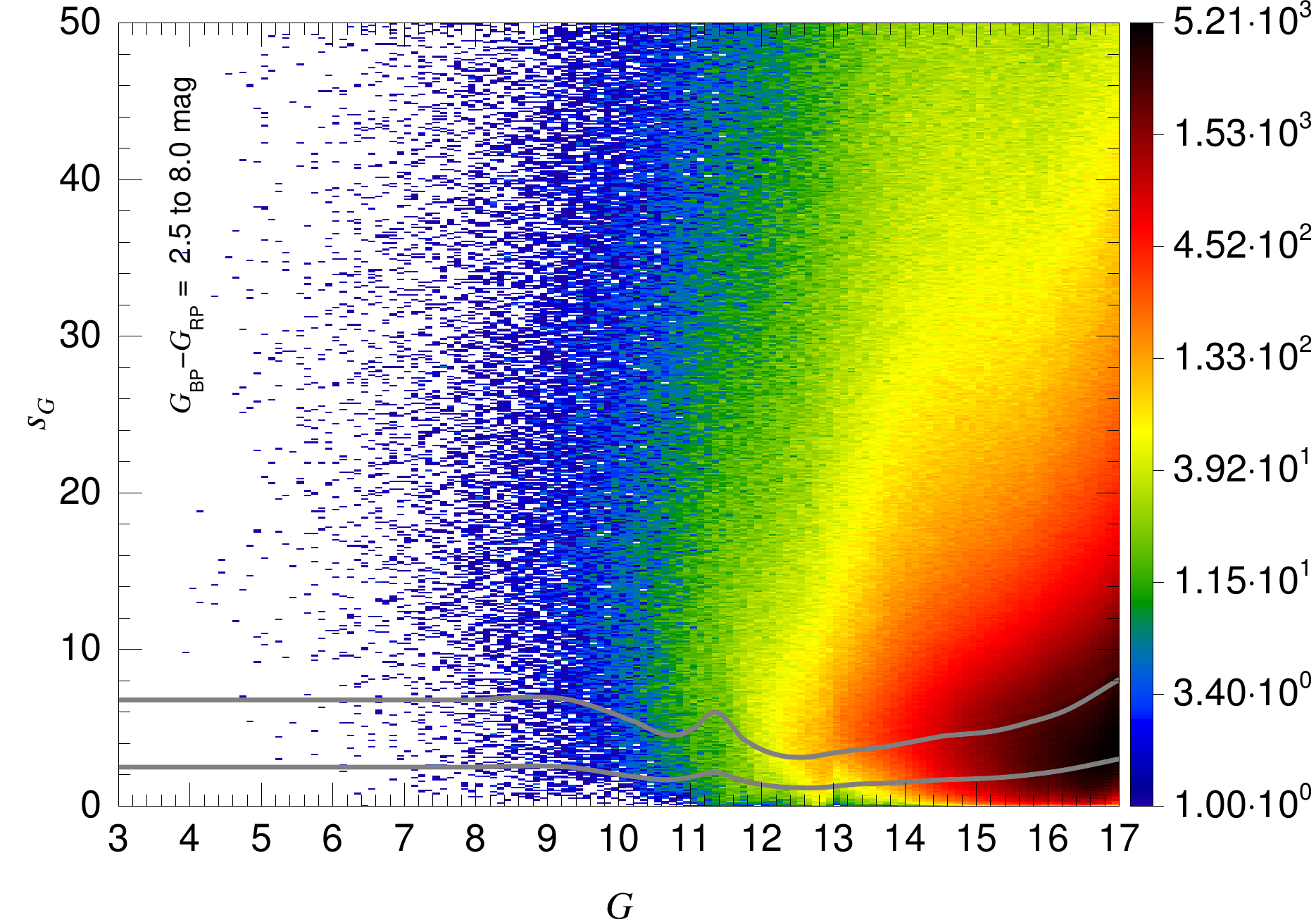}$\!\!\!$
                    \includegraphics[width=0.35\linewidth]{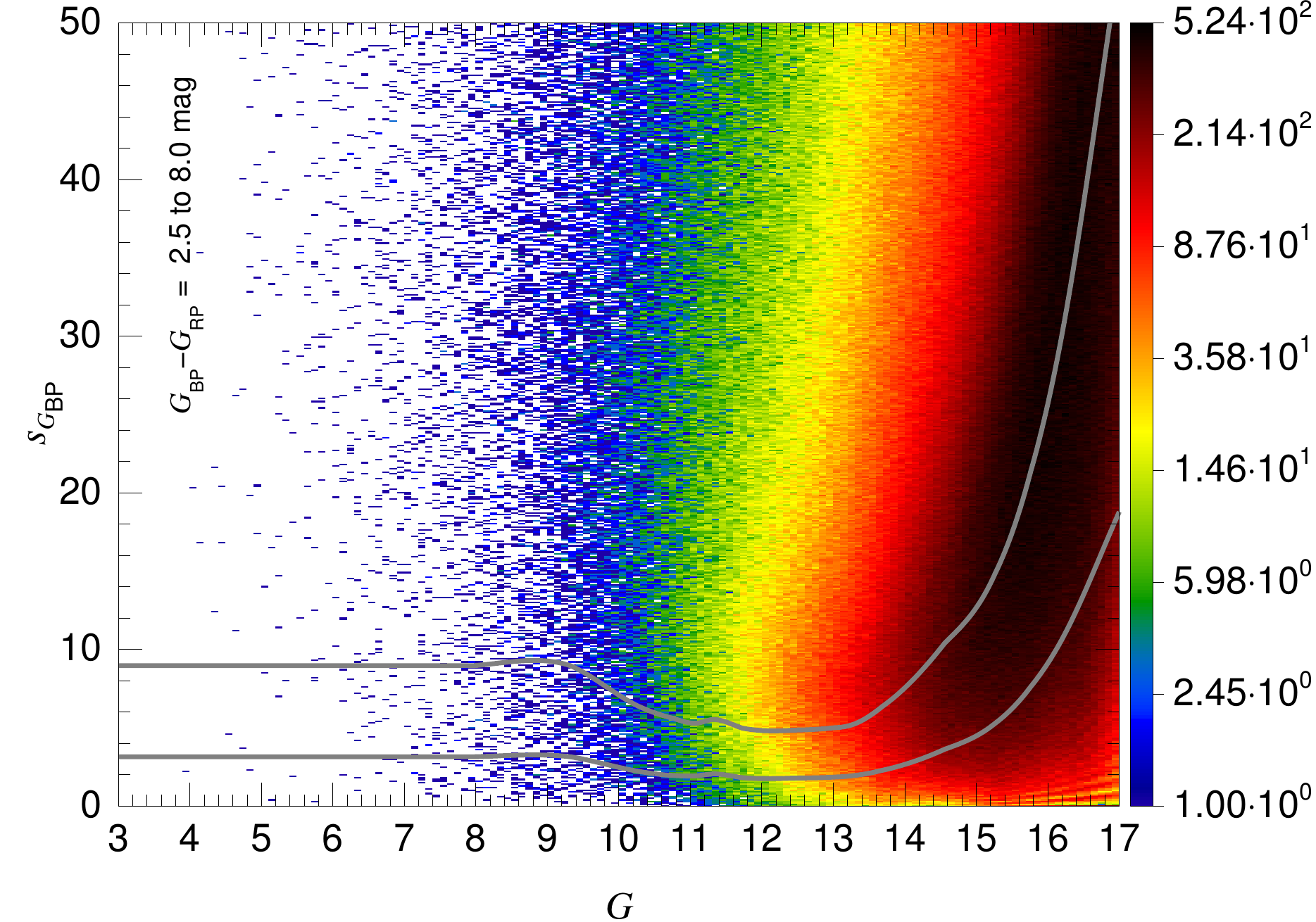}$\!\!\!$
                    \includegraphics[width=0.35\linewidth]{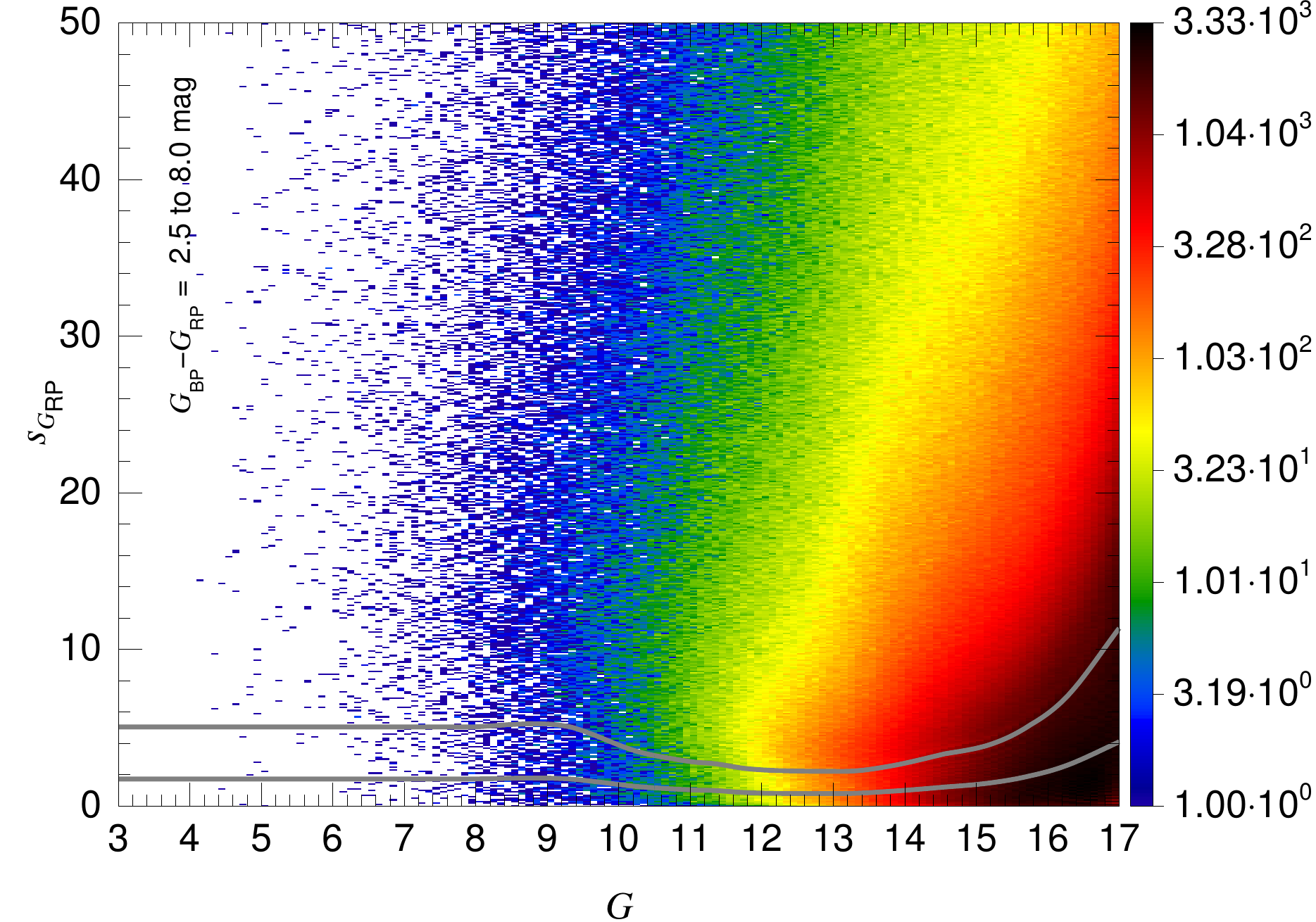}}
\caption{Astrophysical photometric dispersions for \GG\ ({\it left column}), \GBP\ ({\it center column}), and \GRP\ ({\it right column}) 
         for the five \GBPmGRP\ ranges used in this work, from top to bottom: $-$1.0-0.2, 0.2-1.0, 1.0-1.5, 1.5-2.5, and 2.5-8.0 (in 
         mag). The two lines in each plot mark the approximate boundaries between N and M flags (lower one) and between M and V flags 
         (upper one). The horizontal axes are in mag and the vertical ones in mmag.}
\label{hist_mag_sigma}
\end{figure*}

\paragraph{Variability flag.} To facilitate the use of the results, we also provide a three-letter 
variability flag of the type XXX. Each letter corresponds to \GG, \GBP, and \GRP, respectively, and X can be one of four options: 
B(ad), N(on-variable), M(arginal), or V(ariable). A B flag is assigned for the small number of cases where 
$\sXz < \sXi - 3\sigmasXi$ (discussed below). For the rest, an N is assigned if $\sX < \sigmasX$, an M is assigned if 
$\sigmasX \le \sX < 3\sigmasX$, and a V is assigned if $\sX \ge 3\sigmasX$. For example, a star with an NMV flag would be 
non-variable in \GG, marginally variable in \GBP, and variable in \GRP. This flag should be interpreted as a measurement of
variability within the instrumental capability for that magnitude and color. In this way, an object with an observed \sX\ of 
10~mmag would be a V if $\sigmasX = 2$~mmag, an M if $\sigmasX = 5$~mmag, and an N if $\sigmasX = 12$~mmag (all of them reasonable
values for certain points in the magnitude-color space of the sample). This is independent of whether the star is really variable or 
not, as all of the above could correspond to objects with a real astrophysical dispersion of 10~mmag. As a corollary, the fraction of
objects with a V flag is expected to diminish for fainter values of \GG, not because the fraction of real variables is 
smaller\footnote{That could be a secondary factor, as absolute and apparent magnitude are correlated and the fraction of real 
variables is a function of unreddened magnitude and color.}, but because, for magnitudes fainter than $\GG\approx 13$, \sXi\ 
increases with \GG\ and \textit{Gaia} becomes less capable of detecting variability (see 
Figs.~\ref{hist_sigma}~and~\ref{hist_mag_sigma} and next subsection).

\paragraph{Simbad information.} Finally, we cross-matched the sample with the February 2023 version of Simbad by using the
\textit{Gaia}~DR3 identifiers and we found \num{4721984}~matches. % TO BE UPDATED, THERE ARE NEW SOURCES
For each we list the Simbad ID, spectral type, variability flag, and period. The reader should be aware that the cross-match is 
done by the CDS and that Simbad is a live service that changes continuously, so a future cross-match is likely to be different and 
to include more sources. The main purpose of providing the Simbad ID column is to
allow the user to quickly identify already well-known objects, as humans are not at their best when trying to identify objects by
their \textit{Gaia} ``telephone numbers''.  Regarding Simbad spectral types, one should keep in mind that they come from many
sources of diverse quality and include clear mistakes and non-standard classifications. As an example, 
\citet{Maizetal16} list eleven egregious cases of stars of A, F, G, and even K type that were listed in Simbad as being of O type.
All of them have been corrected in Simbad at this point but certainly other similar examples persist.

\subsection{Validation}

\subsubsection{Model}

\begin{figure*}
\centerline{\includegraphics[width=0.49\linewidth]{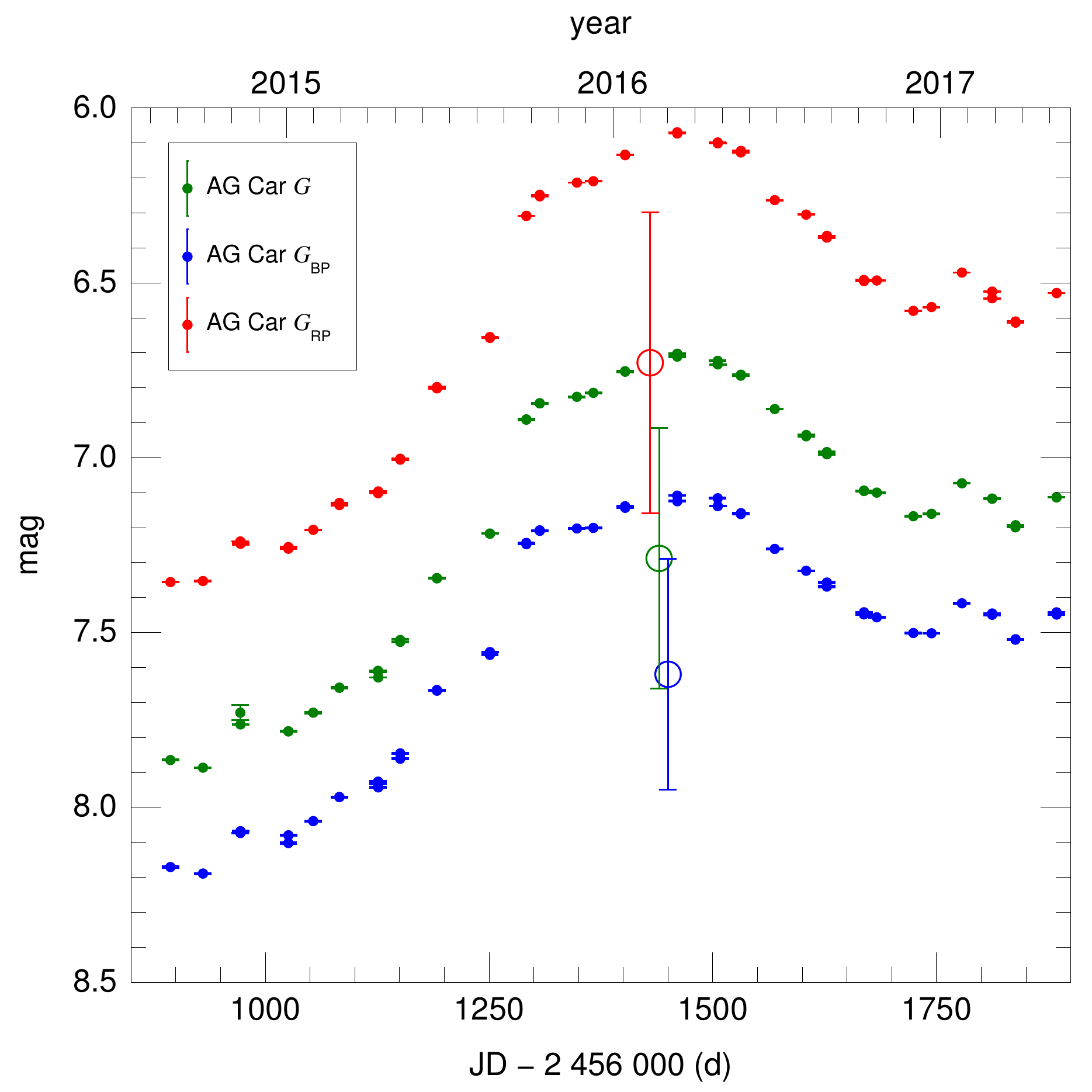} \
            \includegraphics[width=0.49\linewidth]{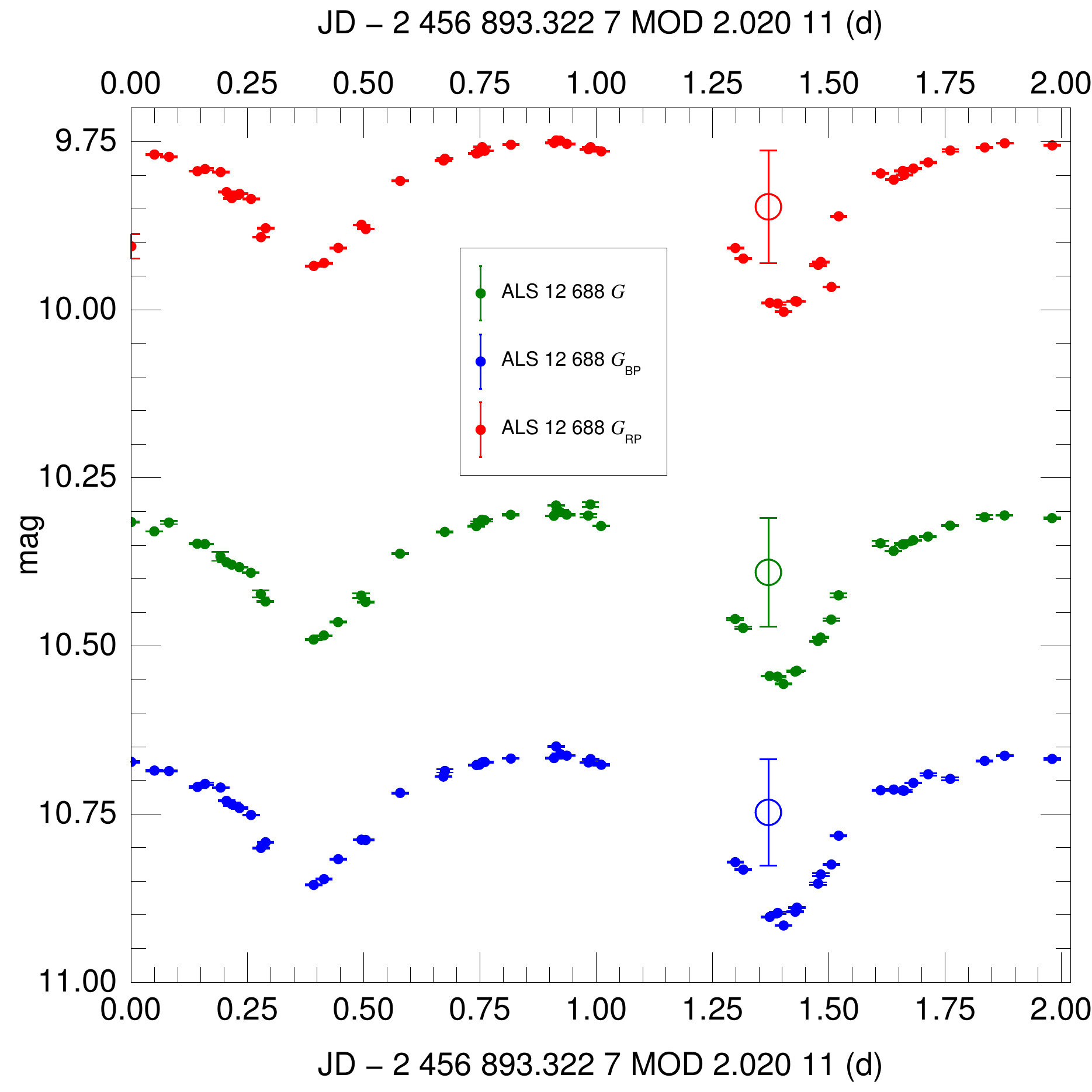}}
\centerline{\includegraphics[width=0.49\linewidth]{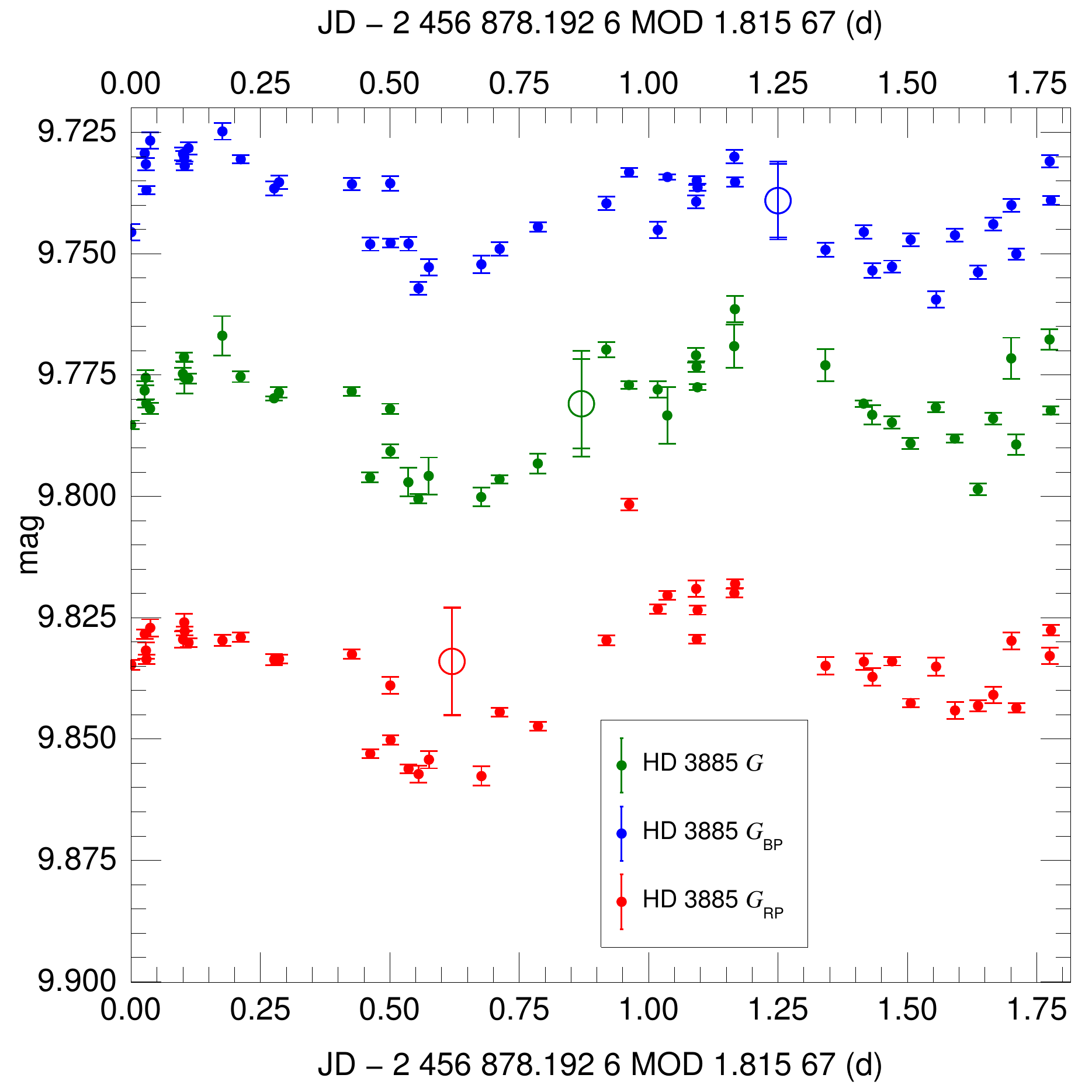} \
            \includegraphics[width=0.49\linewidth]{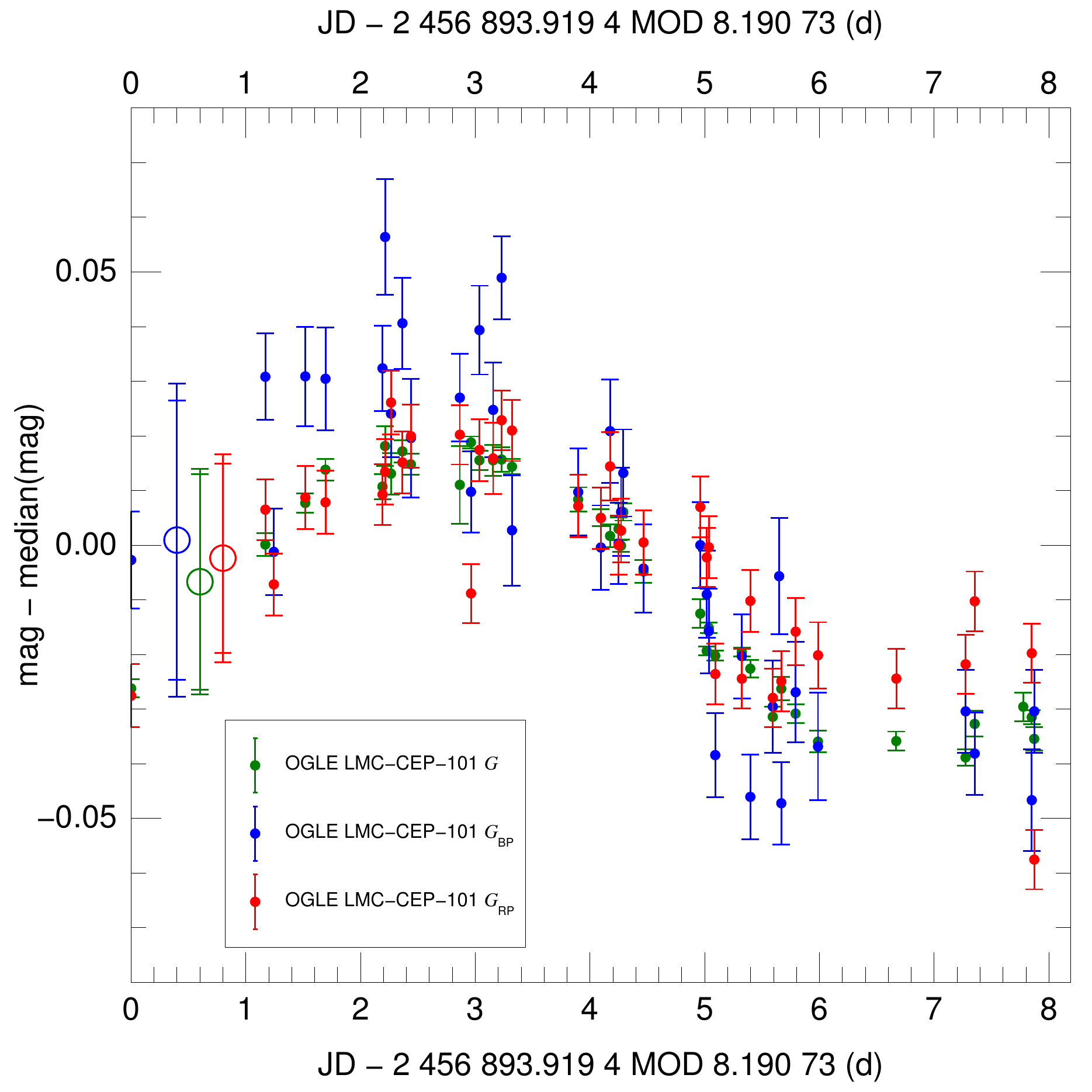}}
 \caption{Light curves for four of the targets with epoch photometry. ({\it Top left}) AG Car, an LBV properly classified in R22. 
         ({\it Top right}) ALS~\num{12688}, not given a classification type there but classified as an eclipsing binary in paper~II of
         this series (Holgado et al. in preparation). ({\it Bottom left}) HD~3885, classified as a slowly pulsating B star in
         R22 and as an $\alpha^2$~CVn variable in Simbad. ({\it Bottom right}) OGLE~LMC-CEP-101, classified as a Cepheid in
         that paper. The last three panels are phased with the best fitting period in each case (\num{2.02011}~d, \num{1.81567}~d,
         and \num{8.19073}~d, respectively). The solid circles with error bars indicate the \textit{Gaia}~DR3 epoch measurements 
         while the open circles with double error bars indicate the average magnitudes and the total and astrophysical dispersions 
         determined in this paper (when both are similar the double error bars overlap). For OGLE~LMC-CEP-101 the median magnitudes 
         are subtracted in the vertical axis to allow for a higher dynamical range to be visualized.}
\label{light_curves}
\end{figure*}

\begin{figure*}[ht!]
\centerline{\includegraphics[width=0.35\linewidth]{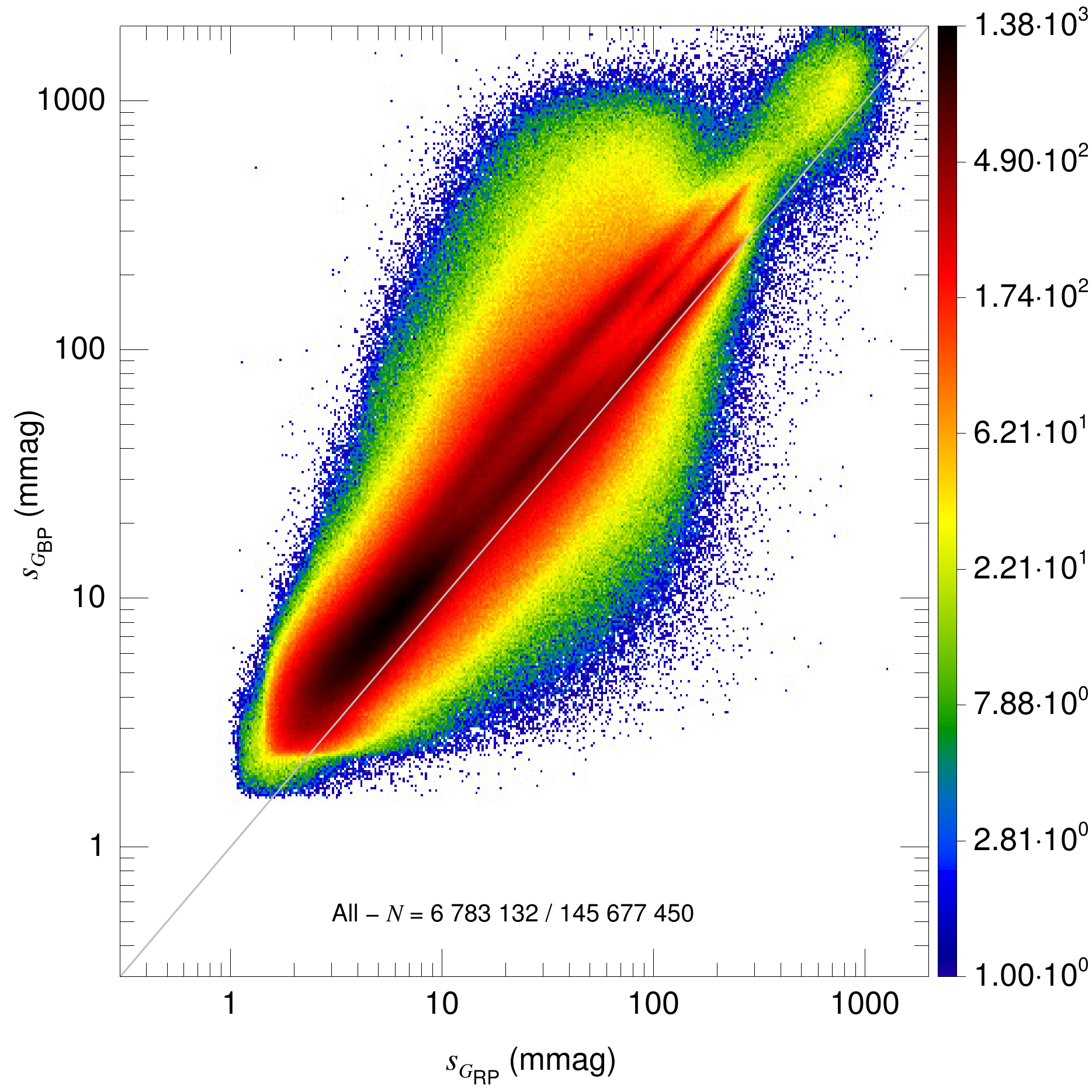}$\!\!\!$
            \includegraphics[width=0.35\linewidth]{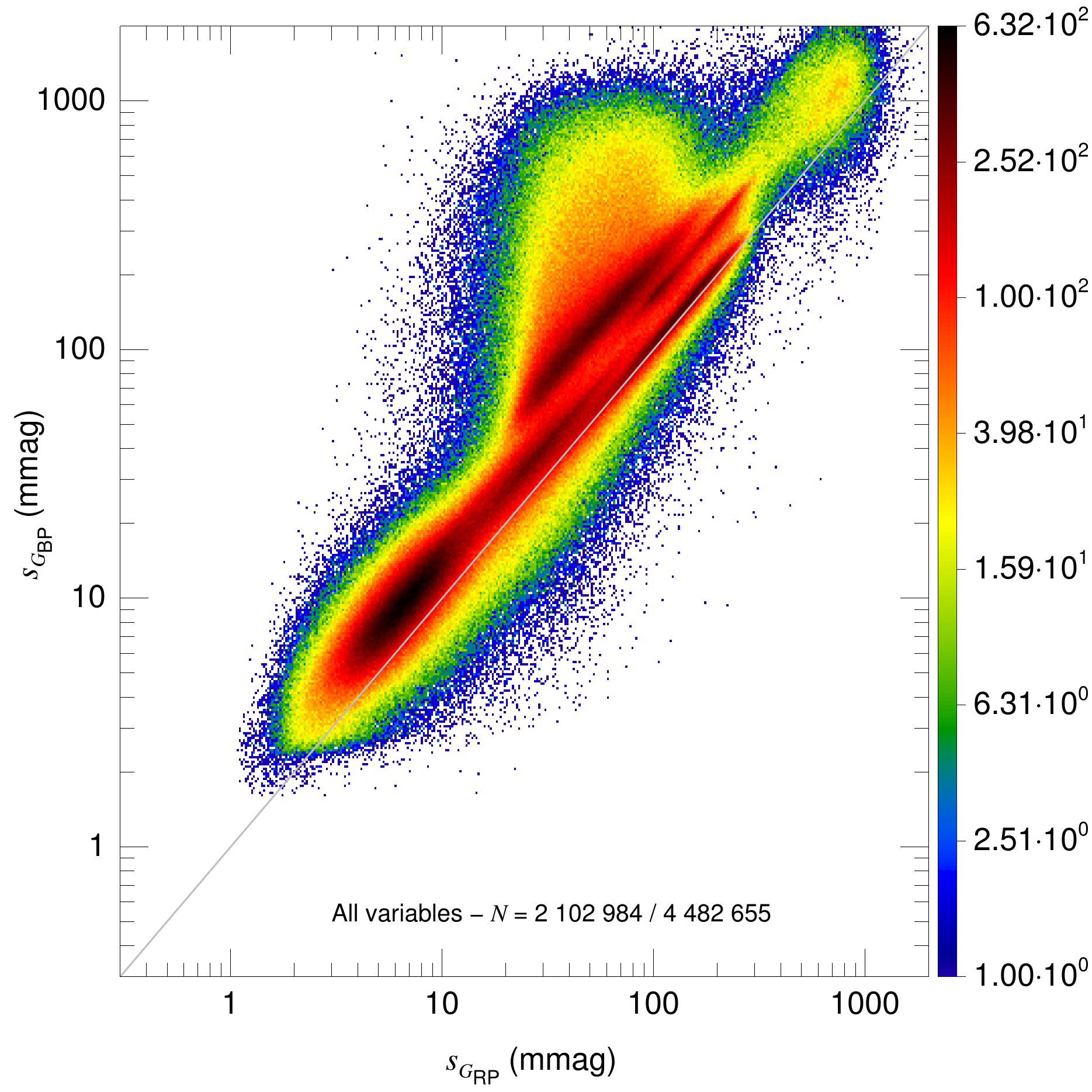}$\!\!\!$
            \includegraphics[width=0.35\linewidth]{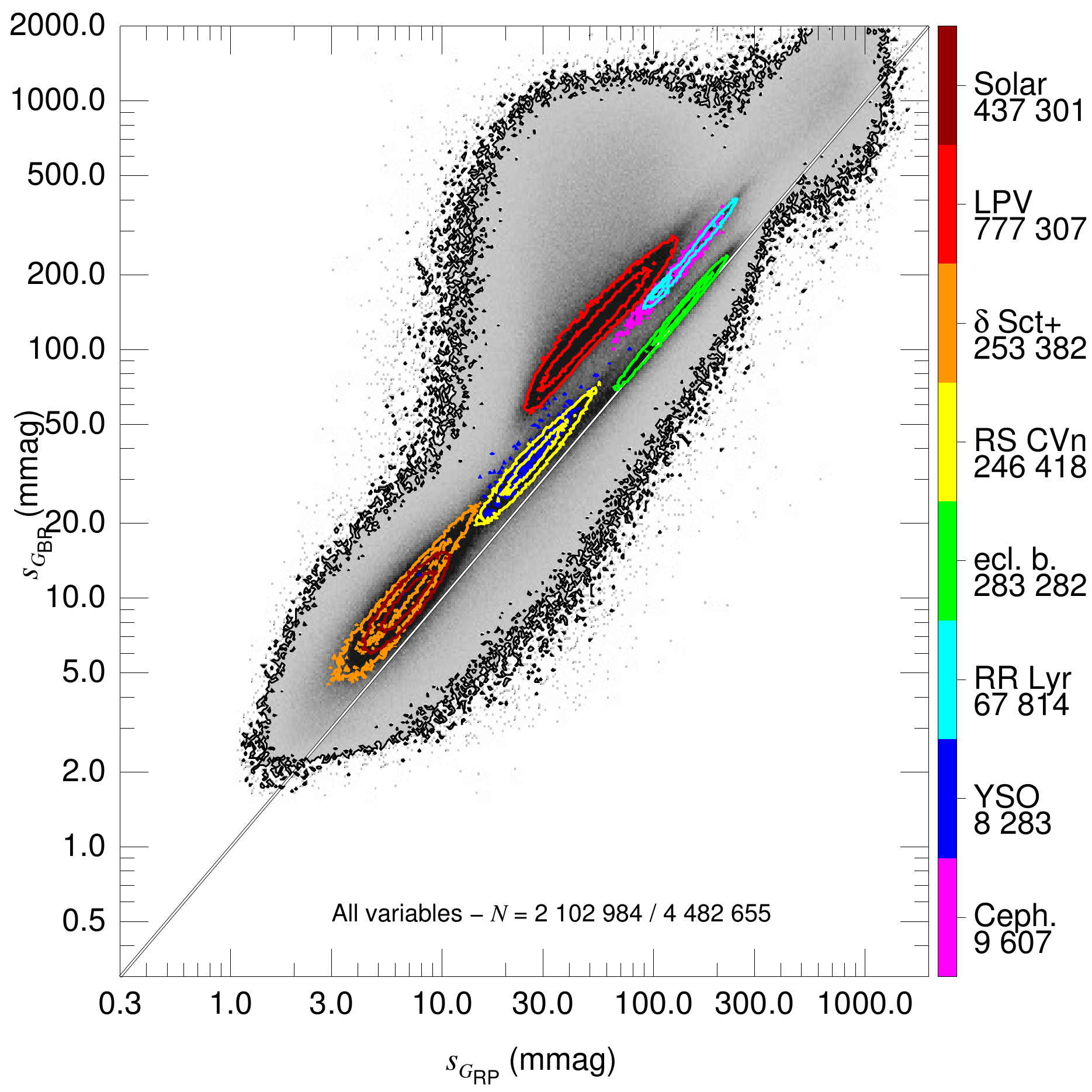}}
\centerline{\includegraphics[width=0.35\linewidth]{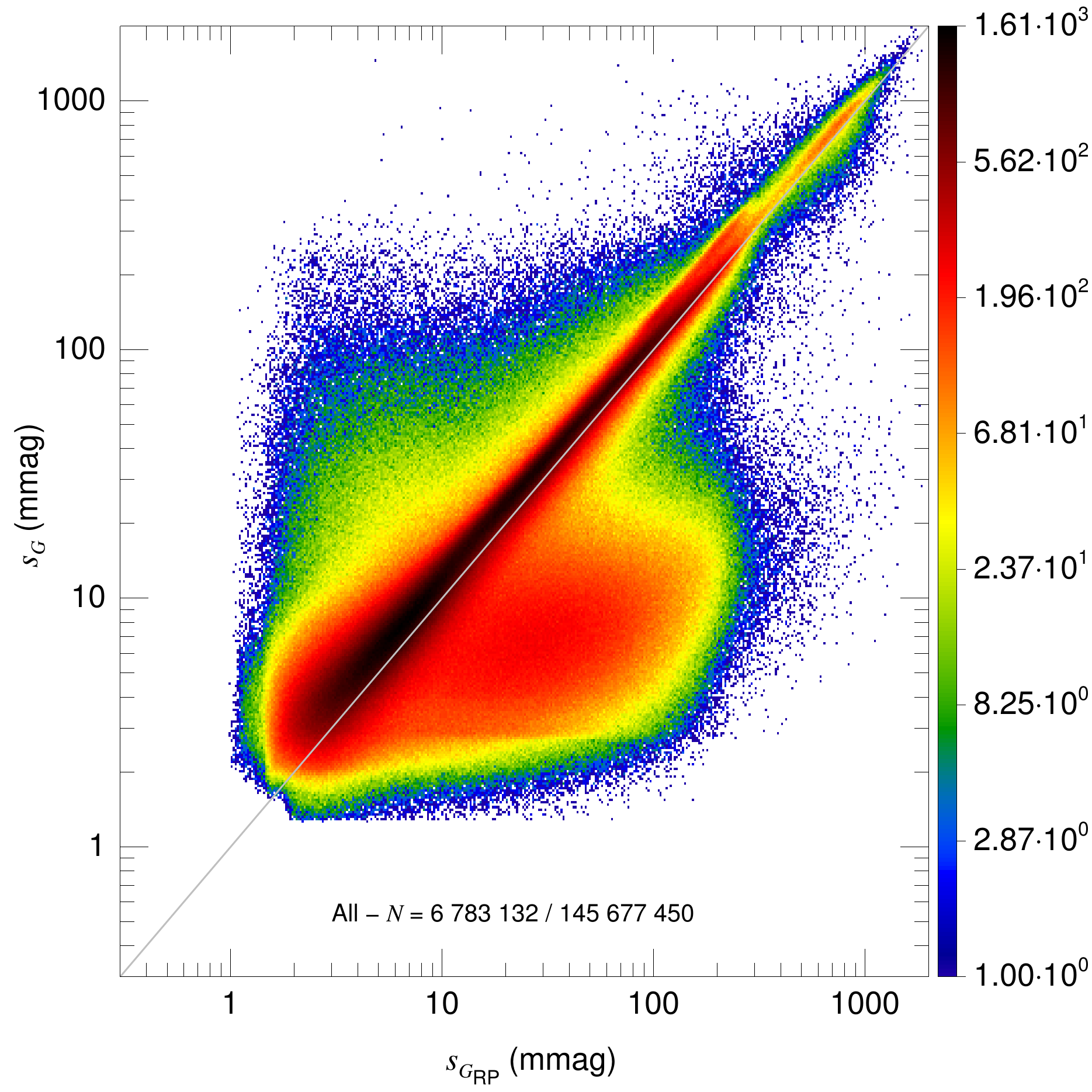}$\!\!\!$
            \includegraphics[width=0.35\linewidth]{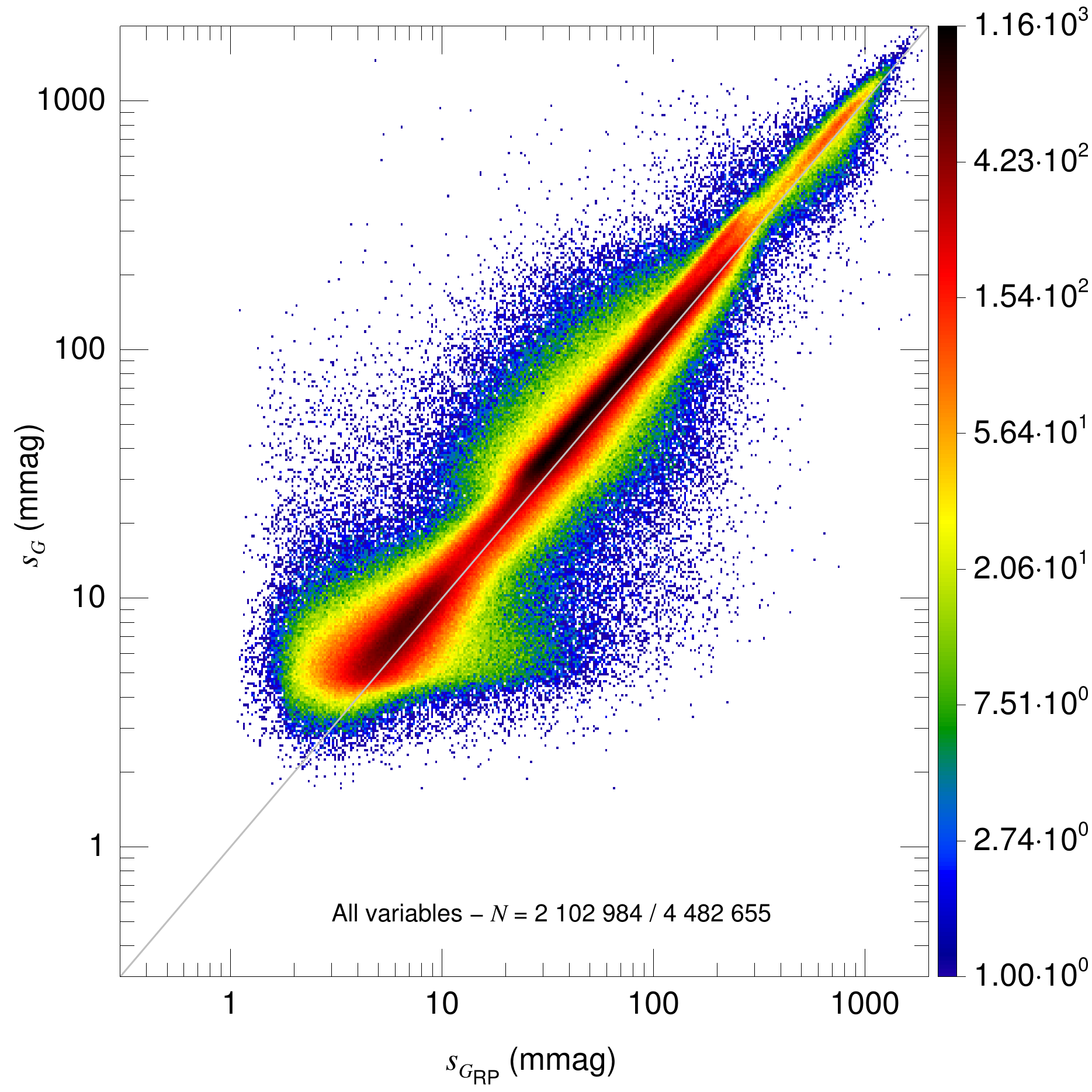}$\!\!\!$
            \includegraphics[width=0.35\linewidth]{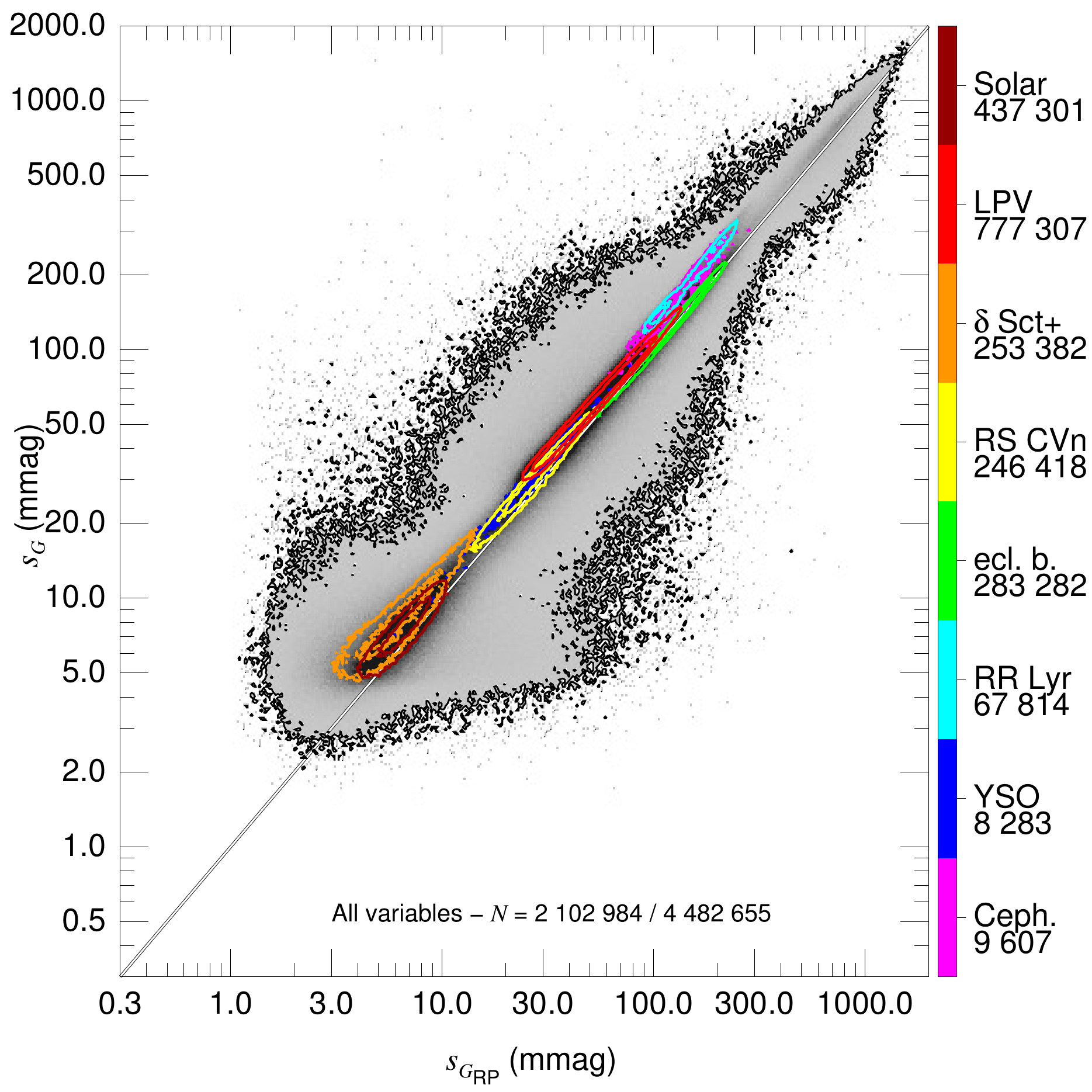}}
\centerline{\includegraphics[width=0.35\linewidth]{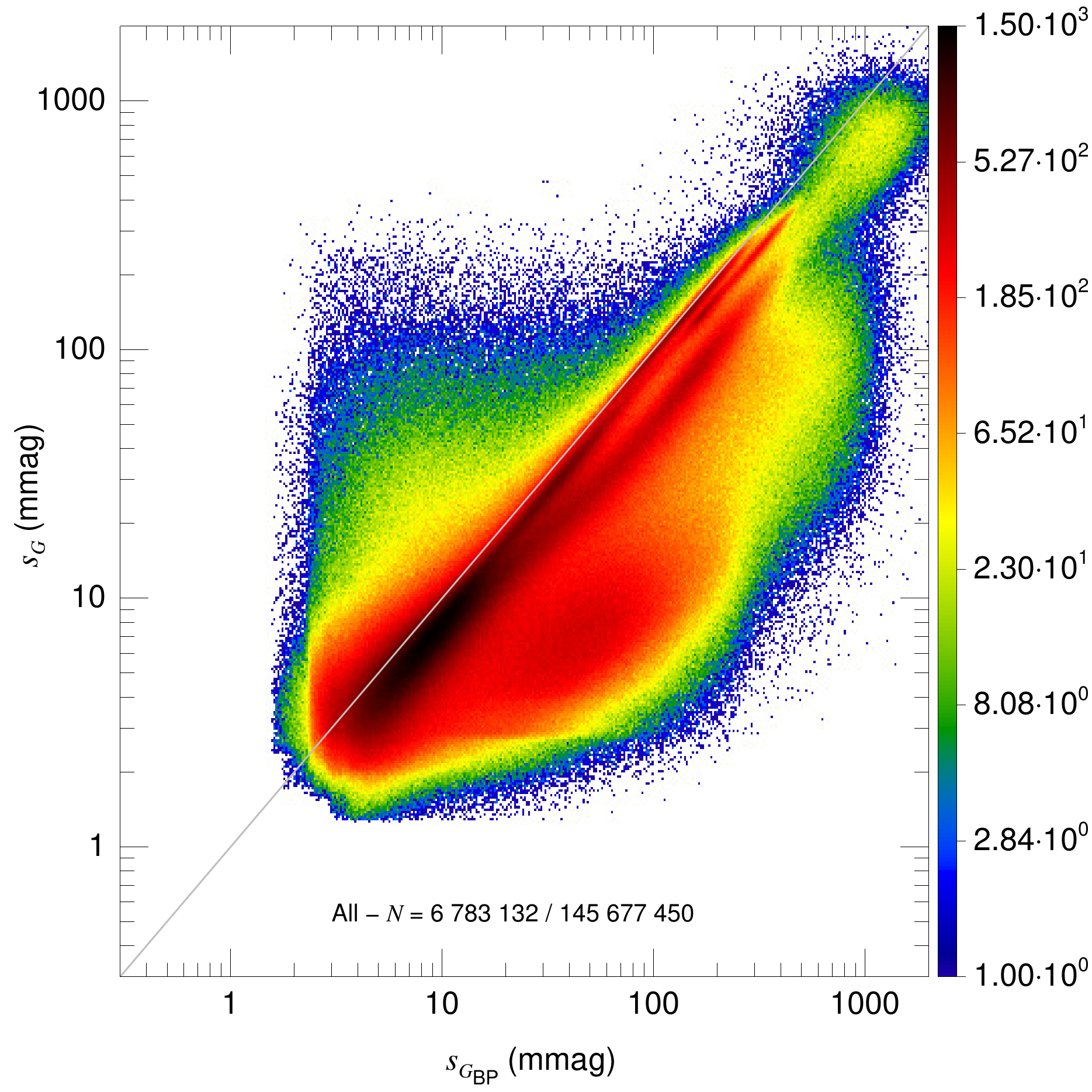}$\!\!\!$
            \includegraphics[width=0.35\linewidth]{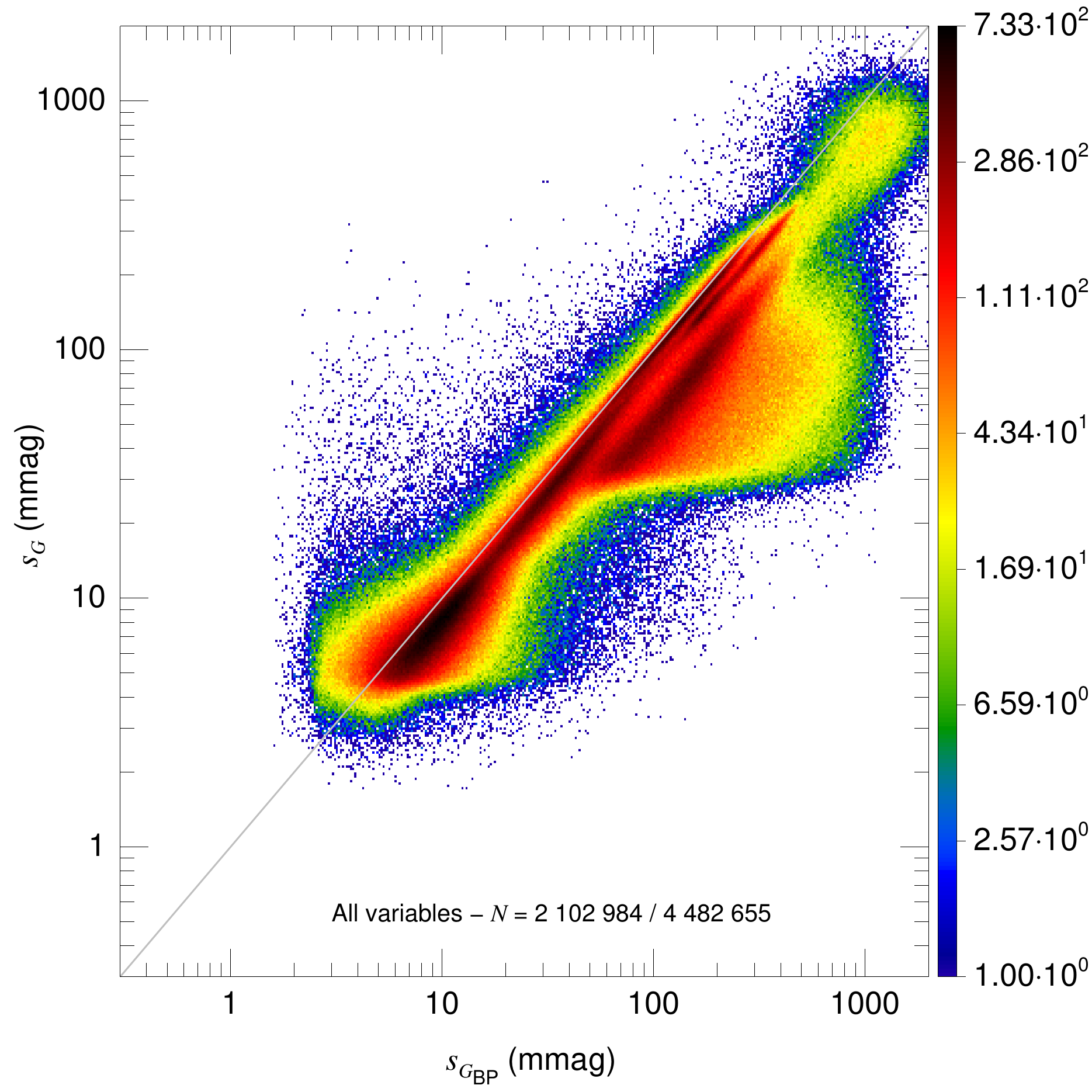}$\!\!\!$
            \includegraphics[width=0.35\linewidth]{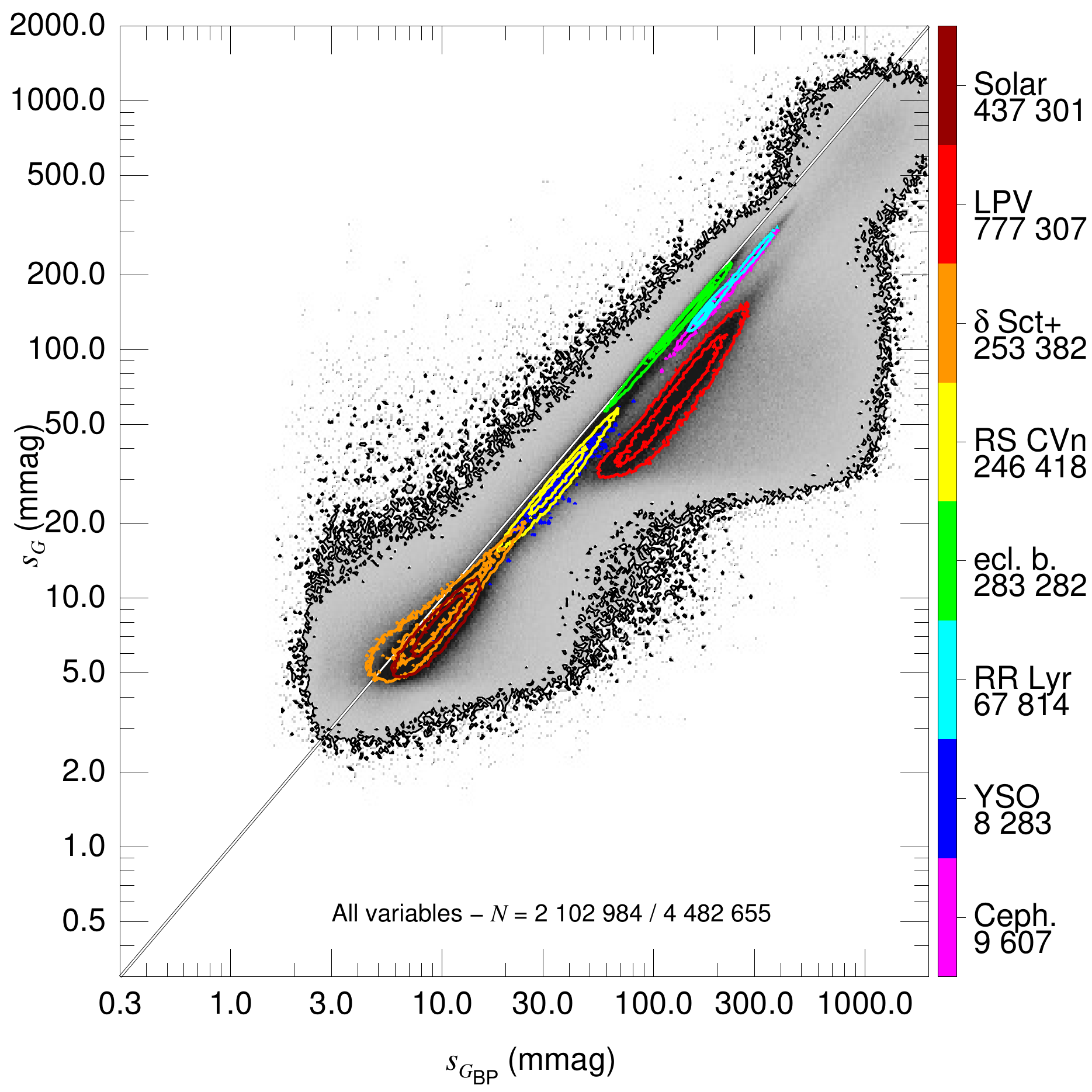}}
\caption{Astrophysical dispersion-dispersion density diagrams for the three color combinations for stars that are classified as VVV 
         here and have dispersions uncertainties \sigmasX\ lower than 1~mmag in the three bands. The left column shows the 
         source density in a colored logarithmic scale for the \num{6979556} objects in the sample that satisfy those conditions. 
         The center column is the equivalent for the \num{2057225} objects that, in addition, have variable types in 
         R22. The right column is a version of the center column with the source density in a gray linear scale and 
         with the addition of a black contour at its lowest level (1 object per bin) and two colored contours (at the 40\% and 70\% 
         of the maximum) for eight of the most common types of variables. See Fig.~\ref{varhist_types} for plots of the ratios of 
         the astrophysical dispersion for the different variable types.}
\label{varhist_All}
\end{figure*}

$\,\!$\indent In this subsection we address several issues to validate the results from our methodology. The first one is the
quality of our model: how well is the distribution of astrophysical dispersions characterized by Eqn.~\ref{fsX} and how well is the
instrumental dispersion described by a Gaussian of mean \sXi\ and width \sigmasXi? To answer those questions, we provide three types
of plots: 

\begin{enumerate}
 \item Figure~\ref{hist_sigma0} (already discussed above) are histograms plotted in a uniform log-log scale for each combination of 20 
       magnitude bins, five color bins, and three filters and they show the total dispersion histograms for (a) the full sample in that
       magnitude-filter bin and (b) the subsample with R22 variability types, plotting the three most common types of variables and the
       rest of the types in a cumulative manner. In addition, we also plot the fitted total dispersion and the corresponding 
       astrophysical dispersion. For large values of \sX\ (or \sXz), Eqn.~\ref{fsX} becomes a power law, which is a straight line in 
       in a log-log plot.  
 \item Figure~\ref{hist_sigma} are selected histograms for the calculated astrophysical dispersion histograms (as opposed to the 
       total dispersion) for three representative ranges of \GG\ (low, intermediate, and high) and the three filters \GG, \GBP, and 
       \GRP. They also show the dispersion ranges that correspond to the different values of the variability flag. As a comparison, 
       the corresponding total dispersion histogram and the fits to both are also shown. In this case a linear scale is used for the 
       horizontal axis to show the behavior near zero astrophysical dispersion. Note that the B-flag histogram consists of a single
       (leftmost) bin, when present.
 \item Figure~\ref{hist_mag_sigma} is the equivalent to Fig.~\ref{hist_mag_sigma0} for the astrophysical dispersion.
\end{enumerate}

The overall result is that the fitted functions work well. In Fig.~\ref{hist_sigma0} both the instrumental peak at low values
of \sXz\ and the power-law behavior at large values are well reproduced for the total dispersion in most panels. In addition, the 
observed distribution in astrophysical dispersions to the left of the instrumental peak is relatively well reproduced by the fitted 
function. The discrepancies are minor and can be classified into three types. 

First, at large values of \sXz, some magnitude-color bins in Fig.~\ref{hist_sigma0} show deviations from the power-law
behavior. In general, when the population is dominated
by a single type of variable star (as determined by the subsample from R22), the fit is better than when the
population is more evenly distributed among variable types. This can be seen by looking at the reddest bins (\GBPmGRP\ of 2.5-8.0,
lower row in Fig.~\ref{hist_sigma0}): the brightest magnitudes are strongly dominated by luminous red stars, of which
the variable stars are classified as LPVs (99\% or more of all detected variables) and those bins have excellent fits. However, as 
we move to fainter magnitudes, populations of lower-luminosity stars appear (as evidenced by the YSOs and solar-like and ellipsoidal
variables) and the diversity of subtypes among the LPVs increases.
The mixture of populations makes the fits worse as the distribution
deviates from a power law. Nevertheless, what should be surprising is that the power-law approximation works so well in general, as
it indicates that enough different types of variables behave in a way similar enough to allow for the fits to be acceptable for very
different parts of the \textit{Gaia} color-magnitude diagram. The power-law description is not always valid, though, as some types of
variables clearly deviate from that behavior. One example are the RR~Lyr in the last page of Fig.~\ref{hist_sigma0} and more are 
discussed later on.

The second discrepancy arises at the left of the instrumental peak: 265 objects in \GG, \num{21852} in \GBP, and \num{18940} in 
\GRP\ are more than 3\sigmasXi\ lower than \sXi\ and are hence given B flags. The number of such targets is much smaller in 
\GG\ than in \GBP\ or \GRP, as seen in %the relative sizes of the orange histograms in Fig.~\ref{hist_sigma} and 
in the differences between the histograms of \sGBPz\ and \sGRPz\ and the fitted functions in either Fig.~\ref{hist_sigma} or 
Fig.~\ref{hist_sigma0}. There is an excess of low \sGBPz\ and \sGRPz\ values for faint magnitudes which is not seen in \sGz.
The effect also appears as a blue region in the lower right corner of the \GBP\ and \GRP\ panels of Fig.~\ref{hist_mag_sigma0}. We 
have looked at the characteristics of those sources and the most likely source of this effect is the low number of observations, 
which apparently distorts the computation of the published magnitude uncertainties. Hence, the label B(ad) in the variability flag to
indicate that those values should not be used.

For the third discrepancy, we look at Fig.~\ref{hist_mag_sigma}. For \GBP\ and \GRP\ (center and right columns), the astrophysical
photometric dispersions follow a smooth distribution where the main trend is the expected one, a relatively flat behavior at bright
and intermediate magnitudes and an increasing trend with magnitude beyond $\GG = 13$. For \GG\ the overall trend is similar but there
are significant residual structures, especially for the bluer color bins. They are not too large but may be introducing systematic 
effects in \sG\ for some specific \GG\ magnitudes at the level of 1~mmag for red stars and 2~mmag for blue ones. This effect is a 
remainder of the existence of uncorrected effects in the \GG\ photometry due to the use of electronic gates.

In summary, our model provides a fit accurate enough for our purposes. However, one should not 
insist in the exactness of Eqn.~\ref{fsX}, which is an empirical approximation. At large values of \sX\ because of the existence of
different types of variables, each one with its distribution, and at low values because slight deviations from the flat character of
$f(s_{\rm X})$ to the left of both \sXi\ and \sXc\ are hard to detect.

\subsubsection{On the use of the variability flag}

$\,\!$\indent We have already introduced the meaning of the variability flag. Figures~\ref{hist_sigma}~and~~\ref{hist_mag_sigma} can
be used to better interpret it. Depending on the \GG\ magnitude of the target, one band or another has a better sensitivity in terms
of being able to measure its astrophysical dispersion. For the brightest stars \GBP\ is the best band, for intermediate ones \GRP\ is
the option of choice, and for the faintest ones \GG\ is the preferred one. Taking as an example the magnitude and color ranges in
Fig.~\ref{hist_sigma}, for $\GG\sim 5$~mag the three-sigma limit defined by the V flag is $\sim 4$~mmag for \GBP, $\sim 6$~mmag for 
\GG, and $\sim 7$~mmag for \GRP. Those values improve for $\GG\sim 10$~mag, with $\sim 1.2$~mmag for \GRP, $\sim 1.8$~mmag for \GBP,
and $\sim 2.5$~mmag for \GG. When we go to fainter stars ($\GG\sim 17$~mag) the three-sigma limit becomes significantly higher: 
$\sim 7$~mmag for \GG, $\sim 17$~mmag for \GRP, and $\sim 23$~mmag for \GBP. Therefore, the best capabilities of \textit{Gaia} for
determining variability take place at intermediate magnitudes. 

Which band is the best one for determining variability? In principle, there is information in all of them but, as we have just seen,
the three sigma limit for \GBP\ and \GRP\ for faint stars (the majority of the sample) becomes significantly worse than the equivalent
for \GG. Therefore, \sG\ is the more uniform of the three bands and the first of the three letters in the variability flag is the one
that is likely to be able to identify more variables. One can take the safe approach of considering only stars with a VVV flag as
objects that are certainly variable but in doing so one would discard a relatively large number of faint objects with a VMM flag that 
have values of \sX\ in the 10-20~mmag range in all three bands. On the other hand, one should be aware of the differences between 
\GG\ and \GBP+\GRP\ that appear in Fig.~\ref{hist_mag_sigma0} that are propagated to Fig.~\ref{hist_mag_sigma}. The \sXz\ and \sX\
distributions as a function of \GG\ for \GBP\ and \GRP\ are relatively smooth but those for \GG\ show discontinuities caused by the
onboard data processing. Therefore, one should expect an increase in the number of discrepant sources arising from \GG\ due to this
effect. Below we address this issue by analyzing the existence of outliers in the dispersion-dispersion planes, which adds 
information to the cases with variability flags where the three letters are not the same.

\subsubsection{Observed dispersion}

$\,\!$\indent A third question is whether the value of \sXz\ determined from $\sigma_X$ corresponds to the real photometric
dispersion. The question is analyzed in the second paper of this series by comparing the values calculated here with those determined
from the sample with light curves. In summary, the agreement is quite good but there are some details about which points are
selected and which are rejected from the light curve and the calculation of final weighted mean magnitude by the \textit{Gaia} team, 
as otherwise disagreements can appear in the \GG\ band. In Fig.~\ref{light_curves} we show some examples of light curves built from 
\textit{Gaia}~EDR3 epoch photometry and the corresponding (independently obtained) values of \sXz\ and \sX.

\subsubsection{Outliers in the dispersion-dispersion planes}

$\,\!$\indent A fourth question is the existence of outliers in the dispersion-dispersion planes. In Fig.~\ref{varhist_All} we show
the density diagrams for stars that are clearly variables 
in the three color combinations and in Fig.~\ref{varhist_types} we show
similar plots for the subsamples with R22 variability classifications but using as $y$ axis the ratio of the previous 
$y$ to $x$ axes. Those figures are discussed below, as the structures seen in them are useful for the identification of different
types of variable stars. Here we discuss a structure that appears in them and that is of instrumental origin. Comparing the left and
center panels of the two bottom rows in Fig.~\ref{varhist_All}, we see that in the left panels a region around (50~mmag,~10~mmag) is 
significantly more populated in the left panels compared to the center ones. In other words, there is an excess of sources with 
significantly larger \sGRP\ and \sGBP\ with respect to \sG\ when comparing the global sample of this paper with the subsample from 
R22. As no equivalent structure is seen in the top row, the phenomenon appears to affect \sGRP\ and \sGBP\ 
simultaneously with respect to \sG. We have verified this by selecting that the targets that simultaneously satisfy having the
variability flag as VVV, $\sG < 20$~mmag, $\log\sG < \log\sGRP - 0.3$, and $\log\sG < \log\sGBP - 0.3$ are \num{5060097} and the 
ones that satisfy only the first three conditions are \num{5327967}, that is, the first number is 95\% of the second one.

We have investigated this issue and discovered that the majority of those stars have companions 1-2\arcsec\ away with a relatively
small magnitude difference. This can be seen more clearly in the (better studied) bright subsample, as most of those objects have
entries in the Washington Double Star catalog (WDS, \citealt{Masoetal01}). Therefore, the likely explanation is that in those cases
\GBP\ and \GRP\ photometry, which is obtained from the collapse of slitless spectrophotometry in pseudowavelength, is contaminated to 
different degrees in different epochs depending on the orientation of the CCDs. Hence, in those cases only \sG\ can be trusted but 
not \sGBP\ or \sGRP.
%Gaia DR3 383445785309696896
%HD 724 A, HD 709 A

In addition, an even smaller number of sources have the opposite problem: large values of \sG\ in comparison with those of \sGBP\ and
\sGRP. We suspect that some of those are also caused by close visual binaries but others by the discontinuities in \sG\ seen in
Fig.~\ref{hist_mag_sigma}. In general, we recommend that those cases for which \sG\ is very different from \sGBP\ or \sGRP be treated 
with caution. The analysis of different types of variables below can serve as a guide for the expected ranges.
%HD 90263, 2MASS J05162172+3042398

\subsubsection{Validation summary}

$\,\!$\indent The provided astrophysical dispersions are overall a good description of reality but,
as with all massive amounts of data, they have to be used with caution. Be aware of the limitations indicated by the variability flag
and on the lookout for cases where one of the dispersions is very different from the other two, as in those cases it is likely that
the higher value(s) are not real. 

\section{Results}

\subsection{The dispersion-dispersion planes}

\begin{table*}
\caption{Observed properties for different types of variables selecting stars classified as VVV and with dispersions uncertainties \sigmasX\ lower than 1~mmag in the three bands.}
\addtolength{\tabcolsep}{-1mm}
\renewcommand{\arraystretch}{1.4}
\centerline{
\begin{tabular}{lrrr@{}lr@{}lr@{}lr@{}lr@{}lr@{}lr@{}lr@{}l}
\hline
Variability type   & \mci{Number} & \mci{Perc.} & \mcii{\GG}   & \mcii{\GBPmGRP} & \mcii{\sG}    & \mcii{\sGBP}  & \mcii{\sGRP}  & \mcii{\sGBP/\sGRP} & \mcii{\sG/\sGRP}   & \mcii{\sG/\sGBP}   \\
                   & \mci{used}   & \mci{used}  & \mcii{(mag)} & \mcii{(mag)}    & \mcii{(mmag)} & \mcii{(mmag)} & \mcii{(mmag)} & \mcii{}            & \mcii{}            & \mcii{}            \\
\hline
Solar-like         & \num{437301} &  26.3 & 13.8&$_{-1.1}^{+0.7}$ &    1.08&$_{-0.16}^{+0.25}$ &   7.4&$_{-   2.0}^{+   3.0}$ &   9.9&$_{-   3.1}^{+   4.7}$ &   6.8&$_{-   2.2}^{+   3.3}$ & 1.434&$_{-0.284}^{+0.393}$ & 1.118&$_{-0.236}^{+0.252}$ & 0.778&$_{-0.173}^{+0.173}$ \\
LPV                & \num{777307} &  85.9 & 14.3&$_{-1.7}^{+1.2}$ &    4.43&$_{-1.34}^{+1.18}$ &  66.7&$_{-  28.1}^{+  65.8}$ & 167.8&$_{-  81.4}^{+ 220.9}$ &  57.4&$_{-  25.4}^{+  59.1}$ & 2.460&$_{-0.624}^{+2.319}$ & 1.150&$_{-0.123}^{+0.162}$ & 0.475&$_{-0.222}^{+0.222}$ \\
$\delta$ Sct+      & \num{253382} &  36.9 & 13.4&$_{-1.3}^{+1.0}$ &    0.63&$_{-0.15}^{+0.19}$ &   8.2&$_{-   2.7}^{+   7.0}$ &   9.3&$_{-   4.0}^{+   9.6}$ &   6.8&$_{-   3.0}^{+   7.2}$ & 1.458&$_{-0.435}^{+0.332}$ & 1.284&$_{-0.351}^{+0.427}$ & 0.876&$_{-0.168}^{+0.168}$ \\
RS CVn             & \num{246418} &  45.7 & 14.8&$_{-1.1}^{+0.7}$ &    1.29&$_{-0.24}^{+0.28}$ &  28.3&$_{-  10.3}^{+  19.0}$ &  36.0&$_{-  13.7}^{+  23.9}$ &  27.2&$_{-  10.3}^{+  18.9}$ & 1.318&$_{-0.223}^{+0.255}$ & 1.070&$_{-0.162}^{+0.115}$ & 0.816&$_{-0.122}^{+0.122}$ \\
Eclipsing binary   & \num{283282} &  55.2 & 15.1&$_{-1.6}^{+1.0}$ &    1.02&$_{-0.30}^{+0.33}$ &  73.9&$_{-  50.2}^{+  79.7}$ &  82.4&$_{-  55.9}^{+  83.4}$ &  78.2&$_{-  52.1}^{+  77.9}$ & 1.059&$_{-0.158}^{+0.191}$ & 0.990&$_{-0.161}^{+0.096}$ & 0.940&$_{-0.151}^{+0.151}$ \\
RR Lyr             & \num{ 67814} &  94.4 & 15.9&$_{-1.0}^{+0.7}$ &    0.76&$_{-0.18}^{+0.35}$ & 176.6&$_{-  51.0}^{+  92.5}$ & 217.9&$_{-  61.1}^{+ 110.8}$ & 150.1&$_{-  49.1}^{+  72.6}$ & 1.561&$_{-0.262}^{+0.130}$ & 1.250&$_{-0.180}^{+0.102}$ & 0.805&$_{-0.056}^{+0.056}$ \\
YSO                & \num{  8283} &  21.6 & 13.9&$_{-1.4}^{+1.1}$ &    1.99&$_{-0.49}^{+0.68}$ &  26.3&$_{-  16.6}^{+  41.5}$ &  38.4&$_{-  23.4}^{+  66.9}$ &  24.0&$_{-  15.3}^{+  38.3}$ & 1.521&$_{-0.274}^{+0.696}$ & 1.135&$_{-0.173}^{+0.204}$ & 0.745&$_{-0.218}^{+0.218}$ \\
Ellipsoidal        & \num{  3010} &  11.2 & 15.1&$_{-0.6}^{+0.5}$ &    2.05&$_{-0.42}^{+0.50}$ &  17.2&$_{-   6.0}^{+  12.6}$ &  46.0&$_{-  18.0}^{+  34.6}$ &  24.6&$_{-   9.5}^{+  17.4}$ & 1.814&$_{-0.559}^{+1.205}$ & 0.854&$_{-0.431}^{+0.274}$ & 0.431&$_{-0.238}^{+0.238}$ \\
Cepheid            & \num{  9607} &  89.9 & 15.5&$_{-2.1}^{+0.7}$ &    0.93&$_{-0.21}^{+1.11}$ & 156.3&$_{-  64.8}^{+  91.4}$ & 200.7&$_{-  81.5}^{+ 112.4}$ & 130.3&$_{-  54.3}^{+  72.1}$ & 1.576&$_{-0.231}^{+0.159}$ & 1.216&$_{-0.158}^{+0.140}$ & 0.782&$_{-0.061}^{+0.061}$ \\
$\alpha^2$ CVn+    & \num{  6540} &  71.9 &  9.3&$_{-1.0}^{+0.5}$ &    0.25&$_{-0.23}^{+0.17}$ &   7.3&$_{-   1.9}^{+   5.6}$ &   6.6&$_{-   2.8}^{+   6.5}$ &   5.2&$_{-   2.5}^{+   8.0}$ & 1.248&$_{-0.436}^{+0.490}$ & 1.368&$_{-0.563}^{+1.167}$ & 1.065&$_{-0.278}^{+0.278}$ \\
Be+                & \num{  4022} &  52.9 & 14.9&$_{-4.2}^{+0.9}$ &    0.06&$_{-0.19}^{+0.67}$ &  37.1&$_{-  20.6}^{+  44.9}$ &  37.3&$_{-  20.9}^{+  40.3}$ &  49.6&$_{-  32.2}^{+  64.1}$ & 0.783&$_{-0.242}^{+0.378}$ & 0.811&$_{-0.203}^{+0.292}$ & 1.051&$_{-0.219}^{+0.219}$ \\
Short timescale    & \num{   997} &  23.8 & 16.1&$_{-1.2}^{+0.5}$ &    1.06&$_{-0.20}^{+0.31}$ & 111.4&$_{-  43.4}^{+  61.5}$ & 120.8&$_{-  74.5}^{+  82.2}$ & 107.0&$_{-  69.4}^{+  73.4}$ & 1.098&$_{-0.173}^{+0.295}$ & 1.208&$_{-0.604}^{+0.982}$ & 1.090&$_{-0.589}^{+0.589}$ \\
AGN                & \num{   994} &  27.8 & 16.0&$_{-0.7}^{+0.5}$ &    0.69&$_{-0.22}^{+0.24}$ & 104.9&$_{-  53.6}^{+ 101.6}$ & 111.1&$_{-  50.5}^{+ 103.4}$ &  89.1&$_{-  42.3}^{+  84.7}$ & 1.246&$_{-0.223}^{+0.300}$ & 1.140&$_{-0.189}^{+0.276}$ & 0.930&$_{-0.141}^{+0.141}$ \\
$\beta$ Cep        & \num{  1271} &  93.4 & 10.3&$_{-1.1}^{+1.3}$ &    0.36&$_{-0.31}^{+0.58}$ &   9.7&$_{-   3.1}^{+   7.7}$ &   8.6&$_{-   3.8}^{+   8.5}$ &   7.7&$_{-   3.6}^{+   7.8}$ & 1.154&$_{-0.198}^{+0.214}$ & 1.237&$_{-0.217}^{+0.591}$ & 1.075&$_{-0.168}^{+0.168}$ \\
Slowly pulsating B & \num{  1116} &  96.8 &  8.6&$_{-1.5}^{+1.6}$ & $-$0.05&$_{-0.09}^{+0.08}$ &  11.7&$_{-   2.7}^{+   4.9}$ &  11.9&$_{-   3.4}^{+   5.5}$ &  10.9&$_{-   3.4}^{+   5.6}$ & 1.145&$_{-0.304}^{+0.201}$ & 1.132&$_{-0.265}^{+0.198}$ & 0.988&$_{-0.113}^{+0.113}$ \\
sdB                & \num{    17} &   1.9 & 13.3&$_{-0.5}^{+0.6}$ & $-$0.39&$_{-0.07}^{+0.11}$ &  11.9&$_{-   1.6}^{+   6.3}$ &  11.5&$_{-   4.5}^{+   9.6}$ &   8.3&$_{-   4.2}^{+   8.9}$ & 1.237&$_{-0.120}^{+0.656}$ & 1.390&$_{-0.363}^{+1.530}$ & 1.033&$_{-0.170}^{+0.170}$ \\
CV                 & \num{   601} &  92.3 & 15.9&$_{-1.4}^{+0.8}$ &    0.53&$_{-0.30}^{+0.49}$ & 265.1&$_{- 119.8}^{+ 351.4}$ & 294.5&$_{- 132.8}^{+ 381.2}$ & 247.8&$_{- 107.7}^{+ 289.8}$ & 1.198&$_{-0.180}^{+0.271}$ & 1.102&$_{-0.146}^{+0.174}$ & 0.916&$_{-0.104}^{+0.104}$ \\
WD                 & \num{    52} &   9.0 & 15.1&$_{-1.6}^{+0.7}$ & $-$0.07&$_{-0.41}^{+0.19}$ &  33.3&$_{-  14.4}^{+  21.5}$ &  35.3&$_{-  18.2}^{+  20.9}$ &  31.1&$_{-  14.4}^{+  34.2}$ & 1.064&$_{-0.338}^{+0.415}$ & 0.987&$_{-0.261}^{+0.419}$ & 0.955&$_{-0.167}^{+0.167}$ \\
Symbiotic          & \num{   495} &  94.1 & 14.6&$_{-2.4}^{+0.8}$ &    3.70&$_{-1.23}^{+0.47}$ &  96.1&$_{-  30.1}^{+  49.7}$ & 213.4&$_{-  92.5}^{+ 100.5}$ &  81.9&$_{-  29.2}^{+  49.9}$ & 2.590&$_{-1.065}^{+0.863}$ & 1.178&$_{-0.143}^{+0.187}$ & 0.465&$_{-0.102}^{+0.102}$ \\
$\alpha$ Cyg       & \num{   299} &  94.0 &  7.4&$_{-1.3}^{+3.8}$ &    0.52&$_{-0.36}^{+0.50}$ &  16.1&$_{-   5.2}^{+   6.7}$ &  16.6&$_{-   5.5}^{+   6.1}$ &  18.0&$_{-   7.0}^{+  10.9}$ & 0.943&$_{-0.255}^{+0.171}$ & 0.944&$_{-0.235}^{+0.157}$ & 1.000&$_{-0.107}^{+0.107}$ \\
Exoplanet transit  & \num{    83} &  39.5 & 12.3&$_{-1.5}^{+1.0}$ &    0.86&$_{-0.18}^{+0.35}$ &   5.1&$_{-   1.7}^{+   3.5}$ &   5.3&$_{-   2.4}^{+   3.6}$ &   3.9&$_{-   1.3}^{+   3.3}$ & 1.254&$_{-0.371}^{+0.317}$ & 1.209&$_{-0.333}^{+0.628}$ & 1.000&$_{-0.214}^{+0.214}$ \\
R CrB              & \num{    87} & 100.0 & 14.4&$_{-2.4}^{+1.5}$ &    2.40&$_{-0.90}^{+0.94}$ & 602.9&$_{- 444.1}^{+1255.1}$ & 671.2&$_{- 466.7}^{+ 780.2}$ & 580.7&$_{- 441.7}^{+ 997.3}$ & 1.205&$_{-0.320}^{+0.316}$ & 1.100&$_{-0.130}^{+0.179}$ & 0.911&$_{-0.174}^{+0.174}$ \\
Microlensing       & \num{     6} &  17.6 & 15.6&$_{-0.7}^{+0.7}$ &    1.91&$_{-0.69}^{+0.83}$ &  48.7&$_{-  27.5}^{+ 120.3}$ & 106.6&$_{-  28.5}^{+ 123.5}$ &  85.6&$_{-  66.0}^{+  89.7}$ & 1.678&$_{-0.715}^{+3.349}$ & 0.966&$_{-0.472}^{+0.179}$ & 0.352&$_{-0.138}^{+0.138}$ \\
\hline
\end{tabular}
\addtolength{\tabcolsep}{1mm}
\renewcommand{\arraystretch}{1.0}
}
\label{variable_types_results}
\end{table*}

$\,\!$\indent We start by looking at the distribution of the sample in the dispersion-dispersion planes, which
we already introduced in the previous section to mention the existence of outliers there caused by instrumental issues. The
distribution is shown in Fig.~\ref{varhist_All} for the stars that are clearly variable, which we define as those classified as VVV
and with dispersion uncertainties lower than 1~mmag in the three bands. The first column shows the density for all stars that
satisfy those conditions and the second one the density for the subsample with R22 variability classifications. The
third column is a gray-scale version of the second one with colors used to indicate the density distribution of eight of the most
common types of variables, which together comprise almost 99\% of all of the R22 sample analyzed here (see 
Table~\ref{vartypestats}). In addition, in Fig.~\ref{varhist_types} we show equivalent plots for each of the variable types in 
R22 but in this case plotting one astrophysical dispersion in the horizontal axis and the ratio of two in the vertical
axis. Finally, Table~\ref{variable_types_results} lists the properties extracted from Fig.~\ref{varhist_types}.

\begin{figure*}
\centerline{\includegraphics[width=0.49\linewidth]{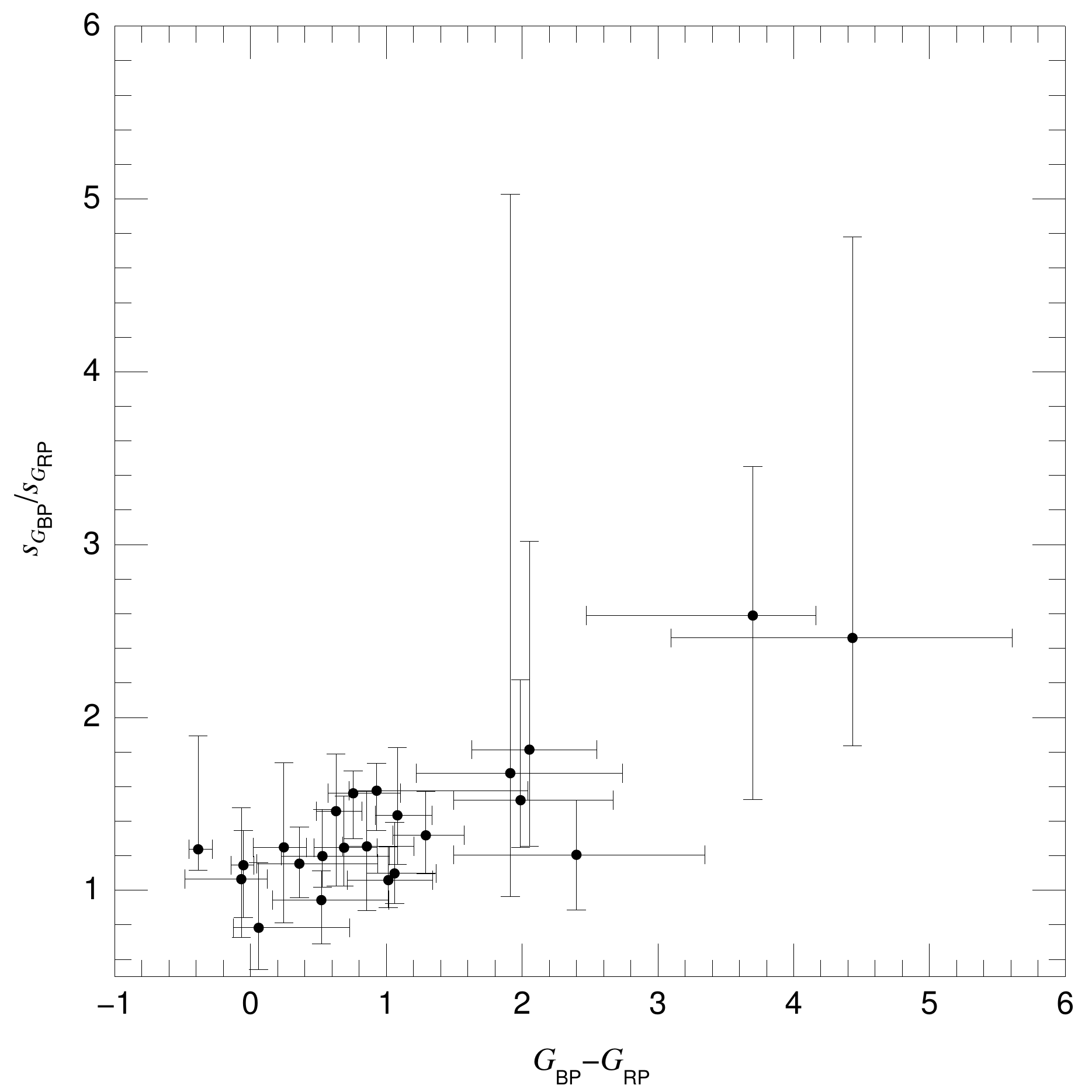} \
            \includegraphics[width=0.49\linewidth]{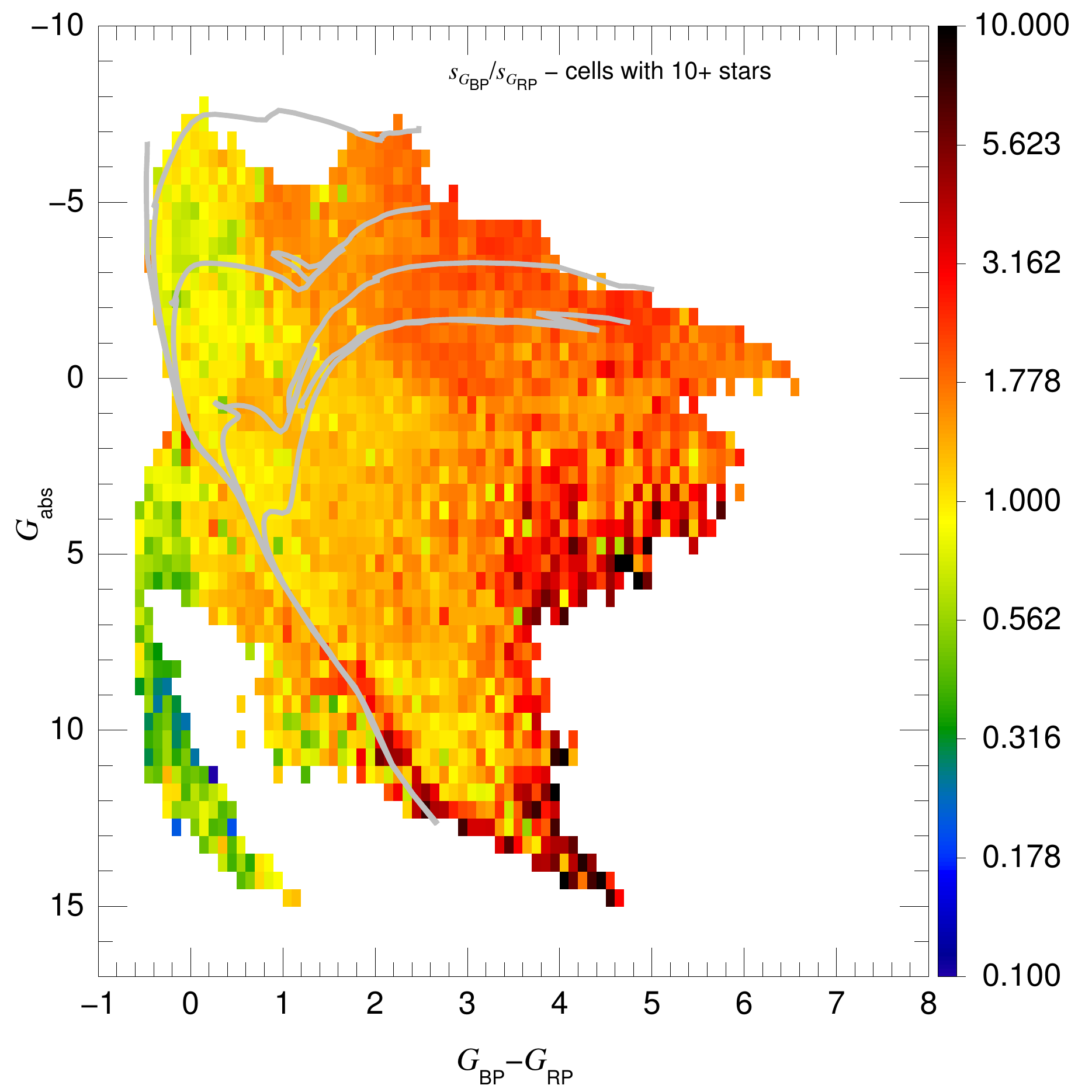}}
\caption{({\it Left}) \sGBP/\sGRP\ as a function of \GBPmGRP\ from the data in Table~\ref{variable_types_results}. The plotted 
         values are the median and the one-sigma equivalents derived from the distribution of each sample, which is done to reflect 
         some of the significant asymmetries in them. ({\it Right}) Average \sGBP/\sGRP\ across the \textit{Gaia} CAMD for 
         color-magnitude cells with 10 or more targets.}
\label{sBPsRP}
\end{figure*}

Figure~5 shows that the three dispersion-dispersion pairs are highly correlated, as expected since stars vary in
similar ways in all optical bands to a first approximation. However, there is significant structure within each diagram. For
the middle row (\sG\ vs. \sGRP), the correlation between the two dispersions is quite tight with the only exception of the
instrumental effect described in the previous section caused by the different nature of the \GG\ and \GRP\ photometry. For the other
two rows where \sGBP\ is involved the correlation is not as tight and at least three trends are seen: one close to the
diagonal line (with \sGBP\ slightly larger than \sGRP\ or \sG) that runs for most of the extension of the diagram, a second one with
\sGBP\ significantly larger than the other dispersion, and a third one in between the two that is seen only at large values of \sGBP\
(200~to~300~mmag). The distinction between the three trends is seen in both the top and bottom rows but is more clear in the top one
(\sGBP\ vs. \sGRP), as there is a better separation between the two passbands. The right column
indicates that different trends correspond to different types of variables: the second trend is dominated by 
LPVs, the third one by RR Lyr stars and Cepheids, and the first one by the other five types. In addition, different types have
significantly different ranges in dispersion (with some biases that we discuss below). Therefore, our first result is that
\textbf{the location of the targets in the dispersion-dispersion planes, most notable on \sGBP\ vs. \sGRP, can be used to constrain 
their variability type.}

Before proceeding further with that idea, there are two types of biases that have to be considered. The first one is 
introduced by our criterion that only VVV objects with \sigmasX\ smaller than 1~mmag are considered. That criterion eliminates
low-dispersion targets, especially faint ones. It can be determined from Table~\ref{variable_types_results}, where we list the
percentage of stars of each type in the R22 sample that pass the criterion. Types with high typical values of \sX\ are
little affected by the criterion, with R~CrB stars, the one with the highest values, not having a single target excluded. The 
exception there are short timescale variables, microlensing events, and AGNs, for which many targets are excluded for being too faint. 
In general, low-dispersion variables such as solar-like and $\delta$~Sct+ stars have a large percentage excluded, indicating that 
their median values of \sX\ in Table~\ref{variable_types_results} are underestimated. The second type of bias is the selection
criterion used by R22 to include stars in their sample. Such a bias is harder to determine but it is present, as seen
by comparing the left and center columns in Fig.~\ref{varhist_All}: the second trend mentioned above associated to LPV variables is
continuous in the top and bottom panels of the first column but is interrupted around $\sGBP\sim 60$~mmag in their equivalents in the
center column. This indicates that LPVs in the R22 sample are biased towards those with large amplitudes, an effect
that is also seen in the lower rows of Fig.~\ref{hist_sigma0}. Therefore, both biases should be at work in 
Table~\ref{variable_types_results} and one would expect that the values of \sX\ should be smaller in general.

Looking into the specific variable-type categories in Table~\ref{variable_types_results} and Fig.~\ref{varhist_types}, there are
several interesting results:

\begin{itemize}
 \item Eclipsing binaries have values of nearly 1.0 in the three median dispersion ratios with little scatter. Ideal eclipsing 
       binaries of identical effective temperature for each component should indeed have ratios of 1.0 and if the temperature
       difference is small they should not stray far away from that. Therefore, this result validates our technique.
 \item In general, the redder the target class, the larger \sGBP/\sGRP\ becomes (Fig.~\ref{sBPsRP}). This is a expected 
       behavior for intrinsically (not orbitally) variable stars, as in a cool star a small change in temperature has a larger flux 
       effect in the blue than in the red, and in a hot star the effect is reversed. However, the correlation is not perfect due to 
       the presence of the biases above, the presence of extinction confounding intrinsic and observed colors, and the existence of 
       different origins and details for the variability of different types (right panel of Fig.~\ref{sBPsRP}). For example, M stars 
       located near the ZAMS have significantly larger values of \sGBP/\sGRP\ than those that have experienced some evolution. This
       is consistent with young cool photospheres being more active at shorter wavelengths than older ones.
 \item As a corollary of the previous point, the only two types with median \sGBP/\sGRP\ lower than 1.0 are Be+ and $\alpha$~Cyg
       stars, which include most of the massive early-type stars analyzed by R22. They are not the ones with lower
       value of \GBPmGRP\ in Table~\ref{variable_types_results} due to their higher average extinction (despite the presence of 
       many LMC+SMC stars among them, see below) but they are likely the two types with the lower unreddened median values of 
       \GBPmGRP.
 \item Cepheids and RR~Lyr stars occupy a relatively distinct (but common for the two types) parameter space around \sX\ of 
       100-200~mmag and \sGBP/\sGRP\ around 1.6.
 \item Some types have complex distributions in Fig.~\ref{varhist_types}, a likely consequence of the fact that they are a mixture of
       different subtypes. For example, most of the $\delta$~Sct+ stars have values of \sX\ lower than 10~mmag but two tails extend
       towards higher values with different \sGBP/\sGRP: one around 1.0 and another around 1.3. LPVs have one major concentration
       around \sGRP\ of 30-100~mmag and \sGBP/\sGRP\ of 2.0-2.5 (truncated towards lower values of \sGRP, see above), but with a tail 
       extending towards much larger values of \sGBP/\sGRP\ and with a distinct group around \sGRP\ of 500-1000~mmag. The latter 
       are likely highly variable AGB stars.
 \item The most variable type by far is that of R~CrB stars. In addition to the already mentioned RR Lyr, Cepheid, and LPV types, two
       others that also have large values of \sX\ are CVs and symbiotic stars.
\end{itemize}

\subsection{Color-absolute magnitude diagrams}

\begin{figure*}
\centerline{\includegraphics[width=0.49\linewidth]{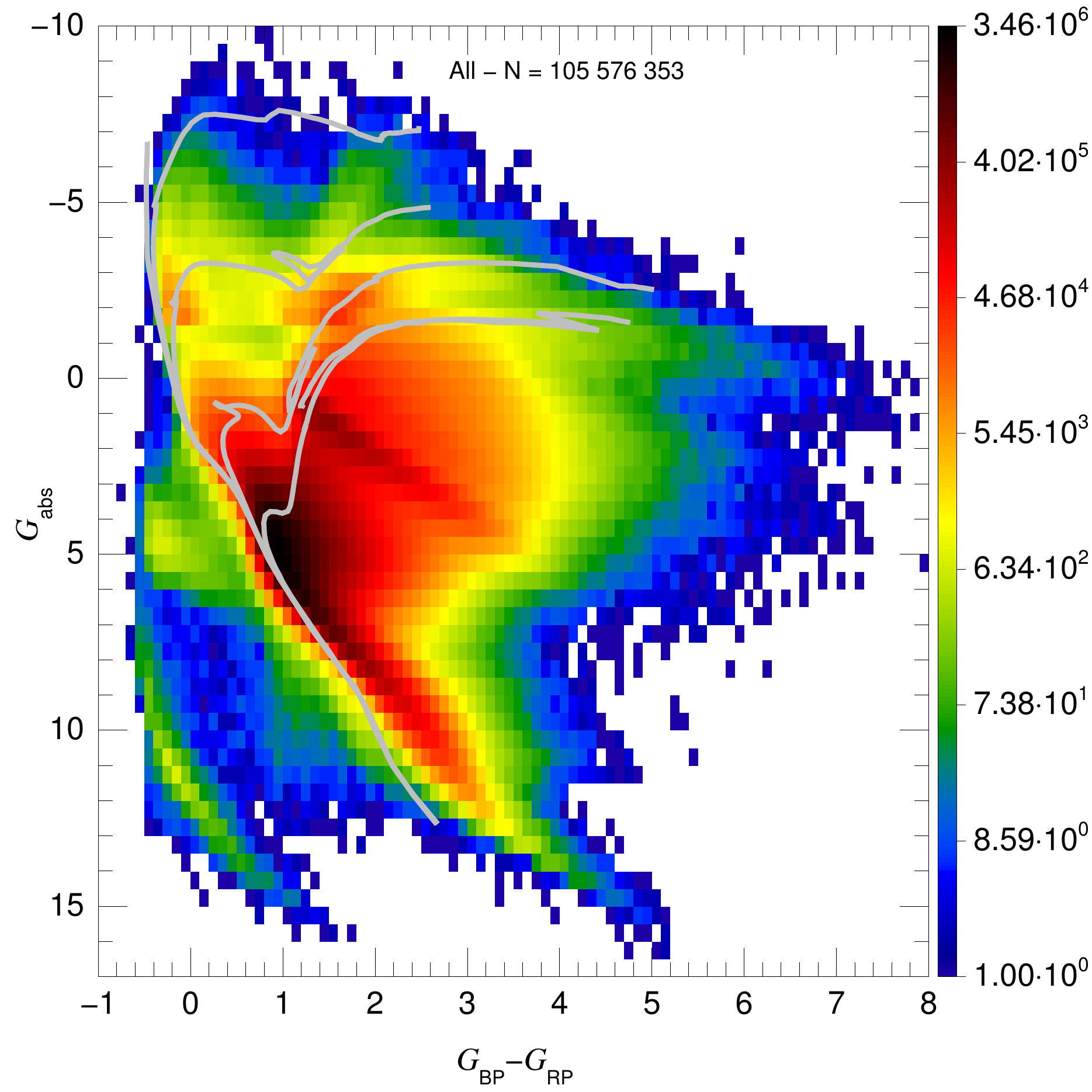} \
            \includegraphics[width=0.49\linewidth]{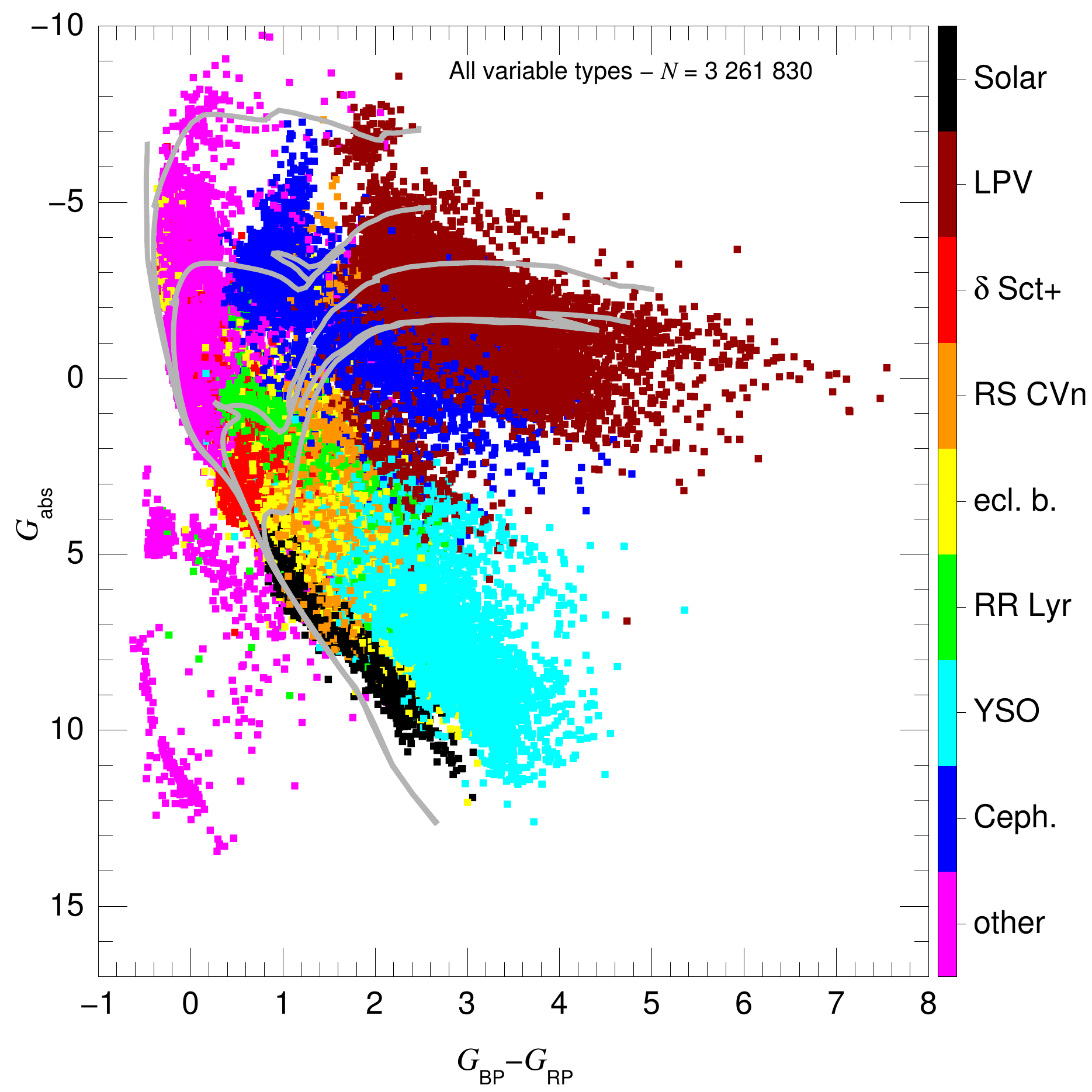}}
\centerline{\includegraphics[width=0.49\linewidth]{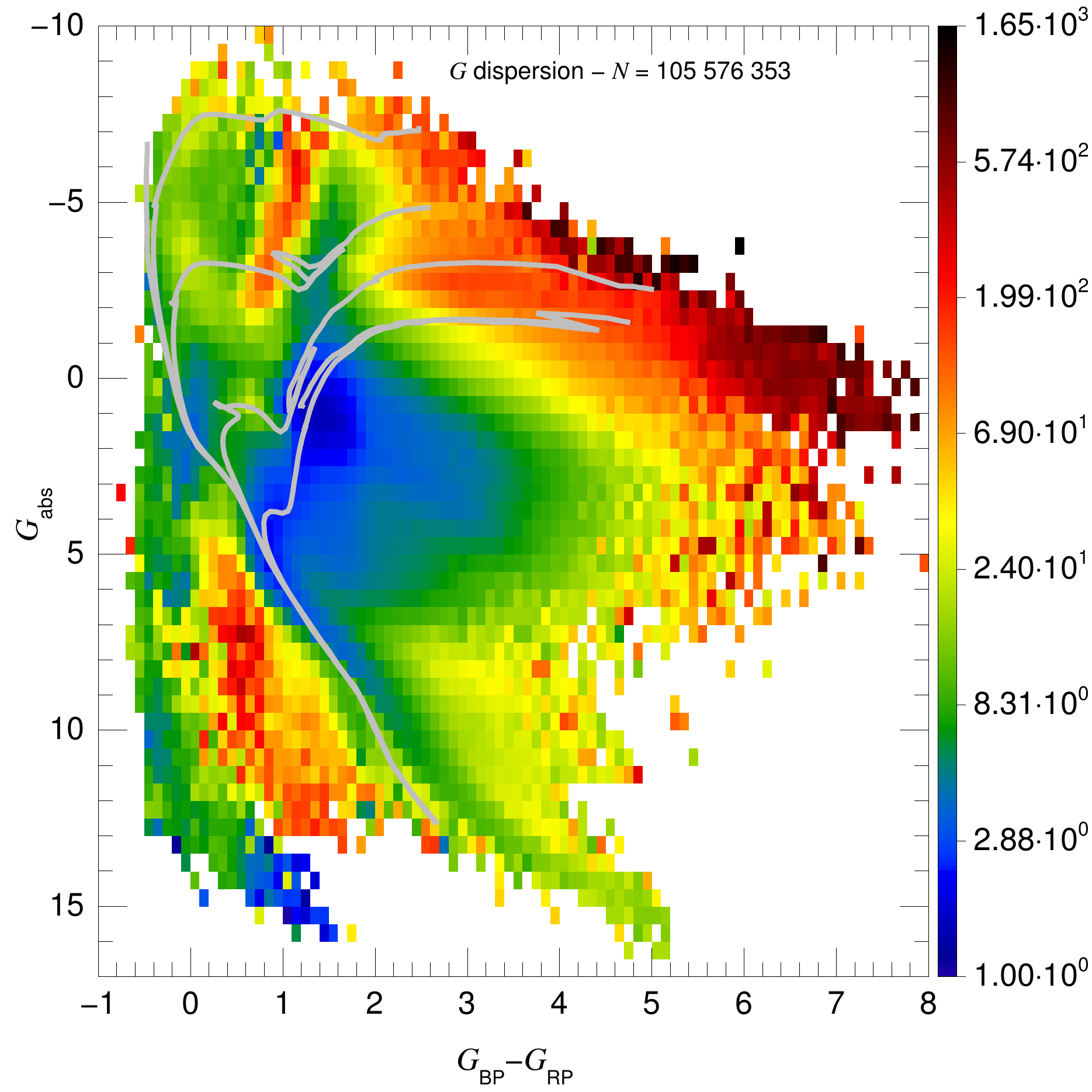} \
            \includegraphics[width=0.49\linewidth]{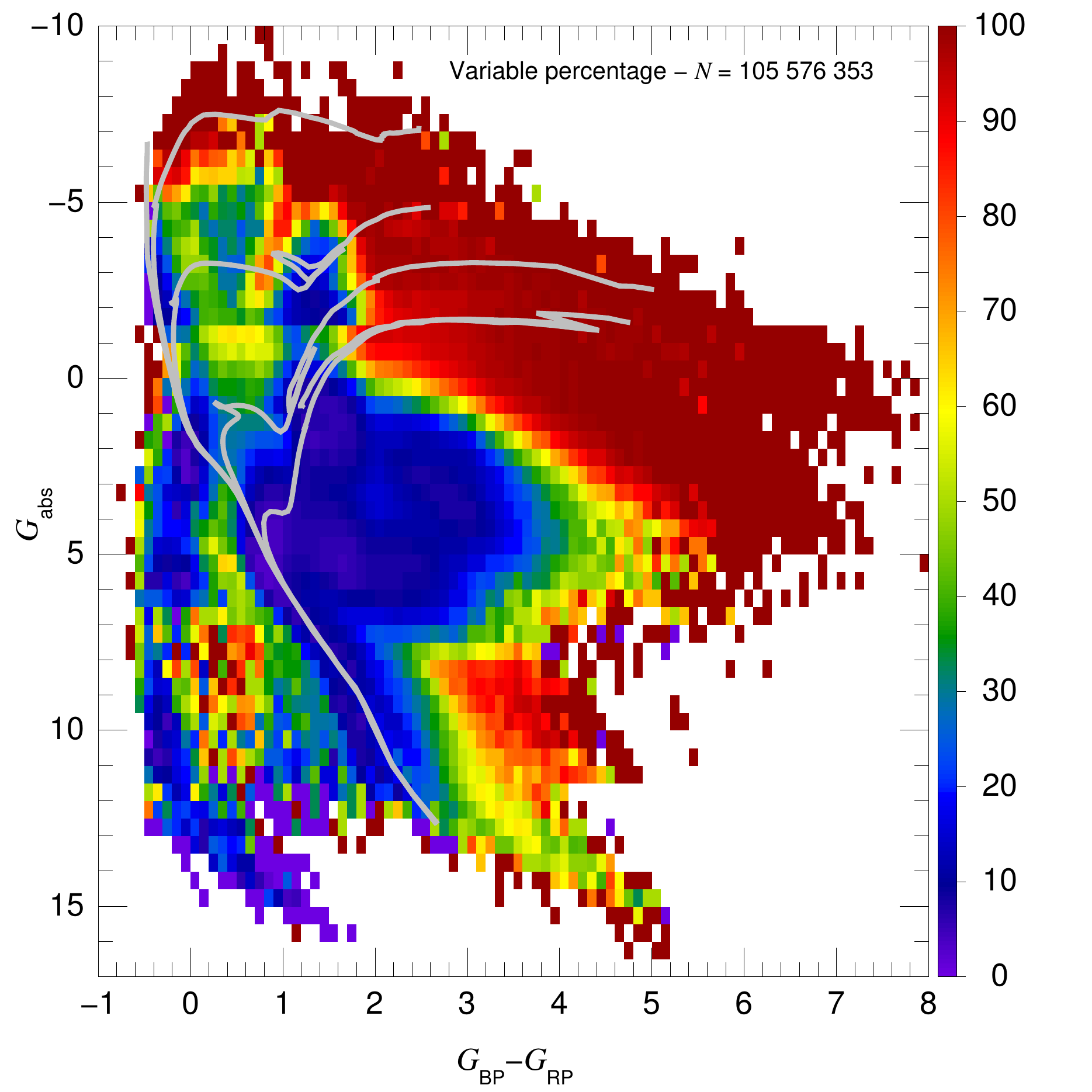}}
\caption{Color-absolute magnitude style diagrams for the sample in this paper with accurate distances. ({\it top left}) Source 
         density. ({\it top right}) Variable types from the R22 subsample with classifications. ({\it bottom left}) Average \GG\ 
         astrophysical dispersion. ({\it bottom right}) Percentage of objects with VVV variability flag. Isochrones for 1~Ma, 10~Ma, 
         100~Ma, 1~Ga, and 10~Ga (in all cases for solar metallicity) are plotted in grey, with the isochrones being a combination
         of Geneva \citep{LejeScha01} for high-mass stars and Padova \citep{Giraetal00,Salaetal00} for low-mass stars (see
         \citealt{Maiz13a}). The two black dotted 
         vertical lines in the first panel show the location of the color limits for the calculation of the instrumental dispersion. 
         The discontinuity seen around $G_{\rm abs} = -1.5$~mag is caused by the inclusion of the LMC (and, to a lesser degree, the 
         SMC at difference of half a magnitude) stars, which dominate the {\it Gaia} luminous star population if one does not
         correct for extinction (compare with Fig.~\ref{CAMD_MW}). In the top right panel the number of plotted sources per variable
         type is capped at \num{10000} to allow all types to be seen better.}
\label{CAMD_all}
\end{figure*}

\begin{figure*}
\centerline{\includegraphics[width=0.49\linewidth]{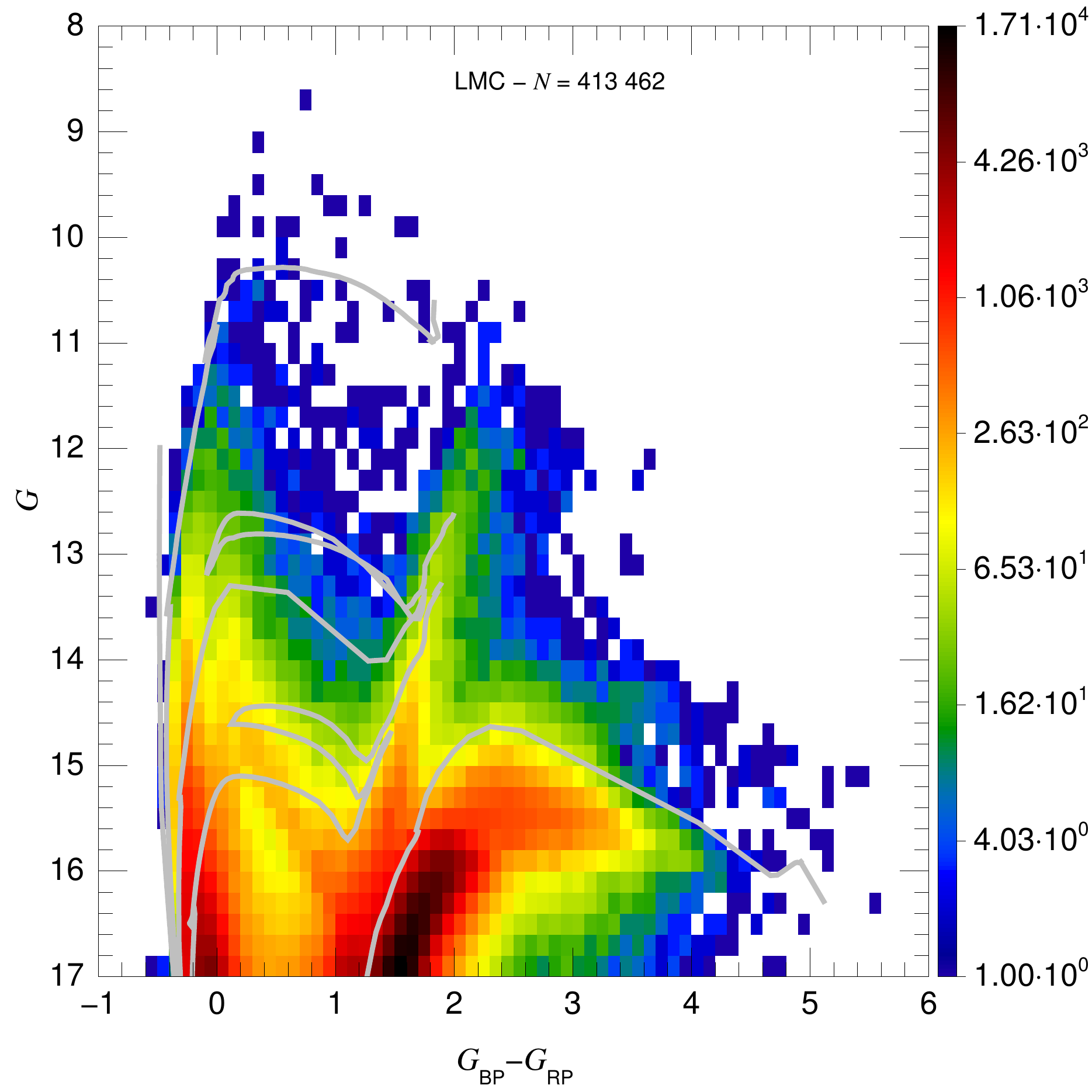} \
            \includegraphics[width=0.49\linewidth]{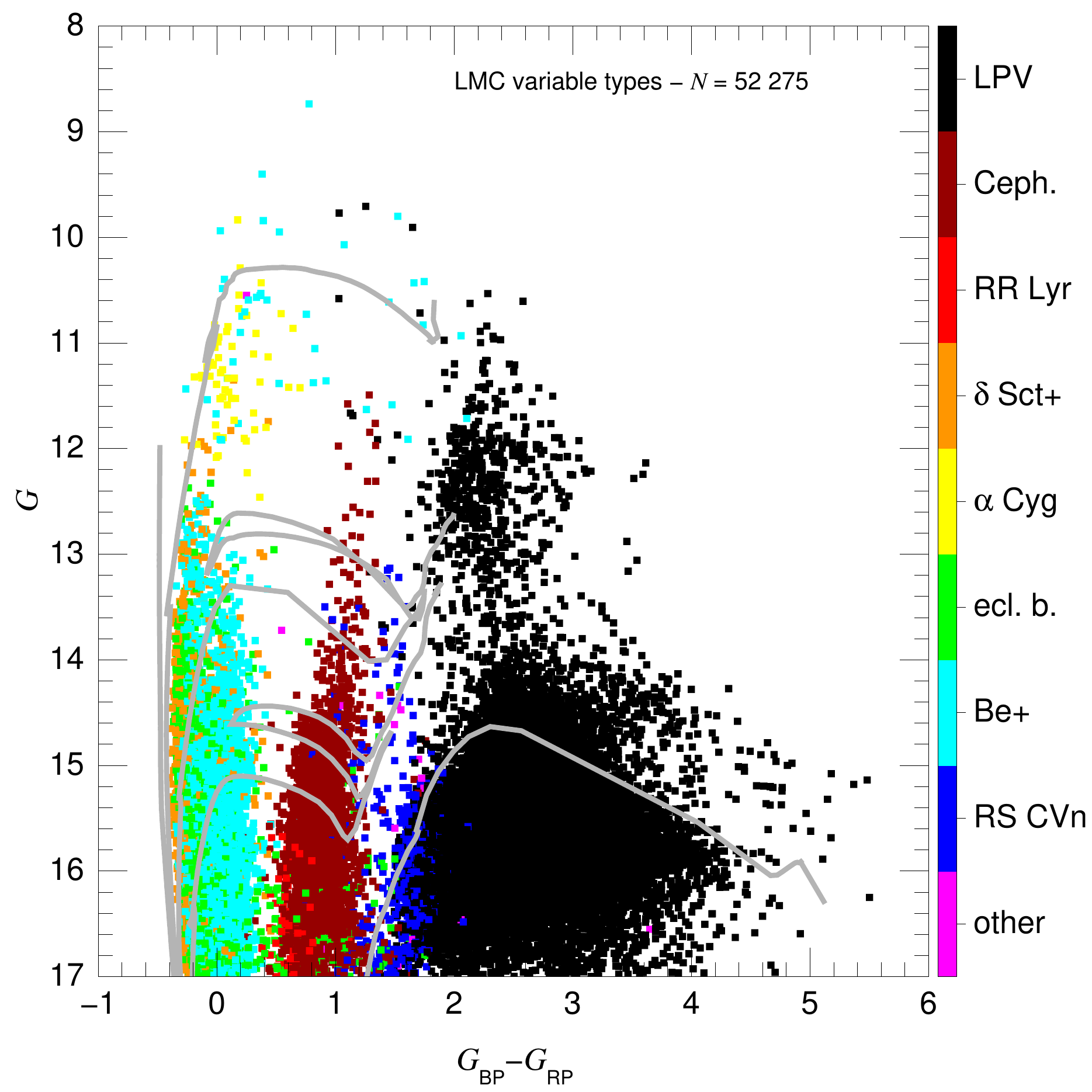}}
\centerline{\includegraphics[width=0.49\linewidth]{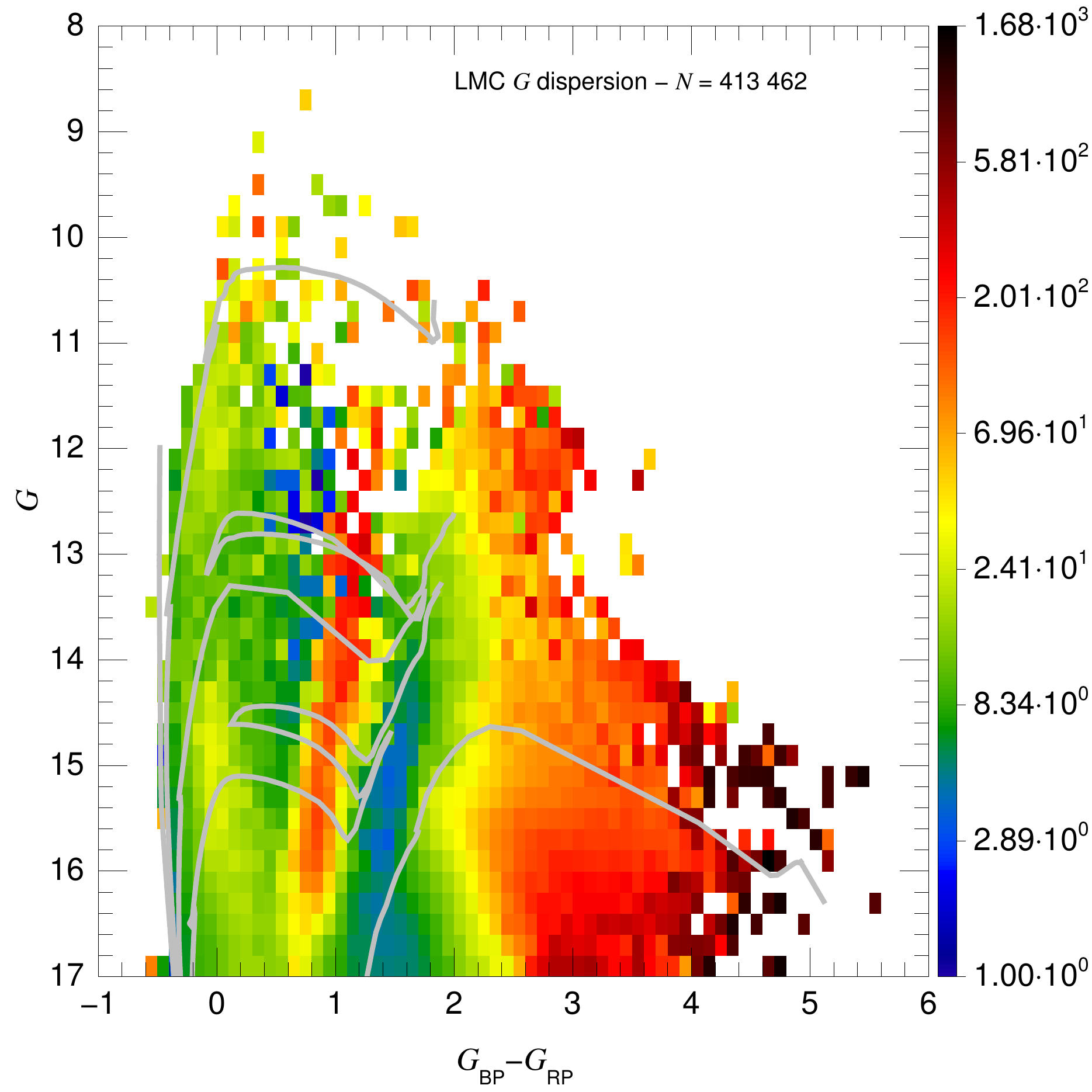} \
            \includegraphics[width=0.49\linewidth]{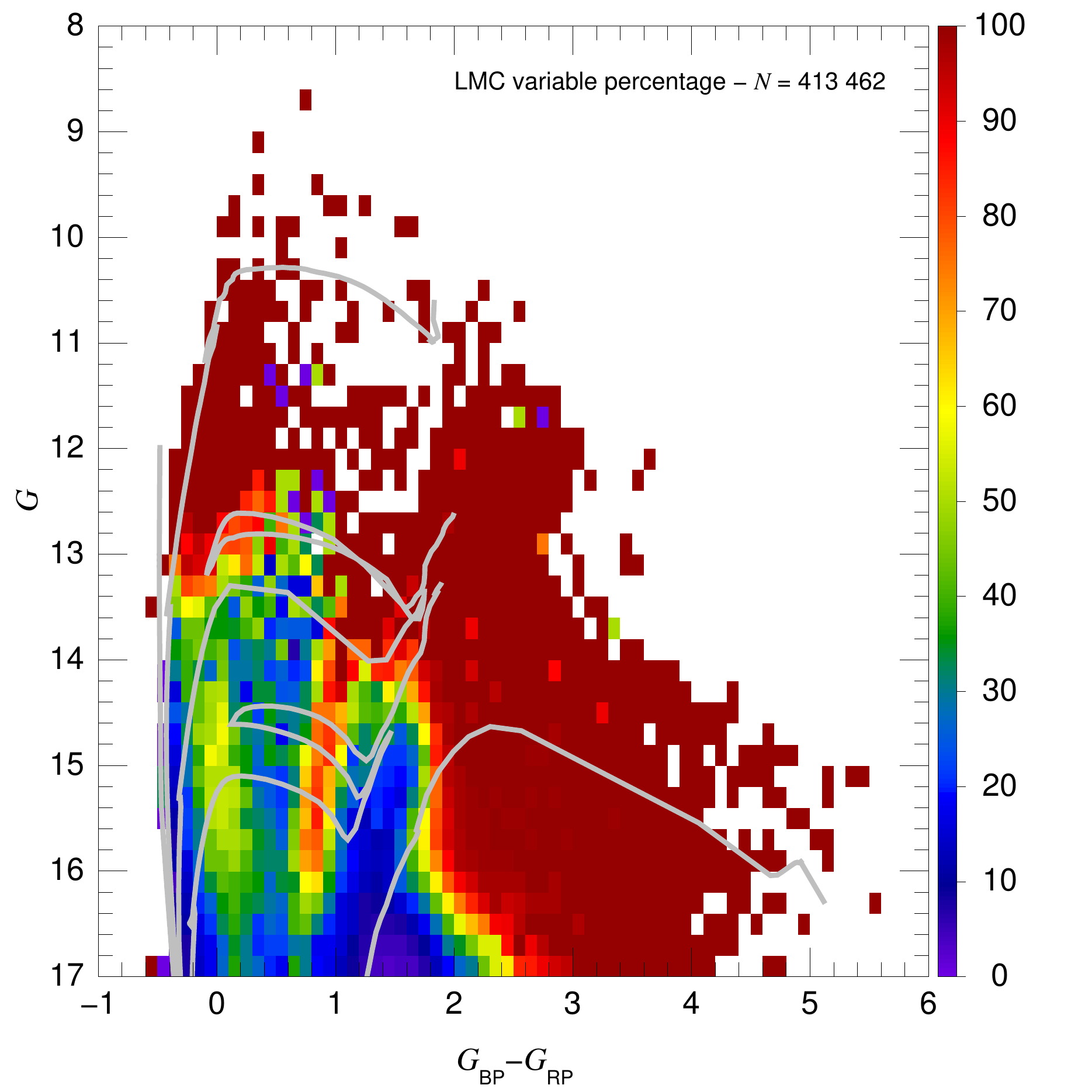}}
\caption{Same as Fig.~\ref{CAMD_all} but only for LMC targets, in the form of color-magnitude diagrams ($m-M$~=~18.473~mag)
         and showing the isochrones for LMC metallicity for 1~Ma, 10~Ma, 32~Ma, 100~Ma, and 1~Ga.}
\label{CMD_LMC}
\end{figure*}

\begin{figure*}
\centerline{\includegraphics[width=0.49\linewidth]{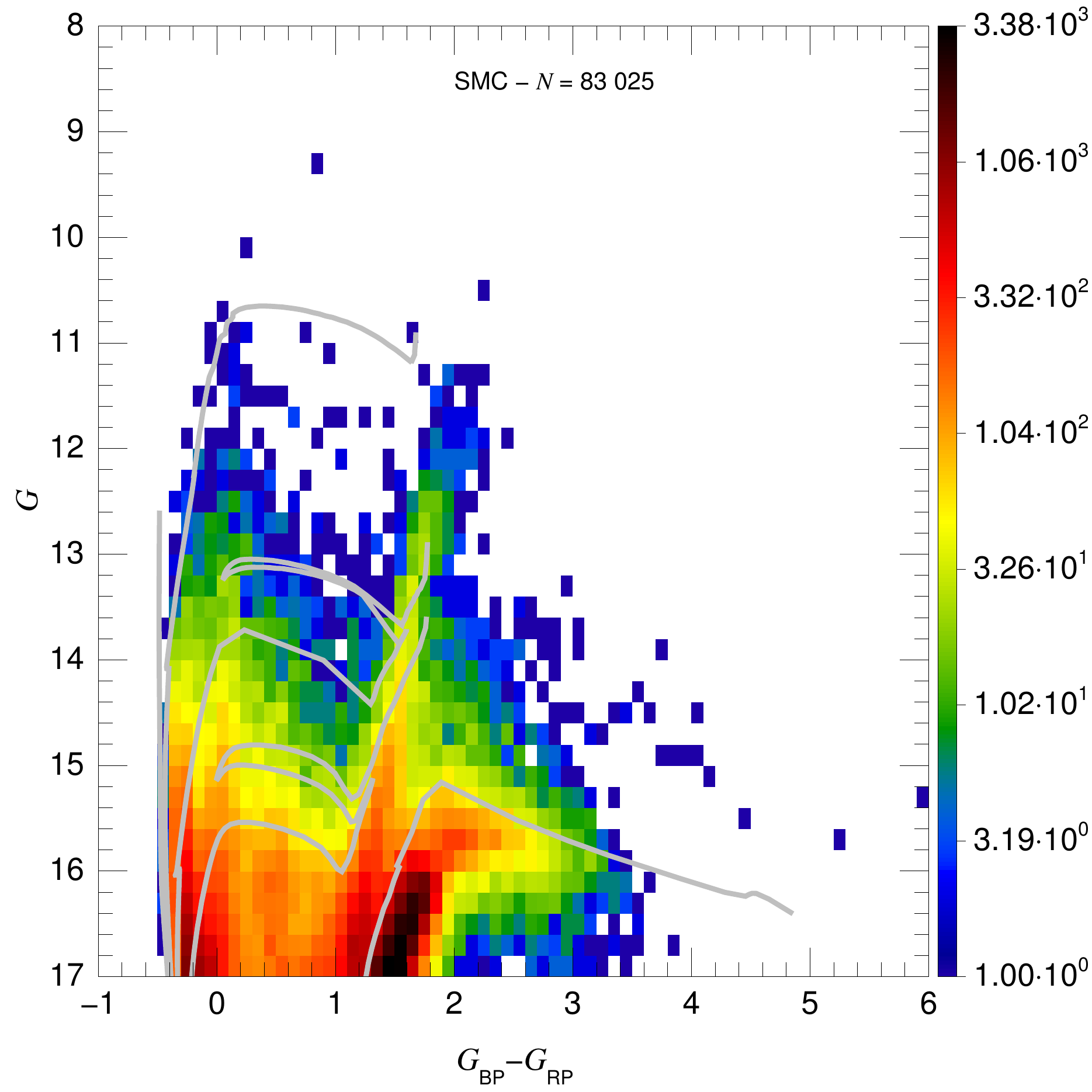} \
            \includegraphics[width=0.49\linewidth]{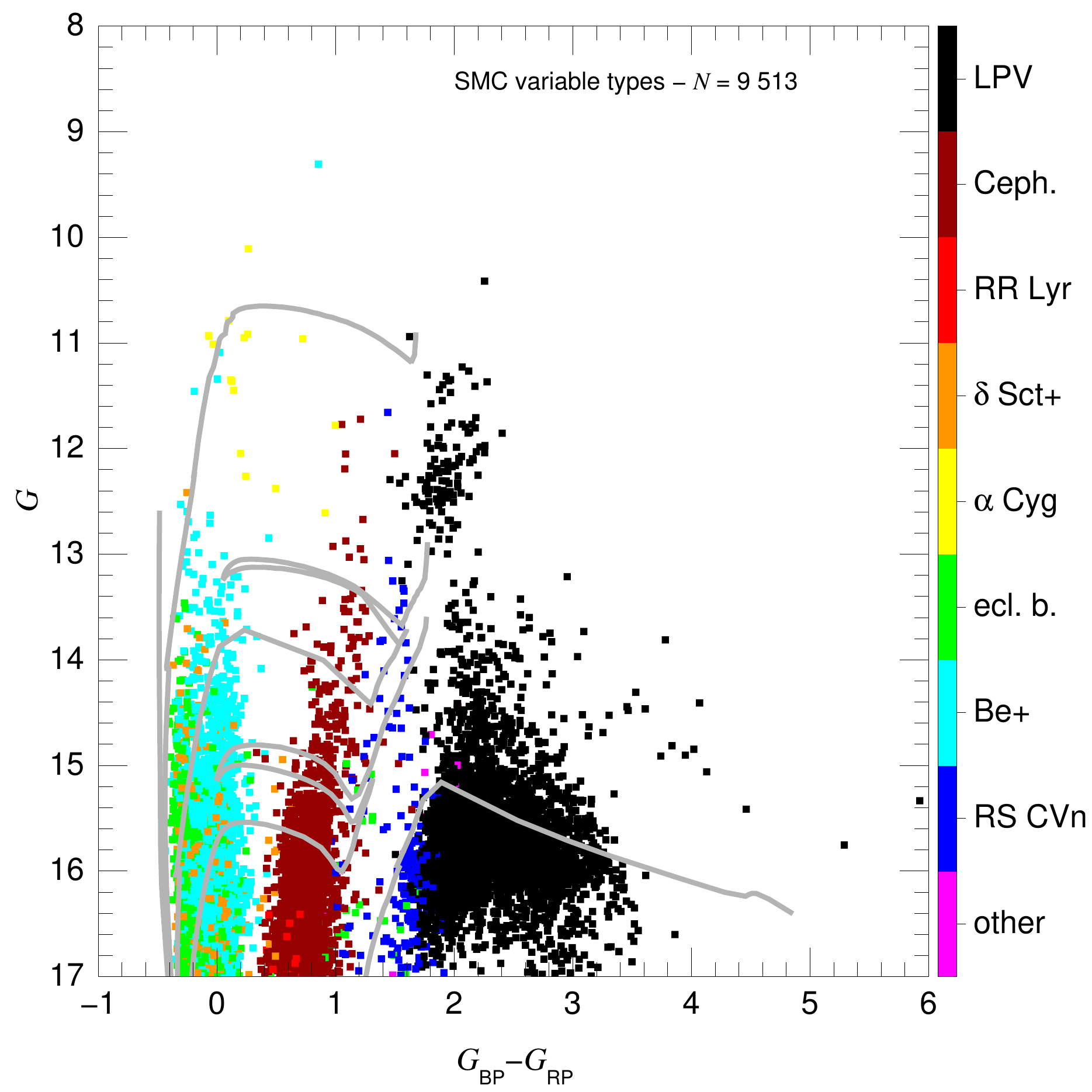}}
\centerline{\includegraphics[width=0.49\linewidth]{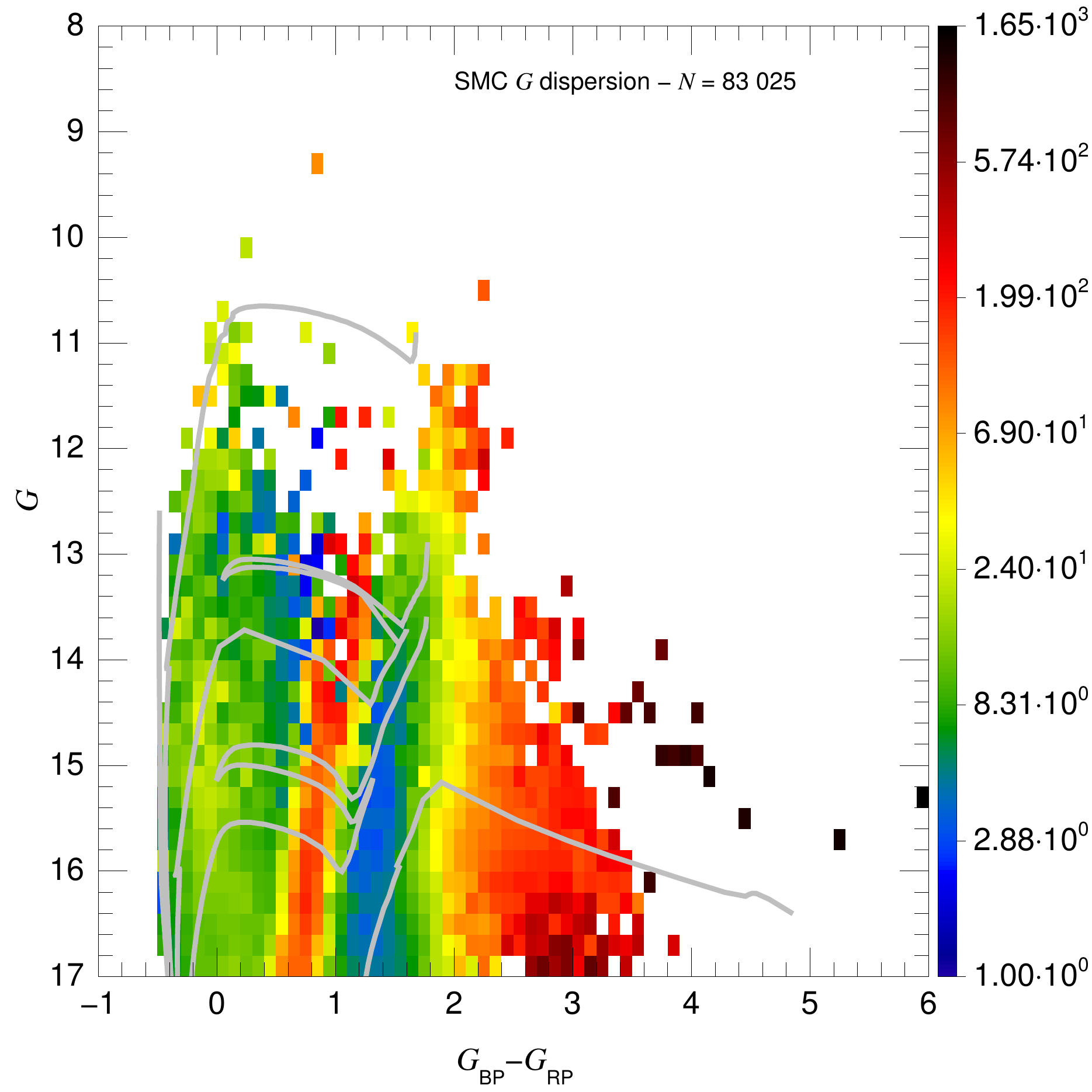} \
            \includegraphics[width=0.49\linewidth]{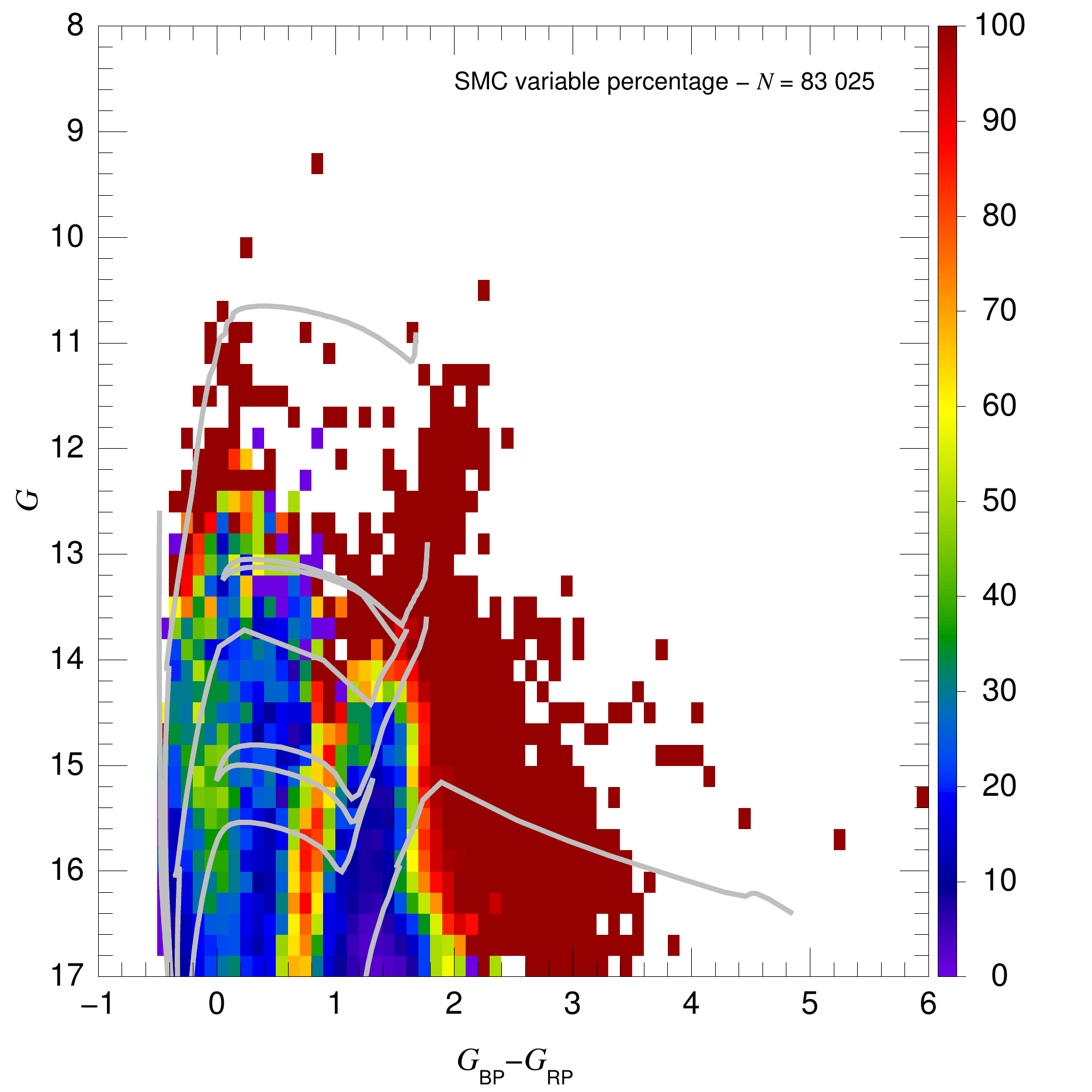}}
\caption{Same as Fig.~\ref{CAMD_all} but only for SMC targets, in the form of color-magnitude diagrams ($m-M$~=~18.993~mag)
         and showing the isochrones for SMC metallicity for 1~Ma, 10~Ma, 32~Ma, 100~Ma, and 1~Ga.}
\label{CMD_SMC}
\end{figure*}

\begin{figure*}
\centerline{\includegraphics[width=0.49\linewidth]{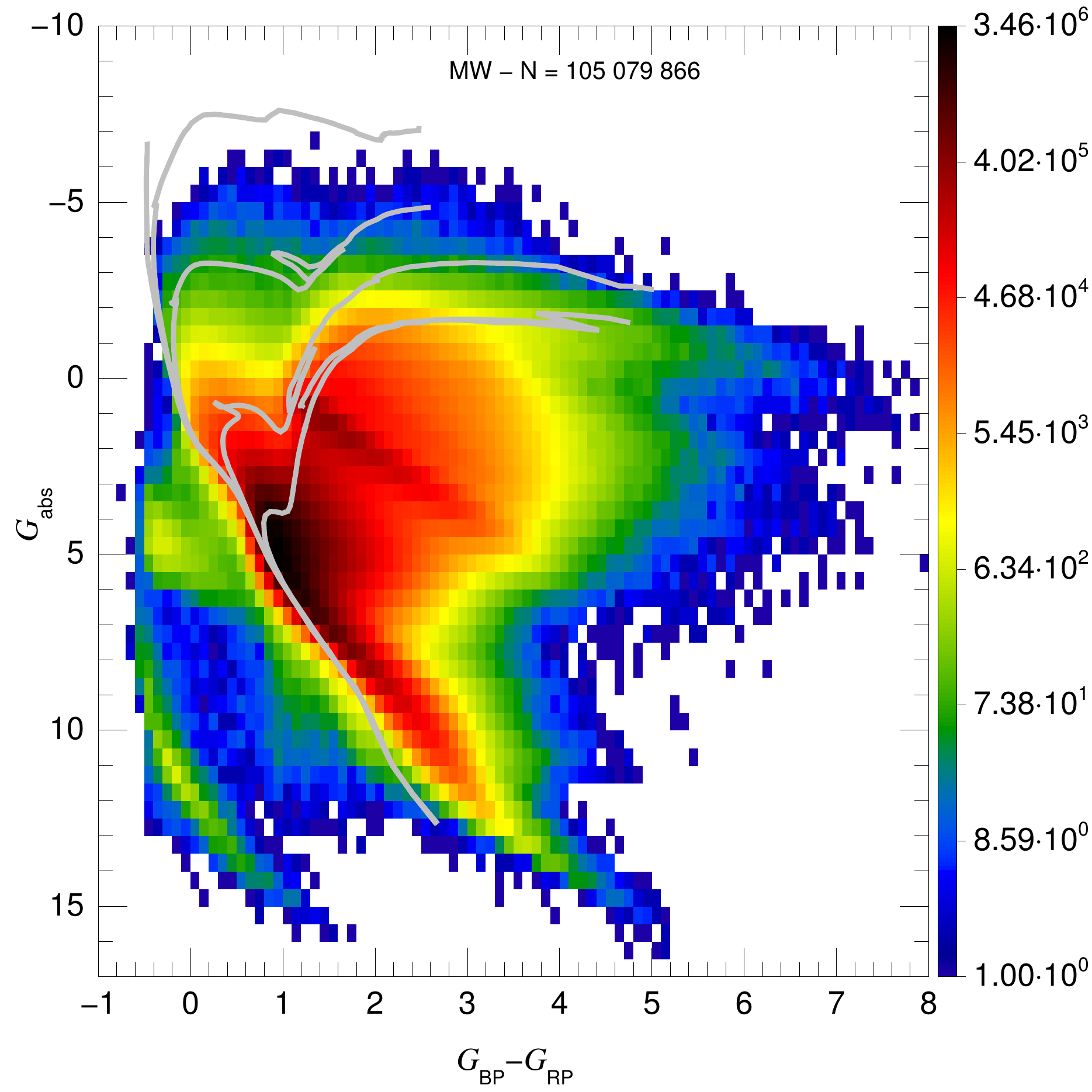} \
            \includegraphics[width=0.49\linewidth]{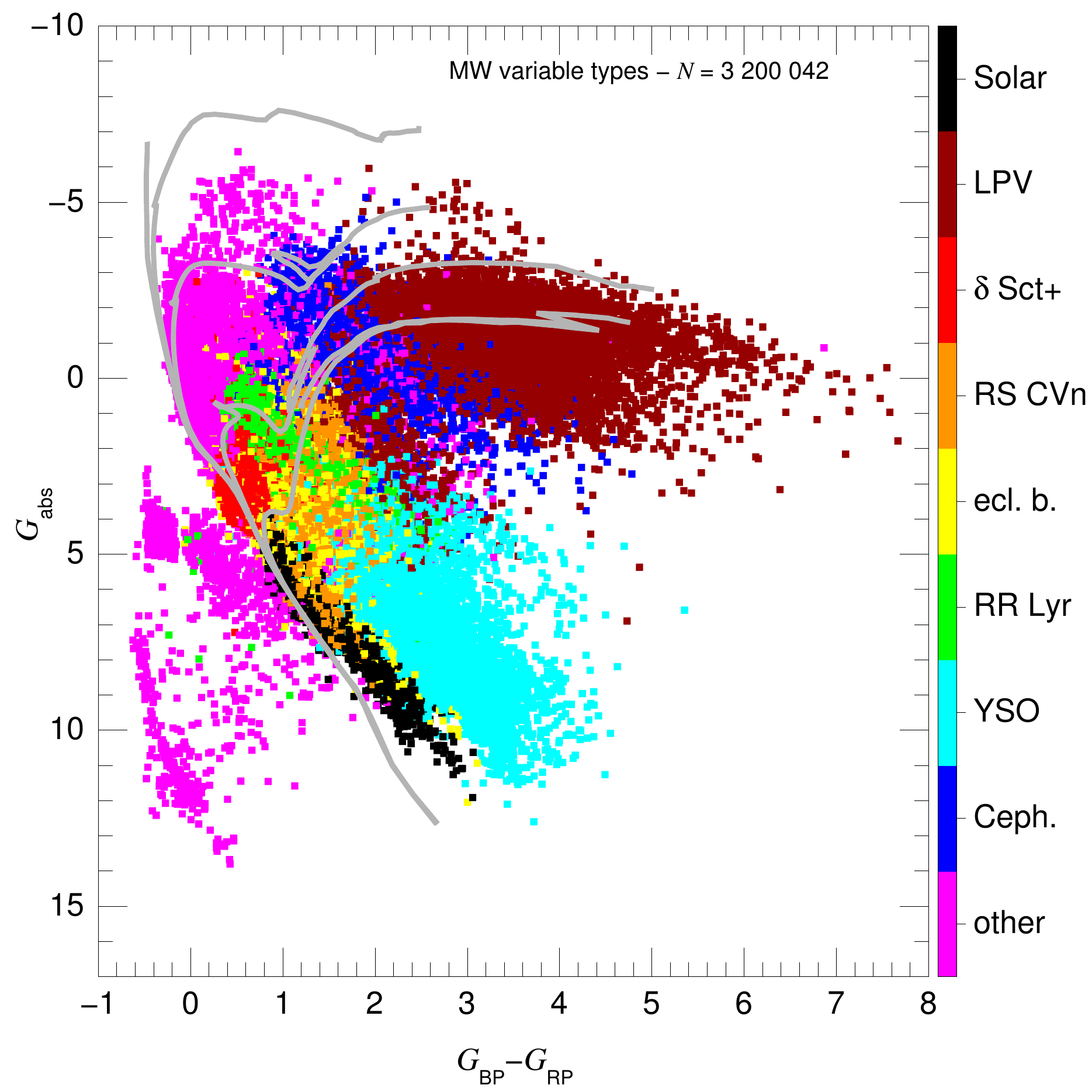}}
\centerline{\includegraphics[width=0.49\linewidth]{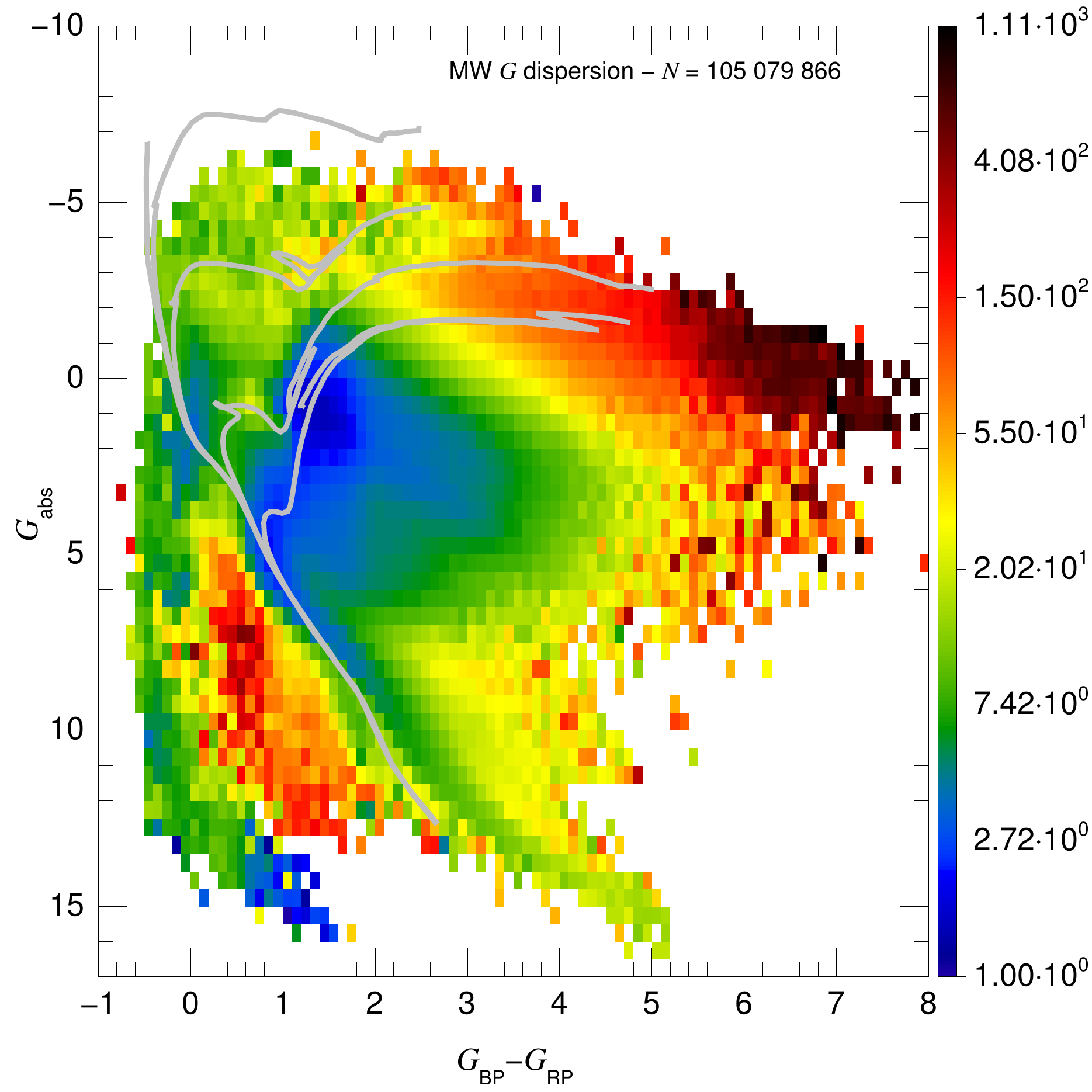} \
            \includegraphics[width=0.49\linewidth]{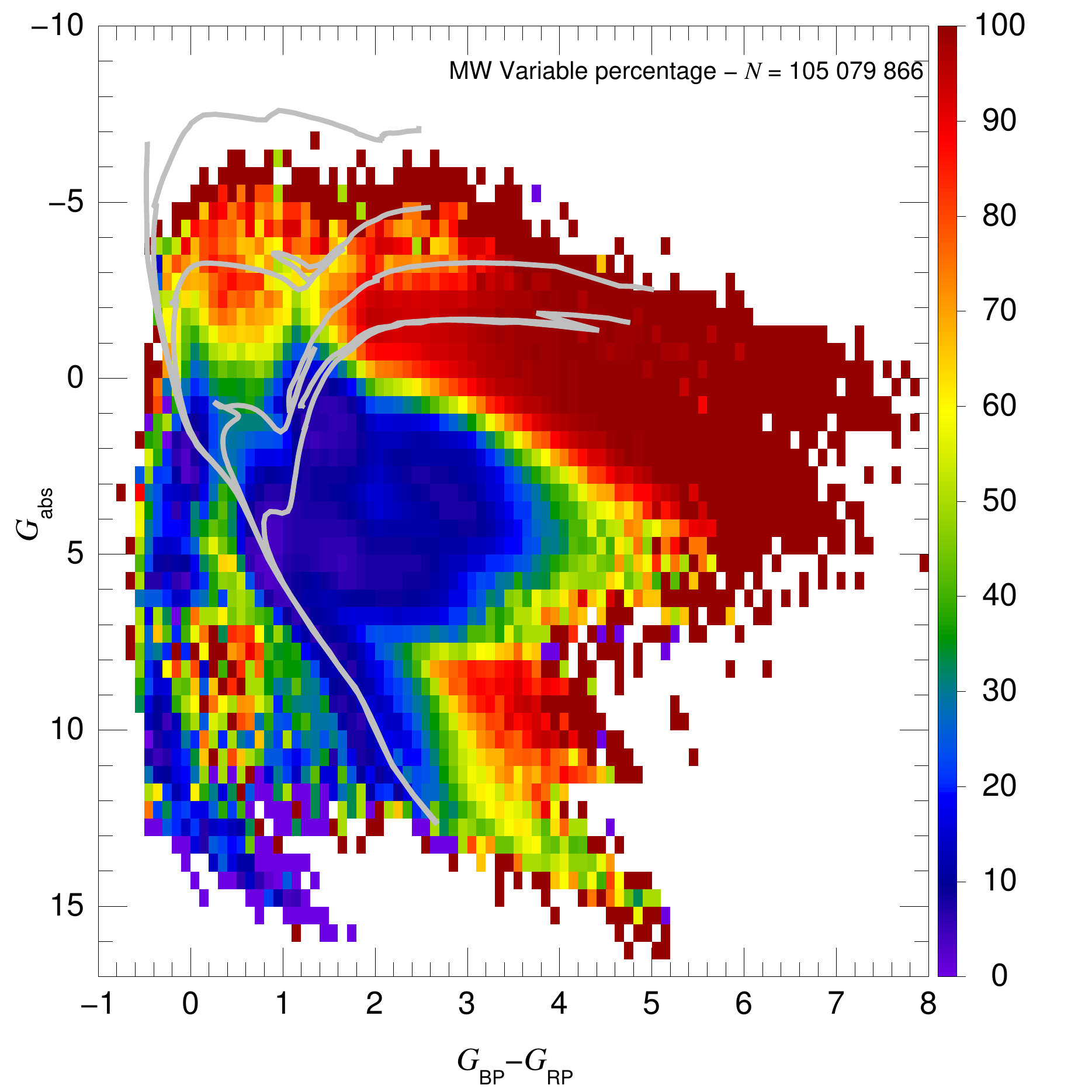}}
 \caption{Same as Fig.~\ref{CAMD_all} but the MW subsample, i.e. excluding targets in the LMC and SMC.}
\label{CAMD_MW}
\end{figure*}

$\,\!$\indent In this subsection we analyze the behavior of \textit{Gaia} photometric variability across the HR diagram. First, we
select the sample of stars with $\pic/\spic > 5$ described above, which automatically excludes QSOs and a small number of luminous 
Local Group sources not in the LMC and SMC. The sample of \num{105576353} stars is subdivided into Galactic (MW), LMC, and SMC
samples. The full sample and the three subsamples are shown in Figs.~\ref{CAMD_all},~\ref{CMD_LMC},~\ref{CMD_SMC},~and~\ref{CAMD_MW},
respectively. Each of those figures has four panels: [a] a CAMD (CMD for the LMC and SMC) density plot, [b] the location of the
R22 variables (capped at \num{10000} stars per type and with the less numerous types grouped in an ``other''
category), [c] the average \sG\ as a function of color of magnitude for all stars independent of the variability flag, and [d] the
percentage of stars with a variability flag of VVV as a function of color and magnitude.

We provide four different figures because each shows different characteristics. The LMC and SMC subsamples have considerably
less objects than the Galactic one and they only cover the brightest part of the Hertzsprung-Russell diagram (HRD), 
but they have two important advantages: their
extinction is negligible compared to the Galaxy (hence allowing a straightforward comparison between the CAMD/CMD and the HRD) and 
the inclusion of two other galaxies allows for the analysis of potential metallicity effects. A comparison between 
Figs.~\ref{CAMD_all}~and~\ref{CAMD_MW} reveals that most stars in the full sample with \Gabs\ brighter than $-$1.5~mag are in the
Magellanic Clouds because most of the bright OB stars in the Milky Way have extinctions that move them downward and to the right in
the CAMD\footnote{A few others do not appear because, even though their extinction is small, they are too bright to be in 
\textit{Gaia}~DR3. $\zeta$~Ori~Aa,Ab, $\beta$~CMa, and $\delta$~Sco are some examples.}. Massive-star astronomers know well that OB 
stars (even those within a few kpc) are a needle in a Galactic haystack. 

\subsubsection{The LMC}

We start discussing the LMC with Fig.~\ref{CMD_LMC}. The CMD there is a version of the (top part of the) left panel of Fig.~2 in
\citet{Lurietal21}, which we briefly summarize here. The two most obvious density concentrations are the nearly vertical one on the
left (the hot-star MS) and the curved one on the left composed by RGB stars in its nearly vertical part and AGB stars in its nearly
horizontal one (see \citealt{Lurietal21}). In addition, there are other interesting structures that deserve attention:

\begin{itemize}
 \item The area termed BL (blue loop) by \citet{Lurietal21} is encompassed by two overdensities that originate at 
       $\GBPmGRP\sim 1.2$, $\GG\sim 16.5$ and extend towards the top. The two overdensities are the two extremes of the blue
       loop characteristic of 3-10~M$_\odot$ stars, as seen in the 32~Ma and 100~Ma isochrones which indeed span the range between
       the two overdensities, small for 3~M$_\odot$ stars and increasingly larger with mass. This agreement validates the Padova
       (100~Ma) and Geneva (32~Ma) isochrones employed. The right overdensity corresponds to the bright red giant (BRG, in luminosity 
       class terms) or BRG but extends to the red supergiants (RSGs), as it includes stars above the 32~Ma isochrone\footnote{There
       is confusion in the literature about the nomenclature for luminous red stars, with a mix based on luminosity classes
       (supergiants or I, bright giants or II, and giants or III) and on evolutionary phases (RGB, RC, AGB, and RSG). Here we
       call BRGs the stars in the same branch in the LMC CMD as RSGs (massive stars) but below them i.e. of 
       intermediate mass, independently of whether they are in that region on their first pass (before the blue loop) or the second
       one (after the blue loop).}. The left overdensity starts also at a \GBPmGRP\ color that corresponds to the K spectral type but 
       moves towards earlier types as it increases in luminosity until it reaches the A supergiant region.
 \item Crossing the BL region nearly vertically around $\GBPmGRP\sim 1.0$~mag there is a band with large values of \sG\ and a
       corresponding high percentage of VVV objects. The top right panel indicates that this band is populated mostly by Cepheids, 
       with some RR Lyr stars near the bottom (but note that \citealt{Lurietal21} put RR Lyr stars at fainter magnitudes). The
       location of this instability strip does not correspond to a source overdensity except at its lower end around $\GG\sim 16$~mag.
       See below for a further analysis of the LMC instability strip.
 \item A similar but less marked vertical structure is seen in the bottom panels of Fig.~\ref{CMD_LMC}, with also larger values of
       \sG\ and more VVV objects. In this case, it also appears to follow an overdensity, as the MS splits into a ``blue branch'' and
       a ``red branch''. The blue branch is the one with the lower values of \sG\ and the red one the one with the higher values.
       The top right panel of Fig.~\ref{CMD_LMC} indicates that the variable stars in the red branch are classified in
       R22 as Be+ stars, dominated by Be stars but also including WRs, LBVs (or S Dor stars), and $\gamma$~Cas stars.
       Given the location in the HRD, we suspect that most of these objects are indeed Be stars.
 \item To the right of the instability strip we find a region of low variability that indicates that 3-5~M$_\odot$ stars are
       photometrically stable during their BRG phases. However, variability increases as we move towards higher luminosities and,
       especially, towards redder colors. The first trend shows that the RSGs are more variable than BRGs (but not extremely so in
       comparison with other luminous red stars) and that low mass stars are more variable during their AGB phase than in the 
       immediately previous phases. Indeed, the highest values of \sG\ are reached for the reddest stars in Fig.~\ref{CMD_LMC}, that 
       is, at the end of their AGB phase. The transition to higher values of \sG\ takes place around $\GBPmGRP = 2.2$ nearly
       independently of luminosity, pointing towards a temperature effect as extinction should be of little importance.
 \item As we climb the MS following the real (blue) branch we see an apparently contradictory phenomenon. The average \sG\ 
       remains approximately constant around 10-20~mmag but the percentage of VVV sources increases as we go from B to O stars. This 
       is likely an instrumental effect, as for brighter stars it is easier to classify a target as VVV (Fig.~\ref{hist_sigma}).
 \item There are few supergiants to the left of the instability strip, with those of B type appearing to be more variable than those
       of A and F type. This is likely related to the mass-loss bi-stability jump \citep{Benaetal07}.
\end{itemize}

\subsubsection{The SMC}

$\,\!$\indent Figure~\ref{CMD_SMC} shows the SMC equivalents to Fig.~\ref{CMD_LMC}. Similar patterns are seen there but we point out
some differences likely caused by metallicity effects.

\begin{itemize}
  \item The left overdensity of the blue loop is apparently missing. Instead, an overdensity is seen at lower luminosities and 
        $\GBPmGRP\sim 1.0$. This could mean that the blue loop is present at SMC metallicities only for lower masses than for the LMC
        but we also have to consider that the SMC subsample has significantly less stars than the LMC one.
  \item The separation between the blue (real) and red (Be) branches of the MS is more marked than for the LMC.
  \item The transition to higher values of \sG\ appears at a slightly lower value of \GBPmGRP\ (2.0 versus 2.0 for the LMC) but
        AGB stars do not reach colors as red as in the LMC.
  \item Similarly the BRGs and RSGs in the SMC are bluer for the same \Gabs\ \citep{Dordetal16b}.
\end{itemize}

\subsubsection{The Milky Way}

\begin{figure}
\centerline{\includegraphics[width=\linewidth]{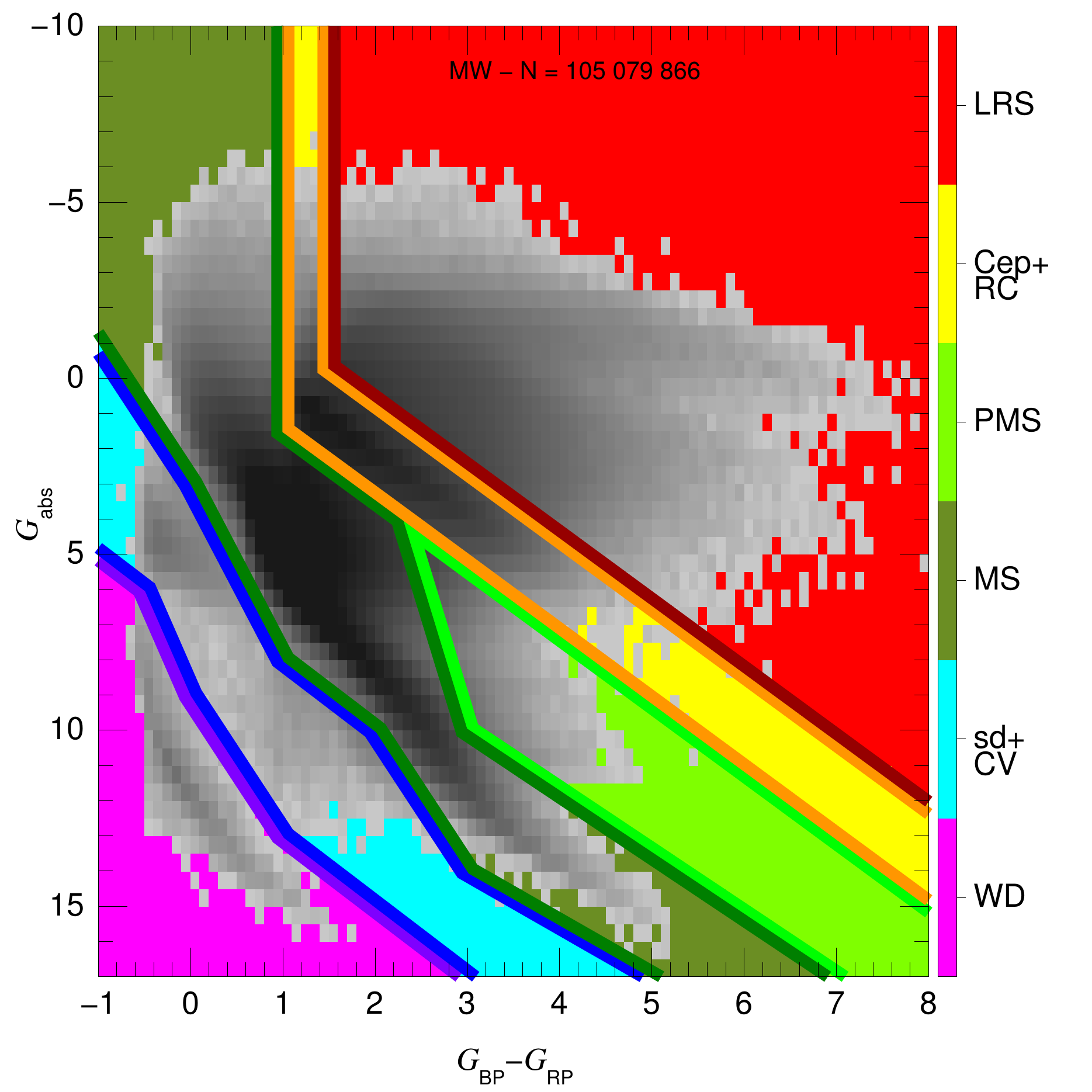}}
\caption{Color-absolute magnitude source density diagram for the Milky Way subsample with the six different regions drawn. The gray
         scale uses the same values as the upper left panel of Fig.~\ref{CAMD_MW} and the background colors are applied only where
         no sources are found.}
\label{CAMD_MW_BW}
\end{figure}

\begin{table*}
\caption{Points used to define the limits for the six CAMD regions of the Milky Way subsample.}
\label{CAMD_regions}
\centerline{
\begin{tabular}{lcrrrrrrr}
\hline
Limit      & Coord.   &     \#1 &    \#2 &  \#3 &  \#4 &  \#5 &  \#6 &  \#7 \\
\hline
WD-sd+CV   & \GBPmGRP &  $-$1.0 & $-$0.5 &  0.2 &  1.0 &  3.0 &      &      \\
           & \Gabs    &     5.0 &    6.0 & 11.5 & 13.0 & 17.0 &      &      \\
sd+CV-MS   & \GBPmGRP &  $-$1.0 &    0.0 &  0.2 &  0.8 &  1.0 &  2.0 &  5.0 \\
           & \Gabs    &  $-$1.0 &    3.0 &  4.0 &  7.0 &  9.5 & 12.0 & 17.0 \\
MS-PMS     & \GBPmGRP &     2.3 &    3.0 &  7.0 &      &      &      &      \\
           & \Gabs    &     4.0 &   10.0 & 17.0 &      &      &      &      \\
MS-Cep+RC  & \GBPmGRP &     1.0 &    1.0 &  2.3 &      &      &      &      \\
           & \Gabs    & $-$10.0 &    1.5 &  4.0 &      &      &      &      \\
PMS-Cep+RC & \GBPmGRP &     2.3 &    8.0 &      &      &      &      &      \\
           & \Gabs    &     4.0 &   15.0 &      &      &      &      &      \\
Cep+RC-LRS & \GBPmGRP &     1.5 &    1.5 &  8.0 &      &      &      &      \\
           & \Gabs    & $-$10.0 & $-$0.3 & 12.2 &      &      &      &      \\
\hline
\end{tabular}
}
\end{table*}

\begin{table*}
\caption{Number and percentage by R22 variable type for the six MW CAMD regions. We exclude three targets misclassified as AGNs.}
\addtolength{\tabcolsep}{-1mm}
\centerline{
\begin{tabular}{lrrrrrrrrrrrr}
\hline
Variability type   & \mcii{WD}                     & \mcii{sd+CV}                  & \mcii{MS}                     & \mcii{PMS}                    & \mcii{Cep+RC}                 & \mcii{LRS}                    \\
                   & \mci{\#}       & \mci{Perc.}  & \mci{\#}       & \mci{Perc.}  & \mci{\#}       & \mci{Perc.}  & \mci{\#}       & \mci{Perc.}  & \mci{\#}       & \mci{Perc.}  & \mci{\#}       & \mci{Perc.}  \\
\hline
Solar-like         & \num{       1} & \num{  0.20} & \num{       3} & \num{  0.20} & \num{ 1637959} & \num{ 53.65} & \num{       8} & \num{  0.08} & \num{       2} & \num{  0.00} &            --- &          --- \\
LPV                &            --- &          --- &            --- &          --- & \num{     171} & \num{  0.01} & \num{     127} & \num{  1.34} & \num{    5172} & \num{  8.53} & \num{   72803} & \num{ 97.11} \\
$\delta$ Sct+      &            --- &          --- & \num{      24} & \num{  1.59} & \num{  601354} & \num{ 19.70} &            --- &          --- & \num{    1144} & \num{  1.89} &            --- &          --- \\
RS CVn             &            --- &          --- & \num{       3} & \num{  0.20} & \num{  403713} & \num{ 13.22} & \num{      66} & \num{  0.70} & \num{   41945} & \num{ 69.18} & \num{      95} & \num{  0.13} \\
Eclipsing binary   &            --- &          --- & \num{     121} & \num{  8.02} & \num{  364766} & \num{ 11.95} & \num{     605} & \num{  6.39} & \num{    8270} & \num{ 13.64} & \num{     291} & \num{  0.39} \\
RR Lyr             &            --- &          --- & \num{      14} & \num{  0.93} & \num{   10235} & \num{  0.34} & \num{       7} & \num{  0.07} & \num{     638} & \num{  1.05} &            --- &          --- \\
YSO                &            --- &          --- & \num{       1} & \num{  0.07} & \num{   21671} & \num{  0.71} & \num{    8636} & \num{ 91.15} & \num{    1546} & \num{  2.55} & \num{      25} & \num{  0.03} \\
Ellipsoidal        &            --- &          --- &            --- &          --- & \num{    1080} & \num{  0.04} & \num{      23} & \num{  0.24} & \num{    1067} & \num{  1.76} & \num{     288} & \num{  0.38} \\
Cepheid            &            --- &          --- &            --- &          --- & \num{     161} & \num{  0.01} &            --- &          --- & \num{     513} & \num{  0.85} & \num{    1050} & \num{  1.40} \\
$\alpha^2$ CVn+    &            --- &          --- &            --- &          --- & \num{    8124} & \num{  0.27} &            --- &          --- &            --- &          --- &            --- &          --- \\
Be+                &            --- &          --- &            --- &          --- & \num{     715} & \num{  0.02} &            --- &          --- & \num{     136} & \num{  0.22} & \num{     340} & \num{  0.45} \\
Short timescale    &            --- &          --- & \num{      41} & \num{  2.72} & \num{     295} & \num{  0.01} & \num{       1} & \num{  0.01} &            --- &          --- &            --- &          --- \\
$\beta$ Cep        &            --- &          --- &            --- &          --- & \num{     999} & \num{  0.03} &            --- &          --- & \num{     139} & \num{  0.23} & \num{      19} & \num{  0.03} \\
Slowly pulsating B &            --- &          --- &            --- &          --- & \num{    1116} & \num{  0.04} &            --- &          --- &            --- &          --- &            --- &          --- \\
sdB                &            --- &          --- & \num{     873} & \num{ 57.89} & \num{       3} & \num{  0.00} &            --- &          --- &            --- &          --- &            --- &          --- \\
CV                 & \num{       1} & \num{  0.20} & \num{     367} & \num{ 24.34} & \num{     245} & \num{  0.01} &            --- &          --- &            --- &          --- &            --- &          --- \\
WD                 & \num{     504} & \num{ 99.60} & \num{      61} & \num{  4.05} & \num{       5} & \num{  0.00} &            --- &          --- &            --- &          --- &            --- &          --- \\
Symbiotic          &            --- &          --- &            --- &          --- & \num{       2} & \num{  0.00} &            --- &          --- & \num{       6} & \num{  0.01} & \num{      47} & \num{  0.06} \\
$\alpha$ Cyg       &            --- &          --- &            --- &          --- & \num{     124} & \num{  0.00} &            --- &          --- & \num{      45} & \num{  0.07} & \num{       2} & \num{  0.00} \\
Exoplanet transit  &            --- &          --- &            --- &          --- & \num{     208} & \num{  0.01} &            --- &          --- &            --- &          --- &            --- &          --- \\
R CrB              &            --- &          --- &            --- &          --- & \num{       3} & \num{  0.00} &            --- &          --- & \num{       6} & \num{  0.01} & \num{      10} & \num{  0.01} \\
Microlensing       &            --- &          --- &            --- &          --- & \num{       3} & \num{  0.00} & \num{       1} & \num{  0.01} &            --- &          --- &            --- &          --- \\
\hline
Total              & \num{     506} & \num{100.00} & \num{    1508} & \num{100.00} & \num{ 3052952} & \num{100.00} & \num{    9474} & \num{100.00} & \num{   60629} & \num{100.00} & \num{   74970} & \num{100.00} \\
\hline
\end{tabular}
\addtolength{\tabcolsep}{1mm}
}
\label{variables_by_region}
\end{table*}

\begin{figure*}[ht!]
\centerline{\includegraphics[width=0.35\linewidth]{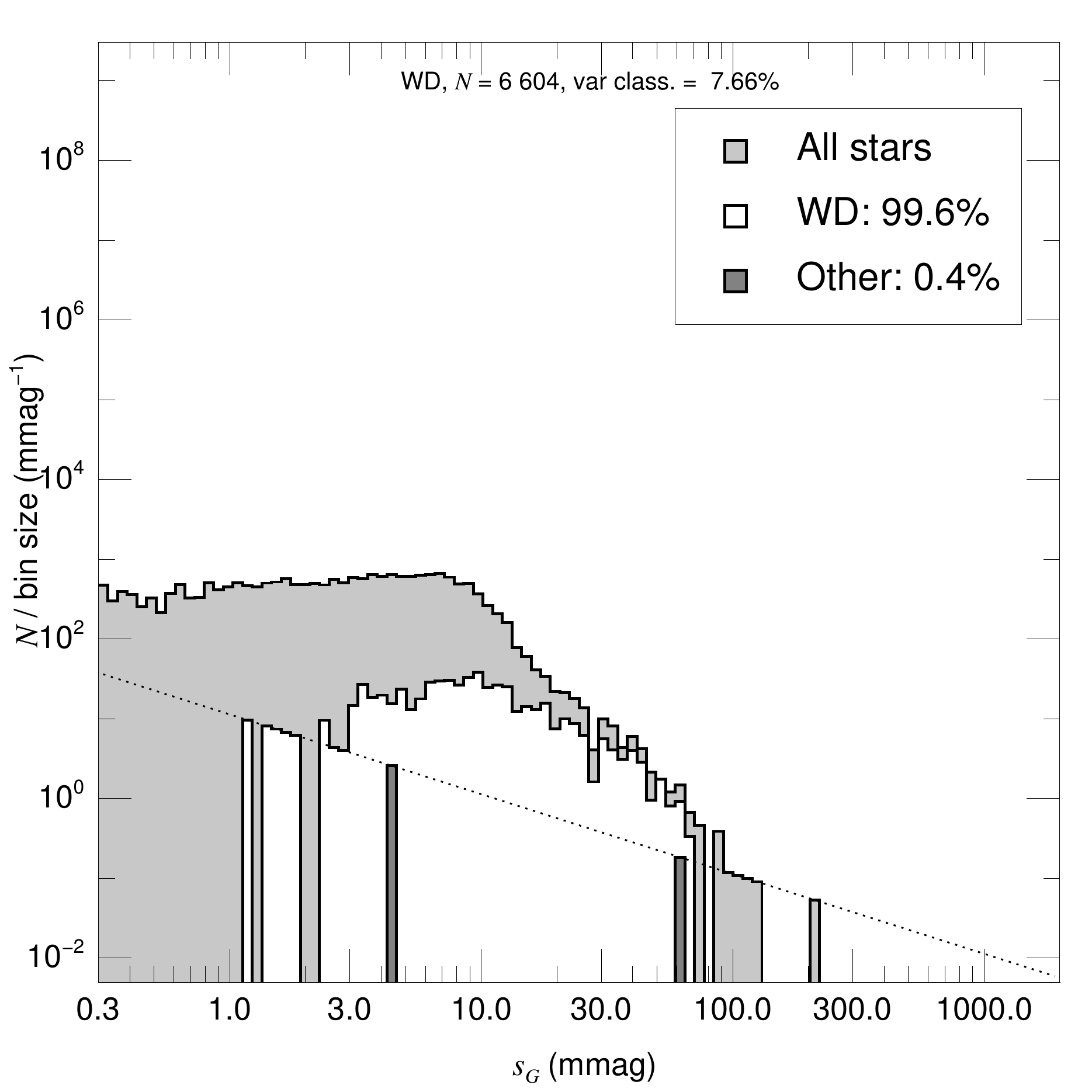}$\!\!\!$
            \includegraphics[width=0.35\linewidth]{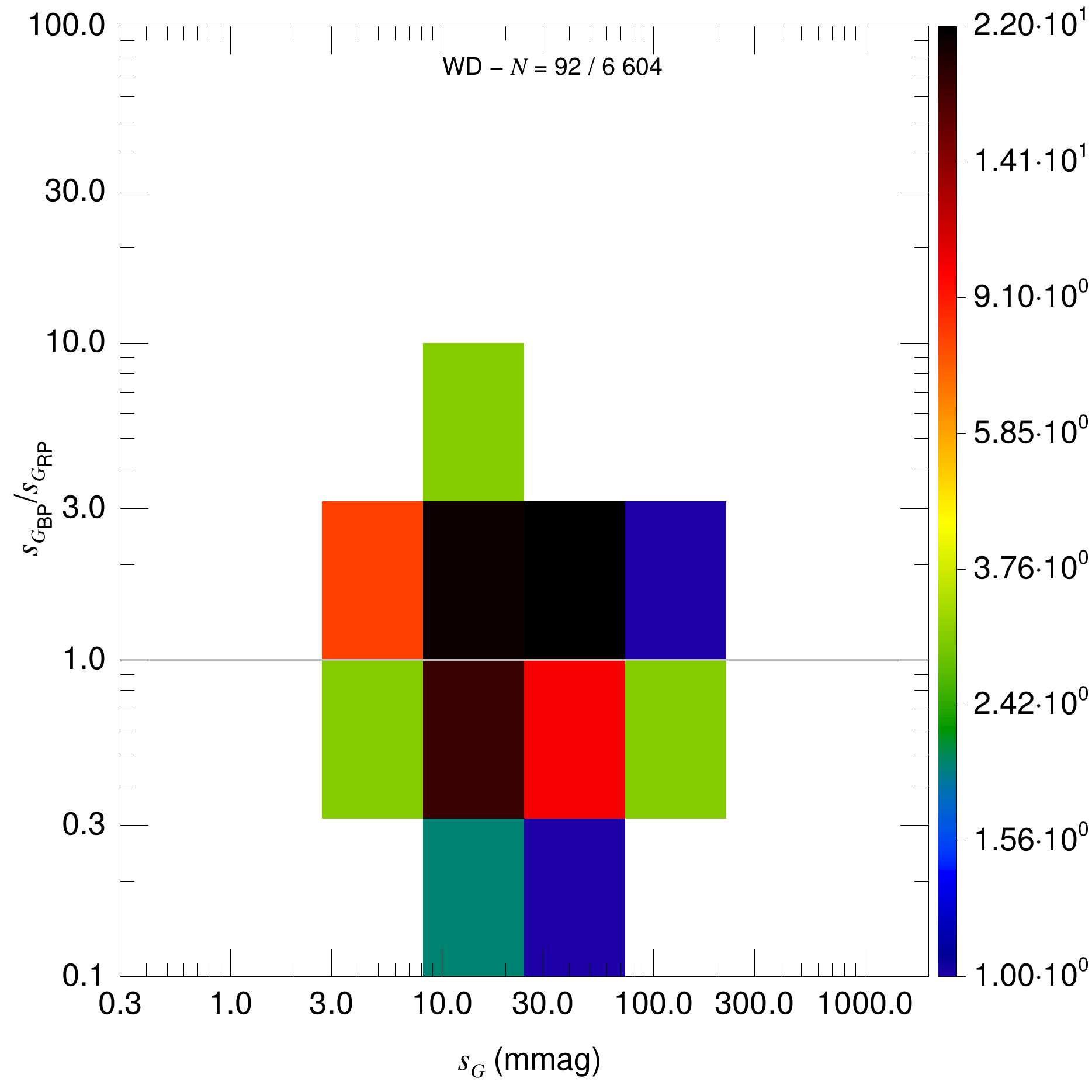}$\!\!\!$
            \includegraphics[width=0.35\linewidth]{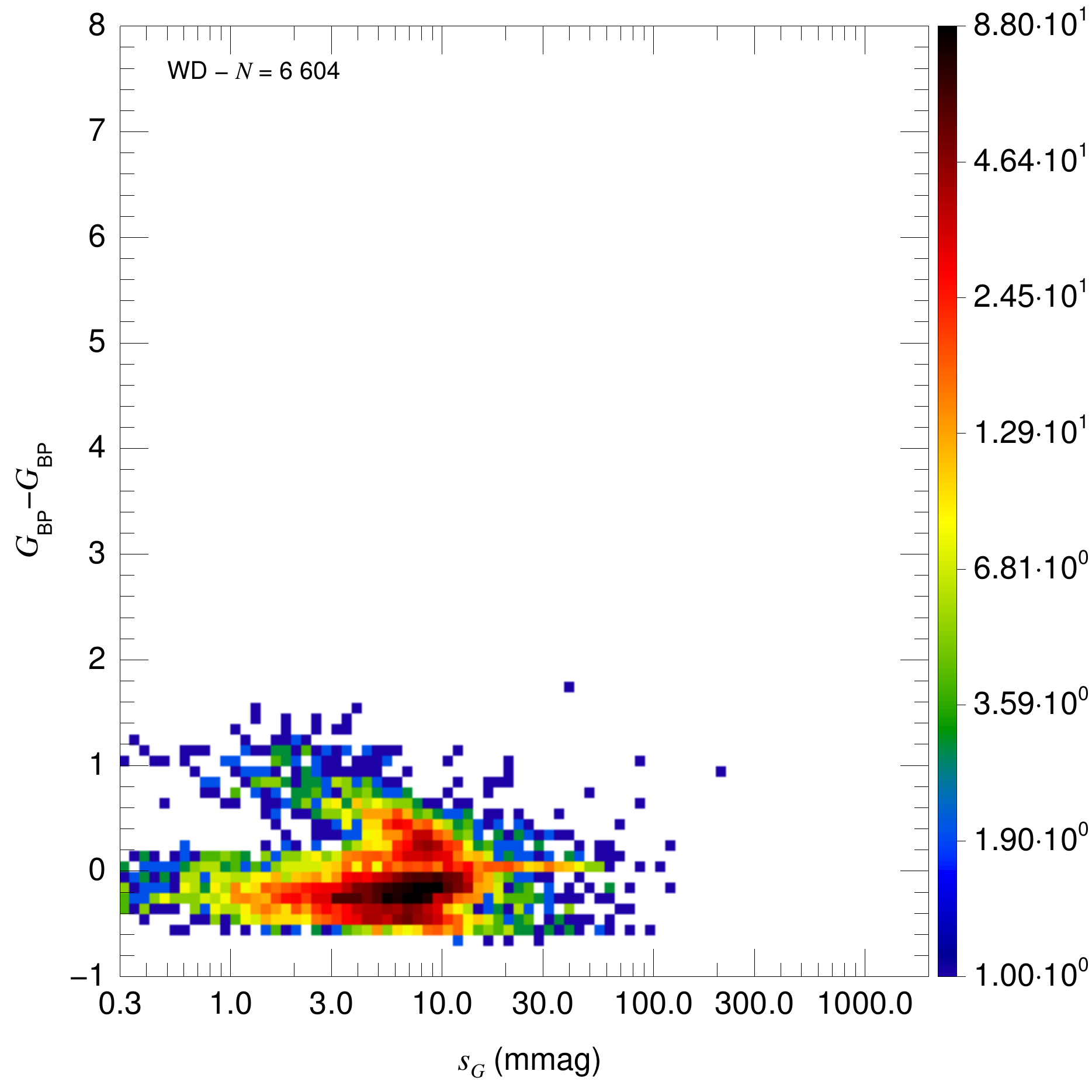}}
\centerline{\includegraphics[width=0.35\linewidth]{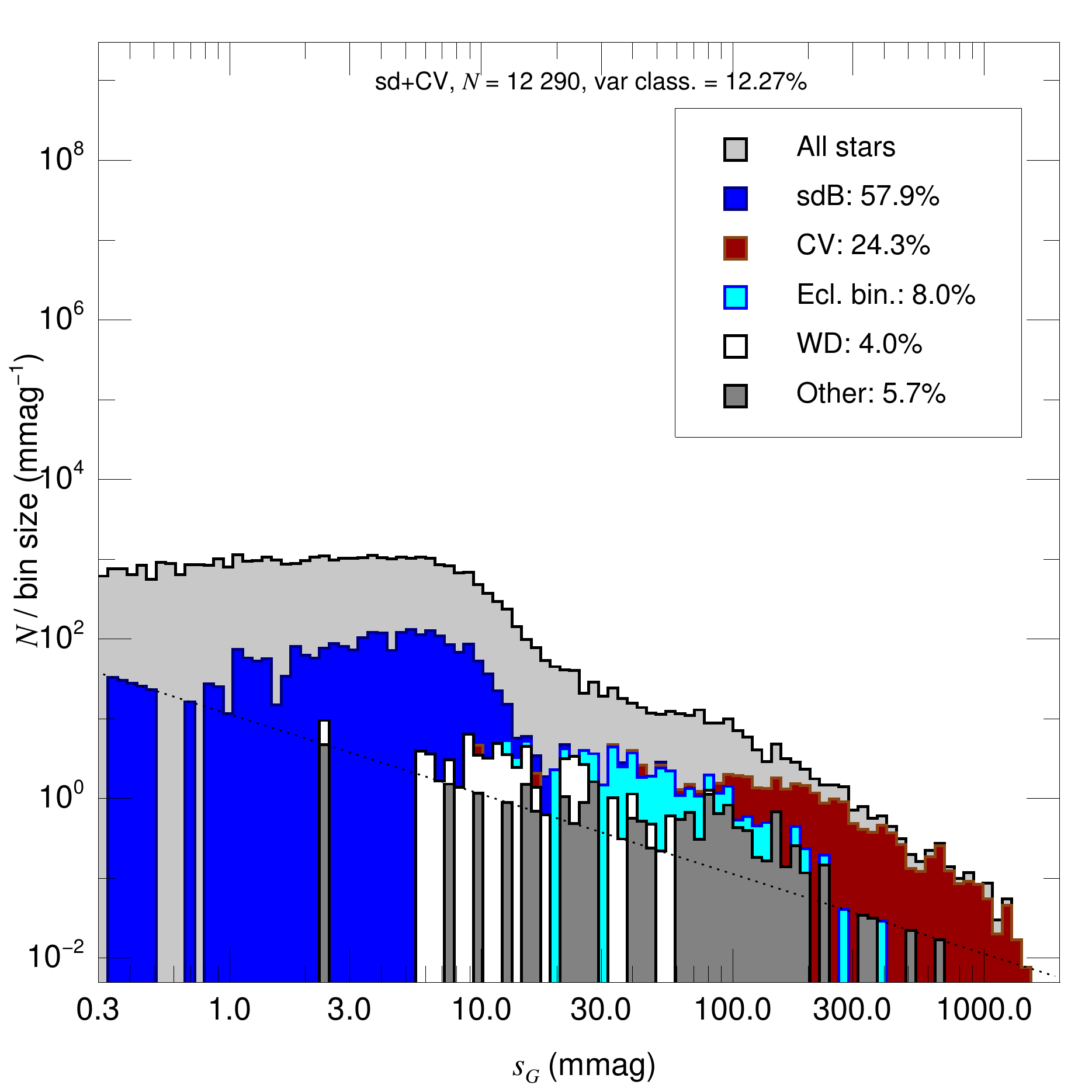}$\!\!\!$
            \includegraphics[width=0.35\linewidth]{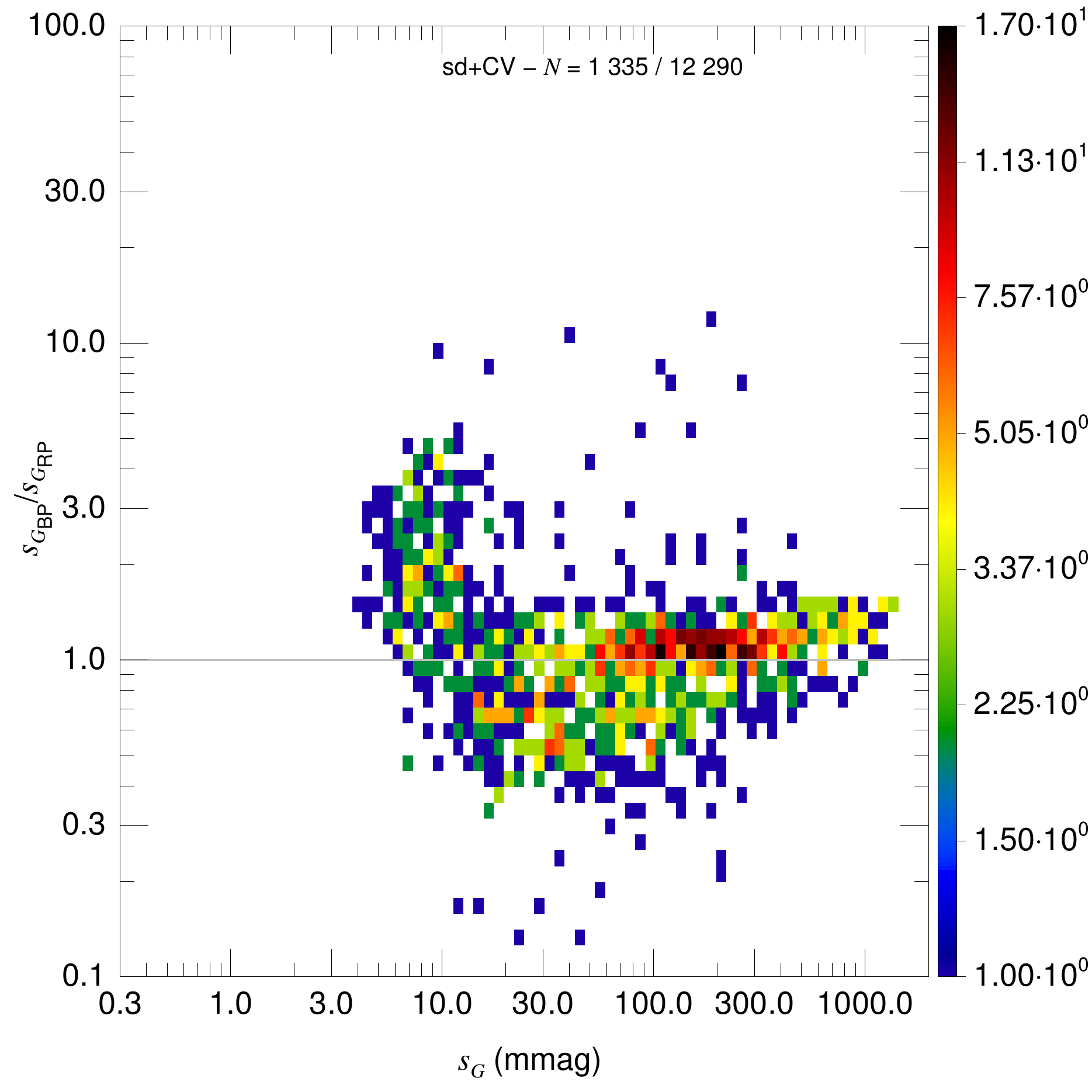}$\!\!\!$
            \includegraphics[width=0.35\linewidth]{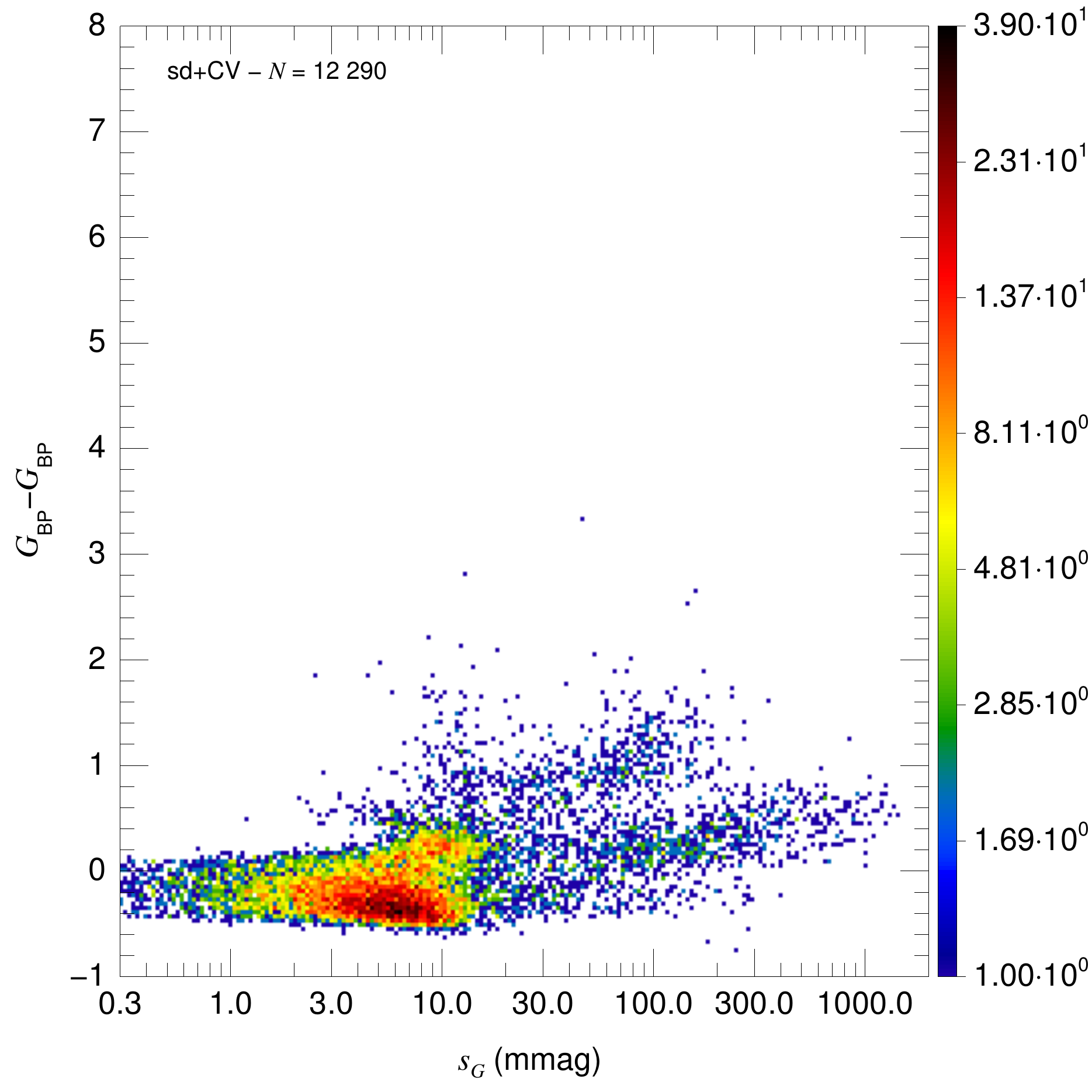}}
\centerline{\includegraphics[width=0.35\linewidth]{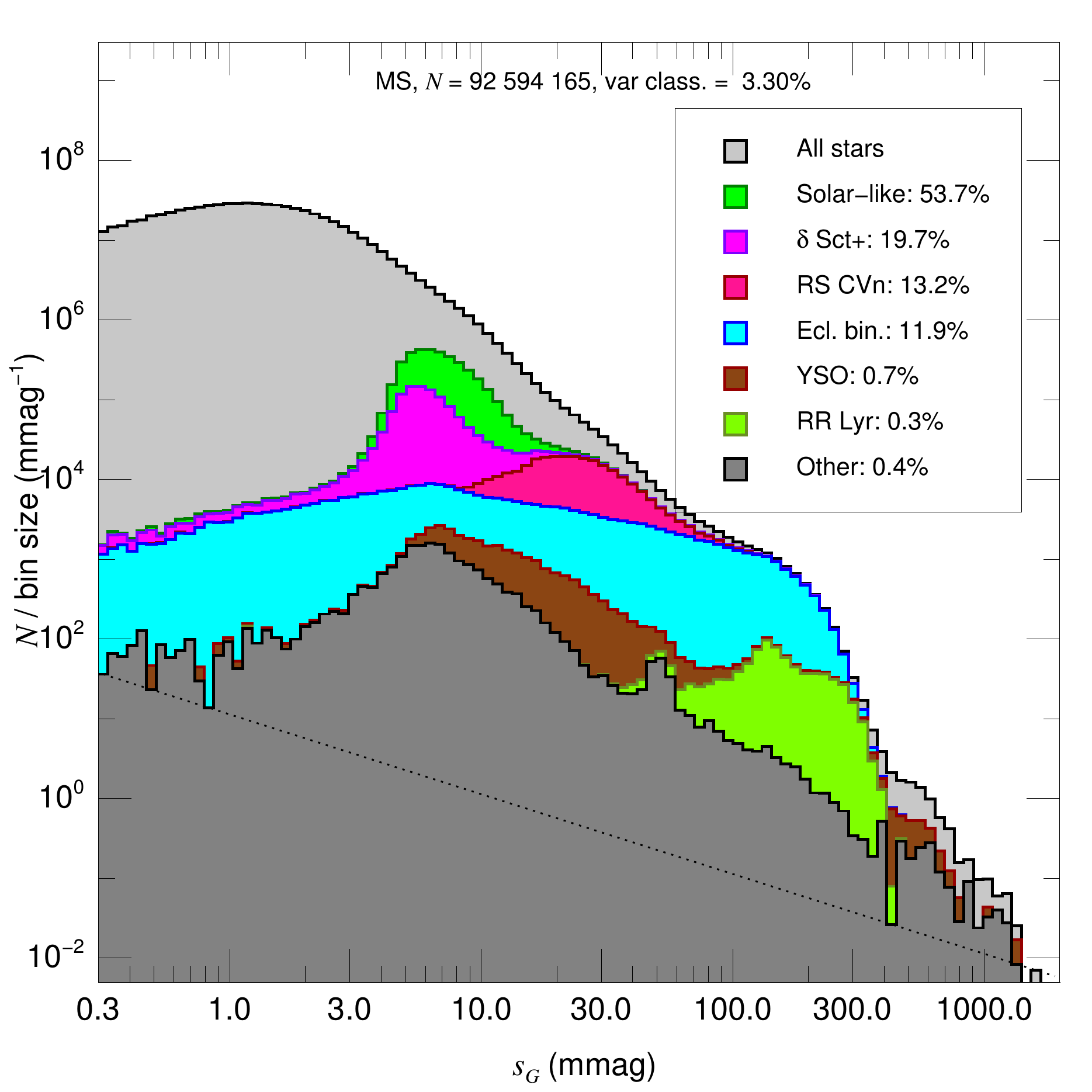}$\!\!\!$
            \includegraphics[width=0.35\linewidth]{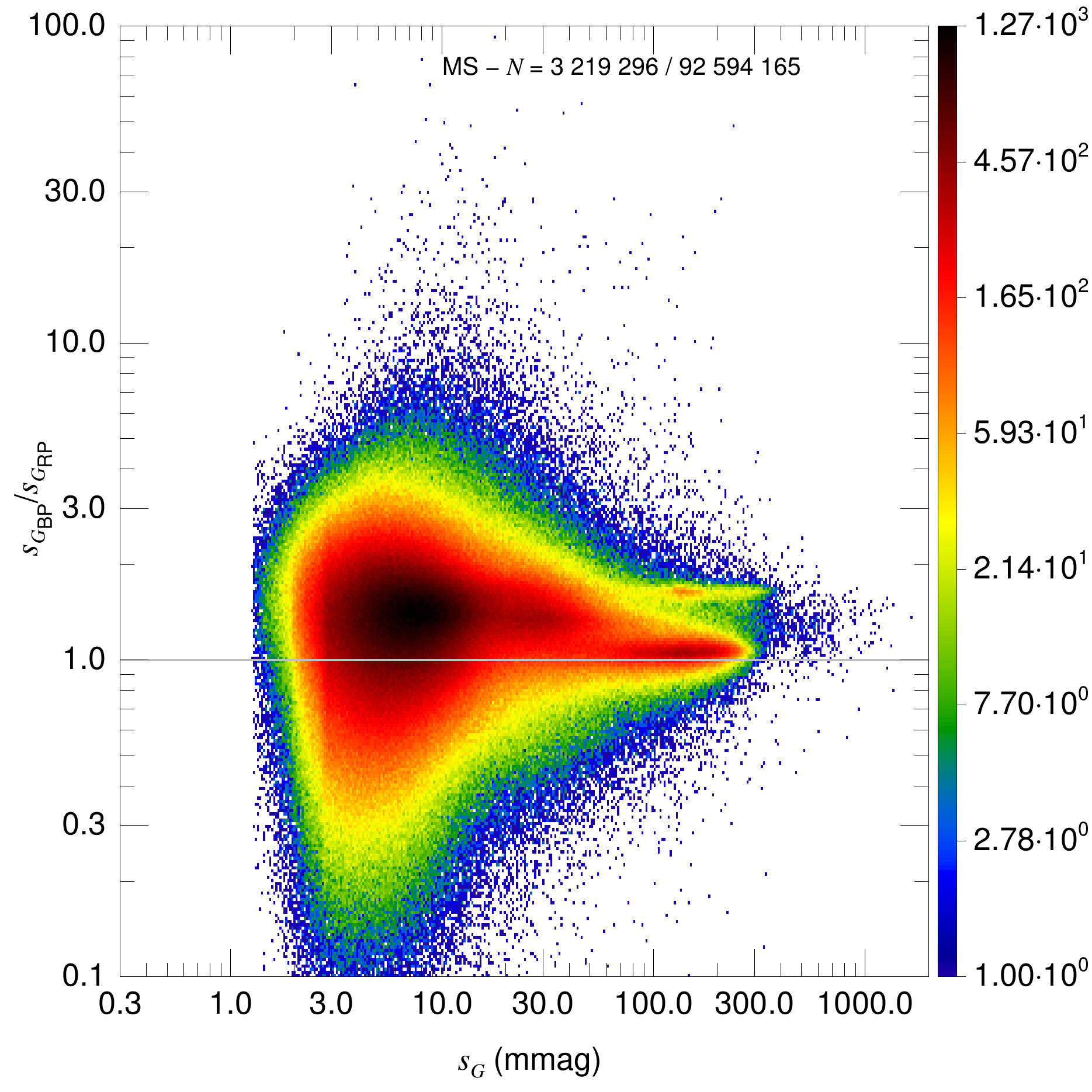}$\!\!\!$
            \includegraphics[width=0.35\linewidth]{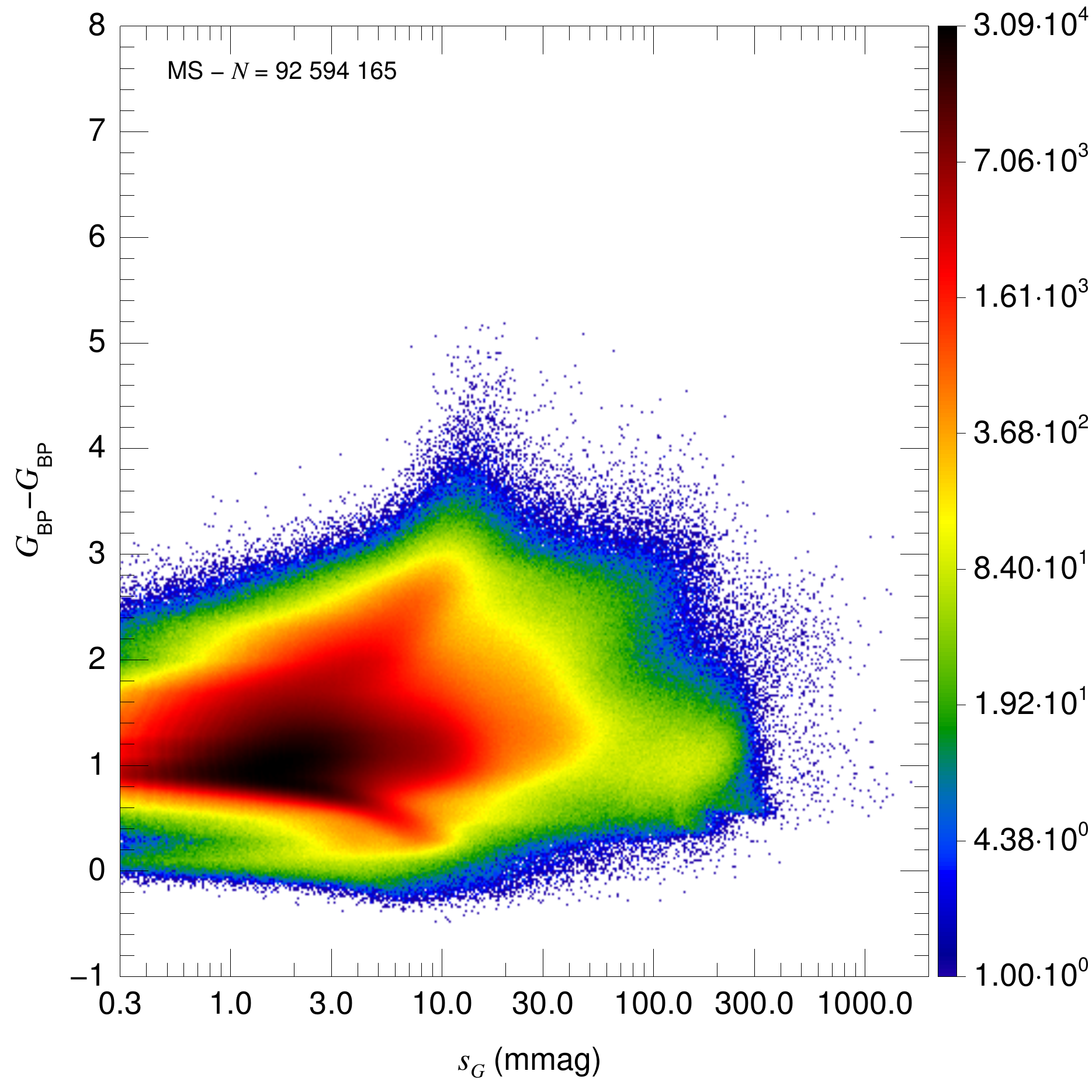}}
 \caption{Results for the first three of the six MW CAMD regions. (left column) Astrophysical \GG\ dispersion histograms. The 
          histograms have a uniform bin size but both the horizontal and vertical scales are logarithmic and the same for all plots 
          to allow for better comparisons. The grey histogram shows the total sample and each panel also shows the histograms for 
          the most common types of variable stars from the R22 data for that CAMD region (each in a different color 
          that is maintained the same throughout this Figure and the next for a given variable type) as well as another histogram 
          with the rest of the R22 variables. The histograms built from R22 data are cumulative, so the
          top colored line represents all of the variables. The text at the top of each panel gives the number of stars and the
          percentage of stars with a variability classification from R22. (center column) Astrophysical \GG\ 
          dispersion - astrophysical \GBP\ to \GRP\ dispersion ratio density diagrams selecting stars that have uncertainties in the 
          $y$ axis lower than 50\%. The number of selected stars and the total number are given at the top of each panel. The bin 
          size is adjusted for each region depending on the number of sources. (right column) Astrophysical \GG\ dispersion - 
          \GBPmGRP\ density diagrams. Note that the left and right columns include all stars in the regions while the center column
          only has a selection.}
\label{results_by_region_1}
\end{figure*}

\begin{figure*}[ht!]
\centerline{\includegraphics[width=0.35\linewidth]{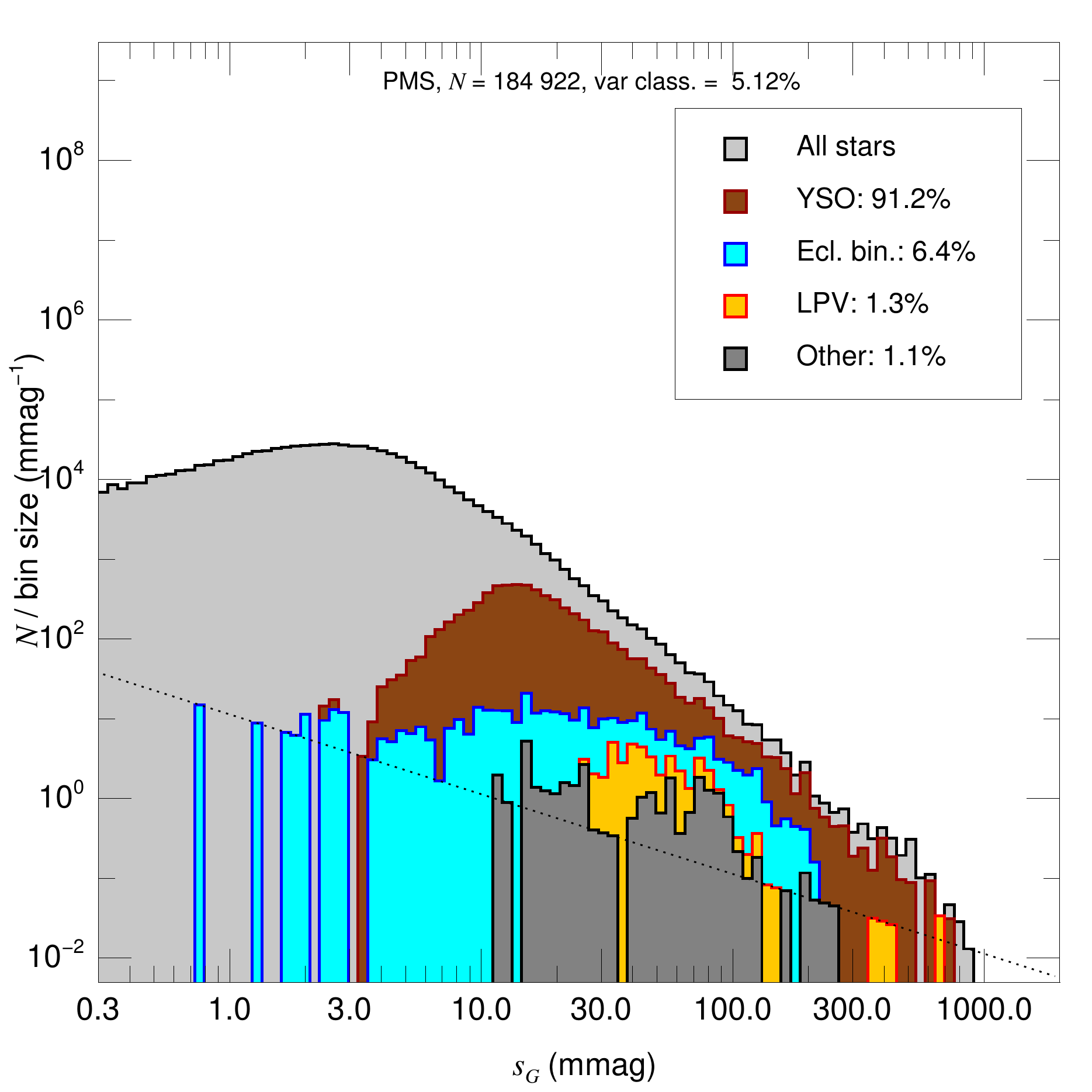}$\!\!\!$
            \includegraphics[width=0.35\linewidth]{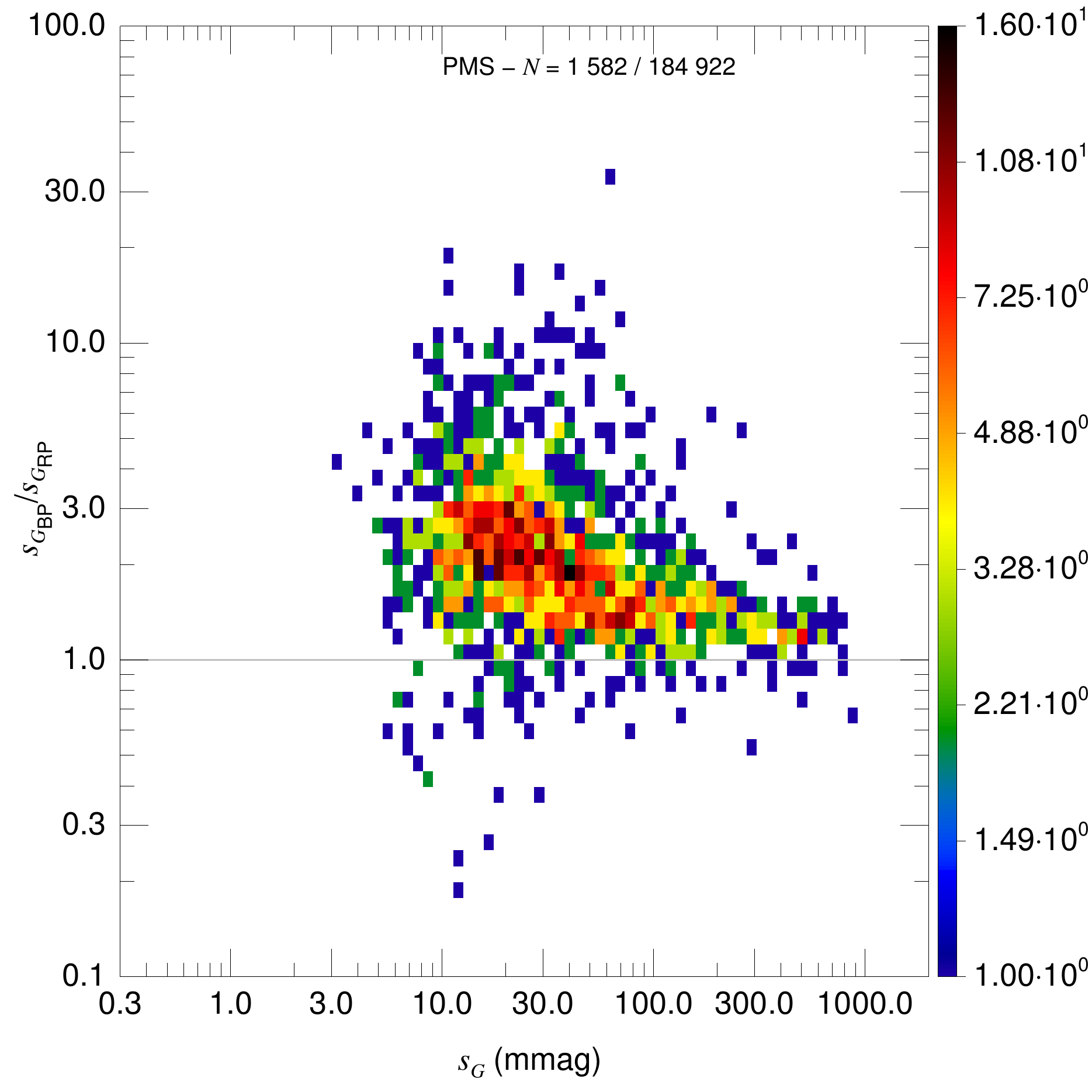}$\!\!\!$
            \includegraphics[width=0.35\linewidth]{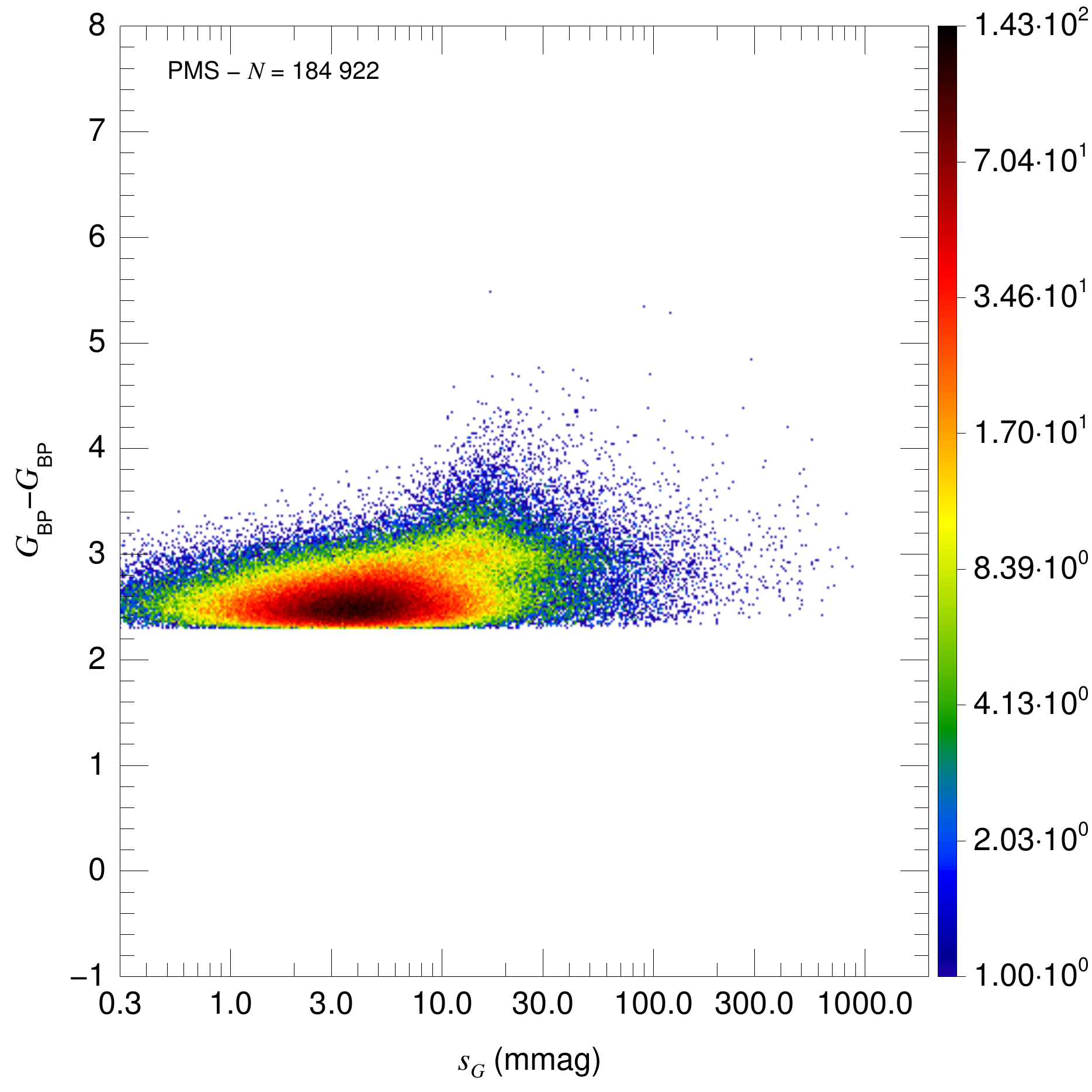}}
\centerline{\includegraphics[width=0.35\linewidth]{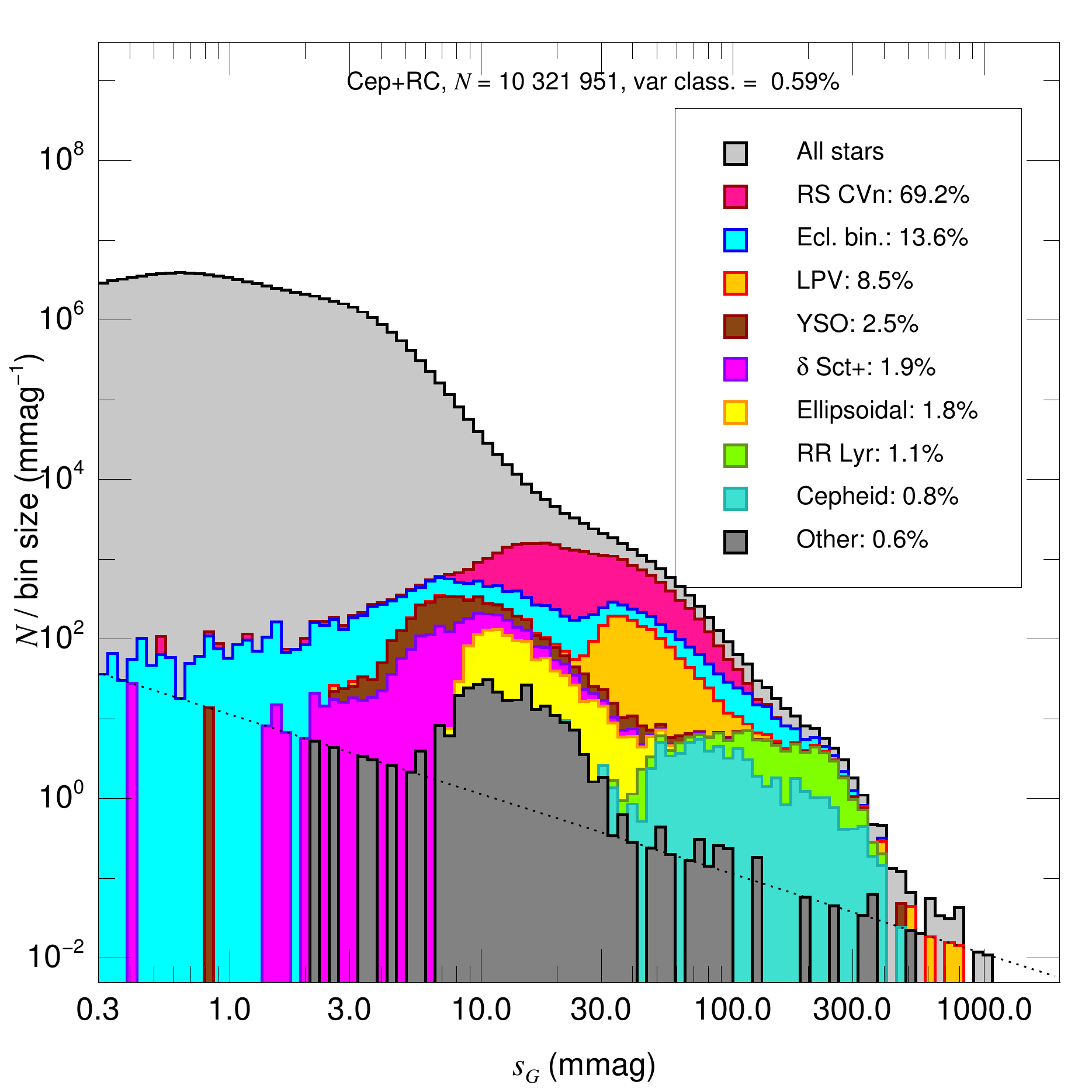}$\!\!\!$
            \includegraphics[width=0.35\linewidth]{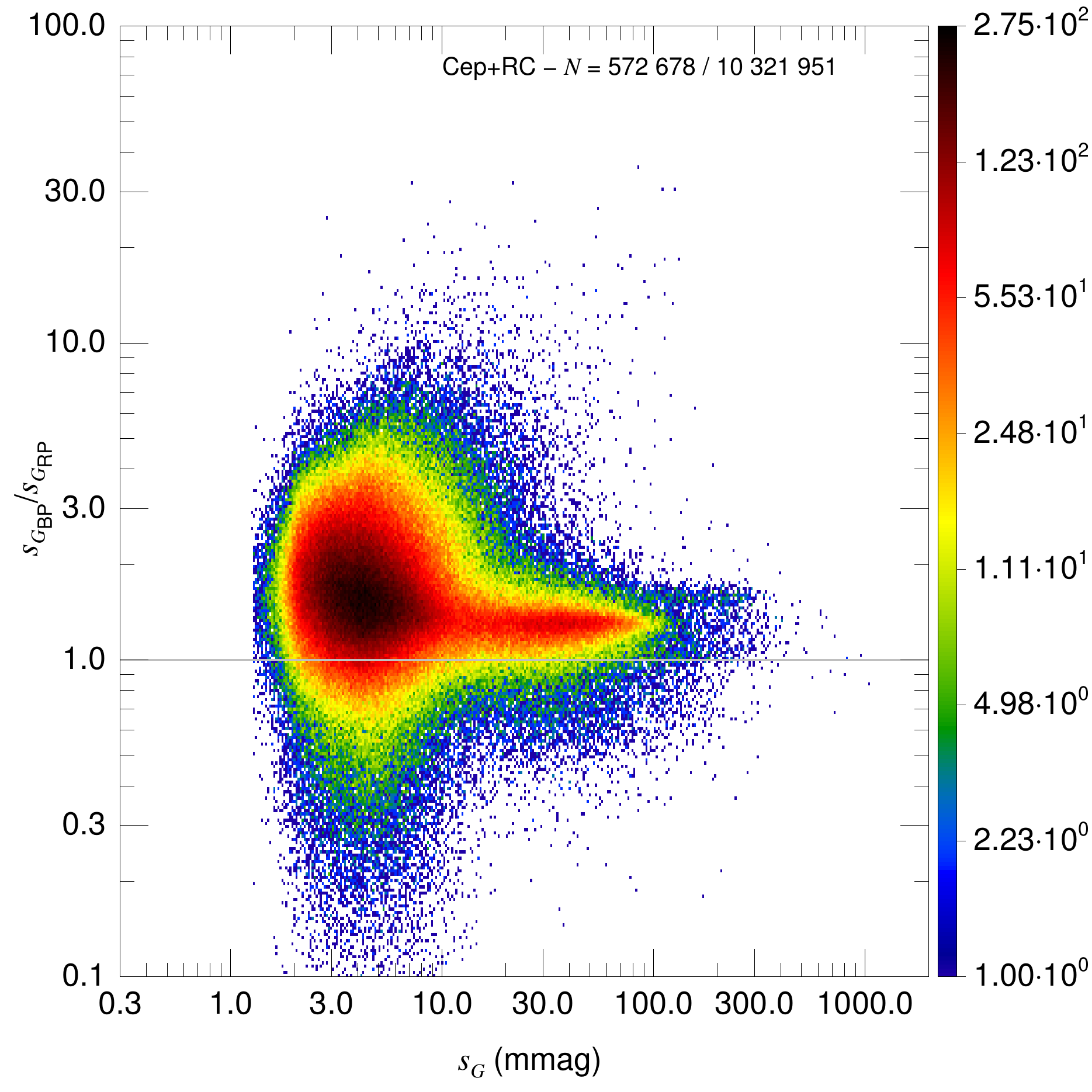}$\!\!\!$
            \includegraphics[width=0.35\linewidth]{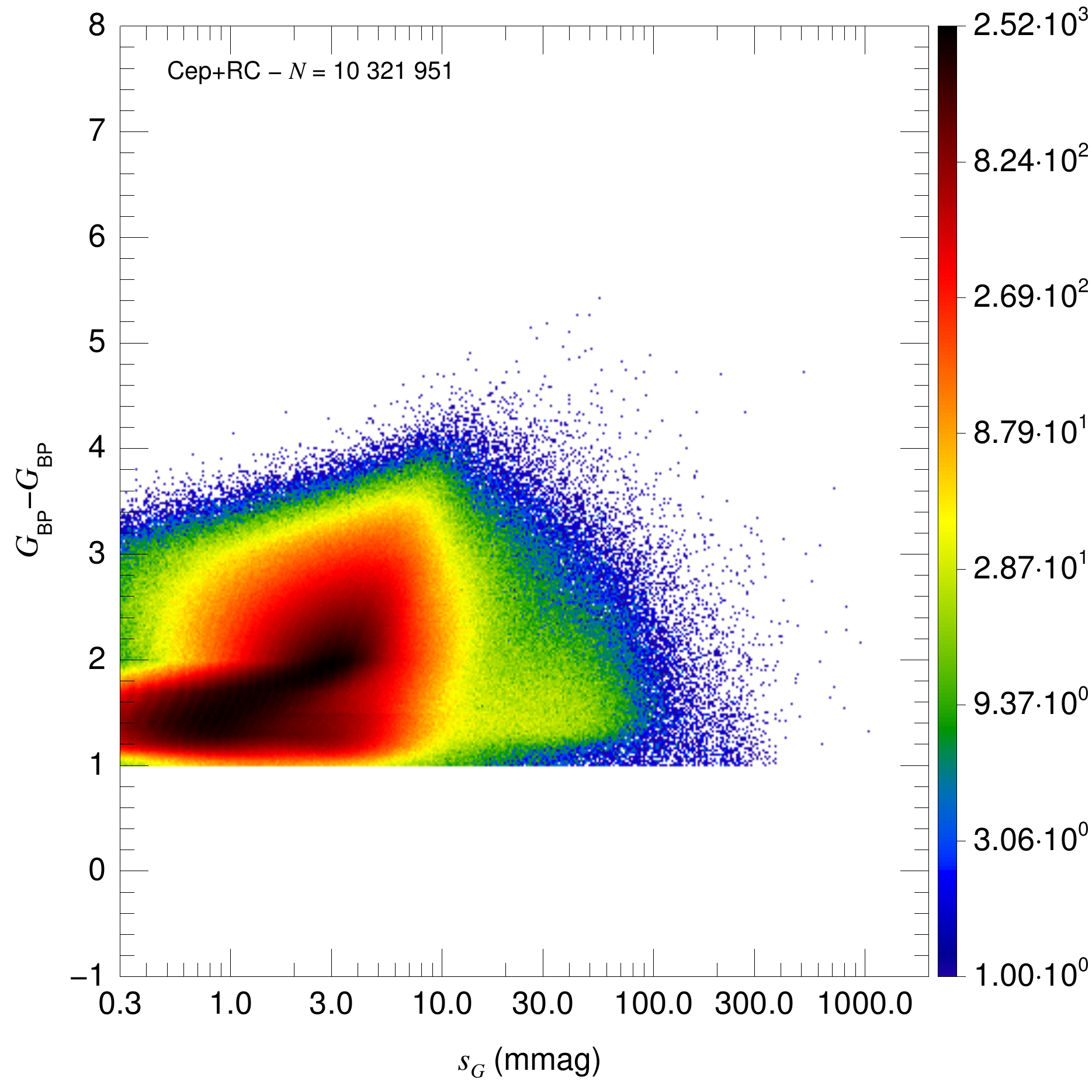}}
\centerline{\includegraphics[width=0.35\linewidth]{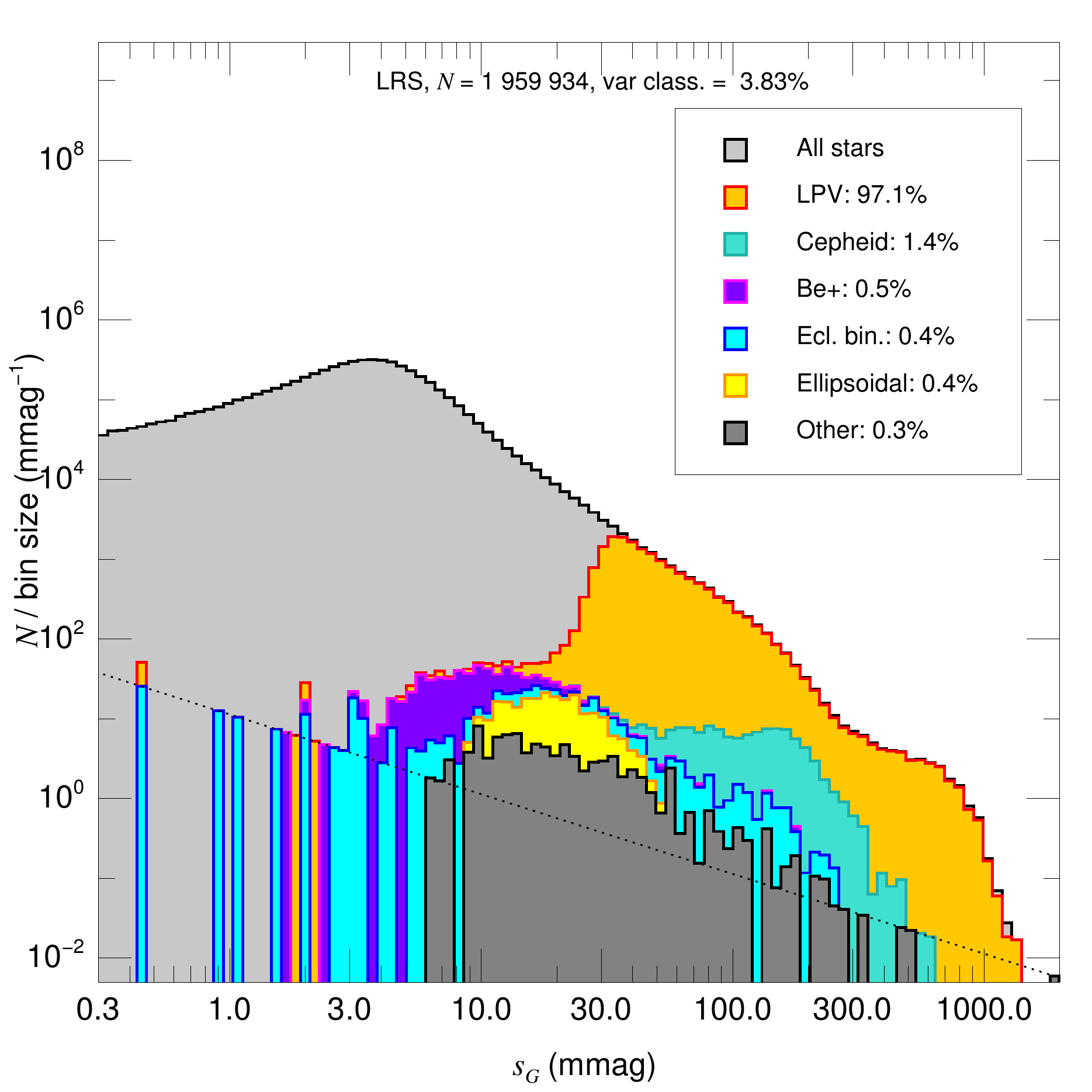}$\!\!\!$
            \includegraphics[width=0.35\linewidth]{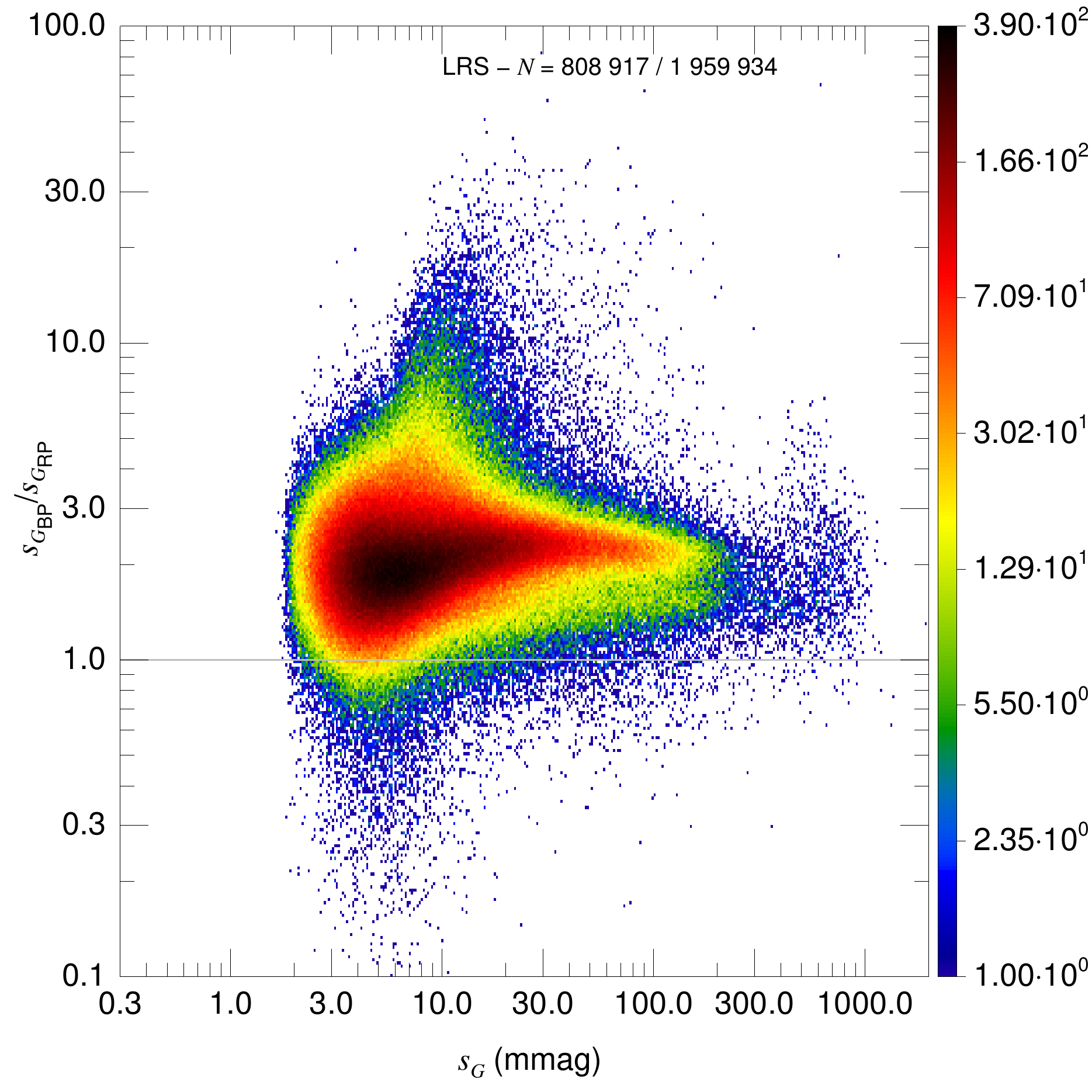}$\!\!\!$
            \includegraphics[width=0.35\linewidth]{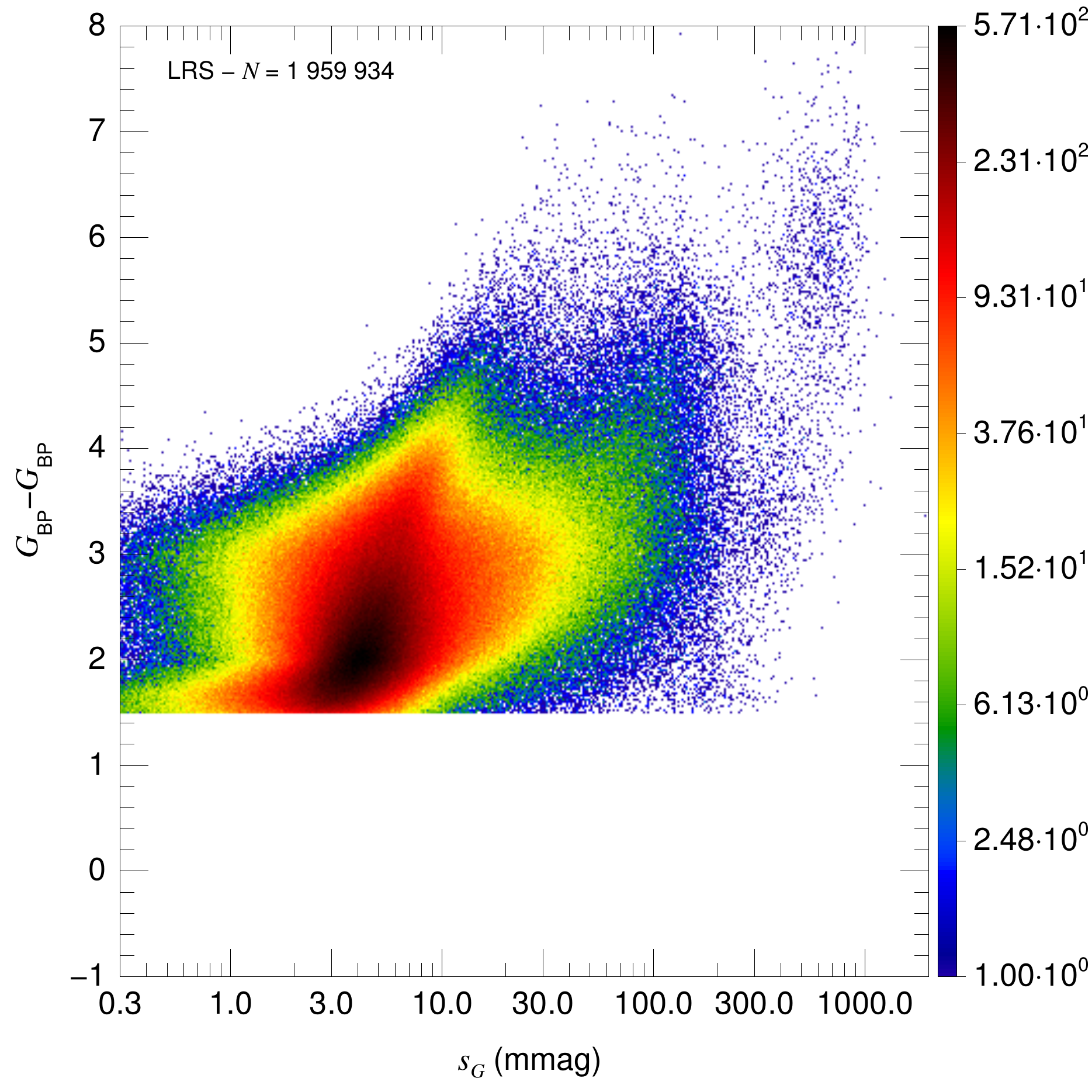}}
 \caption{Same as Fig.~\ref{results_by_region_1} for the last three of the six MW regions.}
\label{results_by_region_2}
\end{figure*}

$\,\!$\indent The Milky Way subsample is more than two orders of magnitude larger than those of the Magellanic Clouds combined and
covers a much larger range in \Gabs. At the same time, the Galactic \textit{Gaia} CAMD is smeared by the effect of extinction, which
moves stars downwards and to the right following curved trajectories (see e.g. \citealt{Berletal23}). As an example, the 
RC stars would all concentrate in a small region of the CAMD if it were not for extinction but in the upper left panel of 
Fig.~\ref{CAMD_MW} they are spread in a trajectory determined by extinction (Nogueras~Lara et al. in prep.). 

To better analyze the different populations seen in Fig.~\ref{CAMD_MW}, we divide the Galactic CAMD in six different regions: white
dwarfs (WD), subdwarfs and cataclysmic variables (sd+CV), main sequence (MS), pre-main sequence (PMS), Cepheids and red clump
(Cep+RC), and luminous red stars (LRS) according to the values given in Table~\ref{CAMD_regions} and shown in Fig.~\ref{CAMD_MW_BW}.
The name of each region describes its dominant population(s) (with LRS objects encompassing RGBs, AGBs, BRGs, and RSGs) but 
other populations may be also present. For example, MS stars of different temperatures may be found in the PMS,
Cep+RC, or LRS regions due to extinction. Below we analyze each of those six CAMD regions with the help of 
Figs.~\ref{CAMD_MW},~\ref{results_by_region_1},~and~\ref{results_by_region_2}. In the last two figures we plot for each region:
(a) its \sG\ histogram combining the information from Figs.~\ref{hist_sigma}~and~\ref{hist_sigma0}, (b) its \sG\ - \sGBP/\sGRP\
density diagram (to be compared with Fig.~\ref{varhist_types}), and (c) its \GBPmGRP\ - \sGBP/\sGRP\ density diagram (to be compared
with the lower left plot of Fig.~\ref{CAMD_MW}).

\paragraph{White dwarfs.} This is the region less likely to be contaminated with sources that do not belong to the dominant 
population, as extinction can only move objects out but not in. The vast majority (99.6\%) of the stars given a variable type by R22 
receive the correct assignment as WD. {\bf White dwarfs have a high degree of variability in comparison to MS or RC stars} (to be
analyzed later on) with a nearly flat distribution for $\sG < 10$~mmag and a power law with a slope of $\sim 3.0$ for $\sG < 10$~mmag 
(top left panel of Fig.~\ref{results_by_region_1}).  Most of the sample consists of blue (young) WDs with $\GBPmGRP < 0.2$, for which 
there is a relatively broad distribution in \sG\ (top right panel of Fig.~\ref{results_by_region_1}). {\bf The reddest (i.e. oldest) 
and less abundant WDs become progressively less variable} starting at $\GBPmGRP\sim 0.2$ (bottom left panel of Fig.~\ref{CAMD_MW} and 
upper right panel of Fig.~\ref{results_by_region_1}). R22 selected objects in this region with a bias for the most variable sources 
but not a strong one. Only a few hundred white dwarfs were previously known to be variable \citep{Cors20} and, to our knowledge, the 
flat distribution in \sG\ (especially among young WDs) we see here was previously unknown.

\paragraph{Subdwarfs and cataclysmic variables.} These two populations are clearly distinguished. {\bf sdBs have \sG\ distributions 
very similar to that of (young) white dwarfs}, with a flat distribution for $\sG < 10$~mmag and a decline at higher values.
Our sample includes $\sim$8 times more stars than that of R22 but the two distributions are not too different.
{\bf CVs, on the other hand, have much higher values of \sG\ (reaching $\sim$1~mag) and a narrow distribution of
\sGBP/\sGRP\ around 1.0 or slightly higher}. A minority of objects in this part of the \sG-\sGBP/\sGRP\ plane are identified as
eclipsing binaries. The majority of objects with \sG\ larger than 30~mmag are likely to be CVs but are not identified as
such in R22. This is the region with the highest fraction of targets with R22 variability classification (12.27\%). 

\paragraph{Main sequence.} This is the region that includes the majority of the stars in the sample and, as a consequence, it contains
a complex mixture of populations in terms of variables. Regarding the R22 sample, a little over half are solar-like (the most common
type overall as well), whose distribution peaks around $\sG\sim 6$~mmag. The second most abundant type, $\delta$~Sct+, follows a
similar distribution with a slightly lower value of the peak in \sG. Overall, {\bf the main-sequence region is dominated by 
low-variability stars}, with a
peak in the lower left panel of Fig.~\ref{results_by_region_1} around 1-2~mmag. The real peak could have an even 
lower value of \sG, as such targets have large uncertainties and, hence, the distribution is not well constrained near zero. The 
third most common type of variables in R22, RS~CVn, have typical values of \sG\ around 20-30~mmag and of \sGBP/\sGRP\ around 1.3. 
Between \sG\ of 3~mmag and 200~mmag the distribution can be approximated by a power law with $\alpha$ around 2.5 (with deviations 
associated to the different types of variables). At the high end the distribution has a cutoff and the population there is dominated 
by eclipsing binaries with a contribution from RR~Lyr and a small one from YSOs and CVs at the extreme end. The double population of 
eclipsing binaries and RR~Lyr at high values of \sG\ is seen in the lower center panel of Fig.~\ref{results_by_region_1} as the two 
structures at $\sGBP/\sGRP\sim 1.0$ and $\sim 1.6$, respectively.

\paragraph{Pre-main sequence.} Contrary to the previous region, this one is heavily dominated by YSOs (91.2\%) in the R22
population, with eclipsing binaries a distant second at 6.4\%. The distribution in \sG\ is very well fit by a power law with 
$\alpha = 2.75$ that extends from 7~mmag to its high end. {\bf There are very few objects with low variability in the pre-main
sequence region}, as expected in a population dominated by YSOs.
A comparison between the total distribution and that of YSOs from R22 reveals their selection
bias towards high values of \sG, both in the slope and the cutoff of the distribution. Together with the white dwarfs region, they
appear to be the two regions with the cleanest samples i.e. dominated by a single population. 

\paragraph{Cepheids and red clump.} This is a rather heterogeneous region defined by the (approximately vertical in the CAMD) Cepheid 
instability strip at zero extinction\footnote{Most Galactic Cepheids actually fall outside the region due to extinction.} 
and by the (diagonal in the CAMD) red-clump extinction track, where the majority of the sample is actually located. Its most
outstanding characteristic in terms of variability are its low values: the global distribution peaks around 1~mmag and only 0.59\% of
the targets have a variability class in R22. {\bf As this region is clearly dominated by RC stars (as shown by the density
distribution in the CAMD), the conclusion is that those objects form one of the most stable categories among stellar types, another
reason besides their well-defined magnitudes and color why they make excellent standard candles}. The variable minority is dominated by
RS~CVn stars (69.2\% of R22 targets) but other populations are also present. At high values of \sG\ there are also objects classified
by R22 as eclipsing binaries, LPVs, YSOs, RR Lyr, and Cepheids. 
%An effect that appears in the center right panel of 
%Fig.~\ref{results_by_region_1} (and also in the lower left panel that corresponds to LRSs and, to a lesser extent, in the lower left
%panel of the previous figure that corresponds to main sequence stars) appears to be of instrumental origin: the average \sG\ increases
%for red stars. This is caused by the fact that \GBPmGRP\ differences in this population are caused mainly by extinction and fainter
%stars with low \sG\ can appear to have an artificially increased value due to the larger uncertainties associated with high values of
%\GG\ (Fig.~\ref{hist_mag_sigma}). We have verified that such an effect does not happen for other regions: more specifically, for
%the white dwarfs and subdwarfs and CVs regions \sG\ does not depend on \GG.

\paragraph{Luminous red stars.} This last region is occupied by luminous stars that appear as red, in most cases because they are
intrinsically so (RGB, AGB, BRG, RSG) but in a few cases due to extinction, a phenomenon that significatively affects the majority of 
the sample. {\bf The luminous red stars and pre-main sequence regions are the two with the highest degrees of variability} and in both 
cases they can be characterized by a power law that extends from $\sG \sim 7$~mmag and a decline at lower values. However, there are some
differences in the distributions: the power-law fit here is not as good, the measured $\alpha$ is lower ($\sim 2.2$), and the decline
at low dispersions is not as marked. The R22 sample is heavily dominated by LPVs (97.1\%), which is actually a catch-all category for
variable LRSs. R22 did a relatively complete job in catching the most variable sources: only three of the 40 targets with highest \sG\
are missing in their list. Most of the stars with very high values of \sG\ appear in the AGB catalog of \citet{Suh22}. 

\section{Example applications}

\begin{figure}
\centerline{\includegraphics[width=\linewidth]{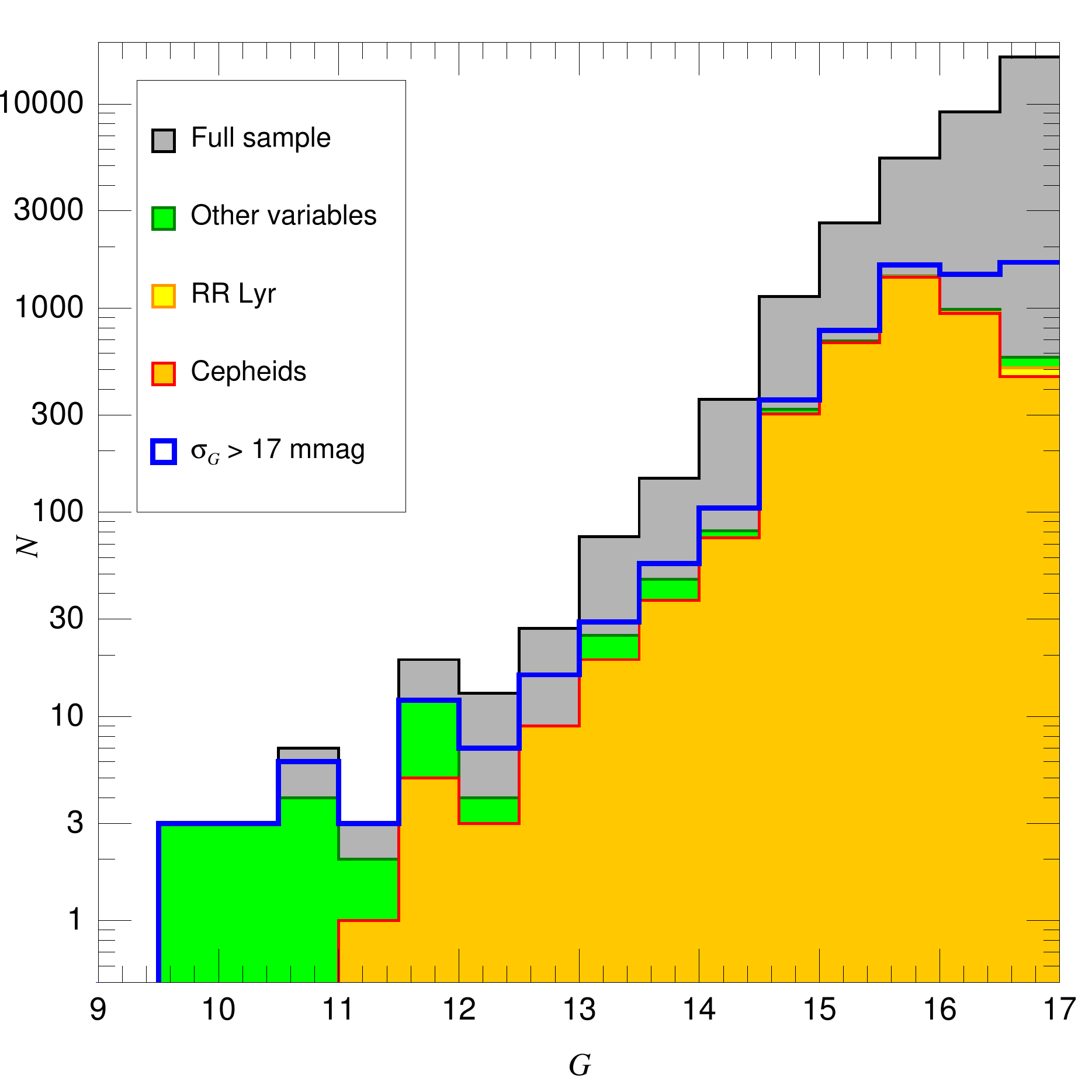}}
\caption{\GG\ histograms for the stars in the LMC instability strip. The filled polygons are cumulative histograms for the full
         sample, non-RR~Lyr non-Cepheid variables from R22, RR~Lyr from R22, and Cepheids from R22. The blue unfilled line is an 
         independent histogram containing all stars with $\sG > 17$~mmag.}
\label{strip_LMC_Ghisto}    
\end{figure}

\begin{table*}
\caption{Stars in the LMC instability strip with $\sG > 17$~mmag and $\GG < 12.5$~mag.}
\centerline{
\begin{tabular}{llrcrrll}
\hline
\textit{Gaia} DR3 ID & Simbad ID & \mci{\GG}   & \mci{\GBPmGRP} & \mci{\sG}    & \mci{\sGBP/\sGRP} & \mci{Variability code} & Simbad   \\
                     &           & \mci{(mag)} & \mci{(mag)}    & \mci{(mmag)} &                   &                        & sp. type \\
\hline
\num{4658617199111762560} & HD 269723                 &  9.707 & 1.254 &  35.7 &  1.643 & LPV                   & G4 0       \\
\num{4655460776104657408} & HD 268757                 &  9.801 & 1.524 &  53.9 &  1.191 & BE$|$GCAS$|$SDOR$|$WR & G8 0       \\
\num{4660603466846646528} & HD 271191A                &  9.906 & 1.649 &  45.9 &  8.431 & LPV                   & M0/1I      \\
\num{4657630731032546176} & HD 270046                 & 10.070 & 1.072 &  48.7 &  1.409 & BE$|$GCAS$|$SDOR$|$WR & G5I        \\
\num{4655456756015488000} & SP77 31$-$16              & 10.418 & 1.747 &  74.7 &  1.256 & BE$|$GCAS$|$SDOR$|$WR & K/M        \\
\num{4657658321911656192} & [MG73] 59                 & 10.431 & 1.661 & 141.8 &  1.314 & BE$|$GCAS$|$SDOR$|$WR & G1Ia       \\
\num{4652214880388720000} & HD 269110                 & 10.580 & 1.028 &  63.9 &  1.369 & LPV                   & G0I:       \\
\num{4658056168985153152} & SK $-$69 148              & 10.612 & 1.449 & 103.5 &  1.178 & BE$|$GCAS$|$SDOR$|$WR & G2Ia       \\
\num{4660276052943084928} & HD 269879                 & 10.617 & 1.117 &  18.4 &  2.223 & ---                   & G2Ia       \\
\num{4660207333449354112} & [W60] D17                 & 10.716 & 1.711 &  19.3 &  1.431 & LPV                   & K5I        \\
\num{4651769406346606976} & SP77 48$-$6               & 10.830 & 1.738 &  75.0 &  1.279 & BE$|$GCAS$|$SDOR$|$WR & M          \\
\num{4655114945354538368} & SP77 31$-$36              & 10.858 & 1.139 &  87.7 &  0.972 & ---                   & M1:I:e     \\
\num{4657672306328206208} & HD 38489                  & 11.358 & 0.919 &  29.1 &  3.500 & BE$|$GCAS$|$SDOR$|$WR & B[e]       \\
\num{4657674814588916864} & W61 8$-$14                & 11.462 & 1.392 &  20.4 &  1.215 & ---                   & ---        \\
\num{4662155148333196800} & SV* HV 5497               & 11.492 & 1.284 &  34.5 &  2.774 & CEP                   & G2Ib       \\
\num{4659422866270932352} & HD 270100                 & 11.577 & 1.103 & 139.5 &  2.319 & CEP                   & G2:Ia      \\
\num{4658684608603152384} & SK $-$67 114              & 11.584 & 1.476 &  37.1 &  0.937 & BE$|$GCAS$|$SDOR$|$WR & OB         \\
\num{4657625263503324544} & CPD$-$69 502              & 11.630 & 1.260 &  53.1 &  0.745 & BE$|$GCAS$|$SDOR$|$WR & B6I        \\
\num{4658745498377157248} & HD 269374                 & 11.664 & 1.124 &  38.6 &  1.469 & LPV                   & M0         \\
\num{4657277066293133056} & ASAS J053625$-$6941.5     & 11.688 & 1.146 &  53.7 &  1.605 & LPV                   & K2III      \\
\num{4658831844404939776} & SP77 37$-$33              & 11.724 & 1.657 &  21.2 &  1.320 & ---                   & K5-M3      \\
\num{4658325789850642048} & SV* HV 2447               & 11.761 & 1.335 & 156.2 &  1.633 & CEP                   & G1Ia?      \\
\num{4664612492399408128} & NSV 1789                  & 11.846 & 1.288 &  62.1 & 10.688 & CEP                   & ---        \\
\num{4652070290243179776} & SK $-$70 80               & 11.871 & 1.522 &  38.6 &  1.090 & LPV                   & B8Iab      \\
\num{4660117306608575488} & [BE74] 356                & 11.916 & 1.354 &  34.7 &  1.524 & LPV                   & ---        \\
\num{4661326189635227648} & SV* HV 883                & 11.971 & 1.338 & 355.0 &  1.476 & CEP                   & F8/G0Ia    \\
\num{4661266399338865408} & V* TT Dor                 & 11.976 & 1.024 &  98.9 &  1.869 & CEP                   & ---        \\
\num{4659110463203521280} & MACHO 32.11203.12         & 12.083 & 1.593 &  17.1 &  1.443 & ---                   & ---        \\
\num{4655463864161679104} & LHA 120$-$S 71            & 12.108 & 1.474 &  43.0 &  1.730 & LPV                   & Be         \\
\num{4659851572721436416} & SV* HV 2827               & 12.156 & 1.374 & 107.9 &  1.457 & ---                   & M          \\
\num{4658656609733883648} & SV* HV 953                & 12.167 & 1.108 & 274.6 &  1.791 & CEP                   & ---        \\
\num{4660251348288638848} & UCAC2 2515786             & 12.309 & 1.339 &  90.0 &  1.679 & CEP                   & K4III      \\
\num{4658929254268139904} & SV* HV 2369               & 12.310 & 1.269 & 218.3 &  3.446 & CEP                   & F5?I?      \\
\num{4657679487513112960} & HTR 13                    & 12.477 & 0.880 &  19.9 &  1.000 & ---                   & BN6Iap     \\
\hline
\end{tabular}
}
\label{strip_LMC}
\end{table*}

$\,\!$\indent This paper is intended as a first step in the study of stellar variability using \textit{Gaia}~DR3 data. There is
just too much information about different types of variables for a single paper. For that reason, here we just analyze three
examples of what these results can be used for and leave further studies for future papers.

\subsection{Cepheids in the LMC}

$\,\!$\indent As a first application, we analyze the objects in the instability strip in the LMC, which we define based on 
Fig.~\ref{CMD_LMC} as the region with:

\begin{gather}   
(G-17) \ge 11.25\,(\GBPmGRP-0.4)\phantom{.} \\
(G-17) \le 11.25\,(\GBPmGRP-1.2). 
\label{instability_strip_LMC}
\end{gather}   

The instability strip contains \num{35956} stars from our sample, of which R22 classified \num{3947} (10.97\%) as Cepheids, 75 
(0.21\%) as RR~Lyr, and 181 (0.50\%) as other types of variables. For the whole sample the median \sG\ is 4.4~mmag but for the
Cepheids it is 145.8~mmag (minimum of 17.4~mmag) and for the RR~Lyr it is 191.0~mmag (minimum of 53.1~mmag). Therefore, the Cepheids
and RR~Lyr can be distinguished from other stars in principle from their dispersions. 

One use of the data in this paper is to search for LMC Cepheids not included in the R22 sample. We do that by selecting all stars in
the instability strip with $\sG > 17$~mmag (value selected from the minimum one for a Cepheid in the R22 LMC sample), computing its 
\GG-magnitude histogram, and comparing it with those of the different variable types (Fig.~\ref{strip_LMC_Ghisto}). We see there 
that R22 classified most of the stars in the LMC instability strip as Cepheids but that other variability types dominate at the 
brighter end and that RR~Lyr start to make their appearance at the faintest end. A thorough search for Cepheids in the LMC would 
be the subject of a different paper but here we concentrate just on the brightest stars ($\GG < 12.5$~mag), which we list in 
Table~\ref{strip_LMC}, as a proof of concept. There we find 34 stars, of which:

\begin{itemize}
 \item Nine are classified as LPV by R22 so they are expected to be late-type stars. The four brightest have indeed Simbad spectral 
       classifications of yellow or red supergiants. The next two have classifications of M0 without a luminosity class and of K2~III.
       Of the last three, one has no spectral classification but is listed as a long-period variable candidate, another one appears as
       a B8~Iab but has alternative classifications as a late-type object and an early+late binary, and the last one is listed as Be 
       but has an alternative classification as C: using \textit{Gaia} data. 
 \item Another nine are classified as Be+ by R22 so they are expected to be early-type stars. However, the six brightest appear in
       Simbad as yellow or red supergiants. Only the last three are classified as being of early type.
 \item Another nine are classified as Cepheids by R22, of which six have spectral types in Simbad. All six appear as FGK giants or
       supergiants, as expected for Cepheids. The four objects with the highest values of \sG\ among the list of 34 are listed as 
       Cepheids by R22, indicating that high variability can be a discerning sign for them. Five of the nine have values of 
       \sGBP/\sGRP\ between 1.4 and 1.9 (that is close to the 1.6 value discussed above) but the other four have significantly higher 
       ones. 
 \item The remaining seven are not in the R22 sample. Of those, the one that looks more promising as a Cepheid from our data is 
       SV*~HV~2827 in terms of its large value of \sG\ and the closeness of its \sGBP/\sGRP\ to 1.6. As it turns out, it is a known 
       Cepheid \citep{MartWarr79}, thus showing that these data can find at least some Cepheids.
\end{itemize}

\subsection{PMS and MS stars in four Villafranca clusters}

\begin{figure*}
\centerline{\includegraphics[width=0.49\linewidth]{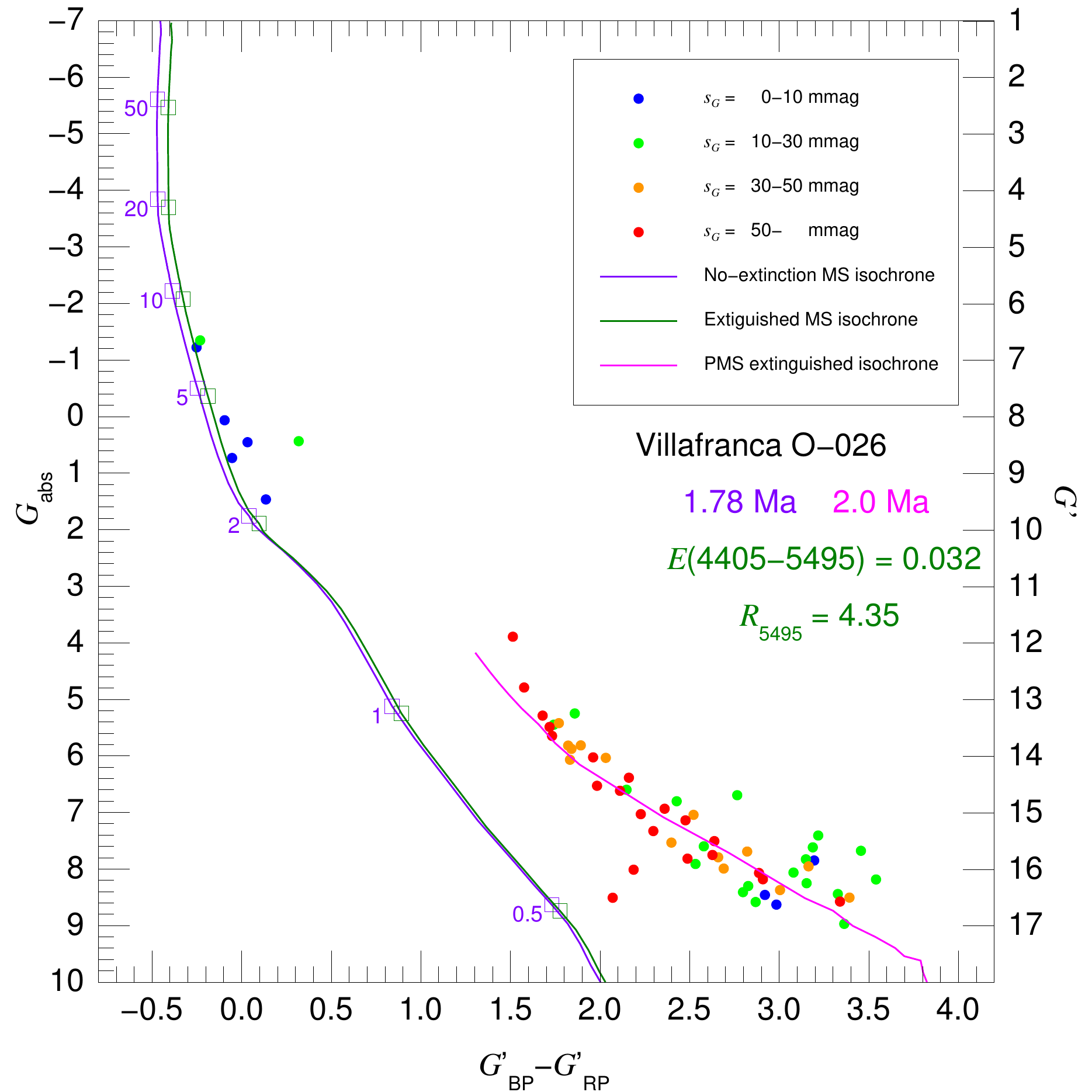} \
            \includegraphics[width=0.49\linewidth]{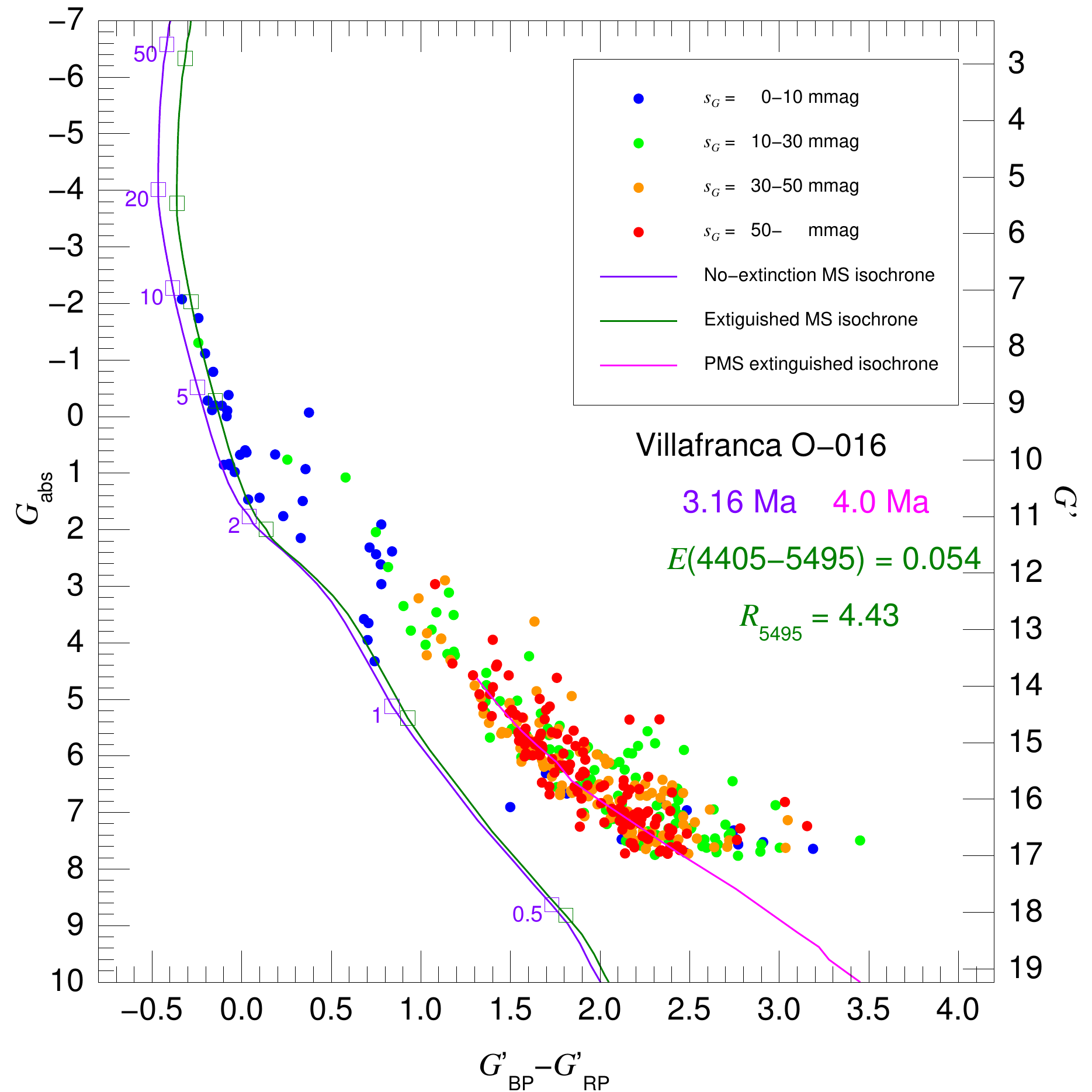}}
\centerline{\includegraphics[width=0.49\linewidth]{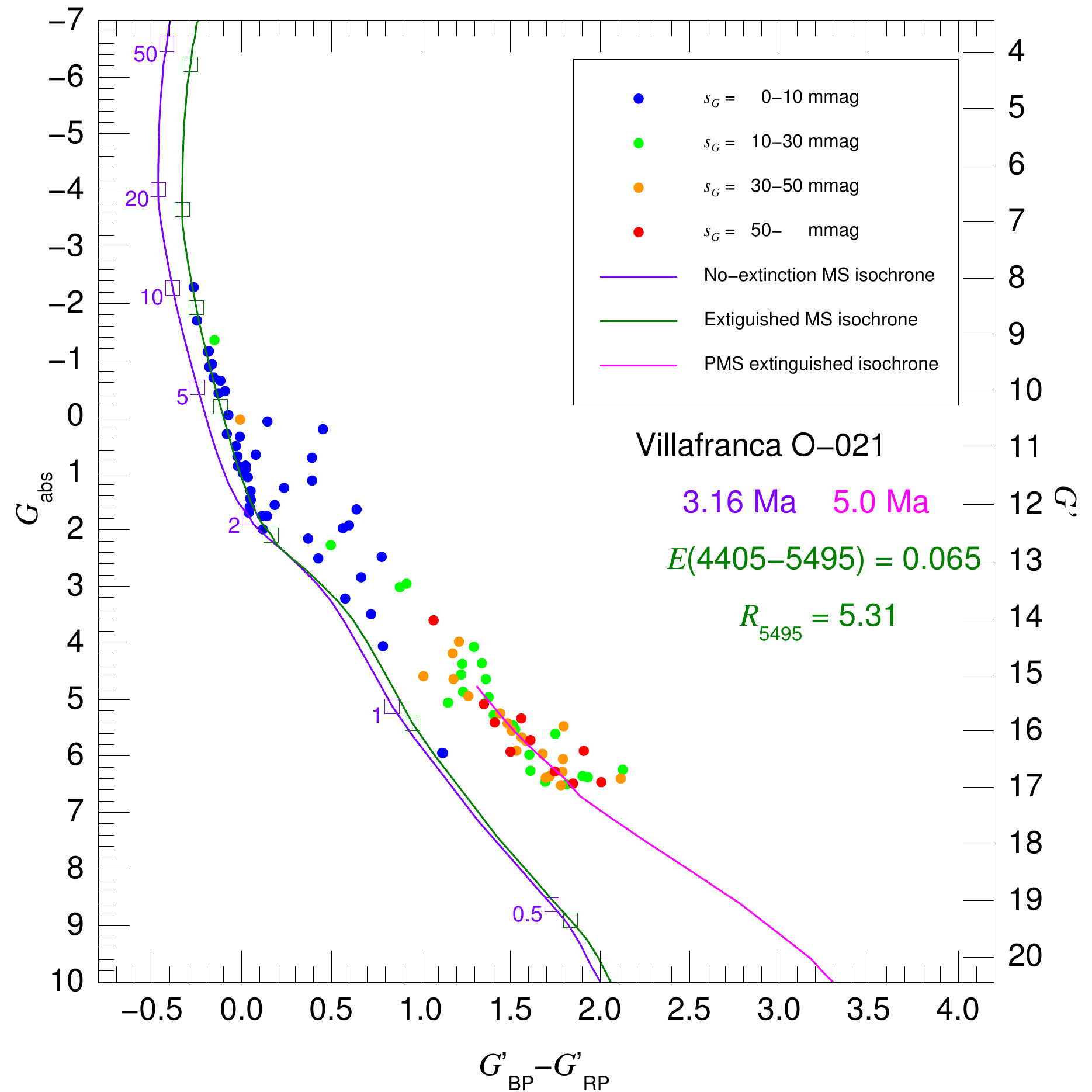} \
            \includegraphics[width=0.49\linewidth]{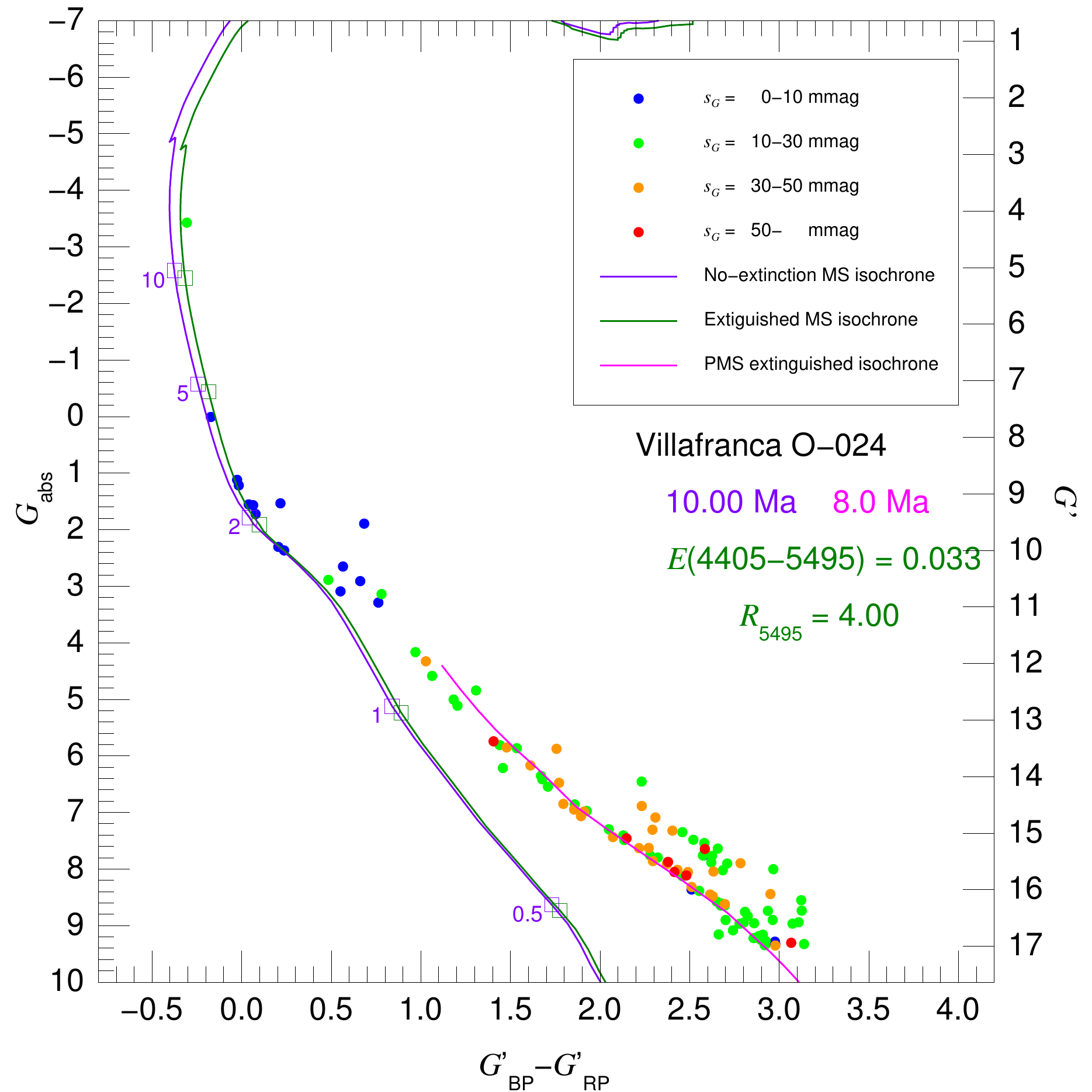}}
\caption{{\it Gaia}~EDR3 CAMDs/CMDs for four Villafranca clusters: \VO{026} ($\sigma$ Orionis cluster, {\it top left}), \VO{016} 
         (NGC~2264, {\it top right}), \VO{021} (NGC~2362, {\it bottom left}), and \VO{024} ($\gamma$~Velorum cluster, {\it bottom 
         right}) sorted from youngest to oldest. 
         In each case we use the sample from Fig.~6 in \citet{Maizetal22a} with $\GG \le 17$~mag. The purple and green lines show MS 
         isochrones with no extinction and with an average extinction appropriate to the cluster, respectively. Initial masses (in 
         solar units) are labelled along the isochrones and the purple and green text at the right side of each panel gives the age 
         and extinction parameters of the isochrones. The magenta line shows the extinguished PMS isochrone from \citet{Baraetal15} of
         the age indicated by the text of the same color at the right side of the panel. The PMS isochrones reach up to
         1.4~M$_\odot$.}
\label{CAMD_Villafranca}    
\end{figure*}

$\,\!$\indent The Villafranca project \citep{Maizetal20b,Maizetal22a} is combining \textit{Gaia} astrometry and photometry with 
data from ground-based spectroscopic \citep{Maizetal11,Maizetal19a} and photometric \citep{Maizetal21d} surveys to analyze Galactic
stellar groups (clusters and subassociations) with massive stars. In \citet{Maizetal22a} we presented distances and memberships for 
26 groups with O stars. Of those, we analyzed four low-foreground-extinction clusters in more detail and we selected the 
\textit{Gaia}~DR3 objects that have a high probability of membership. As a second application of the results in this paper, we 
cross-matched the stellar sample for those four clusters with the results here, discarding those objects fainter than \GG~=~17~mag or 
with six-parameter astrometric solutions.  The resulting CAMDs/CMDs are plotted in Fig.~\ref{CAMD_Villafranca}, which is a new version 
of Fig.~1 in \citet{Maizetal22a} but now using a color code to place each target in one of four \sG\ bins.

Results show a consistent pattern: a large majority of the stars that have already reached the MS are in the first \sG\
bin (0-10~mmag) and most of the rest are in the second bin (10-30~mmag), with just one single object in the four clusters in the third
bin (30-50~mmag). On the other hand, stars that are clearly in the PMS are concentrated in the last three bins, with only a few stars
in the first bin. Furthermore, the fraction of objects in the third and fourth ($\sG > 50$~mmag) bins is higher in the two youngest
clusters (\VO{026} and \VO{016}, PMS ages of $\sim$2~Ma and $\sim$4~Ma, respectively) than in the two oldest clusters (\VO{021} and 
\VO{024}, PMS ages of $\sim$5~Ma and $\sim$8~Ma, respectively), indicating an age effect in the sense that variability decreases with
age (see also \citealt{BarbMann23}). 

As the four clusters have similar low extinctions and the transition from high to low variability appears to take place at a similar
color ($\GBPmGRP\sim 0.8$~mag), it is possible that the effective temperature of the stars is also playing a role, with hotter PMS stars
being less variable than cooler ones. The competition between the two effects could be tested by analyzing massive PMS stars. However, we
point out that the two stars with masses below 1~M$_\odot$ in \VO{016} and \VO{021} that are closer to the MS isochrone (either because 
they are somewhat older or because they are cluster interlopers) have low values of \sG, indicating that an age effect has to be
present, in line with our comparison above between the MS and PMS regions. Therefore, the results in this paper can be used to 
characterize PMS populations in clusters.

\subsection{The most variable stars in \textit{Gaia}~DR3}

\begin{table*}
\caption{Sample stars with very high values of \sG.}
\centerline{
\begin{tabular}{llrrrrrrll}
\hline
\textit{Gaia} DR3 ID & Simbad ID & \mci{\GG}   & \mci{\Gabs} & \mci{\GBPmGRP} & \mci{\sG}    & \mci{\sGBP}  & \mci{\sGRP}  & \mci{Var.} & \mci{Gal.} \\
                     &           & \mci{(mag)} & \mci{(mag)} & \mci{(mag)}    & \mci{(mmag)} & \mci{(mmag)} & \mci{(mmag)} & \mci{code} &            \\
\hline
\num{4089520833939676544} & V* V3816 Sgr              & 16.124 &    2.425 & 2.859 & 5160 &  956 &  755 & RCB & MW  \\
\num{4506868269285941504} & IRAS 18592$+$1451         & 16.236 &      --- & 2.272 & 3239 & 2640 & 3102 & RCB & MW  \\
\num{5958230391179294208} & V* V653 Sco               & 15.885 &      --- & 2.565 & 3035 & 2269 & 2717 & RCB & MW  \\
\num{4068751643411385344} & ATO J266.3809$-$23.5402   & 15.479 &      --- & 2.255 & 3025 & 2123 & 2743 & RCB & MW  \\
\num{4239325689754161024} & V* ES Aql                 & 14.373 &    1.072 & 2.790 & 2895 & 3038 & 2046 & RCB & MW  \\
\num{5836539158653589376} & 2MASS J16020546$-$5527418 & 15.796 &      --- & 2.267 & 2624 & 1691 & 2344 & RCB & MW  \\
\num{4168623896624089088} & IRAS 17349$-$0726         & 15.478 &      --- & 3.124 & 2603 & 1453 & 1581 & RCB & MW  \\
\num{2019421879469272576} & IRAS 19282$+$2253         & 16.379 &      --- & 4.928 & 2549 & 1169 & 2180 & LPV & MW  \\
\num{5982321339413825536} & V* V409 Nor               & 13.873 &    1.713 & 3.189 & 2465 & 3264 & 2470 & RCB & MW  \\
\num{2066869246454772224} & V* V2492 Cyg              & 16.033 &    6.571 & 2.188 & 2403 & 1829 & 2025 & YSO & MW  \\
\num{5874882324544560512} & ---                       & 15.524 &      --- & 5.997 & 1740 &  811 &  509 & --- & MW  \\
\num{4146094937780229632} & ---                       & 14.773 &    7.329 & 0.857 & 1661 & 1969 & 1440 & --- & MW  \\
\num{4090114093469071744} & ---                       & 15.987 &      --- & 0.077 & 1377 & 1542 &  970 & --- & MW  \\
\num{3481965141177021568} & TWA 30                    & 15.239 &   11.865 & 3.179 & 1356 & 1117 & 1389 & --- & MW  \\
\num{3371277026437339776} & IRAS 06303$+$1819         & 14.382 &      --- & 3.126 & 1323 & 1405 & 1364 & --- & MW  \\
\num{4255191573882029696} & ---                       & 16.471 &      --- & 5.200 & 1321 &  381 & 1376 & --- & MW  \\
\num{5932094003397889280} & V* HP Nor                 & 15.198 &    6.718 & 0.934 & 1287 & 1580 & 1008 & --- & MW  \\
\num{5201350088612791168} & Ass Cha T 1$-$23          & 14.955 &    8.553 & 2.665 & 1287 & 1153 & 1207 & --- & MW  \\
\num{5903939358802958720} & ---                       & 15.300 &    5.550 & 0.794 & 1260 & 1488 &  960 & --- & MW  \\
\num{4748477123328077696} & V* R Hor                  &  6.117 & $-$0.677 & 5.600 & 1236 & 1213 &  603 & --- & MW  \\
\num{4655541487140819584} & IRAS 04498$-$6842         & 15.881 & $-$2.592 & 4.655 & 1678 & 1333 & 1535 & LPV & LMC \\
\num{5278704373764216320} & CRTS J062446.0$-$700113   & 15.807 & $-$2.666 & 4.377 & 1644 & 2037 & 1245 & LPV & LMC \\
\num{4655510632093225984} & IRAS 04544$-$6849         & 15.743 & $-$2.730 & 4.072 & 1483 & 1903 & 1194 & LPV & LMC \\
\num{4658367472584132736} & SHV 0513471$-$684007      & 16.959 & $-$1.514 & 4.705 & 1461 & 1751 & 1256 & LPV & LMC \\
\num{4655077768111946624} & WORC 82                   & 15.529 & $-$2.944 & 3.506 & 1460 & 2208 & 1220 & LPV & LMC \\
\num{4655482525823008768} & 2MASS J04534486$-$6857593 & 15.138 & $-$3.336 & 5.481 & 1404 & 1466 & 1295 & LPV & LMC \\
\num{4650475349856505344} & SHV 0535302$-$730413      & 16.327 & $-$2.146 & 4.741 & 1402 & 1753 & 1182 & LPV & LMC \\
\num{4658266867298645248} & WOH S 183                 & 15.810 & $-$2.663 & 4.277 & 1322 & 1679 & 1126 & LPV & LMC \\
\num{4653285598494818176} & WOH S 24                  & 15.373 & $-$3.101 & 4.758 & 1322 & 2148 & 1062 & LPV & LMC \\
\num{4652142346930346240} & SHV 0516511$-$700222      & 16.434 & $-$2.039 & 4.356 & 1320 & 1412 & 1175 & LPV & LMC \\
\num{4685923742225050752} & SV* HV 11417              & 15.334 & $-$3.659 & 5.924 & 1653 &  701 & 1432 & LPV & SMC \\
\num{4690479156274391936} & [MH95] 580                & 16.374 & $-$2.619 & 0.891 & 1558 & 1359 & 1320 & RCB & SMC \\
\num{4688960902568226432} & IRAS F00486$-$7308        & 15.754 & $-$3.239 & 5.288 & 1431 & 1495 & 1225 & LPV & SMC \\
\num{4689002306059371520} & SV* HV 12149              & 15.060 & $-$3.933 & 4.128 & 1370 & 2748 & 1226 & LPV & SMC \\
\num{4688923656590953472} & SV* HV 1349               & 15.414 & $-$3.579 & 4.462 & 1294 & 1564 & 1087 & LPV & SMC \\
\num{4685915392806928896} & SV* HV 1719               & 14.902 & $-$4.091 & 3.948 & 1195 & 2340 &  917 & LPV & SMC \\
\num{4685938821845281280} & BMB$-$B 75                & 14.911 & $-$4.082 & 3.741 & 1092 & 1788 &  841 & LPV & SMC \\
\num{4688924549945469184} & SV* HV 1366               & 16.041 & $-$2.952 & 3.615 & 1031 & 1769 &  750 & LPV & SMC \\
\num{4687441308747959936} & SV* HV 859                & 14.408 & $-$4.585 & 4.068 & 1010 & 2529 &  836 & LPV & SMC \\
\num{4689063844337635712} & SV* HV 12122              & 14.814 & $-$4.179 & 3.836 &  994 & 1557 &  850 & LPV & SMC \\
\hline
\end{tabular}
}
\label{high_variability}
\end{table*}

$\,\!$\indent As a final application of these results, we list in Table~\ref{high_variability} a total of 40 targets: the ten targets 
each with the highest values of \sG\ (a) overall, (b) without a variability classification in R22, (c) in then LMC, and (d) in the 
SMC.

The seven targets with the highest values of \sG\ are all identified in R22 as R~CrB variables, a rare kind of variable that has the
highest average values in all three bands in Table~\ref{variable_types_results}. If we discard those, the highest-variability sample
is dominated by LPVs. Both types are LRSs and the majority of cases are in the AGB phase. The ten stars with the highest values of 
\sG\ are in the Milky Way but this is possibly a sample size effect (the LMC+SMC LRS samples are much smaller). LMC and SMC stars can 
also reach high values of \sG\ but there are two significant differences between the Magellanic and Galactic samples in 
Table~\ref{high_variability}: (a) the first ones have extremely red colors (with the exception of one R~CrB variable, the rest have 
all $\GBPmGRP > 3.5$) while the second one is a mixed bag in \GBPmGRP\ and (b) stars in the LMC and SMC have brighter \Gabs\ (though
extinction possibly plays a secondary role in that). As Magellanic interstellar extinction is generally low, this means that its 
highest variability sample is dominated by extreme AGB stars (see Figs.~\ref{CMD_LMC}~and~Figs.\ref{CMD_SMC}). The Galactic
equivalent is more of a mixture.

R22 did a good job in selecting the targets with the highest variability but a significant number is not in their sample,
including the second group of ten stars with \sG\ larger than 1200~mmag in Table~\ref{high_variability}. In particular we point out 
the case of R~Hor (=HD~\num{18242}), a relatively well-known Mira variable of sixth magnitude in \GG\ that is missing in the R22 
sample. Most of that second group have red colors and are likely to be LPVs but some have $\GBPmGRP < 1$~mag. One of those, HP~Nor, is
classified as a cataclysmic variable in Simbad. Therefore, the results in this paper can be used to select high-variability targets
that are not included in R22.

\section{Summary}

$\,\!$\indent In this paper we have presented the \textit{Gaia}~DR3 three-band photometric dispersions for the 145 million sources
brighter than $\GG=17$~mag, normal colors, and 5-parameter astrometric solutions. This work is intended to serve as a complement to
R22: the information provided per star is smaller as we do not use epoch photometry but in exchange we use a much larger and unbiased
sample. The results are made available so they can be used for future papers or for the preparation of DR4 by \textit{Gaia}~CU7.

Our main scientific results are:

\begin{itemize}
 \item A number of interesting results are observed for the Magellanic Clouds: the two regions in their CMDs with the highest 
       variability are the Cepheid-RR Lyr instability strip and the coolest AGB stars, Be stars form a distinct branch with respect to
       the real main sequence in the CMD and show higher variability, the two ends of the blue loop are relatively quiet phases in
       terms of variability, and \sG\ increases as we ascend in luminosity through the BRG-RSG branch without reaching the extreme
       values of the AGB stars.
 \item Metallicity effects account for differences in some of the structures seen in the LMC and SMC CMDs, with a general displacement
       towards bluer colors for the position of red stars in the SMC with respect to the LMC, but the overall values of \sG\ do not 
       change much.
 \item White dwarfs and B-type subdwarfs are significantly more variable than main-sequence stars. Most of the exceptions correspond to 
       old WDs, indicating that the cause of the phenomenon becomes less important with age.
 \item Main-sequence stars (and objects that fall in the same region of the CAMD) can show many types of variability but, overall, 
       form one of the most stable regions in the HRD, with a peak in the \sG\ distribution at just 1-2~mmag. This is an important
       consequence for astrobiology in terms of the stability of the habitable zone in short time scales.
 \item We confirm that young stellar objects, on the other hand, are significantly more variable that MS stars in the optical in 
       time scales of months, that is, what is sampled in \textit{Gaia}~DR3.
 \item Red clump stars are also very stable, even more so than main-sequence stars. This is excellent news for their use as standard
       candles.
 \item The LRS region of the \textit{Gaia} CAMD is the most variable of all. The most variable objects are the 
       extreme but uncommon R~CrB variables and the more common AGB stars but other types of LRSs also contribute to the high
       variability. 
 \item There is a strong correlation between \sG\ and position in the HRD in the four Villafranca clusters we have analyzed. Stars still 
       in the PMS are significantly more variable than those that are already in the MS.
\end{itemize}

\begin{acknowledgements}
We thank Michael Weiler for the inspiration for this paper and useful comments and Sergio Sim\'on-D\'{\i}az, Danny Lennon, 
Xavier Luri, and Lee Patrick for additional comments. 
 J.~M.~A., G.~H., and M.~P.~G. acknowledge support from the Spanish Government Ministerio de Ciencia e 
Innovaci\'on and Agencia Estatal de Investigaci\'on (\num{10.13039}/\num{501100011033}) through grant PGC2018-0\num{95049}-B-C22 and 
from the Consejo Superior de Investigaciones Cient\'ificas (CSIC) through grant 2022-AEP~005. 
J.~A.~C. acknowledges support from the Spanish Government Ministerio de Ciencia e 
Innovaci\'on and Agencia Estatal de Investigaci\'on (\num{10.13039}/\num{501100011033}) through grant PID2019-\num{109522}-GB-C51.
This work has made use of data from the European Space Agency (ESA) mission {\it Gaia} 
(\url{https://www.cosmos.esa.int/gaia}), processed by the {\it Gaia} Data Processing and Analysis Consortium (DPAC, 
\url{https://www.cosmos.esa.int/web/gaia/dpac/consortium}).  Funding for the DPAC has been provided by national institutions, 
in particular the institutions participating in the {\it Gaia} Multilateral Agreement. The {\it Gaia} data are processed with 
the computer resources at Mare Nostrum and the technical support provided by BSC-CNS.
This research has made extensive use of the \href{http://simbad.u-strasbg.fr/simbad/}{SIMBAD} and 
\href{https://vizier.u-strasbg.fr/viz-bin/VizieR}{VizieR} databases, operated at 
\href{https://cds.u-strasbg.fr}{CDS}, Strasbourg, France. 
\end{acknowledgements}

%------------------------------------------------------------------

%\vfill
%
%\eject

\bibliographystyle{aa} % style aa.bst
\bibliography{general} % your references references.bib

%\vfill
%
%\eject

%$\,\!$
%
%\vfill
%
%\eject

\begin{appendix}

\section{Glossary}

$\,\!$\indent We provide a list of acronyms and terms used in this paper. In addition, acronyms for variable types are listed in 
Table~\ref{vartypestats}.

\begin{itemize}
 \item $\alpha$: (Modified) power-law slope, see Eqn.~\ref{fsX}.
 \item AGB: Asymptotic Giant Branch.
 \item BRG: Bright Red Giant.
 \item CAMD: Color-Absolute Magnitude Diagram.
 \item CCD: Charged-Coupled Device.
 \item CDS: Centre de Donn\'ees astronomiques de Strasbourg.
 \item CMD: Color-Magnitude Diagram.
 \item CSIC: Consejo Superior de Investigaciones Cient{\'\i}ficas.
 \item CV: Cataclysmic Variable.
 \item ESA: European Space Agency.
 \item \textit{Gaia} CU7: \textit{Gaia} \href{https://www.cosmos.esa.int/web/gaia/dpac}{Coordination Unit 7} (Variability).
 \item \textit{Gaia} DPAC: \textit{Gaia} \href{https://www.cosmos.esa.int/web/gaia/dpac}{Data Processing and Analysis Consortium}.
 \item \textit{Gaia} DR3: Third \textit{Gaia} Data Release.
 \item \textit{Gaia} DR4: Fourth \textit{Gaia} Data Release.
 \item HRD: Hertzsprung-Russell Diagram.
 \item LMC: Large Magellanic Cloud.
 \item LRS: Luminous Red Star.
 \item MS: Main Sequence.
 \item MW: Milky Way.
 \item PMS: Pre-main sequence.
 \item QSO: Quasi-Stellar Object.
 \item R22: \citet{Rimoetal22}.
 \item RC: Red Clump.
 \item RGB: Red Giant Branch. 
 \item RSG: Red SuperGiant. 
 \item sd(B): (B-type) sub dwarf.
 \item SMC: Small Magellanic Cloud.
 \item \sXz: Observed dispersion for band $X$ (\GG, \GBP, or \GRP).
 \item \sX: Astrophysical dispersion for band $X$ (\GG, \GBP, or \GRP), see Eqn.~\ref{sX}.
 \item \sXc: Characteristic dispersion for band $X$ (\GG, \GBP, or \GRP), see Eqn.~\ref{fsX}.
 \item \sXins: Instrumental dispersion for band $X$ (\GG, \GBP, or \GRP), modelled by a Gaussian of average \sXi\ and width \sigmasXi.
 \item \sigmasX: Astrophysical dispersion uncertainty for band $X$ (\GG, \GBP, or \GRP).
 \item WD: White Dwarf.
 \item WDS: Washington Double Star catalog.
 \item YSO: Young Stellar Object.
\end{itemize}

\section{Additional tables and figures}

\begin{landscape}

\begin{table*}
\caption{Fit results for the observed dispersions of \GG.}
\addtolength{\tabcolsep}{-1mm}
\hspace{-8.8cm}\begin{tabular}{r@{-}rc@{\hspace{4mm}}rrrrr@{\hspace{4mm}}rrrrr@{\hspace{4mm}}rrrrr@{\hspace{4mm}}rrrrr@{\hspace{4mm}}rrrrr}
\hline
\mciii{}                  & \mcv{$N$}                                                                          & \mcv{$s_{G{\rm,c}}$}                       & \mcv{$\alpha$}                        & \mcv{$s_{G{\rm,i}}$}                  & \mcv{$\sigma_{G{\rm,i}}$}             \\
\mciii{\GBPmGRPmin}       & $-$1.0         & 0.2            & 1.0            & 1.5            & 2.5            &  $-$1.0& 0.2    & 1.0    & 1.5    & 2.5    & $-$1.0& 0.2   & 1.0   & 1.5   & 2.5   & $-$1.0& 0.2   & 1.0   & 1.5   & 2.5   & $-$1.0& 0.2   & 1.0   & 1.5   & 2.5   \\
\mciii{\GBPmGRPmax}       & 0.2            & 1.0            & 1.5            & 2.5            & 8.0            &  0.2   & 1.0    & 1.5    & 2.5    & 8.0    & 0.2   & 1.0   & 1.5   & 2.5   & 8.0   & 0.2   & 1.0   & 1.5   & 2.5   & 8.0   & 0.2   & 1.0   & 1.5   & 2.5   & 8.0   \\
\mcii{\GG} & \hspace{4mm} & \mcv{}                                                                             & \mcv{}                                     & \mcv{}                                & \mcv{}                                & \mcv{}                                \\
\hline
     3.0 &      5.4 &     & \num{     547} & \num{     410} & \num{     627} & \num{     509} & \num{     168} &  16.01 &  11.36 &  13.20 &  17.50 &  88.08 &  3.32 &  2.85 &  8.12 &  4.31 &  4.87 & 13.17 & 10.95 &  9.19 & 10.74 & 25.74 &  1.21 &  1.25 &  0.79 &  1.15 &  6.94 \\
     5.4 &      5.9 &     & \num{     565} & \num{     481} & \num{     697} & \num{     480} & \num{     171} &   8.64 &   7.93 &   8.52 &  10.60 &  95.86 &  3.03 &  3.17 &  9.90 &  3.10 &  5.96 &  9.32 &  7.43 &  4.98 &  6.63 & 17.45 &  1.23 &  1.01 &  0.10 &  0.93 &  4.09 \\
     5.9 &      6.5 &     & \num{    1088} & \num{    1218} & \num{    1783} & \num{    1138} & \num{     433} &   5.15 &   3.47 &   3.96 &   7.50 &  74.82 &  2.71 &  2.39 &  4.99 &  2.77 &  3.06 &  6.24 &  4.24 &  2.57 &  3.90 & 14.04 &  0.72 &  0.85 &  0.26 &  0.60 &  3.87 \\
     6.5 &      7.7 &     & \num{    4772} & \num{    8400} & \num{    9768} & \num{    6534} & \num{    2309} &   3.81 &   3.64 &   1.11 &   7.30 &  76.70 &  2.57 &  2.63 &  2.46 &  2.96 &  3.24 &  6.22 &  3.74 &  2.74 &  3.69 & 14.35 &  0.71 &  0.52 &  0.29 &  0.48 &  4.01 \\
     7.7 &      8.2 &     & \num{    3460} & \num{    9089} & \num{    9667} & \num{    6240} & \num{    2068} &   3.19 &   3.99 &   1.30 &   6.96 &  61.15 &  2.39 &  2.78 &  2.53 &  2.93 &  2.81 &  6.70 &  4.23 &  3.30 &  4.05 & 13.91 &  0.72 &  0.52 &  0.28 &  0.43 &  3.85 \\
     8.2 &     10.0 &     & \num{   24536} & \num{  141954} & \num{  120807} & \num{   72510} & \num{   22374} &   3.41 &   3.26 &   1.41 &   6.02 &  55.29 &  2.56 &  2.65 &  2.54 &  2.93 &  2.69 &  5.32 &  3.18 &  2.61 &  3.61 & 10.00 &  0.60 &  0.51 &  0.35 &  0.50 &  2.32 \\
    10.0 &     10.5 &     & \num{    8572} & \num{  100923} & \num{   75155} & \num{   46910} & \num{   14698} &   3.39 &   4.33 &   1.96 &   5.26 &  47.56 &  2.61 &  2.92 &  2.66 &  2.88 &  2.58 &  5.88 &  3.38 &  2.91 &  4.06 &  7.94 &  0.68 &  0.59 &  0.46 &  0.68 &  1.44 \\
    10.5 &     11.0 &     & \num{    9370} & \num{  176346} & \num{  122703} & \num{   78646} & \num{   23315} &   2.80 &   3.56 &   2.12 &   4.64 &  39.77 &  2.56 &  2.80 &  2.82 &  2.81 &  2.42 &  5.17 &  2.85 &  2.36 &  3.35 &  6.91 &  0.62 &  0.52 &  0.38 &  0.58 &  1.15 \\
    11.0 &     11.2 &     & \num{    4116} & \num{  105607} & \num{   69959} & \num{   45961} & \num{   13118} &   4.51 &   4.35 &   3.03 &   4.87 &  38.00 &  2.94 &  2.88 &  2.91 &  2.81 &  2.46 &  6.10 &  4.28 &  3.51 &  4.43 &  7.24 &  0.91 &  0.96 &  0.74 &  0.91 &  1.31 \\
    11.2 &     11.4 &     & \num{    3922} & \num{  129650} & \num{   83012} & \num{   55125} & \num{   15286} &   5.13 &   4.58 &   3.68 &   5.59 &  39.05 &  3.06 &  2.84 &  2.96 &  2.92 &  2.49 &  7.21 &  5.30 &  4.27 &  5.14 &  7.94 &  1.27 &  1.41 &  1.10 &  1.21 &  1.63 \\
    11.4 &     12.0 &     & \num{   10564} & \num{  569266} & \num{  354119} & \num{  239657} & \num{   62582} &   6.61 &   5.36 &   4.53 &   5.81 &  31.31 &  3.52 &  3.57 &  3.93 &  3.17 &  2.35 &  4.85 &  3.29 &  2.74 &  3.17 &  5.95 &  0.83 &  0.54 &  0.40 &  0.57 &  1.27 \\
    12.0 &     12.9 &     & \num{   12958} & \num{ 1628024} & \num{ 1019694} & \num{  716203} & \num{  178971} &   4.72 &   2.19 &   2.08 &   2.99 &  21.92 &  2.98 &  2.52 &  2.59 &  2.58 &  2.13 &  4.51 &  3.80 &  3.17 &  3.10 &  5.02 &  0.55 &  0.55 &  0.41 &  0.36 &  0.70 \\
    12.9 &     13.1 &     & \num{    2690} & \num{  535674} & \num{  352726} & \num{  248669} & \num{   62321} &   7.09 &   2.72 &   1.80 &   3.60 &  19.46 &  3.41 &  2.57 &  2.36 &  2.80 &  2.09 &  7.88 &  4.05 &  3.20 &  3.40 &  5.91 &  1.51 &  0.53 &  0.39 &  0.36 &  0.81 \\
    13.1 &     13.3 &     & \num{    2703} & \num{  621253} & \num{  421733} & \num{  294349} & \num{   74788} &   7.79 &   2.91 &   1.30 &   4.30 &  17.60 &  3.64 &  2.62 &  2.17 &  3.10 &  2.05 &  8.86 &  3.38 &  2.67 &  3.17 &  6.42 &  1.08 &  0.40 &  0.23 &  0.35 &  0.82 \\
    13.3 &     14.0 &     & \num{   10571} & \num{ 2966689} & \num{ 2243368} & \num{ 1512763} & \num{  394814} &   4.18 &   2.57 &   1.37 &   4.56 &  13.70 &  2.58 &  2.54 &  2.17 &  3.41 &  2.00 &  9.81 &  3.50 &  2.91 &  3.47 &  6.68 &  1.37 &  0.38 &  0.27 &  0.42 &  0.83 \\
    14.0 &     15.0 &     & \num{   23399} & \num{ 6823247} & \num{ 7130531} & \num{ 4293854} & \num{ 1125231} &   3.24 &   1.82 &   1.62 &   3.75 &   8.14 &  2.12 &  2.32 &  2.19 &  2.90 &  1.95 & 10.64 &  4.45 &  4.08 &  4.80 &  7.57 &  1.39 &  0.52 &  0.52 &  0.71 &  0.94 \\
    15.0 &     15.5 &     & \num{   19666} & \num{ 4692385} & \num{ 6761752} & \num{ 3811212} & \num{  946936} &   4.15 &   1.56 &   1.87 &   3.91 &   6.35 &  2.15 &  2.24 &  2.22 &  2.74 &  2.06 & 11.51 &  5.72 &  5.47 &  6.11 &  8.55 &  1.31 &  0.45 &  0.42 &  0.60 &  0.92 \\
    15.5 &     16.0 &     & \num{   28462} & \num{ 5603789} & \num{ 9992472} & \num{ 5750590} & \num{ 1315612} &   5.03 &   1.56 &   2.38 &   3.90 &   6.14 &  2.29 &  2.23 &  2.32 &  2.58 &  2.20 & 12.54 &  7.11 &  6.90 &  7.54 &  9.69 &  1.35 &  0.60 &  0.58 &  0.72 &  0.97 \\
    16.0 &     16.5 &     & \num{   40273} & \num{ 6536166} & \num{14053493} & \num{ 8269363} & \num{ 1772380} &   6.10 &   1.31 &   2.32 &   3.82 &   6.87 &  2.43 &  2.13 &  2.24 &  2.44 &  2.42 & 14.22 &  9.36 &  9.19 &  9.72 & 11.49 &  1.33 &  0.77 &  0.76 &  0.86 &  1.03 \\
    16.5 &     17.0 &     & \num{   57345} & \num{ 7493508} & \num{19081660} & \num{11543799} & \num{ 2335979} &   7.16 &   1.45 &   2.61 &   3.99 &   7.64 &  2.56 &  2.16 &  2.29 &  2.41 &  2.57 & 16.21 & 11.93 & 11.82 & 12.30 & 13.86 &  1.34 &  1.00 &  0.99 &  1.05 &  1.23 \\
\hline
\end{tabular}
\addtolength{\tabcolsep}{1mm}
\label{fit_results_G}
\end{table*}

\begin{table*}
\caption{Fit results for the observed dispersions of \GBP.}
\addtolength{\tabcolsep}{-1mm}
\hspace{-8.8cm}\begin{tabular}{r@{-}rc@{\hspace{4mm}}rrrrr@{\hspace{4mm}}rrrrr@{\hspace{4mm}}rrrrr@{\hspace{4mm}}rrrrr@{\hspace{4mm}}rrrrr}
\hline
\mciii{}                  & \mcv{$N$}                                                                          & \mcv{$s_{G_{\rm BP}{\rm,c}}$}              & \mcv{$\alpha$}                        & \mcv{$s_{G_{\rm BP}{\rm,i}}$}         & \mcv{$\sigma_{G_{\rm BP}{\rm,i}}$}    \\
\mciii{\GBPmGRPmin}       & $-$1.0         & 0.2            & 1.0            & 1.5            & 2.5            &  $-$1.0& 0.2    & 1.0    & 1.5    & 2.5    & $-$1.0& 0.2   & 1.0   & 1.5   & 2.5   & $-$1.0& 0.2   & 1.0   & 1.5   & 2.5   & $-$1.0& 0.2   & 1.0   & 1.5   & 2.5   \\
\mciii{\GBPmGRPmax}       & 0.2            & 1.0            & 1.5            & 2.5            & 8.0            &  0.2   & 1.0    & 1.5    & 2.5    & 8.0    & 0.2   & 1.0   & 1.5   & 2.5   & 8.0   & 0.2   & 1.0   & 1.5   & 2.5   & 8.0   & 0.2   & 1.0   & 1.5   & 2.5   & 8.0   \\
\mcii{\GG} & \hspace{4mm} & \mcv{}                                                                             & \mcv{}                                     & \mcv{}                                & \mcv{}                                & \mcv{}                                \\
\hline
     3.0 &      5.4 &     & \num{     547} & \num{     410} & \num{     627} & \num{     509} & \num{     168} &   3.60 &   2.41 &   2.95 &   7.97 &  69.46 &  2.22 &  1.99 &  2.74 &  2.18 &  2.39 &  5.55 &  4.56 &  3.78 &  5.15 & 35.17 &  0.95 &  0.86 &  0.60 &  0.86 & 16.68 \\
     5.4 &      5.9 &     & \num{     565} & \num{     481} & \num{     697} & \num{     480} & \num{     171} &   4.10 &   4.00 &   3.93 &   9.66 & 105.49 &  2.52 &  2.55 &  3.38 &  2.39 &  3.05 &  5.70 &  4.70 &  3.84 &  4.99 & 27.18 &  0.85 &  0.89 &  0.60 &  1.03 &  7.99 \\
     5.9 &      6.5 &     & \num{    1088} & \num{    1218} & \num{    1783} & \num{    1138} & \num{     433} &   4.62 &   3.05 &   4.85 &   9.24 & 140.67 &  2.54 &  2.22 &  4.93 &  2.45 &  3.52 &  5.46 &  4.54 &  3.70 &  5.11 & 21.41 &  1.03 &  0.82 &  0.67 &  1.00 &  8.51 \\
     6.5 &      7.7 &     & \num{    4772} & \num{    8400} & \num{    9768} & \num{    6534} & \num{    2309} &   4.05 &   3.12 &   4.06 &   9.23 & 140.41 &  2.58 &  2.36 &  4.65 &  2.72 &  3.50 &  3.72 &  2.75 &  2.51 &  4.31 & 20.33 &  0.79 &  0.71 &  0.56 &  0.94 &  9.48 \\
     7.7 &      8.2 &     & \num{    3460} & \num{    9089} & \num{    9667} & \num{    6240} & \num{    2068} &   2.99 &   3.06 &   3.70 &   8.51 & 111.15 &  2.38 &  2.45 &  4.17 &  2.70 &  2.89 &  3.52 &  2.52 &  2.46 &  4.04 & 18.88 &  0.71 &  0.55 &  0.48 &  0.87 &  9.35 \\
     8.2 &     10.0 &     & \num{   24536} & \num{  141954} & \num{  120807} & \num{   72510} & \num{   22374} &   2.23 &   2.00 &   2.78 &   6.69 & 102.17 &  2.29 &  2.25 &  3.62 &  2.64 &  2.75 &  2.79 &  1.97 &  2.24 &  3.54 & 11.19 &  0.51 &  0.37 &  0.41 &  0.73 &  3.80 \\
    10.0 &     10.5 &     & \num{    8572} & \num{  100923} & \num{   75155} & \num{   46910} & \num{   14698} &   1.82 &   1.39 &   1.39 &   5.55 &  88.13 &  2.26 &  2.11 &  2.47 &  2.58 &  2.61 &  2.61 &  2.16 &  2.46 &  3.21 &  8.52 &  0.46 &  0.45 &  0.48 &  0.59 &  2.47 \\
    10.5 &     11.0 &     & \num{    9370} & \num{  176346} & \num{  122703} & \num{   78646} & \num{   23315} &   1.71 &   1.31 &   1.34 &   5.13 &  73.70 &  2.26 &  2.11 &  2.40 &  2.56 &  2.42 &  3.12 &  2.77 &  2.95 &  3.44 &  7.79 &  0.56 &  0.50 &  0.50 &  0.60 &  1.87 \\
    11.0 &     11.2 &     & \num{    4116} & \num{  105607} & \num{   69959} & \num{   45961} & \num{   13118} &   1.91 &   1.32 &   1.39 &   4.87 &  67.24 &  2.37 &  2.13 &  2.39 &  2.55 &  2.37 &  3.48 &  3.20 &  3.36 &  3.74 &  7.56 &  0.53 &  0.47 &  0.48 &  0.59 &  1.68 \\
    11.2 &     11.4 &     & \num{    3922} & \num{  129650} & \num{   83012} & \num{   55125} & \num{   15286} &   1.78 &   1.35 &   1.44 &   4.74 &  65.31 &  2.30 &  2.15 &  2.38 &  2.56 &  2.33 &  3.62 &  3.33 &  3.47 &  3.83 &  7.63 &  0.55 &  0.47 &  0.49 &  0.59 &  1.75 \\
    11.4 &     12.0 &     & \num{   10564} & \num{  569266} & \num{  354119} & \num{  239657} & \num{   62582} &   2.04 &   1.39 &   1.55 &   4.39 &  51.16 &  2.31 &  2.19 &  2.39 &  2.53 &  2.18 &  3.66 &  3.42 &  3.56 &  3.93 &  7.13 &  0.52 &  0.47 &  0.49 &  0.60 &  1.48 \\
    12.0 &     12.9 &     & \num{   12958} & \num{ 1628024} & \num{ 1019694} & \num{  716203} & \num{  178971} &   2.56 &   1.49 &   1.98 &   4.20 &  37.22 &  2.37 &  2.24 &  2.46 &  2.53 &  2.01 &  3.71 &  3.53 &  3.69 &  4.18 &  7.19 &  0.55 &  0.50 &  0.52 &  0.66 &  1.36 \\
    12.9 &     13.1 &     & \num{    2690} & \num{  535674} & \num{  352726} & \num{  248669} & \num{   62321} &   3.65 &   1.67 &   2.27 &   4.59 &  32.10 &  2.67 &  2.32 &  2.51 &  2.66 &  1.96 &  3.90 &  3.75 &  3.98 &  4.54 &  7.54 &  0.58 &  0.55 &  0.57 &  0.71 &  1.36 \\
    13.1 &     13.3 &     & \num{    2703} & \num{  621253} & \num{  421733} & \num{  294349} & \num{   74788} &   4.08 &   1.79 &   2.46 &   4.84 &  30.89 &  2.72 &  2.36 &  2.53 &  2.72 &  1.94 &  3.99 &  3.92 &  4.18 &  4.79 &  7.82 &  0.57 &  0.57 &  0.61 &  0.75 &  1.34 \\
    13.3 &     14.0 &     & \num{   10571} & \num{ 2966689} & \num{ 2243368} & \num{ 1512763} & \num{  394814} &   4.30 &   2.47 &   3.44 &   5.98 &  27.45 &  2.65 &  2.54 &  2.72 &  2.93 &  1.94 &  4.59 &  4.56 &  4.89 &  5.67 &  9.09 &  0.75 &  0.75 &  0.81 &  0.98 &  1.64 \\
    14.0 &     15.0 &     & \num{   23399} & \num{ 6823247} & \num{ 7130531} & \num{ 4293854} & \num{ 1125231} &   4.19 &   8.39 &   7.63 &  10.74 &  26.99 &  2.22 &  5.03 &  3.47 &  3.31 &  2.03 &  6.94 &  6.22 &  7.22 &  8.43 & 13.62 &  1.43 &  1.13 &  1.47 &  1.74 &  2.91 \\
    15.0 &     15.5 &     & \num{   19666} & \num{ 4692385} & \num{ 6761752} & \num{ 3811212} & \num{  946936} &   6.67 &  10.45 &  10.39 &  16.52 &  33.41 &  2.36 &  4.37 &  3.47 &  3.47 &  2.22 &  9.78 & 10.37 & 11.29 & 13.07 & 20.51 &  1.66 &  1.71 &  1.98 &  2.32 &  3.94 \\
    15.5 &     16.0 &     & \num{   28462} & \num{ 5603789} & \num{ 9992472} & \num{ 5750590} & \num{ 1315612} &  11.28 &  15.56 &  16.16 &  25.49 &  44.10 &  2.74 &  4.77 &  3.88 &  3.89 &  2.42 & 12.36 & 14.32 & 15.33 & 17.90 & 29.05 &  2.03 &  2.41 &  2.68 &  3.22 &  5.87 \\
    16.0 &     16.5 &     & \num{   40273} & \num{ 6536166} & \num{14053493} & \num{ 8269363} & \num{ 1772380} &  18.11 &  22.93 &  25.30 &  38.73 &  61.55 &  3.29 &  4.93 &  4.45 &  4.64 &  2.69 & 16.11 & 20.20 & 21.09 & 24.98 & 42.08 &  2.66 &  3.52 &  3.67 &  4.55 &  8.90 \\
    16.5 &     17.0 &     & \num{   57345} & \num{ 7493508} & \num{19081660} & \num{11543799} & \num{ 2335979} &  27.75 &  33.81 &  39.38 &  58.10 &  91.04 &  3.85 &  4.91 &  5.13 &  5.44 &  3.03 & 21.39 & 29.05 & 29.58 & 35.51 & 61.43 &  3.41 &  5.32 &  5.19 &  6.60 & 13.42 \\
\hline
\end{tabular}
\addtolength{\tabcolsep}{1mm}
\label{fit_results_BP}
\end{table*}

\begin{table*}
\caption{Fit results for the observed dispersions of \GRP.}
\addtolength{\tabcolsep}{-1mm}
\hspace{-8.8cm}\begin{tabular}{r@{-}rc@{\hspace{4mm}}rrrrr@{\hspace{4mm}}rrrrr@{\hspace{4mm}}rrrrr@{\hspace{4mm}}rrrrr@{\hspace{4mm}}rrrrr}
\hline
\mciii{}                  & \mcv{$N$}                                                                          & \mcv{$s_{G_{\rm RP}{\rm,c}}$}              & \mcv{$\alpha$}                        & \mcv{$s_{G_{\rm RP}{\rm,i}}$}         & \mcv{$\sigma_{G_{\rm RP}{\rm,i}}$}    \\
\mciii{\GBPmGRPmin}       & $-$1.0         & 0.2            & 1.0            & 1.5            & 2.5            &  $-$1.0& 0.2    & 1.0    & 1.5    & 2.5    & $-$1.0& 0.2   & 1.0   & 1.5   & 2.5   & $-$1.0& 0.2   & 1.0   & 1.5   & 2.5   & $-$1.0& 0.2   & 1.0   & 1.5   & 2.5   \\
\mciii{\GBPmGRPmax}       & 0.2            & 1.0            & 1.5            & 2.5            & 8.0            &  0.2   & 1.0    & 1.5    & 2.5    & 8.0    & 0.2   & 1.0   & 1.5   & 2.5   & 8.0   & 0.2   & 1.0   & 1.5   & 2.5   & 8.0   & 0.2   & 1.0   & 1.5   & 2.5   & 8.0   \\
\mcii{\GG} & \hspace{4mm} & \mcv{}                                                                             & \mcv{}                                     & \mcv{}                                & \mcv{}                                & \mcv{}                                \\
\hline
     3.0 &      5.4 &     & \num{     547} & \num{     410} & \num{     627} & \num{     509} & \num{     168} &  12.14 &  12.95 &  14.15 &  22.52 &  54.30 &  2.69 &  2.92 &  4.50 &  5.13 &  3.09 &  8.99 &  9.08 & 10.17 & 12.07 & 24.48 &  1.98 &  1.90 &  2.45 &  2.85 &  3.82 \\
     5.4 &      5.9 &     & \num{     565} & \num{     481} & \num{     697} & \num{     480} & \num{     171} &   8.97 &  10.04 &  12.15 &  17.32 & 100.31 &  3.37 &  3.30 &  4.20 &  3.70 &  9.90 &  7.21 &  7.05 &  7.80 &  9.93 & 18.91 &  1.69 &  1.24 &  2.03 &  2.51 &  4.13 \\
     5.9 &      6.5 &     & \num{    1088} & \num{    1218} & \num{    1783} & \num{    1138} & \num{     433} &   4.53 &   2.90 &   6.35 &  12.28 &  72.46 &  2.40 &  2.14 &  3.83 &  3.52 &  3.28 &  4.27 &  4.35 &  3.64 &  4.54 & 17.10 &  1.07 &  1.25 &  0.91 &  0.96 &  5.56 \\
     6.5 &      7.7 &     & \num{    4772} & \num{    8400} & \num{    9768} & \num{    6534} & \num{    2309} &   2.65 &   2.18 &   3.11 &   6.37 &  64.17 &  2.19 &  2.18 &  4.04 &  2.99 &  3.13 &  3.17 &  2.79 &  2.59 &  3.33 & 10.57 &  0.56 &  0.49 &  0.41 &  0.63 &  3.66 \\
     7.7 &      8.2 &     & \num{    3460} & \num{    9089} & \num{    9667} & \num{    6240} & \num{    2068} &   1.93 &   1.67 &   1.96 &   5.25 &  48.48 &  2.12 &  2.14 &  3.24 &  2.85 &  2.65 &  2.75 &  2.27 &  2.10 &  2.67 &  9.78 &  0.46 &  0.42 &  0.35 &  0.47 &  3.78 \\
     8.2 &     10.0 &     & \num{   24536} & \num{  141954} & \num{  120807} & \num{   72510} & \num{   22374} &   1.26 &   0.98 &   0.94 &   3.62 &  42.30 &  2.03 &  2.01 &  2.49 &  2.67 &  2.51 &  2.05 &  1.59 &  1.49 &  1.88 &  5.64 &  0.34 &  0.31 &  0.30 &  0.43 &  2.47 \\
    10.0 &     10.5 &     & \num{    8572} & \num{  100923} & \num{   75155} & \num{   46910} & \num{   14698} &   0.96 &   0.75 &   0.75 &   3.18 &  38.38 &  1.96 &  1.94 &  2.30 &  2.62 &  2.41 &  1.78 &  1.45 &  1.36 &  1.61 &  4.05 &  0.26 &  0.25 &  0.23 &  0.33 &  1.43 \\
    10.5 &     11.0 &     & \num{    9370} & \num{  176346} & \num{  122703} & \num{   78646} & \num{   23315} &   1.02 &   0.81 &   0.83 &   2.98 &  35.33 &  2.01 &  1.99 &  2.29 &  2.59 &  2.40 &  2.01 &  1.81 &  1.73 &  1.85 &  3.79 &  0.32 &  0.32 &  0.33 &  0.37 &  1.07 \\
    11.0 &     11.2 &     & \num{    4116} & \num{  105607} & \num{   69959} & \num{   45961} & \num{   13118} &   1.16 &   0.86 &   0.89 &   2.89 &  34.14 &  2.10 &  2.02 &  2.30 &  2.59 &  2.52 &  2.23 &  2.03 &  1.93 &  2.00 &  3.64 &  0.32 &  0.32 &  0.31 &  0.34 &  0.89 \\
    11.2 &     11.4 &     & \num{    3922} & \num{  129650} & \num{   83012} & \num{   55125} & \num{   15286} &   1.18 &   0.88 &   0.94 &   2.82 &  31.89 &  2.08 &  2.04 &  2.31 &  2.61 &  2.39 &  2.29 &  2.08 &  1.95 &  1.99 &  3.62 &  0.32 &  0.33 &  0.31 &  0.33 &  0.82 \\
    11.4 &     12.0 &     & \num{   10564} & \num{  569266} & \num{  354119} & \num{  239657} & \num{   62582} &   1.28 &   0.95 &   1.02 &   2.53 &  23.18 &  2.07 &  2.08 &  2.31 &  2.55 &  2.21 &  2.40 &  2.12 &  1.93 &  1.92 &  3.20 &  0.35 &  0.34 &  0.32 &  0.32 &  0.72 \\
    12.0 &     12.9 &     & \num{   12958} & \num{ 1628024} & \num{ 1019694} & \num{  716203} & \num{  178971} &   1.86 &   1.12 &   1.25 &   2.28 &  15.34 &  2.17 &  2.16 &  2.32 &  2.48 &  2.00 &  2.65 &  2.35 &  2.08 &  2.00 &  2.93 &  0.40 &  0.40 &  0.36 &  0.34 &  0.63 \\
    12.9 &     13.1 &     & \num{    2690} & \num{  535674} & \num{  352726} & \num{  248669} & \num{   62321} &   2.48 &   1.31 &   1.46 &   2.40 &  12.93 &  2.28 &  2.23 &  2.35 &  2.54 &  1.96 &  3.12 &  2.63 &  2.34 &  2.21 &  2.94 &  0.44 &  0.42 &  0.37 &  0.35 &  0.59 \\
    13.1 &     13.3 &     & \num{    2703} & \num{  621253} & \num{  421733} & \num{  294349} & \num{   74788} &   2.79 &   1.42 &   1.55 &   2.47 &  12.39 &  2.34 &  2.26 &  2.35 &  2.57 &  1.95 &  3.37 &  2.80 &  2.51 &  2.34 &  2.99 &  0.49 &  0.44 &  0.39 &  0.35 &  0.57 \\
    13.3 &     14.0 &     & \num{   10571} & \num{ 2966689} & \num{ 2243368} & \num{ 1512763} & \num{  394814} &   3.25 &   1.77 &   1.97 &   2.70 &  10.13 &  2.38 &  2.34 &  2.41 &  2.62 &  1.91 &  4.21 &  3.37 &  3.01 &  2.78 &  3.27 &  0.74 &  0.58 &  0.51 &  0.45 &  0.59 \\
    14.0 &     15.0 &     & \num{   23399} & \num{ 6823247} & \num{ 7130531} & \num{ 4293854} & \num{ 1125231} &   4.42 &   3.02 &   3.25 &   3.49 &   7.49 &  2.15 &  2.55 &  2.55 &  2.58 &  1.95 &  6.75 &  5.12 &  4.62 &  4.08 &  4.26 &  1.47 &  1.05 &  0.94 &  0.77 &  0.77 \\
    15.0 &     15.5 &     & \num{   19666} & \num{ 4692385} & \num{ 6761752} & \num{ 3811212} & \num{  946936} &   8.14 &   4.07 &   4.64 &   5.13 &   7.02 &  2.30 &  2.61 &  2.63 &  2.63 &  2.08 & 10.47 &  7.93 &  6.87 &  5.79 &  5.53 &  1.90 &  1.41 &  1.19 &  0.90 &  0.86 \\
    15.5 &     16.0 &     & \num{   28462} & \num{ 5603789} & \num{ 9992472} & \num{ 5750590} & \num{ 1315612} &  14.30 &   6.67 &   7.15 &   7.95 &   7.96 &  2.63 &  2.85 &  2.84 &  2.83 &  2.25 & 14.34 & 10.85 &  9.27 &  7.70 &  6.98 &  2.66 &  2.02 &  1.63 &  1.24 &  1.10 \\
    16.0 &     16.5 &     & \num{   40273} & \num{ 6536166} & \num{14053493} & \num{ 8269363} & \num{ 1772380} &  24.40 &  11.92 &  11.88 &  12.16 &  10.72 &  3.10 &  3.28 &  3.25 &  3.23 &  2.57 & 20.15 & 15.20 & 12.70 & 10.47 &  9.10 &  3.58 &  2.94 &  2.29 &  1.77 &  1.48 \\
    16.5 &     17.0 &     & \num{   57345} & \num{ 7493508} & \num{19081660} & \num{11543799} & \num{ 2335979} &  39.98 &  21.04 &  19.70 &  18.07 &  15.76 &  3.66 &  3.87 &  3.85 &  3.80 &  3.02 & 28.30 & 21.57 & 17.79 & 14.77 & 12.29 &  4.81 &  4.18 &  3.23 &  2.64 &  2.14 \\
\hline
\end{tabular}
\addtolength{\tabcolsep}{1mm}
\label{fit_results_RP}
\end{table*}

\end{landscape}

\clearpage

\begin{figure*}
\centerline{$\!\!\!$\includegraphics[width=0.35\linewidth]{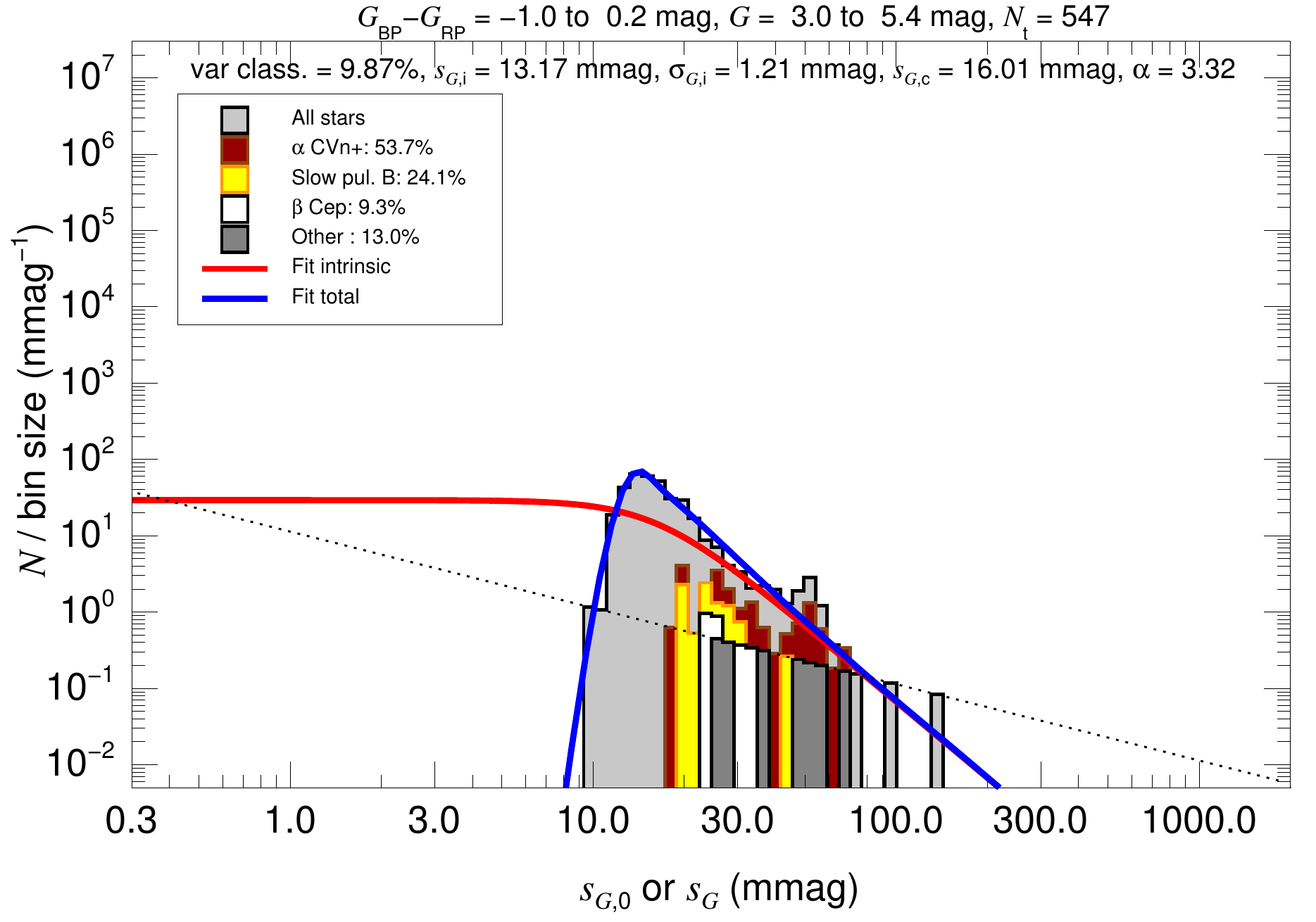}$\!\!\!$
                    \includegraphics[width=0.35\linewidth]{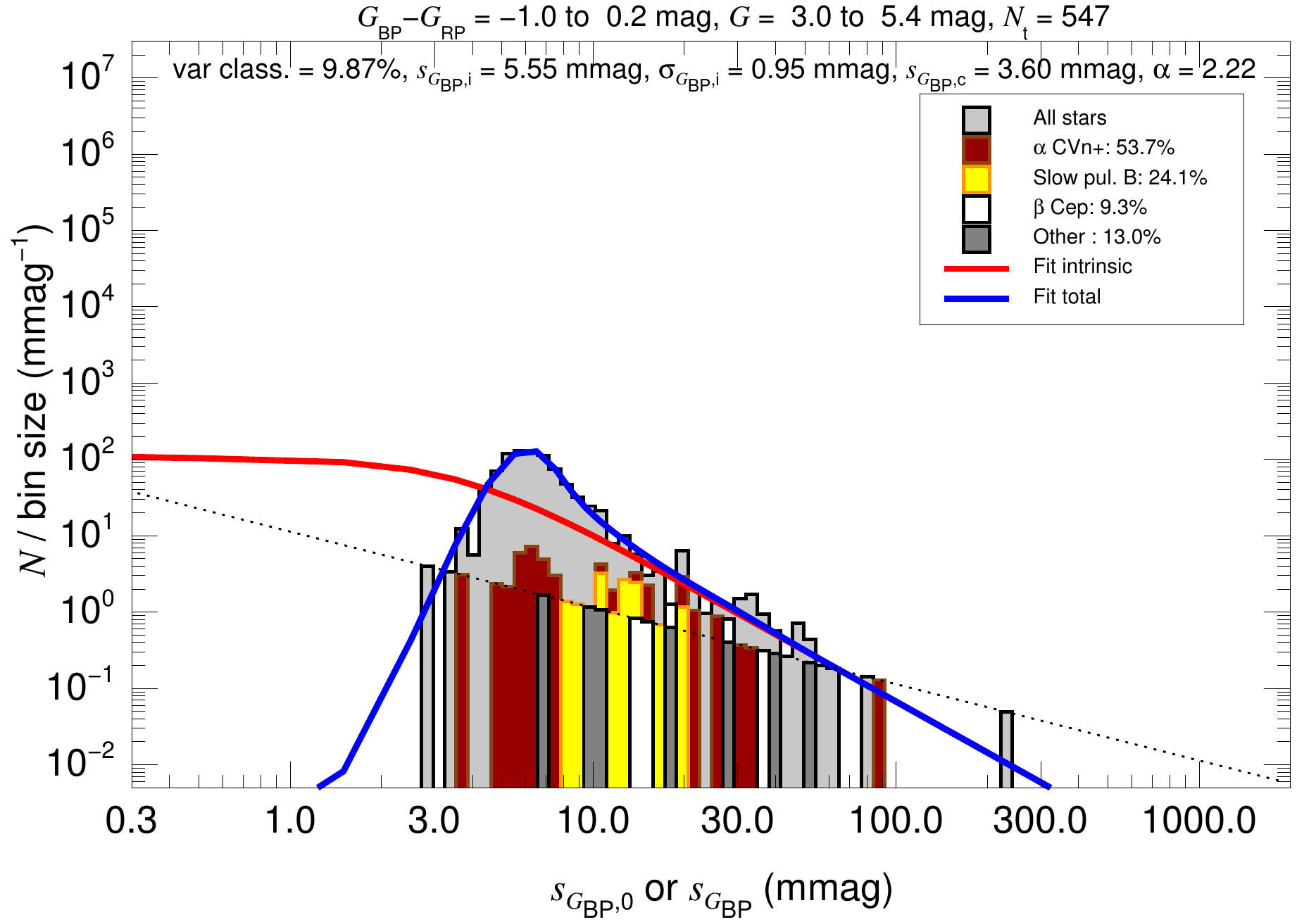}$\!\!\!$
                    \includegraphics[width=0.35\linewidth]{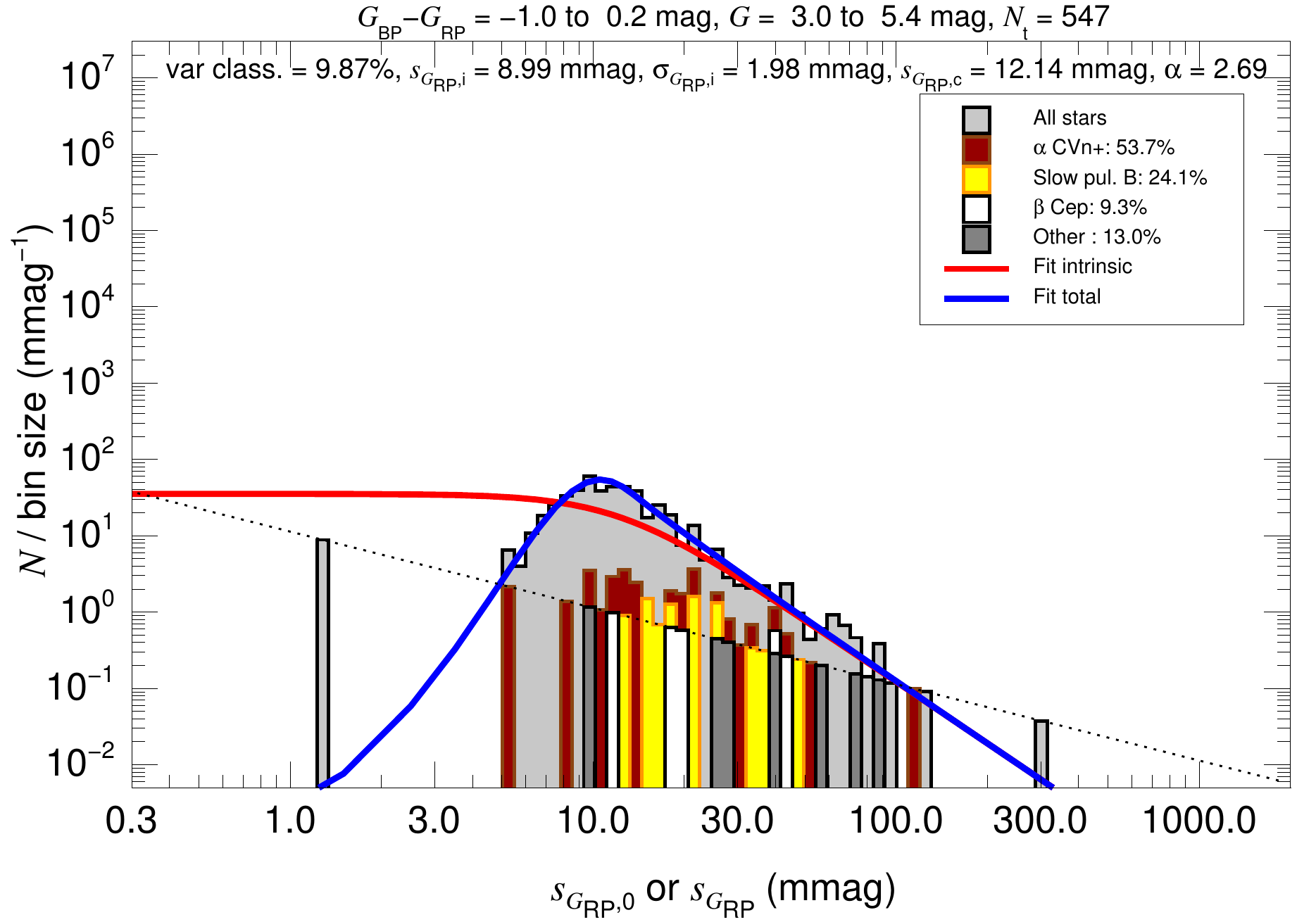}}
\centerline{$\!\!\!$\includegraphics[width=0.35\linewidth]{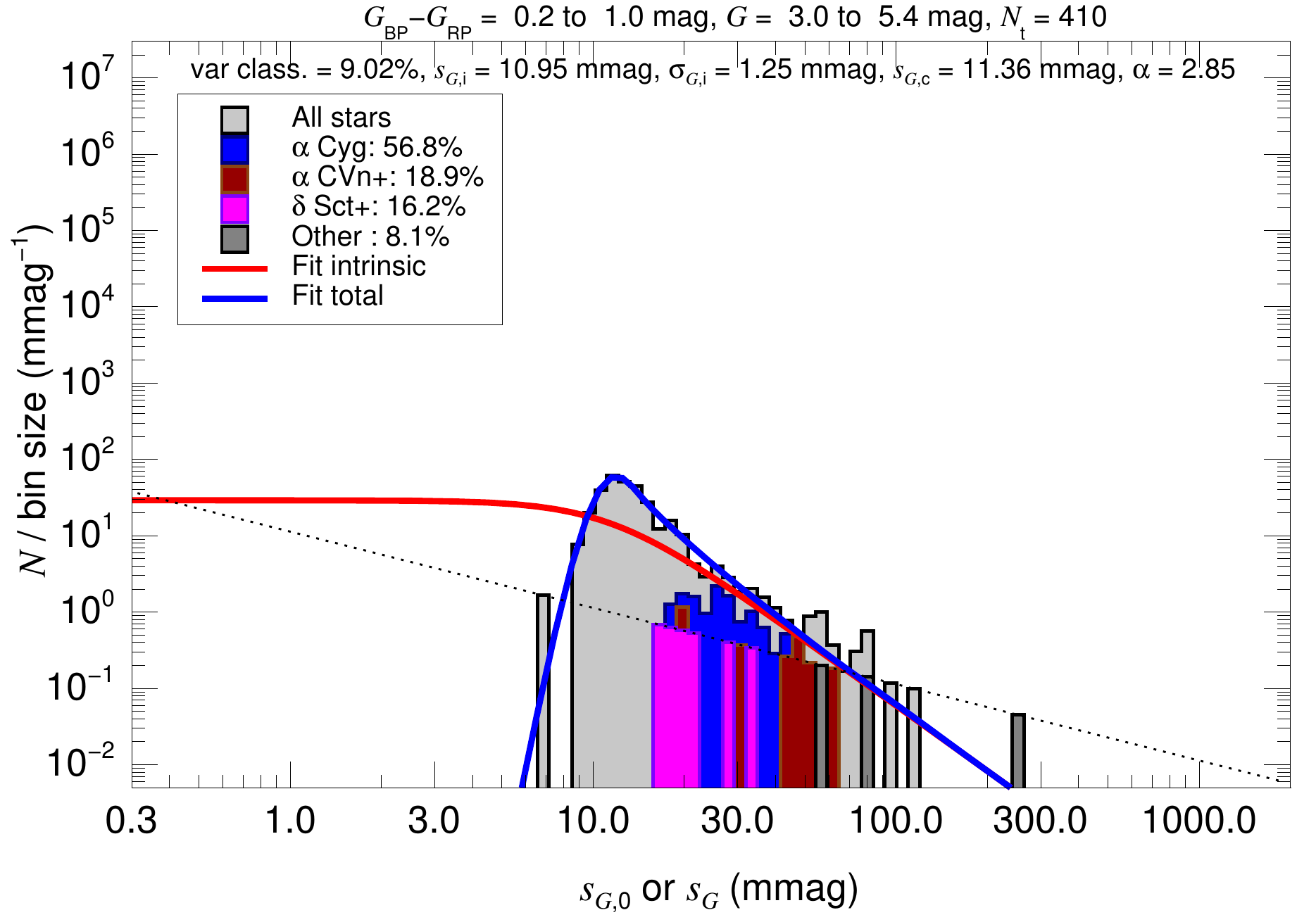}$\!\!\!$
                    \includegraphics[width=0.35\linewidth]{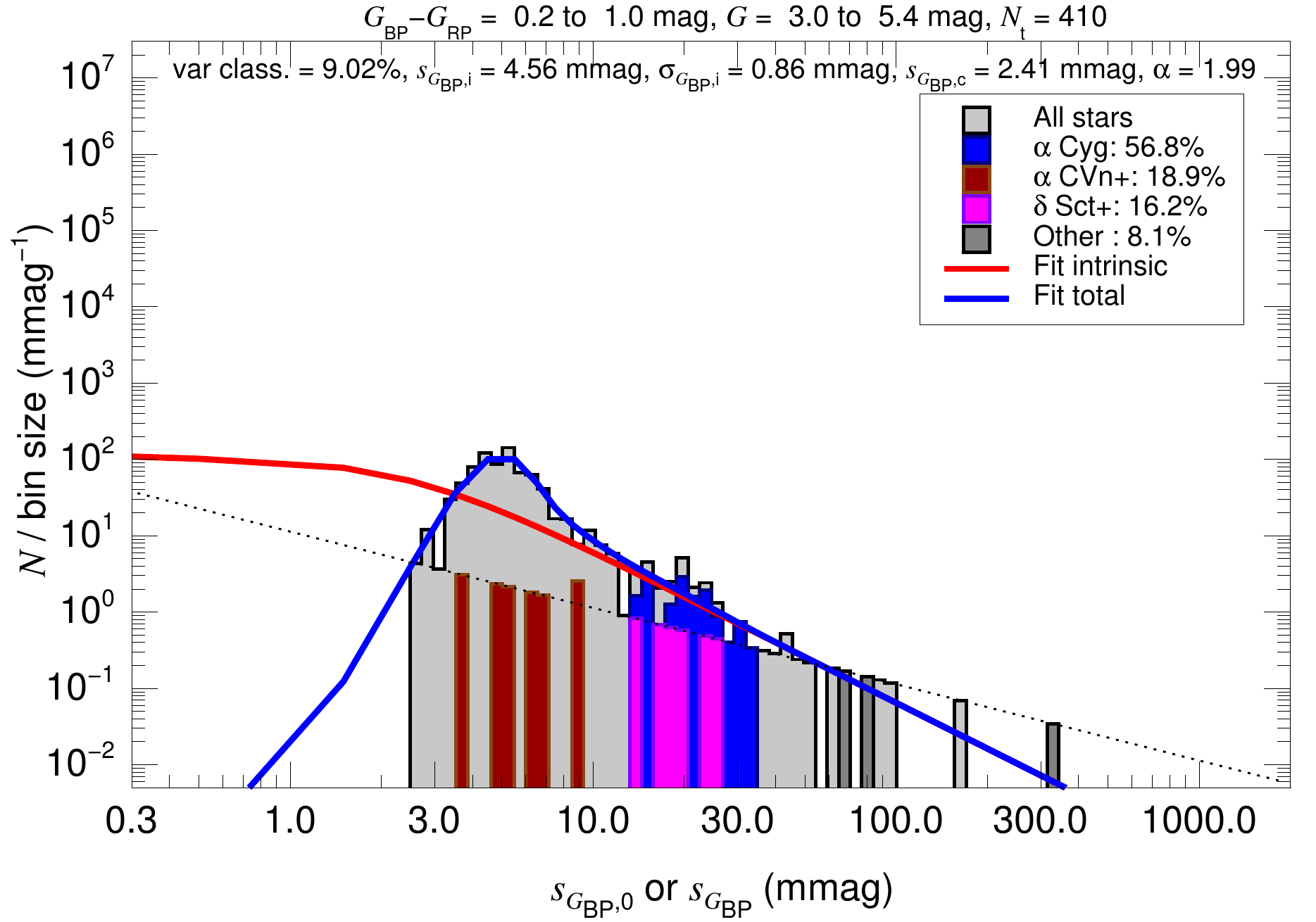}$\!\!\!$
                    \includegraphics[width=0.35\linewidth]{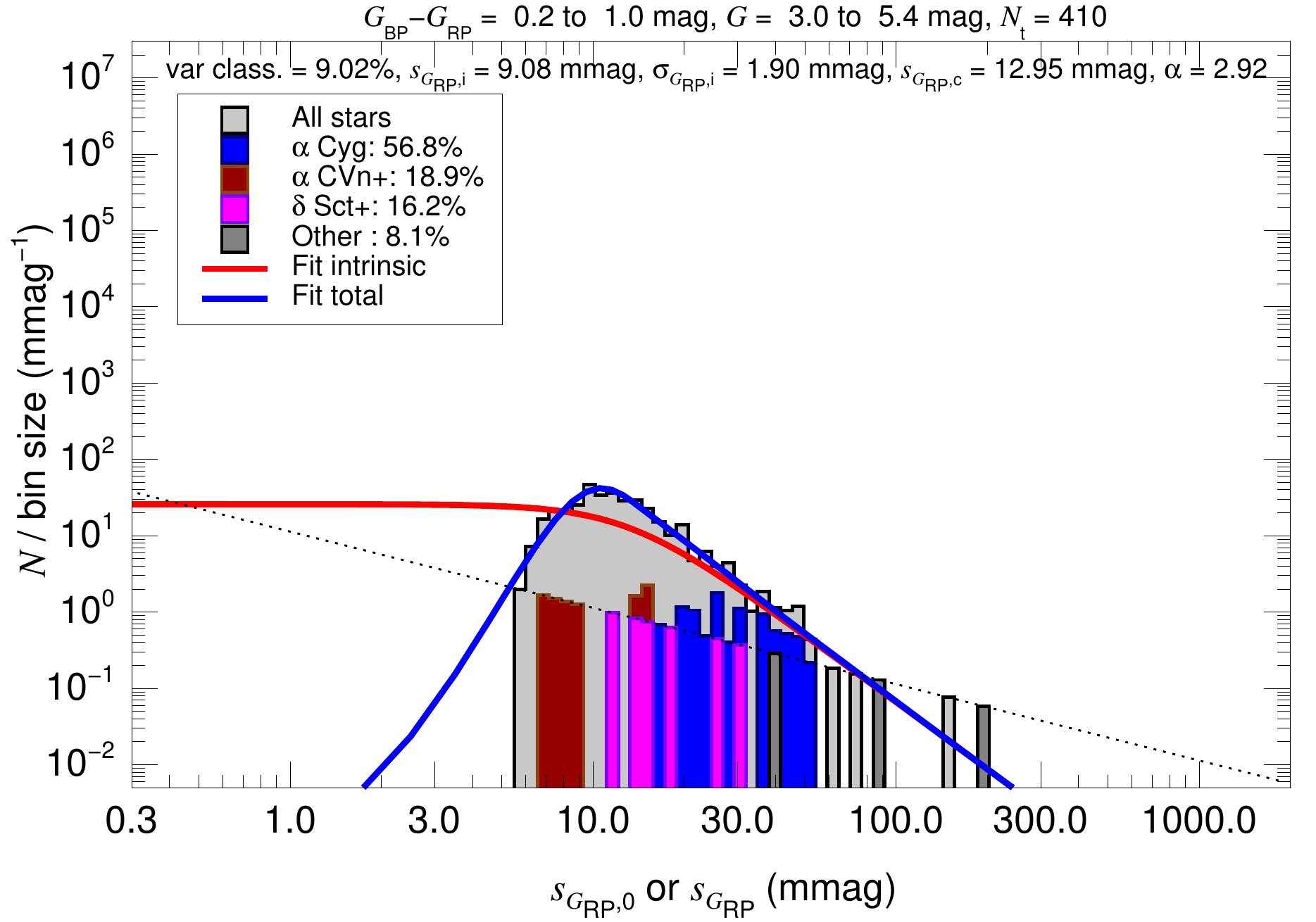}}
\centerline{$\!\!\!$\includegraphics[width=0.35\linewidth]{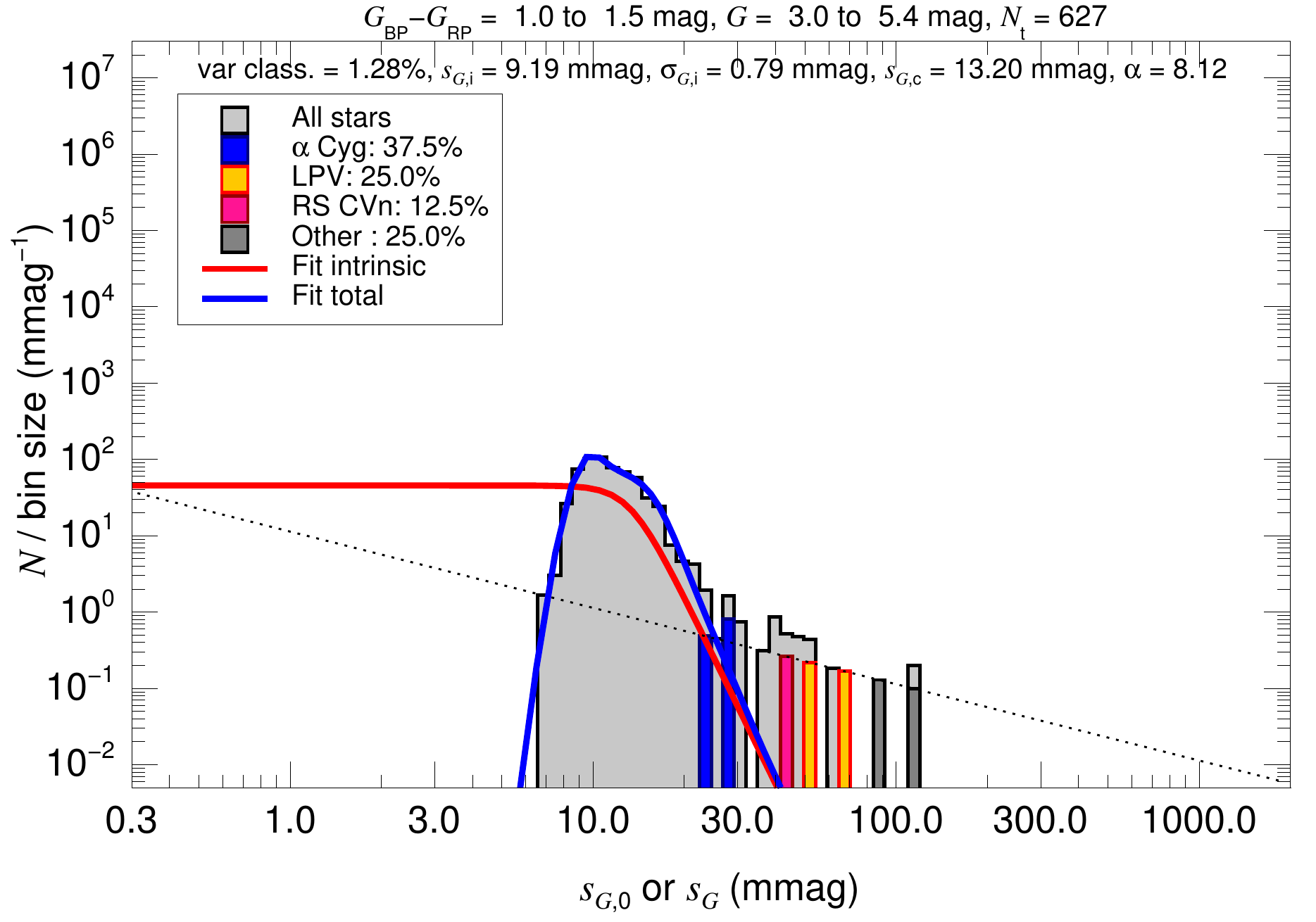}$\!\!\!$
                    \includegraphics[width=0.35\linewidth]{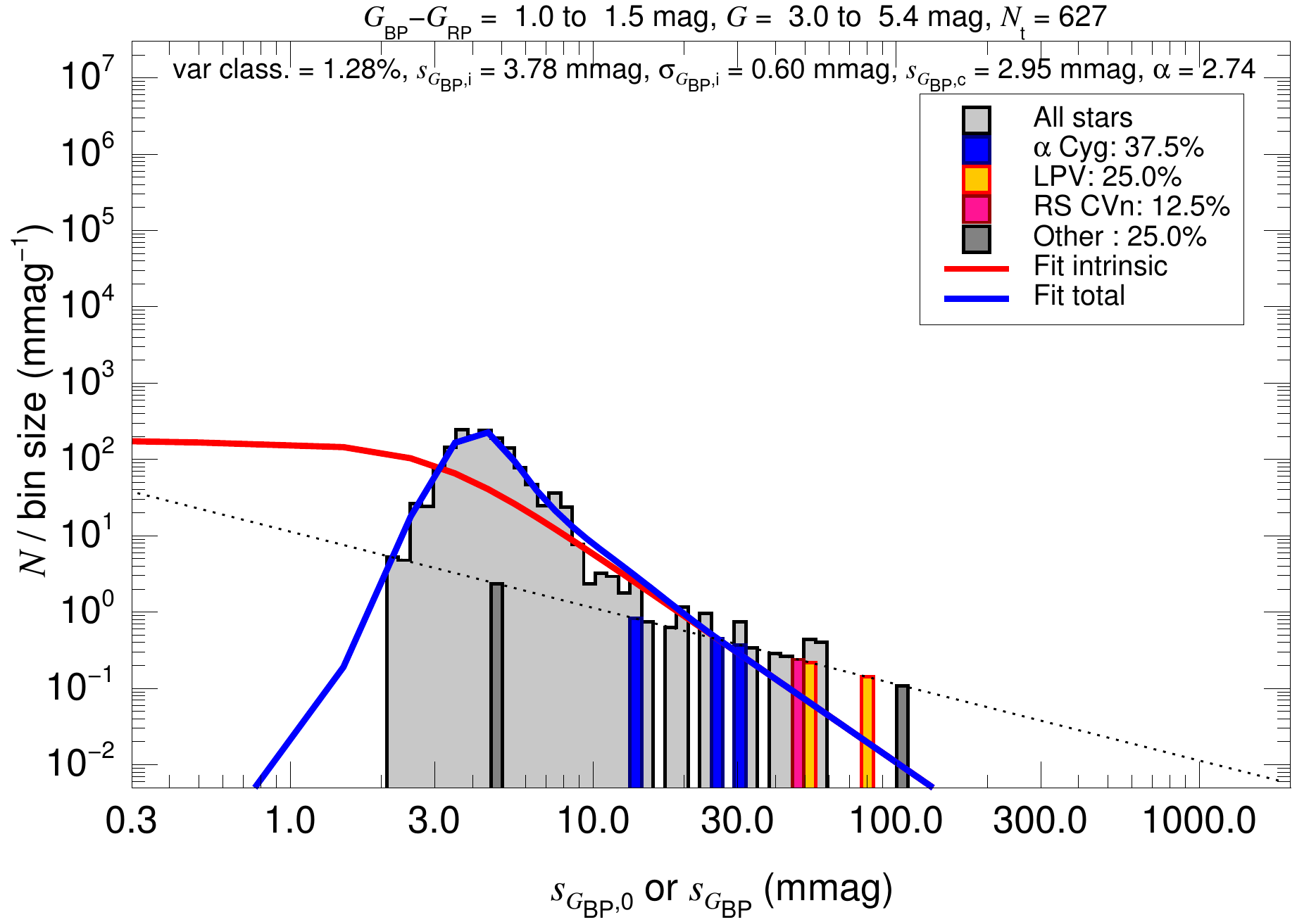}$\!\!\!$
                    \includegraphics[width=0.35\linewidth]{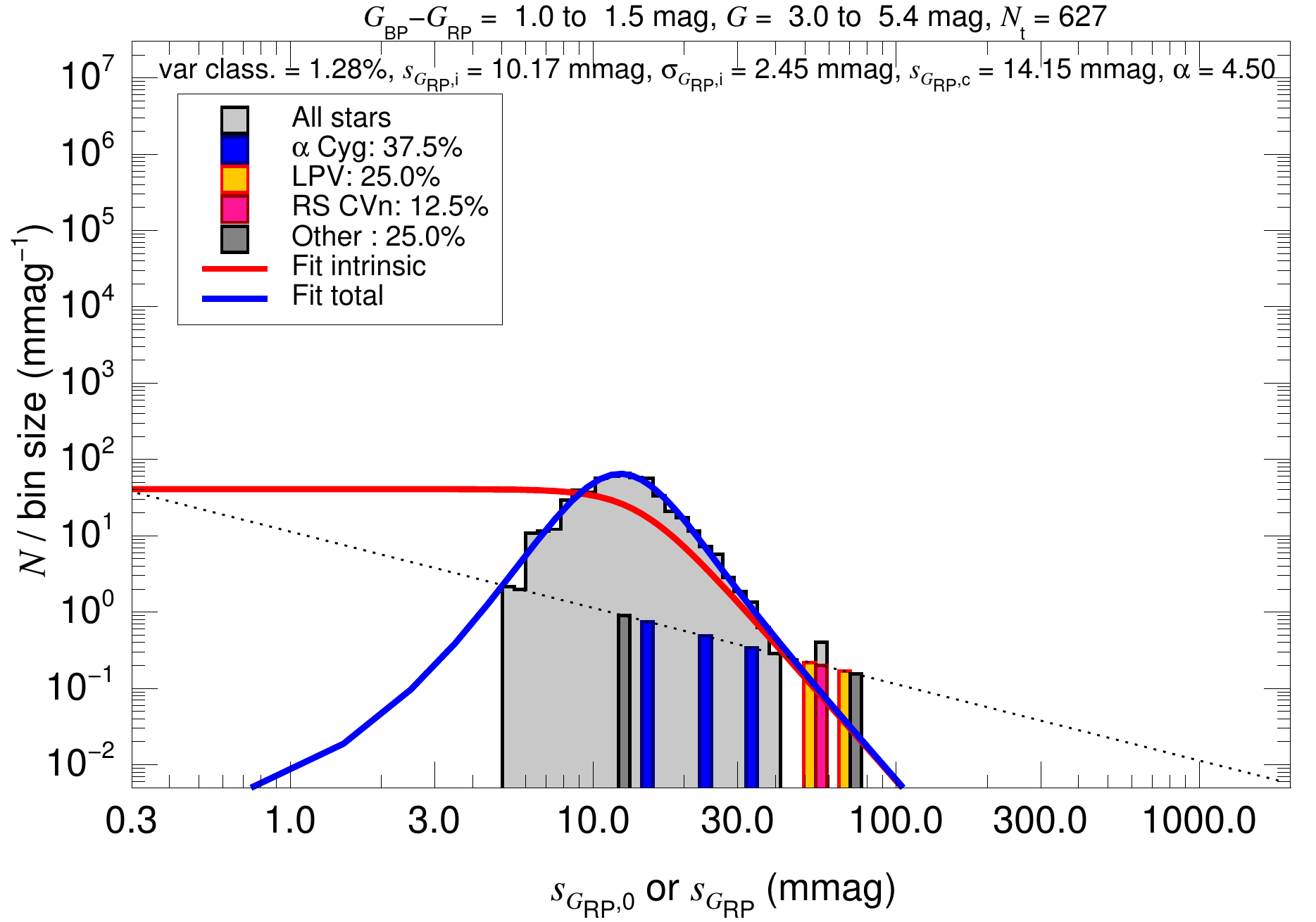}}
\centerline{$\!\!\!$\includegraphics[width=0.35\linewidth]{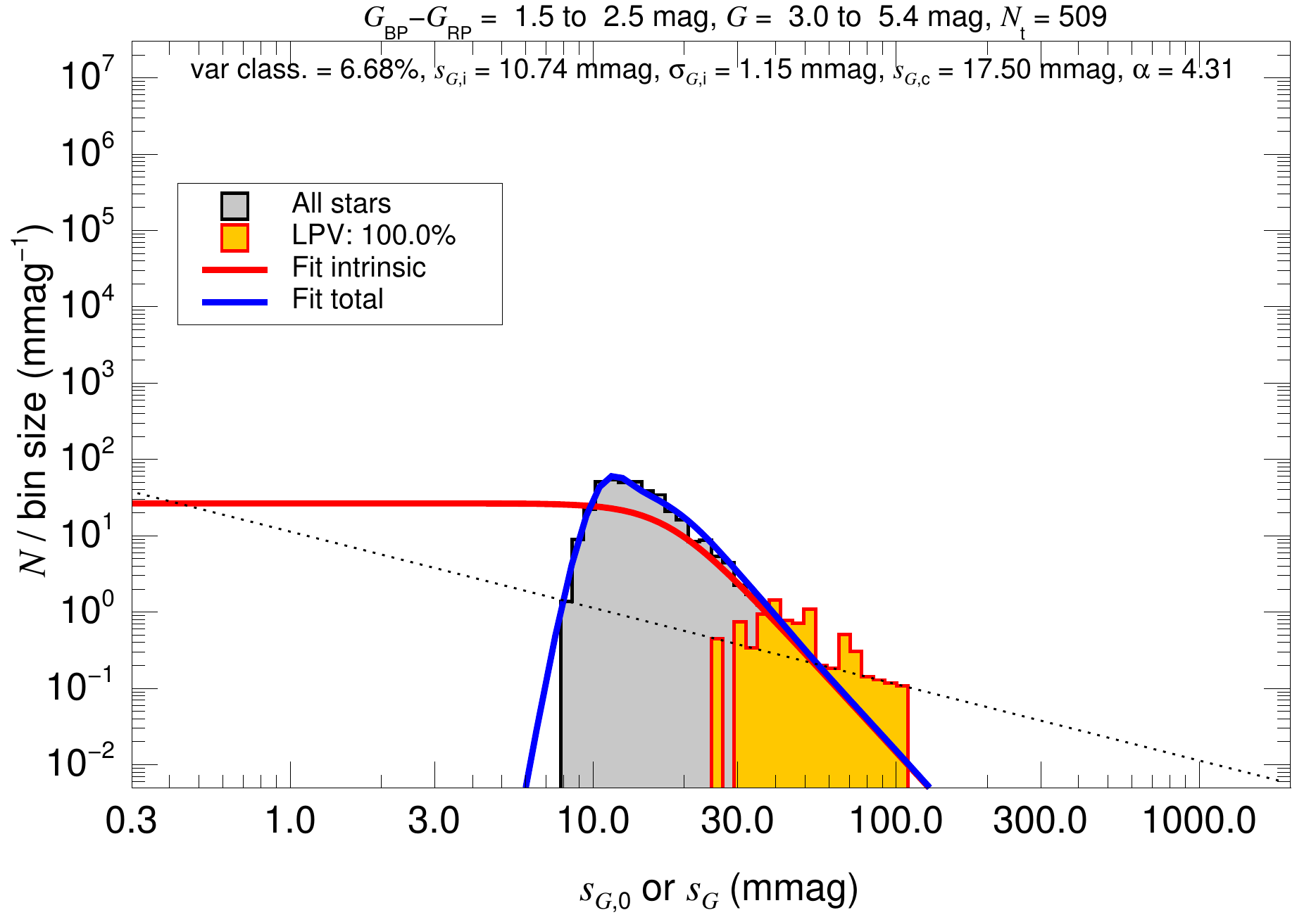}$\!\!\!$
                    \includegraphics[width=0.35\linewidth]{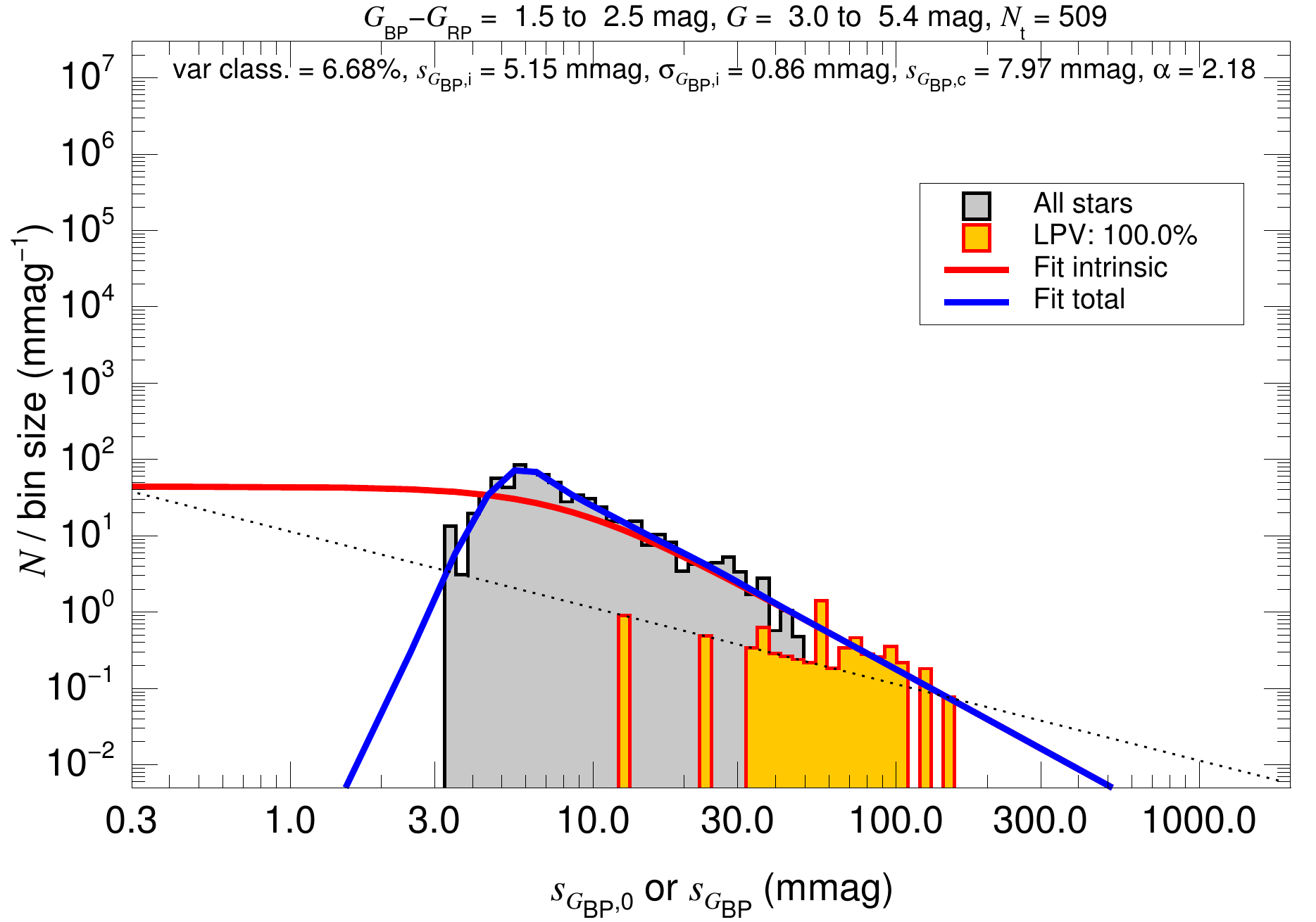}$\!\!\!$
                    \includegraphics[width=0.35\linewidth]{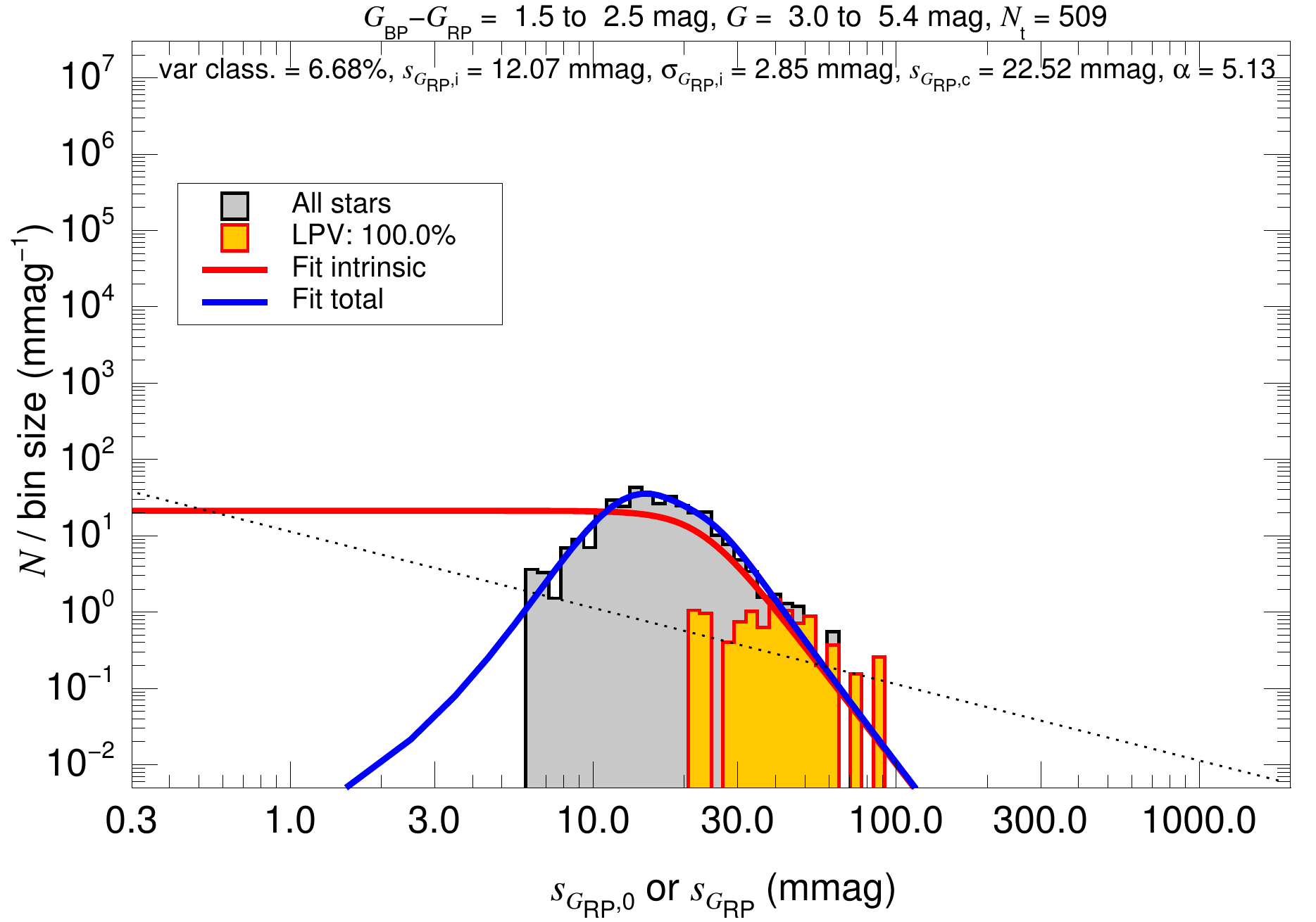}}
\centerline{$\!\!\!$\includegraphics[width=0.35\linewidth]{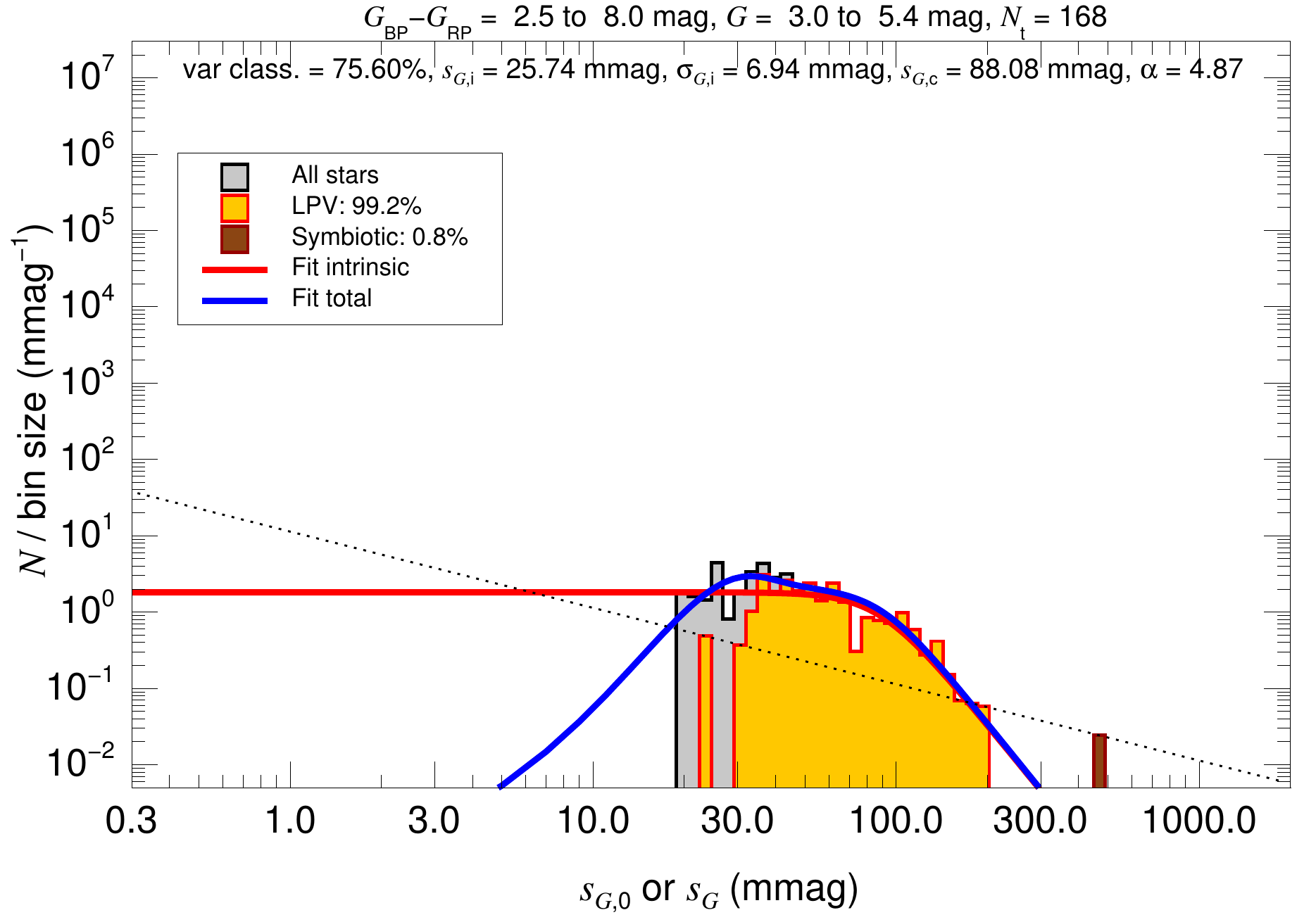}$\!\!\!$
                    \includegraphics[width=0.35\linewidth]{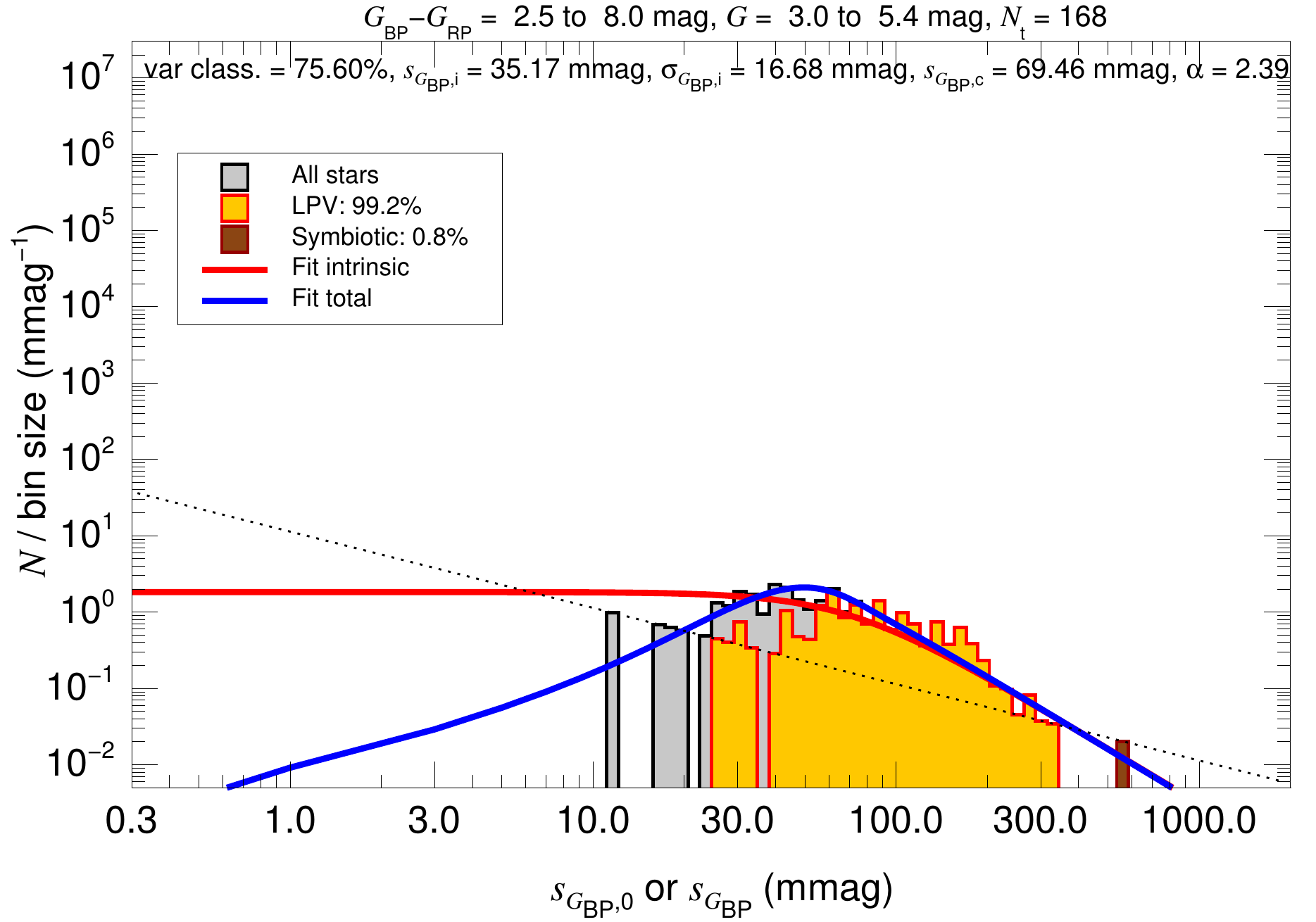}$\!\!\!$
                    \includegraphics[width=0.35\linewidth]{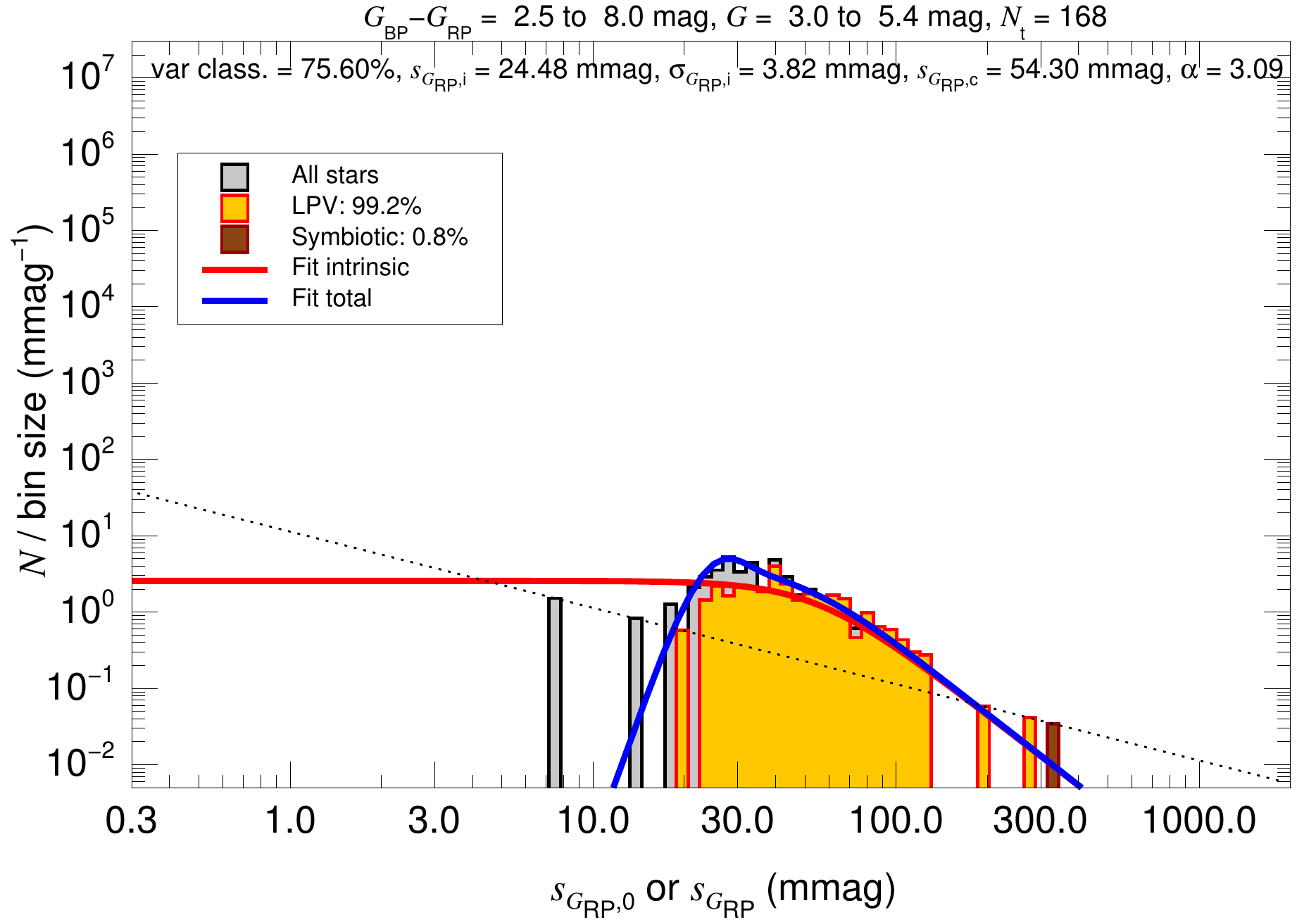}}
\caption{Total dispersion histograms used t1Go calculate the instrumental dispersion as a function of color. Each page shows the 15 
         plots for a given \GG\ magnitude range, starting with 3.0-5.3~mag. The three columns show the histograms for \GG\ (left),
         \GBP\ (center) and \GRP\ (right). The five rows show the \GBPmGRP\ ranges, from top to bottom: $-$1.0-0.2, 0.2-1.0, 1.0-1.5,
         1.5-2.5, and 2.5-8.0 (in mag). The histograms have a uniform bin size in logarithmic units (see the text for the bins used 
         for fitting) and both the horizontal and vertical scales are logarithmic and the same in all plots to allow for better 
         comparisons. The dotted diagonal line indicates the location of one star per bin. The caption for this figure continues on 
         the next page.}
\label{hist_sigma0}
\end{figure*}

\addtocounter{figure}{-1}

\begin{figure*}
\centerline{$\!\!\!$\includegraphics[width=0.35\linewidth]{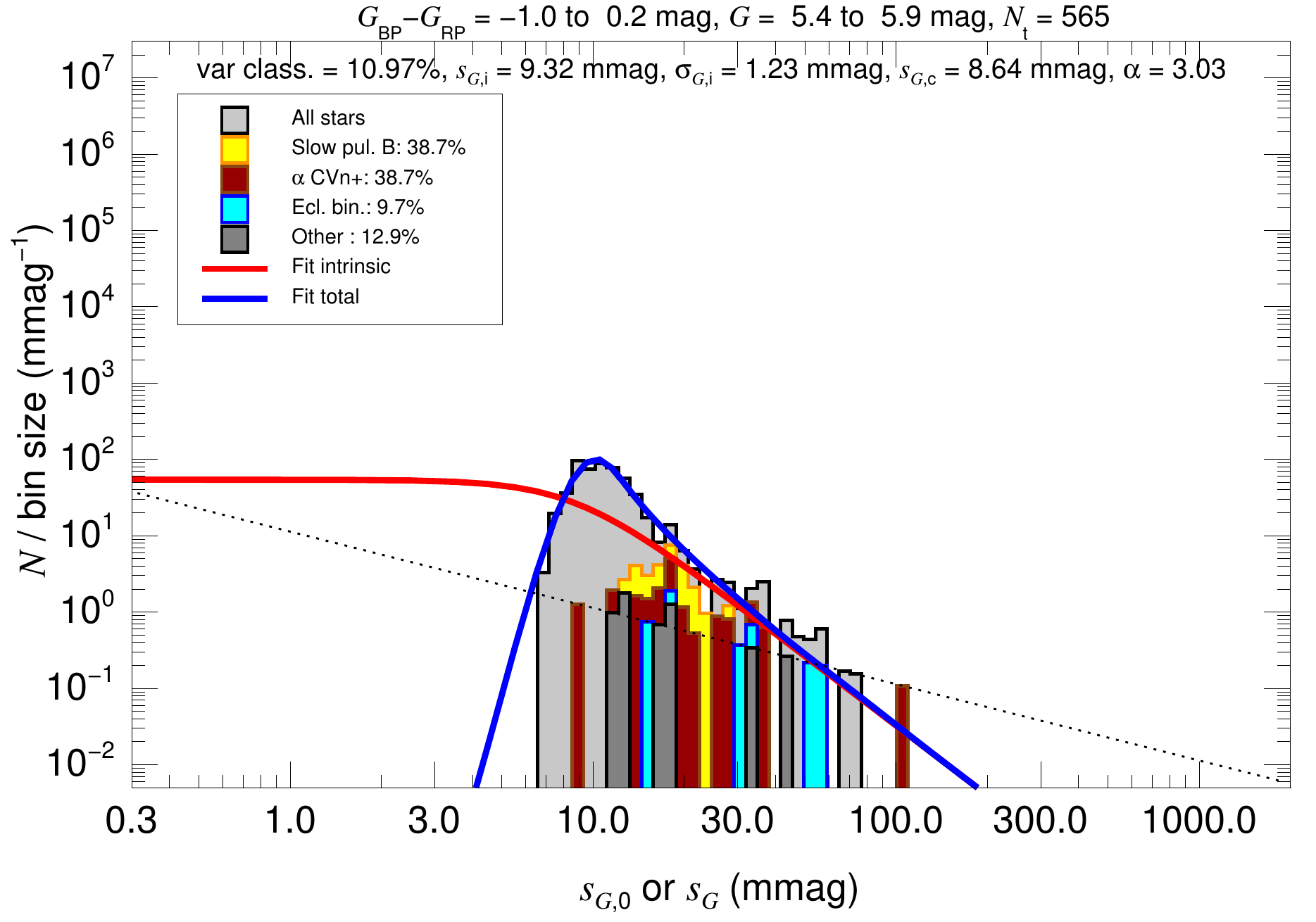}$\!\!\!$
                    \includegraphics[width=0.35\linewidth]{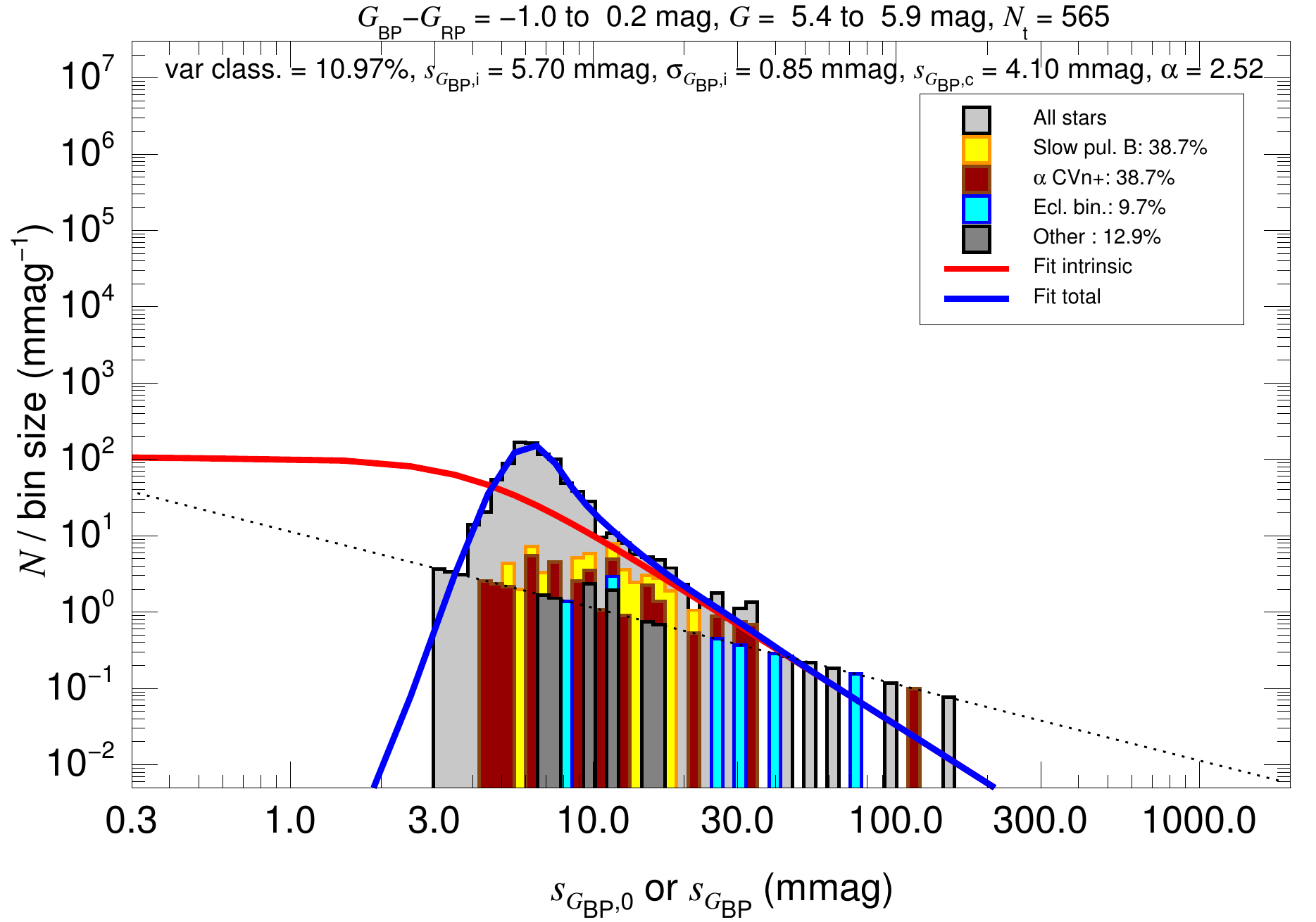}$\!\!\!$
                    \includegraphics[width=0.35\linewidth]{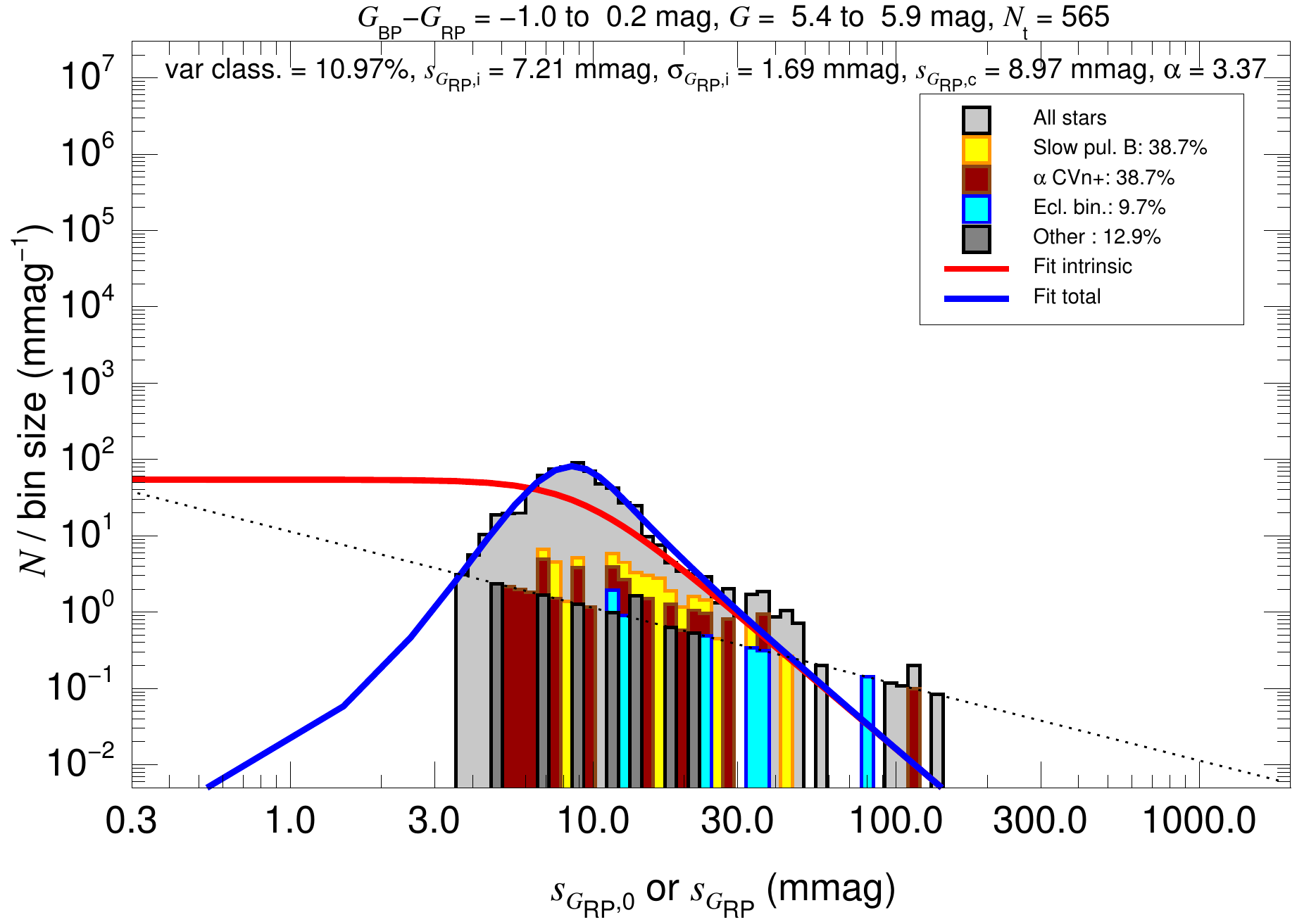}}
\centerline{$\!\!\!$\includegraphics[width=0.35\linewidth]{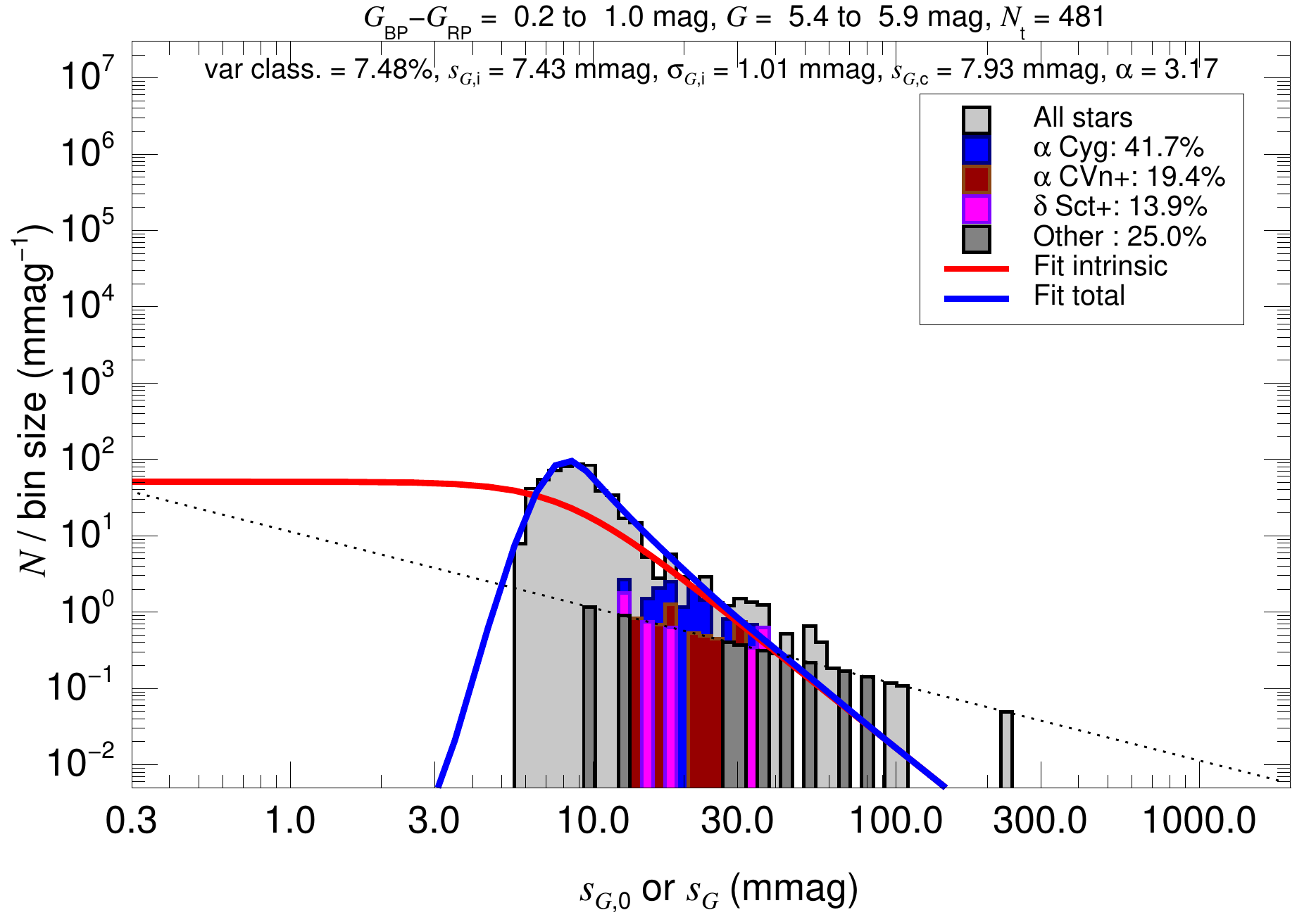}$\!\!\!$
                    \includegraphics[width=0.35\linewidth]{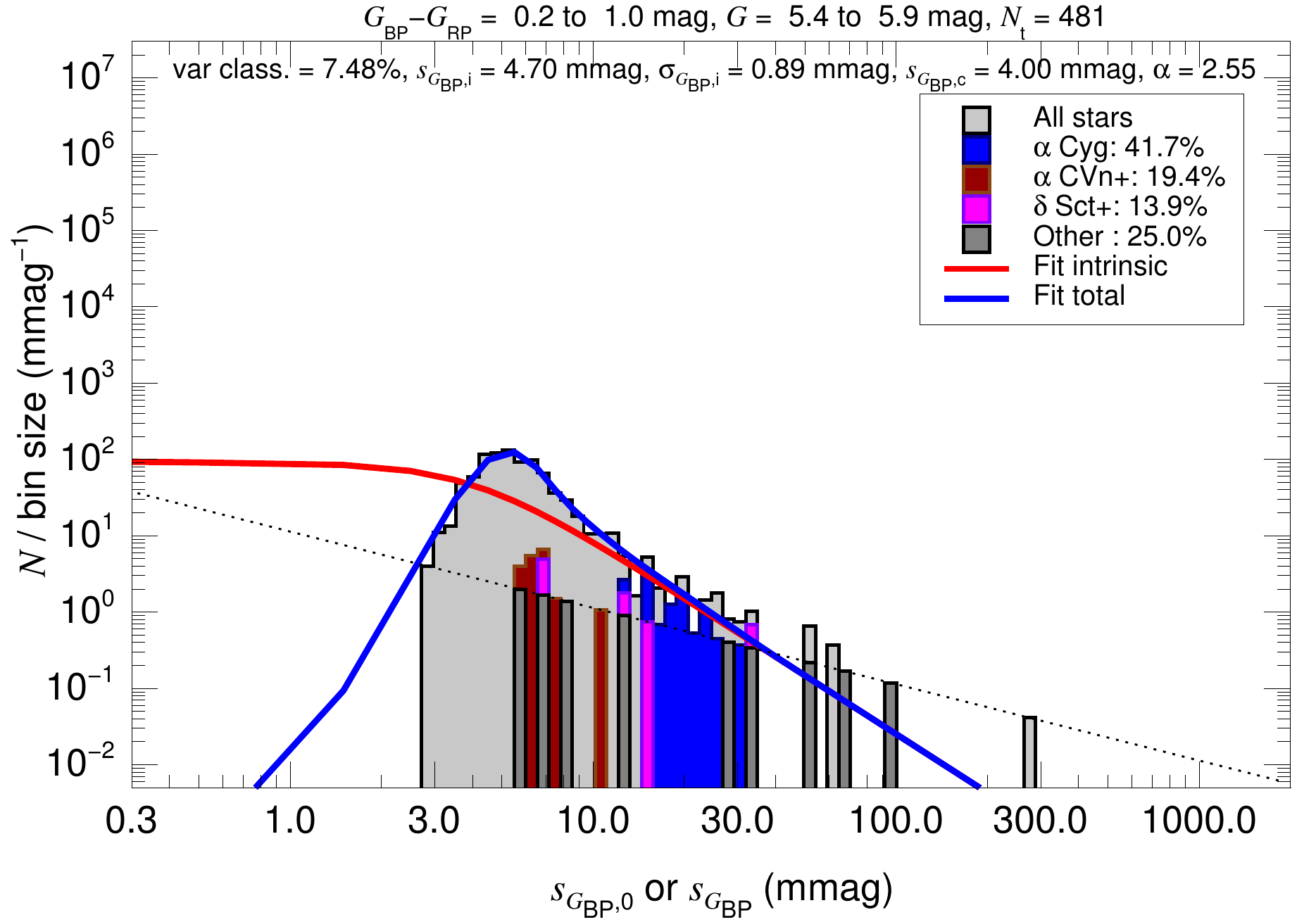}$\!\!\!$
                    \includegraphics[width=0.35\linewidth]{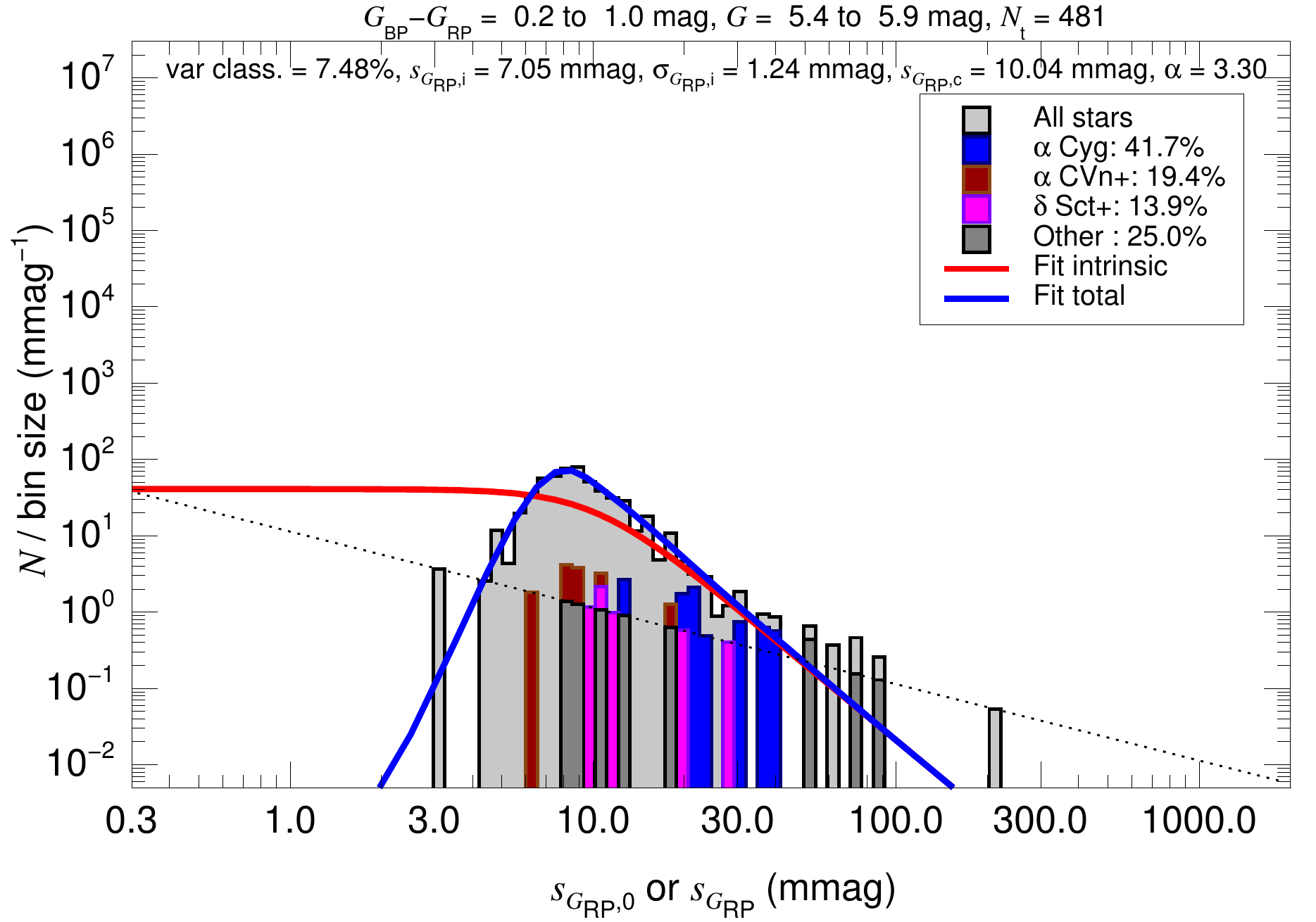}}
\centerline{$\!\!\!$\includegraphics[width=0.35\linewidth]{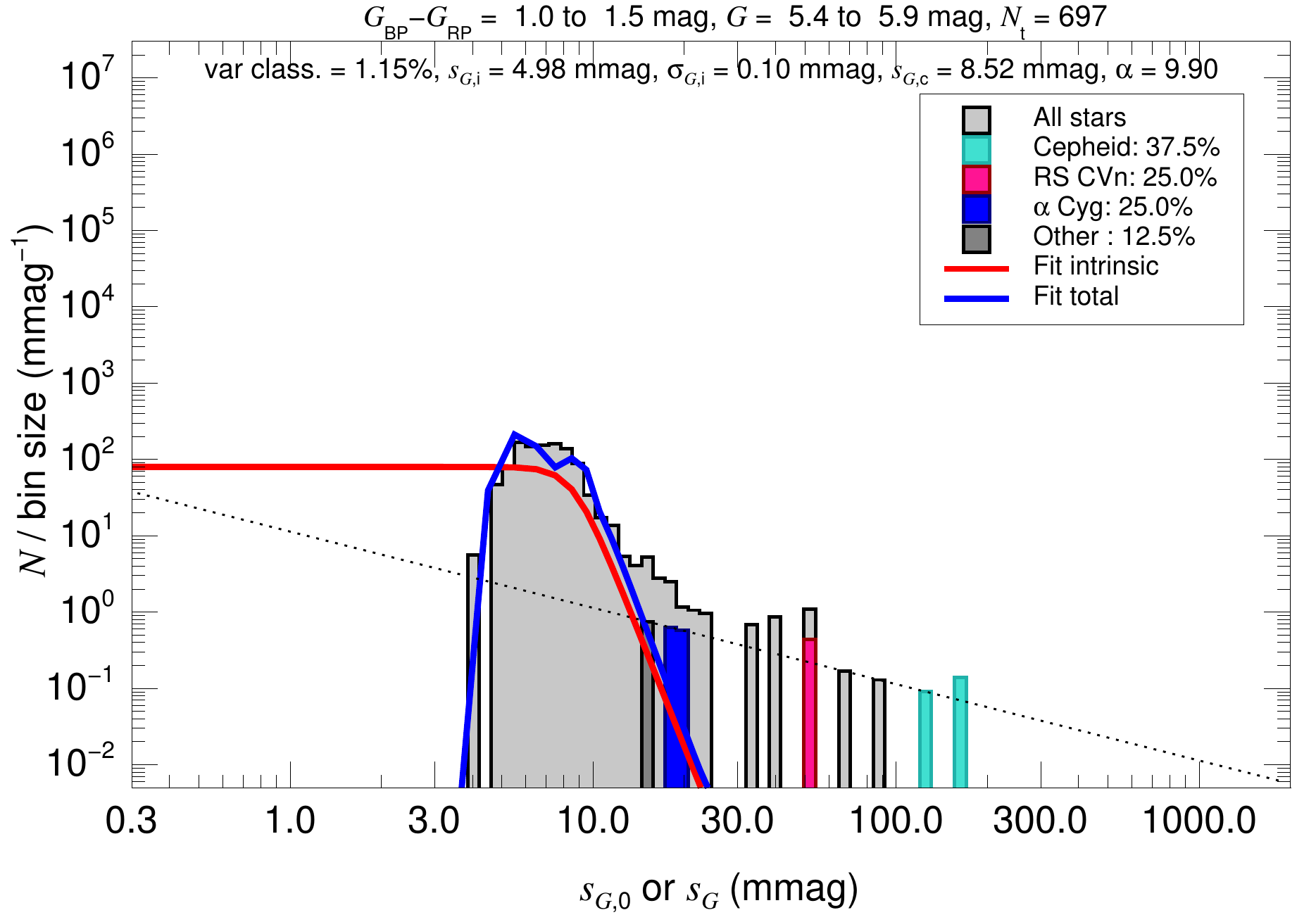}$\!\!\!$
                    \includegraphics[width=0.35\linewidth]{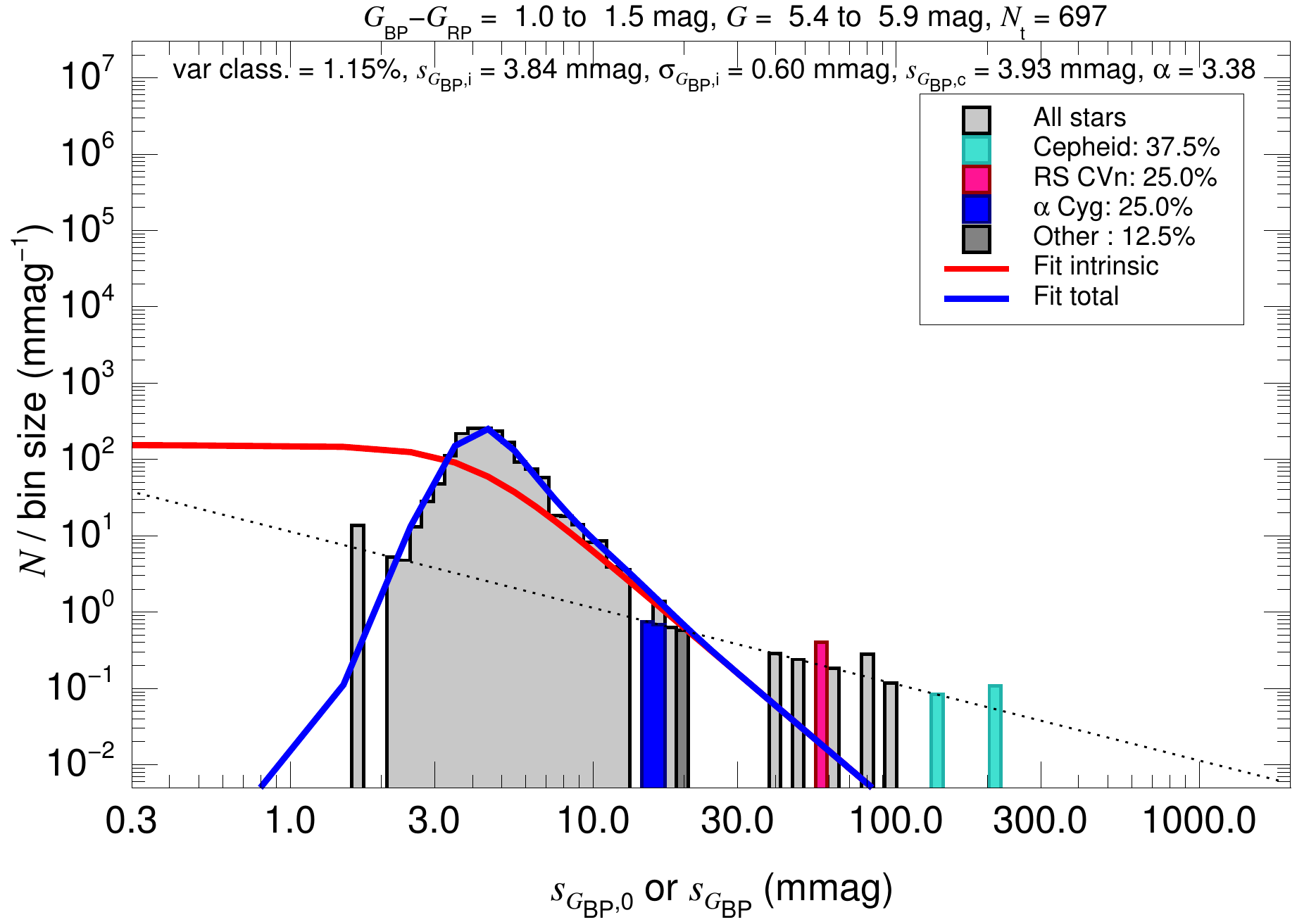}$\!\!\!$
                    \includegraphics[width=0.35\linewidth]{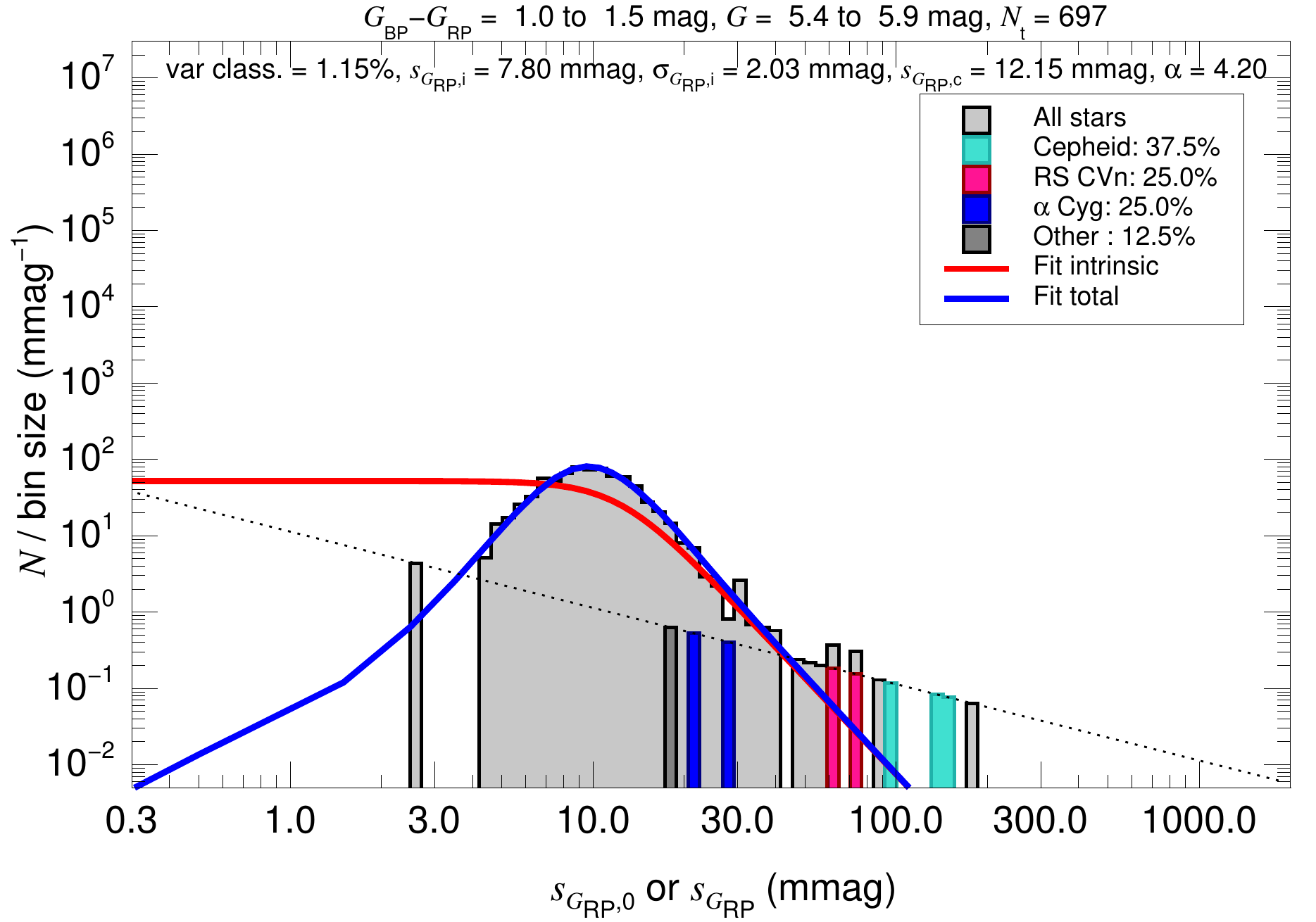}}
\centerline{$\!\!\!$\includegraphics[width=0.35\linewidth]{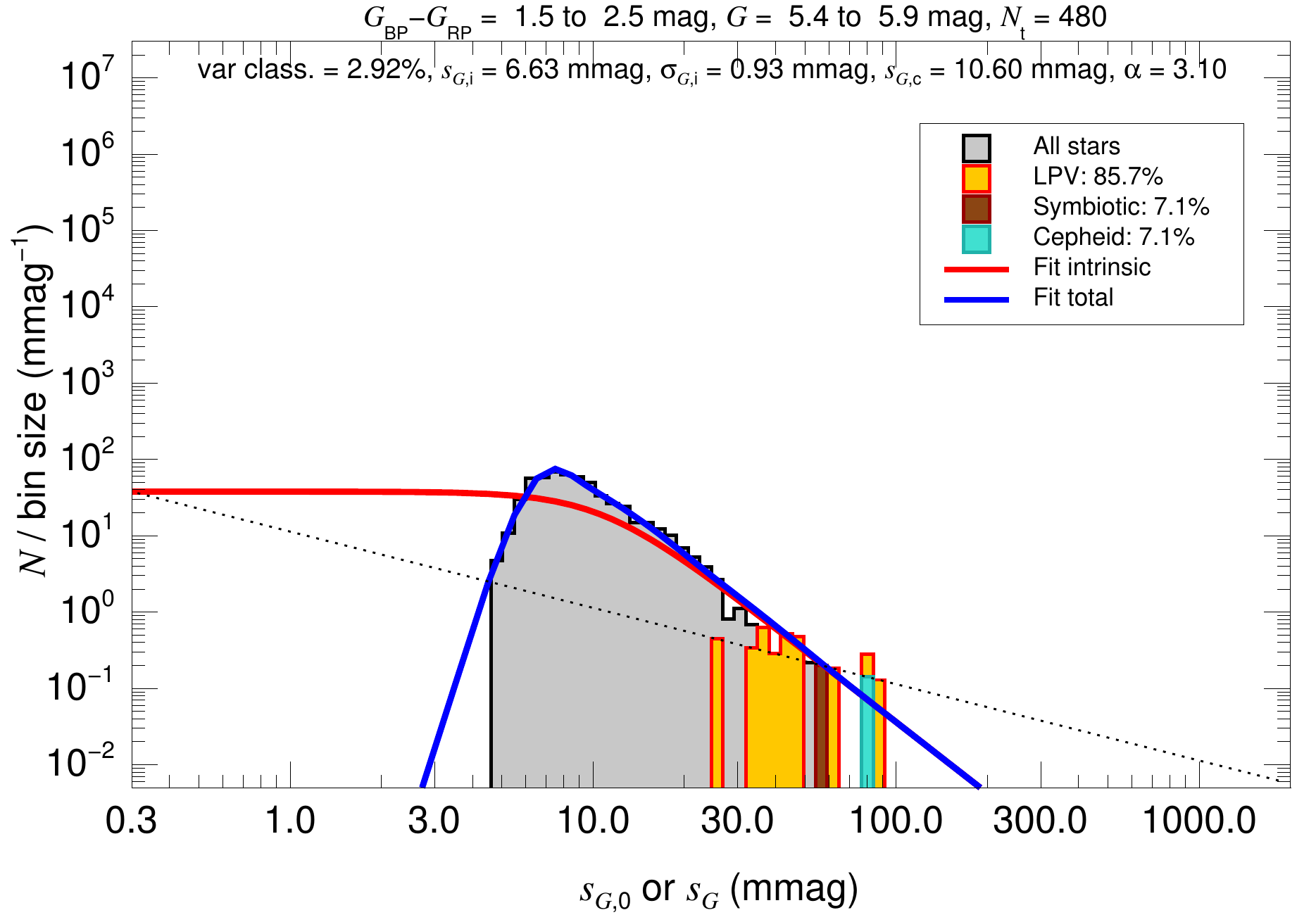}$\!\!\!$
                    \includegraphics[width=0.35\linewidth]{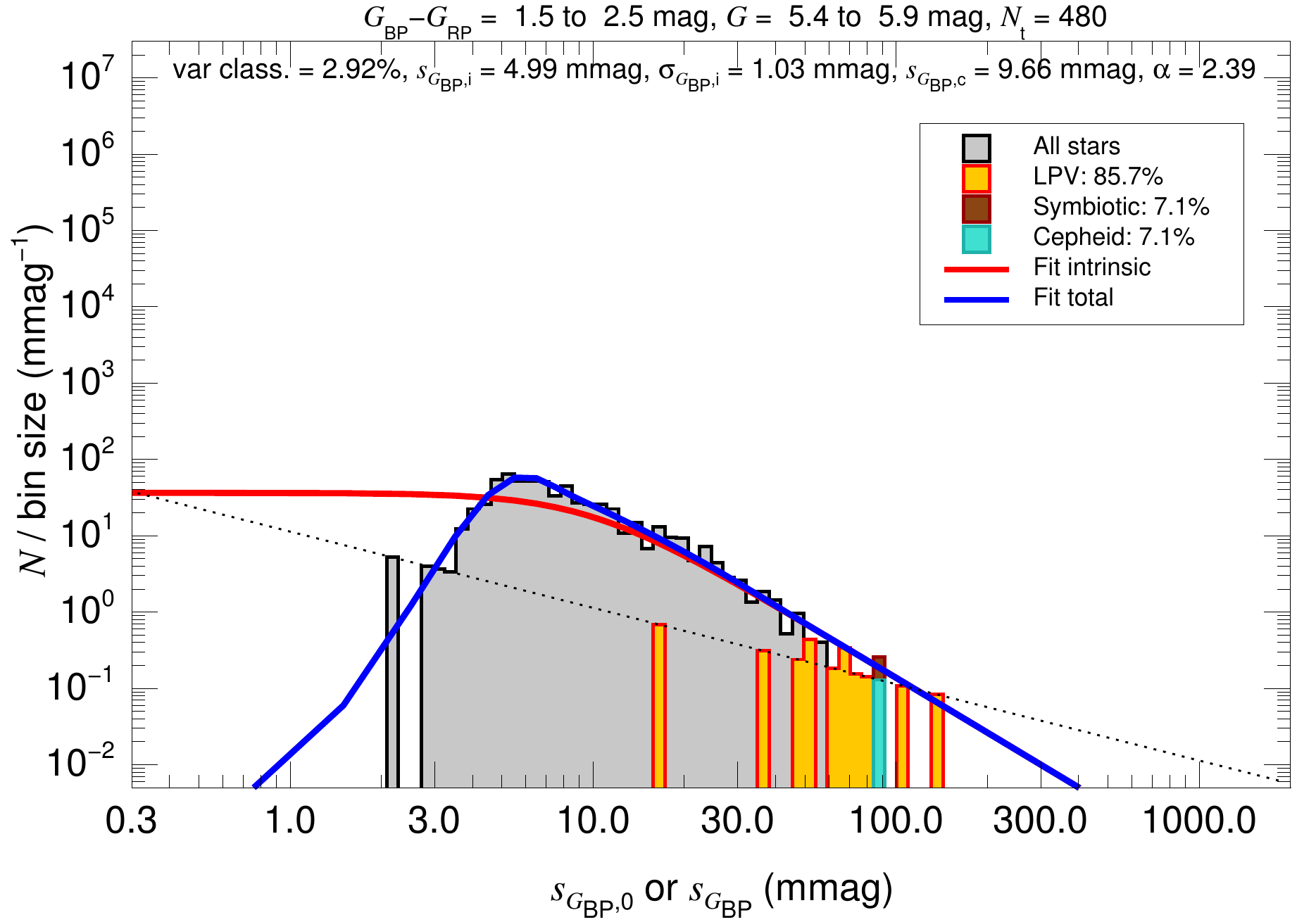}$\!\!\!$
                    \includegraphics[width=0.35\linewidth]{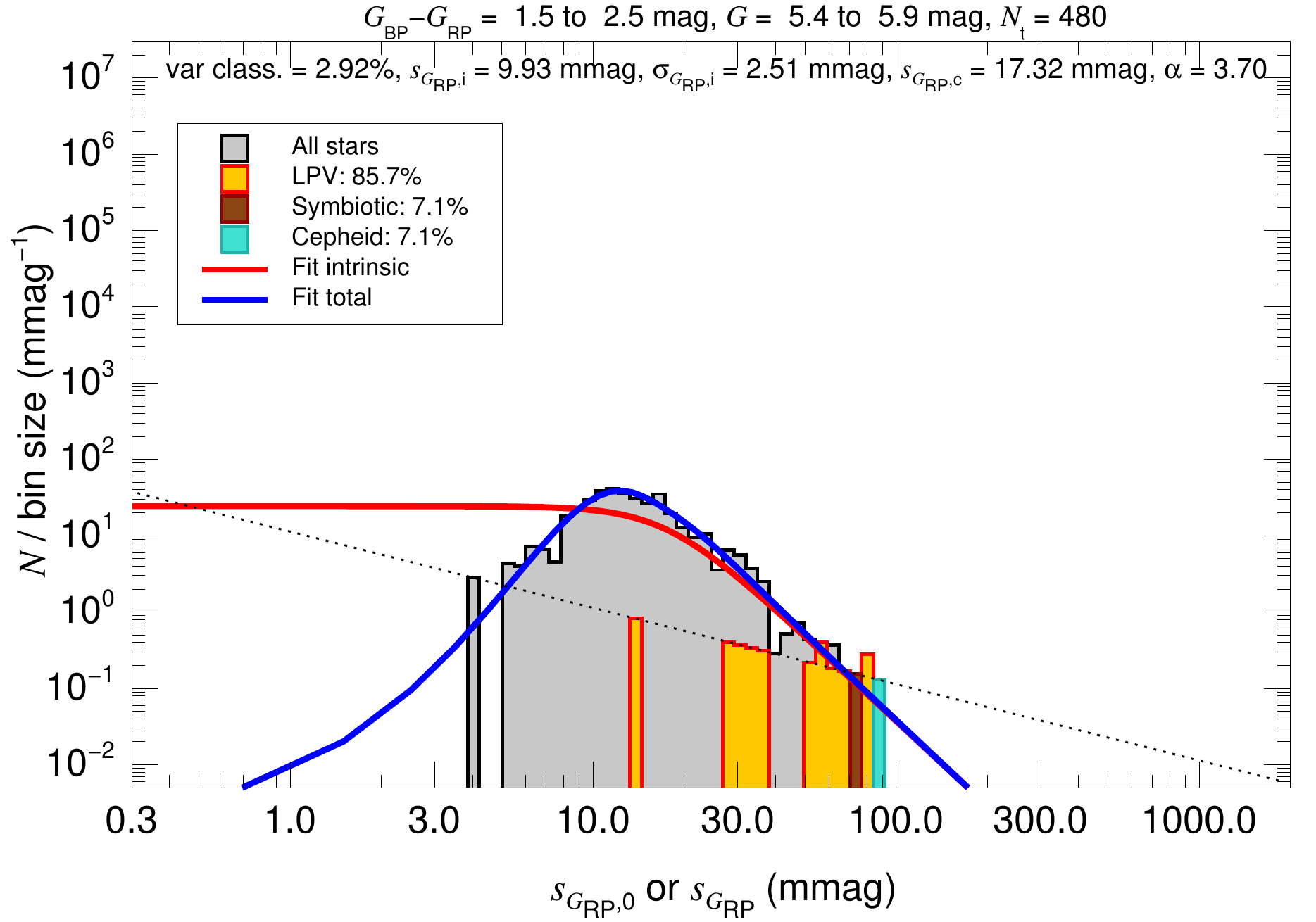}}
\centerline{$\!\!\!$\includegraphics[width=0.35\linewidth]{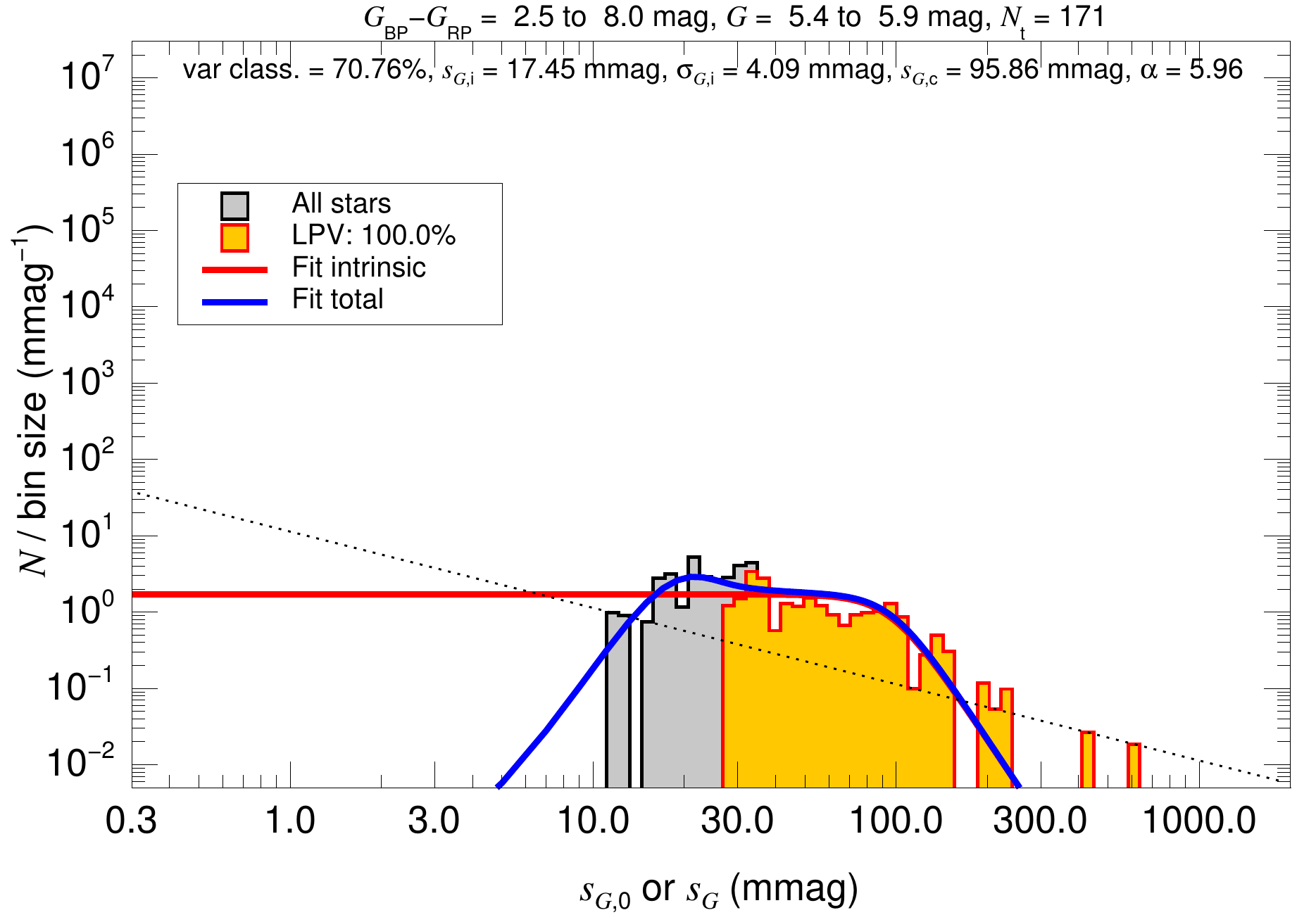}$\!\!\!$
                    \includegraphics[width=0.35\linewidth]{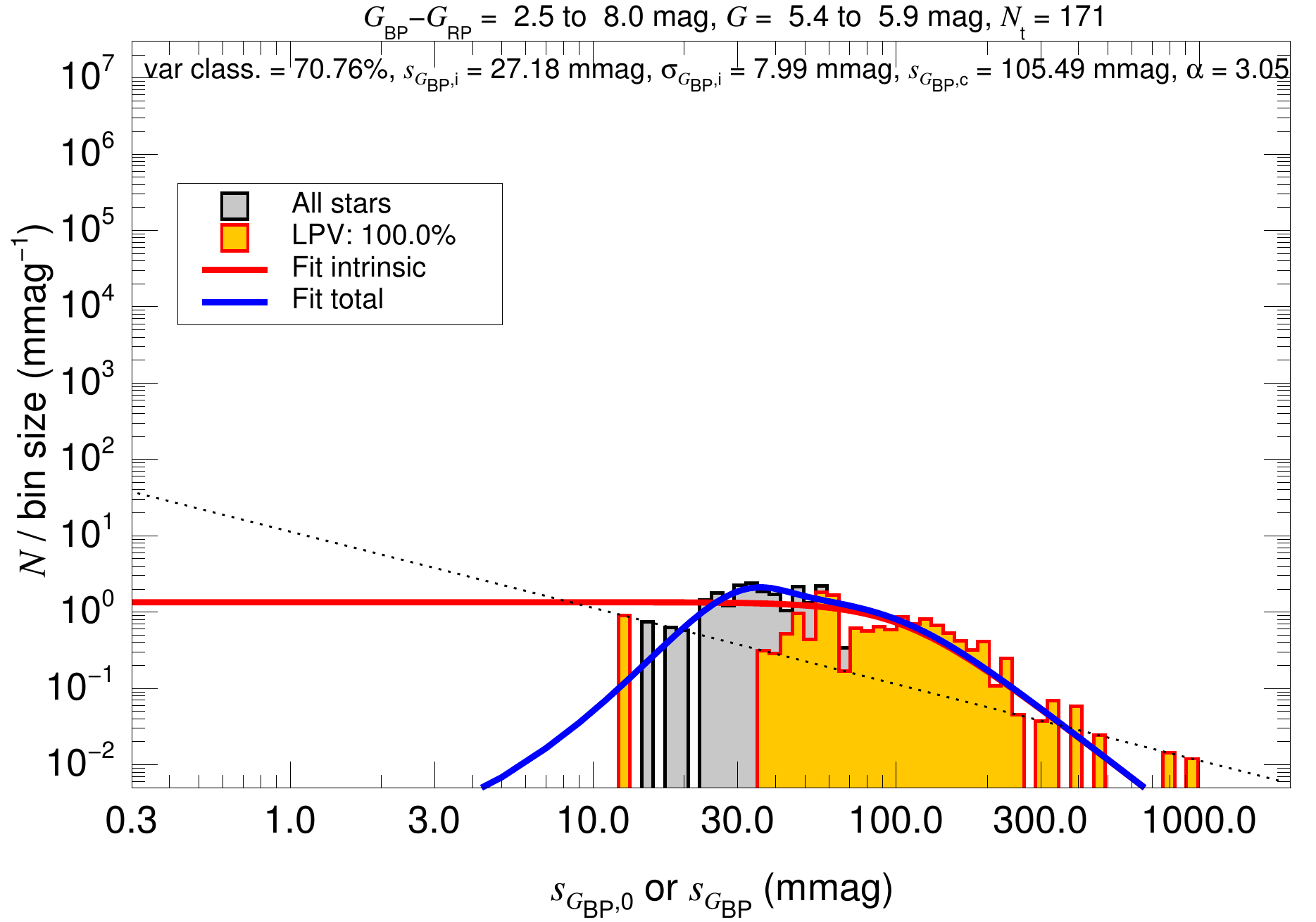}$\!\!\!$
                    \includegraphics[width=0.35\linewidth]{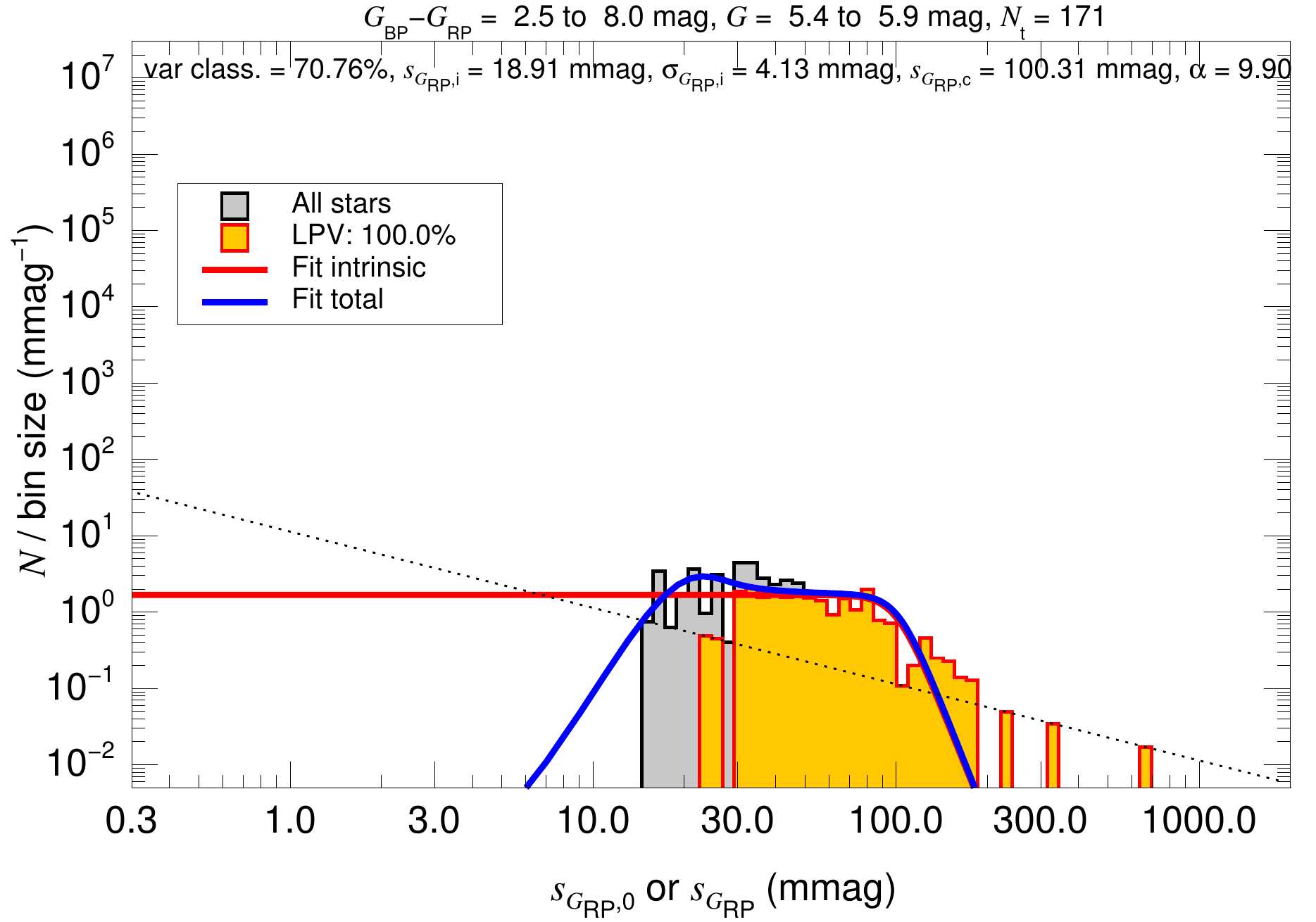}}
\caption{(Continued from previous page). The grey histogram shows the total sample and each panel also shows the histograms for the 
         three most common types of variable stars from the R22 data for that magnitude and color range (each in a different color 
         that is maintained the same throughout the figure for a given variable type) as well as a fourth histogram with the rest of 
         the R22 variables. The four histograms built from R22 data are cumulative, so the top colored line represents all of the 
         variables. The blue line shows the fit for the total dispersion and the red line the corresponding distribution for the 
         astrophysical (intrinsic) dispersion. The text at the top of each panel gives the results of the fits and the percentage of 
         stars with a variability classification from R22..}
\end{figure*}

\addtocounter{figure}{-1}

\begin{figure*}
\centerline{$\!\!\!$\includegraphics[width=0.35\linewidth]{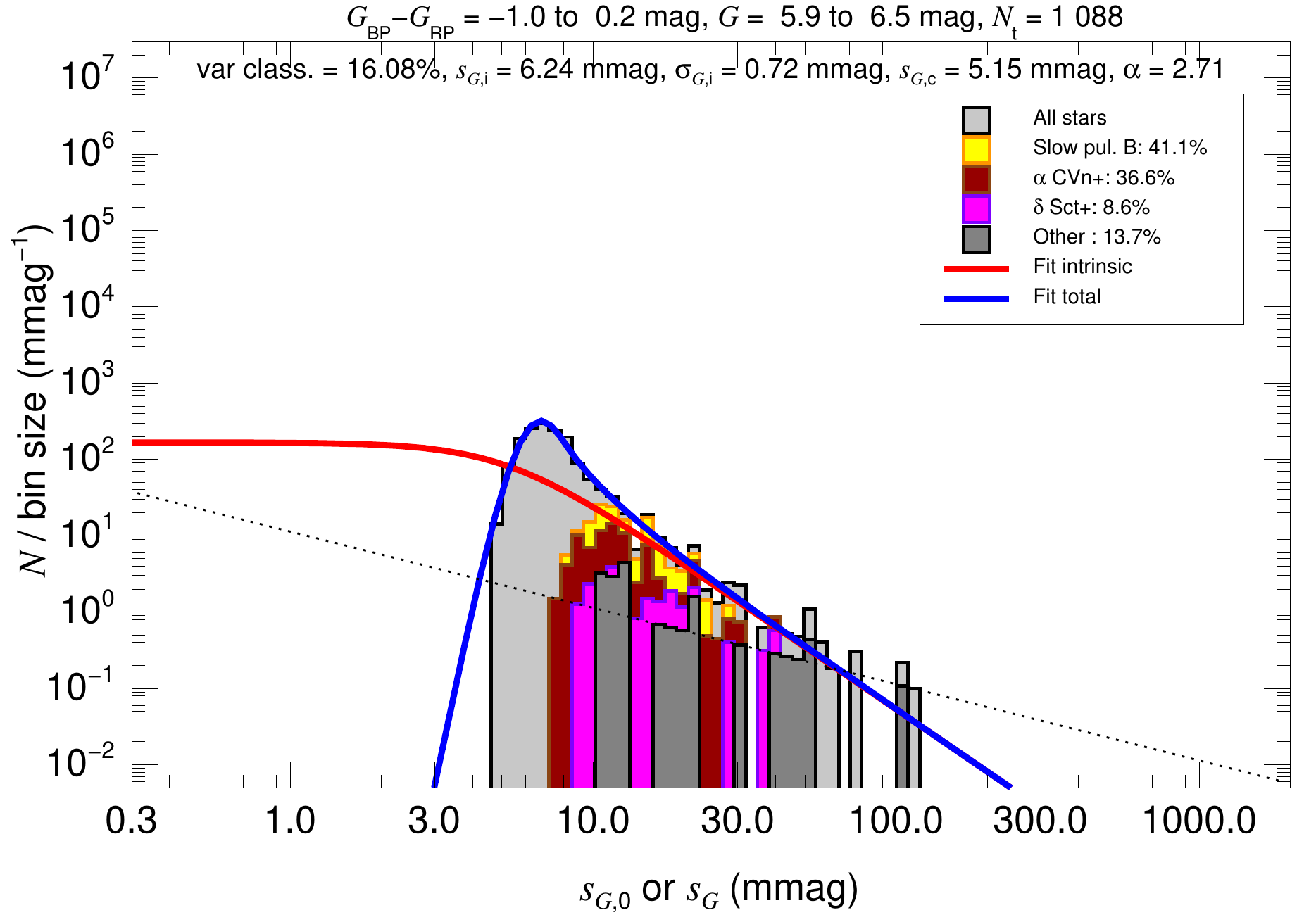}$\!\!\!$
                    \includegraphics[width=0.35\linewidth]{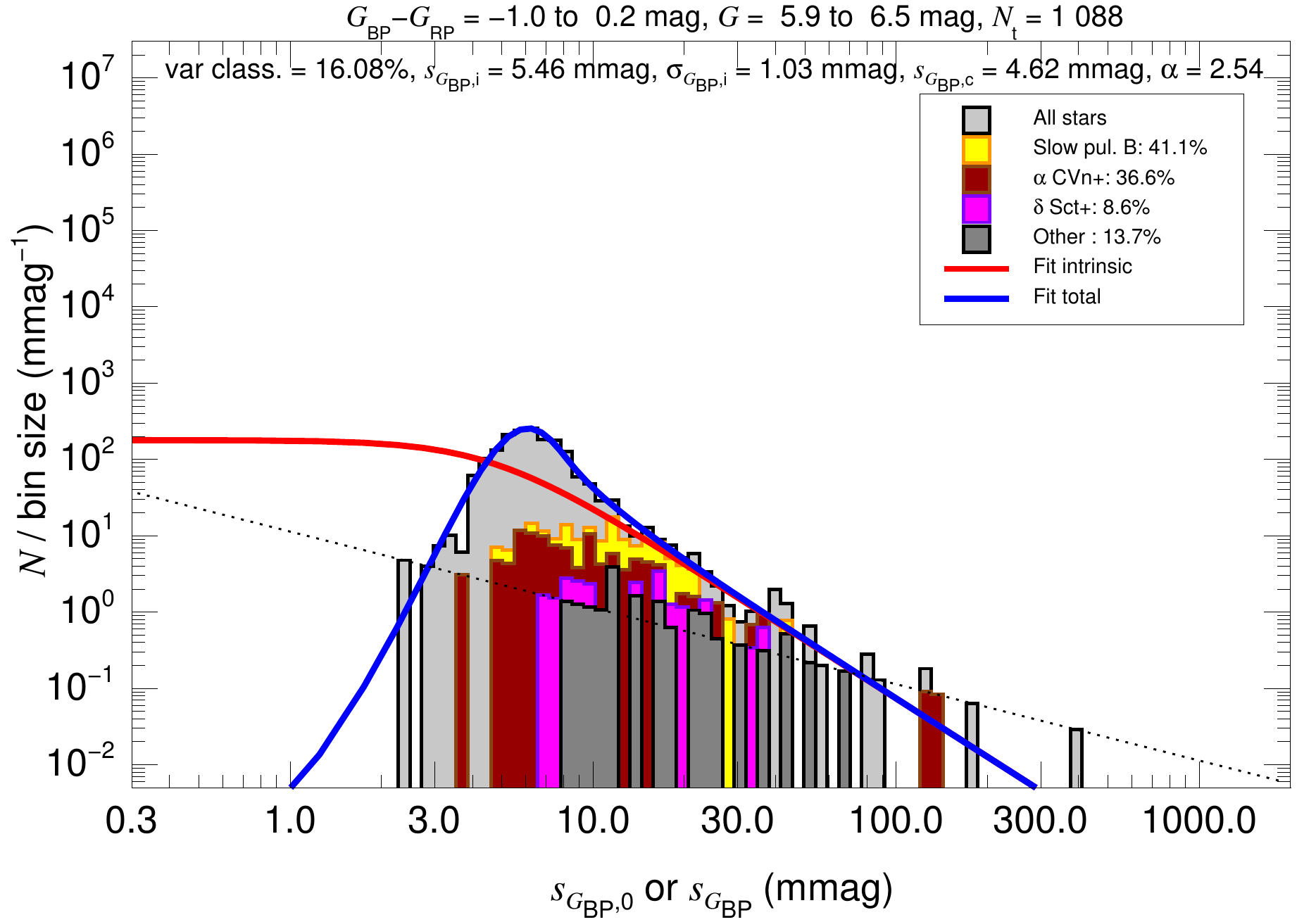}$\!\!\!$
                    \includegraphics[width=0.35\linewidth]{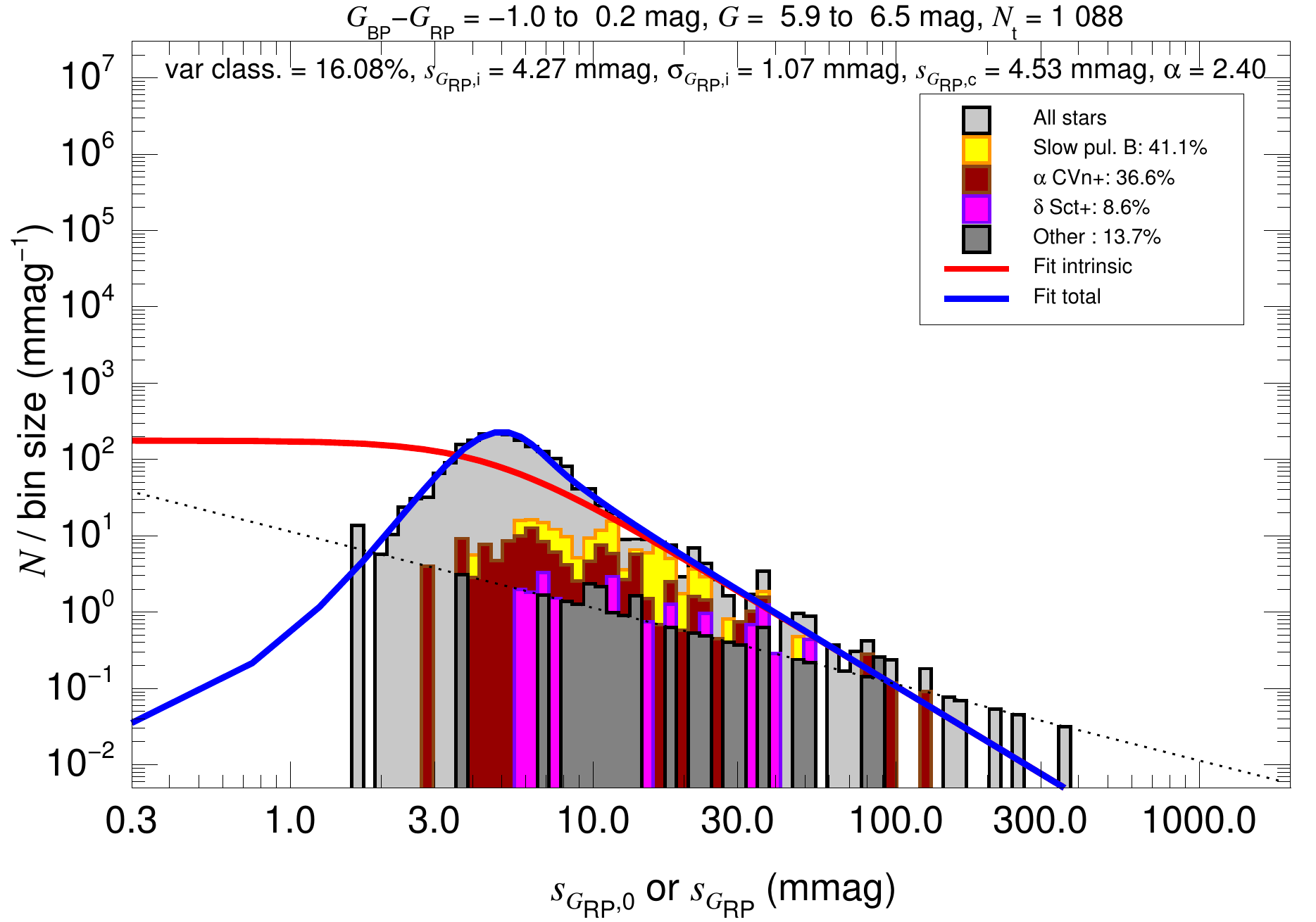}}
\centerline{$\!\!\!$\includegraphics[width=0.35\linewidth]{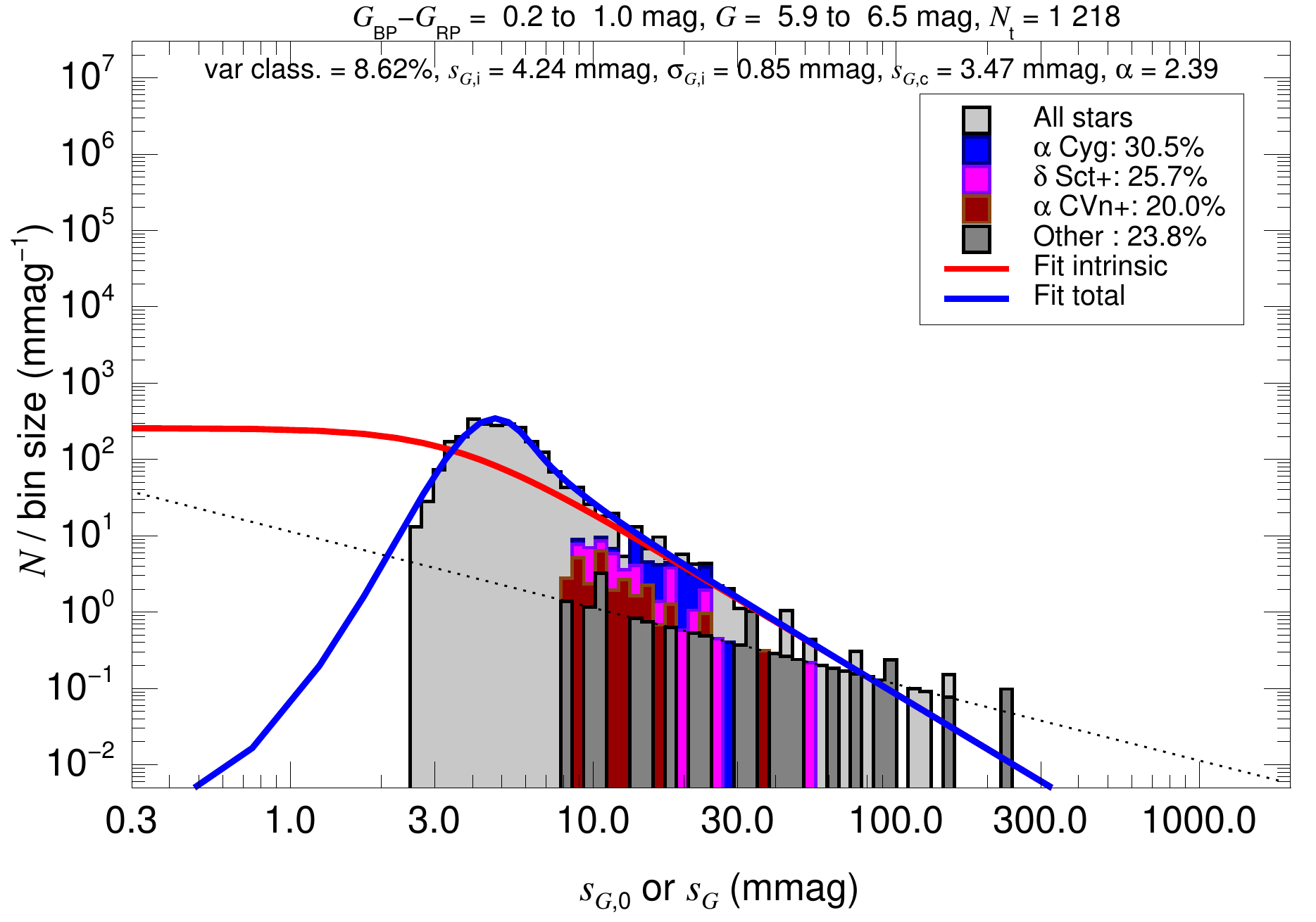}$\!\!\!$
                    \includegraphics[width=0.35\linewidth]{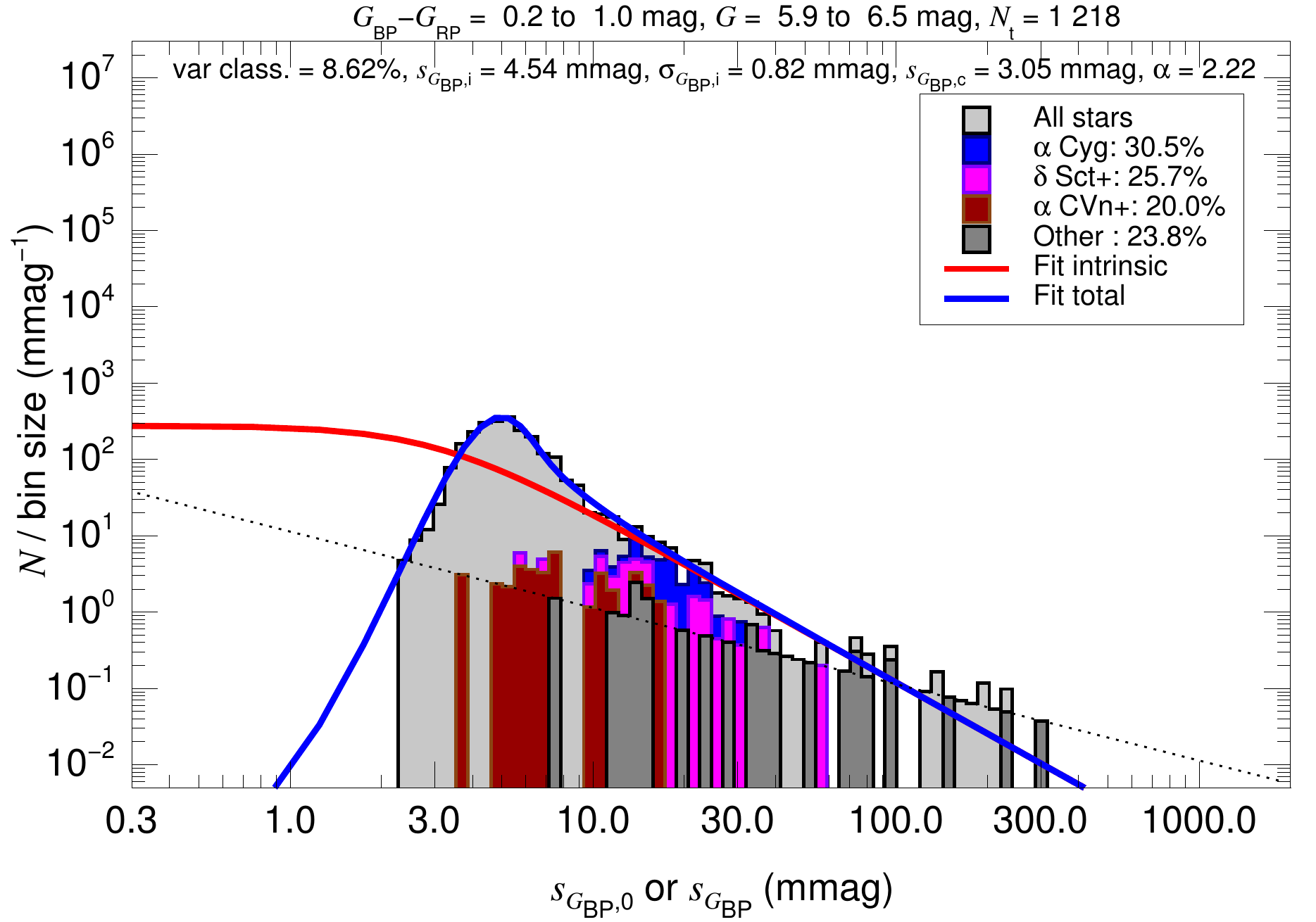}$\!\!\!$
                    \includegraphics[width=0.35\linewidth]{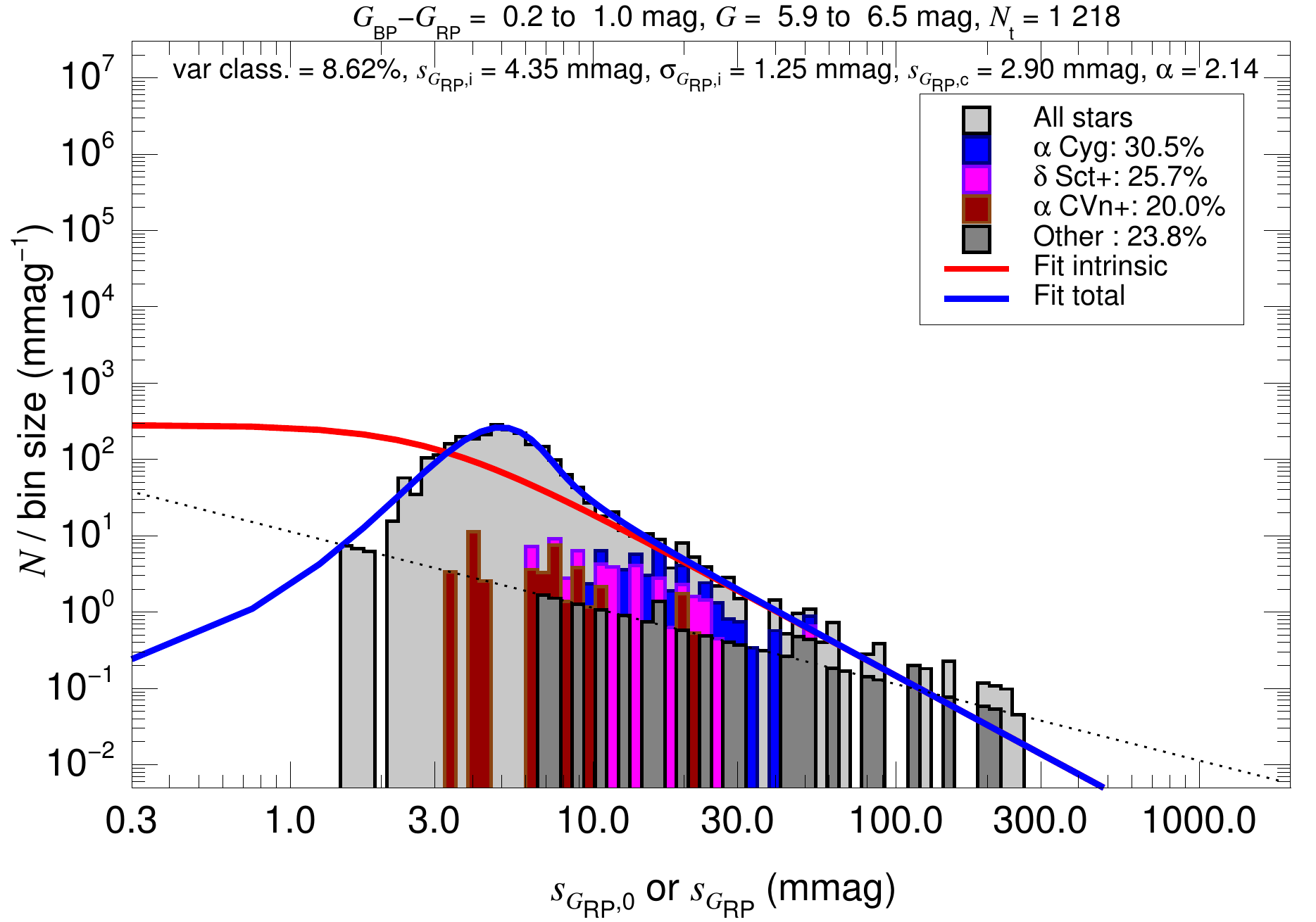}}
\centerline{$\!\!\!$\includegraphics[width=0.35\linewidth]{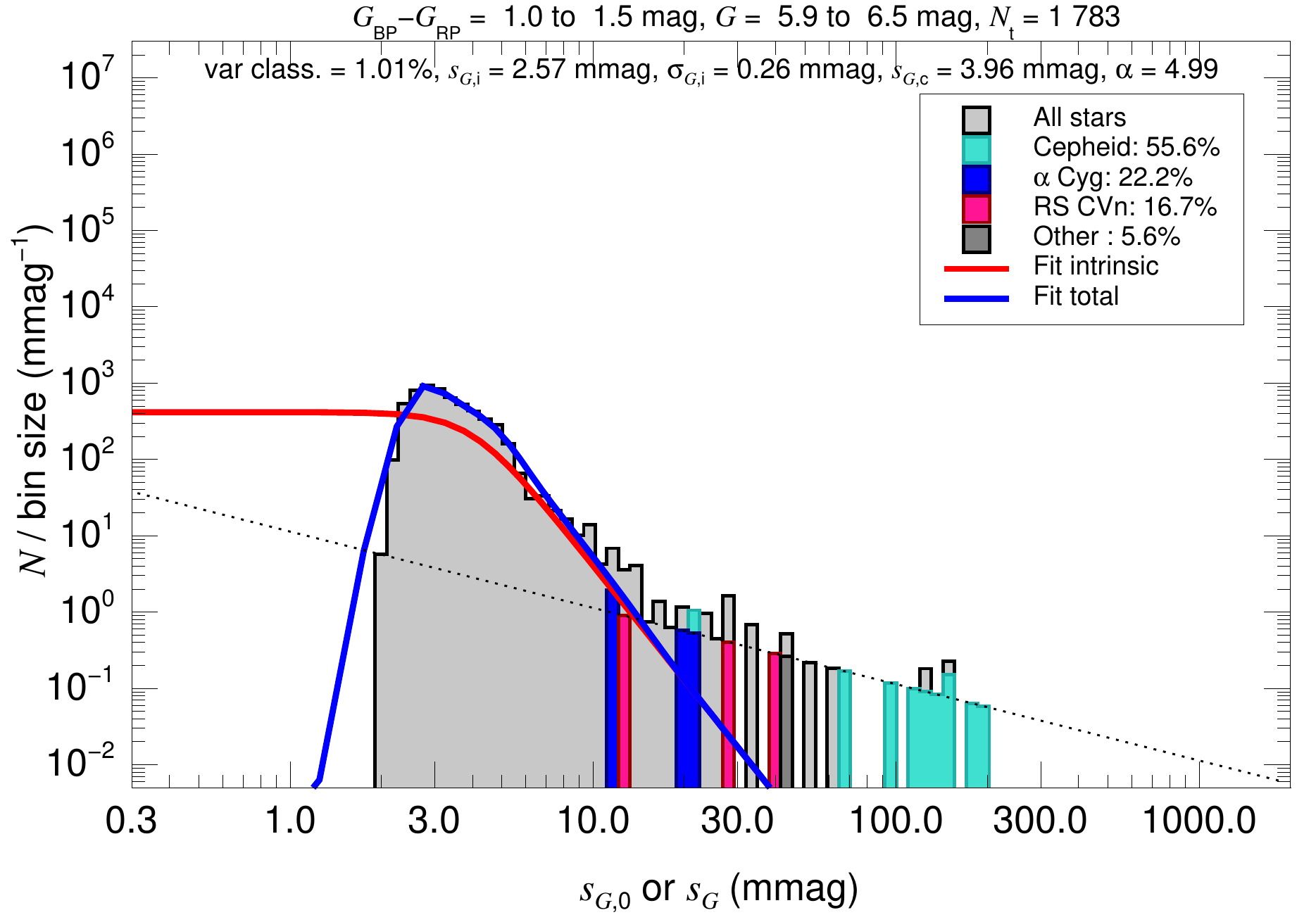}$\!\!\!$
                    \includegraphics[width=0.35\linewidth]{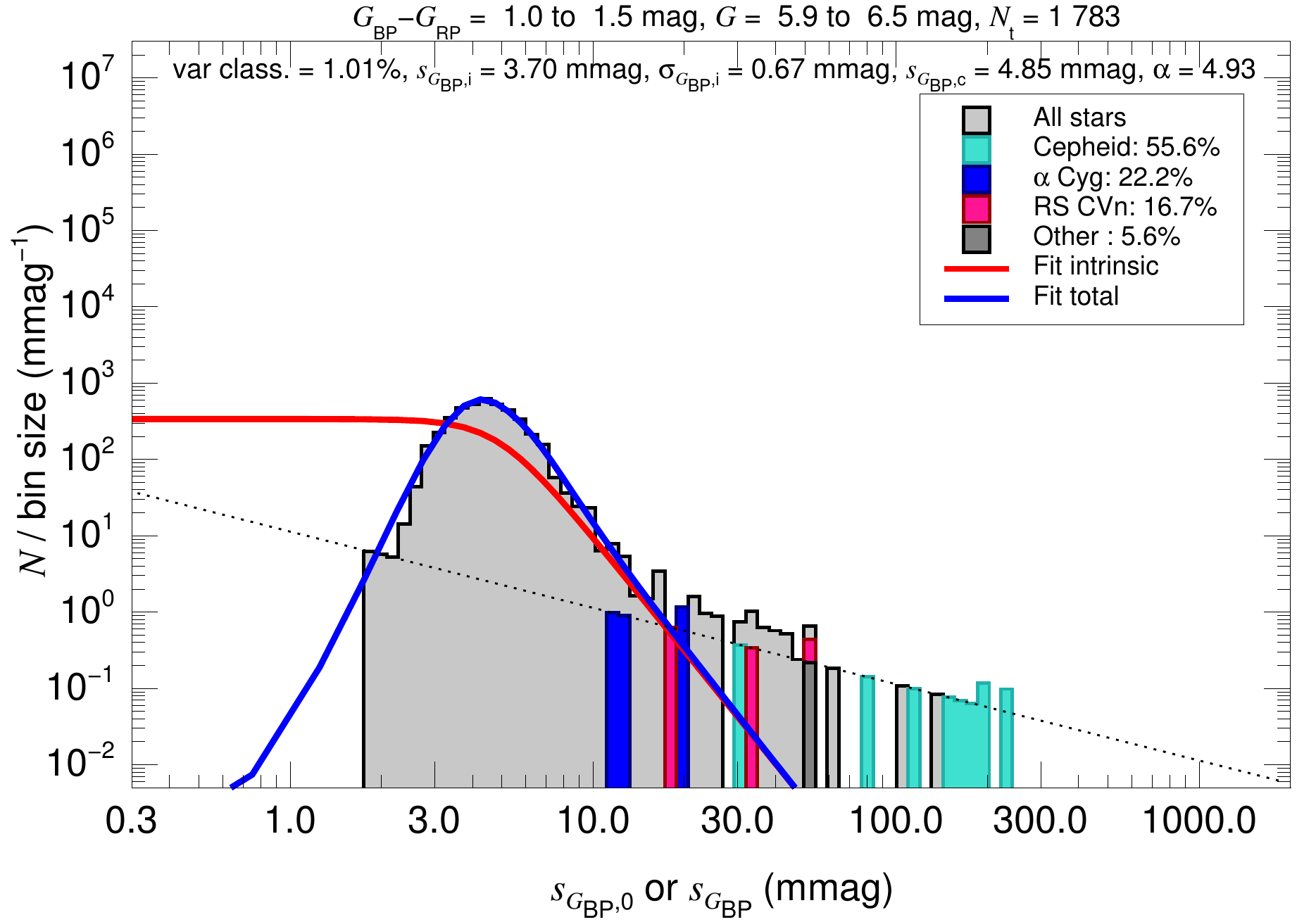}$\!\!\!$
                    \includegraphics[width=0.35\linewidth]{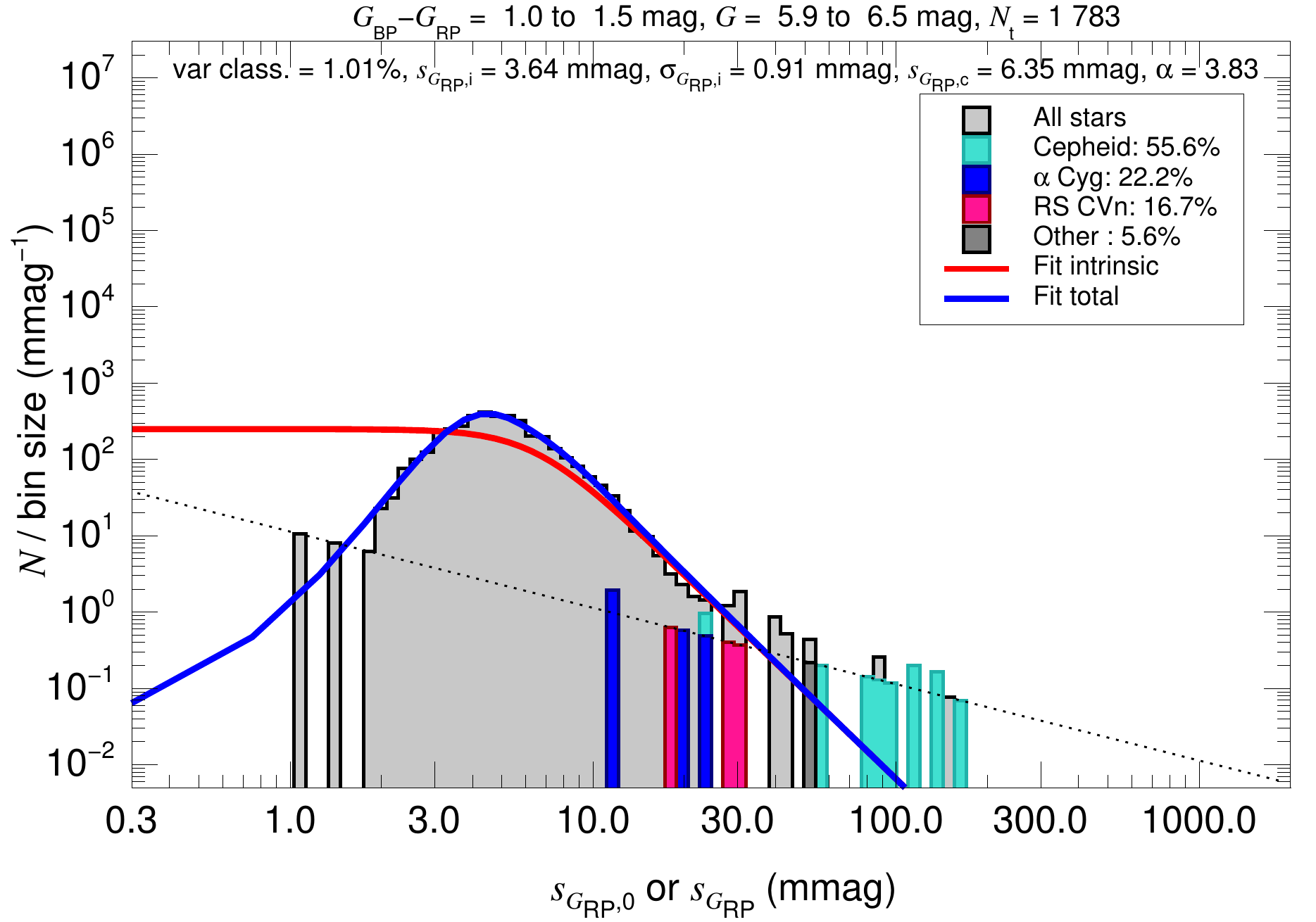}}
\centerline{$\!\!\!$\includegraphics[width=0.35\linewidth]{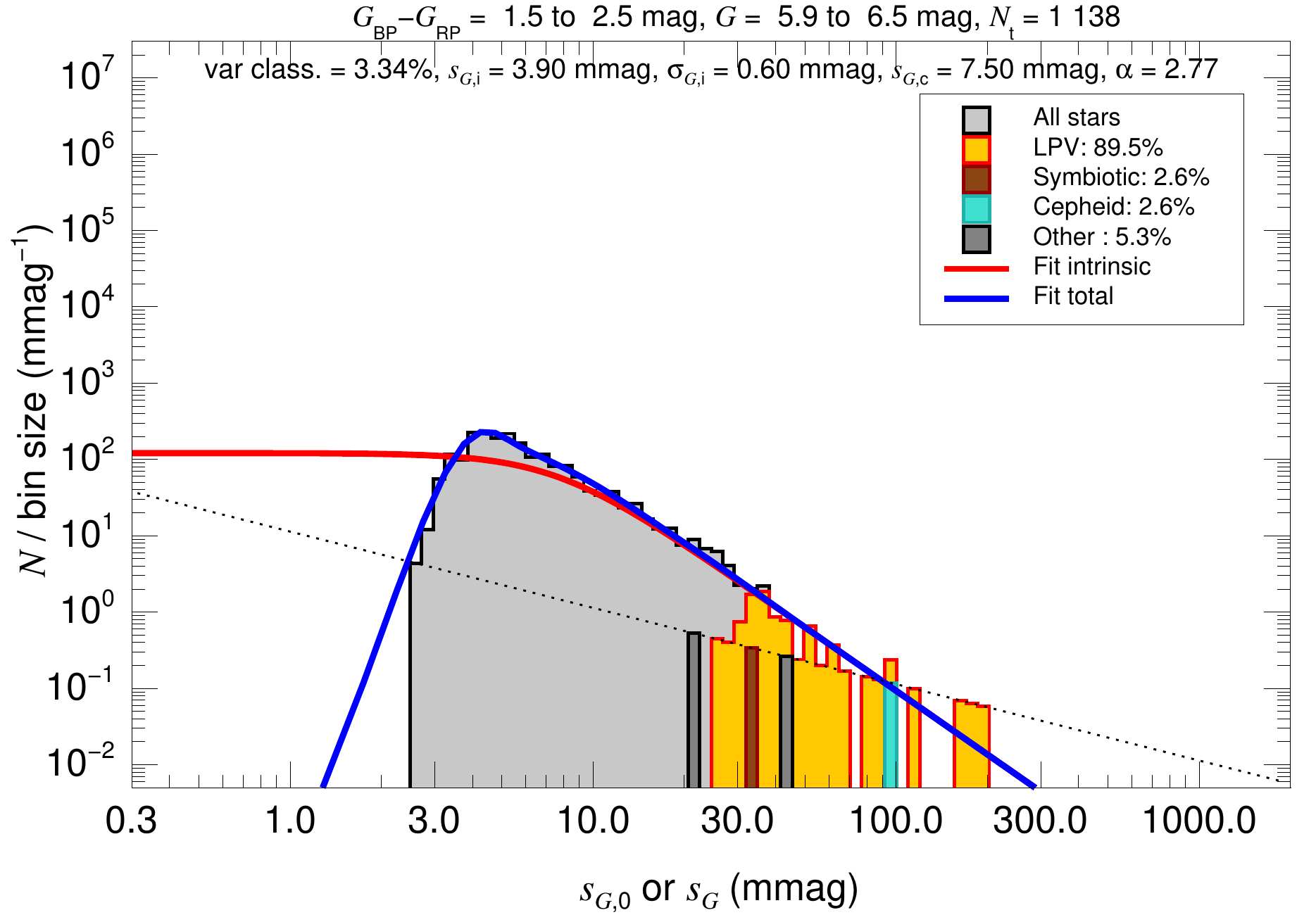}$\!\!\!$
                    \includegraphics[width=0.35\linewidth]{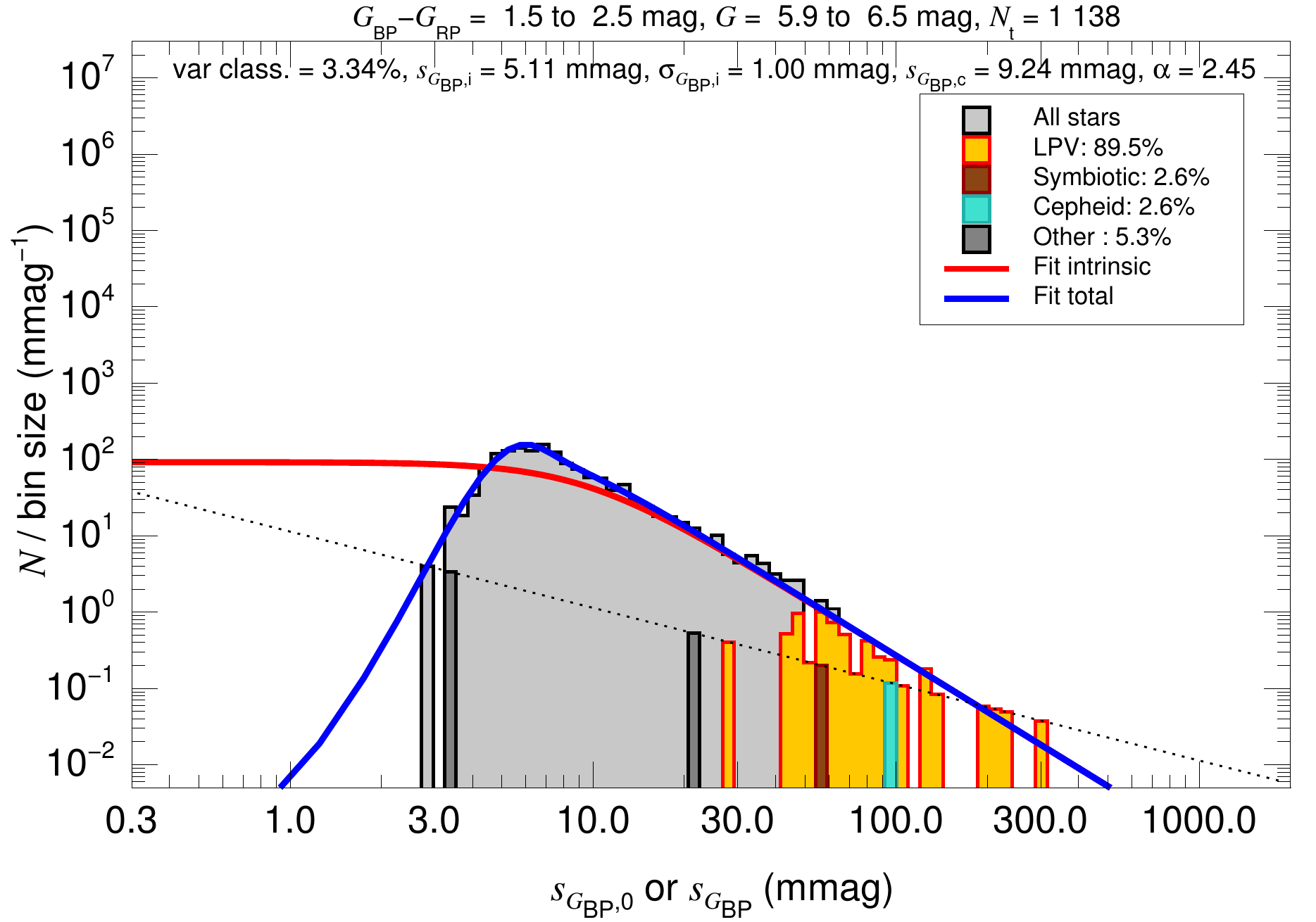}$\!\!\!$
                    \includegraphics[width=0.35\linewidth]{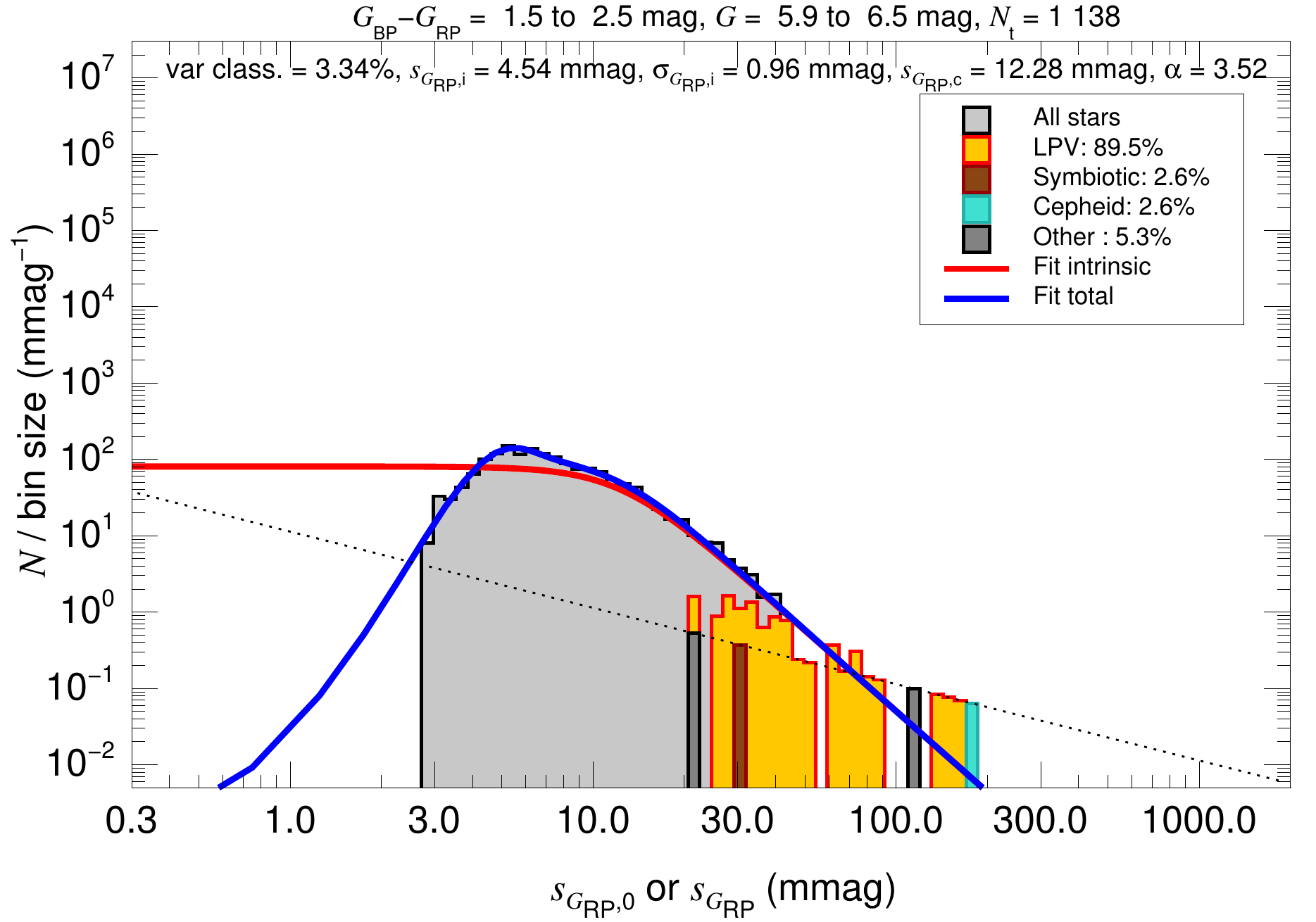}}
\centerline{$\!\!\!$\includegraphics[width=0.35\linewidth]{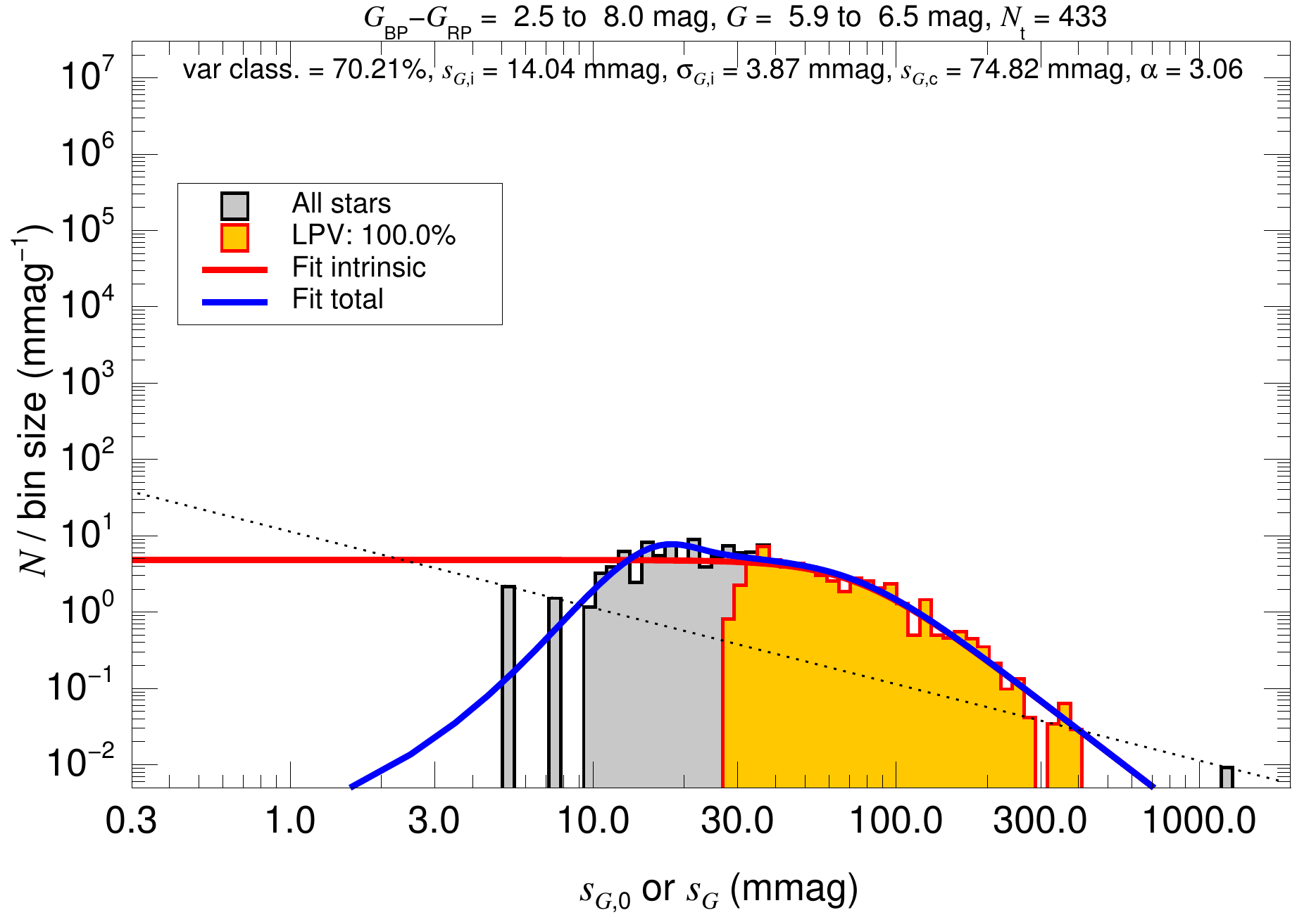}$\!\!\!$
                    \includegraphics[width=0.35\linewidth]{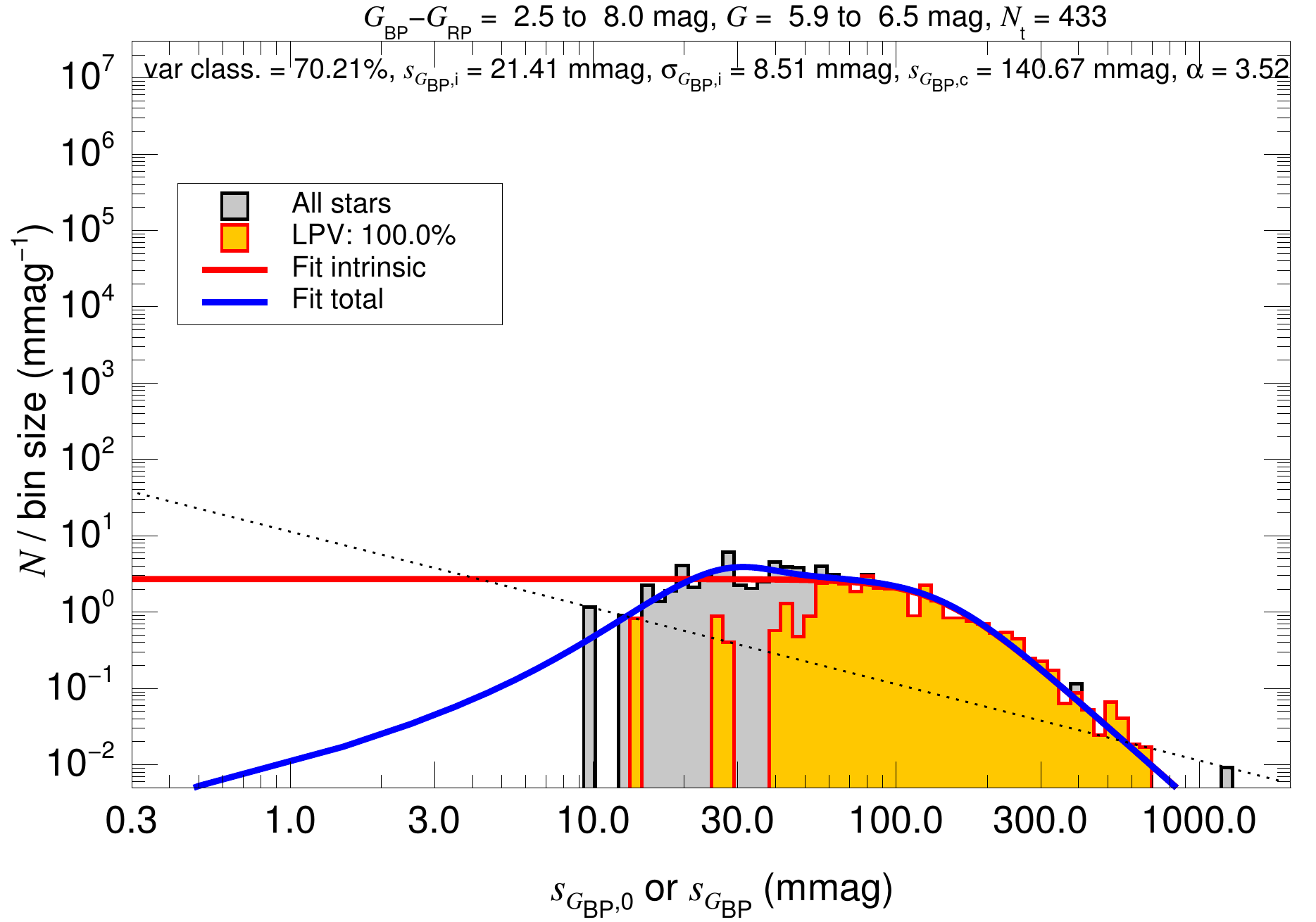}$\!\!\!$
                    \includegraphics[width=0.35\linewidth]{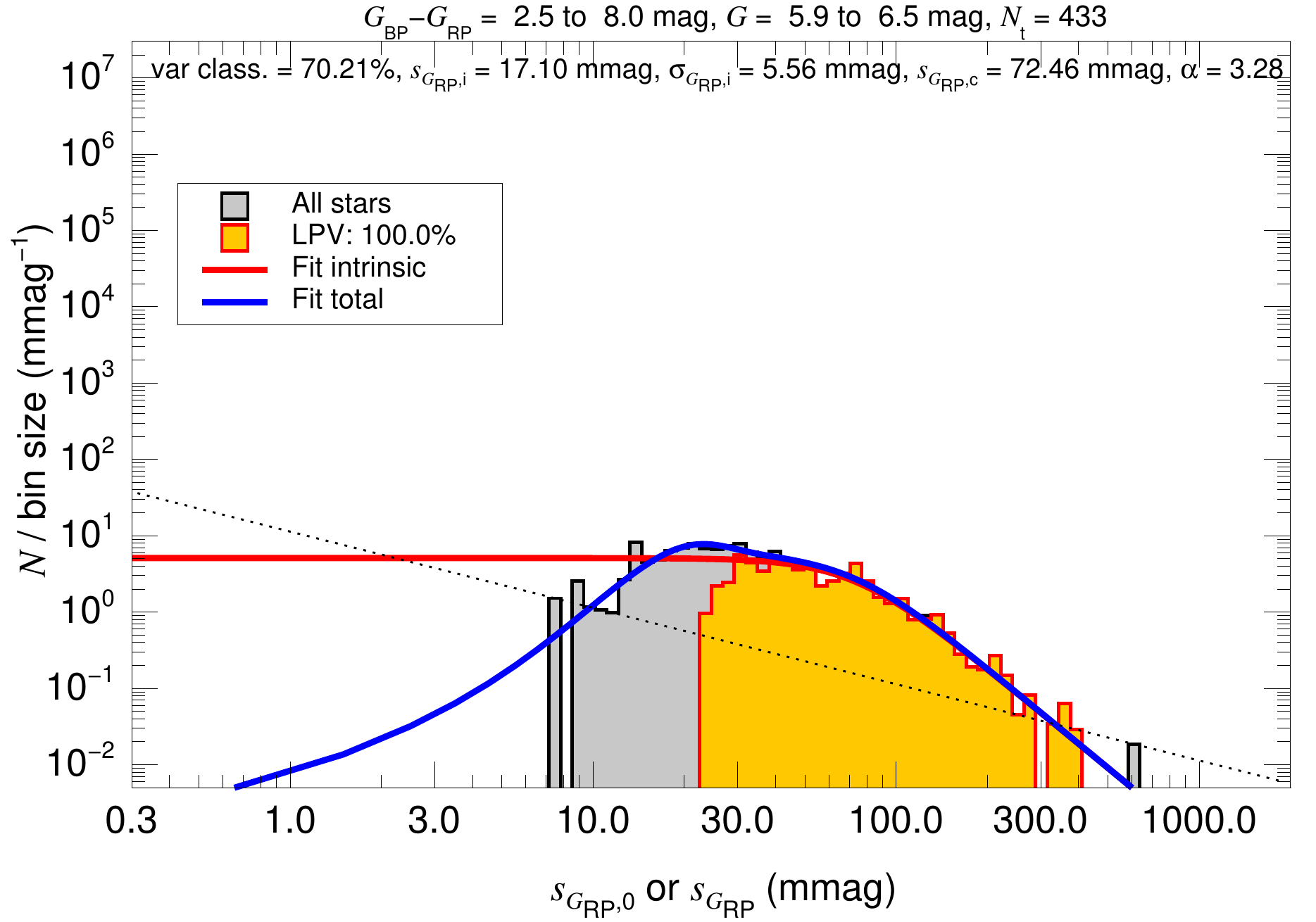}}
\caption{(Continued).}
\end{figure*}

\addtocounter{figure}{-1}

\begin{figure*}
\centerline{$\!\!\!$\includegraphics[width=0.35\linewidth]{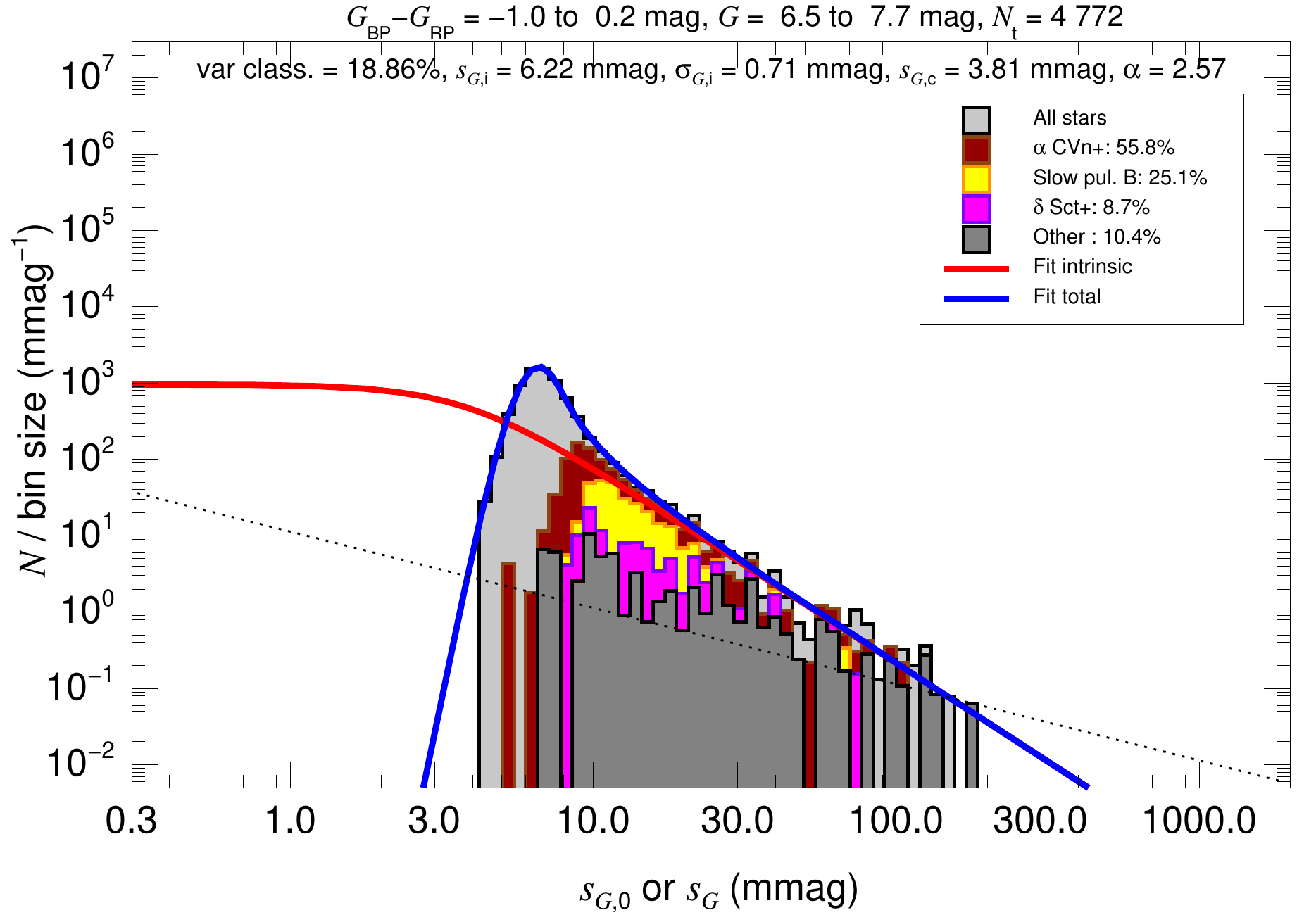}$\!\!\!$
                    \includegraphics[width=0.35\linewidth]{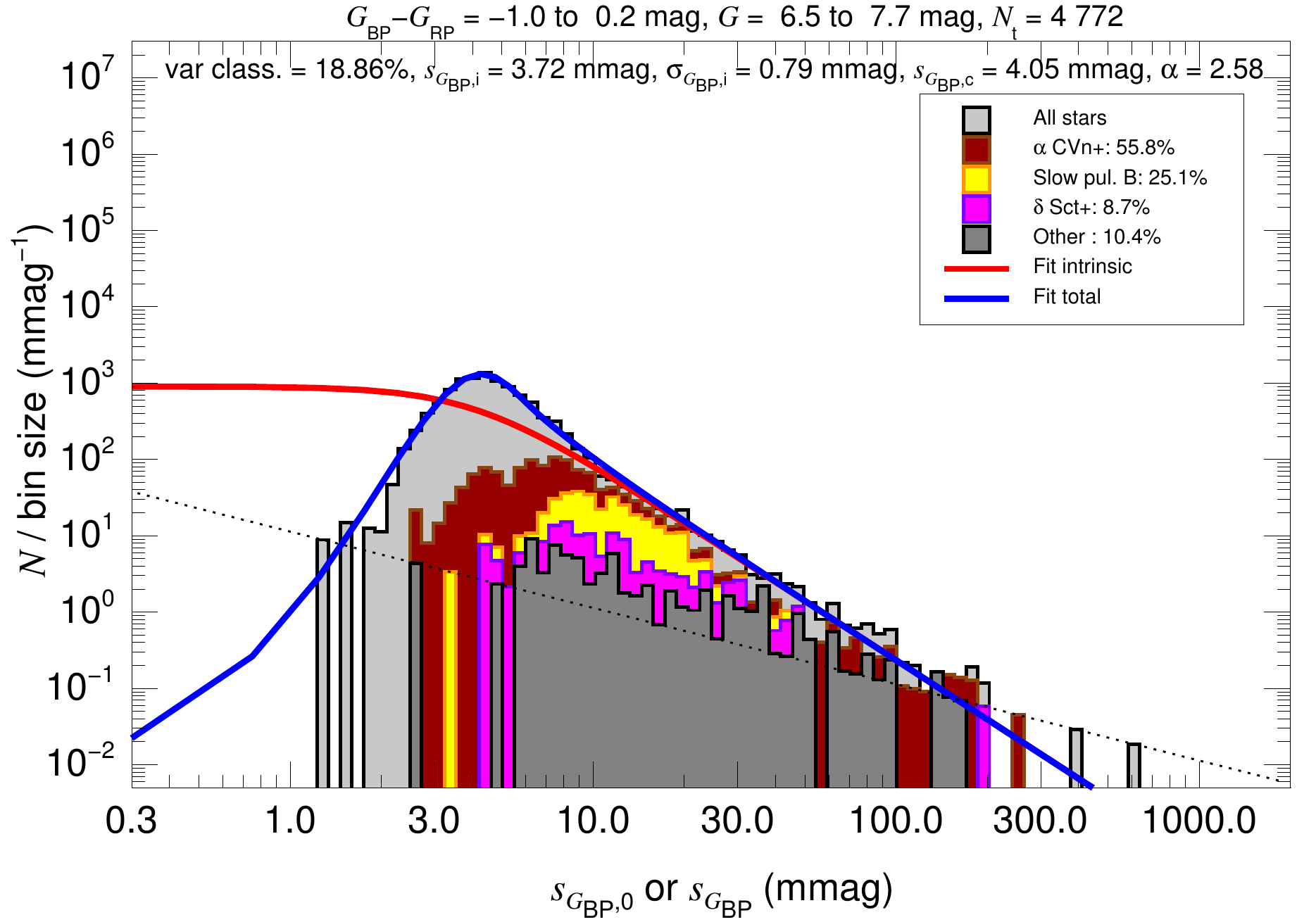}$\!\!\!$
                    \includegraphics[width=0.35\linewidth]{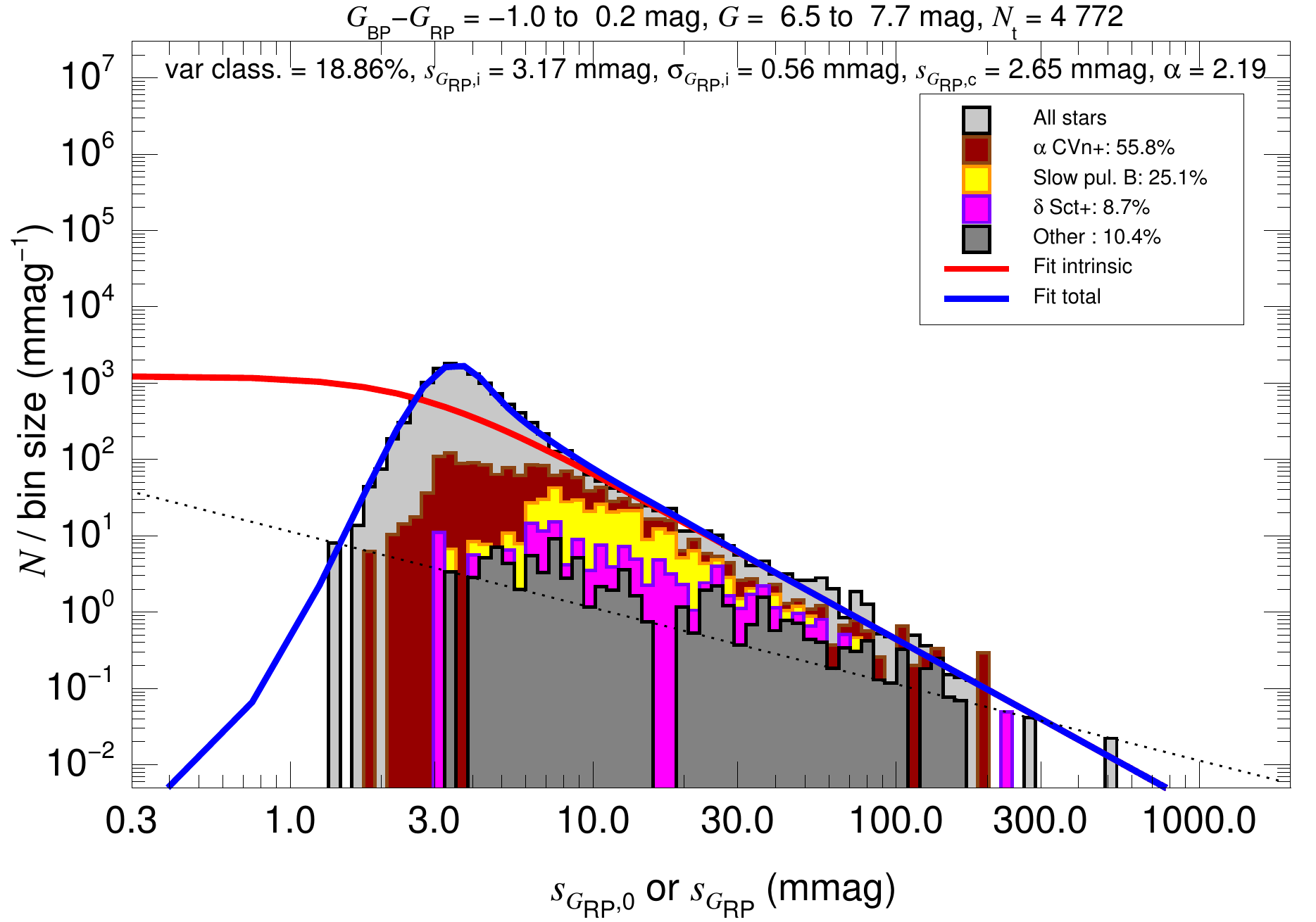}}
\centerline{$\!\!\!$\includegraphics[width=0.35\linewidth]{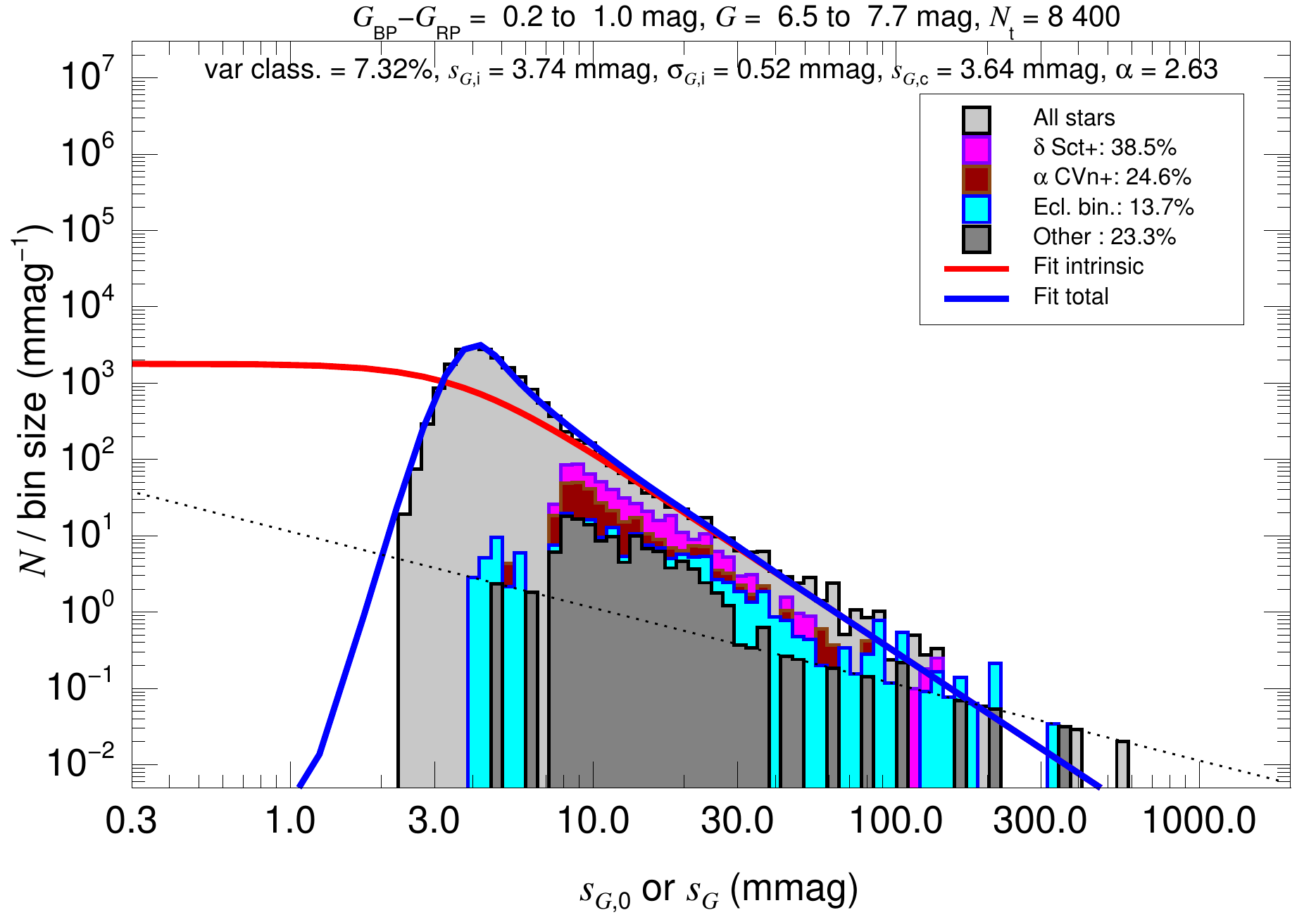}$\!\!\!$
                    \includegraphics[width=0.35\linewidth]{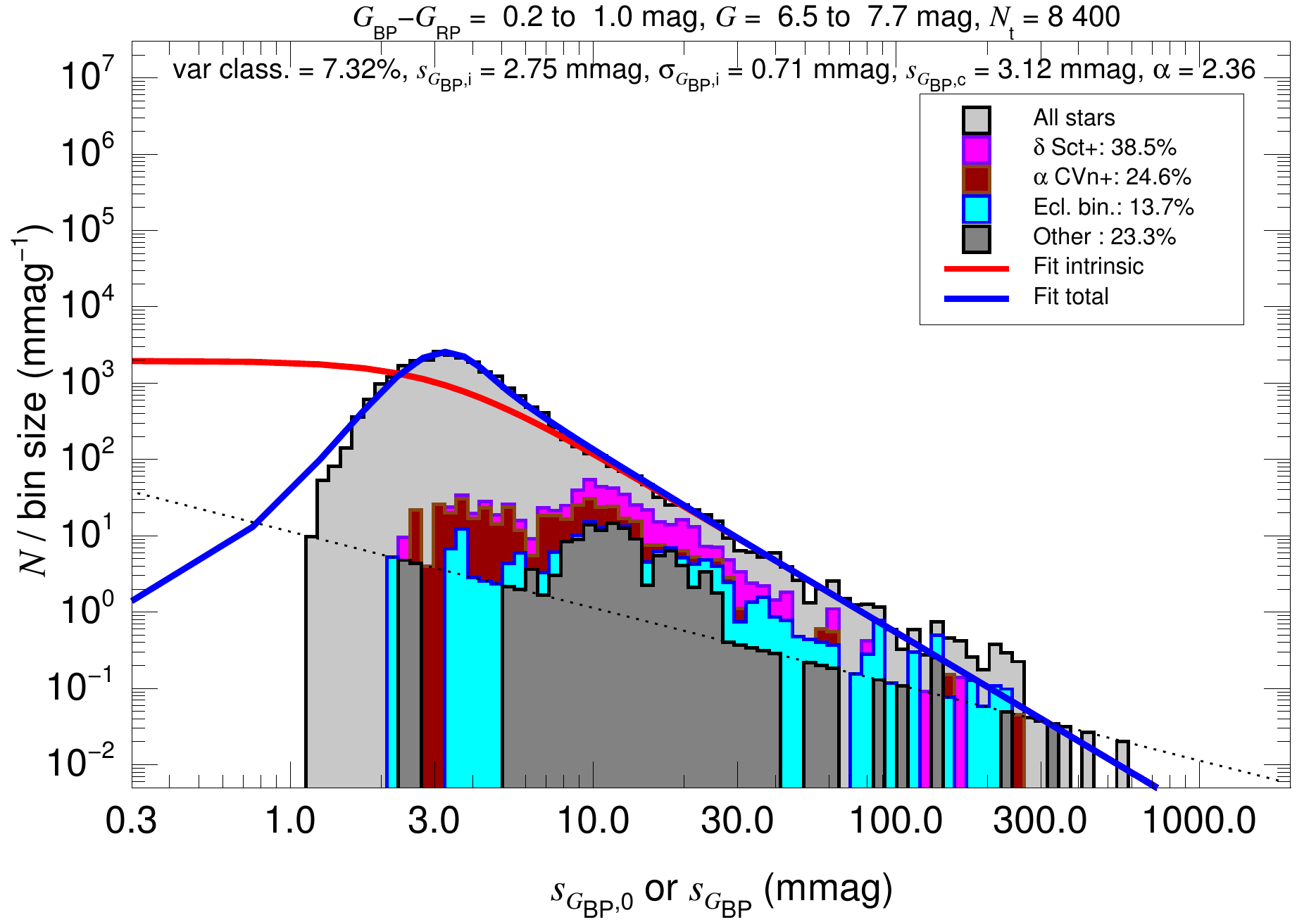}$\!\!\!$
                    \includegraphics[width=0.35\linewidth]{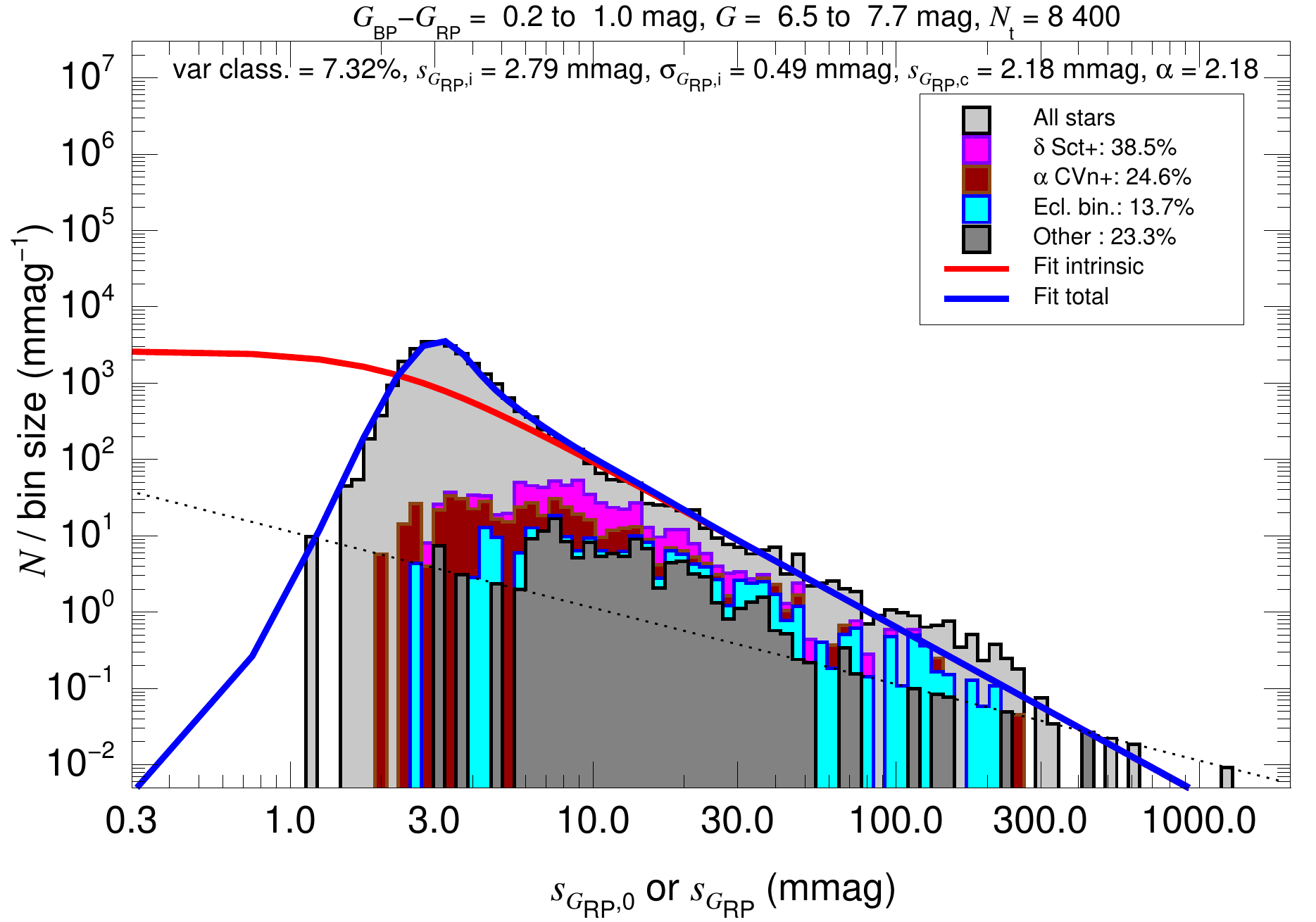}}
\centerline{$\!\!\!$\includegraphics[width=0.35\linewidth]{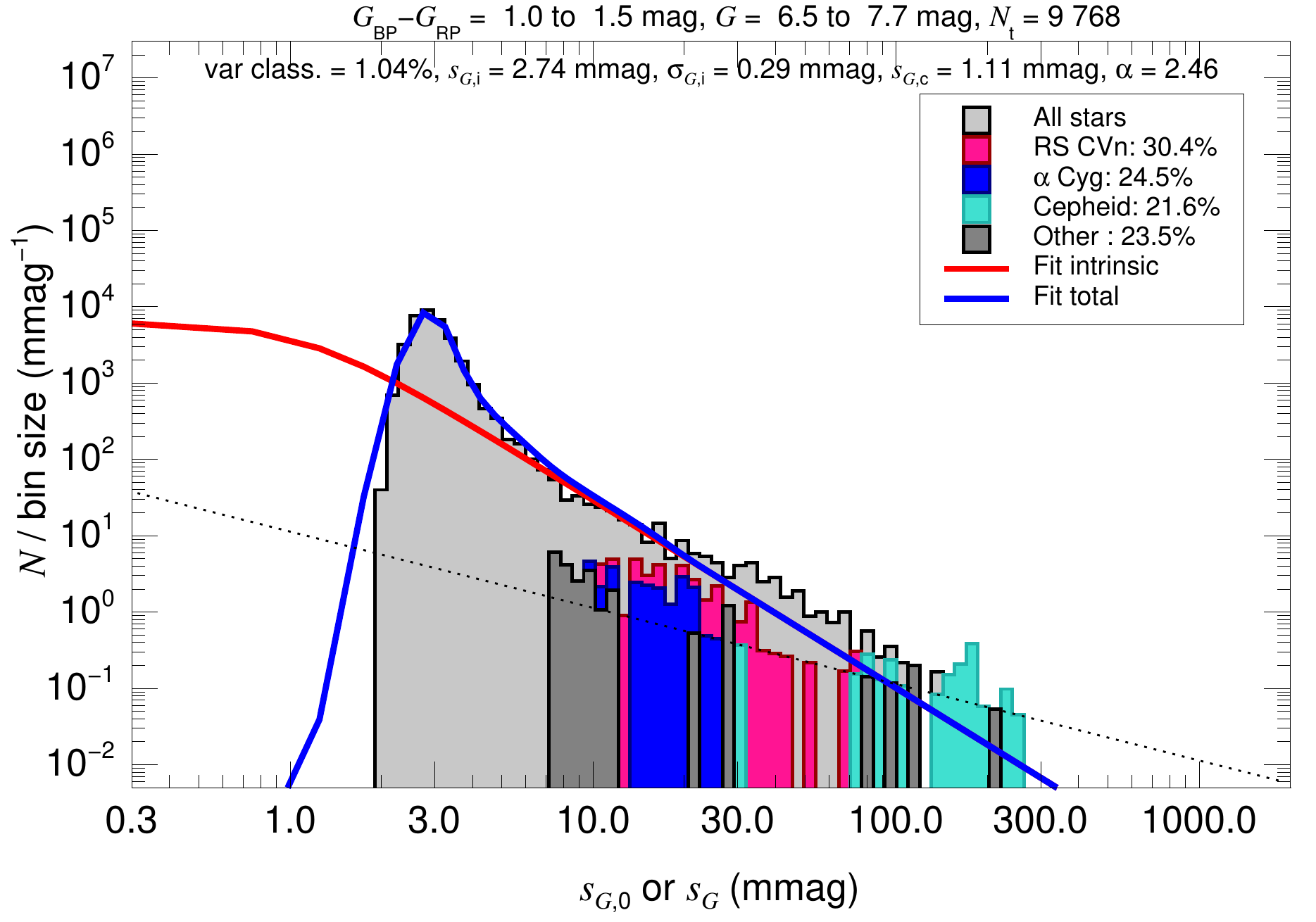}$\!\!\!$
                    \includegraphics[width=0.35\linewidth]{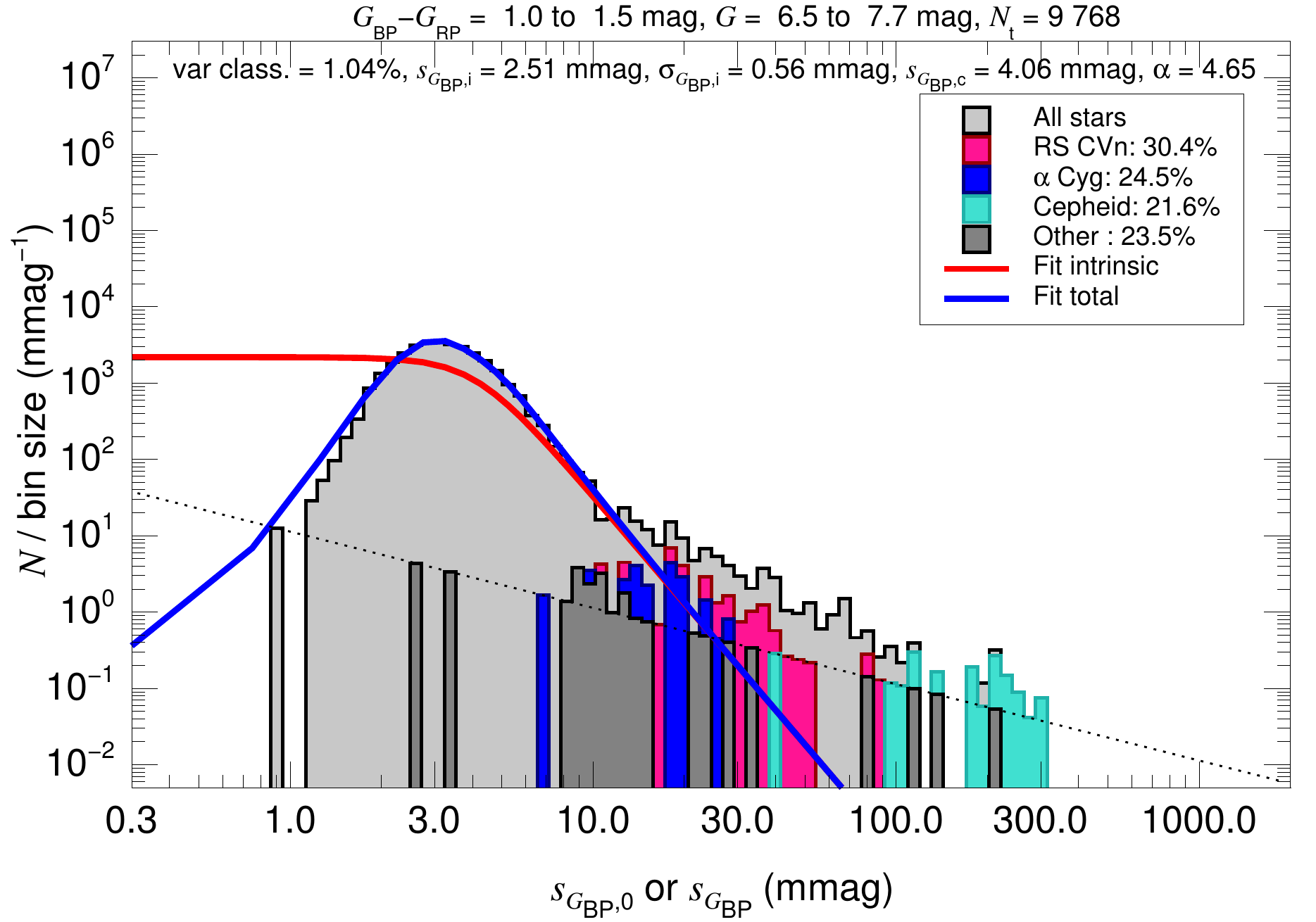}$\!\!\!$
                    \includegraphics[width=0.35\linewidth]{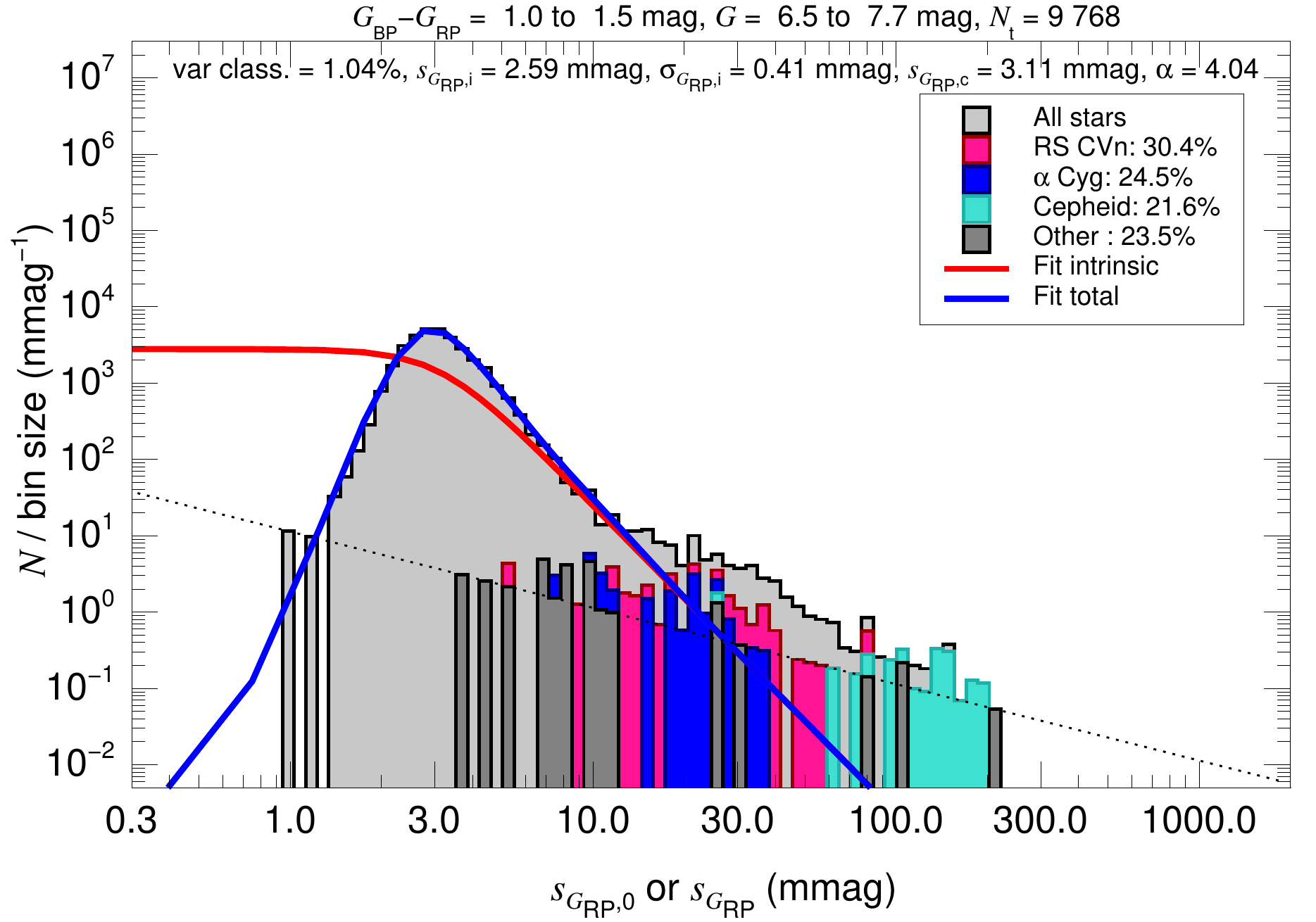}}
\centerline{$\!\!\!$\includegraphics[width=0.35\linewidth]{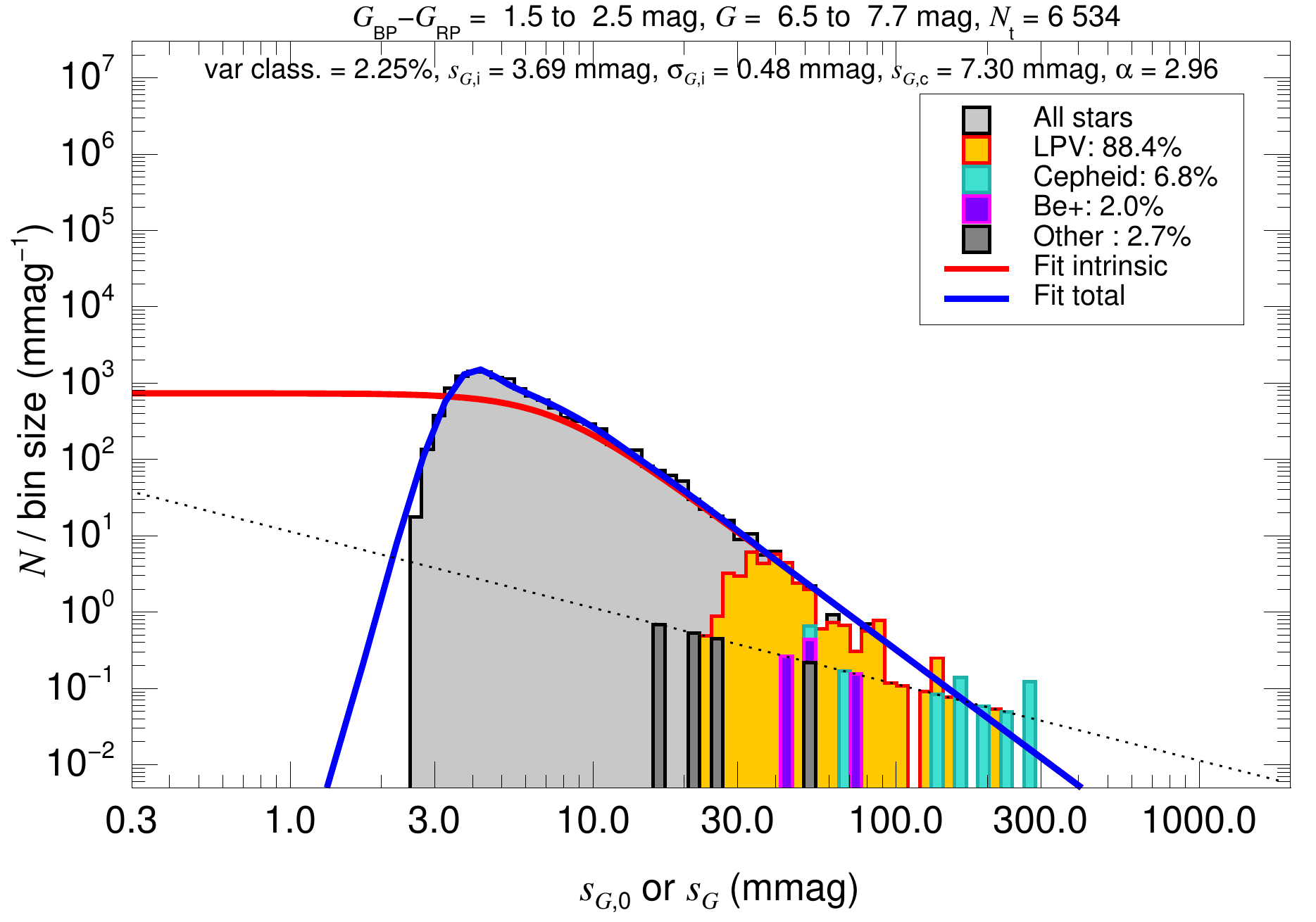}$\!\!\!$
                    \includegraphics[width=0.35\linewidth]{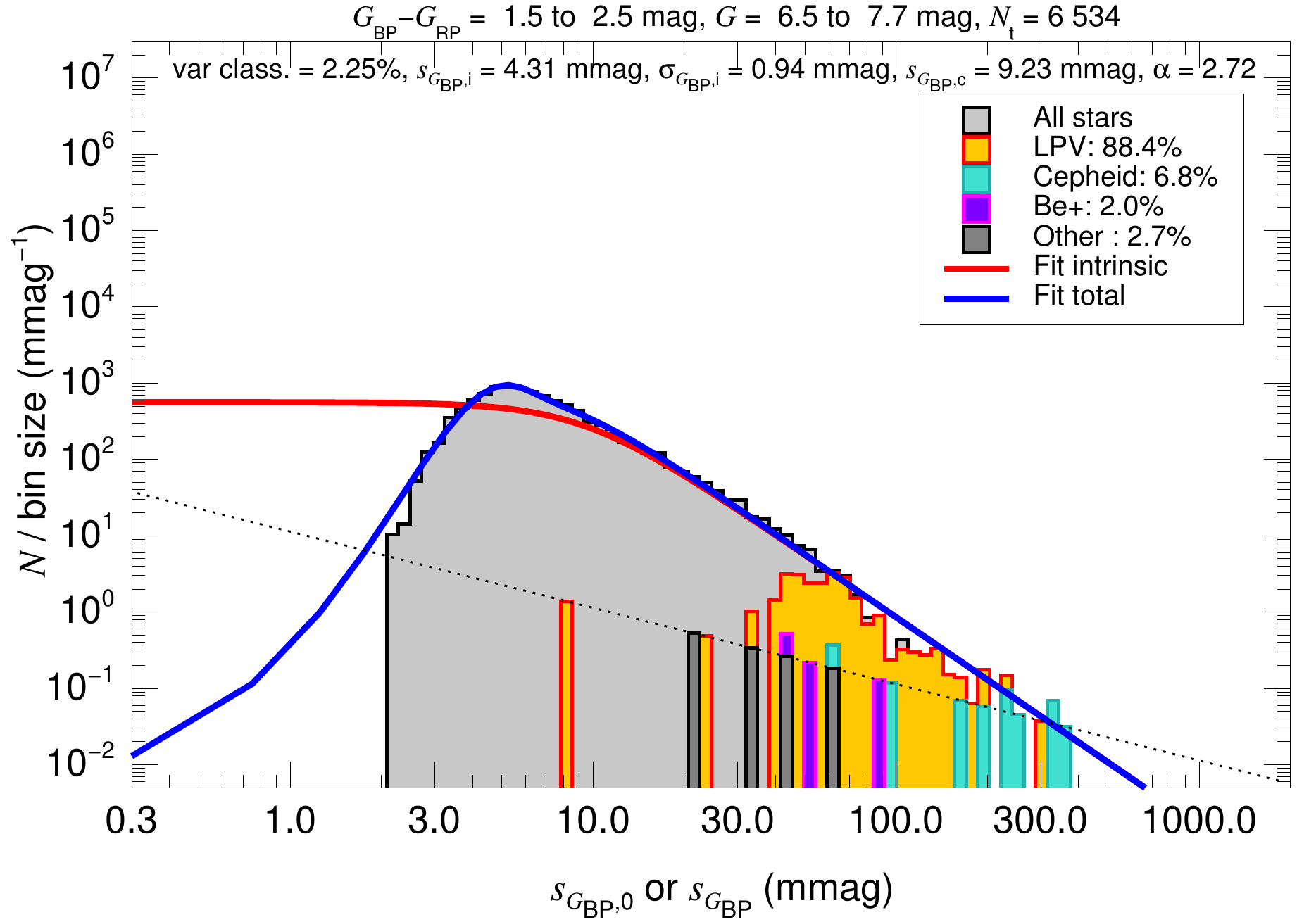}$\!\!\!$
                    \includegraphics[width=0.35\linewidth]{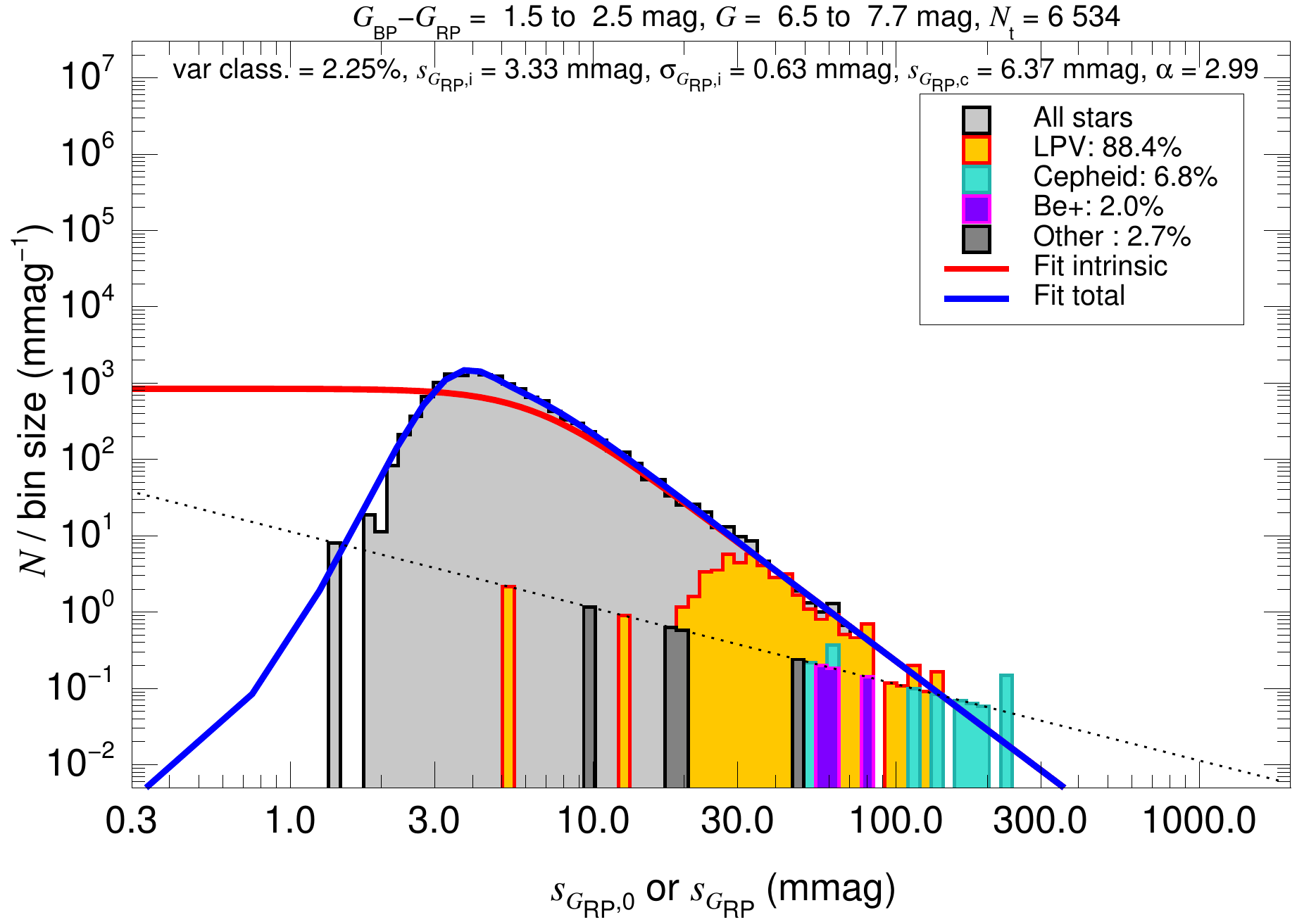}}
\centerline{$\!\!\!$\includegraphics[width=0.35\linewidth]{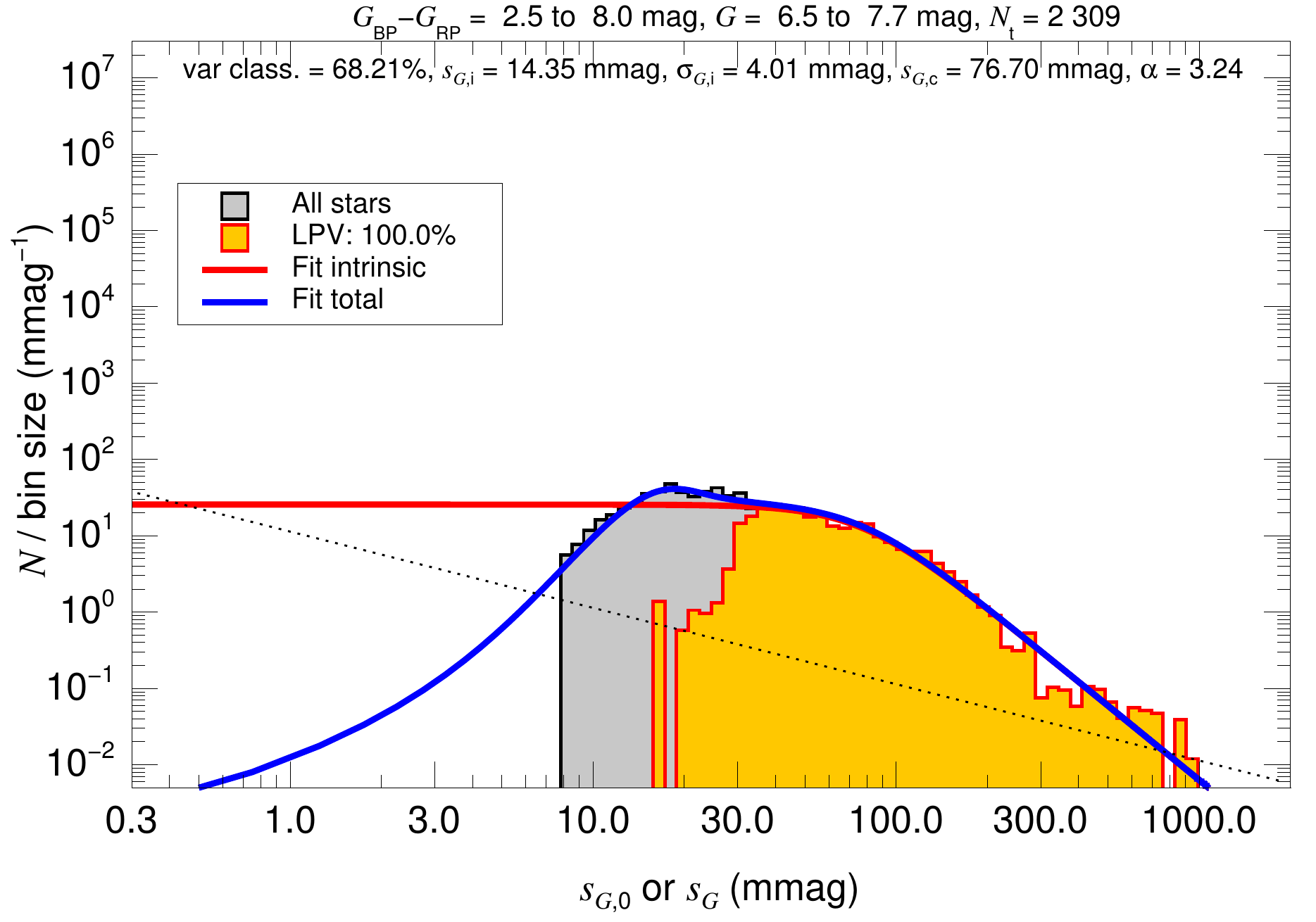}$\!\!\!$
                    \includegraphics[width=0.35\linewidth]{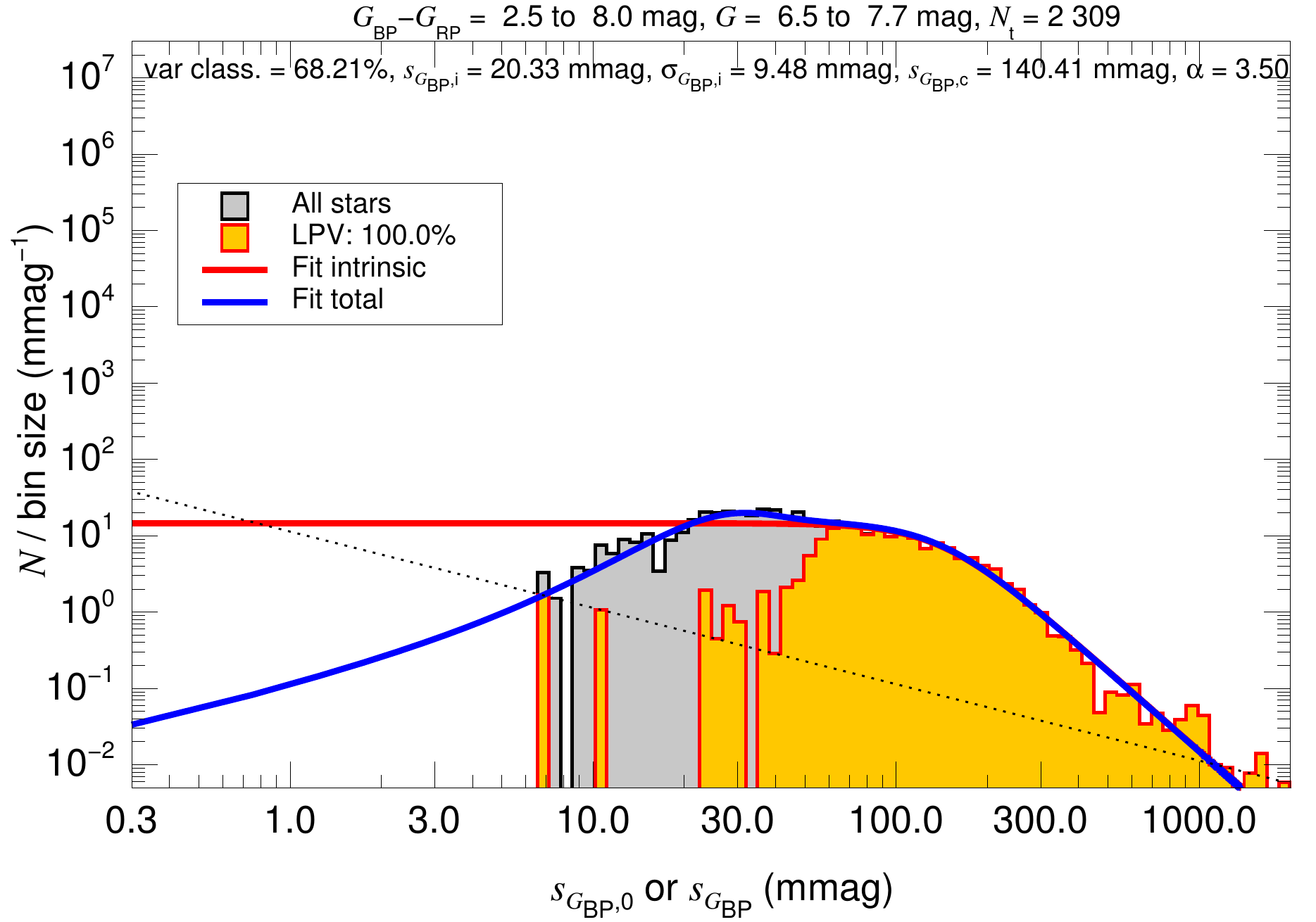}$\!\!\!$
                    \includegraphics[width=0.35\linewidth]{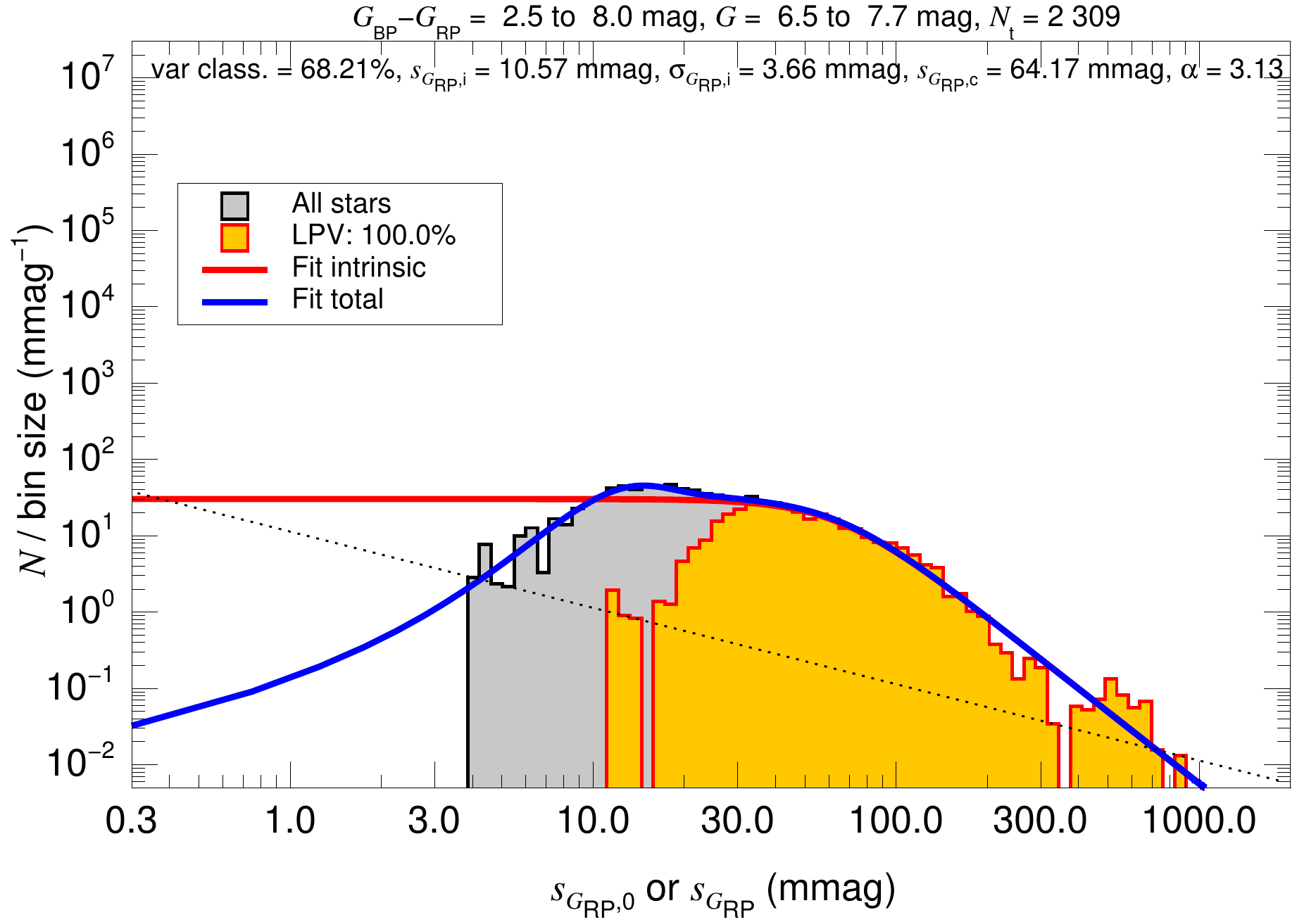}}
\caption{(Continued).}
\end{figure*}

\addtocounter{figure}{-1}

\begin{figure*}
\centerline{$\!\!\!$\includegraphics[width=0.35\linewidth]{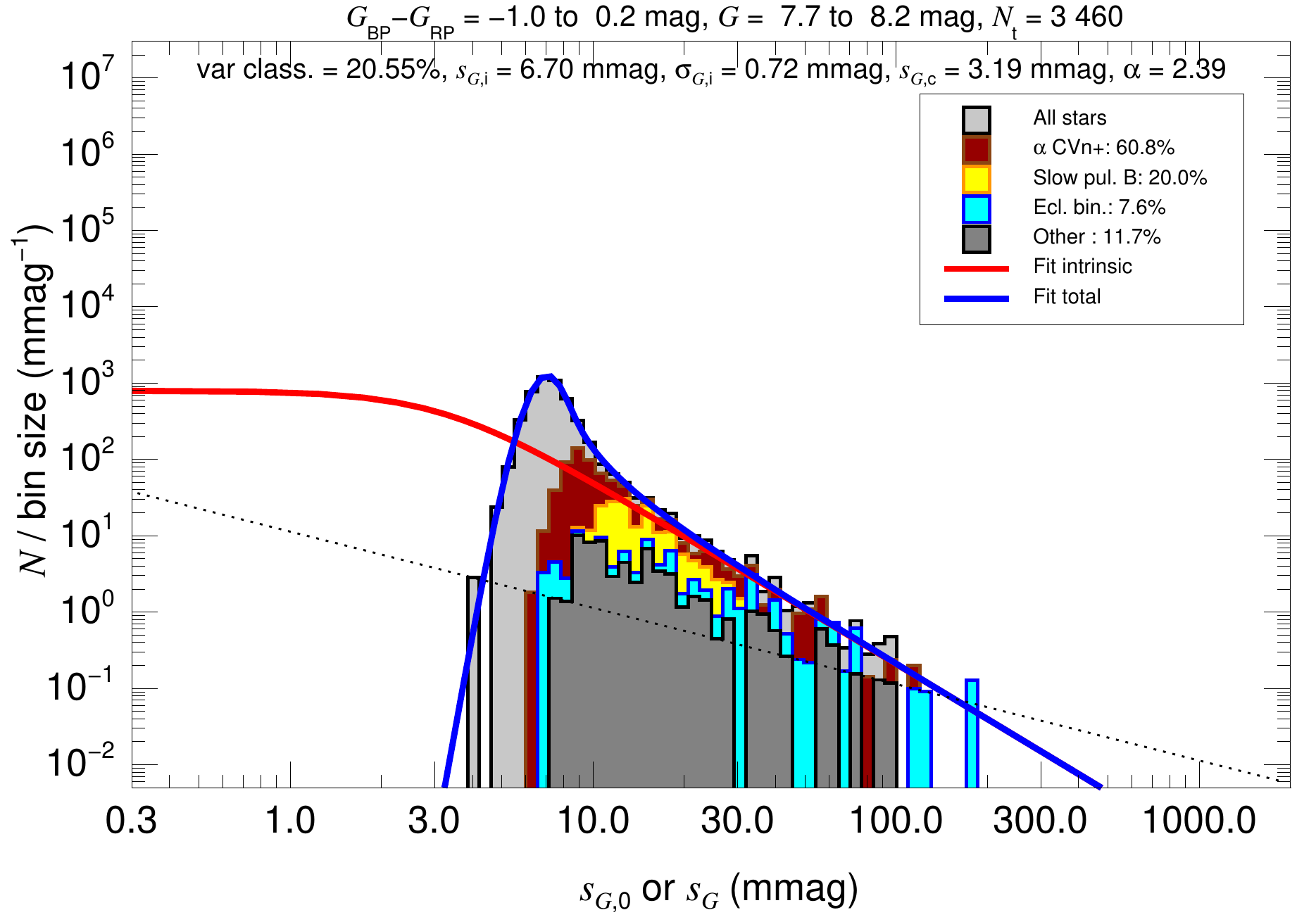}$\!\!\!$
                    \includegraphics[width=0.35\linewidth]{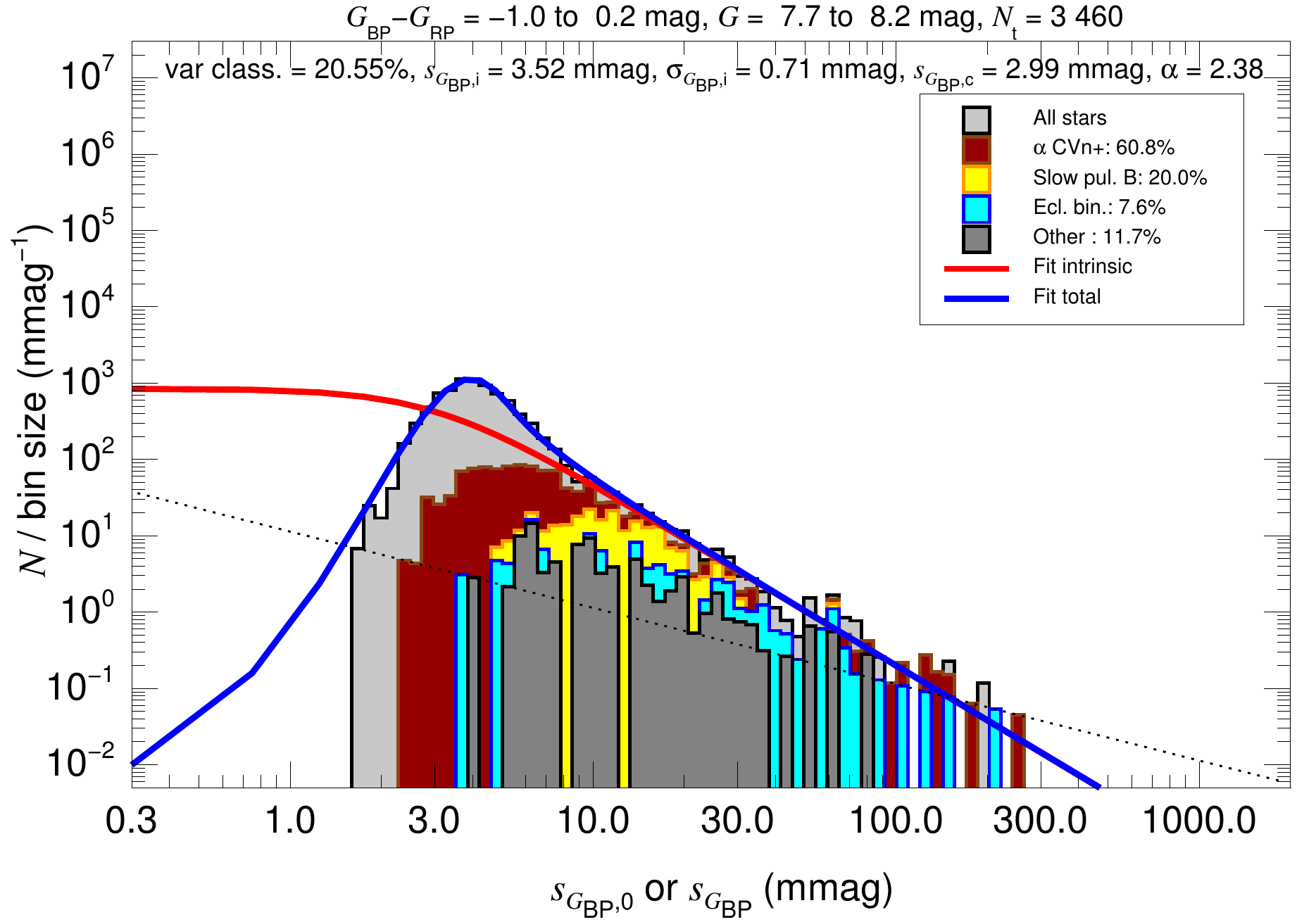}$\!\!\!$
                    \includegraphics[width=0.35\linewidth]{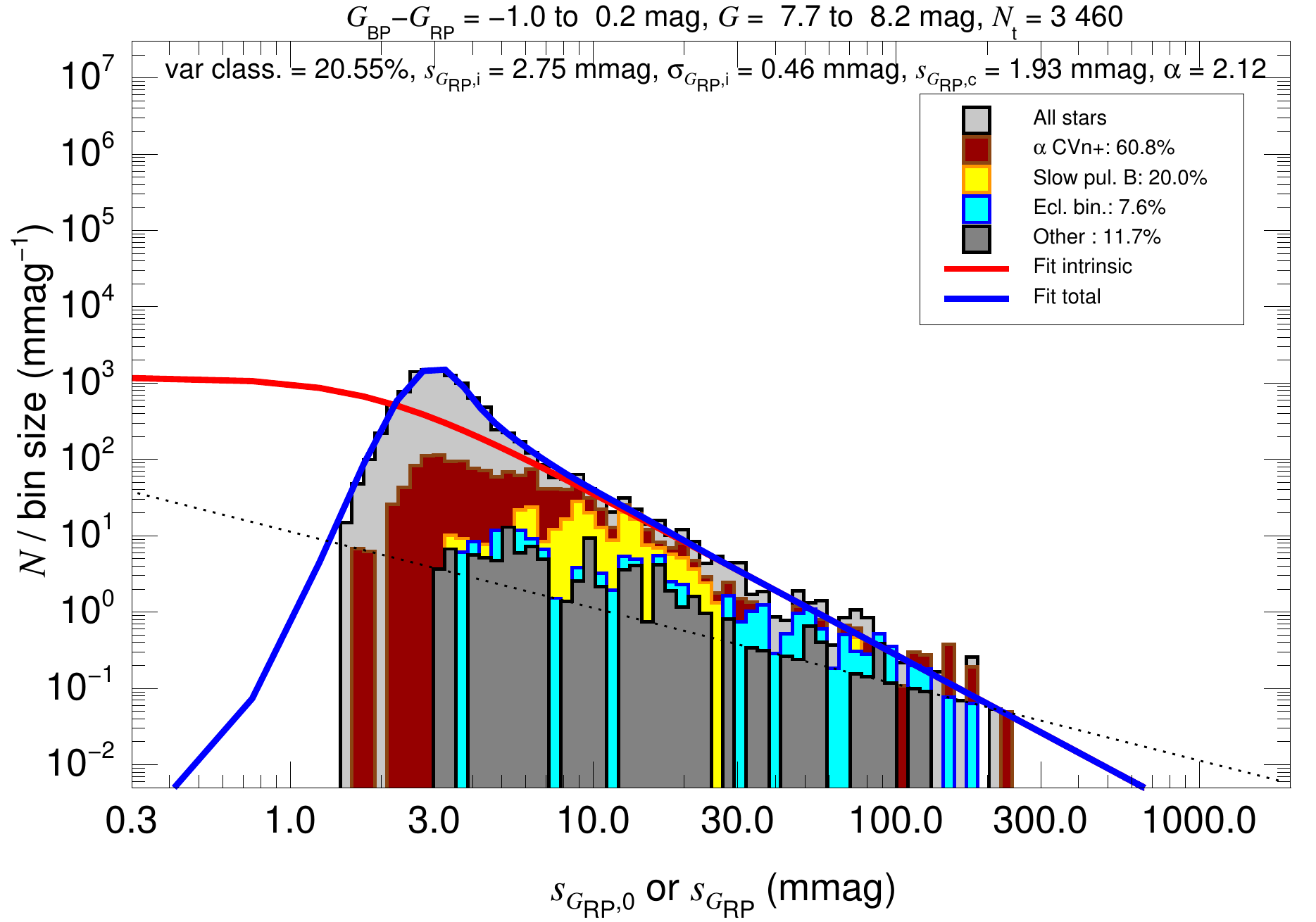}}
\centerline{$\!\!\!$\includegraphics[width=0.35\linewidth]{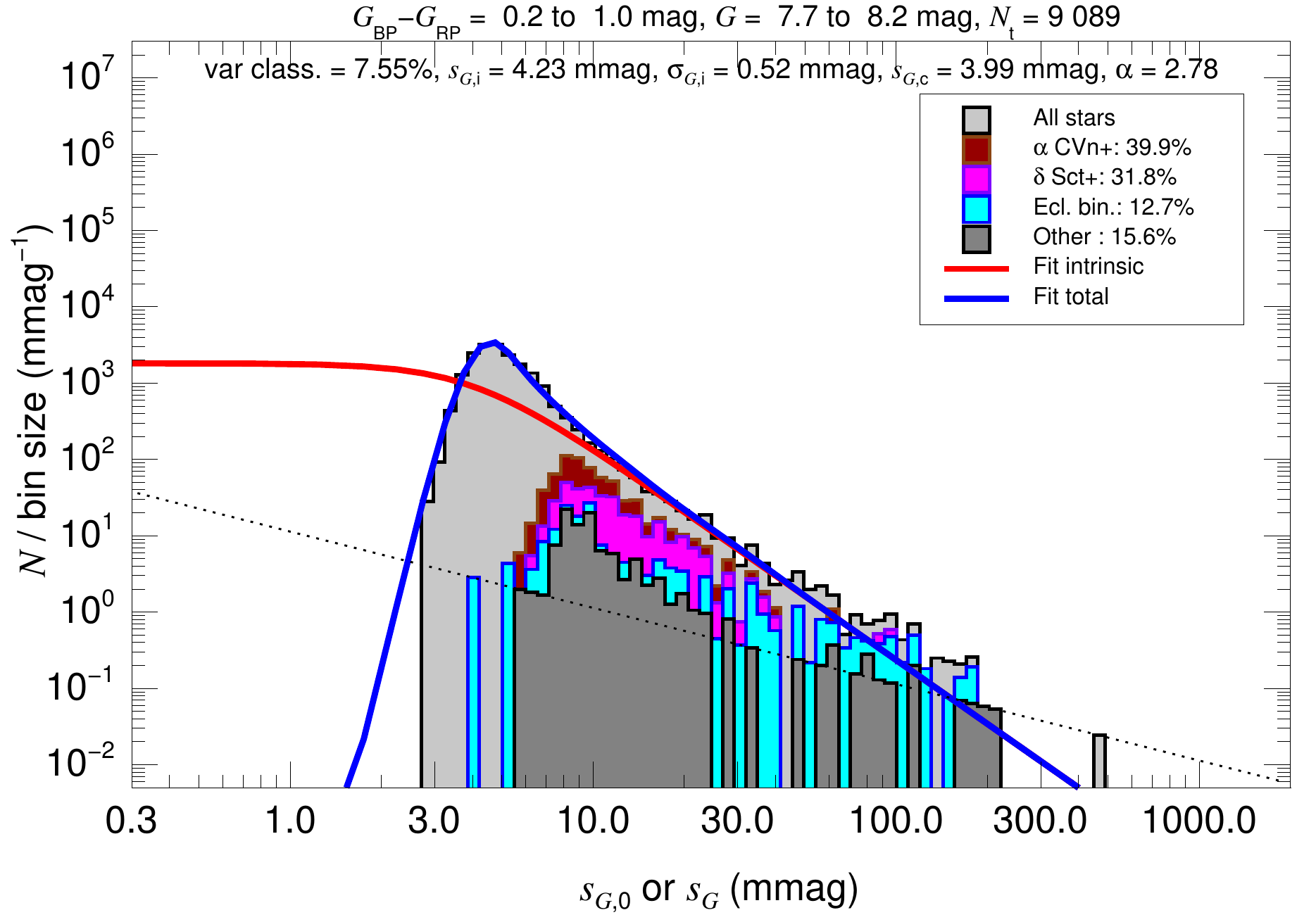}$\!\!\!$
                    \includegraphics[width=0.35\linewidth]{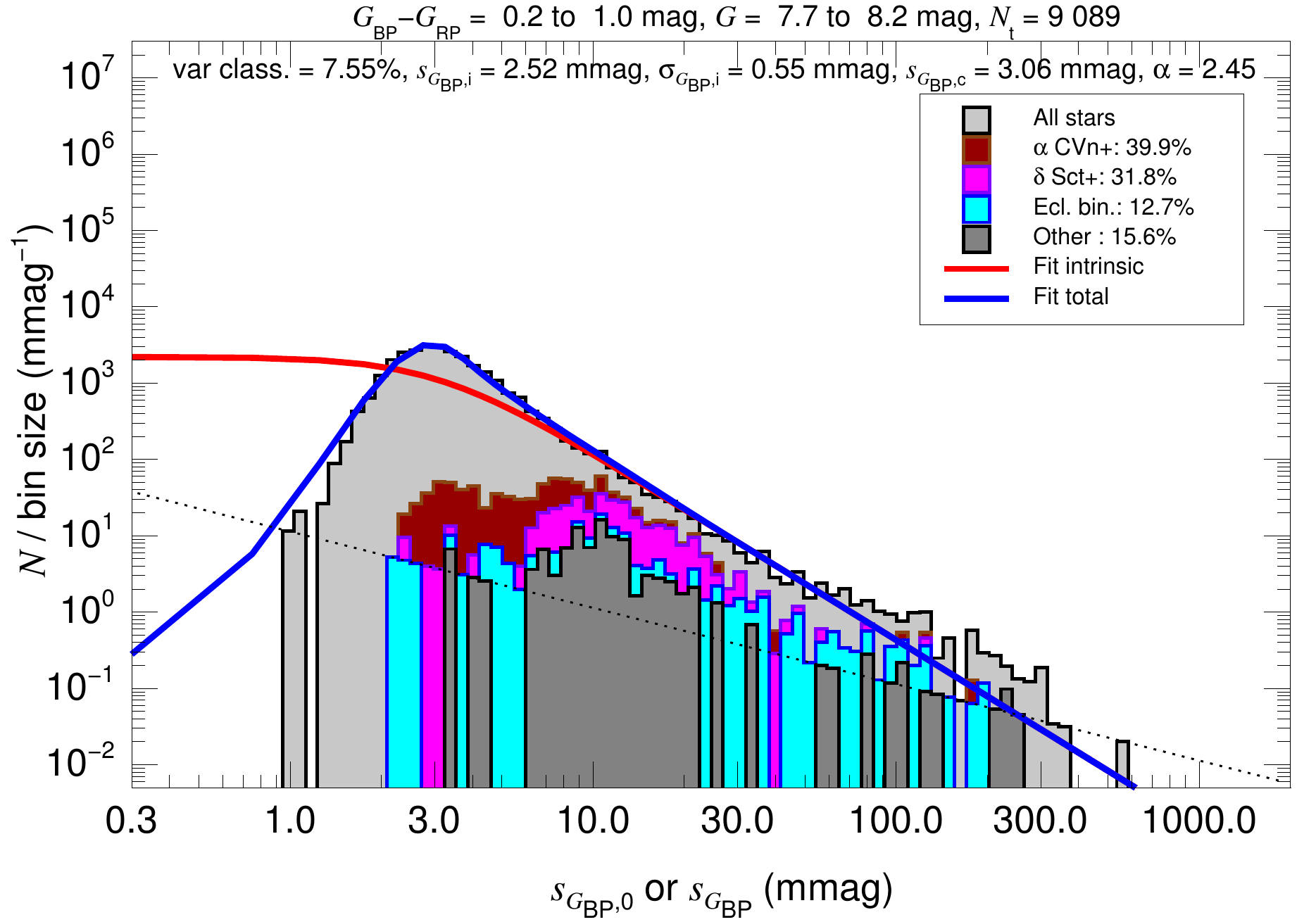}$\!\!\!$
                    \includegraphics[width=0.35\linewidth]{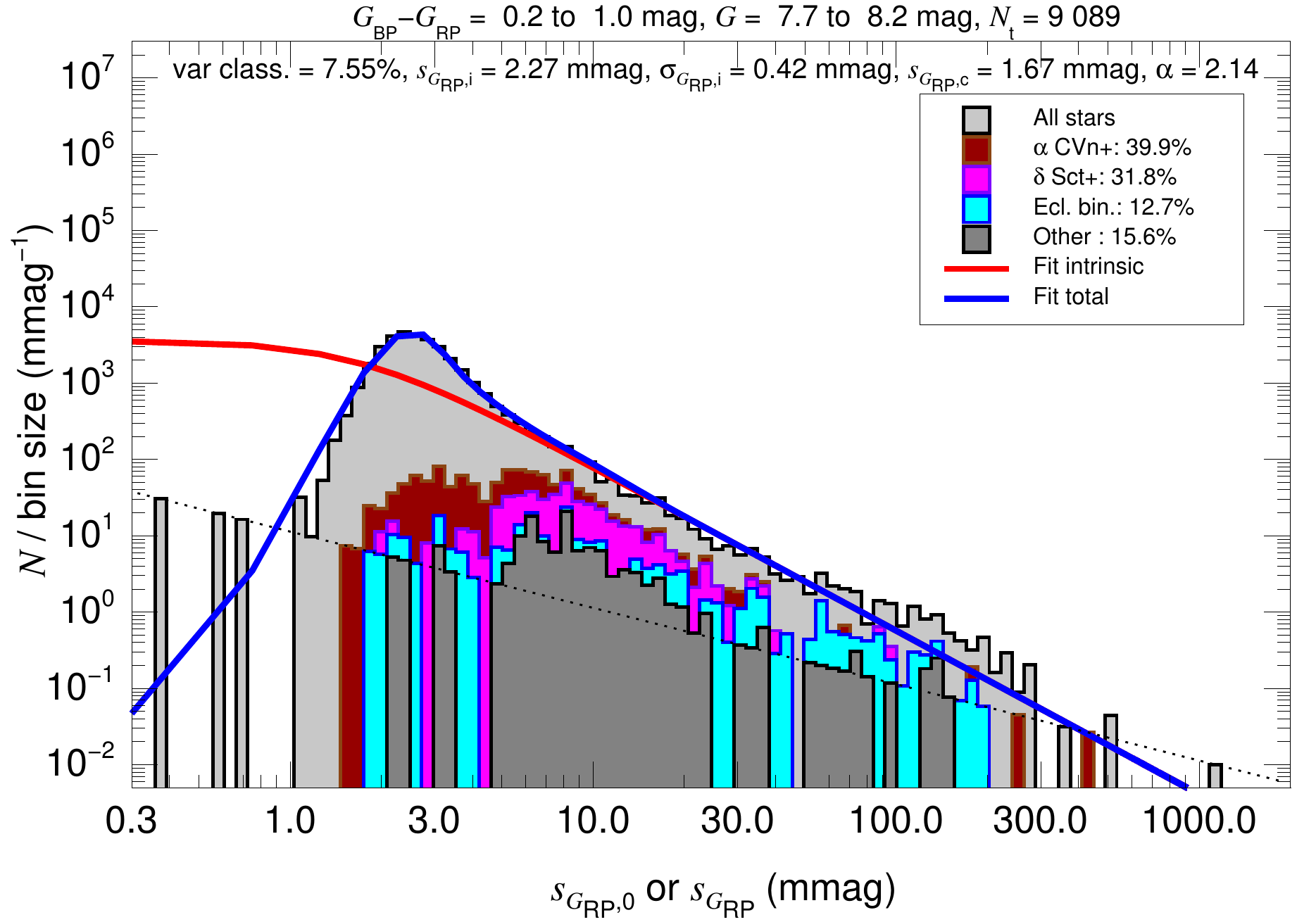}}
\centerline{$\!\!\!$\includegraphics[width=0.35\linewidth]{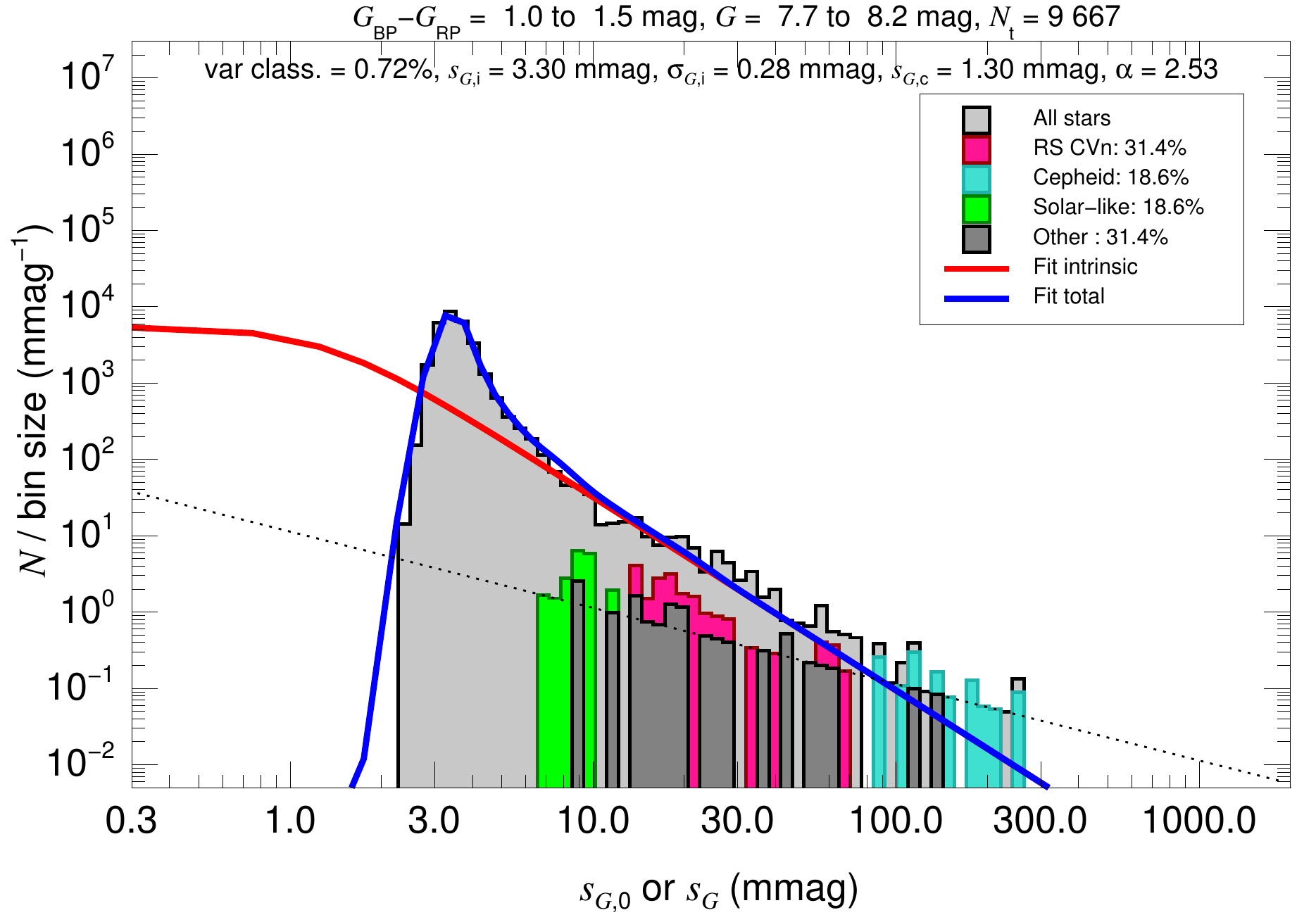}$\!\!\!$
                    \includegraphics[width=0.35\linewidth]{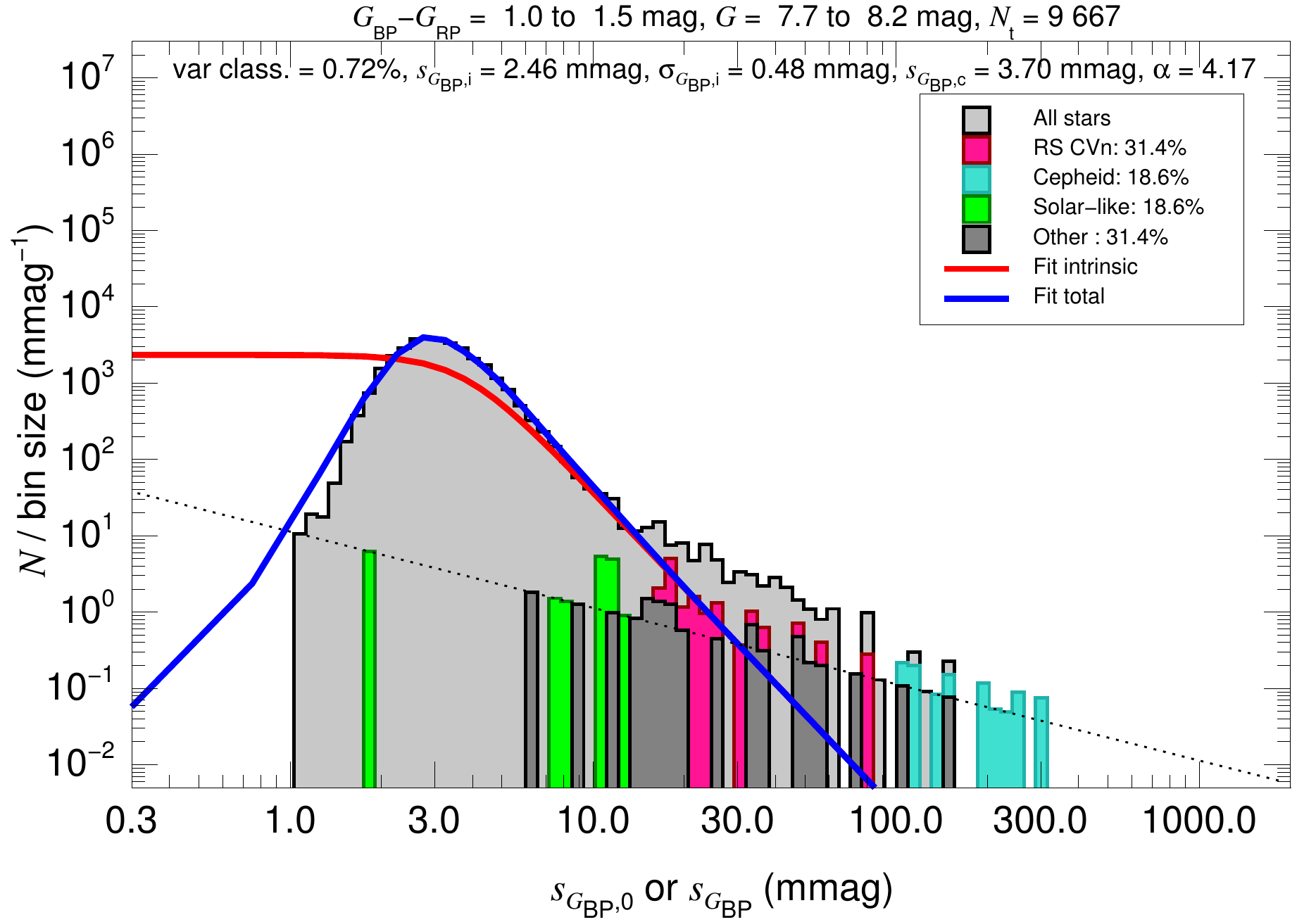}$\!\!\!$
                    \includegraphics[width=0.35\linewidth]{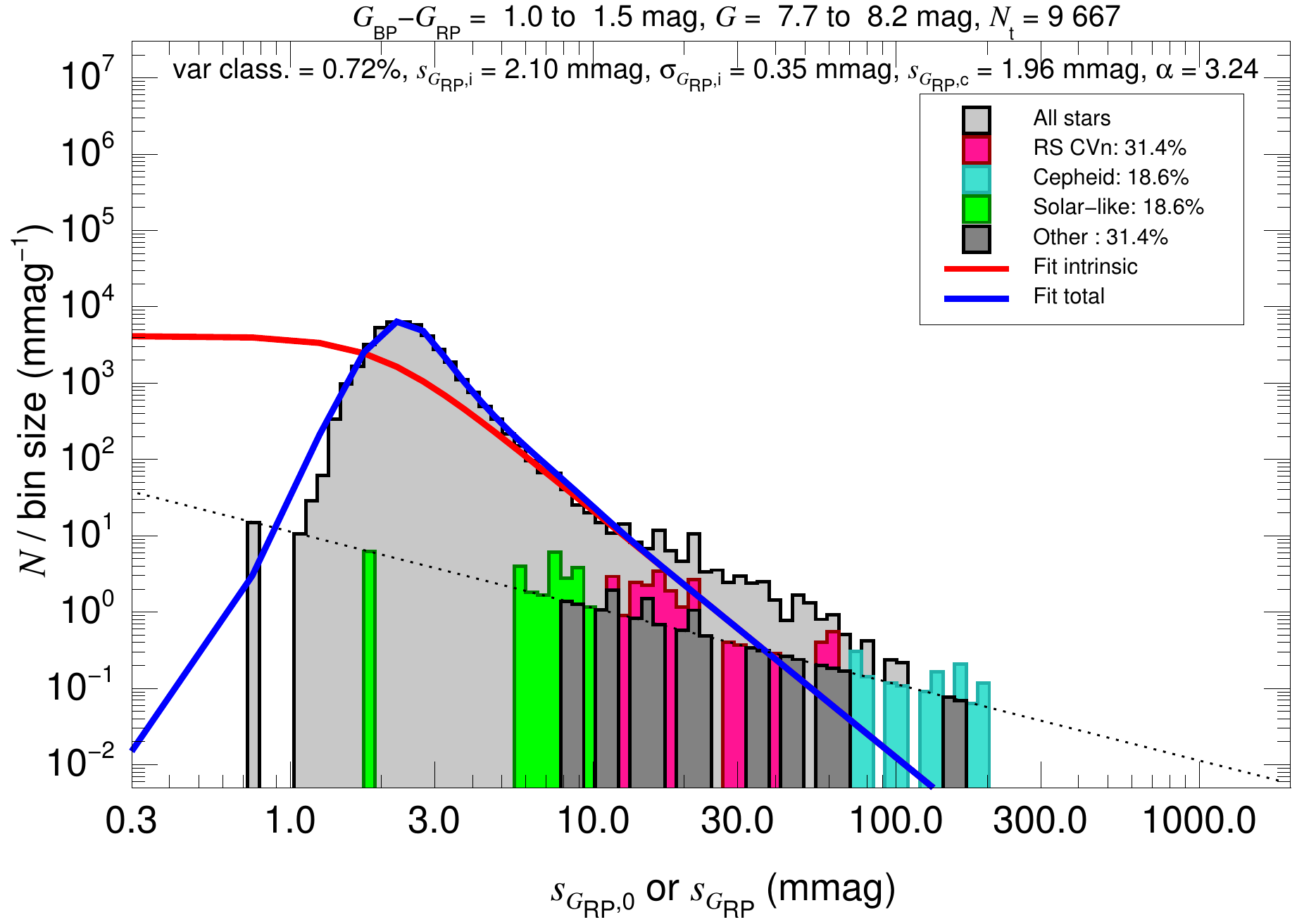}}
\centerline{$\!\!\!$\includegraphics[width=0.35\linewidth]{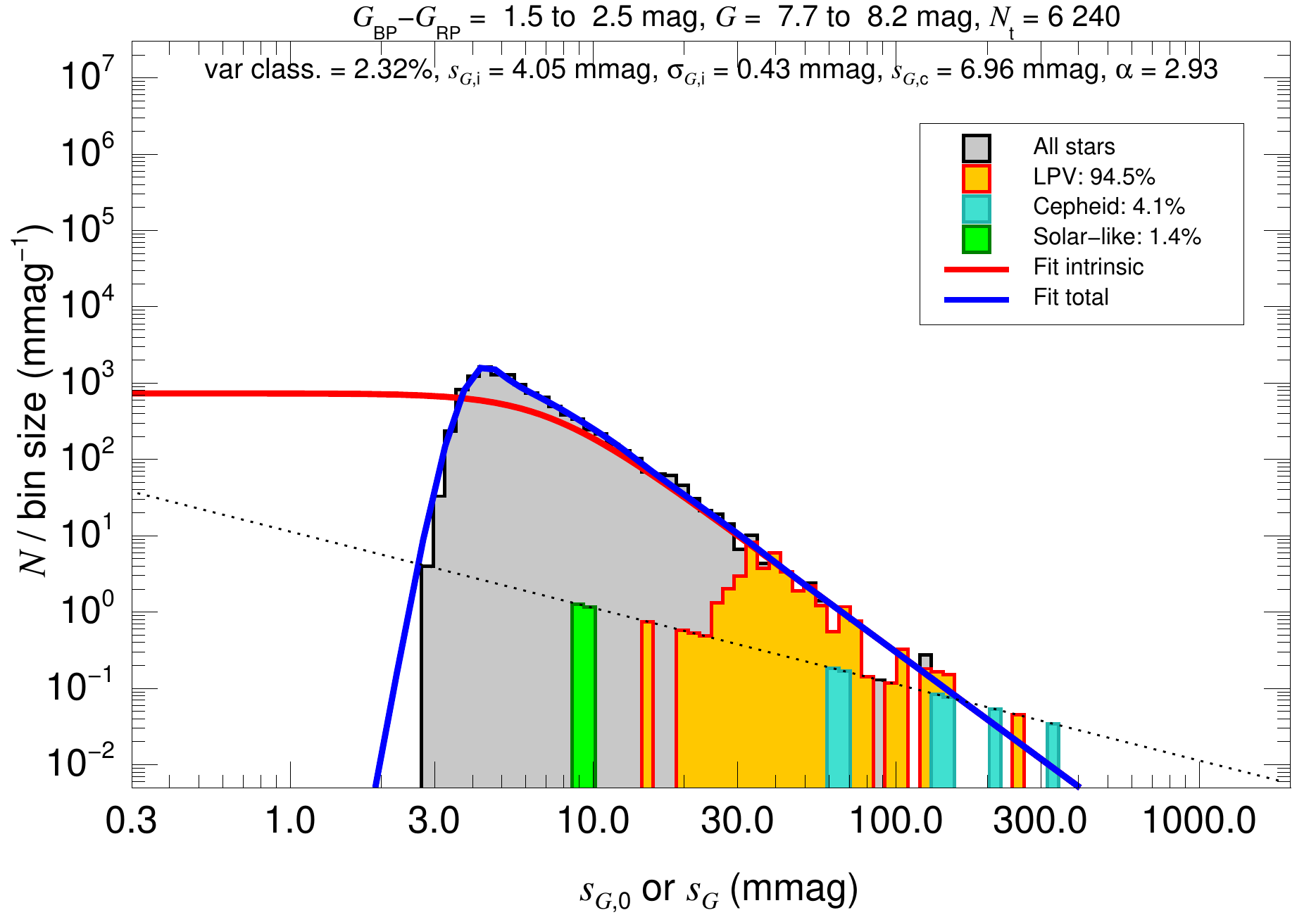}$\!\!\!$
                    \includegraphics[width=0.35\linewidth]{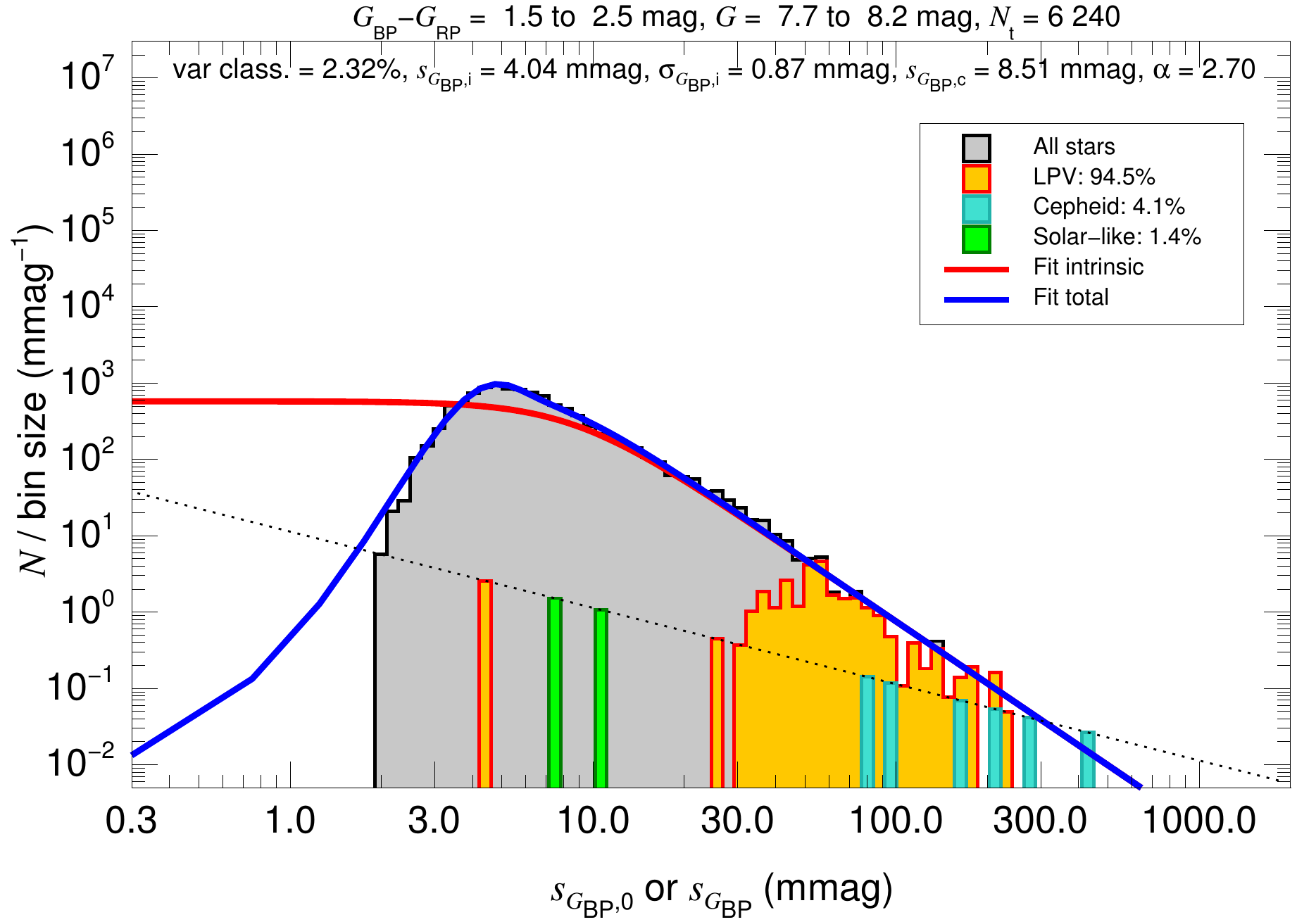}$\!\!\!$
                    \includegraphics[width=0.35\linewidth]{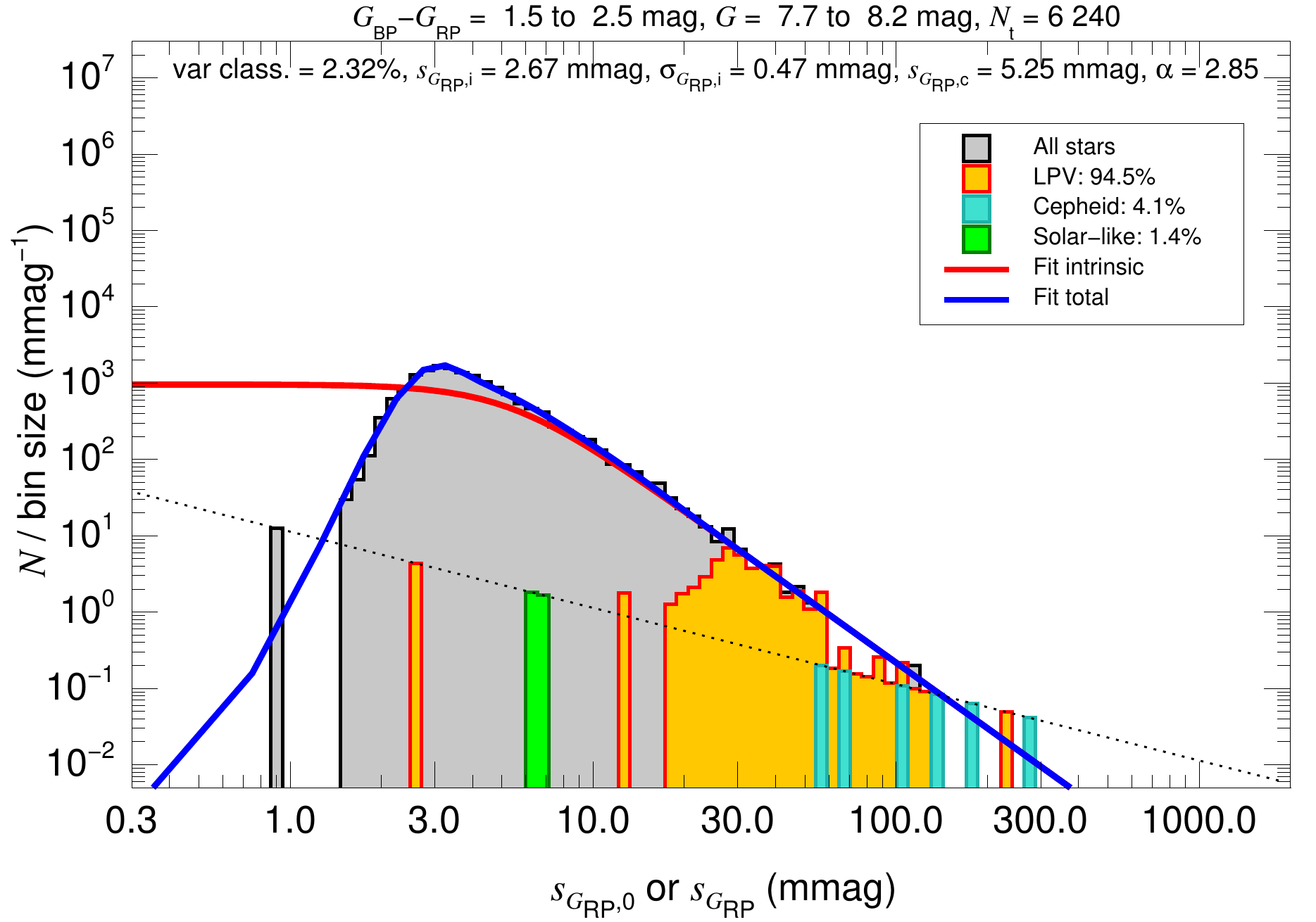}}
\centerline{$\!\!\!$\includegraphics[width=0.35\linewidth]{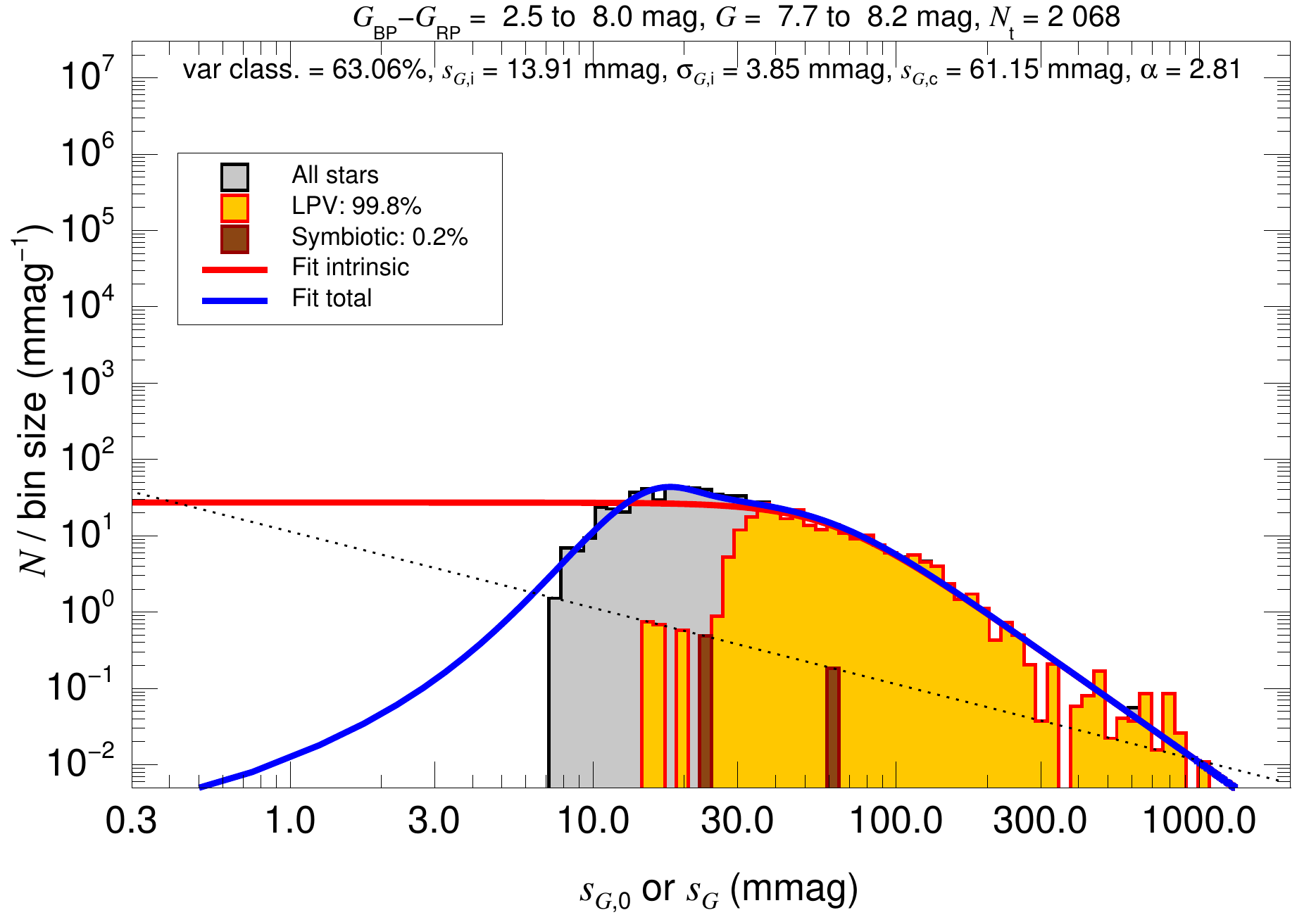}$\!\!\!$
                    \includegraphics[width=0.35\linewidth]{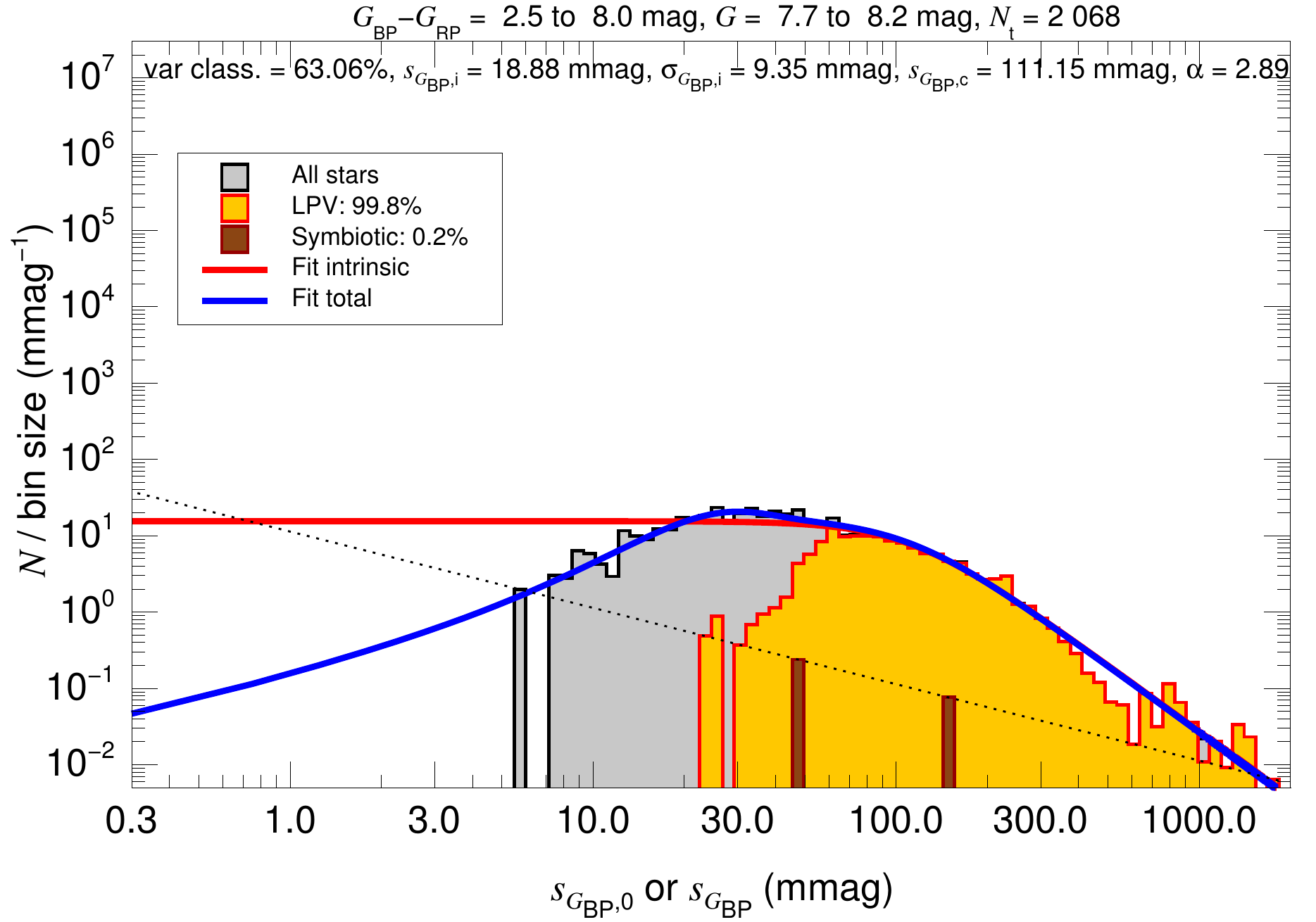}$\!\!\!$
                    \includegraphics[width=0.35\linewidth]{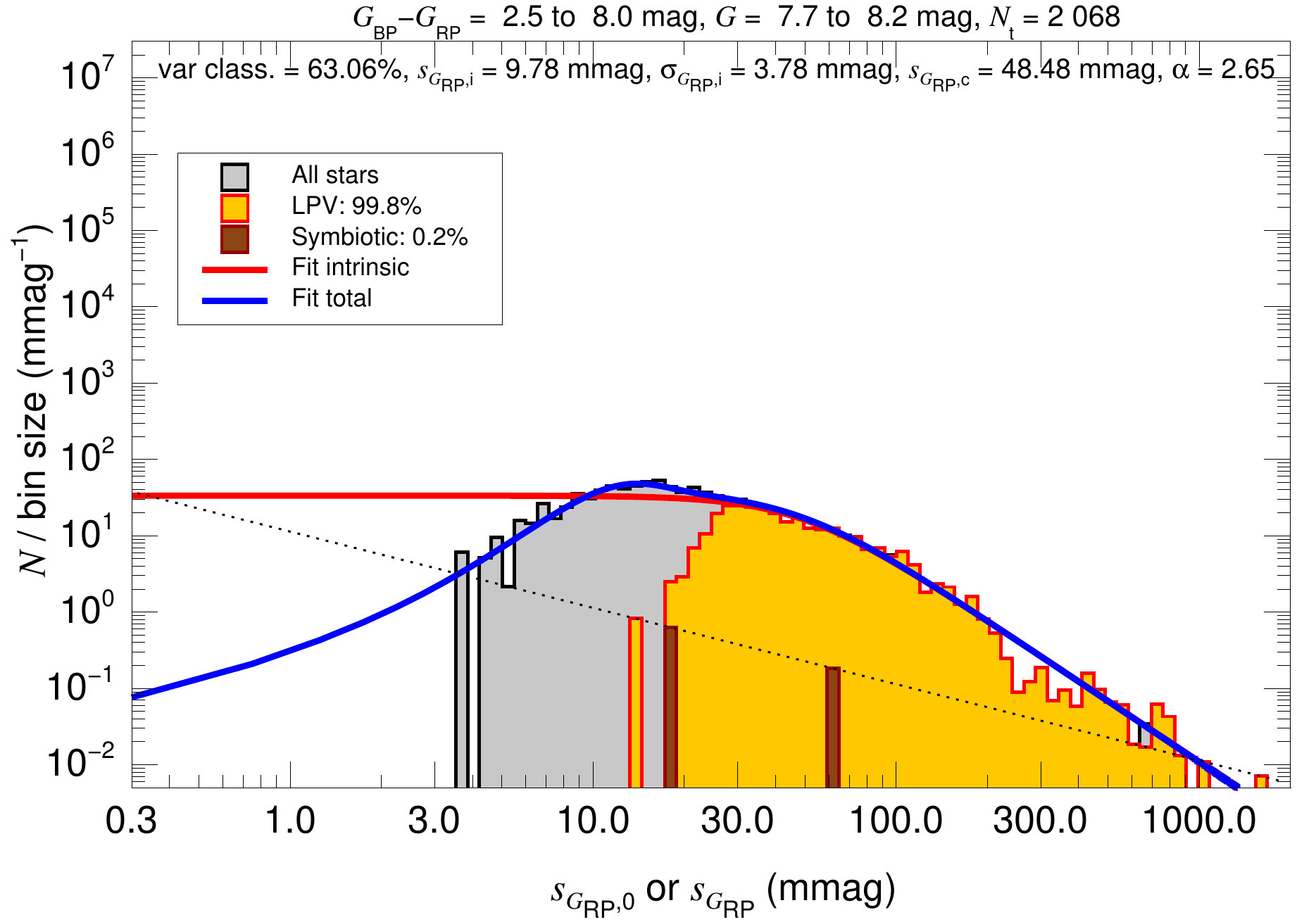}}
\caption{(Continued).}
\end{figure*}

\addtocounter{figure}{-1}

\begin{figure*}
\centerline{$\!\!\!$\includegraphics[width=0.35\linewidth]{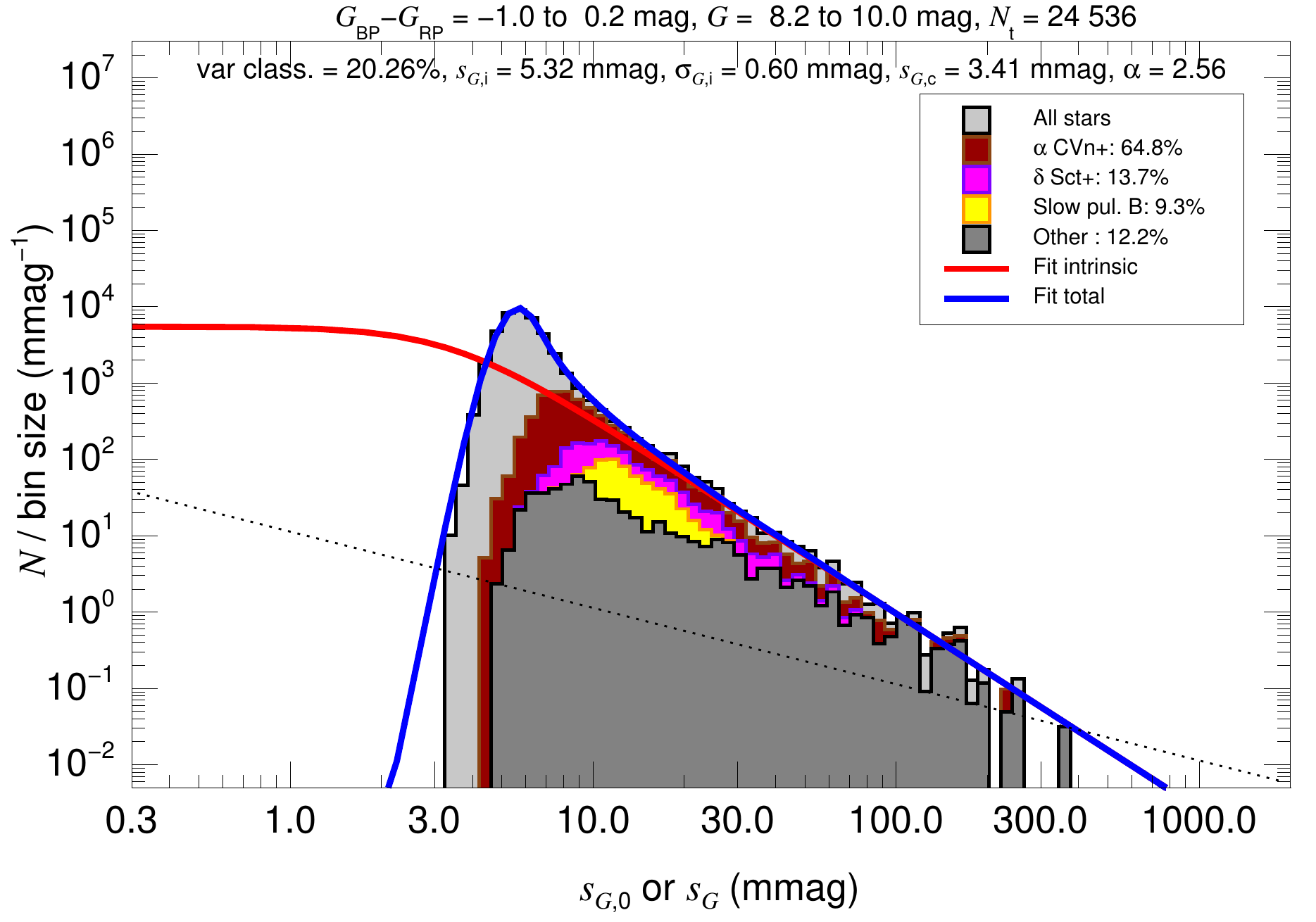}$\!\!\!$
                    \includegraphics[width=0.35\linewidth]{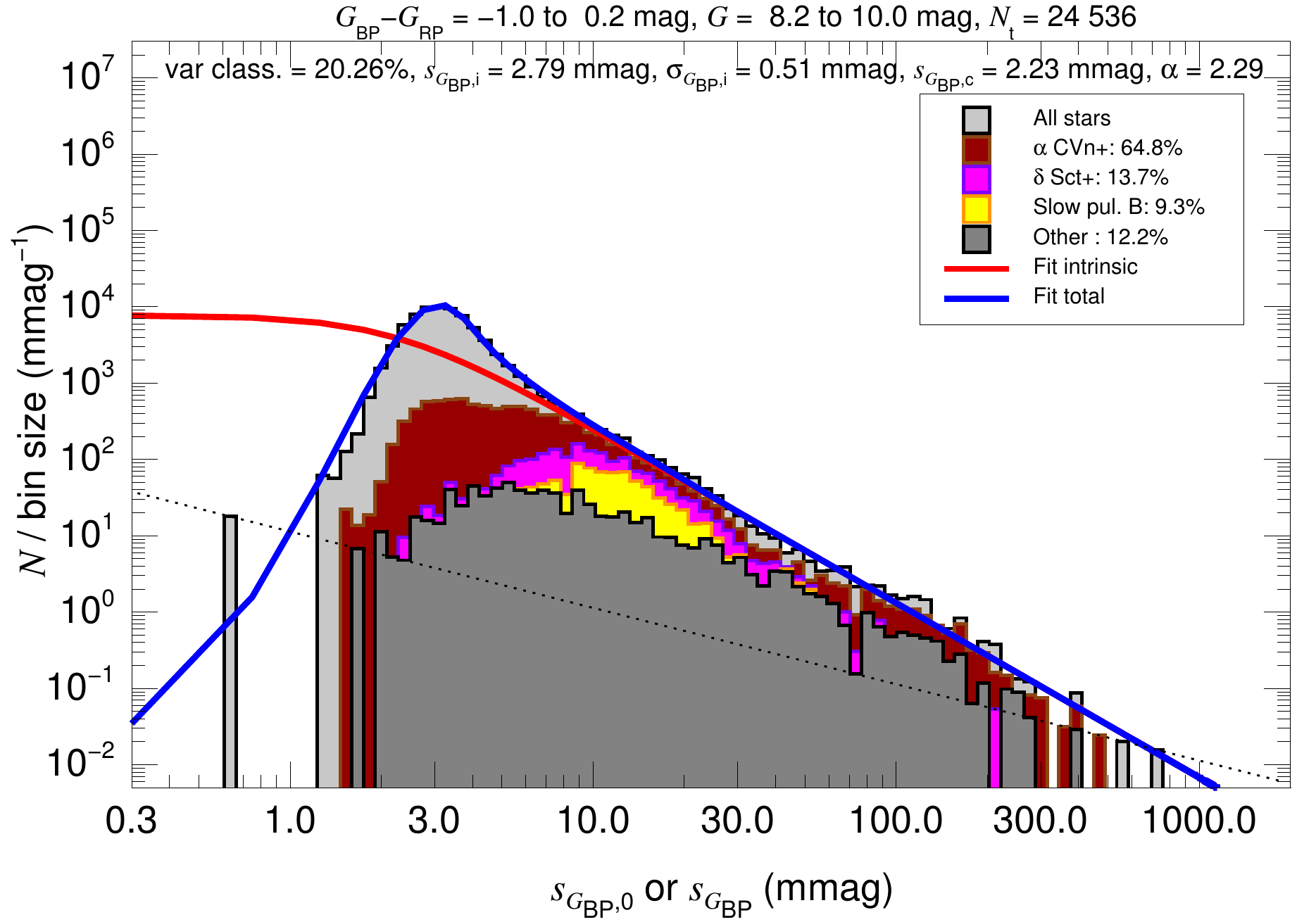}$\!\!\!$
                    \includegraphics[width=0.35\linewidth]{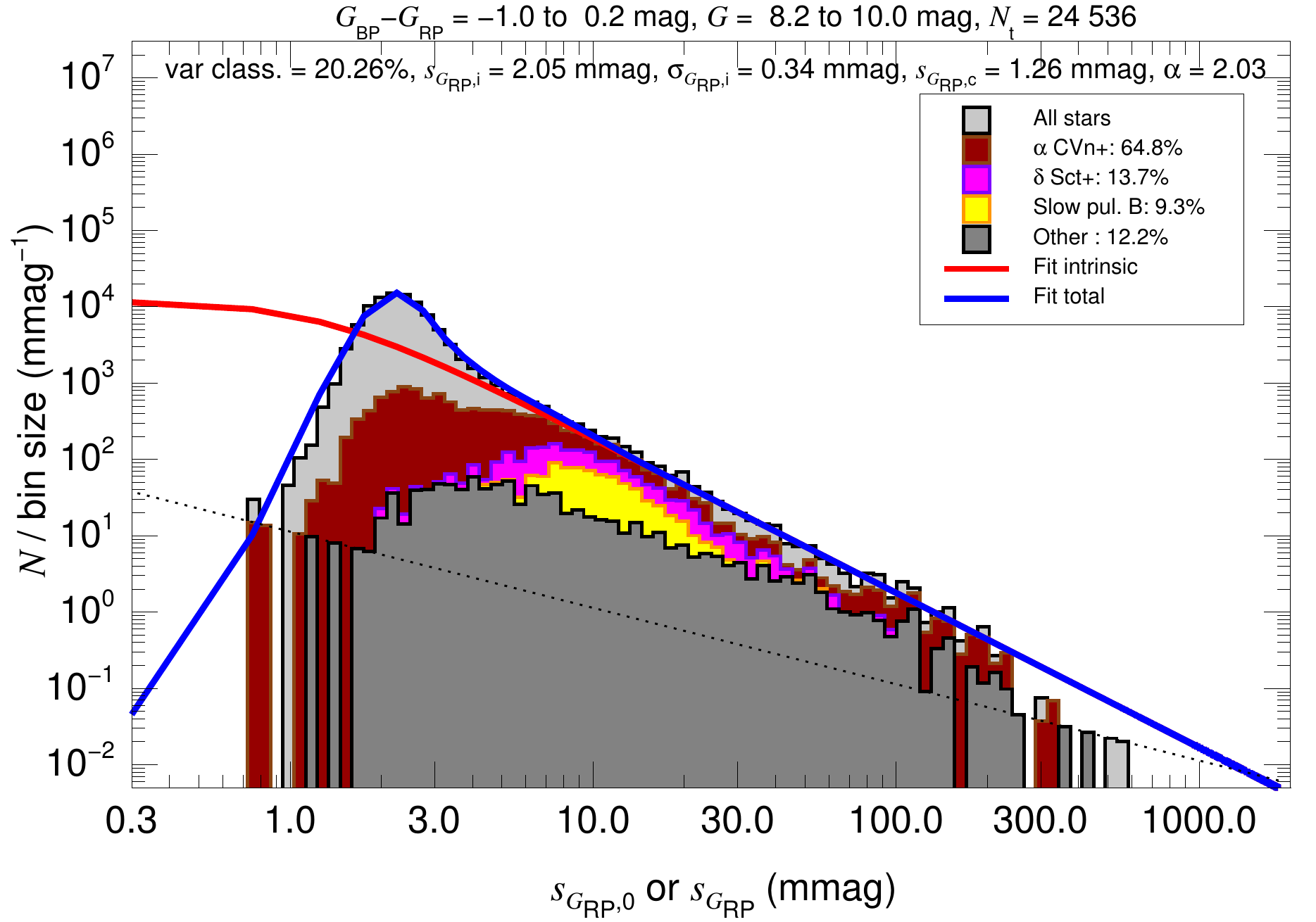}}
\centerline{$\!\!\!$\includegraphics[width=0.35\linewidth]{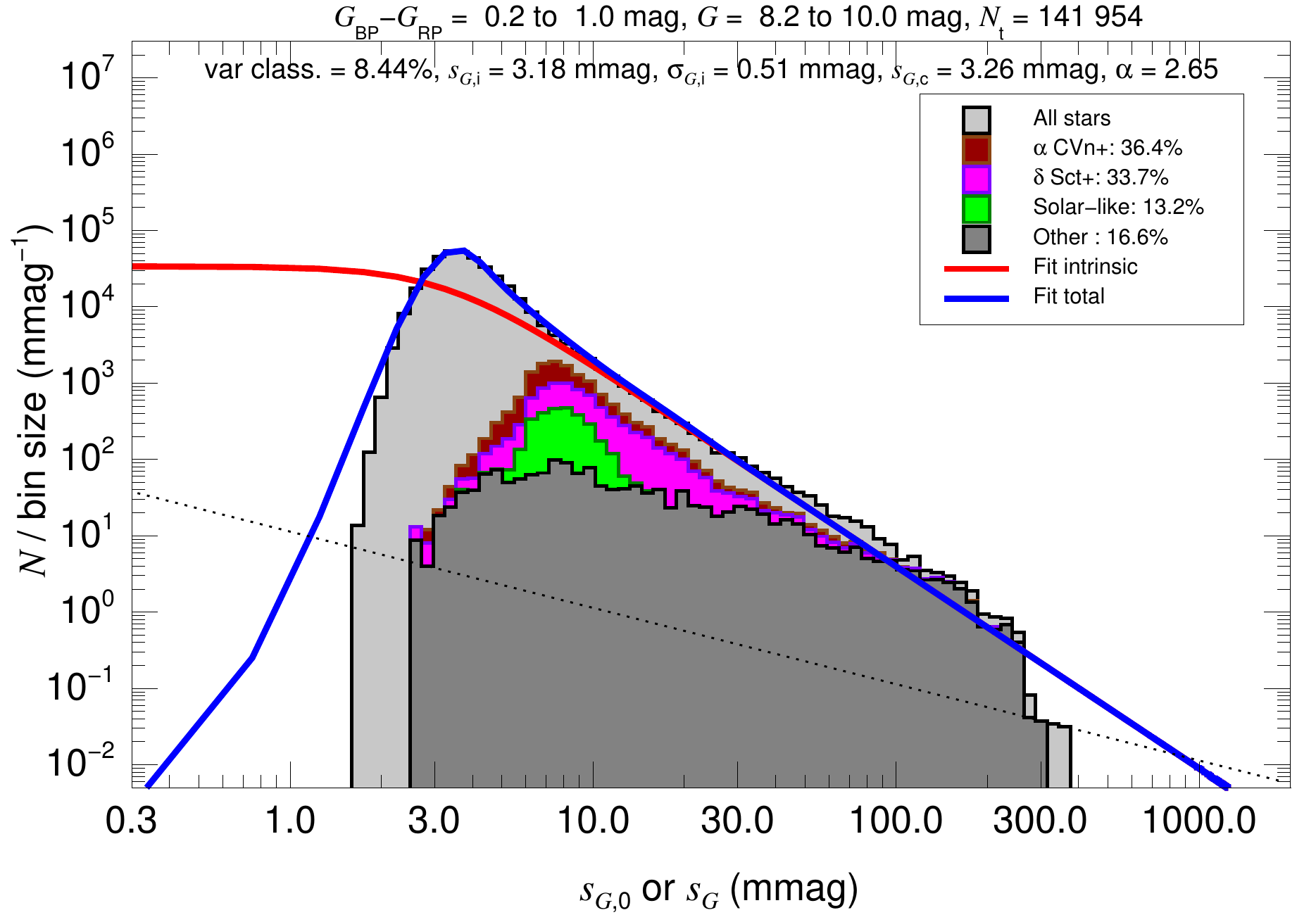}$\!\!\!$
                    \includegraphics[width=0.35\linewidth]{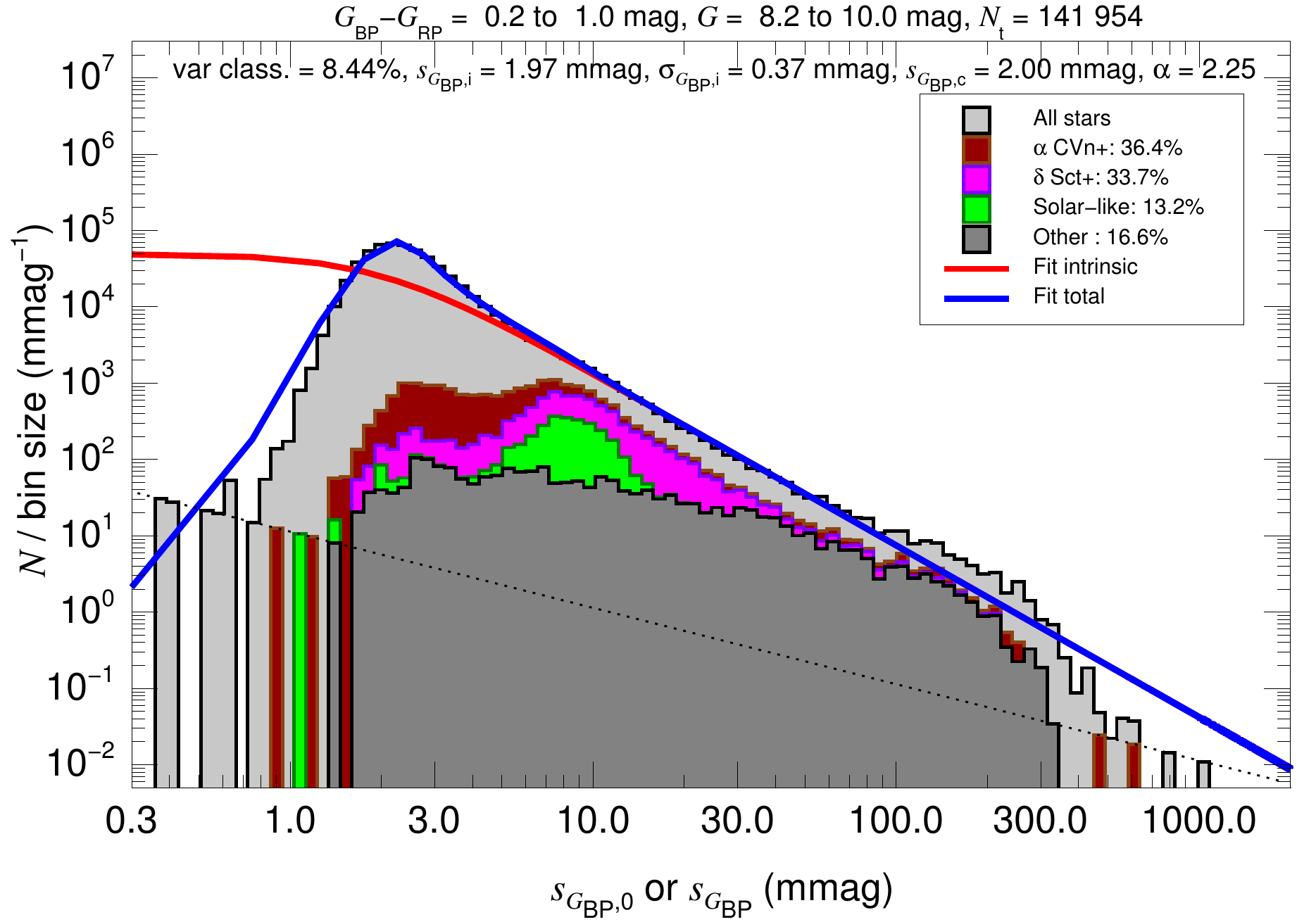}$\!\!\!$
                    \includegraphics[width=0.35\linewidth]{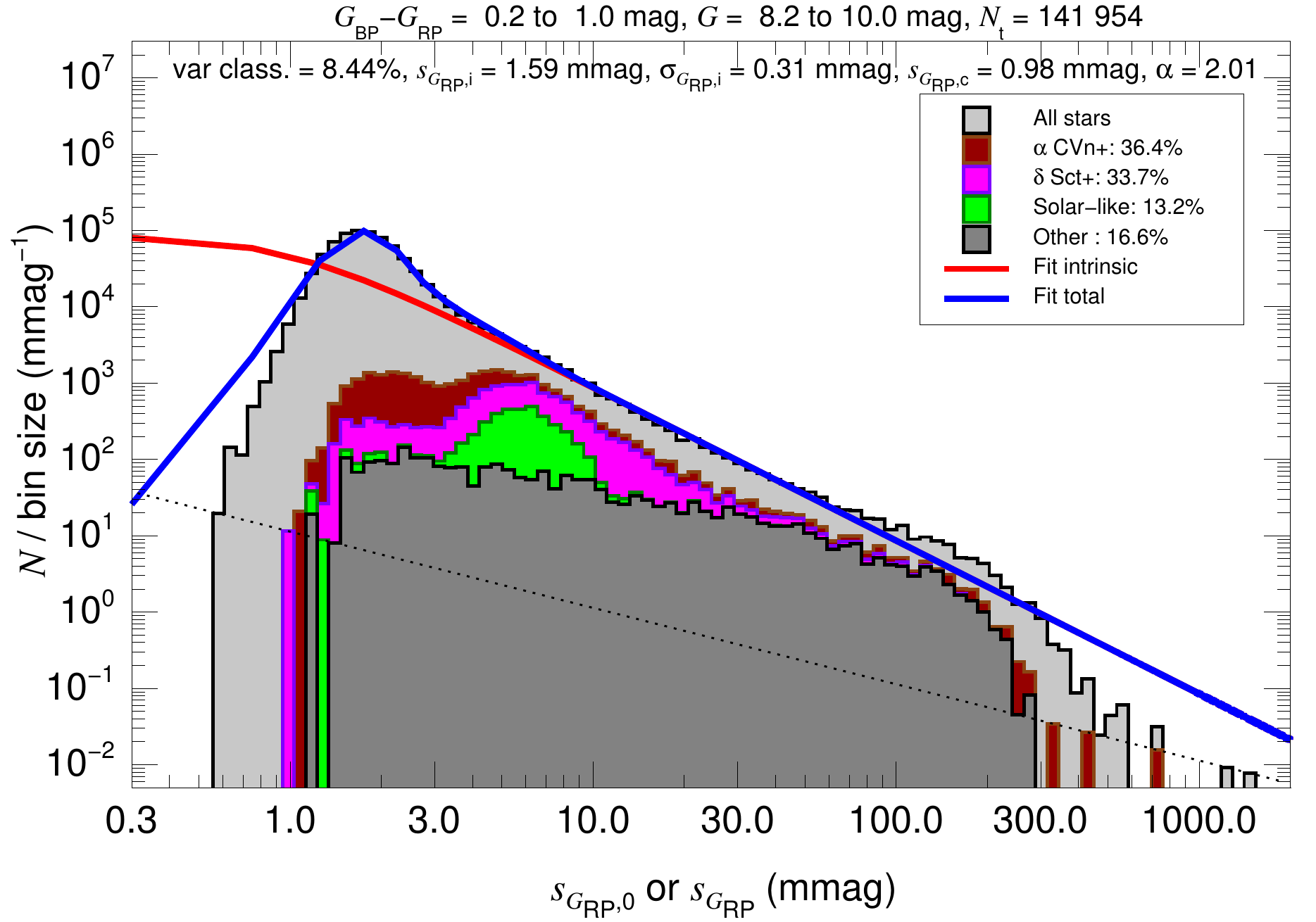}}
\centerline{$\!\!\!$\includegraphics[width=0.35\linewidth]{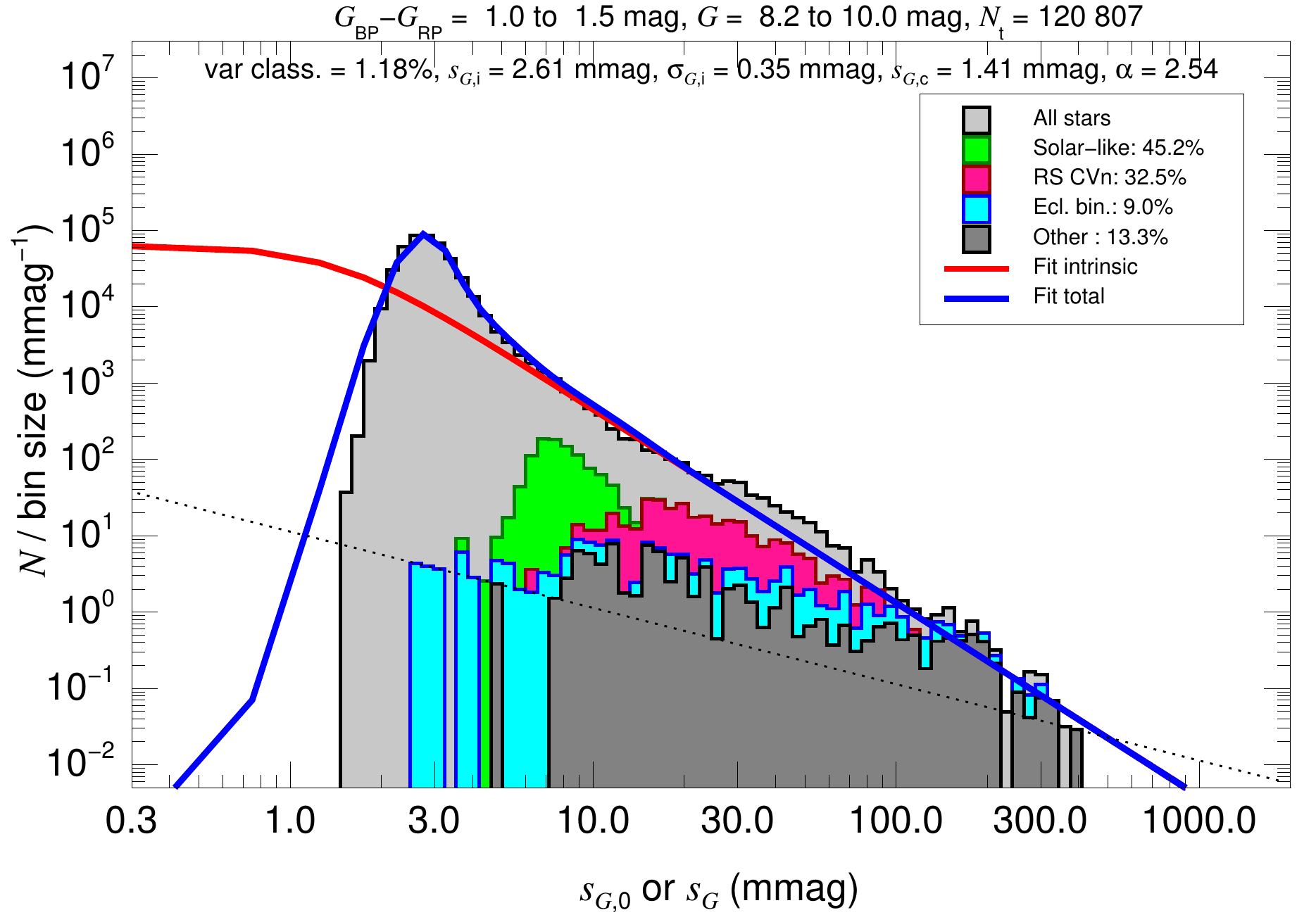}$\!\!\!$
                    \includegraphics[width=0.35\linewidth]{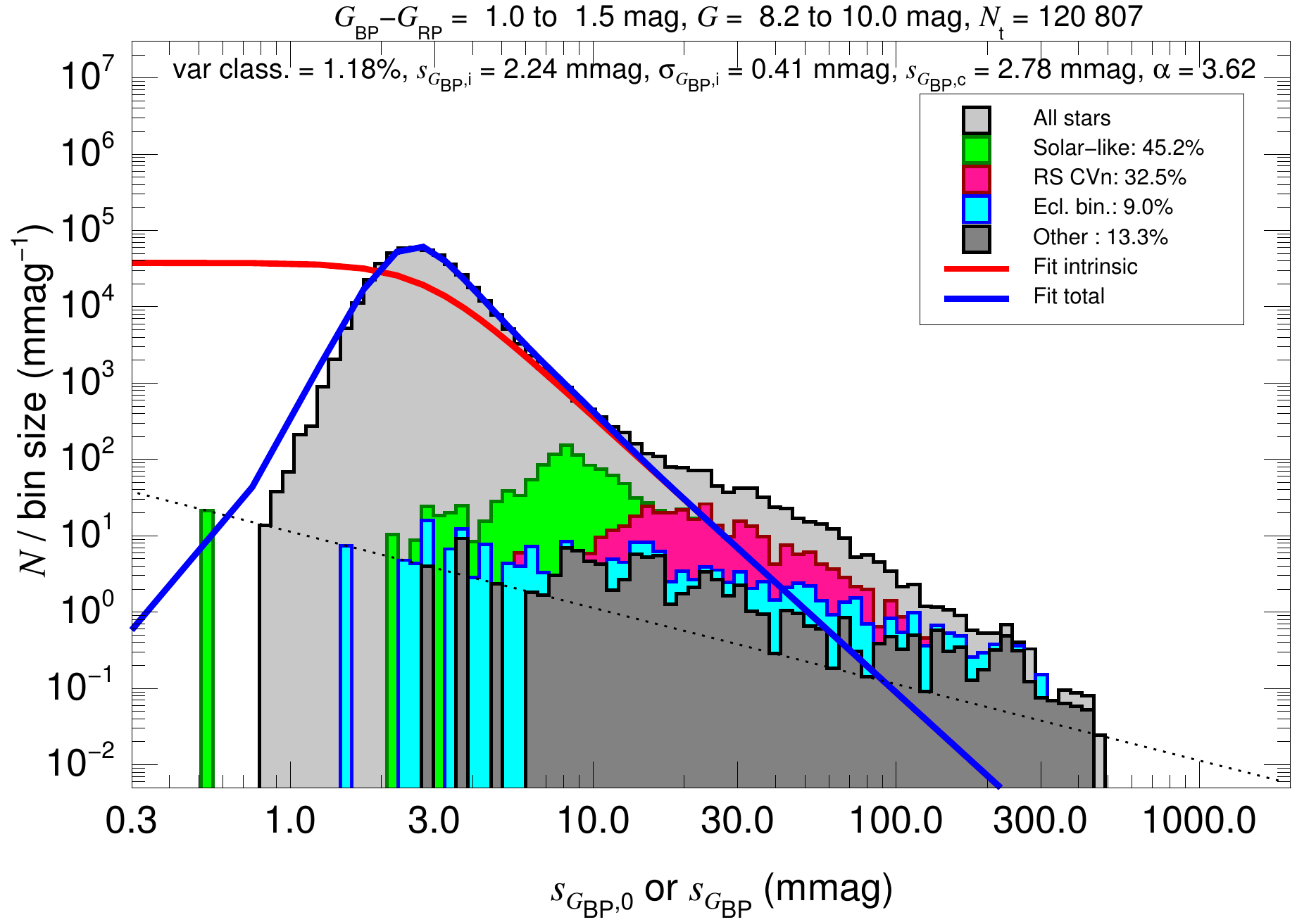}$\!\!\!$
                    \includegraphics[width=0.35\linewidth]{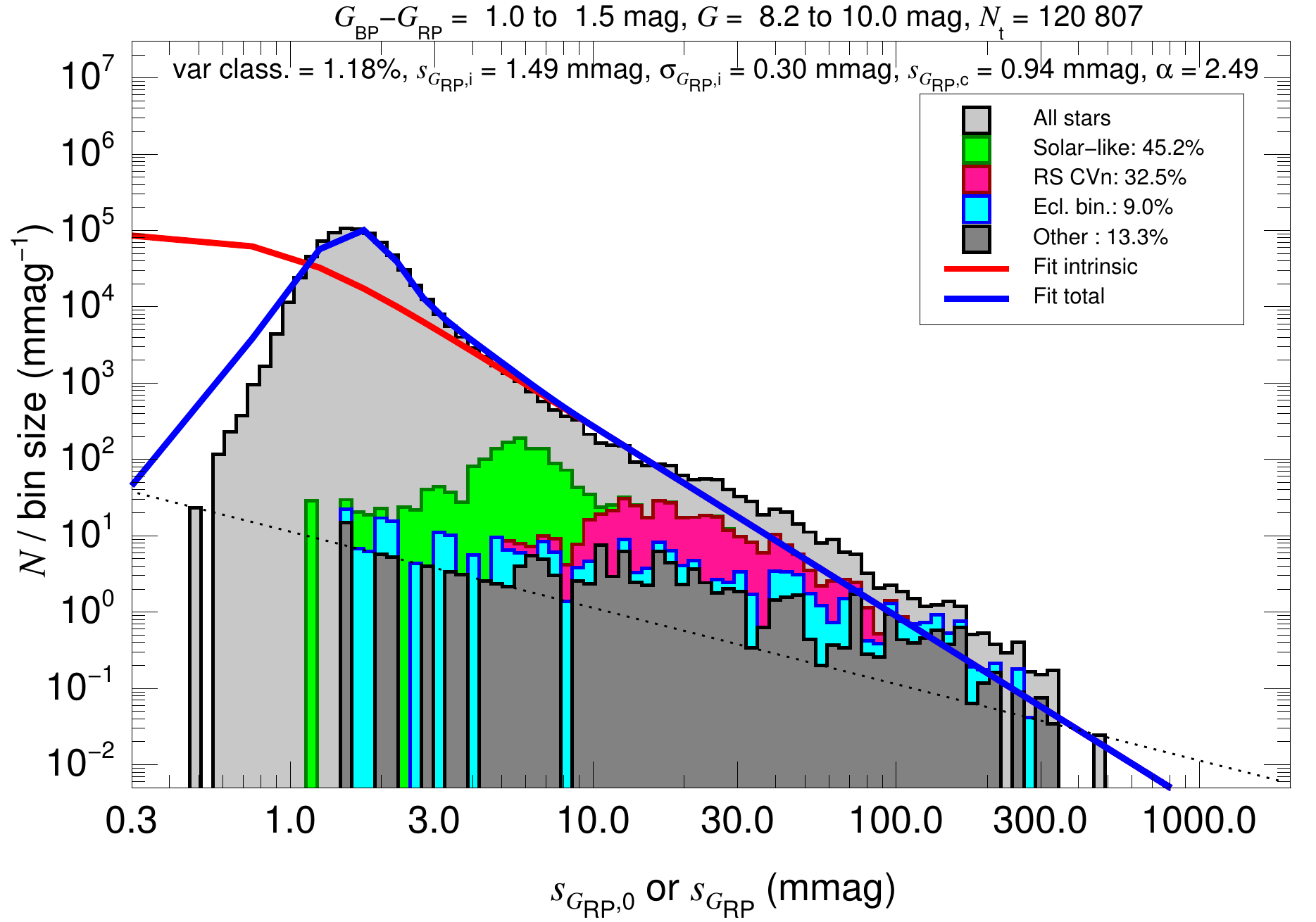}}
\centerline{$\!\!\!$\includegraphics[width=0.35\linewidth]{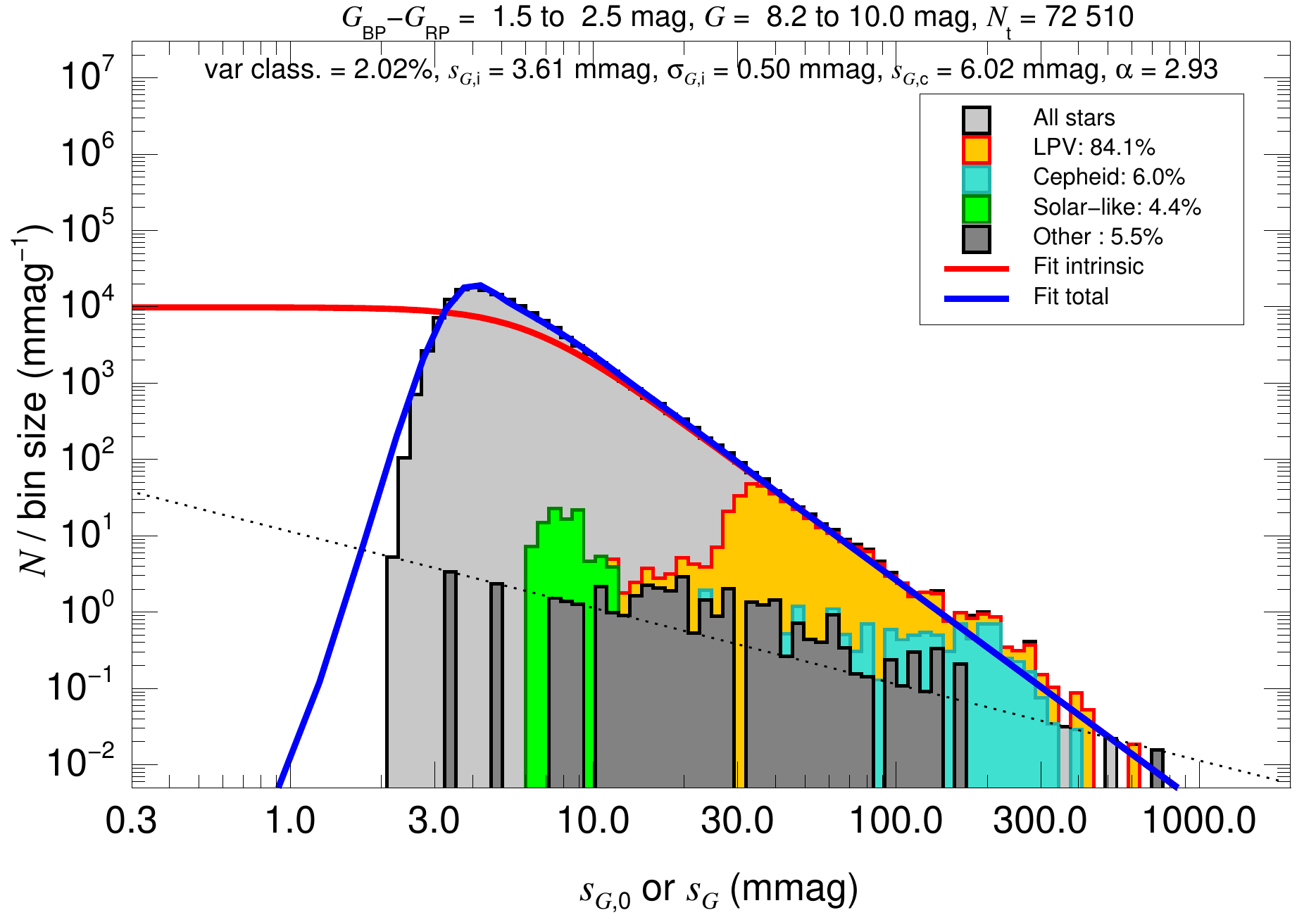}$\!\!\!$
                    \includegraphics[width=0.35\linewidth]{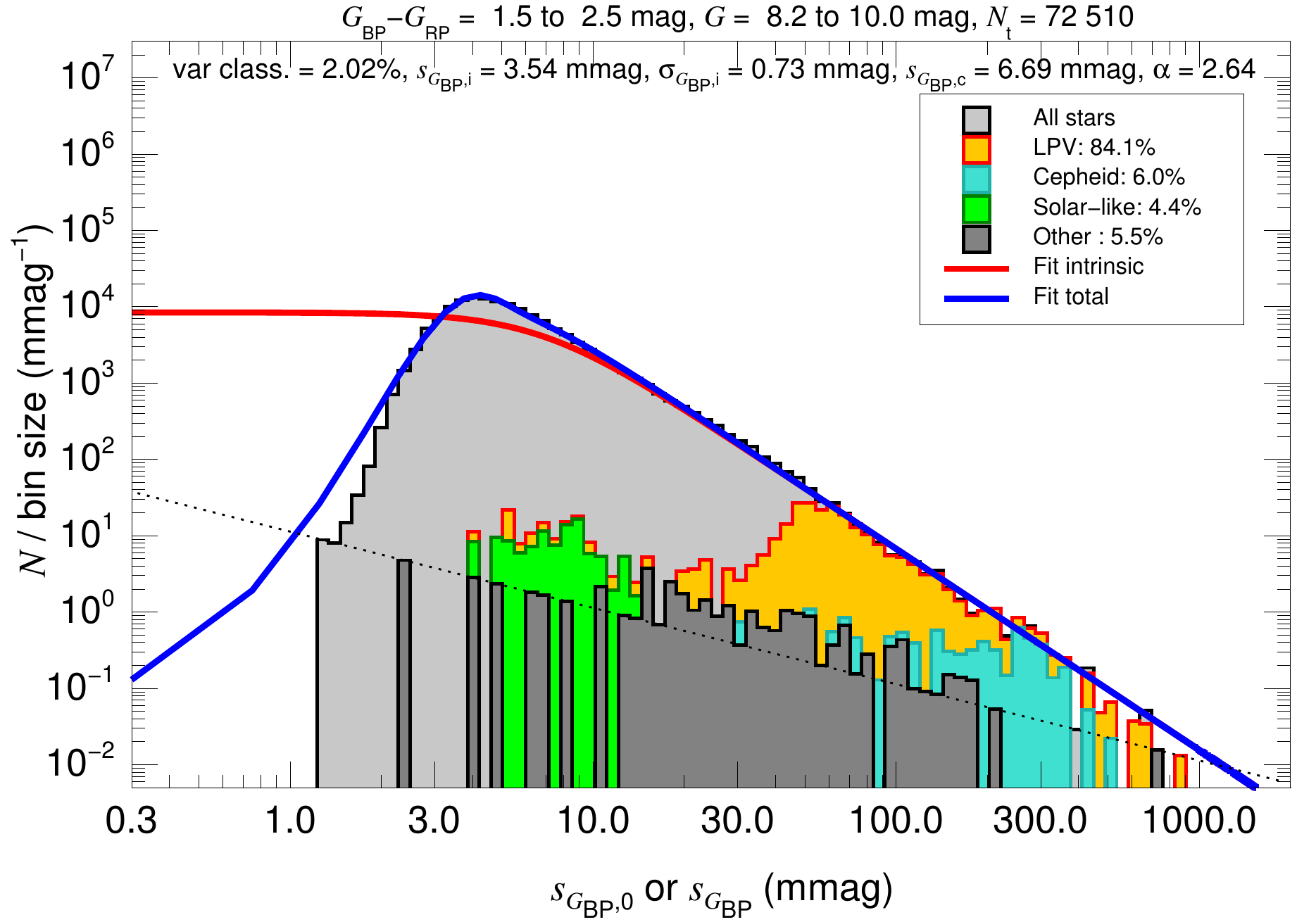}$\!\!\!$
                    \includegraphics[width=0.35\linewidth]{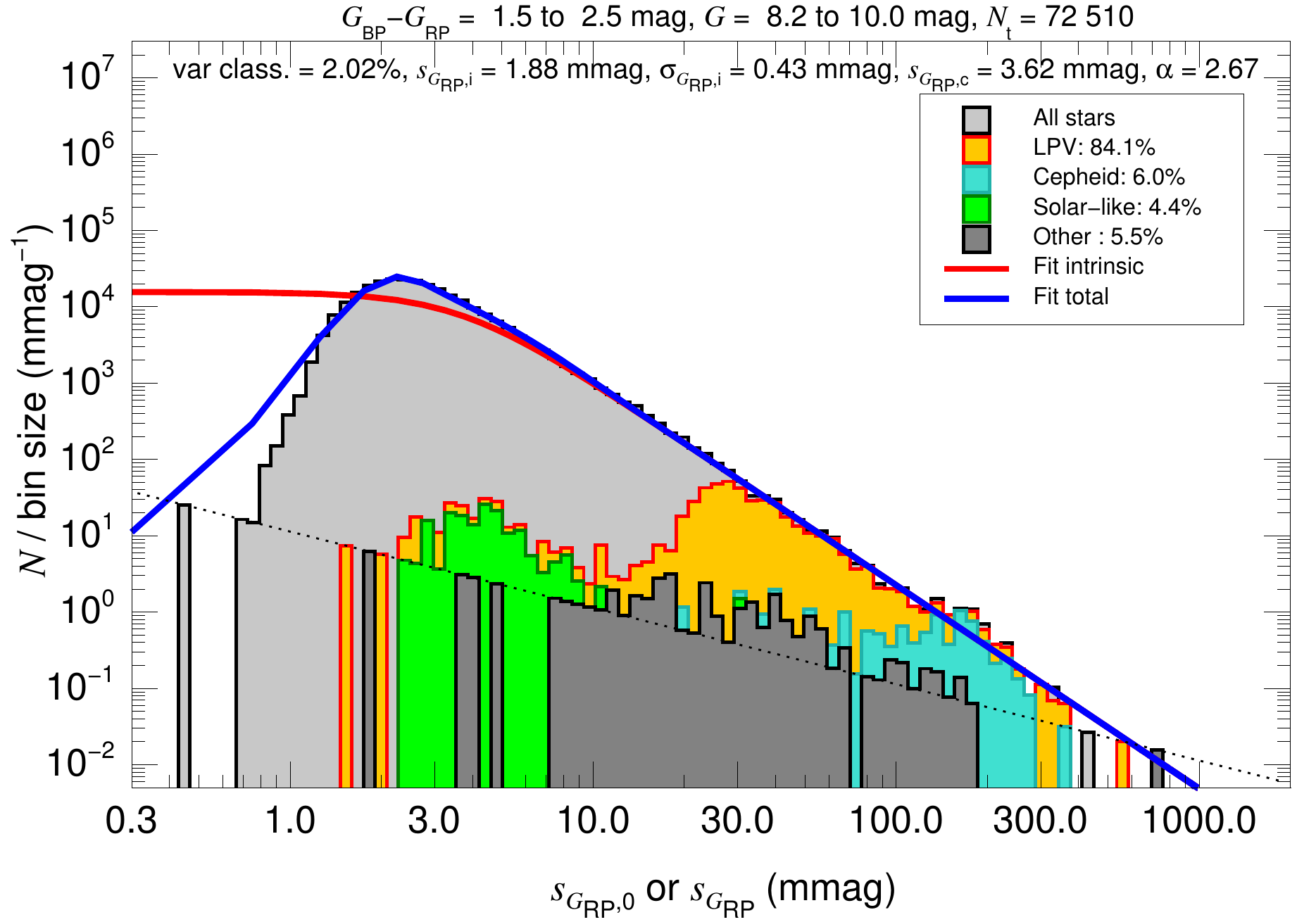}}
\centerline{$\!\!\!$\includegraphics[width=0.35\linewidth]{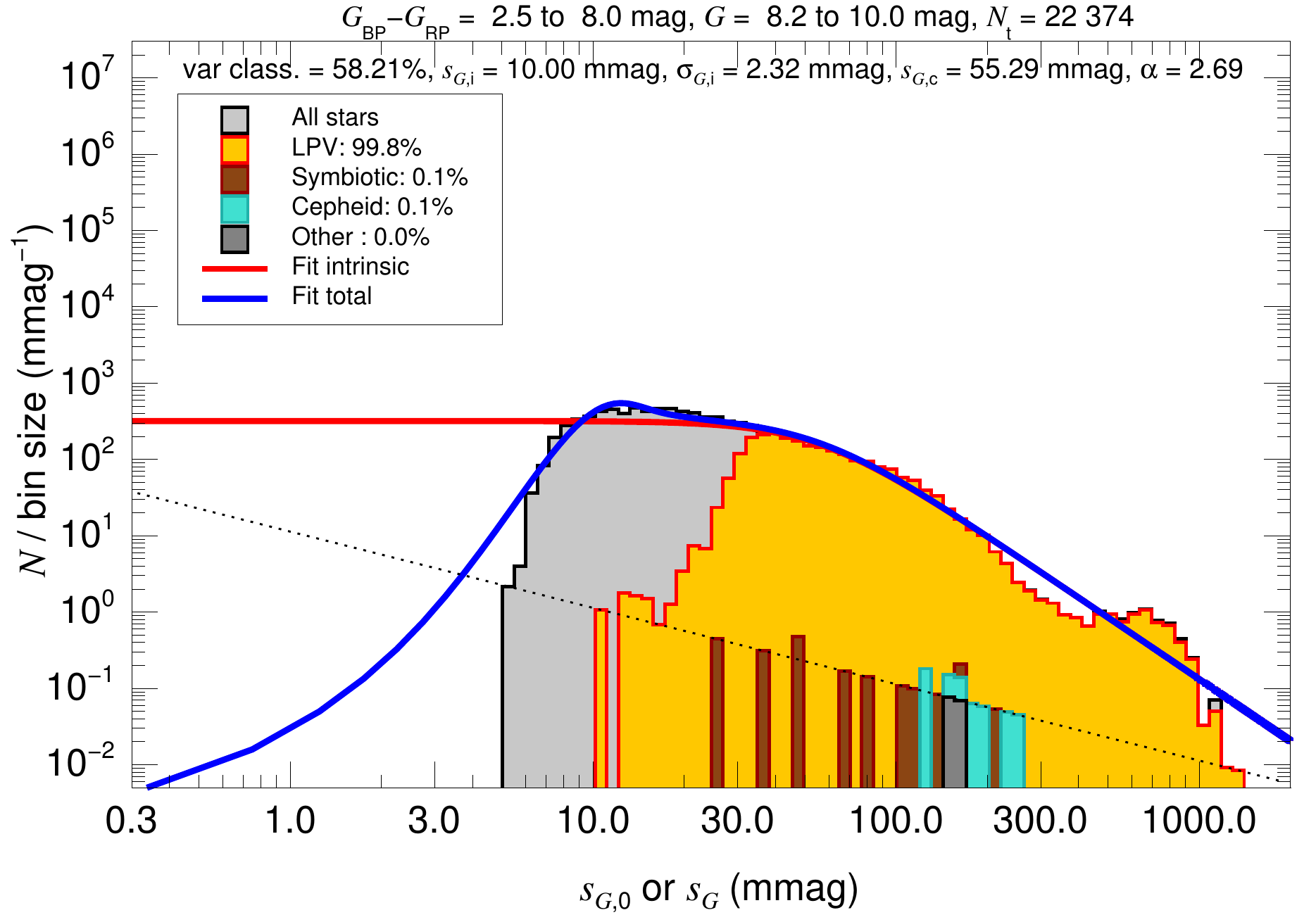}$\!\!\!$
                    \includegraphics[width=0.35\linewidth]{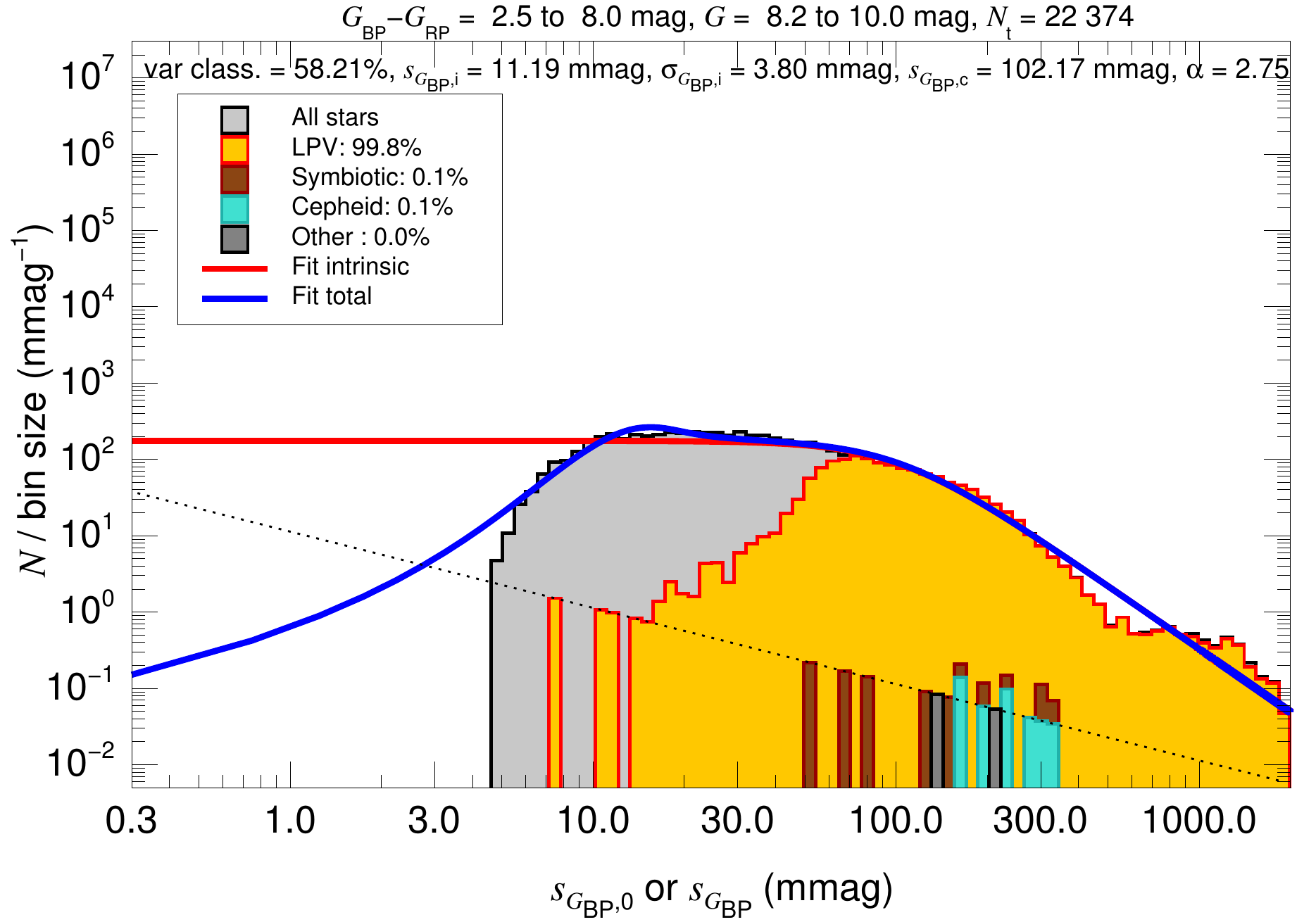}$\!\!\!$
                    \includegraphics[width=0.35\linewidth]{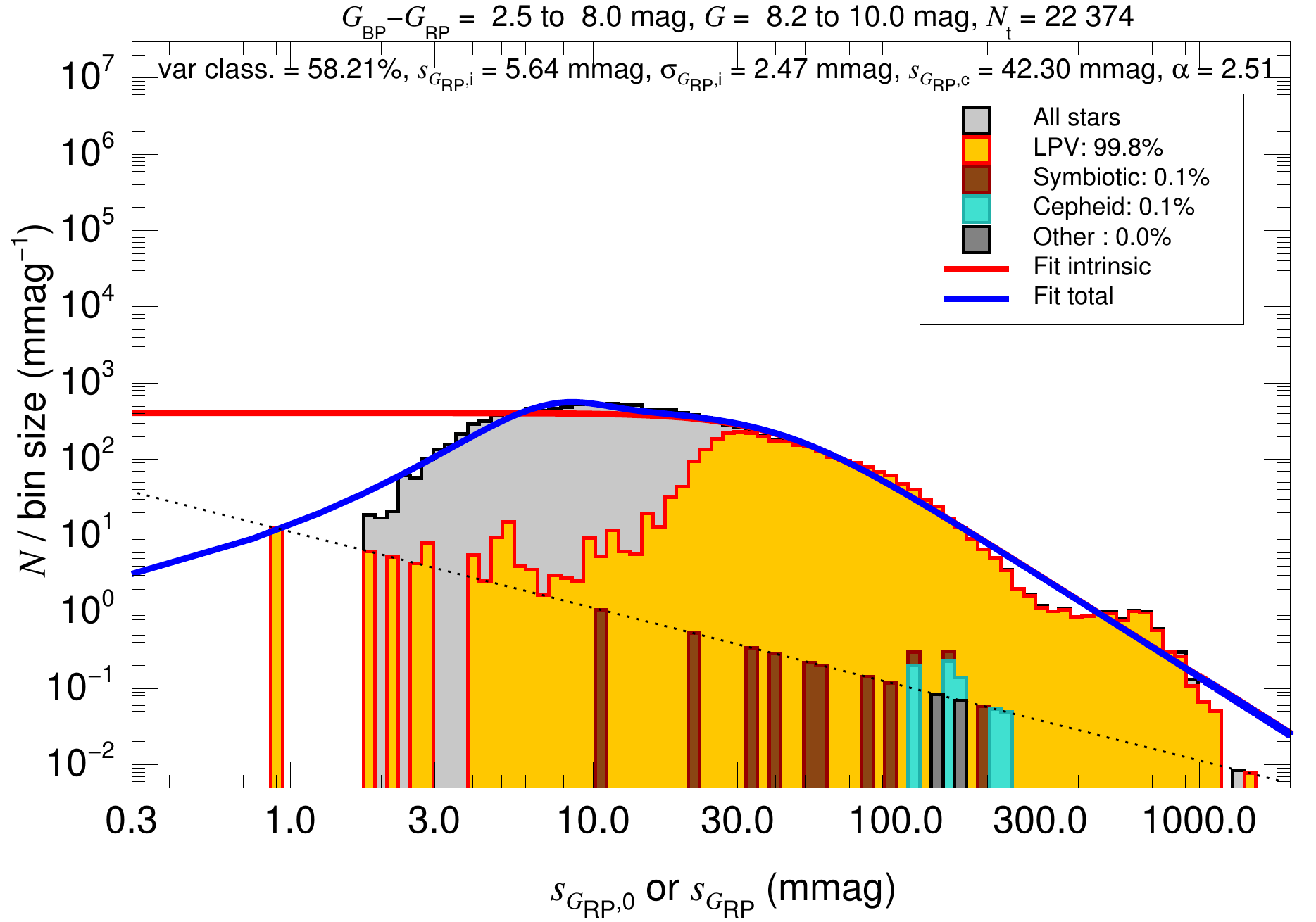}}
\caption{(Continued).}
\end{figure*}

\addtocounter{figure}{-1}

\begin{figure*}
\centerline{$\!\!\!$\includegraphics[width=0.35\linewidth]{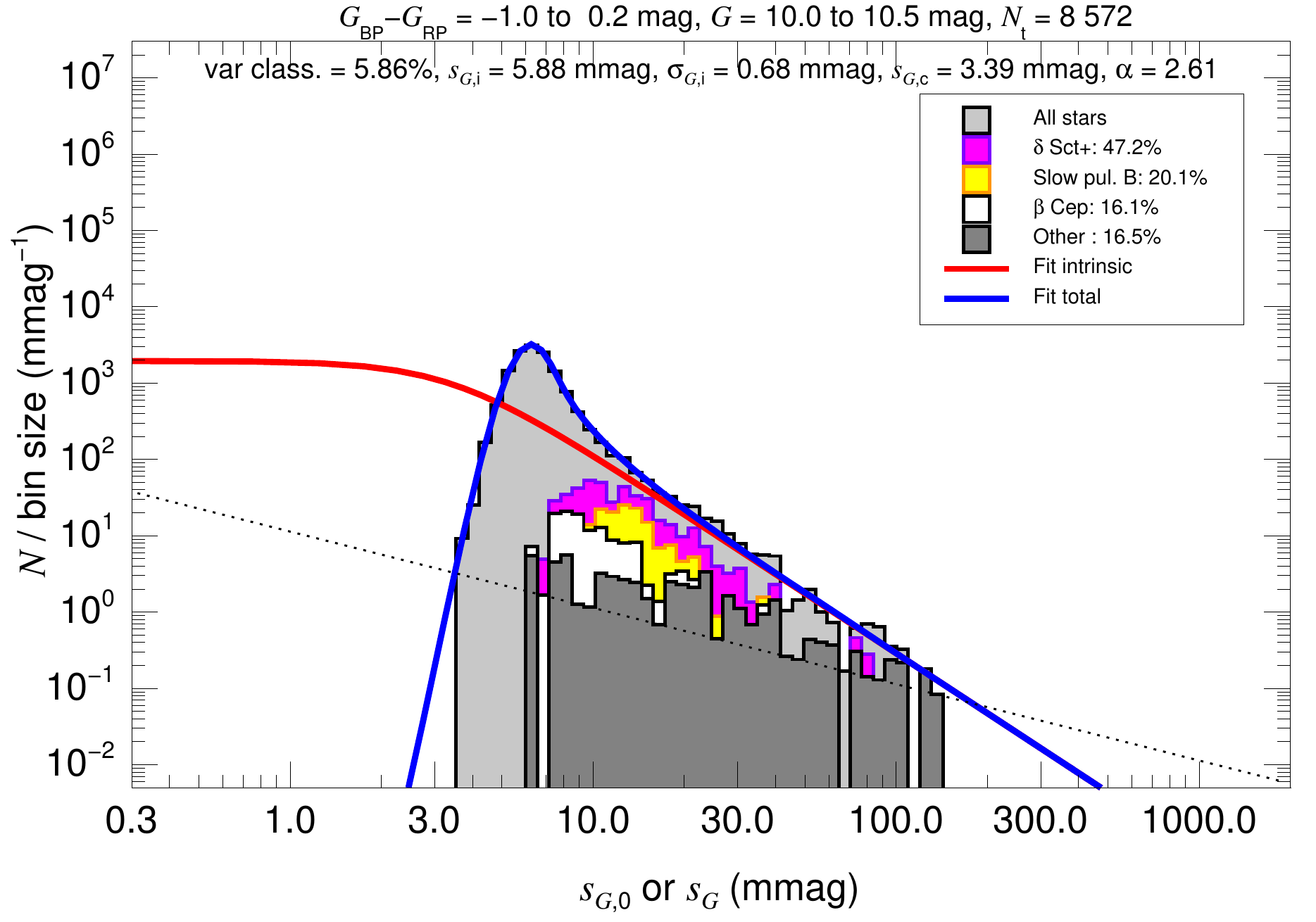}$\!\!\!$
                    \includegraphics[width=0.35\linewidth]{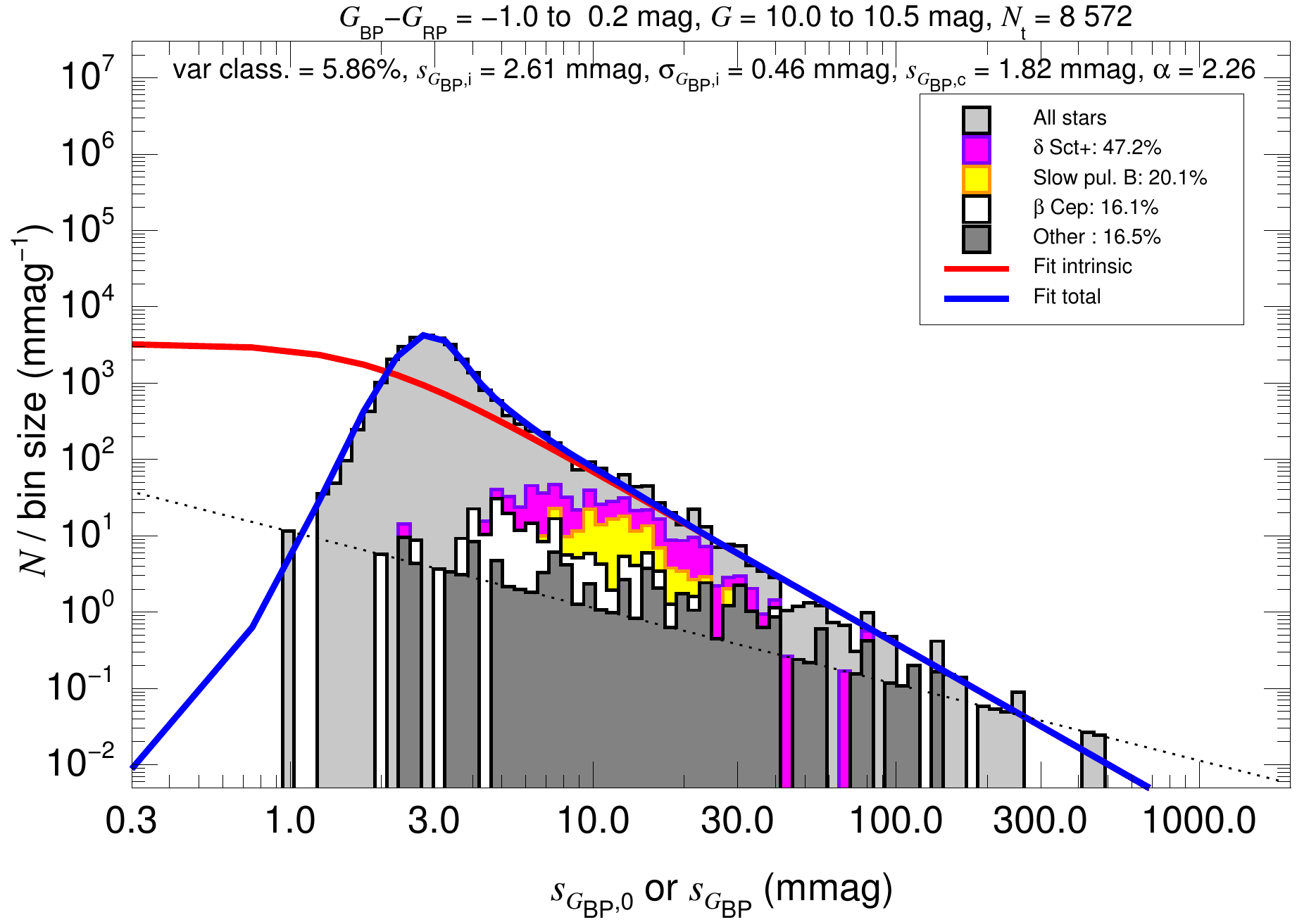}$\!\!\!$
                    \includegraphics[width=0.35\linewidth]{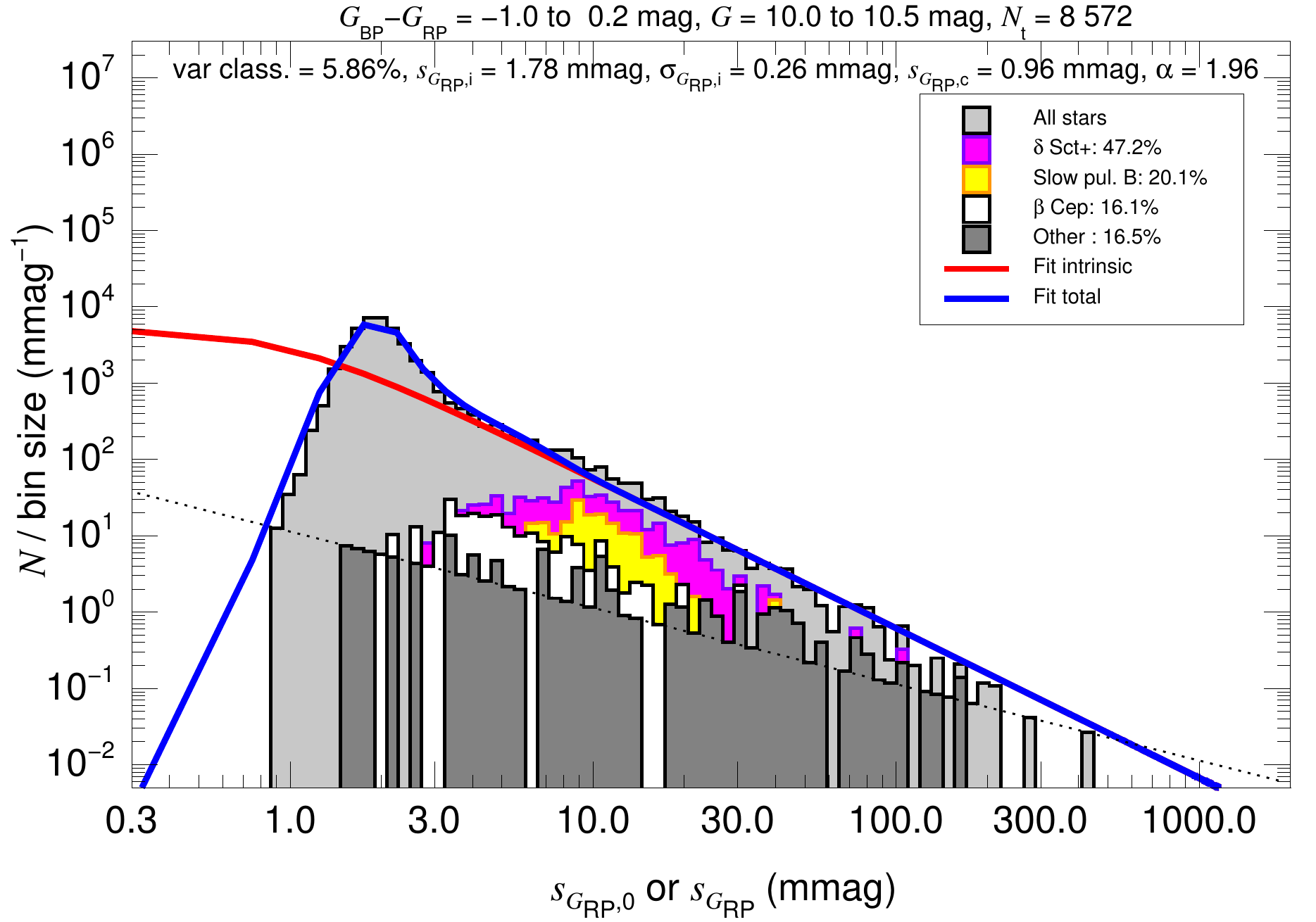}}
\centerline{$\!\!\!$\includegraphics[width=0.35\linewidth]{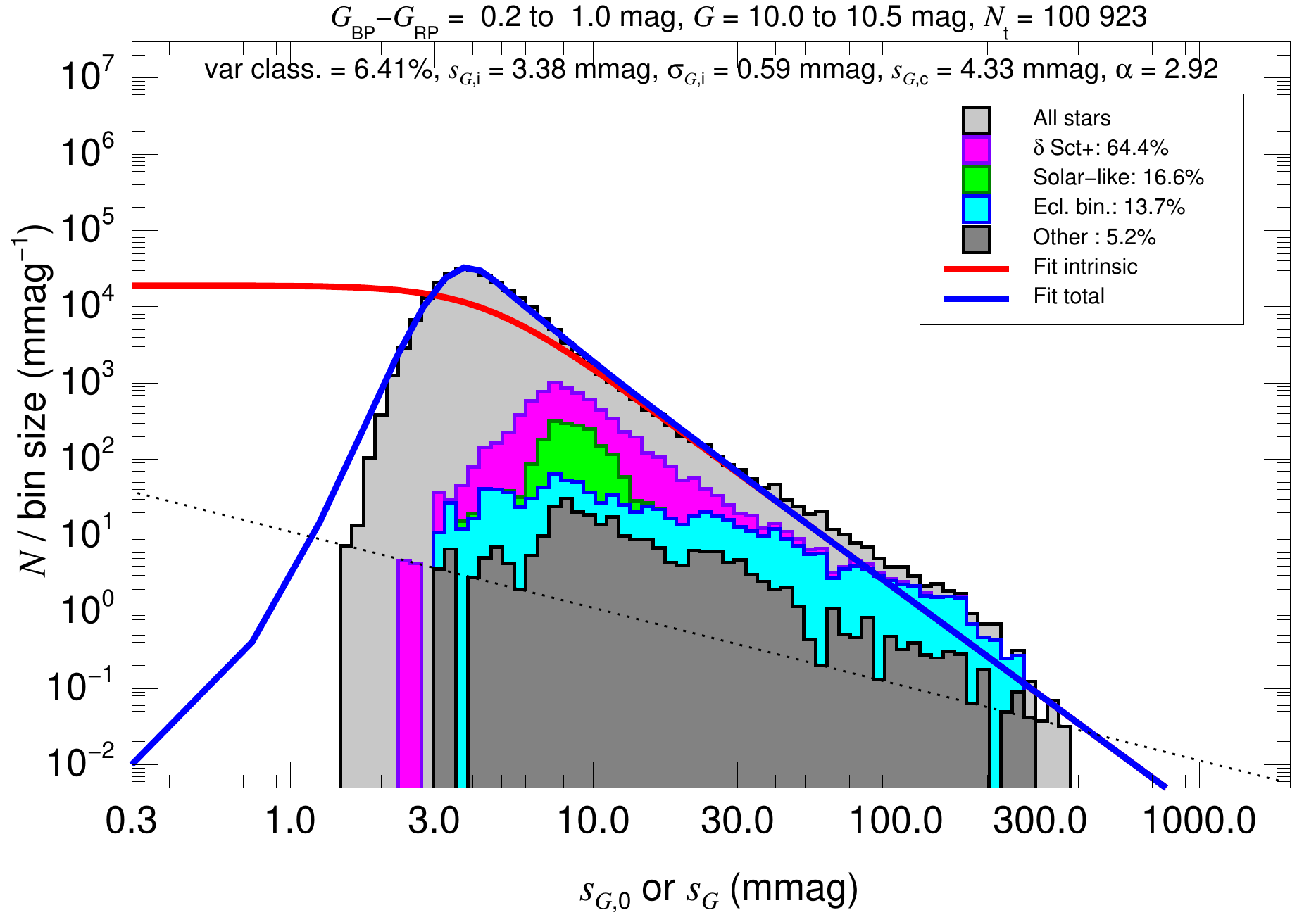}$\!\!\!$
                    \includegraphics[width=0.35\linewidth]{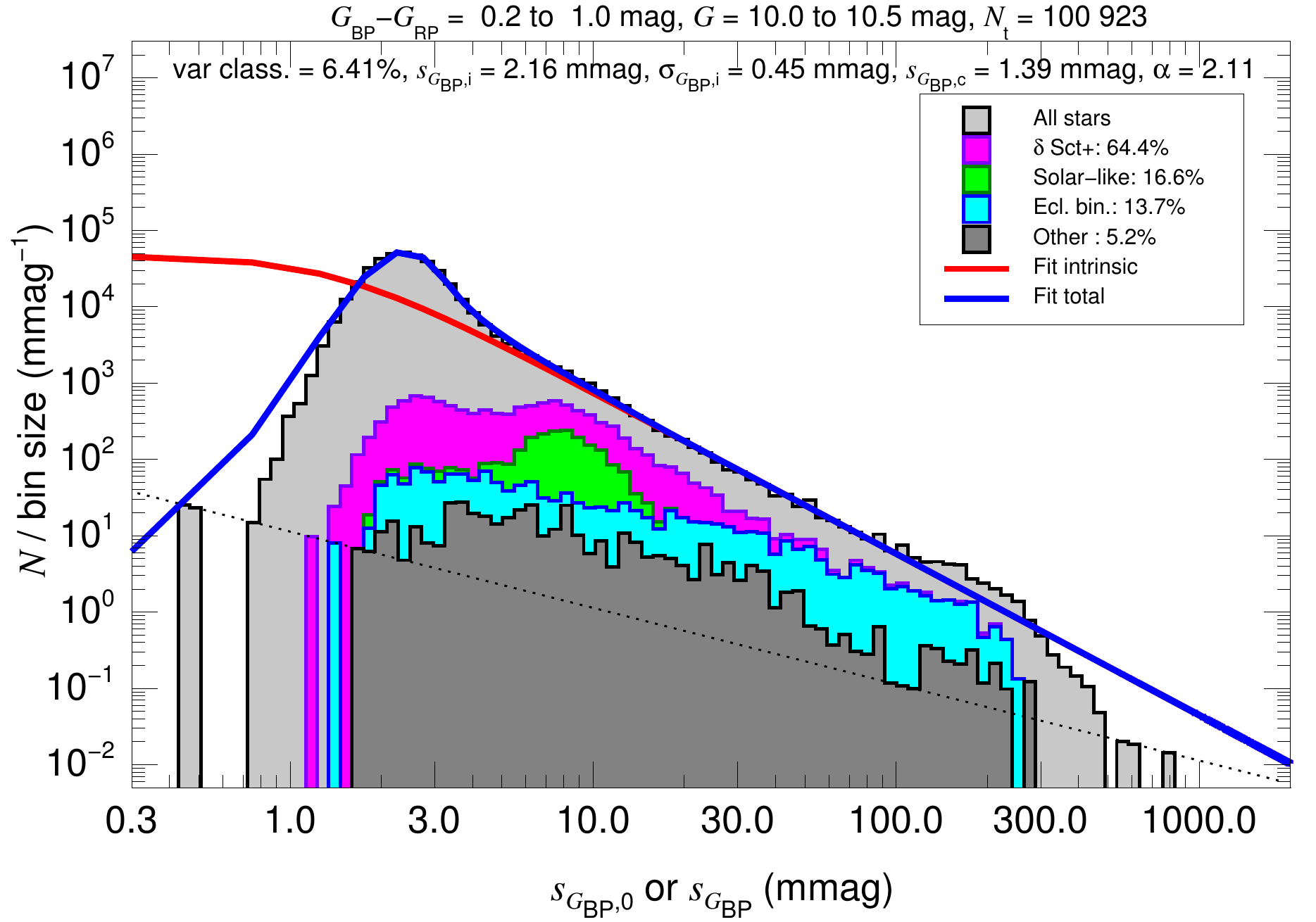}$\!\!\!$
                    \includegraphics[width=0.35\linewidth]{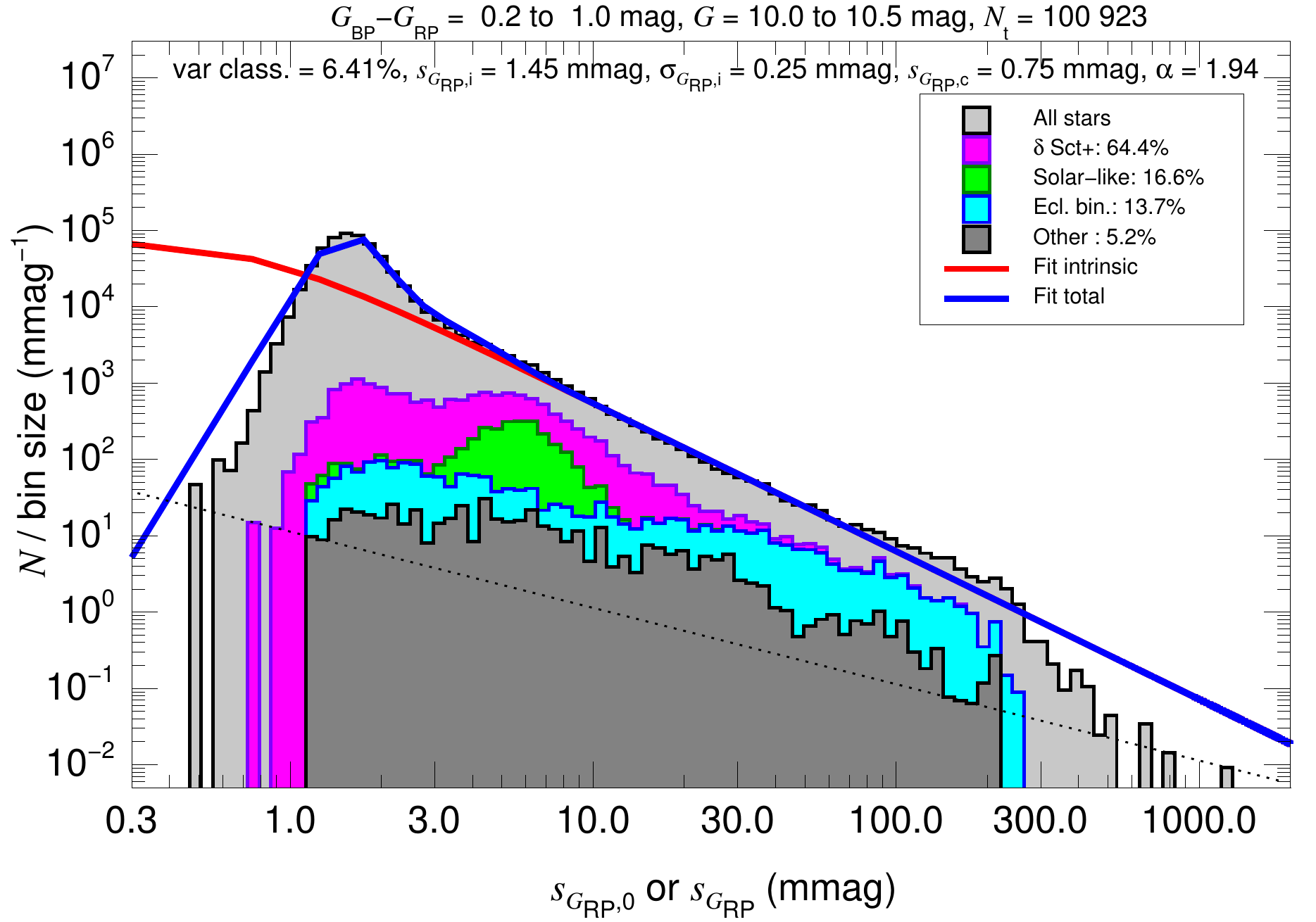}}
\centerline{$\!\!\!$\includegraphics[width=0.35\linewidth]{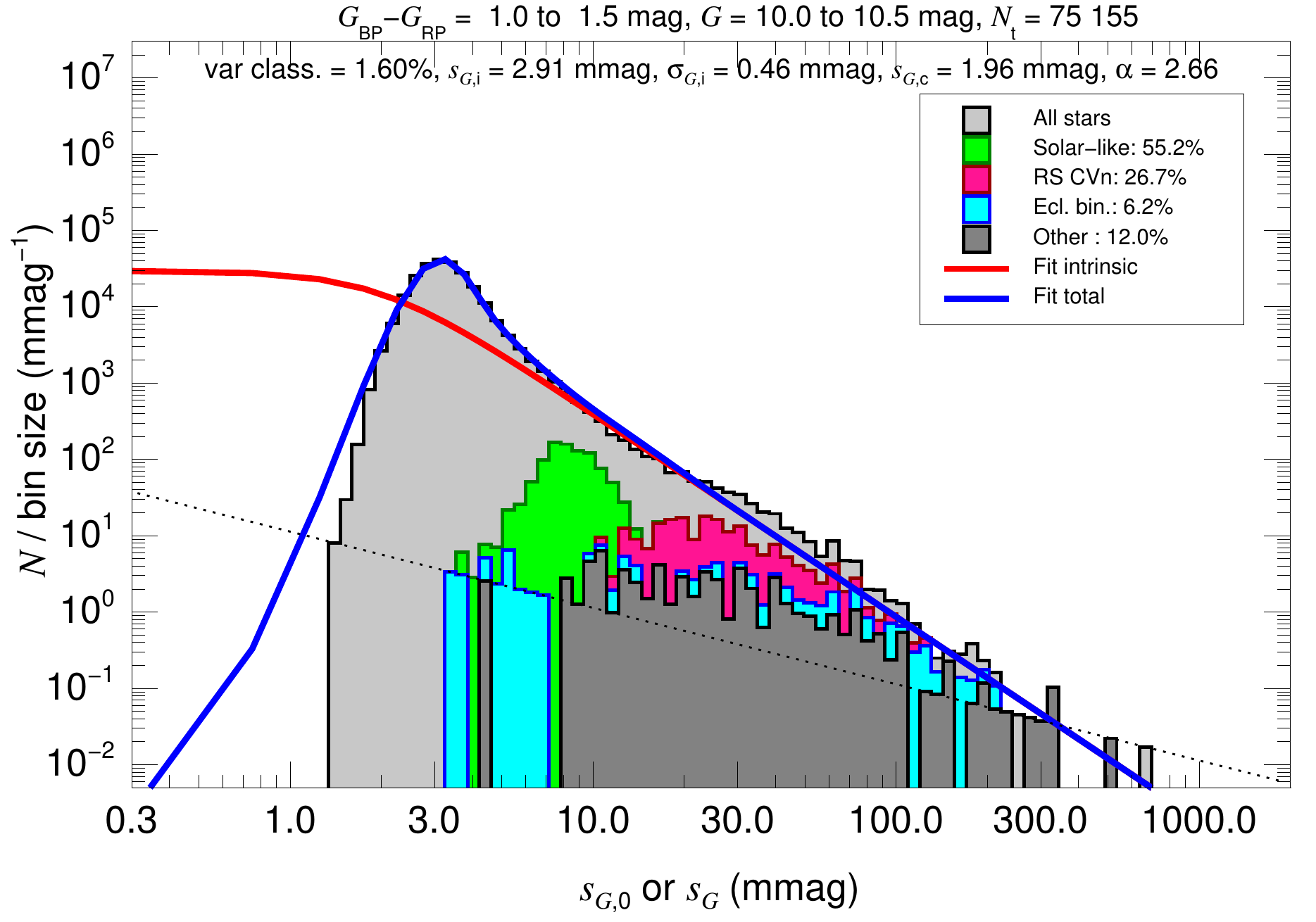}$\!\!\!$
                    \includegraphics[width=0.35\linewidth]{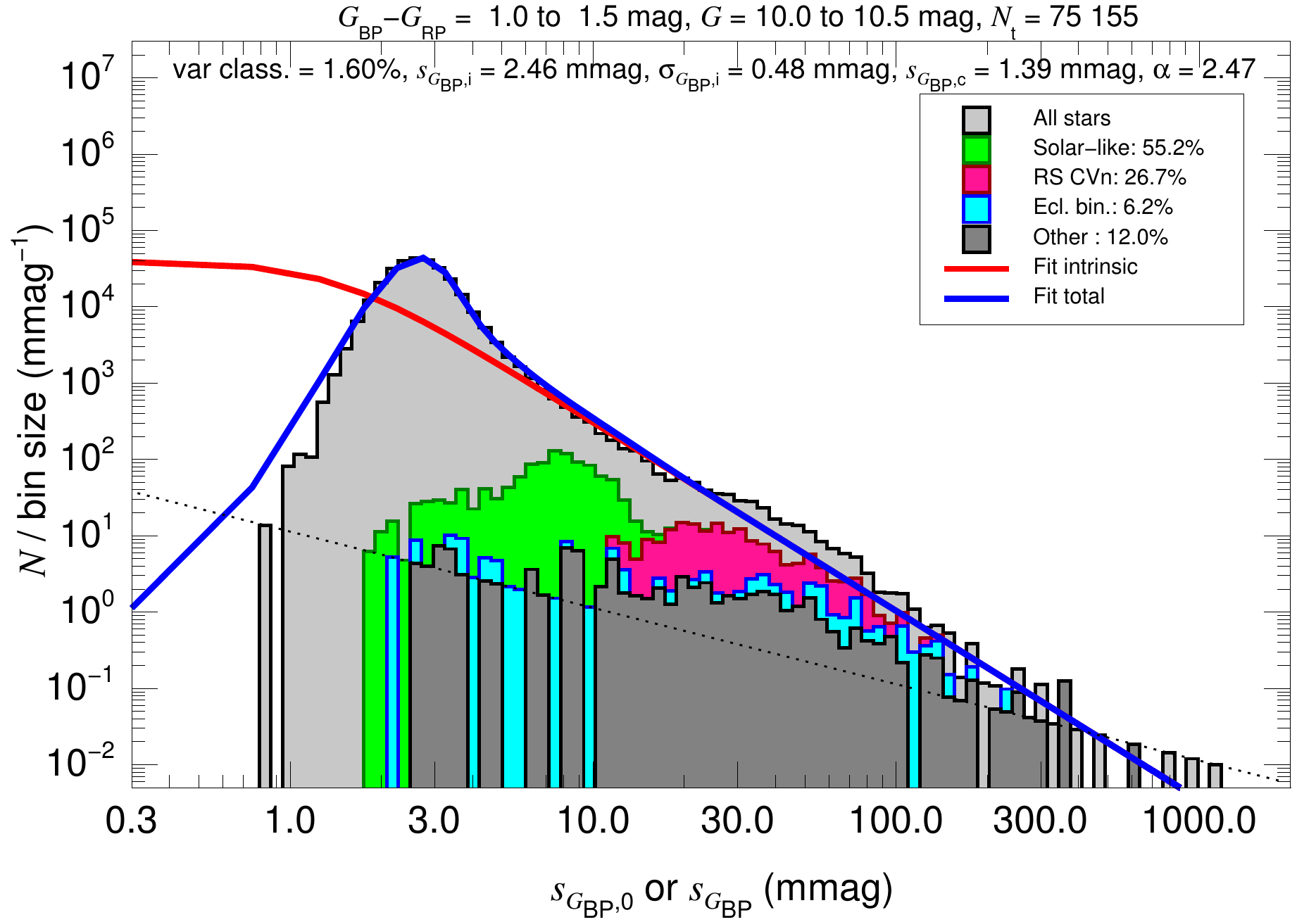}$\!\!\!$
                    \includegraphics[width=0.35\linewidth]{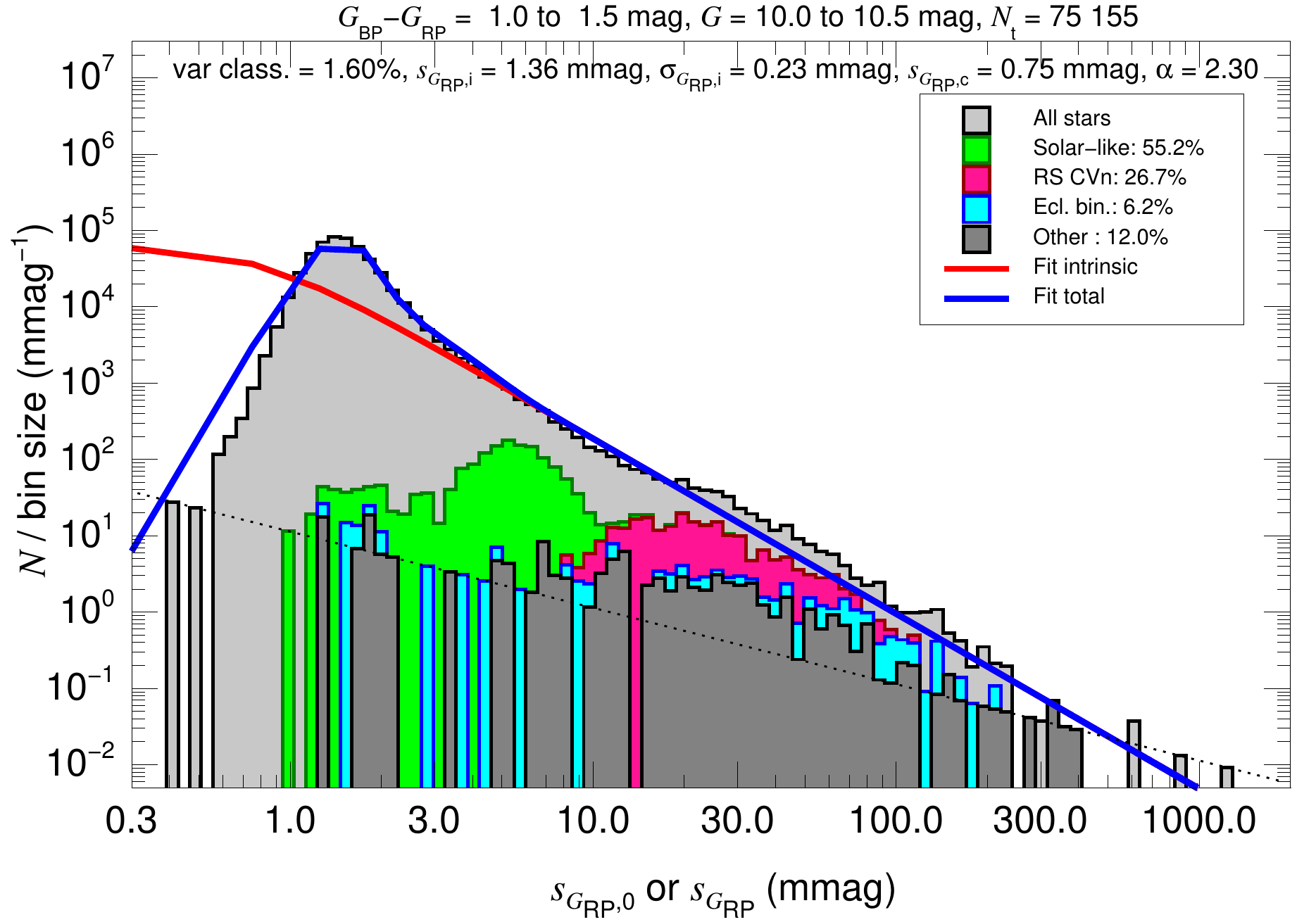}}
\centerline{$\!\!\!$\includegraphics[width=0.35\linewidth]{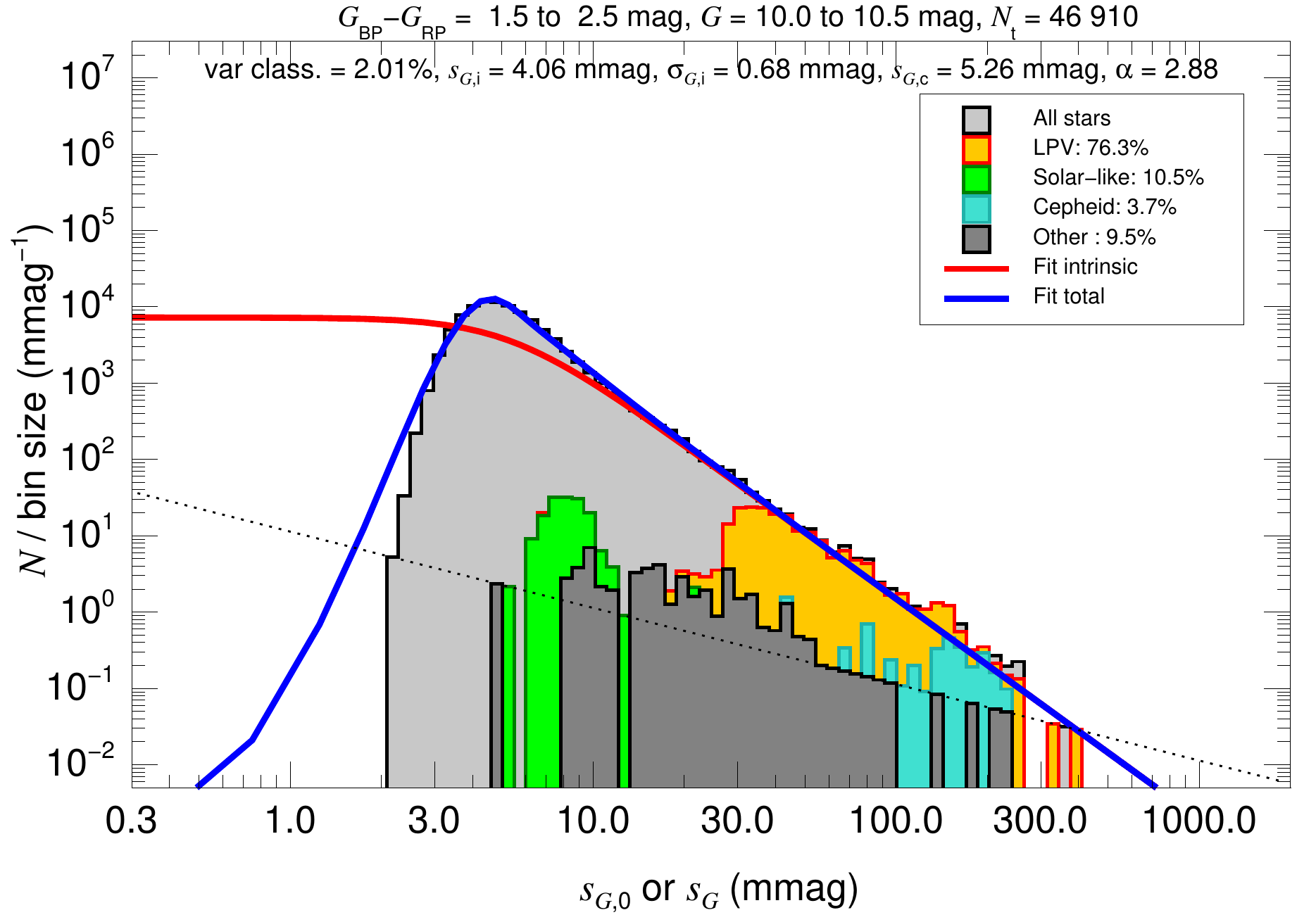}$\!\!\!$
                    \includegraphics[width=0.35\linewidth]{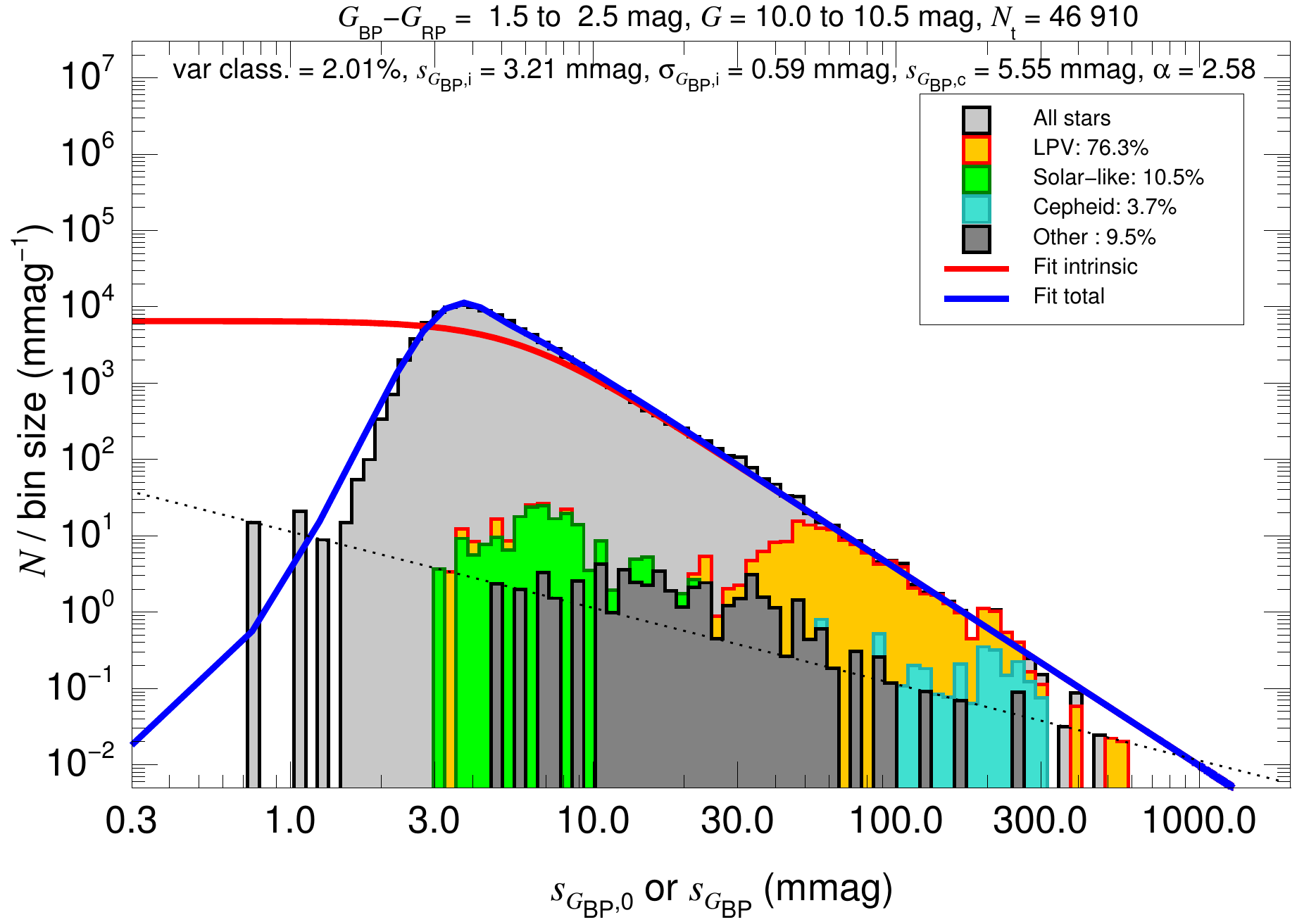}$\!\!\!$
                    \includegraphics[width=0.35\linewidth]{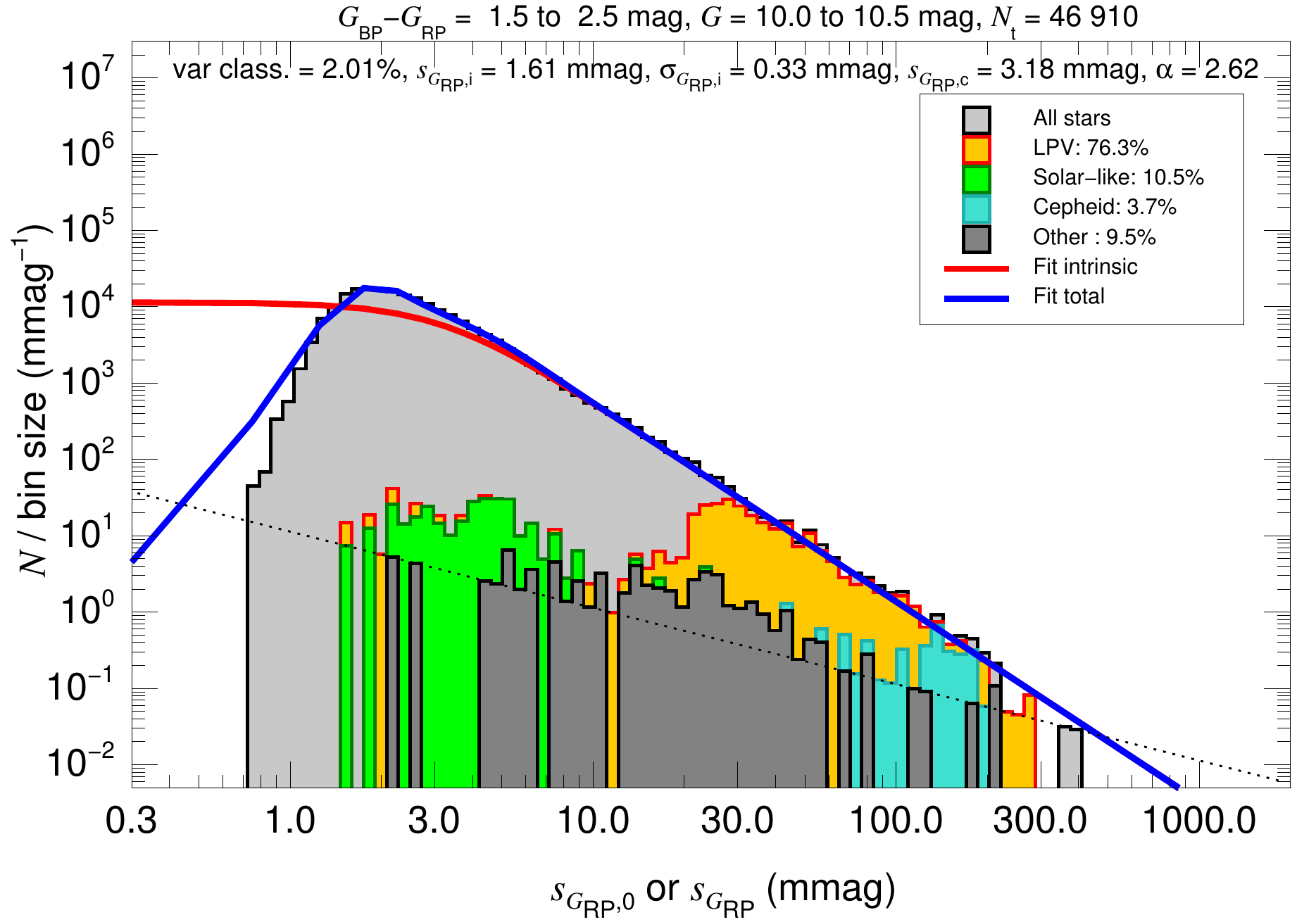}}
\centerline{$\!\!\!$\includegraphics[width=0.35\linewidth]{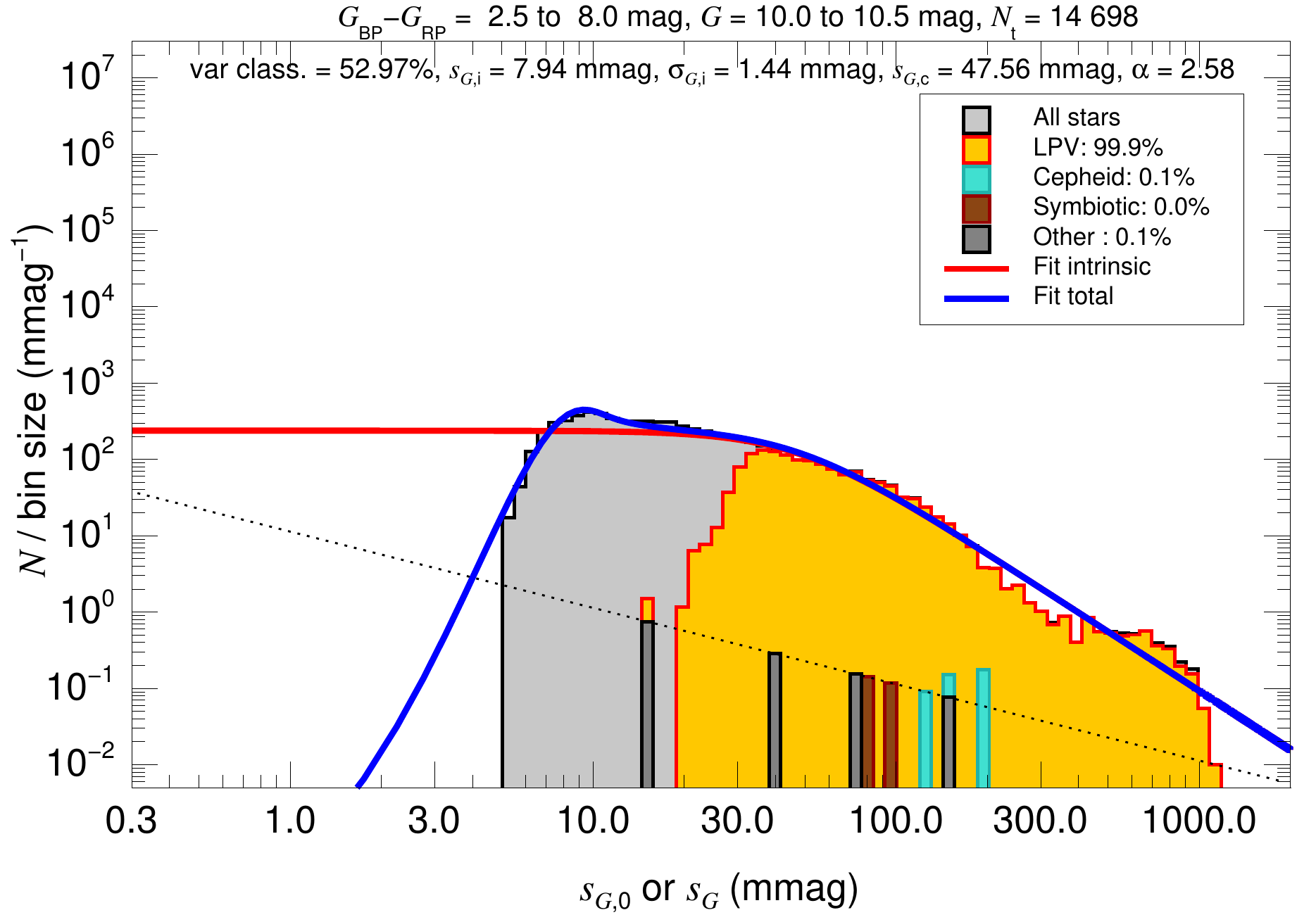}$\!\!\!$
                    \includegraphics[width=0.35\linewidth]{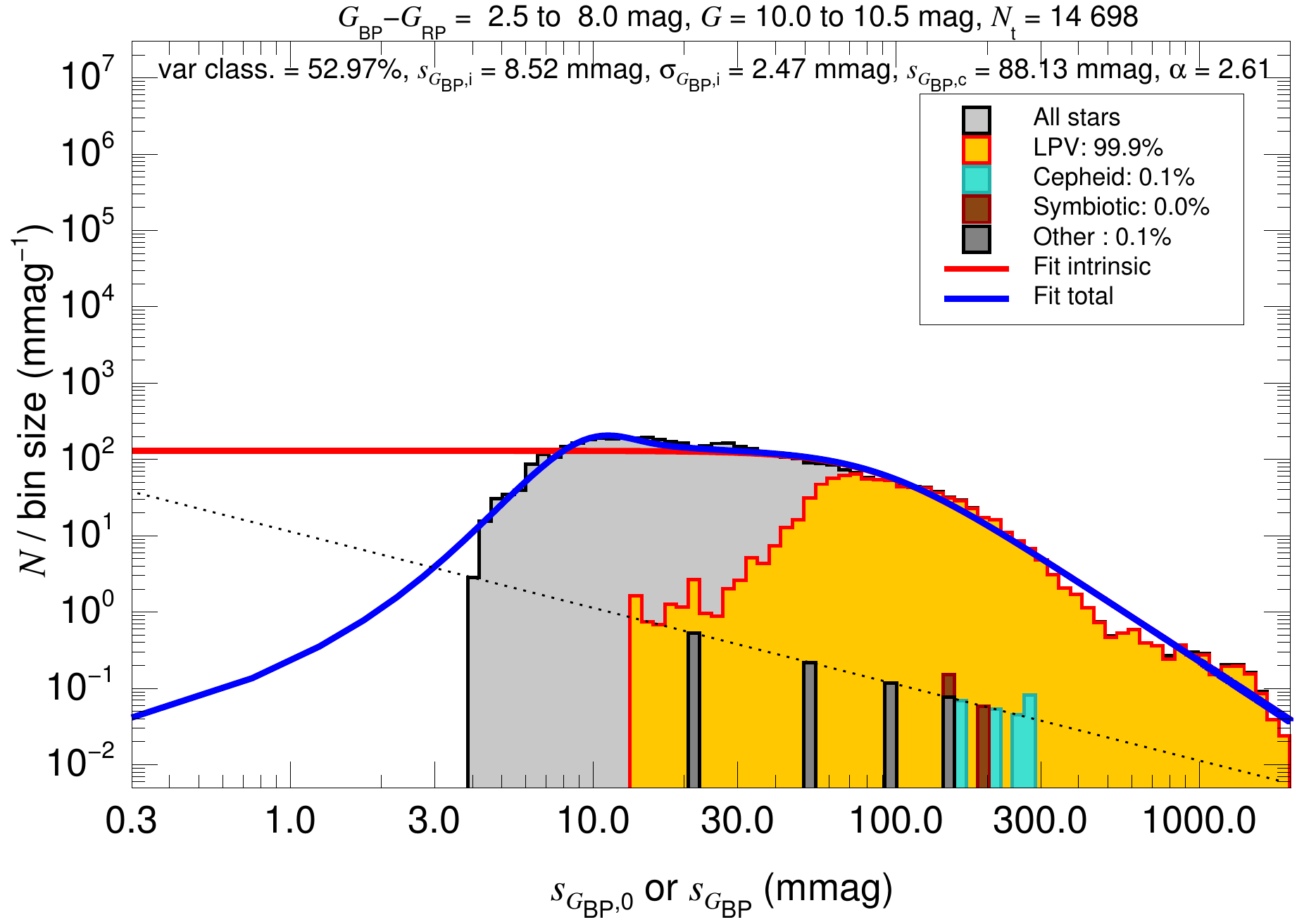}$\!\!\!$
                    \includegraphics[width=0.35\linewidth]{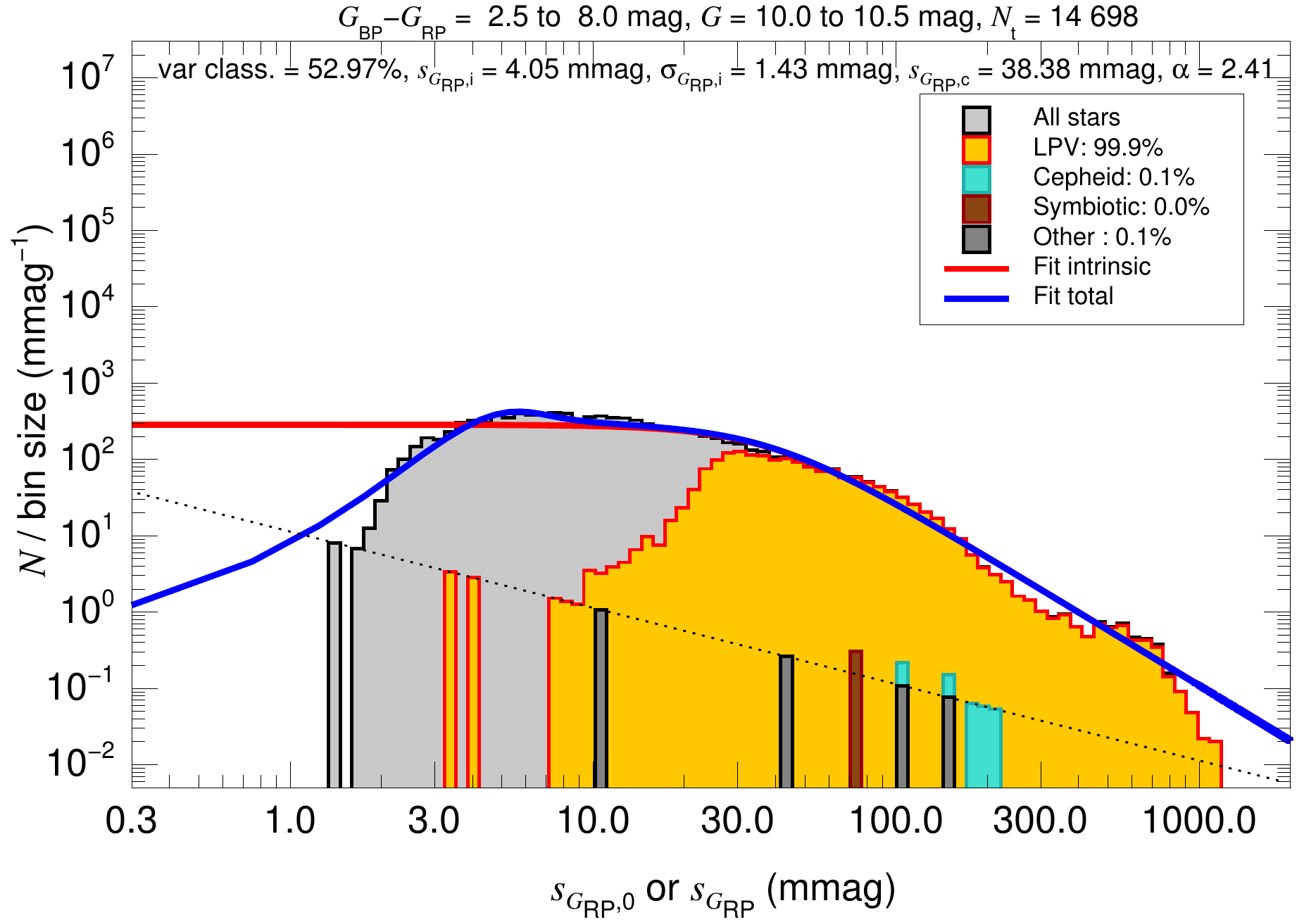}}
\caption{(Continued).}
\end{figure*}

\clearpage

\addtocounter{figure}{-1}

\begin{figure*}
\centerline{$\!\!\!$\includegraphics[width=0.35\linewidth]{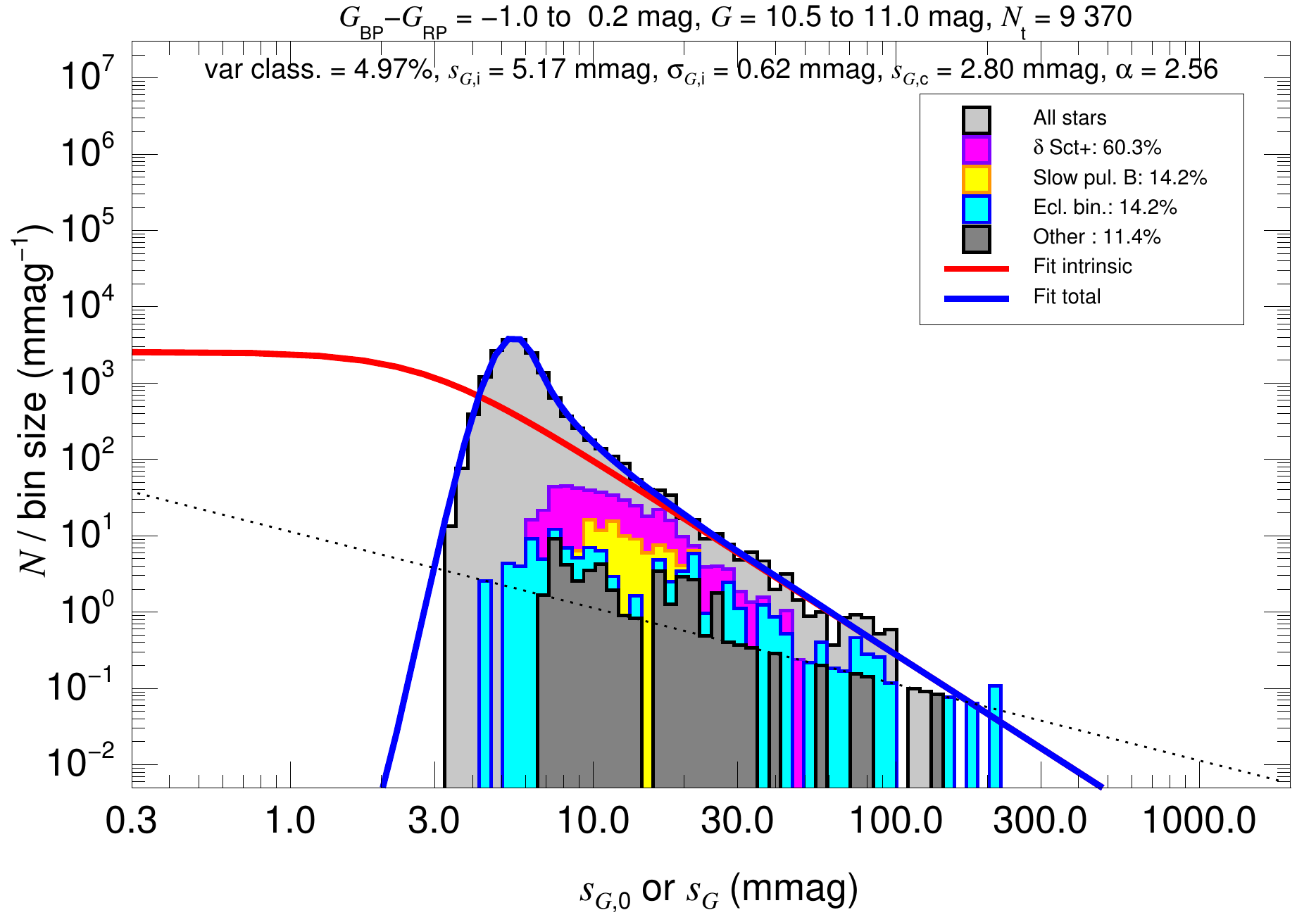}$\!\!\!$
                    \includegraphics[width=0.35\linewidth]{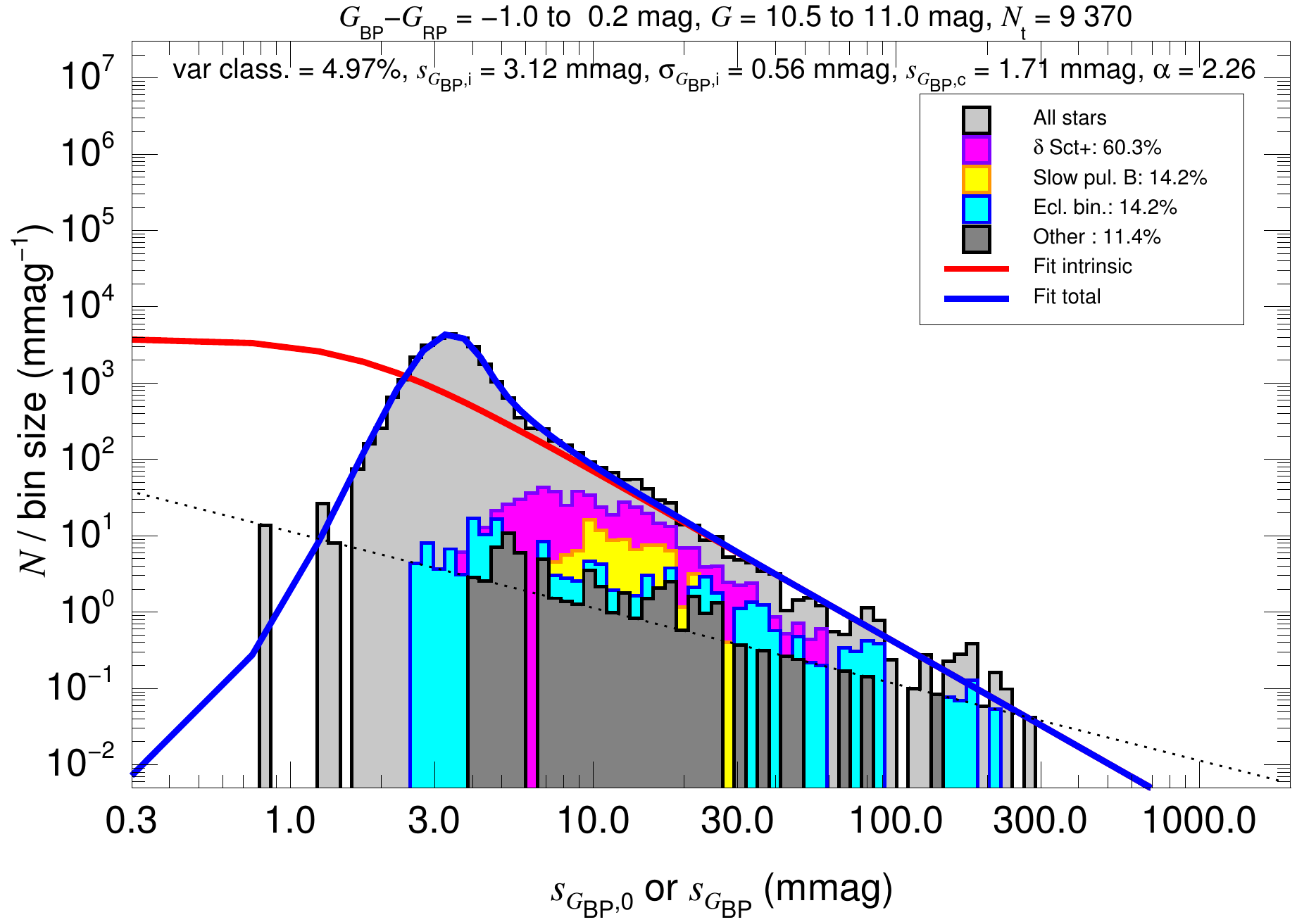}$\!\!\!$
                    \includegraphics[width=0.35\linewidth]{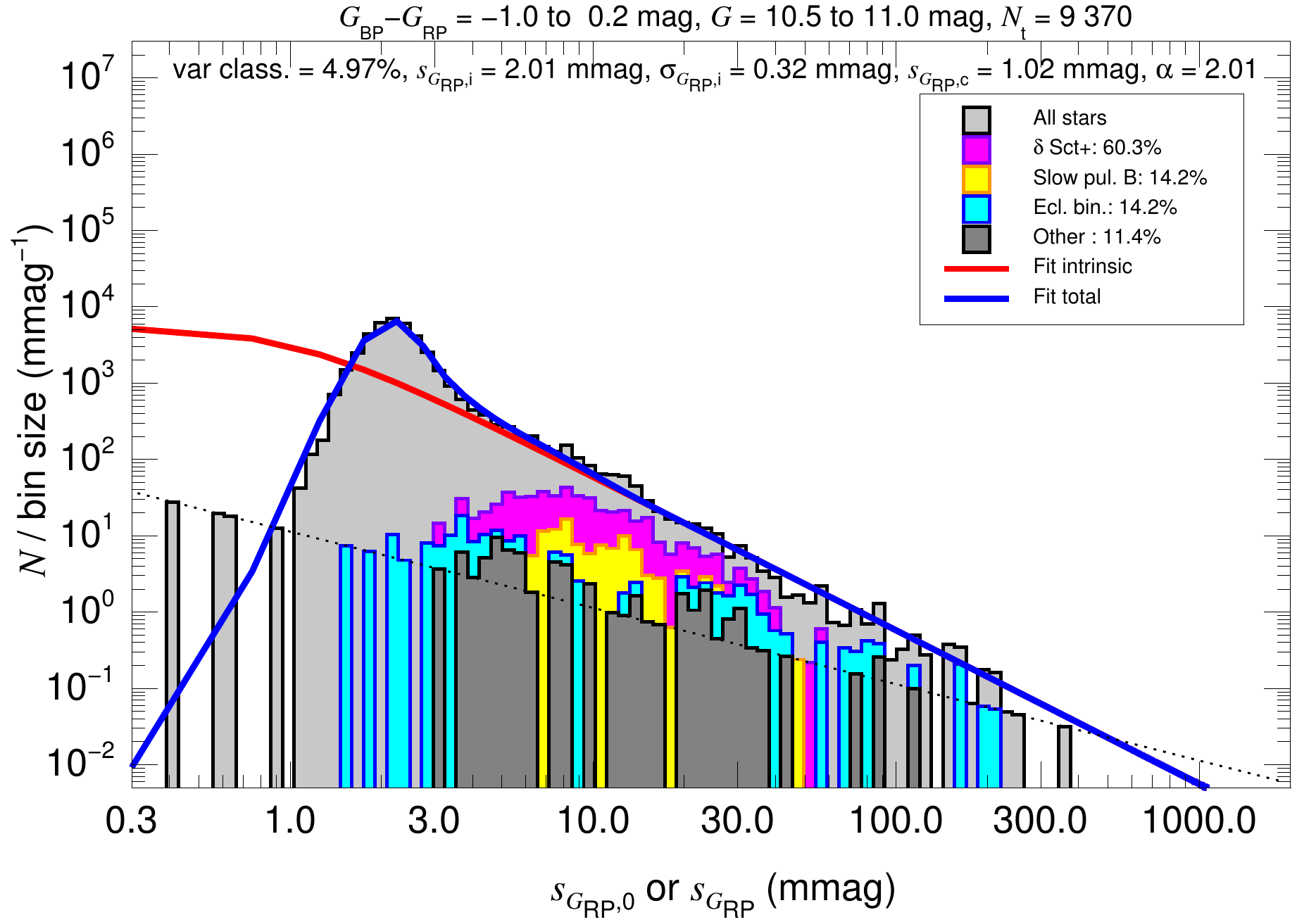}}
\centerline{$\!\!\!$\includegraphics[width=0.35\linewidth]{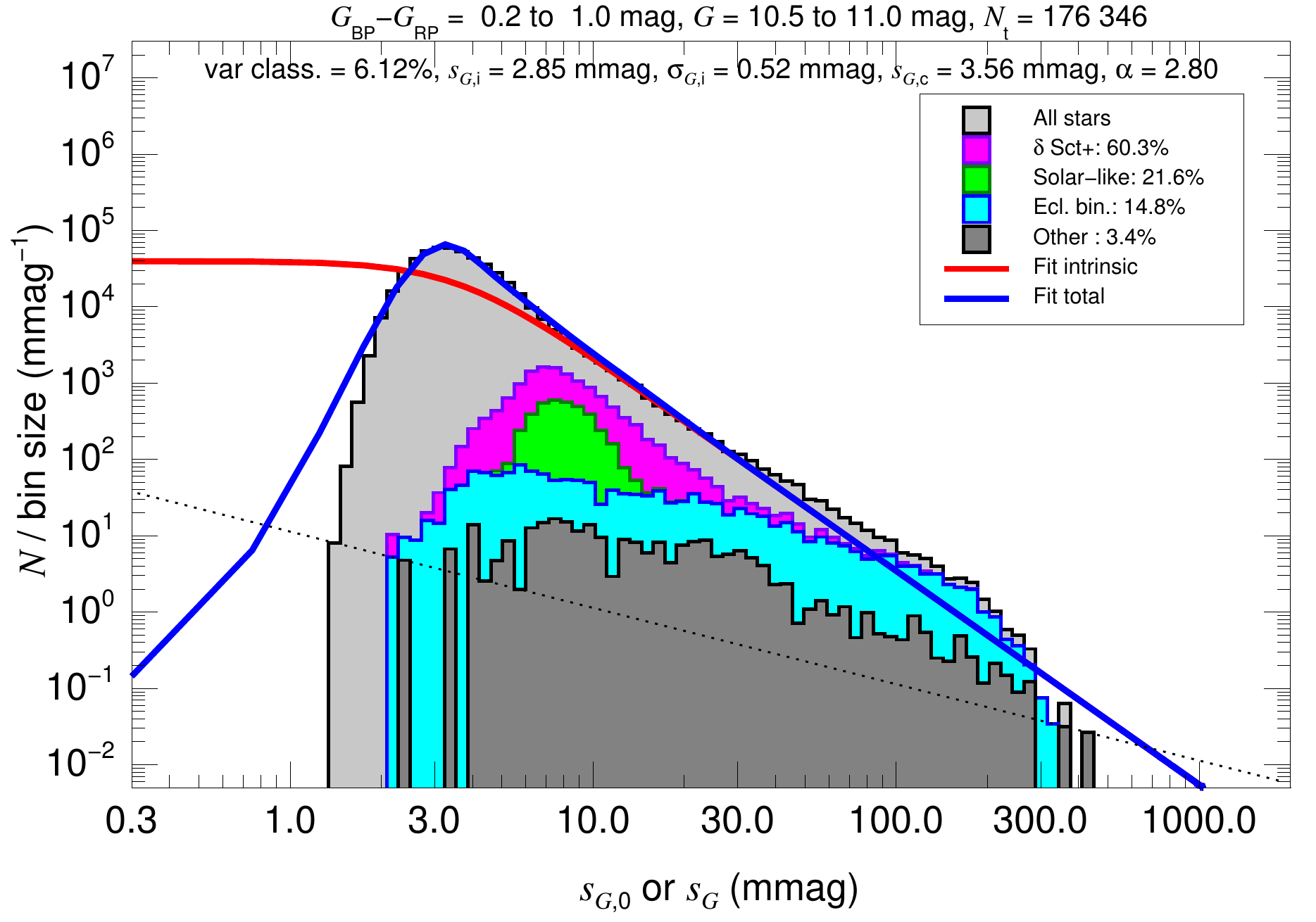}$\!\!\!$
                    \includegraphics[width=0.35\linewidth]{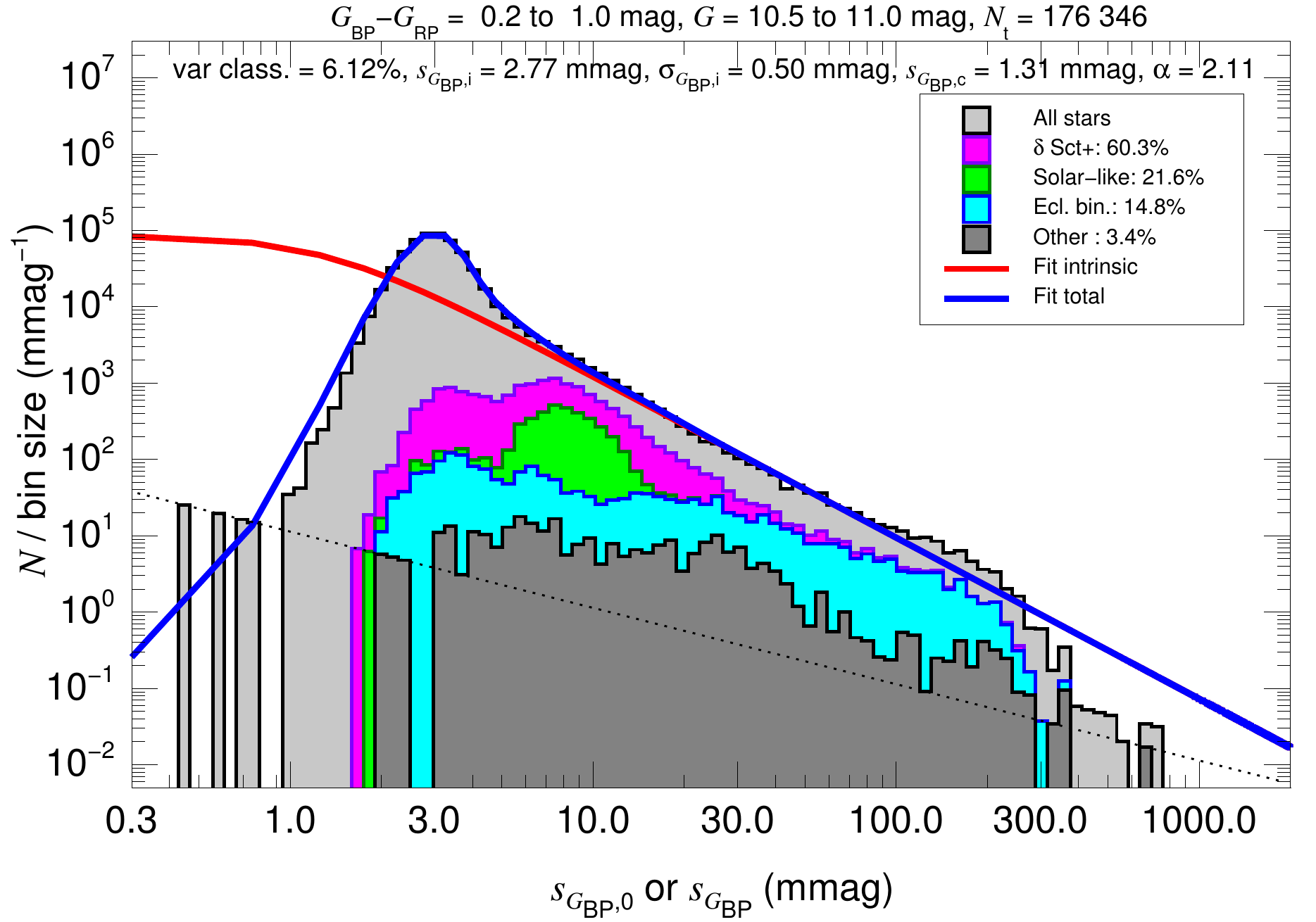}$\!\!\!$
                    \includegraphics[width=0.35\linewidth]{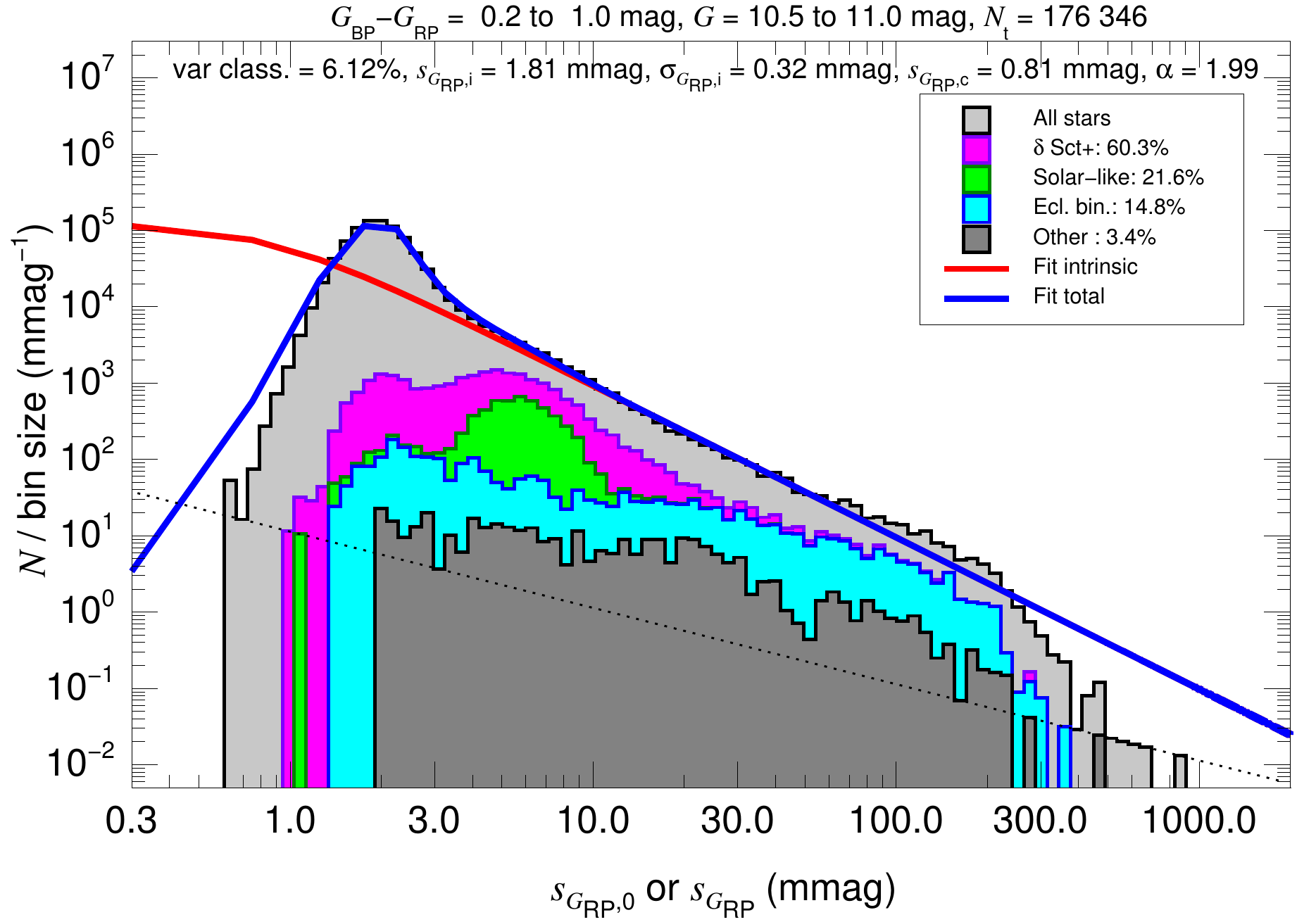}}
\centerline{$\!\!\!$\includegraphics[width=0.35\linewidth]{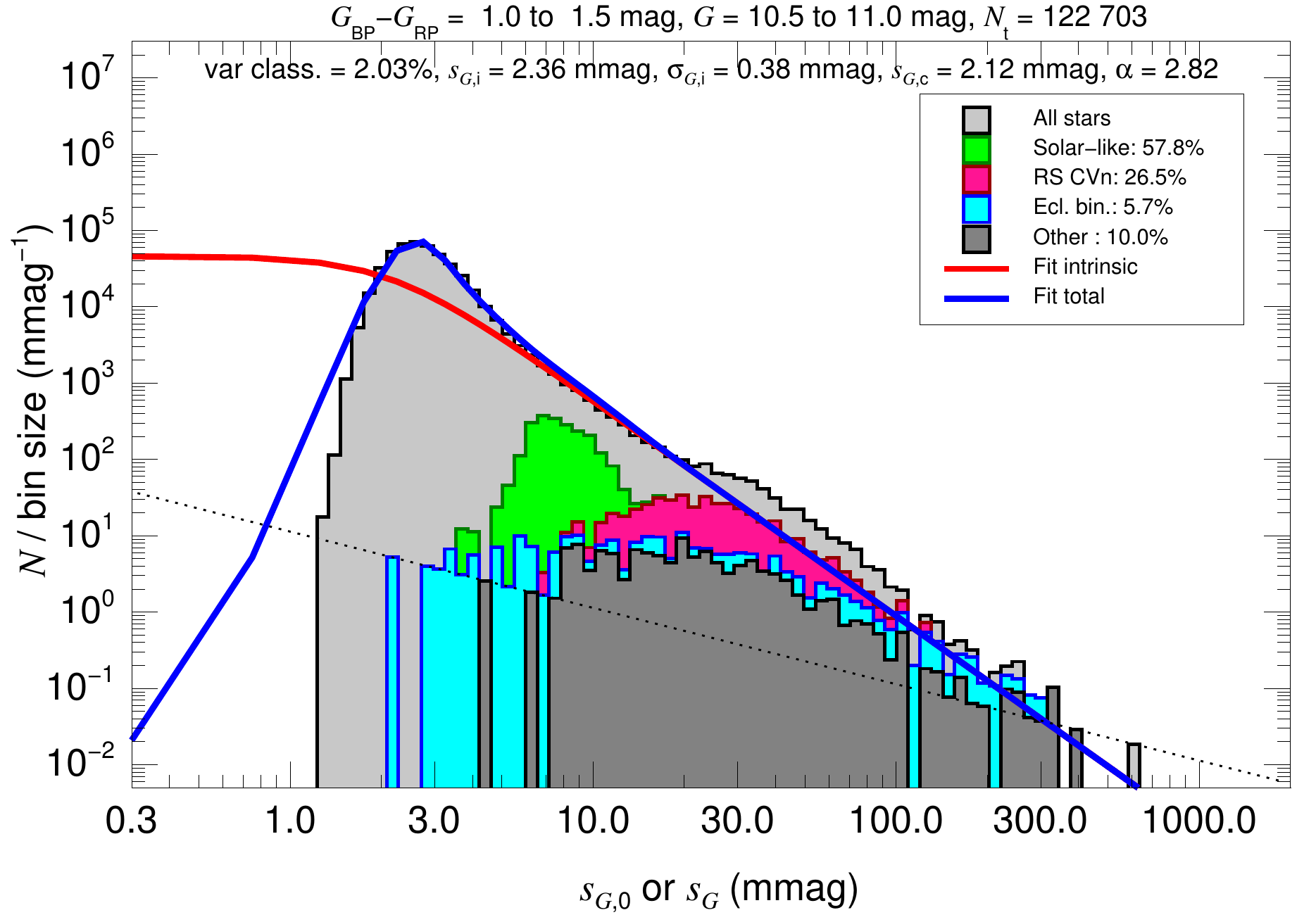}$\!\!\!$
                    \includegraphics[width=0.35\linewidth]{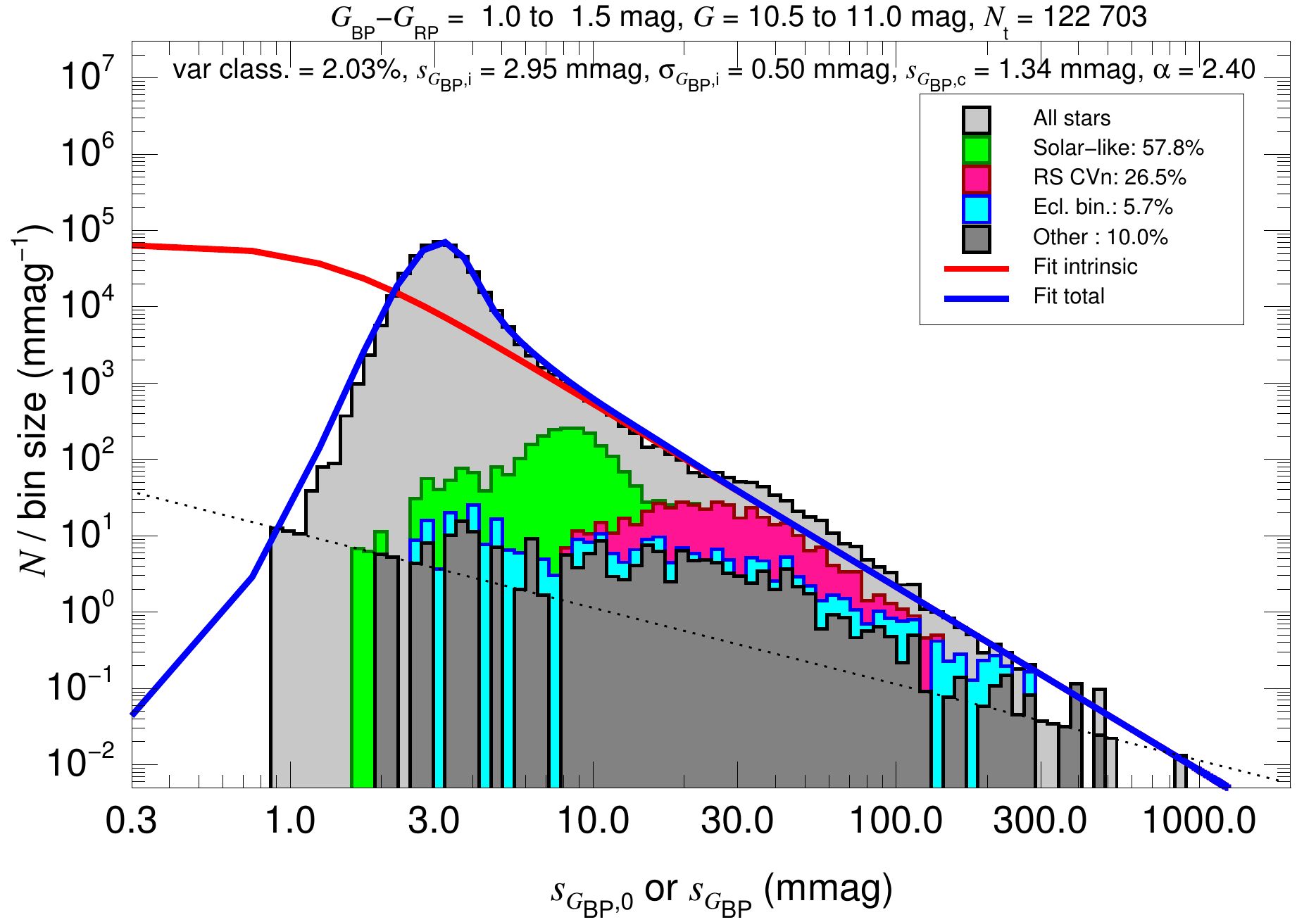}$\!\!\!$
                    \includegraphics[width=0.35\linewidth]{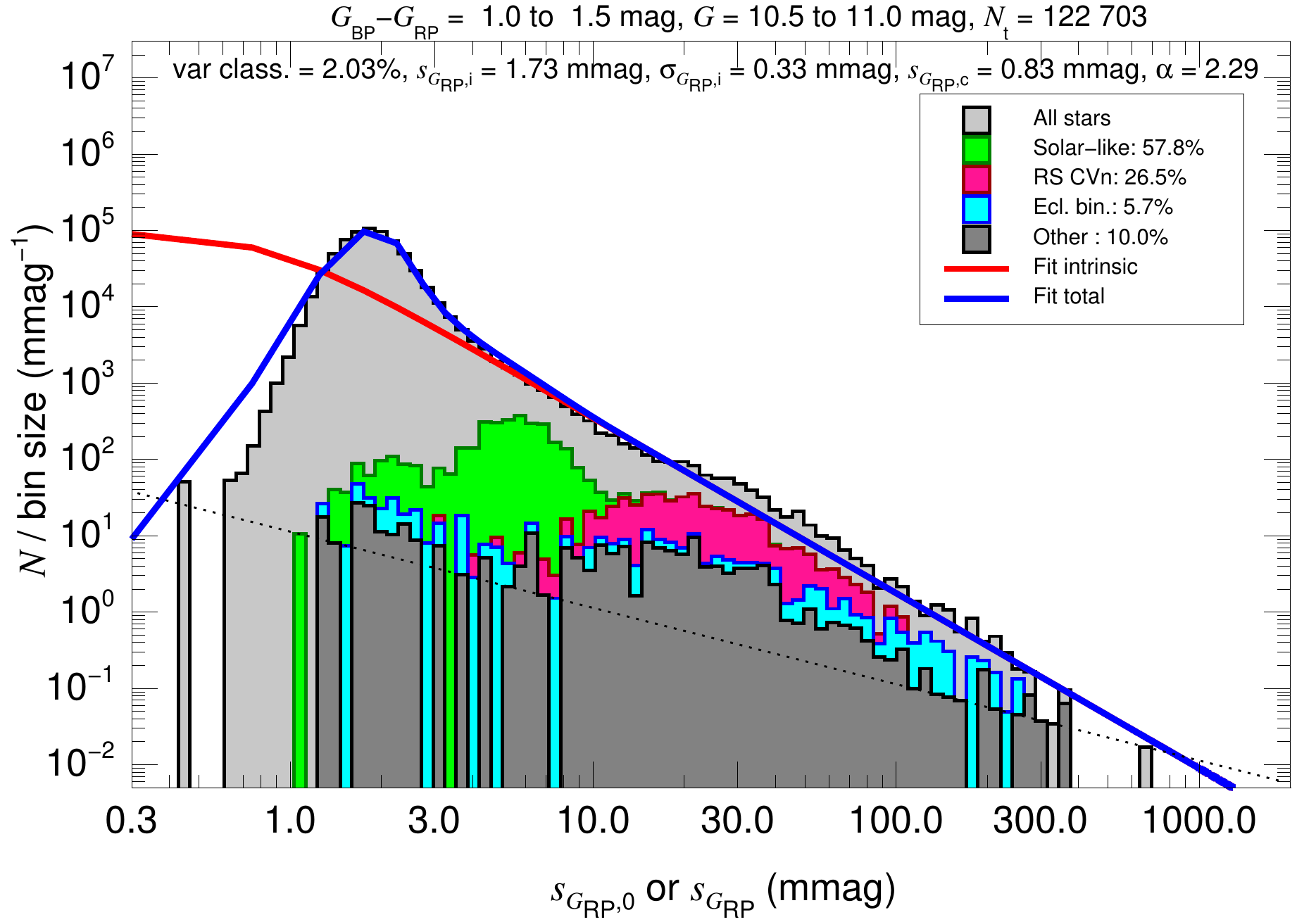}}
\centerline{$\!\!\!$\includegraphics[width=0.35\linewidth]{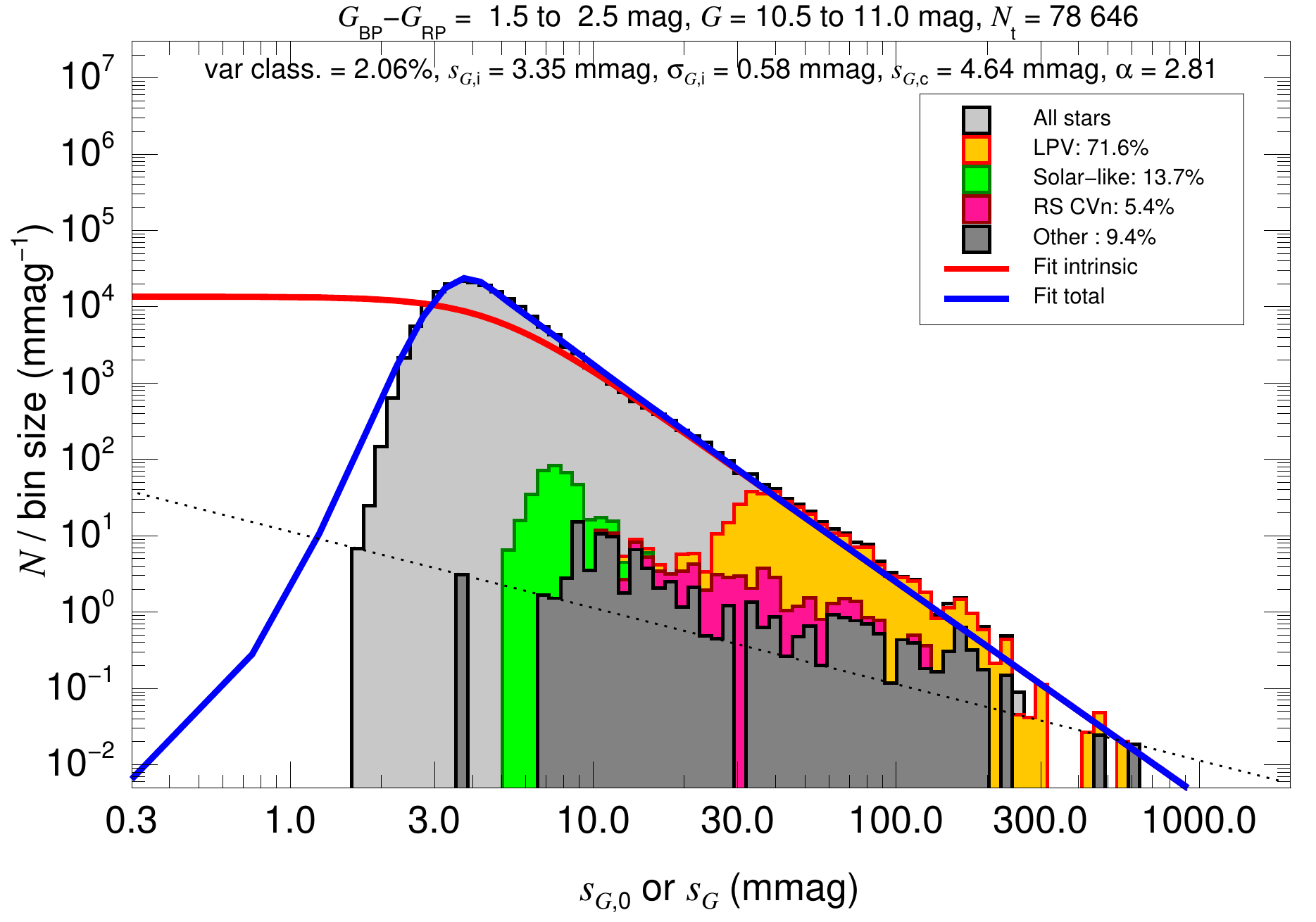}$\!\!\!$
                    \includegraphics[width=0.35\linewidth]{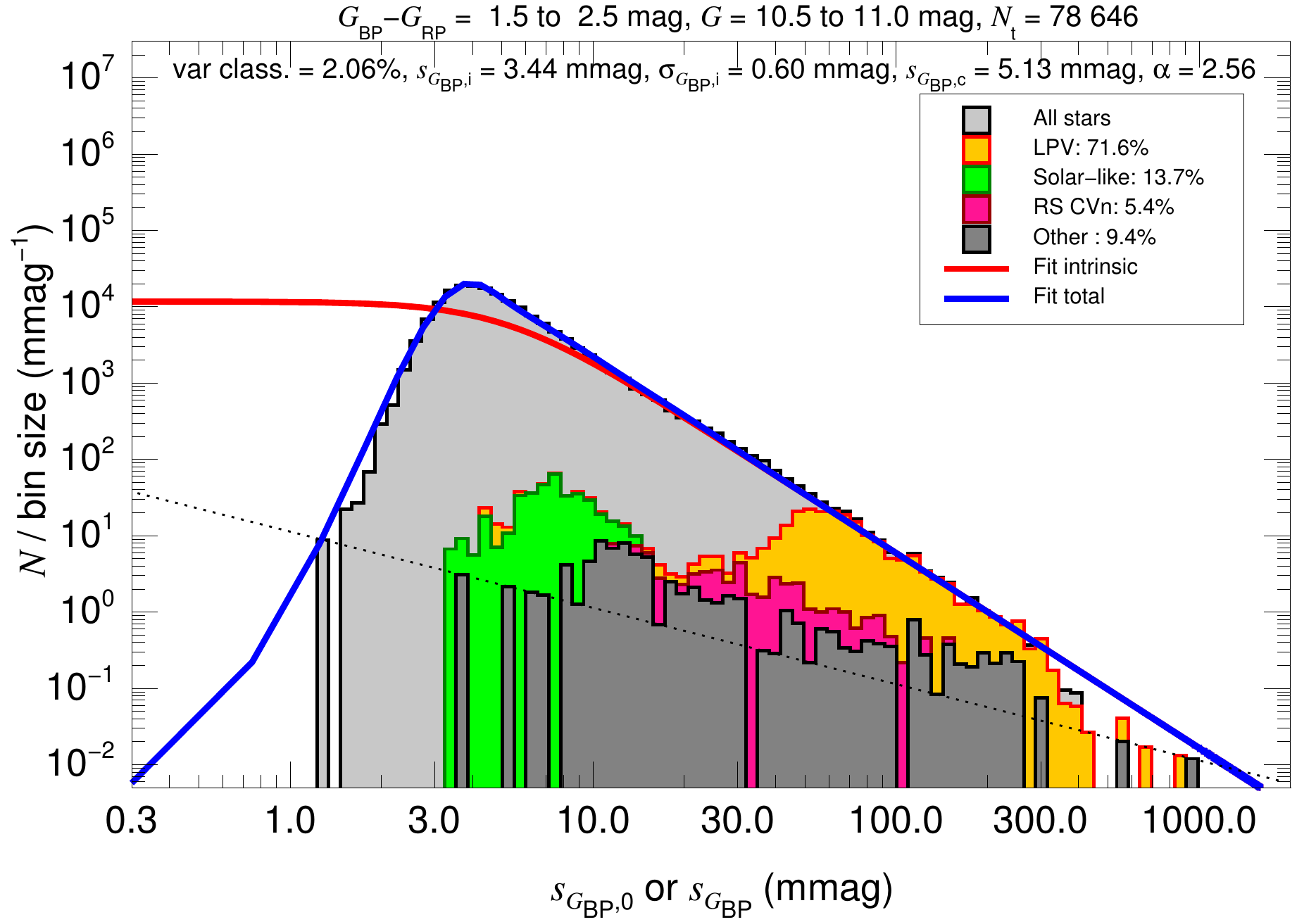}$\!\!\!$
                    \includegraphics[width=0.35\linewidth]{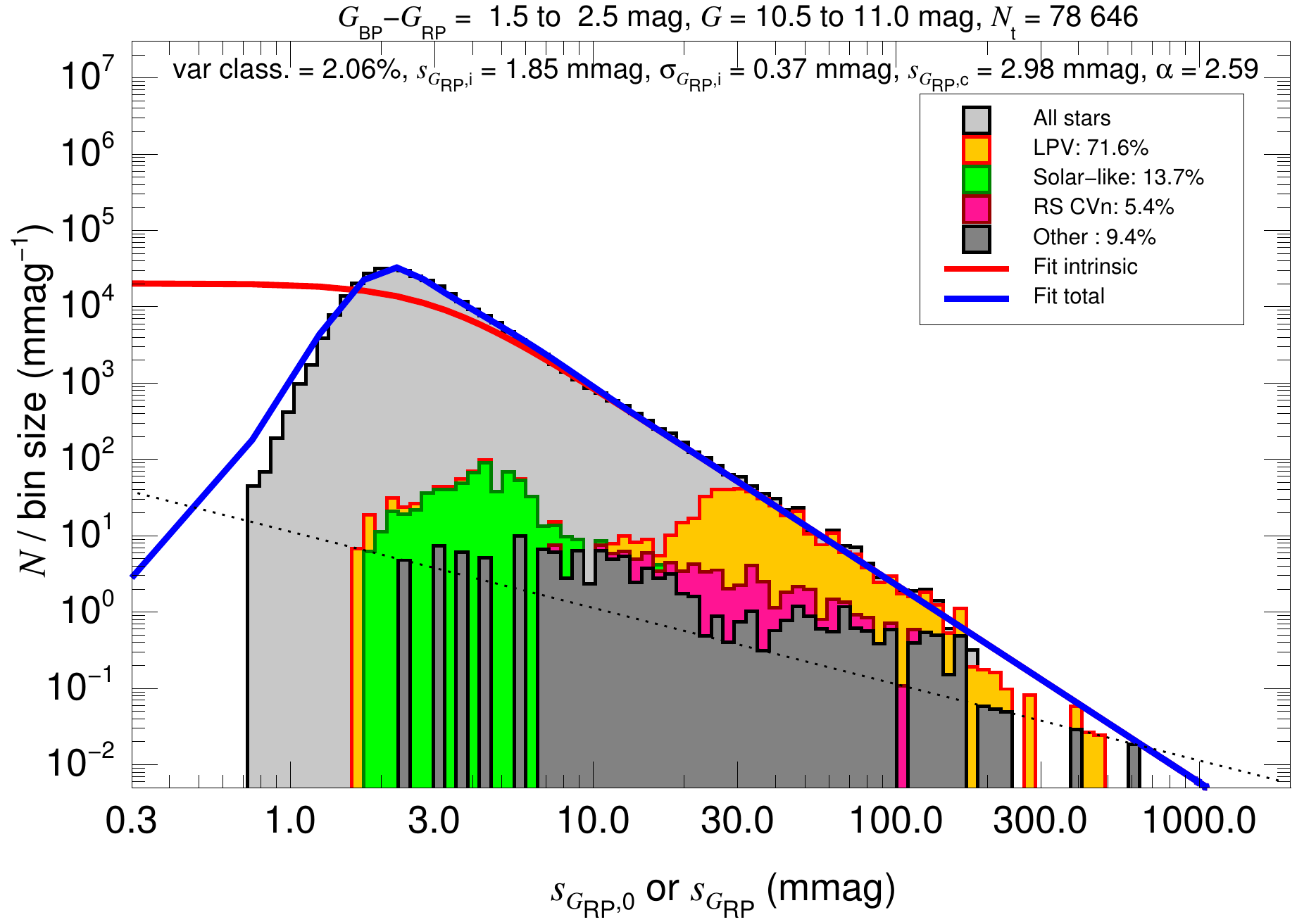}}
\centerline{$\!\!\!$\includegraphics[width=0.35\linewidth]{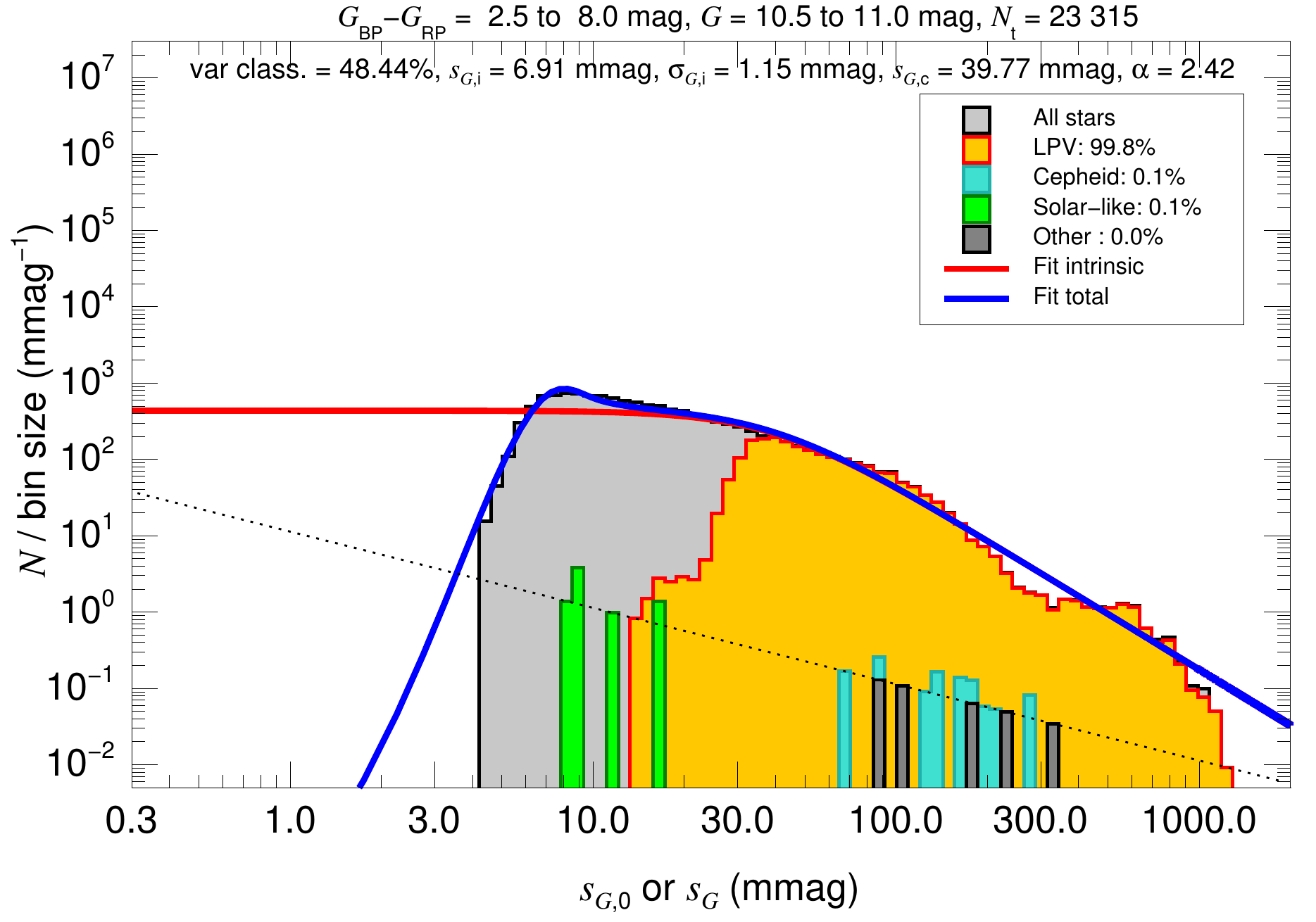}$\!\!\!$
                    \includegraphics[width=0.35\linewidth]{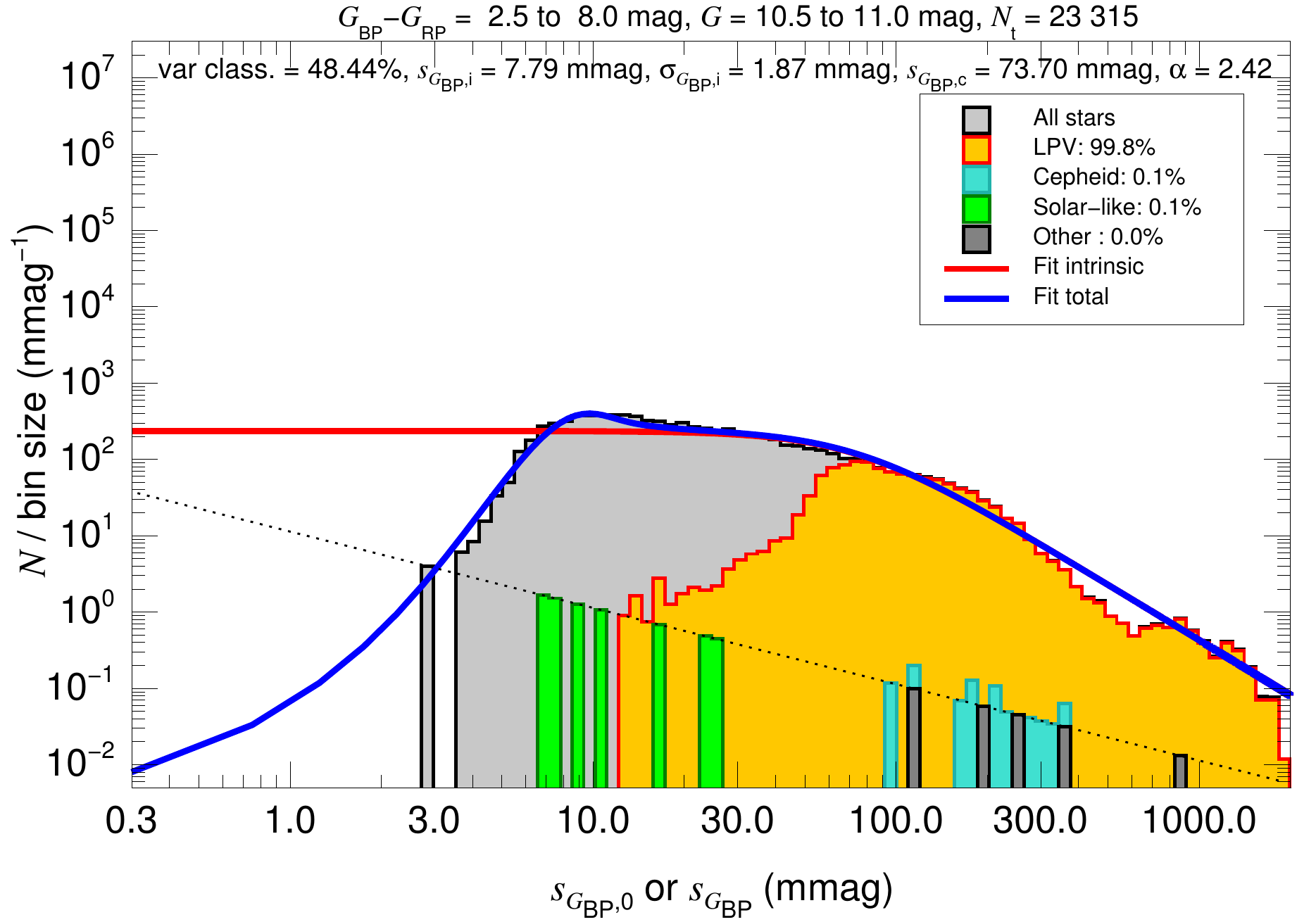}$\!\!\!$
                    \includegraphics[width=0.35\linewidth]{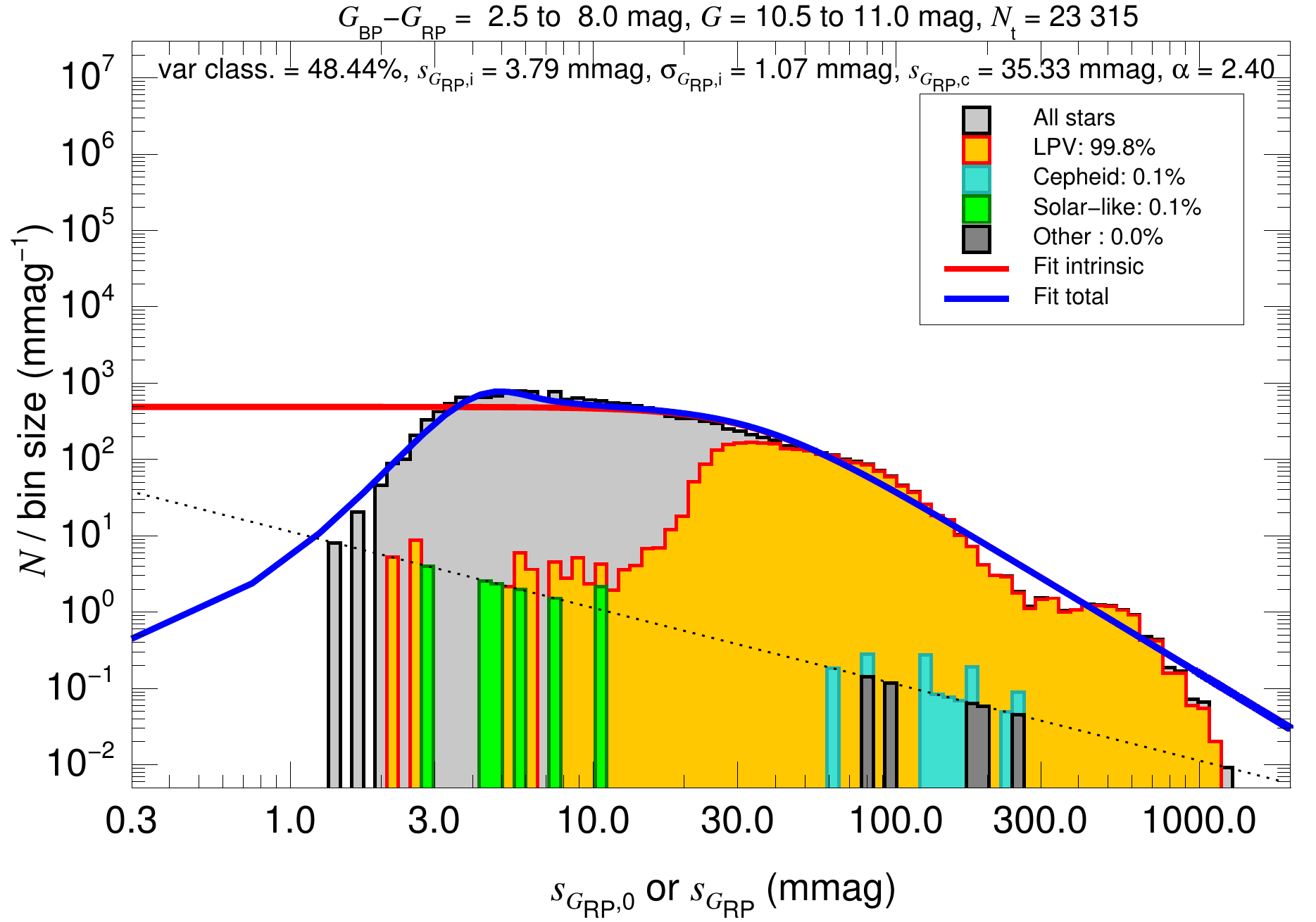}}
\caption{(Continued).}
\end{figure*}

\addtocounter{figure}{-1}

\begin{figure*}
\centerline{$\!\!\!$\includegraphics[width=0.35\linewidth]{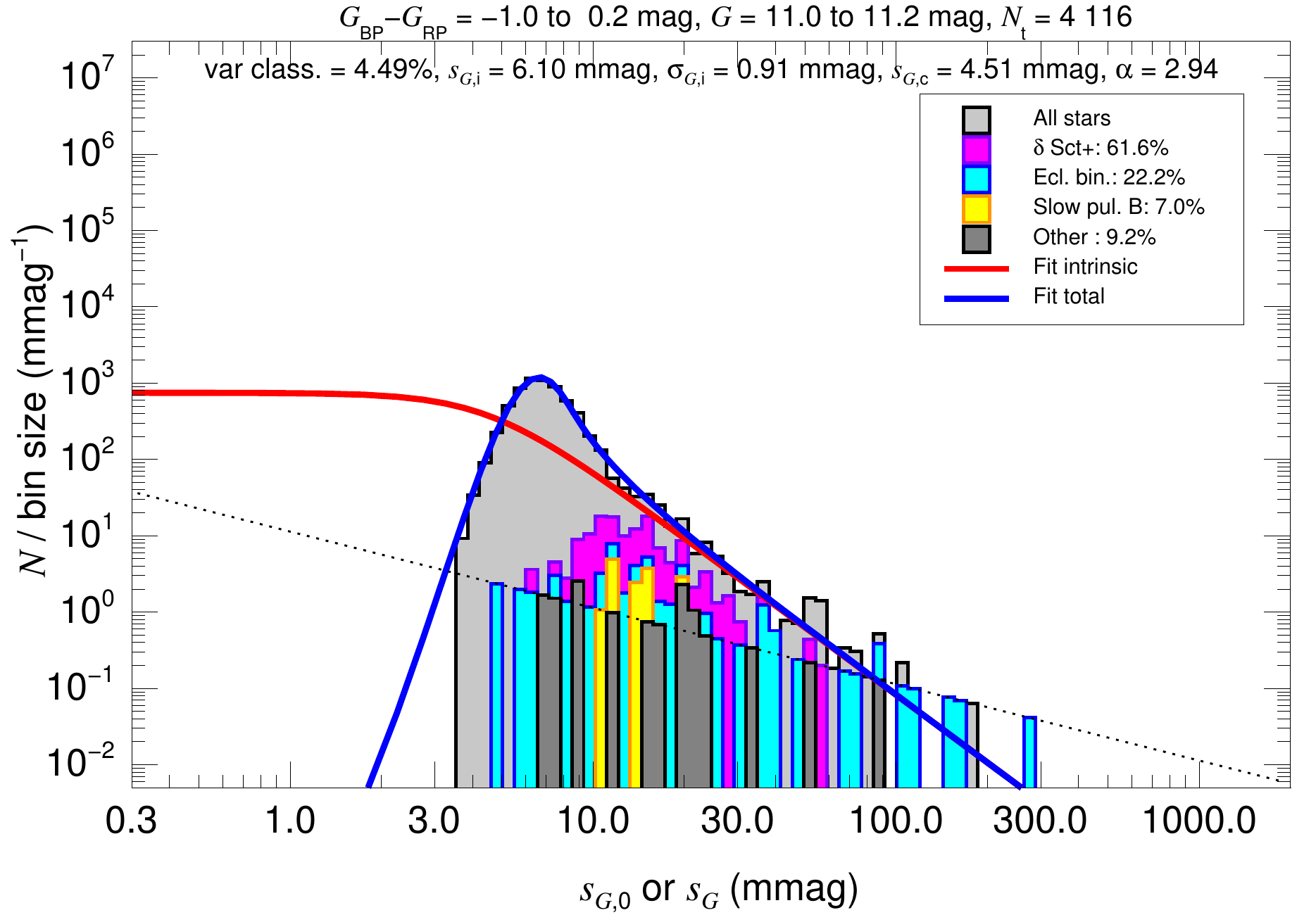}$\!\!\!$
                    \includegraphics[width=0.35\linewidth]{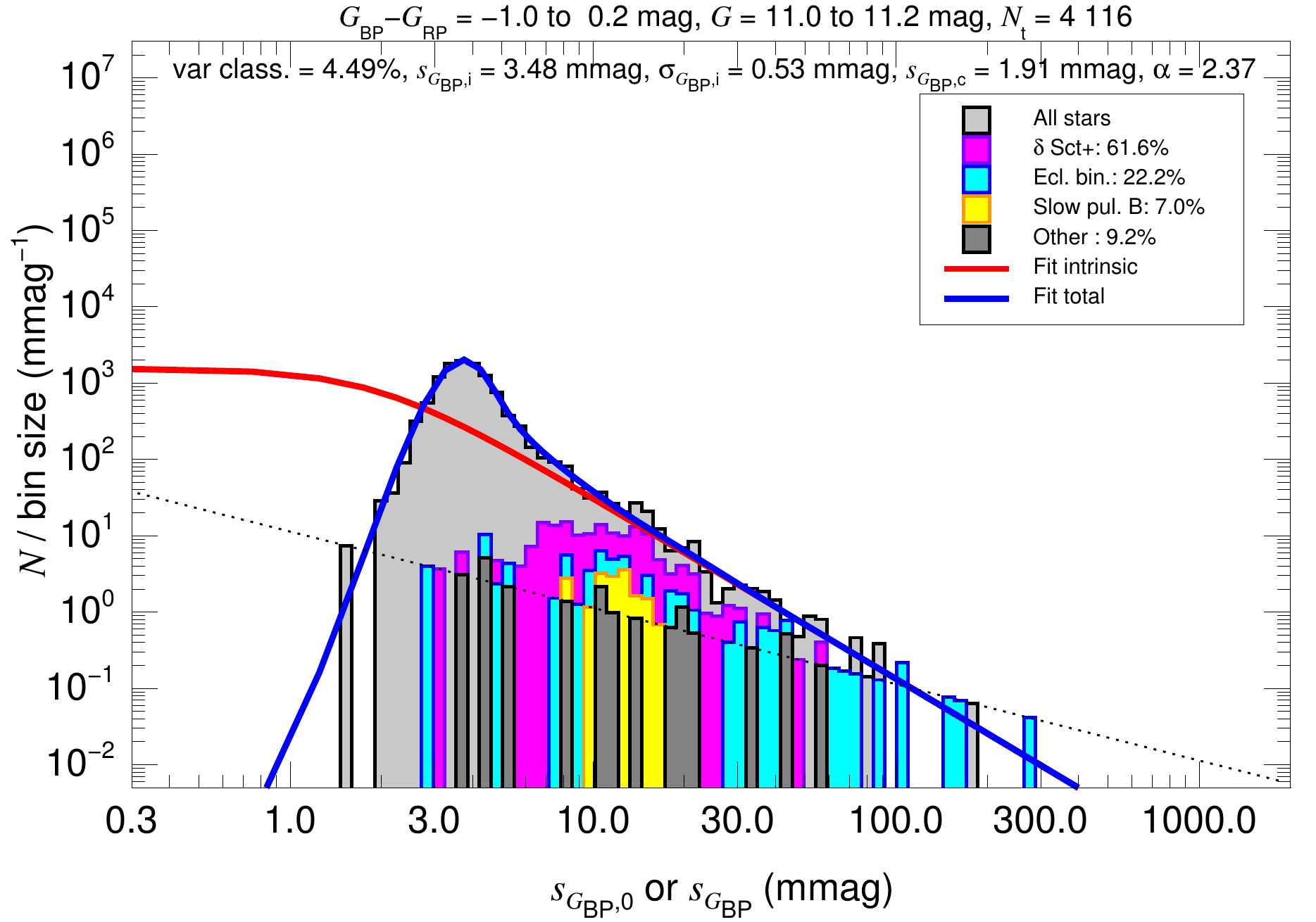}$\!\!\!$
                    \includegraphics[width=0.35\linewidth]{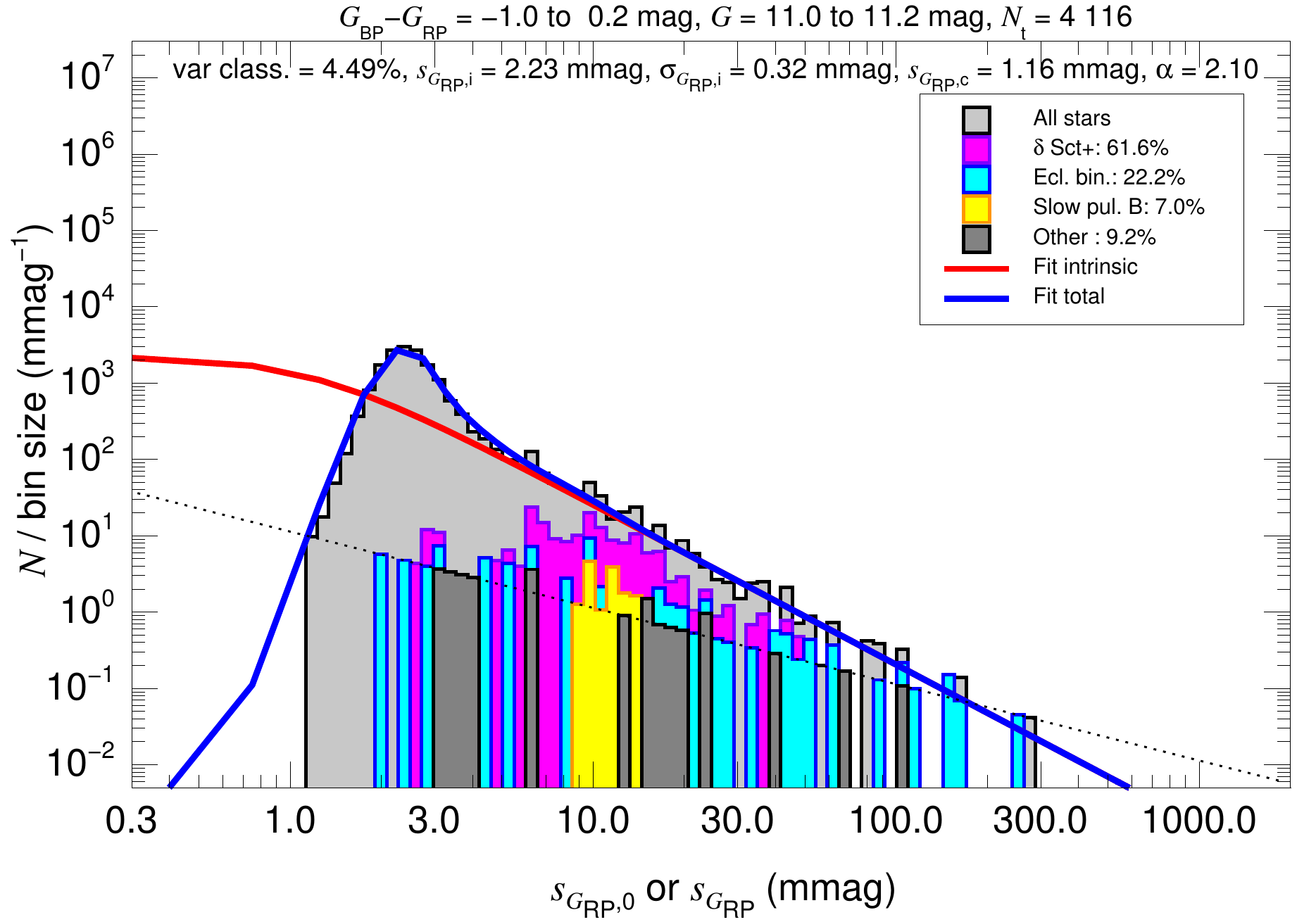}}
\centerline{$\!\!\!$\includegraphics[width=0.35\linewidth]{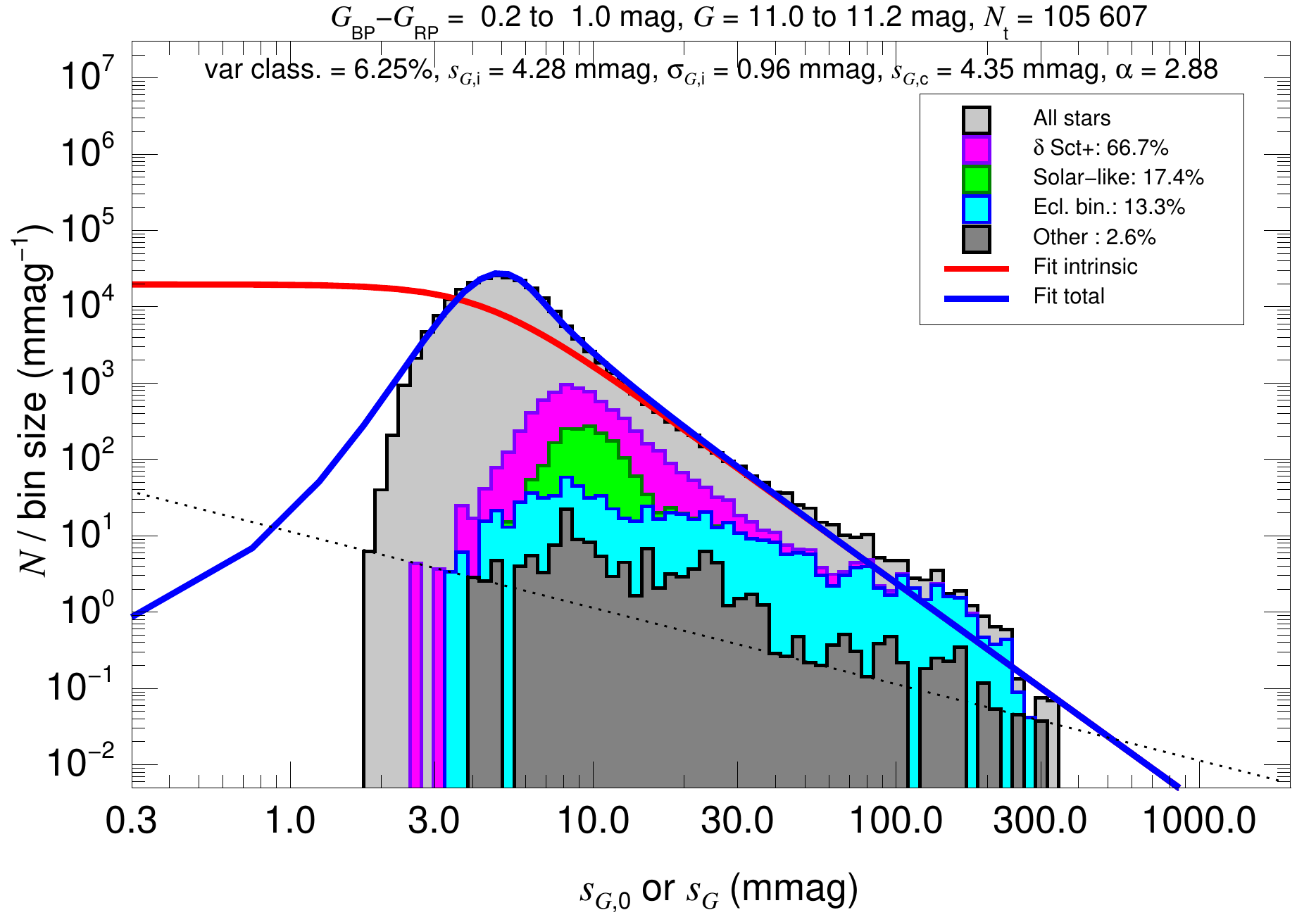}$\!\!\!$
                    \includegraphics[width=0.35\linewidth]{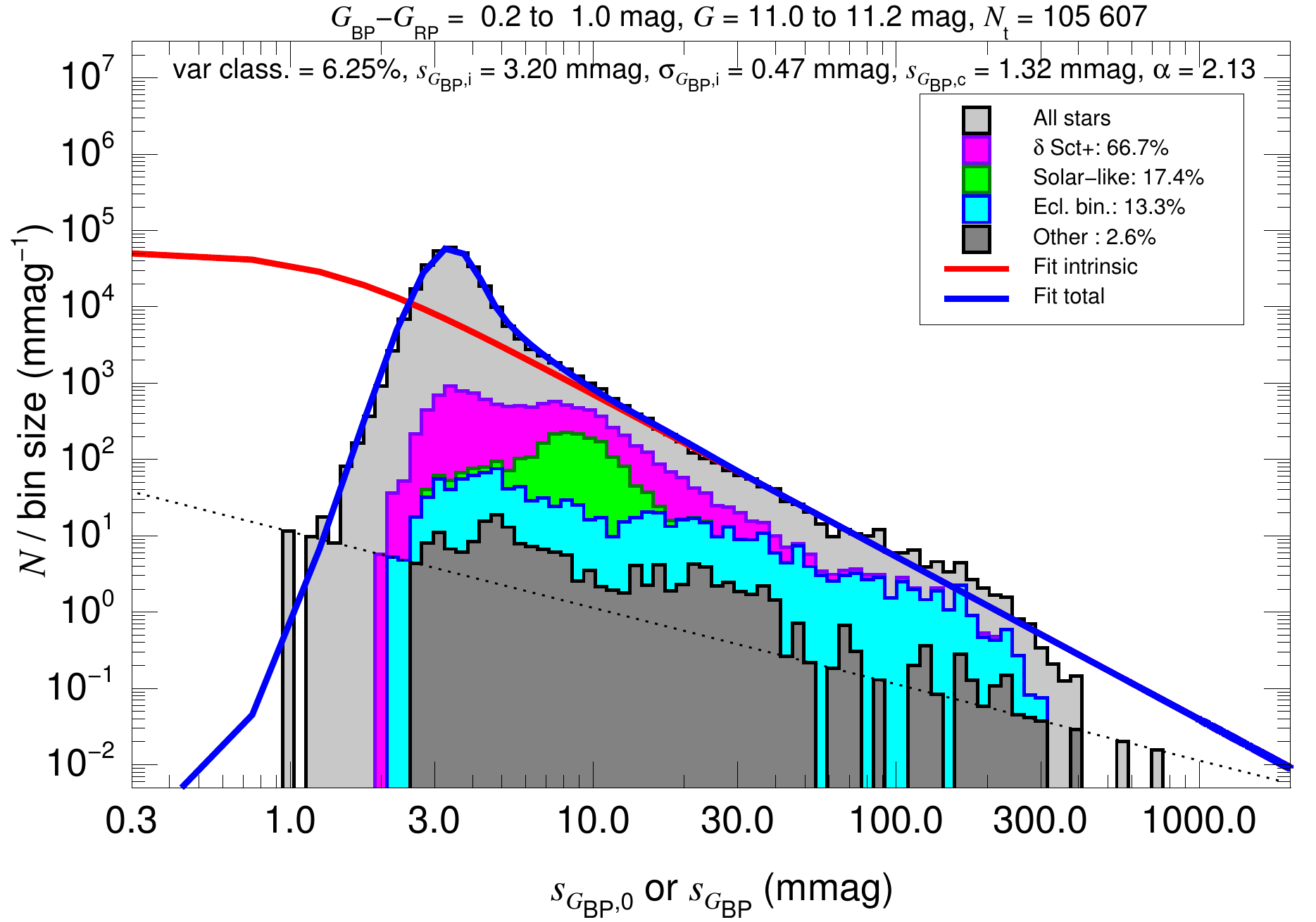}$\!\!\!$
                    \includegraphics[width=0.35\linewidth]{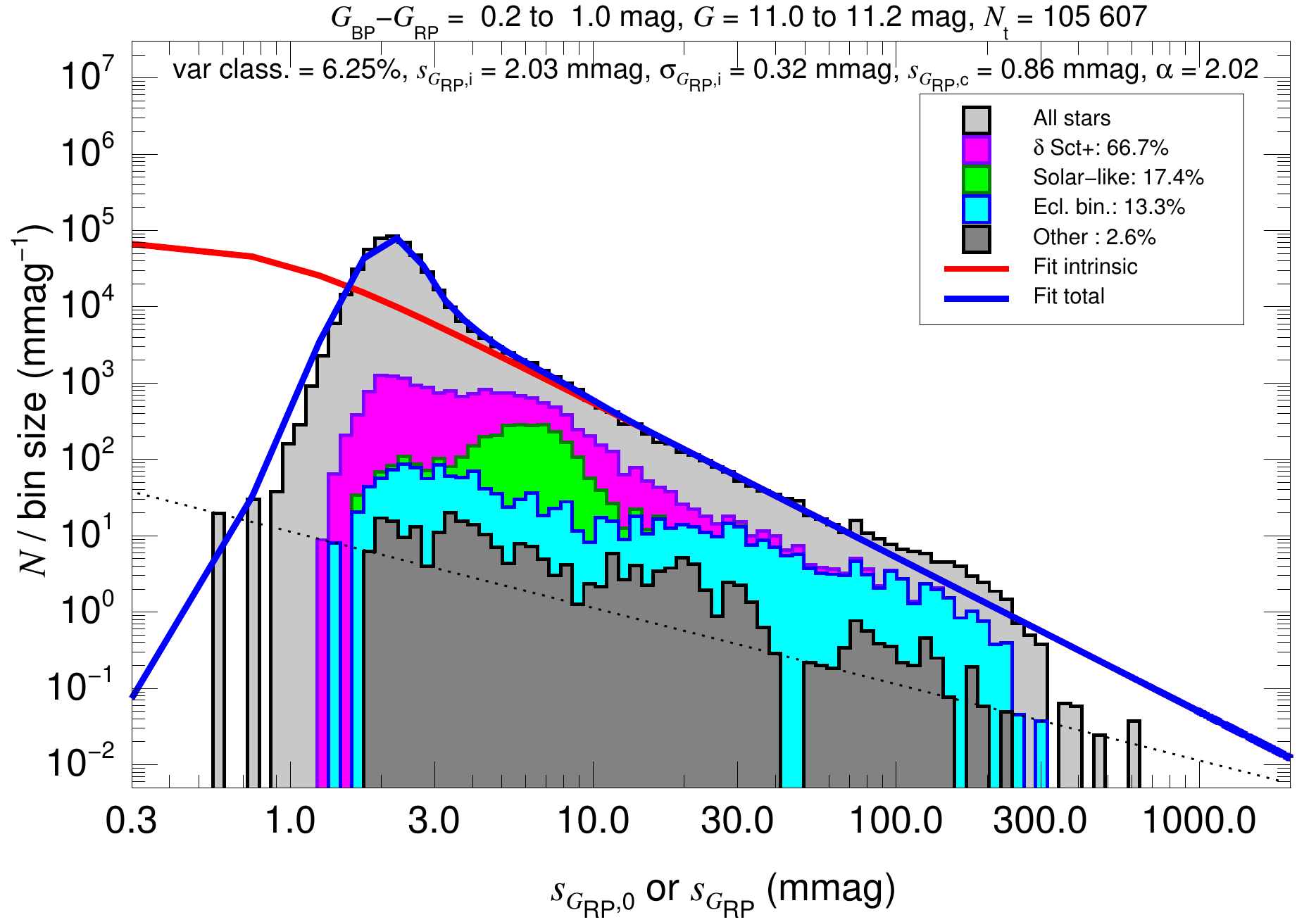}}
\centerline{$\!\!\!$\includegraphics[width=0.35\linewidth]{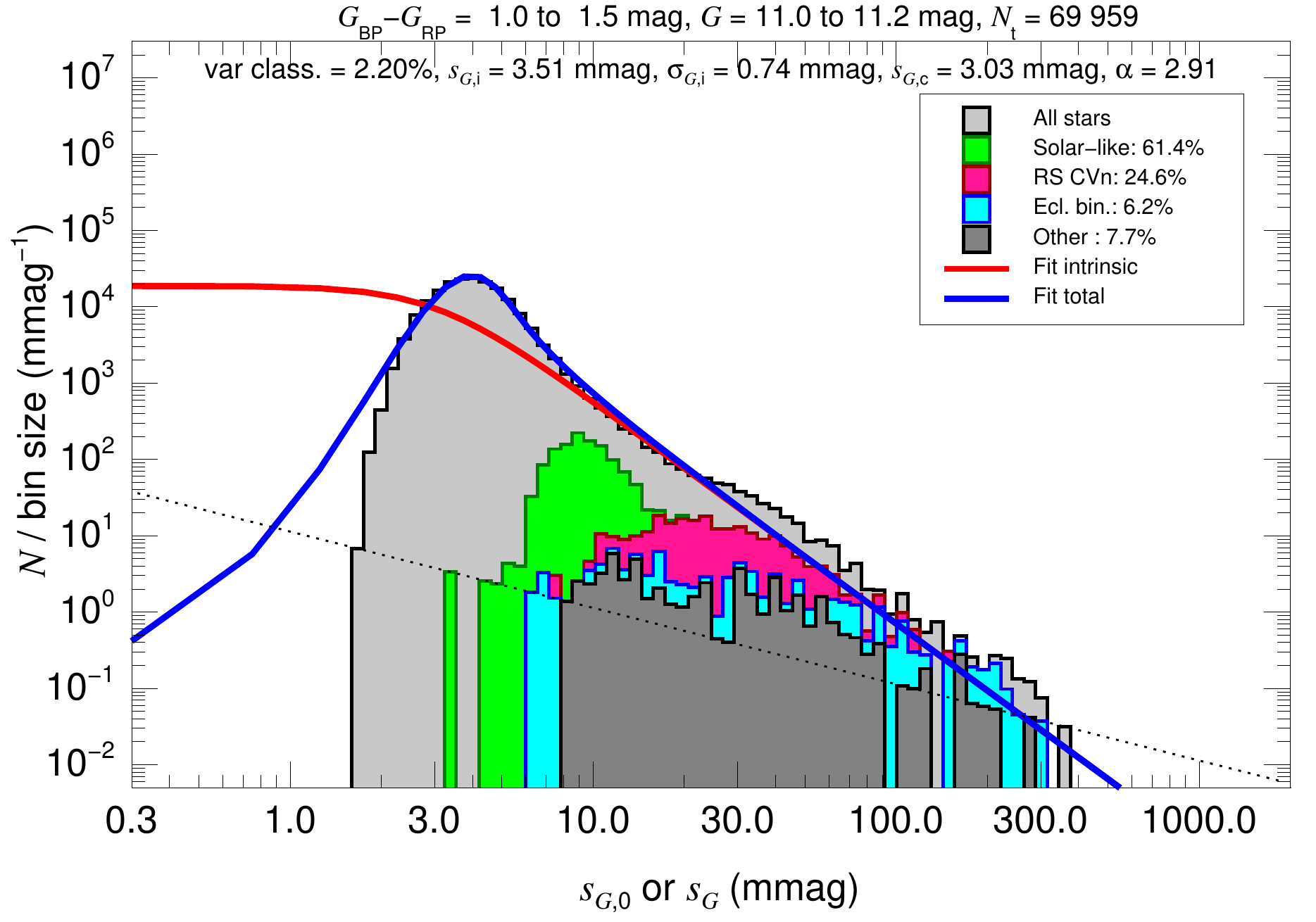}$\!\!\!$
                    \includegraphics[width=0.35\linewidth]{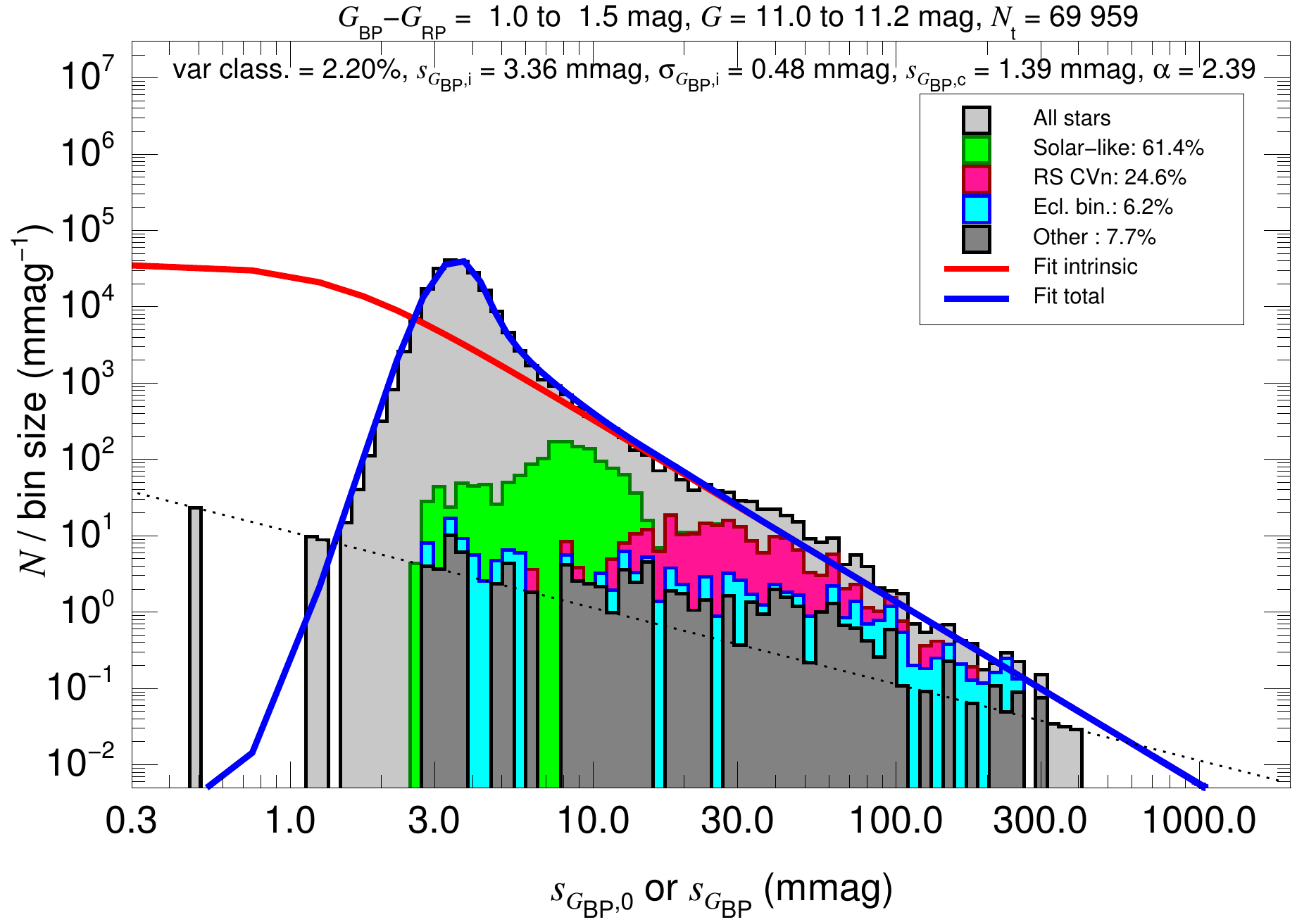}$\!\!\!$
                    \includegraphics[width=0.35\linewidth]{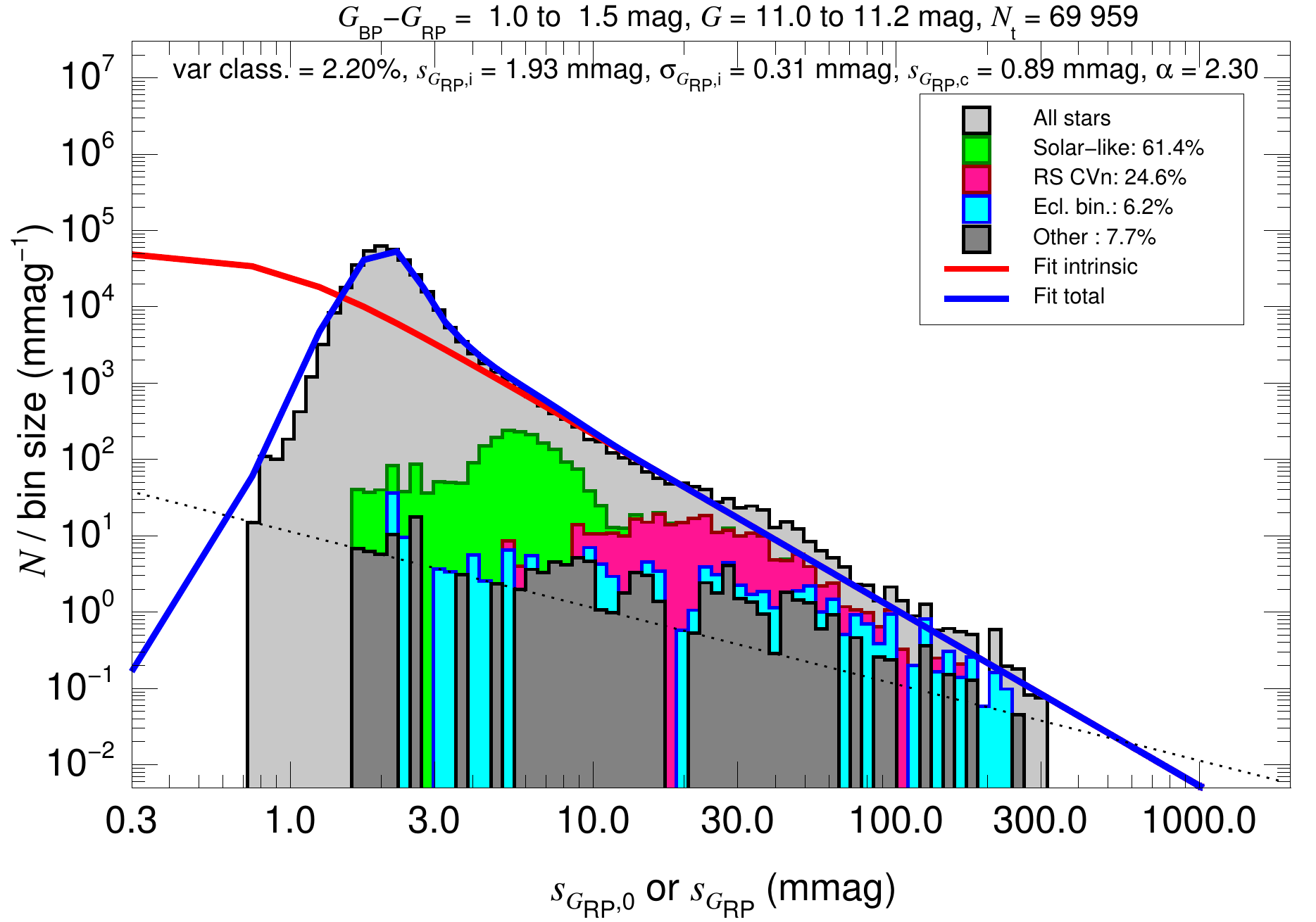}}
\centerline{$\!\!\!$\includegraphics[width=0.35\linewidth]{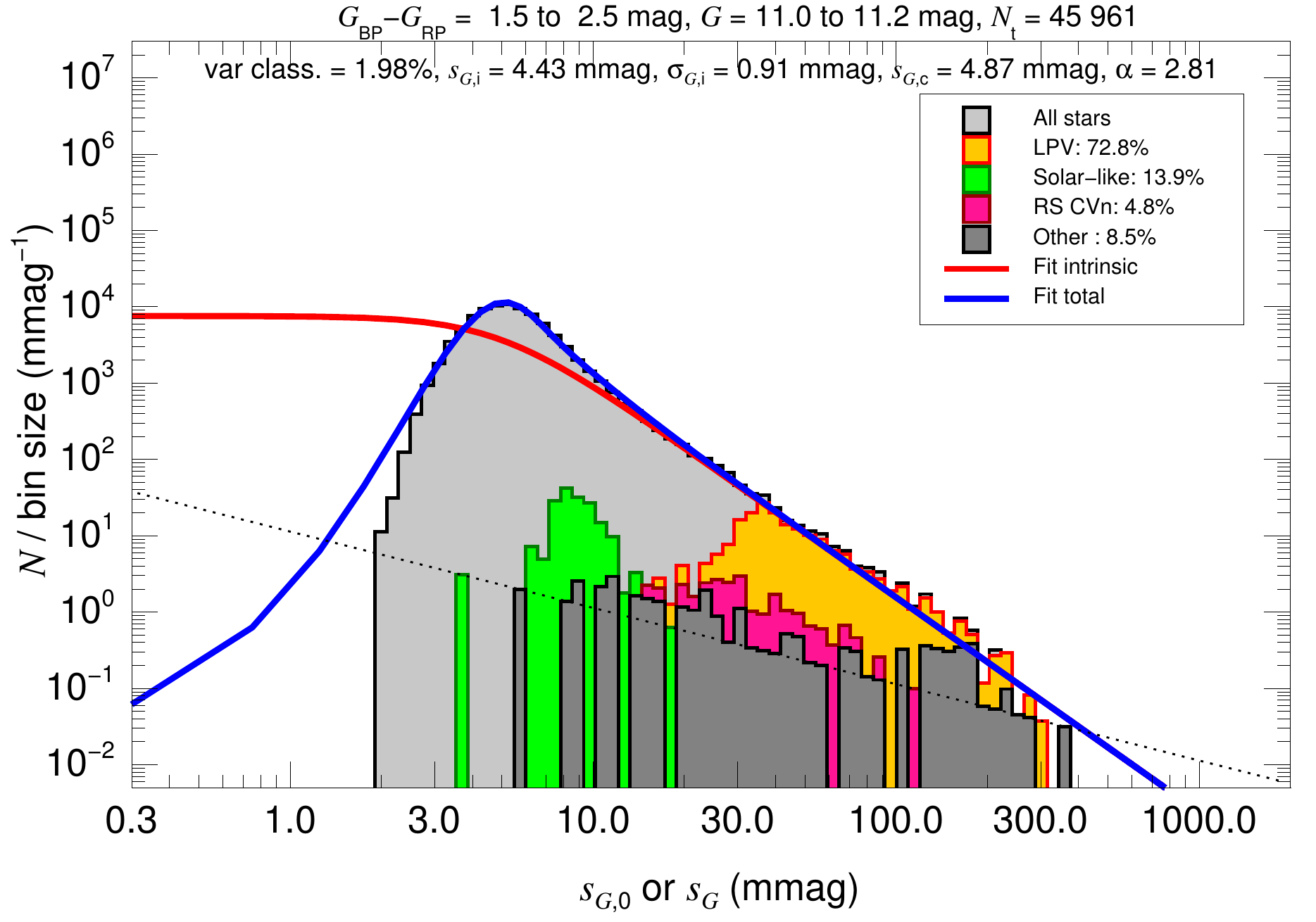}$\!\!\!$
                    \includegraphics[width=0.35\linewidth]{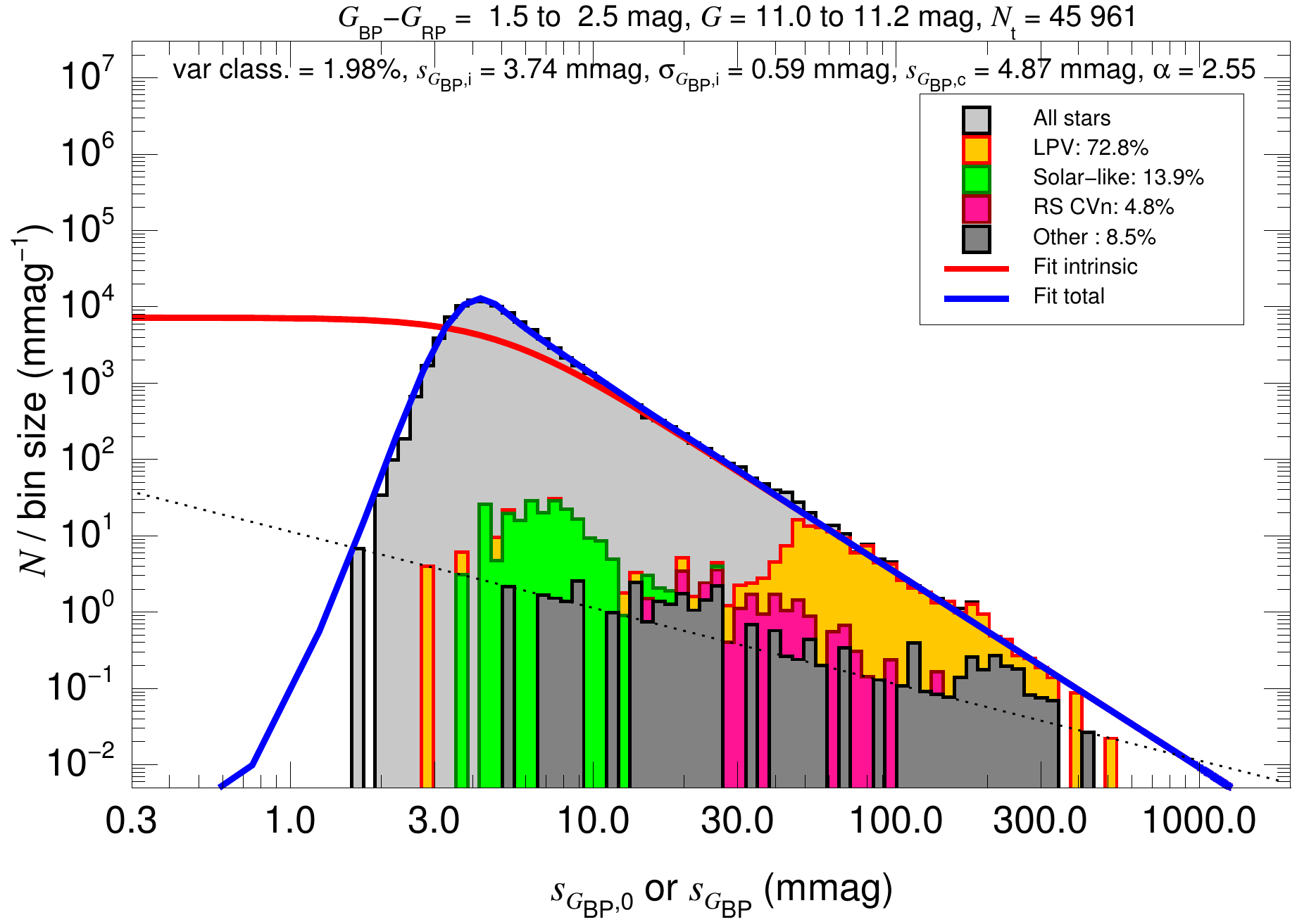}$\!\!\!$
                    \includegraphics[width=0.35\linewidth]{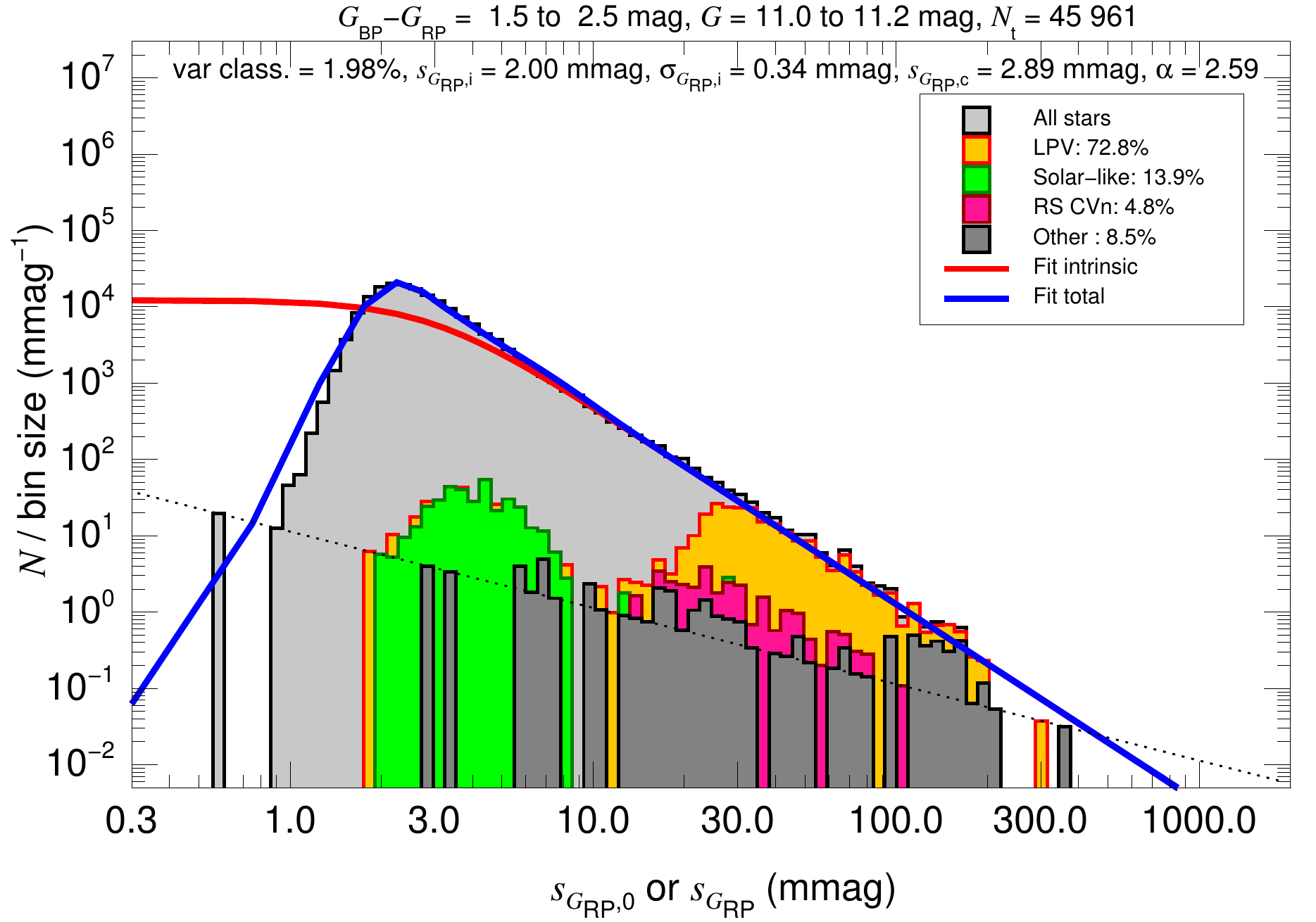}}
\centerline{$\!\!\!$\includegraphics[width=0.35\linewidth]{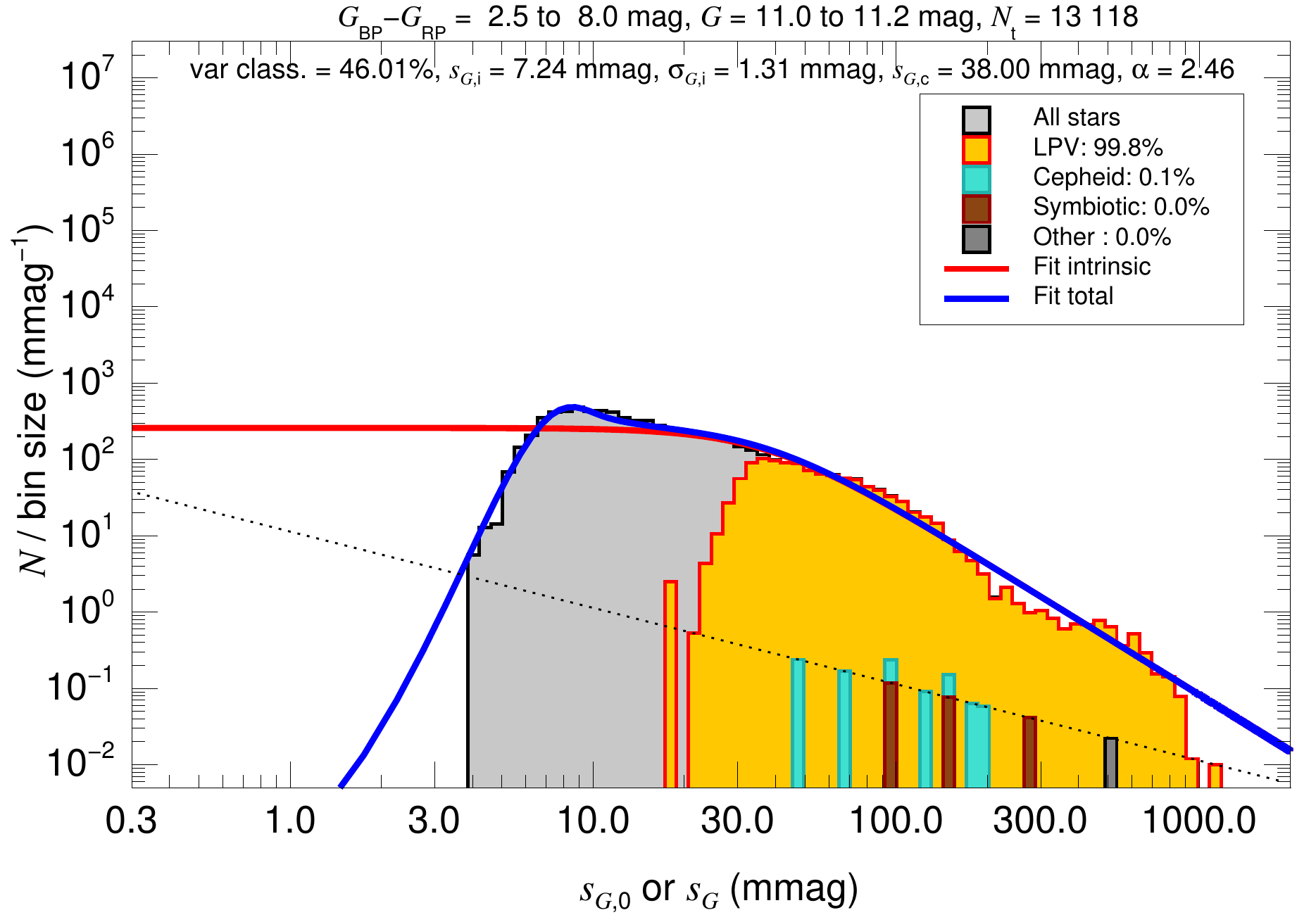}$\!\!\!$
                    \includegraphics[width=0.35\linewidth]{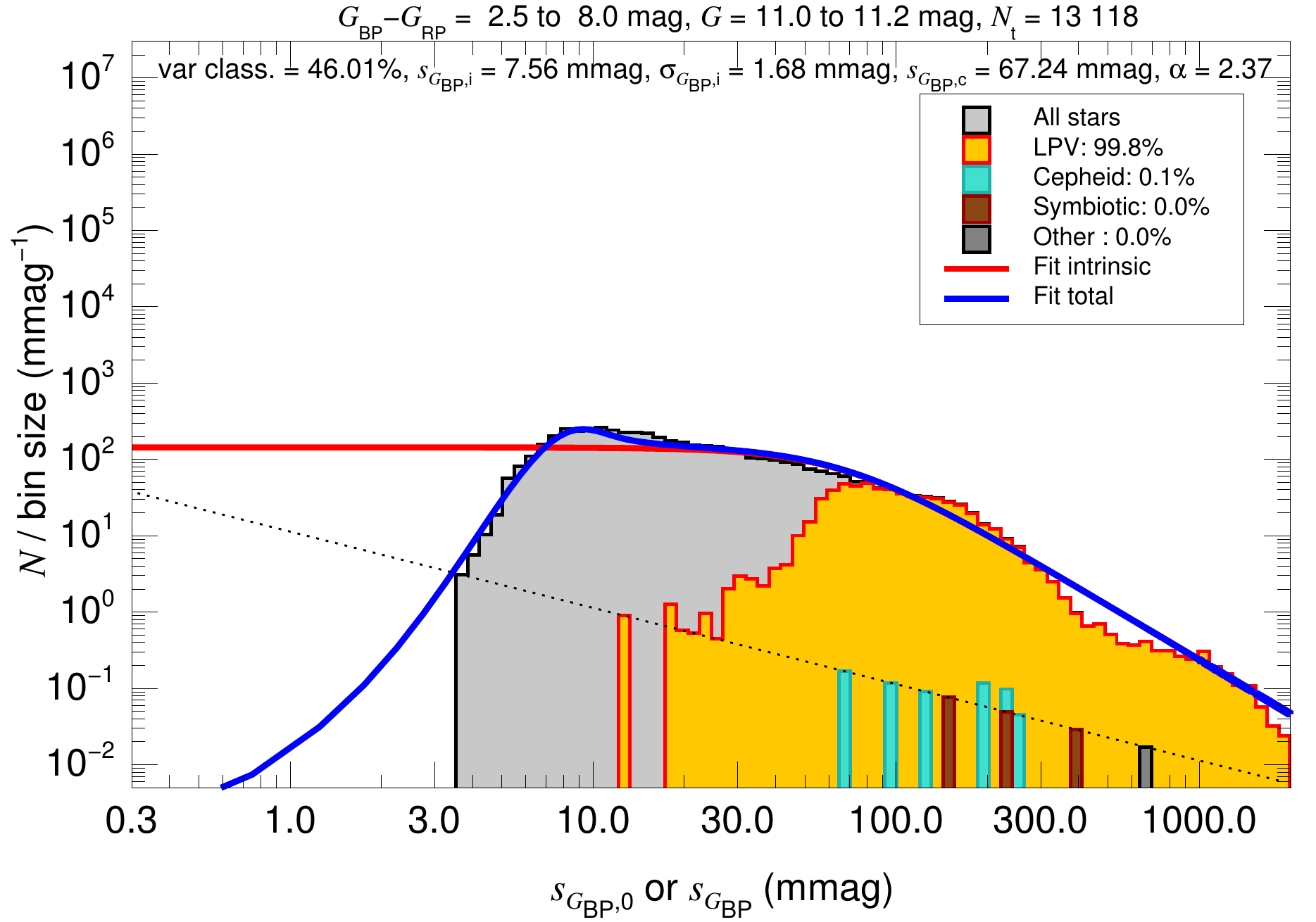}$\!\!\!$
                    \includegraphics[width=0.35\linewidth]{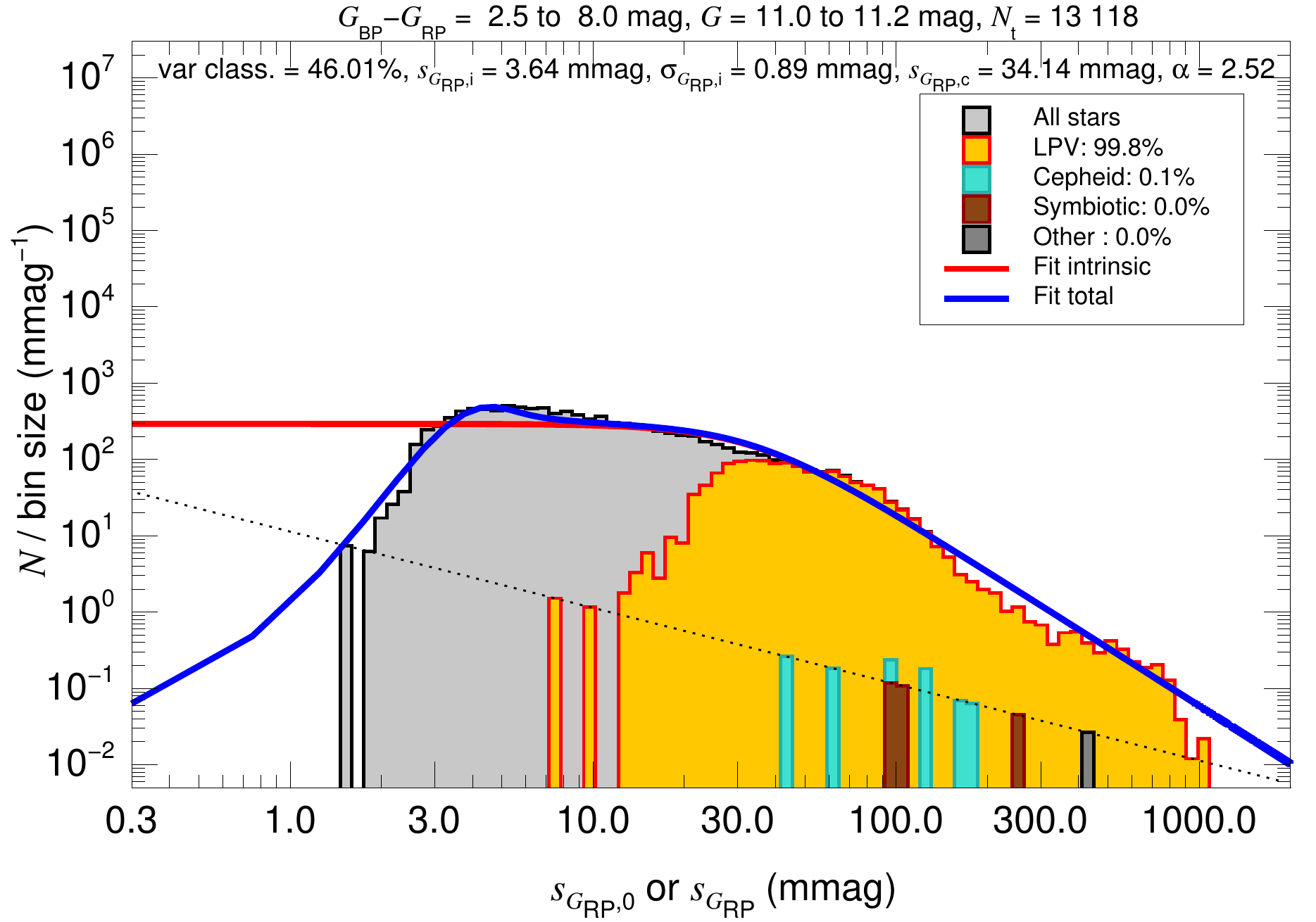}}
\caption{(Continued).}
\end{figure*}

\addtocounter{figure}{-1}

\begin{figure*}
\centerline{$\!\!\!$\includegraphics[width=0.35\linewidth]{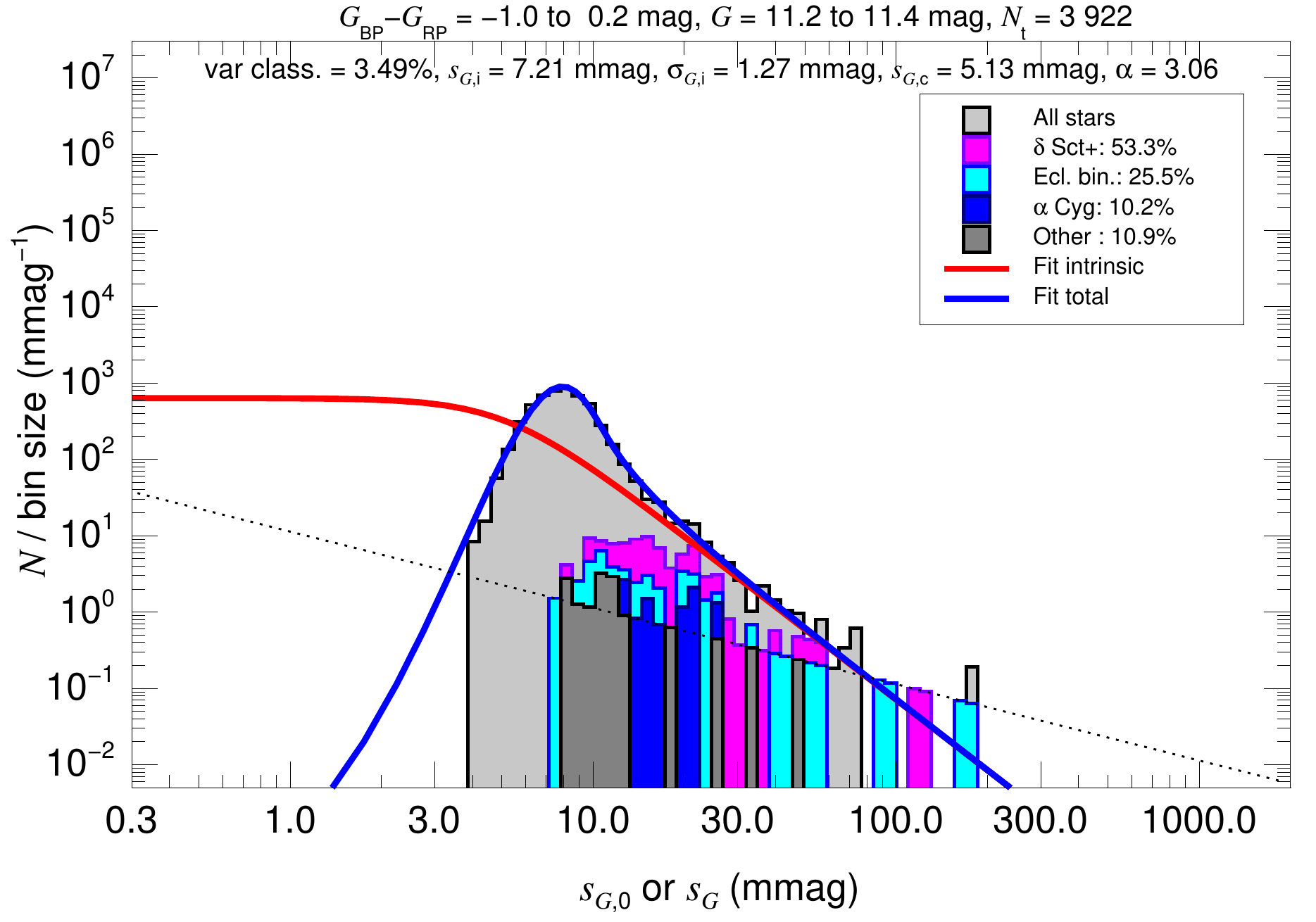}$\!\!\!$
                    \includegraphics[width=0.35\linewidth]{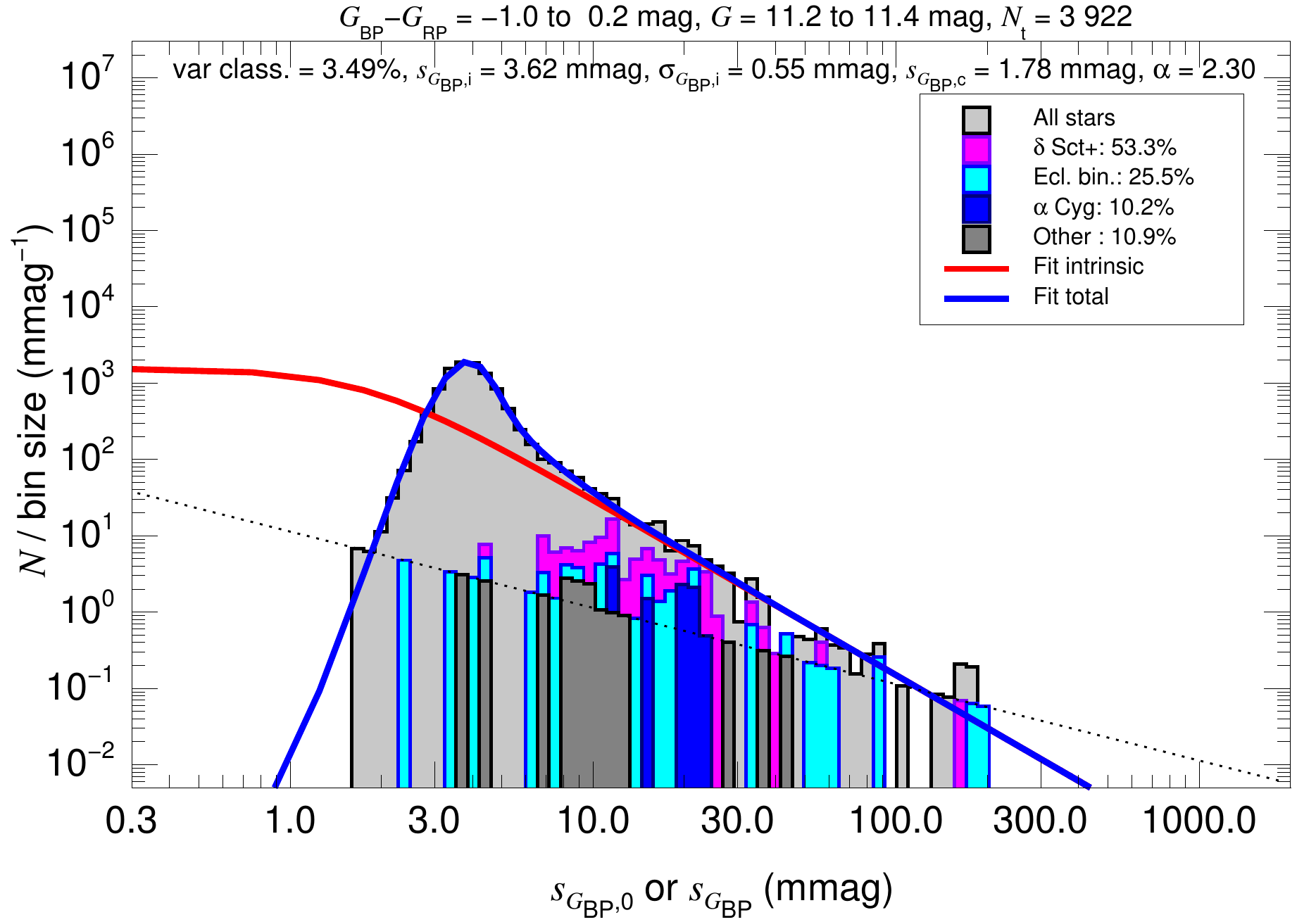}$\!\!\!$
                    \includegraphics[width=0.35\linewidth]{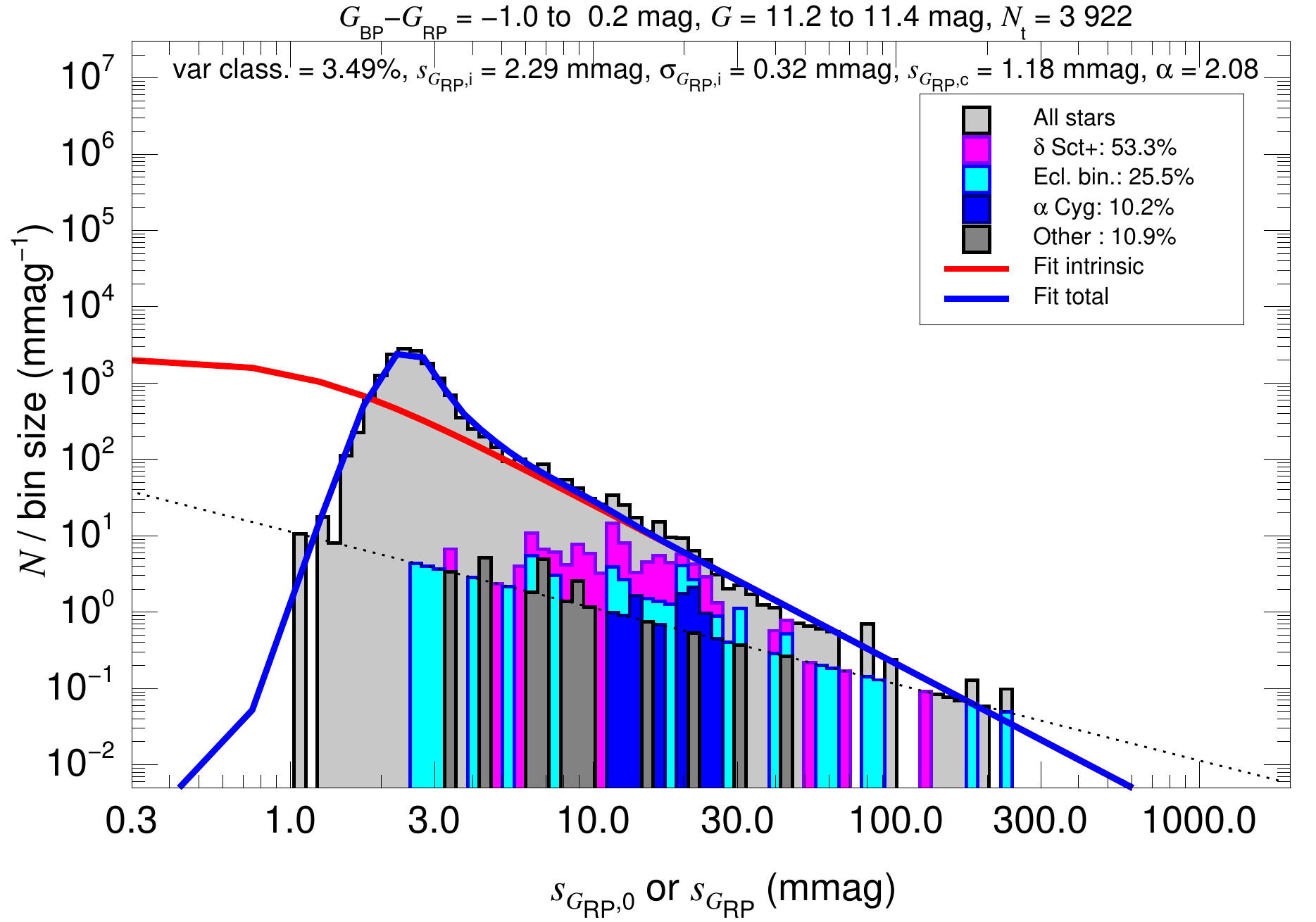}}
\centerline{$\!\!\!$\includegraphics[width=0.35\linewidth]{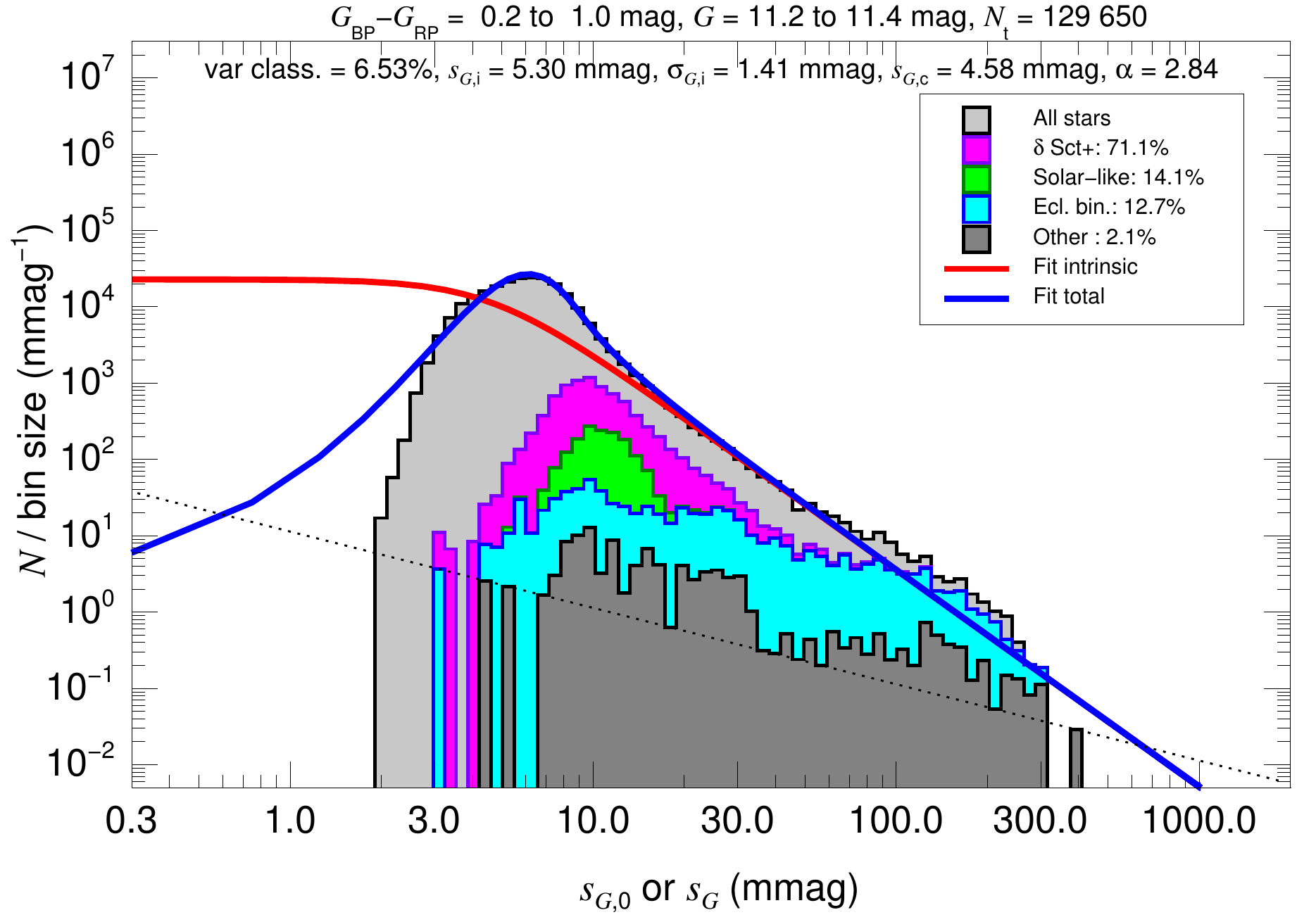}$\!\!\!$
                    \includegraphics[width=0.35\linewidth]{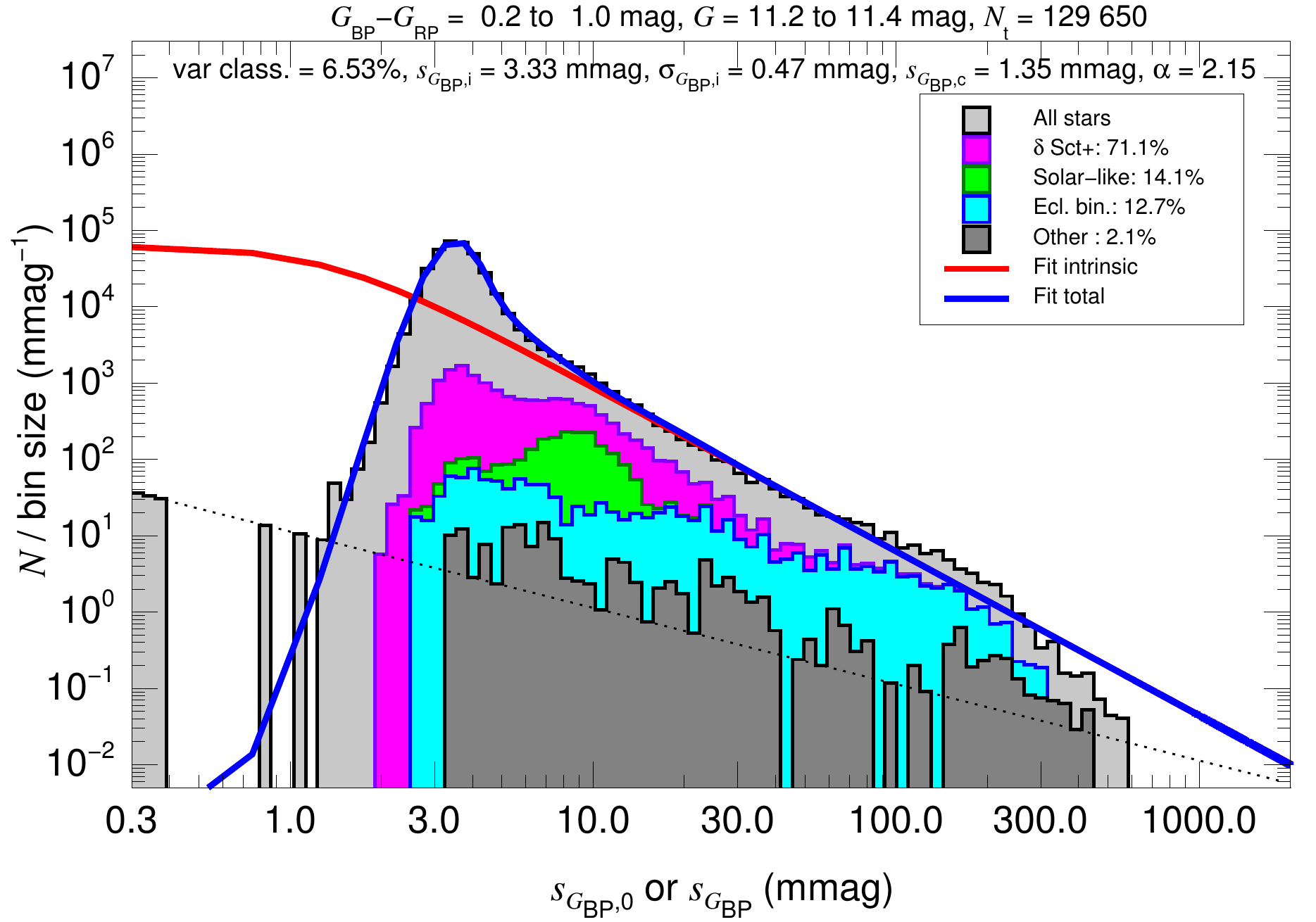}$\!\!\!$
                    \includegraphics[width=0.35\linewidth]{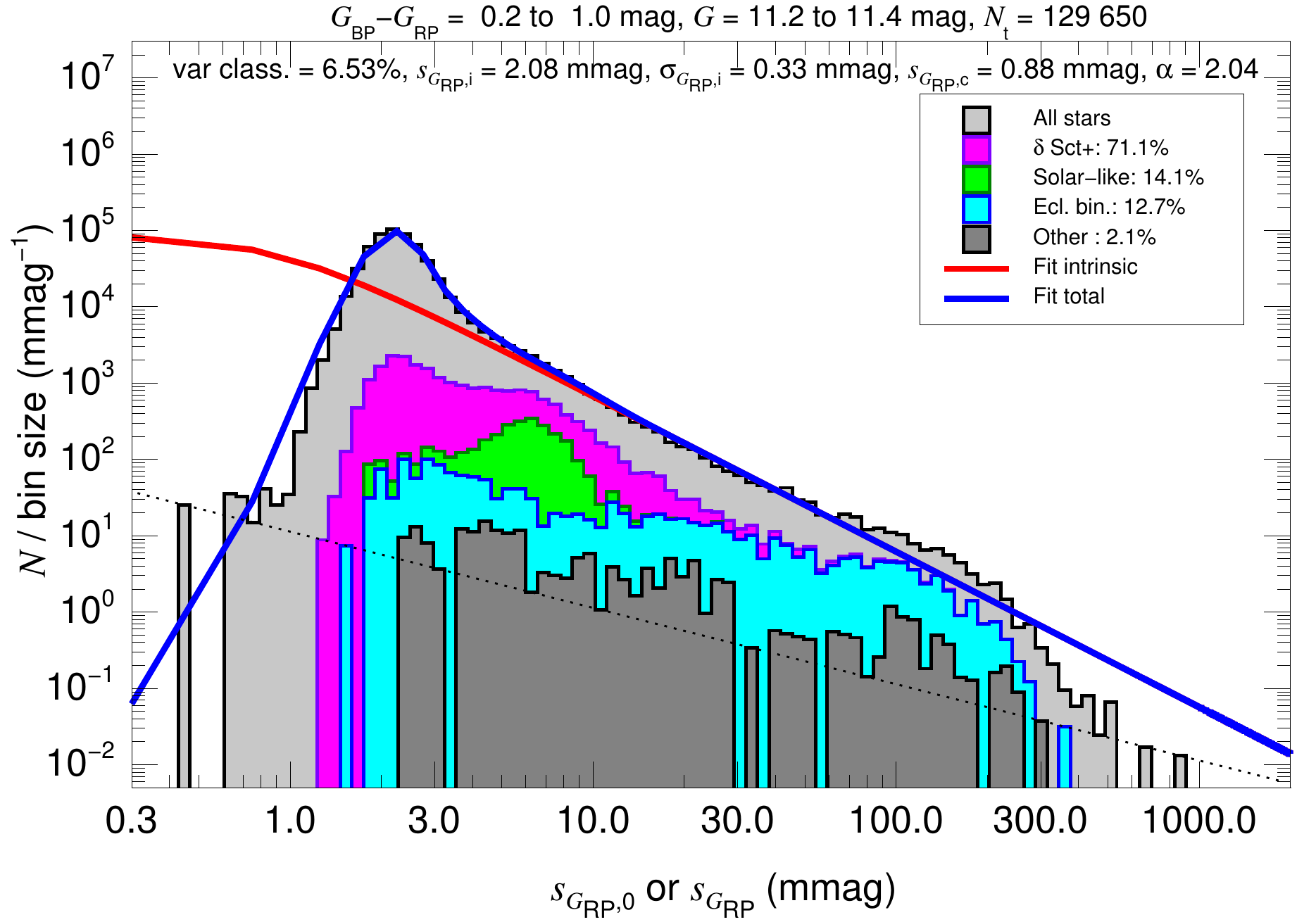}}
\centerline{$\!\!\!$\includegraphics[width=0.35\linewidth]{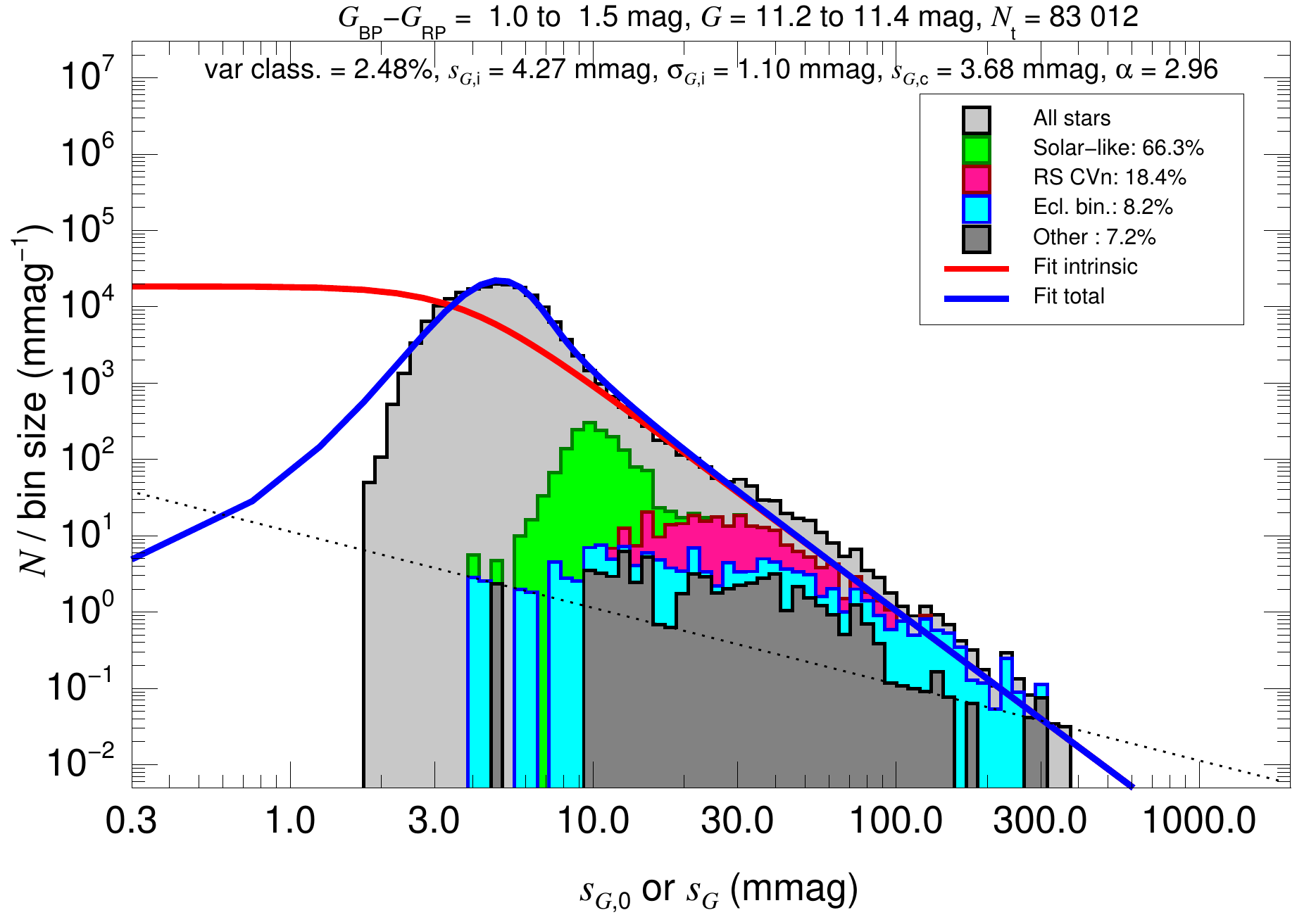}$\!\!\!$
                    \includegraphics[width=0.35\linewidth]{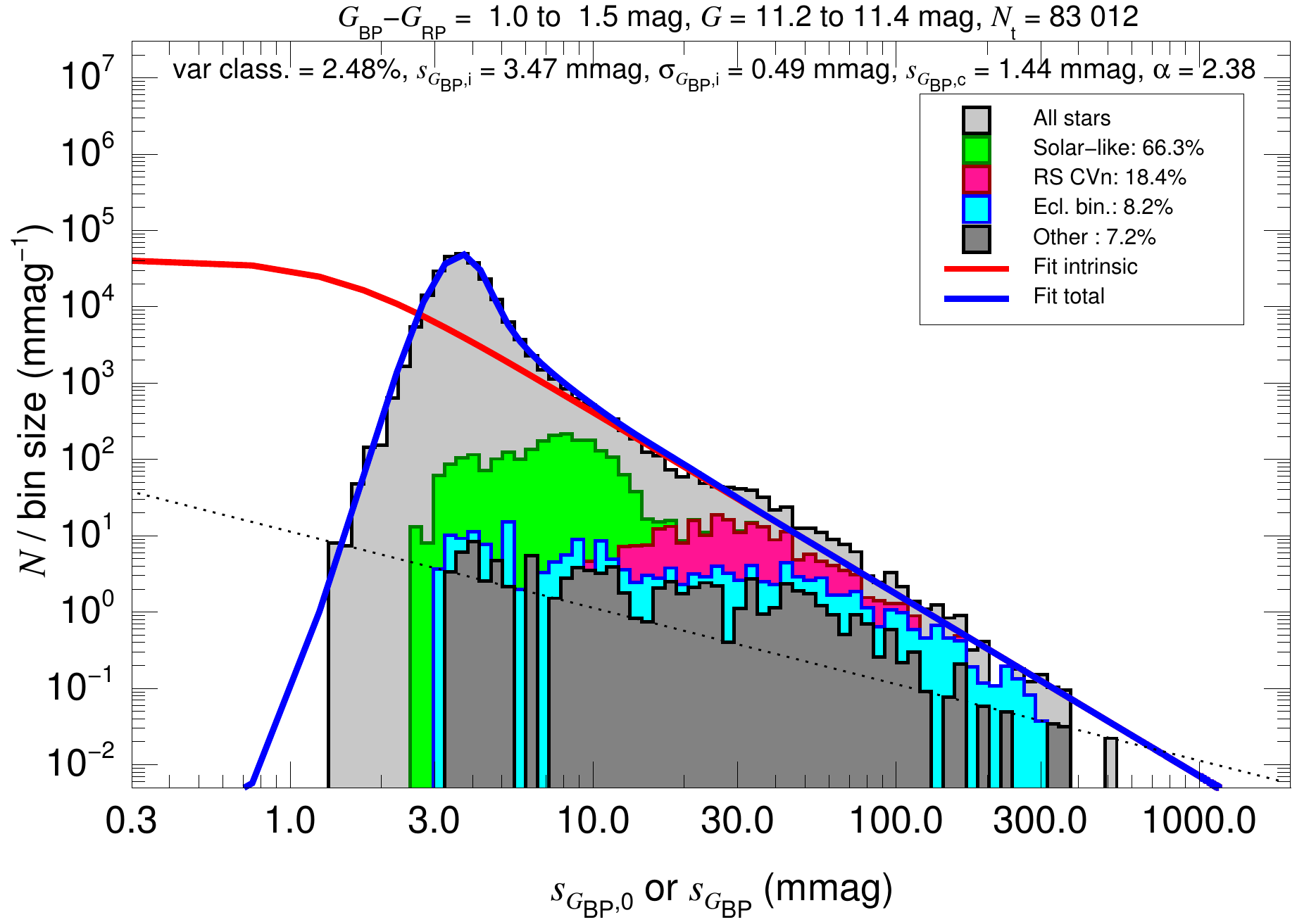}$\!\!\!$
                    \includegraphics[width=0.35\linewidth]{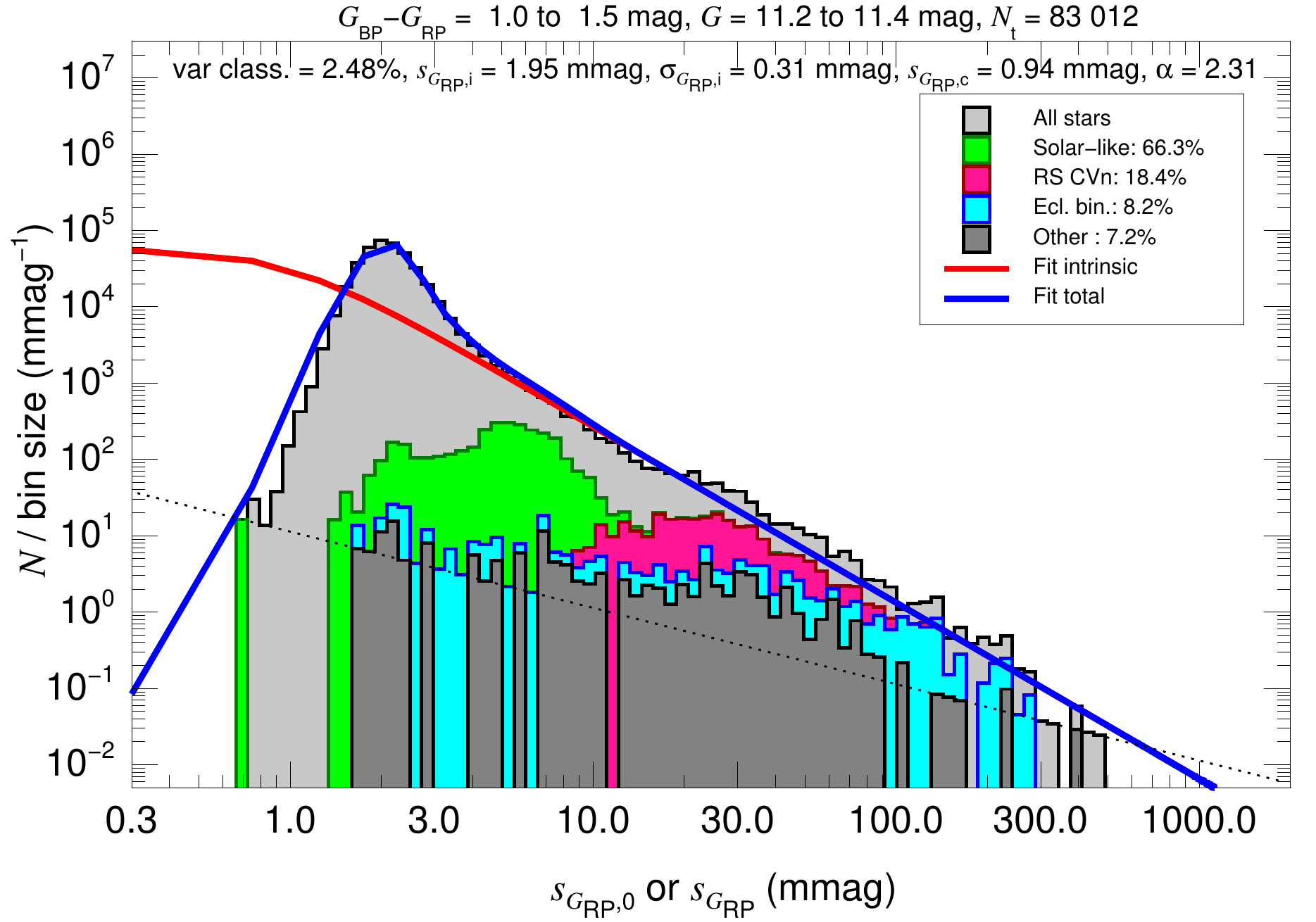}}
\centerline{$\!\!\!$\includegraphics[width=0.35\linewidth]{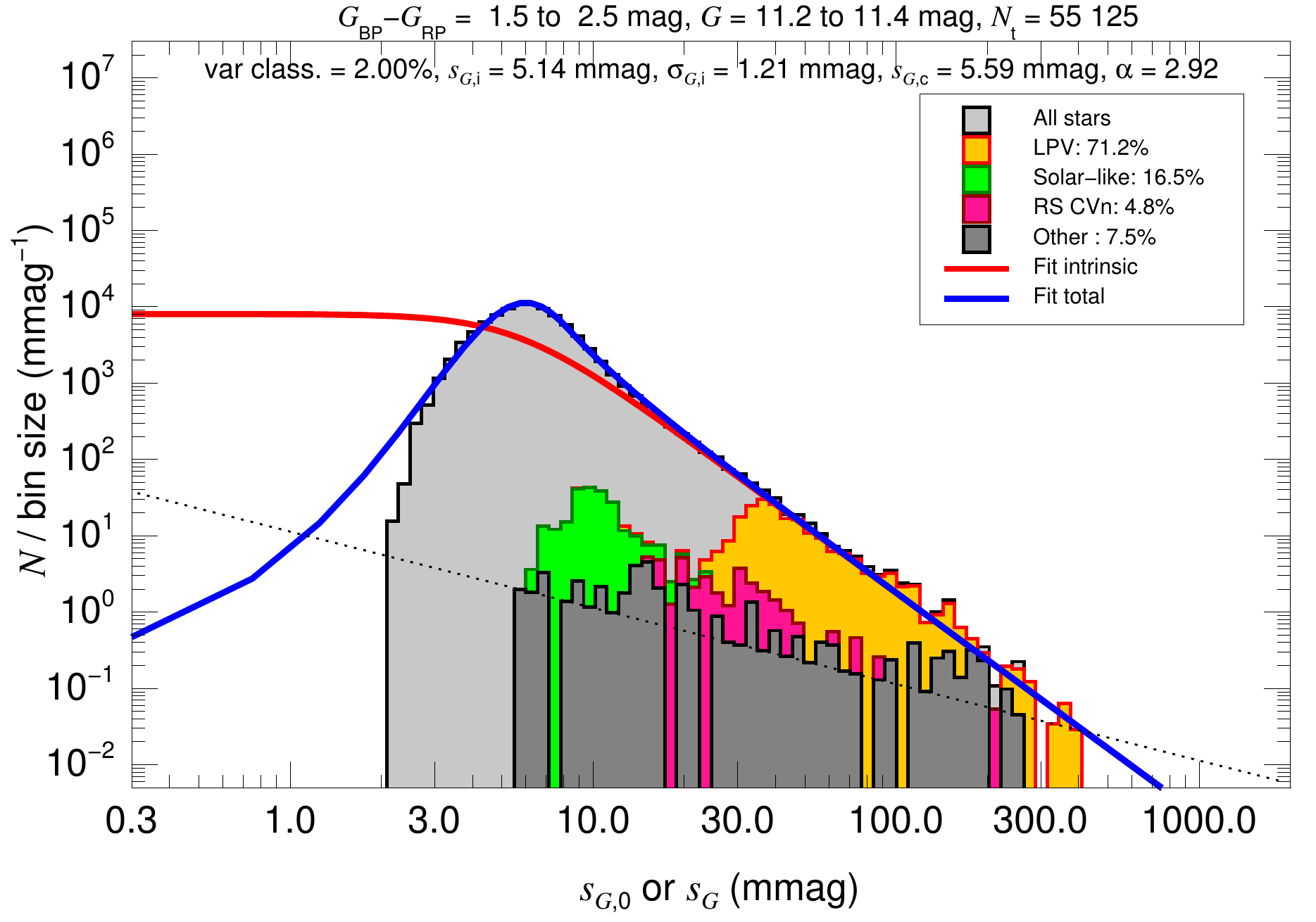}$\!\!\!$
                    \includegraphics[width=0.35\linewidth]{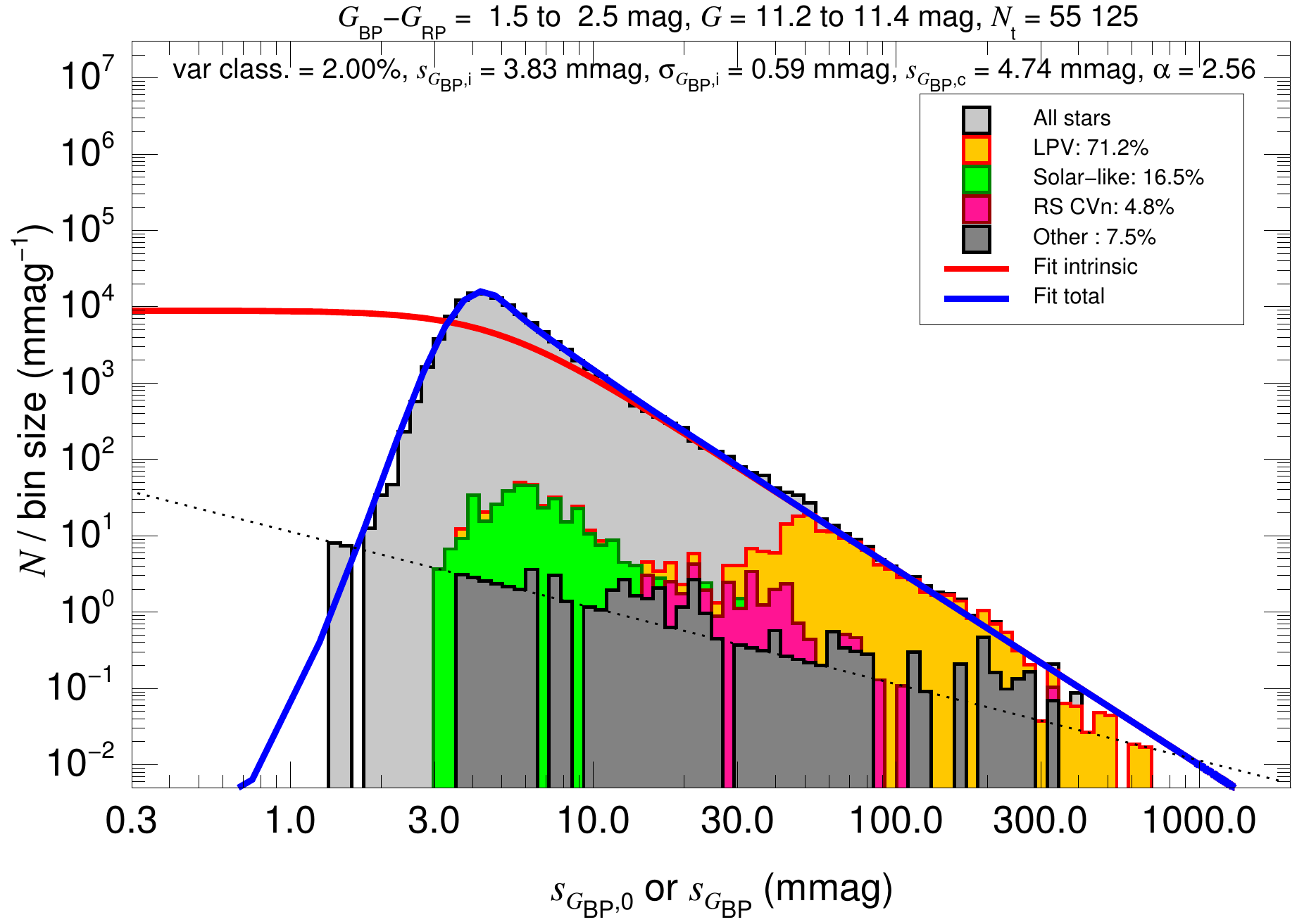}$\!\!\!$
                    \includegraphics[width=0.35\linewidth]{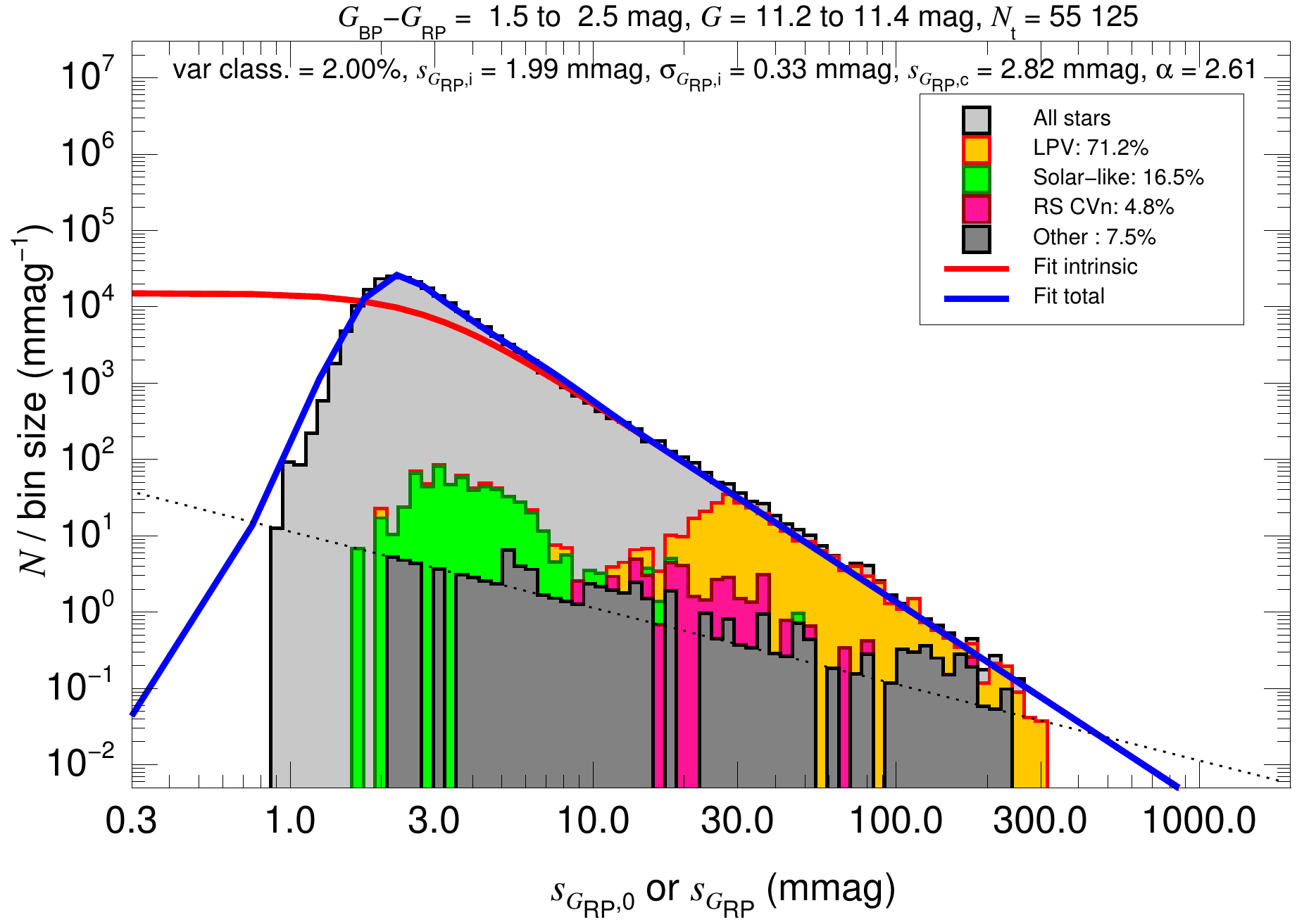}}
\centerline{$\!\!\!$\includegraphics[width=0.35\linewidth]{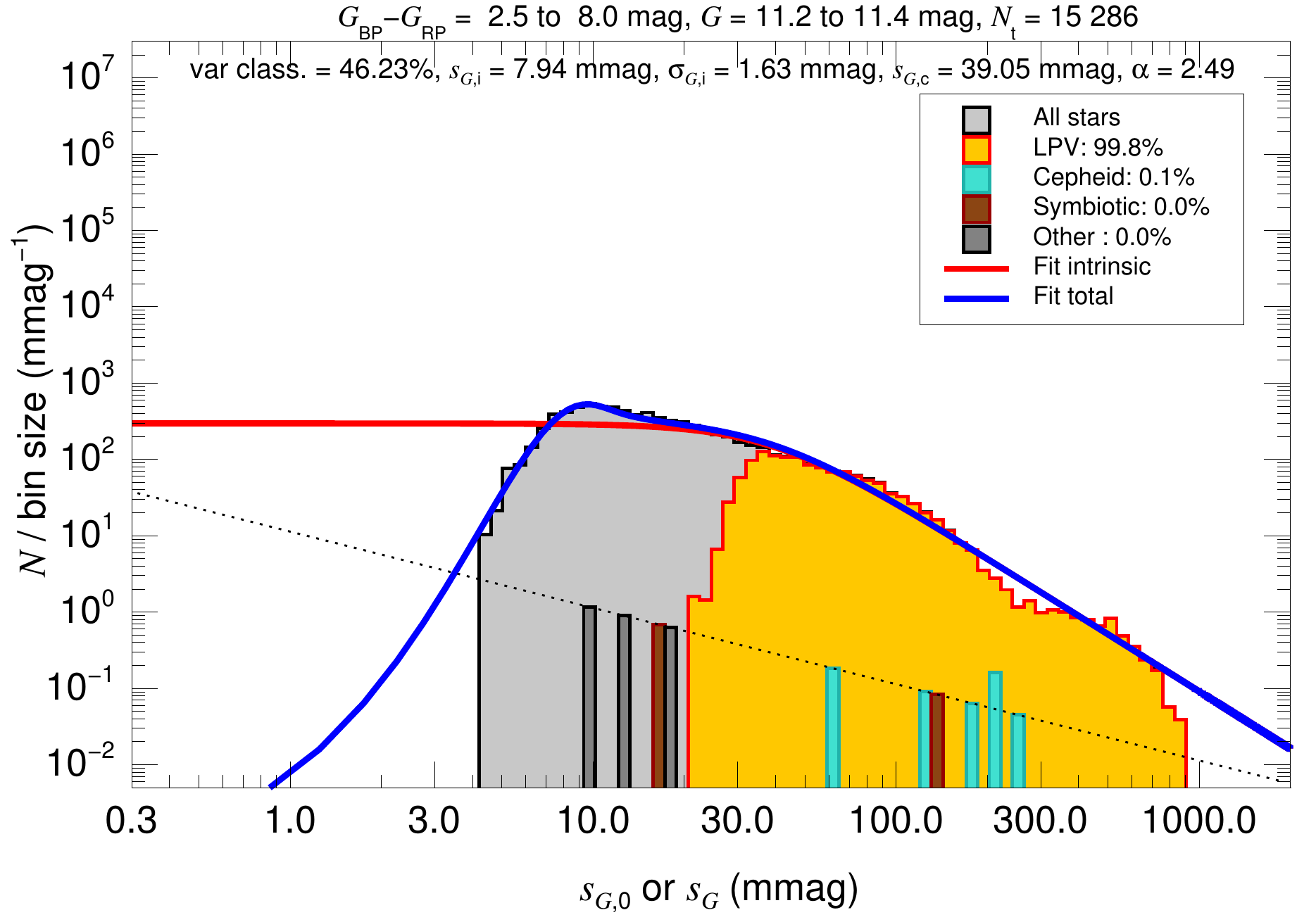}$\!\!\!$
                    \includegraphics[width=0.35\linewidth]{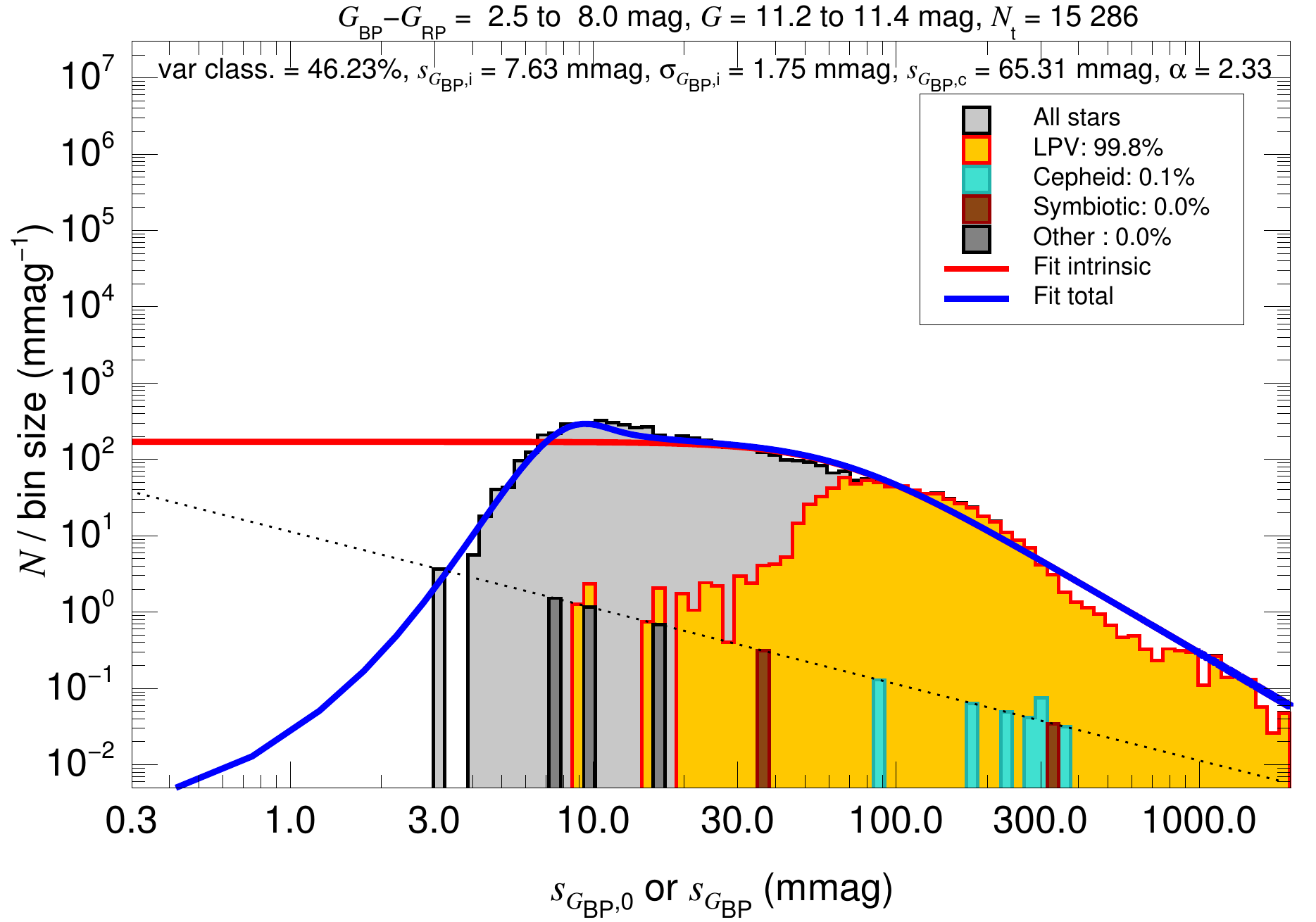}$\!\!\!$
                    \includegraphics[width=0.35\linewidth]{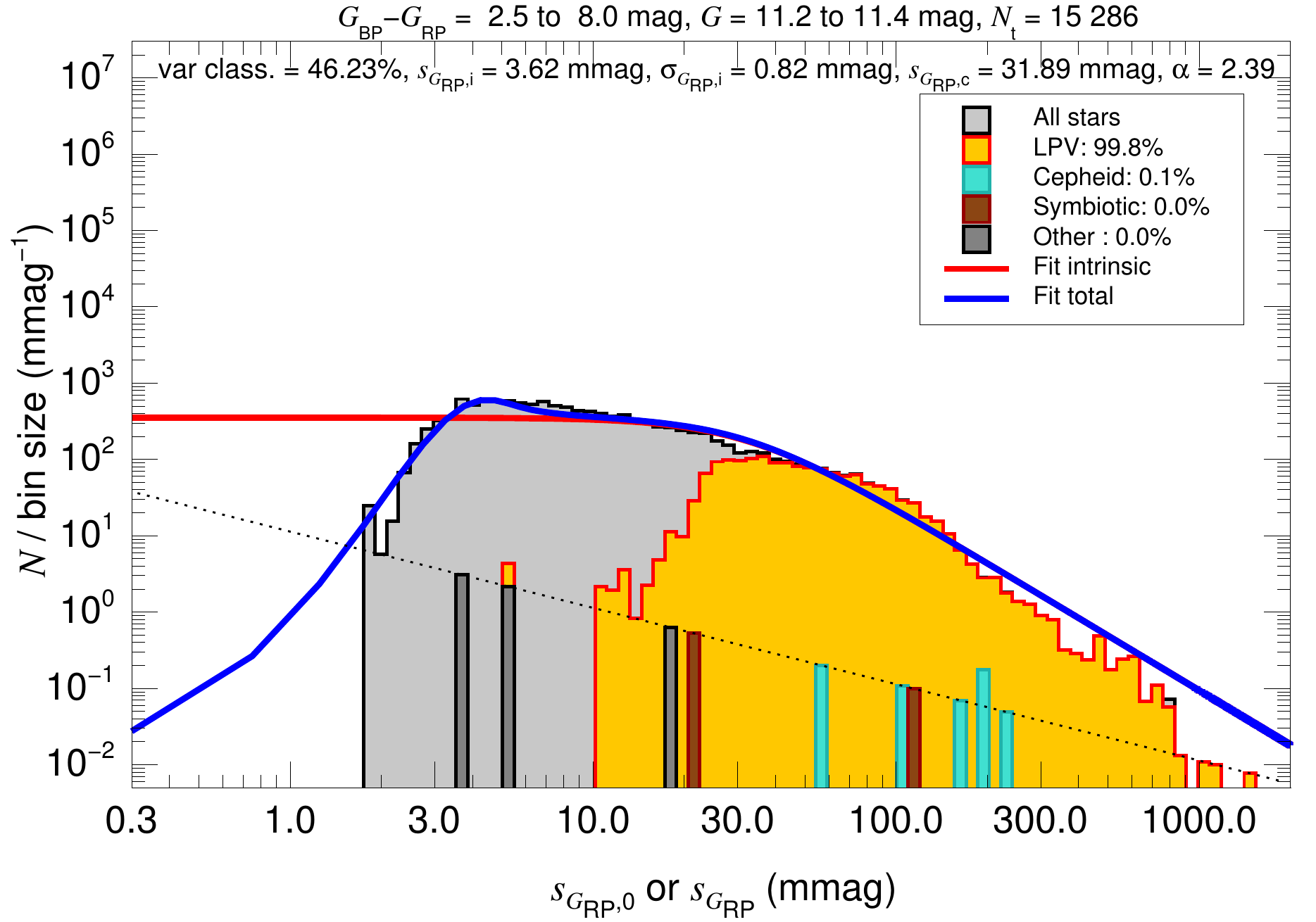}}
\caption{(Continued).}
\end{figure*}

\addtocounter{figure}{-1}

\begin{figure*}
\centerline{$\!\!\!$\includegraphics[width=0.35\linewidth]{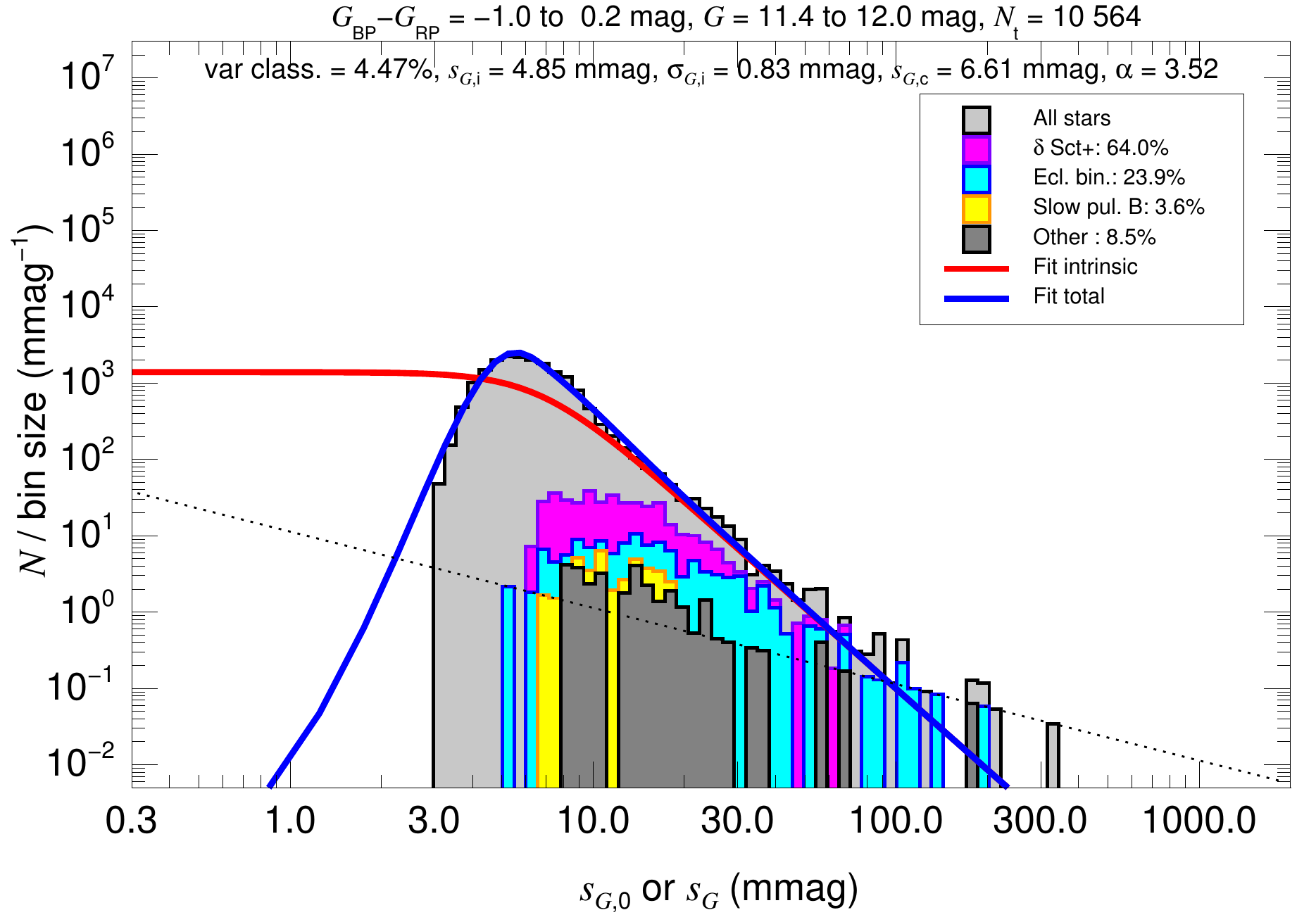}$\!\!\!$
                    \includegraphics[width=0.35\linewidth]{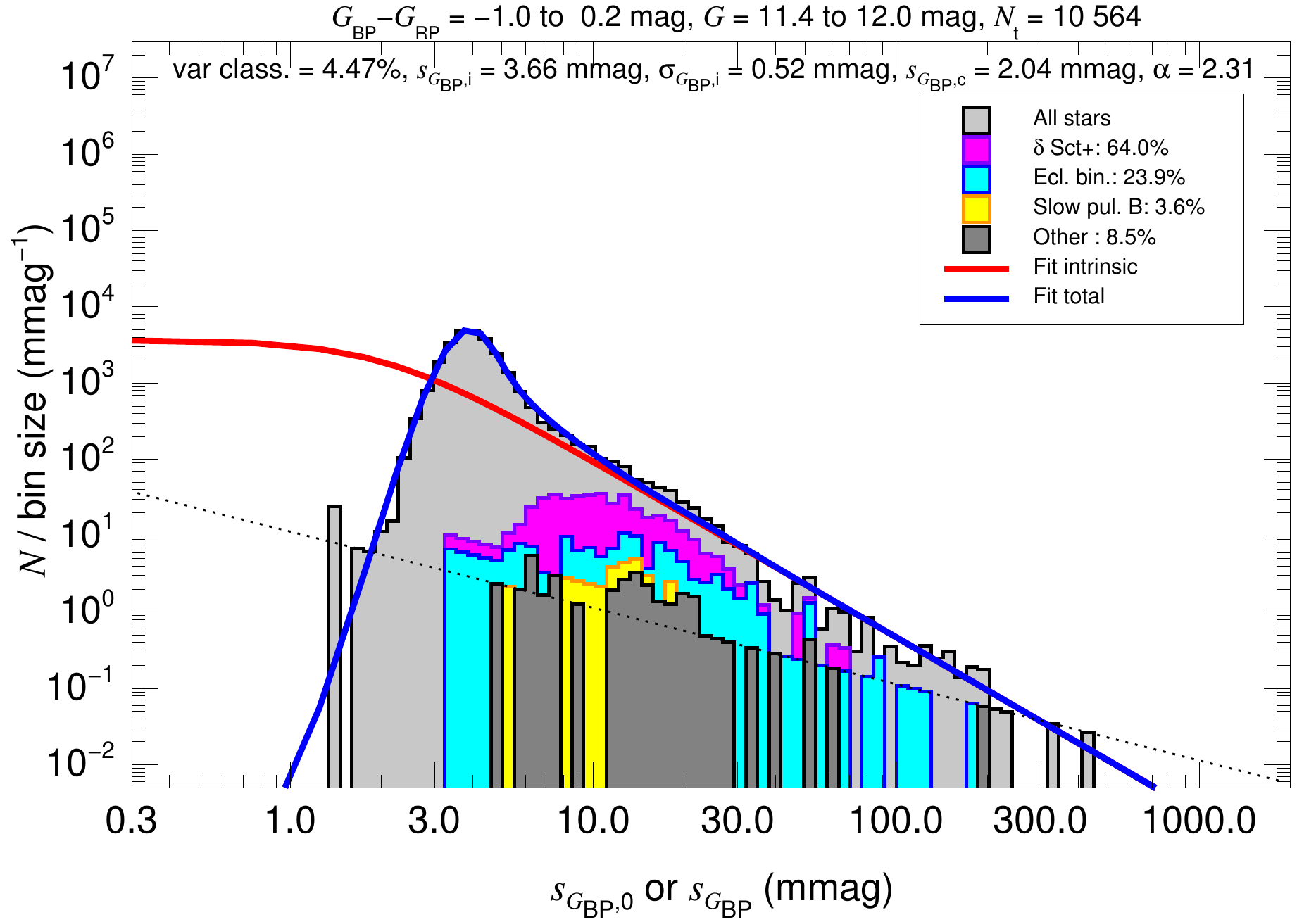}$\!\!\!$
                    \includegraphics[width=0.35\linewidth]{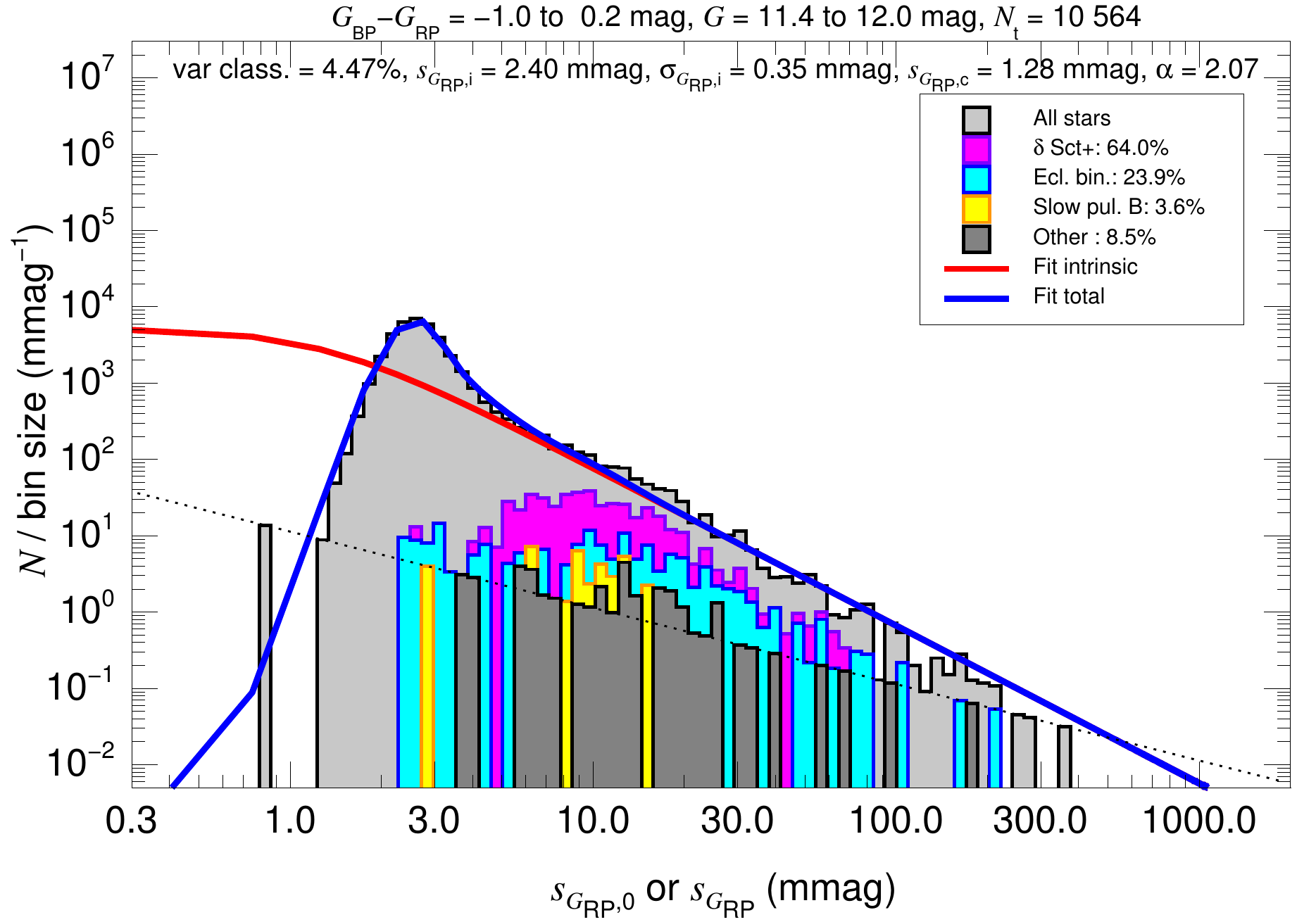}}
\centerline{$\!\!\!$\includegraphics[width=0.35\linewidth]{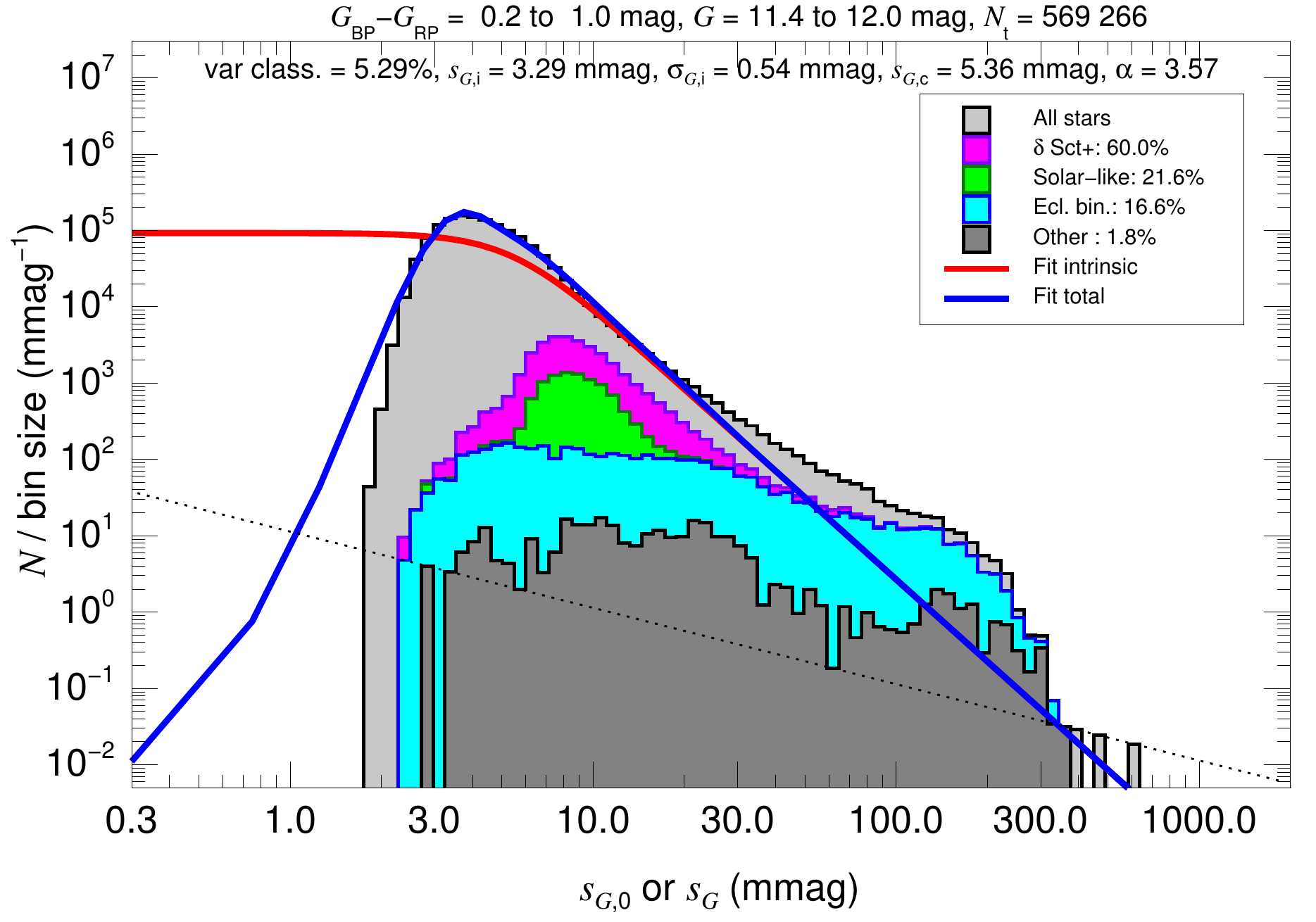}$\!\!\!$
                    \includegraphics[width=0.35\linewidth]{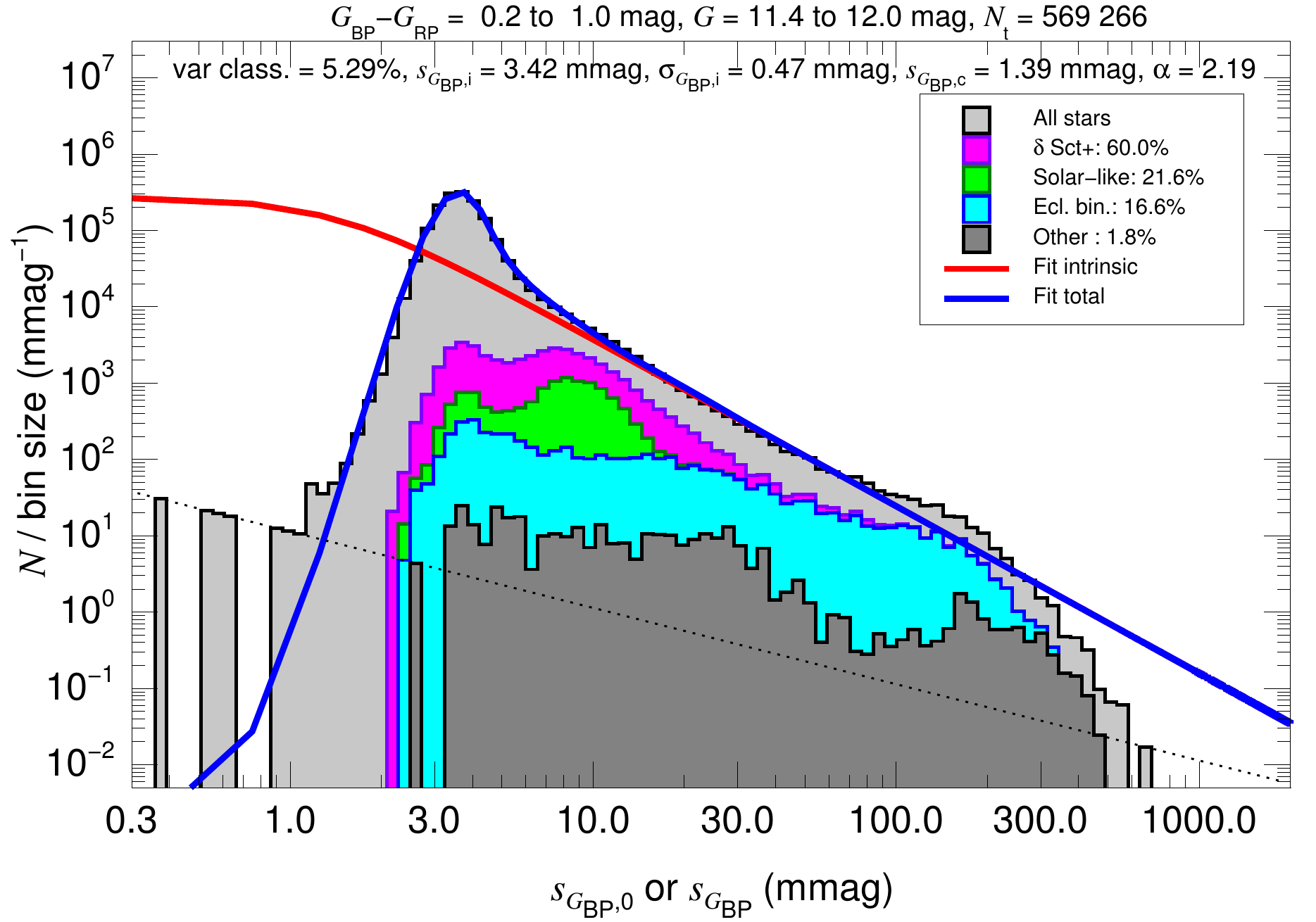}$\!\!\!$
                    \includegraphics[width=0.35\linewidth]{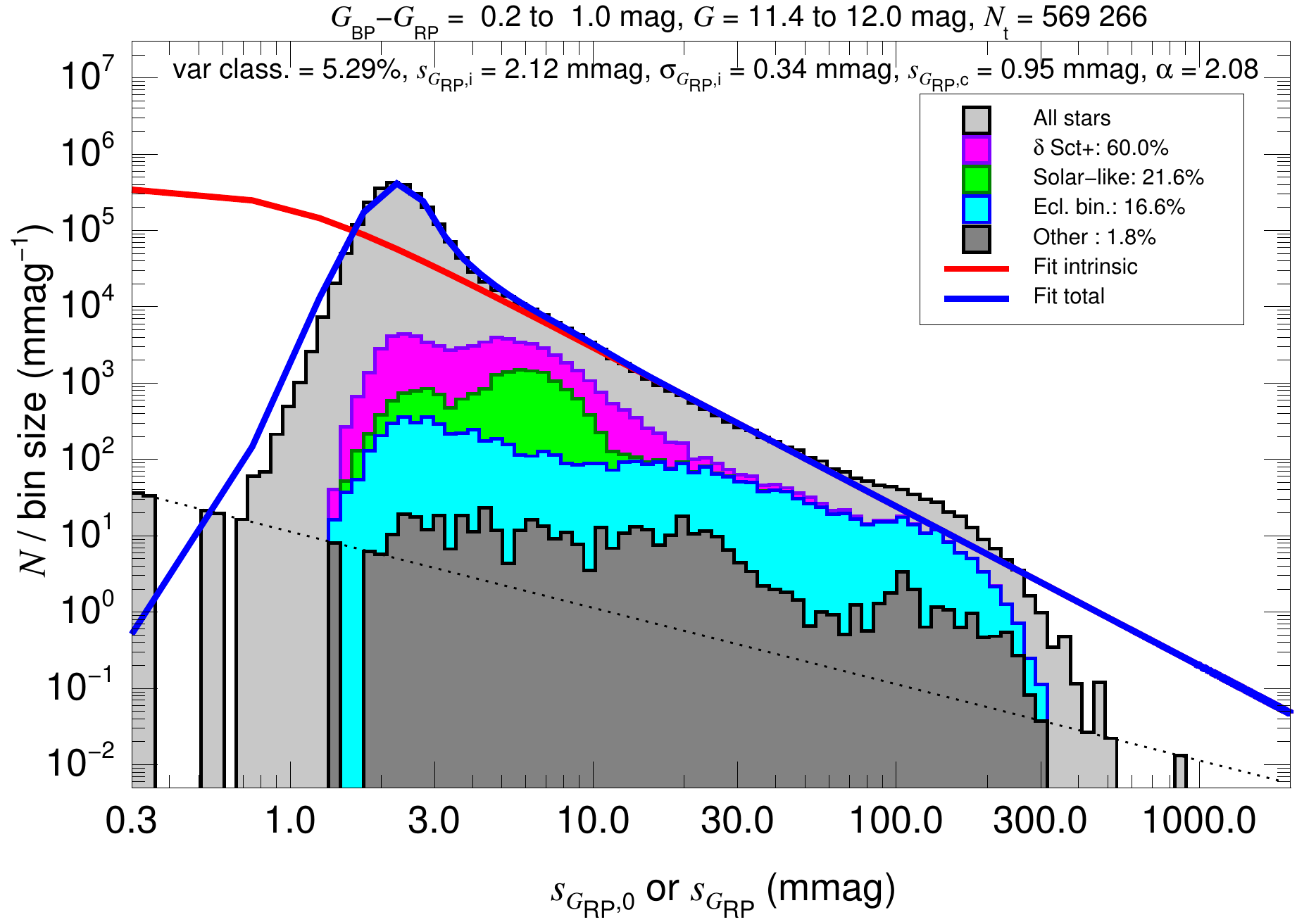}}
\centerline{$\!\!\!$\includegraphics[width=0.35\linewidth]{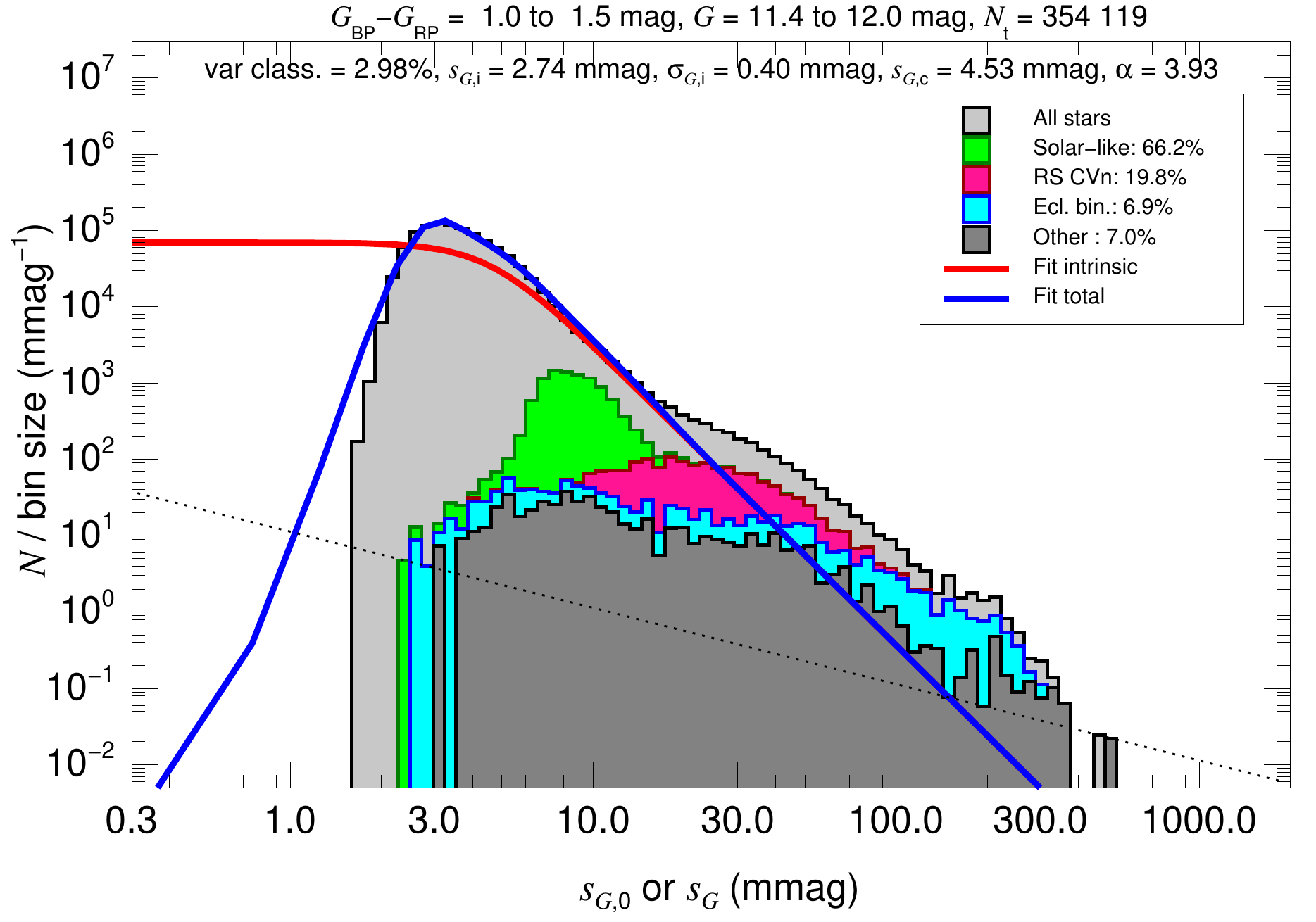}$\!\!\!$
                    \includegraphics[width=0.35\linewidth]{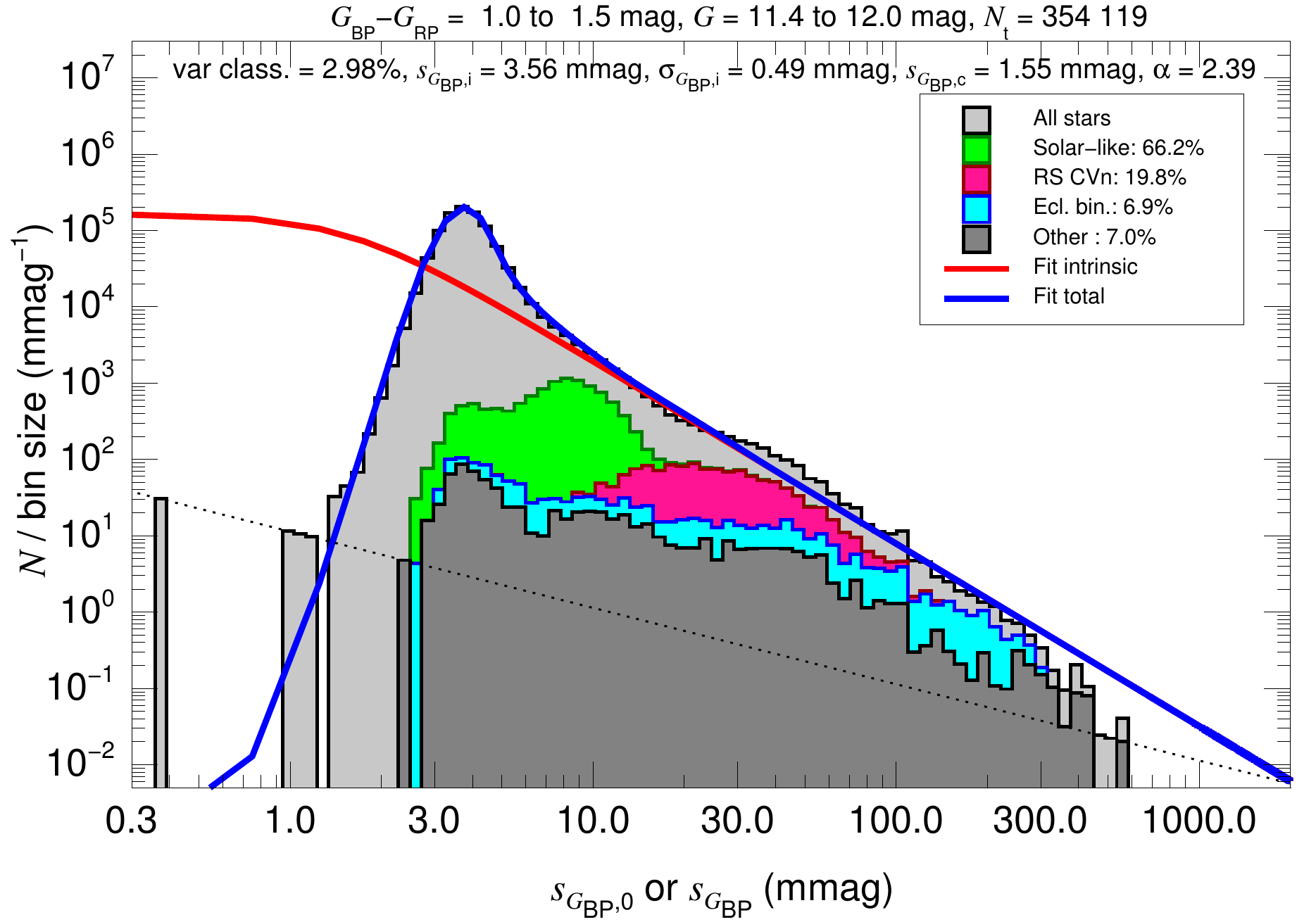}$\!\!\!$
                    \includegraphics[width=0.35\linewidth]{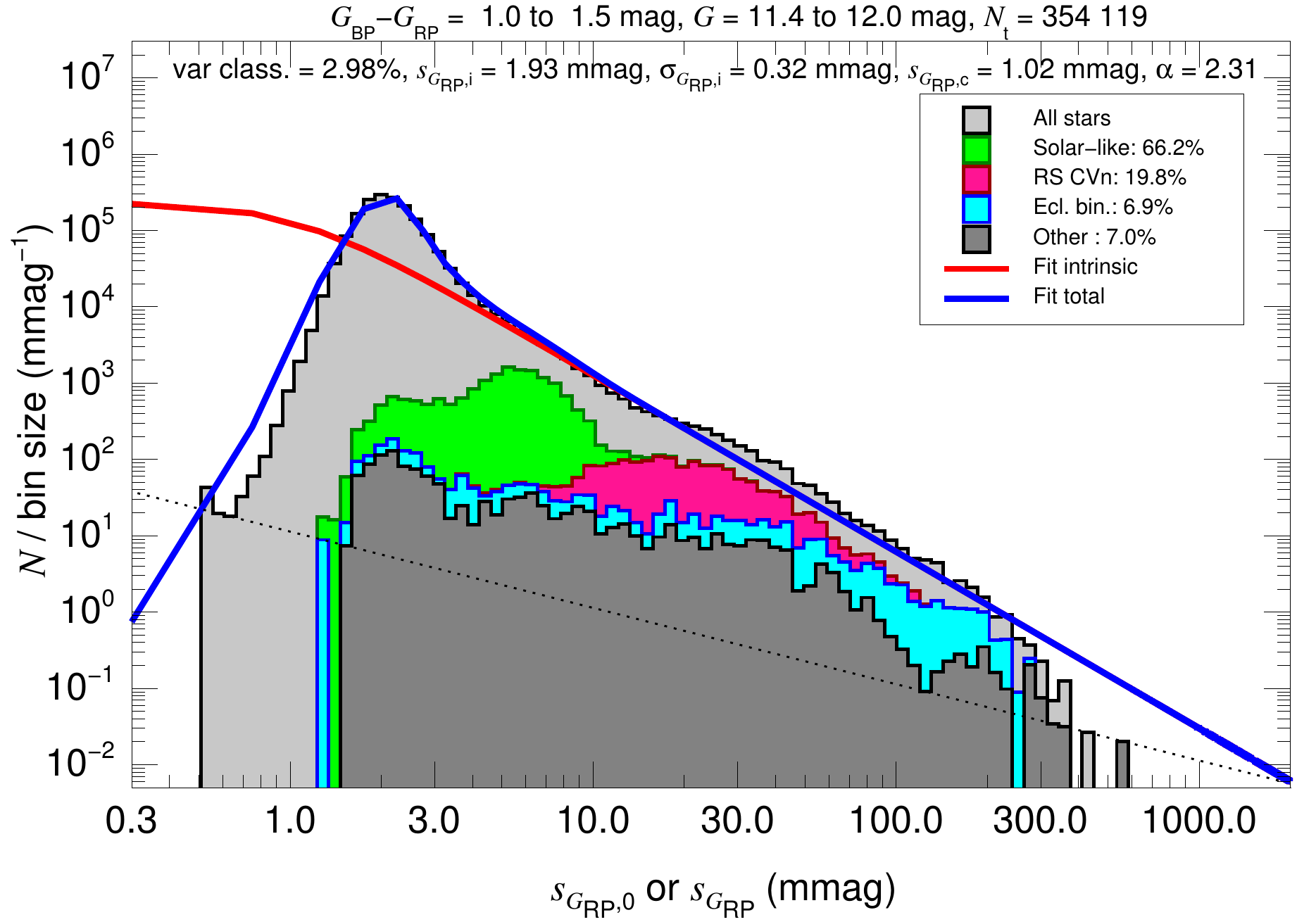}}
\centerline{$\!\!\!$\includegraphics[width=0.35\linewidth]{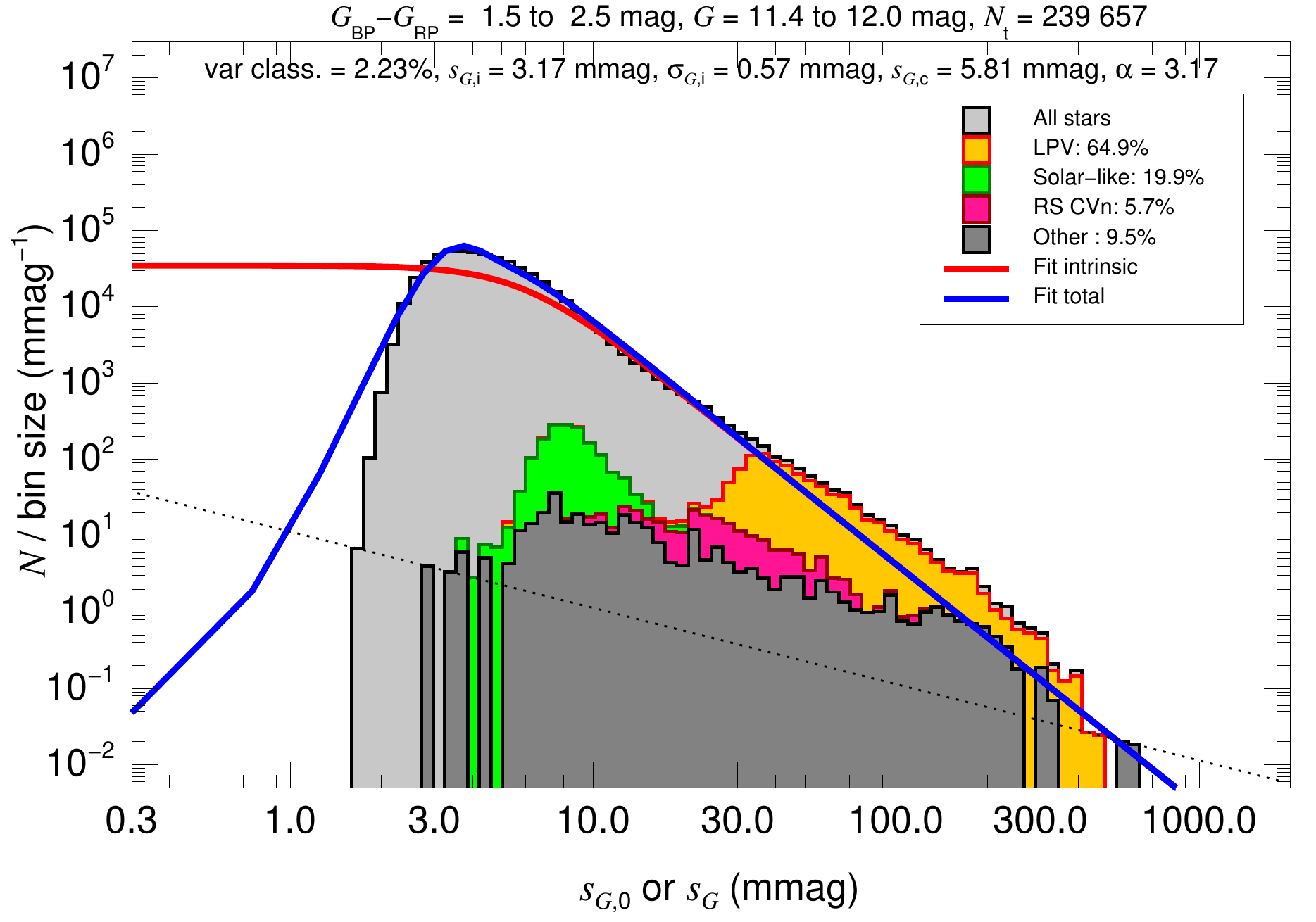}$\!\!\!$
                    \includegraphics[width=0.35\linewidth]{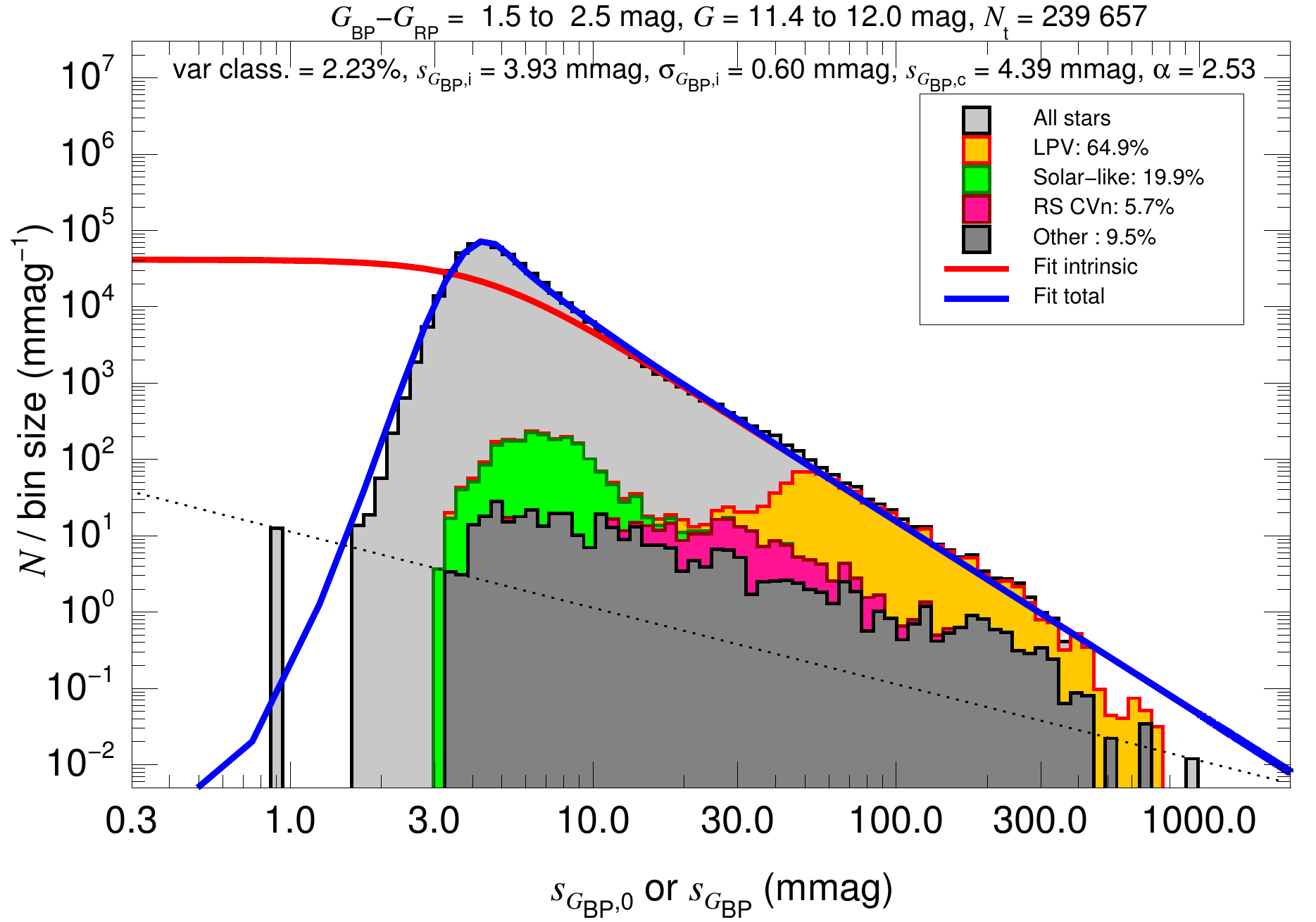}$\!\!\!$
                    \includegraphics[width=0.35\linewidth]{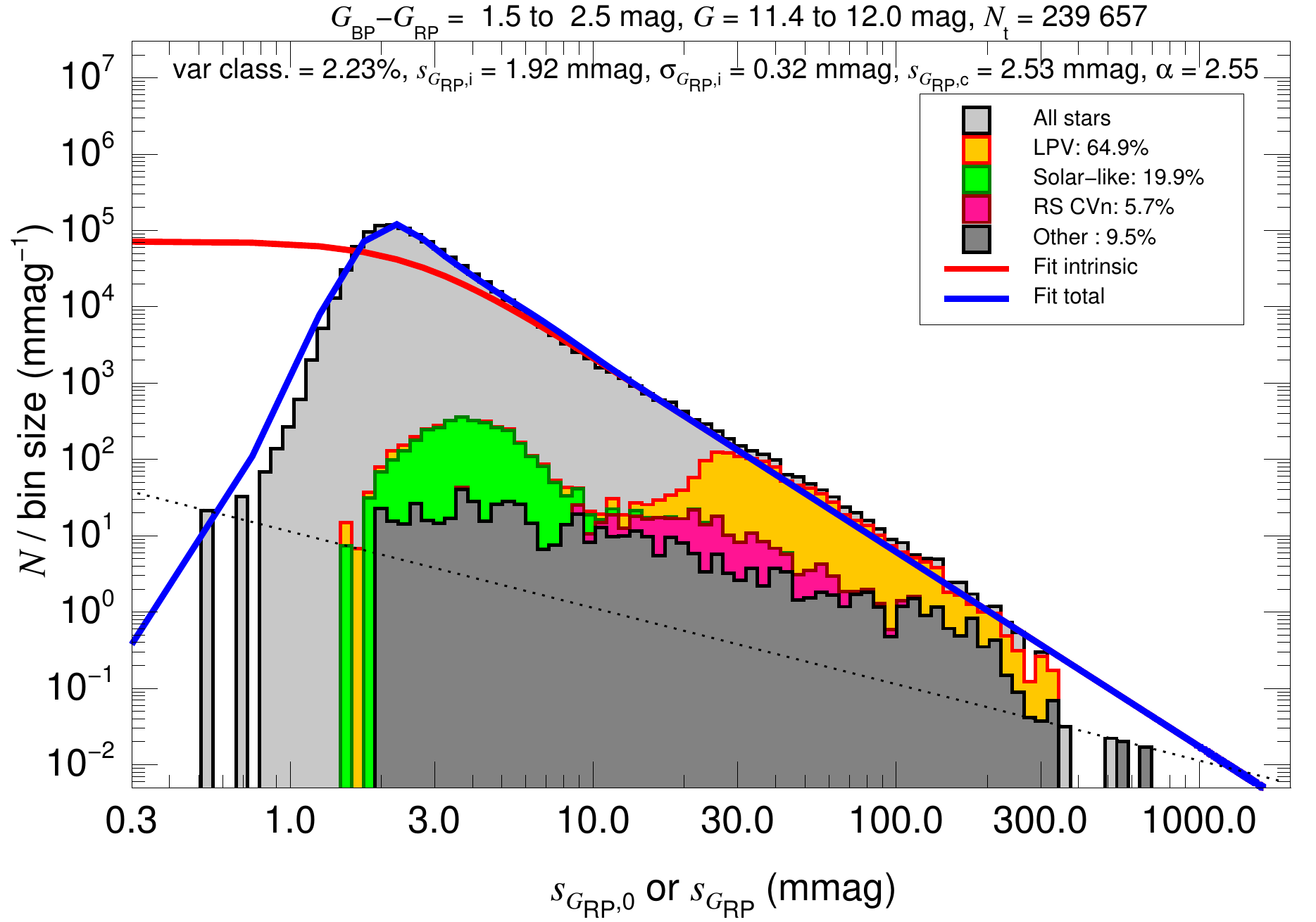}}
\centerline{$\!\!\!$\includegraphics[width=0.35\linewidth]{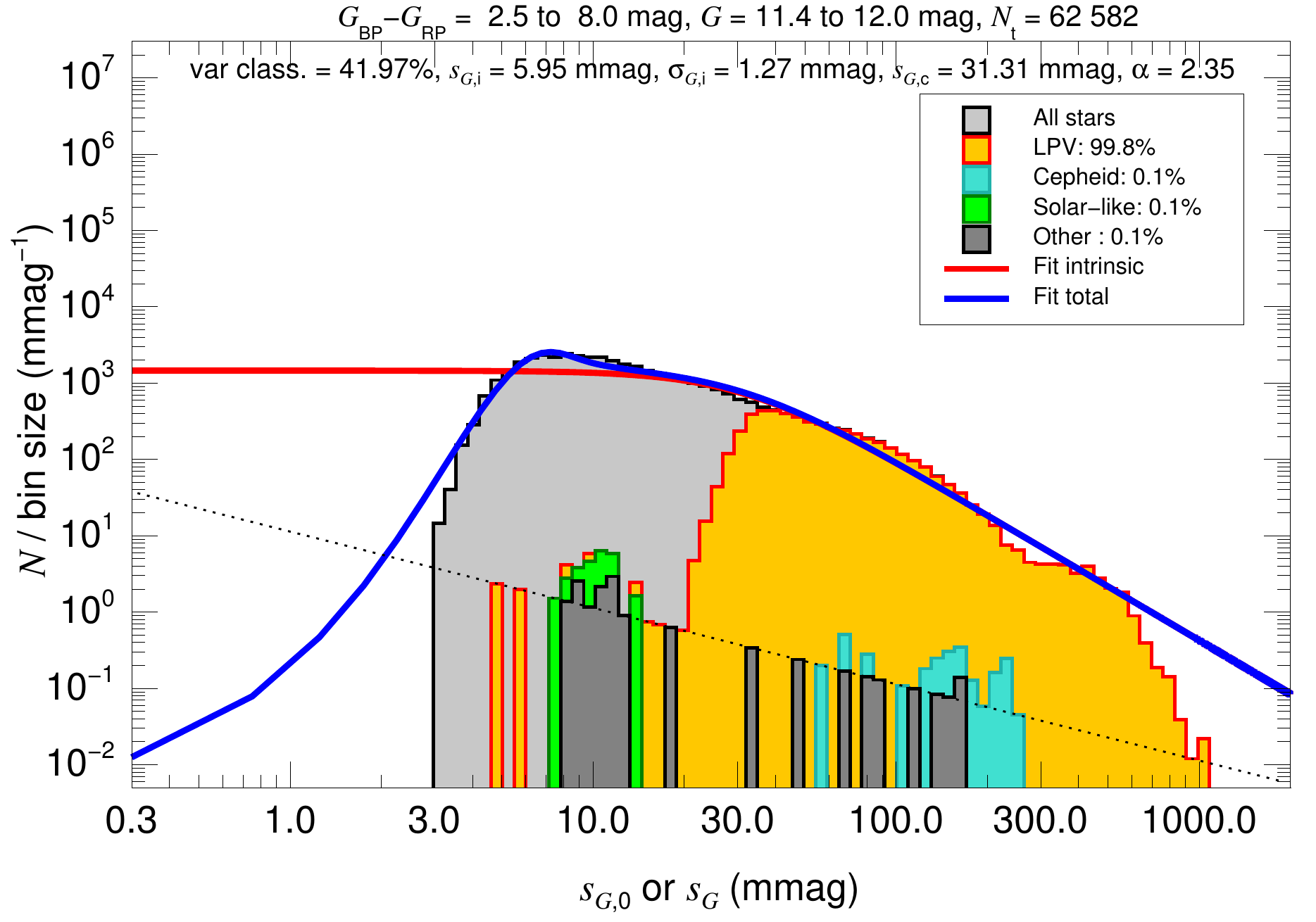}$\!\!\!$
                    \includegraphics[width=0.35\linewidth]{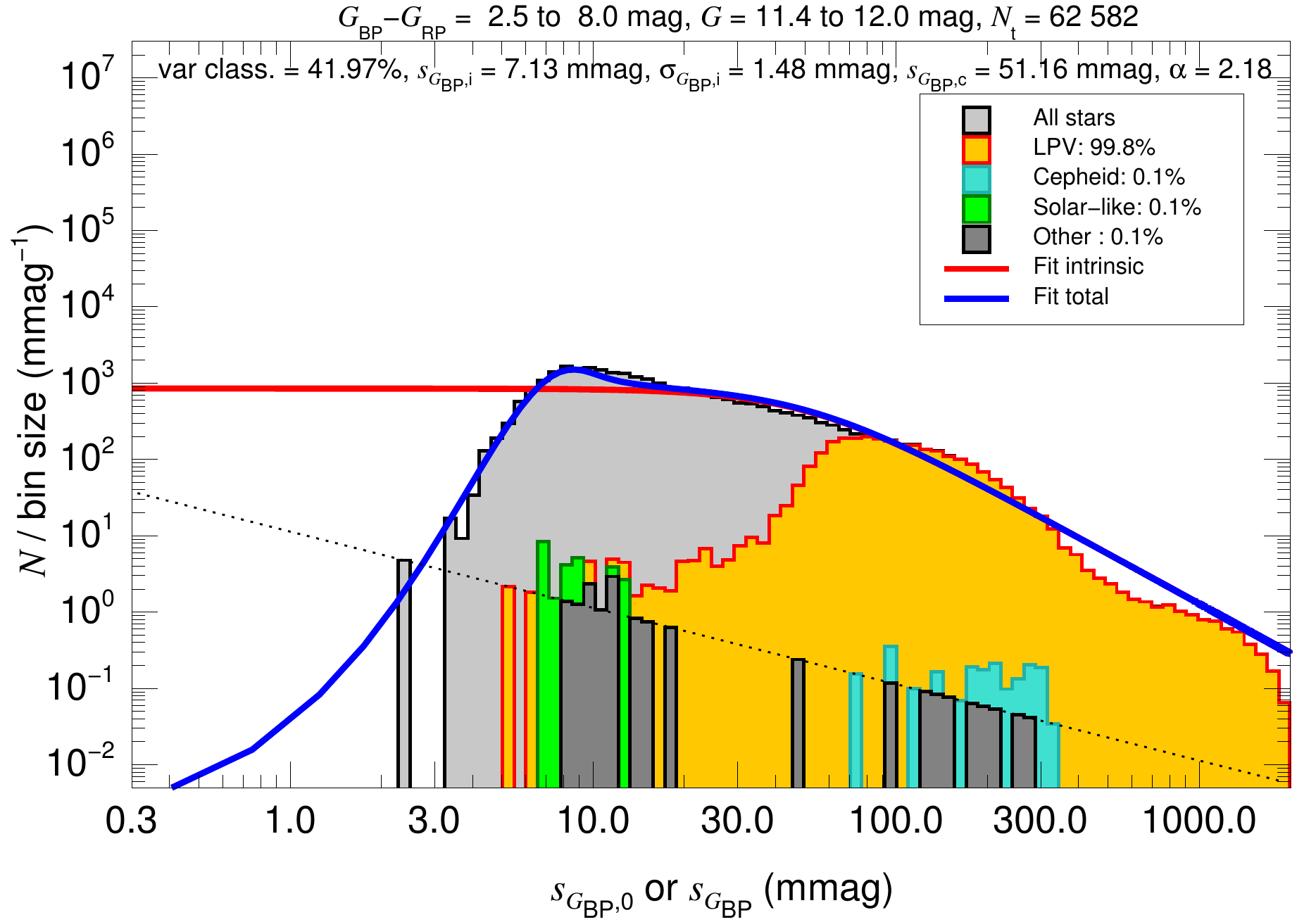}$\!\!\!$
                    \includegraphics[width=0.35\linewidth]{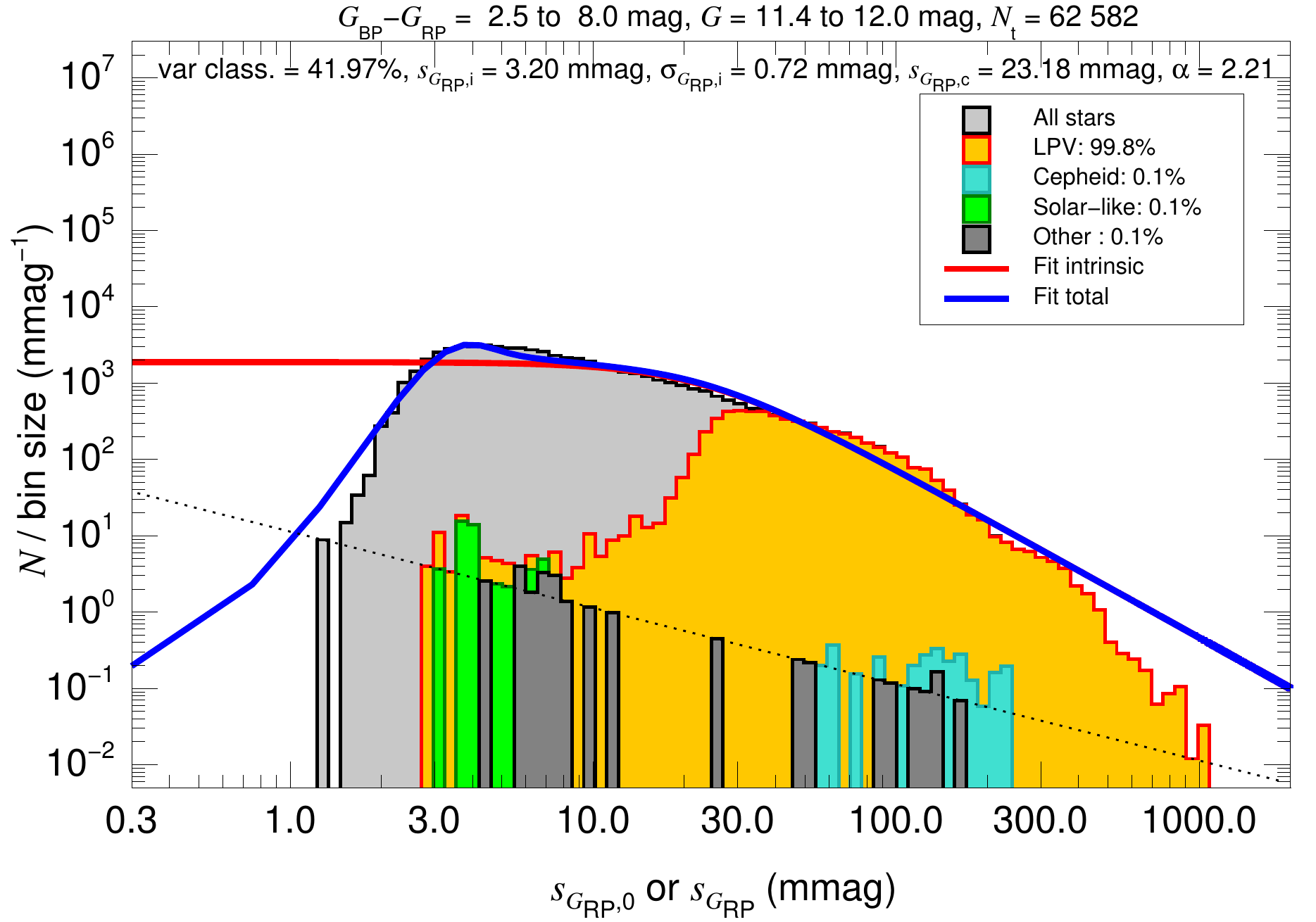}}
\caption{(Continued).}
\end{figure*}

\addtocounter{figure}{-1}

\begin{figure*}
\centerline{$\!\!\!$\includegraphics[width=0.35\linewidth]{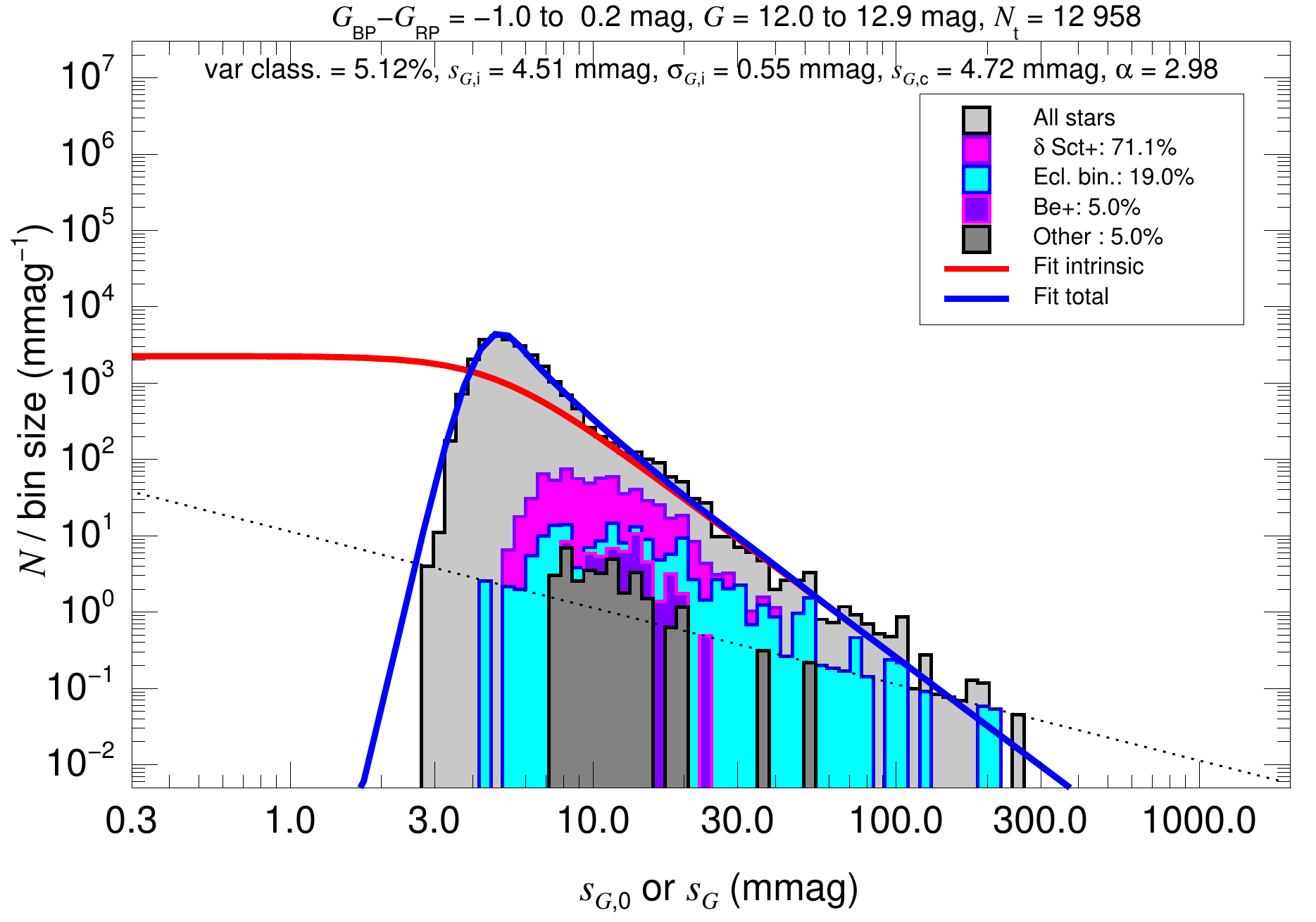}$\!\!\!$
                    \includegraphics[width=0.35\linewidth]{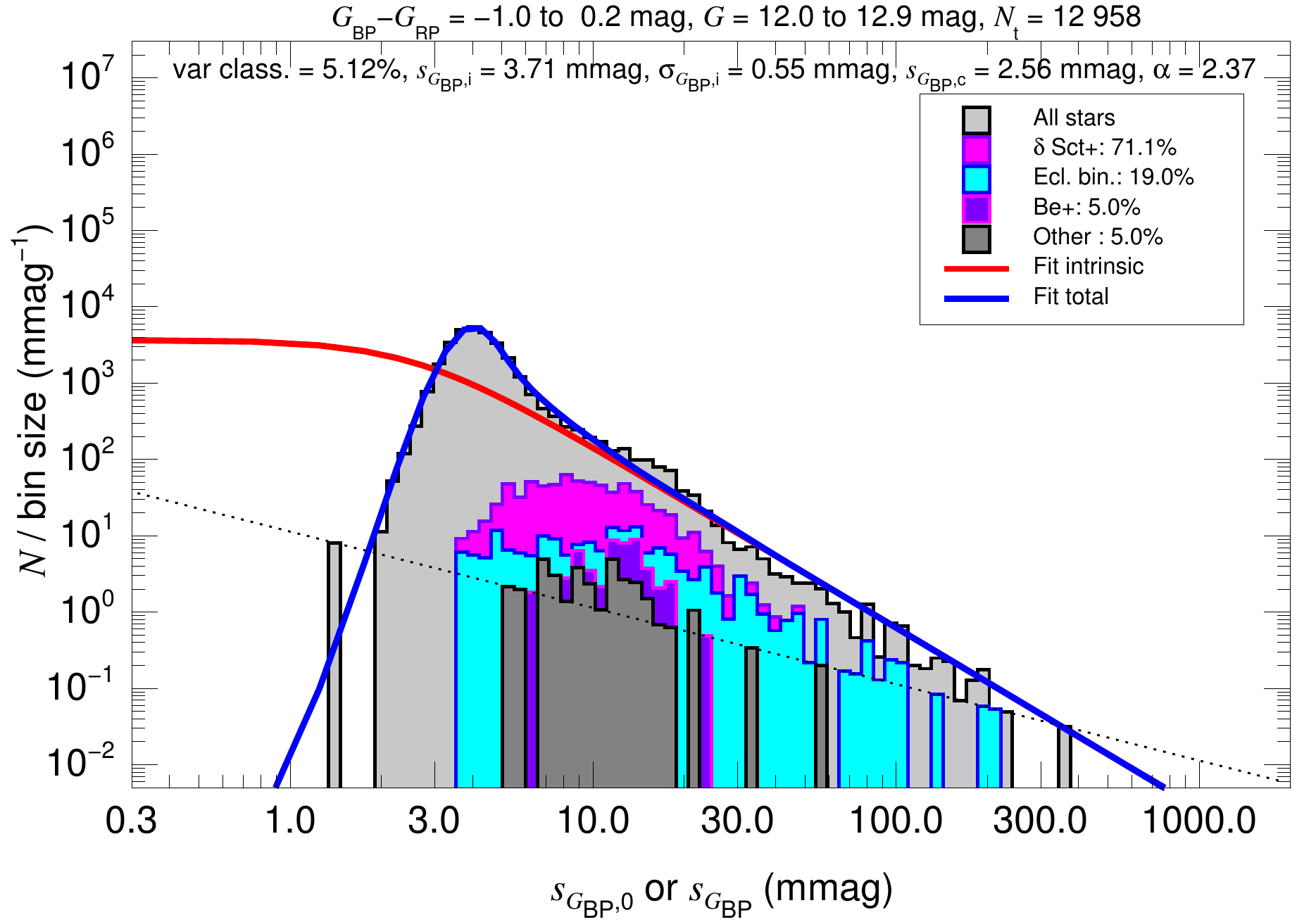}$\!\!\!$
                    \includegraphics[width=0.35\linewidth]{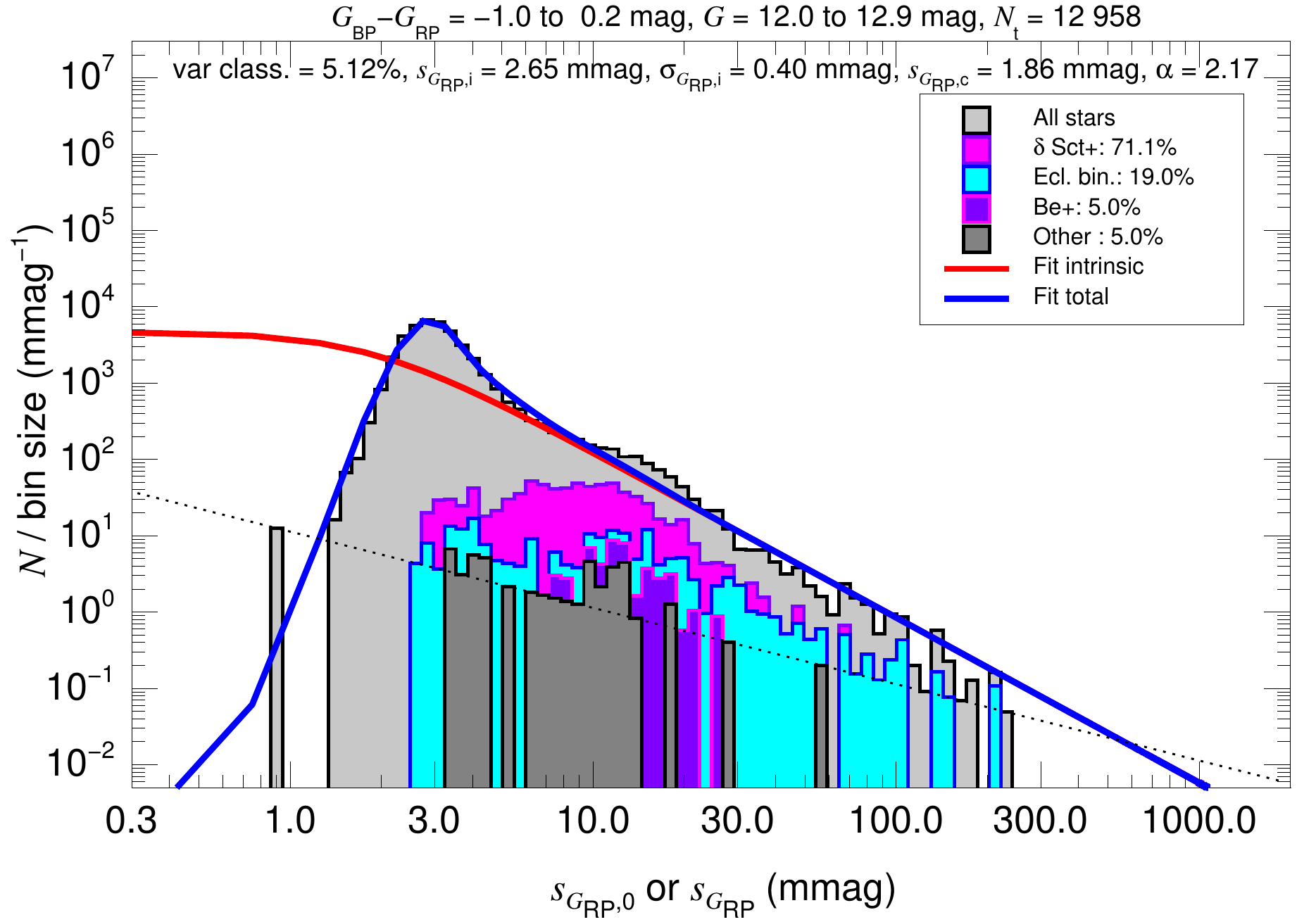}}
\centerline{$\!\!\!$\includegraphics[width=0.35\linewidth]{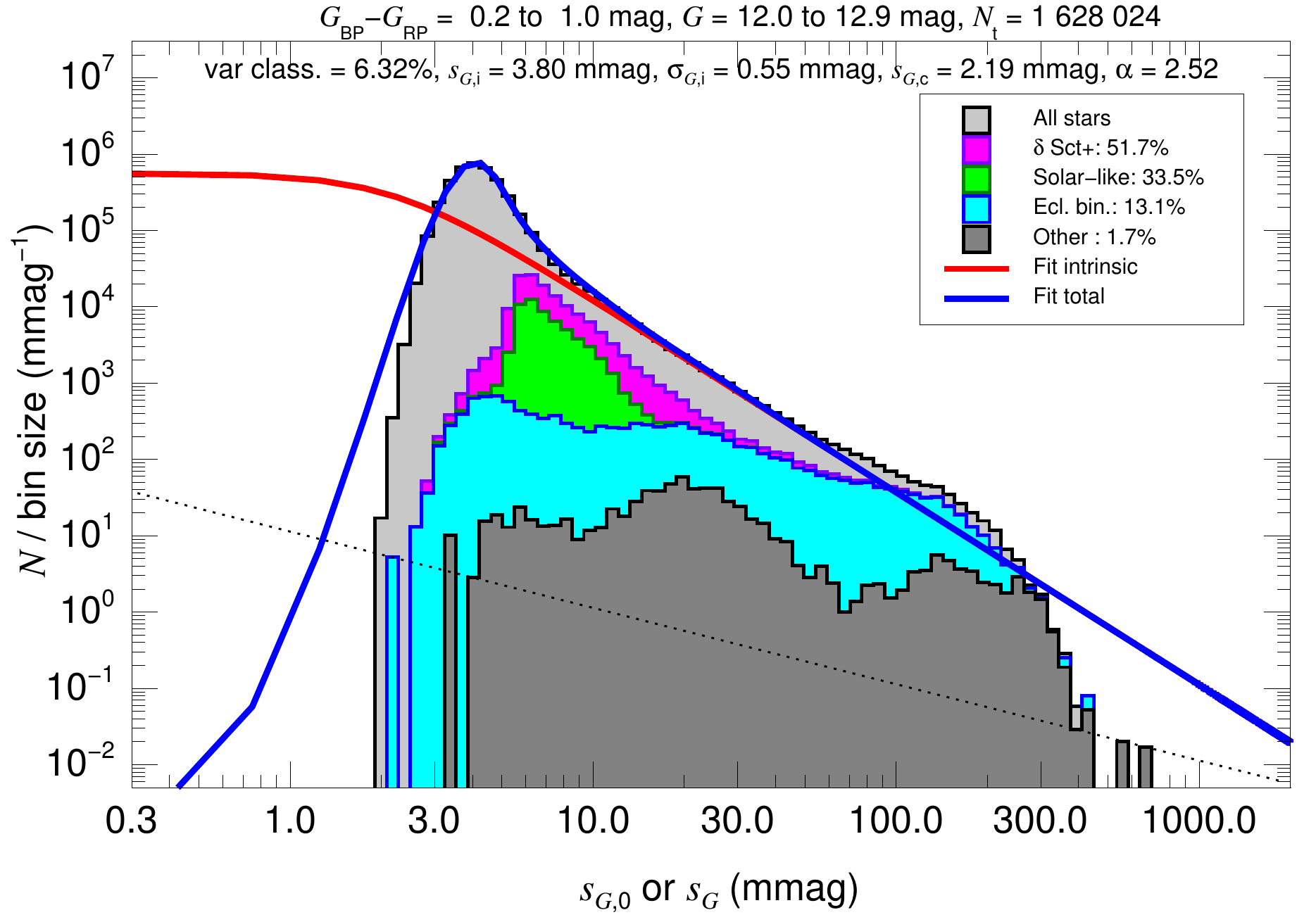}$\!\!\!$
                    \includegraphics[width=0.35\linewidth]{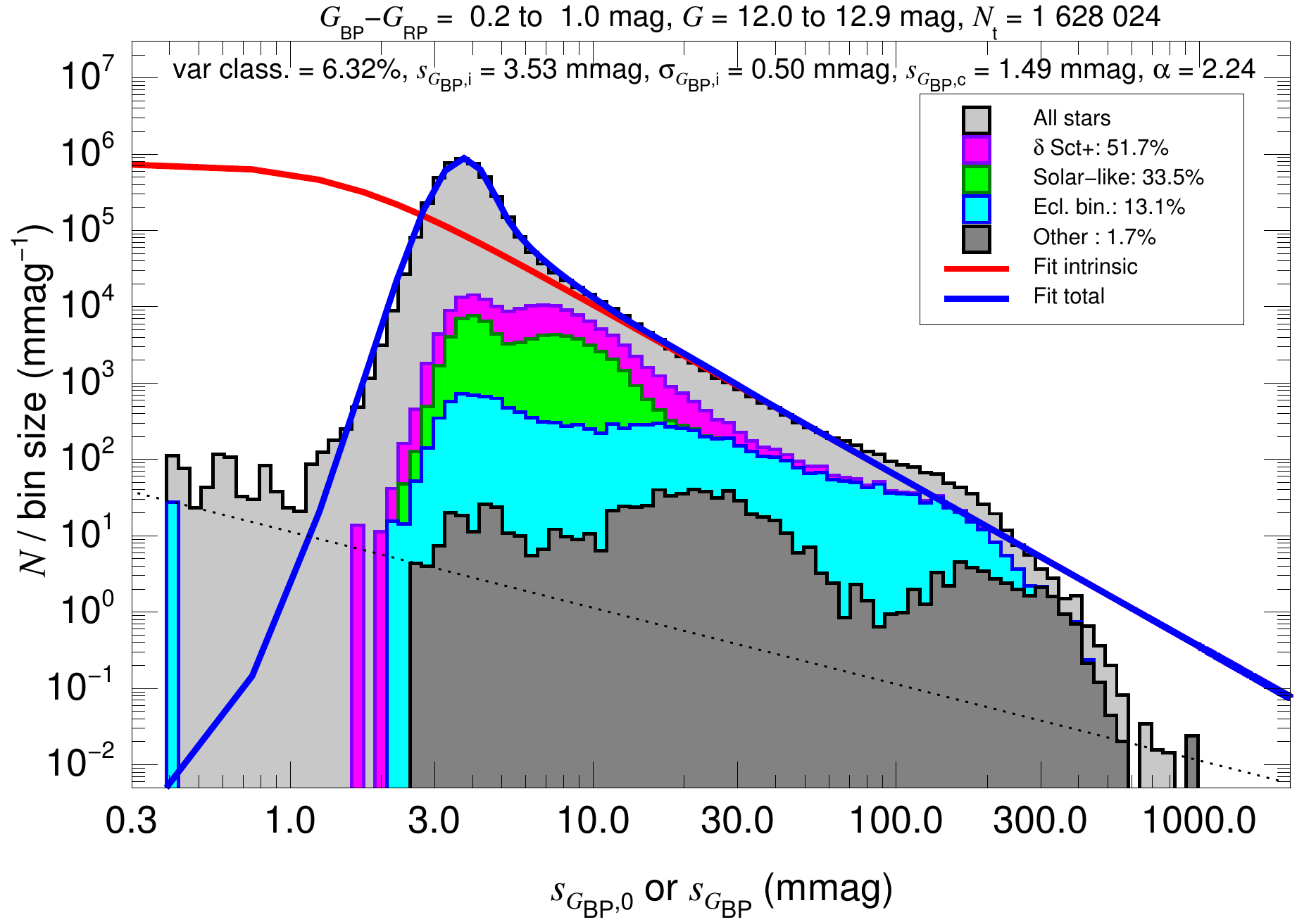}$\!\!\!$
                    \includegraphics[width=0.35\linewidth]{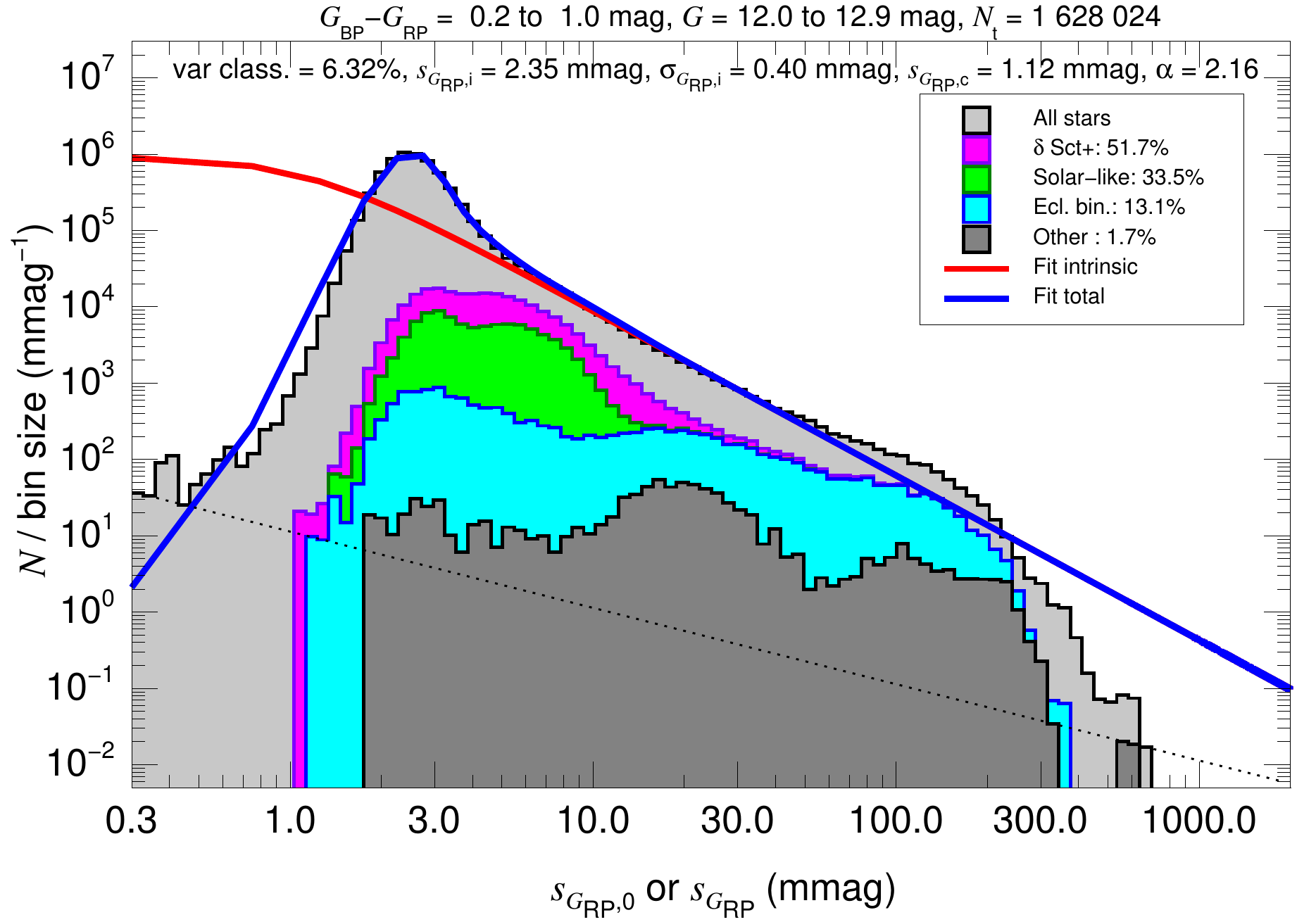}}
\centerline{$\!\!\!$\includegraphics[width=0.35\linewidth]{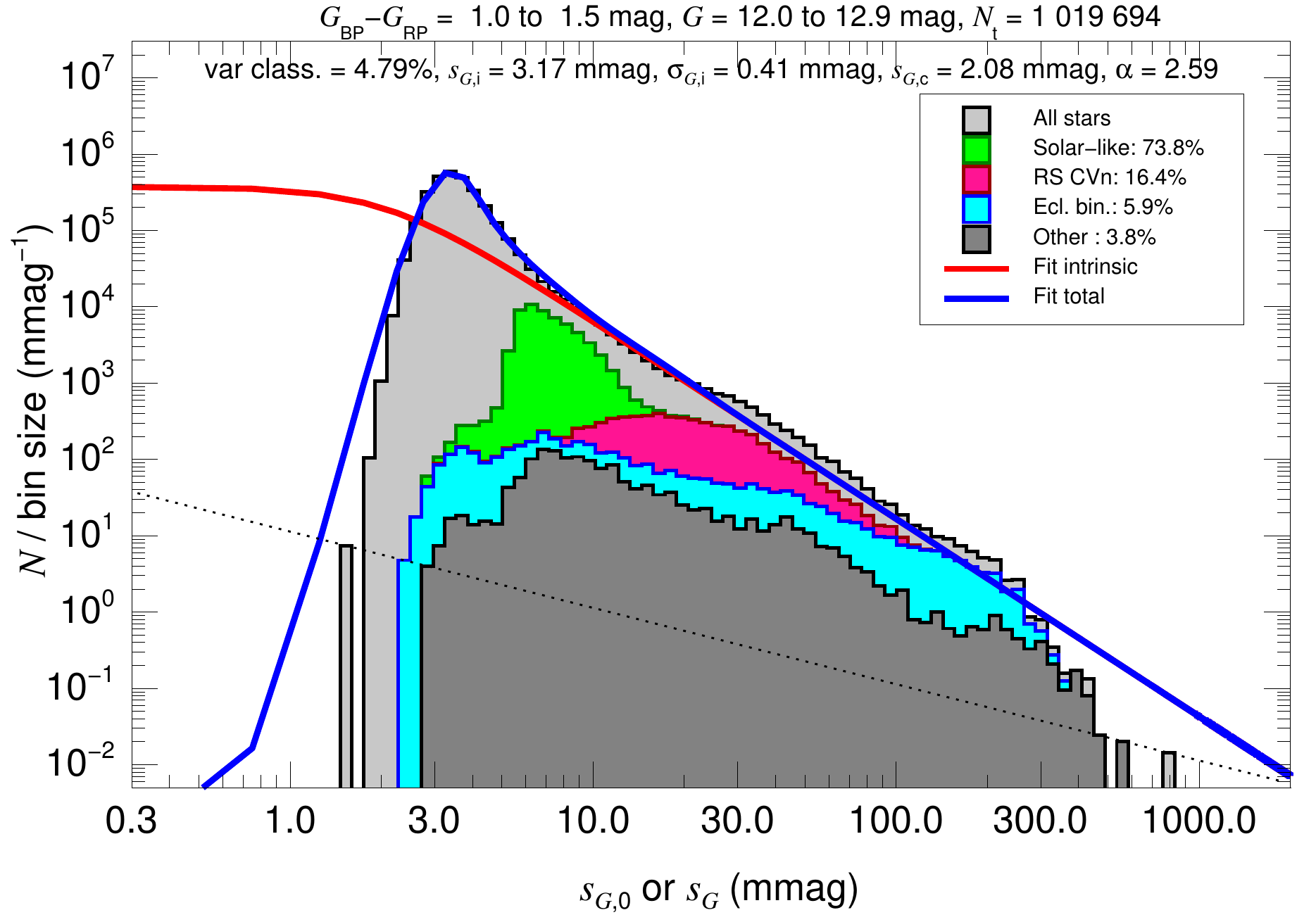}$\!\!\!$
                    \includegraphics[width=0.35\linewidth]{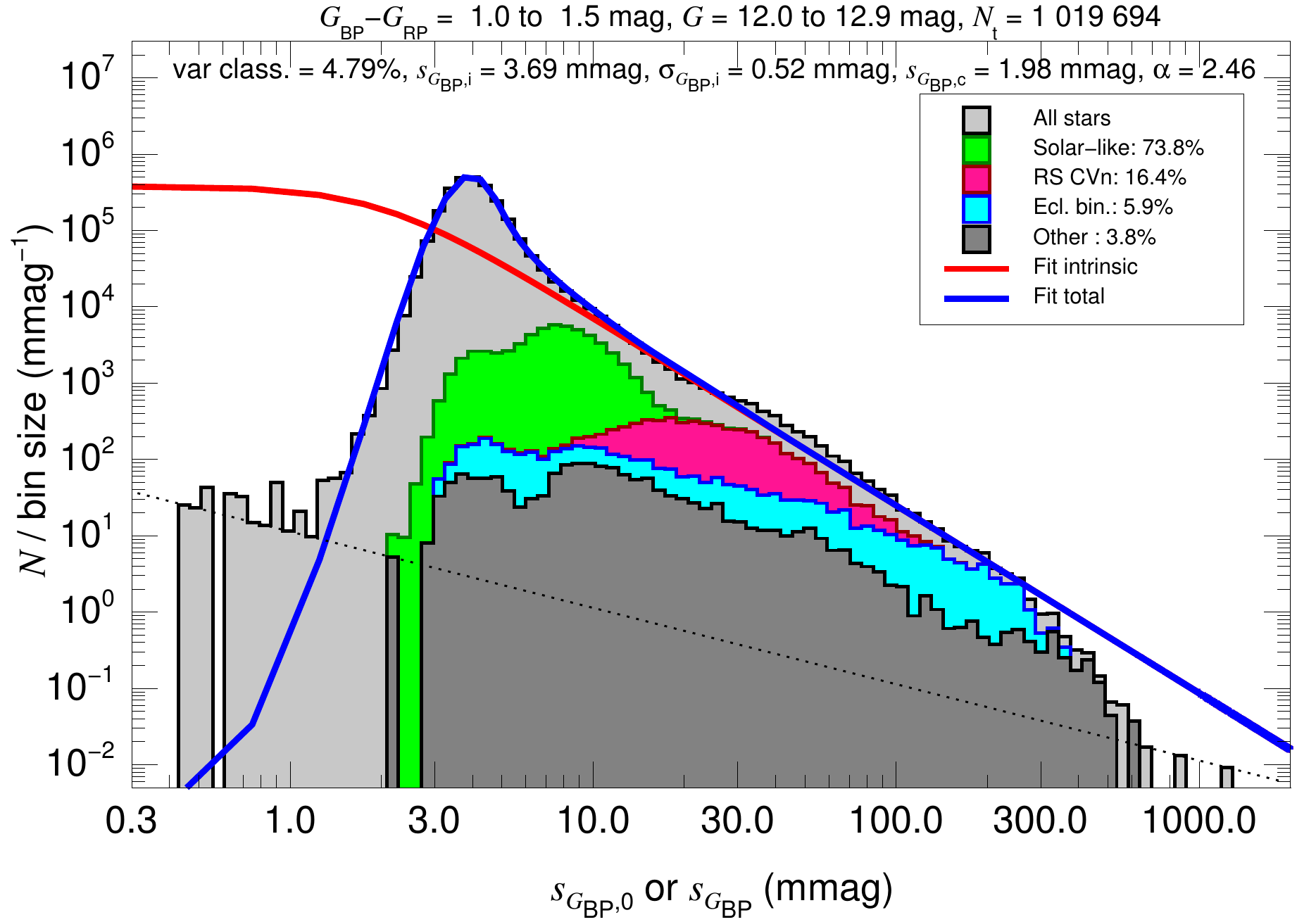}$\!\!\!$
                    \includegraphics[width=0.35\linewidth]{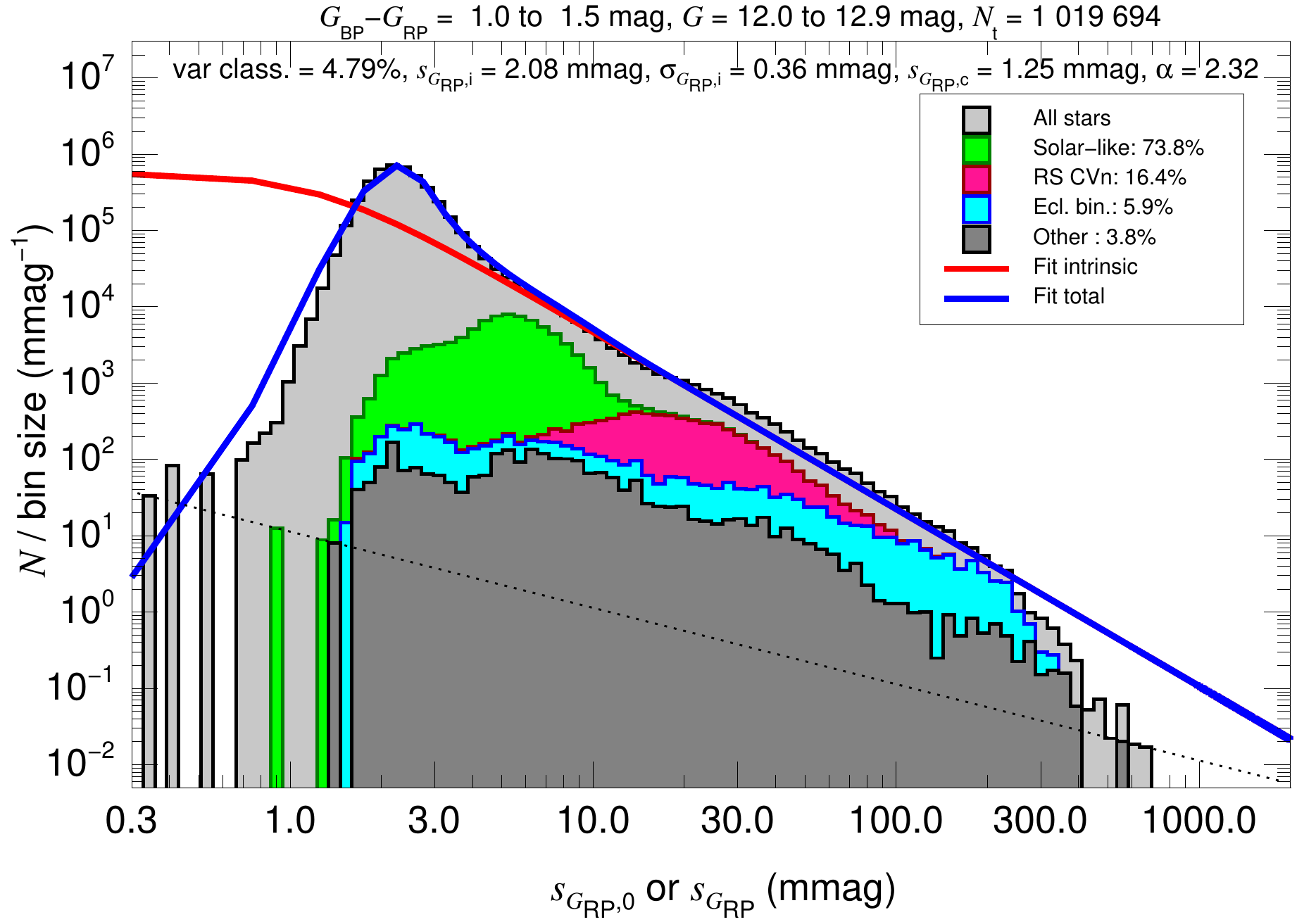}}
\centerline{$\!\!\!$\includegraphics[width=0.35\linewidth]{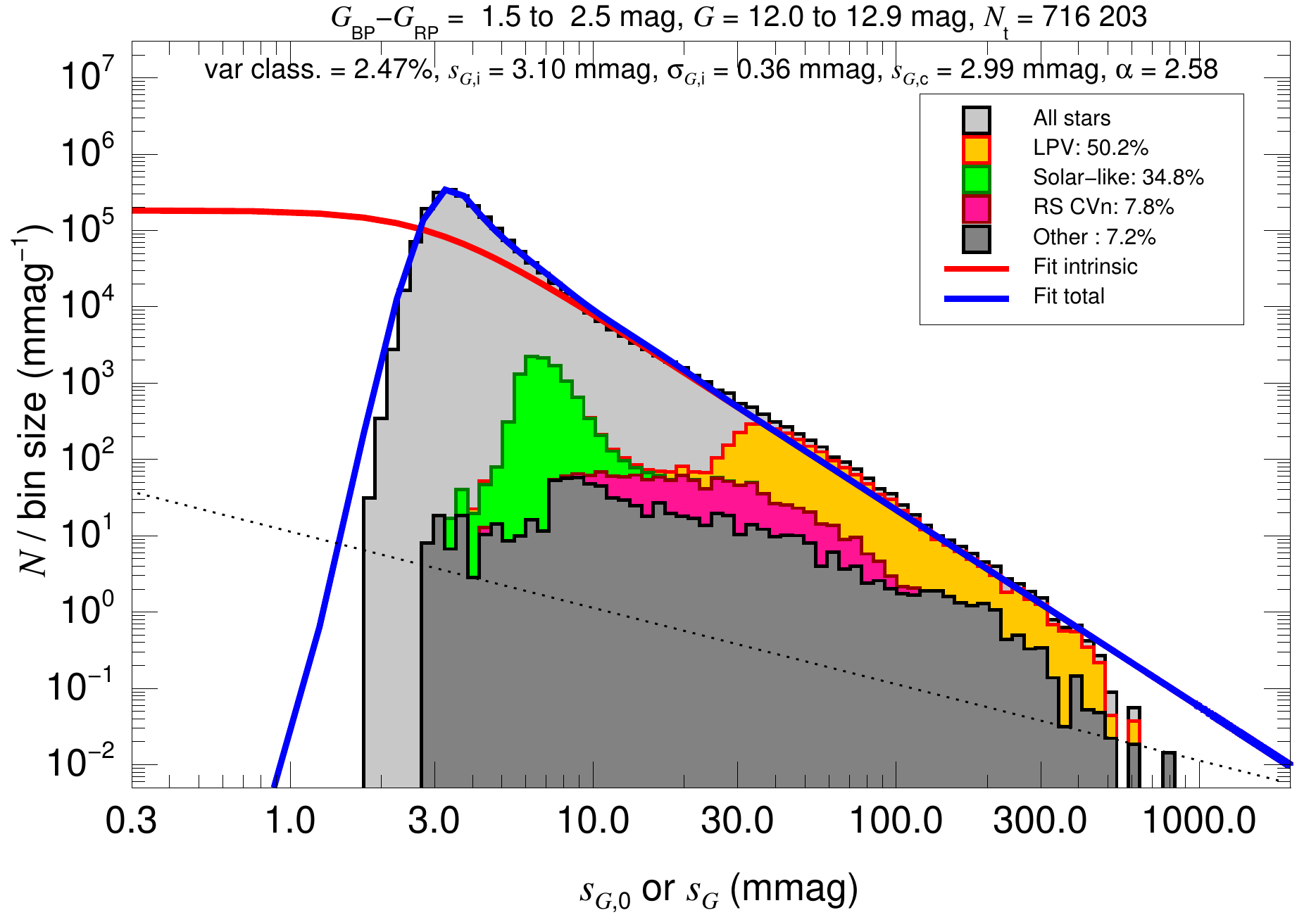}$\!\!\!$
                    \includegraphics[width=0.35\linewidth]{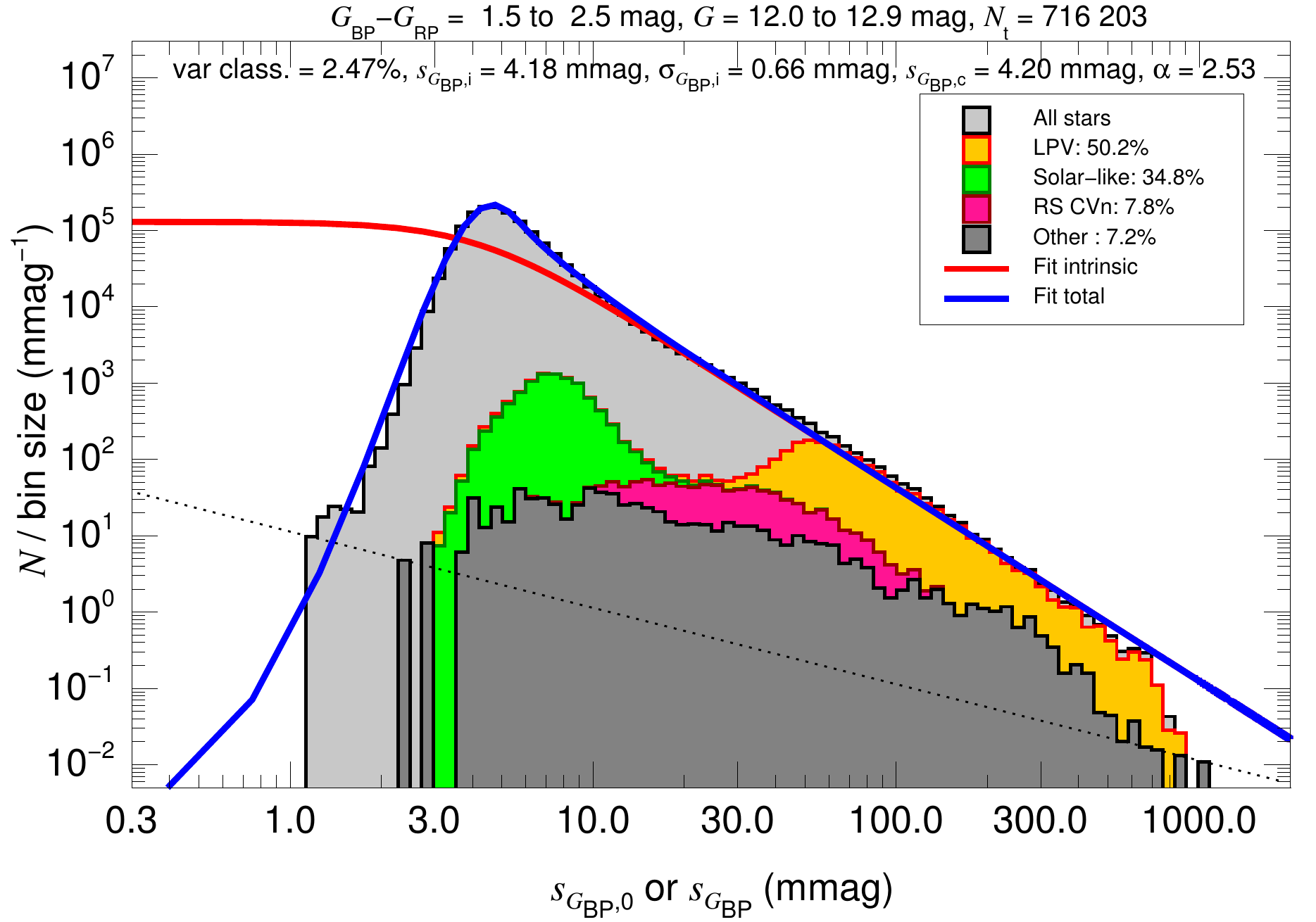}$\!\!\!$
                    \includegraphics[width=0.35\linewidth]{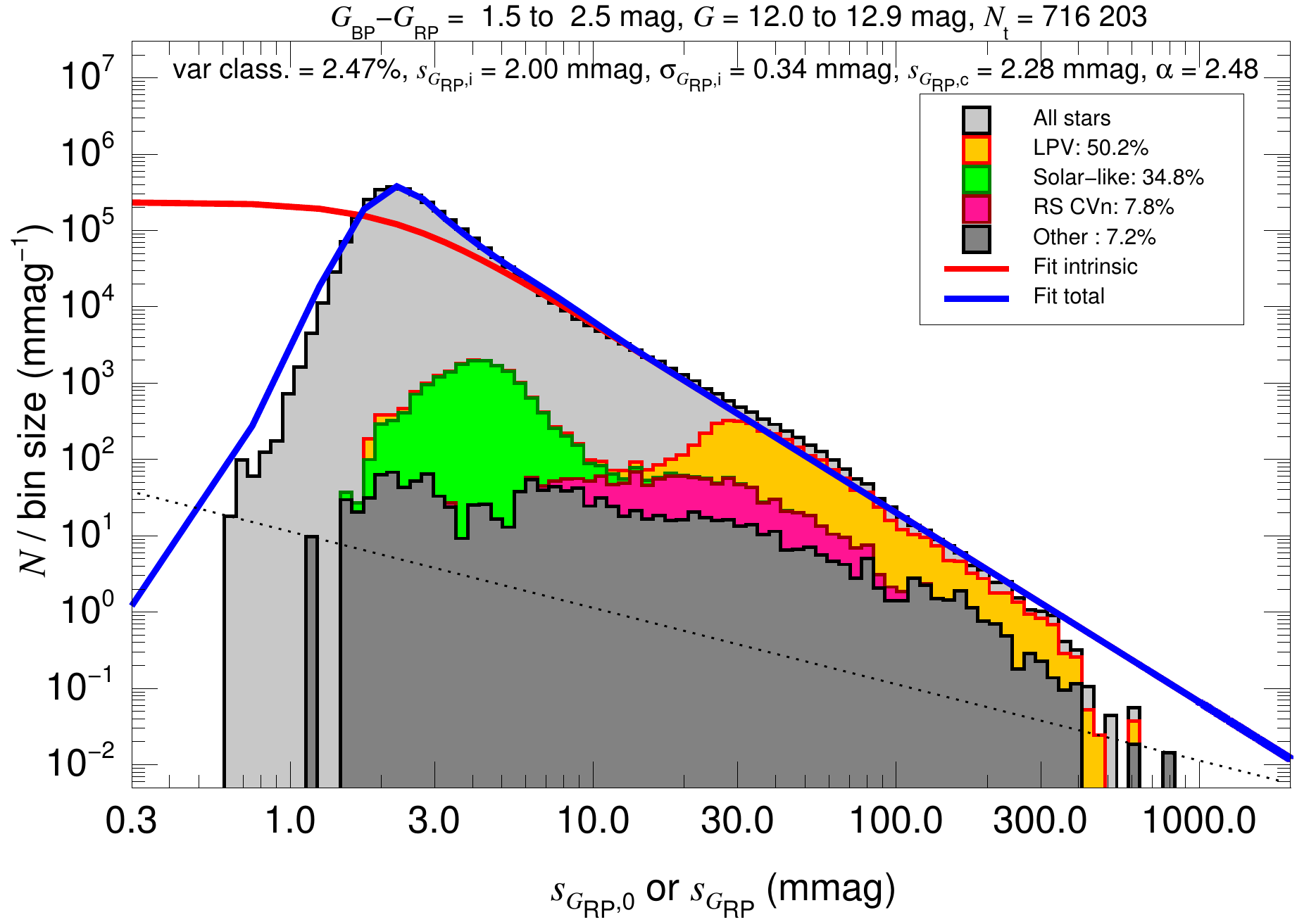}}
\centerline{$\!\!\!$\includegraphics[width=0.35\linewidth]{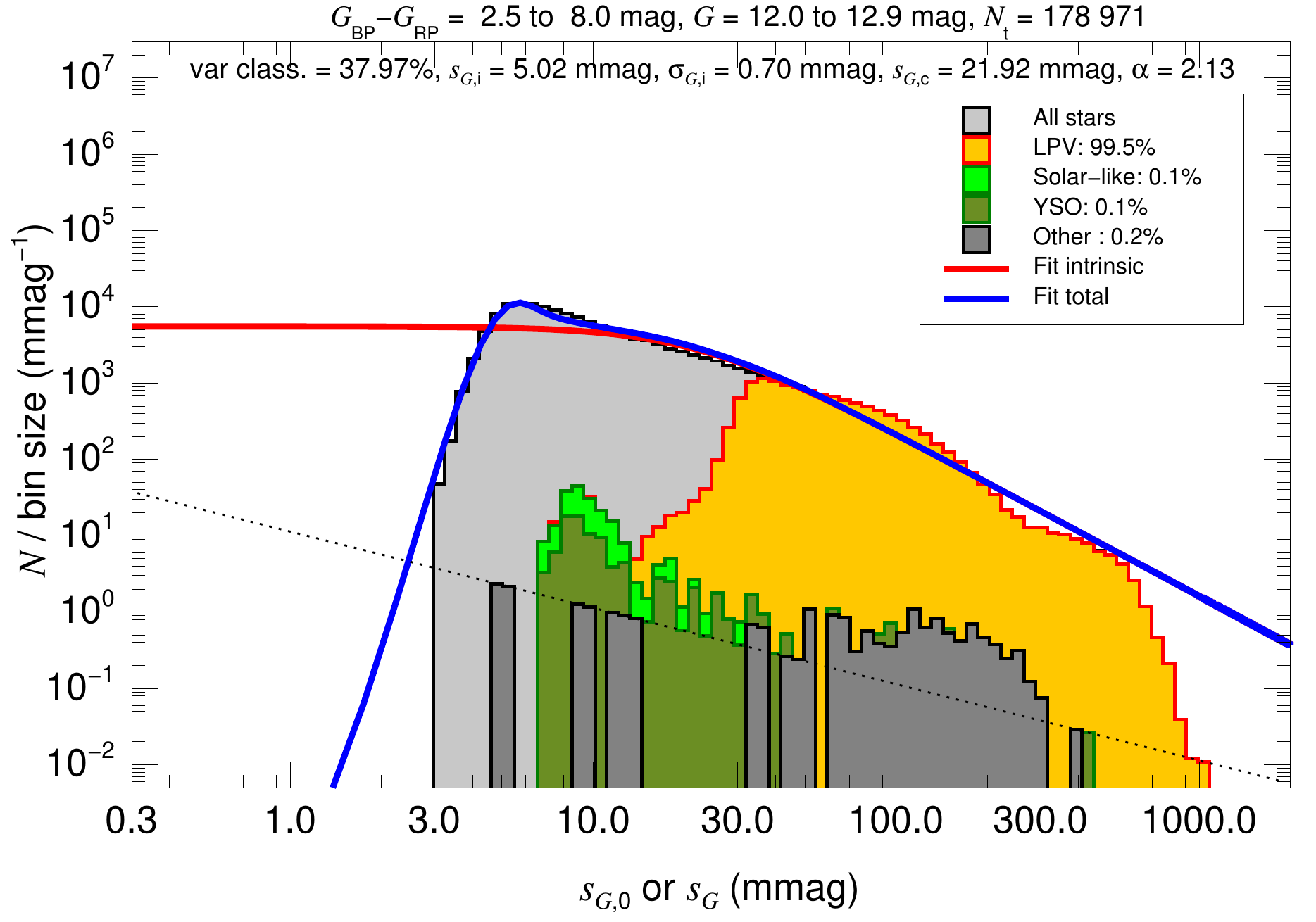}$\!\!\!$
                    \includegraphics[width=0.35\linewidth]{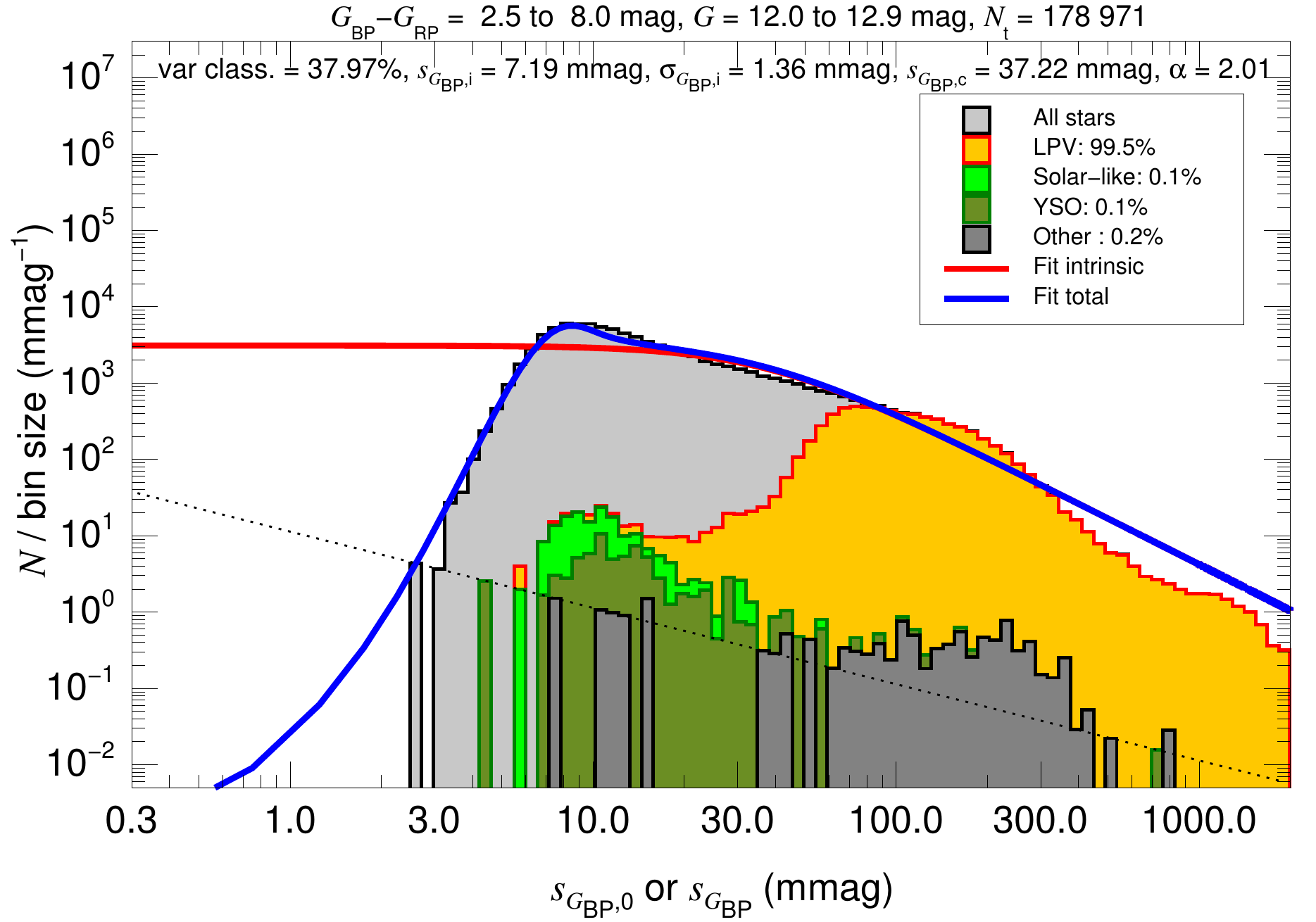}$\!\!\!$
                    \includegraphics[width=0.35\linewidth]{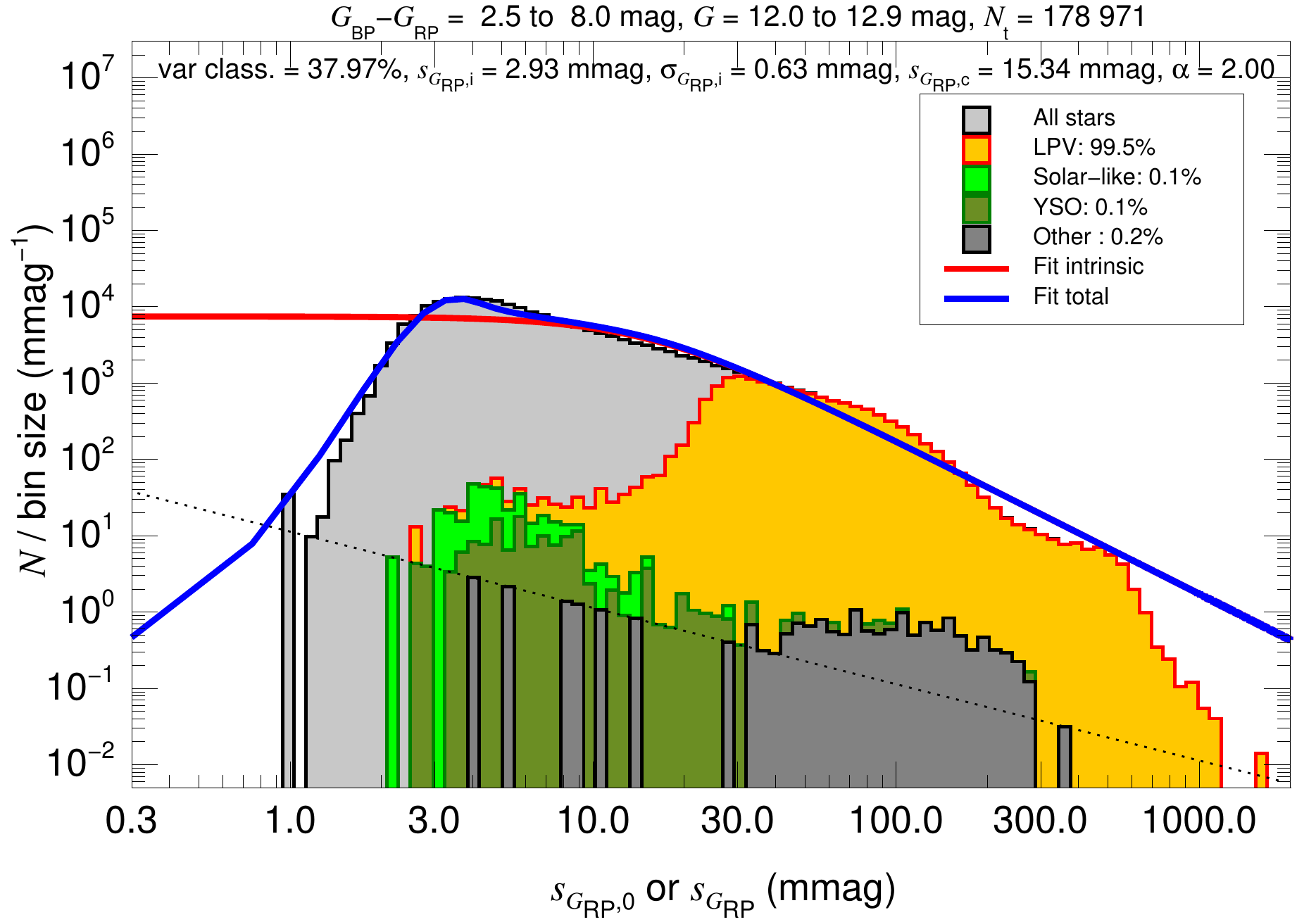}}
\caption{(Continued).}
\end{figure*}

\addtocounter{figure}{-1}

\begin{figure*}
\centerline{$\!\!\!$\includegraphics[width=0.35\linewidth]{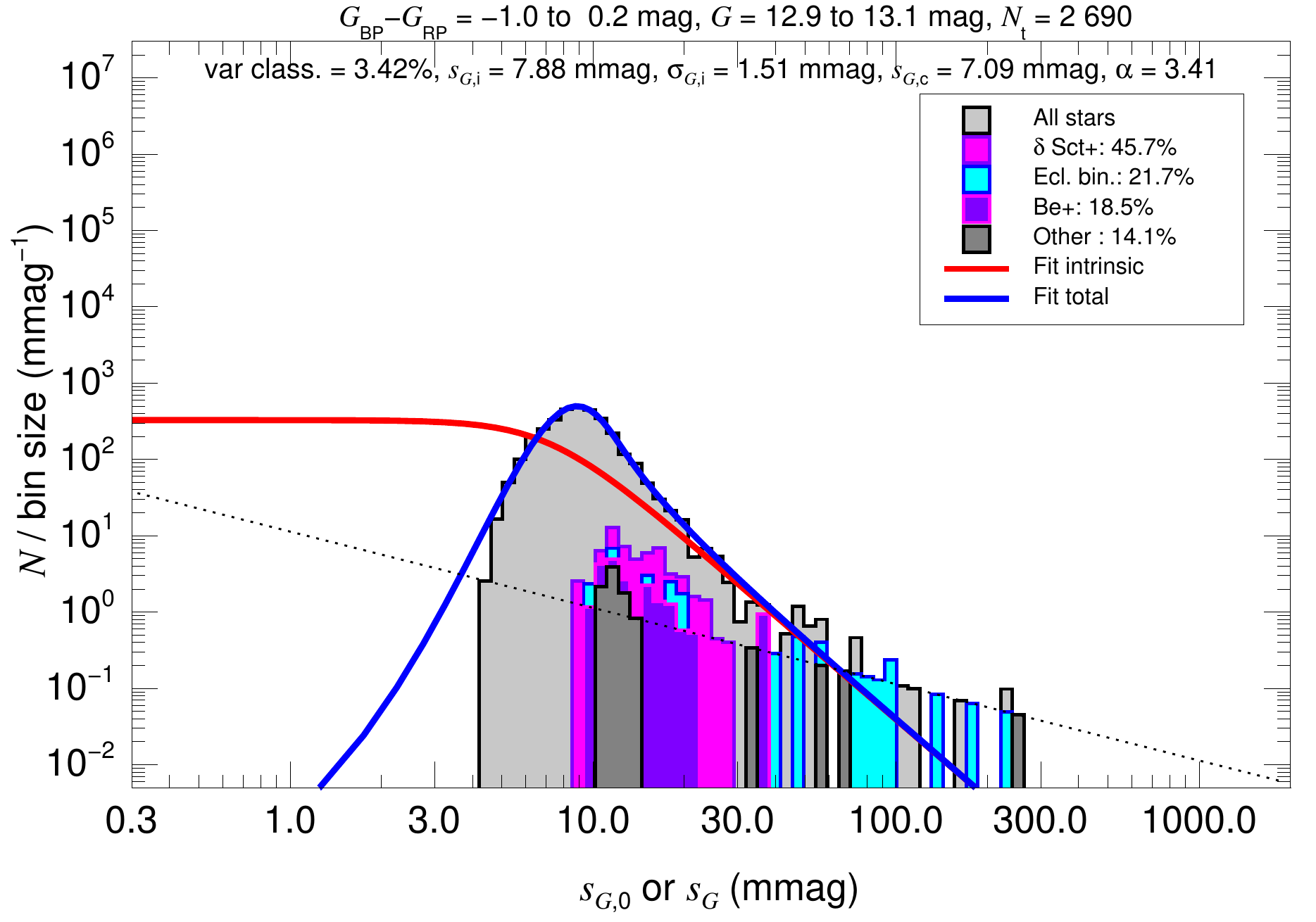}$\!\!\!$
                    \includegraphics[width=0.35\linewidth]{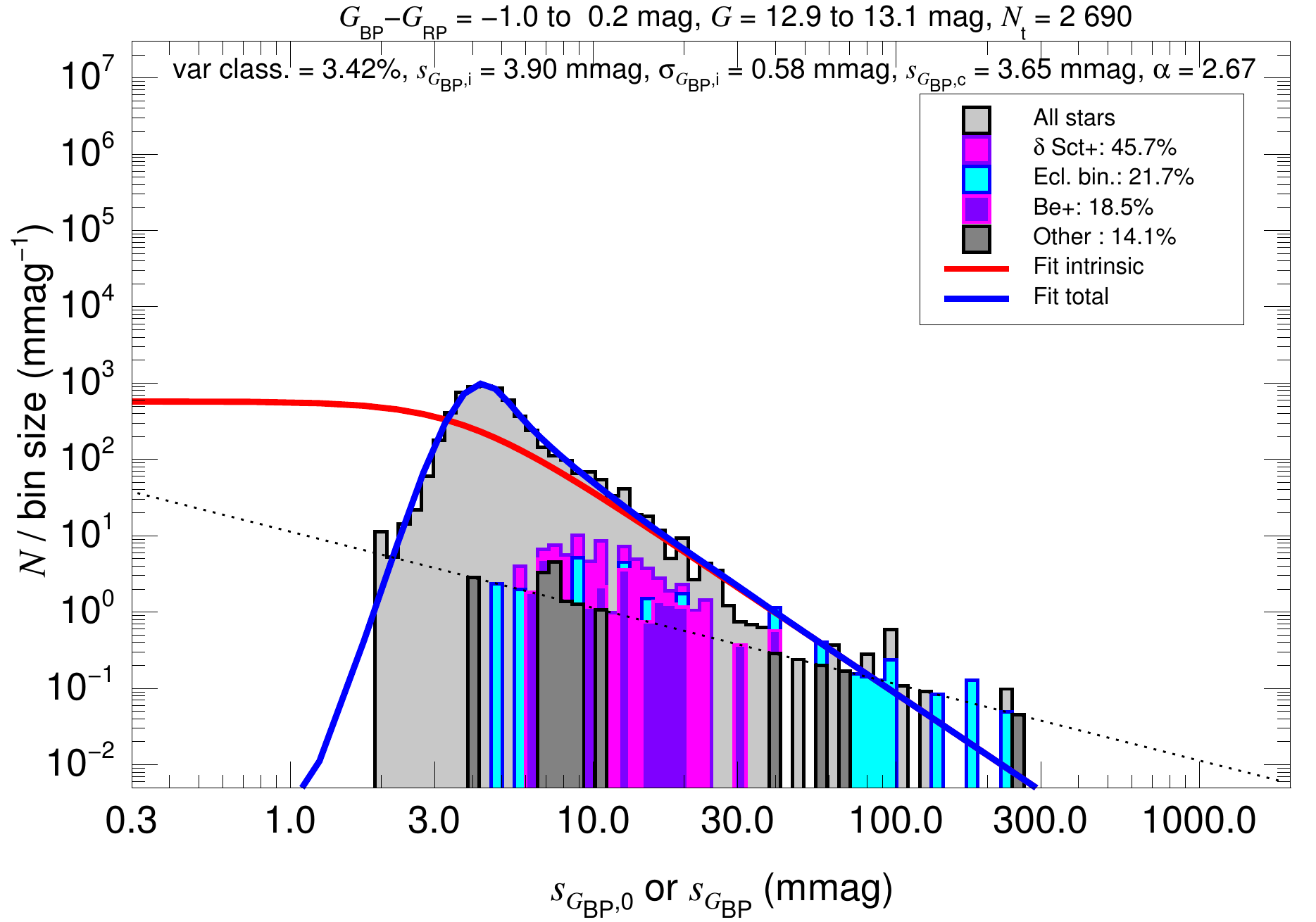}$\!\!\!$
                    \includegraphics[width=0.35\linewidth]{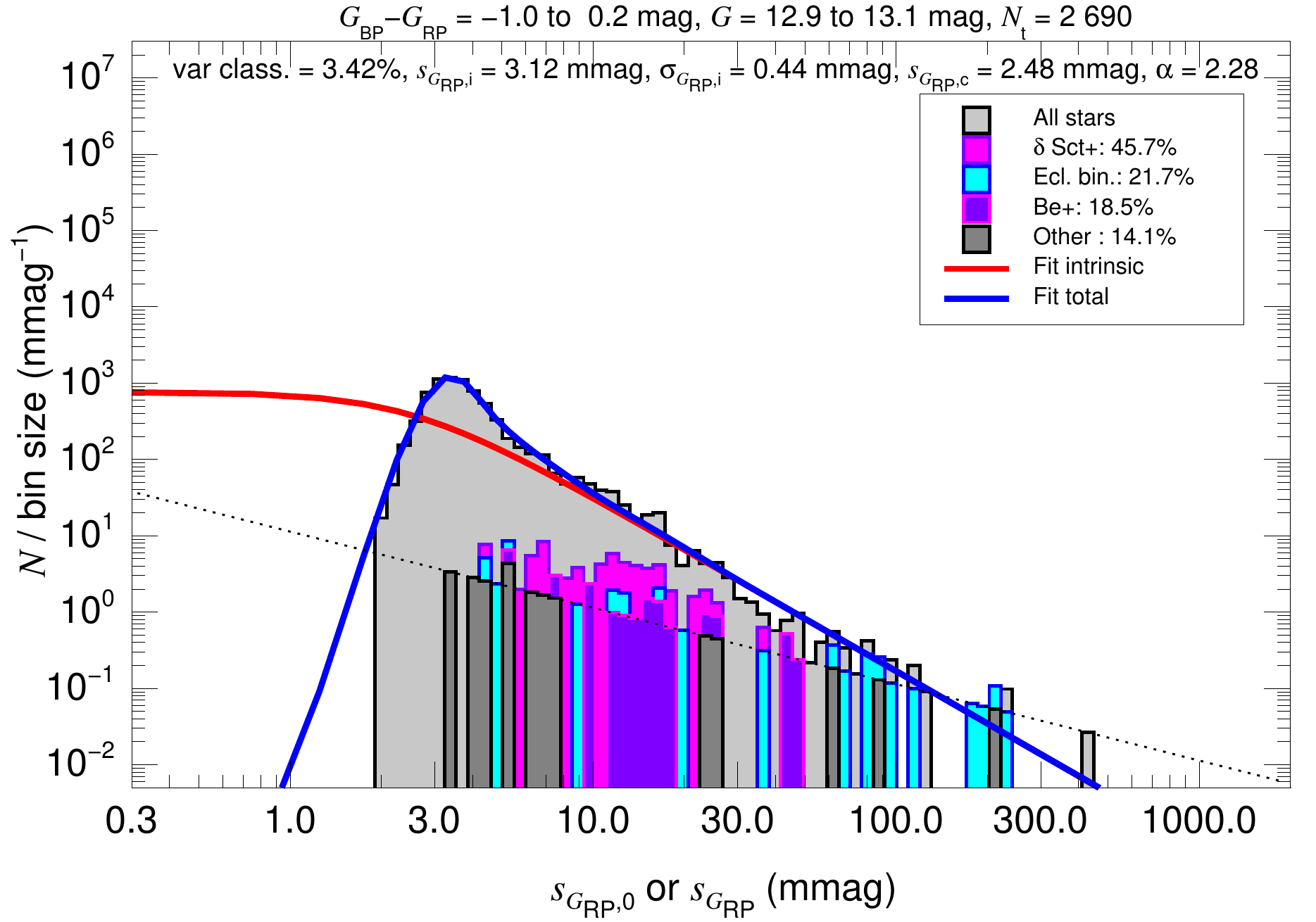}}
\centerline{$\!\!\!$\includegraphics[width=0.35\linewidth]{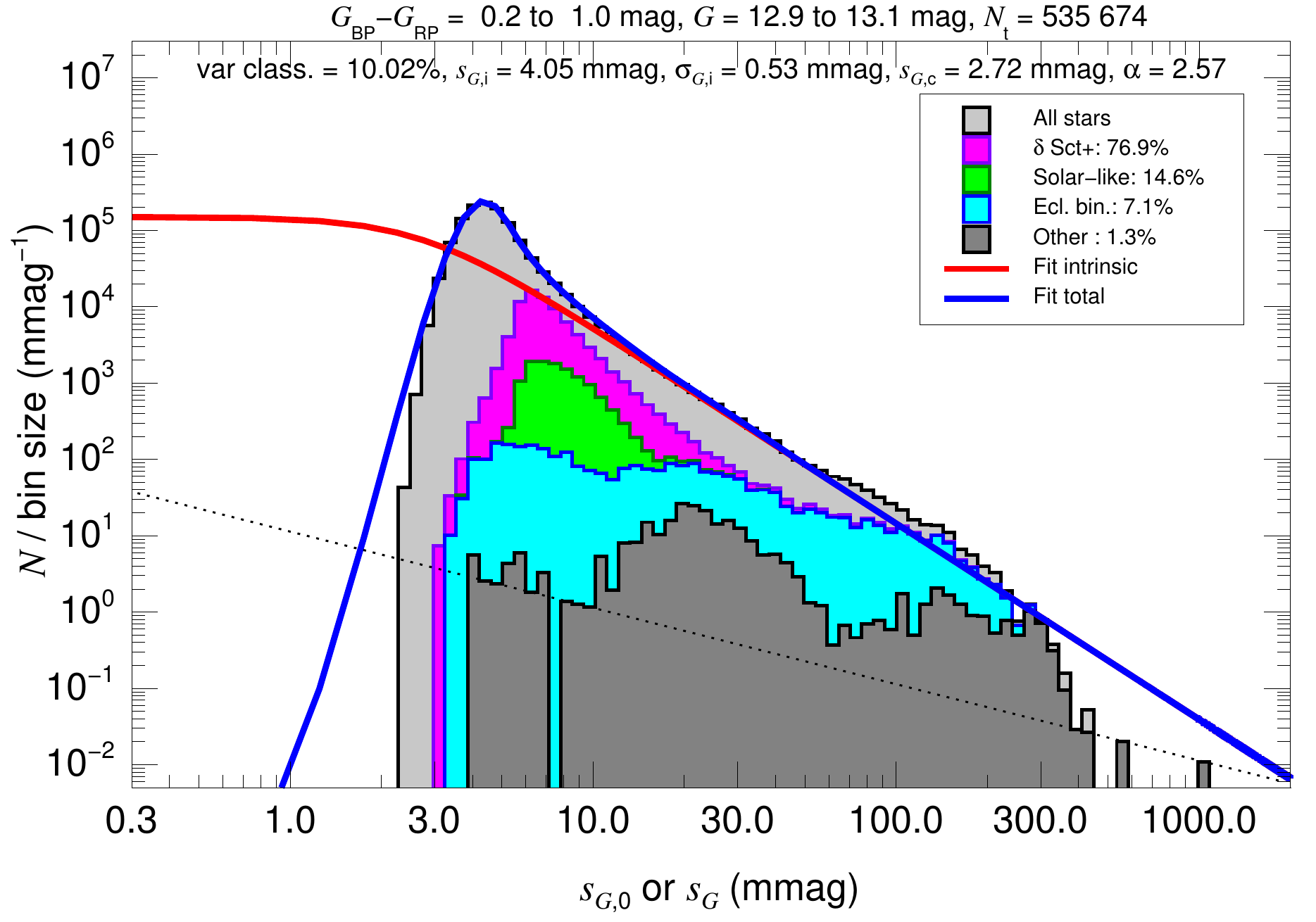}$\!\!\!$
                    \includegraphics[width=0.35\linewidth]{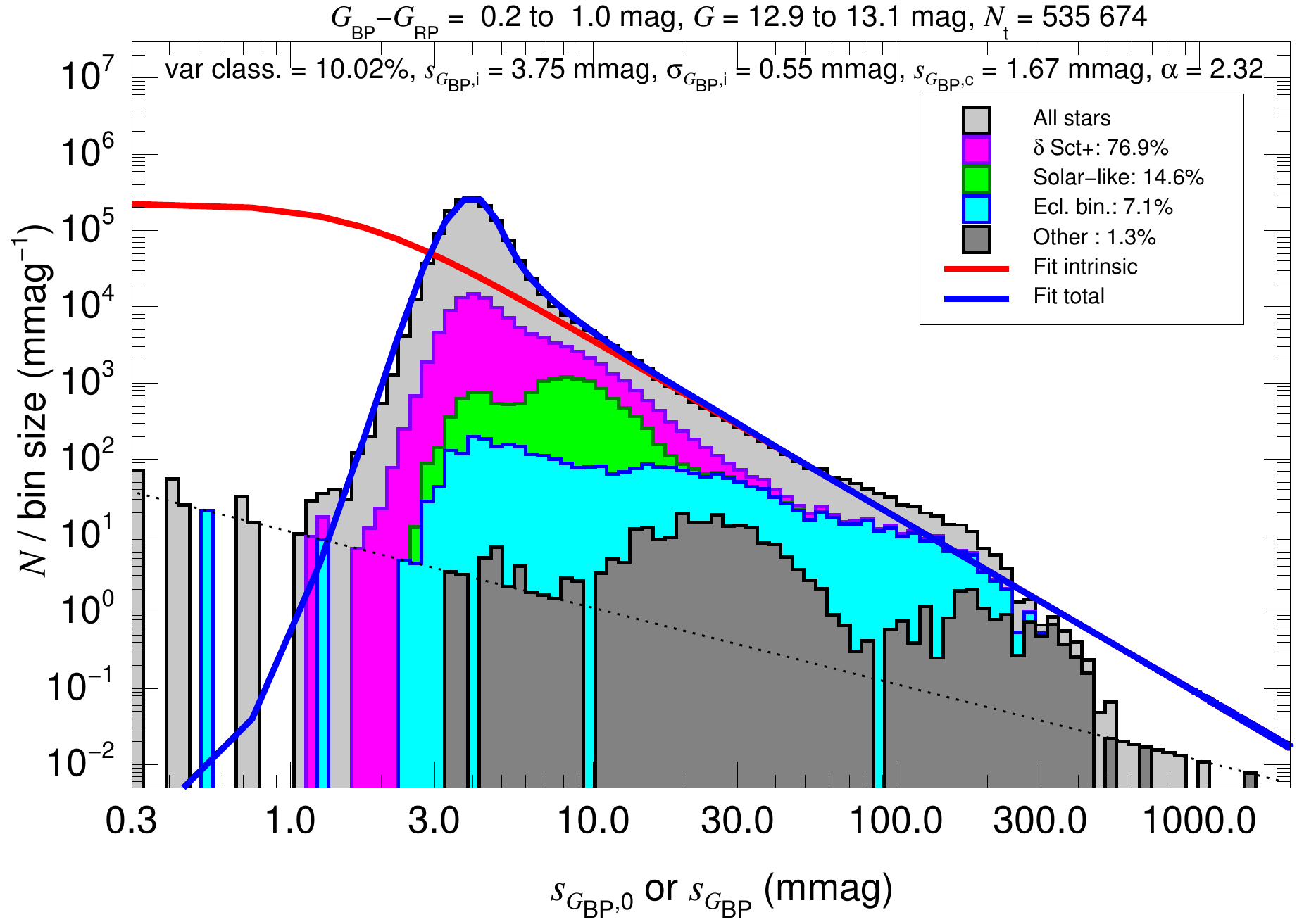}$\!\!\!$
                    \includegraphics[width=0.35\linewidth]{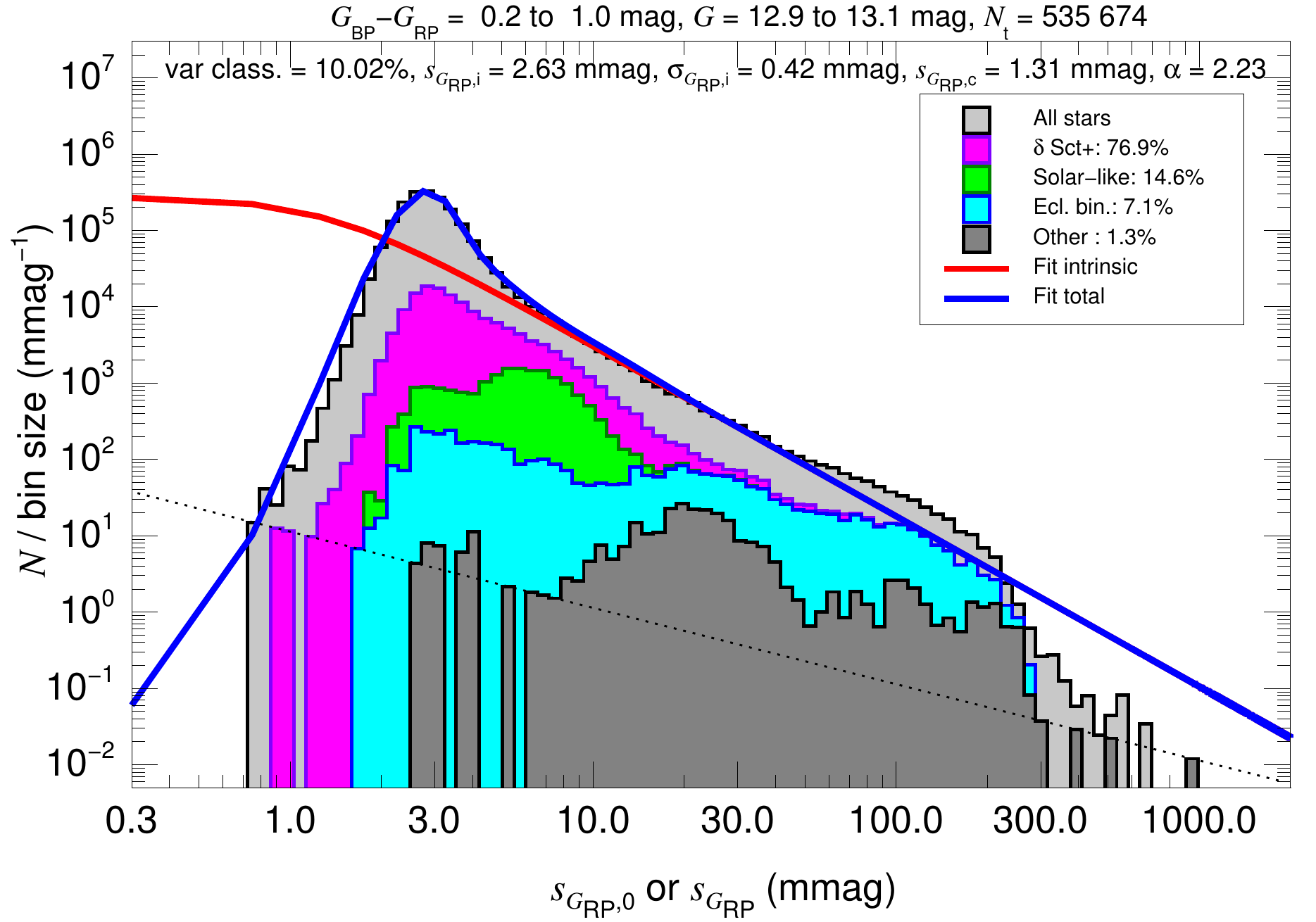}}
\centerline{$\!\!\!$\includegraphics[width=0.35\linewidth]{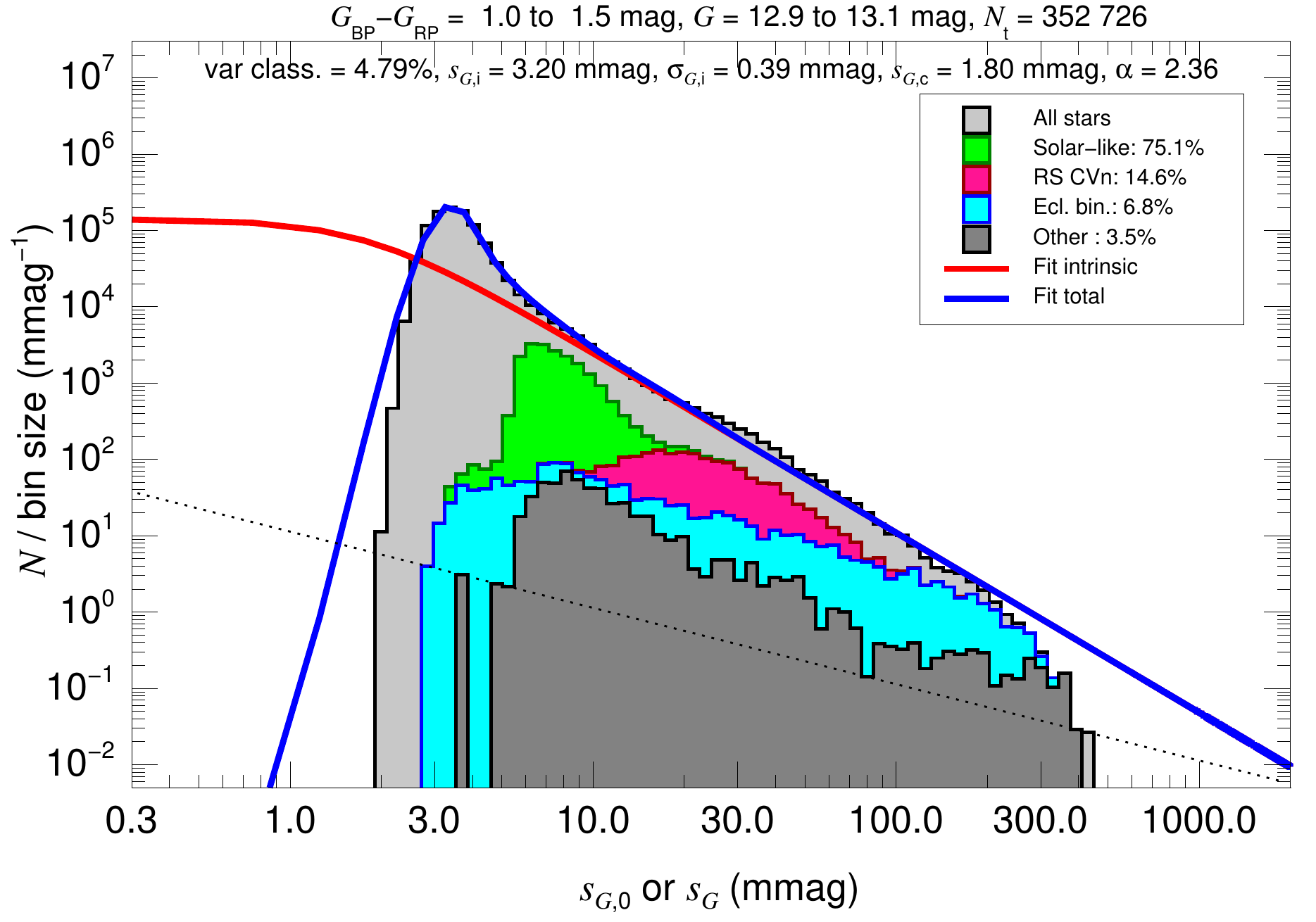}$\!\!\!$
                    \includegraphics[width=0.35\linewidth]{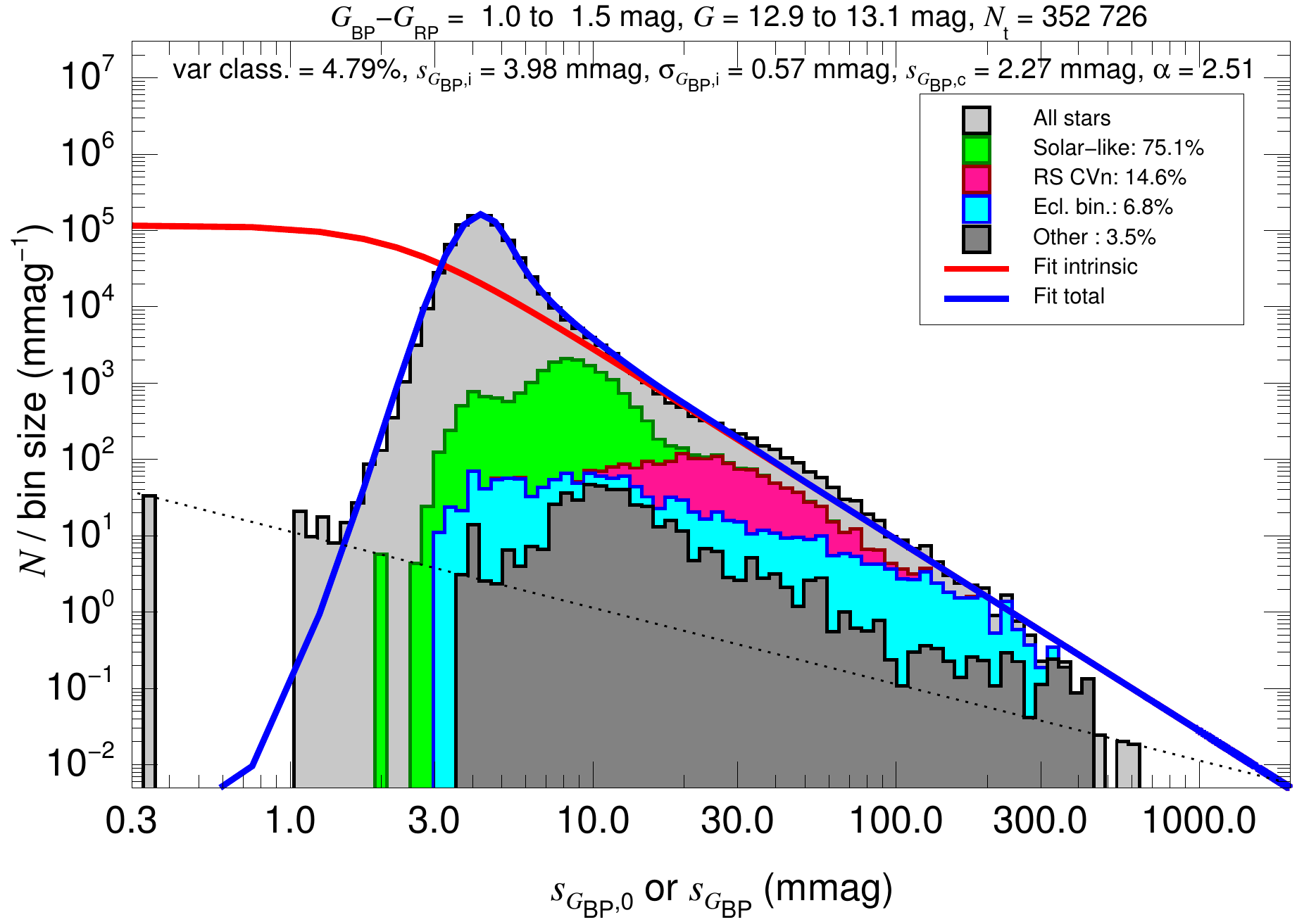}$\!\!\!$
                    \includegraphics[width=0.35\linewidth]{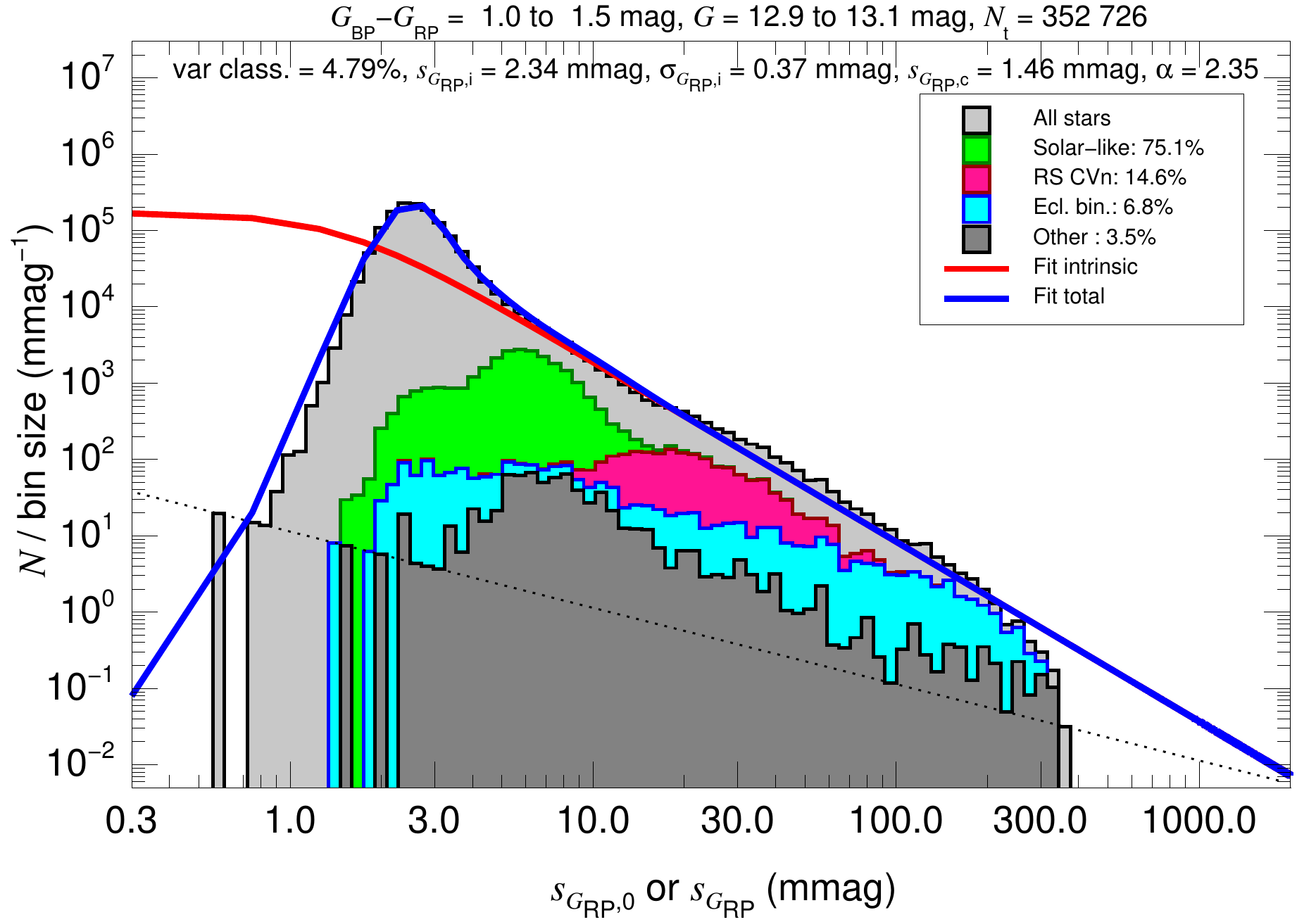}}
\centerline{$\!\!\!$\includegraphics[width=0.35\linewidth]{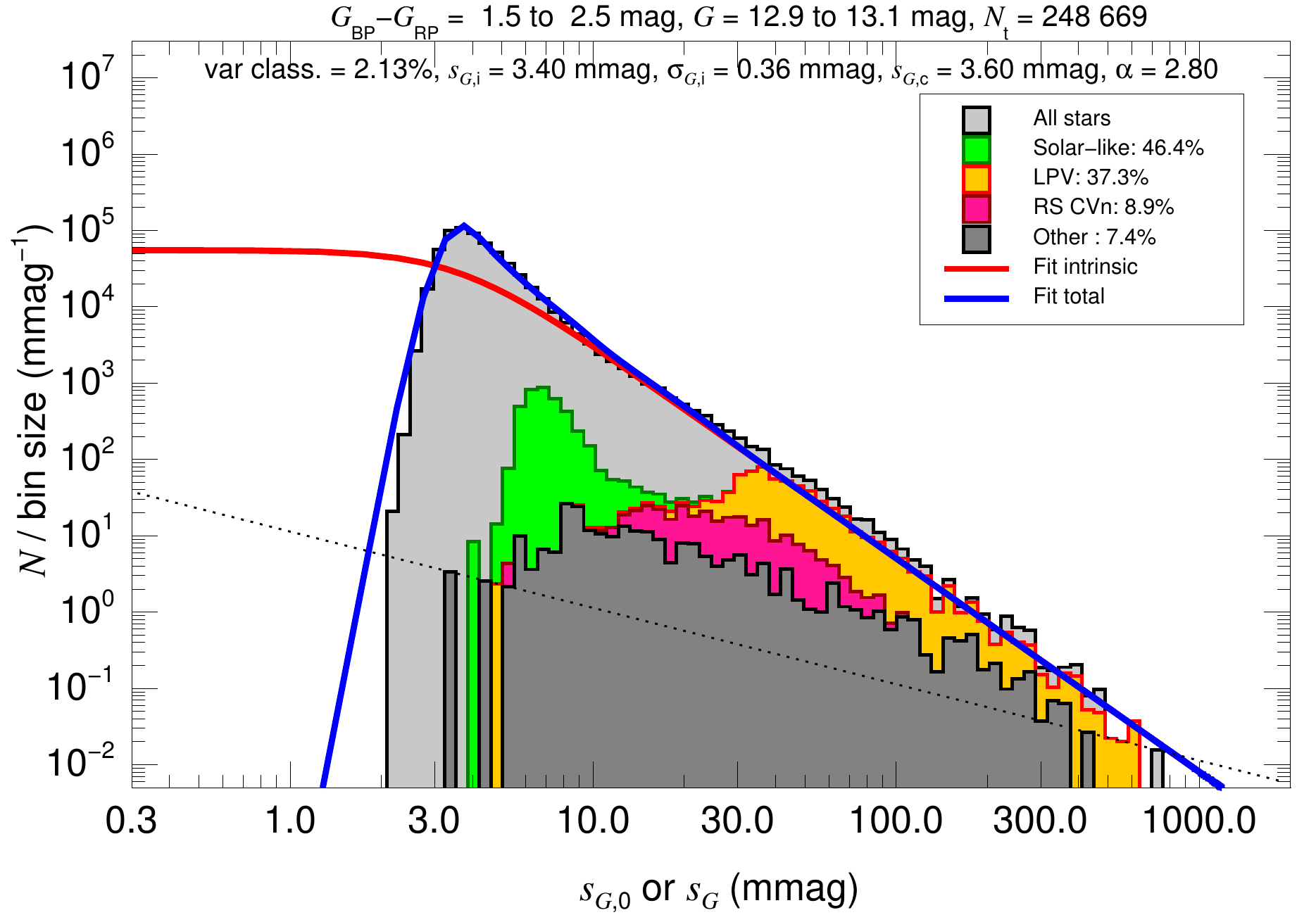}$\!\!\!$
                    \includegraphics[width=0.35\linewidth]{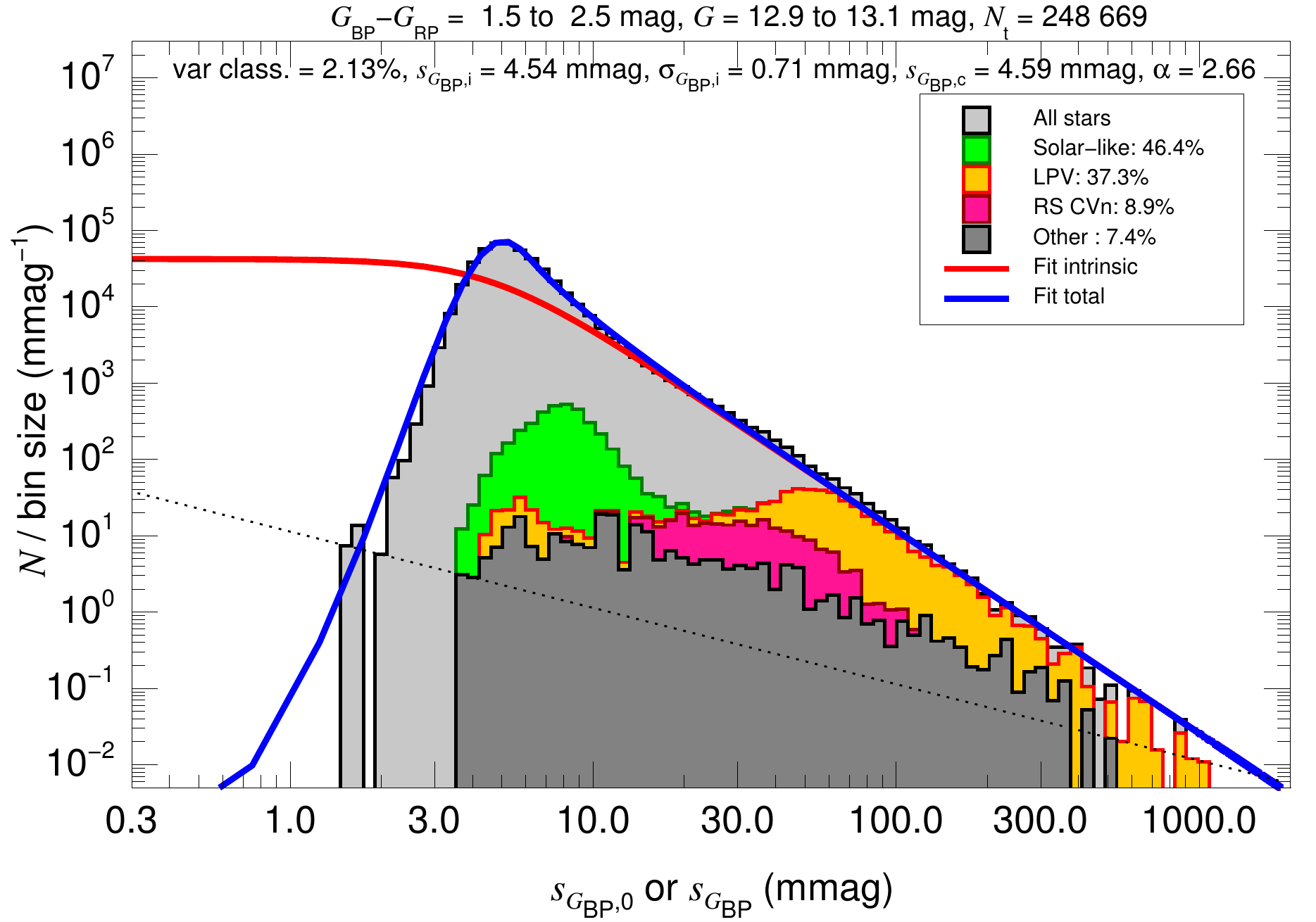}$\!\!\!$
                    \includegraphics[width=0.35\linewidth]{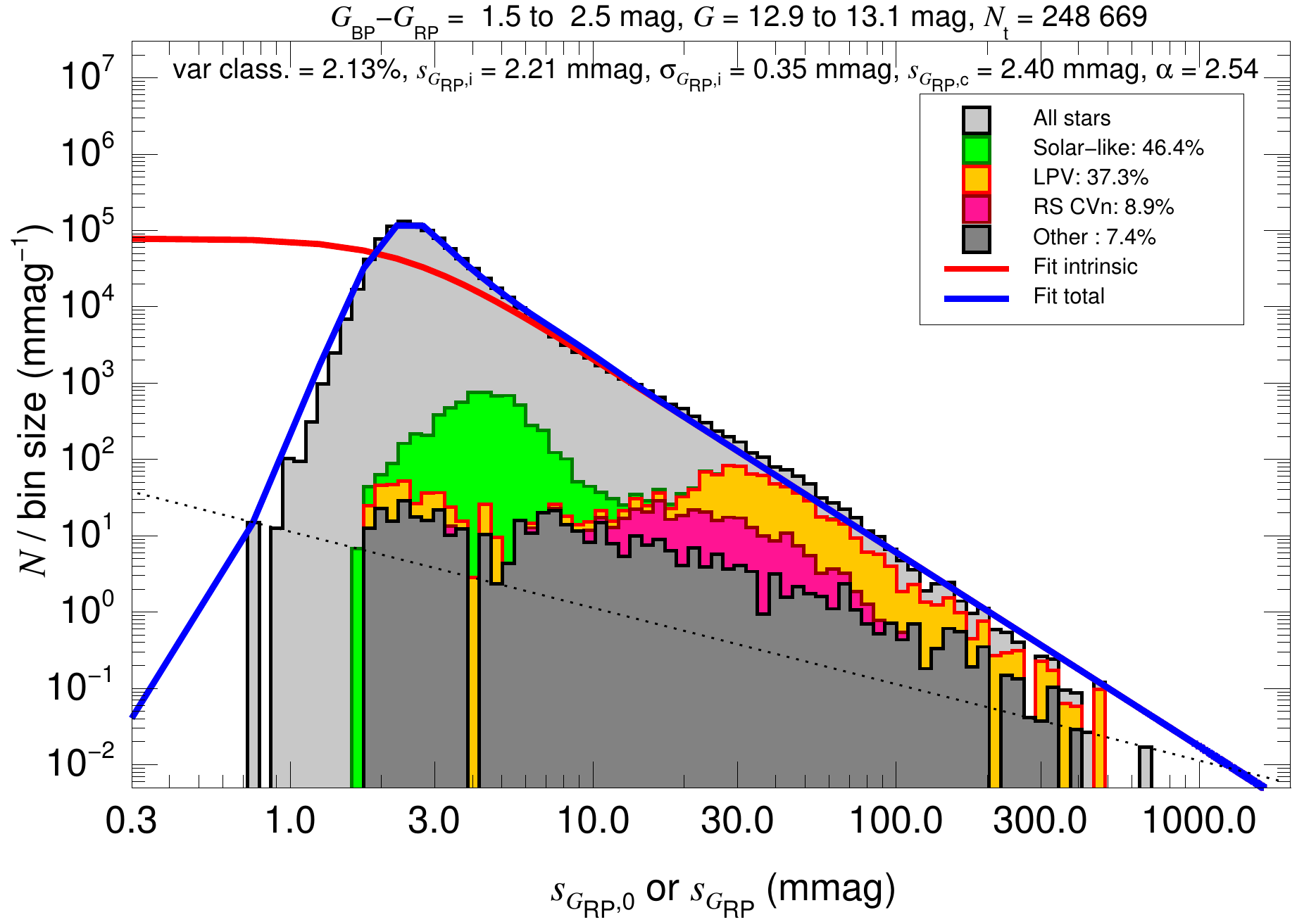}}
\centerline{$\!\!\!$\includegraphics[width=0.35\linewidth]{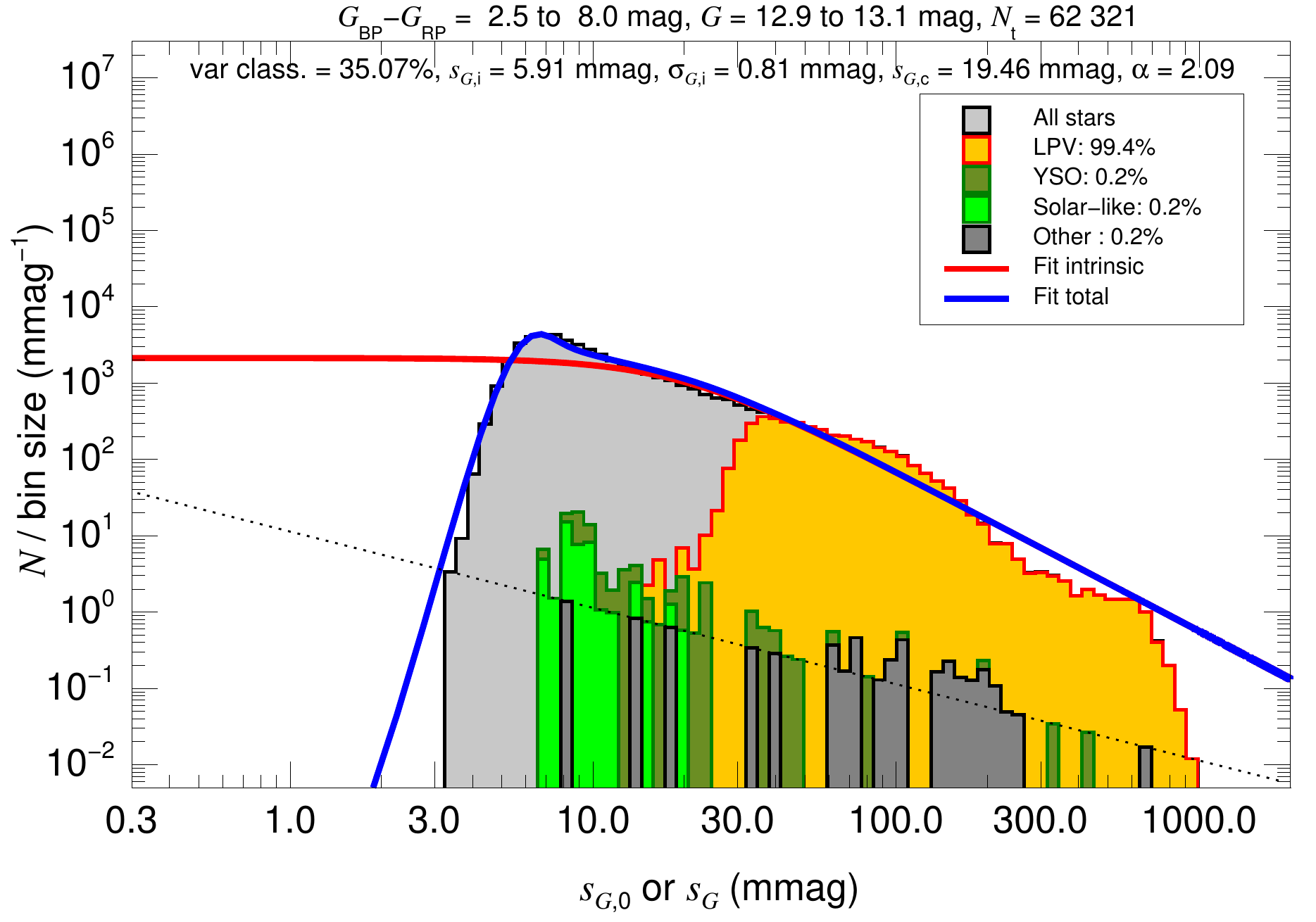}$\!\!\!$
                    \includegraphics[width=0.35\linewidth]{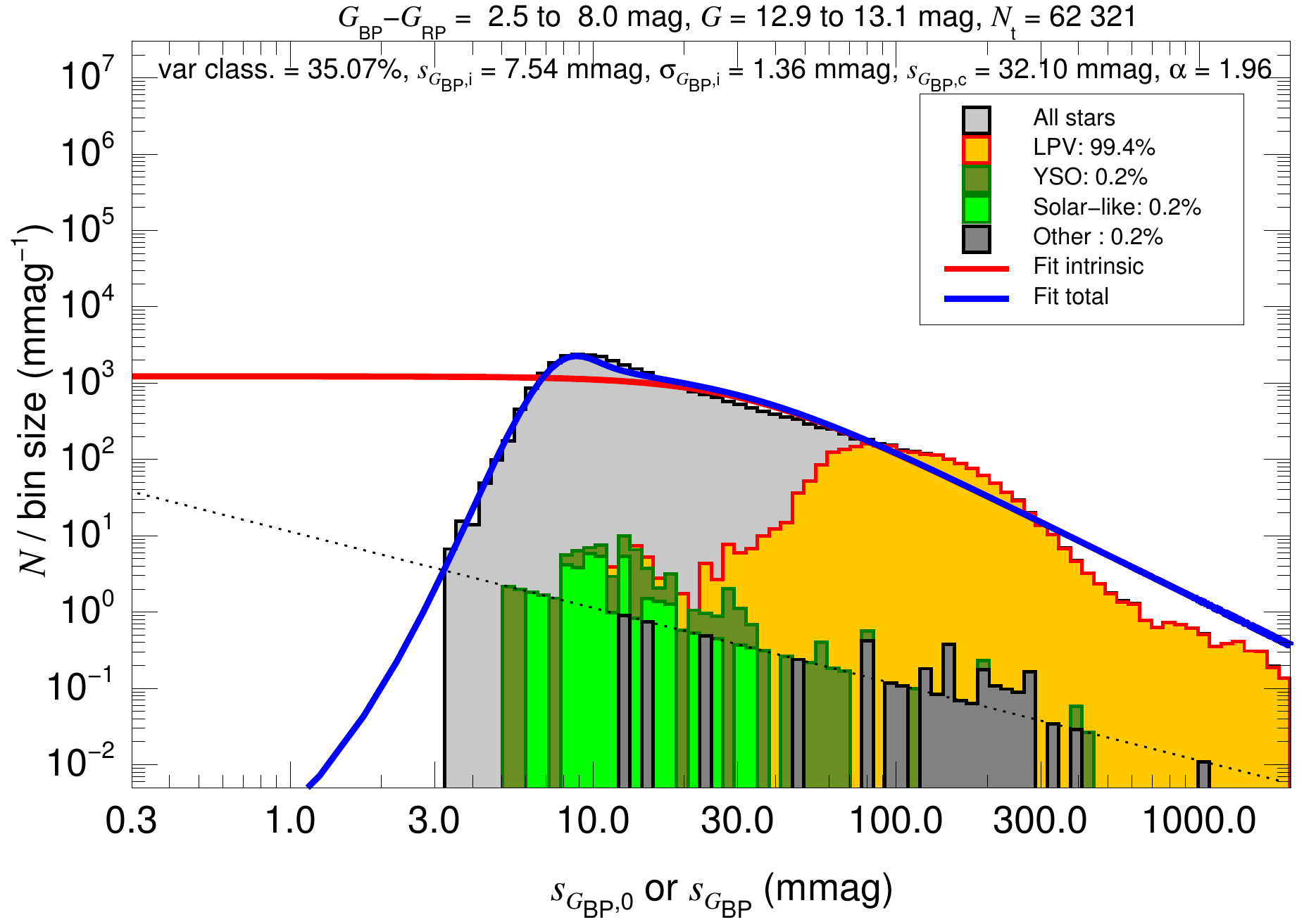}$\!\!\!$
                    \includegraphics[width=0.35\linewidth]{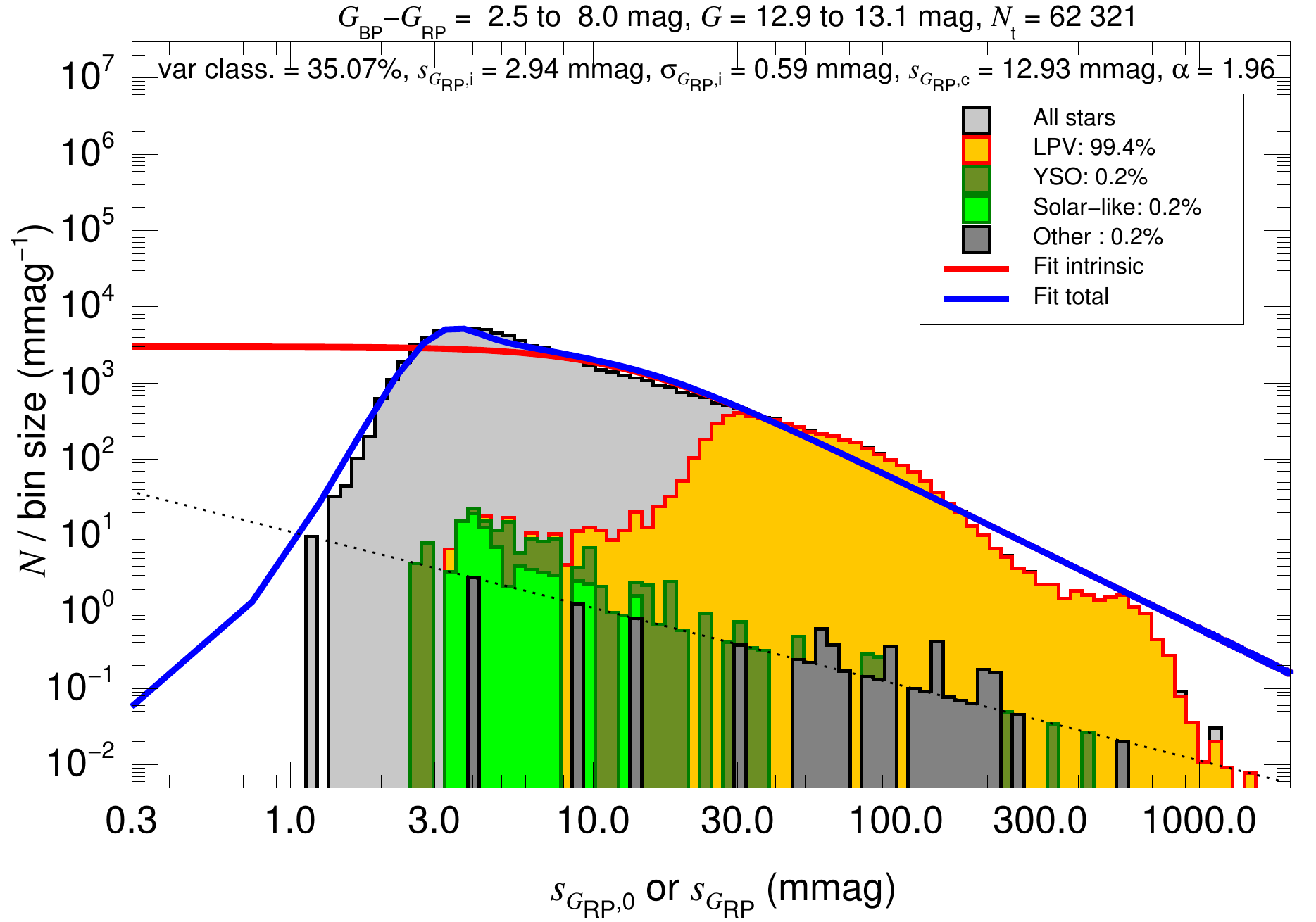}}
\caption{(Continued).}
\end{figure*}

\addtocounter{figure}{-1}

\begin{figure*}
\centerline{$\!\!\!$\includegraphics[width=0.35\linewidth]{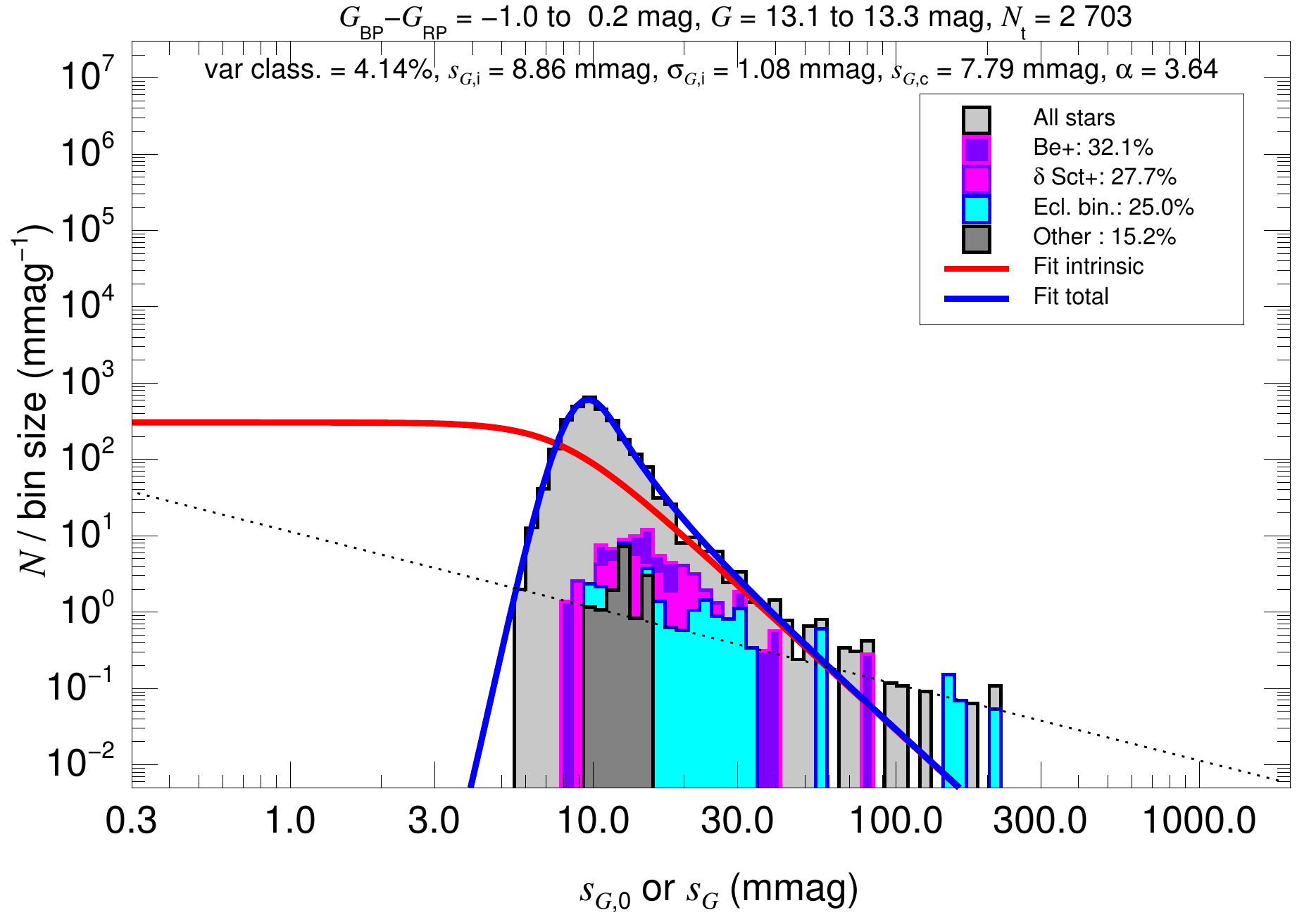}$\!\!\!$
                    \includegraphics[width=0.35\linewidth]{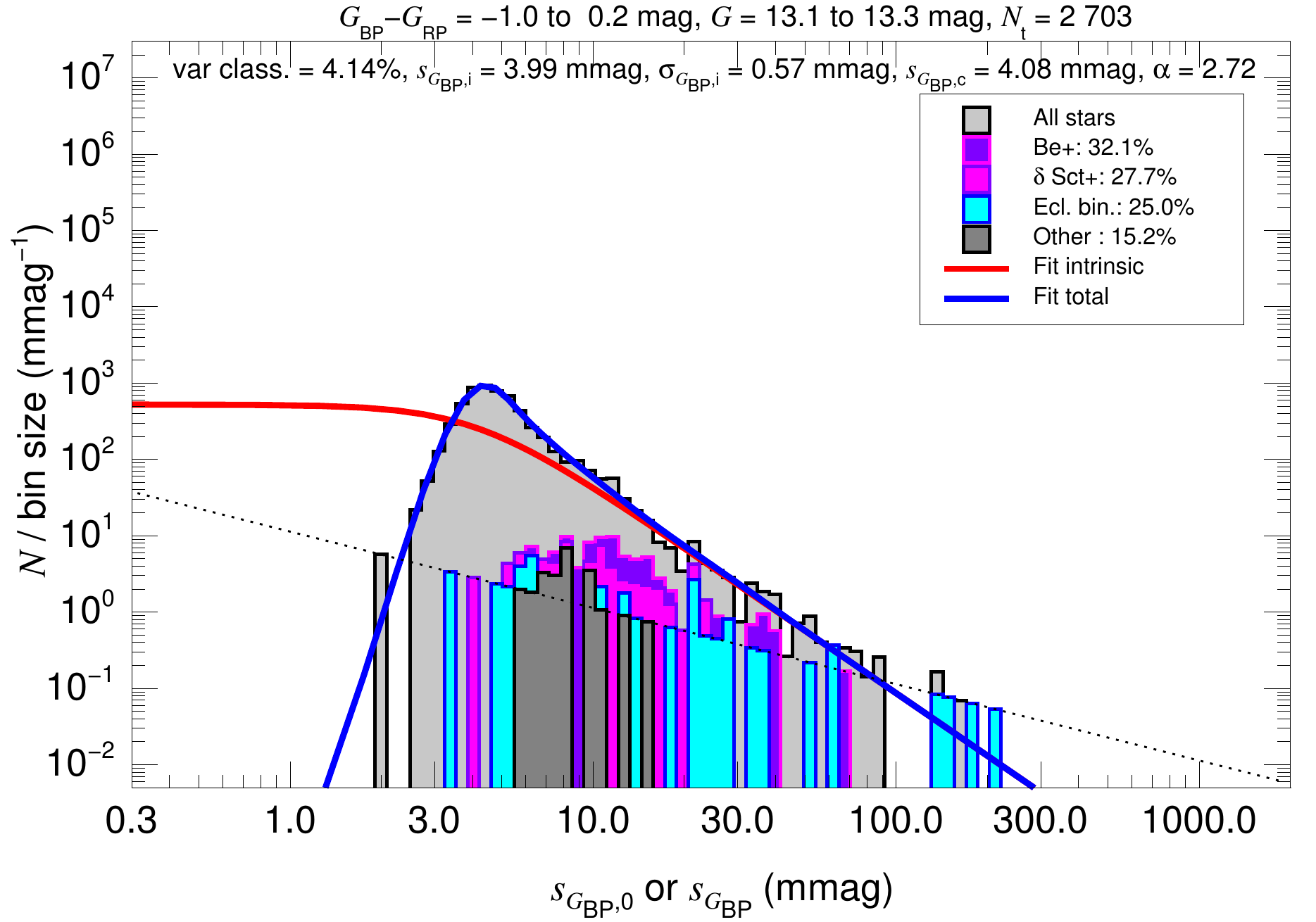}$\!\!\!$
                    \includegraphics[width=0.35\linewidth]{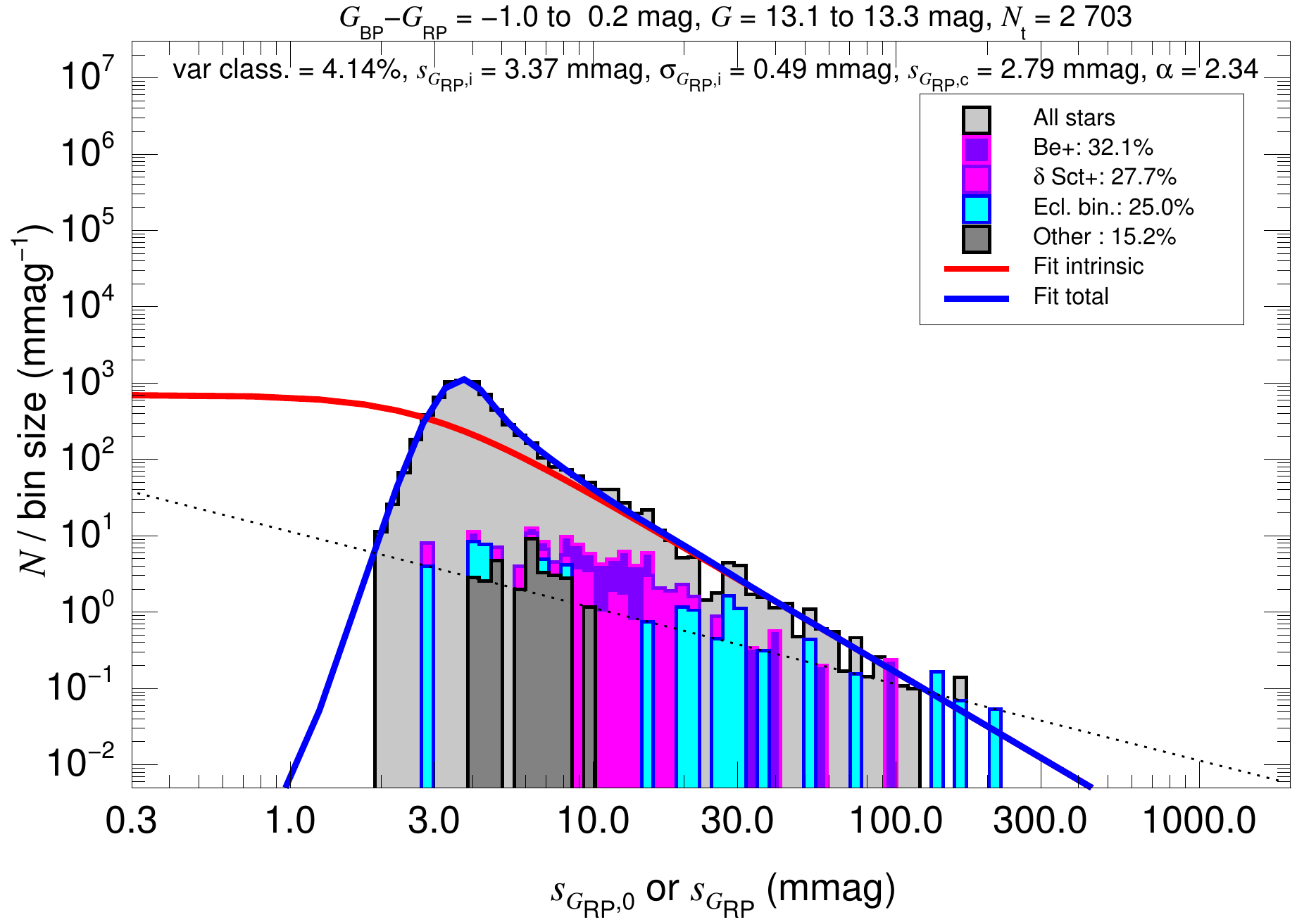}}
\centerline{$\!\!\!$\includegraphics[width=0.35\linewidth]{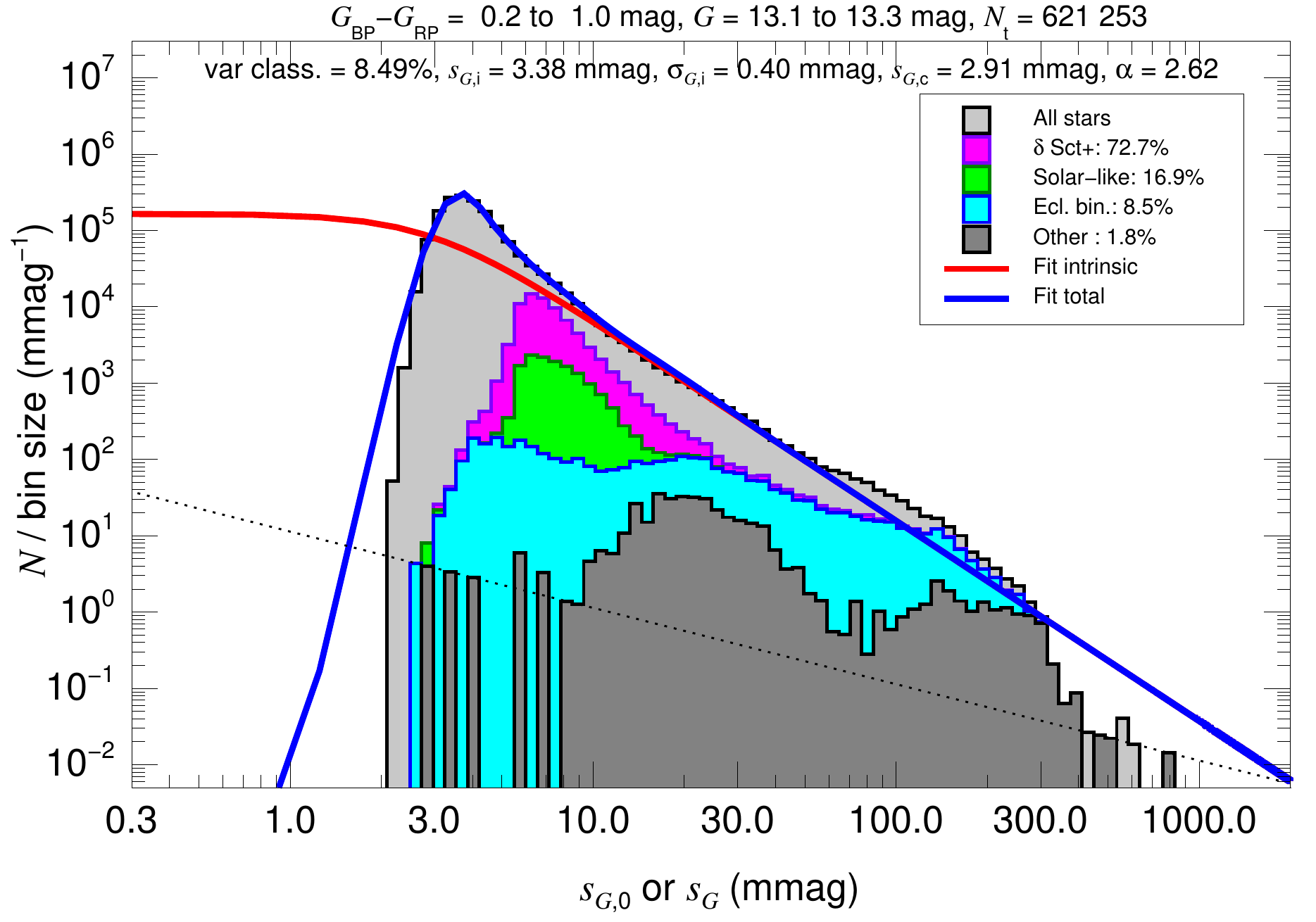}$\!\!\!$
                    \includegraphics[width=0.35\linewidth]{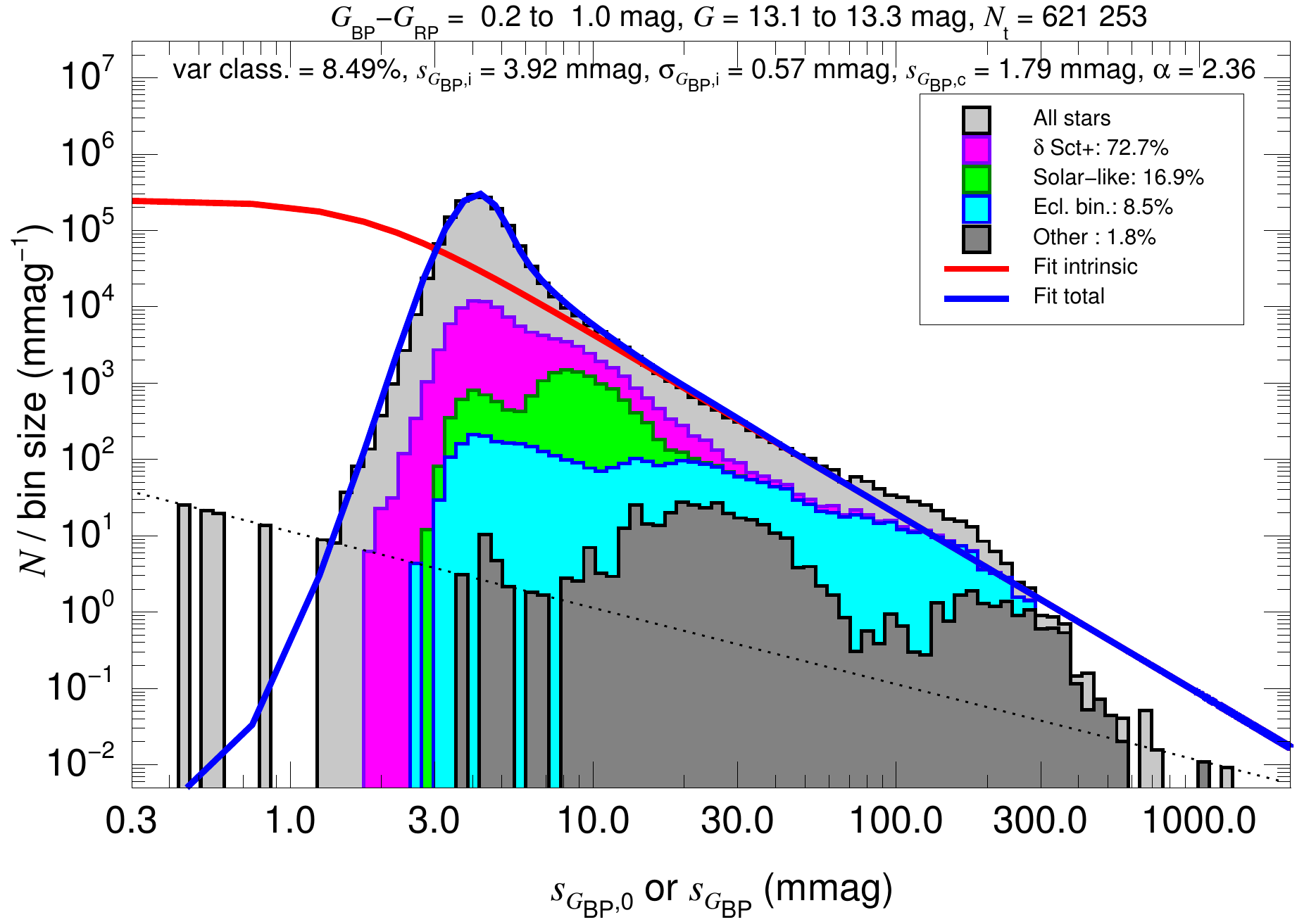}$\!\!\!$
                    \includegraphics[width=0.35\linewidth]{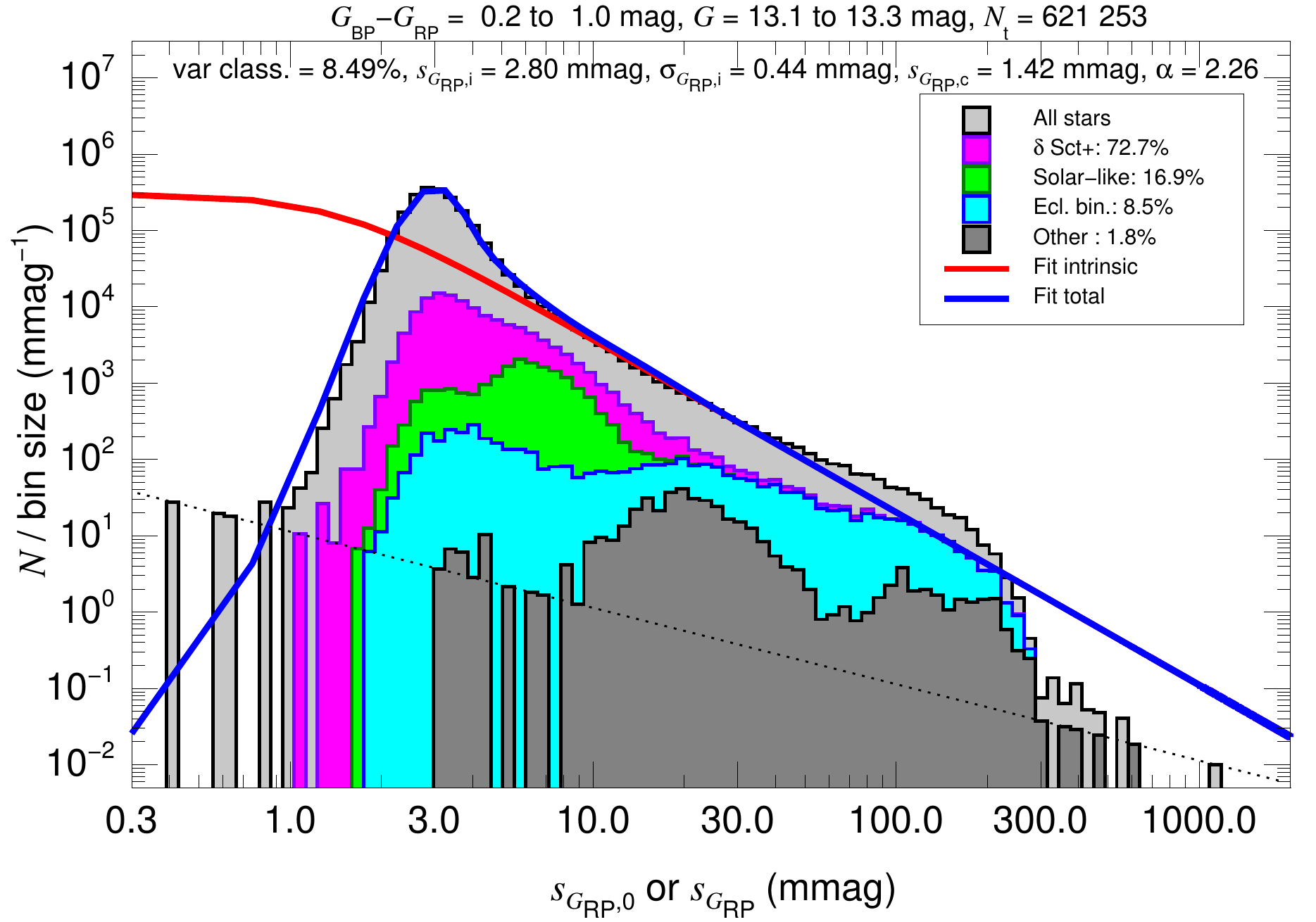}}
\centerline{$\!\!\!$\includegraphics[width=0.35\linewidth]{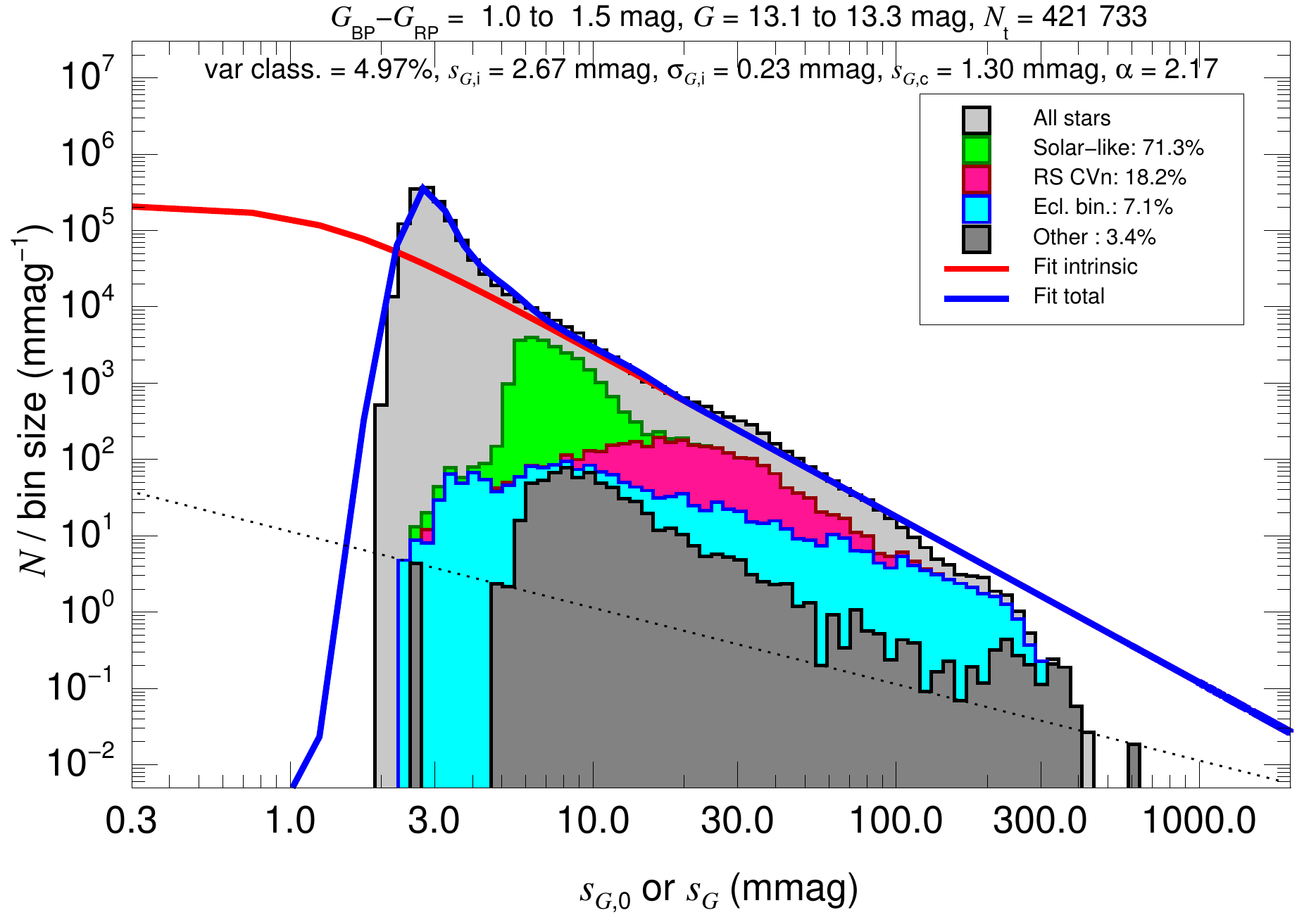}$\!\!\!$
                    \includegraphics[width=0.35\linewidth]{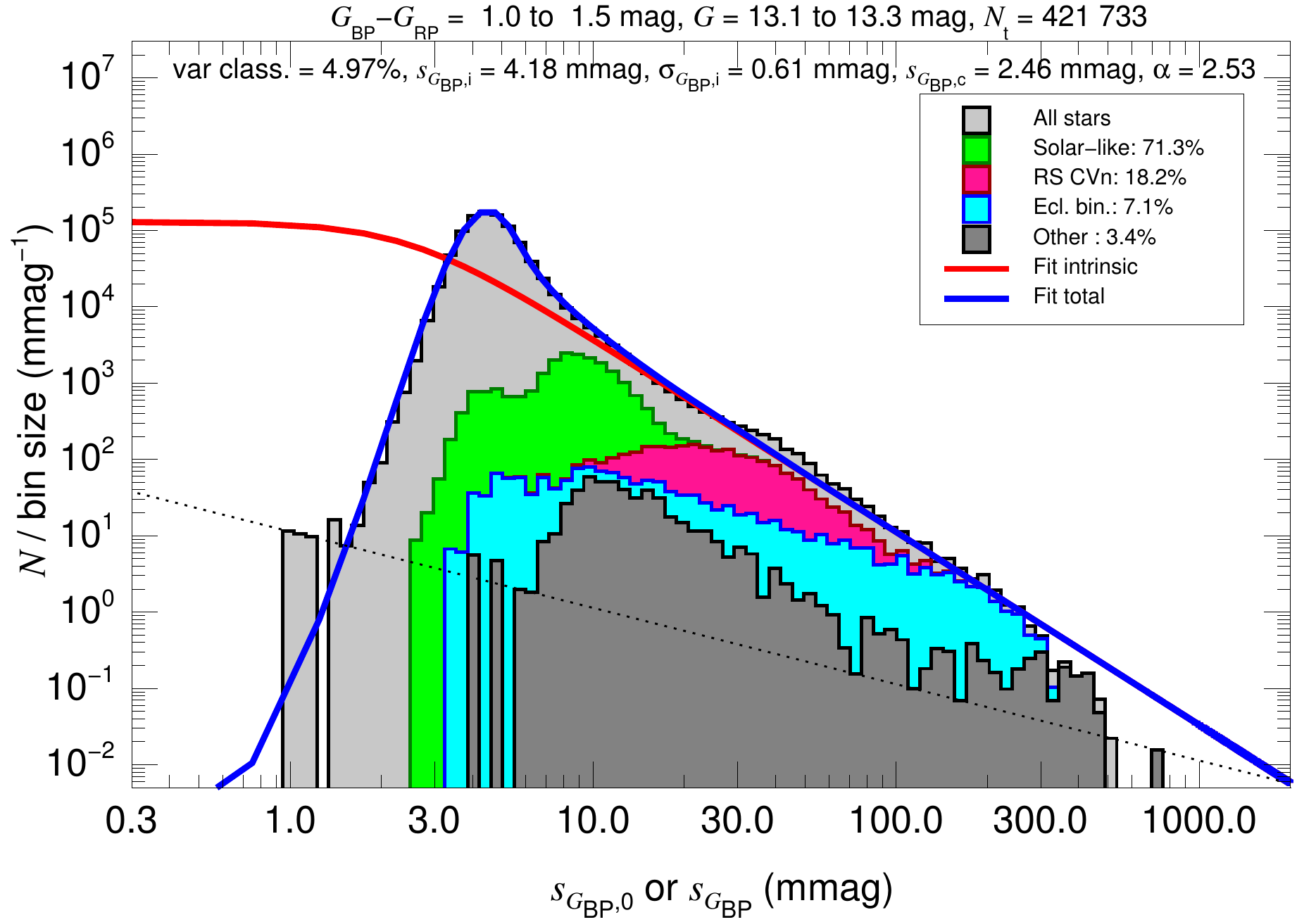}$\!\!\!$
                    \includegraphics[width=0.35\linewidth]{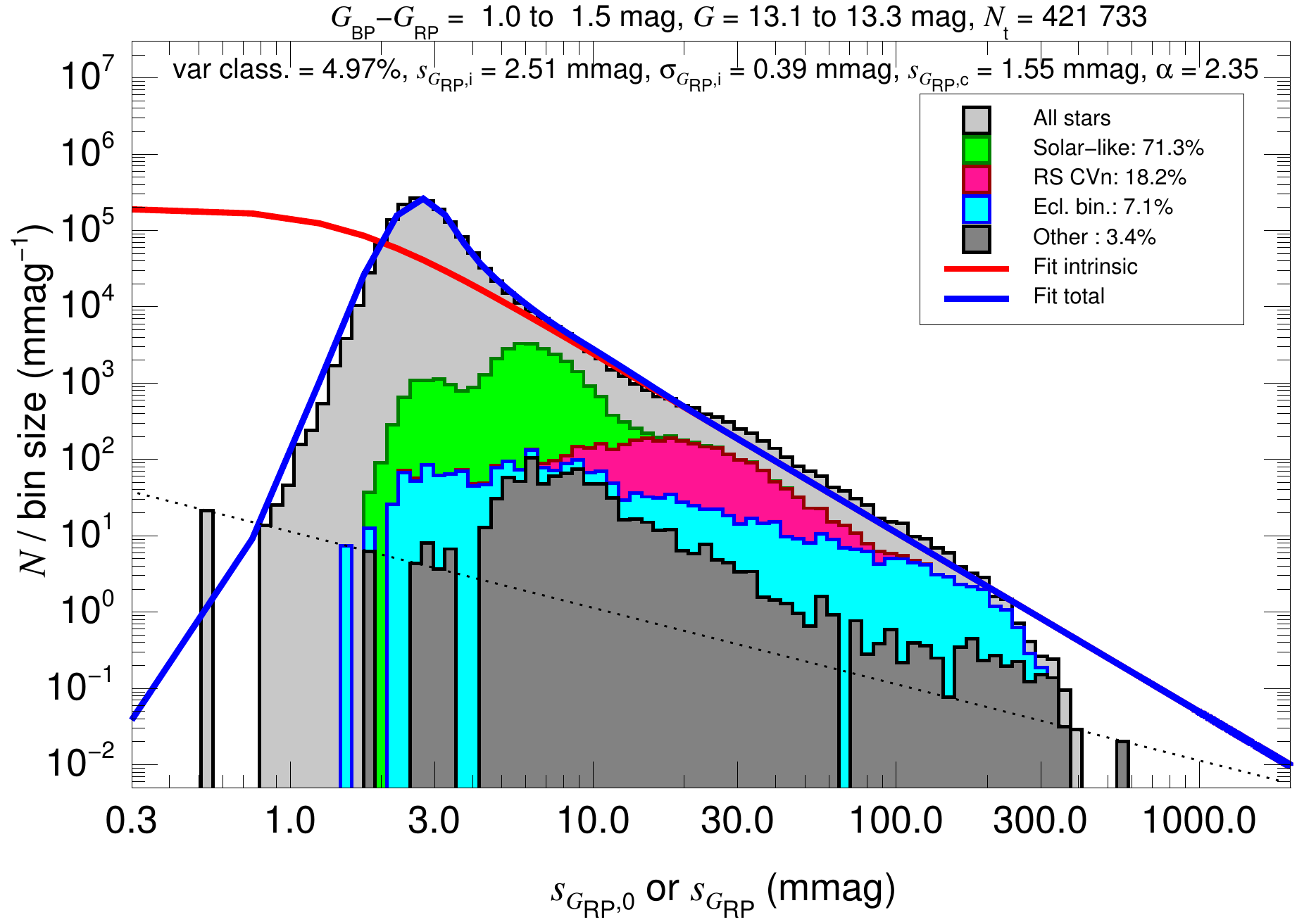}}
\centerline{$\!\!\!$\includegraphics[width=0.35\linewidth]{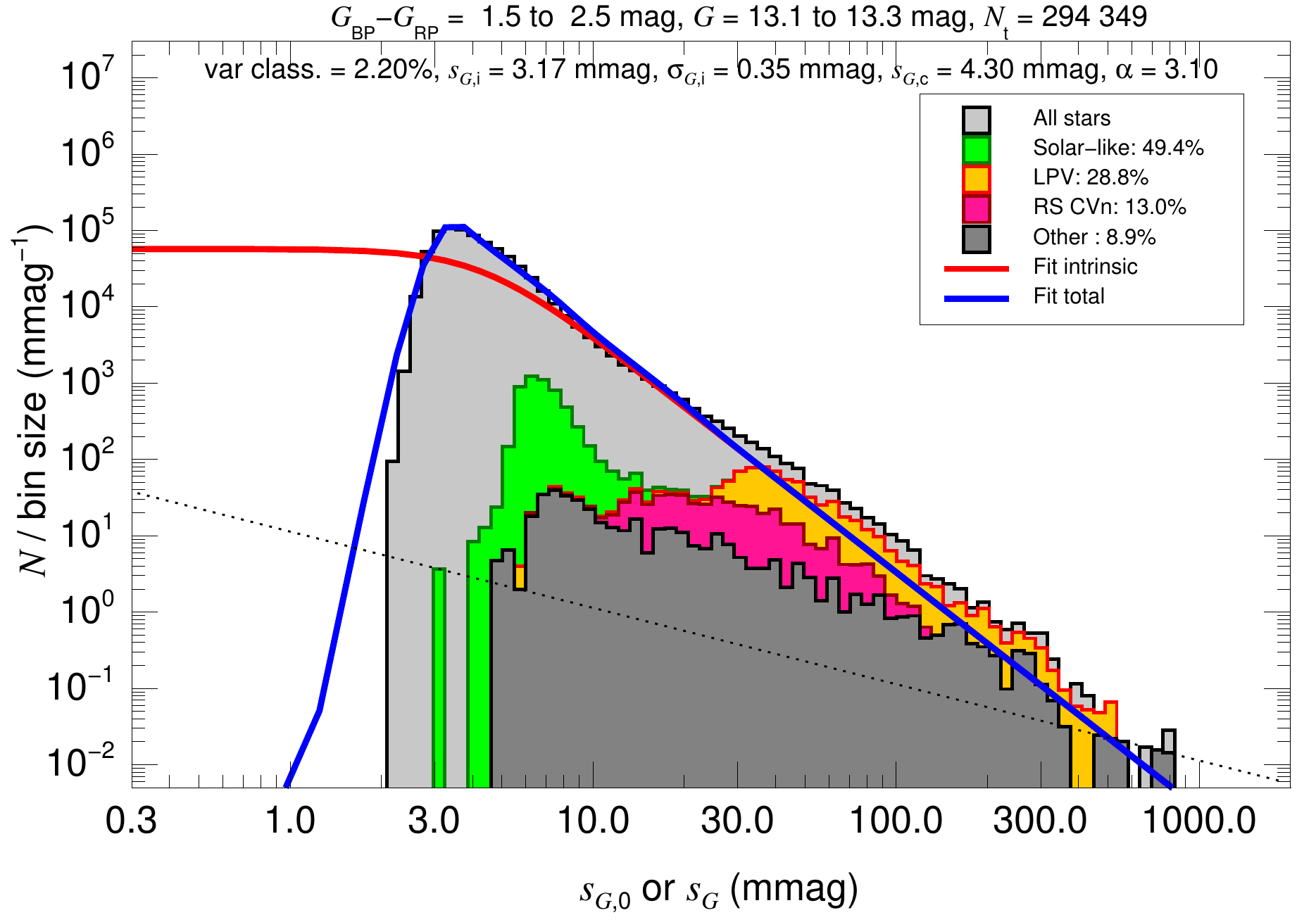}$\!\!\!$
                    \includegraphics[width=0.35\linewidth]{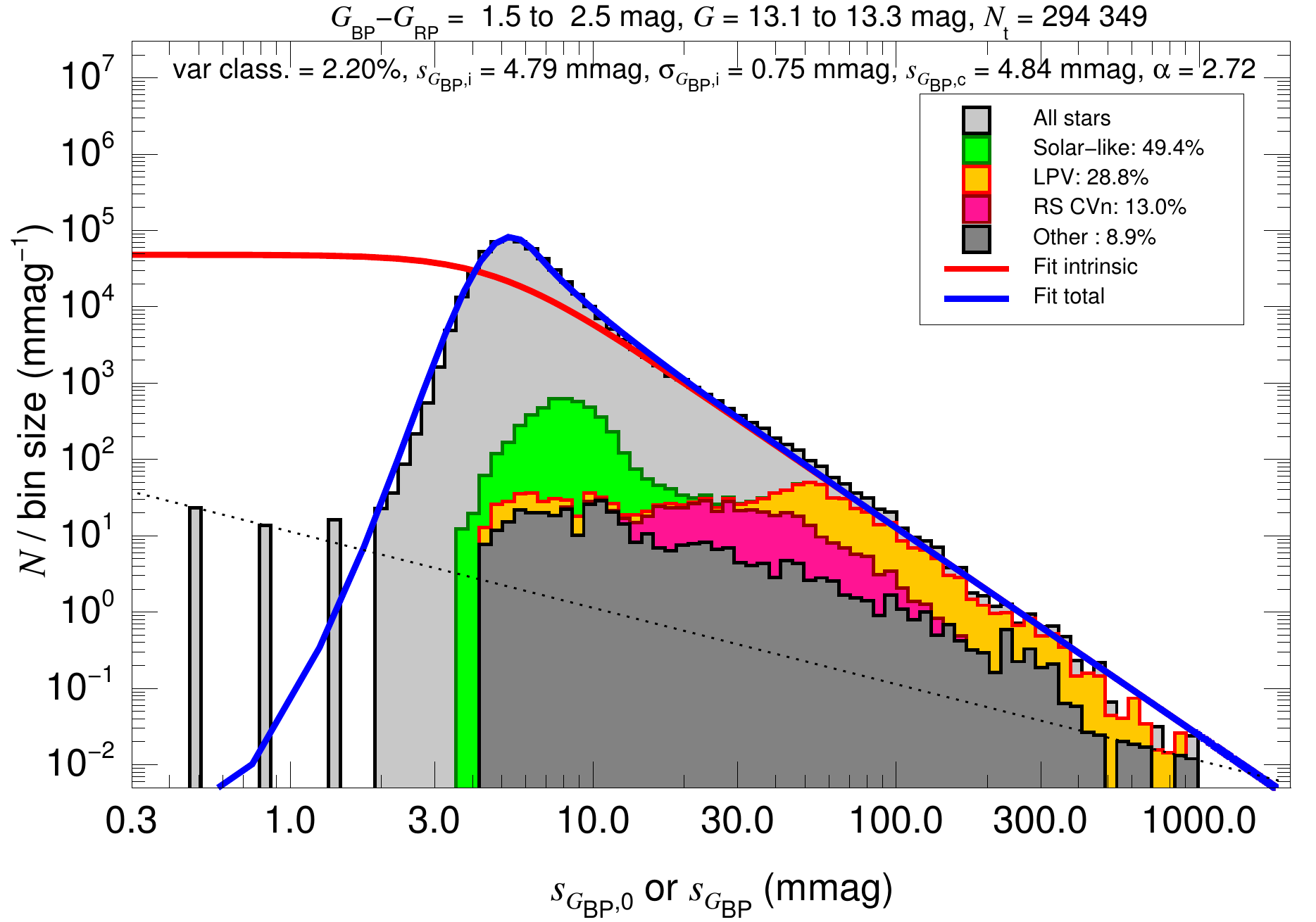}$\!\!\!$
                    \includegraphics[width=0.35\linewidth]{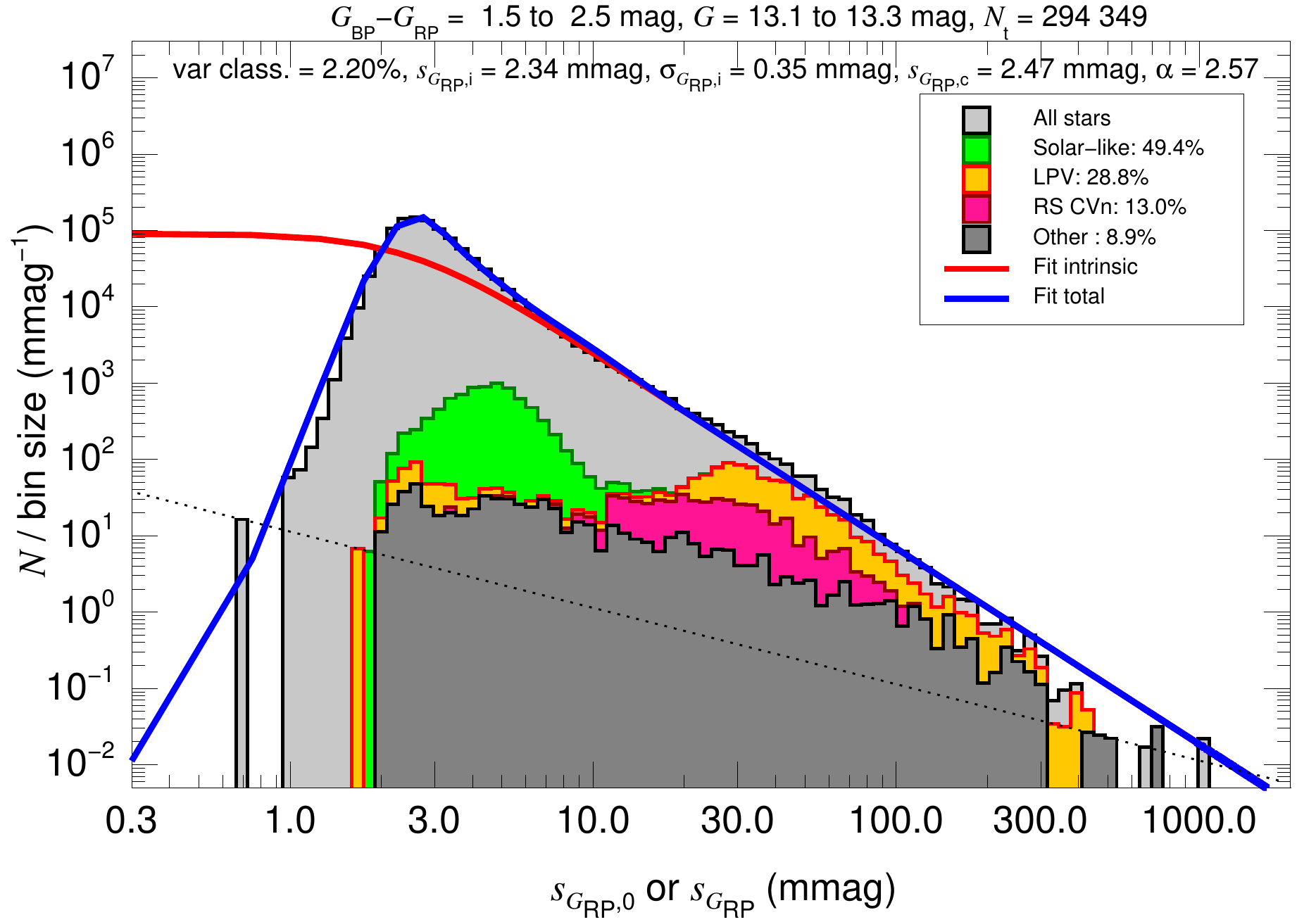}}
\centerline{$\!\!\!$\includegraphics[width=0.35\linewidth]{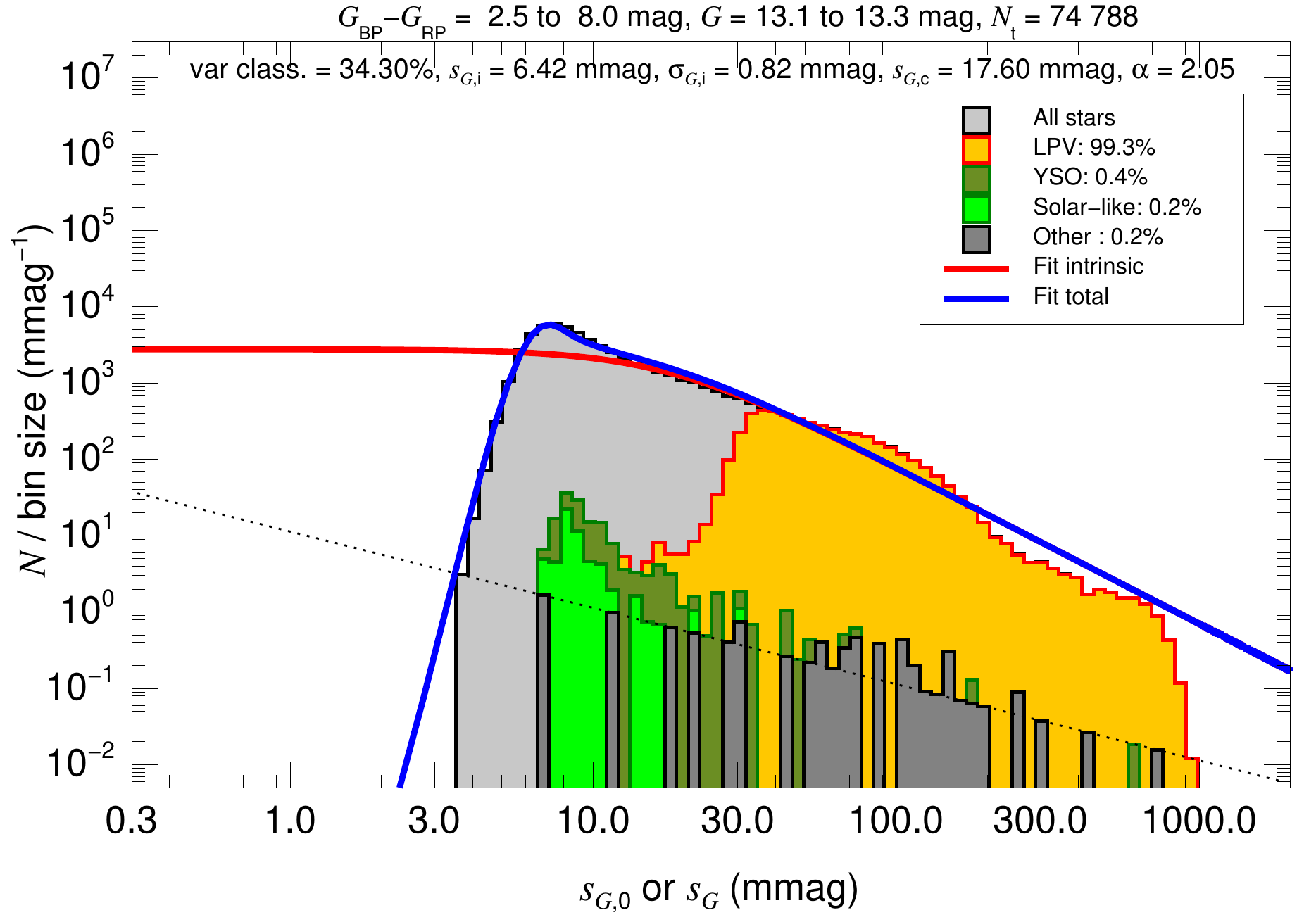}$\!\!\!$
                    \includegraphics[width=0.35\linewidth]{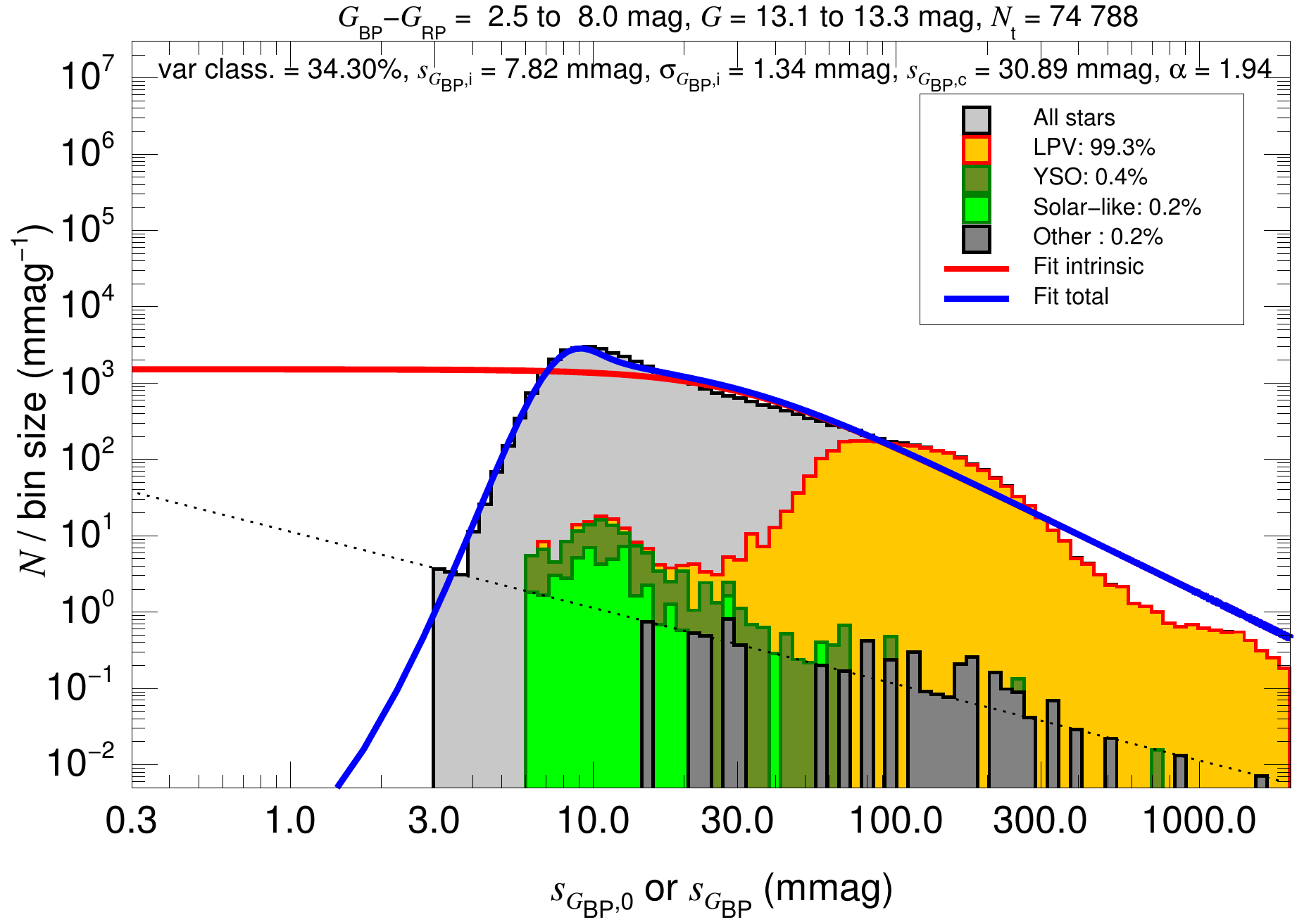}$\!\!\!$
                    \includegraphics[width=0.35\linewidth]{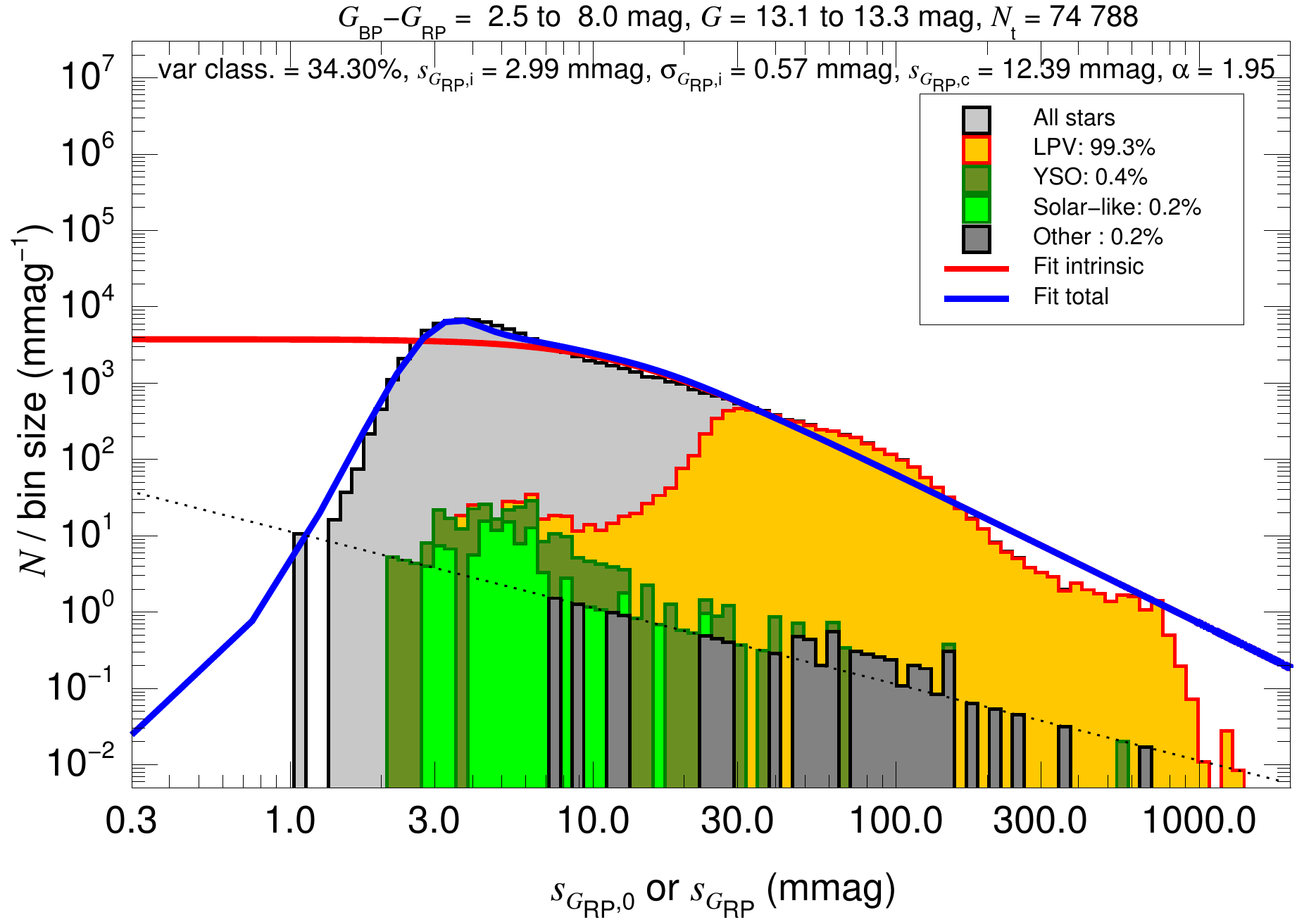}}
\caption{(Continued).}
\end{figure*}

\addtocounter{figure}{-1}

\begin{figure*}
\centerline{$\!\!\!$\includegraphics[width=0.35\linewidth]{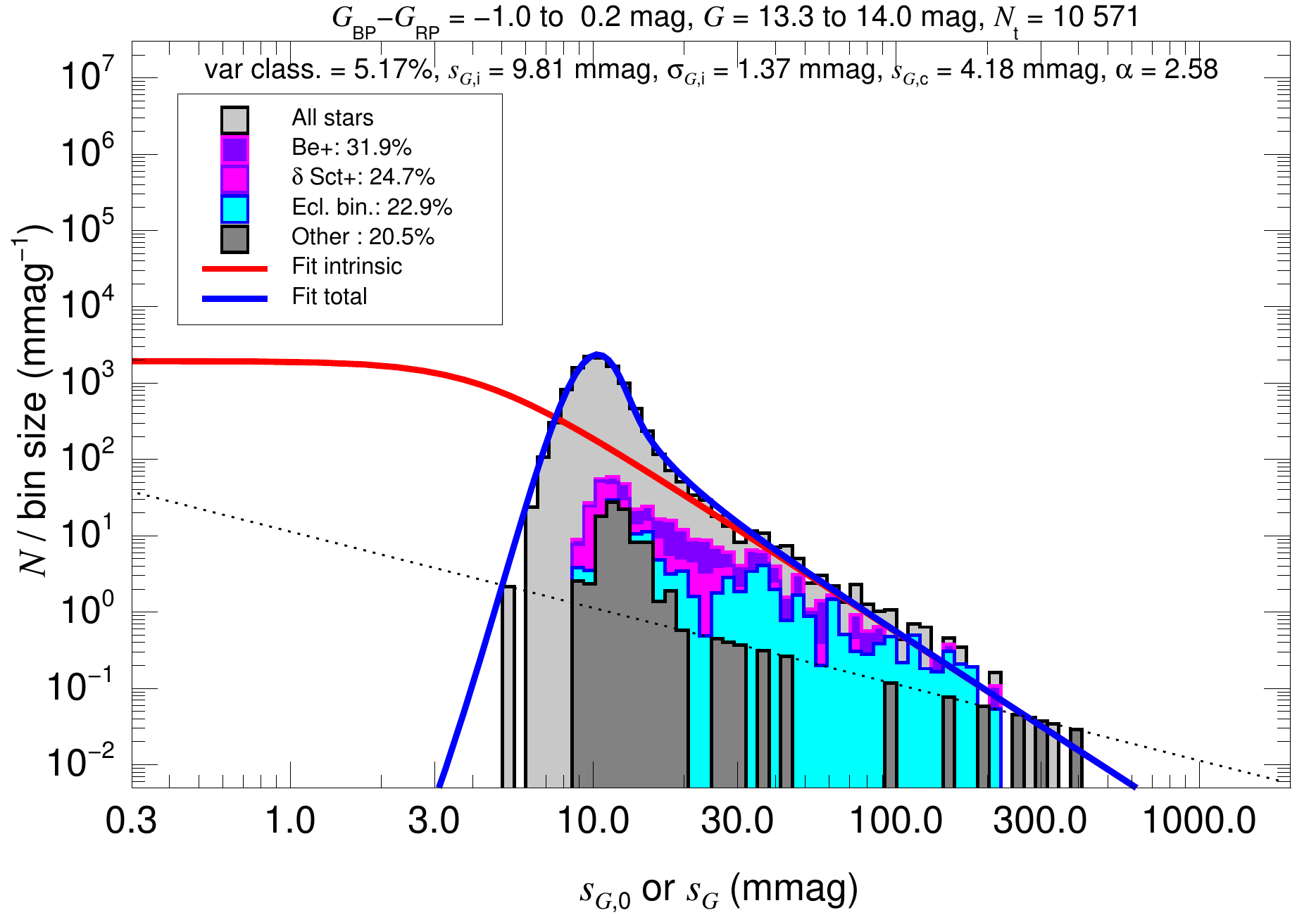}$\!\!\!$
                    \includegraphics[width=0.35\linewidth]{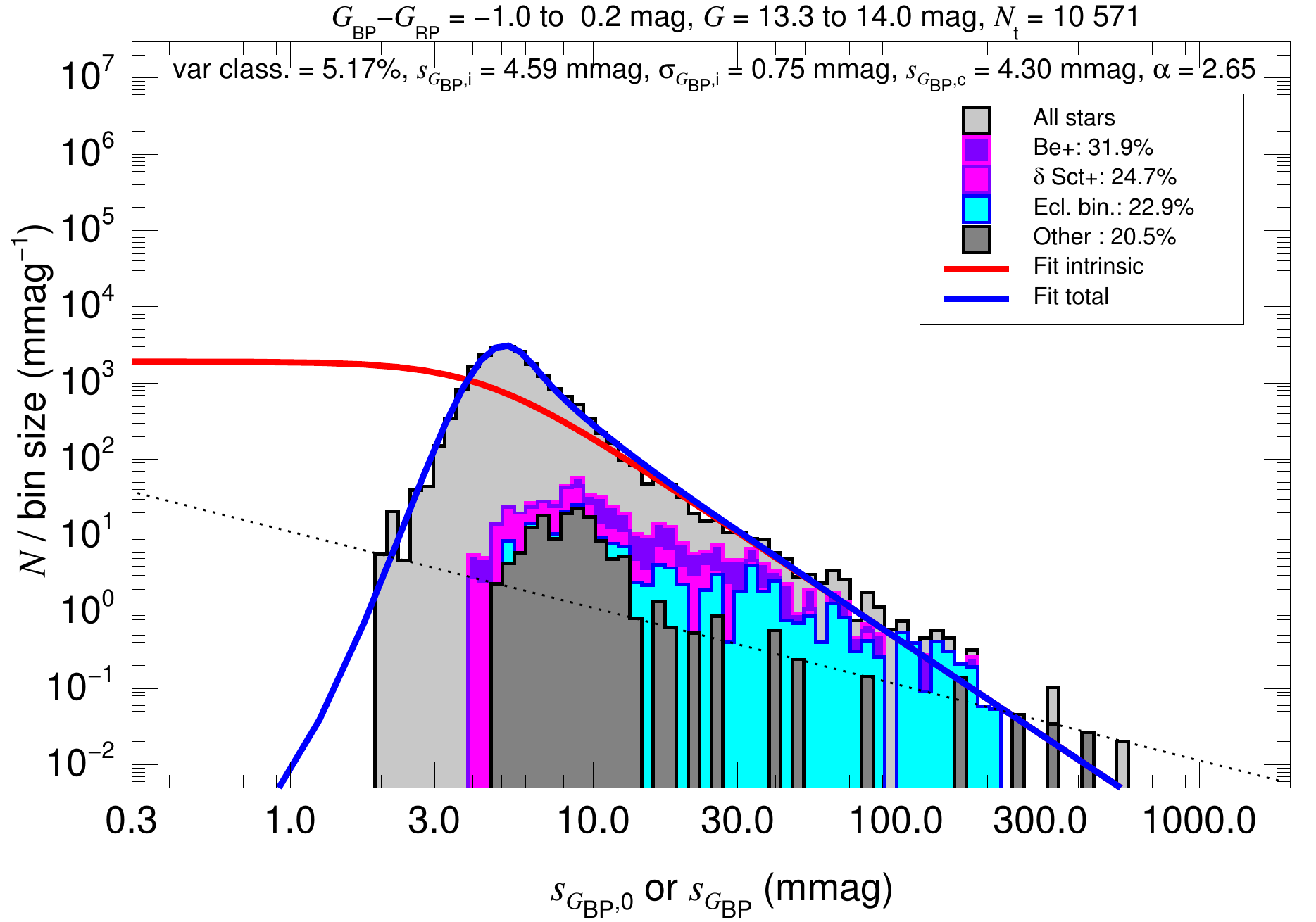}$\!\!\!$
                    \includegraphics[width=0.35\linewidth]{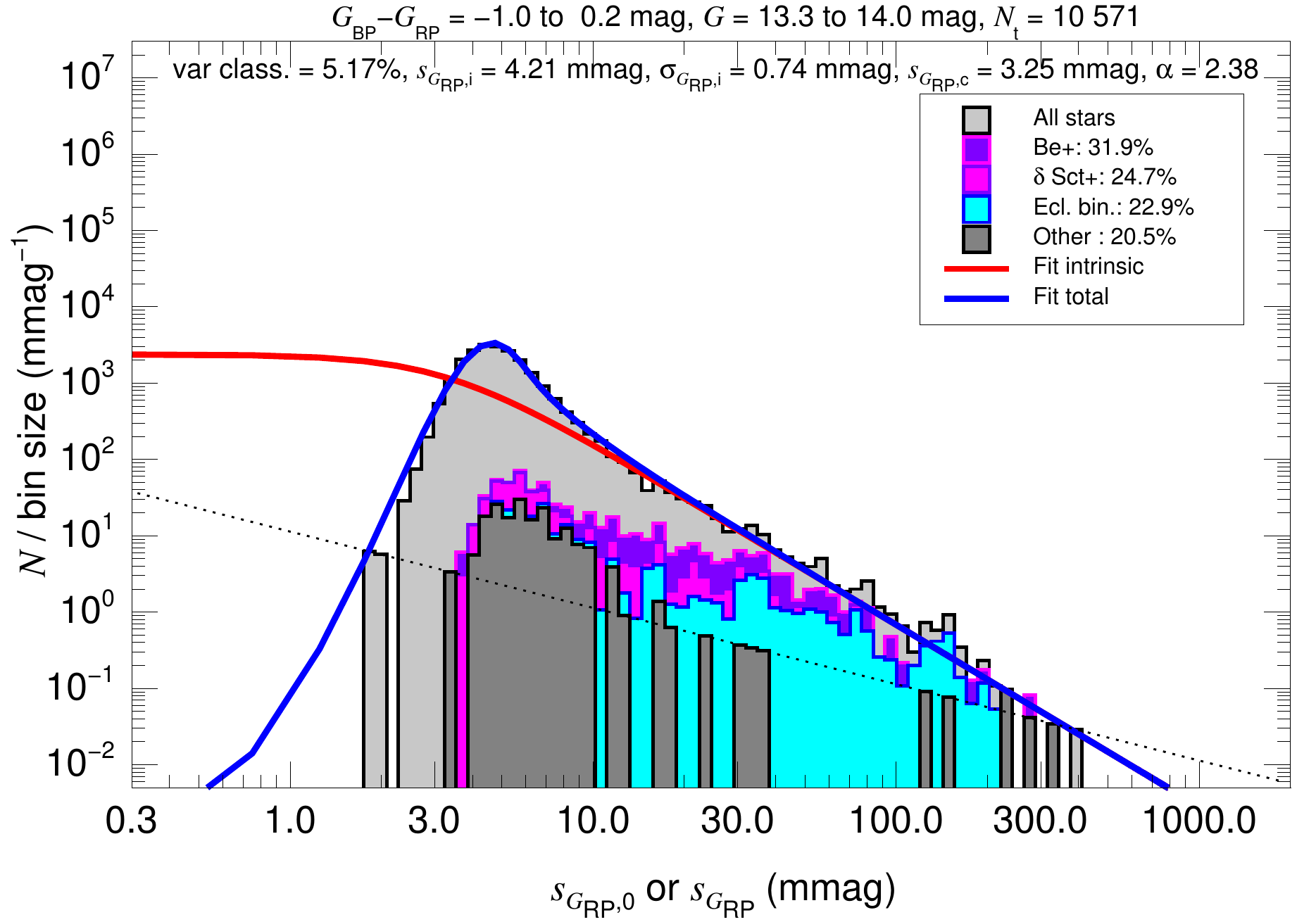}}
\centerline{$\!\!\!$\includegraphics[width=0.35\linewidth]{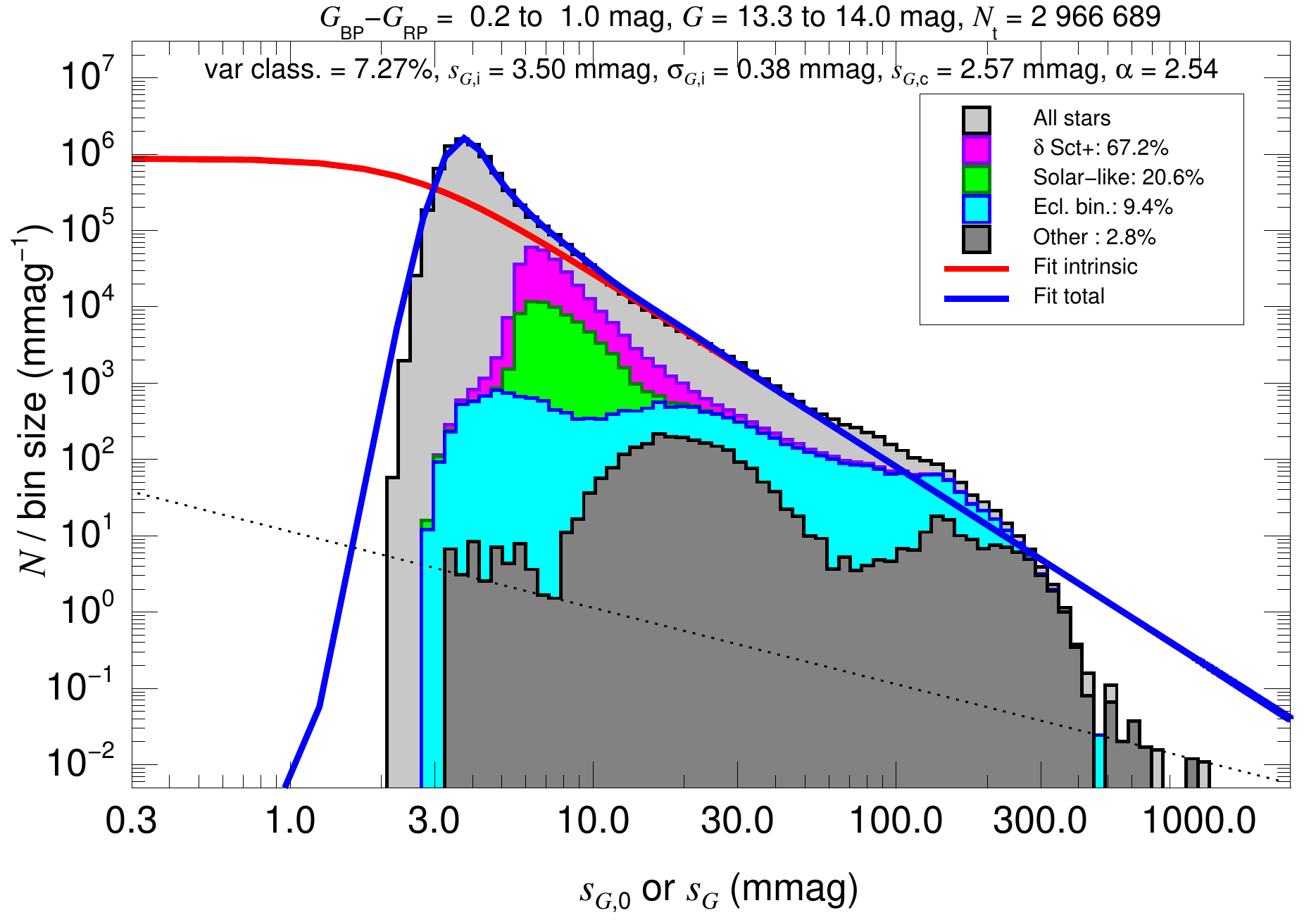}$\!\!\!$
                    \includegraphics[width=0.35\linewidth]{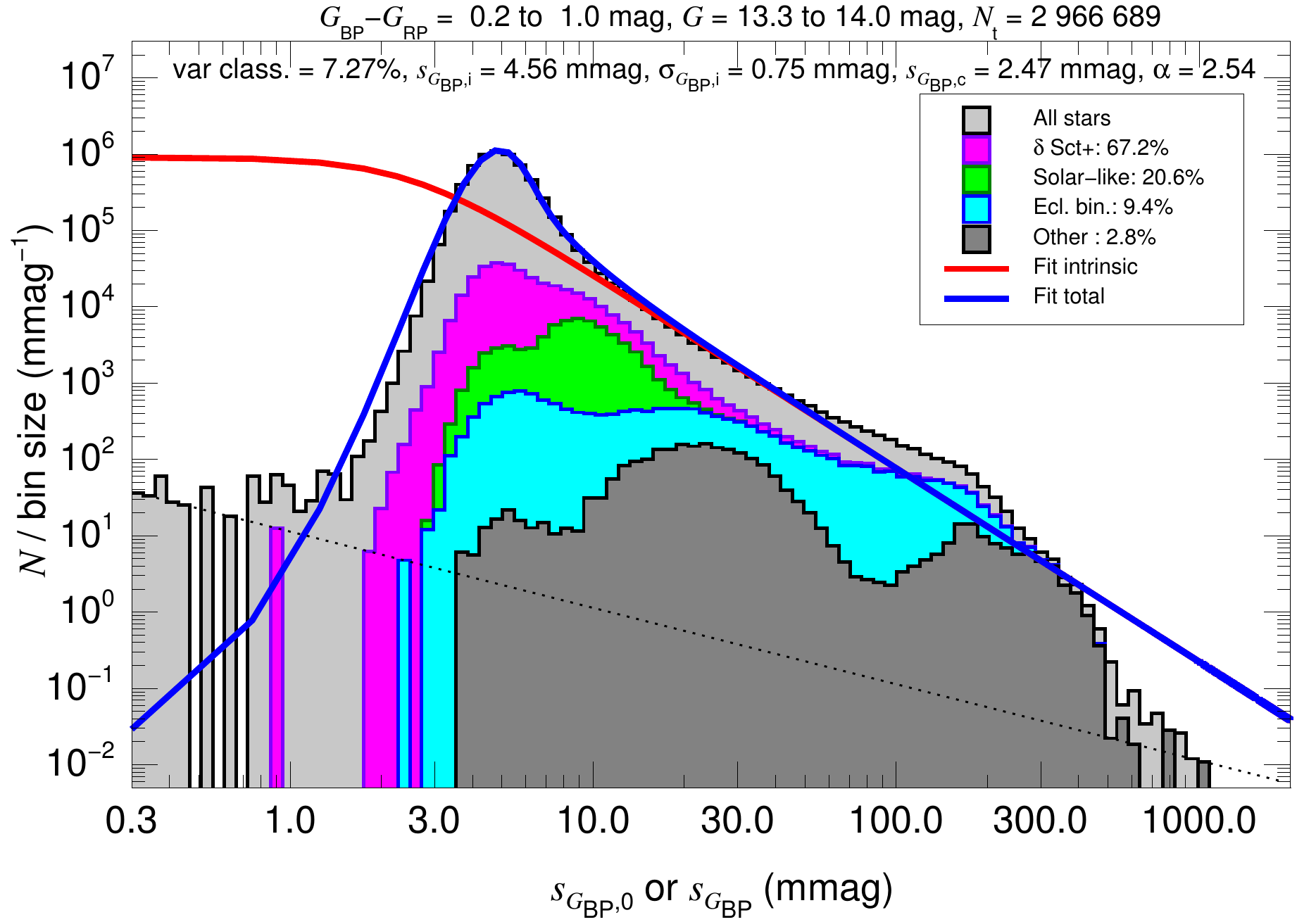}$\!\!\!$
                    \includegraphics[width=0.35\linewidth]{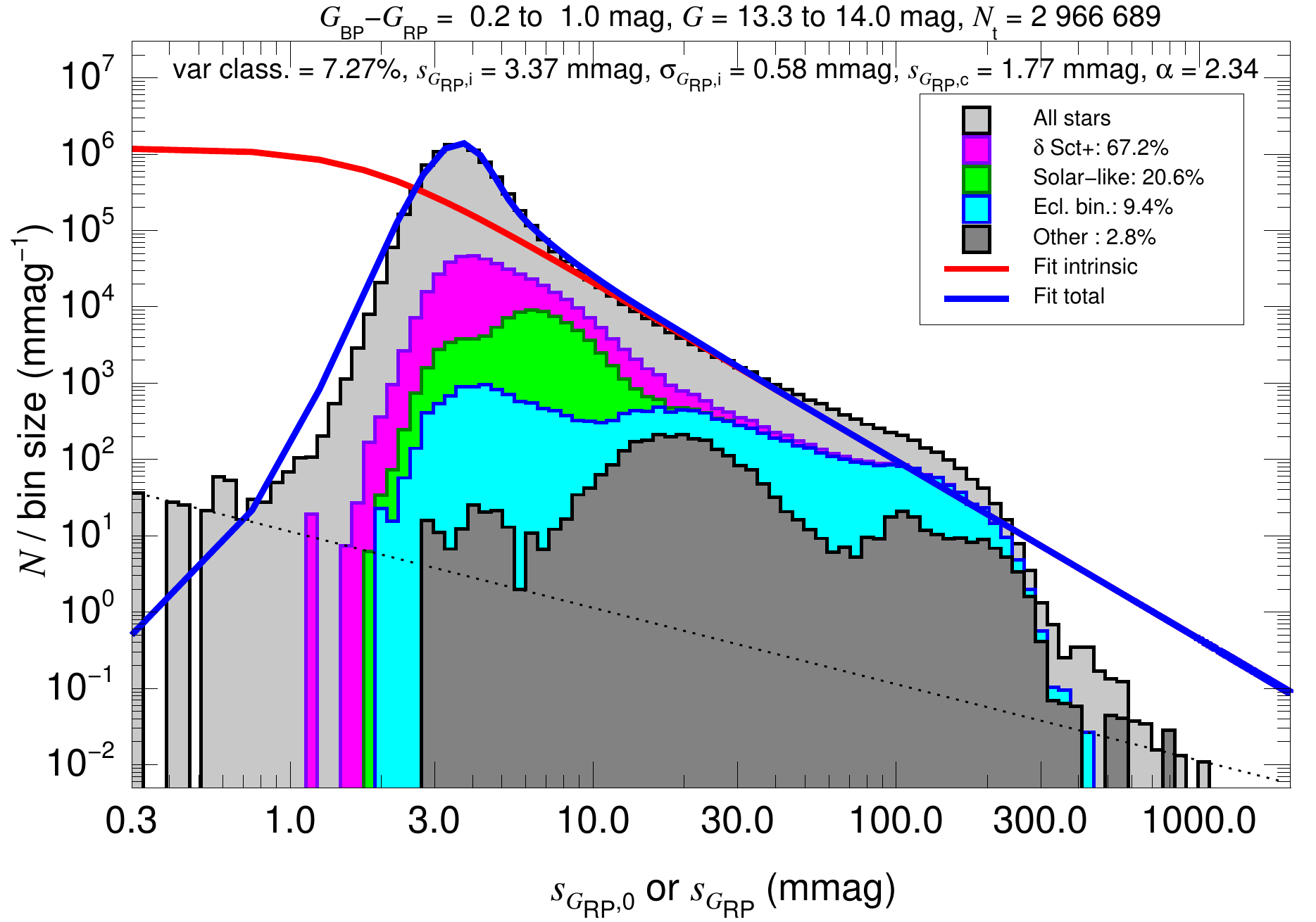}}
\centerline{$\!\!\!$\includegraphics[width=0.35\linewidth]{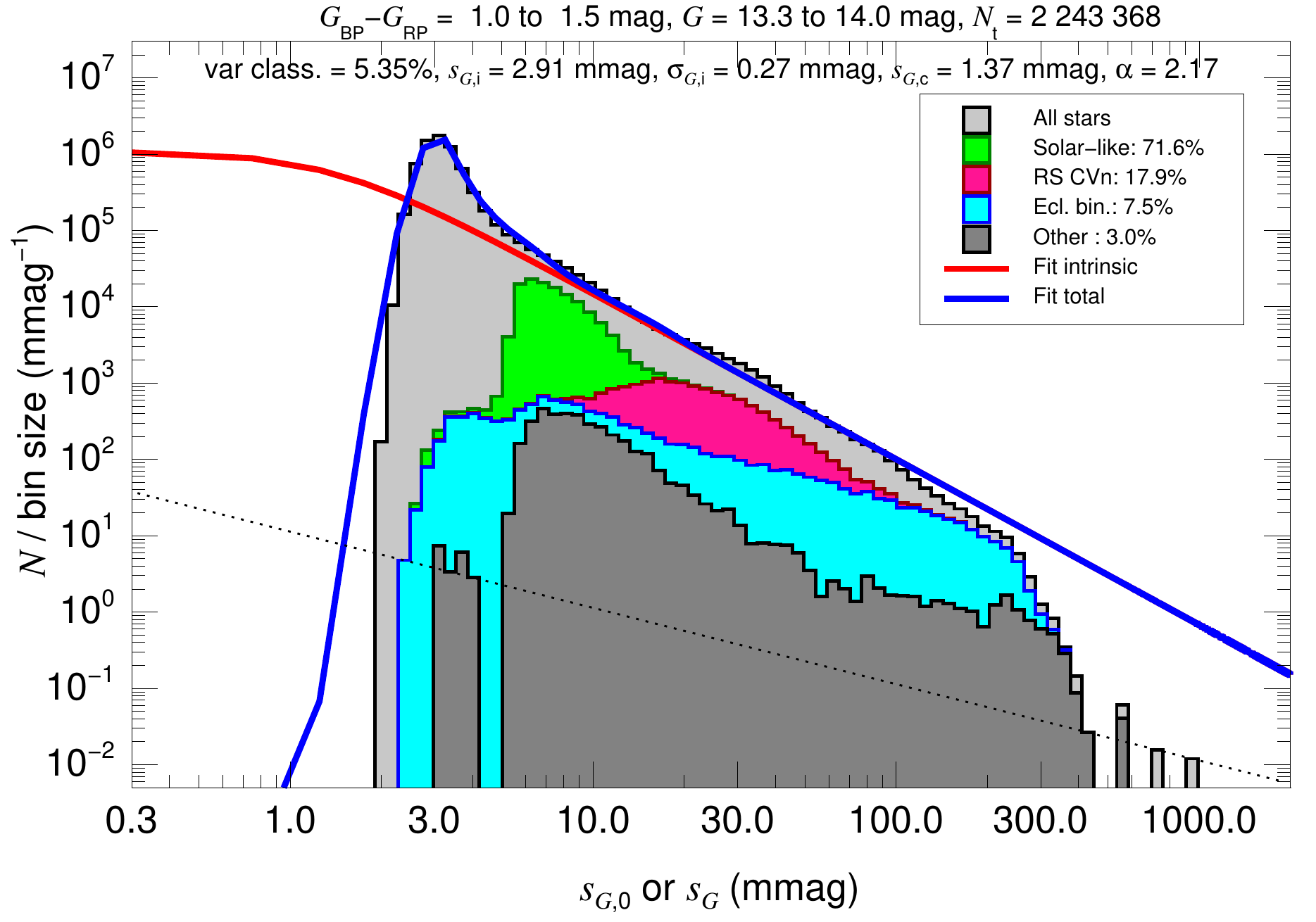}$\!\!\!$
                    \includegraphics[width=0.35\linewidth]{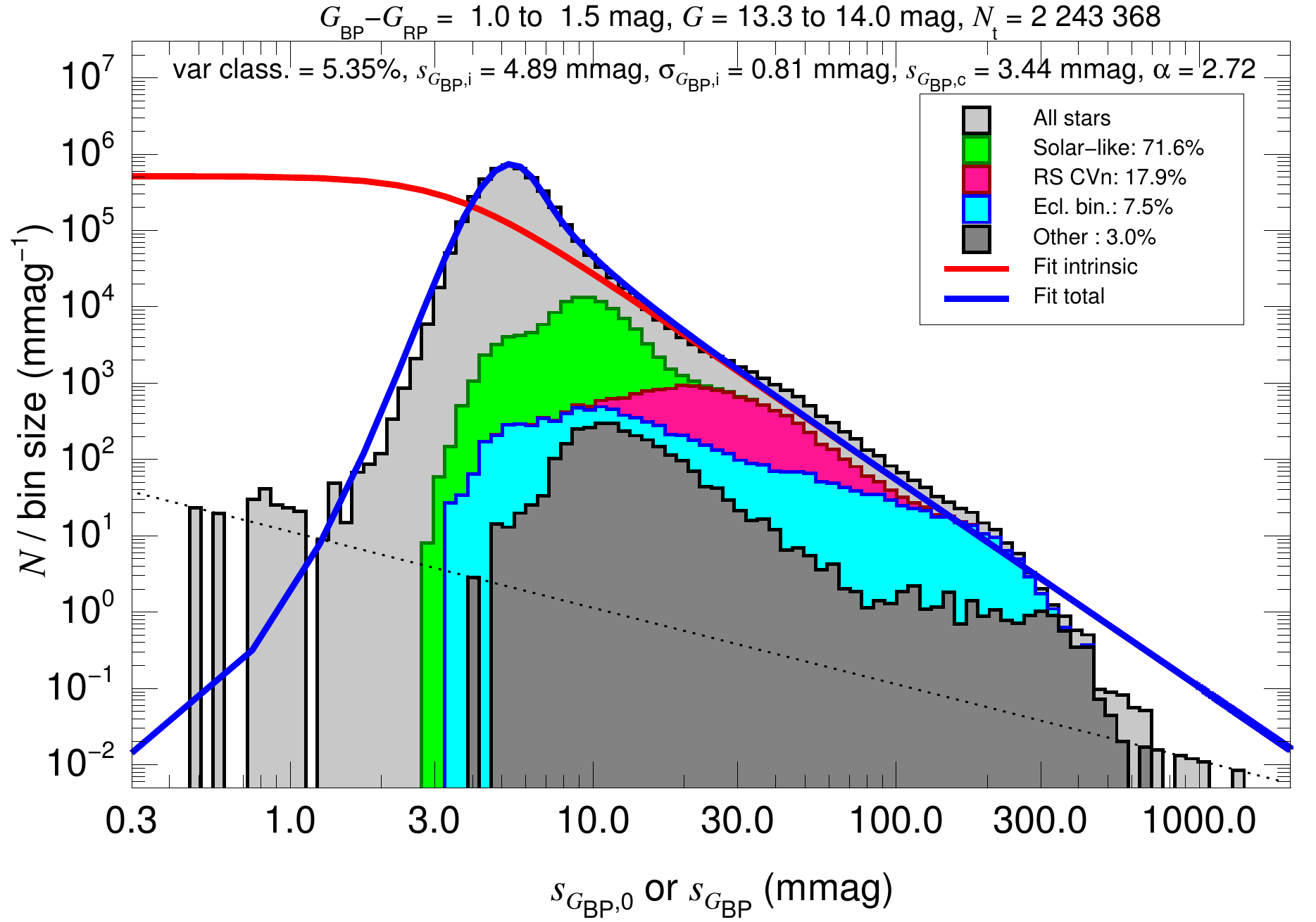}$\!\!\!$
                    \includegraphics[width=0.35\linewidth]{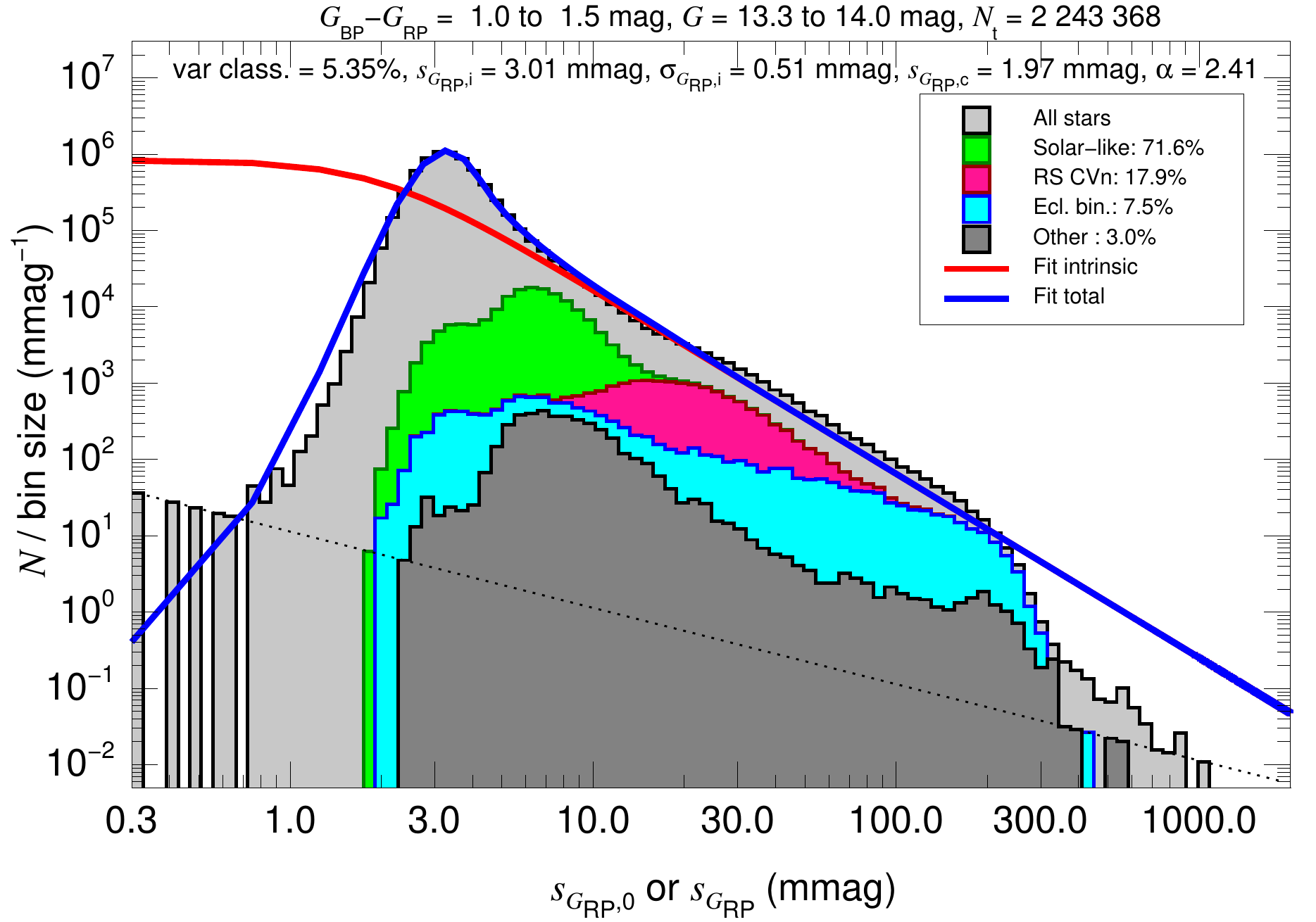}}
\centerline{$\!\!\!$\includegraphics[width=0.35\linewidth]{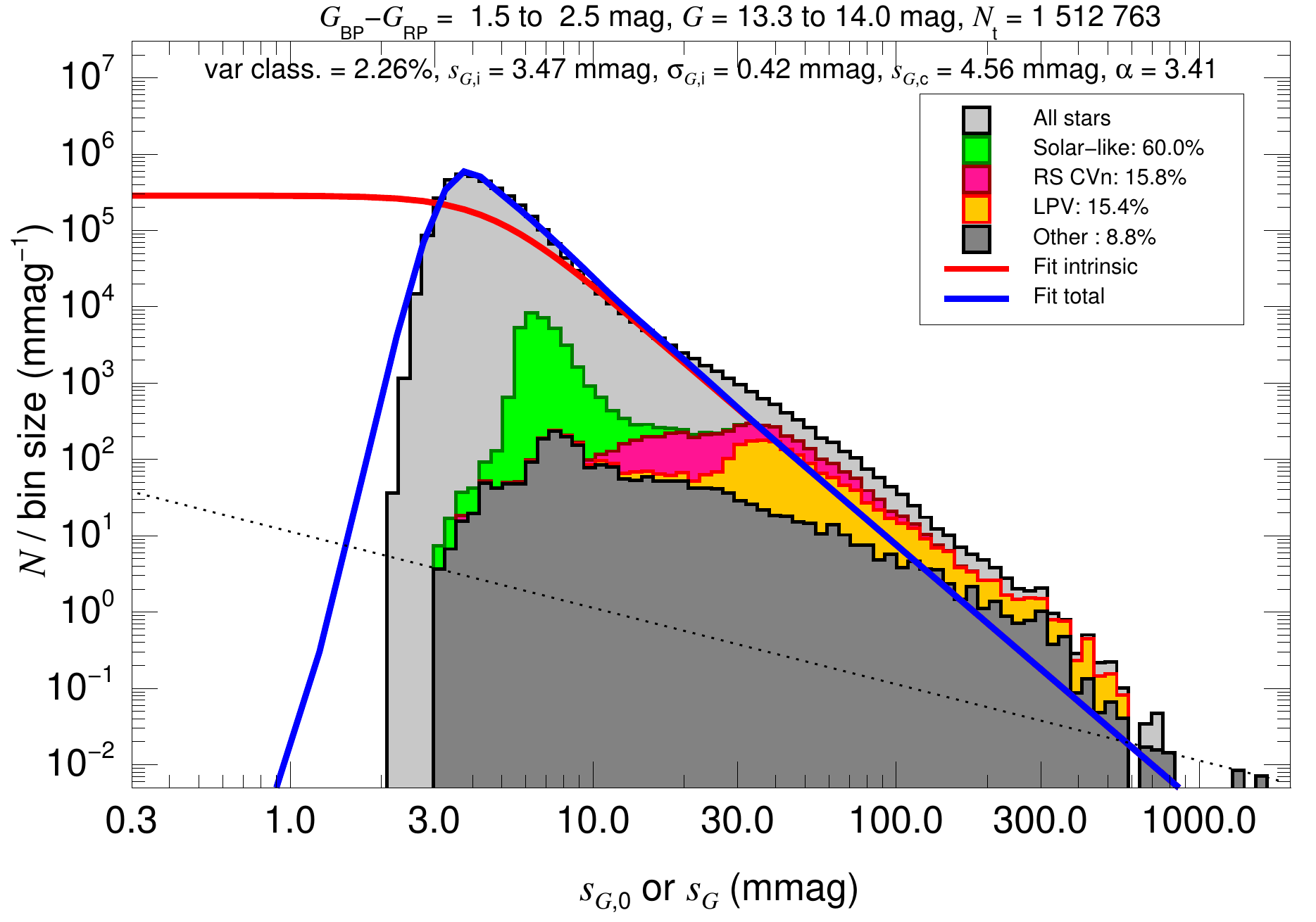}$\!\!\!$
                    \includegraphics[width=0.35\linewidth]{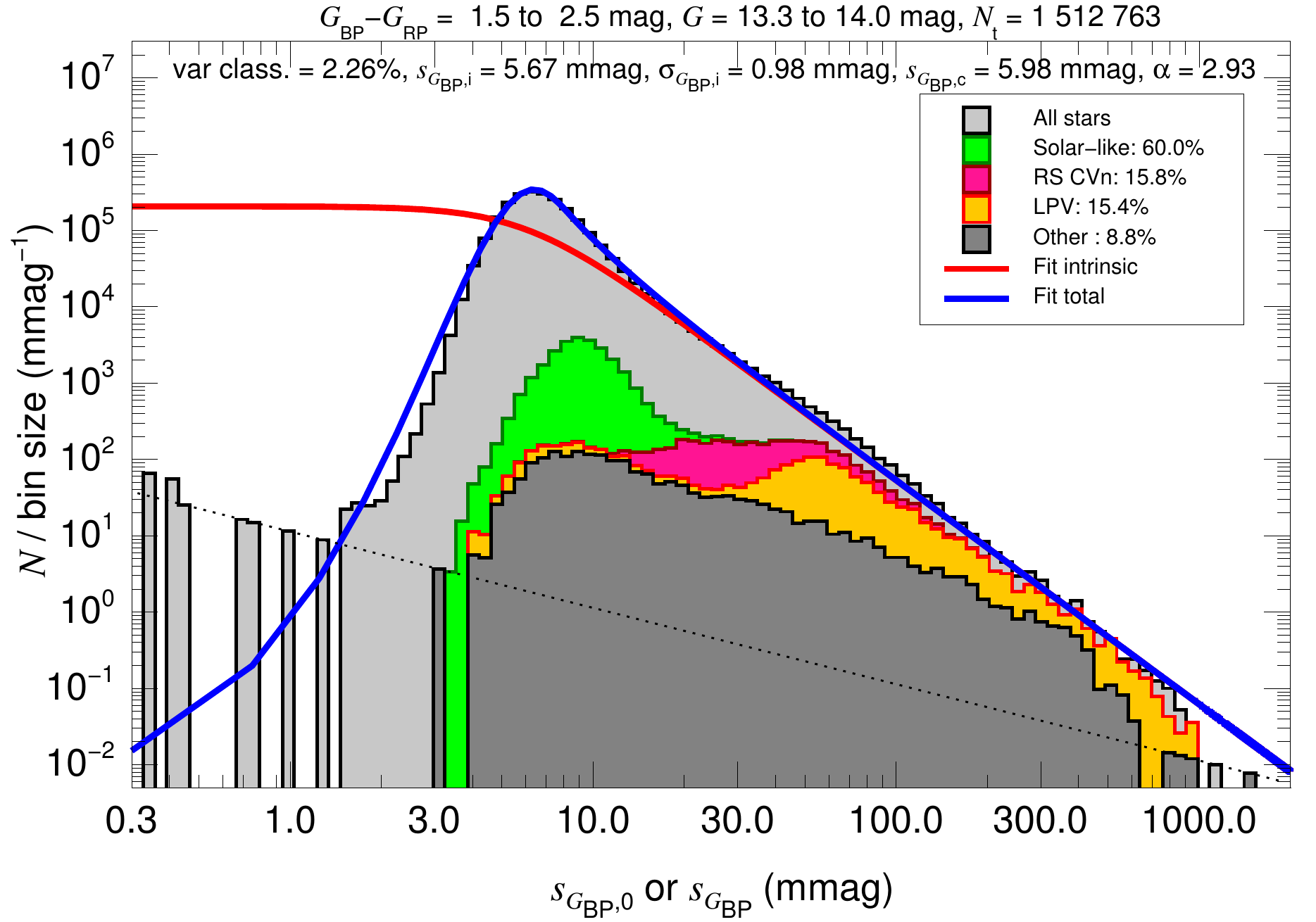}$\!\!\!$
                    \includegraphics[width=0.35\linewidth]{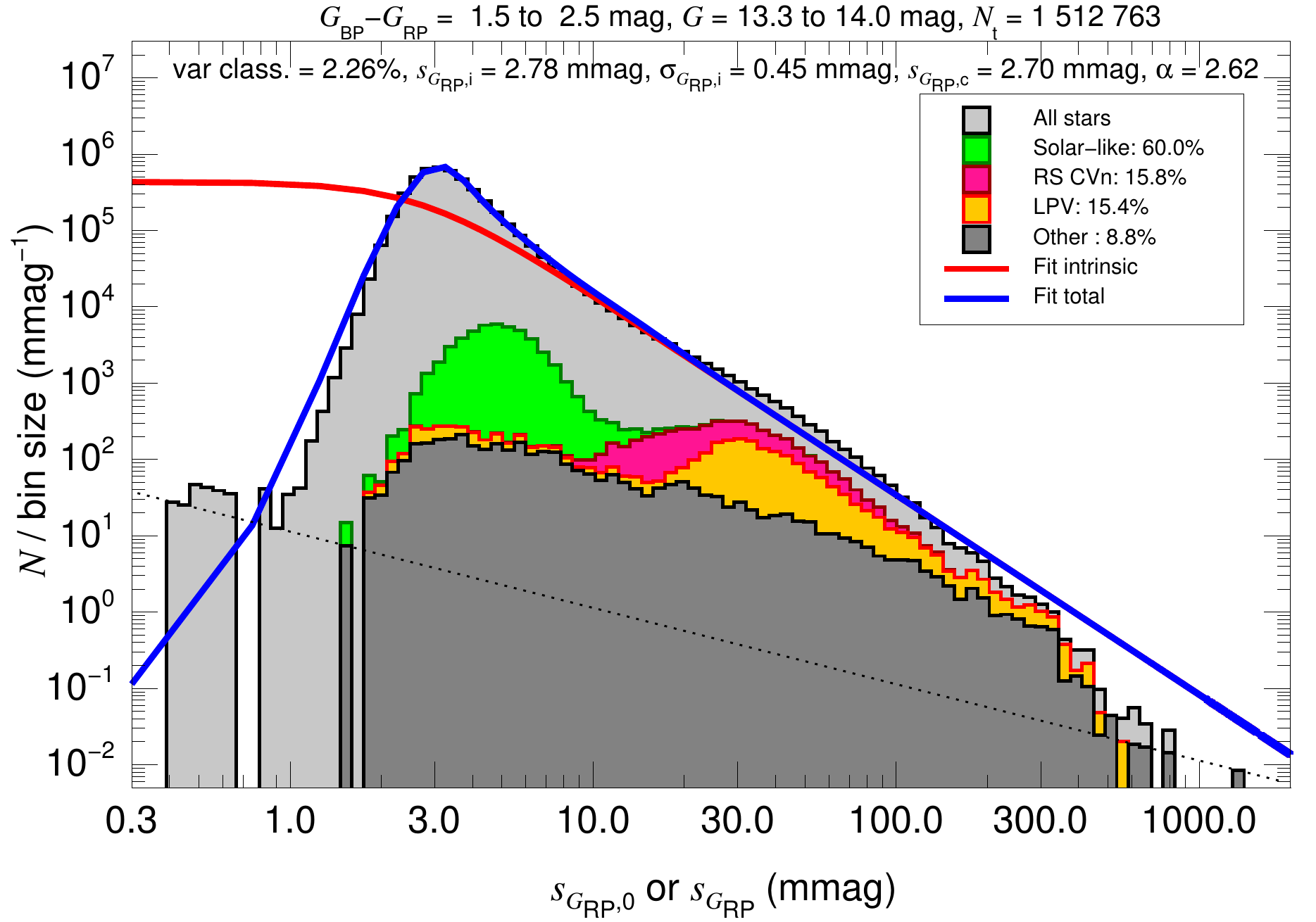}}
\centerline{$\!\!\!$\includegraphics[width=0.35\linewidth]{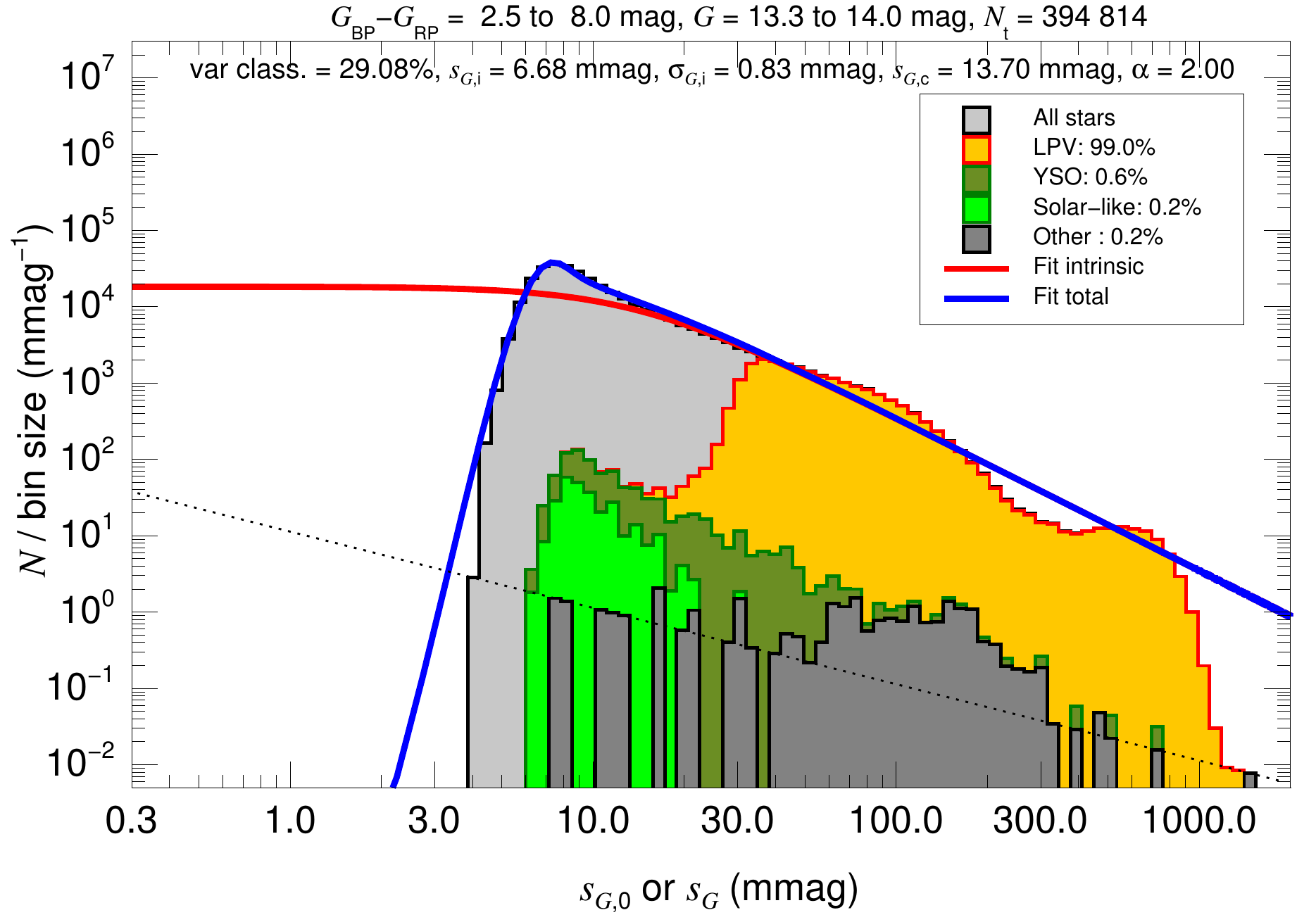}$\!\!\!$
                    \includegraphics[width=0.35\linewidth]{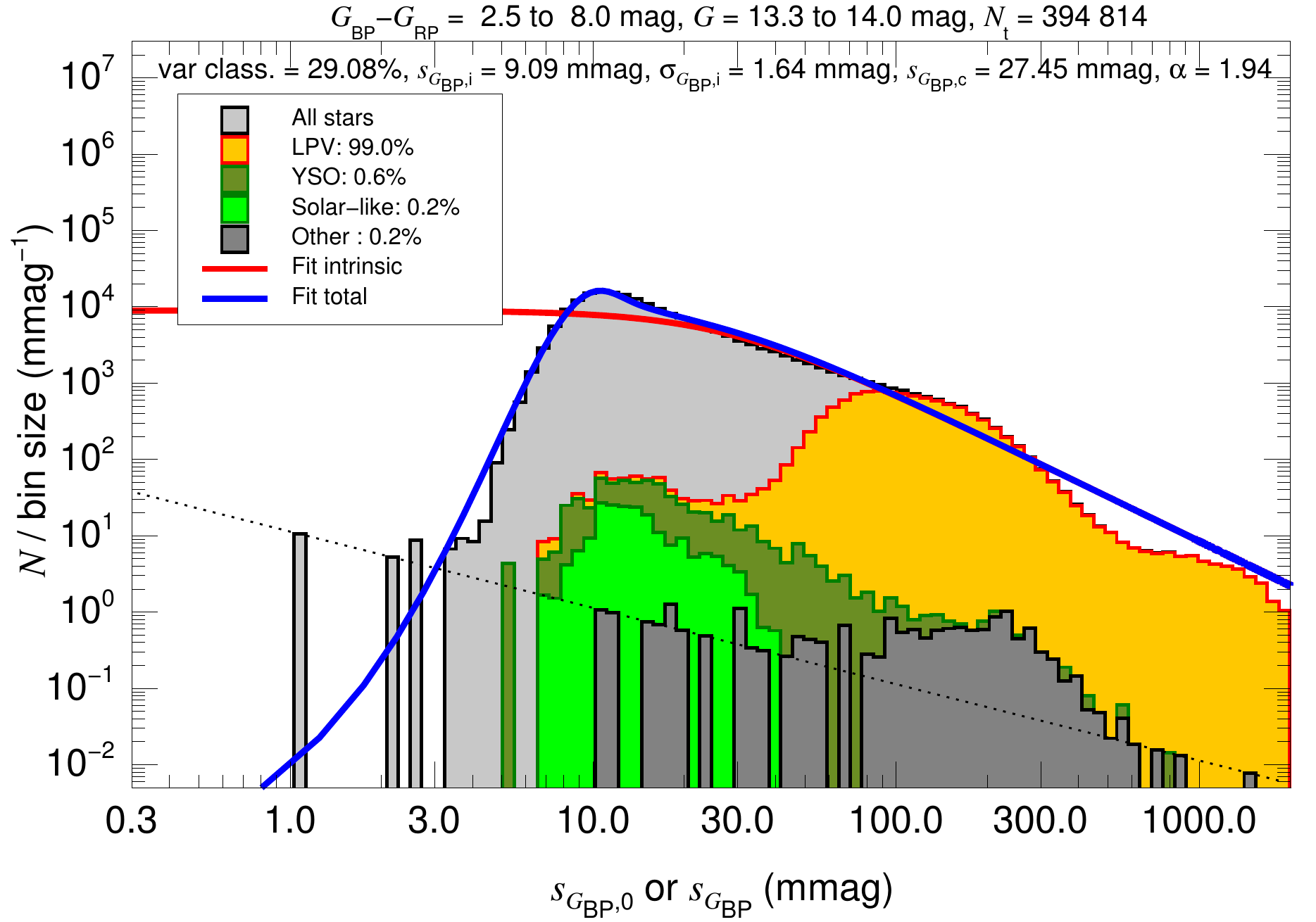}$\!\!\!$
                    \includegraphics[width=0.35\linewidth]{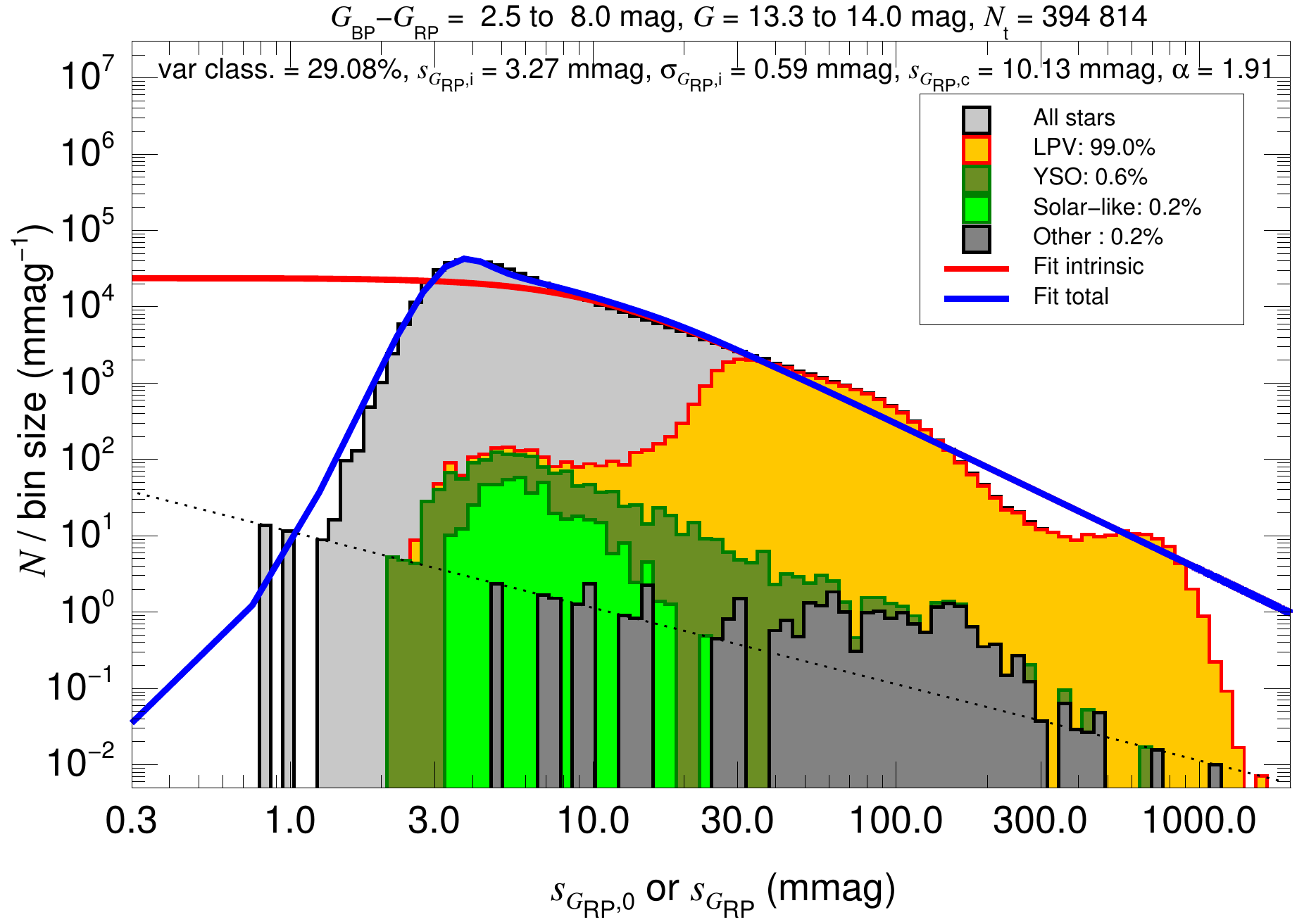}}
\caption{(Continued).}
\end{figure*}

\addtocounter{figure}{-1}

\begin{figure*}
\centerline{$\!\!\!$\includegraphics[width=0.35\linewidth]{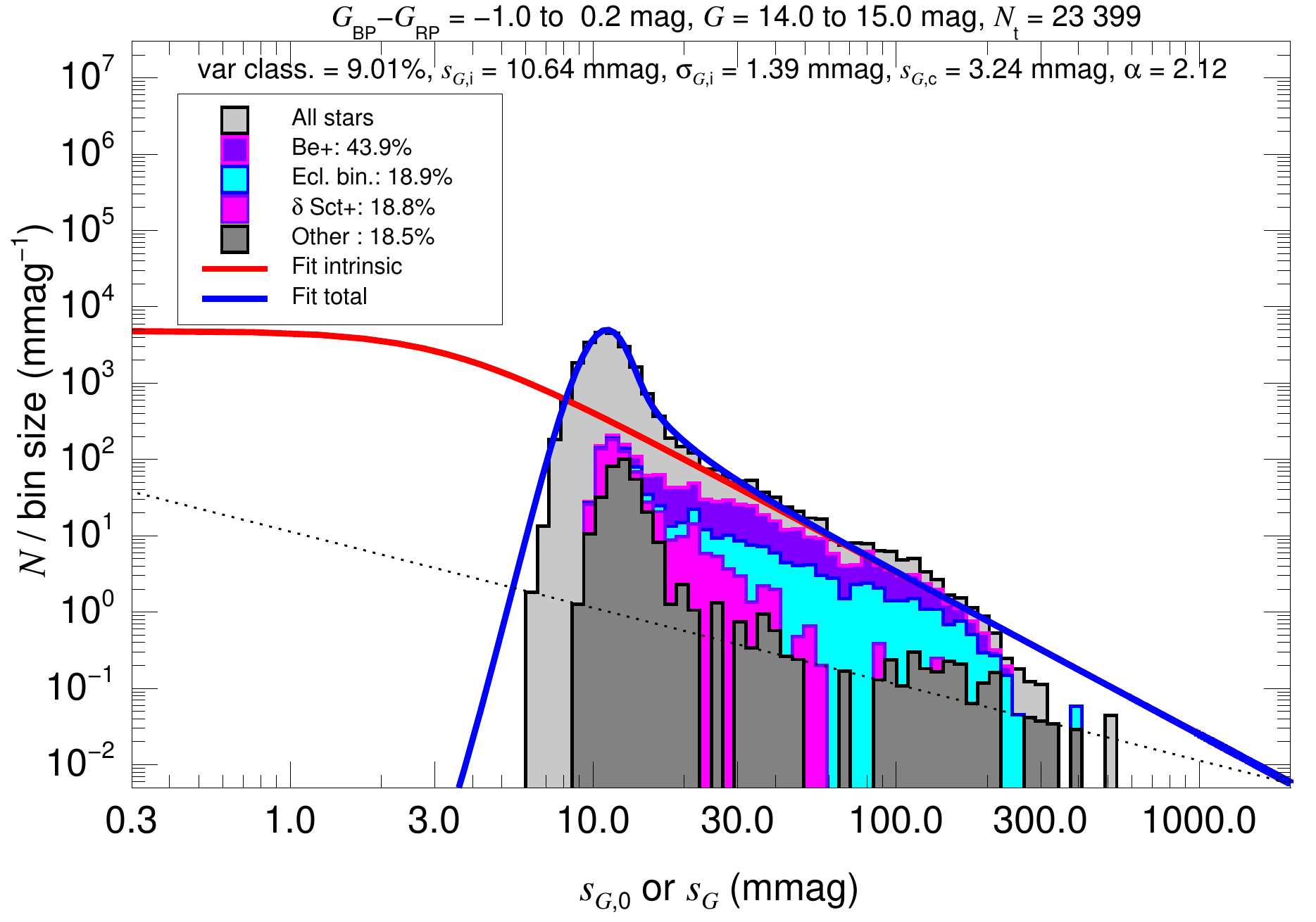}$\!\!\!$
                    \includegraphics[width=0.35\linewidth]{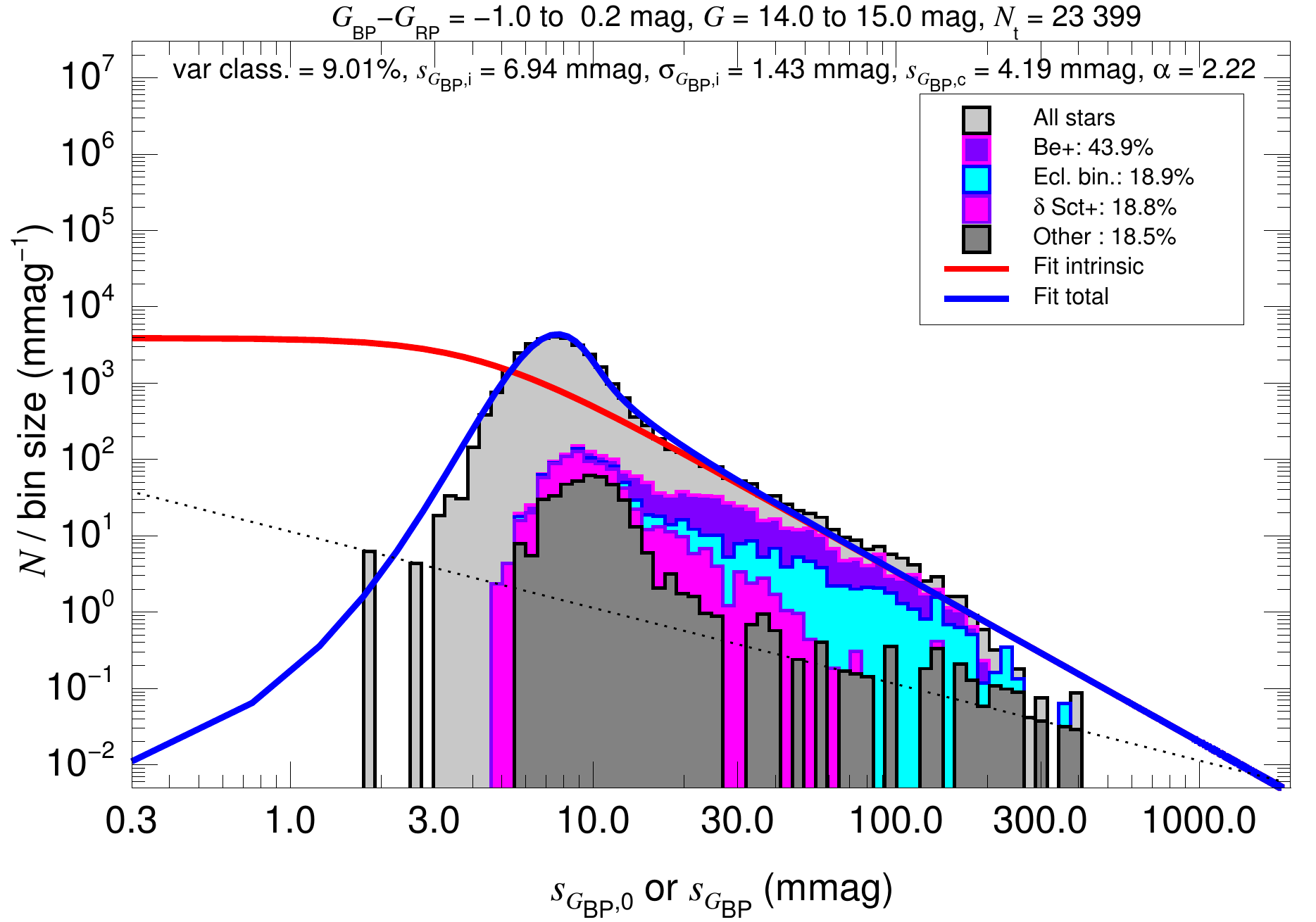}$\!\!\!$
                    \includegraphics[width=0.35\linewidth]{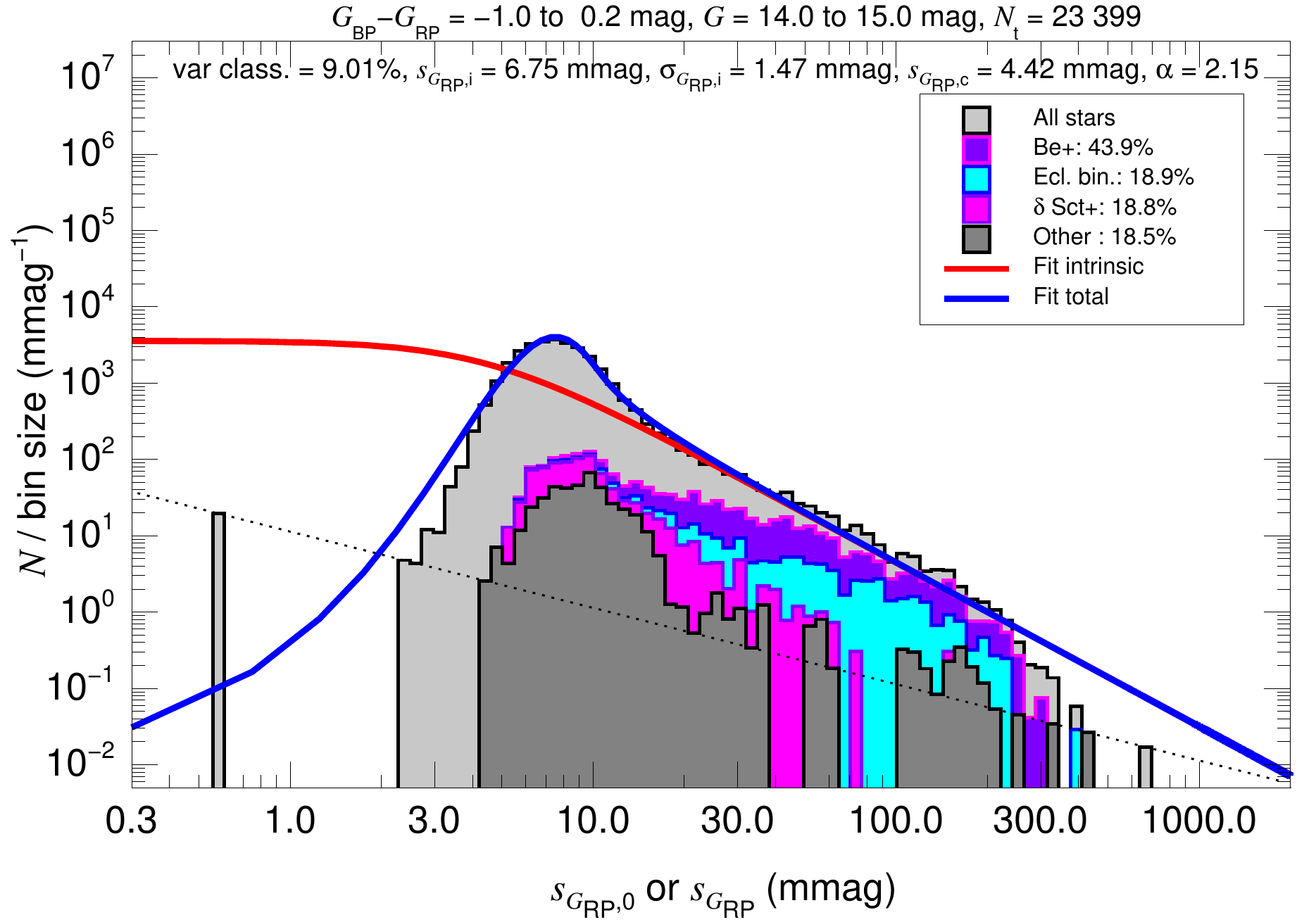}}
\centerline{$\!\!\!$\includegraphics[width=0.35\linewidth]{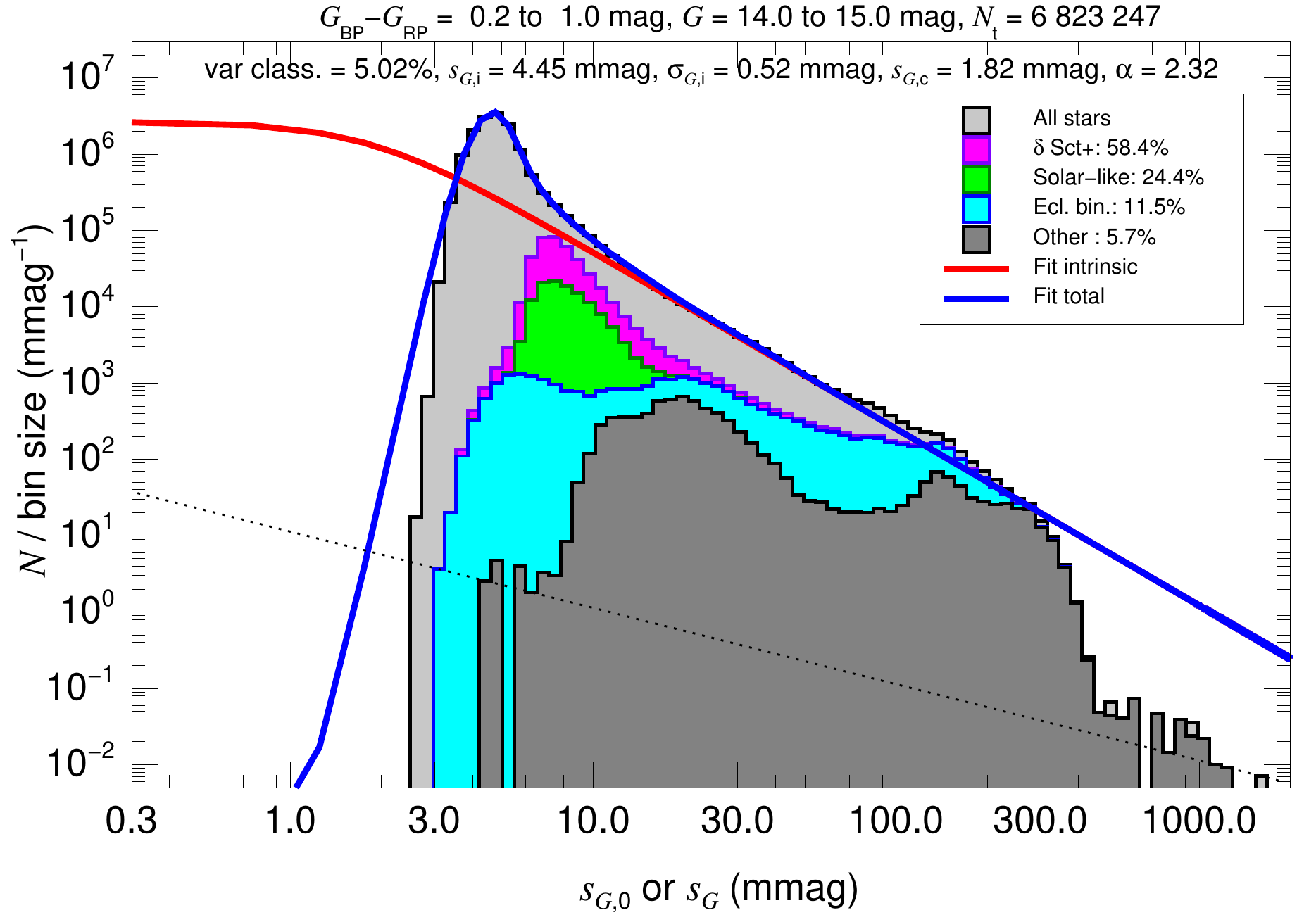}$\!\!\!$
                    \includegraphics[width=0.35\linewidth]{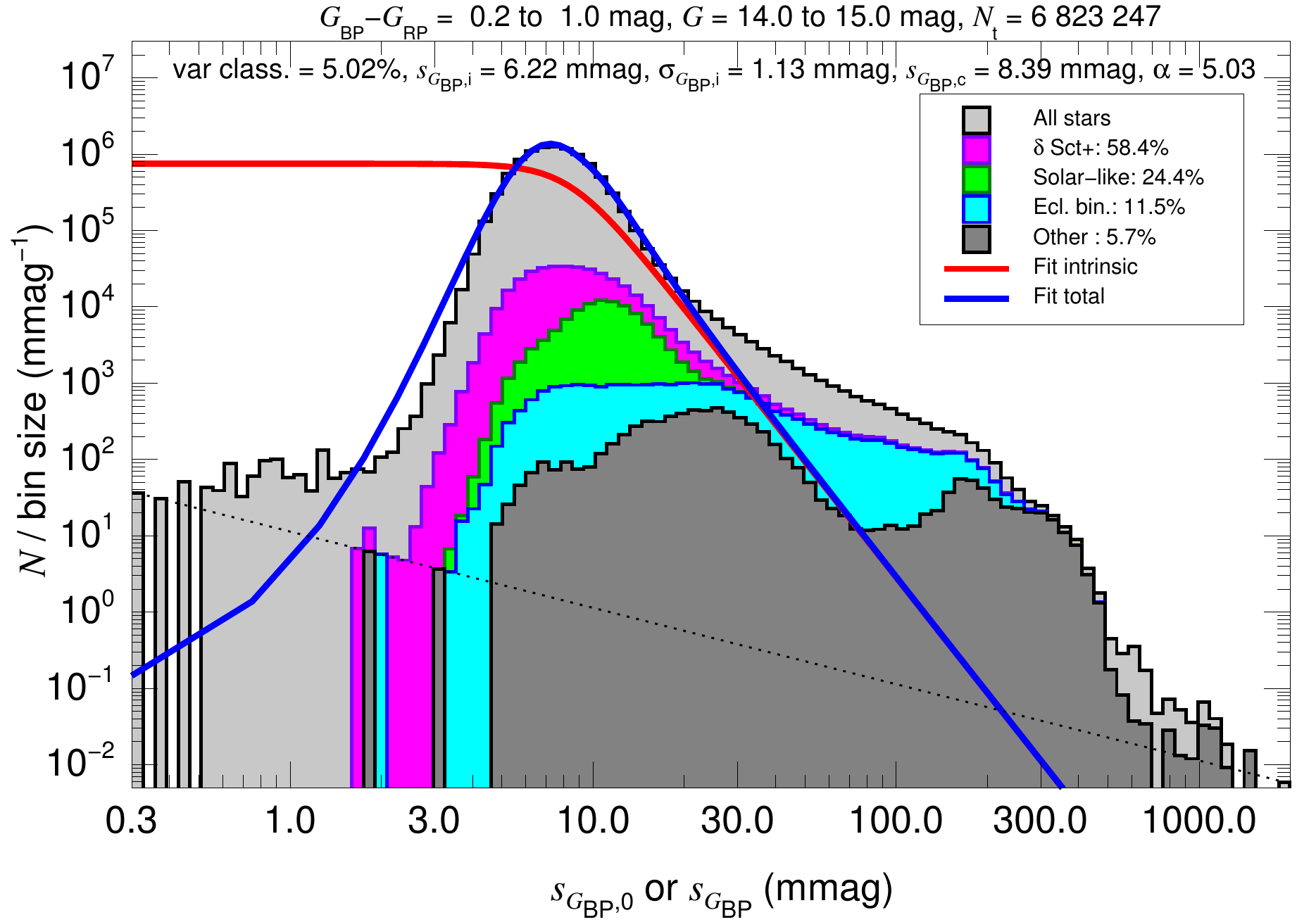}$\!\!\!$
                    \includegraphics[width=0.35\linewidth]{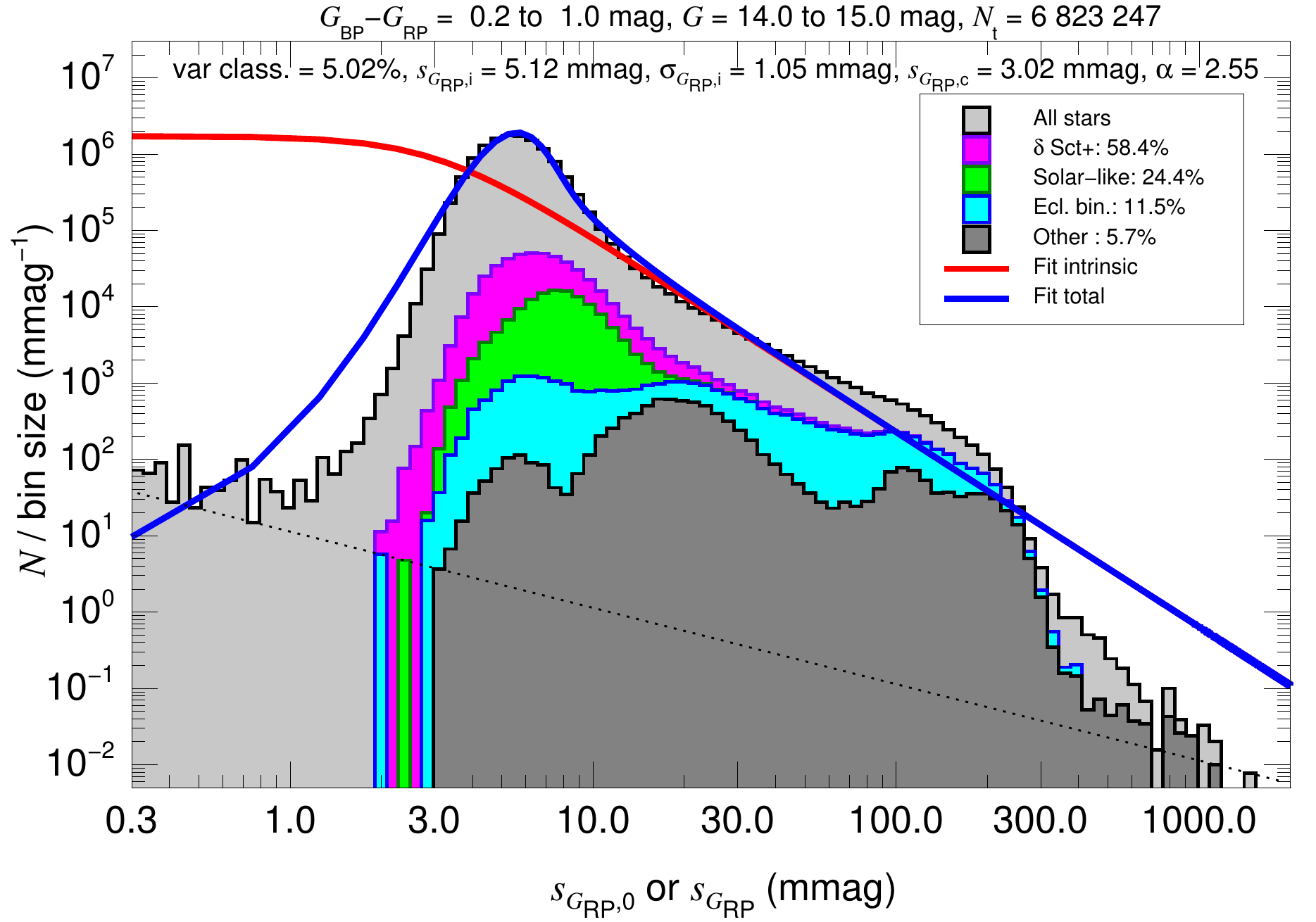}}
\centerline{$\!\!\!$\includegraphics[width=0.35\linewidth]{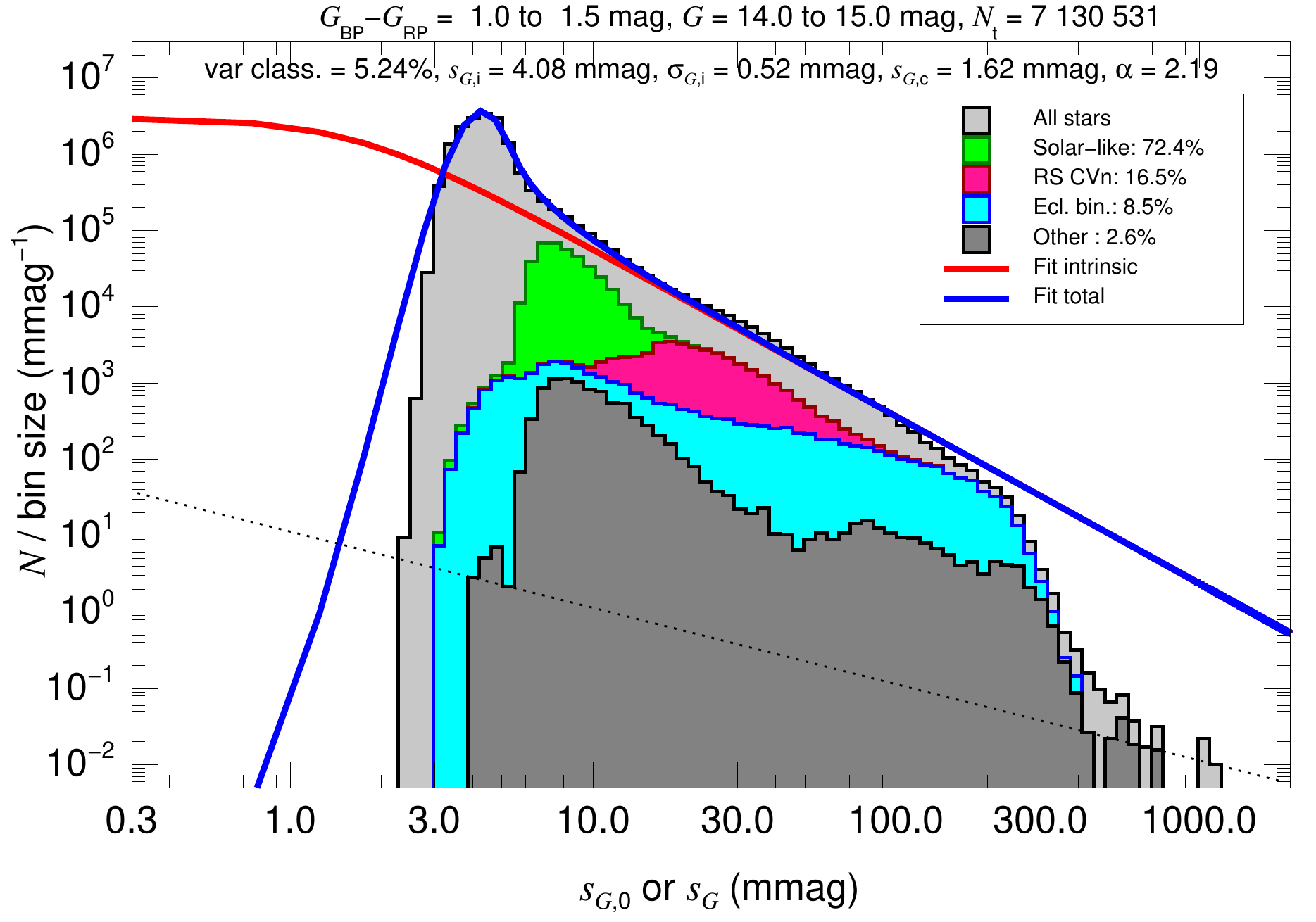}$\!\!\!$
                    \includegraphics[width=0.35\linewidth]{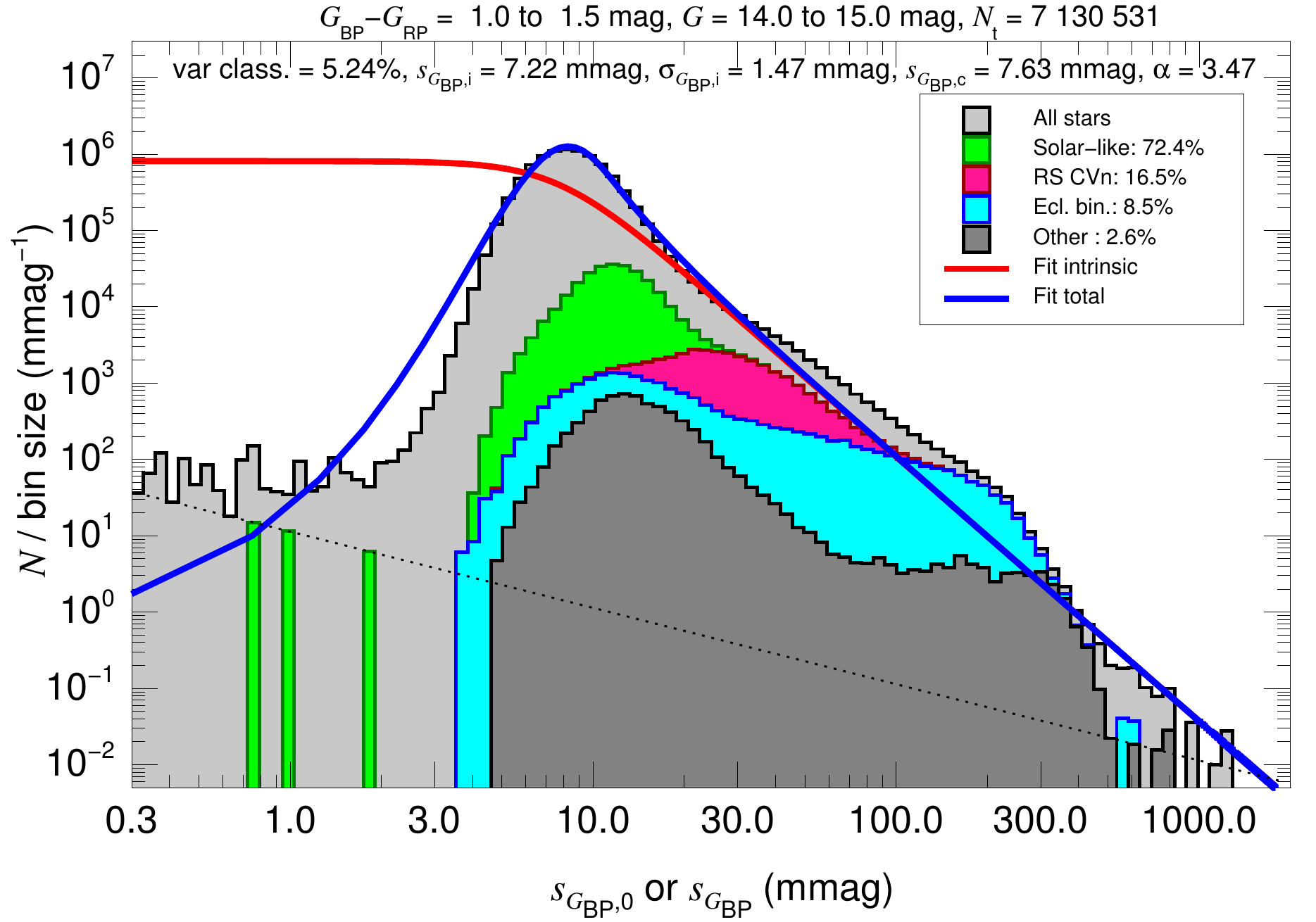}$\!\!\!$
                    \includegraphics[width=0.35\linewidth]{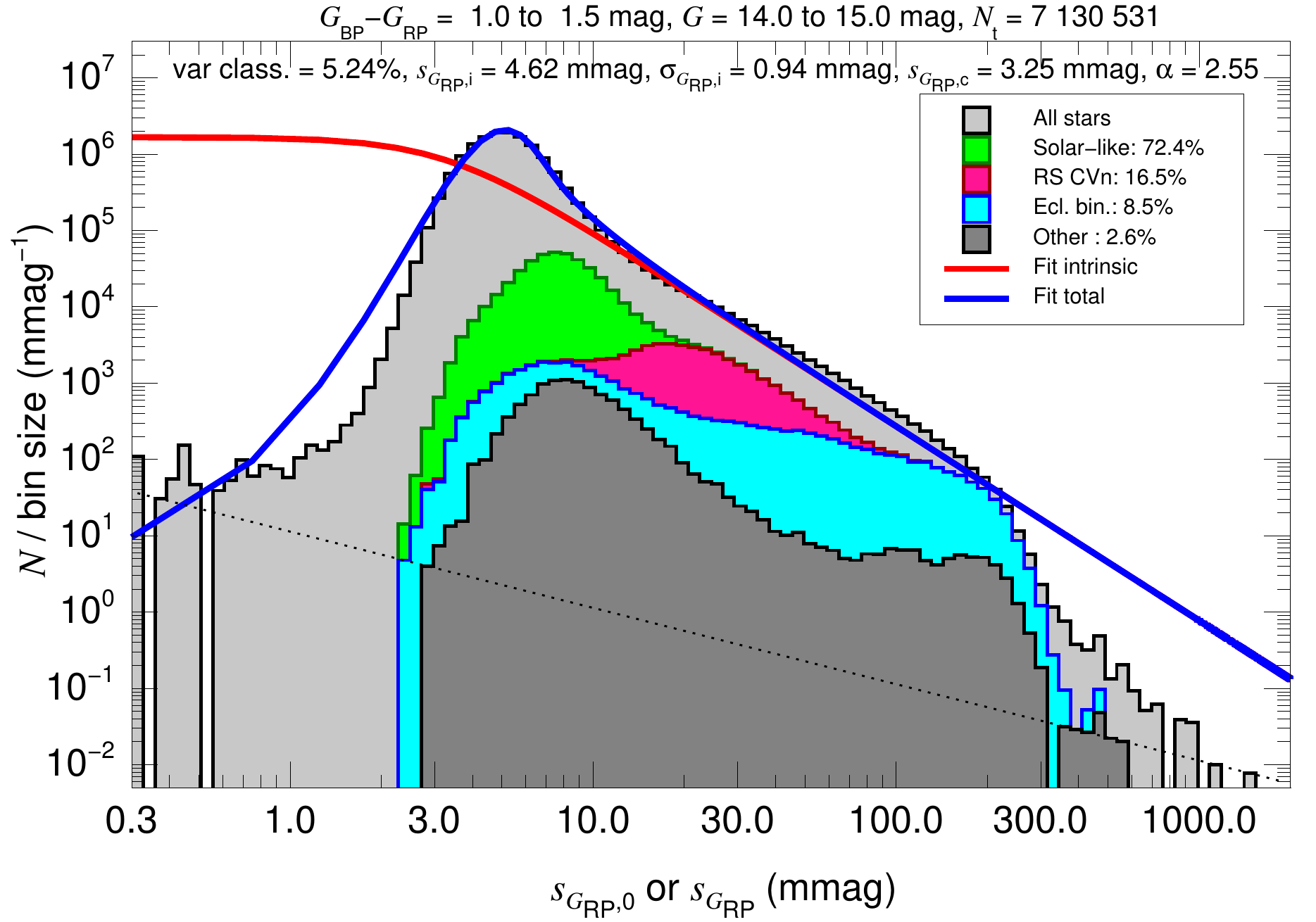}}
\centerline{$\!\!\!$\includegraphics[width=0.35\linewidth]{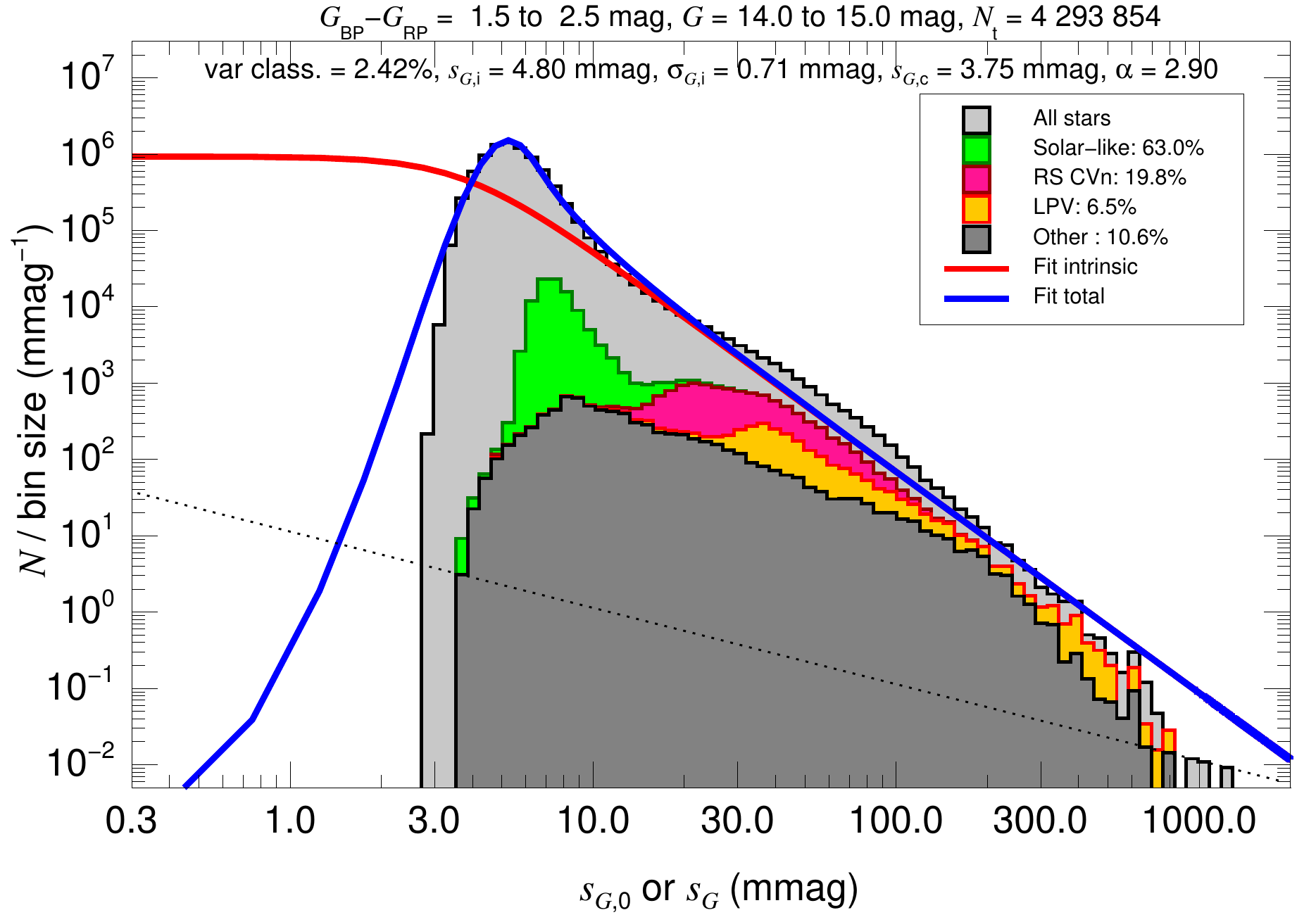}$\!\!\!$
                    \includegraphics[width=0.35\linewidth]{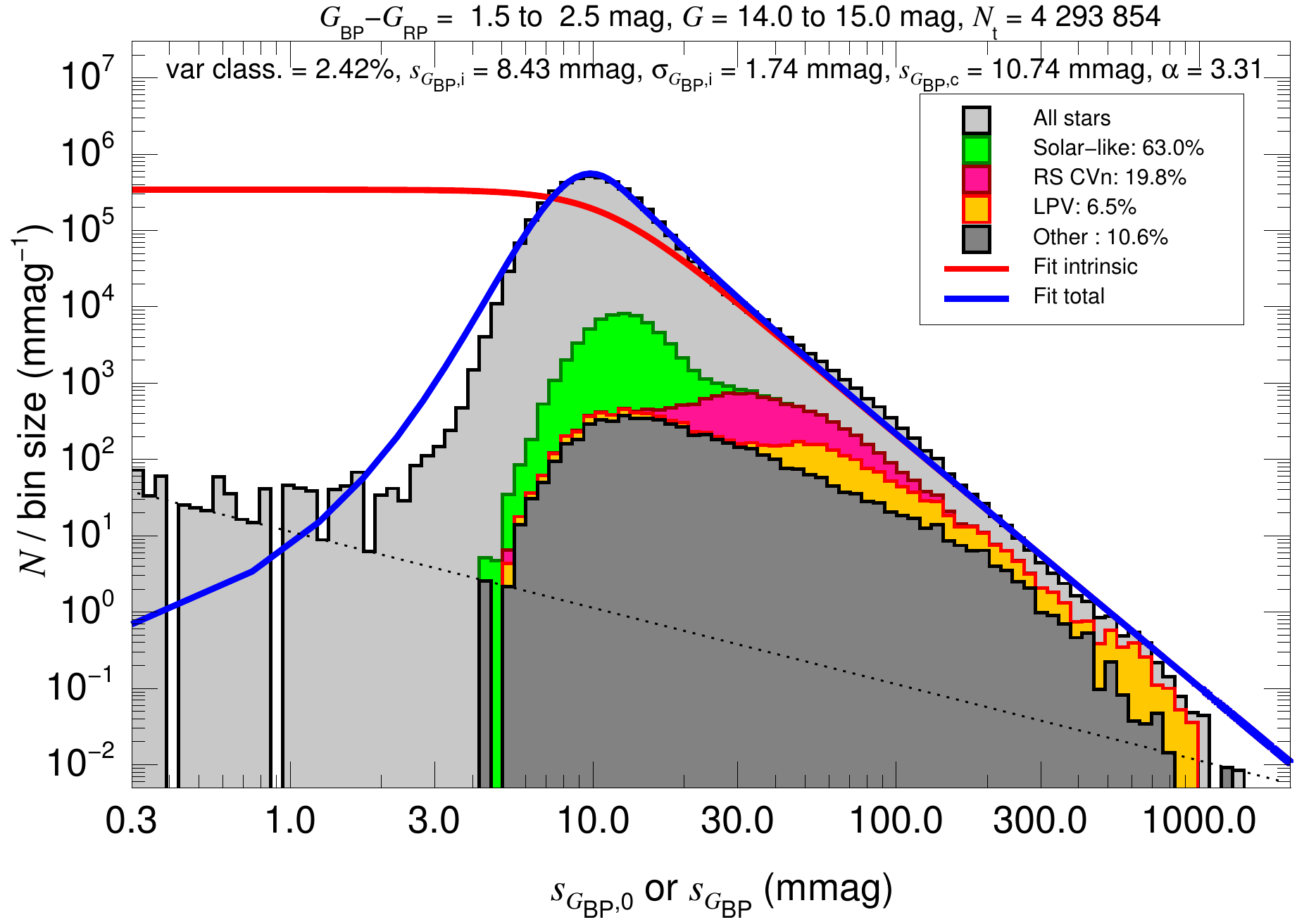}$\!\!\!$
                    \includegraphics[width=0.35\linewidth]{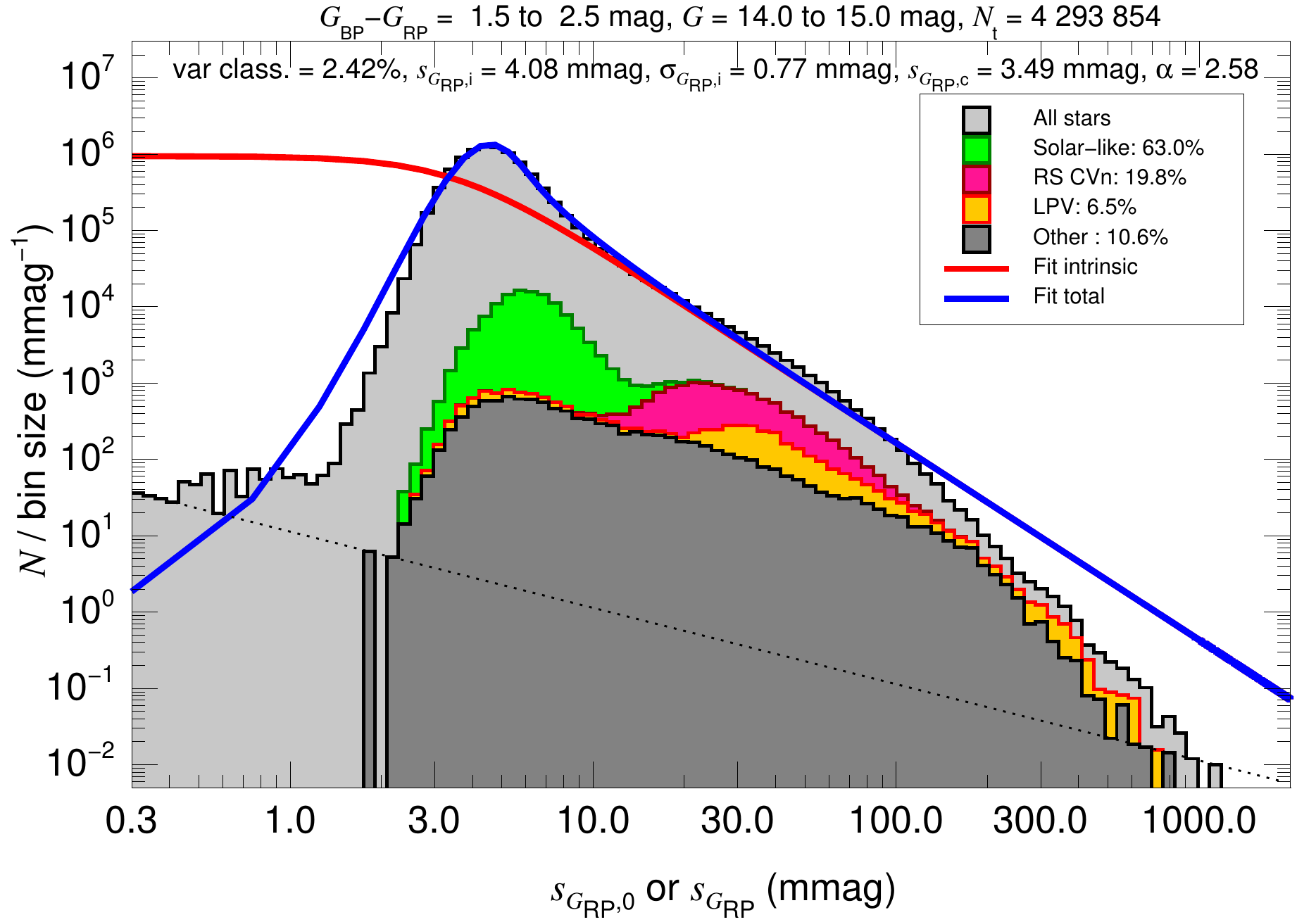}}
\centerline{$\!\!\!$\includegraphics[width=0.35\linewidth]{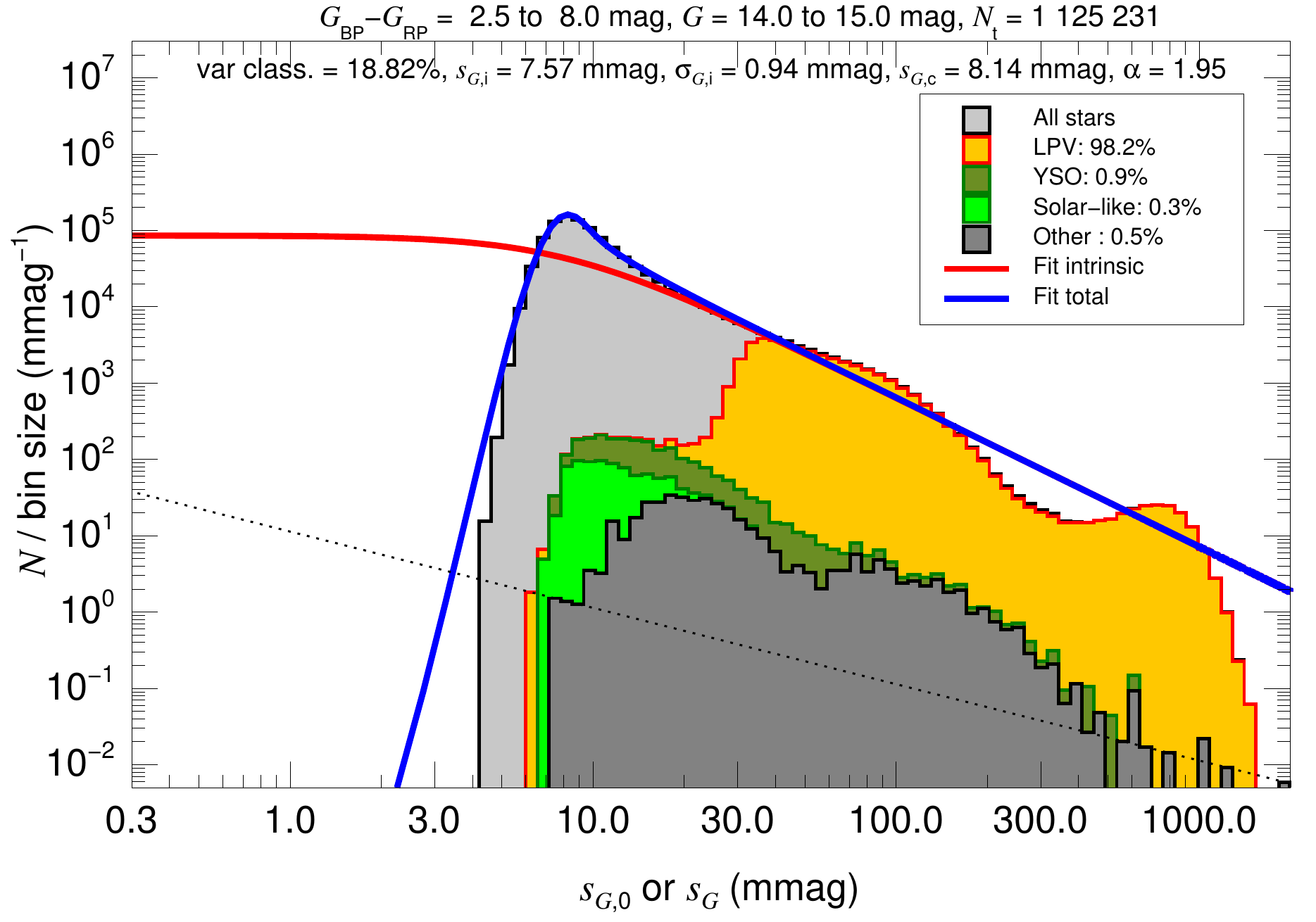}$\!\!\!$
                    \includegraphics[width=0.35\linewidth]{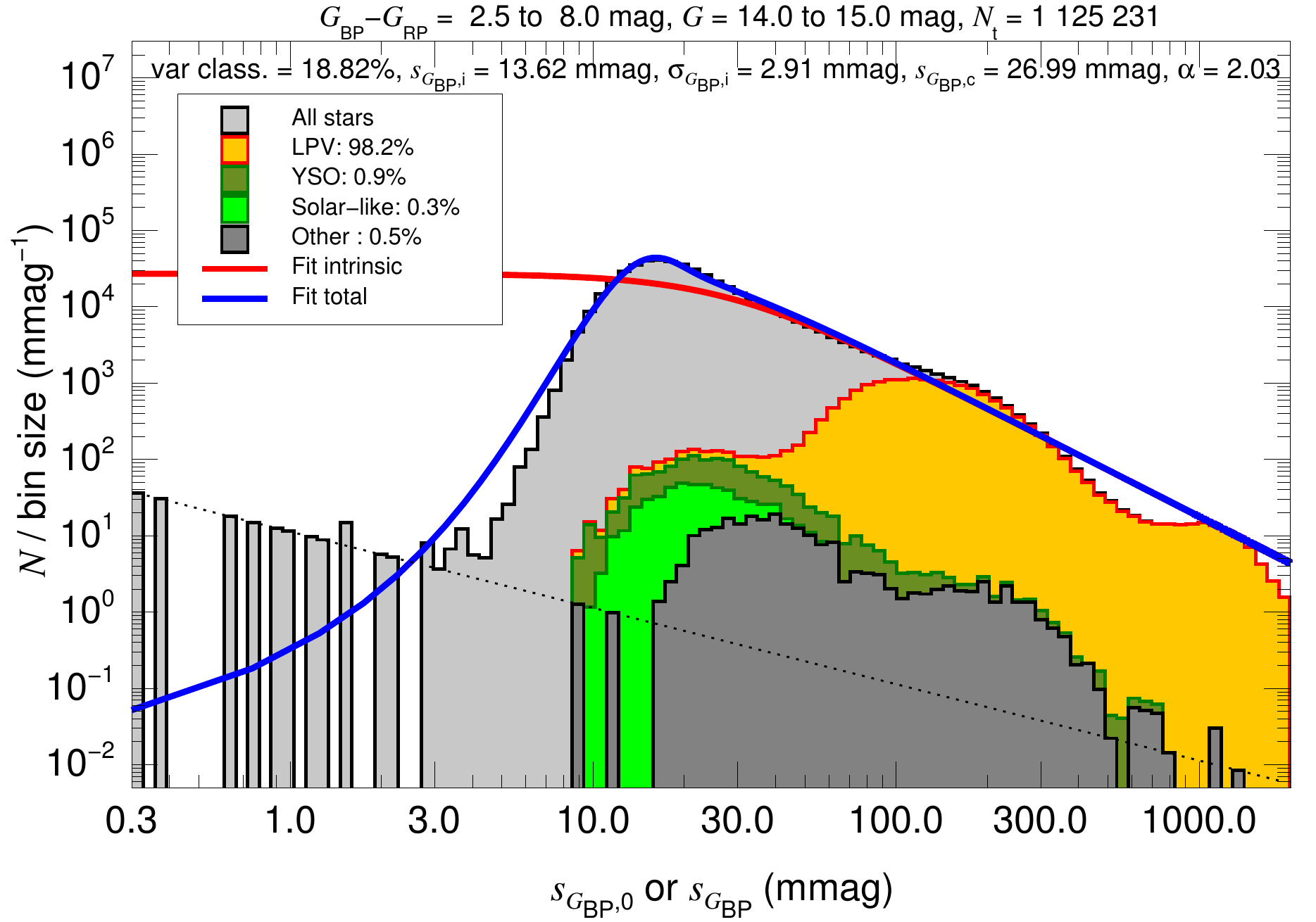}$\!\!\!$
                    \includegraphics[width=0.35\linewidth]{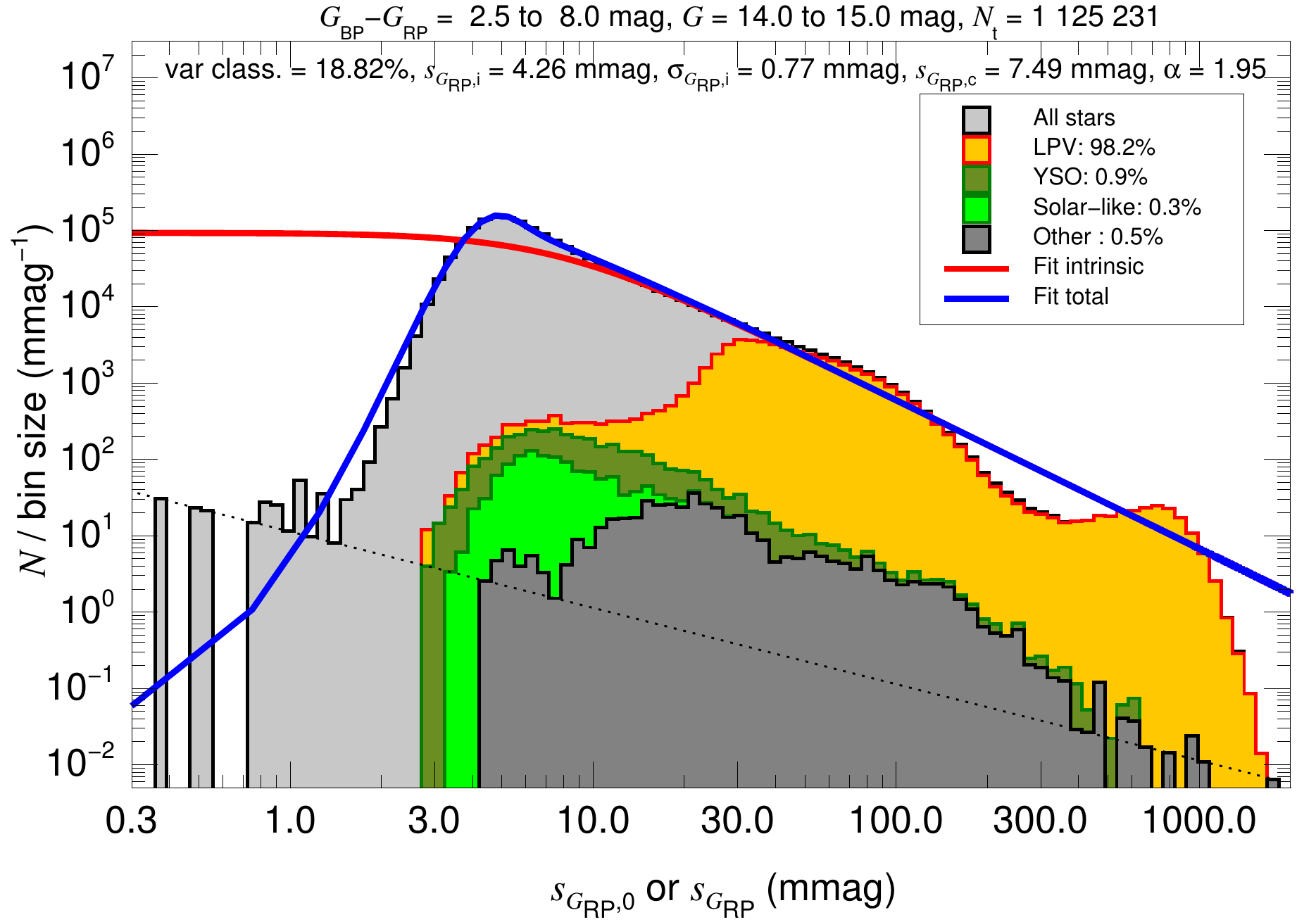}}
\caption{(Continued).}
\end{figure*}

\addtocounter{figure}{-1}

\begin{figure*}
\centerline{$\!\!\!$\includegraphics[width=0.35\linewidth]{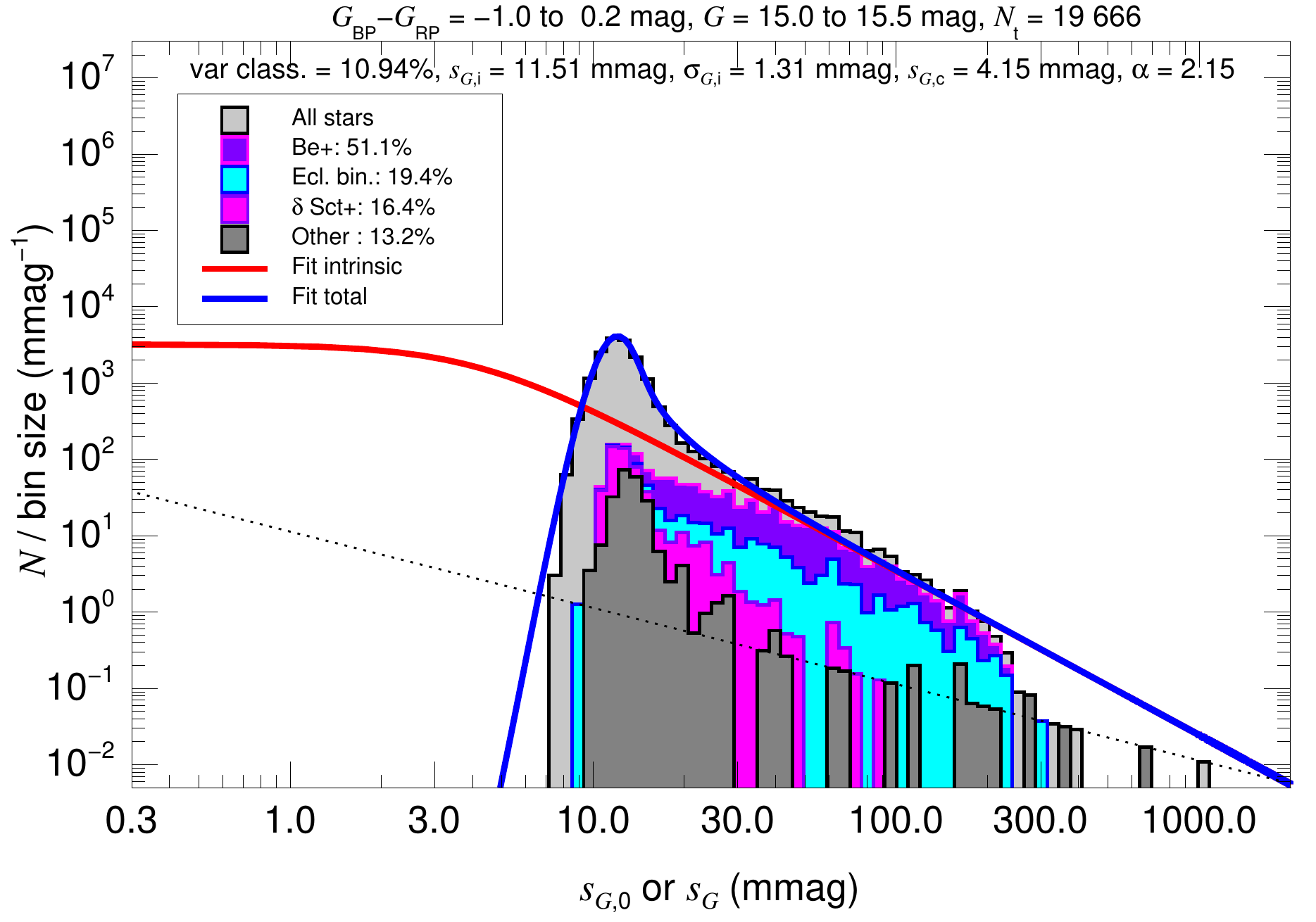}$\!\!\!$
                    \includegraphics[width=0.35\linewidth]{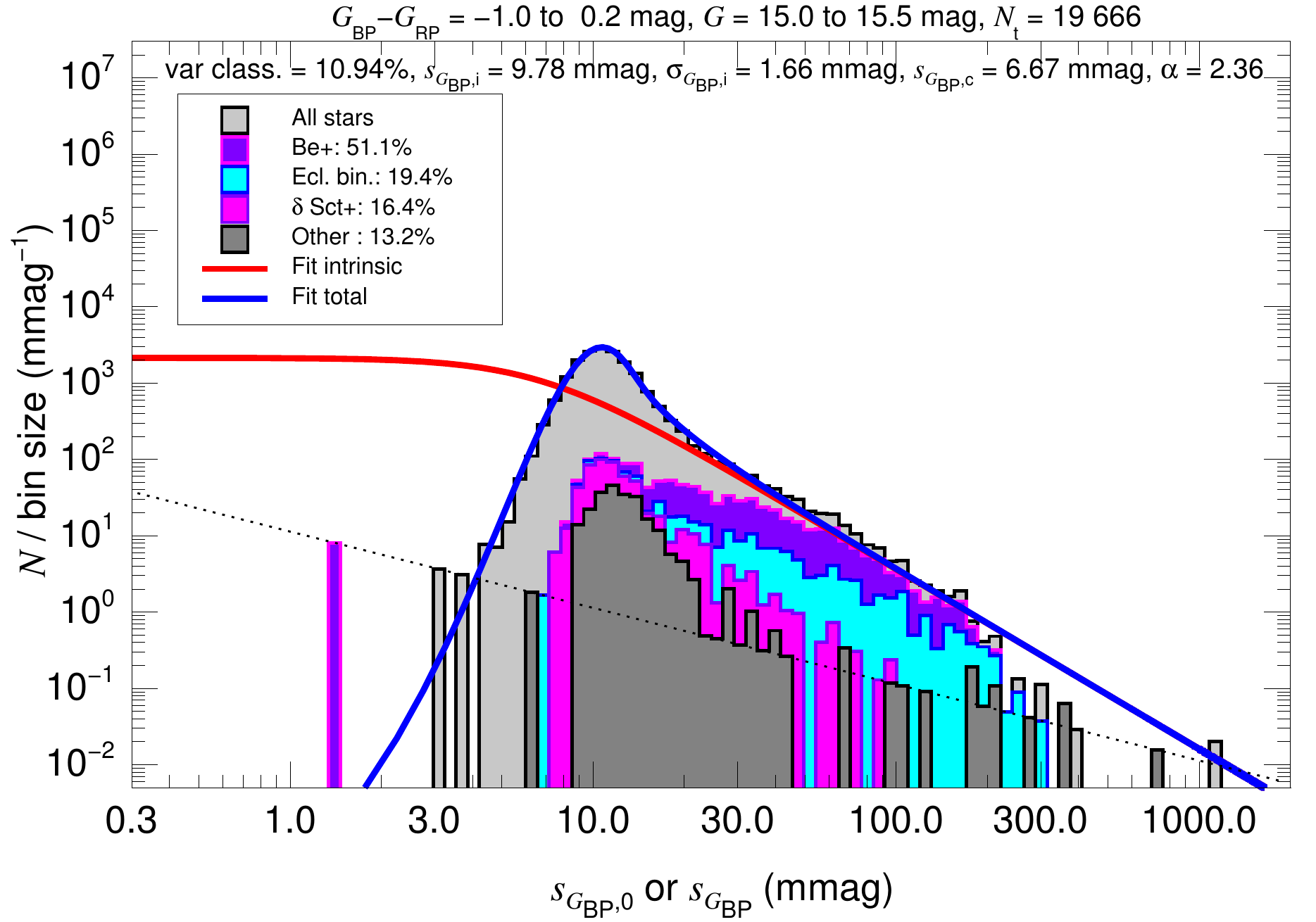}$\!\!\!$
                    \includegraphics[width=0.35\linewidth]{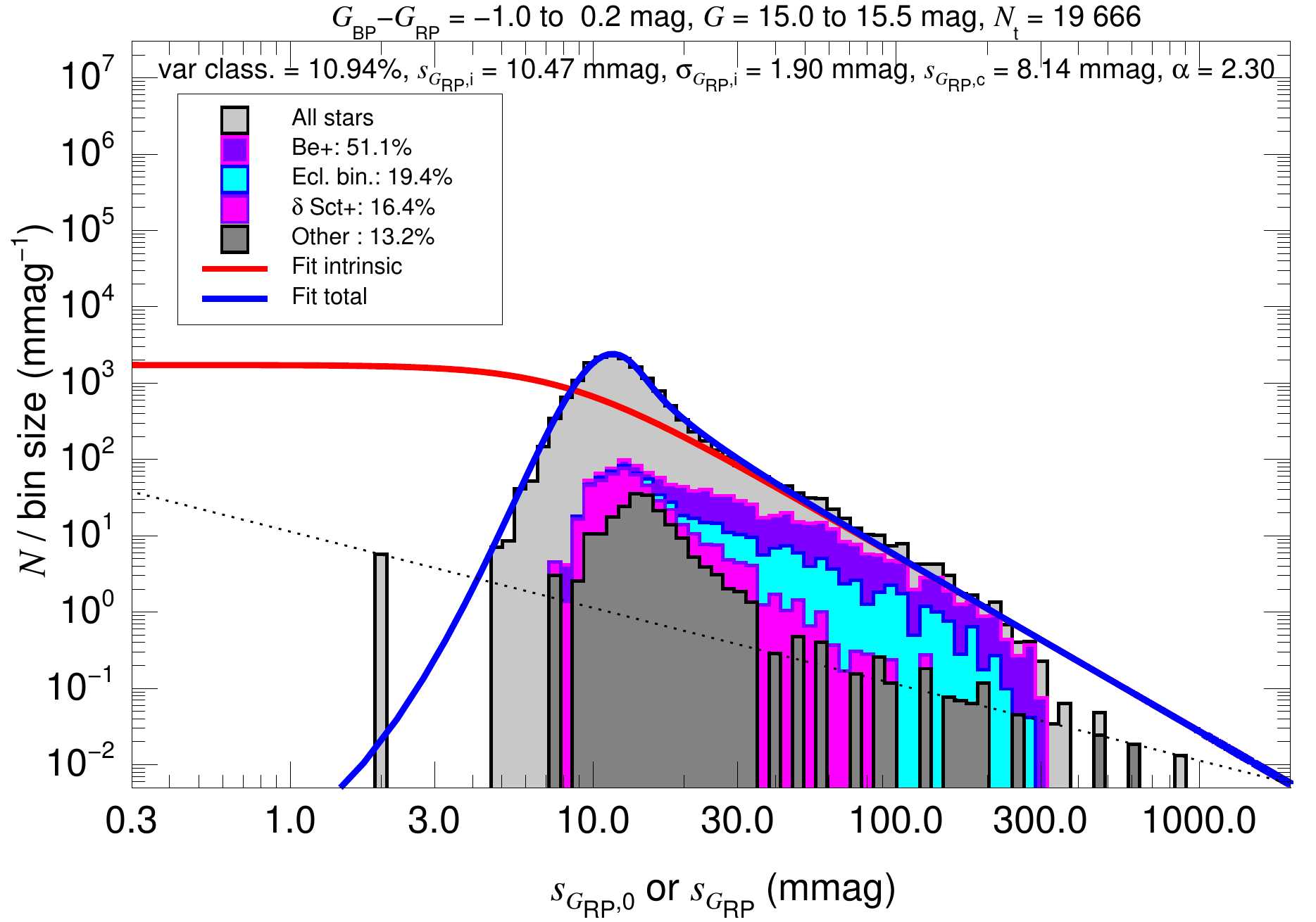}}
\centerline{$\!\!\!$\includegraphics[width=0.35\linewidth]{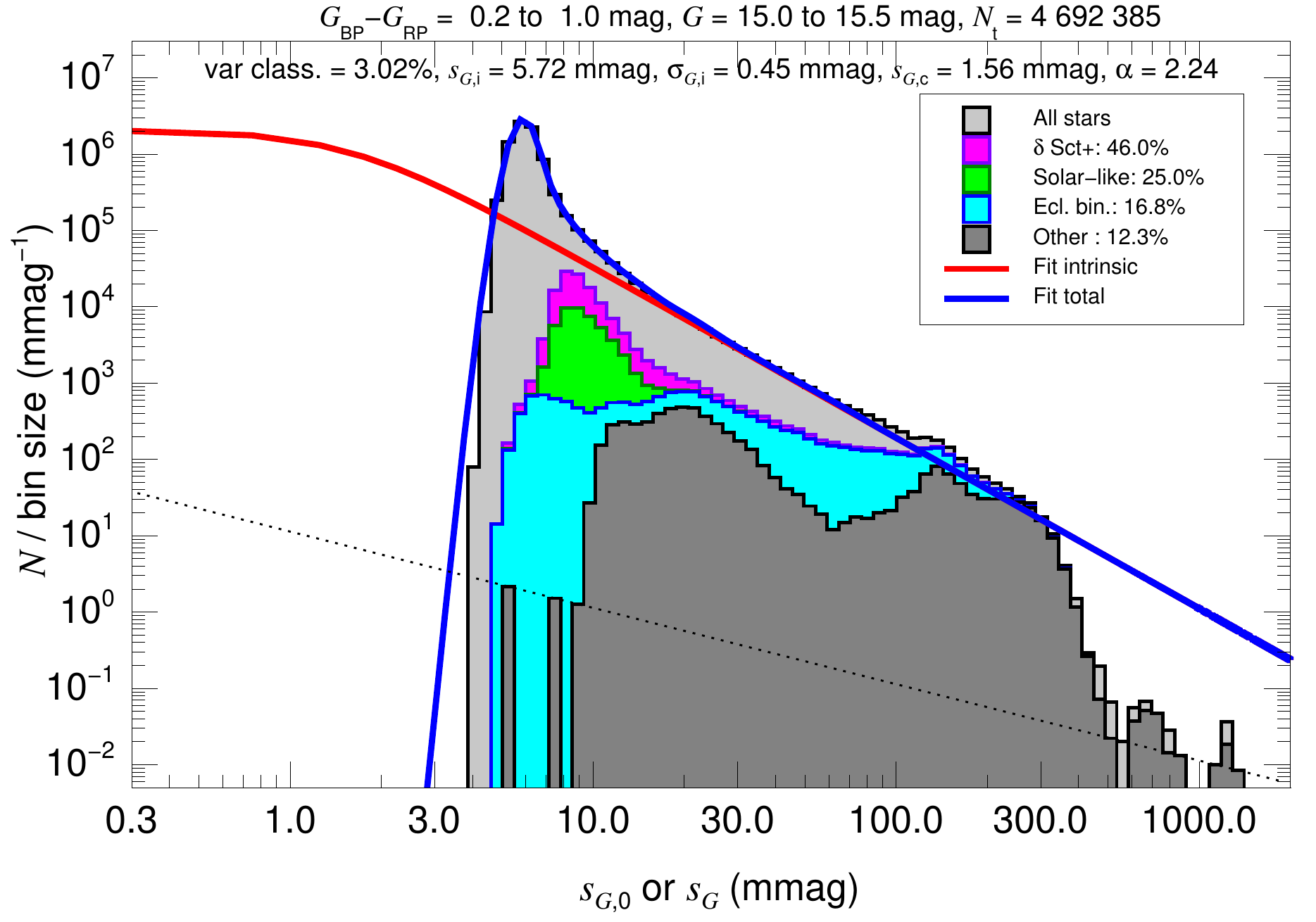}$\!\!\!$
                    \includegraphics[width=0.35\linewidth]{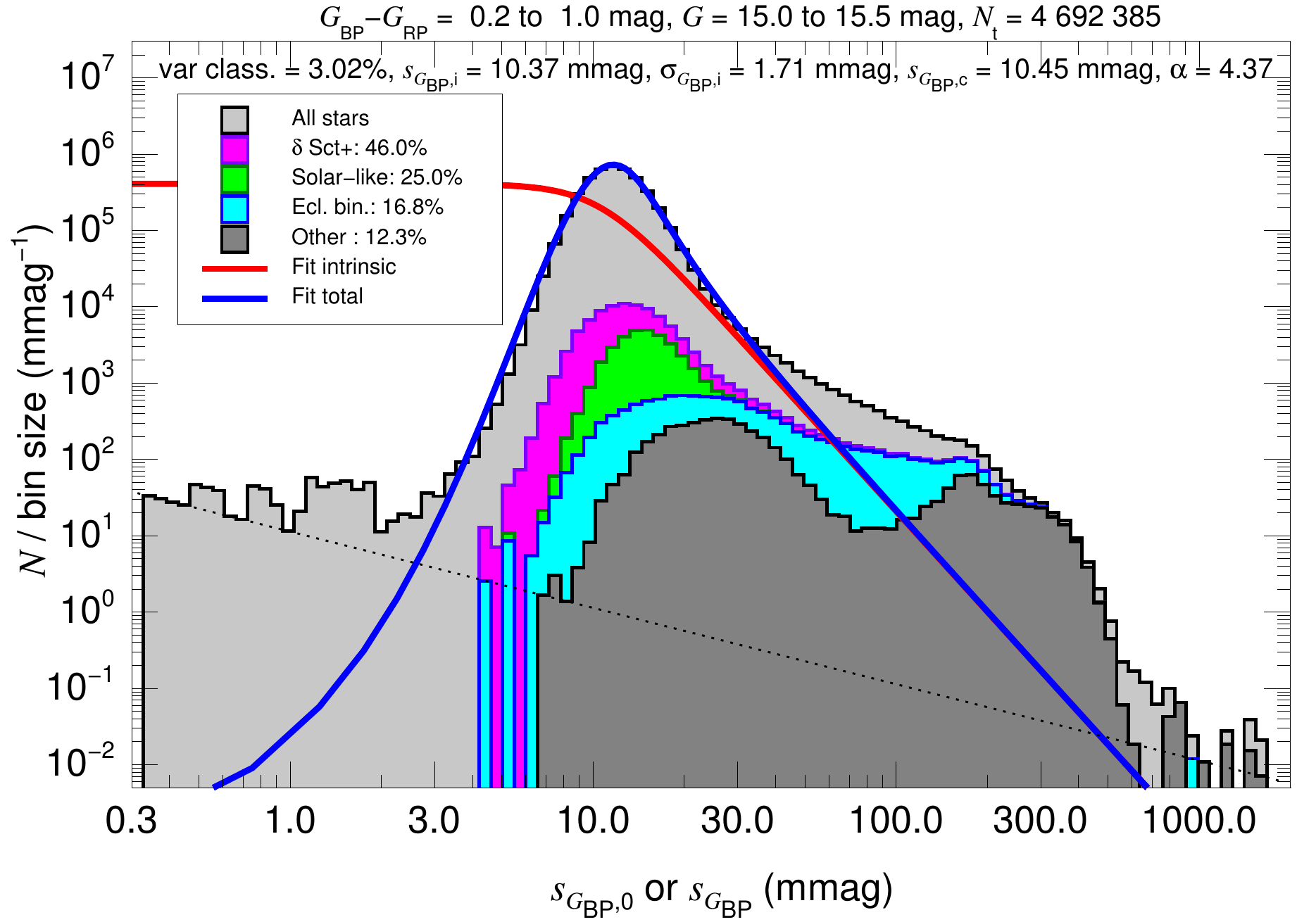}$\!\!\!$
                    \includegraphics[width=0.35\linewidth]{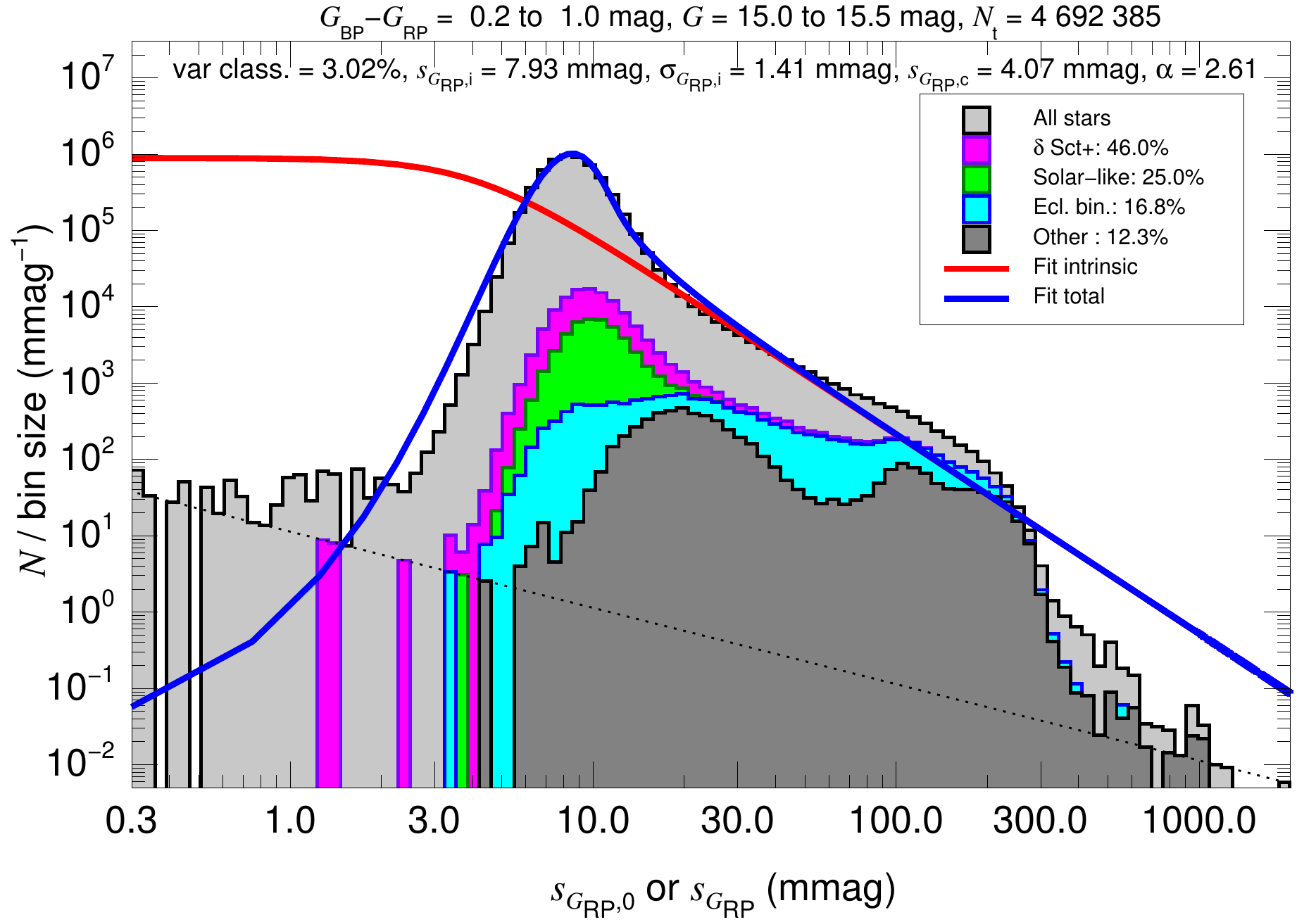}}
\centerline{$\!\!\!$\includegraphics[width=0.35\linewidth]{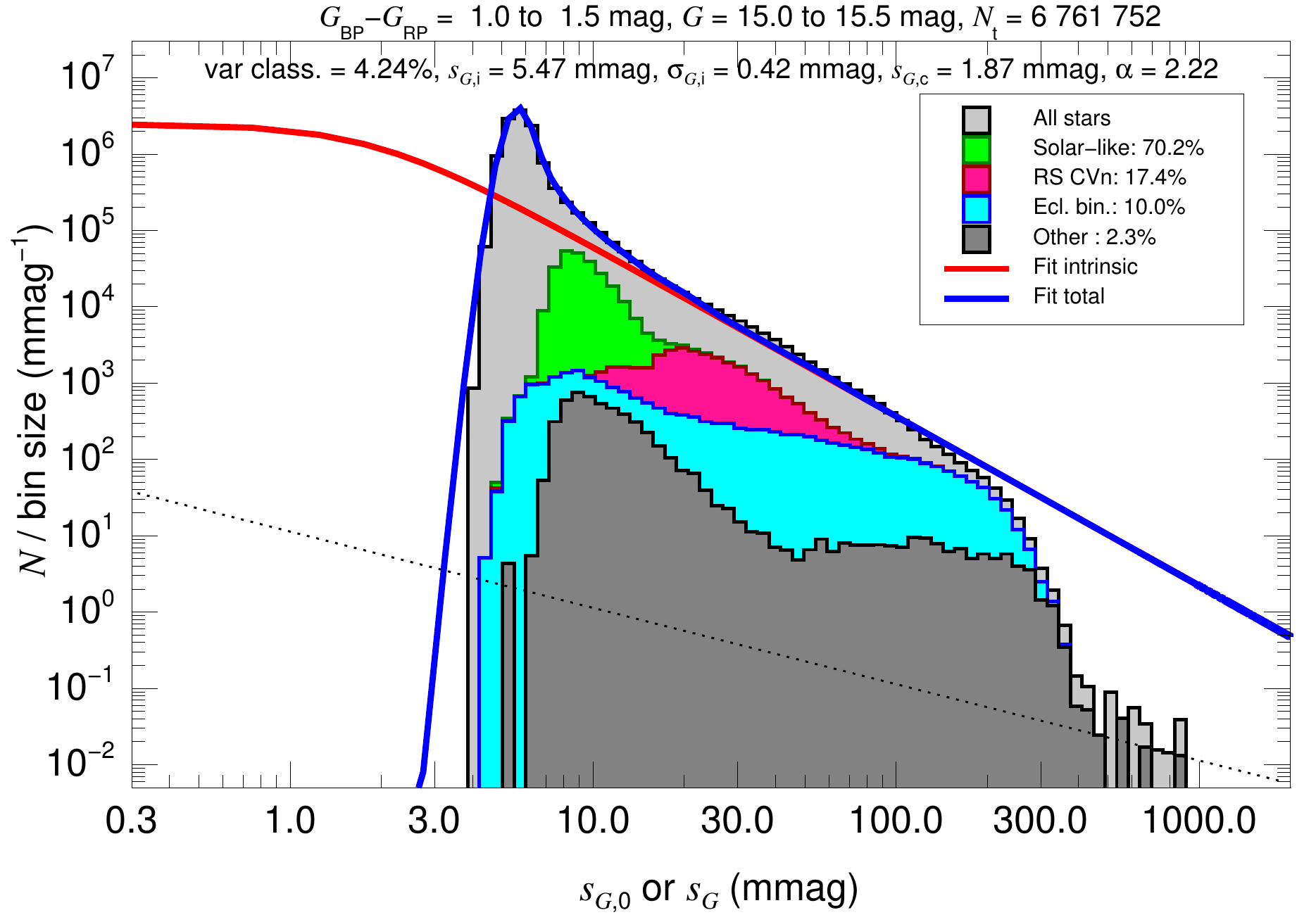}$\!\!\!$
                    \includegraphics[width=0.35\linewidth]{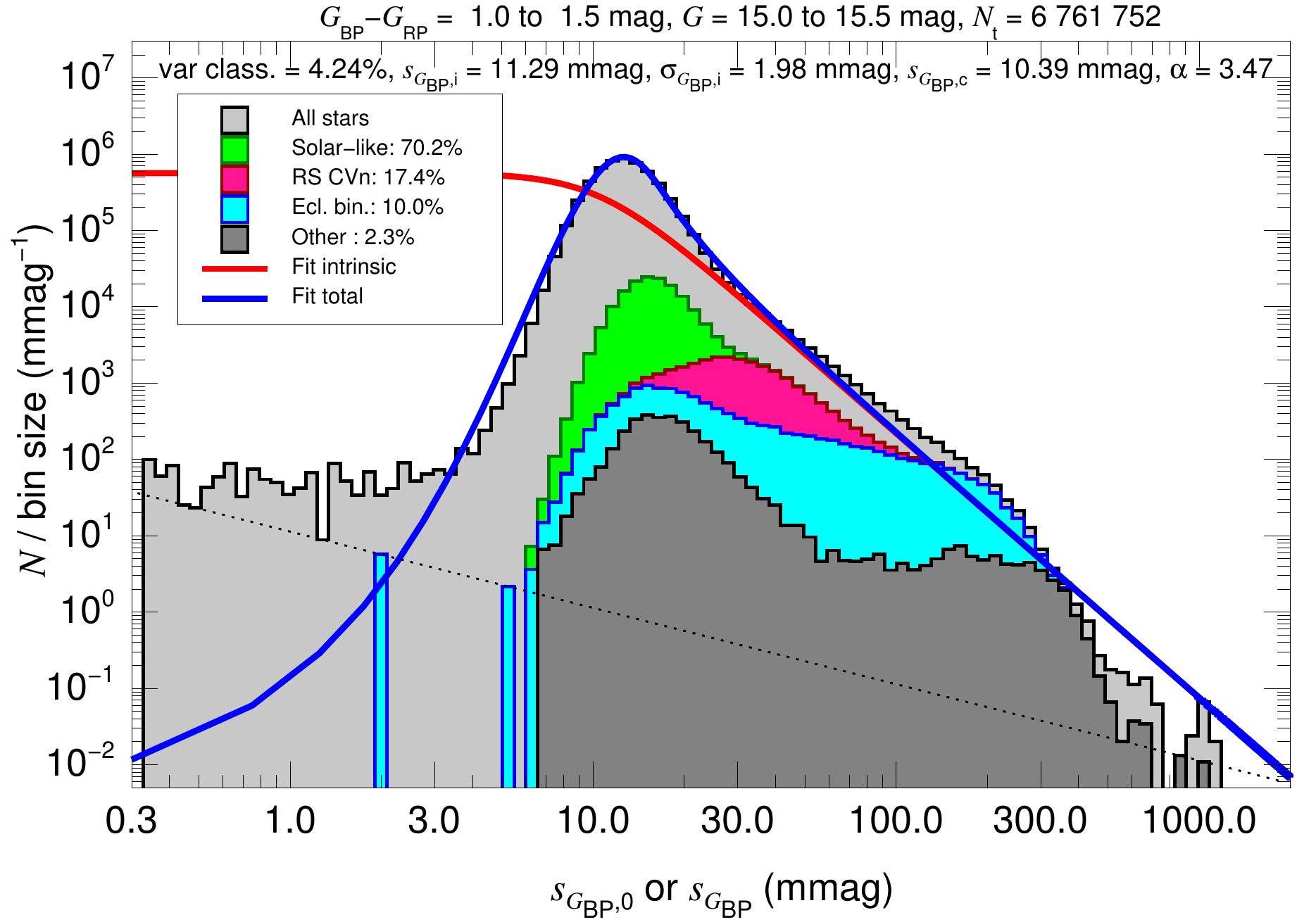}$\!\!\!$
                    \includegraphics[width=0.35\linewidth]{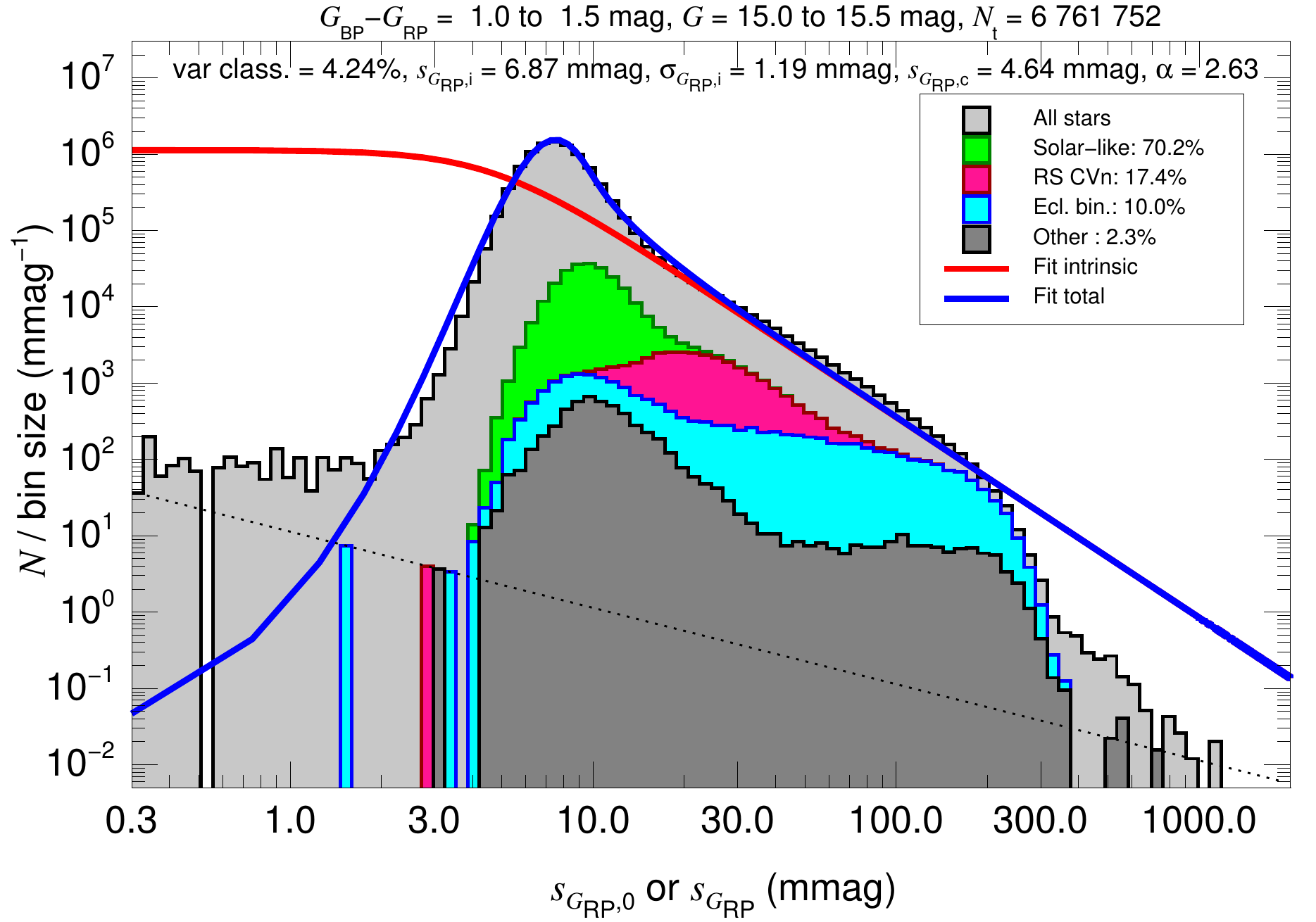}}
\centerline{$\!\!\!$\includegraphics[width=0.35\linewidth]{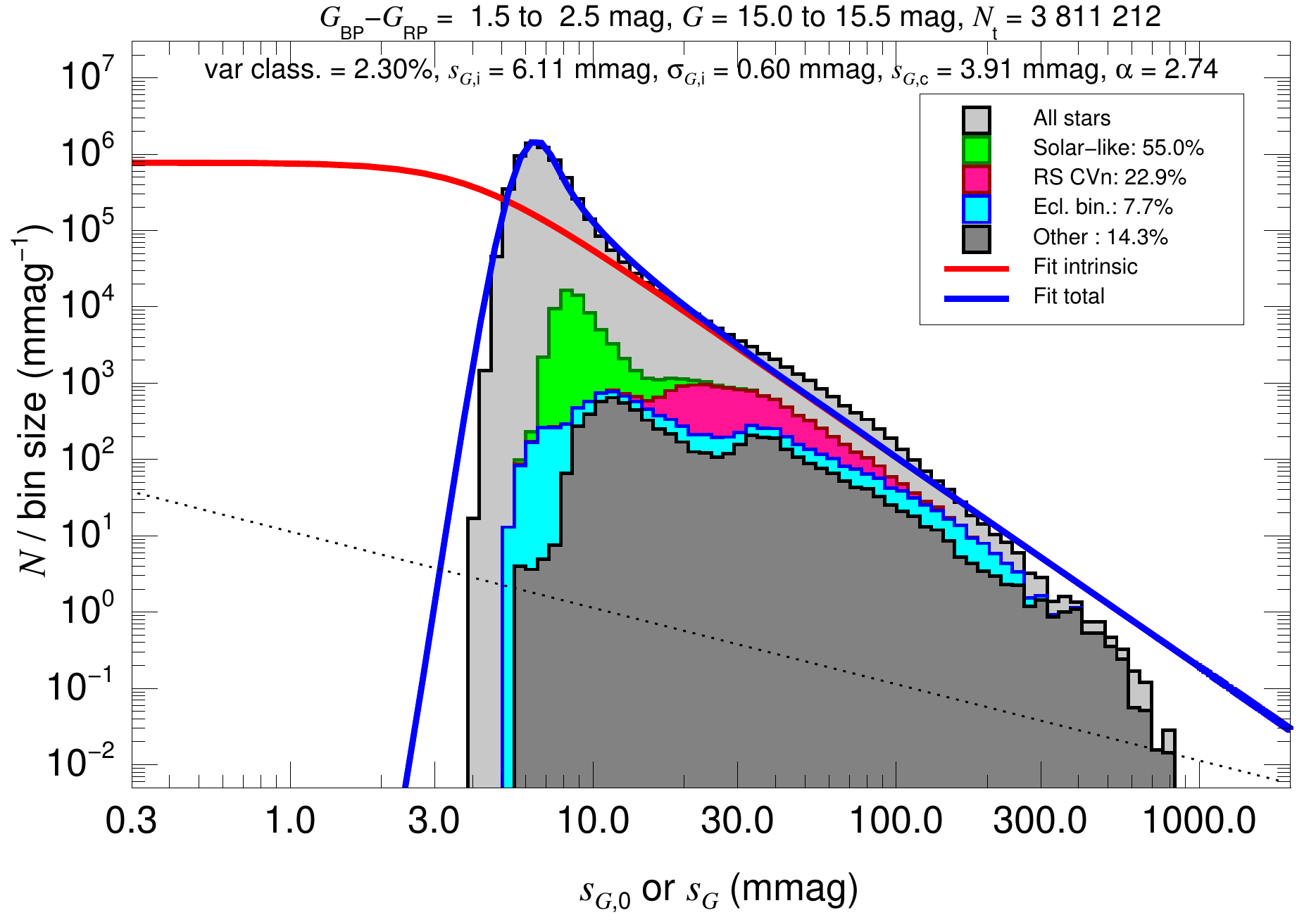}$\!\!\!$
                    \includegraphics[width=0.35\linewidth]{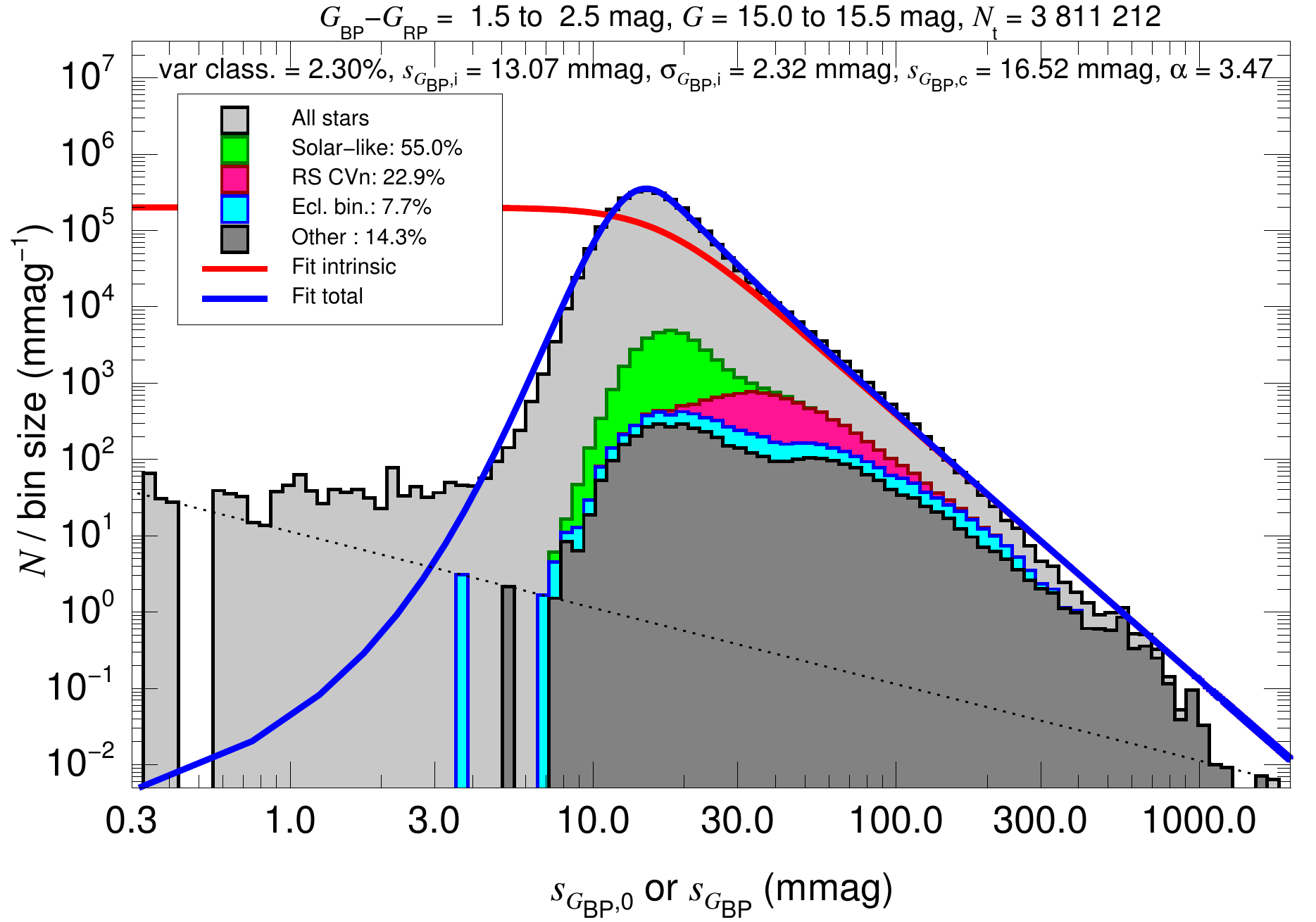}$\!\!\!$
                    \includegraphics[width=0.35\linewidth]{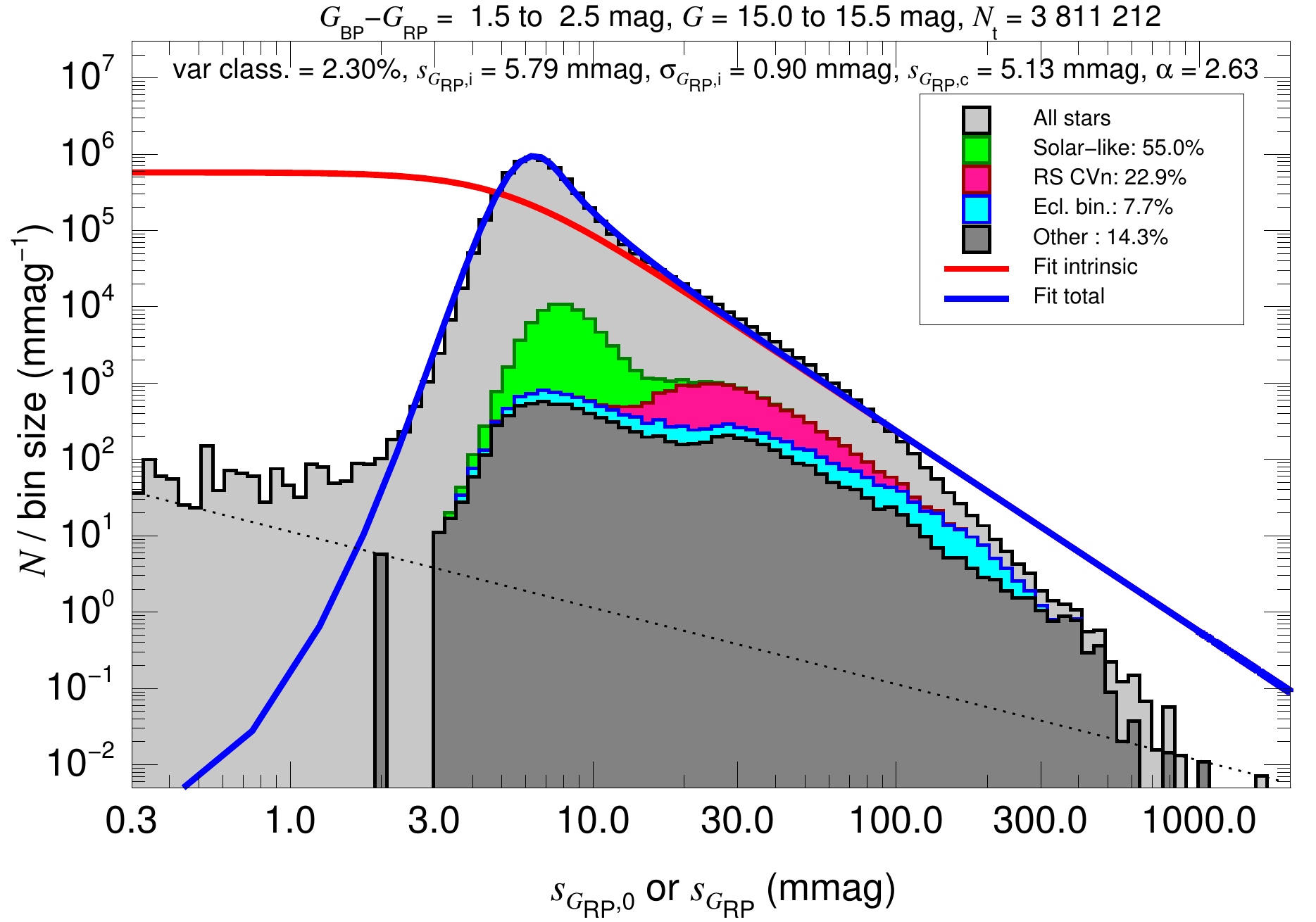}}
\centerline{$\!\!\!$\includegraphics[width=0.35\linewidth]{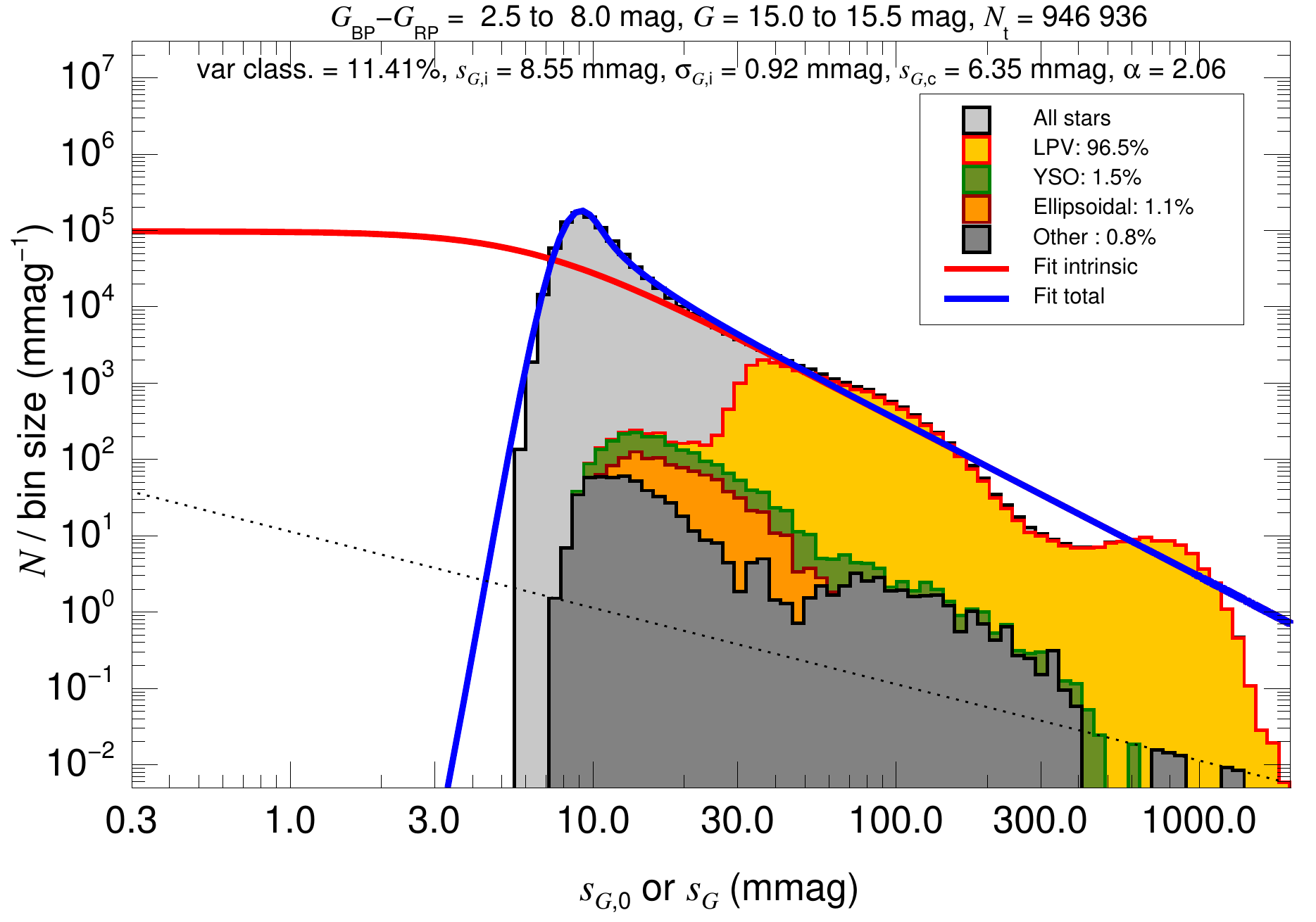}$\!\!\!$
                    \includegraphics[width=0.35\linewidth]{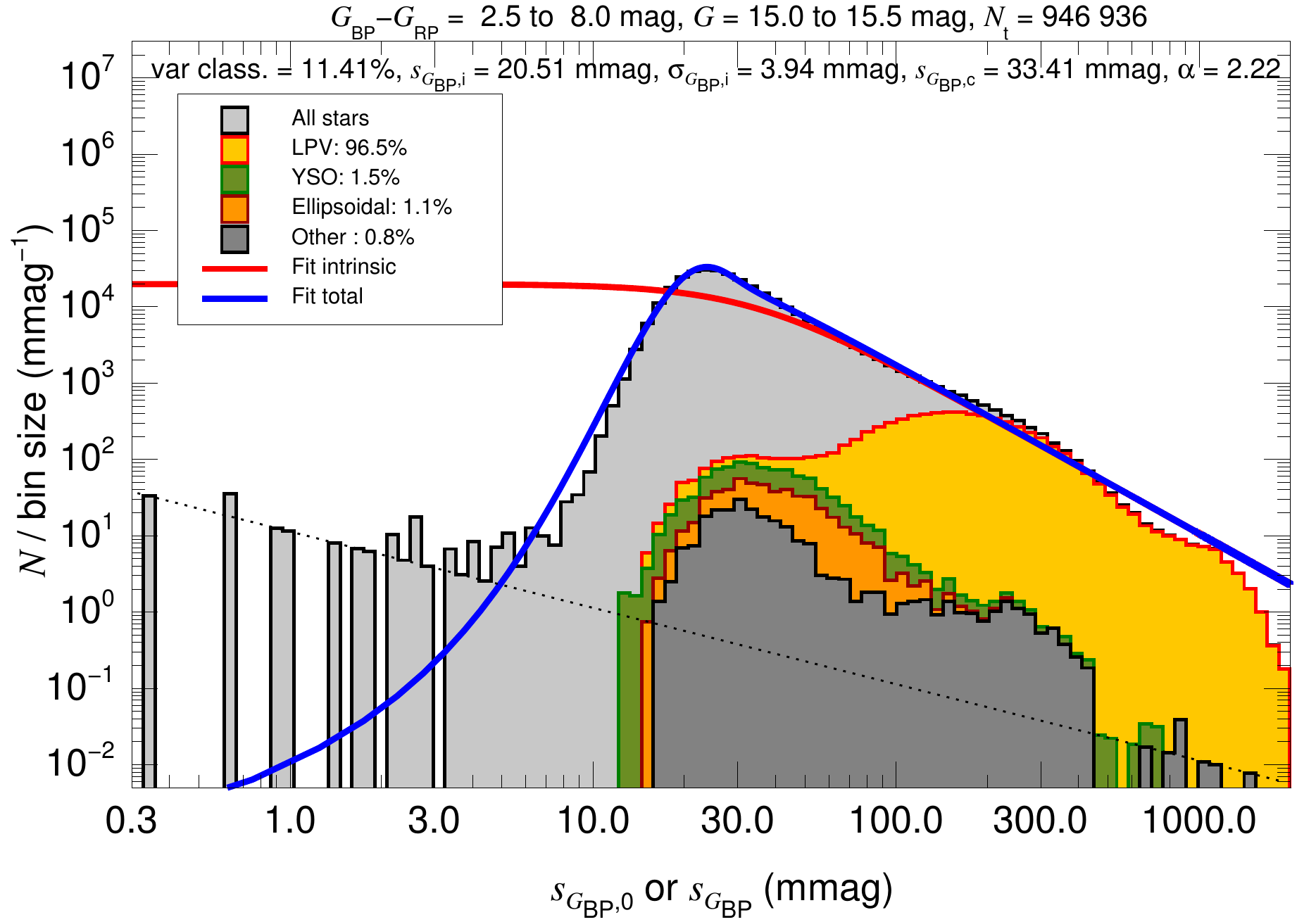}$\!\!\!$
                    \includegraphics[width=0.35\linewidth]{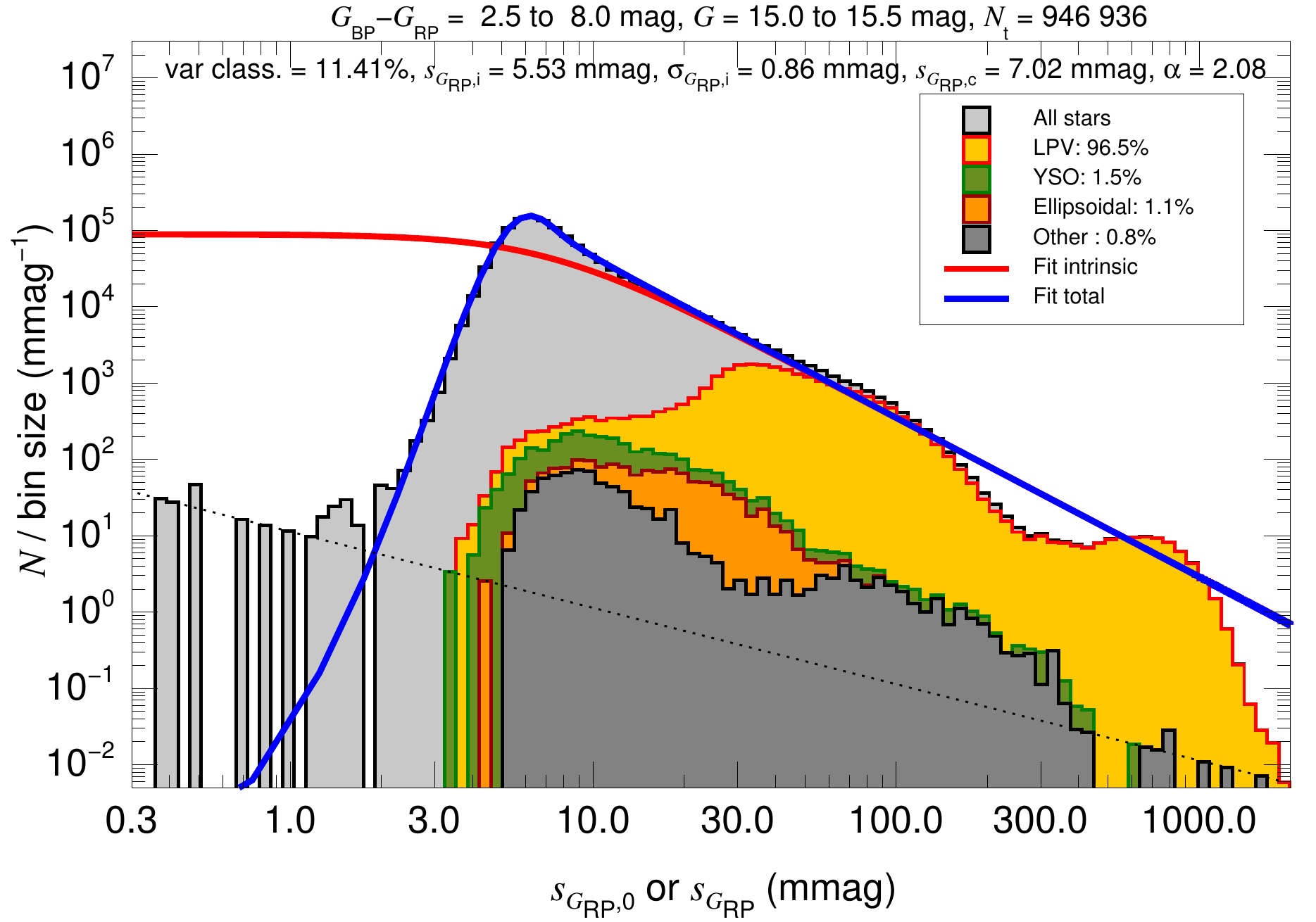}}
\caption{(Continued).}
\end{figure*}

\addtocounter{figure}{-1}

\begin{figure*}
\centerline{$\!\!\!$\includegraphics[width=0.35\linewidth]{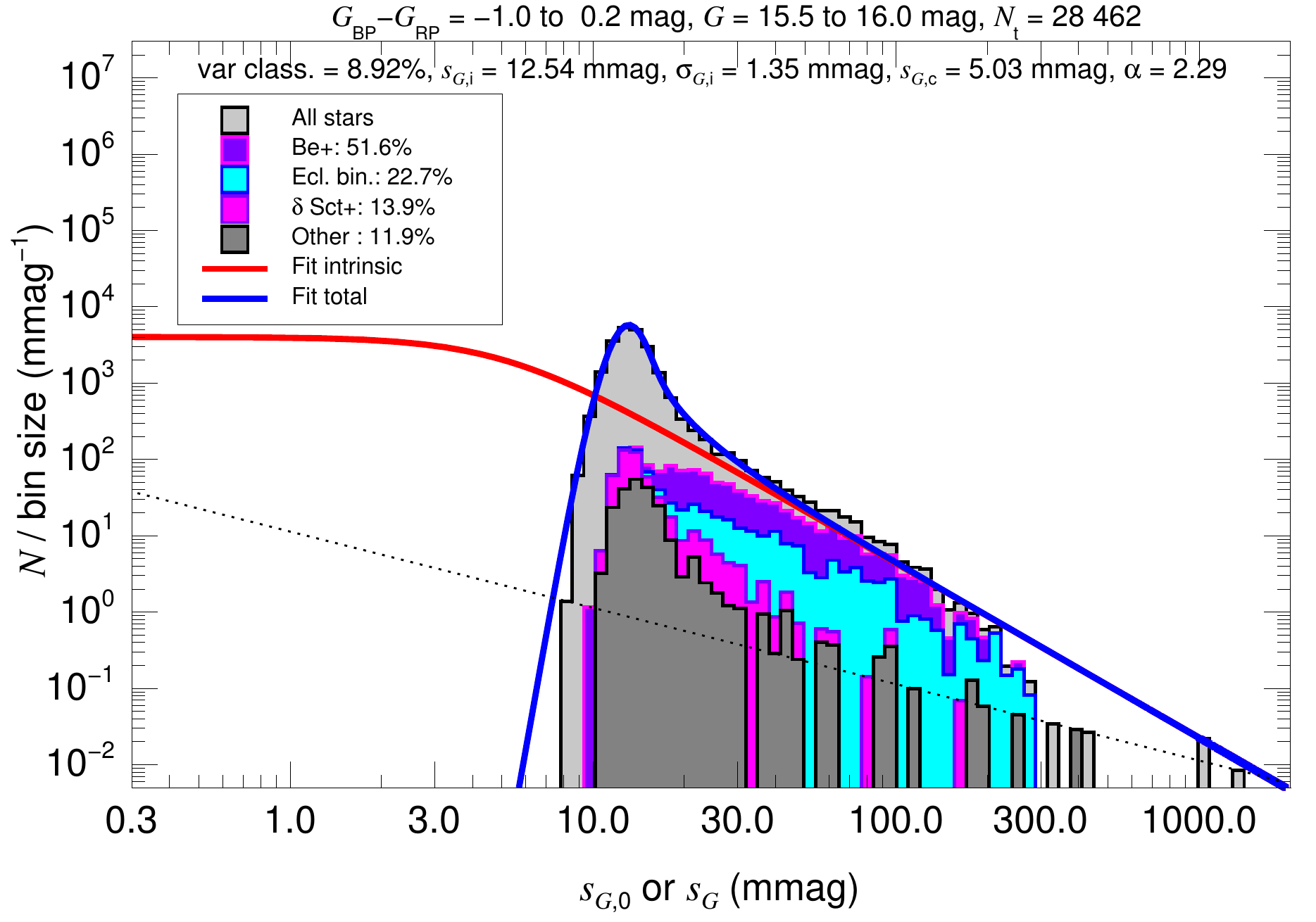}$\!\!\!$
                    \includegraphics[width=0.35\linewidth]{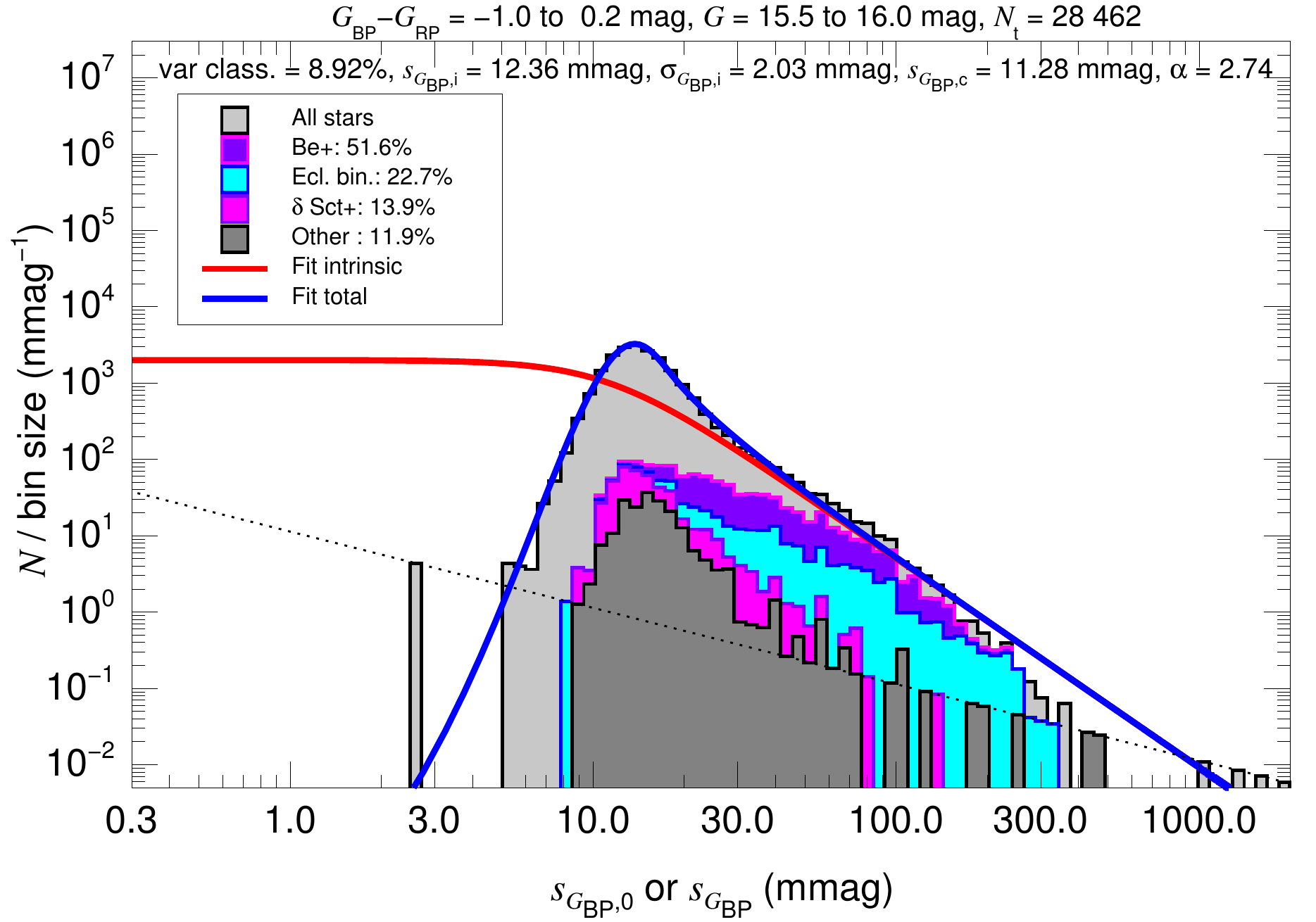}$\!\!\!$
                    \includegraphics[width=0.35\linewidth]{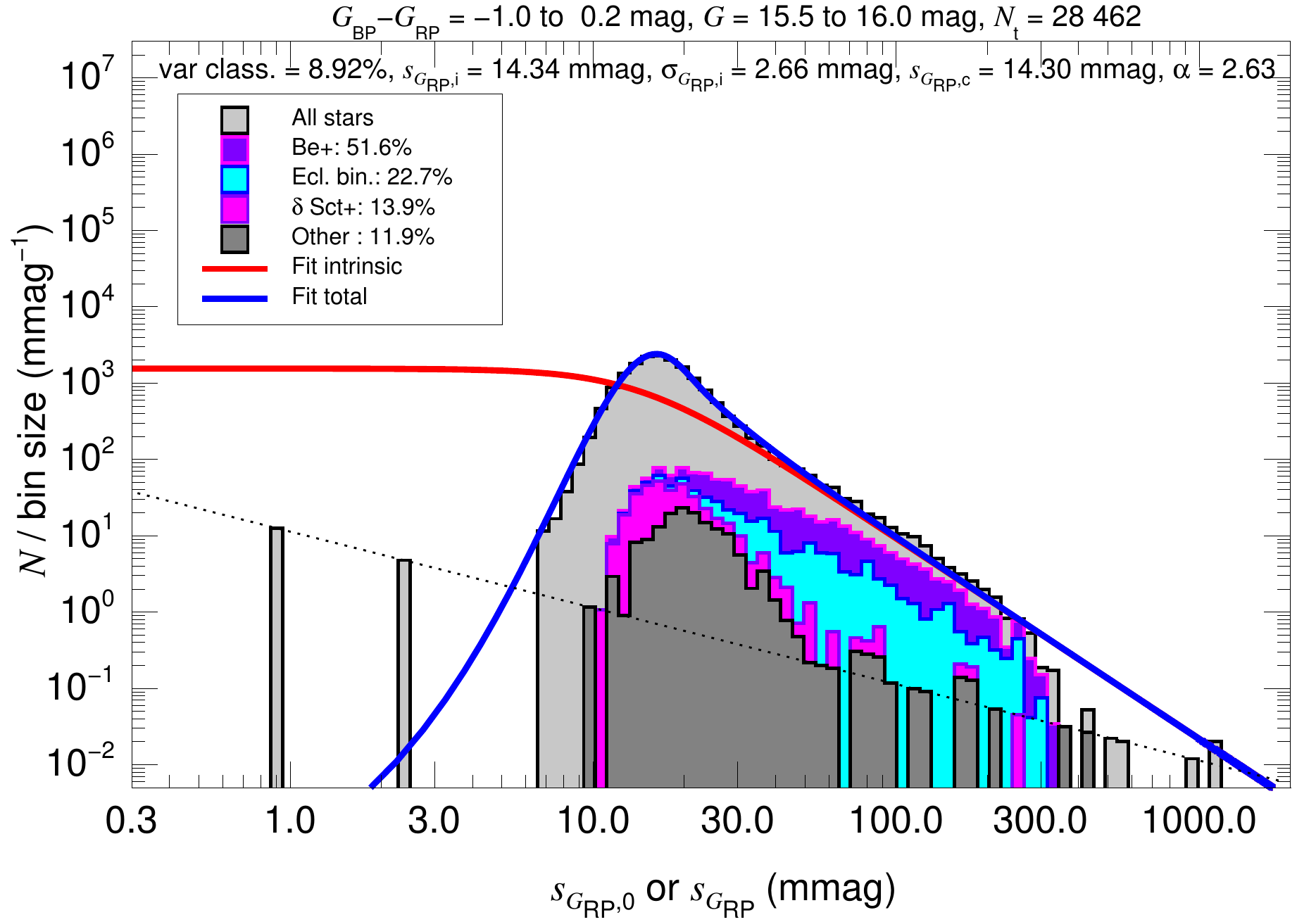}}
\centerline{$\!\!\!$\includegraphics[width=0.35\linewidth]{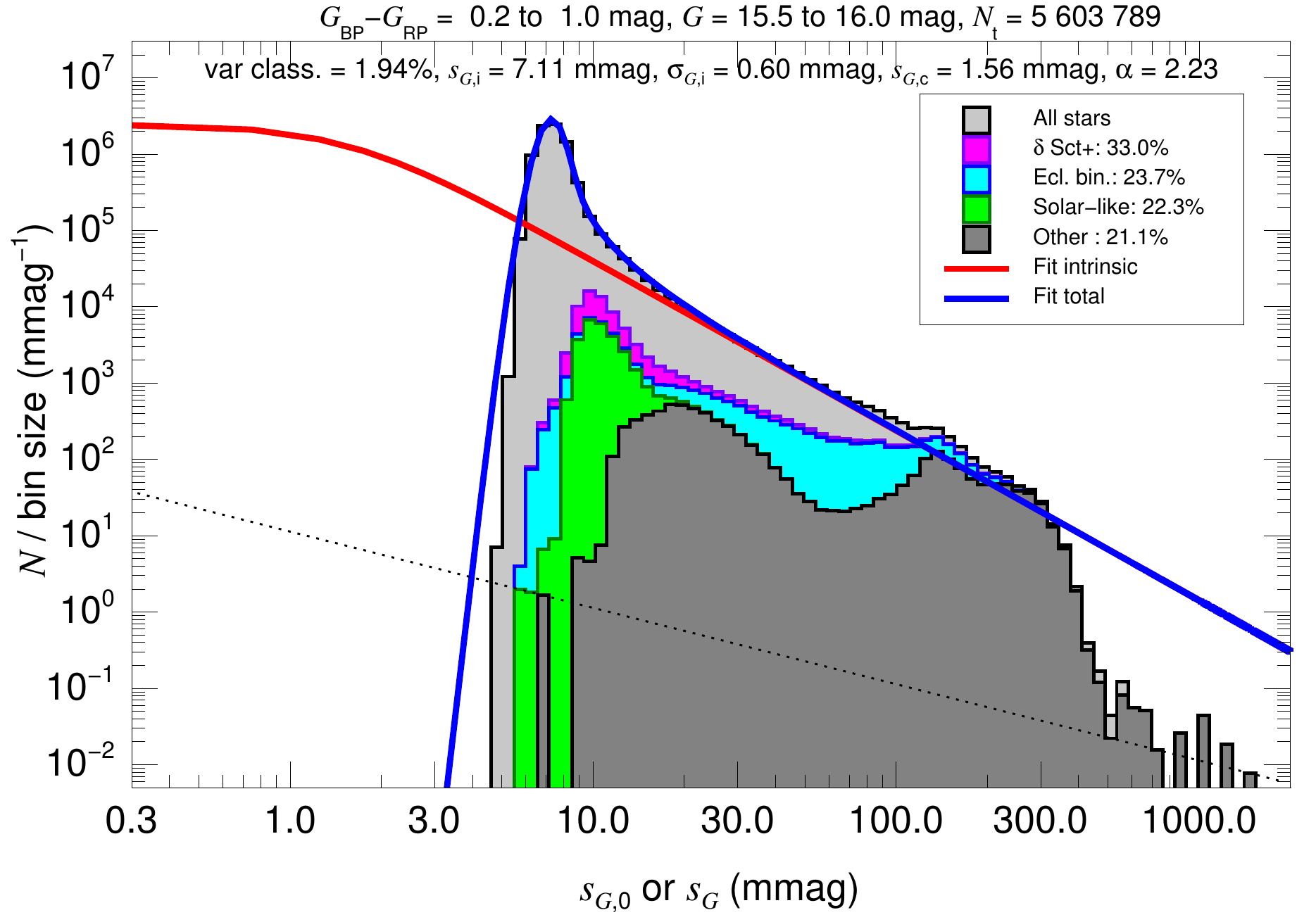}$\!\!\!$
                    \includegraphics[width=0.35\linewidth]{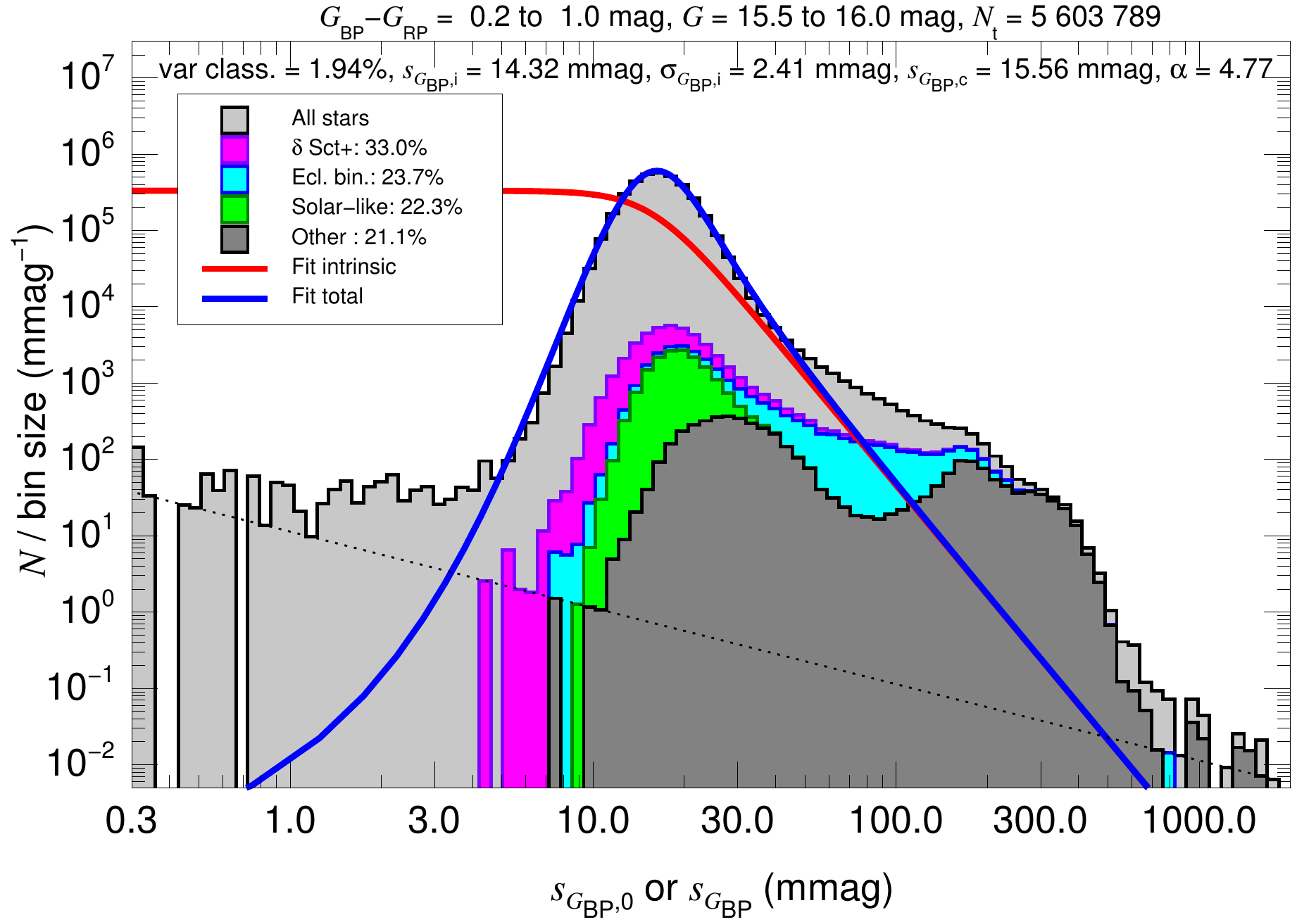}$\!\!\!$
                    \includegraphics[width=0.35\linewidth]{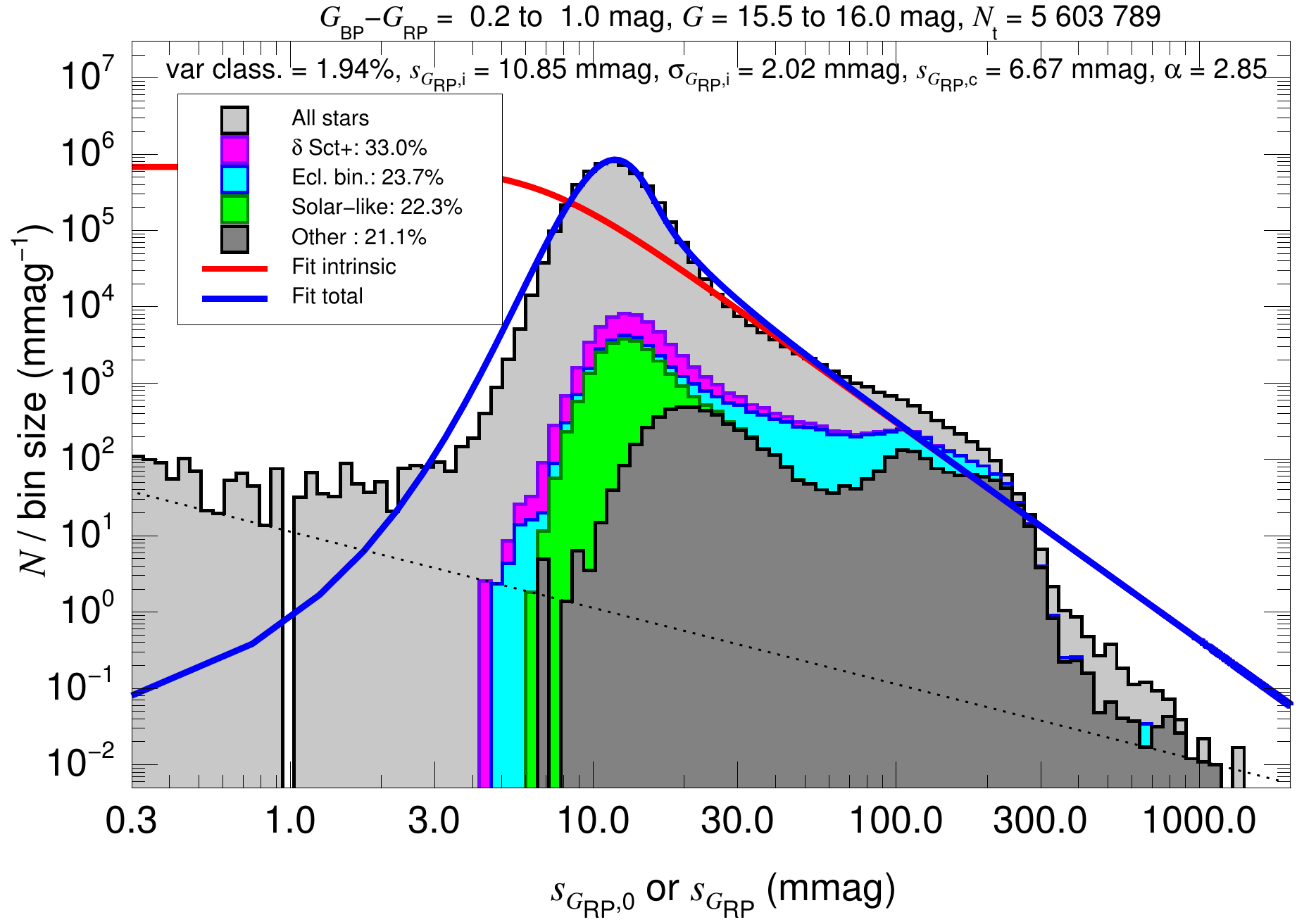}}
\centerline{$\!\!\!$\includegraphics[width=0.35\linewidth]{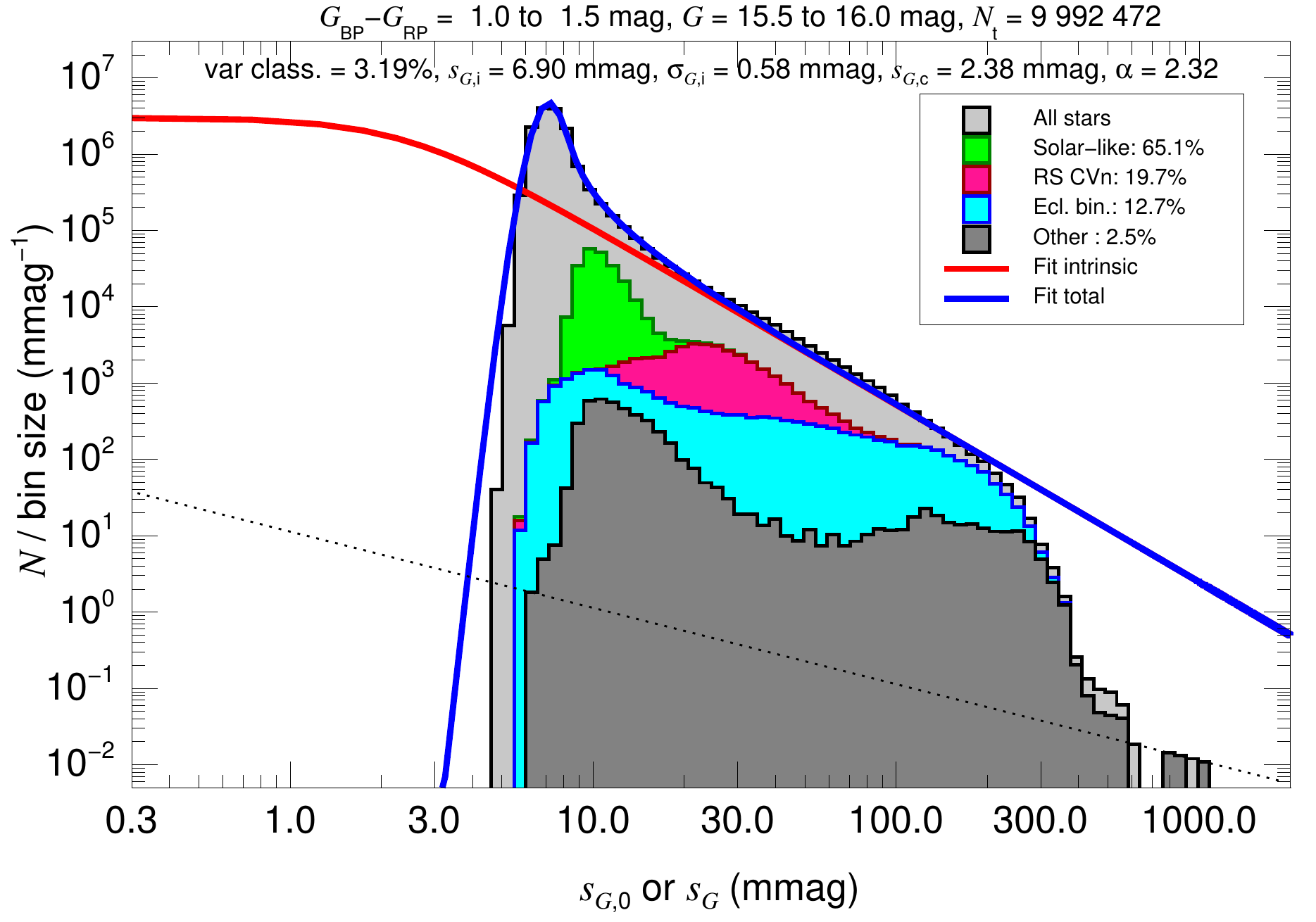}$\!\!\!$
                    \includegraphics[width=0.35\linewidth]{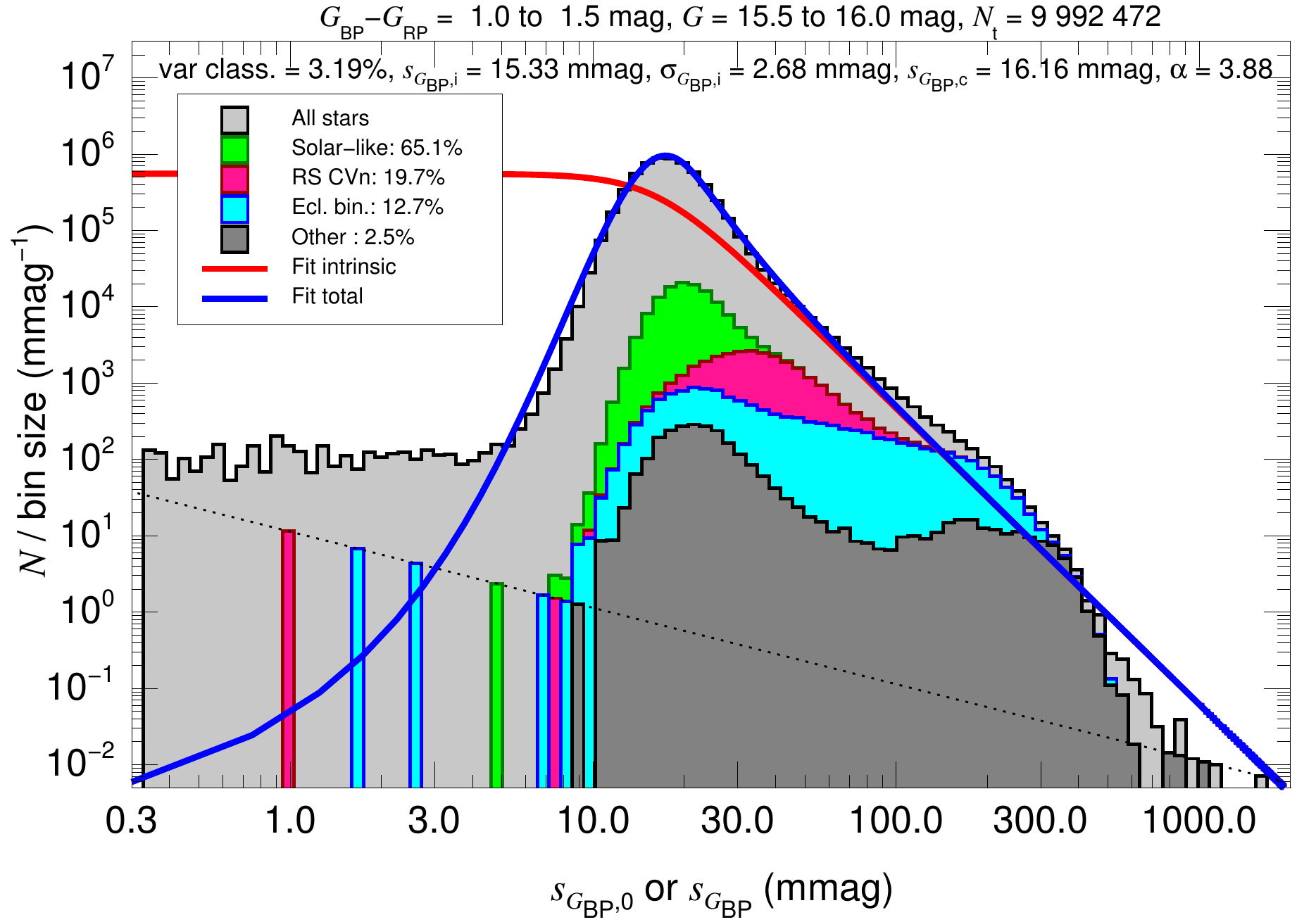}$\!\!\!$
                    \includegraphics[width=0.35\linewidth]{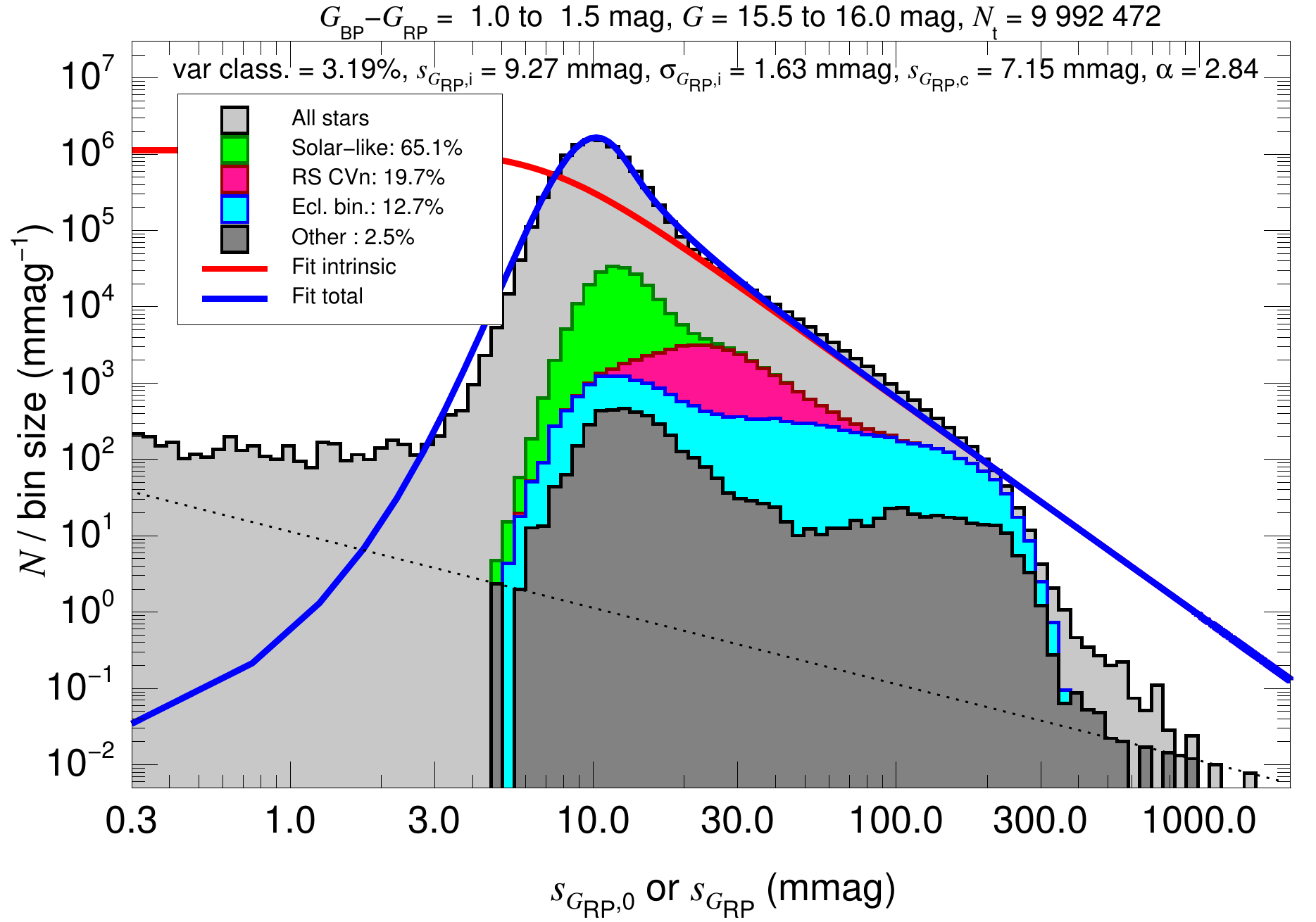}}
\centerline{$\!\!\!$\includegraphics[width=0.35\linewidth]{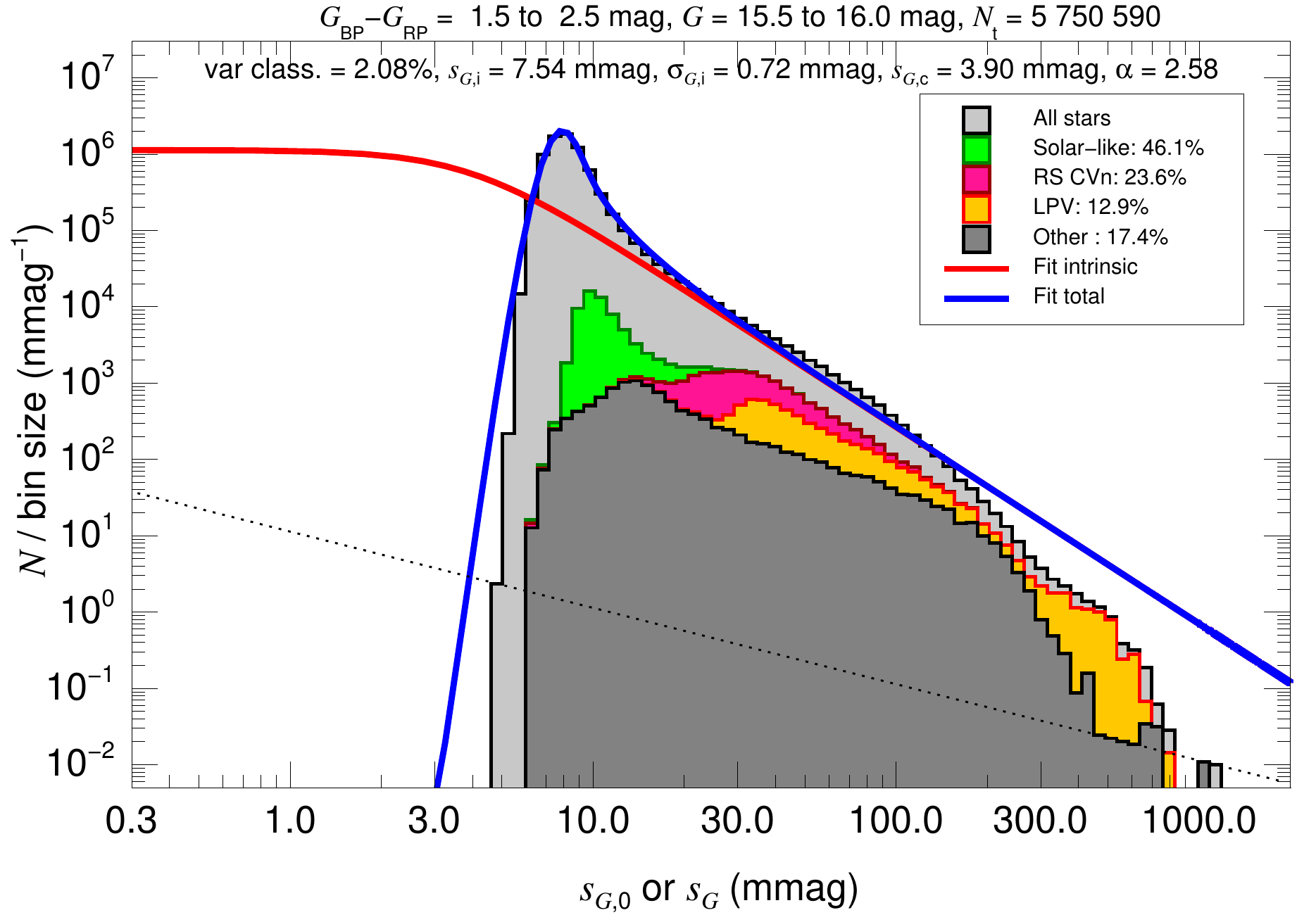}$\!\!\!$
                    \includegraphics[width=0.35\linewidth]{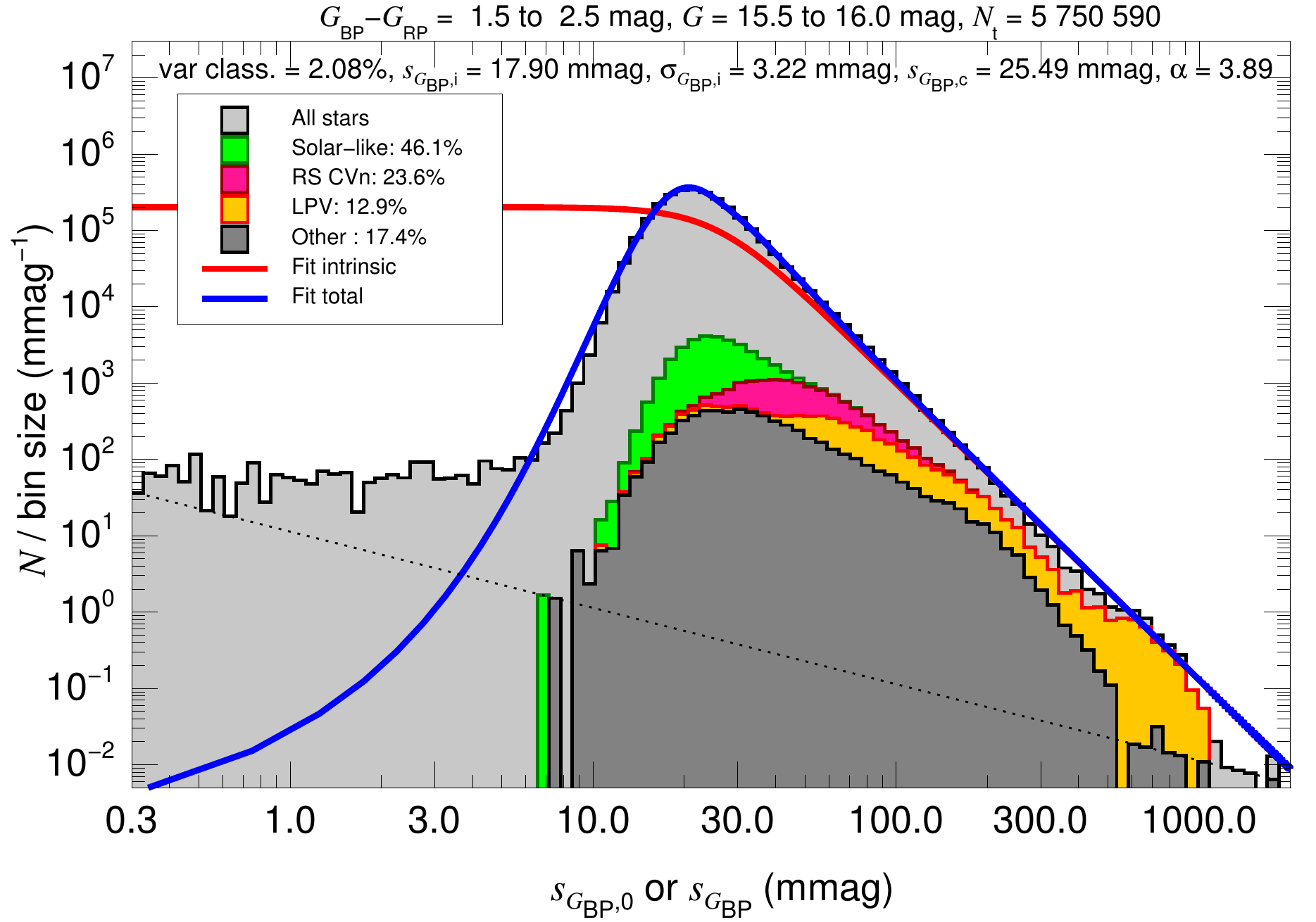}$\!\!\!$
                    \includegraphics[width=0.35\linewidth]{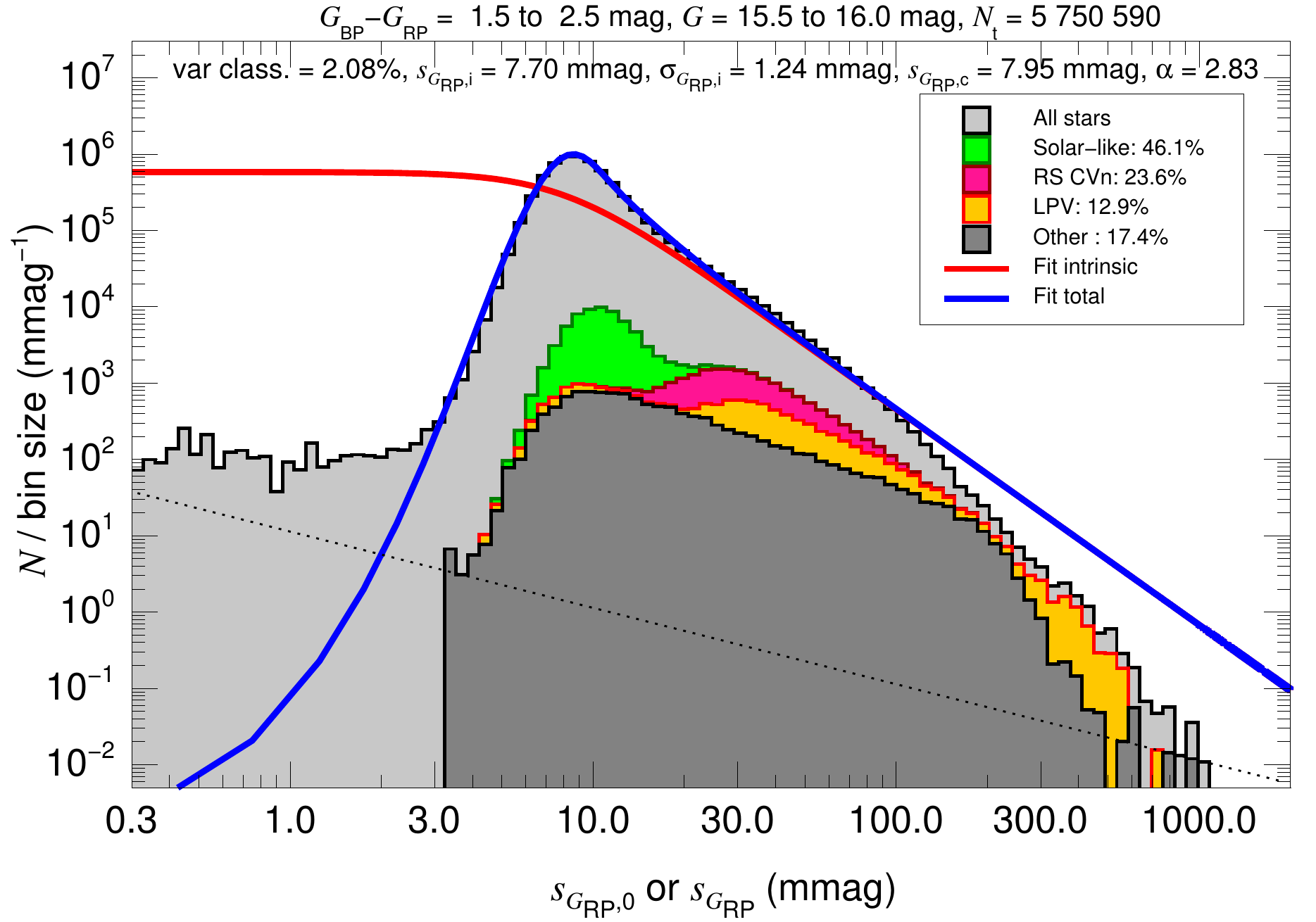}}
\centerline{$\!\!\!$\includegraphics[width=0.35\linewidth]{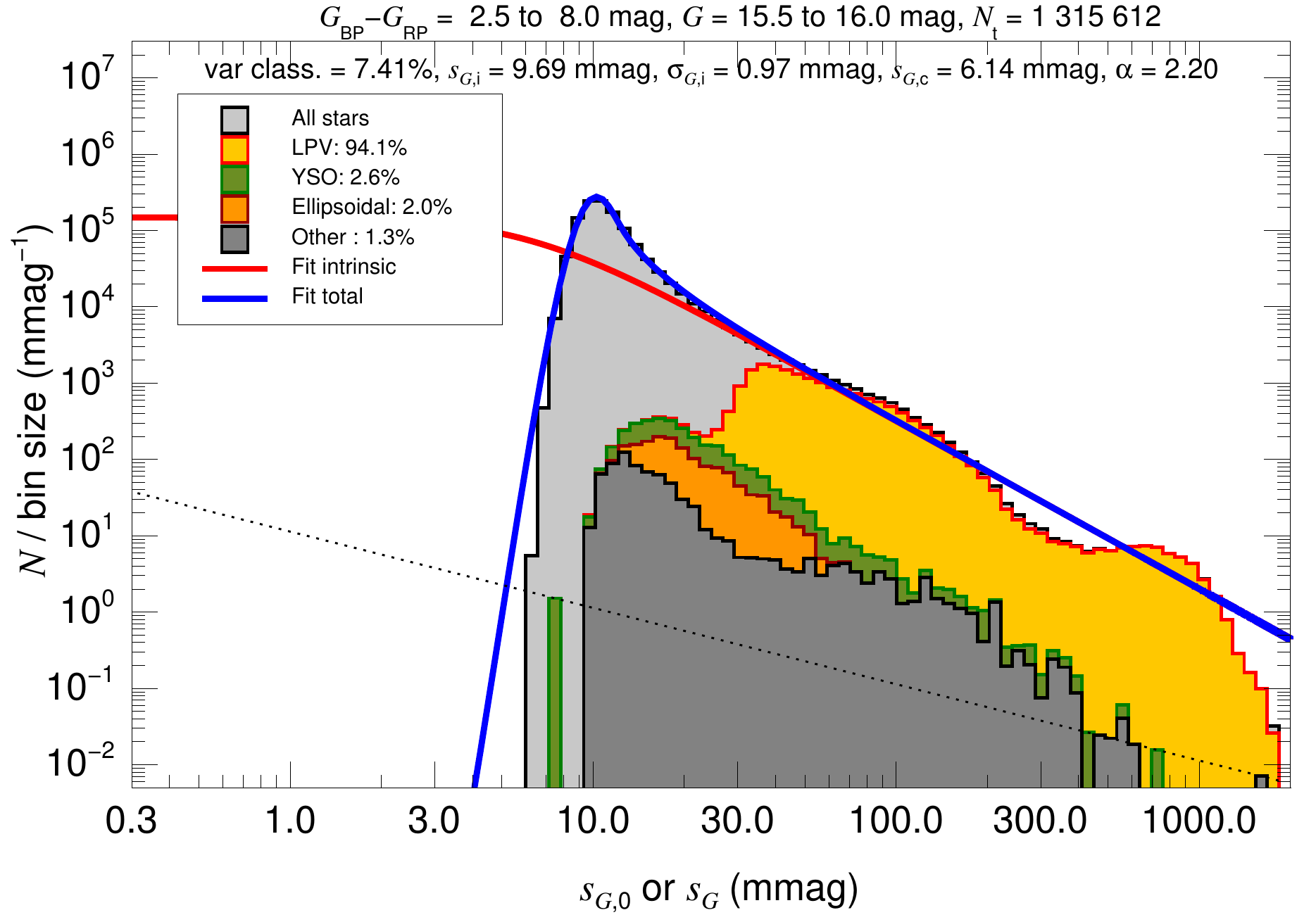}$\!\!\!$
                    \includegraphics[width=0.35\linewidth]{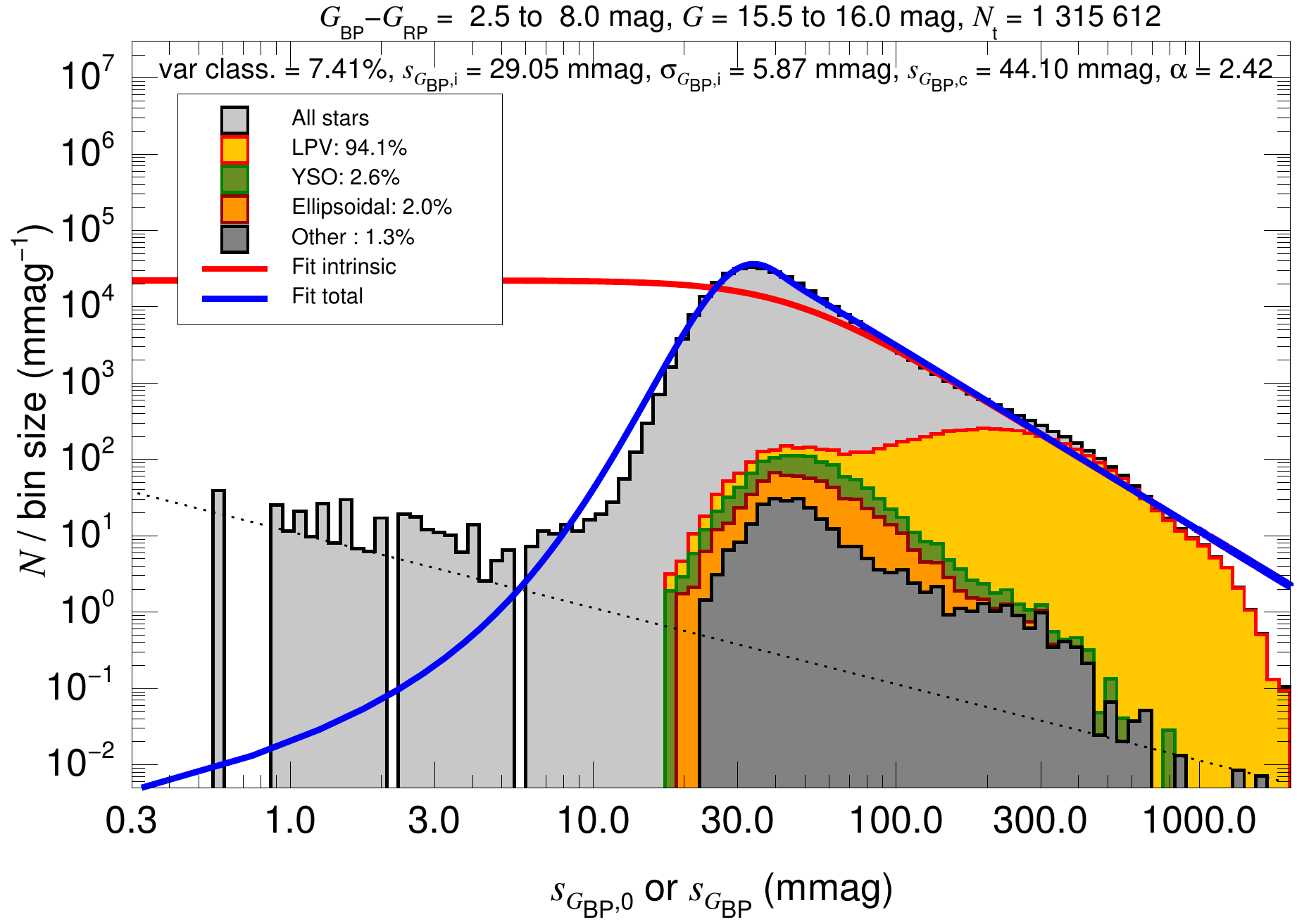}$\!\!\!$
                    \includegraphics[width=0.35\linewidth]{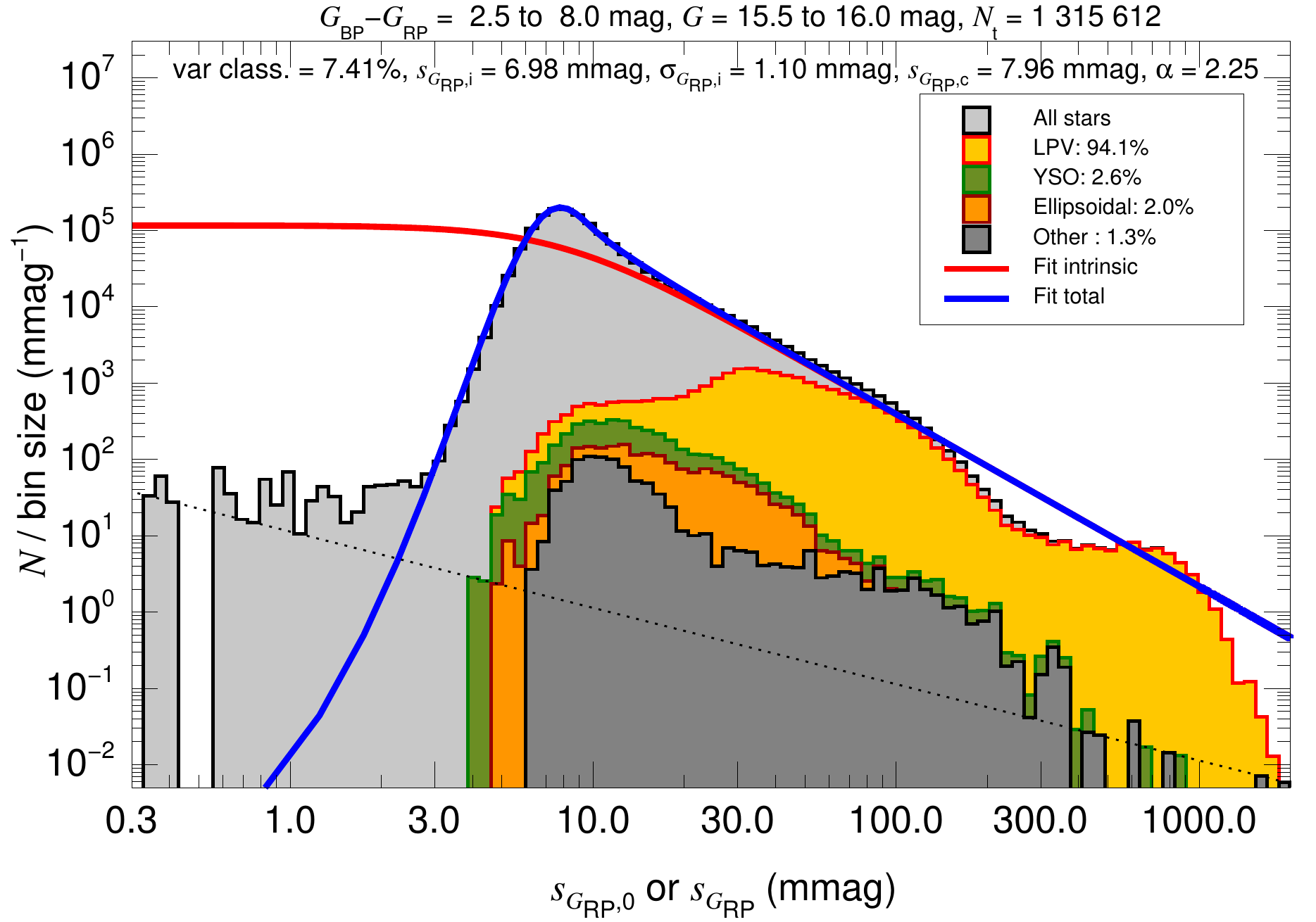}}
\caption{(Continued).}
\end{figure*}

\addtocounter{figure}{-1}

\begin{figure*}
\centerline{$\!\!\!$\includegraphics[width=0.35\linewidth]{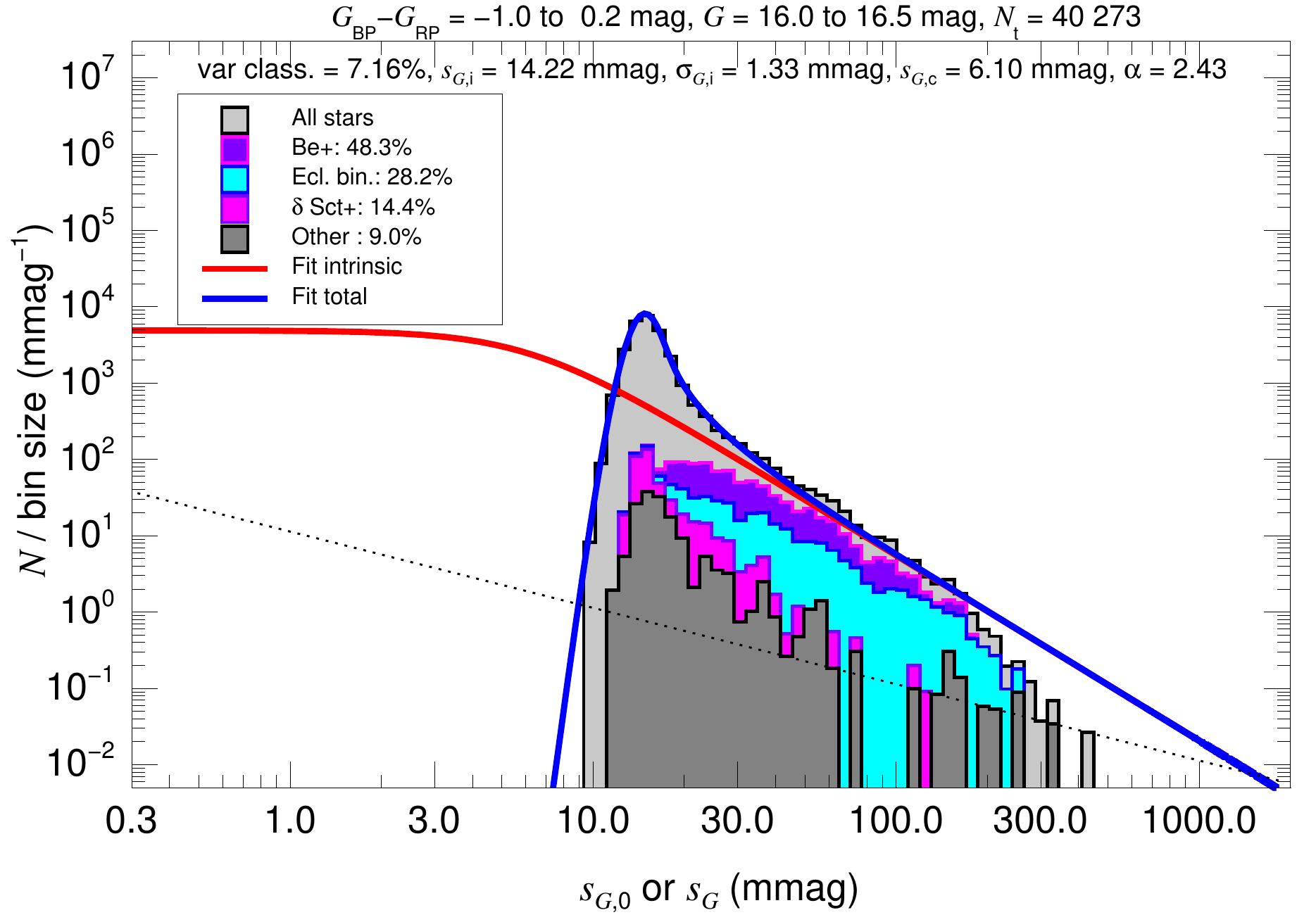}$\!\!\!$
                    \includegraphics[width=0.35\linewidth]{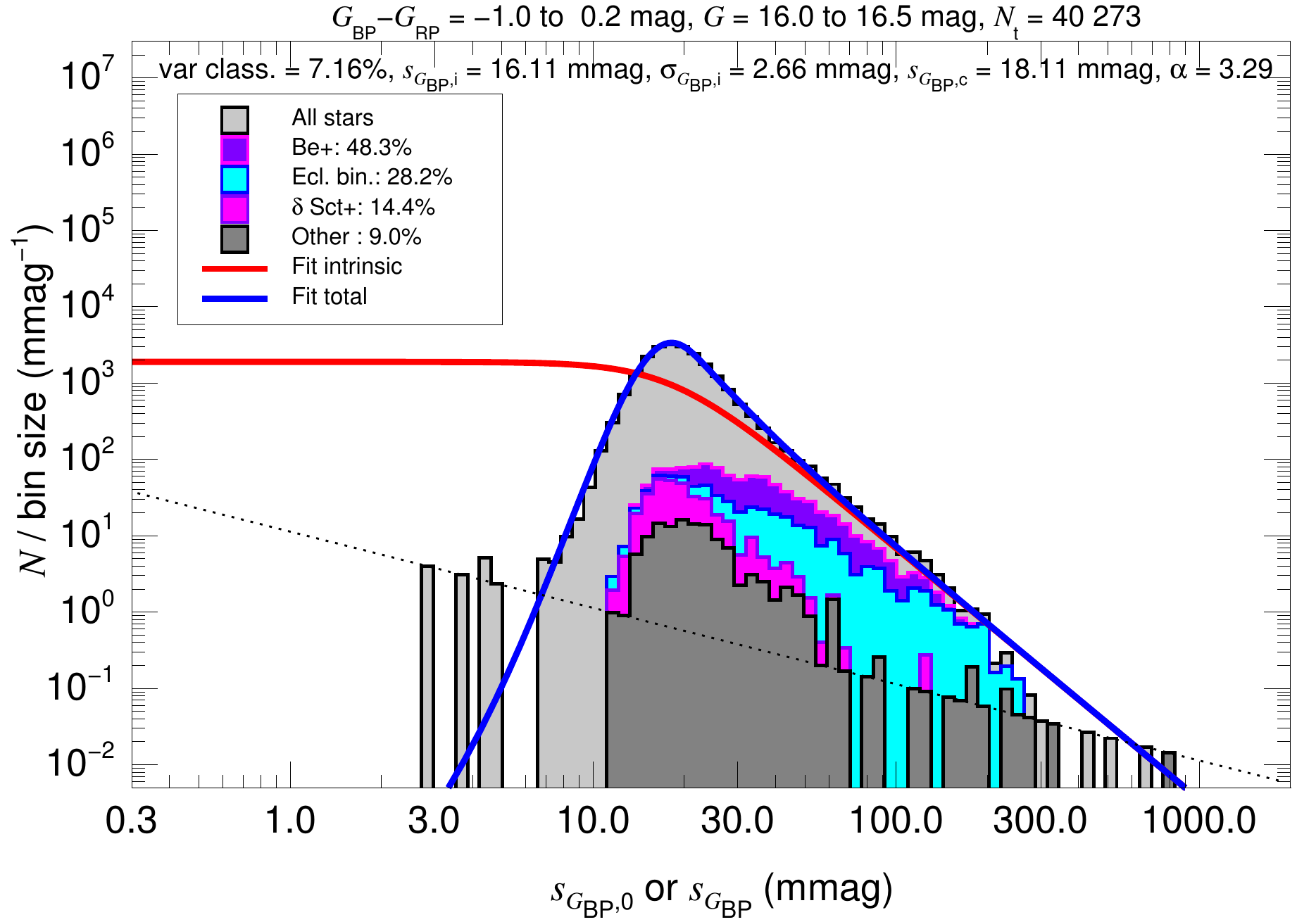}$\!\!\!$
                    \includegraphics[width=0.35\linewidth]{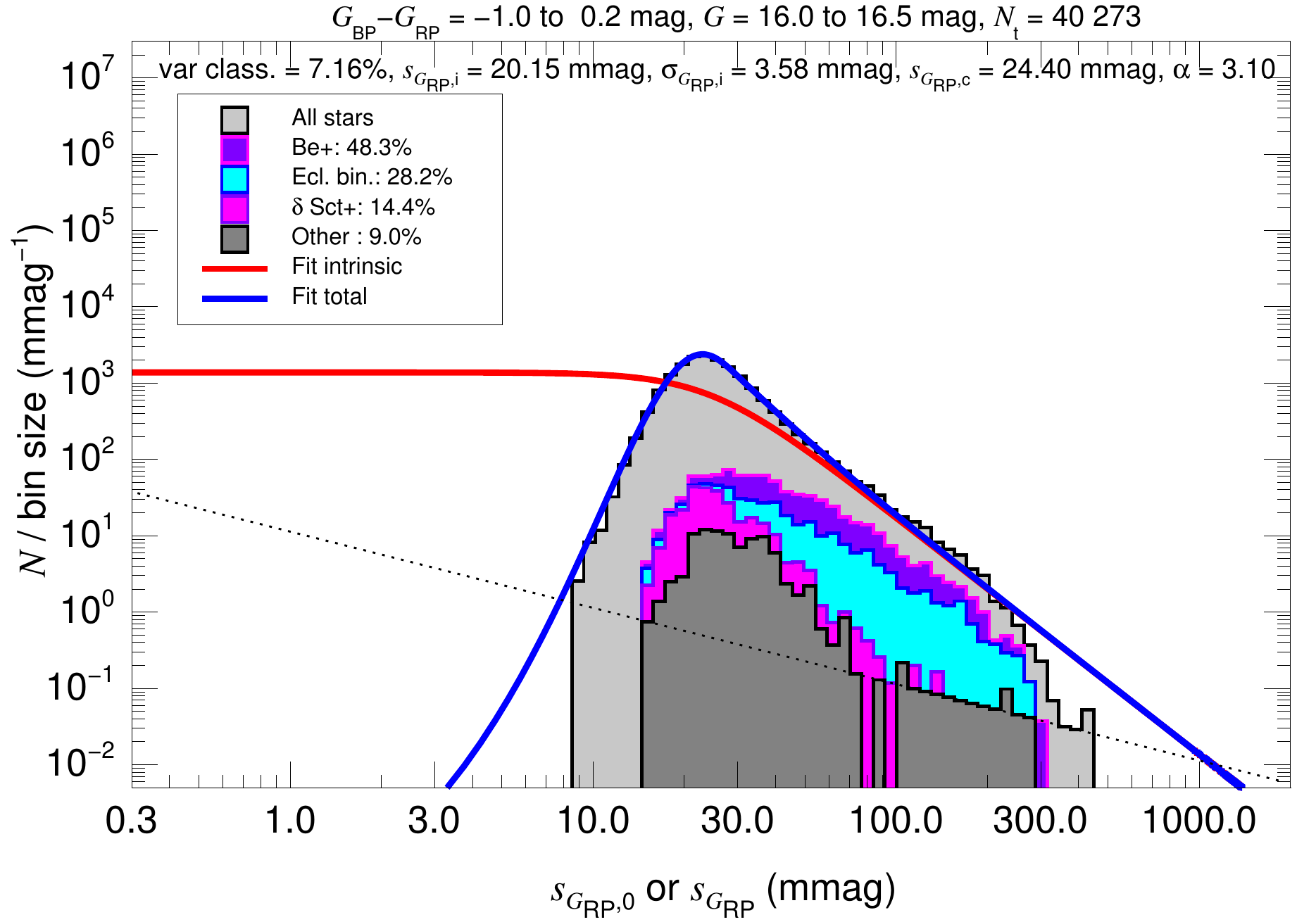}}
\centerline{$\!\!\!$\includegraphics[width=0.35\linewidth]{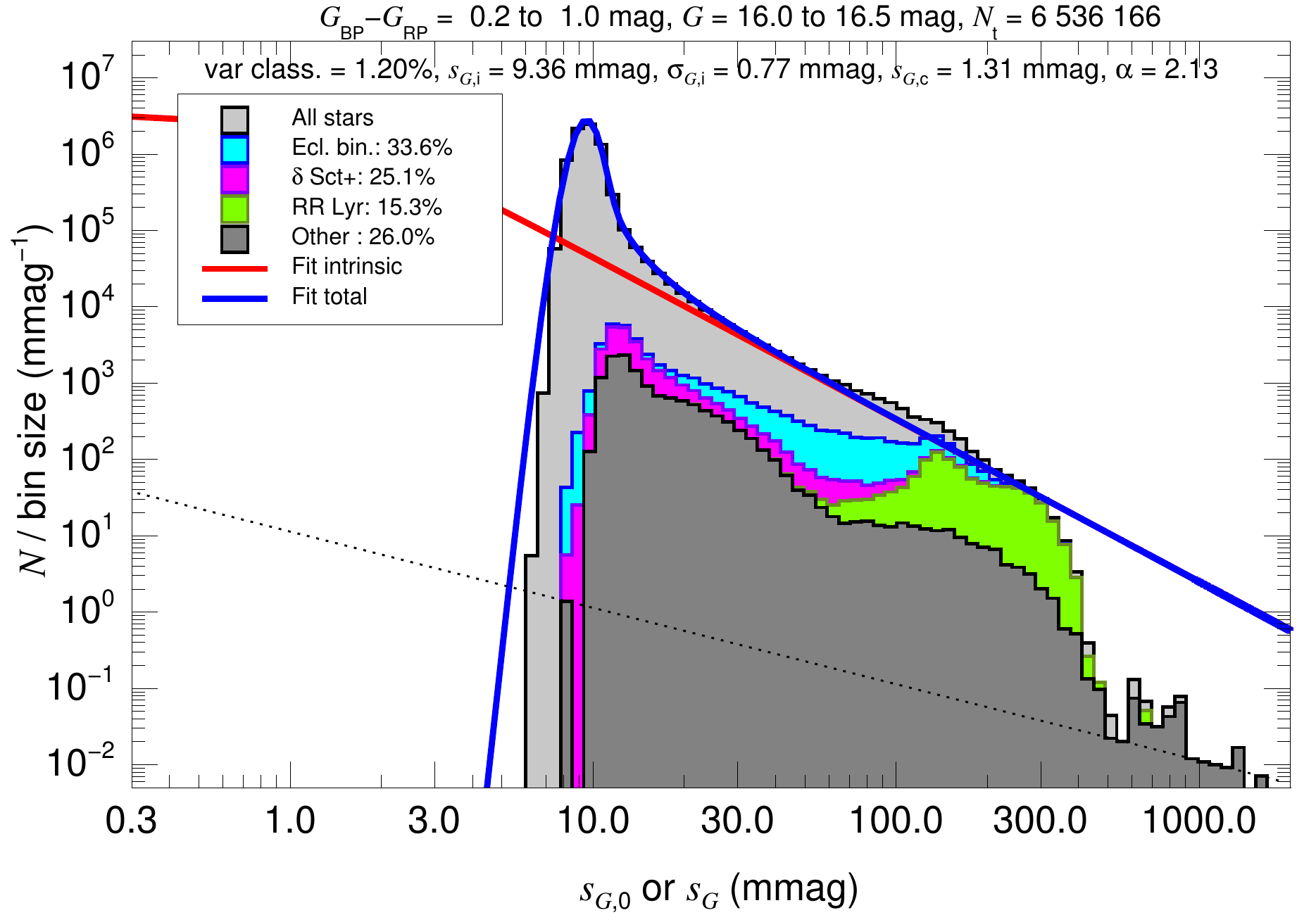}$\!\!\!$
                    \includegraphics[width=0.35\linewidth]{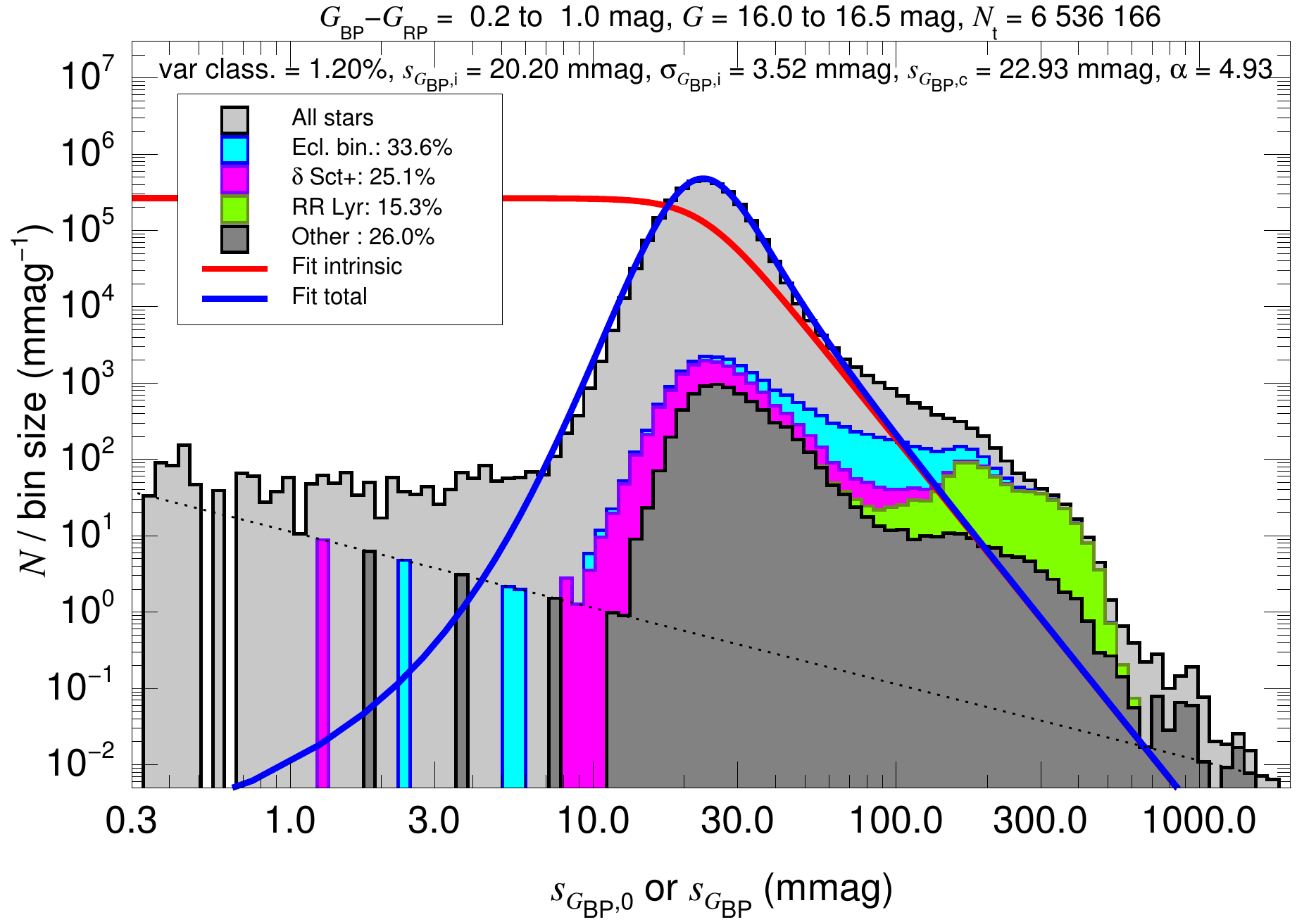}$\!\!\!$
                    \includegraphics[width=0.35\linewidth]{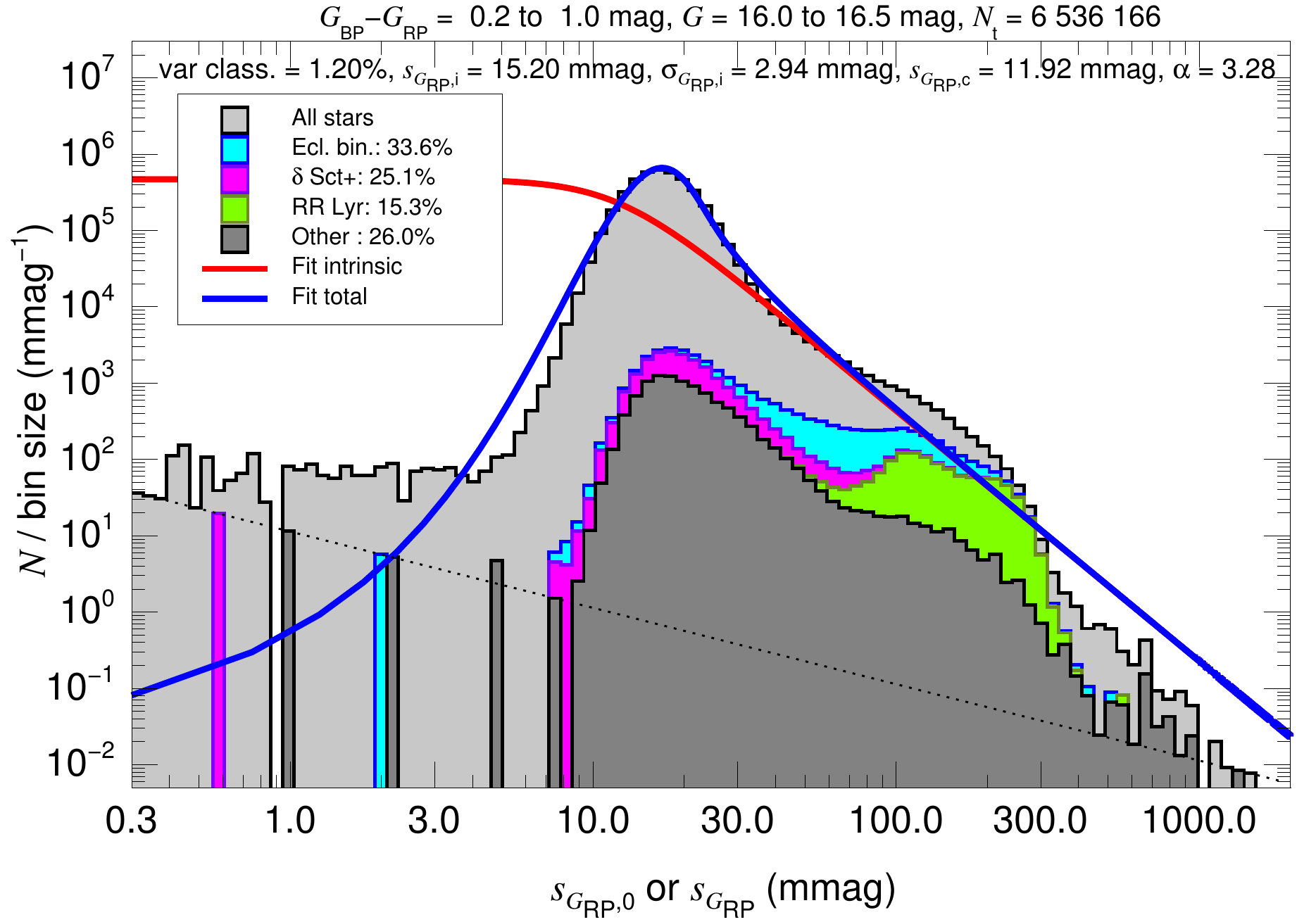}}
\centerline{$\!\!\!$\includegraphics[width=0.35\linewidth]{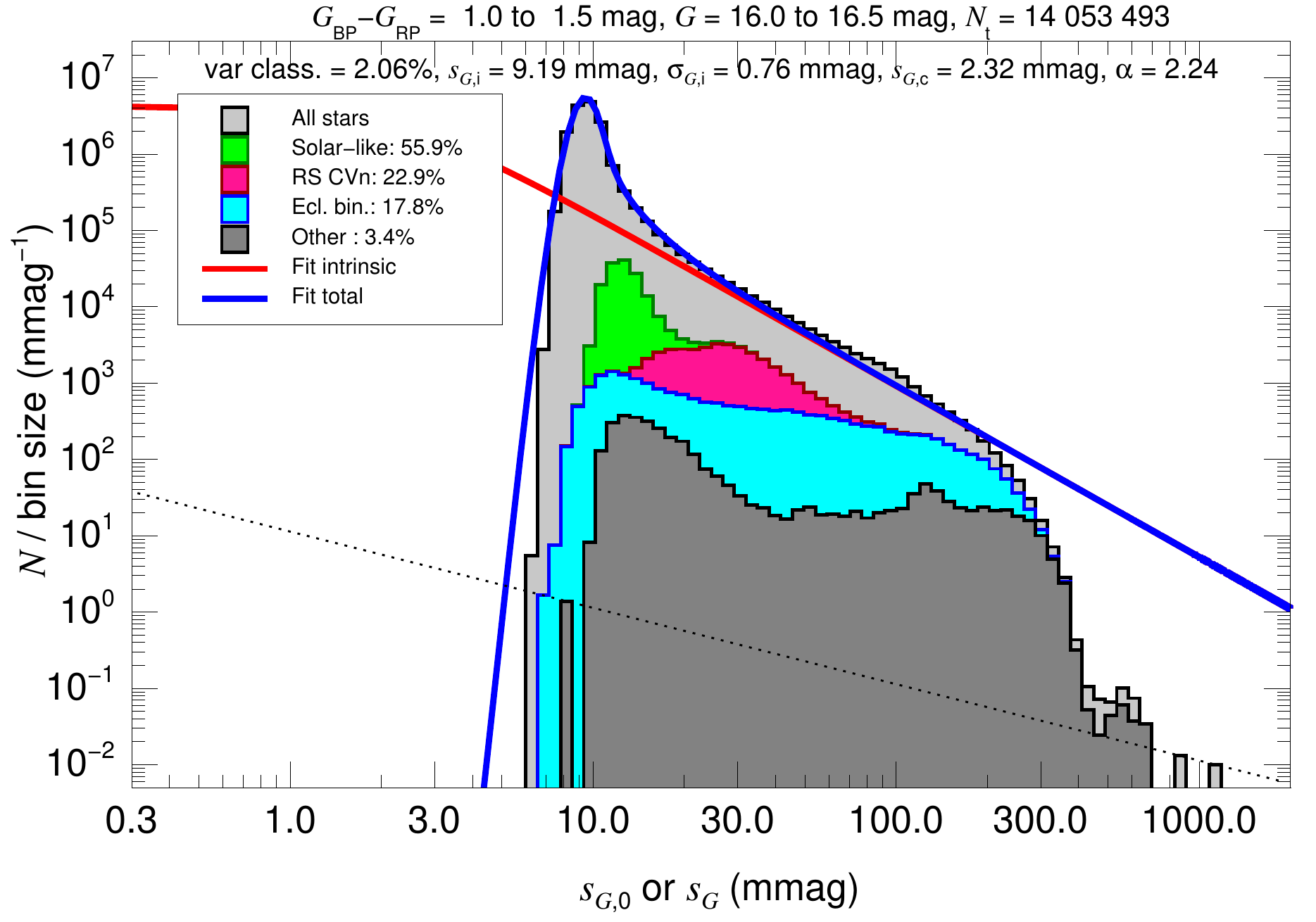}$\!\!\!$
                    \includegraphics[width=0.35\linewidth]{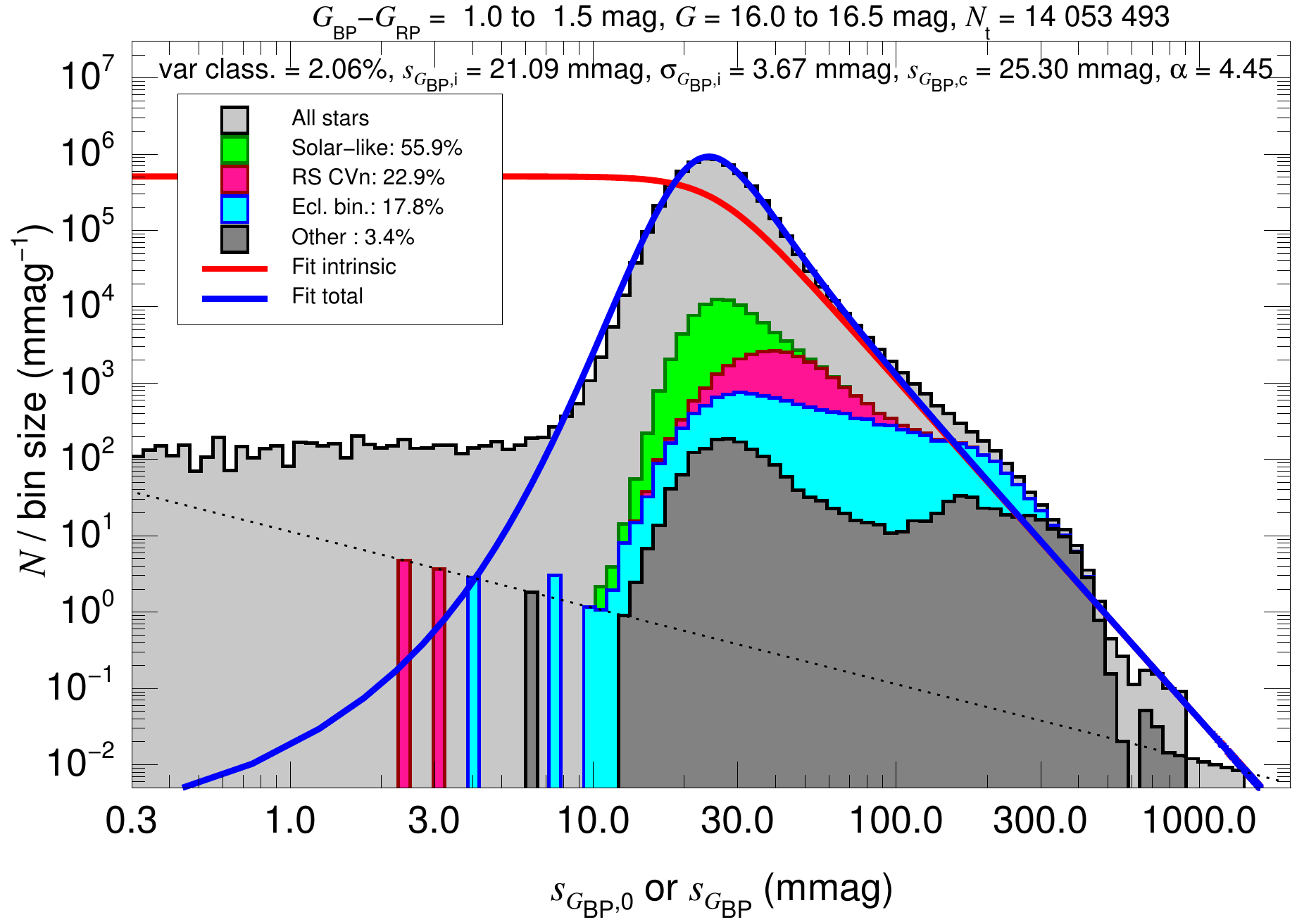}$\!\!\!$
                    \includegraphics[width=0.35\linewidth]{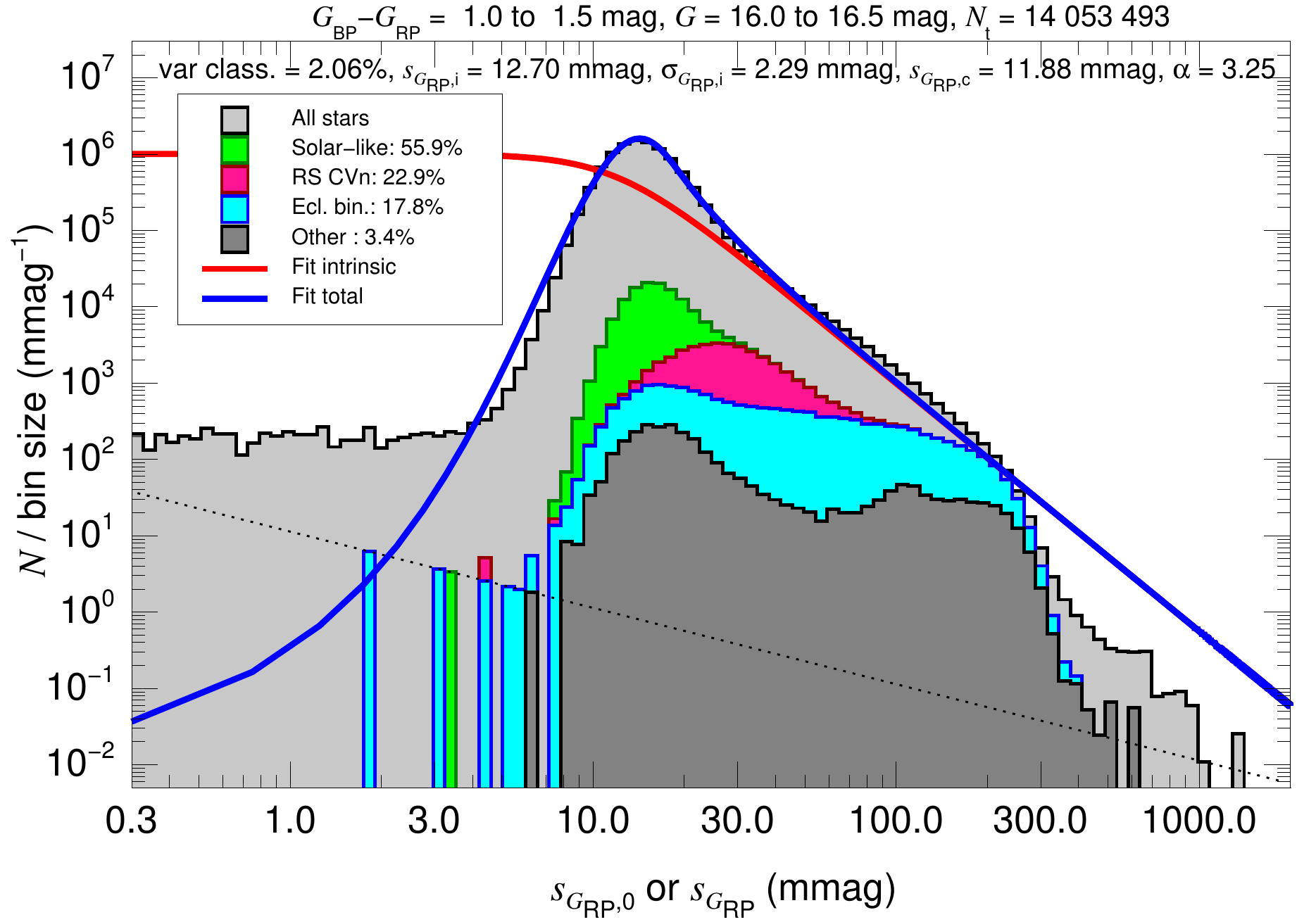}}
\centerline{$\!\!\!$\includegraphics[width=0.35\linewidth]{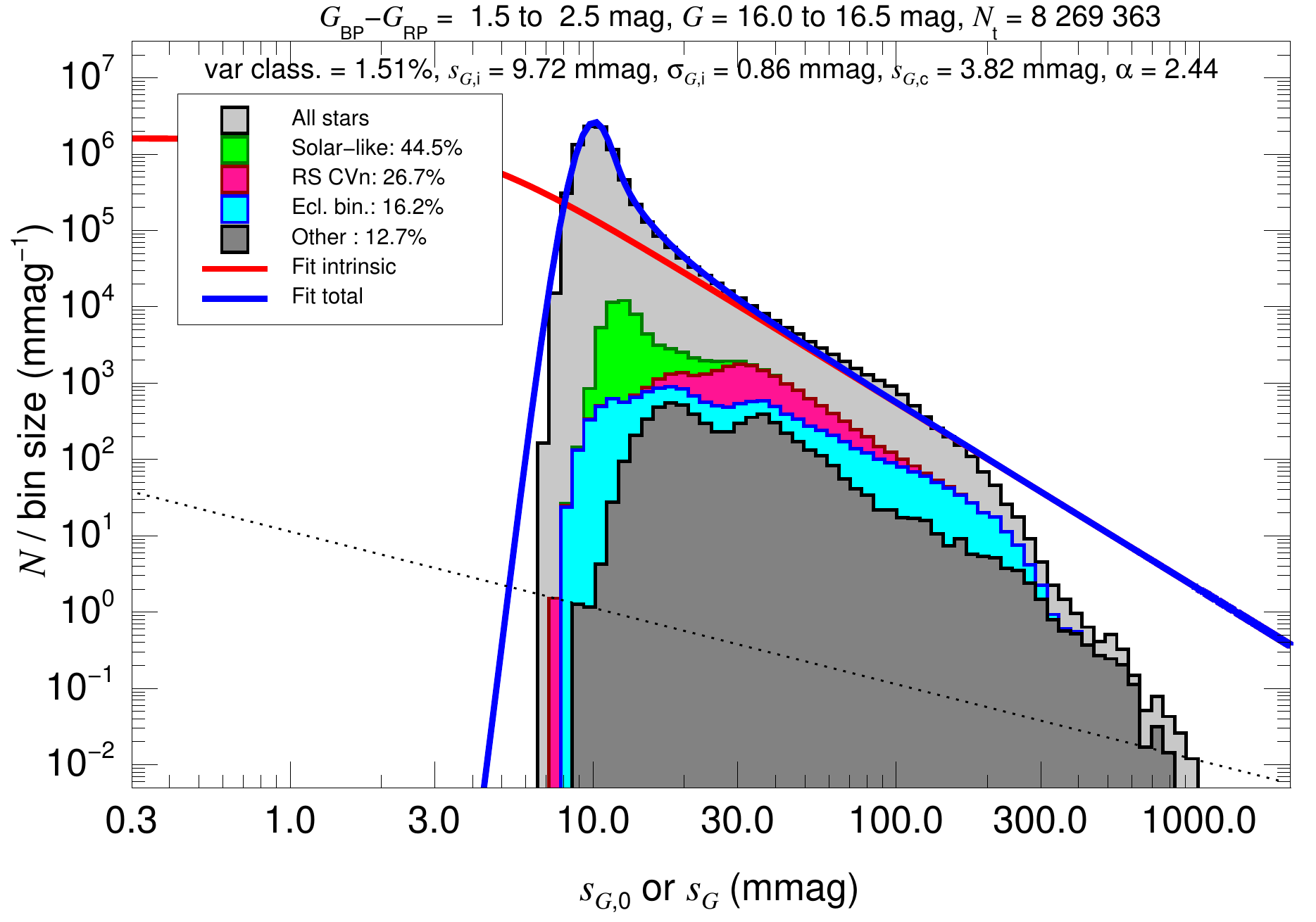}$\!\!\!$
                    \includegraphics[width=0.35\linewidth]{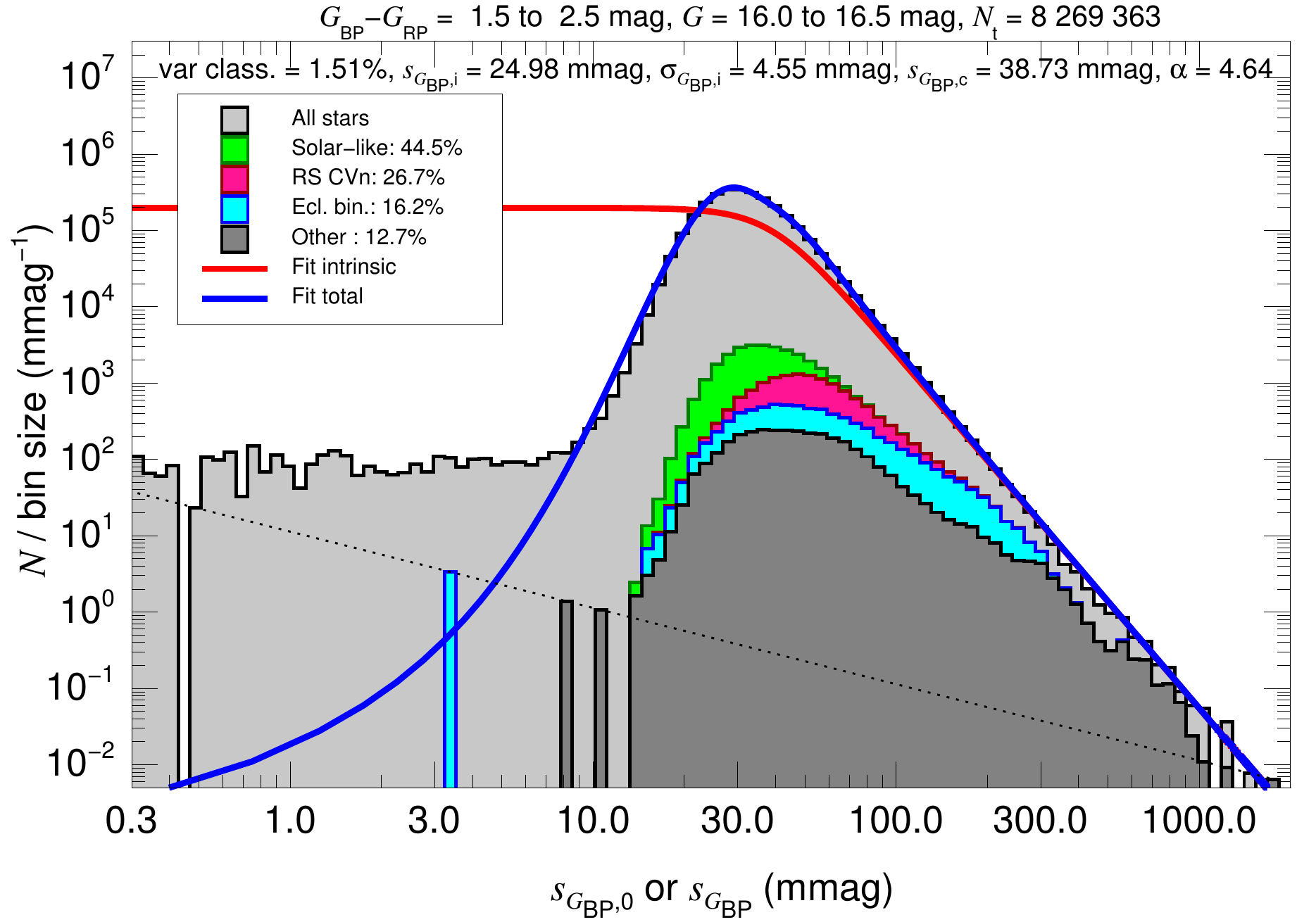}$\!\!\!$
                    \includegraphics[width=0.35\linewidth]{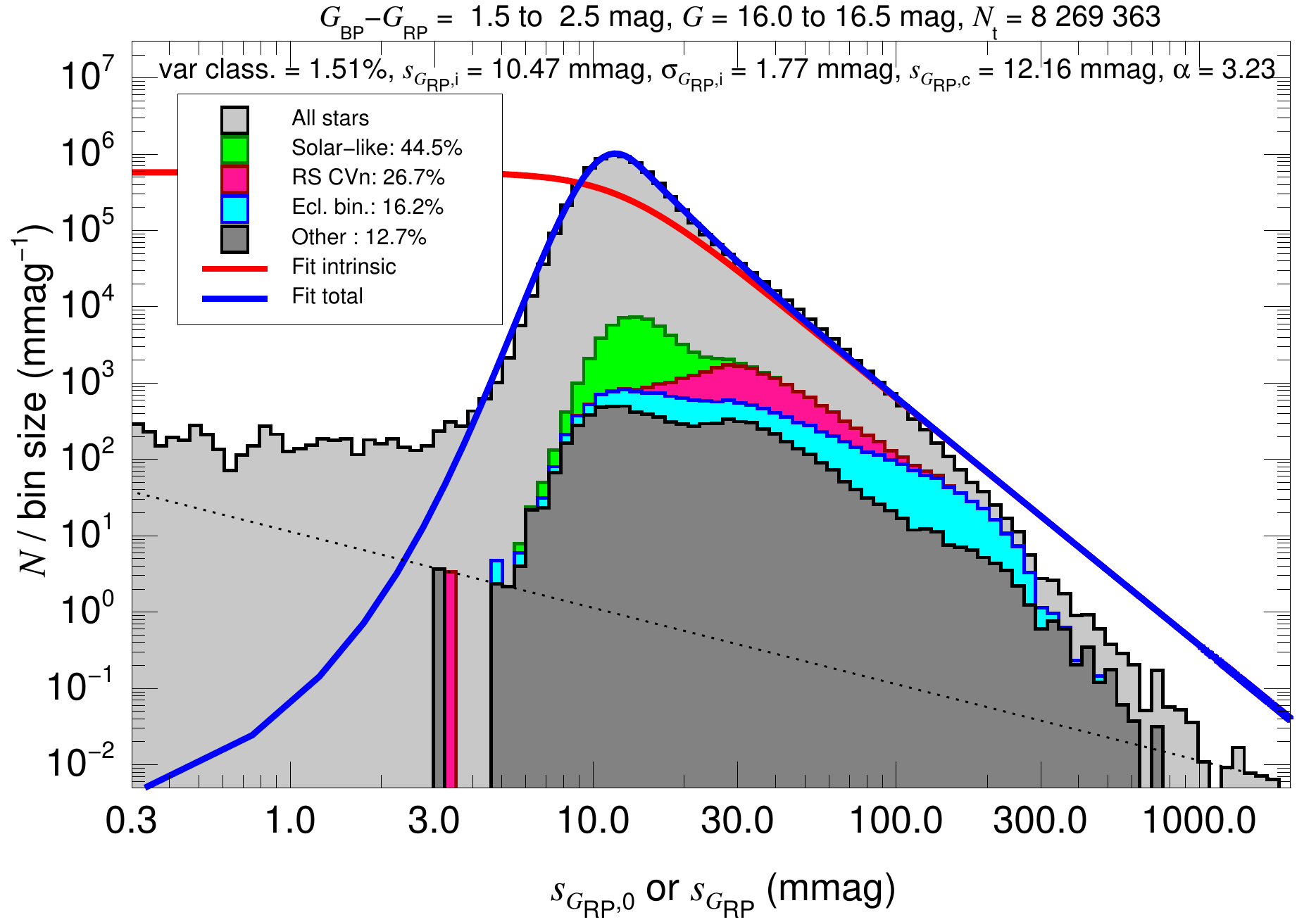}}
\centerline{$\!\!\!$\includegraphics[width=0.35\linewidth]{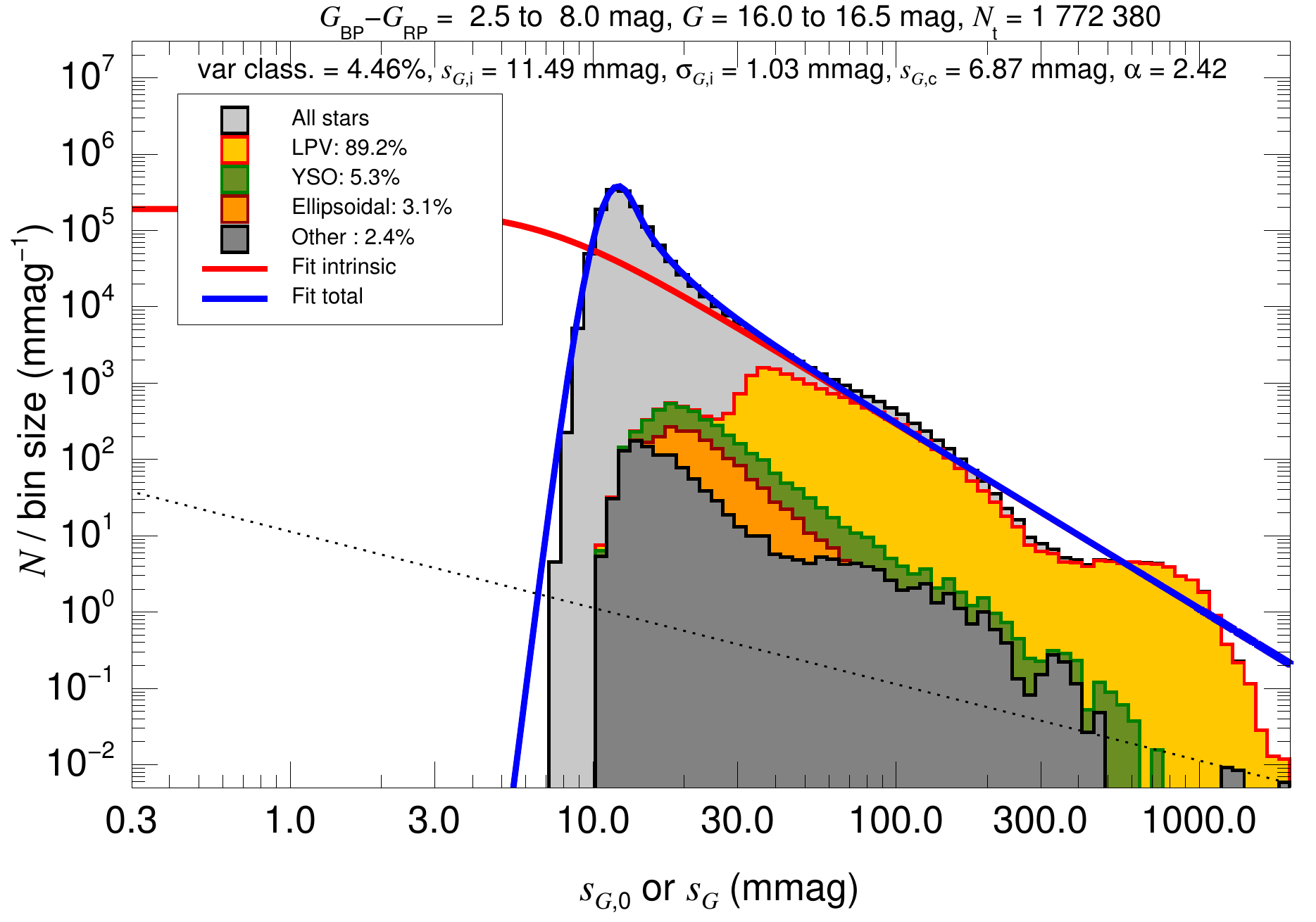}$\!\!\!$
                    \includegraphics[width=0.35\linewidth]{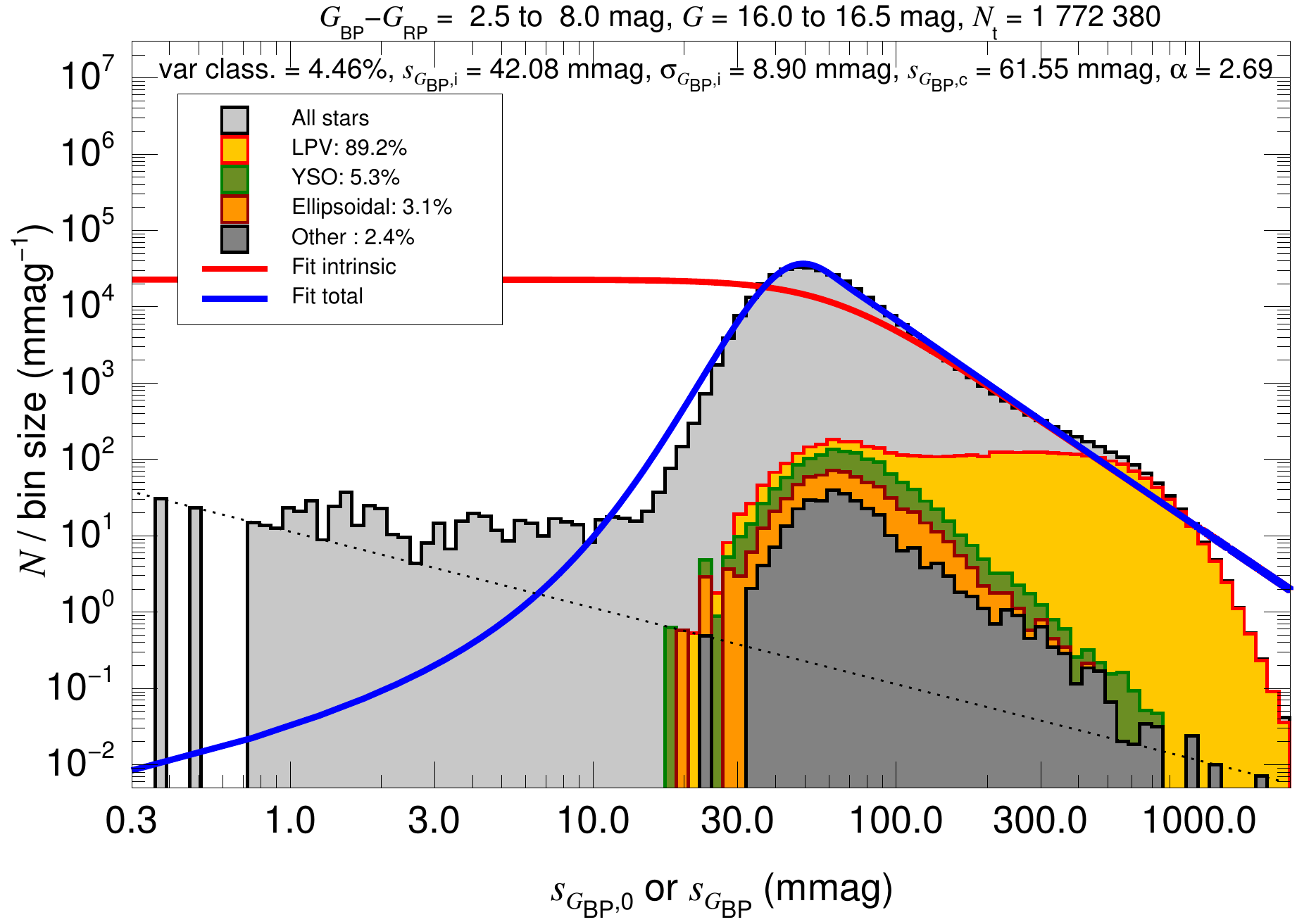}$\!\!\!$
                    \includegraphics[width=0.35\linewidth]{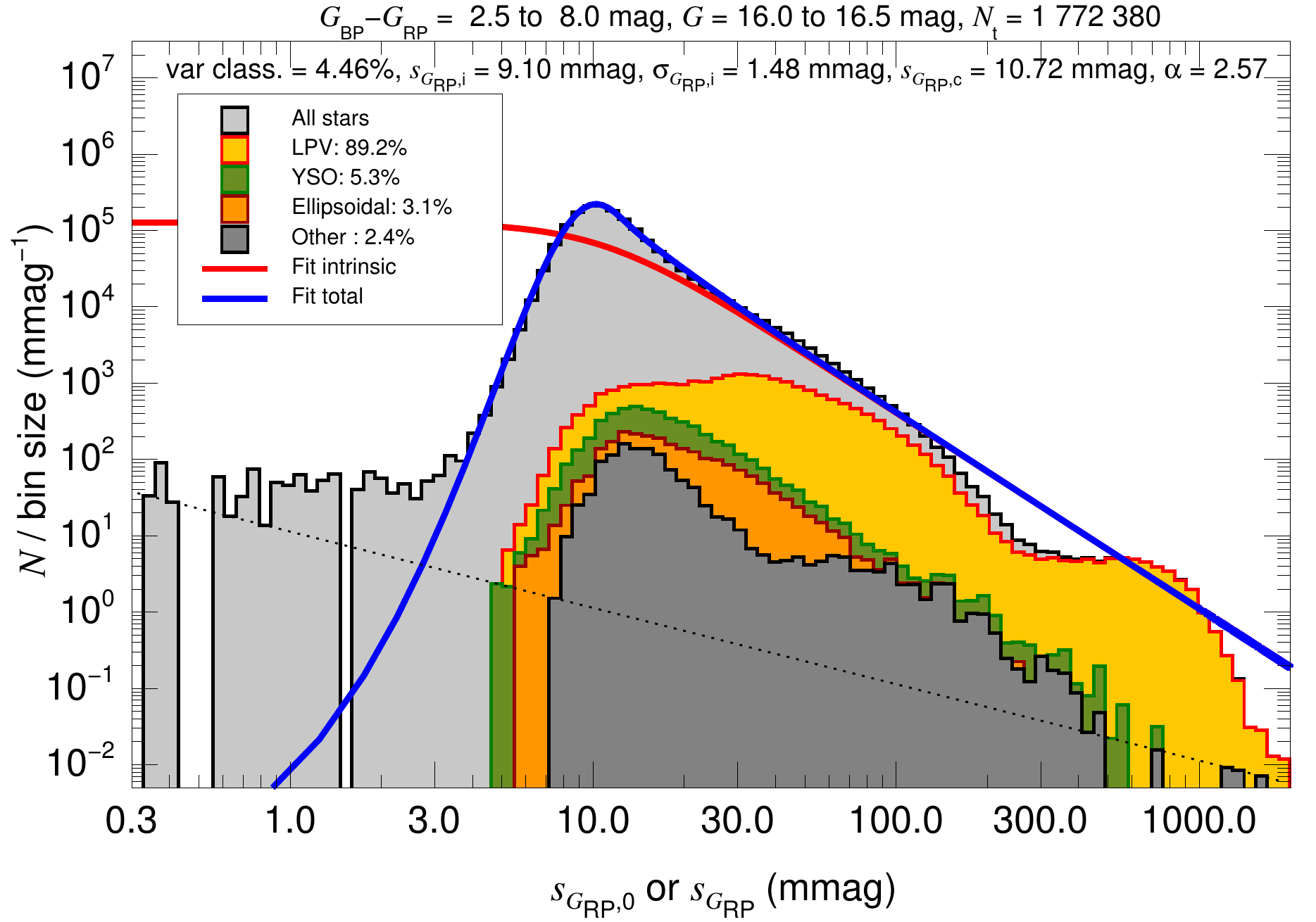}}
\caption{(Continued).}
\end{figure*}

\addtocounter{figure}{-1}

\begin{figure*}
\centerline{$\!\!\!$\includegraphics[width=0.35\linewidth]{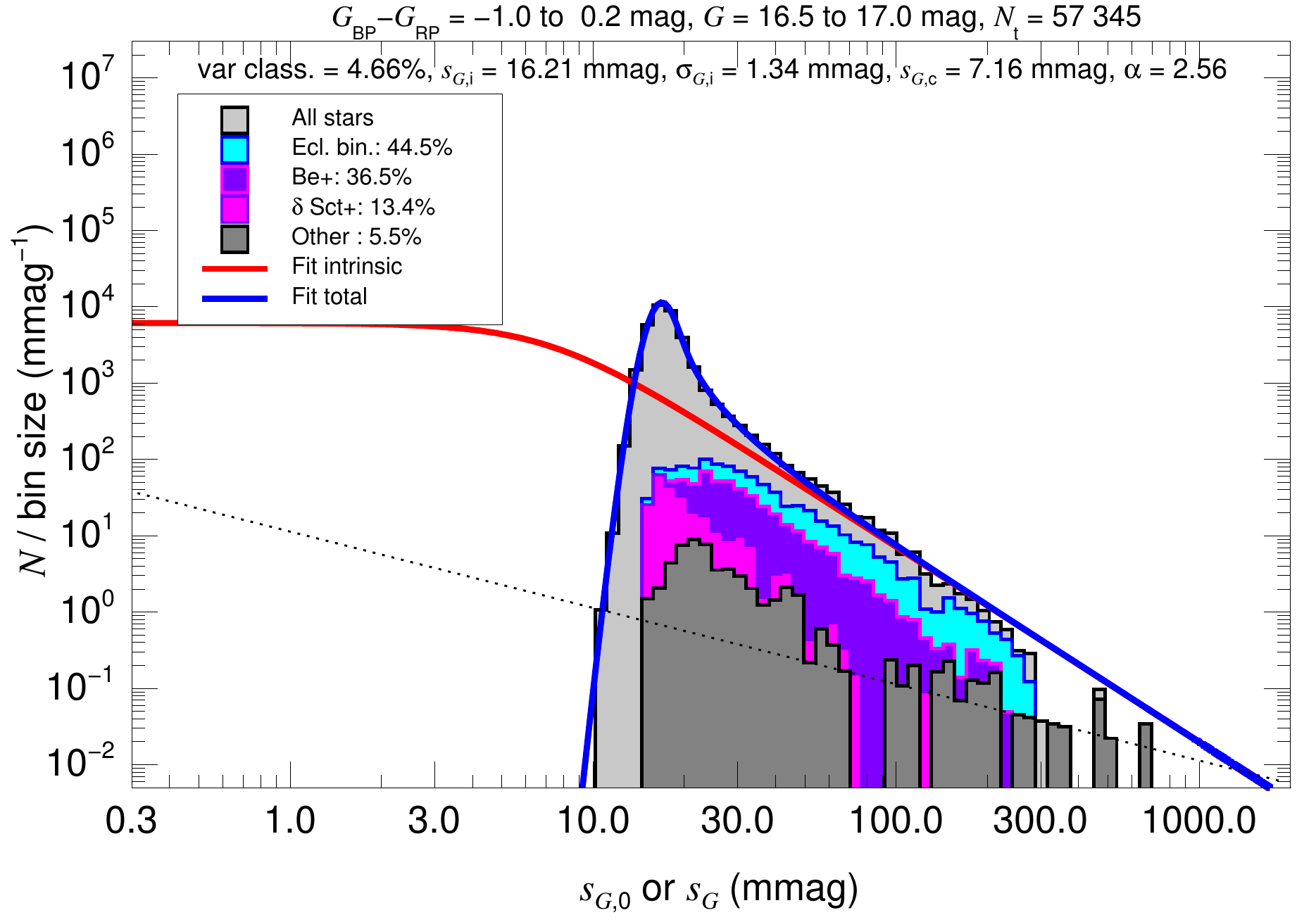}$\!\!\!$
                    \includegraphics[width=0.35\linewidth]{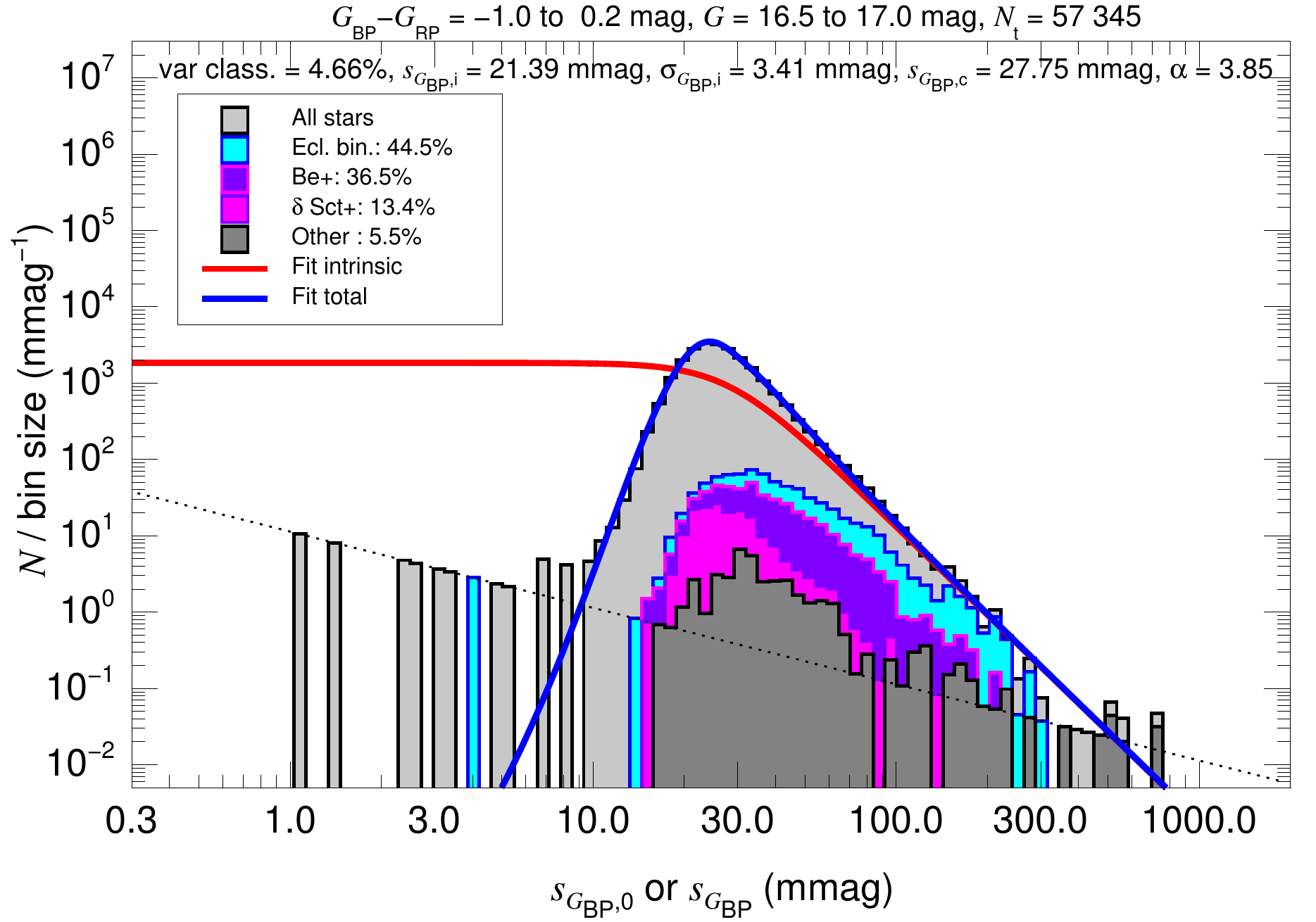}$\!\!\!$
                    \includegraphics[width=0.35\linewidth]{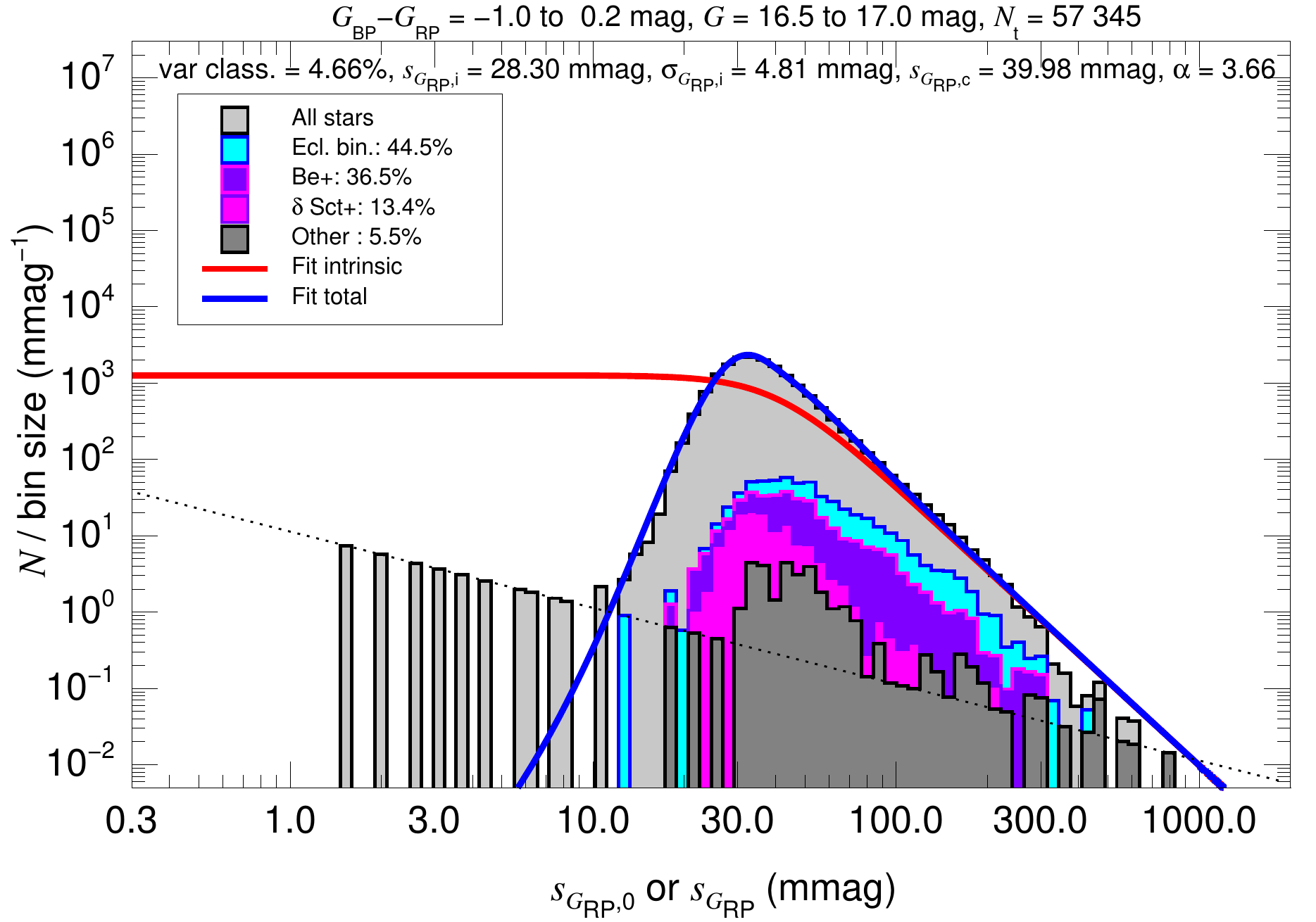}}
\centerline{$\!\!\!$\includegraphics[width=0.35\linewidth]{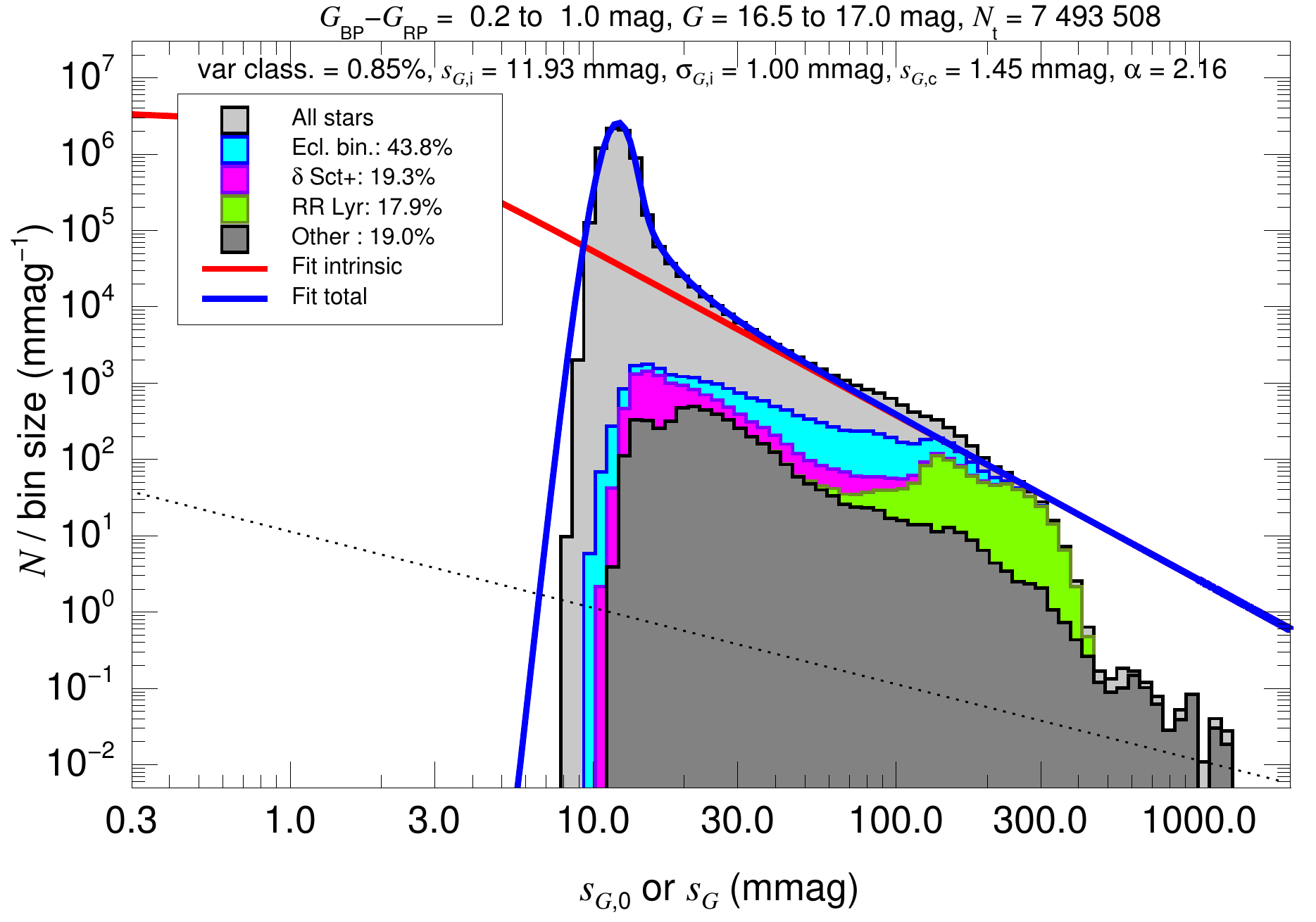}$\!\!\!$
                    \includegraphics[width=0.35\linewidth]{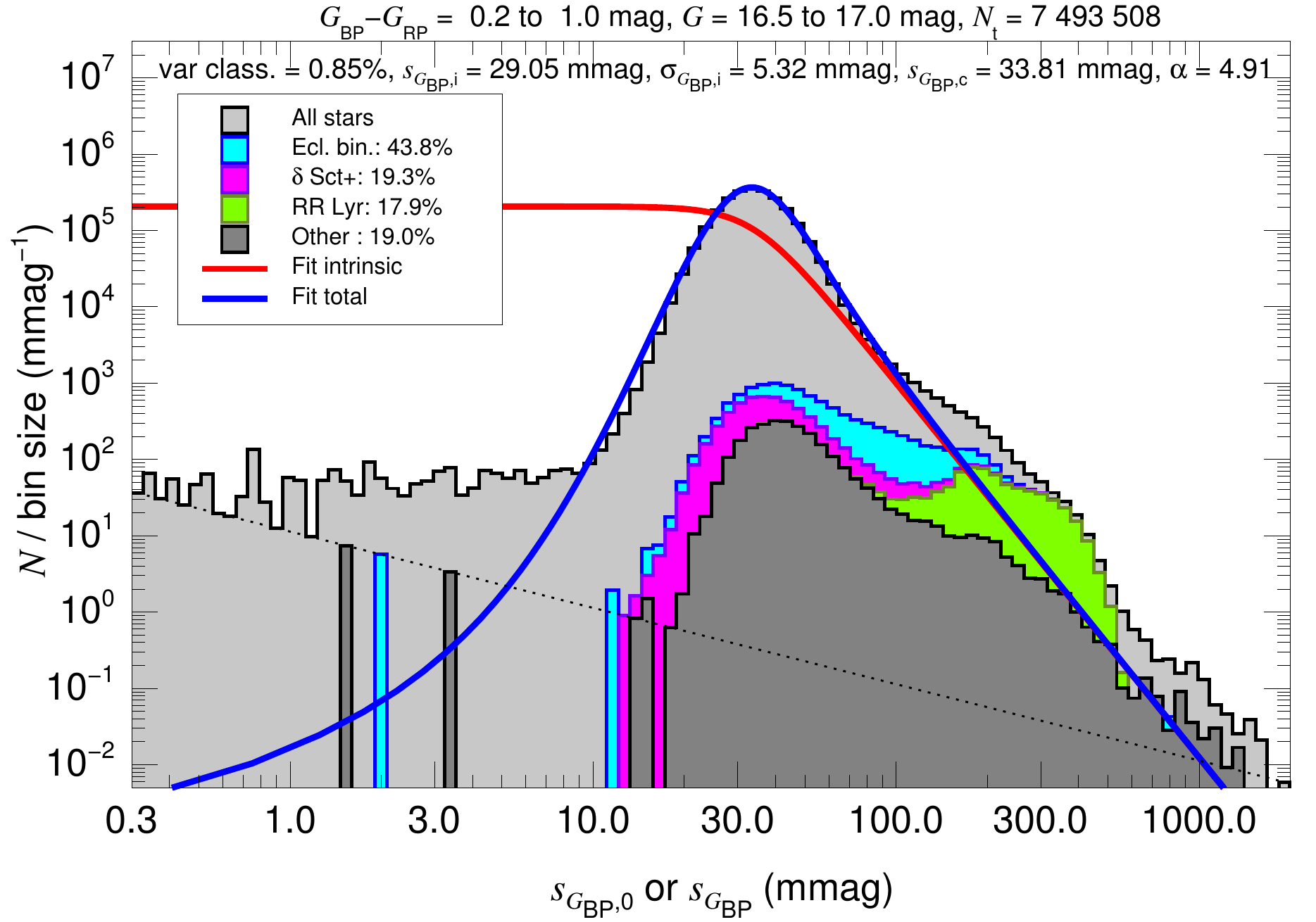}$\!\!\!$
                    \includegraphics[width=0.35\linewidth]{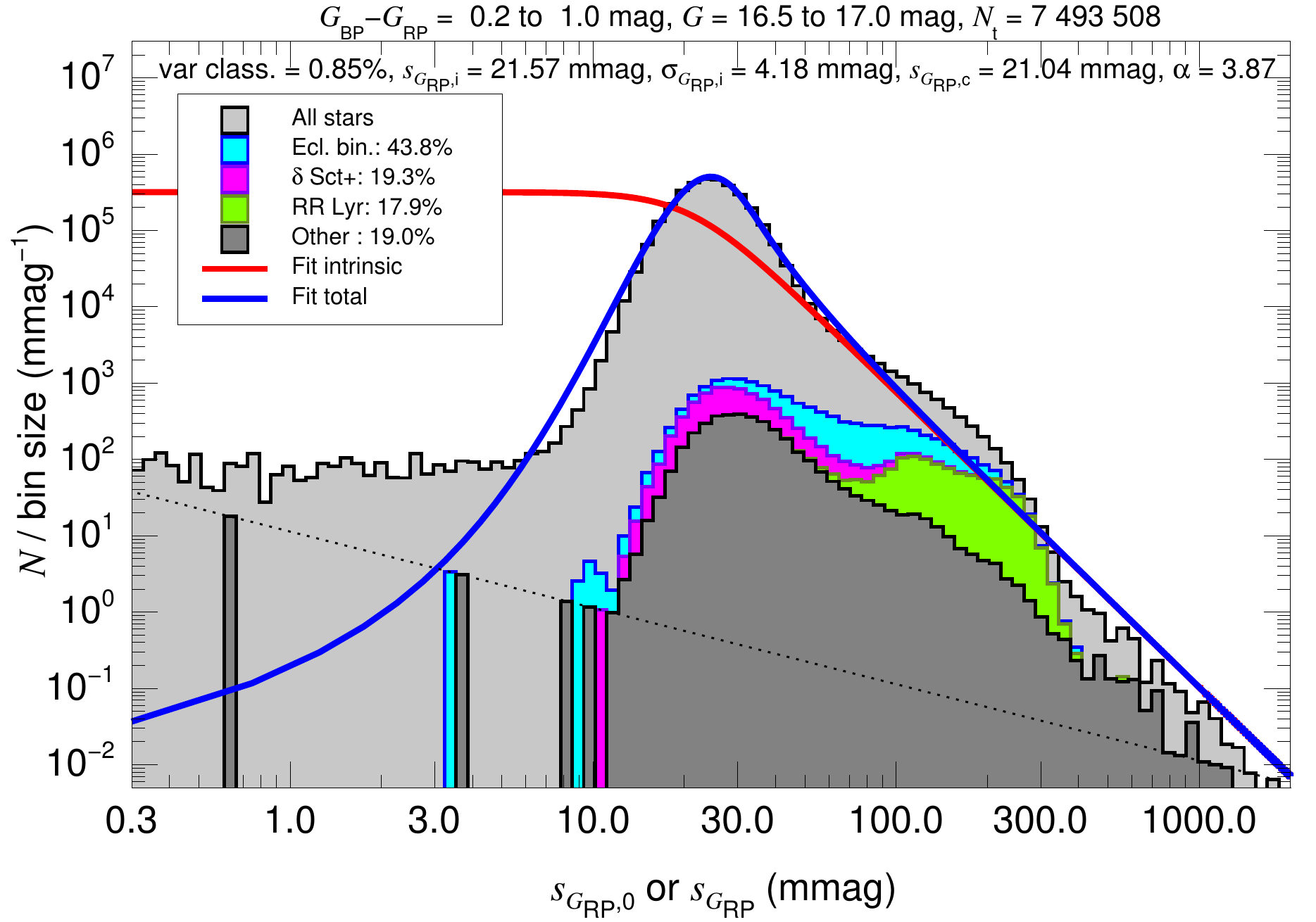}}
\centerline{$\!\!\!$\includegraphics[width=0.35\linewidth]{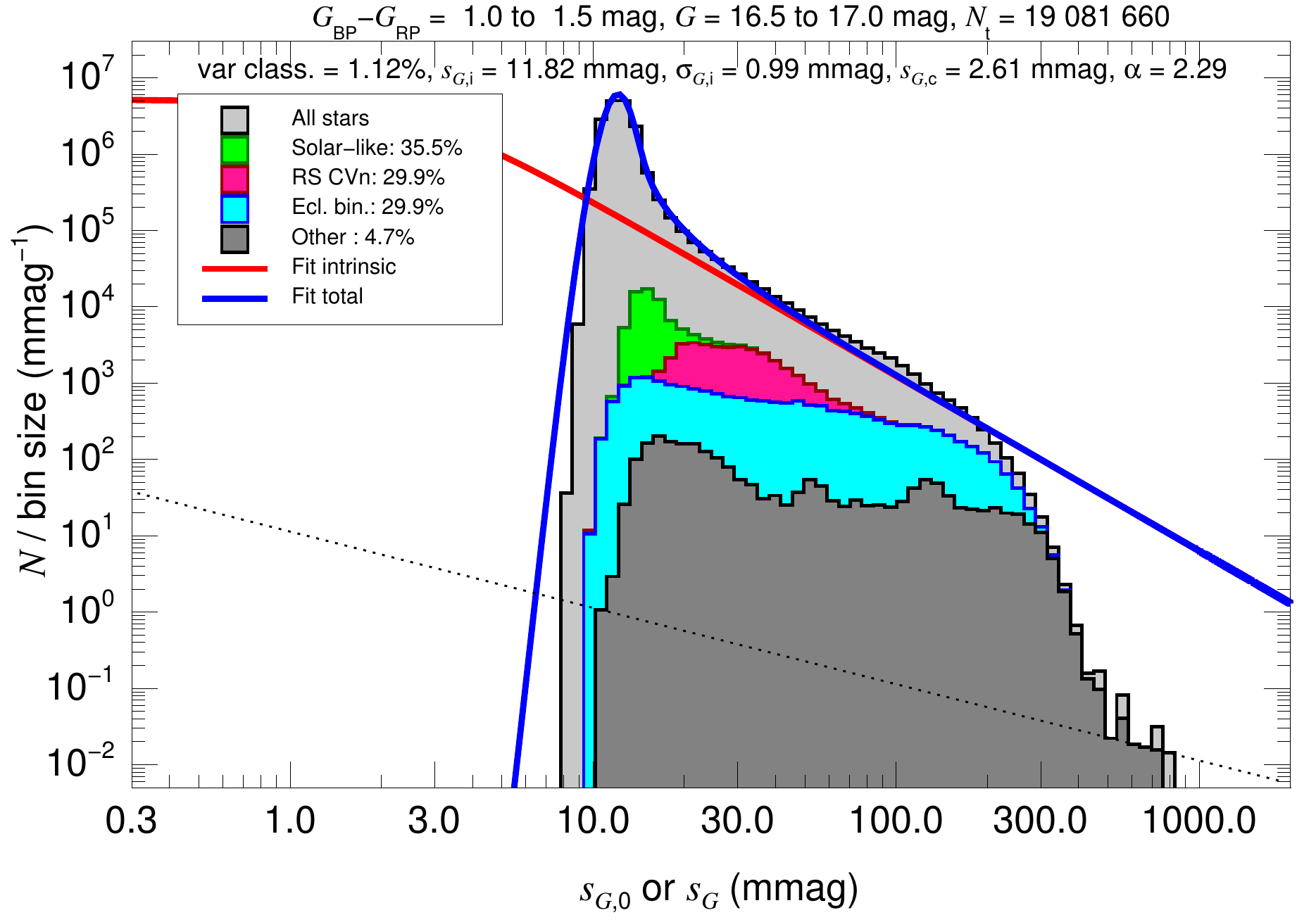}$\!\!\!$
                    \includegraphics[width=0.35\linewidth]{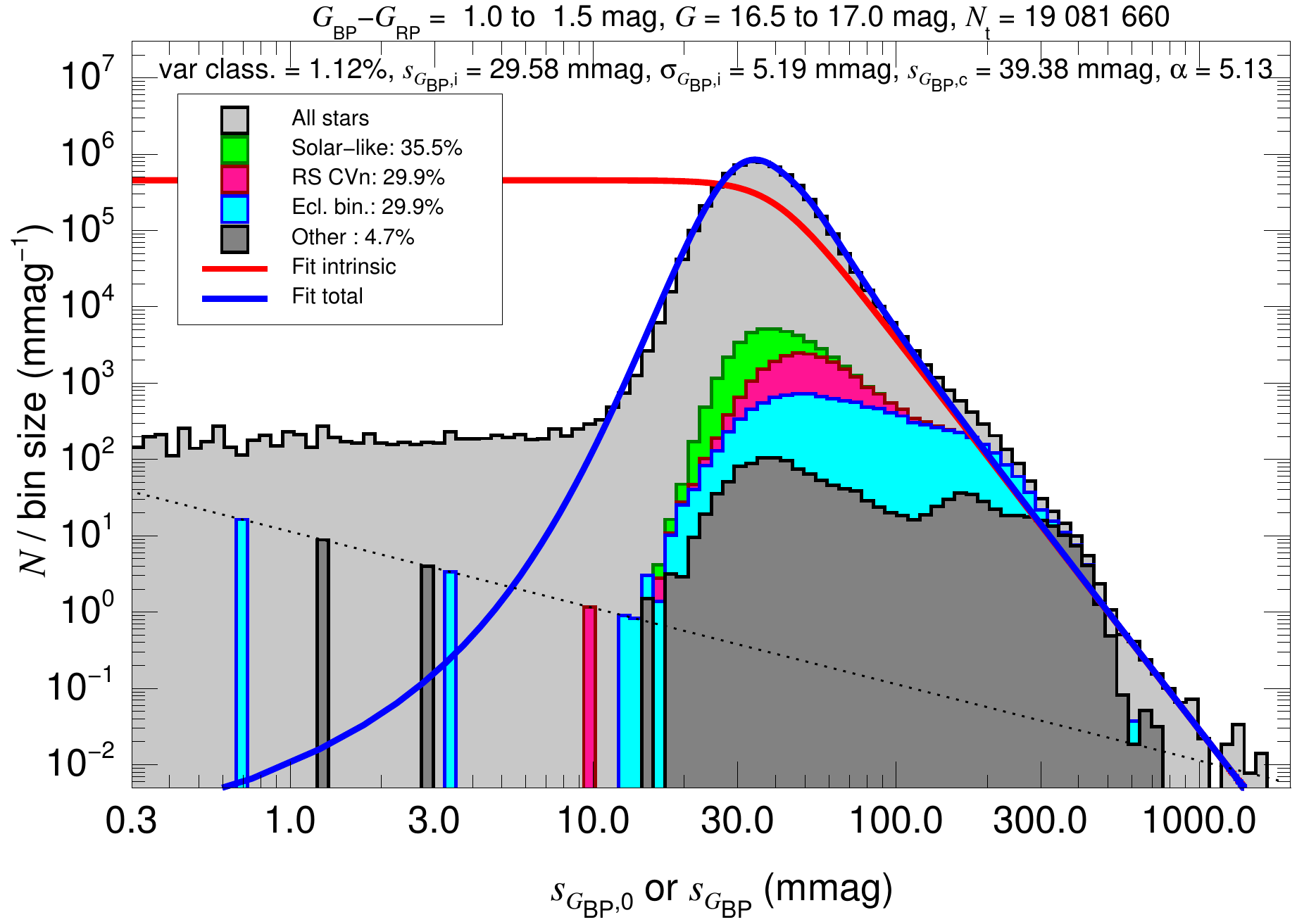}$\!\!\!$
                    \includegraphics[width=0.35\linewidth]{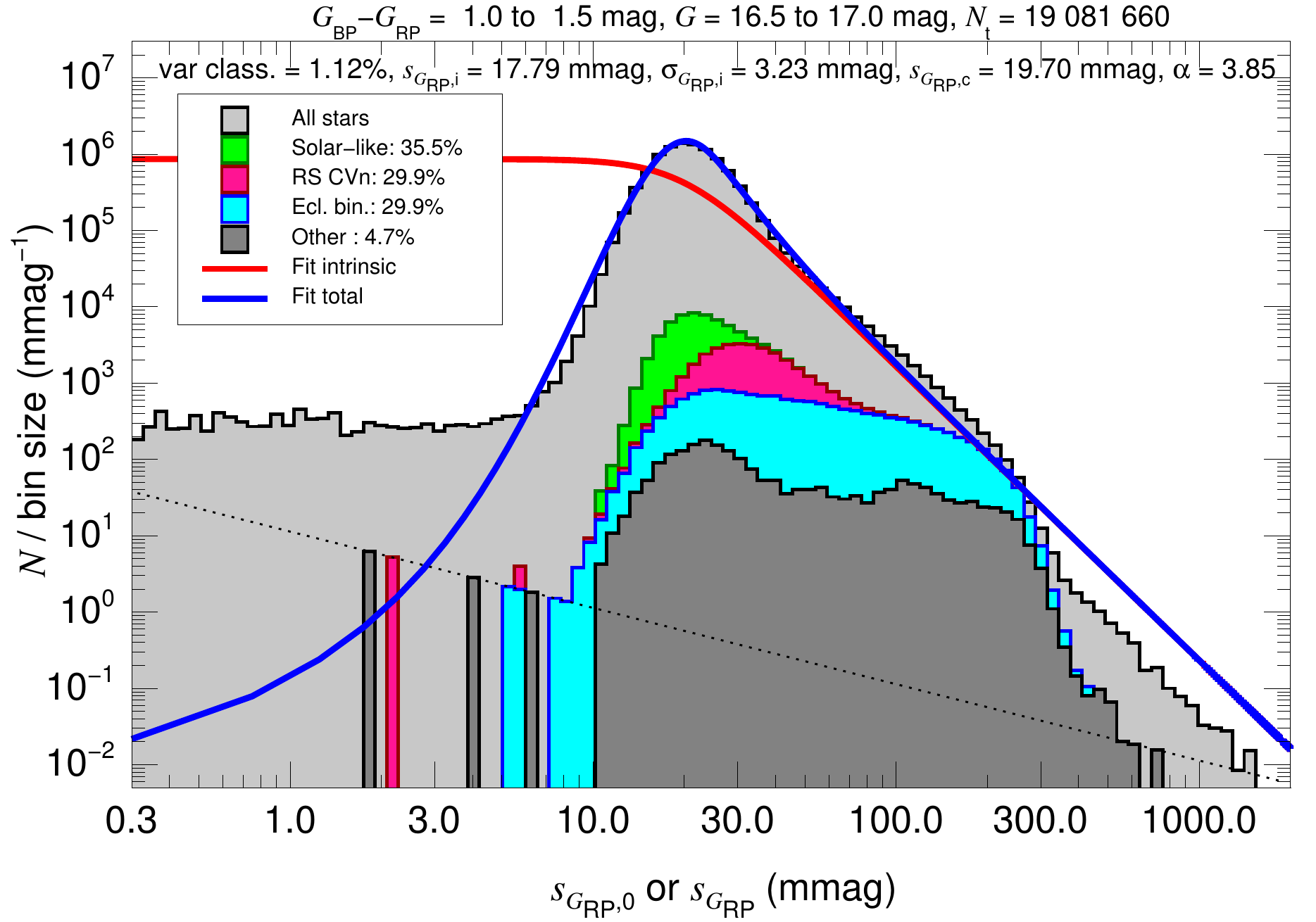}}
\centerline{$\!\!\!$\includegraphics[width=0.35\linewidth]{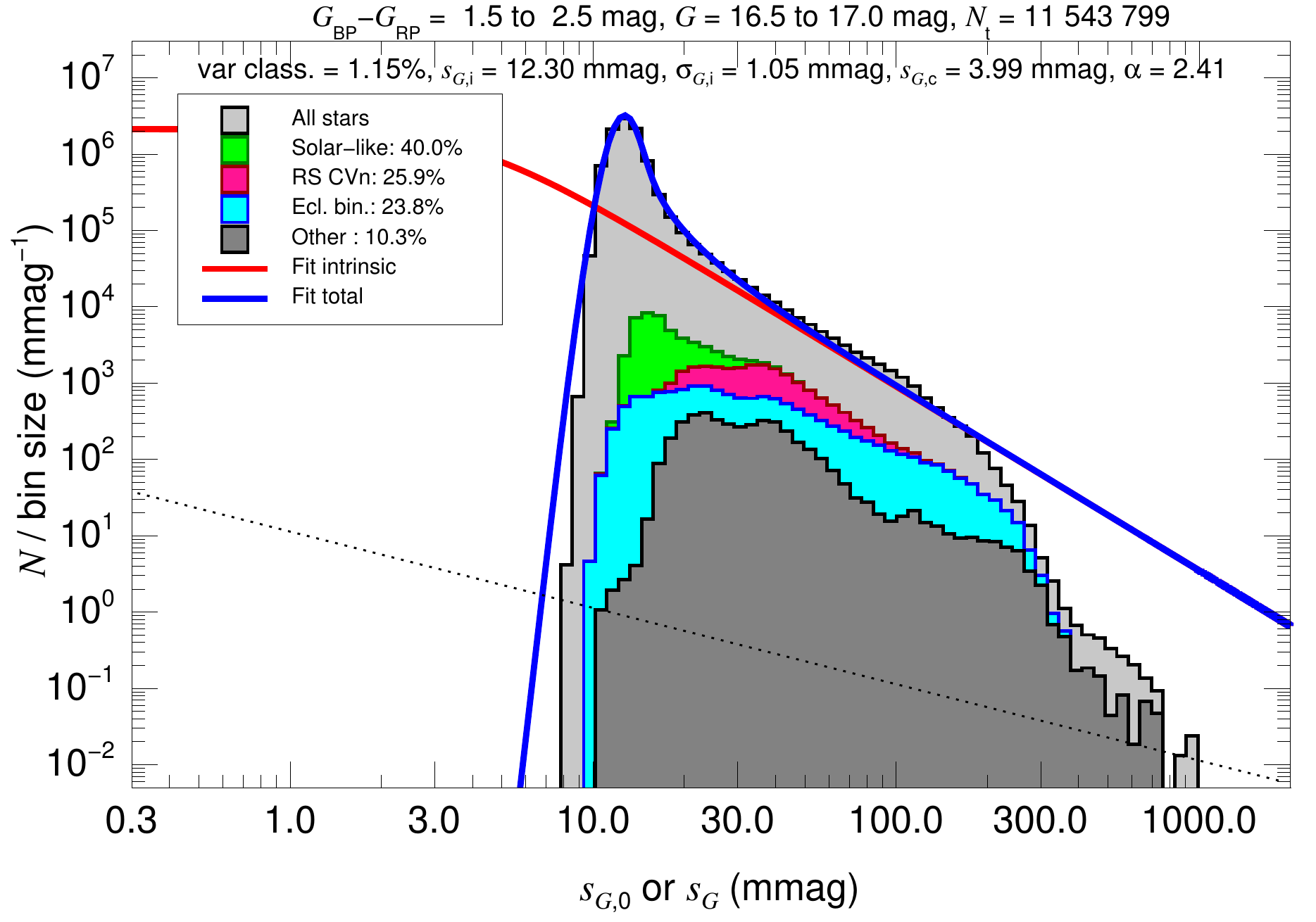}$\!\!\!$
                    \includegraphics[width=0.35\linewidth]{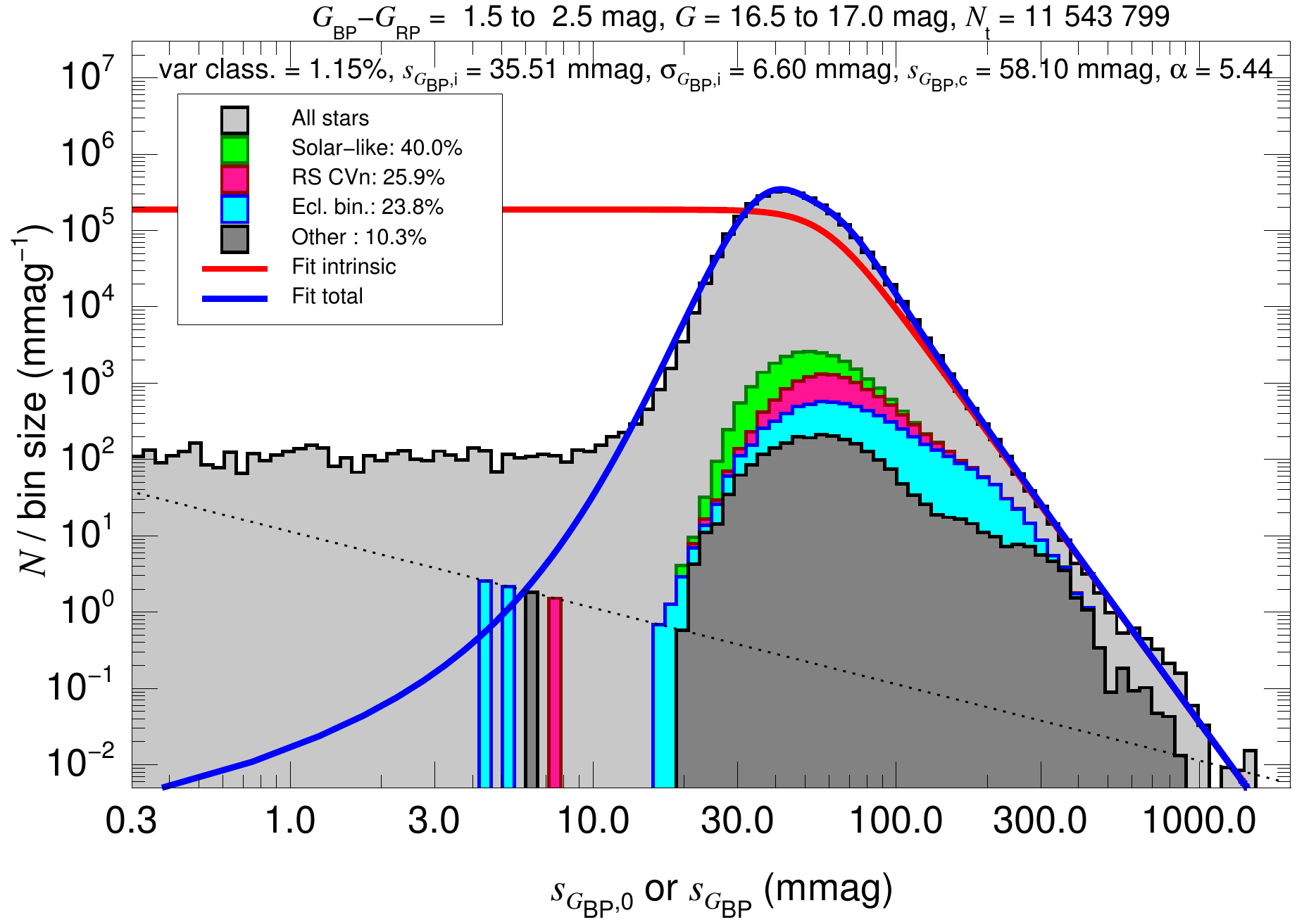}$\!\!\!$
                    \includegraphics[width=0.35\linewidth]{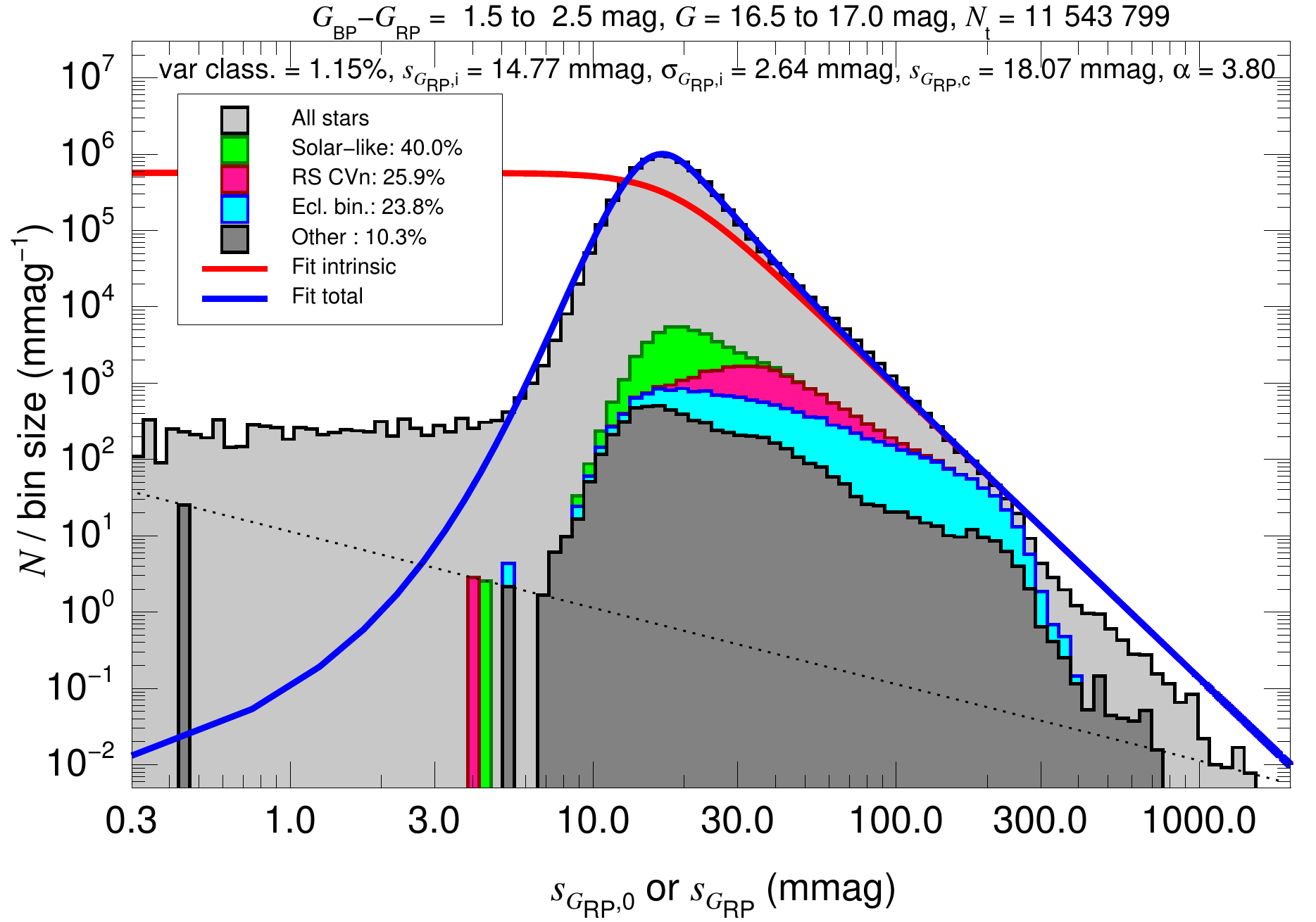}}
\centerline{$\!\!\!$\includegraphics[width=0.35\linewidth]{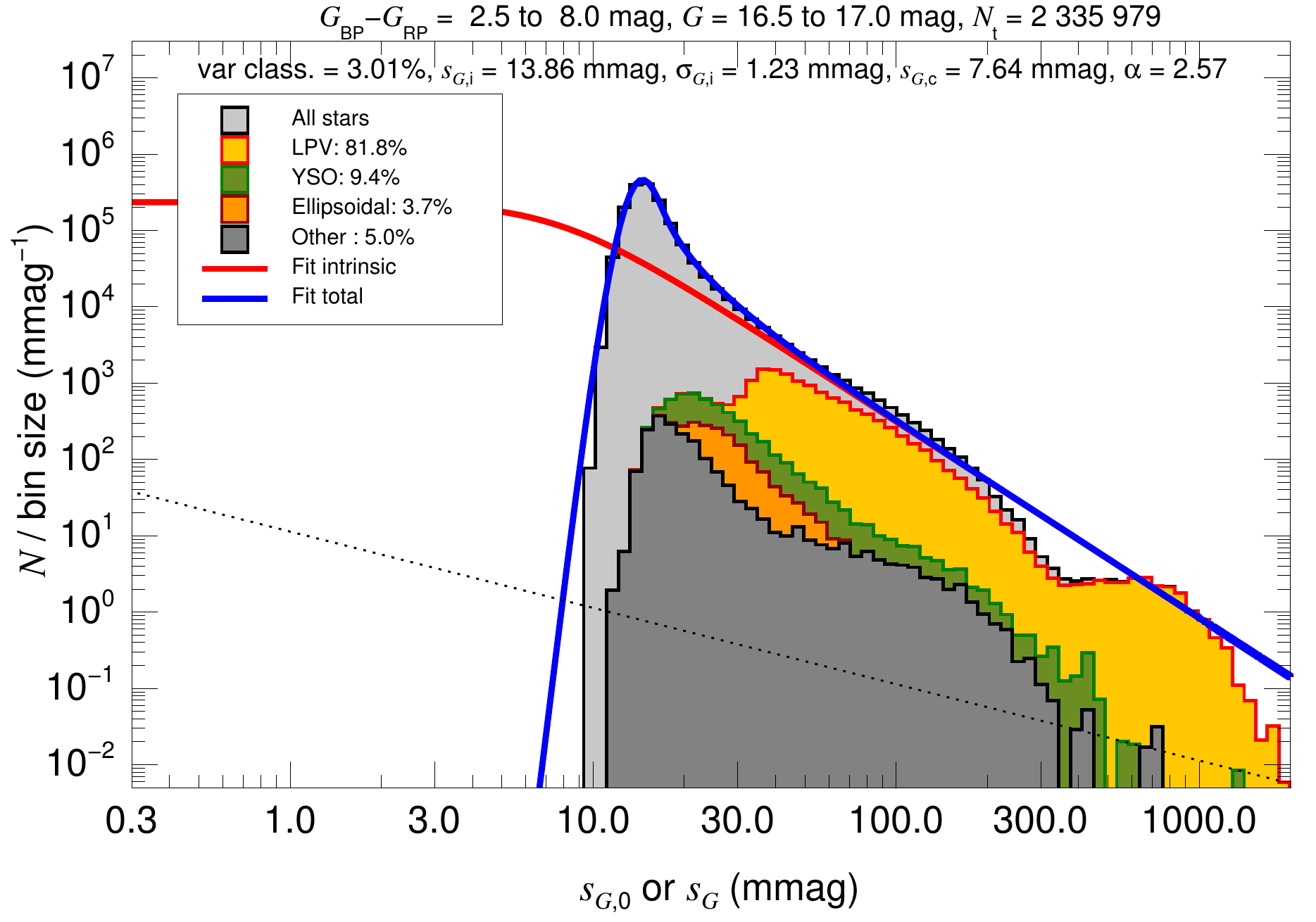}$\!\!\!$
                    \includegraphics[width=0.35\linewidth]{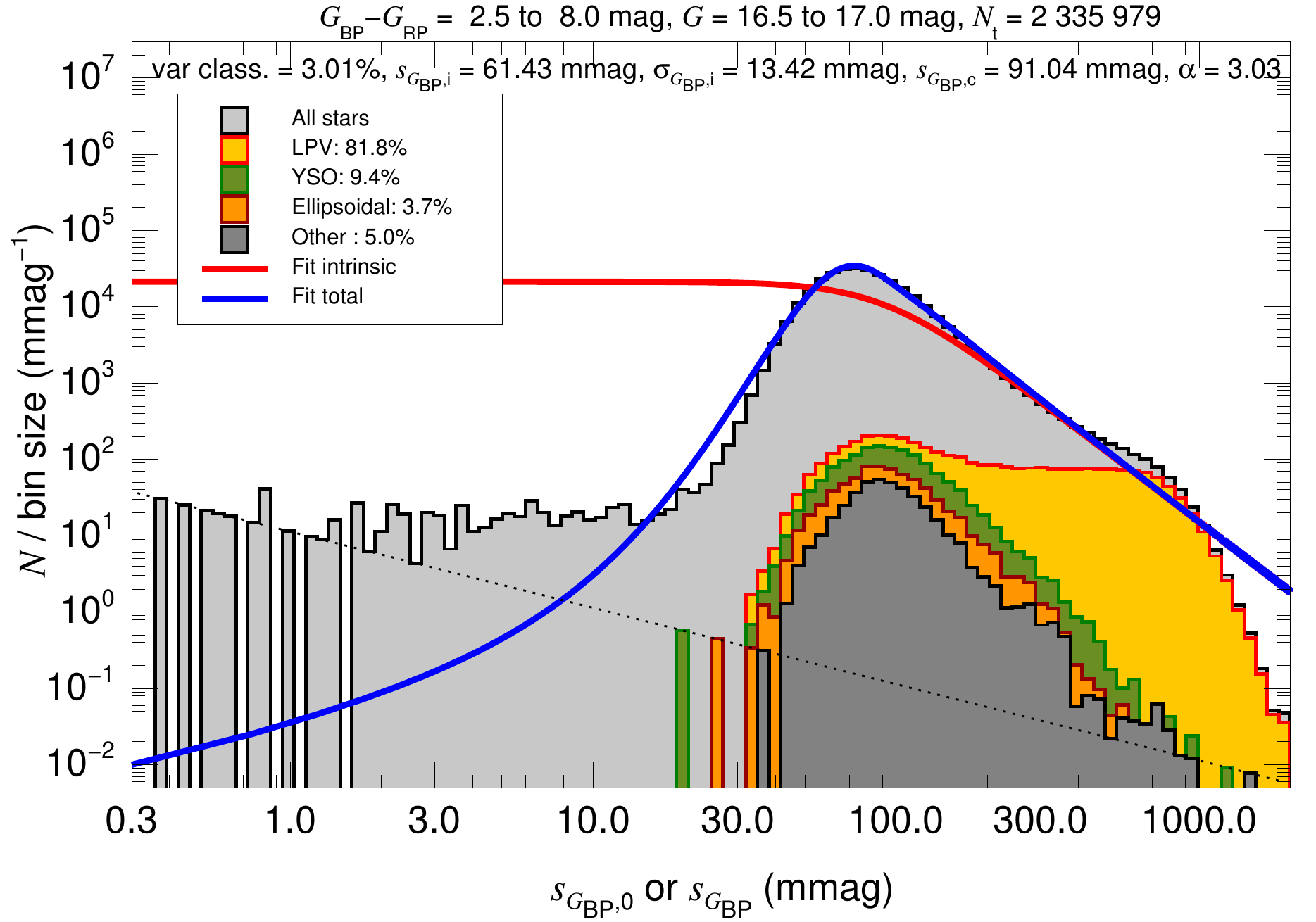}$\!\!\!$
                    \includegraphics[width=0.35\linewidth]{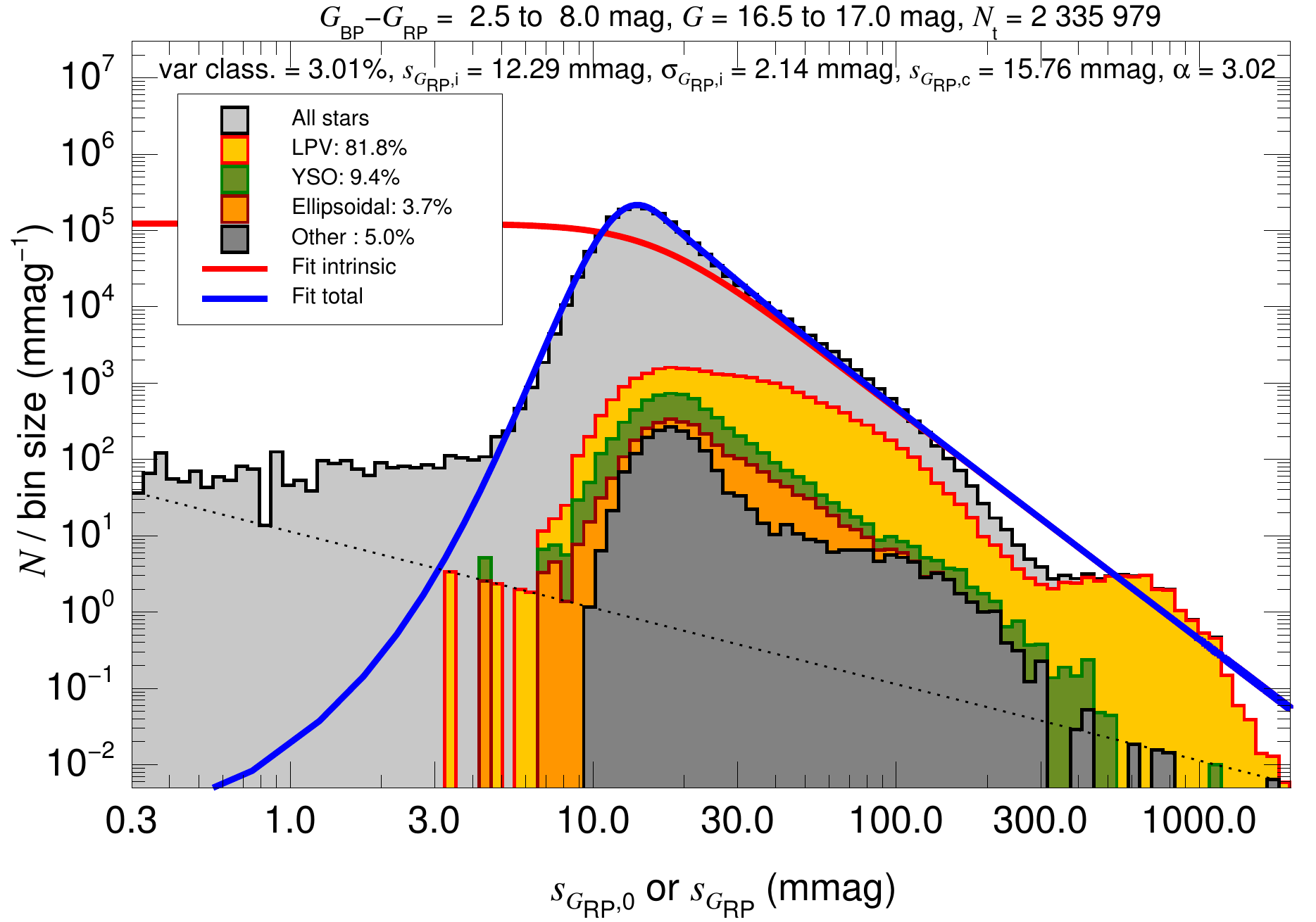}}
\caption{(Continued).}
\end{figure*}

\clearpage

\begin{figure*}
\centerline{\includegraphics[width=0.35\linewidth]{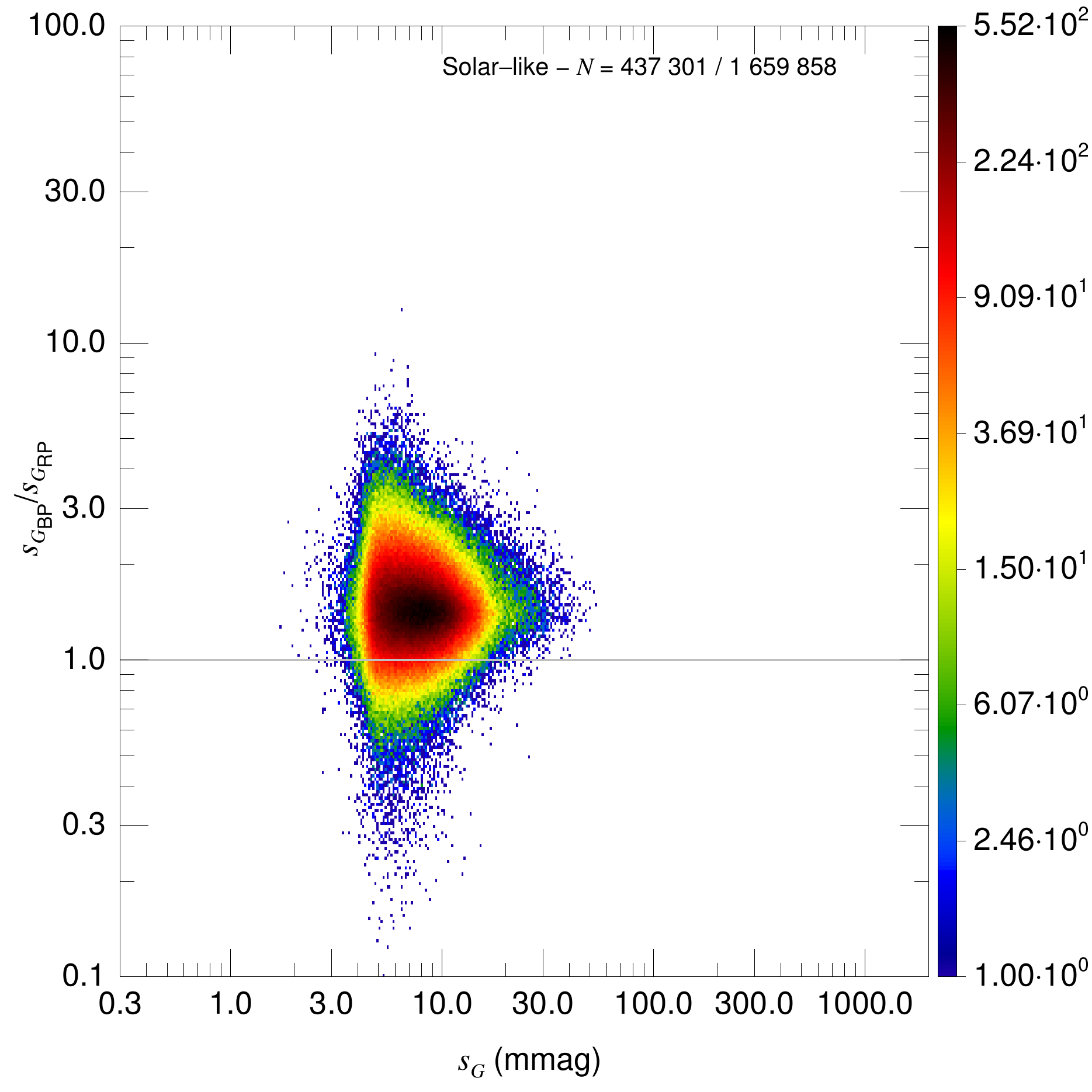}$\!\!\!$
            \includegraphics[width=0.35\linewidth]{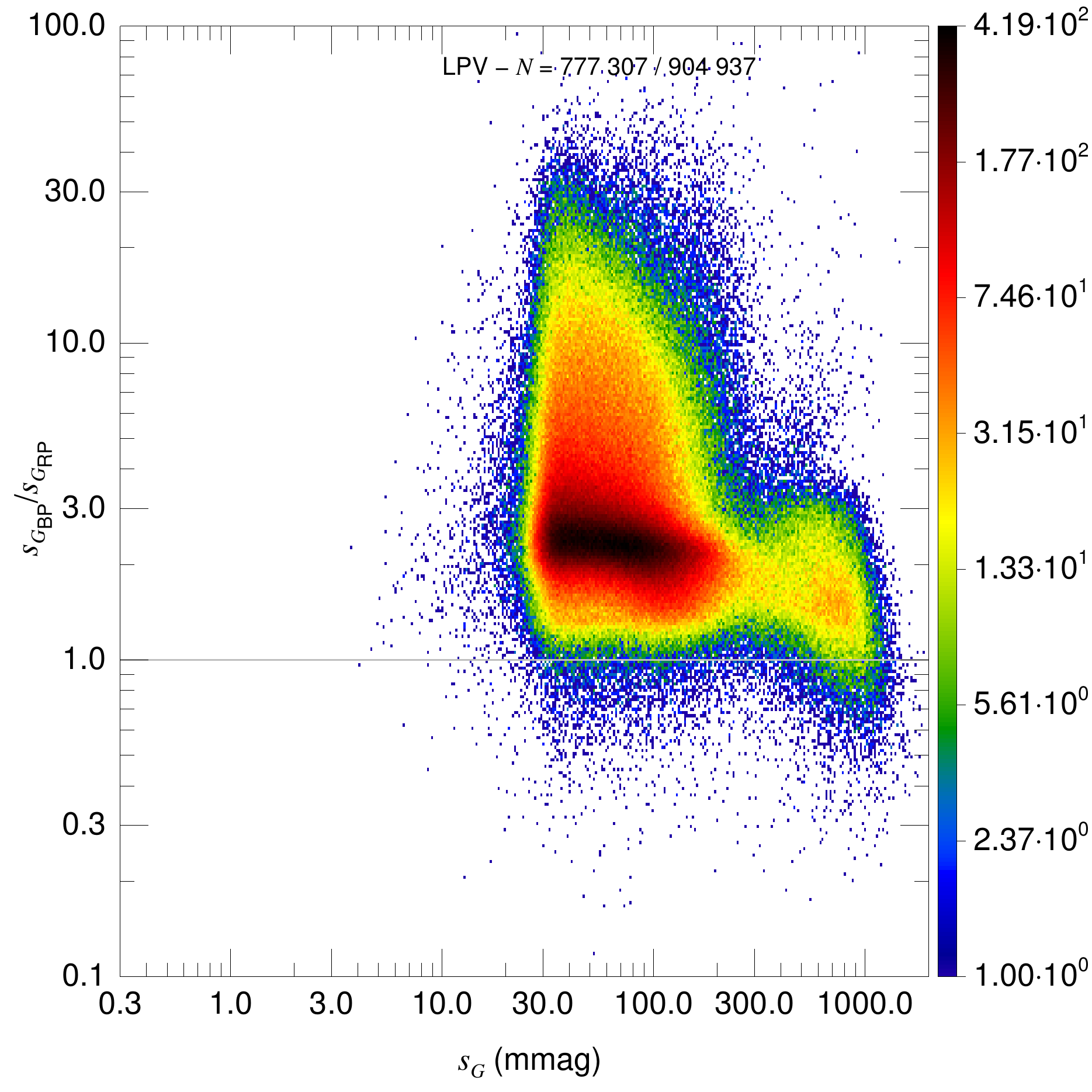}$\!\!\!$
            \includegraphics[width=0.35\linewidth]{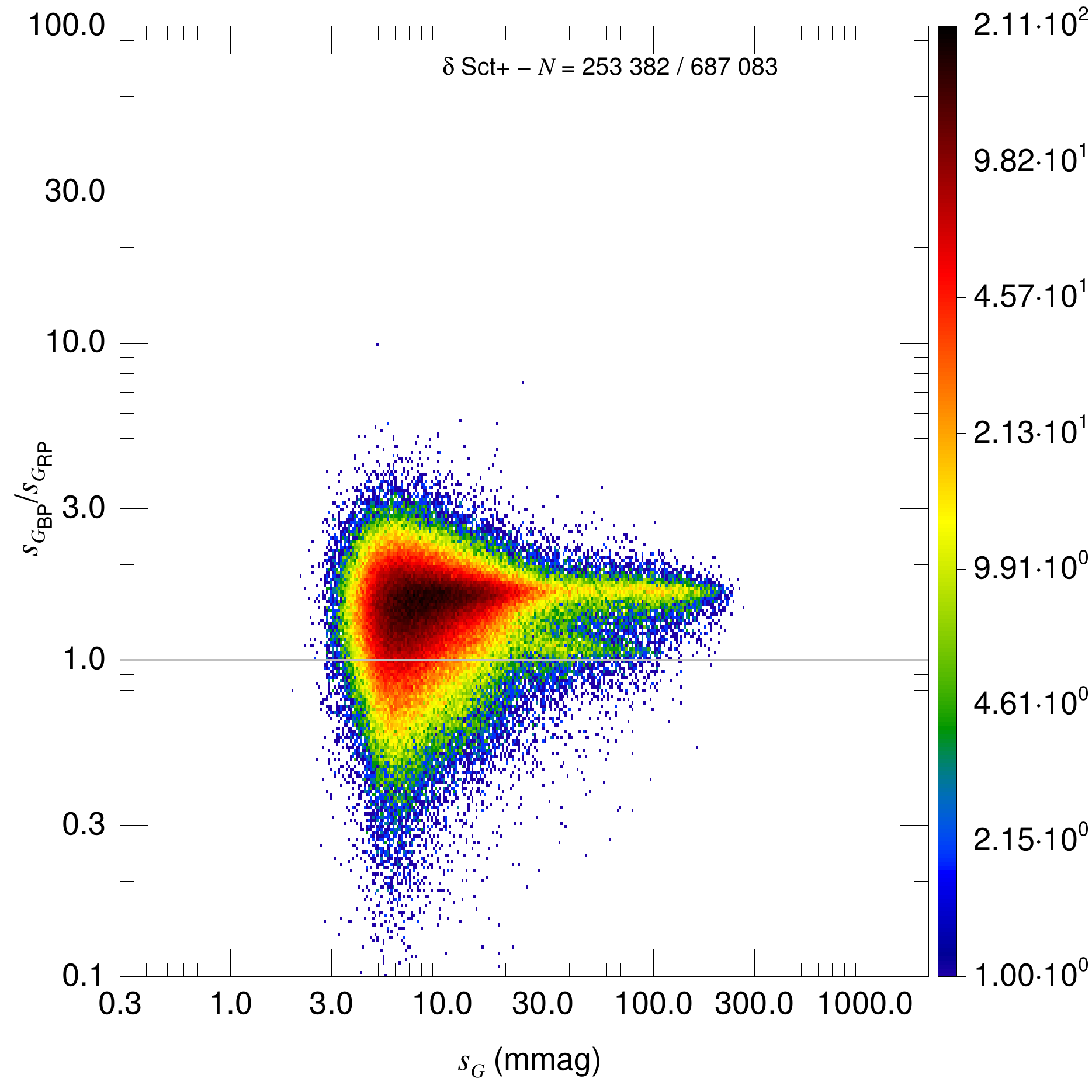}}
\centerline{\includegraphics[width=0.35\linewidth]{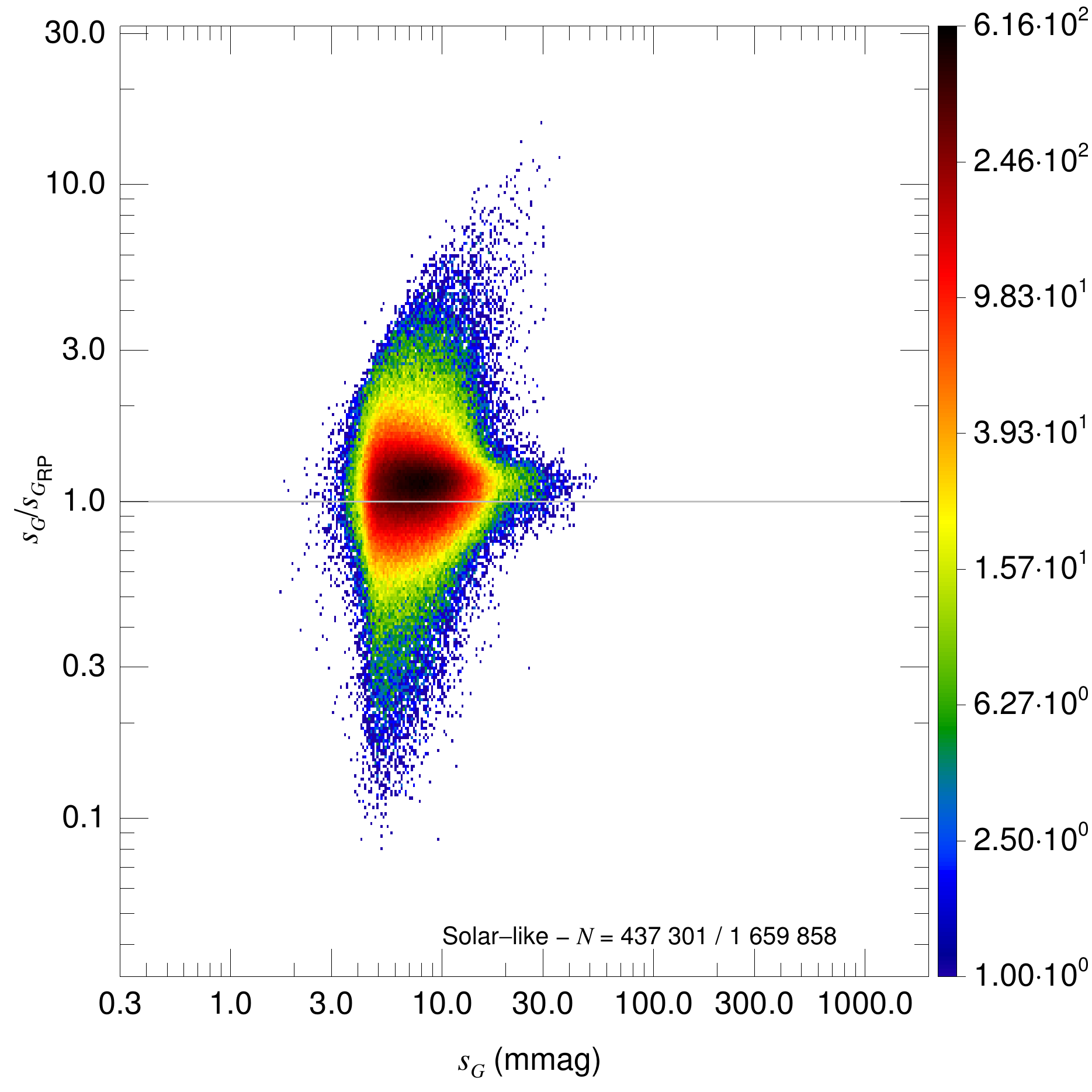}$\!\!\!$
            \includegraphics[width=0.35\linewidth]{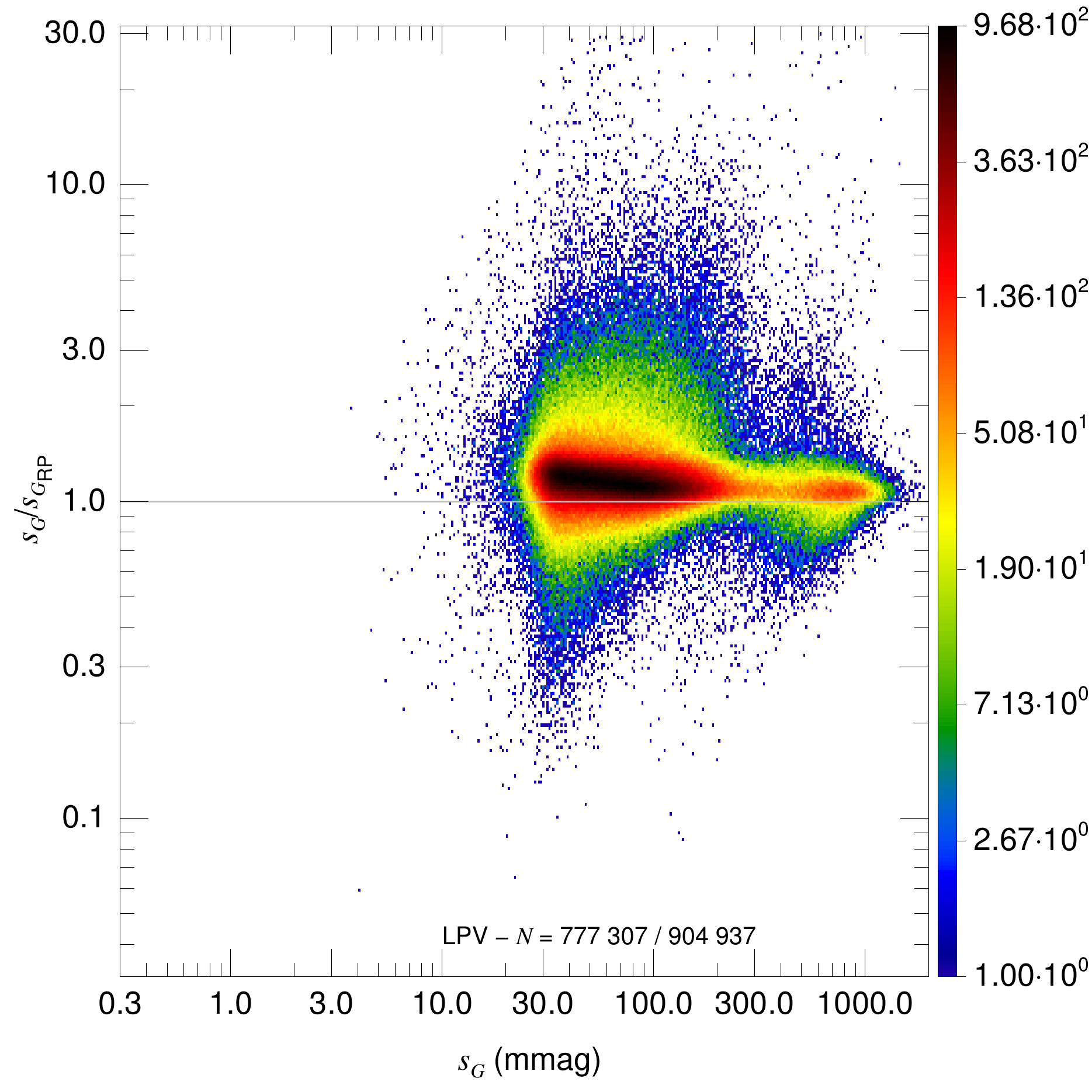}$\!\!\!$
            \includegraphics[width=0.35\linewidth]{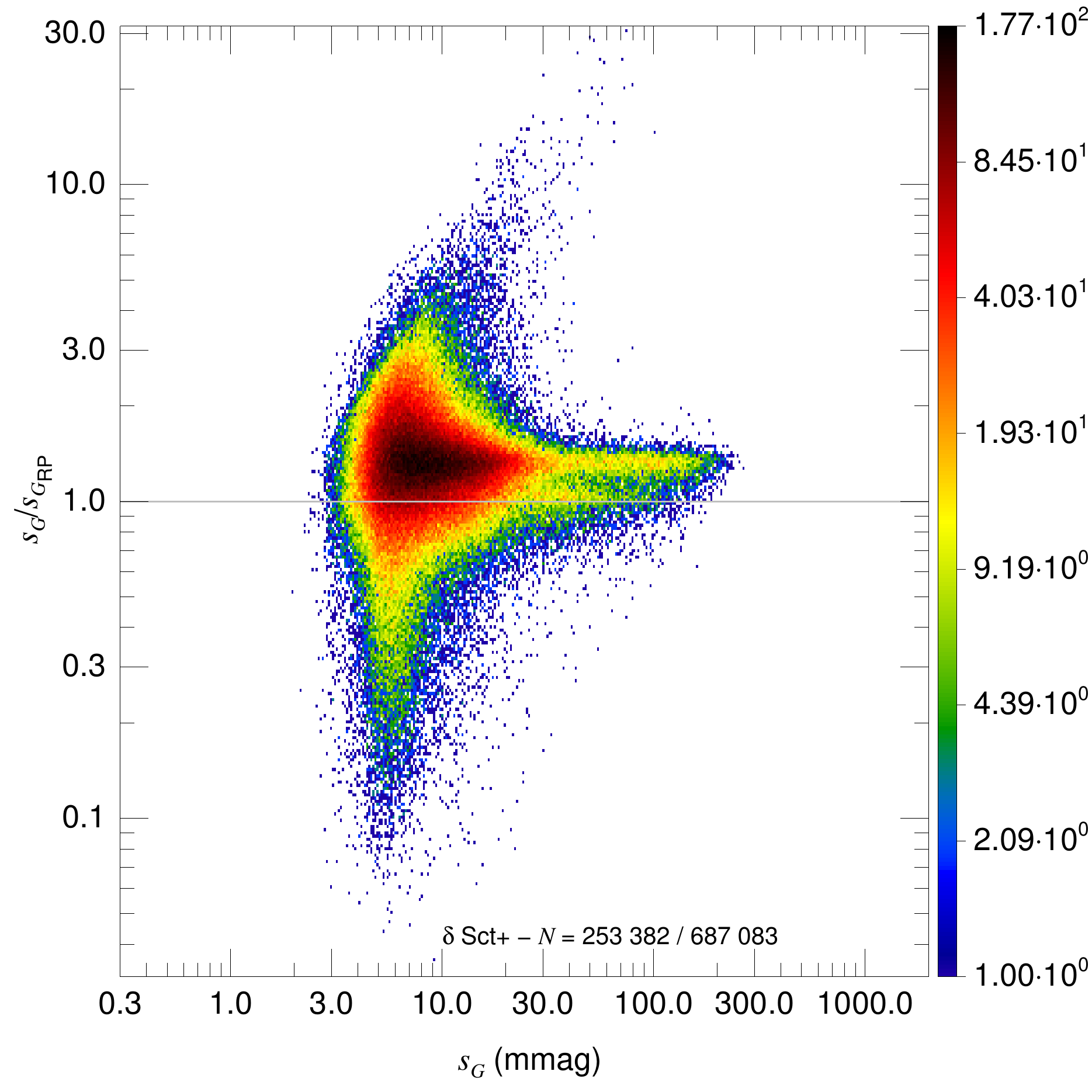}}
\centerline{\includegraphics[width=0.35\linewidth]{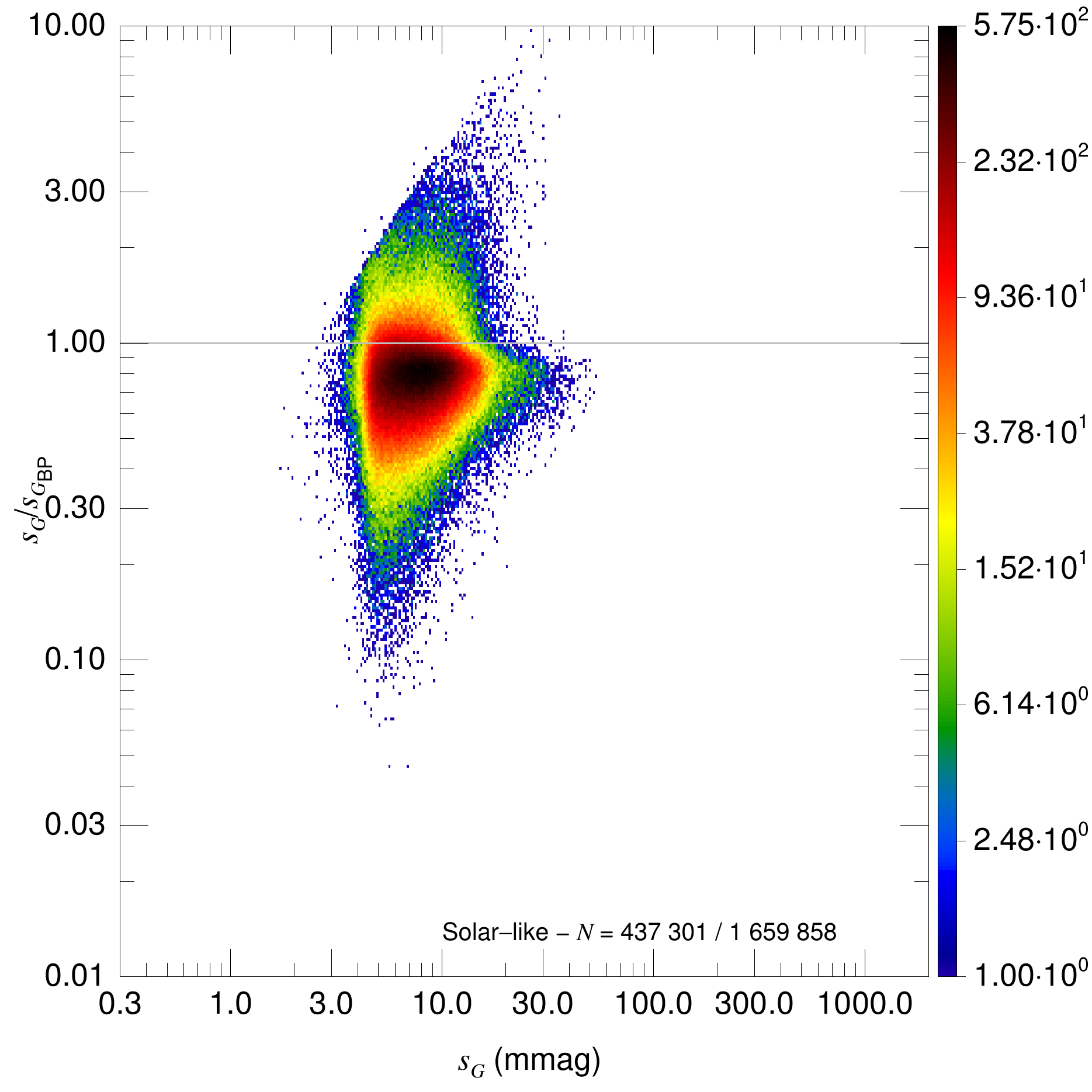}$\!\!\!$
            \includegraphics[width=0.35\linewidth]{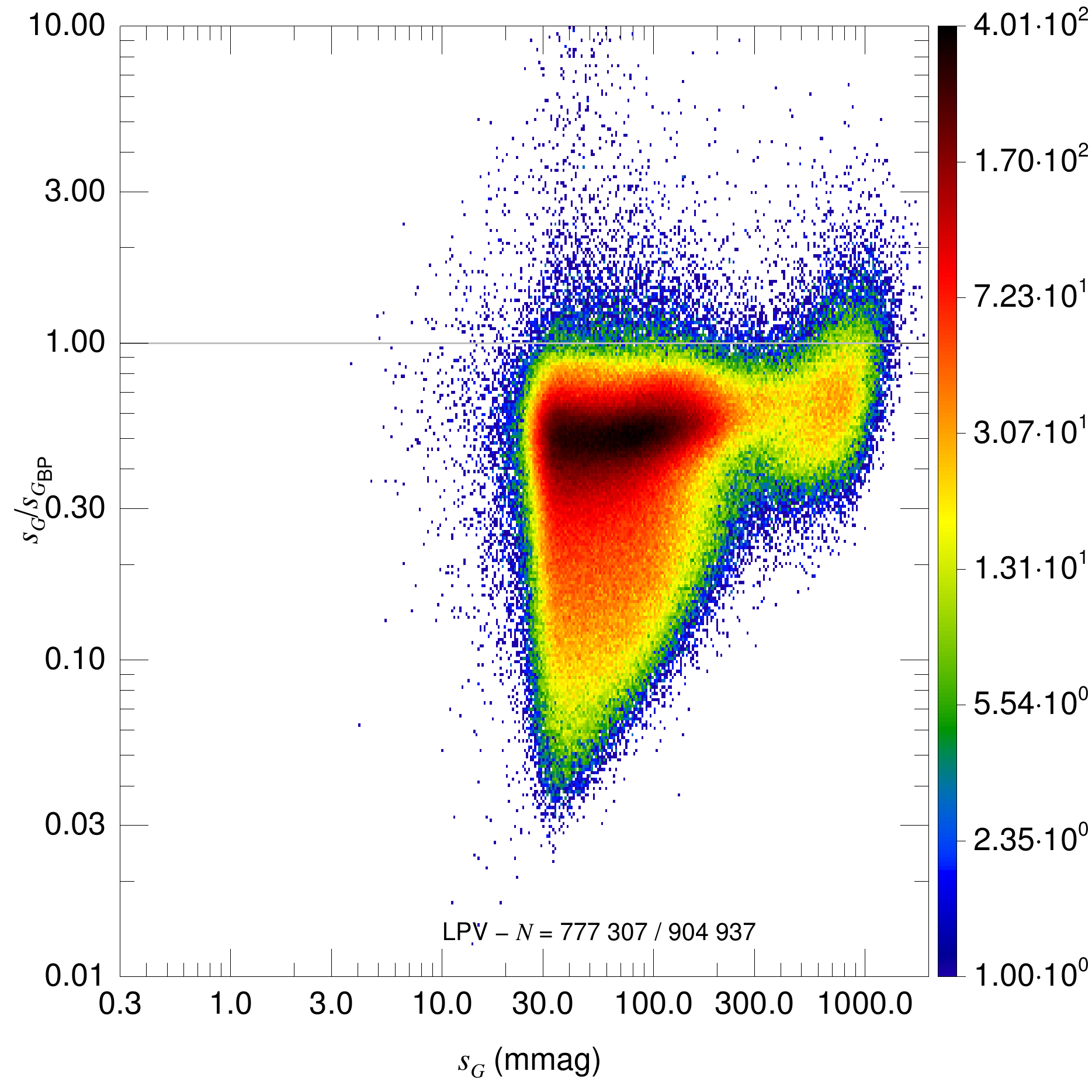}$\!\!\!$
            \includegraphics[width=0.35\linewidth]{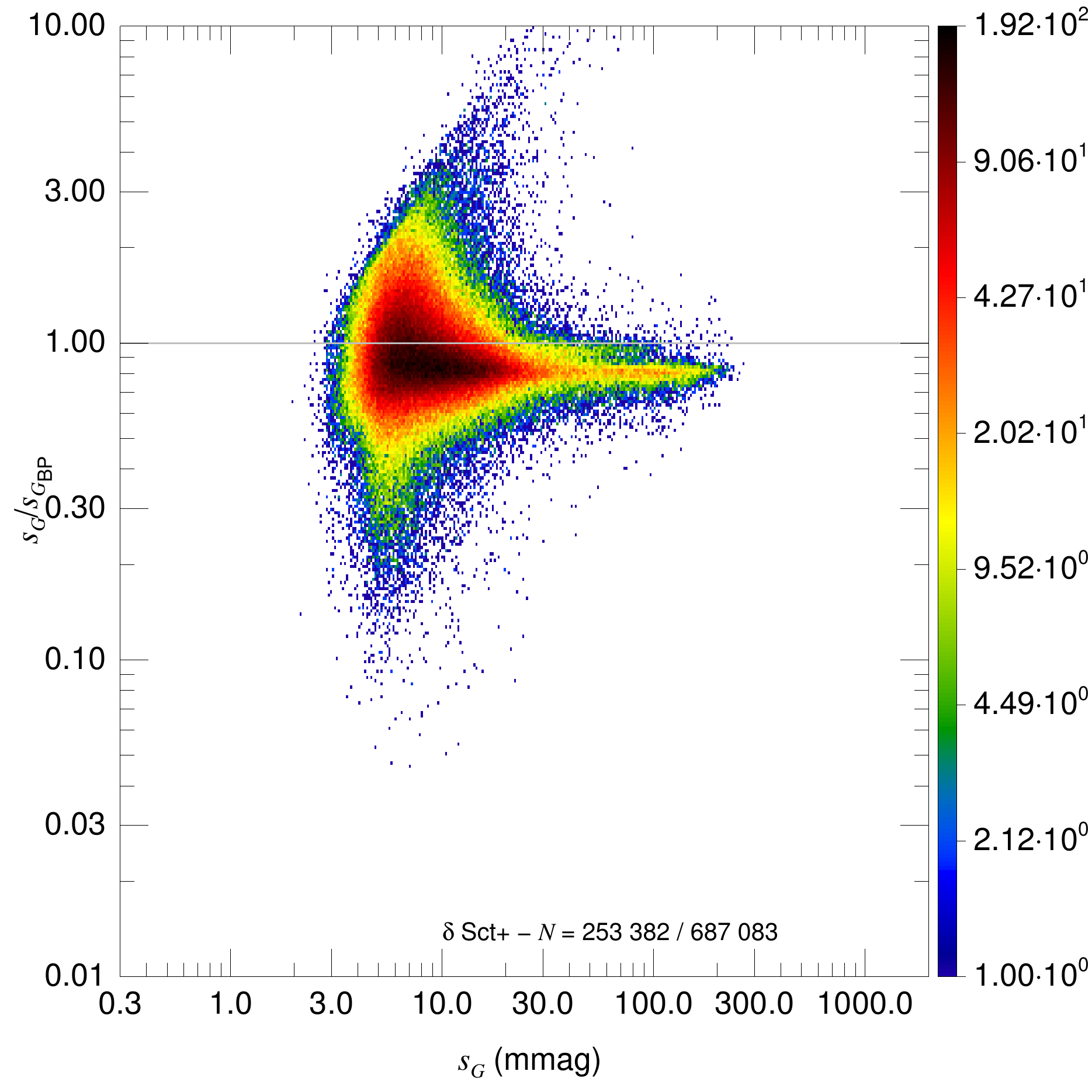}}
\caption{Astrophysical dispersion-dispersion ratio density diagrams for the three color combinations for each of the variable types 
         in R22 selecting stars that are classified as VVV here and have dispersions uncertainties \sigmasX\ lower 
         than 1~mmag in the three bands. Each column shows a different variable type sorted as in Table~\ref{vartypestats} and each 
         row a different color combination. The bin size is adjusted for each variable type depending on the number of sources.}
\label{varhist_types}
\end{figure*}

\addtocounter{figure}{-1}

\begin{figure*}
\centerline{\includegraphics[width=0.35\linewidth]{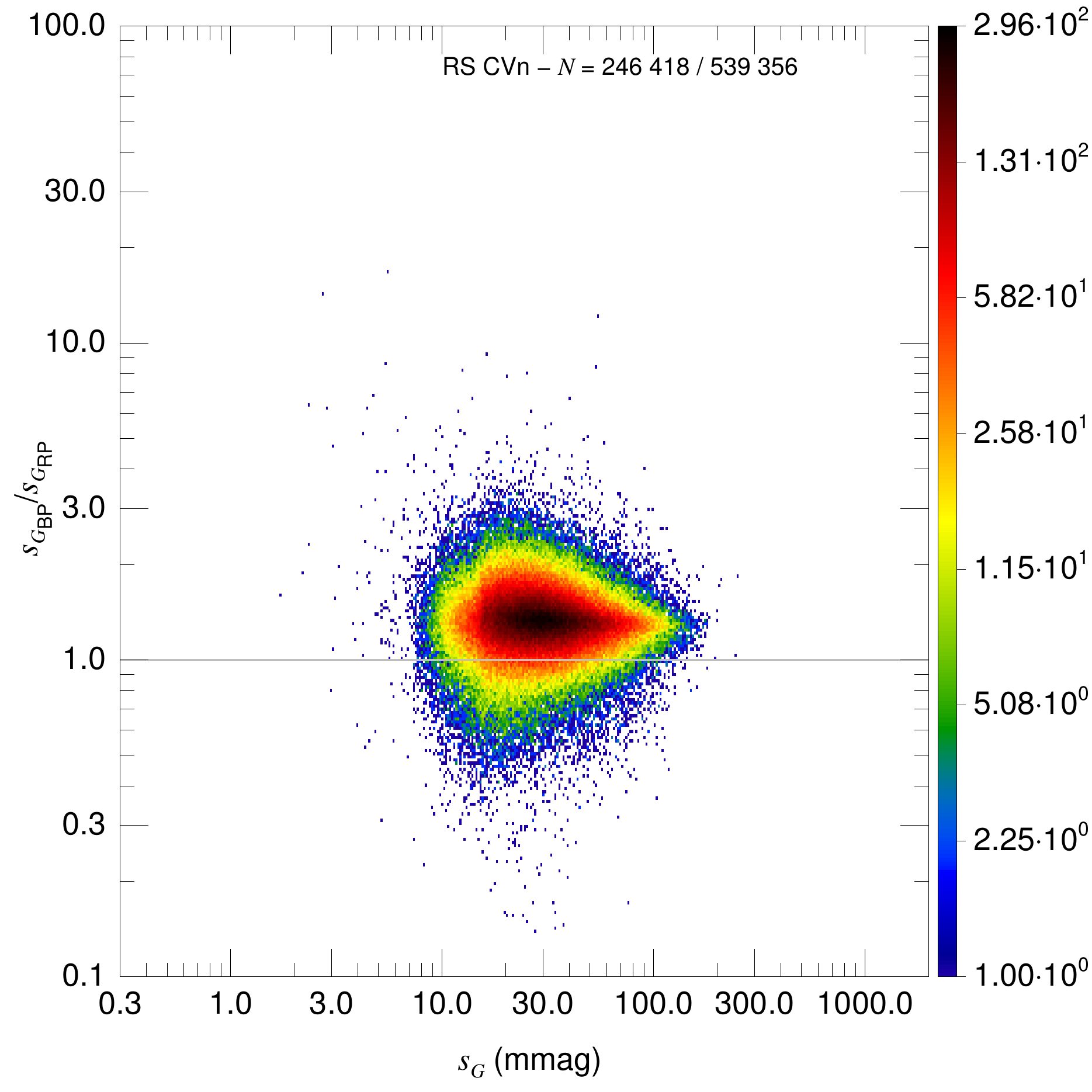}$\!\!\!$
            \includegraphics[width=0.35\linewidth]{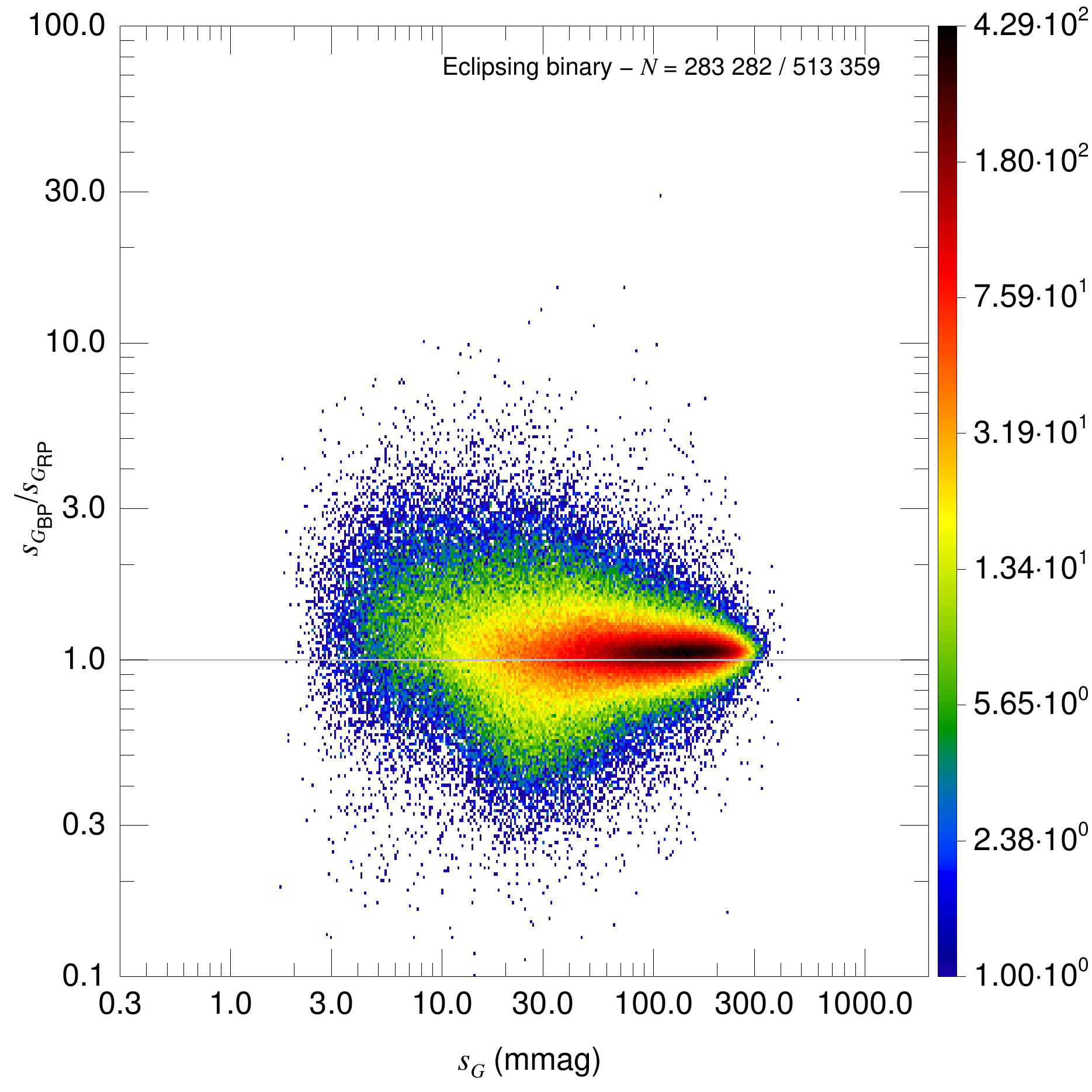}$\!\!\!$
            \includegraphics[width=0.35\linewidth]{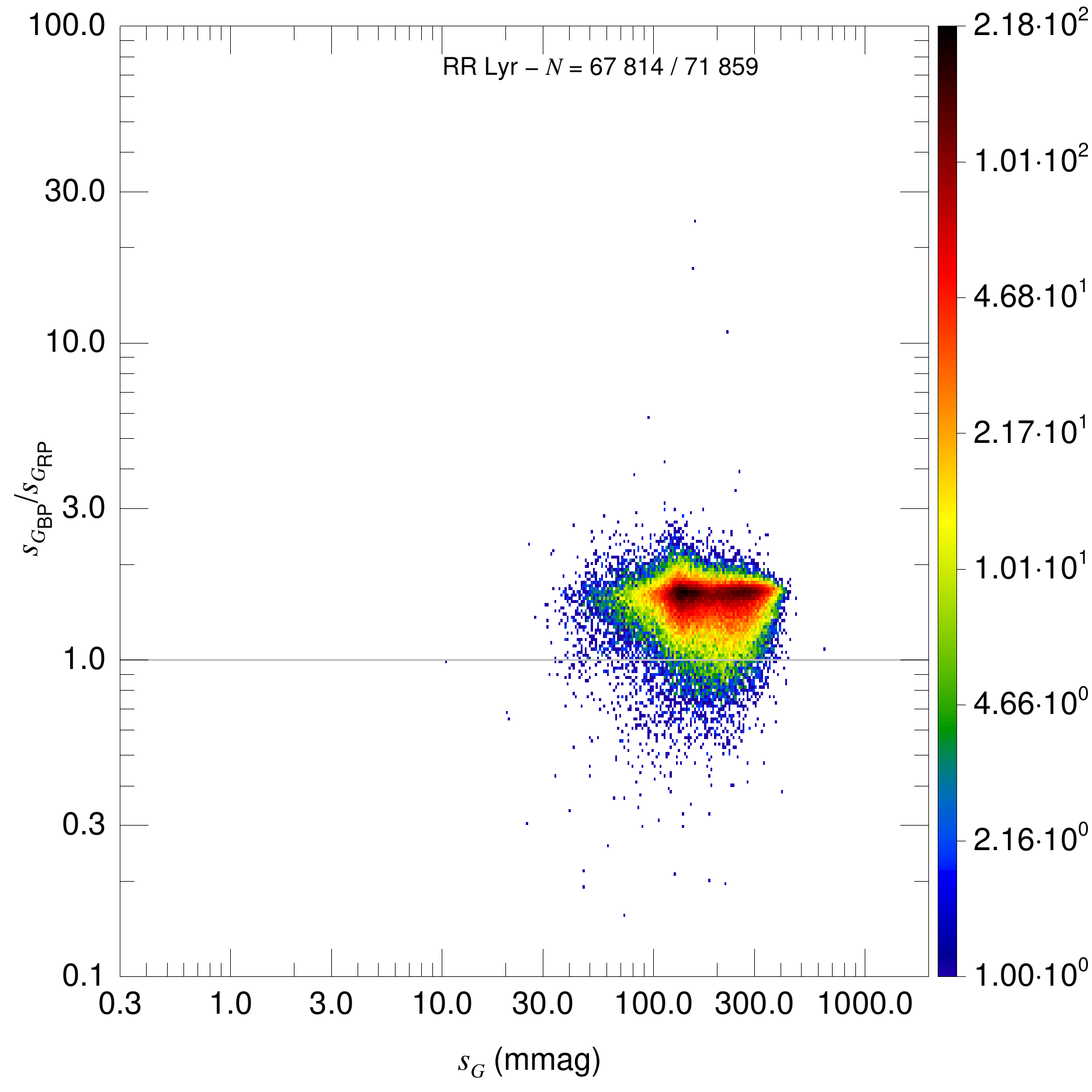}}
\centerline{\includegraphics[width=0.35\linewidth]{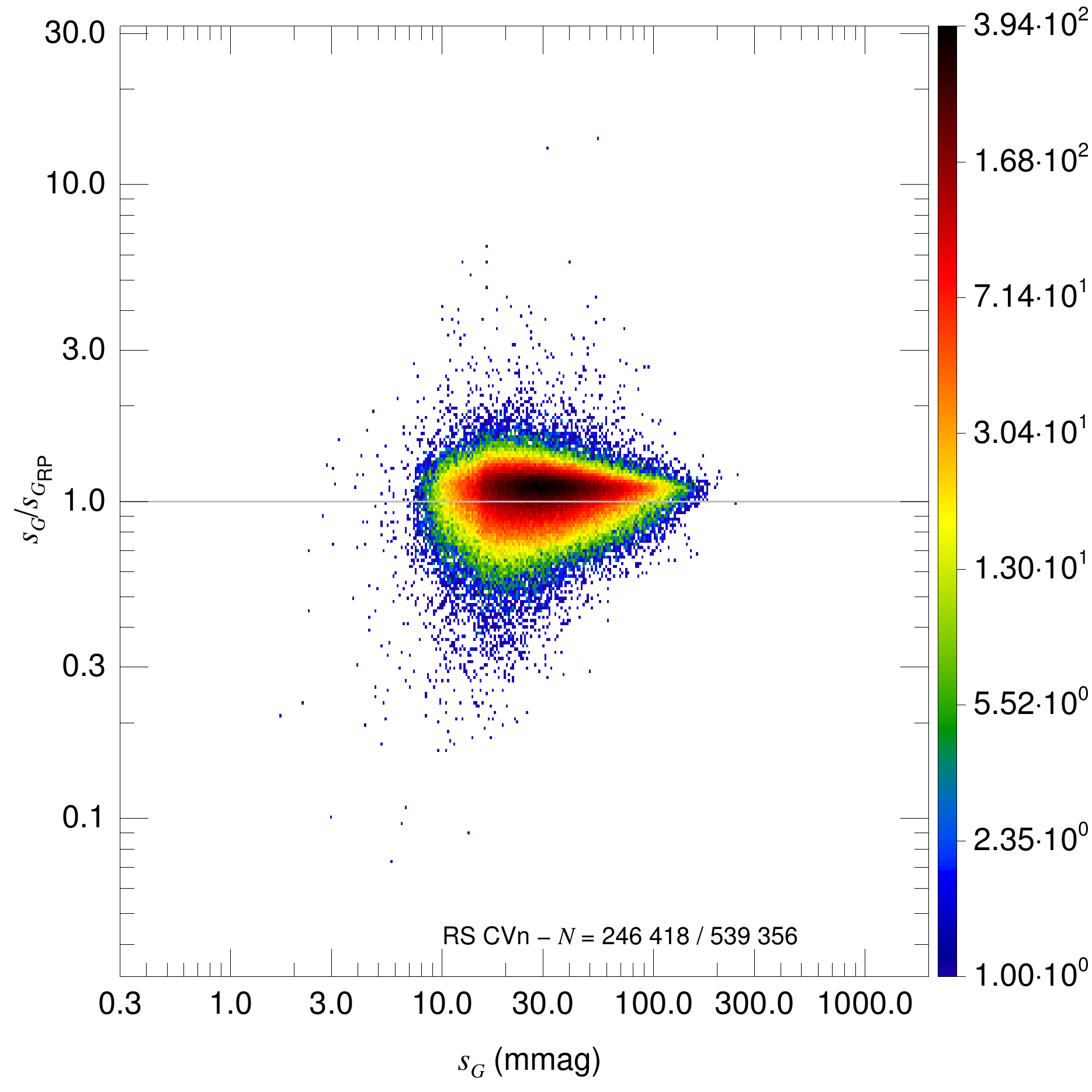}$\!\!\!$
            \includegraphics[width=0.35\linewidth]{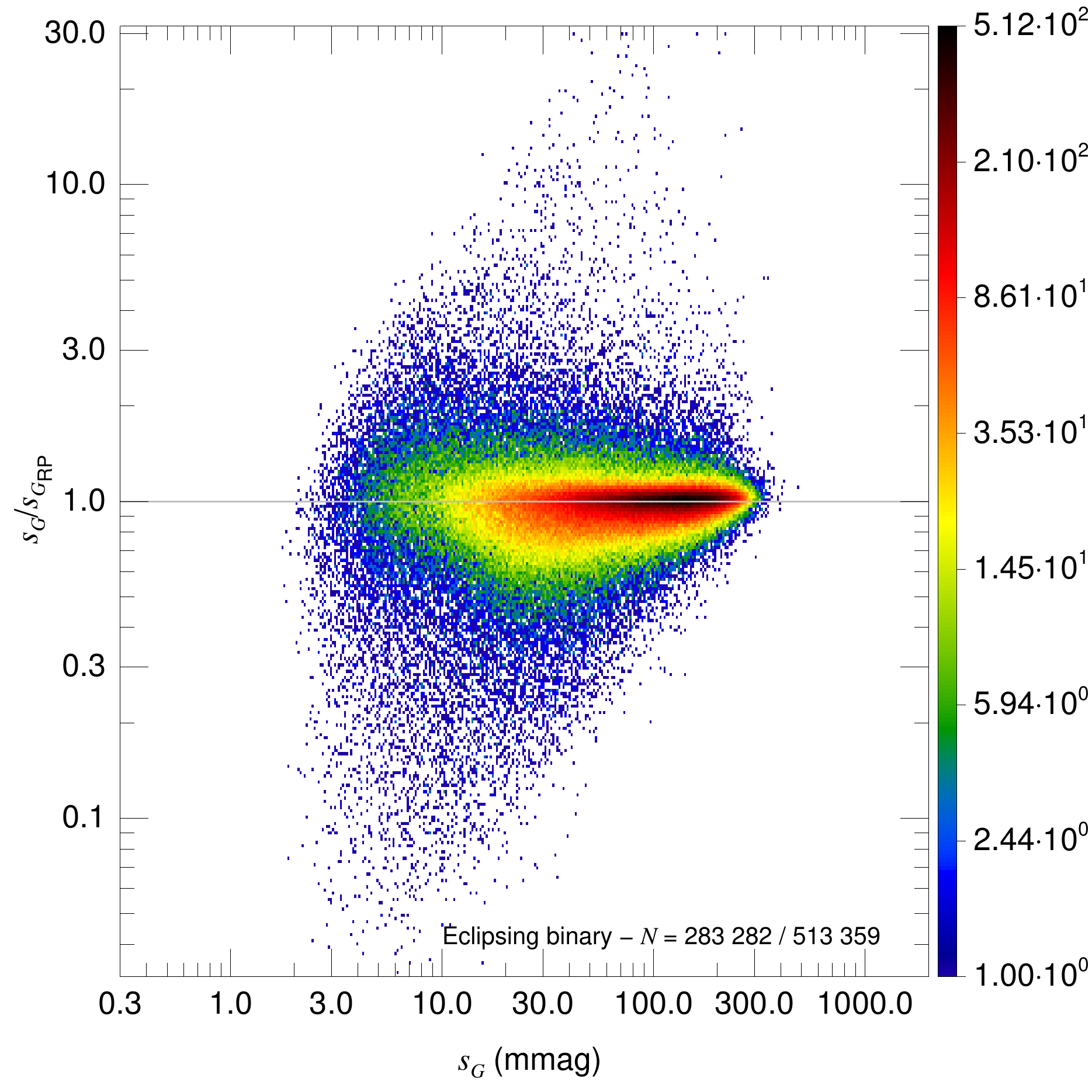}$\!\!\!$
            \includegraphics[width=0.35\linewidth]{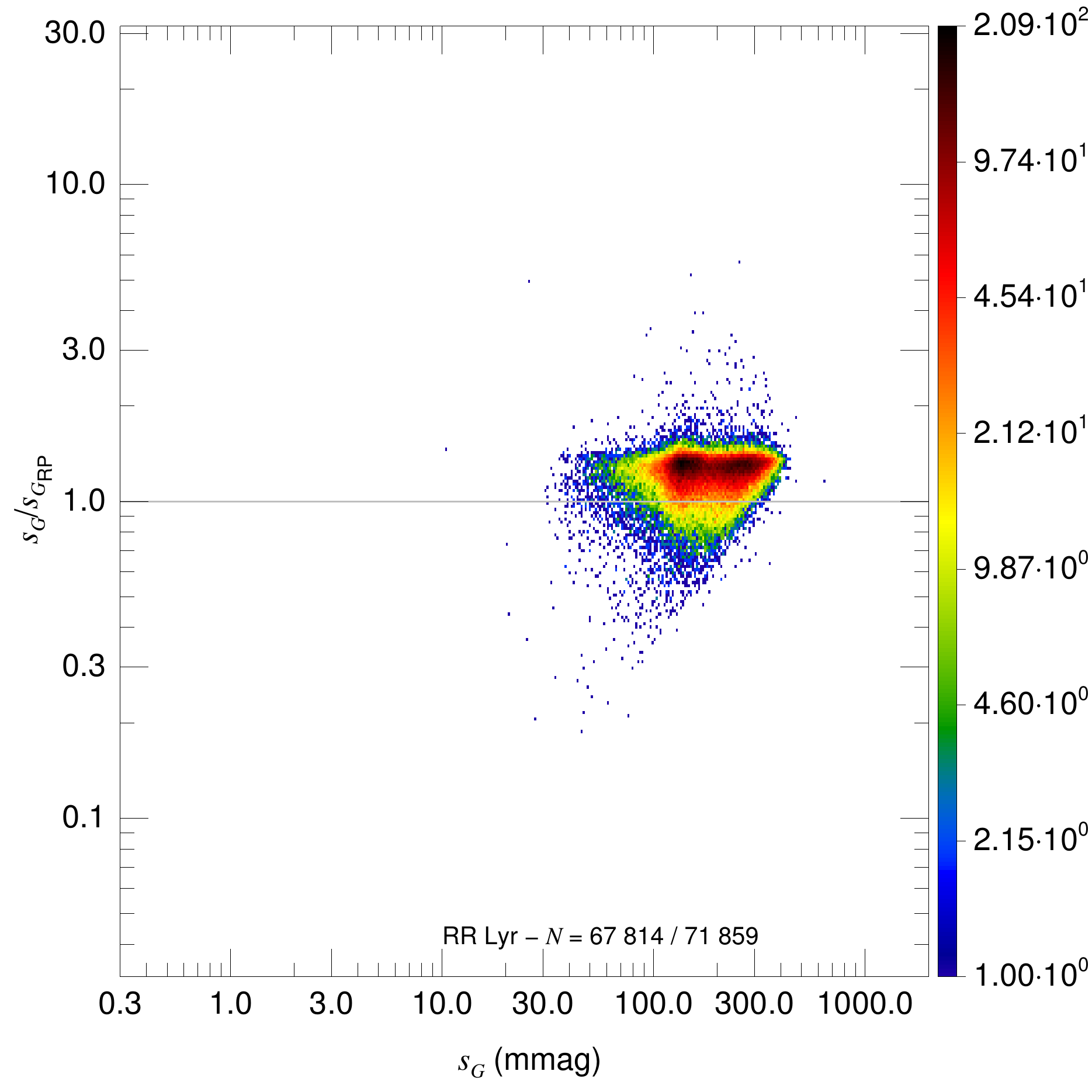}}
\centerline{\includegraphics[width=0.35\linewidth]{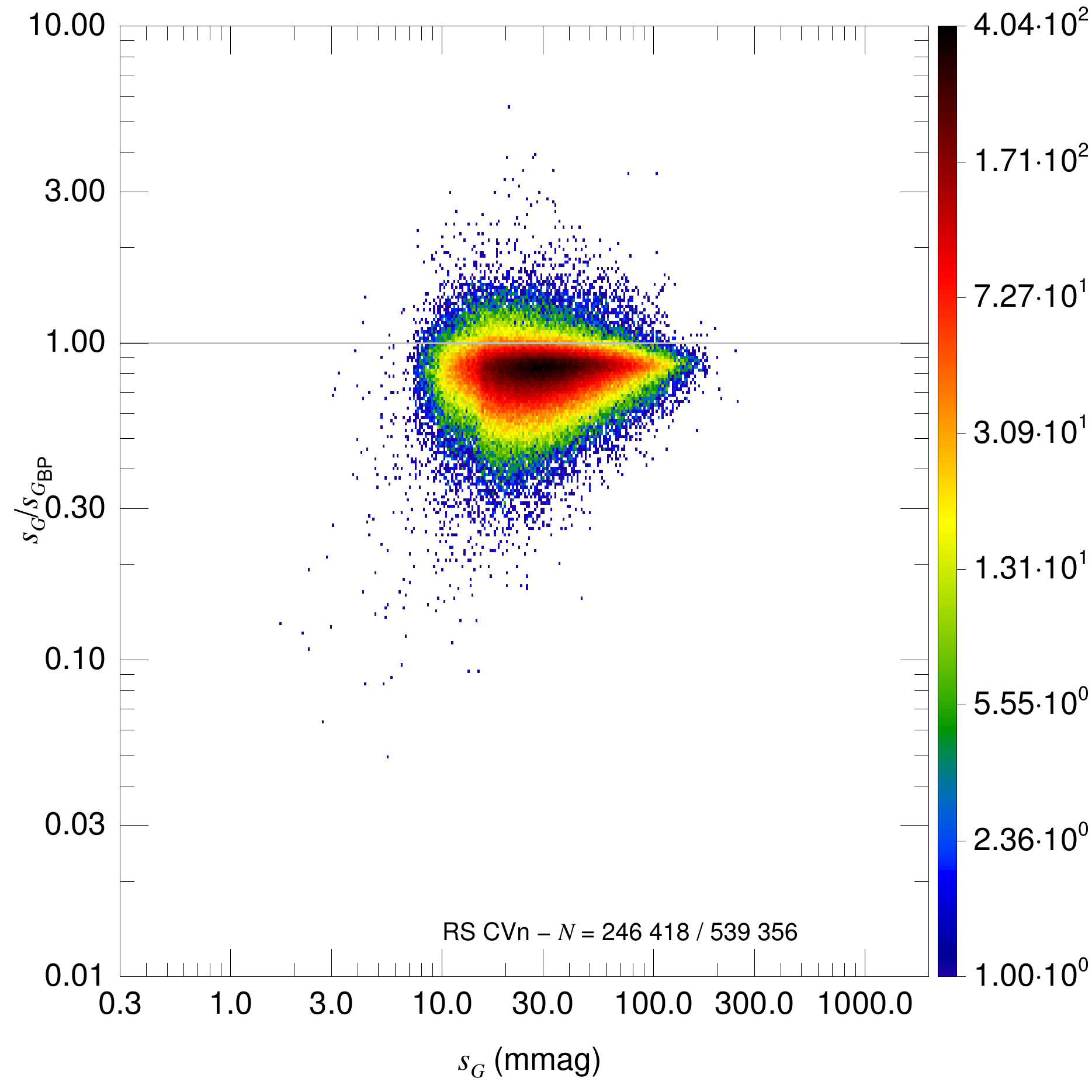}$\!\!\!$
            \includegraphics[width=0.35\linewidth]{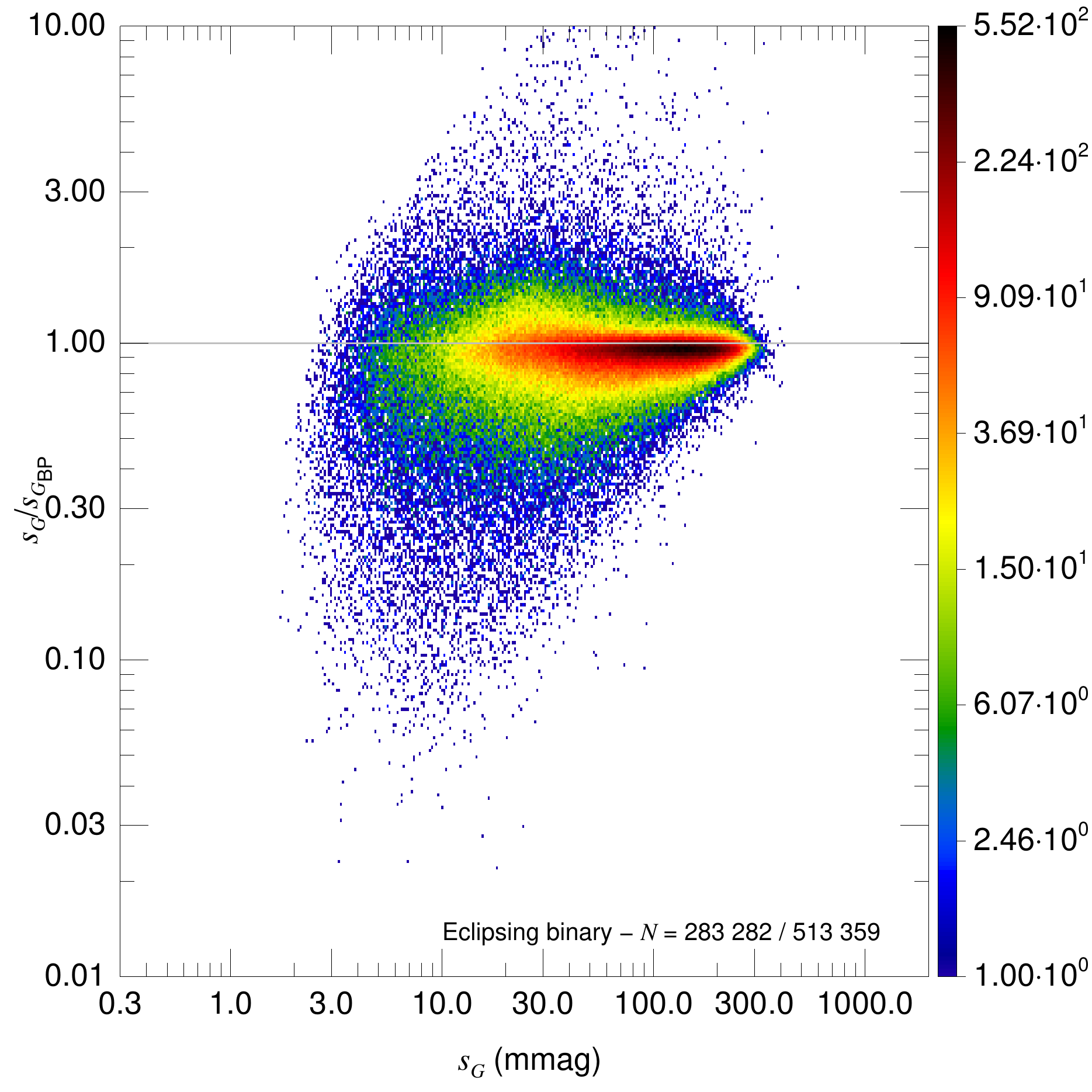}$\!\!\!$
            \includegraphics[width=0.35\linewidth]{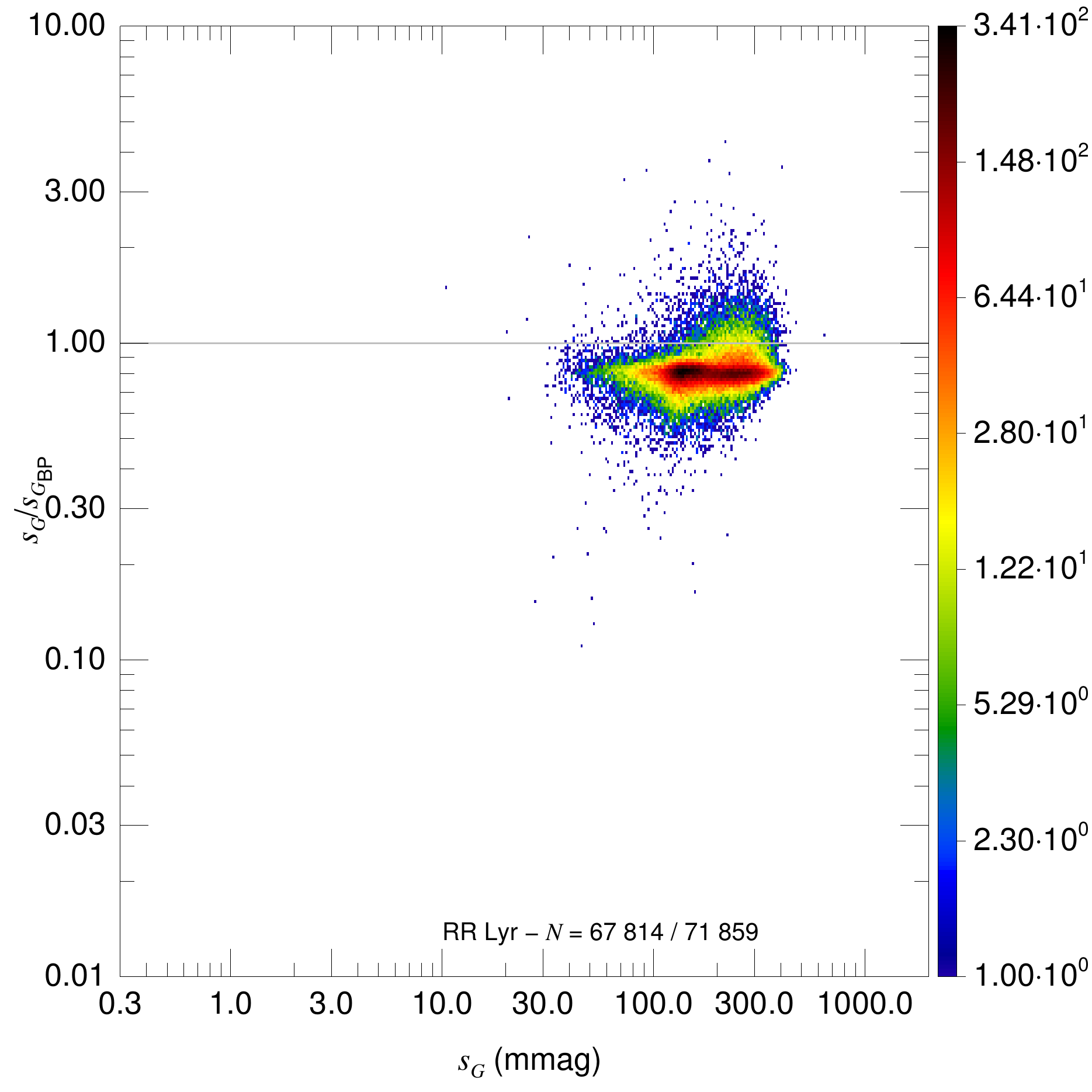}}
\caption{(Continued).}
\end{figure*}

\addtocounter{figure}{-1}

\begin{figure*}
\centerline{\includegraphics[width=0.35\linewidth]{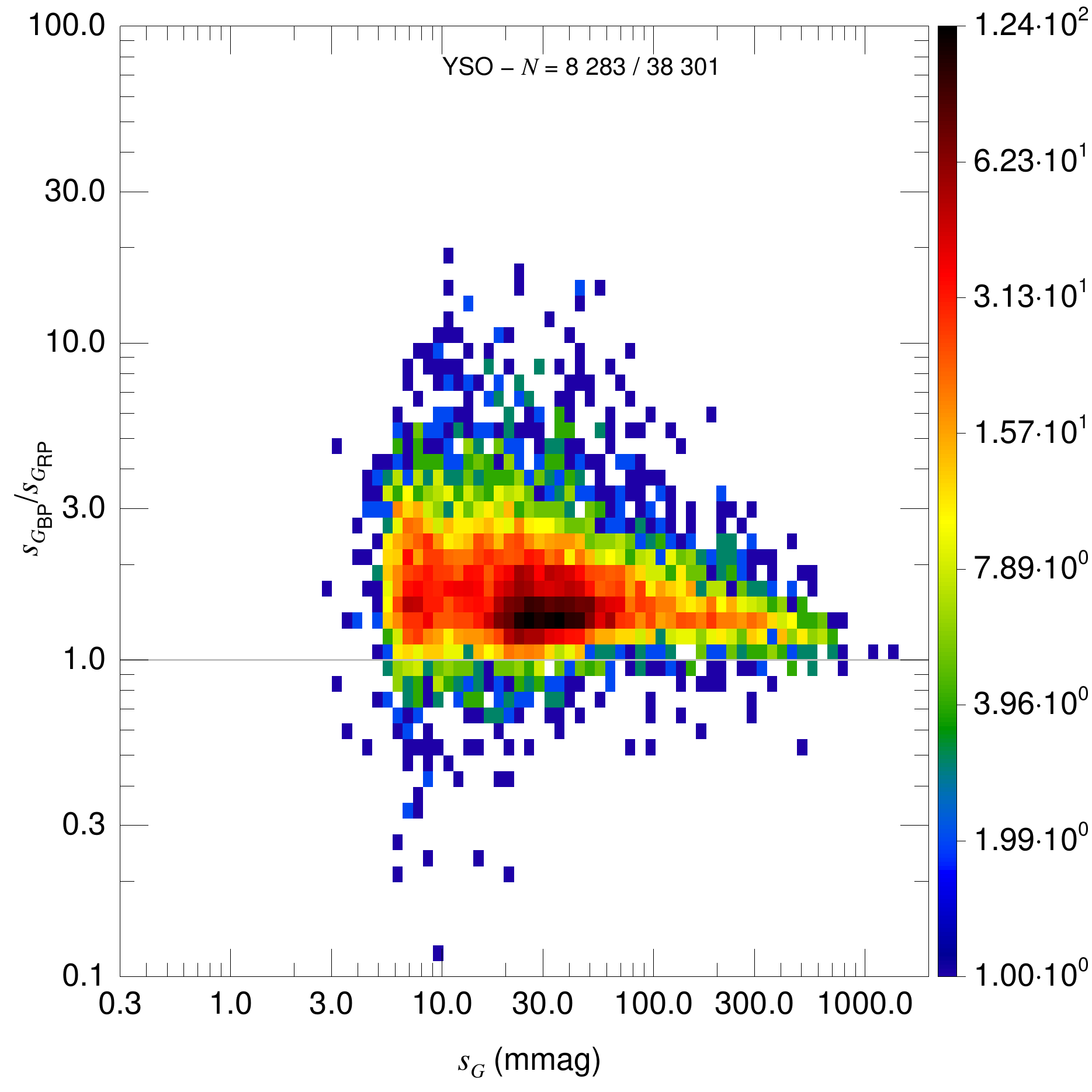}$\!\!\!$
            \includegraphics[width=0.35\linewidth]{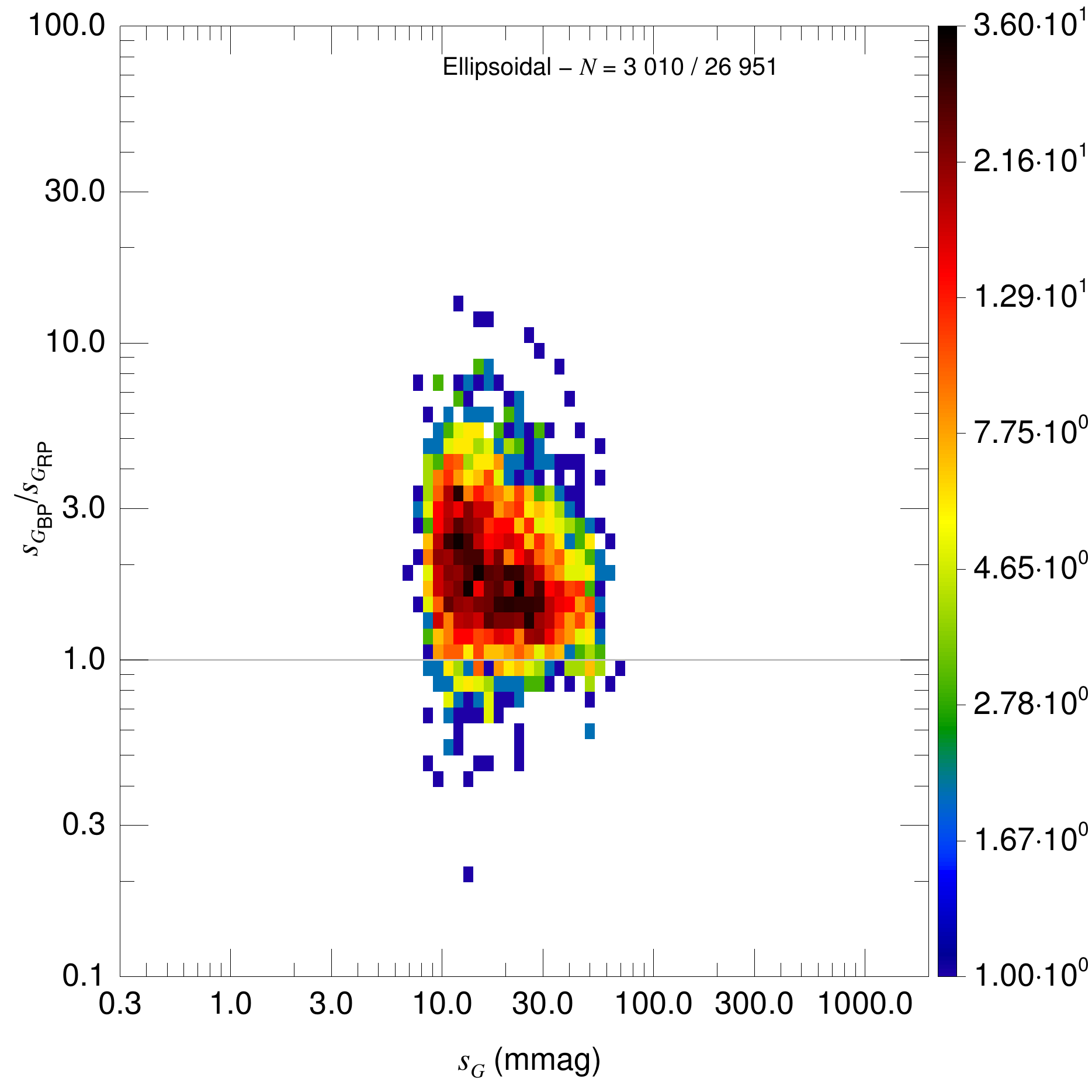}$\!\!\!$
            \includegraphics[width=0.35\linewidth]{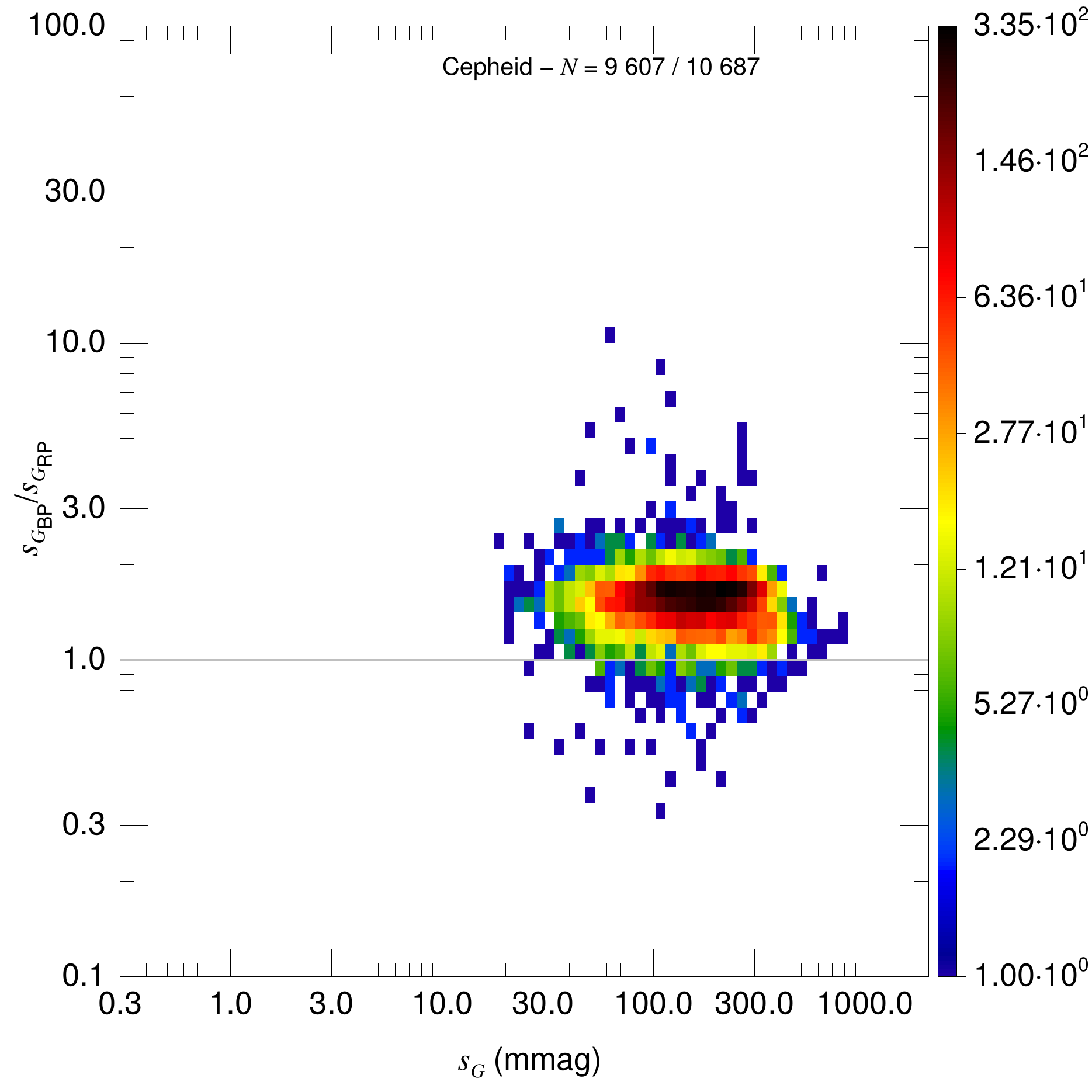}}
\centerline{\includegraphics[width=0.35\linewidth]{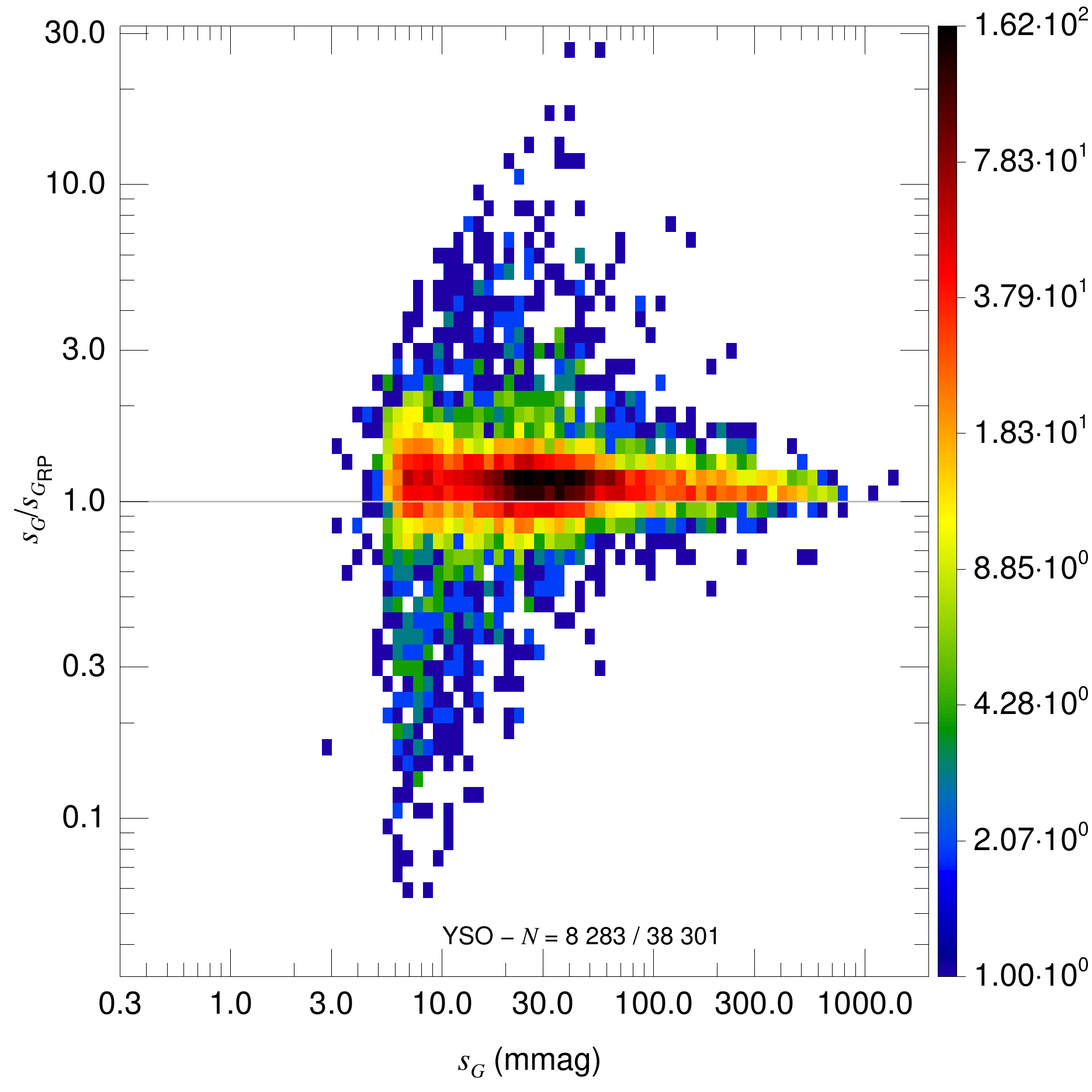}$\!\!\!$
            \includegraphics[width=0.35\linewidth]{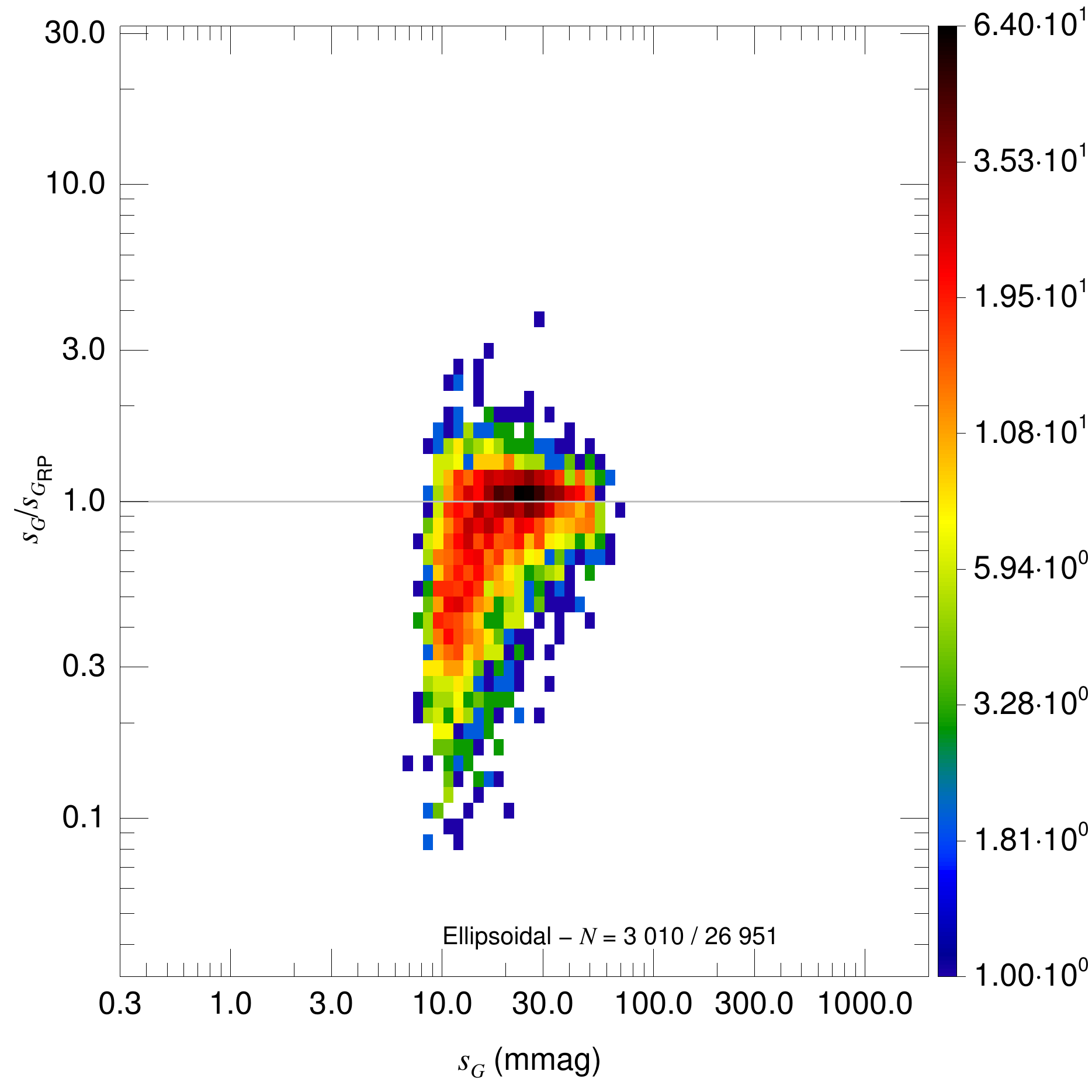}$\!\!\!$
            \includegraphics[width=0.35\linewidth]{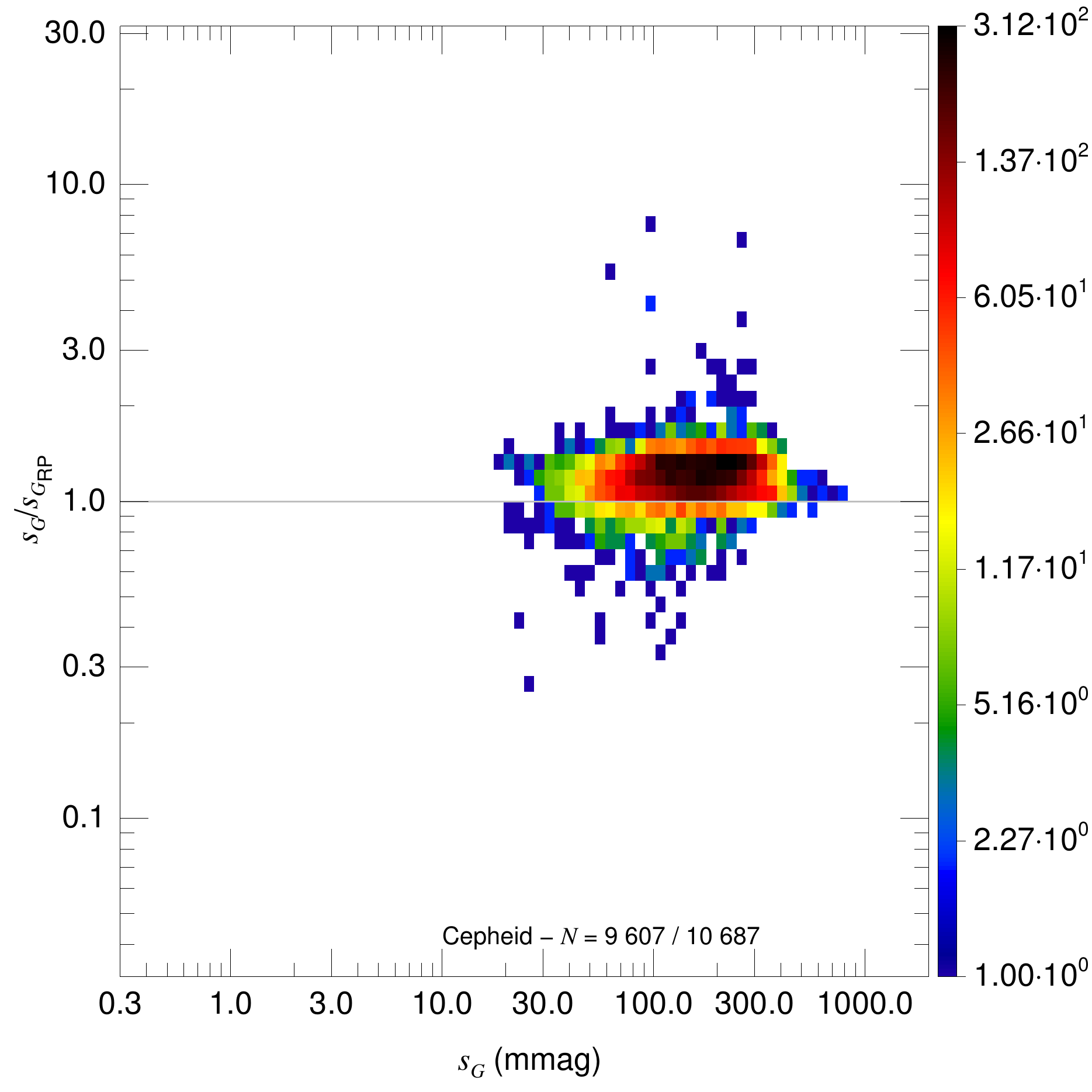}}
\centerline{\includegraphics[width=0.35\linewidth]{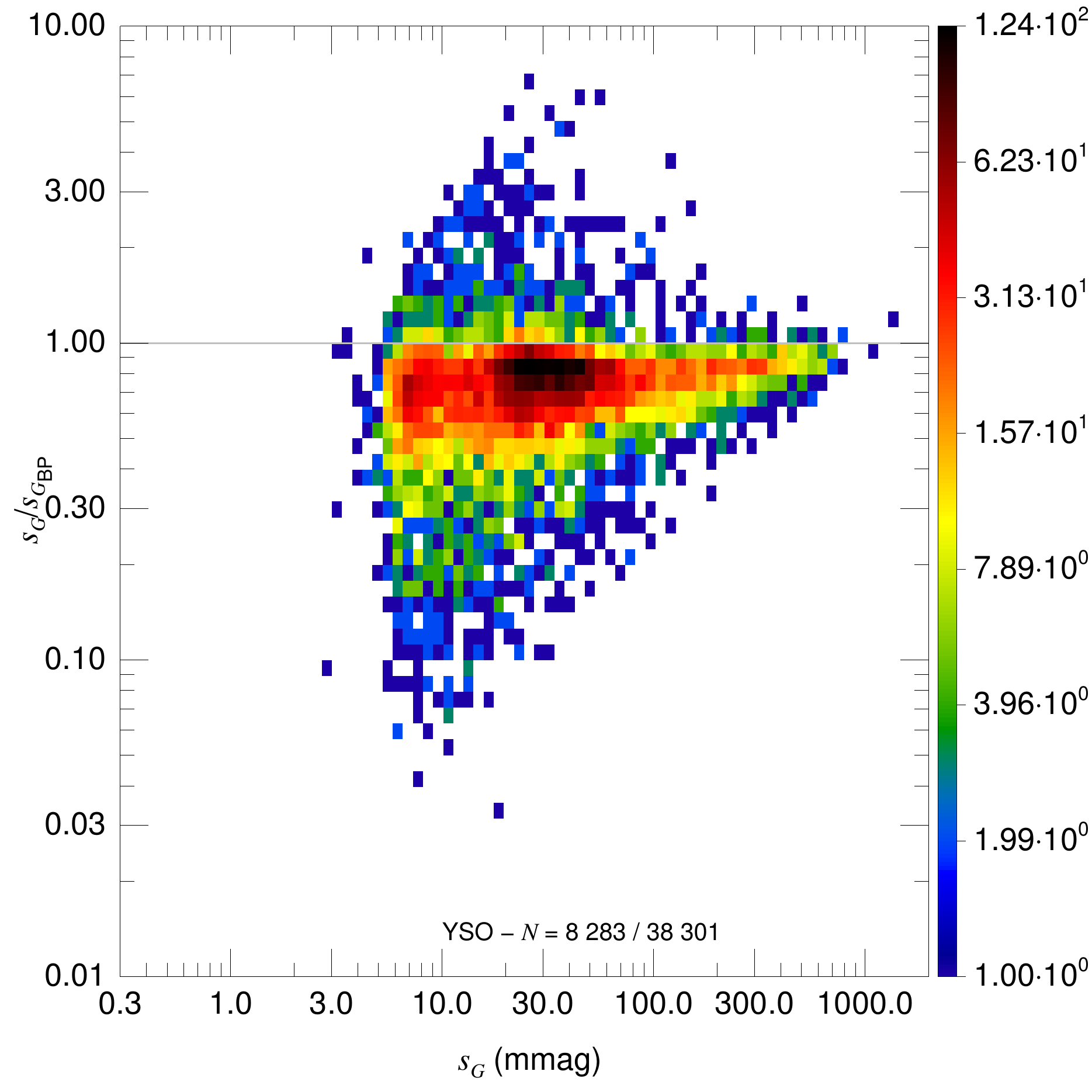}$\!\!\!$
            \includegraphics[width=0.35\linewidth]{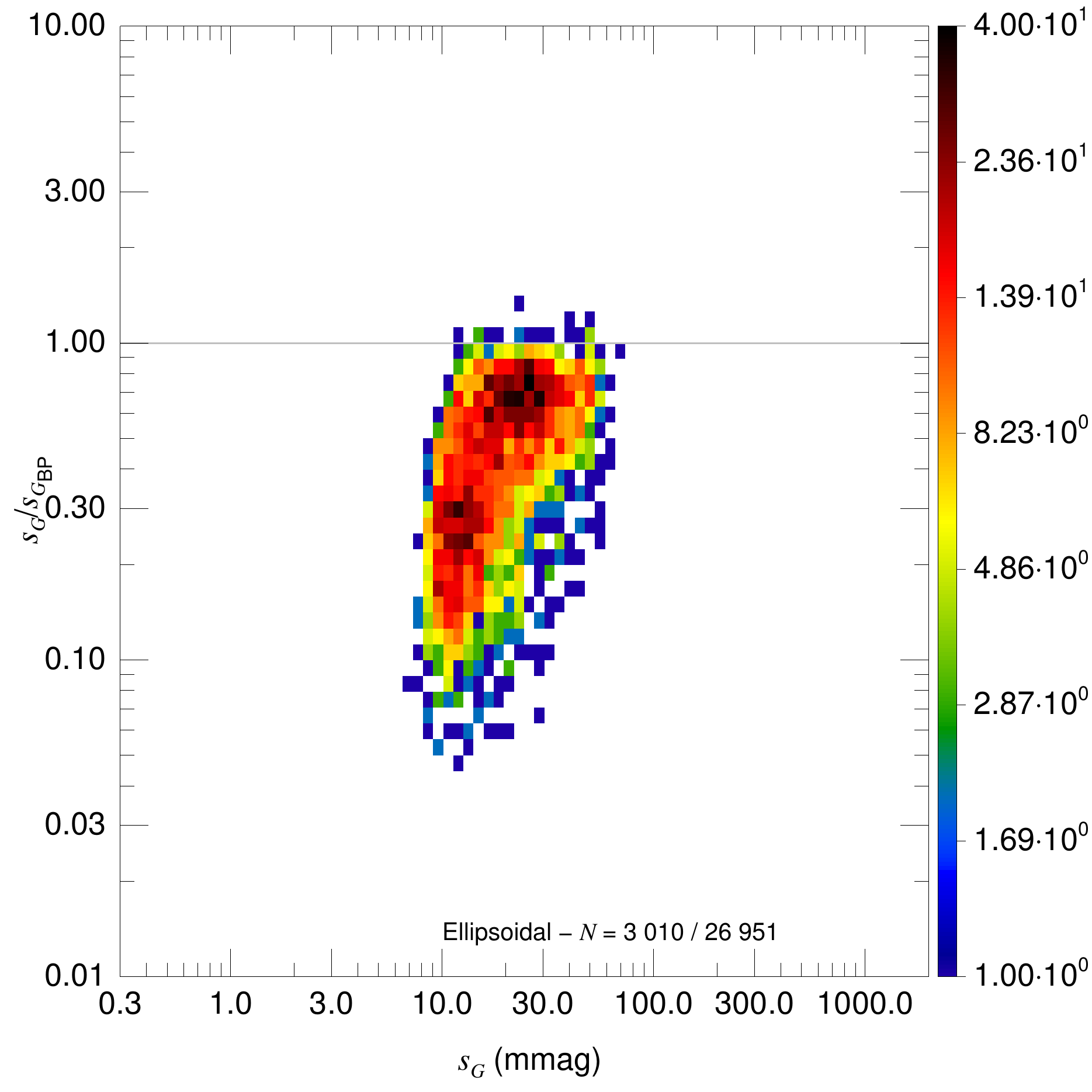}$\!\!\!$
            \includegraphics[width=0.35\linewidth]{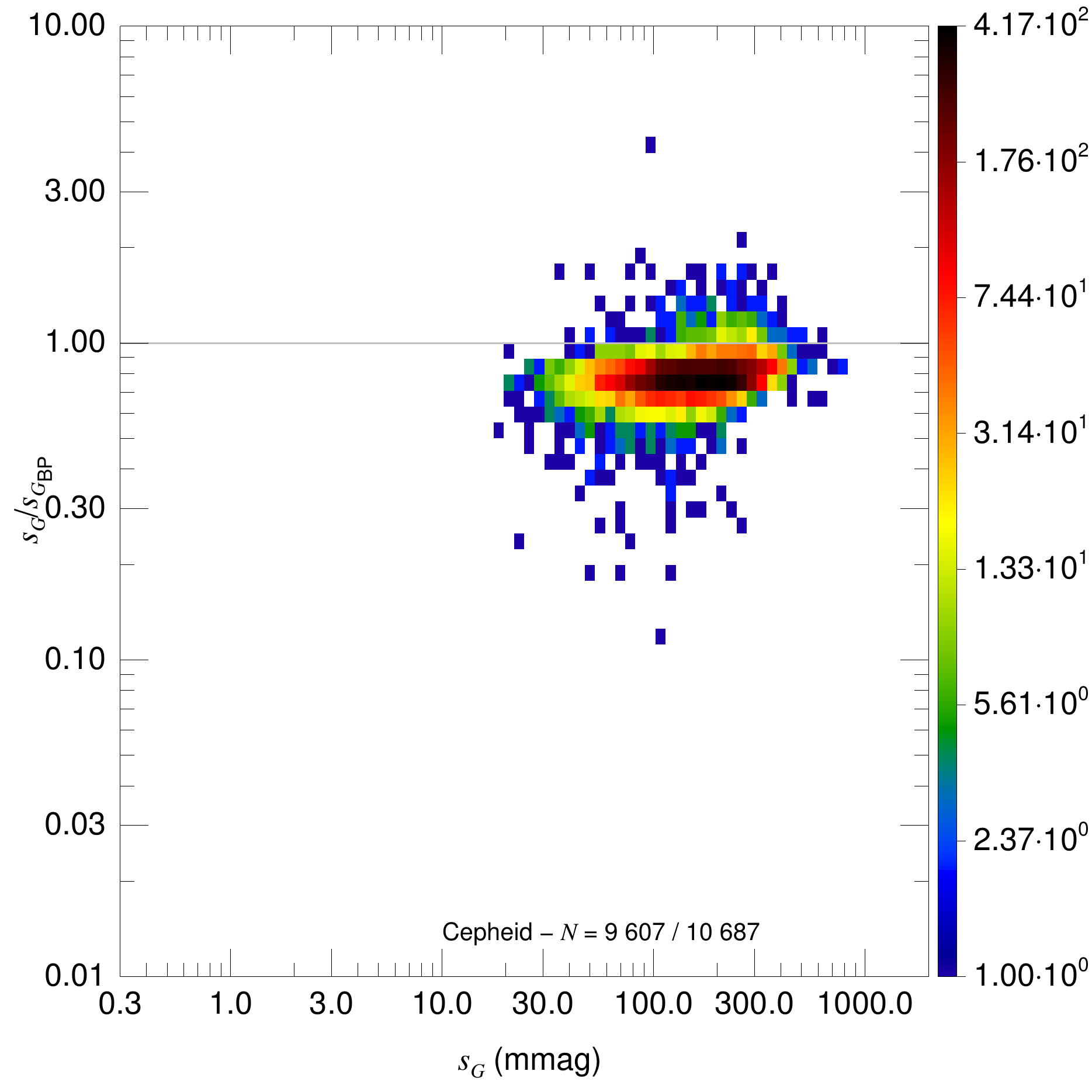}}
\caption{(Continued).}
\end{figure*}

\addtocounter{figure}{-1}

\begin{figure*}
\centerline{\includegraphics[width=0.35\linewidth]{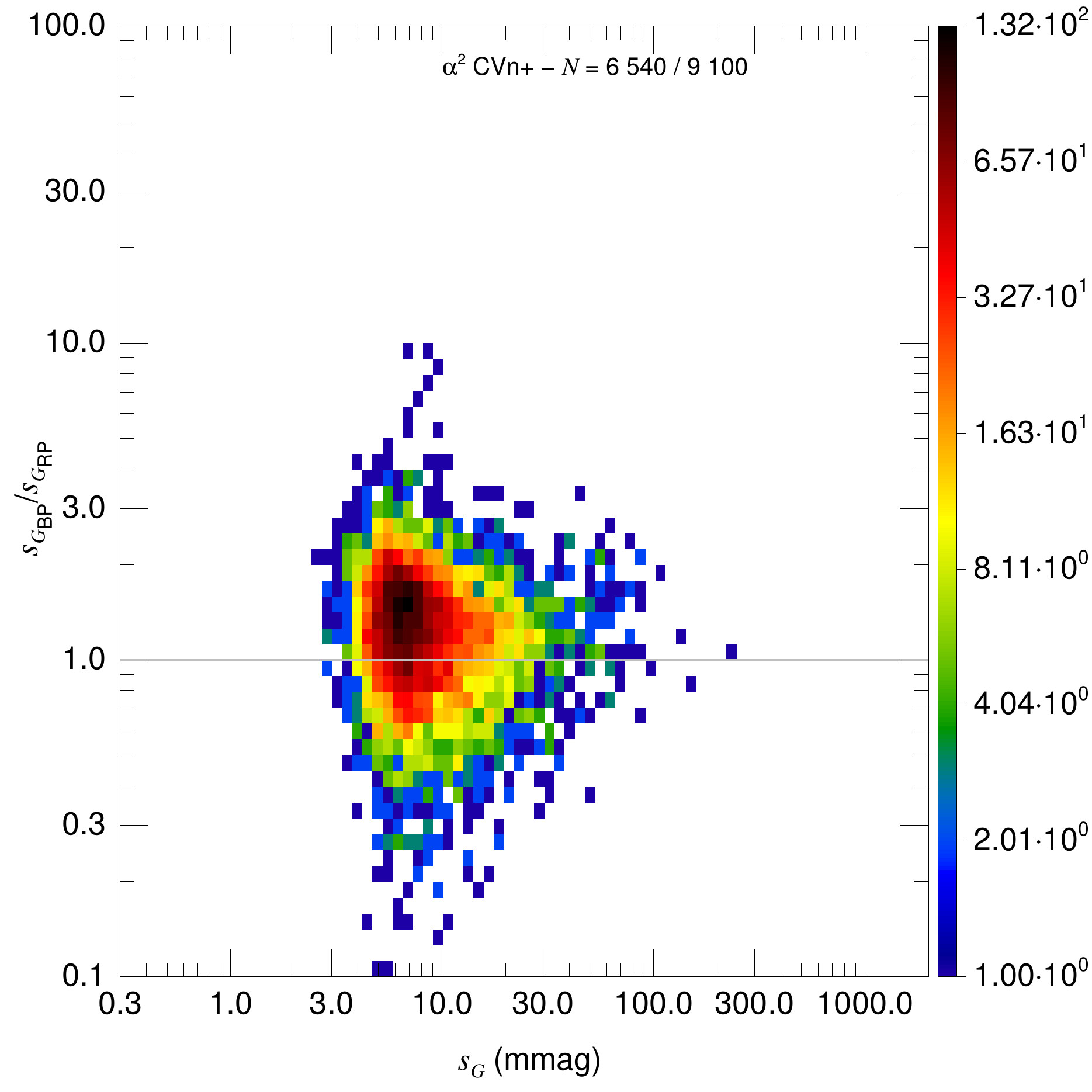}$\!\!\!$
            \includegraphics[width=0.35\linewidth]{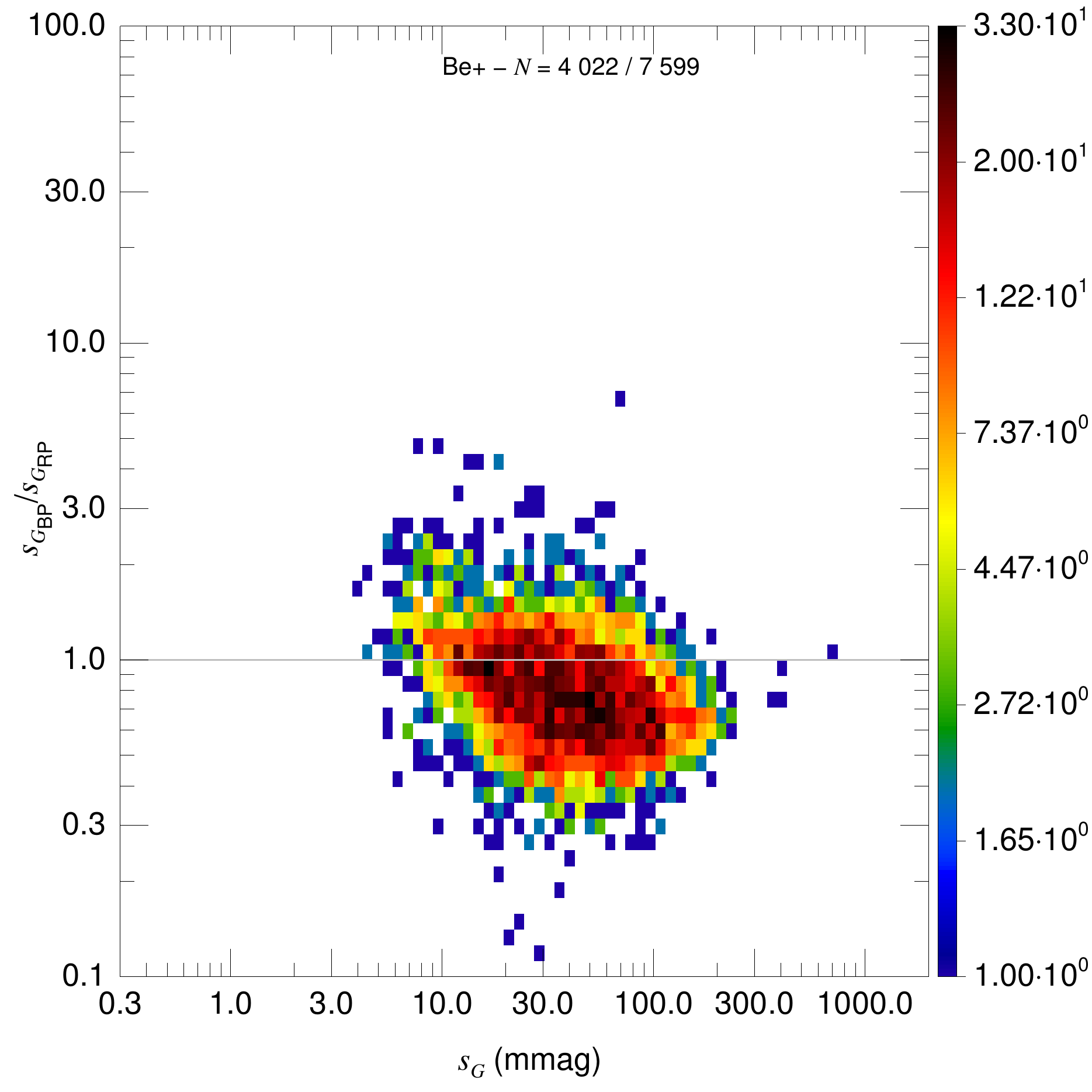}$\!\!\!$
            \includegraphics[width=0.35\linewidth]{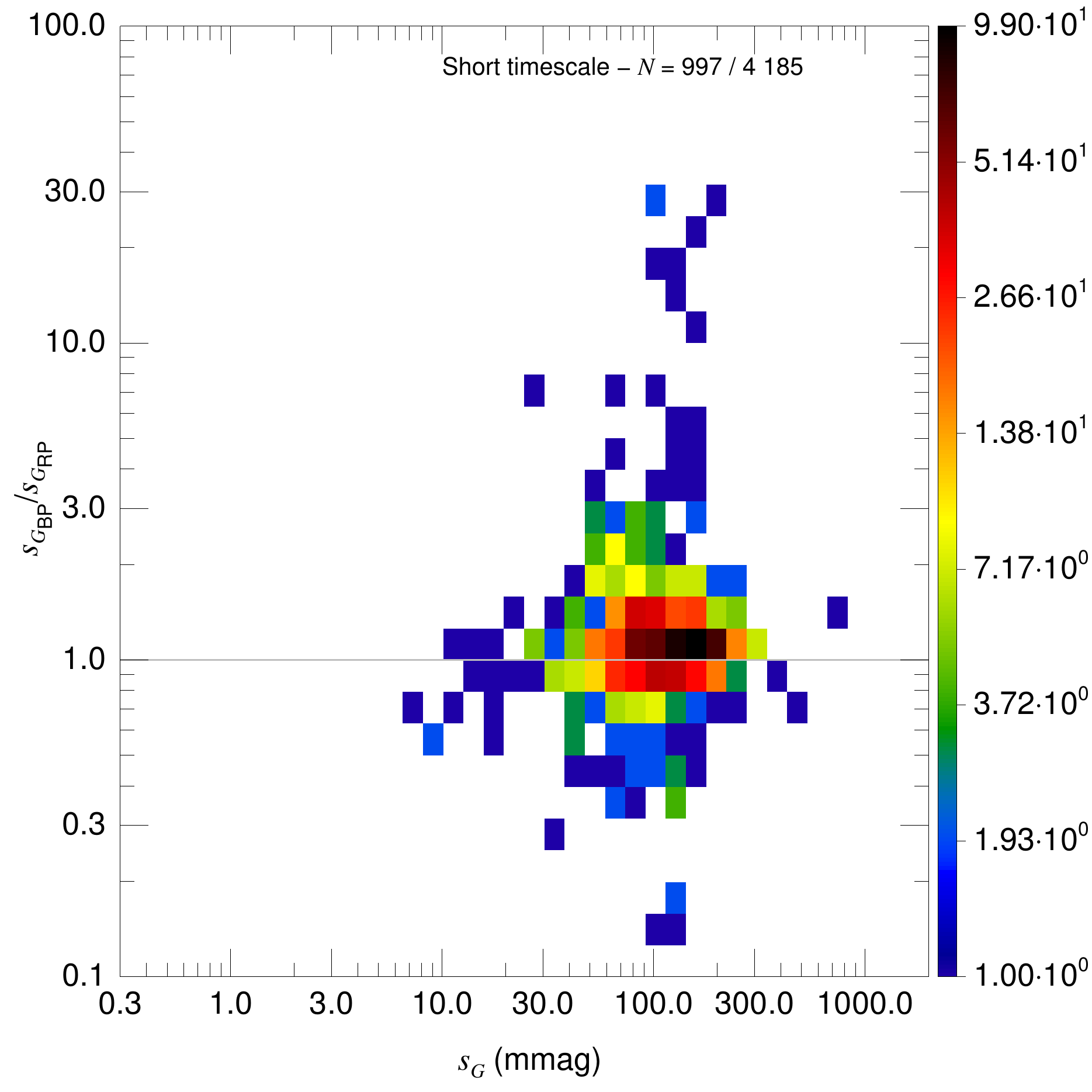}}
\centerline{\includegraphics[width=0.35\linewidth]{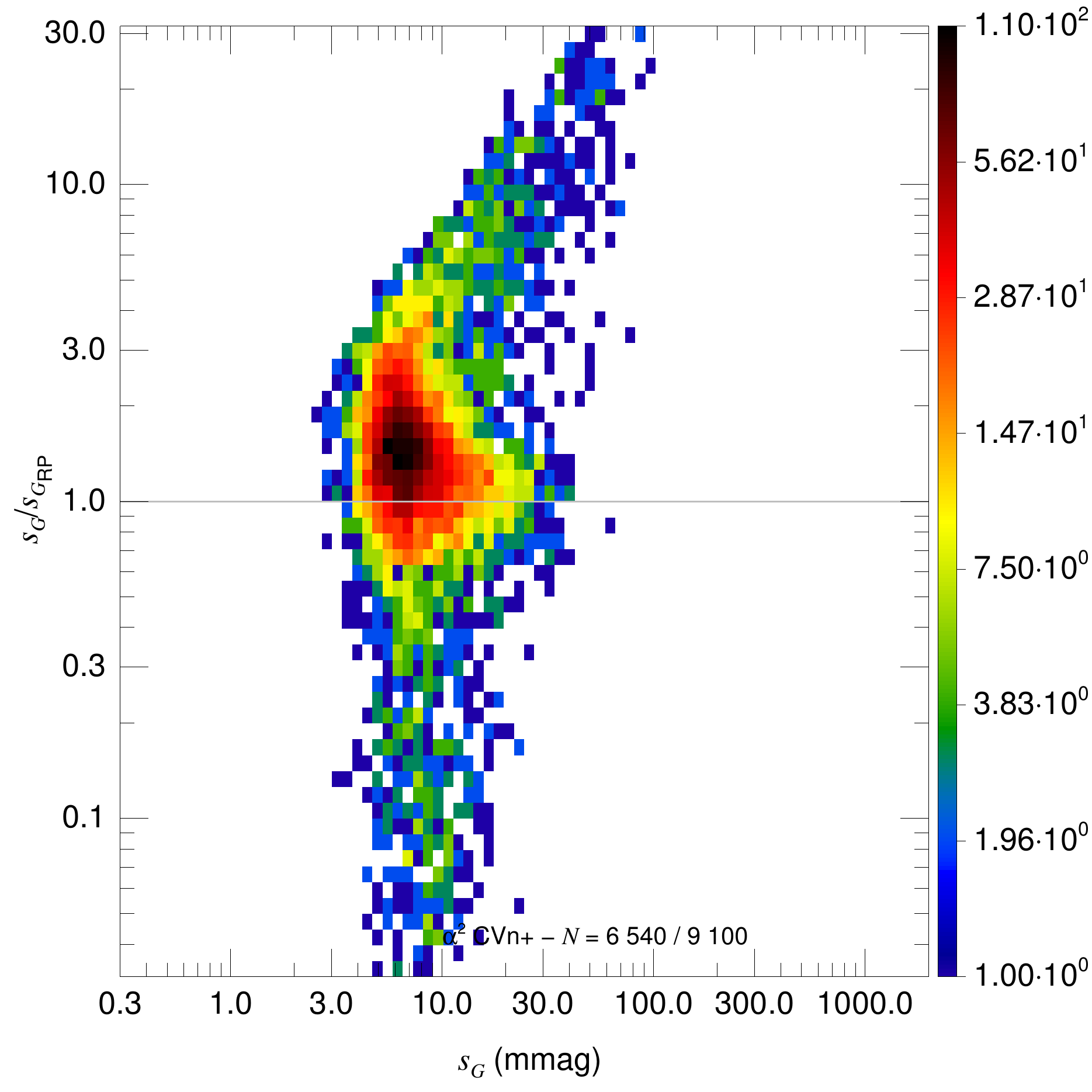}$\!\!\!$
            \includegraphics[width=0.35\linewidth]{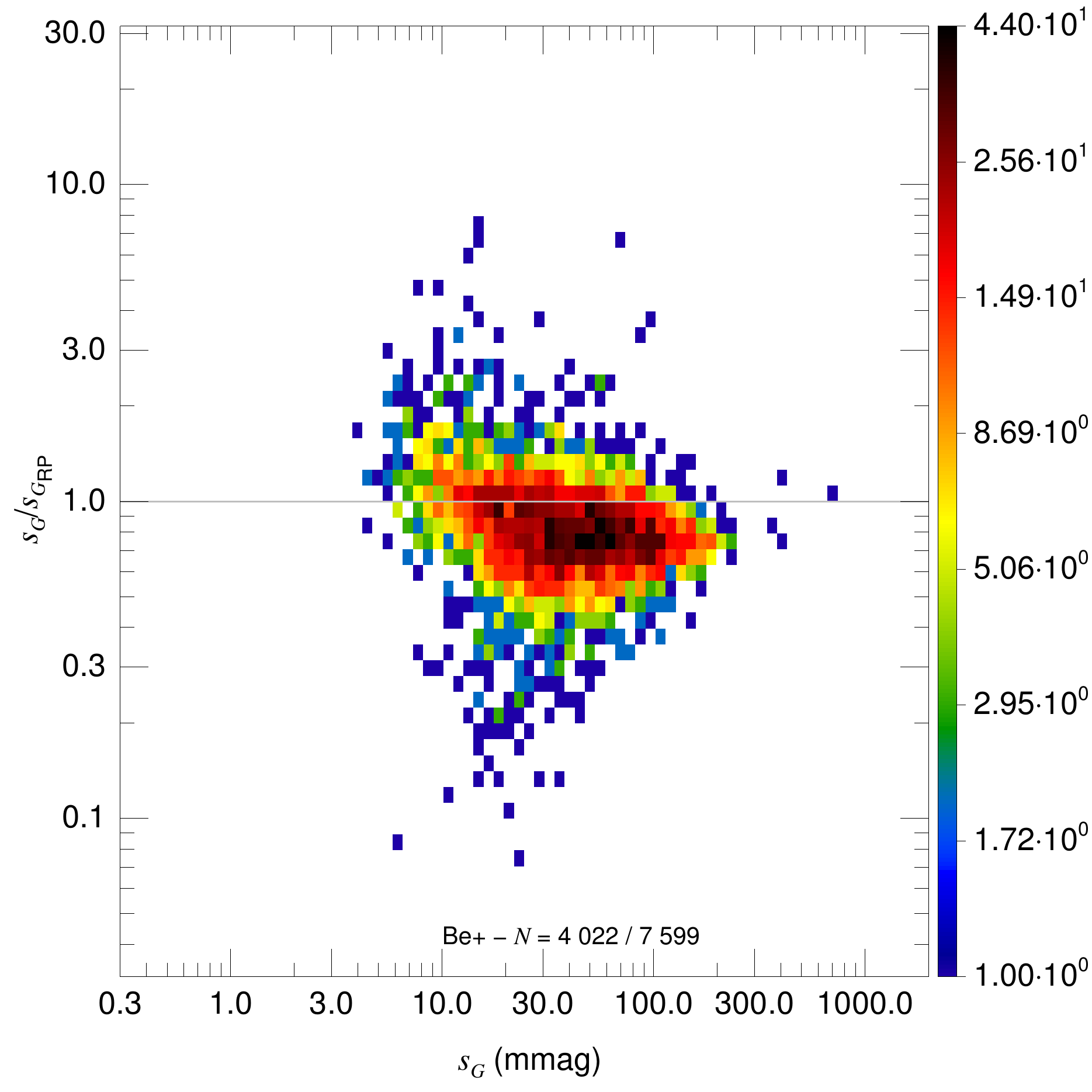}$\!\!\!$
            \includegraphics[width=0.35\linewidth]{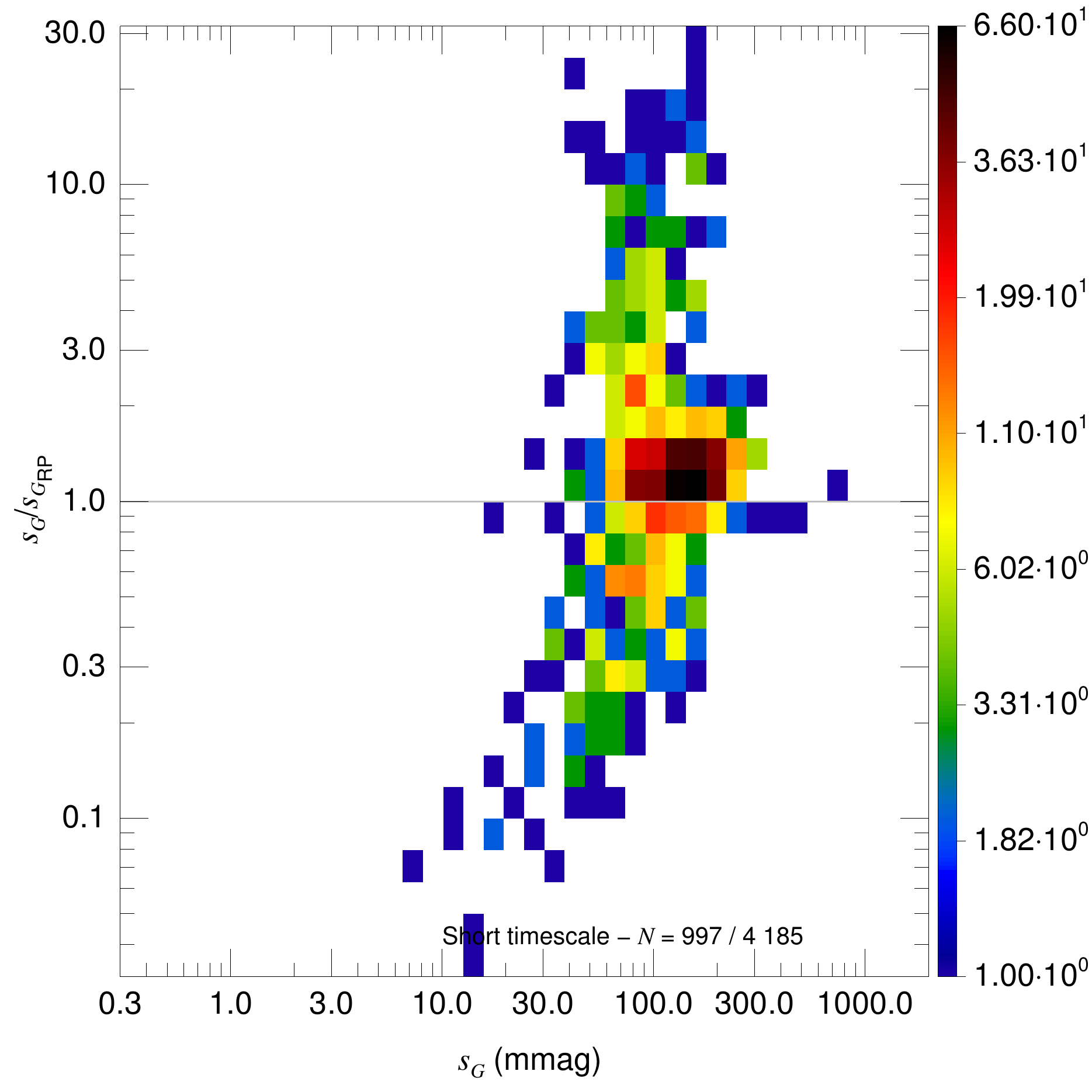}}
\centerline{\includegraphics[width=0.35\linewidth]{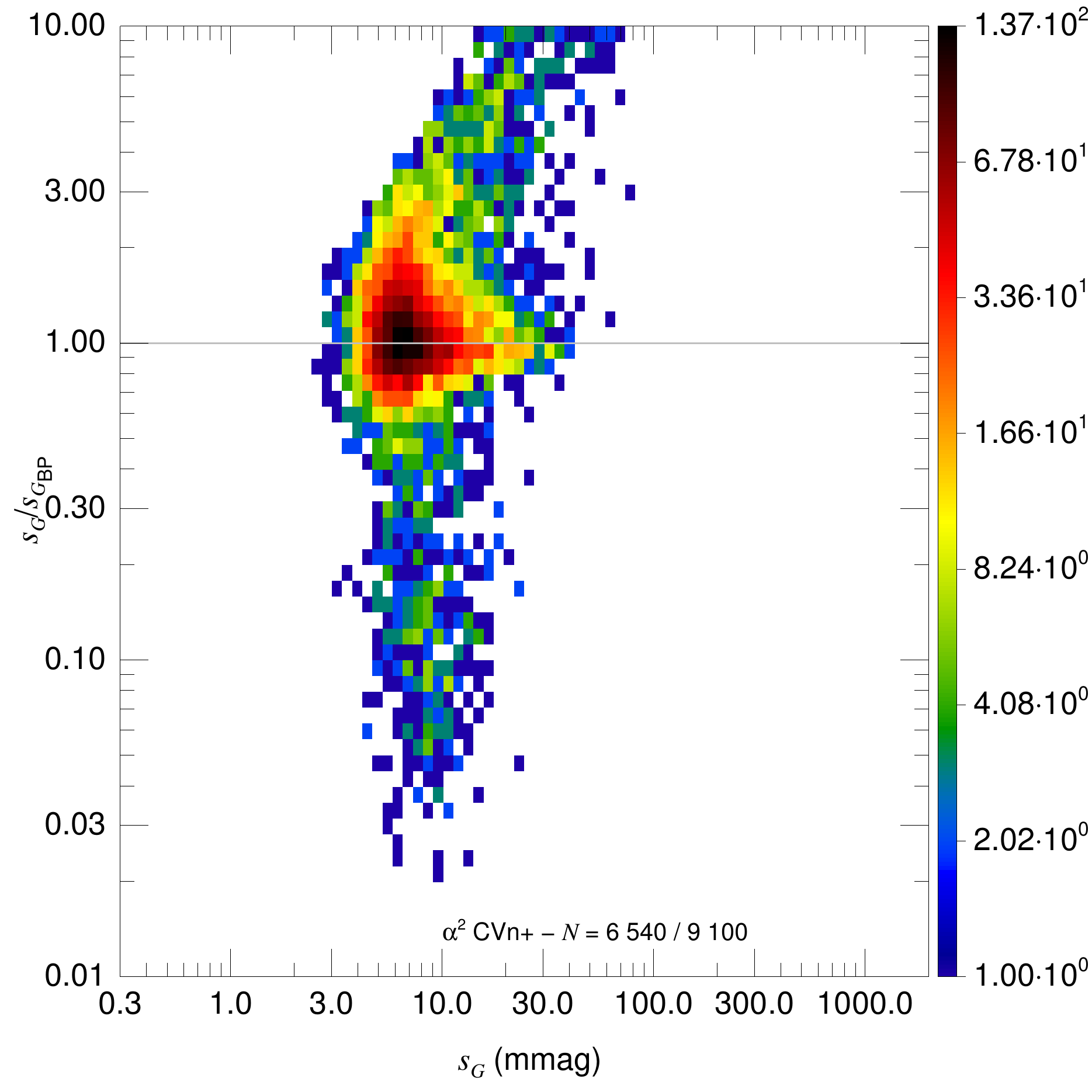}$\!\!\!$
            \includegraphics[width=0.35\linewidth]{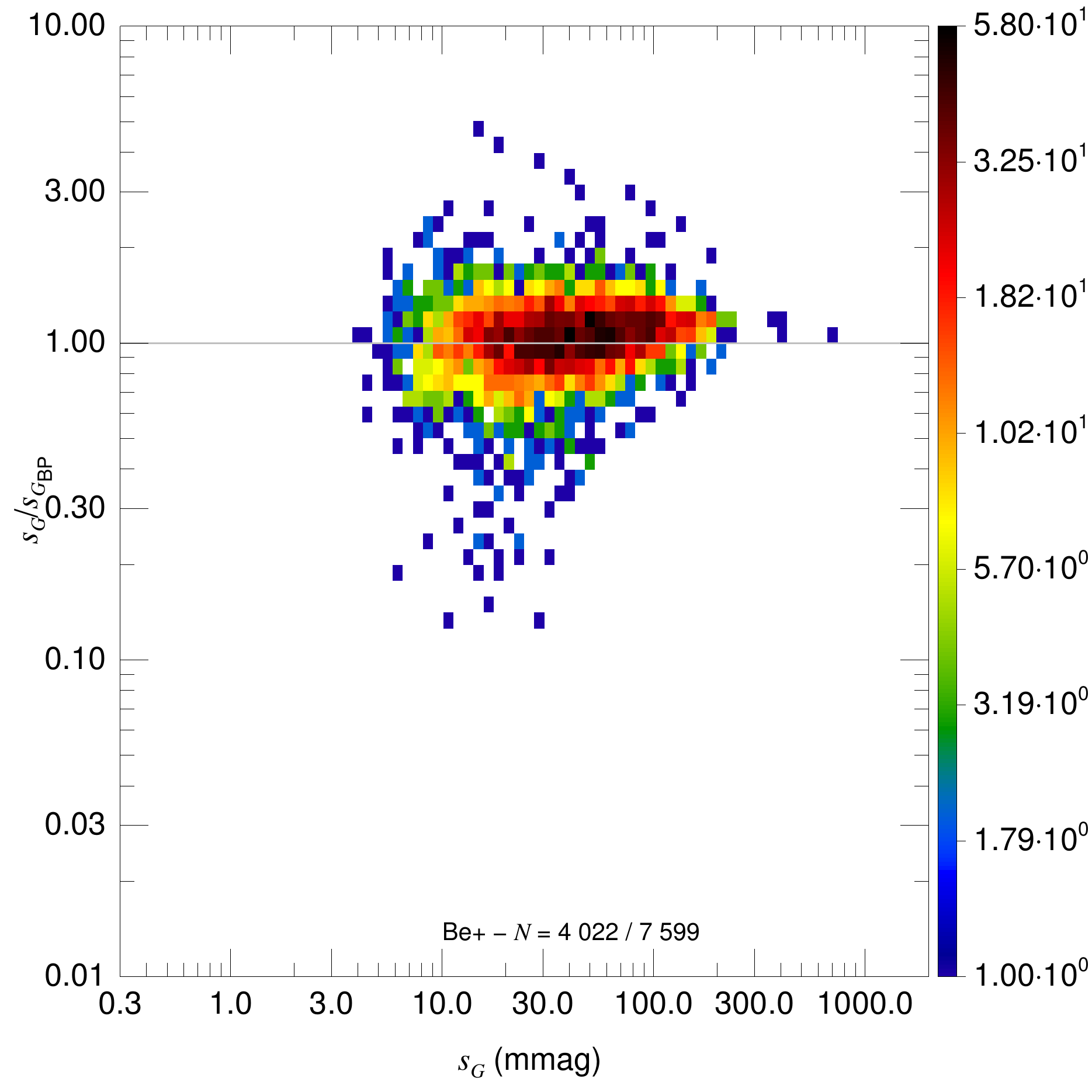}$\!\!\!$
            \includegraphics[width=0.35\linewidth]{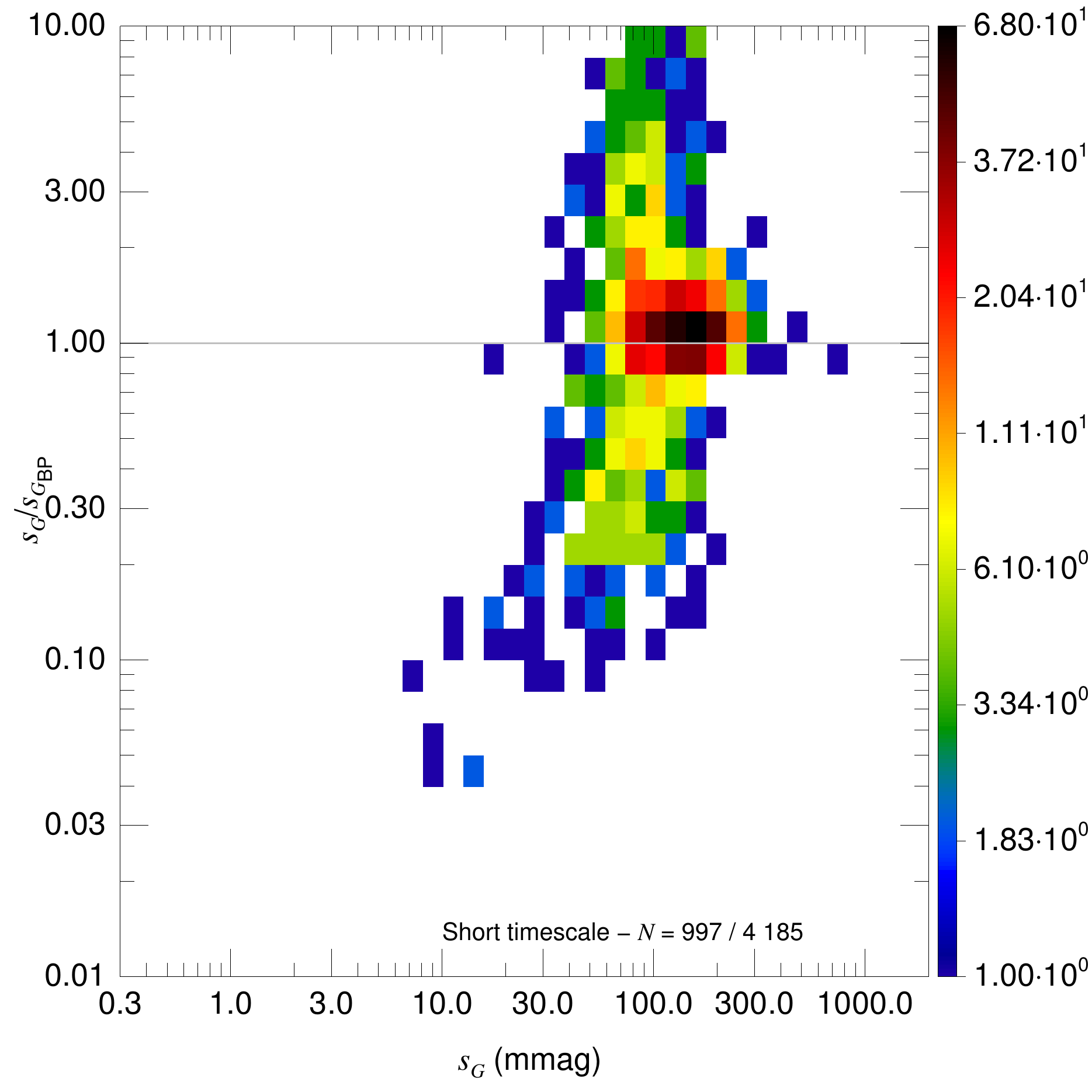}}
\caption{(Continued).}
\end{figure*}

\addtocounter{figure}{-1}

\begin{figure*}
\centerline{\includegraphics[width=0.35\linewidth]{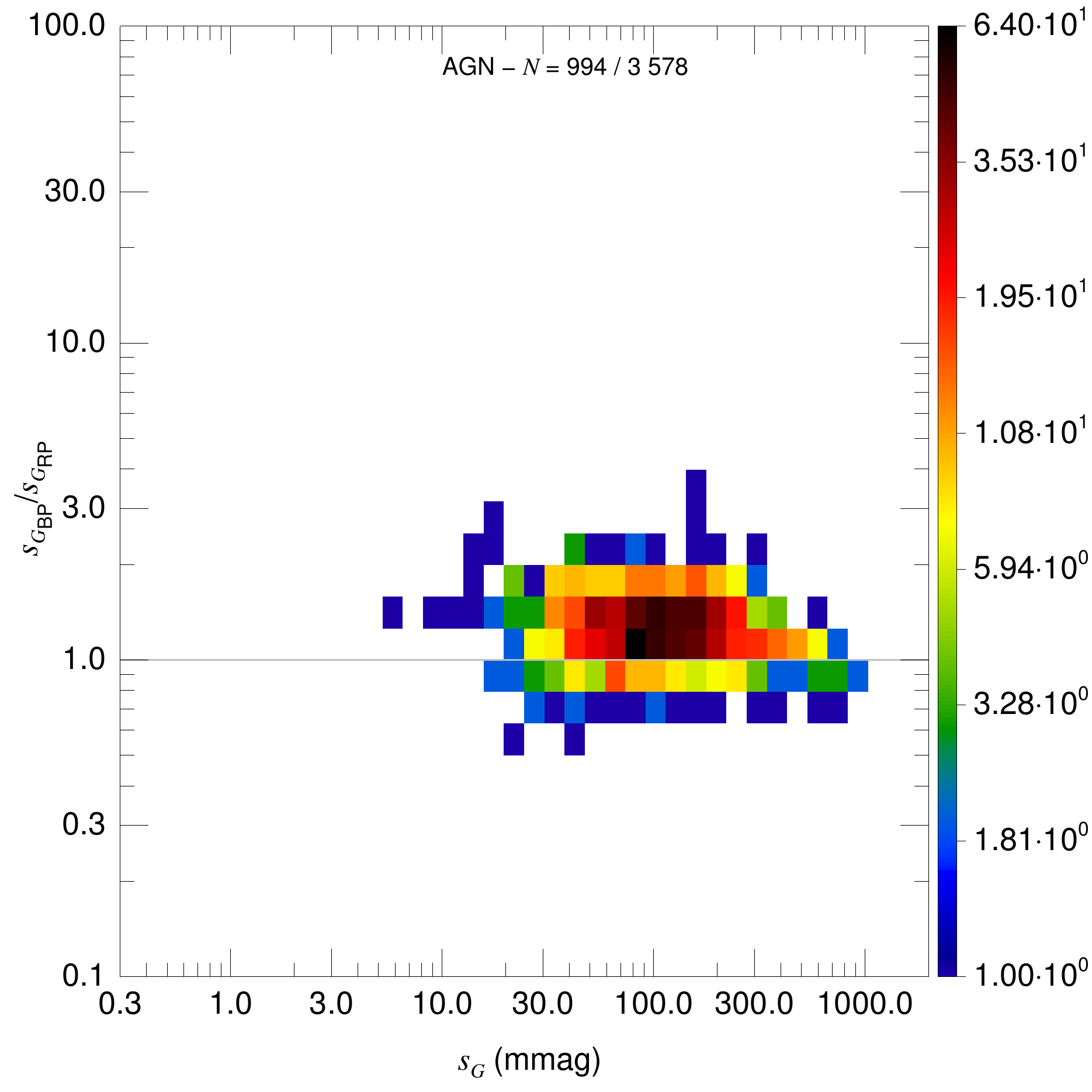}$\!\!\!$
            \includegraphics[width=0.35\linewidth]{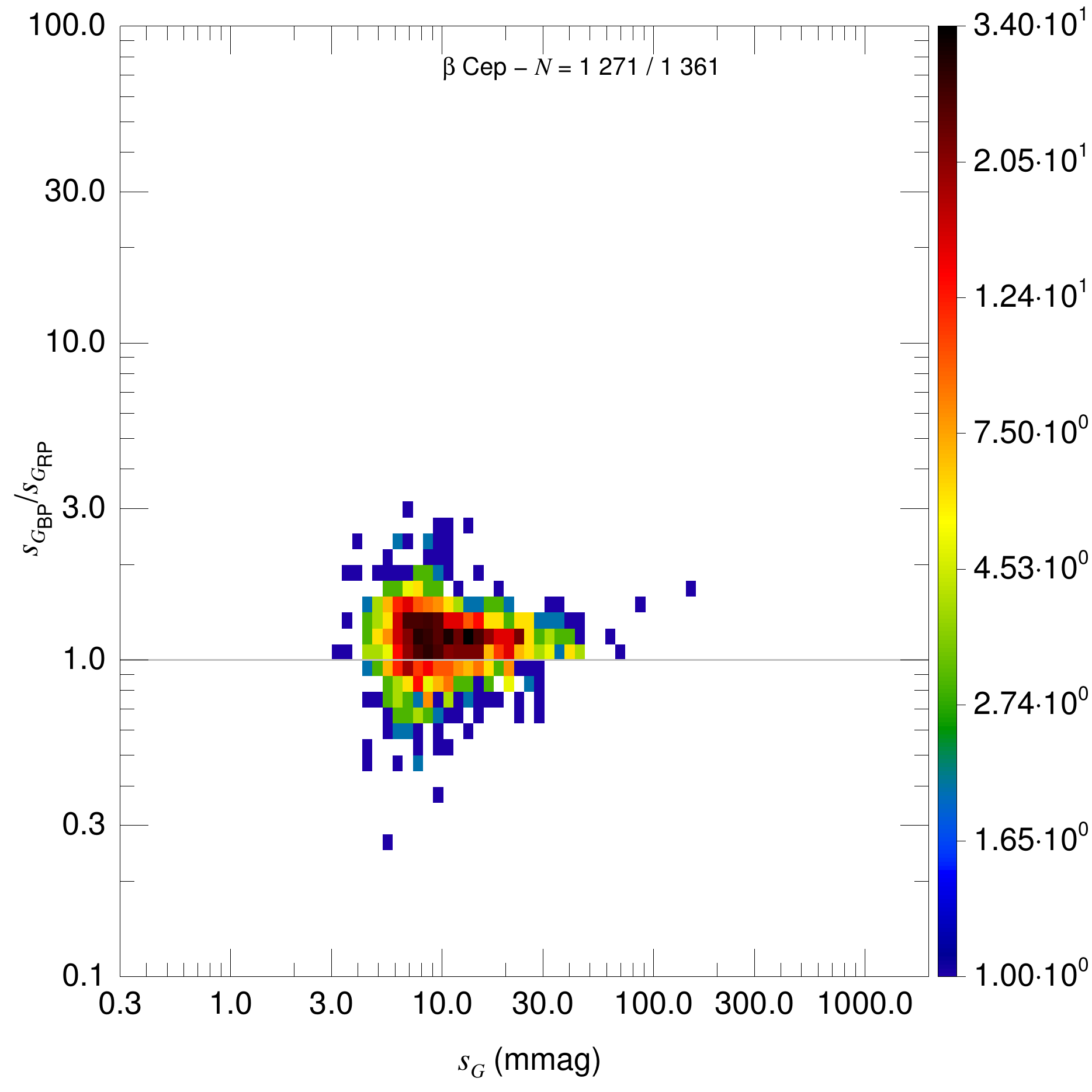}$\!\!\!$
            \includegraphics[width=0.35\linewidth]{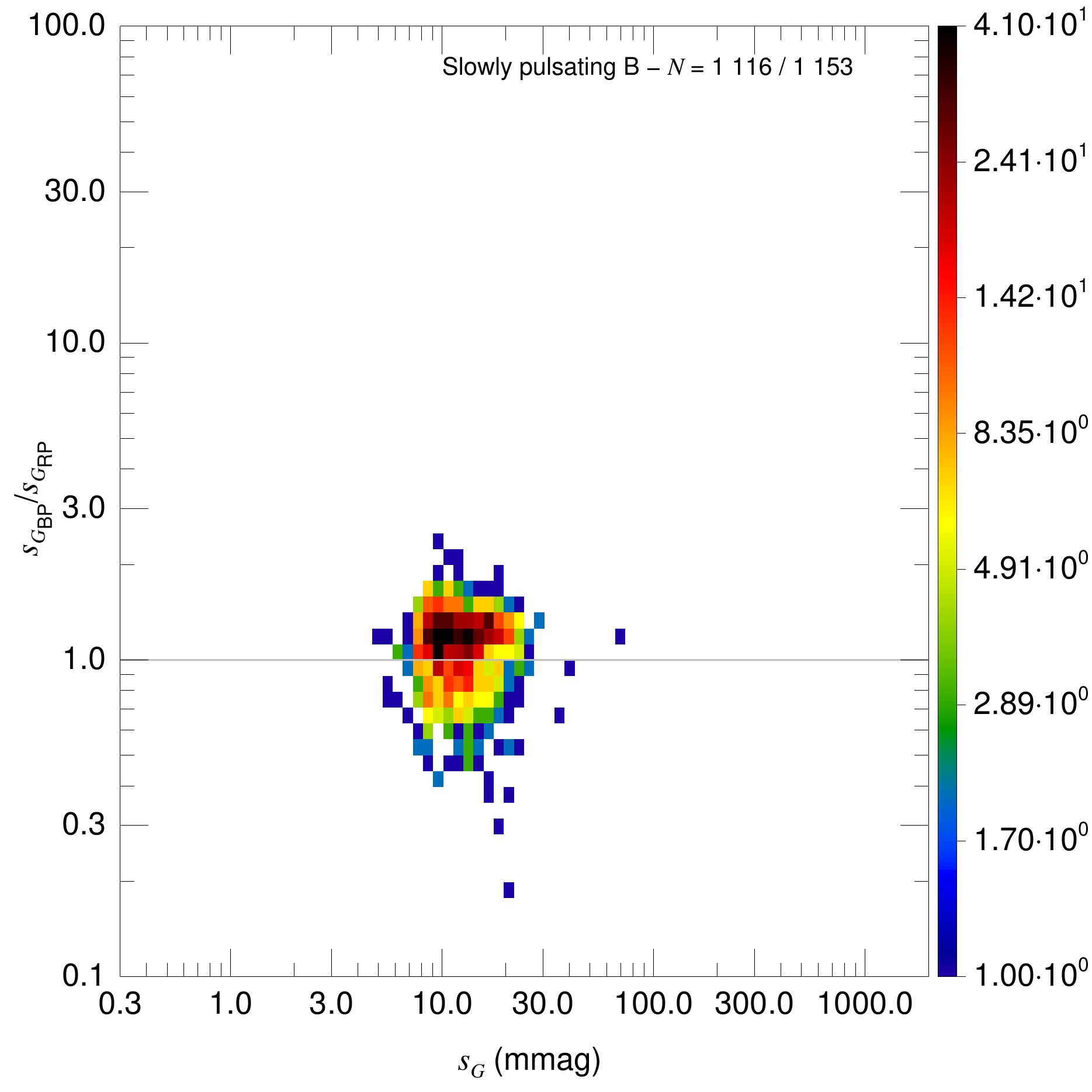}}
\centerline{\includegraphics[width=0.35\linewidth]{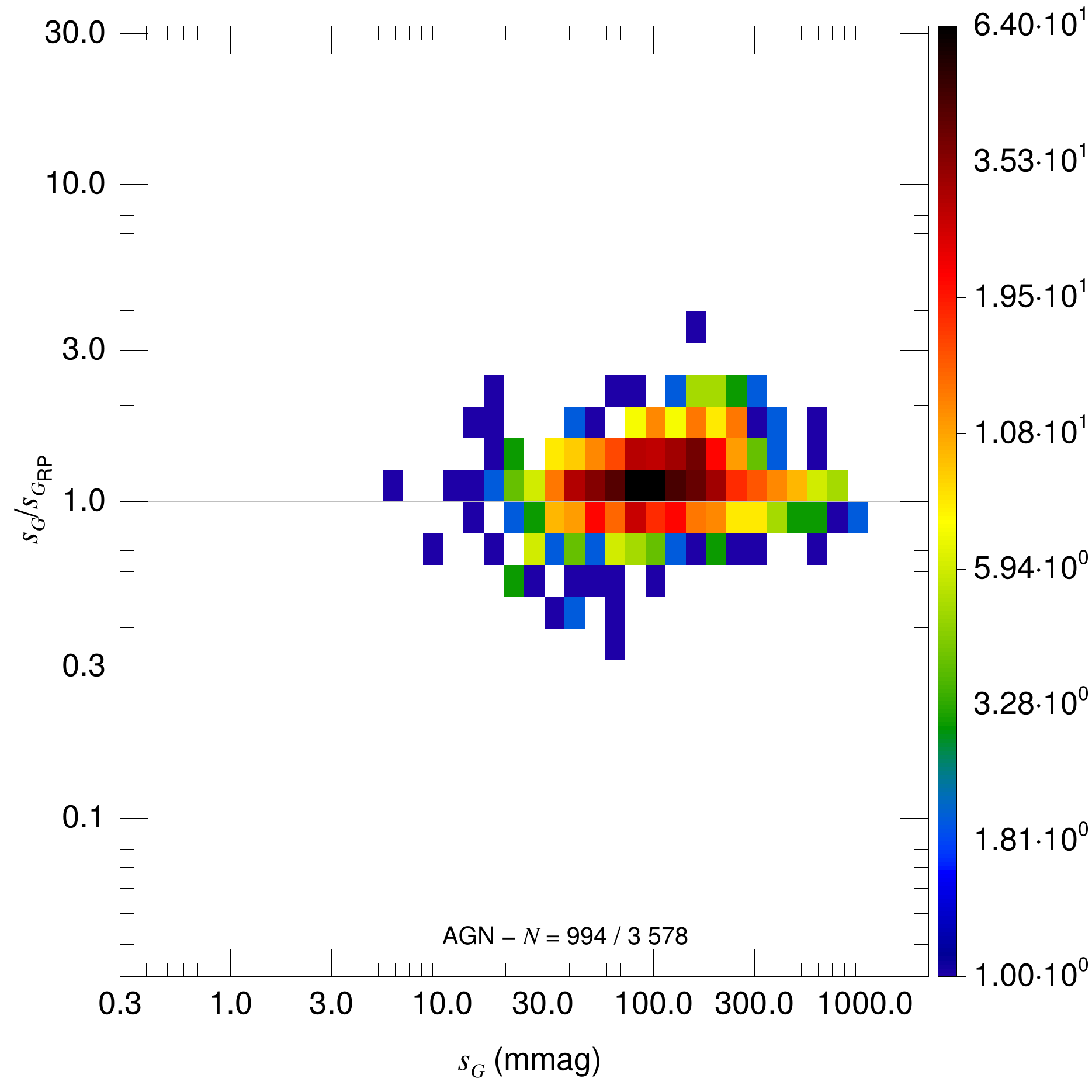}$\!\!\!$
            \includegraphics[width=0.35\linewidth]{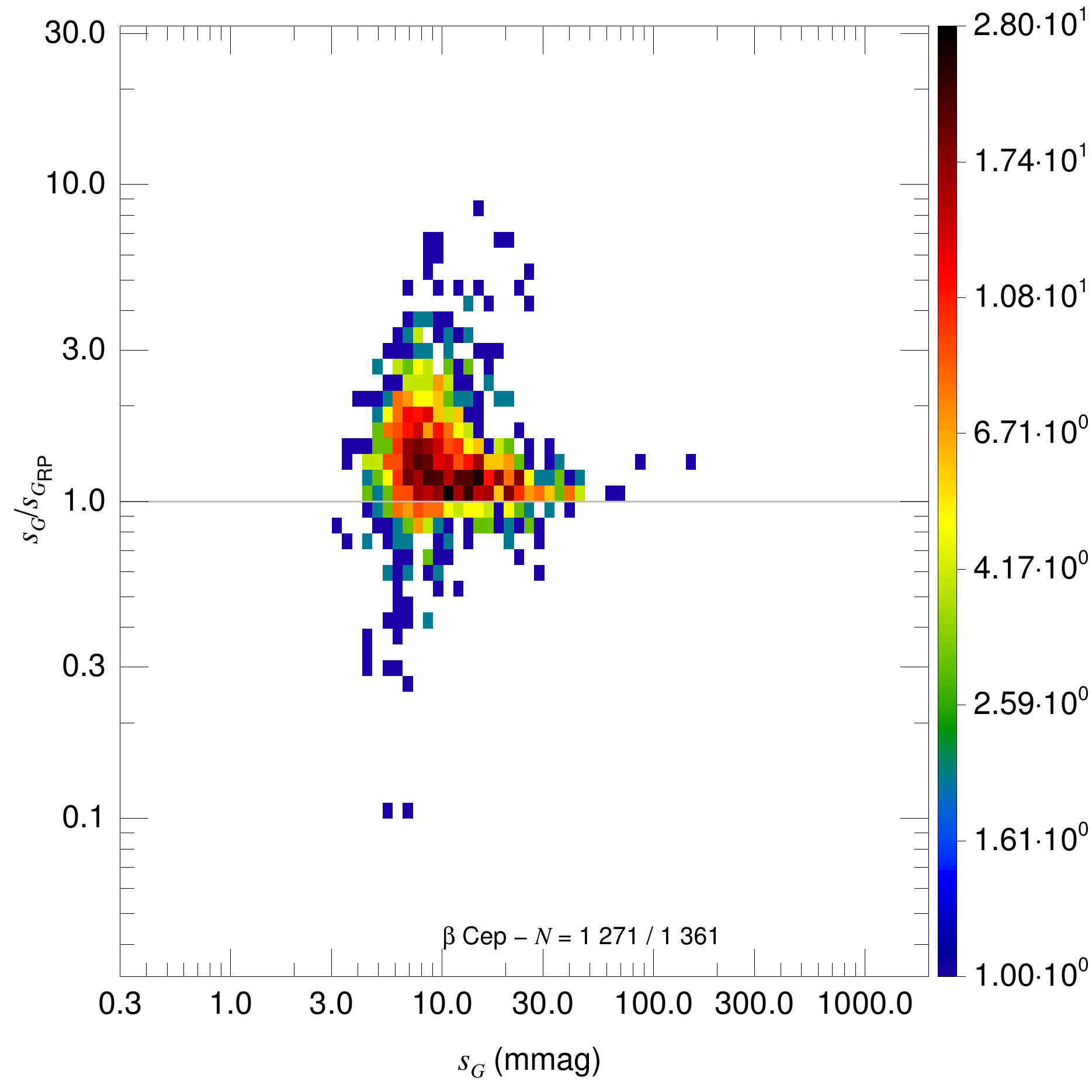}$\!\!\!$
            \includegraphics[width=0.35\linewidth]{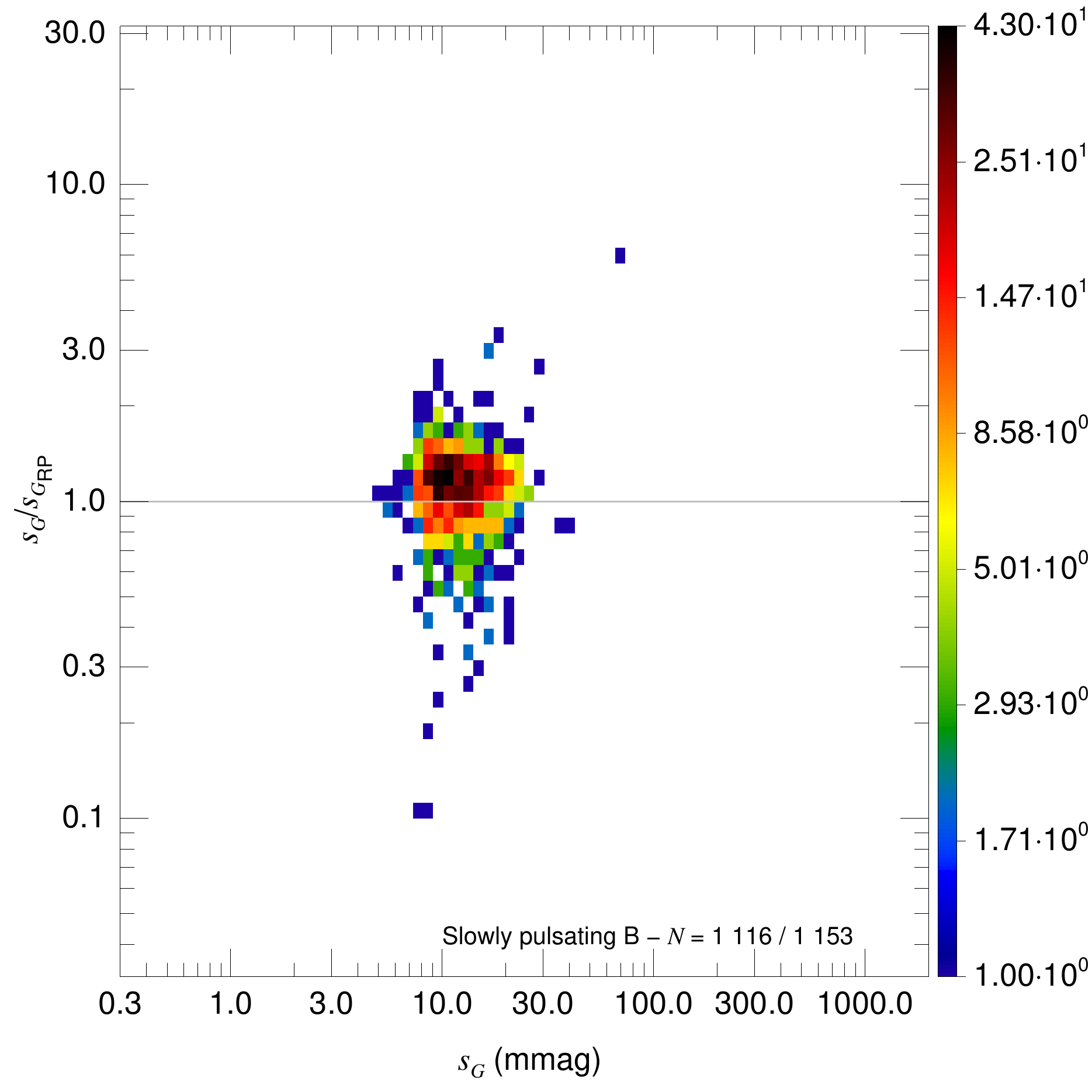}}
\centerline{\includegraphics[width=0.35\linewidth]{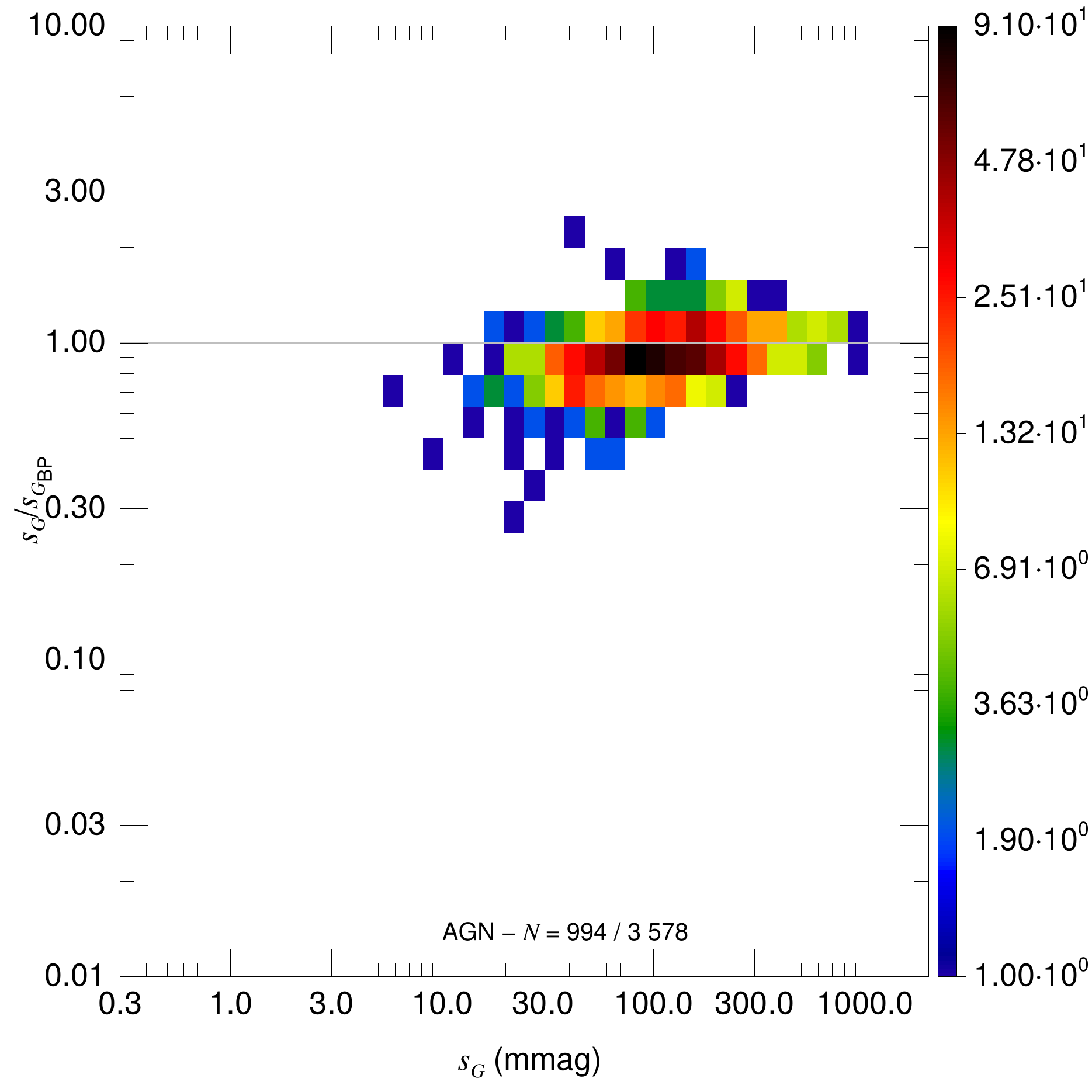}$\!\!\!$
            \includegraphics[width=0.35\linewidth]{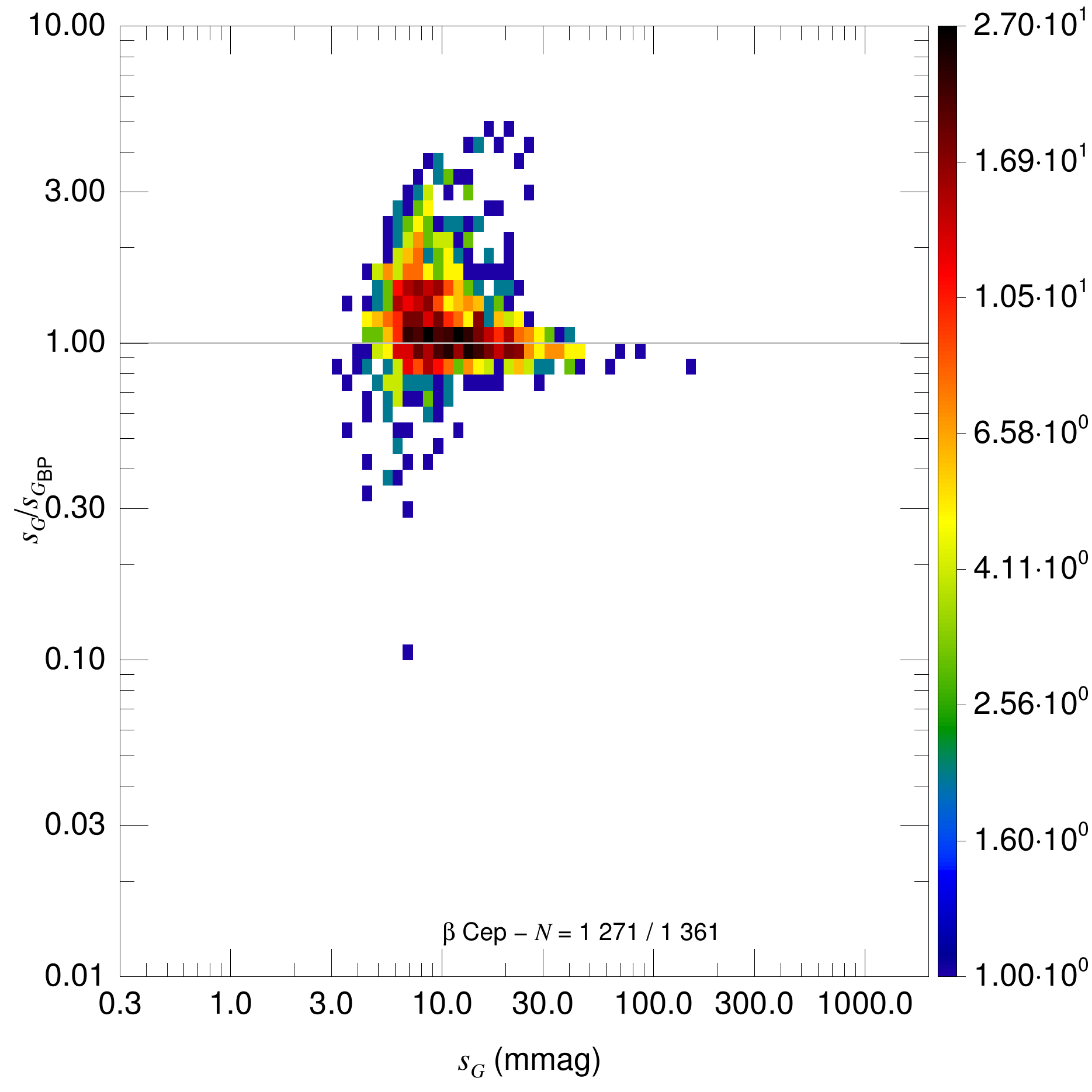}$\!\!\!$
            \includegraphics[width=0.35\linewidth]{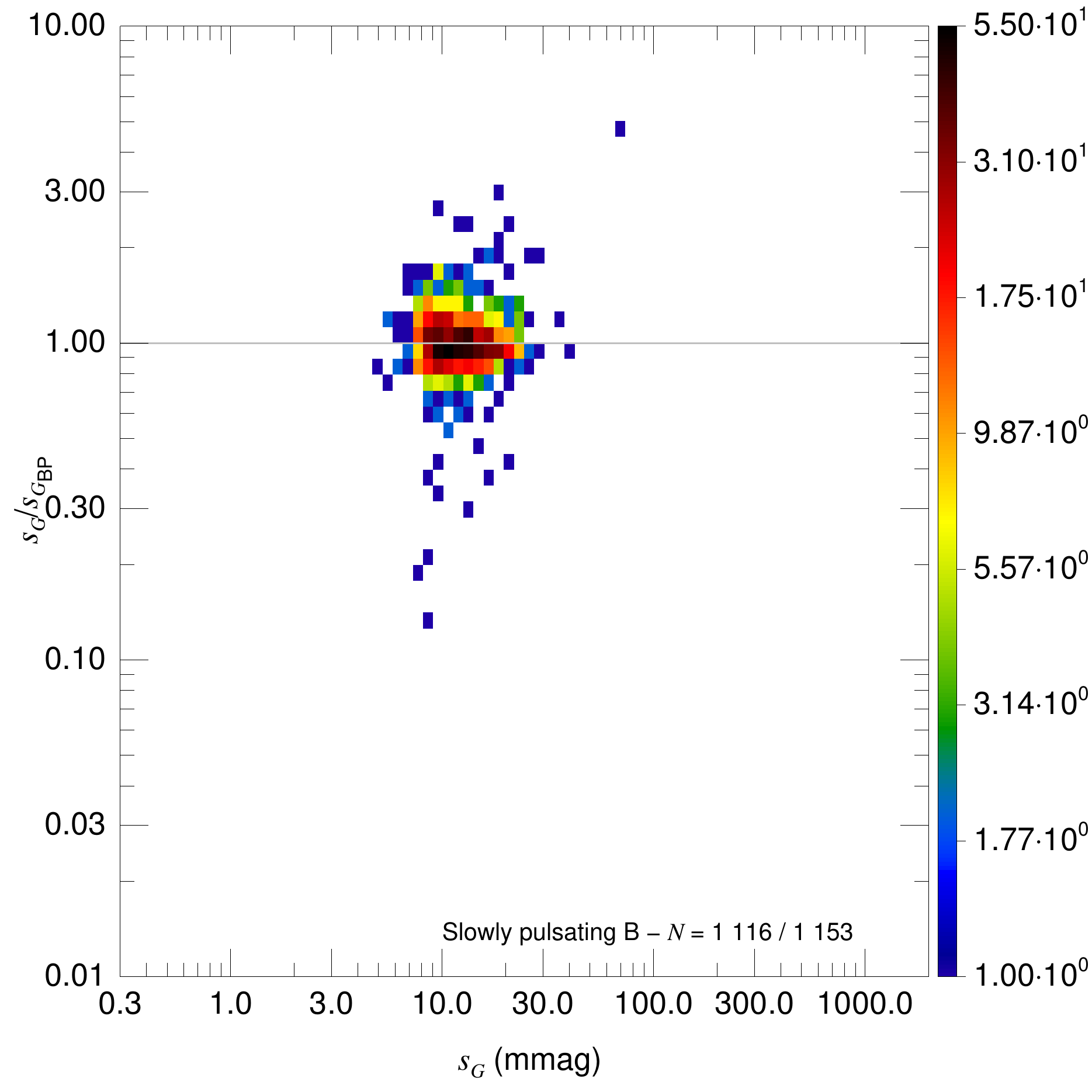}}
\caption{(Continued).}
\end{figure*}

\addtocounter{figure}{-1}

\begin{figure*}
\centerline{\includegraphics[width=0.35\linewidth]{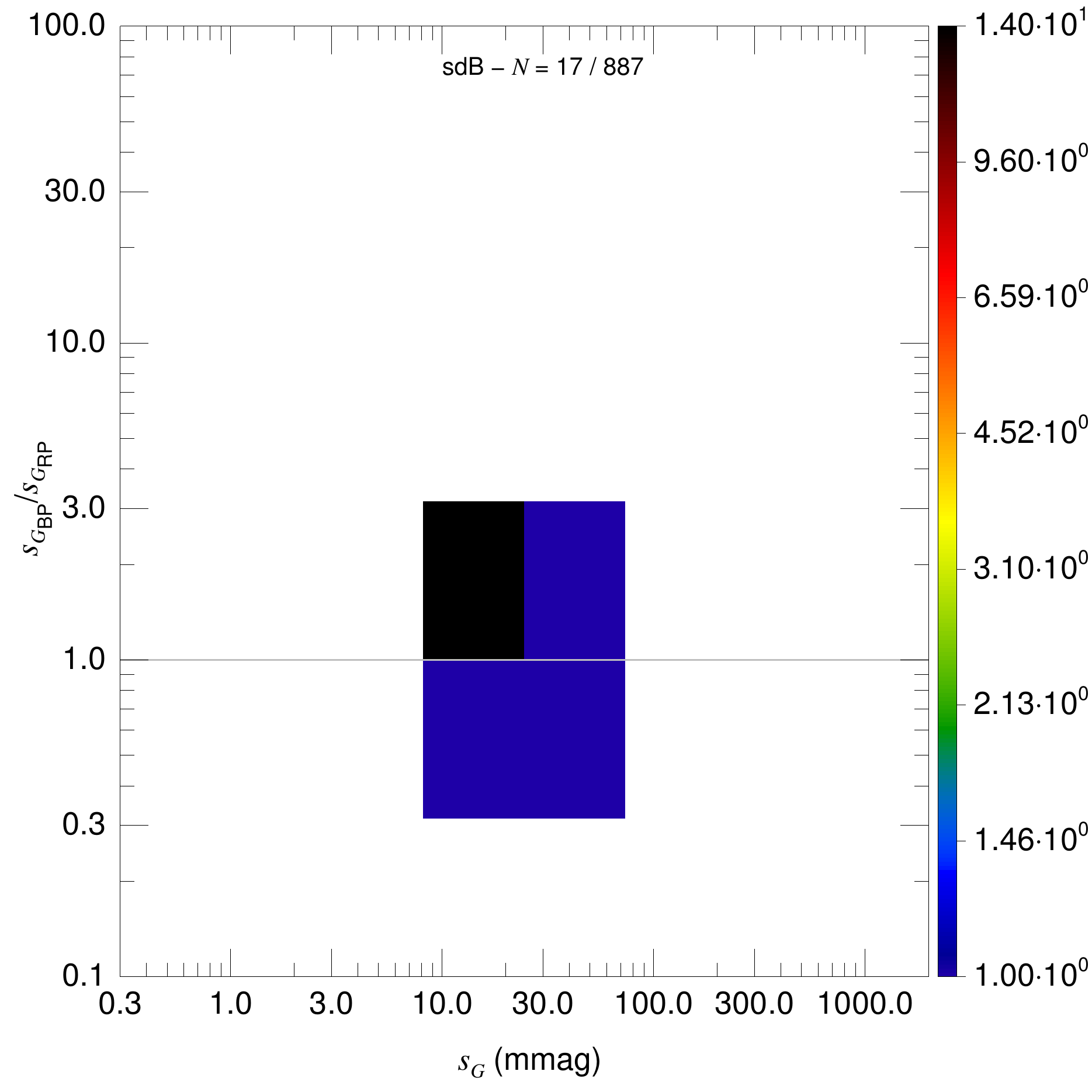}$\!\!\!$
            \includegraphics[width=0.35\linewidth]{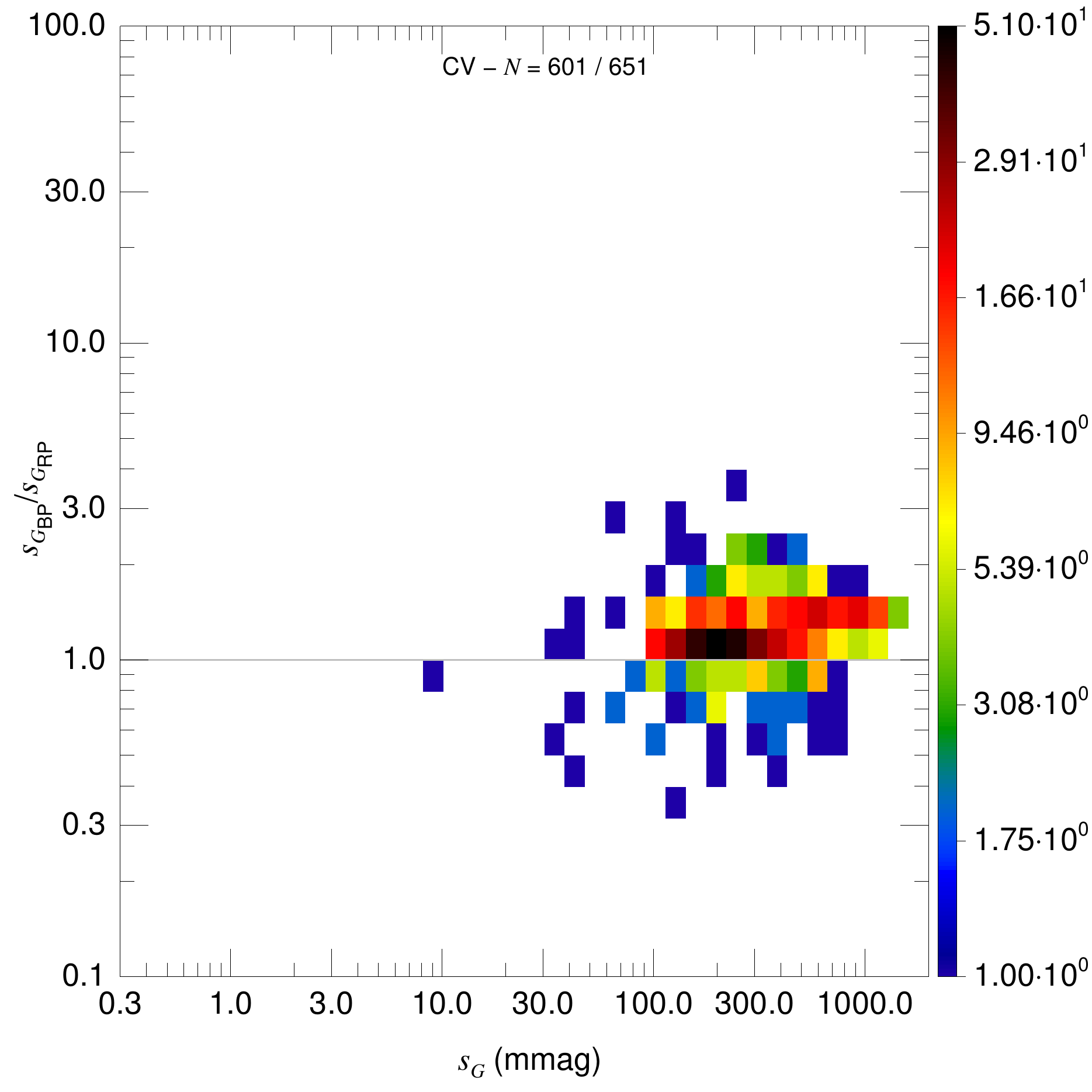}$\!\!\!$
            \includegraphics[width=0.35\linewidth]{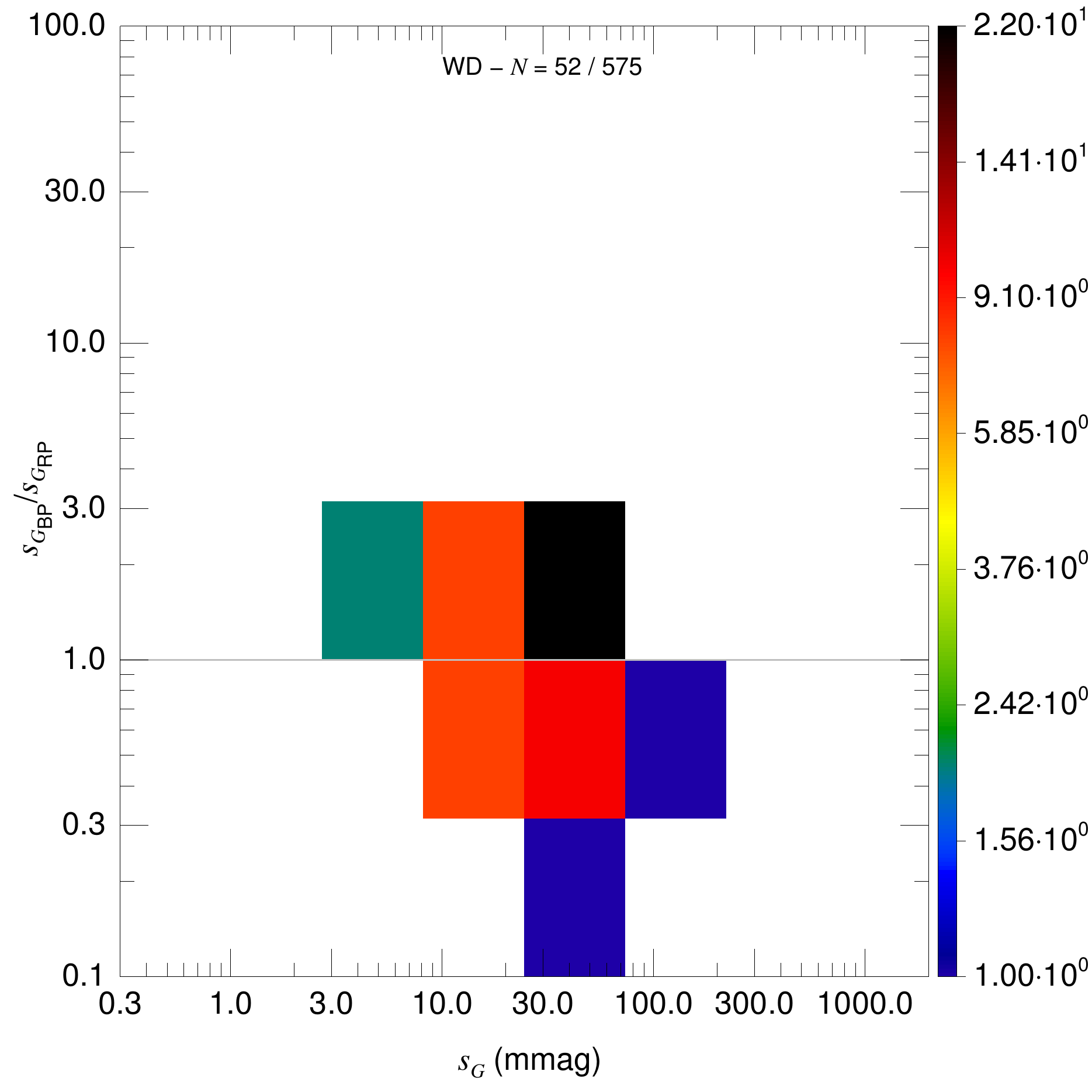}}
\centerline{\includegraphics[width=0.35\linewidth]{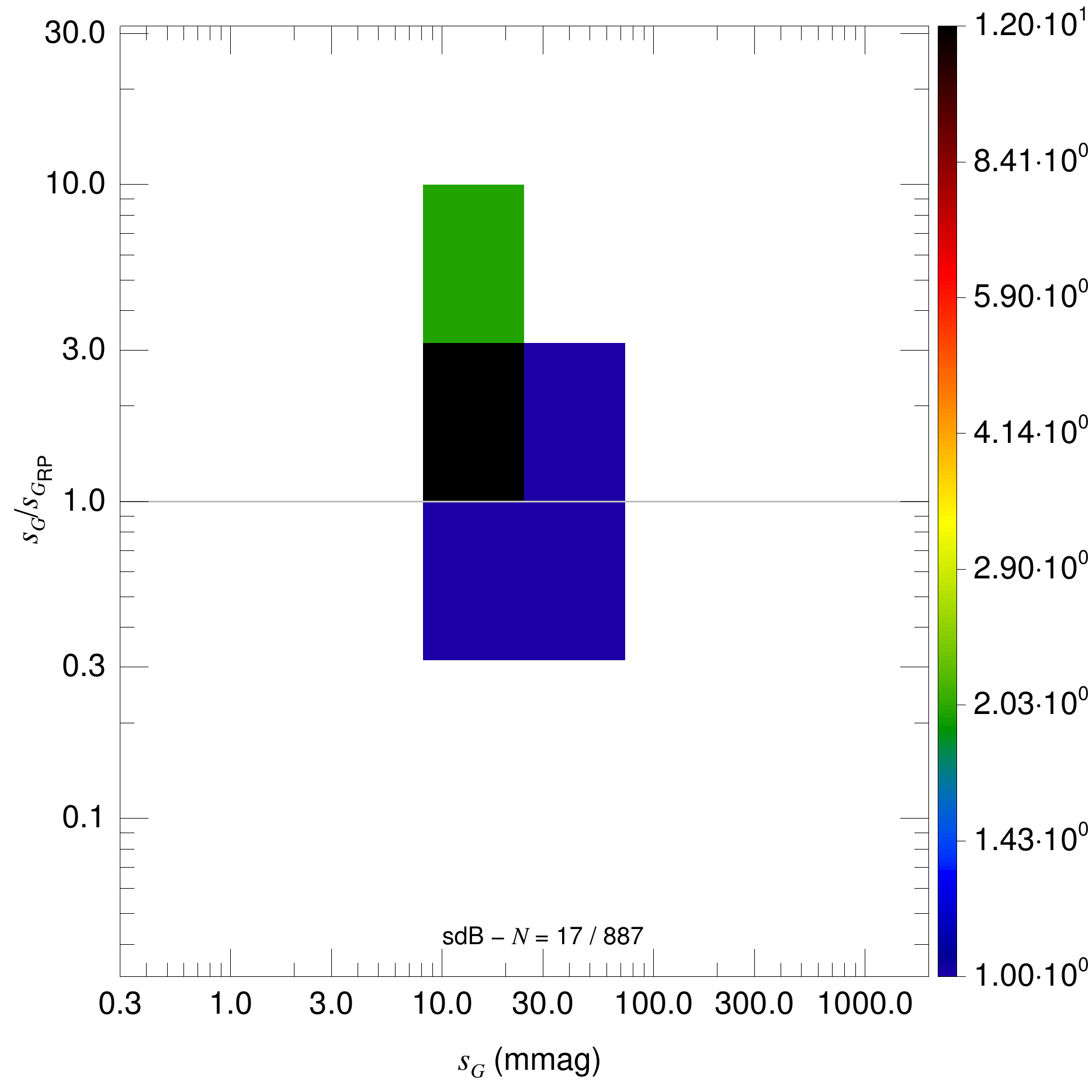}$\!\!\!$
            \includegraphics[width=0.35\linewidth]{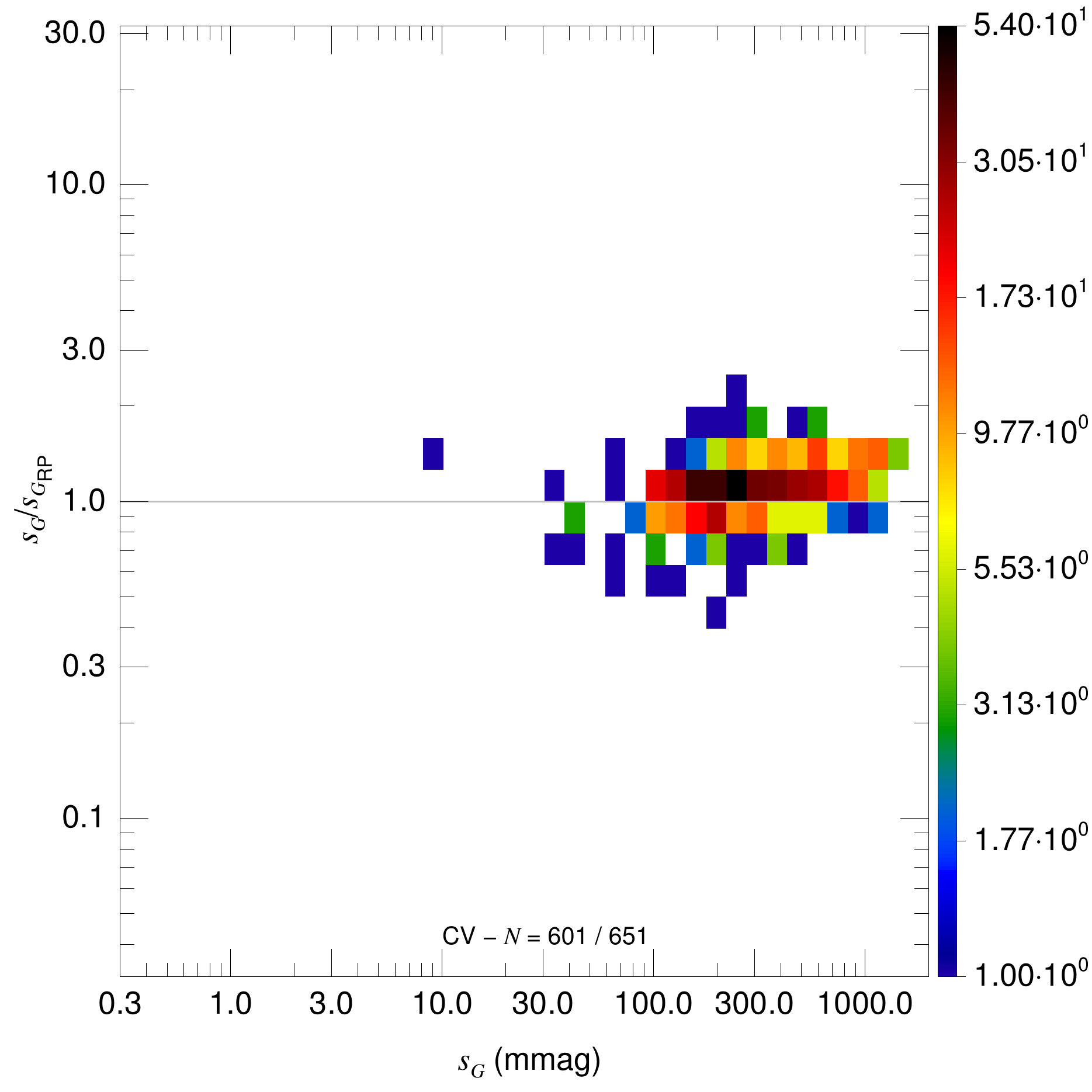}$\!\!\!$
            \includegraphics[width=0.35\linewidth]{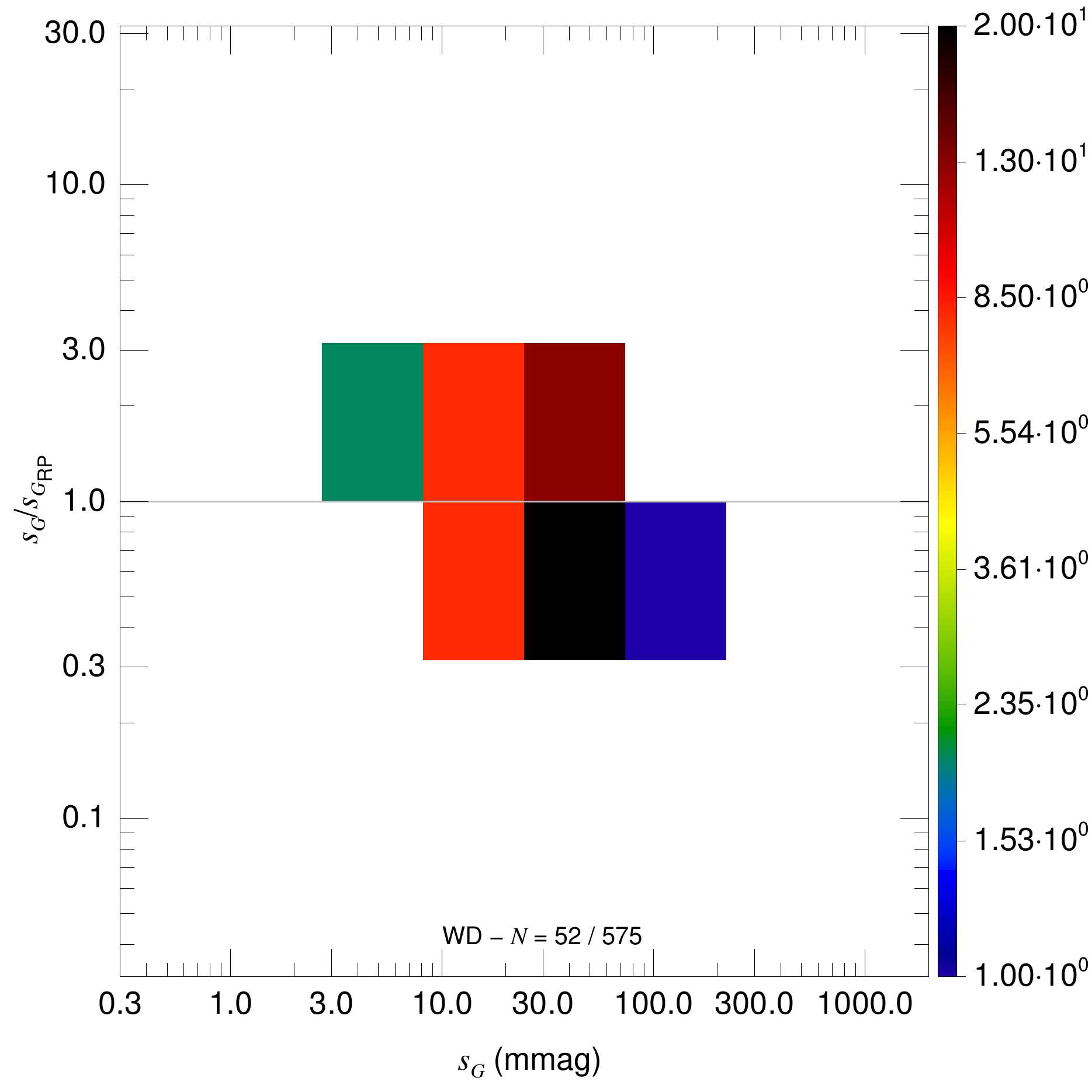}}
\centerline{\includegraphics[width=0.35\linewidth]{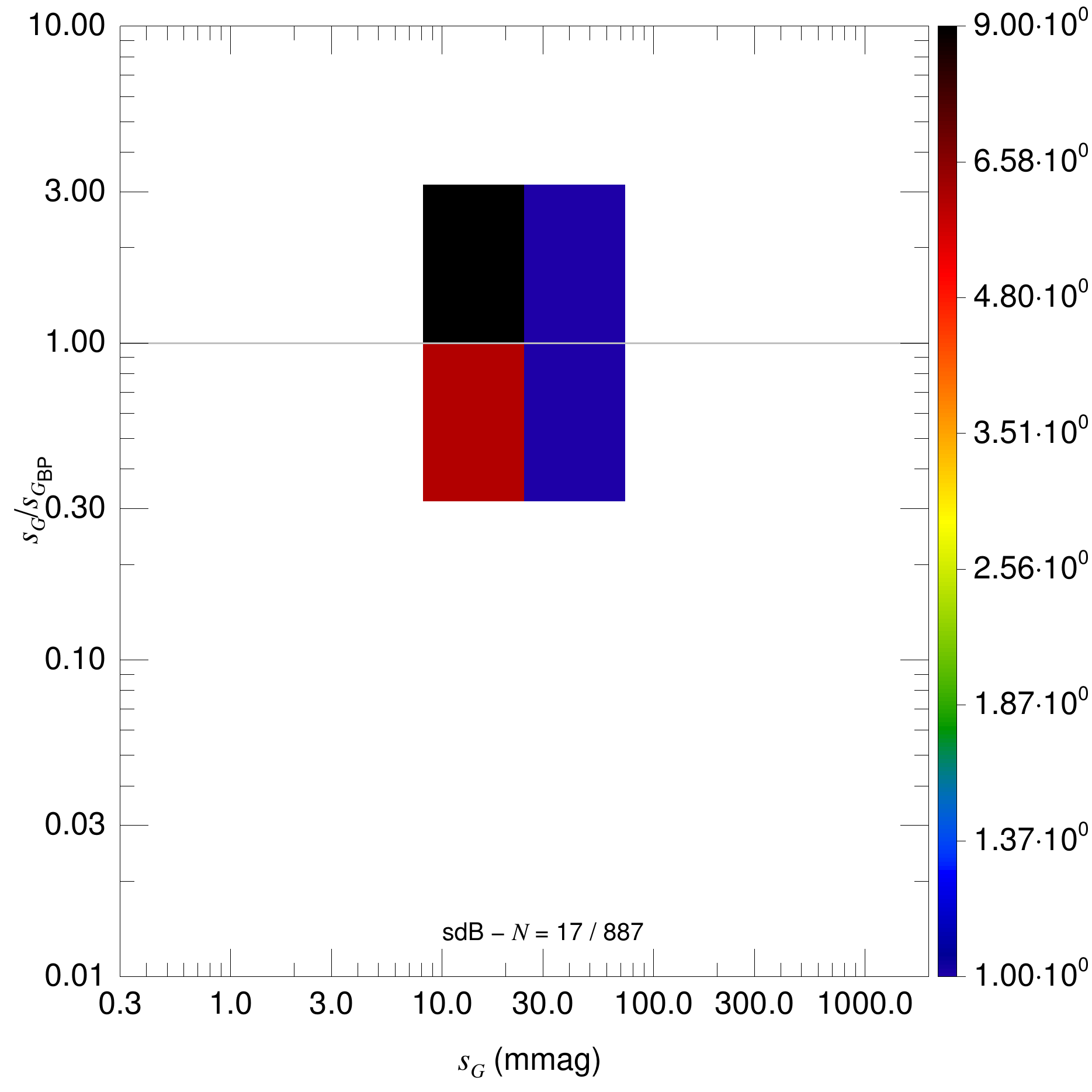}$\!\!\!$
            \includegraphics[width=0.35\linewidth]{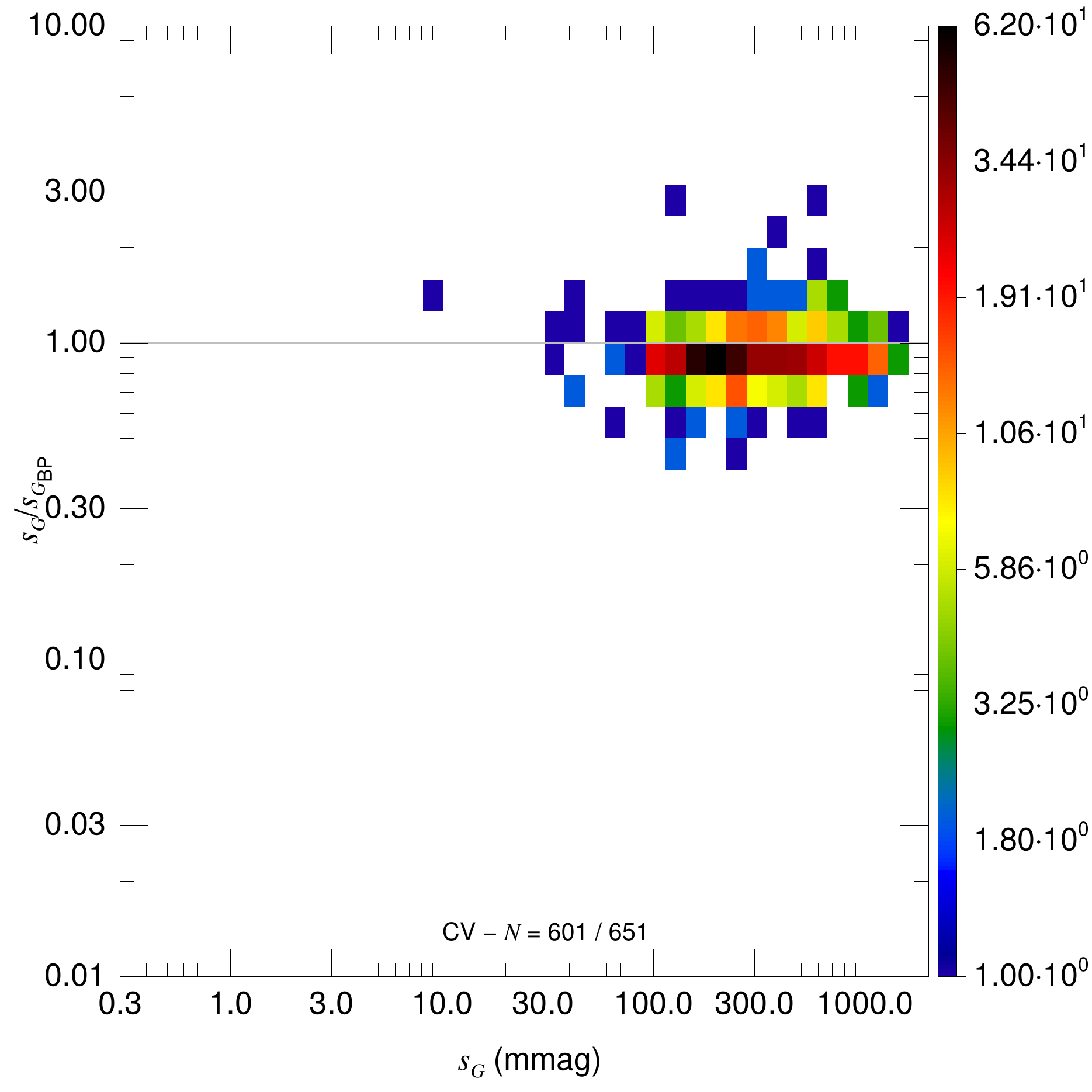}$\!\!\!$
            \includegraphics[width=0.35\linewidth]{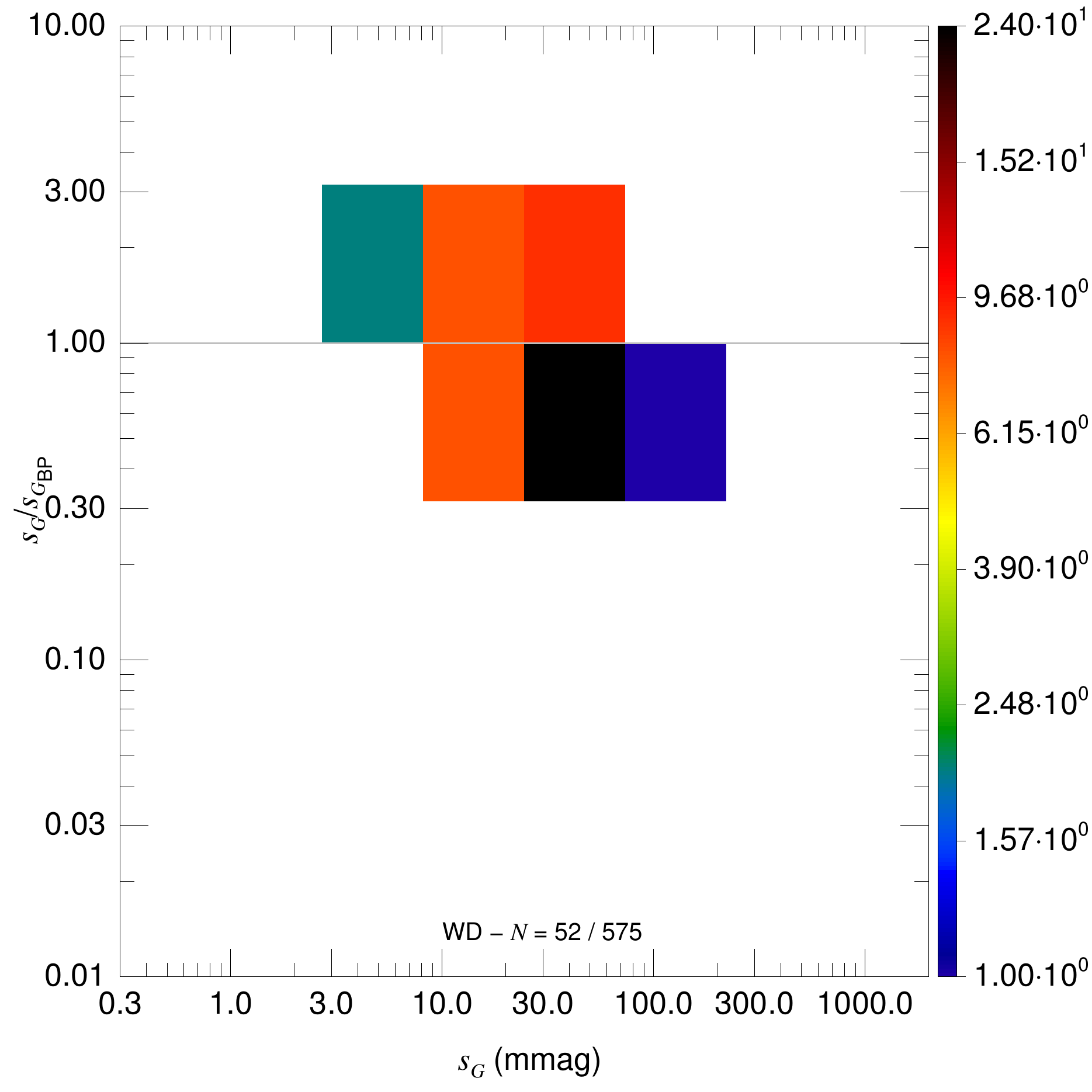}}
\caption{(Continued).}
\end{figure*}

\addtocounter{figure}{-1}

\begin{figure*}
\centerline{\includegraphics[width=0.35\linewidth]{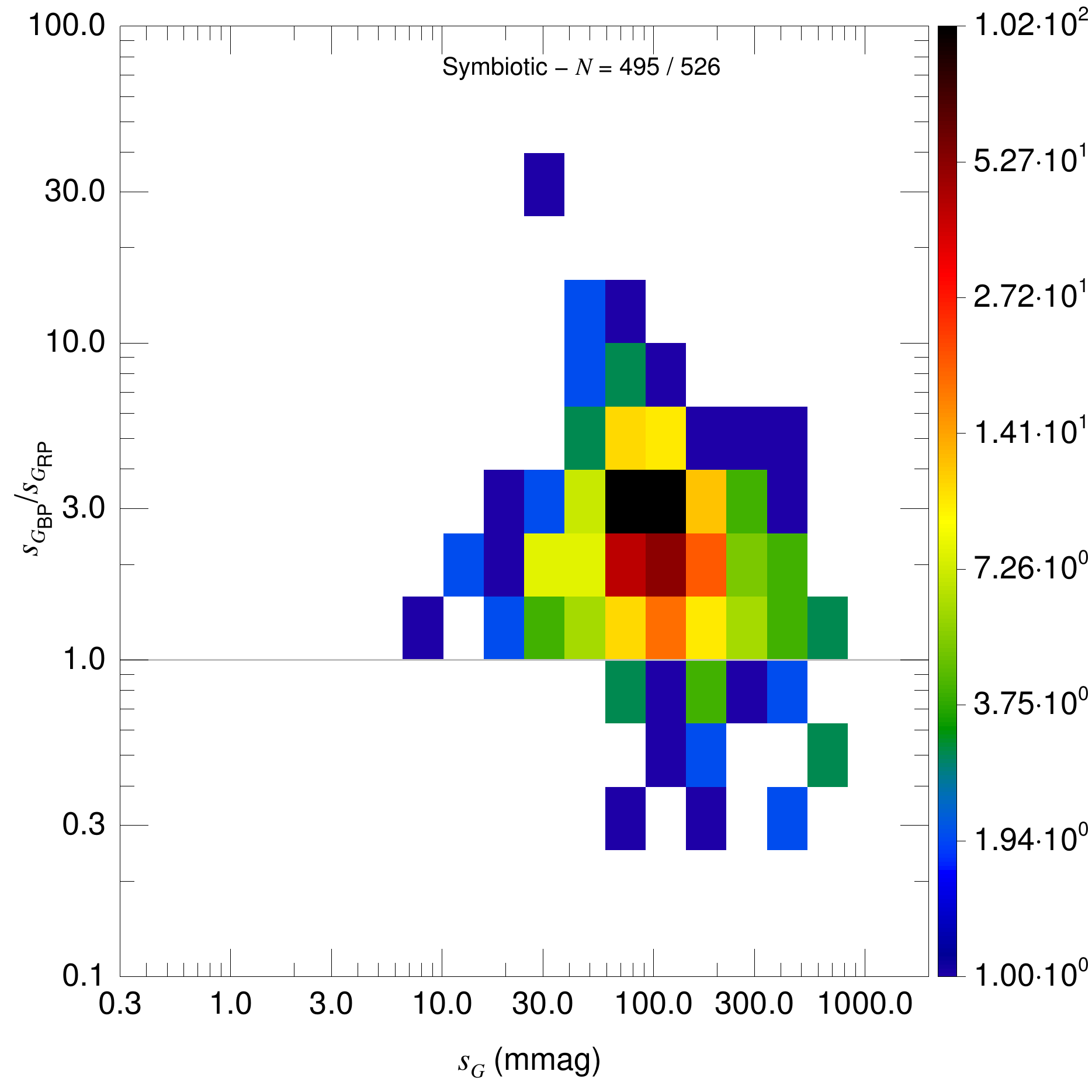}$\!\!\!$
            \includegraphics[width=0.35\linewidth]{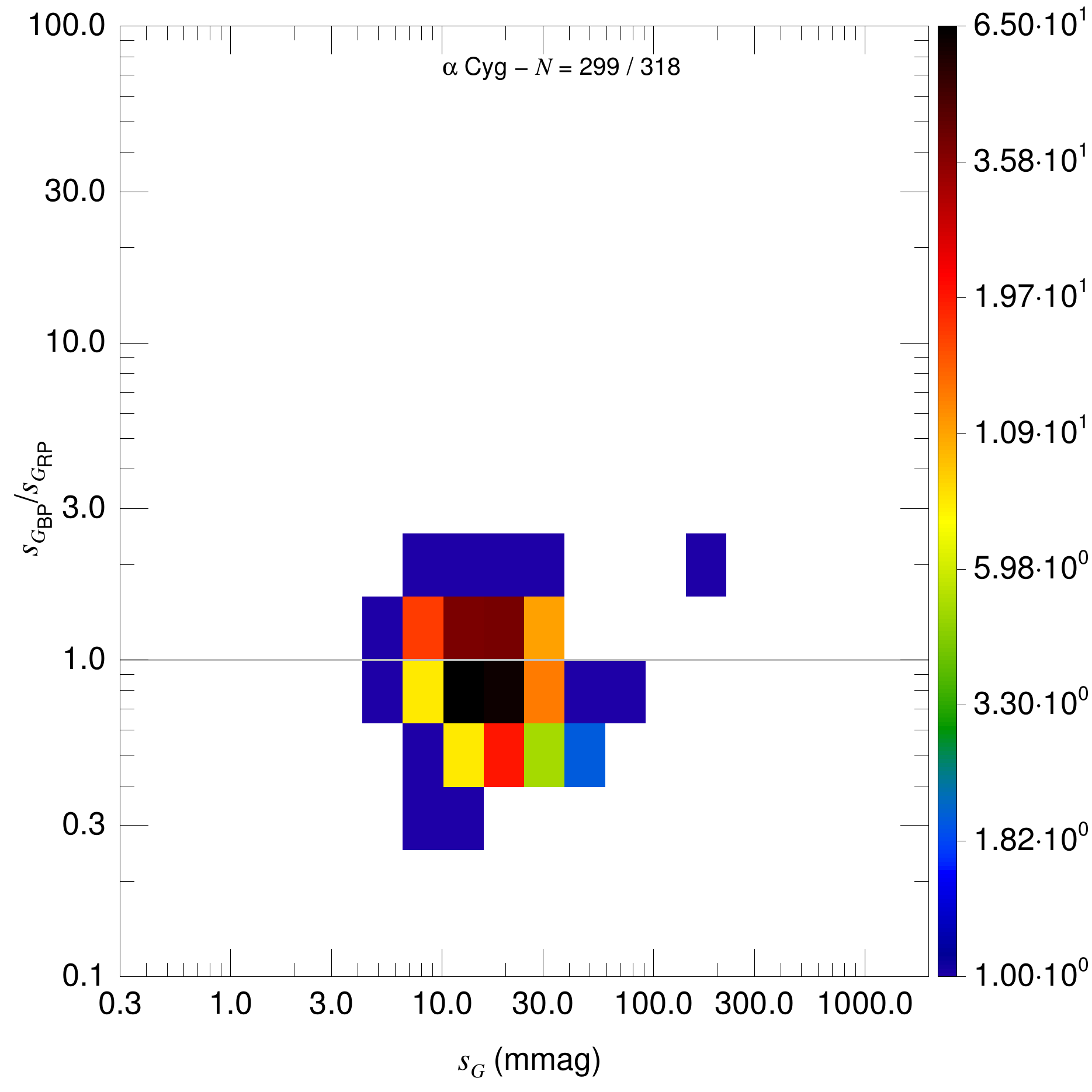}$\!\!\!$
            \includegraphics[width=0.35\linewidth]{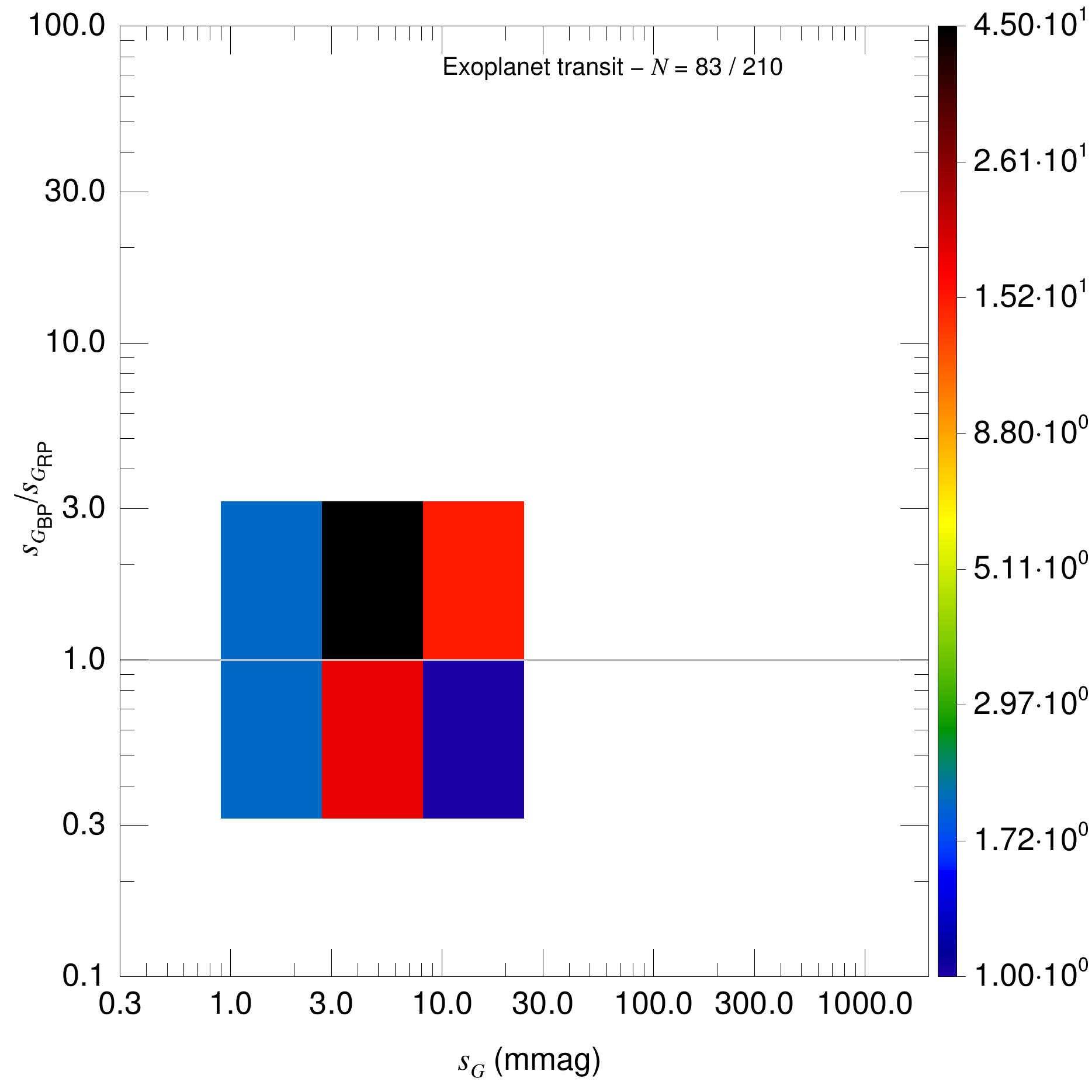}}
\centerline{\includegraphics[width=0.35\linewidth]{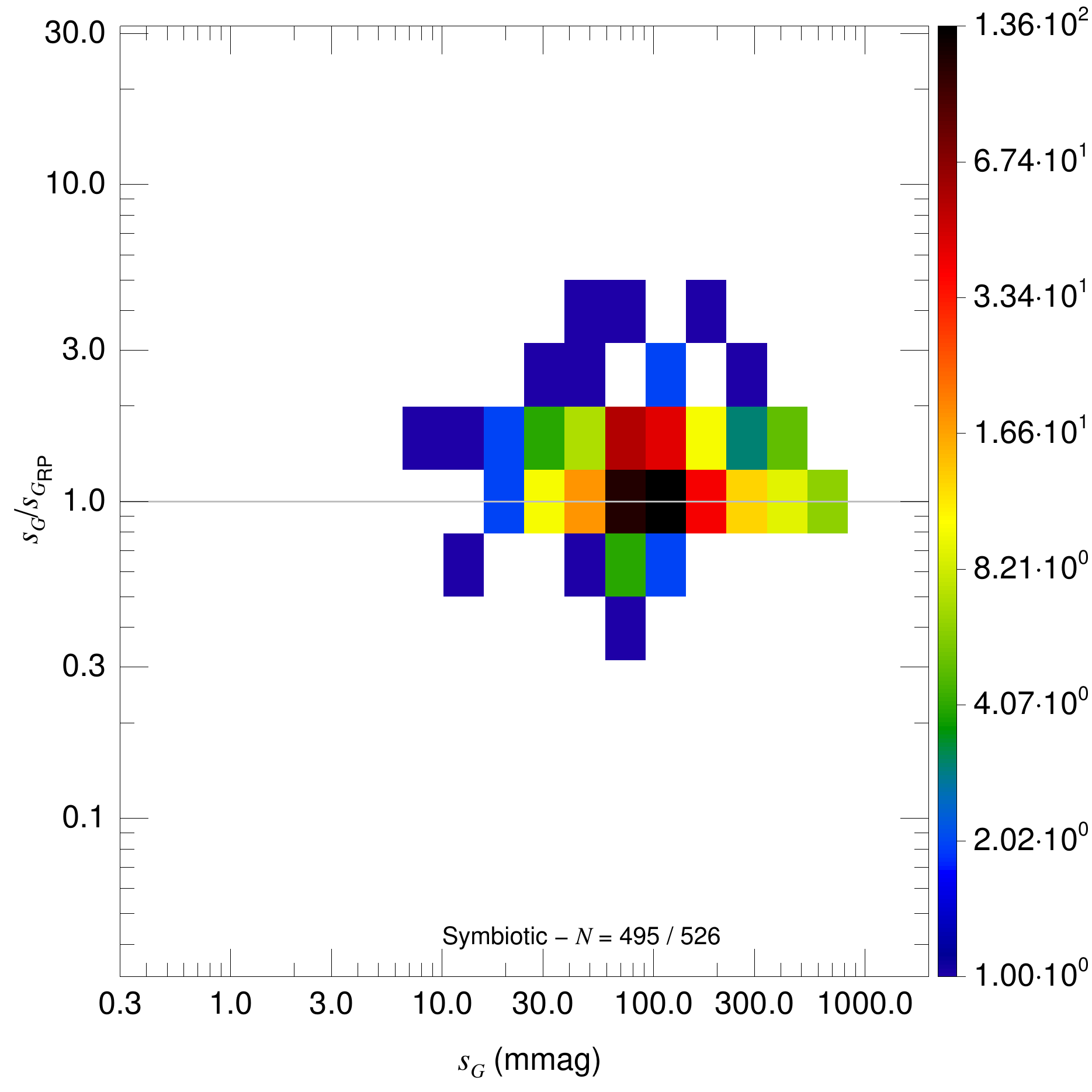}$\!\!\!$
            \includegraphics[width=0.35\linewidth]{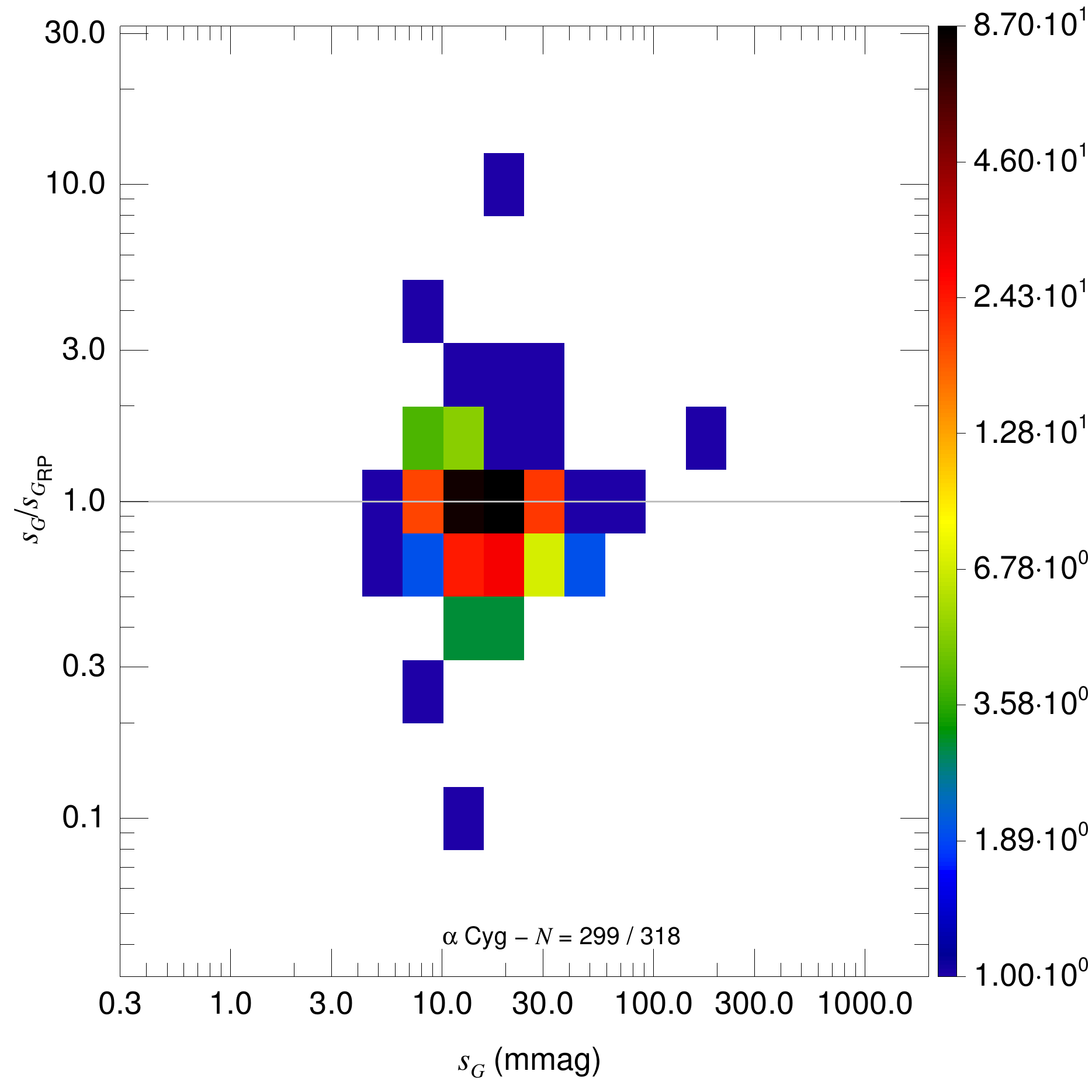}$\!\!\!$
            \includegraphics[width=0.35\linewidth]{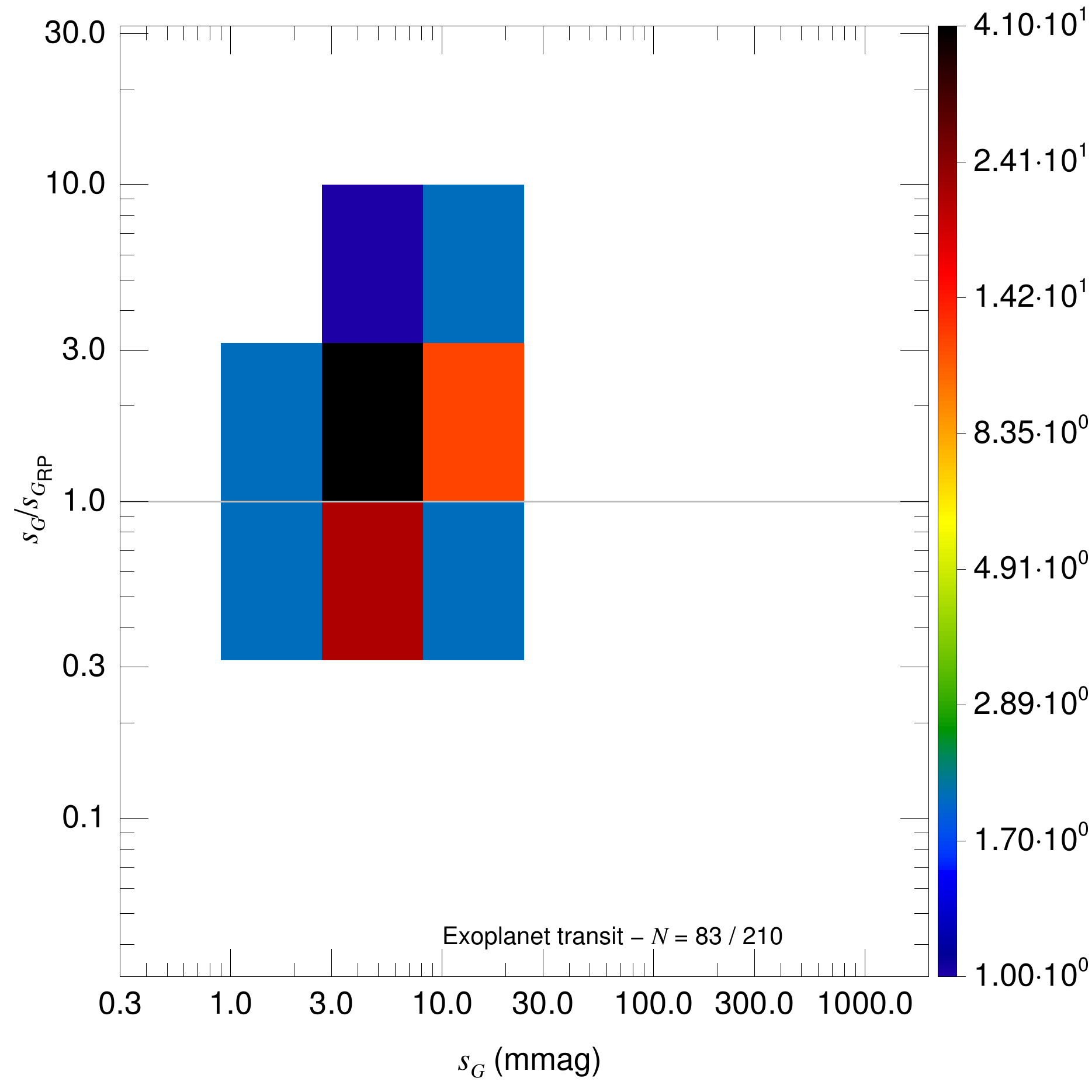}}
\centerline{\includegraphics[width=0.35\linewidth]{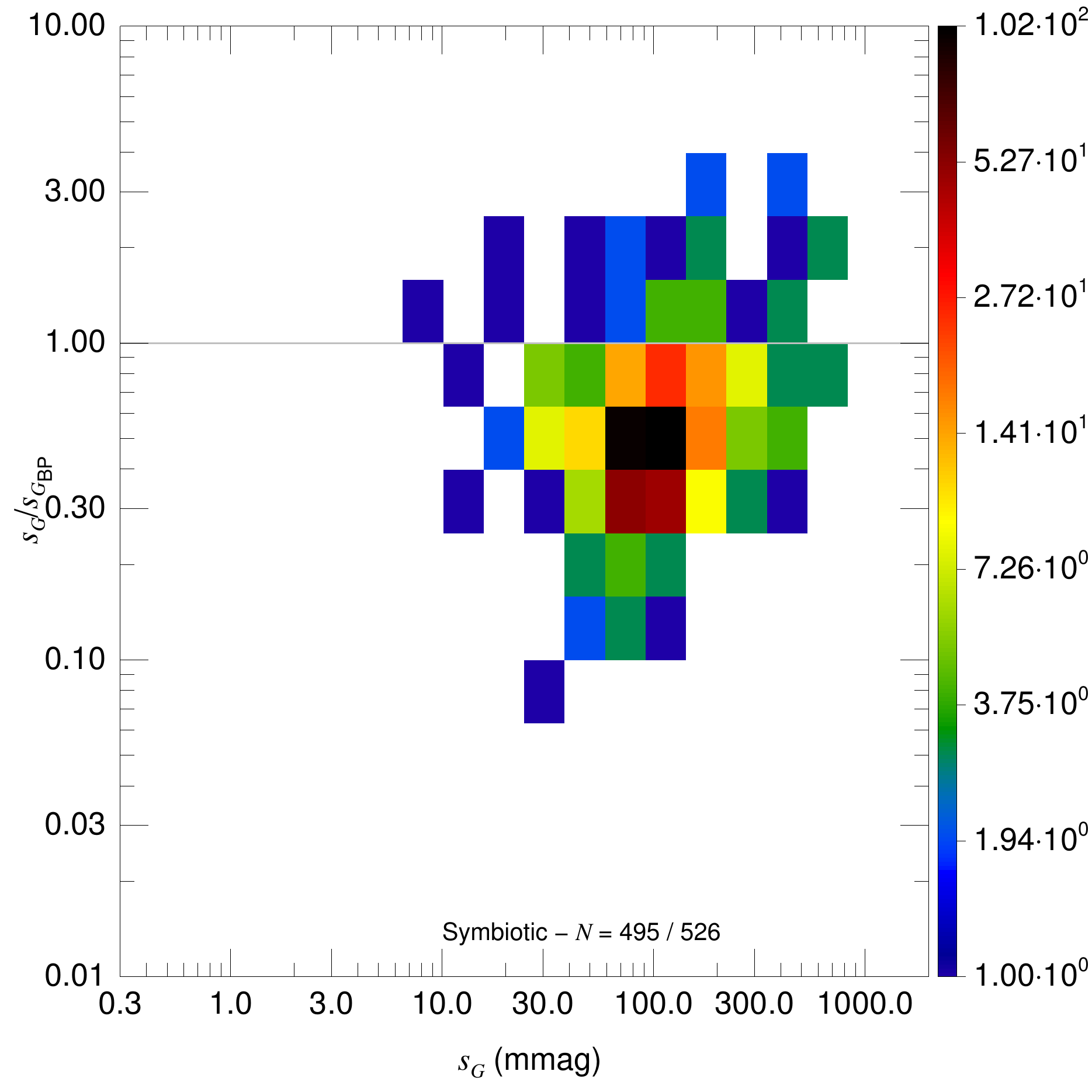}$\!\!\!$
            \includegraphics[width=0.35\linewidth]{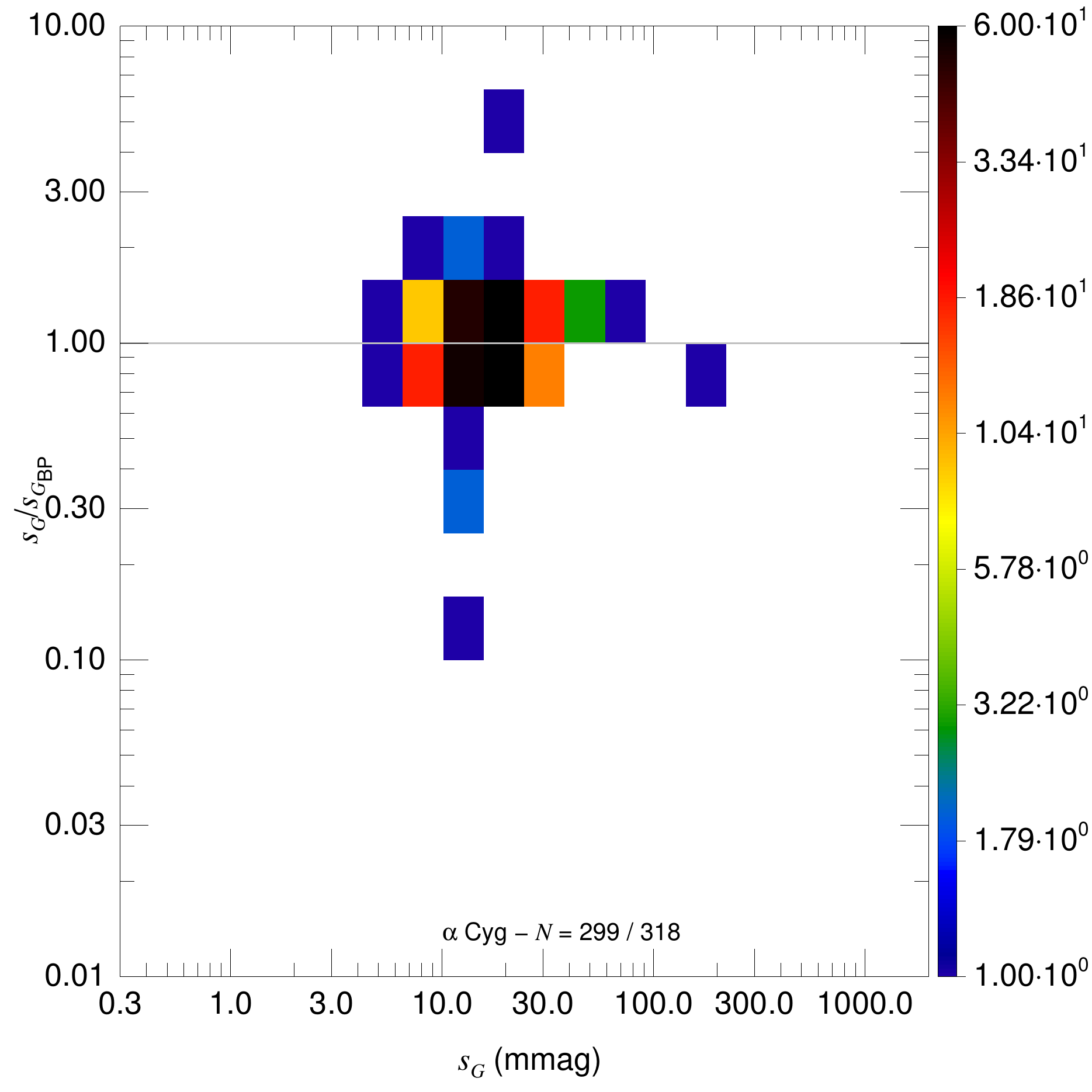}$\!\!\!$
            \includegraphics[width=0.35\linewidth]{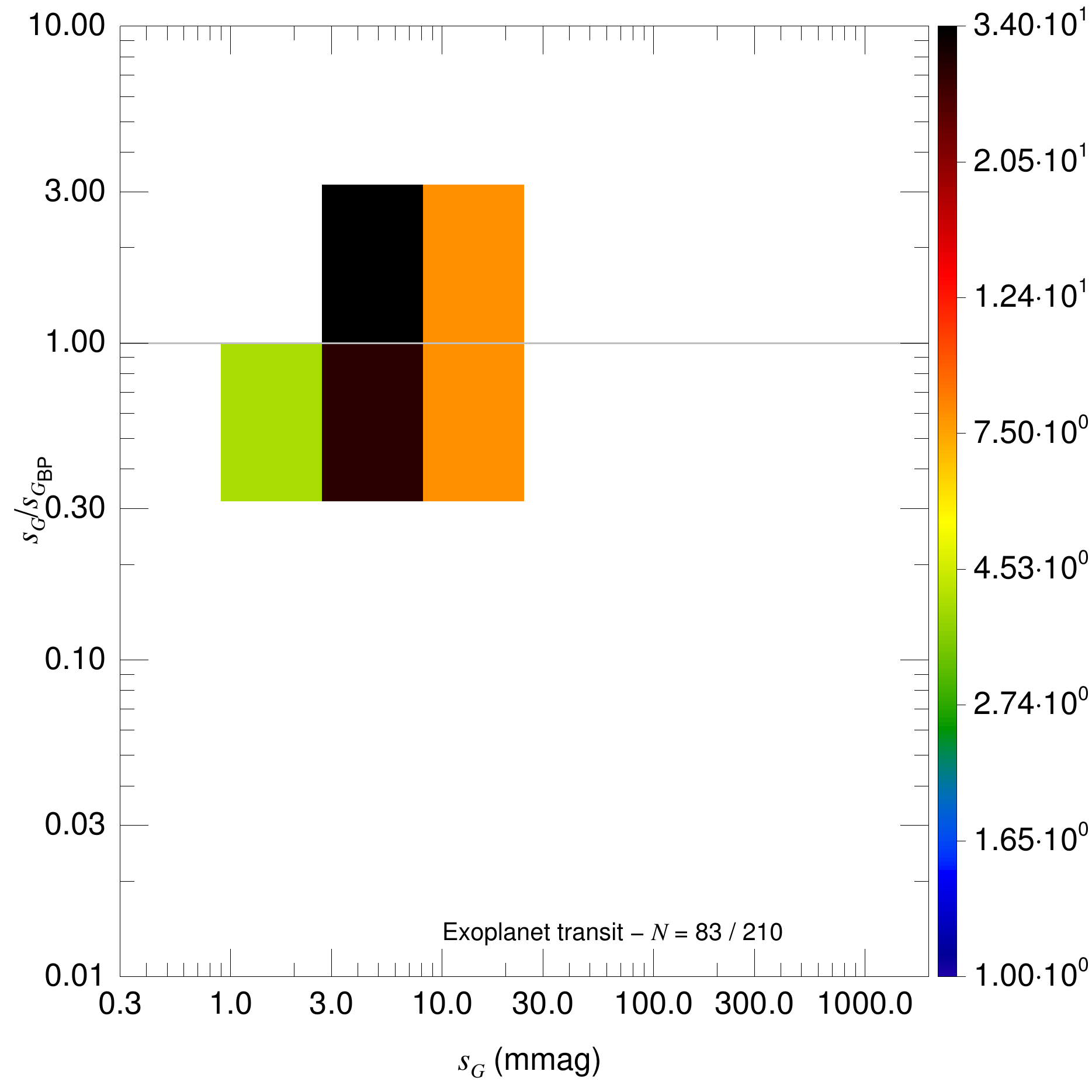}}
\caption{(Continued).}
\end{figure*}

\addtocounter{figure}{-1}

\begin{figure*}
\centerline{\includegraphics[width=0.35\linewidth]{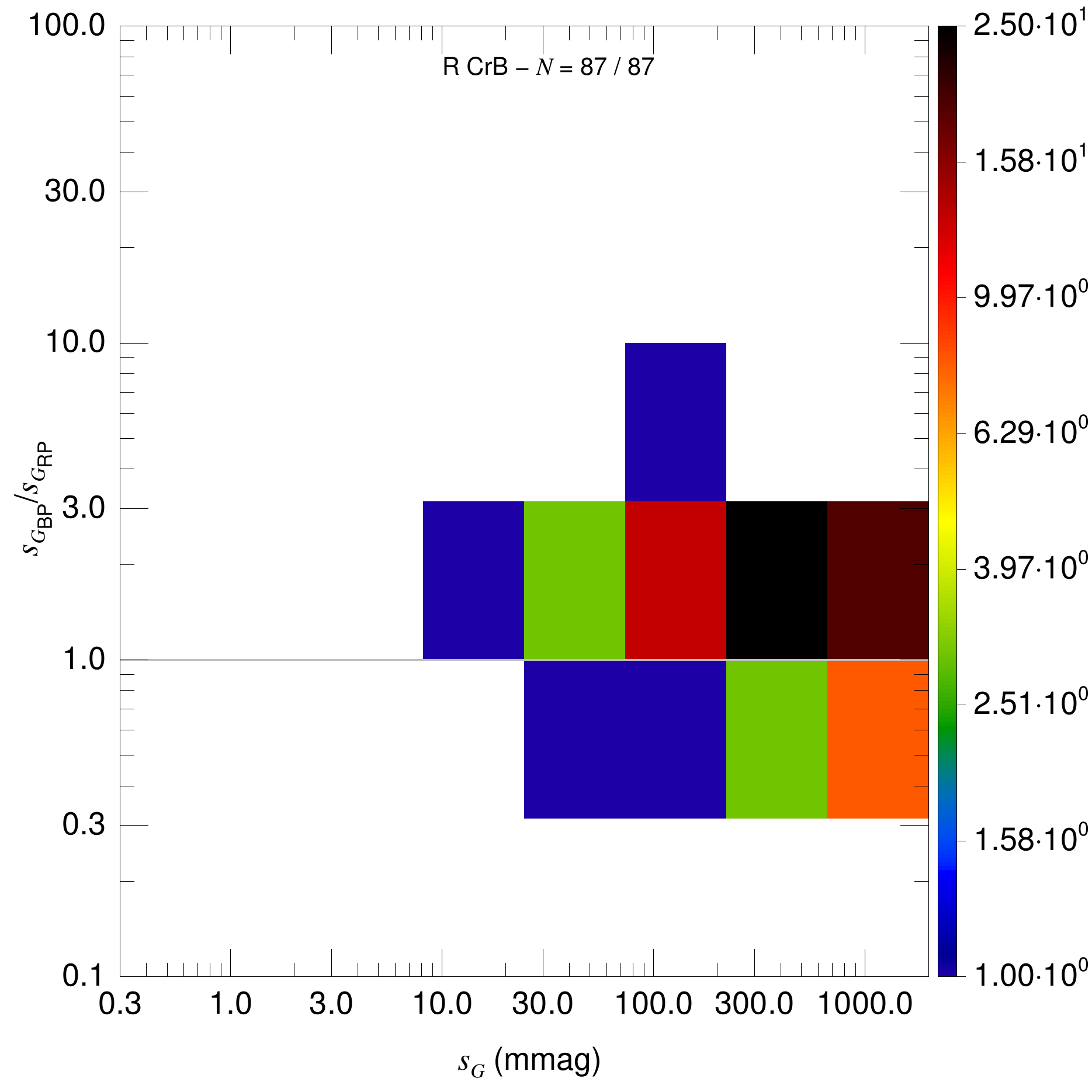}$\!\!\!$
            \includegraphics[width=0.35\linewidth]{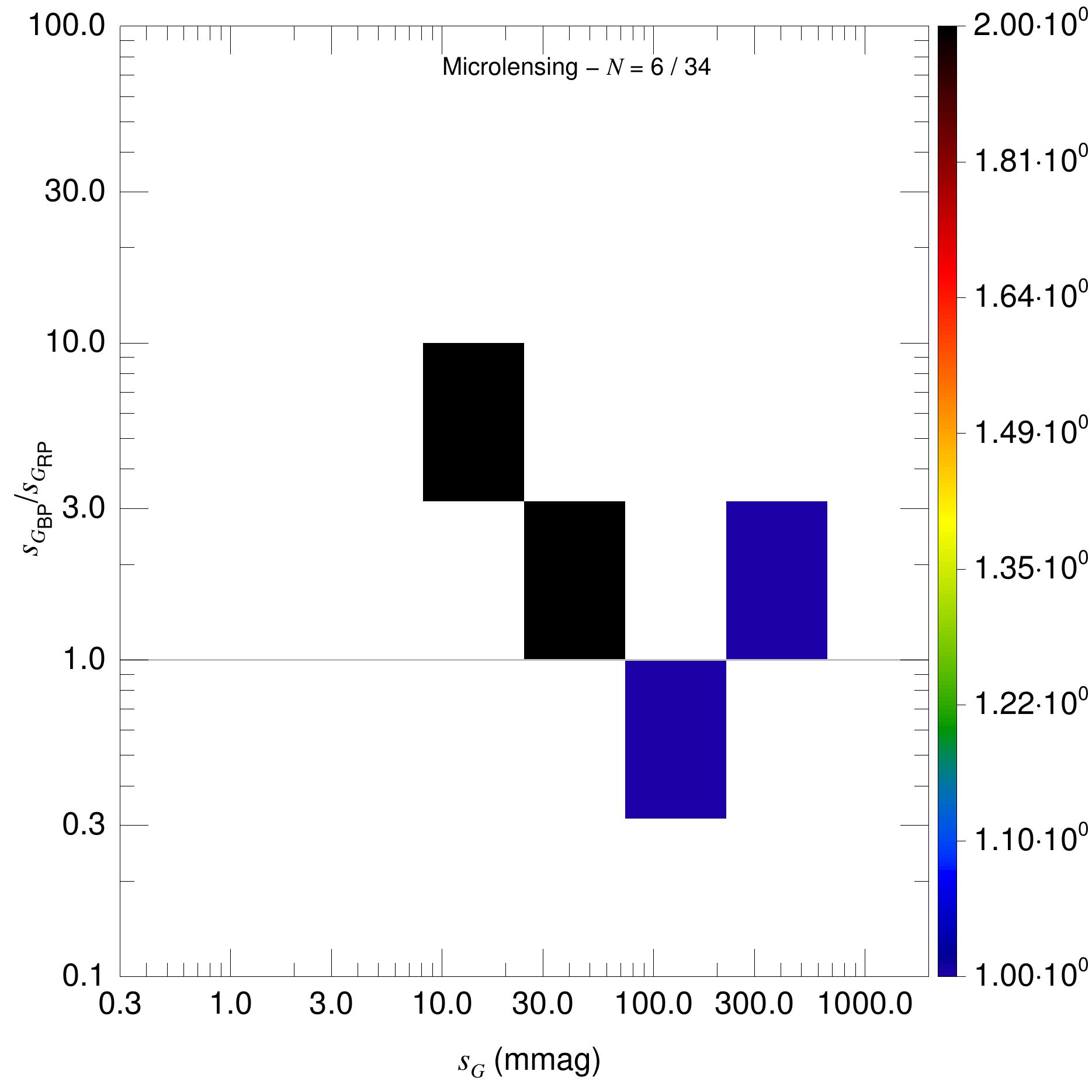}}
\centerline{\includegraphics[width=0.35\linewidth]{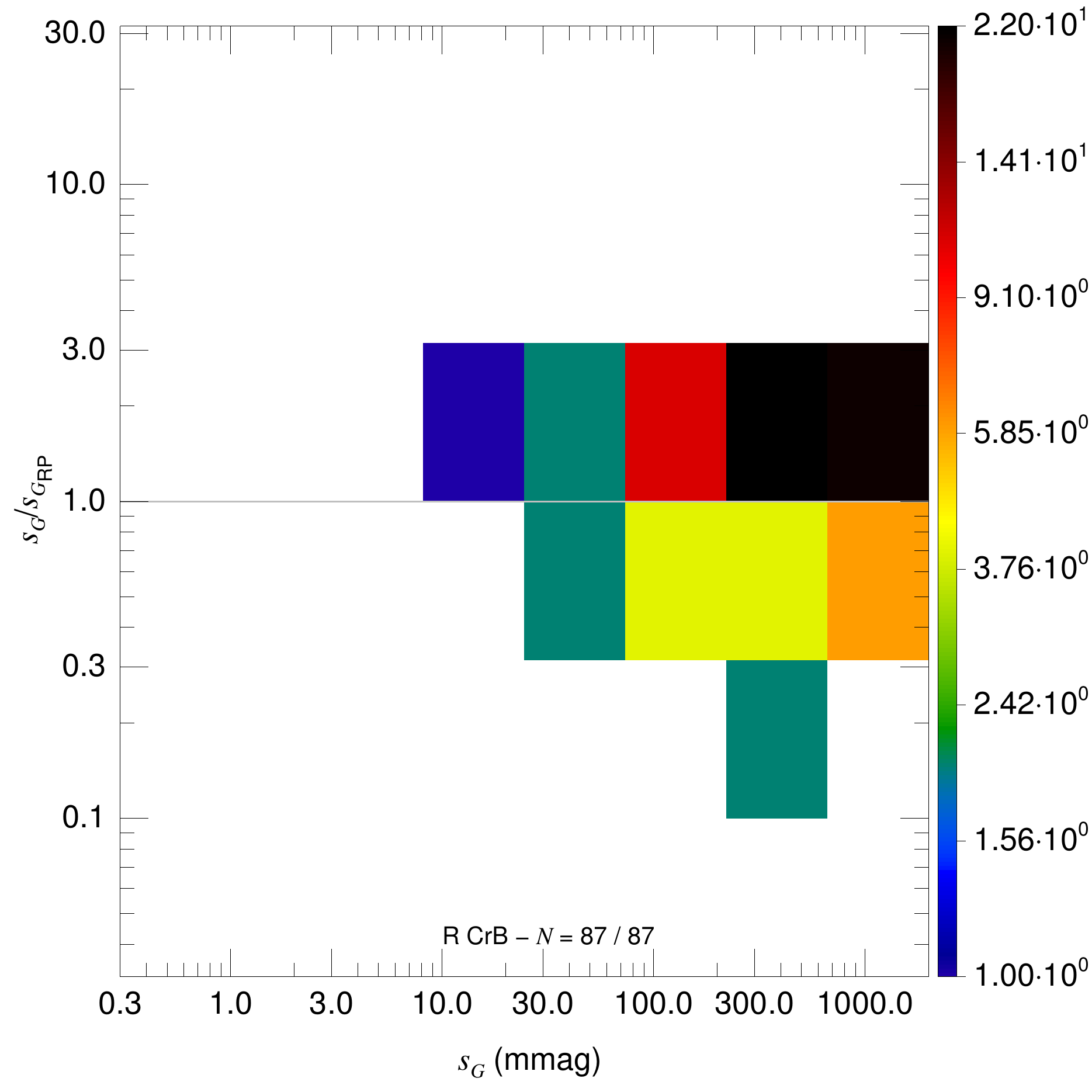}$\!\!\!$
            \includegraphics[width=0.35\linewidth]{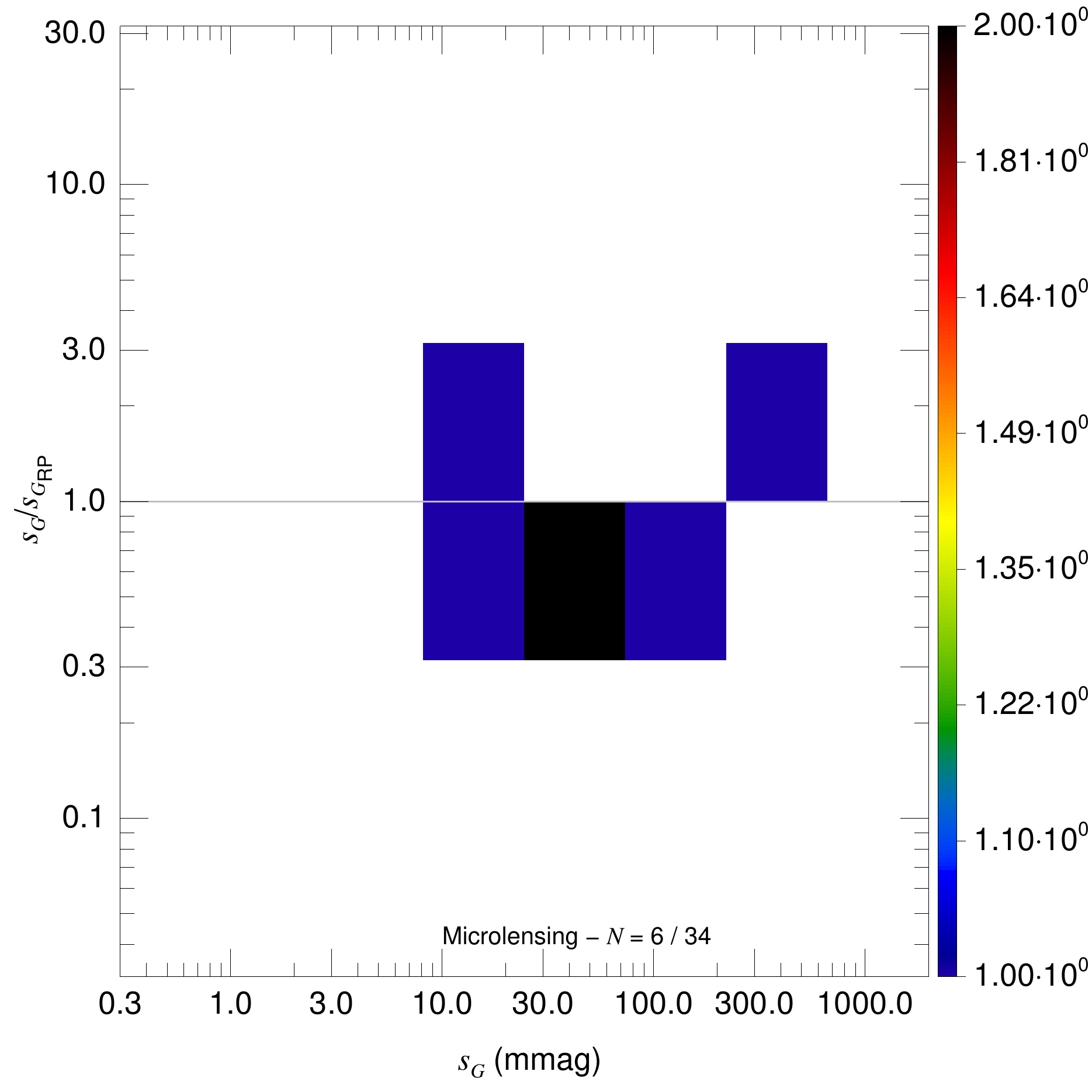}}
\centerline{\includegraphics[width=0.35\linewidth]{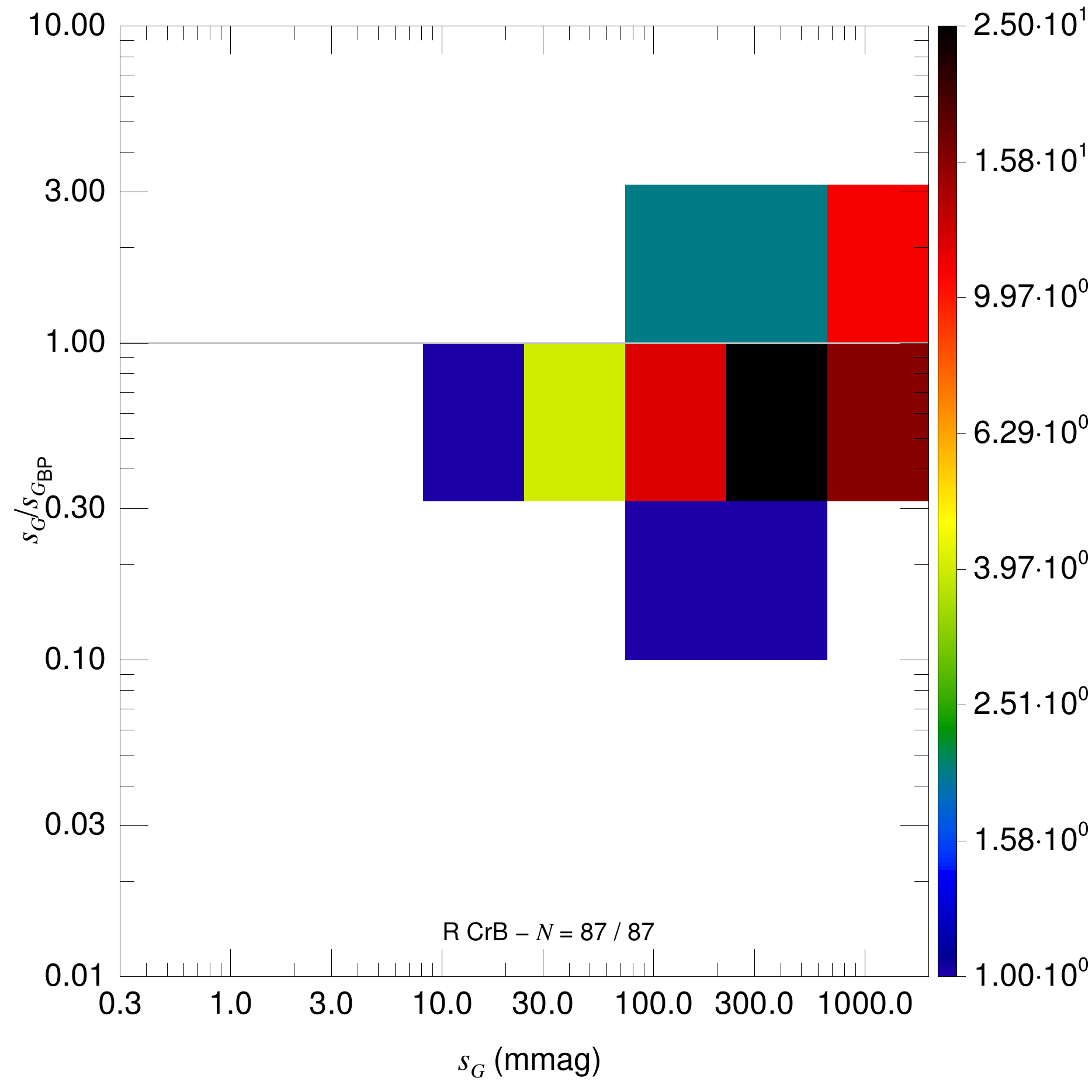}$\!\!\!$
            \includegraphics[width=0.35\linewidth]{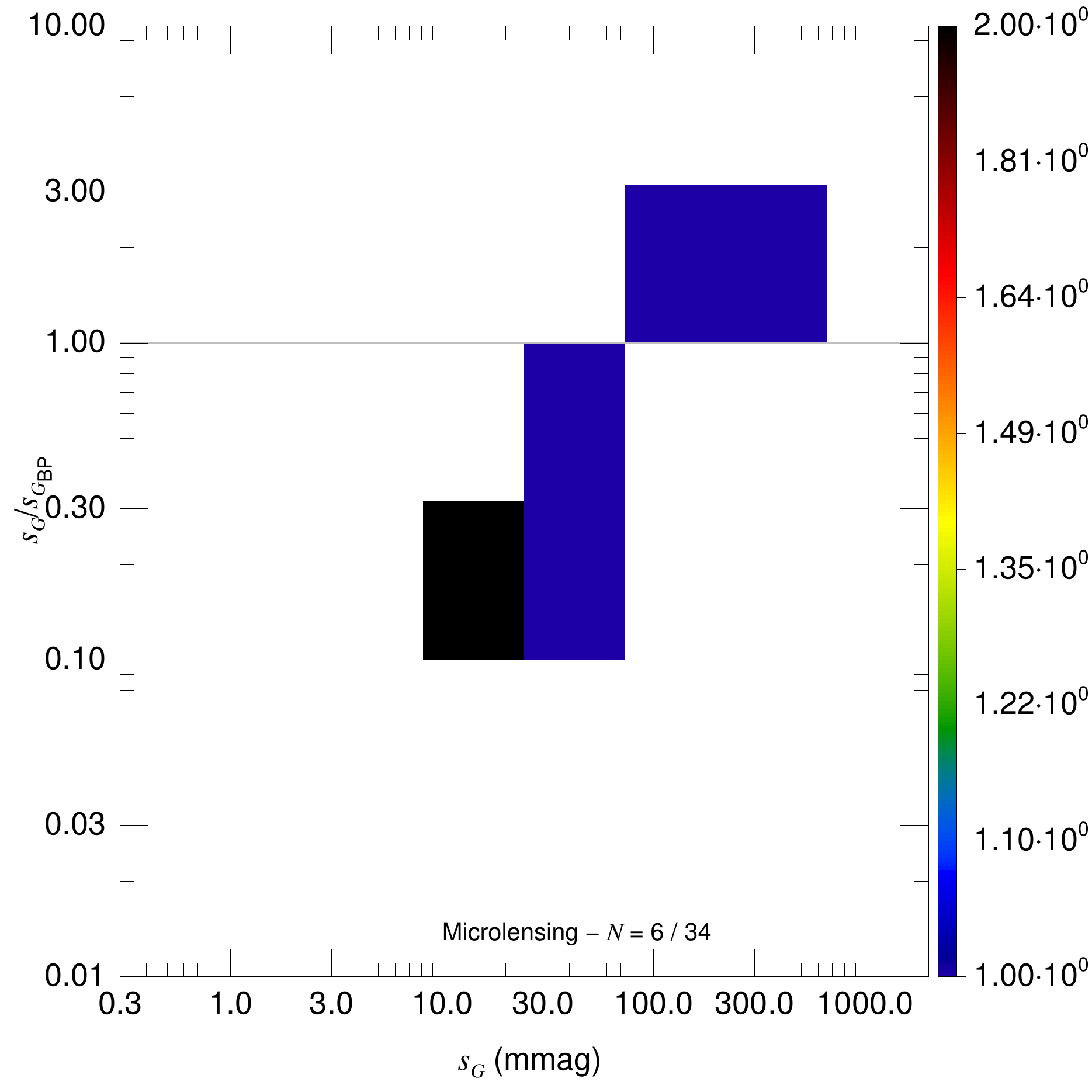}}
\caption{(Continued).}
\end{figure*}

\end{appendix}

\end{document}